\documentclass[11pt,a4paper]{article}
\usepackage{jcappubexpET}
\usepackage{orcidlink}

\pdfoutput=1

\usepackage{amsmath, amssymb, amsfonts, mathtools,bm, graphicx, booktabs, multirow, multicol, url,siunitx,color,appendix,latexsym, epsfig, rotating, makecell,fontawesome}

\usepackage{longtable} 
\usepackage{array} 

\usepackage{lineno}
\usepackage{pdflscape} 
\usepackage{tabularx}
\usepackage{tikz}
\usetikzlibrary{shapes.geometric, arrows}
\tikzstyle{startstop} = [rectangle, rounded corners, minimum width=1cm, minimum height=0.5cm,text centered, draw=black, fill=red!30]
\tikzstyle{io} = [rectangle, minimum width=1cm, minimum height=0.5cm, text centered, draw=black, fill=blue!30]
\tikzstyle{process} = [rectangle, minimum width=1cm, minimum height=0.5cm, text centered, draw=black, fill=orange!30]
\tikzstyle{case} = [rectangle, minimum width=1cm, minimum height=0.5cm, text centered, draw=black]
\tikzstyle{quantity} = [rectangle, minimum width=1cm, minimum height=0.5cm, text centered, draw=black, fill=cyan!30]
\tikzstyle{decision} = [rectangle, minimum width=1cm, minimum height=0.5cm, text centered, draw=black, fill=green!30]
\tikzstyle{arrow} = [thick,->,>=stealth]
\tikzstyle{narrow} = [thick,-,>=stealth]
\tikzstyle{darrow} = [thick,<->,>=stealth]

\usepackage[most]{tcolorbox}
\usepackage{colortbl}  
\newtcolorbox{highlightbox}[1]{breakable,colback=blue!5!white,colframe=blue!70!black,fonttitle=\bfseries,title=#1}

\usepackage{subcaption} 
\usepackage{relsize,nicefrac,physics} 
\usepackage{xcolor}
\usepackage{aas_macros}
\usepackage{pifont}
\usepackage{relsize}
\usepackage{verbatim}
\usepackage{tablefootnote}
\usepackage[acronym]{glossaries} 
\makeglossaries
\newacronym{et}{ET}{Einstein Telescope}
\newacronym{lvc}{LVC}{LIGO-Virgo Collaboration}
\newacronym{lvk}{LVK}{LIGO-Virgo-KAGRA Collaboration}
\newacronym{gw}{GW}{Gravitational Wave}
\newacronym{ns}{NS}{Neutron Star}
\newacronym{pns}{PNS}{Proto-Neutron Star}
\newacronym{snr}{SNR}{Signal-to-Noise Ratio}
\newacronym{em}{EM}{Electromagnetic}
\newacronym{eos}{EoS}{Equation of State}
\newacronym{sasi}{SASI}{Standing Accretion Shock Instability} 
\newacronym{sn}{SN}{Supernova}
\newacronym{cw}{CW}{Continuous Wave}
\newacronym{tcw}{tCW}{Transient Continuous Wave}
\newacronym{cr}{CR}{Critical Ratio}
\newacronym{far}{FAR}{False Alarm Rate}
\newacronym{bbh}{BBH}{Binary Black Hole}
\newacronym{bh}{BH}{Black Hole}
\newacronym{bns}{BNS}{Binary Neutron Star}
\newacronym{nsbh}{NSBH}{Neutron Star-Black Hole}
\newacronym{grb}{GRB}{Gamma Ray Burst}
\newacronym{3g}{3G}{Third Generation}
\newacronym{xdins}{XDINS}{X-Ray Dim Isolated Neutron Star}
\newacronym{ccsn}{CCSN}{Core Collapse Supernova}
\newacronym{pisn}{PISN}{Pair Instability Supernova}
\newacronym{tpagb}{TP-AGB}{Thermally Pulsing Asymptotic Giant Branch}
\newacronym{cowd}{CO-WD}{Carbon-Oxygen White Dwarf}
\newacronym{ecsn}{ECSN}{Electron Capture Supernova}
\newacronym{agb}{AGB}{Asymptotic Giant Branch}
\newacronym{sagb}{SAGB}{Super Asymptotic Giant Branch}
\newacronym{ec}{EC}{Electron Capture}
\newacronym{dc}{DC}{Direct Collapse}
\newacronym{ppisn}{PPISN}{Pulsation Pair Instability Supernova}
\newacronym{lmc}{LMC}{Large Magellanic Cloud}
\newacronym{wr}{WR}{Wolf-Rayet}
\newacronym{rsg}{RSG}{Red Super Giant}
\newacronym{bsg}{BSG}{Blue Super Giant}
\newacronym{lgrb}{LGRB}{Long Gamma Ray Burst}
\newacronym{snii}{SNII}{Type II Supernova}
\newacronym{snib}{SNIb}{Type Ib Supernova}
\newacronym{snic}{SNIc}{Type Ic Supernova}
\newacronym{snia}{SNIa}{Type Ia Supernova}
\newacronym{sniip}{SNII-P}{Type II Plateau Supernova}
\newacronym{sniil}{SNII-L}{Type II Linear Supernova}
\newacronym{sniinp}{SNII-np}{Type II nP Supernova}
\newacronym{csm}{CSM}{Circum Stellar Material}
\newacronym{sniib}{SNIIb}{Type IIb Supernova}
\newacronym{hst}{HST}{Hubble Space Telescope}
\newacronym{jwst}{JWST}{James Webb Space Telescope}
\newacronym{slsn}{SLSN}{Super Luminous Supernova}
\newacronym{tde}{TDE}{Tidal disruption event}
\newacronym{lbv}{LBV}{Luminous Blue Variable}
\newacronym{ilrt}{ILRT}{Intermediate Luminosity Red Transient}
\newacronym{lrn}{LRN}{Luminous Red Novae}
\newacronym{snicbl}{SNIc-BL}{Broad-Line Type Ic Supernova}
\newacronym{atlas}{ATLAS}{Asteroid Terrestrial Impact Last Alert System}
\newacronym{ccsnr}{CCSNR}{Core collapse supernovae rate}
\newacronym{ztf}{ZTF}{Zwicky Transient Facility}
\newacronym{asassn}{ASASSN}{All Sky Automated Survey for SuperNovae}
\newacronym{sfr}{SFR}{Star Formation Rate}
\newacronym{snibc}{SN Ib/c}{Type Ib/Ic Supernova}
\newacronym{imf}{IMF}{Initial Mass Function}
\newacronym{sfd}{SFD}{Star Formation Density}
\newacronym{ilot}{ILOT}{Intermediate Luminosity Optical Transient}
\newacronym{loss}{LOSS}{Lick Observatory Supernova Search}
\newacronym{ulirg}{ULIRG}{Ultra Luminous InfraRed Galaxies}
\newacronym{lirg}{LIRG}{Luminous InfraRed Galaxies}
\newacronym{spirits}{SPIRITS}{Spitzer InfraRed Intensive Transients Survey}
\newacronym{11hugs}{11 HUGS}{11 Mpc H$\alpha$ UV Galaxy Survey}
\newacronym{iifscz}{IIFSCz}{Imperial IRAS Faint Source Catalogue Redshift Database}
\newacronym{mw}{MW}{Milky Way}
\newacronym{tir}{TIR}{Total Infrared}
\newacronym{fuv}{FUV}{Far Ultraviolet}
\newacronym{ir}{IR}{Infrared}
\newacronym{mir}{MIR}{Mid-infrared}
\newacronym{fir}{FIR}{Far Infrared}
\newacronym{uv}{UV}{Ultraviolet}
\newacronym{rrat}{RRAT}{Rotating Radio Transient}
\newacronym{panstarrs}{PANSTARRS}{}

\glsdisablehyper
\definecolor{cyan1}{RGB}{52,177,201} 
\definecolor{lightgray}{gray}{0.9}

\DeclareSIUnit{\msun}{M_{\odot}}
\DeclareSIUnit{\year}{\text{yr}}
\DeclareSIUnit{\radiant}{\text{rad}}
\DeclareSIUnit{\degree}{\text{deg}}
\DeclareSIUnit{\parsec}{\text{pc}}
\DeclareSIUnit \day{day}
\DeclareSIUnit \hour{hr}
\DeclareSIUnit \radiant{rad}
\DeclareSIUnit \degfull{deg}
\DeclareSIUnit \erg {erg}
\DeclareSIUnit \Lsun {L_{\odot}}
\DeclareSIUnit \AstroUnit {au}
\DeclareSIUnit \steradian {sr}

\newcommand{\Msun}{M_{\odot}}

\newcommand{\msol}[0]{M_{\odot}}
\newcommand{\Rsun}{R_{\odot}}
\newcommand{\Lsun}{L_{\odot}}

\newcommand{\hubble}{H_0}

\newcommand{\snr}{\rm SNR}
\newcommand{\hplus}{h_{+}}
\newcommand{\hcross}{h_{\times}}
\newcommand{\Fplus}{F_{+}}
\newcommand{\Fcross}{F_{\times}}
\newcommand{\anglesavg}[1]{\langle #1 \rangle}

\newcommand{\approptoinn}[2]{\mathrel{\vcenter{
  \offinterlineskip\halign{\hfil$##$\cr
    #1\propto\cr\noalign{\kern1pt}#1\sim\cr\noalign{\kern-2pt}}}}}
\newcommand{\appropto}{\mathpalette\approptoinn\relax}

\def\ga{\,\,\raise0.14em\hbox{$>$}\kern-0.76em\lower0.28em\hbox
{$\sim$}\,\,}
\def\la{\,\,\raise0.14em\hbox{$<$}\kern-0.76em\lower0.28em\hbox
{$\sim$}\,\,}

\newcommand{\TEOBResumS}{\texttt{TEOBResumS}}

\newcommand{\eps}{\epsilon}

\newcommand{\ssim}{\mathchar"5218\relax\,}

\newcommand\eq[1]{eq.~(\ref{#1})}
\newcommand\eqs[2]{eqs.~(\ref{#1}) and (\ref{#2})}

\newcommand\eqsss[4]{eqs.~(\ref{#1}), (\ref{#2}),  (\ref{#3}) and (\ref{#4})}
\newcommand\eqst[2]{eqs.~(\ref{#1})--(\ref{#2})}
\newcommand\Eq[1]{Equation~(\ref{#1})}

\newcommand\be{\begin{equation}}
\newcommand\ee{\end{equation}}
\newcommand\bees{\begin{eqnarray}}
\newcommand\ees{\end{eqnarray}}
\newcommand{\hatn}{\hat{\bf n}}
\renewcommand\[{\left[}
\renewcommand\]{\right]}
\renewcommand\({\left(}
\renewcommand\){\right)}
\newcommand{\nn}{\nonumber}
\newcommand{\nev}{N_{\rm ev}}
\newcommand{\snrth}{{\rm SNR}_{\rm th}}
\newcommand{\snrobs}{{\rm SNR}^{\rm obs}}

\newcommand{\dgw}{d_L^{\,\rm gw}}
\newcommand{\dem}{d_L^{\,\rm em}}
\newcommand{\rde}{\rho_{\rm DE}}
\newcommand{\ode}{\Omega_{\rm DE}}
\newcommand{\oma}{\Omega_{M}}

\newcommand{\ra}{\rightarrow}

\newcommand{\dtens}{\{\Delta t_{i}\}}
\def\lsim{\raise 0.4ex\hbox{$<$}\kern -0.8em\lower 0.62
ex\hbox{$\sim$}}
\def\gsim{\raise 0.4ex\hbox{$>$}\kern -0.7em\lower 0.62
ex\hbox{$\sim$}}

\newcounter{myparagraph}[subsubsection] 
\renewcommand{\themyparagraph}{\thesubsubsection.\arabic{myparagraph}}

\let\oldparagraph\paragraph 
\renewcommand{\paragraph}[1]{%
    \refstepcounter{myparagraph}
    \oldparagraph{\themyparagraph\ #1}
}

\numberwithin{equation}{section}
\numberwithin{figure}{section}
\numberwithin{table}{section}

\definecolor{MyBlue}{rgb}{0.15, 0.15, 0.70}
\hypersetup{
colorlinks,
citecolor=MyBlue,
linkcolor=MyBlue,
urlcolor=MyBlue
}
\definecolor{lightgray}{gray}{0.9}

\title{The Science of the Einstein Telescope}

\collaboration{Einstein Telescope collaboration}
\affiliationAdd{\Byurakan}{Byurakan Astrophysical Observatory, National Academy of Sciences, Byurakan 0213, Armenia}
\affiliationAdd{\UYerevan}{Institute of Physics, Yerevan State University, Yerevan 0025, Armenia}

\affiliationAdd{\ANU}{OzGrav, Australian National University, Canberra, Australian Capital Territory 0200, Australia}

\affiliationAdd{\swin}{Centre for Astrophysics and Supercomputing, Swinburne University of Technology, Melbourne, Australia, VIC 3122}

\affiliationAdd{\AustrianAcademy}{Affiliation: Institute of High Energy Physics - Austrian Academy of Sciences, 1010 Vienna, Austria}

\affiliationAdd{\UAntwerpen}{Universiteit Antwerpen, Prinsstraat 13, 2000 Antwerpen, Belgium}
\affiliationAdd{\ULB}{Institute of Astronomy and Astrophysics, Université Libre de Bruxelles, CP 226, 1050 Brussels, Belgium}
\affiliationAdd{\VUB}{56 Department of Physics \& Astronomy, Vrije Universiteit Brussel (VUB), Pleinlaan 2, 1050 Brussels, Belgium}

\affiliationAdd{\BelObs}{Royal Observatory of Belgium, Avenue Circulaire, 3, 1180 Uccle, Belgium}

\affiliationAdd{\KULeuvenA}{Institute of Astronomy, KU Leuven, Celestijnenlaan 200D, B-3001 Leuven, Belgium}
\affiliationAdd{\KULeuvenI}{Leuven Gravity Institute, KU Leuven, Celestijnenlaan 200D box 2415, 3001 Leuven, Belgium}
\affiliationAdd{\KULeuvenD}{Department of Physics and Astronomy, Laboratory for
Semiconductor Physics, KU Leuven, B-3001 Leuven, Belgium}
\affiliationAdd{\KULeuvenDD}{KU Leuven, Department of Electrical Engineering (ESAT), STADIUS Center for Dynamical Systems, Signal Processing and Data Analytics, B-3001 Leuven, Belgium}
\affiliationAdd{\UniLouvain}{Centre for Cosmology, Particle Physics and Phenomenology - CP3, Universit\'e Catholique de Louvain, Louvain-La-Neuve, B-1348, Belgium}
\affiliationAdd{\CurlLouvain}{Cosmology, Universe and Relativity at Louvain (CURL), Institute of Mathematics and Physics, University of Louvain, 2 Chemin du Cyclotron, 1348 Louvain-la-Neuve, Belgium}
\affiliationAdd{\STARInstitute}{STAR Institute, B\^atiment B5, Universit\'e de Li\`ege, Sart Tilman B4000 Li\`ege, Belgium}

\affiliationAdd{\ifufrj}{Instituto de Física, Universidade Federal do Rio de Janeiro, 21941-972, Rio de Janeiro, RJ, Brazil}
\affiliationAdd{\ovufrj}{Observatório do Valongo, Universidade Federal do Rio de Janeiro, 20080-090, Rio de Janeiro, RJ, Brazil}
\affiliationAdd{\UFF}{Instituto de F\'isica, Universidade Federal Fluminense, Niter\'oi, Rio de Janeiro, 24210-346, Brazil.}
\affiliationAdd{\ifusp}{Instituto de Física, Universidade de São Paulo, Rua do Matão 1371, 05508-090, São Paulo, SP, Brazil}
\affiliationAdd{\IFTUNESP}{Instituto de Física Teórica, UNESP - Universidade Estadual Paulista, São Paulo 01140-070, SP, Brazil}
\affiliationAdd{\ICTPSAIFR}{International Institute of Theoretical Physics - South American Institute for Fundamental Research, São Paulo 01140-070, SP, Brazil}
\affiliationAdd{\ppgcosmo}{PPGCosmo, Universidade Federal do Espírito Santo, 29075-910, Vitória, ES, Brazil}

\affiliationAdd{\UniSofia}{Department of Theoretical Physics, Faculty of Physics, Sofia University, Sofia 1164, Bulgaria}
\affiliationAdd{\IMISofia}{Institute of Mathematics and Informatics, Bulgarian Academy of Sciences, Acad. G. Bonchev St. 8, Sofia 1113, Bulgaria}

\affiliationAdd{\ASU}{Astronomical Institute of the Czech Academy of Sciences,
Bo\v{c}n\'{i} II 1401/1a, CZ-141 00 Prague, Czech Republic}
\affiliationAdd{\Charles}{Institute of Theoretical Physics, Faculty of Mathematics and Physics, Charles University, V Hole\v{s}ovi\v{c}k\'{a}ch 2, 180 00 Prague, Czech Republic}
\affiliationAdd{\Opava}{Institute of Physics, Silesian University in Opava, Bezručovo nám. 13, CZ-746-01 Opava, Czech Republic}

\affiliationAdd{\AndresBello}{Universidad Andres Bello, Facultad de Ciencias Exactas, Departamento de Fisica, Instituto de Astrofisica, Fernandez Concha 700, Las Condes, Santiago RM, Chile}

\affiliationAdd{\KIAA}{Kavli Institute for Astronomy and Astrophysics, Peking University, Beijing 100871, China}
\affiliationAdd{\NAOC}{National Astronomical Observatories, Chinese Academy of Sciences, Beijing 100012, China}
\affiliationAdd{\SMU}{Department of Mathematics and Physics, School of Biomedical Engineering, Southern Medical University, Guangzhou 510515, China}

\affiliationAdd{\BNU}{Institute for Frontier in Astronomy and Astrophysics, Department of Astronomy, Beijing Normal University, Beijing, People's Republic of China}

\affiliationAdd{\Aalborg}{Dept. Materials and Production, Aalborg University, 9220 Aalborg, Denmark}
\affiliationAdd{\NBI}{Niels Bohr Institute, University of Copenhagen, Blegdamsvej 17, 2100 Copenhagen, Denmark}

\affiliationAdd{\KBFITalinn}{Keemilise ja Bioloogilise Füüsika Instituut, Rävala pst. 10, 10143 Tallinn, Estonia}

\affiliationAdd{\Meudon}{Laboratoire Univers et Théories, Observatoire de Paris, Université PSL, Université Paris Cité, CNRS, F-92190 Meudon, France}
\affiliationAdd{\mrs}{Aix-Marseille Universit\'e, Universit\'e de Toulon, CNRS, CPT, Marseille, France}
\affiliationAdd{\OBaS}{Observatoire astronomique de Strasbourg, CNRS, Université de Strasbourg, 11 rue de l'Université, 67000 Strasbourg, France}
\affiliationAdd{\OBSPM}{Observatoire de Paris, CNRS, Université PSL, 5 place Jules Janssen, 92915 Meudon, France}
\affiliationAdd{\GANIL}{Grand Accélérateur National d'Ions Lourds (GANIL), CEA/DRF – CNRS/IN2P3, Boulevard Henri Becquerel, 14076 Caen, France}
\affiliationAdd{\IAP}{Sorbonne Universit\'e, CNRS, UMR 7095, Institut d’Astrophysique de Paris (IAP), 98 bis boulevard Arago, 75014 Paris, France}
\affiliationAdd{\IUF}{Institut Universitaire de France, Minist\'ere de l’Enseignement Sup\'erieur et de la Recherche, 1 rue Descartes, 75231 Paris Cedex F-05, France}
\affiliationAdd{\UniParisSaclay}{Université Paris-Saclay, CNRS/IN2P3, IJCLab, 91405 Orsay, France}
\affiliationAdd{\LPENS}{Laboratoire de Physique de l’\'Ecole Normale Sup\'erieure, ENS, CNRS, Universit\'e PSL, Sorbonne Universit\'e, Universit\'e Paris Cit\'e, F-75005 Paris, France}
\affiliationAdd{\LtwoIT}{Laboratoire des 2 Infinis - Toulouse (L2IT-IN2P3), Universit\'e de Toulouse, CNRS, UPS, F-31062 Toulouse Cedex 9, France}
\affiliationAdd{\OCA}{Universit\'e C\^ote d'Azur, Observatoire de la C\^ote d'Azur, CNRS, Laboratoire Lagrange, Bd de l'Observatoire, CS 34229, 06304 Nice cedex 4, France}
\affiliationAdd{\Artemis}{Universit\'e C\^ote d'Azur, Observatoire de la C\^ote d'Azur, CNRS, Artemis, F-06304 Nice, France}
\affiliationAdd{\UniParis}{Université Paris Cité, CNRS, Astroparticule et Cosmologie, F-75013 Paris, France}
\affiliationAdd{\LPC}{Laboratoire de Physique Corpusculaire (LPC-Caen), CNRS/IN2P3, 6 Bvd du maréchal Juin, 14050 Caen, France}
\affiliationAdd{\Annecy}{Univ.~Savoie Mont Blanc, CNRS, Laboratoire d’Annecy de Physique des Particules - IN2P3, 74940 Annecy, France}

\affiliationAdd{\zah}{Institut f{\"u}r Theoretische Astrophysik, Zentrum f\"ur Astronomie Heidelberg (ZAH), Universit{\"a}t Heidelberg, Albert-Ueberle-Stra{\ss}e 2, 69120, Heidelberg, Germany}
\affiliationAdd{\IWRHeidelberg}{Interdisziplinäres Zentrum für Wissenschaftliches Rechnen (IWR), Universität Heidelberg, Im Neuenheimer Feld 225, 69120 Heidelberg, Germany}

\affiliationAdd{\HITS}{Heidelberger Institut f{\"u}r Theoretische Studien, Schloss-Wolfsbrunnenweg 35, 69118 Heidelberg, Germany}
\affiliationAdd{\DESY}{Deutsches Elektronen-Synchrotron DESY, Notkestra{\ss}e 85, 22607 Hamburg, Germany}
\affiliationAdd{\UniHamburg}{II. Institute of Theoretical Physics, Universit\"{a}t  Hamburg, 22761, Hamburg, Germany}
\affiliationAdd{\UniPotsdam}{Institut f\"ur Physik und Astronomie, Universit\"at Potsdam, Haus 28, Karl-Liebknecht-Str. 24/25, 14476, Potsdam, Germany}
\affiliationAdd{\AEI}{Max Planck Institute for Gravitational Physics (Albert Einstein Institute), 
Am M\"{u}hlenberg 1, D-14476 Potsdam, Germany}
\affiliationAdd{\aeihannover}{Max Planck Institute for Gravitational Physics (Albert Einstein Institute), 30167 Hannover, Germany}
\affiliationAdd{\leibnitzuni}{Leibniz Universität Hannover, 30167 Hannover, Germany}
\affiliationAdd{\LUH}{Institute for Theoretical Physics, Leibniz University Hannover, Appelstraße 2, 30167 Hannover, Germany}
\affiliationAdd{\GSI}{GSI Helmholtzzentrum f\"ur Schwerionenforschung, Planckstra{\ss}e 1, 64291 Darmstadt, Germany} 
\affiliationAdd{\IKPDarmstadt}{Institut {f\"ur} Kernphysik (Theoriezentrum), Technische
Universit{\"a}t Darmstadt, Schlossgartenstra{\ss}e 2, D-64289 Darmstadt,
Germany}
\affiliationAdd{\HFHF}{Helmholtz Forschungsakademie Hessen f\"ur FAIR (HFHF), GSI
   Helmholtzzentrum f\"ur Schwerionenforschung, Planckstra{\ss}e~1, 64291 Darmstadt, Germany}
\affiliationAdd{\UniJena}{Theoretisch-Physikalisches Institut, Friedrich-Schiller-Universität Jena, 07743, Jena, Germany}
\affiliationAdd{\UniTuebingen}{Theoretical Astrophysics, IAAT, Eberhard Karls University of T\"{u}bingen, T\"{u}bingen 72076, Germany}
\affiliationAdd{\FIAS}{Frankfurt Institute for Advanced Studies, D-60438 Frankfurt am Main, Germany}
\affiliationAdd{\MPBonn}{Max Planck Institut für Radioastronomie, Auf dem H\"ugel 69, 53121, Bonn, Germany}
\affiliationAdd{\ESO}{European Southern Observatory, Karl-Schwarzschild-Str. 2, 85748 Garching-bei-M"unchen, Germany}
\affiliationAdd{\KR}{Institute for Astroparticle Physics, Karlsruhe Institute of Technology, Karlsruhe, Germany}
\affiliationAdd{\MPA}{Max Planck Institute for Astrophysics, Karl-Schwarzschild-Str. 1, 85748 Garching, Germany}

\affiliationAdd{\UniThessaloniki}{Department of Physics, Aristotle University of Thessaloniki, 54124 Thessaloniki, Greece}
\affiliationAdd{\Heraklion}{Institute of Astrophysics, Foundation for Research and Technology - Hellas, N. Plastira 100, 70013, Heraklion, Greece} 

\affiliationAdd{\HUNWigner}{HUN-REN Wigner Research Centre for Physics, Konkoly-Thege Miklós út 29-33, 1121 Budapest, Hungary}
\affiliationAdd{\Wigner}{Department of Theoretical Physics, HUN-REN Wigner Research Centre for Physics, Konkoly-Thege Mikl\'{o}s \'{u}t 29-33, 1121 Budapest, Hungary}
\affiliationAdd{\SzegedU}{Department of Theoretical Physics, University of Szeged, Tisza Lajos krt. 84-86, 6720 Szeged, Hungary}

\affiliationAdd{\TorVergata}{Dipartimento di Fisica, Università degli Studi di Roma Tor Vergata, Via della Ricerca Scientifica 1, 00133, Roma, Italy}
\affiliationAdd{\UniBo}{Dipartimento di Fisica e Astronomia "Augusto Righi", Alma Mater Studiorum, University of Bologna
Viale Berti Pichat 6/2, Bologna, Italy}
\affiliationAdd{\INFNBologna}{INFN–Bologna, Viale C. Berti Pichat, 6/2 – 40127 Bologna, Italy}
\affiliationAdd{\INAFOAS}{INAF - Osservatorio di Astrofisica e Scienza dello Spazio di Bologna, via Piero Gobetti 93/3, I-40129 Bologna, Italy}
\affiliationAdd{\UniPadova}{Physics and Astronomy Department ``Galileo Galilei'', University of Padova, Via Marzolo 8, 35131 Padova, Italy}
\affiliationAdd{\INFNPadova}{INFN–Padova, Via Marzolo 8, 35131 Padova, Italy}
\affiliationAdd{\INAFPadova}{ INAF-Osservatorio Astronomico di Padova,  Vicolo dell'Osservatorio, 5, I-35122 Padova (PD), Italy}
\affiliationAdd{\gssi}{Gran Sasso Science Institute, L'Aquila (AQ), I-67100, Italy}
\affiliationAdd{\infnlngs}{INFN - Laboratori Nazionali del Gran Sasso, Assergi, I-67100, Italy}
\affiliationAdd{\infnlabs}{INFN, Laboratori Nazionali del Gran Sasso, Via Giovanni Acitelli 22, I-67100 Assergi, Italy}
\affiliationAdd{\sapienza}{Dipartimento di Fisica, Sapienza Università di Roma, Piazzale Aldo Moro 5, 00185, Roma, Italy}
\affiliationAdd{\INFNRoma}{INFN, Sezione di Roma, Piazzale Aldo Moro 2, 00185, Roma, Italy}
\affiliationAdd{\INFNFerrara}{INFN Sezione di Ferrara, Via Saragat 1, 44122 Ferrara, Italy}
\affiliationAdd{\INFNFirenze}{INFN Sezione di Firenze, Sesto Fiorentino, I-50019 , Firenze, Italy}
\affiliationAdd{\INFNCatania}{INFN Sezione di Catania, Via S. Sofia 64, 95123 Catania, Italy}
\affiliationAdd{\UniMilanoB}{Dipartimento di Fisica G. Occhialini, Università di Milano-Bicocca, Piazza della Scienza 3, I-20126 Milano, Italy}
\affiliationAdd{\INFNMilanoB}{INFN, Sezione di Milano-Bicocca, Piazza della Scienza 3, I-20126 Milano, Italy}
\affiliationAdd{\INAFMilano}{INAF, Osservatorio Astronomico di Brera, Via E. Bianchi 46, I-23807 Merate, Italy}
\affiliationAdd{\UniMilanoS}{Dipartimento di Fisica Aldo Pontremoli, Università degli Studi di Milano, Via Celoria 16, 20133 Milano, Italy}
\affiliationAdd{\INFNMilanoS}{INFN, Sezione di Milano, Via Celoria 16, 20133 Milano, Italy}
\affiliationAdd{\IASFMilano}{INAF, Istituto di Astrofisica Spaziale e Fisica Cosmica, Via Alfonso Corti 12, I-20133, Milano, Italy}
\affiliationAdd{\INAFNapoli}{INAF, Osservatorio Astronomico di Capodimonte, Salita Moiariello 16, I-80131, Naples, Italy}
\affiliationAdd{\UniNa}{Department of Physics ``Ettore Pancini”, Università degli Studi di Napoli Federico II, Via Cinthia, 21, I-80126 Napoli, Italy} 
\affiliationAdd{\INAFCagliari}{INAF-Osservatorio Astronomico di Cagliari, via della Scienza 5, 09047 Selargius (CA), Italy}
\affiliationAdd{\UniCagliari}{Dipartimento di Fisica, Università degli Studi di Cagliari, SP Monserrato--Sestu km 0.7, 09042 Monserrato, Italy.}
\affiliationAdd{\UniTrento}{Dipartimento di Fisica, Università di Trento, Via Sommarive 14, 38123 Trento, Italy}
\affiliationAdd{\INFNTrento}{INFN-TIFPA, Trento Institute for Fundamental Physics and Applications, via Sommarive 14, I-38123 Trento, Italy}
\affiliationAdd{\UniPisa}{Dipartimento di Fisica “E. Fermi”, Università di Pisa, Largo B. Pontecorvo 3, I-56127 Pisa, Italy}
\affiliationAdd{\InfnPisa}{INFN-Pisa, Largo B. Pontecorvo 3, I-56127 Pisa, Italy}
\affiliationAdd{\INFNPerugia}{INFN–Perugia, Via A.Pascoli snc, 06123 Perugia, Italy}
\affiliationAdd{\UniUrbino}{Universit\`a degli Studi di Urbino ``Carlo Bo'', I-61029, Urbino, Italy}
\affiliationAdd{\unisa}{Dipartimento di Fisica E.R. Caianiello, Universit\`a di Salerno, Italy}
\affiliationAdd{\INFNSa}{INFN - Gruppo Collegato di Salerno, Italy}
\affiliationAdd{\CNRSa}{Istituto SPIN-CNR, Salerno Italy}
\affiliationAdd{\unibas}{Dipartimento di Ingegneria, Universit\`a della Basilicata, Italy}
\affiliationAdd{\uniRomaTre}{Dipartimento di Ingegneria Industriale Elettronica e Meccanica, Universit\`a di Roma Tre, Italy}
\affiliationAdd{\SISSA}{SISSA, Via Bonomea 265, 34136 Trieste, Italy \&  INFN Sezione di Trieste}
\affiliationAdd{\INAFTeramo}{INAF–Osservatorio Astronomico d’Abruzzo, Teramo, Italy}
\affiliationAdd{\INAFRoma}{INAF–Osservatorio Astronomico di Roma, via Frascati 33, 00078 Monte Porzio Catone (RM), Italy}

\affiliationAdd{\ifpu}{IFPU - Institute for Fundamental Physics of the Universe, Via Beirut 2, 34014 Trieste, Italy}
\affiliationAdd{\TOINFN}{INFN sezione di Torino, Torino, 10125, Italy}

\affiliationAdd{\IPMU}{Kavli IPMU (WPI), UTIAS, The University of Tokyo, Kashiwa, Chiba 277-8583, Japan}

\affiliationAdd{\ITPUtrecht}{Institute for Theoretical Physics, Utrecht University, Princetonplein 5, 3584 CC Utrecht, The Netherlands}
\affiliationAdd{\GRASP}{Institute for Gravitational and Subatomic Physics (GRASP), Utrecht University, Princetonplein 1, 3584 CC Utrecht, The Netherlands}

\affiliationAdd{\Nikhef}{Nikhef, Science Park 105, 1098 XG Amsterdam, The Netherlands}
\affiliationAdd{\GRAPPA}{GRAPPA, Anton-Pannekoek Institute for Astronomy, Institute of Physics, Universiteit van Amsterdam, Amsterdam, The Netherlands}
\affiliationAdd{\UniAmsterdam}{Anton Pannekoek Institute for Astronomy, University of Amsterdam, 1090 GE Amsterdam, The Netherlands}
\affiliationAdd{\Radboud}{Department of Astrophysics/IMAPP, Radboud University, P.O. Box 9010, NL-6500 GL Nijmegen, The Netherlands}
\affiliationAdd{\SRON}{SRON, Netherlands Institute for Space Research, Niels Bohrweg 4, NL-2333 CA Leiden, The Netherlands}
\affiliationAdd{\VSI}{Van Swinderen Institute for Particle Physics and Gravity, University of Groningen, Nijenborgh 3, 9747 AG Groningen, The Netherlands}

\affiliationAdd{\UniWroclaw}{Institute of Theoretical Physics, University of Wroc{\l}aw, Pl. M. Borna 9, 50-204 Wroc\l{}aw, Poland}
\affiliationAdd{\CAMK}{Nicolaus Copernicus Astronomical Center of the Polish Academy of Sciences, ul. Bartycka 18, 00-716 Warszawa, Poland}
\affiliationAdd{\WroclawUniScienceTech}{Institute of Theoretical Physics, Wroc{\l}aw University of Science and Technology, Wybrzeże Wyspiańskiego 27, 50-370 Wroc\l{}aw, Poland}

\affiliationAdd{\OAUW}{Astronomical Observatory, University of Warsaw, al. Ujazdowskie 4,
00-478 Warsaw, Poland}

\affiliationAdd{\UniCoimbra}{CFisUC, Department of Physics, University of Coimbra, Rua Larga P-3004-516, Coimbra, Portugal}
\affiliationAdd{\centra}{CENTRA, Departamento de F\'{\i}sica, Instituto Superior T\'ecnico -- IST, Universidade de Lisboa -- UL, Avenida Rovisco Pais 1, 1049-001 Lisboa, Portugal}

\affiliationAdd{\IFIN}{National Institute for Physics and Nuclear Engineering (IFIN-HH), RO-077125 Bucharest, Romania}

\affiliationAdd{\BIST}{Institut de F\'isica d'Altes Energies (IFAE), The Barcelona Institute of Science and Technology, UAB Campus, E-08193 Barcelona, Spain}

\affiliationAdd{\ICE}{Institute of Space Sciences (ICE, CSIC), Campus UAB, Carrer de Magrans, 08193 Barcelona, Spain}
\affiliationAdd{\IEEC}{Institut d’Estudis Espacials de Catalunya (IEEC), Edifici RDIT, Campus UPC, 08860 Castelldefels (Barcelona), Spain}
\affiliationAdd{\UPVEHU}{Department of Physics, University of the Basque Country UPV/EHU, 48080, Bilbao, Spain}
\affiliationAdd{\EHUQC}{EHU Quantum Center, University of the Basque Country UPV/EHU, Bilbao, Spain}
\affiliationAdd{\IKERBASQUE}{IKERBASQUE, Basque Foundation for Science, 48011, Bilbao, Spain}
\affiliationAdd{\ULE}{Department of Mathematics, University of León, Campus de Vegazana, s/n 24071 León, Spain}
\affiliationAdd{\UCM}{Dept. F{\'i}sica Te{\'o}rica and IPARCOS, Univ. Computense de Madrid, Spain}
\affiliationAdd{\IFTUAM}{Instituto de F\'isica Te\'orica, Universidad Aut\'onoma de Madrid, Cantoblanco, 28049 Madrid, Spain}
\affiliationAdd{\iem}{Instituto de Estructura de la Materia, CSIC, Serrano 121, 28006 Madrid, Spain}
\affiliationAdd{\ific}{Instituto de F\'isica Corpuscular (IFIC), Universitat de Val\`{e}ncia-CSIC,\\C/ Catedrático José Beltrán 2, E-46980, Paterna, Spain}
\affiliationAdd{\DFTUSalamanca}{Department of Fundamental Physics and IUFFyM, University of Salamanca, Plaza de la Merced S/N E-37008, Salamanca, Spain}
\affiliationAdd{\USALDfp}{Department of Fundamental Physics, University of Salamanca, Plaza de la Merced s/n E-37008 Salamanca, Spain}
\affiliationAdd{\DAAValencia}{Departament d'Astronomia i Astrofísica, Universitat de València, E-46100 Burjassot, València, Spain}
\affiliationAdd{\ObsValencia}{Observatori Astronòmic, Universitat de València, E-46980 Paterna, València, Spain}
\affiliationAdd{\IFFMadrid}{Instituto de F\'isica Fundamental, Consejo Superior de Investigaciones Cient\'ificas, c/. Serrano 121, E-28006, Madrid, Spain}
\affiliationAdd{\ICBarcelona}{
Institut de Ciències del Cosmos Universitat de Barcelona (ICCUB), c. Martí i
Franquès, 1, 08028 Barcelona, Spain}

\affiliationAdd{\UIB}{Institut d’Aplicacions Computacionals i de Codi Comunitari (IAC3) - IEEC, Universitat de les Illes Balears, Crta. Valldemossa km 7.5, E-07122 Palma, Spain}

\affiliationAdd{\DPTunige}{D\'epartement de Physique Th\'eorique, Universit\'e de Gen\`eve, 24 quai Ernest Ansermet, 1211 Gen\`eve 4, Switzerland}
\affiliationAdd{\AstroGeneve}{D\'epartement d’Astronomie, Universit\'e de Gen\`eve, Chemin Pegasi 51, CH-1290 Versoix, Switzerland}

\affiliationAdd{\gwsc}{Gravitational Wave Science Center (GWSC), Universit\'e de Gen\`eve, CH-1211 Geneva, Switzerland}
\affiliationAdd{\CERN}{CERN, Theoretical Physics Department, Esplanade des Particules 1, Gen\`eve 1211, Switzerland}
\affiliationAdd{\UZurich}{Universit\"{a}t Z\"{u}rich, Winterthurerstrasse 190, 8057 Z\"{u}rich, Switzerland}

\affiliationAdd{\UniBirmingham}{School of Physics and Astronomy and Institute for Gravitational Wave Astronomy, University of Birmingham, Edgbaston, Birmingham, B15 2TT, United Kingdom}

\affiliationAdd{\Birmingham}{School of Physics and Astronomy and Institute for Gravitational Wave Astronomy, University of Birmingham, Edgbaston, Birmingham, B15 2TT, United Kingdom}

\affiliationAdd{\Cambridge}{Institute of Astronomy, University of Cambridge, Madingley Road, Cambridge, CB3 0HA, United Kingdom}
\affiliationAdd{\damtp}{DAMTP, Centre for Mathematical Sciences, University of Cambridge, Wilberforce Road, Cambridge CB3 0WA, United Kingdom}
\affiliationAdd{\KavliCambridge}{
Kavli Institute for Cosmology, University of Cambridge, Madingley Road, Cambridge CB3 0HA, United Kingdom}

\affiliationAdd{\UniCardiff}{Gravity Exploration Institute, School of Physics and Astronomy, Cardiff University, Cardiff, CF24 3AA, United Kingdom}

\affiliationAdd{\Cardiff}{School of Physics and Astronomy, Cardiff University, Cardiff, CF24 3AA, United Kingdom}

\affiliationAdd{\UniGlas}{Institute for Gravitational Research, School of Physics and Astronomy, University of Glasgow, Glasgow G12 8QQ, Scotland, United Kingdom}

\affiliationAdd{\ICL}{Blackett Laboratory, Imperial College London, SW7 2AZ, United Kingdom}

\affiliationAdd{\UCL}{Department of Physics and Astronomy, University College London, London WC1E 6BT, United Kingdom}
\affiliationAdd{\smsqm}{School of Mathematical Sciences, Queen Mary University of London, Mile End Road, London, E1 4NS, United Kingdom}

\affiliationAdd{\NotthinghamCG}{Nottingham Centre of Gravity, Nottingham NG7 2RD, United Kingdom}
\affiliationAdd{\UniNotthingham}{School of Mathematical Sciences \& School of Physics and Astronomy, University of Nottingham, University Park, NG7 2RD, Nottingham, UK}

\affiliationAdd{\icgport}{Institute of Cosmology and Gravitation, University of Portsmouth, Burnaby Road, Portsmouth PO1 3FX, United Kingdom}

\affiliationAdd{\STAG}{Mathematical Sciences and STAG Research Centre, University of Southampton, Southampton SO17 1BJ, United Kingdom}

\affiliationAdd{\Warwick}{Department of Physics, University of Warwick, Coventry, CV4 7AL, United Kingdom}

\affiliationAdd{\Oxford}{Astrophysics, Department of Physics, University of Oxford, Keble Road, Oxford, OX1 3RH, United Kingdom}

\affiliationAdd{\Belfast}{Astrophysics Research Centre, School of Mathematics and Physics, Queen’s University Belfast, Belfast BT7 1NN, United Kingdom}

\affiliationAdd{\Swansea}{Physics Department, Swansea University, SA2 8PP, United Kingdom}

\affiliationAdd{\USanDiego}{Department of Astronomy and Astrophysics, University of California, San Diego, La Jolla, CA 92093, USA}

\affiliationAdd{\princeton}{Department of Astrophysical Sciences, Princeton University, 4 Ivy Lane, Princeton, NJ 08544, USA}

\affiliationAdd{\IAS}{School of Natural Sciences, Institute for Advanced Study, Princeton, NJ 08540, USA}

\affiliationAdd{\flatiron}{Center for Computational Astrophysics, Flatiron Institute, 162 Fifth Avenue, New York, NY 10010, USA}

\affiliationAdd{\JohnsHopkins}{William H. Miller III Department of Physics and Astronomy, Johns Hopkins University, Baltimore, Maryland 21218, USA}

\affiliationAdd{\UMD}{Department of Physics, University of Maryland, College Park, MD 20742, USA}

\affiliationAdd{\UPenn}{Center for Particle Cosmology, Department of Physics and Astronomy, University of Pennsylvania 209 South 33rd Street, Philadelphia, Pennsylvania 19104, USA}

\affiliationAdd{\PennState}{Institute for Gravitation and the Cosmos, Department of Physics, Pennsylvania State University, University Park, PA 16802, USA}

\affiliationAdd{\PennStateAA}{Department of Astronomy and Astrophysics, The Pennsylvania State University, University Park, PA 16802, USA}

\affiliationAdd{\NCarolina}{Department of Physics and Astronomy, University of North Carolina at Chapel Hill, Chapel Hill, NC 27599-3255, USA}
\affiliationAdd{\JSSI}{Joint Space-Science Institute, University of Maryland, College Park, MD 20742, USA}

\affiliationAdd{\GWU}{Department of Physics, The George Washington University, 725 21st St NW, Washington, DC 20052, USA}

\affiliationAdd{\Berkeley}{Department of Physics, University of California, Berkeley, CA 94720, USA}

\affiliationAdd{\UMassD}{Department of Mathematics, Center for Scientific Computing and Data Science Research, University of Massachusetts, Dartmouth, MA 02747, USA}

\affiliationAdd{\UIUC}{The Grainger College of Engineering,
Department of Physics \& Illinois Center for Advanced Studies of the Universe, University of Illinois Urbana-Champaign, Urbana, Illinois 61801, USA}

\affiliationAdd{\UNewHampshire}{Department of Physics \& Astronomy, University of New Hampshire, 9 Library Way, Durham, NH 03824, USA}

\affiliationAdd{\kuwait}{Department of Physics, College of Science, Kuwait University, Sabah Al Salem University City, P.O. Box 2544, Safat 1320, Kuwait}

\affiliationAdd{\KHU}{Kyung Hee University, 1732, Deogyeong-daero, Giheung-gu, Yongin-si, Gyeonggi-do 17104, Republic of Korea}
\affiliationAdd{\KASI}{Korea Astronomy and Space Science Institute, 776 Daedeok-daero, Yuseong-gu, Daejeon 34055, Republic of Korea}
\affiliationAdd{\EWU}{Ewha Womans University, 52, Ewhayeodae-gil, Seodaemun-gu, Seoul 03760, Republic of Korea}
\affiliationAdd{\YonseiUniversity}{Yonsei University, 50 Yonsei-ro, Seodaemun-gu, Seoul, Republic of Korea}

\affiliationAdd{\MaastrichtUni}{Maastricht University, 6200 MD Maastricht, Netherlands}
\affiliationAdd{\VUAmsterdam}{Vrije Universiteit Amsterdam, De Boelelaan 1105, 1081 HV Amsterdam, Netherlands}

\affiliationAdd{\IPARCOSUCM}{Instituto de F\'isica de Part\'iculas y del Cosmos, Universidad Complutense de Madrid, Avda. de S\'eneca, 2 Ciudad Universitaria 28040 MADRID, Spain}

\affiliationAdd{\CSIC}{Consejo Superior de Investigaciones Cient\'ificas, Calle de Serrano, 117, 28006 Madrid, Spain}
\affiliationAdd{\ICREA}{Institucio Catalana de Recerca i Estudis Avan\c{c}ats, Passeig de Llu\'is Companys, 23, 08010 Barcelona, Spain}

\affiliationAdd{\PhysBRWTH}{3. Physikalisches Institut B, RWTH Aachen University, Physikzentrum, Otto-Blumenthal-Stra{\ss}e 26, 52074 Aachen, Germany}
\affiliationAdd{\ILT}{Fraunhofer Institute for Laser Technology ILT, Steinbachstra{\ss}e 15, 52074 Aachen, Germany}
\affiliationAdd{\UHAM}{Universit\"at Hamburg, Mittelweg 177, 20148 Hamburg, Germany}
\affiliationAdd{\ECAP}{Erlangen Centre for Astroparticle Physics, Erlangen Centre for Astroparticle Physics (ECAP), Friedrich-Alexander-Universit\"at Erlangen-N\"urnberg, Nikolas Fiebinger Strasse 2, 91058 Erlangen, Germany}
\affiliationAdd{\TUBS}{Technische Universit\"at Braunschweig, Institut f\"ur Halbleitertechnik (IHT) Hans-Sommer-Str. 66 38106 Braunschweig , Germany}
\affiliationAdd{\InstitutfurPhysikundAstronomieUniversitatPotsdam}{Institut f\"ur Physik und Astronomie, Universit\"at Potsdam, Haus 28, Karl-Liebknecht-Str. 24/25, Potsdam, Germany}
\affiliationAdd{\HeidelbergUniversity}{Heidelberg University, Grabengasse 1, 69117 Heidelberg, Germany}
\affiliationAdd{\HSU}{Helmut Schmidt Universit\"at, Holstenhofweg 85, 22043 Hamburg, Germany}
\affiliationAdd{\PhysARWTH}{3. Physikalisches Institut A, RWTH Aachen University, Physikzentrum, Otto-Blumenthal-Stra{\ss}e 26, 52074 Aachen, Germany}
\affiliationAdd{\LUHIG}{Institut f\"ur Gravitationsphysik, Institut f\"ur Gravitationsphysik Leibniz Universit\"at Hannover Callinstr. 38 30167 Hannover, Germany}

\affiliationAdd{\KITETP}{Karlsruhe Institute of Technology - Institute of Experimental Particle Physics, 76021 Karlsruhe, Germany}

\affiliationAdd{\SNS}{Scuola Normale Superiore, Piazza dei Cavalieri 7, 56126, Pisa}

\affiliationAdd{\LNFINFN}{Laboratori Nazionali di Frascati dell’Istituto Nazionale di Fisica Nucleare, Via Enrico Fermi 54 00044 Frascati (Rome), Italy}
\affiliationAdd{\SSM}{Scuola Superiore Meridionale, Largo San Marcellino 10, 80138 Napoli, Italy}
\affiliationAdd{\INFNCagliari}{Istituto Nazionale di Fisica Nucleare - Sezione di Cagliari, Dipartimento di Fisica, Complesso Universitario di Monserrato, Strada provinciale per Sestu, Km 1, 09042 Monserrato CA, Italy}
\affiliationAdd{\INGVPisa}{Istituto Nazionale di Geofisica e Vulcanologia, Sezione di Pisa, Via Cesare Battisti 53, 56125 Pisa, Italy}
\affiliationAdd{\INAFOATo}{National Institue for Astrophysics - Astrophysical Observatory of Turin, Via Osservatorio 20, 10095- Pino Torinese, Italy}
\affiliationAdd{\UNITOMathematicsDept}{University of Turin, Department  of Mathematics "G. Peano”, Via Carlo Alberto 10 - 10123 Torino, Italy}
\affiliationAdd{\UniSS}{Universit\`a degli Studi di Sassari, Department of Chemical, Physical, Mathematical and Natural Science, Via Vienna 2, 07100 Sassari, Italy}
\affiliationAdd{\INFNLNS}{INFN Laboratori Nazionali del Sud, via S.Sofia 62, 95123 Catania, Italy}
\affiliationAdd{\UniSannio}{Universit\`a del Sannio, I-82100 Benevento, Italy}
\affiliationAdd{\INFNNA}{Istituto Nazionale di Fisica Nucleare - Sezione di Napoli, Complesso Universitario di Monte S.Angelo, via Cintia - 80126 Napoli, Italy}

\affiliationAdd{\INGVBologna}{Istituto Nazionale di Geofisica e Vulcanologia, Sezione di Bologna, Viale Berti Pichat 6/2, 40127 Bologna, Italy}
\affiliationAdd{\UniCA}{Universit\`a degli Studi di Cagliari, Via Universit\`a 40, 09124 Cagliari, Italy}
\affiliationAdd{\EGO}{European Gravitational Observatory, I-56021 Cascina, Pisa, Italy}
\affiliationAdd{\UniPDDFA}{Universit\`a di Padova, Dipartimento di Fisica e Astronomia, via Marzolo, 8, I-35131 Padova, Italy}
\affiliationAdd{\INGVRomaone}{Istituto Nazionale di Geofisica e Vulcanologia, Sezione di Roma 1, Via di Vigna Murata 605, 00143 Roma, Italy}
\affiliationAdd{\UNITOPhysicsDept}{University of Turin, Department of Physics, Via P. Giuria, 1 - 10125 Torino, Italy}
\affiliationAdd{\INFNTrieste}{Istituto Nazionale di Fisica Nucleare - Sezione di Trieste, via Valerio 2 - 34127 Trieste, Italy}
\affiliationAdd{\UniNA}{Universit\`a di Napoli ``Federico II", Complesso Universitario di Monte S.Angelo, via Cintia, 80126 Napoli, Italy}
\affiliationAdd{\UniFerrara}{Dipartimento di Fisica e Scienze della Terra, Università di Ferrara, Via Saragat 1 - 44122 Ferrara, Italy}

\affiliationAdd{\INFNRomaTorVergata}{INFN, Sezione di Roma Tor Vergata, Via della Ricerca Scientifica 1, I-00133, Italy}

\affiliationAdd{\KCL}{King's College London, Physics Department, Strand, London WC2R 2LS, United Kingdom}

\affiliationAdd{\UGent}{Universiteit Gent, B-9000 Gent, Belgium}
\affiliationAdd{\UCLouvain}{Universit\'e catholique de Louvain, B-1348 Louvain-la-Neuve, Belgium}

\affiliationAdd{\IPtwoILyonINtwoPthree}{Institut de Physique des 2 Infinis de Lyon - IN2P3, Universit\'e Claude Bernard Lyon 1, CNRS, 4 rue Enrico Fermi, 69622 Villeurbanne Cedex, France}
\affiliationAdd{\LMAINtwoPthree}{Laboratoire des Mat\'eriaux Avanc\'es, IP2I Lyon - IN2P3, Universit\'e Claude Bernard Lyon 1, CNRS, B\^atiment Virgo, 7 Av. Pierre de Coubertin, 69100 Villeurbanne, France}

\affiliationAdd{\CFisUCDeptPhysicsUniCoimbra}{CFisUC, Department of Physics, University of Coimbra, Rua Larga, 3004-516 Coimbra, Portugal}

\affiliationAdd{\UniGeDepartmentofNuclearandParticlePhysics}{Department of Nuclear and Particle Physics, Universit\'e de Gen\`eve, quai Ansermet 24, 1205 Gen\`eve, Switzerland}

\affiliationAdd{\UFESVitoria}{Universidade Federal do Esp\'irito Santo, Av. Fernando Ferrari, 514 - Goiabeiras, Vit\'oria - ES, 29075-910, Brazil}

\affiliationAdd{\HUNRENCSFKKonkolyObservatory}{Konkoly Observatory, HUN-REN Research Centre for Astronomy and Earth Sciences, H-1121 Budapest, Konkoly Thege Mikl\'os \'ut 15-17., Hungary}

\affiliationAdd{\HUNRENCSFK}{HUN-REN CSFK, MTA Centre of Excellence, H-1121, Budapest, Konkoly Thege Mikl\'os \'ut 15-17, Hungary}

\affiliationAdd{\ELTE}{ELTE E\"otv\"os Lor\'and University, Institute of Physics and Astronomy, H-1117, Budapest, Pázmány Péter sétány 1/a, Hungary}

\affiliationAdd{\Stavanger}{Institute of Mathematics and Physics, University of Stavanger, NO-4036, Stavanger, Norway}

\author{Adrian Abac\orcidlink{0000-0003-4786-2698},}
\affiliation{\AEI}

\author{Raul Abramo\orcidlink{0000-0001-8295-7022},}
\affiliation{\ifusp}

\author{Simone Albanesi \orcidlink{0000-0001-7345-4415},}
\affiliation{\UniJena}
\affiliation{\TOINFN}

\author{Angelica Albertini \orcidlink{0000-0002-9556-1323},}
\affiliation{\ASU}
\affiliation{\Charles}

\author{Alessandro Agapito\orcidlink{0009-0005-9004-3163},} 
\affiliation{\sapienza} 
\affiliation{\INFNRoma}
\affiliation{\mrs}

\author{Michalis Agathos\orcidlink{0000-0002-9072-1121},}
\affiliation{\damtp}
\affiliation{\smsqm}

\author{Conrado Albertus\orcidlink{0000-0002-0248-8260},}
\affiliation{\DFTUSalamanca}

\author{Nils Andersson\orcidlink{0000-0001-8550-3843},}
\affiliation{\STAG}

\author{Tomás Andrade\orcidlink{0000-0001-6000-1302},}
\affiliation{\ICBarcelona}

\author{Igor Andreoni\orcidlink{0000-0002-8977-1498},}
\affiliation{\NCarolina}
\affiliation{\JSSI}

\author{Federico Angeloni\orcidlink{0009-0007-8669-6393},}
\affiliation{\sapienza}
\affiliation{\TorVergata}
\affiliation{\INFNRoma}
\affiliation{\INAFRoma}

\author{Marco Antonelli\orcidlink{0000-0002-5470-4308},}
\affiliation{\LPC}

\author{John Antoniadis\orcidlink{0000-0003-4453-3776},}
\affiliation{\Heraklion}
\affiliation{\MPBonn}

\author{Fabio Antonini\orcidlink{0000-0003-3138-6199},}
\affiliation{\UniCardiff}

\author{Manuel Arca Sedda\orcidlink{0000-0002-3987-0519},}
\affiliation{\gssi}
\affiliation{\infnlngs}
\affiliation{\INAFTeramo}
\affiliation{\UniPadova}

\author{M. Celeste Artale \orcidlink{0000-0003-0570-785X},}
\affiliation{\AndresBello}
\affiliation{\zah}

\author{Stefano Ascenzi\orcidlink{0000-0001-5116-6789},}
\affiliation{\gssi}

\author{Pierre Auclair\orcidlink{0000-0002-4814-1406},}
\affiliation{\CurlLouvain}

\author{Matteo Bachetti\orcidlink{0000-0002-4576-9337},}
\affiliation{\INAFCagliari}

\author{Charles Badger\orcidlink{0000-0002-5238-5622},}
\affiliation{\KCL}

\author{Biswajit Banerjee\orcidlink{0000-0002-8008-2485},}
\affiliation{\gssi}

\author{David Barba-González\orcidlink{0000-0002-1518-5584},}
\affiliation{\DFTUSalamanca}

\author{Dániel Barta\orcidlink{0000-0001-6841-550X},}
\affiliation{\HUNWigner}

\author{Nicola Bartolo\orcidlink{0000-0001-8584-6037},}
\affiliation{\UniPadova}
\affiliation{\INFNPadova}
\affiliation{\INAFPadova}

\author{Andreas Bauswein,}
\affiliation{\GSI}

\author{Andrea Begnoni\orcidlink{0000-0003-4603-6992},}
\affiliation{\UniPadova}
\affiliation{\INFNPadova}

\author{Freija Beirnaert\orcidlink{0000-0002-4003-7233},}
\affiliation{\UGent}

\author{Micha{\l} Bejger\orcidlink{0000-0002-4991-8213},}
\affiliation{\INFNFerrara}
\affiliation{\CAMK}

\author{Enis Belgacem\orcidlink{0000-0003-4920-0911},}
\affiliation{\DPTunige}
\affiliation{\gwsc}

\author{Nicola Bellomo\orcidlink{0000-0002-4375-705X},}
\affiliation{\UniPadova}
\affiliation{\INFNPadova}
\affiliation{\INAFPadova}

\author{Laura Bernard \orcidlink{0000-0003-2856-1662},}
\affiliation{\Meudon}

\author{Maria Grazia Bernardini\orcidlink{0000-0001-6106-3046},}
\affiliation{\INAFMilano}

\author{Sebastiano Bernuzzi\orcidlink{0000-0002-2334-0935},}
\affiliation{\UniJena}

\author{Christopher P. L. Berry\orcidlink{0000-0003-3870-7215},}
\affiliation{\UniGlas}

\author{Emanuele Berti\orcidlink{0000-0003-0751-5130},}
\affiliation{\JohnsHopkins}

\author{Gianfranco Bertone\orcidlink{0000-0002-6191-1487},}
\affiliation{\GRAPPA}

\author{Dario Bettoni\orcidlink{0000-0002-0176-5537},}
\affiliation{\ULE}
\affiliation{\DFTUSalamanca}

\author{Miguel Bezares\orcidlink{0000-0002-3739-6980},}
\affiliation{\NotthinghamCG}
\affiliation{\UniNotthingham}

\author{Swetha Bhagwat\orcidlink{0000-0003-4700-5274},}
\affiliation{\UniBirmingham}

\author{Sofia Bisero\orcidlink{0009-0005-6643-1473},}
\affiliation{\OBSPM}

\author{Marie Anne Bizouard\orcidlink{0000-0002-4618-1674},}
\affiliation{\Artemis}

\author{Jose J. Blanco-Pillado\orcidlink{0000-0003-2260-9047},}
\affiliation{\UPVEHU}
\affiliation{\EHUQC}
\affiliation{\IKERBASQUE}

\author{Simone Blasi\orcidlink{0000-0002-9578-624X},}
\affiliation{\DESY}
\affiliation{\VUB}

\author{Alice Bonino\orcidlink{0000-0001-6502-284X},}
\affiliation{\UniBirmingham}

\author{Alice Borghese\orcidlink{0000-0001-8785-5922},}
\affiliation{\INAFRoma}

\author{Nicola Borghi\orcidlink{0000-0002-2889-8997},}
\affiliation{\UniBo}
\affiliation{\INAFOAS}

\author{Ssohrab Borhanian\orcidlink{0000-0003-0161-6109},}
\affiliation{\UniJena}
\affiliation{\UniMilanoB}
\affiliation{\INFNMilanoB}

\author{Elisa Bortolas\orcidlink{0000-0001-9458-821X},}
\affiliation{\INAFPadova}
\affiliation{\UniMilanoB}
\affiliation{\INFNMilanoB}

\author{Maria Teresa Botticella\orcidlink{0000-0002-3938-692X},}
\affiliation{\INAFNapoli}

\author{Marica Branchesi\orcidlink{0000-0003-1643-0526},}
\emailAdd{marica.branchesi@gssi.it}
\affiliation{\gssi}
\affiliation{\infnlngs}

\author{Matteo Breschi \orcidlink{0000-0002-3327-3676},}
\affiliation{\UniJena}

\author{Richard Brito\orcidlink{0000-0002-7807-3053},} 
\affiliation{\centra} 

\author{Enzo Brocato\orcidlink{0000-0001-7988-8177},}
\affiliation{\INAFTeramo}
\affiliation{\INAFRoma}

\author{Floor S. Broekgaarden\orcidlink{0000-0002-4421-4962},}
\affiliation{\USanDiego}

\author{Tomasz Bulik\orcidlink{0000-0003-2045-4803},}
\affiliation{\OAUW}

\author{Alessandra Buonanno \orcidlink{0000-0002-5433-1409},}
\affiliation{\AEI}
\affiliation{\UMD}

\author{Fiorella Burgio\orcidlink{0000-0003-2195-5693},}
\affiliation{\INFNCatania}

\author{Adam Burrows\orcidlink{0000-0002-3099-5024},}
\affiliation{\princeton}

\author{Gianluca Calcagni\orcidlink{0000-0003-2631-4588},}
\affiliation{\iem}

\author{Sofia Canevarolo\orcidlink{0000-0001-5091-0780},}
\affiliation{\ITPUtrecht}

\author{Enrico Cappellaro\orcidlink{0000-0001-5008-8619},} 
\affiliation{\INAFPadova}

\author{Giulia Capurri\orcidlink{0000-0003-0889-1015},}
\affiliation{\UniPisa}
\affiliation{\InfnPisa}

\author{Carmelita Carbone \orcidlink{0000-0003-0125-3563},}
\affiliation{\IASFMilano}

\author{Roberto Casadio\orcidlink{0000-0002-1330-778},}
\affiliation{\UniBo}
\affiliation{\INFNBologna}

\author{Ramiro Cayuso\orcidlink{0000-0001-6271-0233},}
\affiliation{\SISSA}
\affiliation{\ifpu}

\author{Pablo Cerdá-Durán\orcidlink{0000-0003-4293-340X},}
\affiliation{\DAAValencia}
\affiliation{\ObsValencia}

\author{Prasanta Char\orcidlink{0000-0001-6592-6590},}
\affiliation{\DFTUSalamanca}

\author{Sylvain Chaty\orcidlink{0000-0002-5769-8601},}
\affiliation{\UniParis}

\author{Tommaso Chiarusi\orcidlink{0000-0001-8454-8644},}
\affiliation{\INFNBologna}

\author{Martyna Chruslinska\orcidlink{0000-0002-8901-6994},}
\affiliation{\ESO}
\affiliation{\MPA}

\author{Francesco Cireddu \orcidlink{0009-0002-7074-4278},}
\affiliation{\KULeuvenI}
\affiliation{\KULeuvenD}
\affiliation{\UniPisa}

\author{Philippa Cole\orcidlink{0000-0001-6045-6358},}
\affiliation{\UniMilanoB}
\affiliation{\INFNMilanoB}

\author{Alberto Colombo\orcidlink{0000-0002-7439-4773},}
\affiliation{\INAFMilano}
\affiliation{\INFNMilanoB}

\author{Monica Colpi\orcidlink{0000-0002-3370-6152},}
\affiliation{\UniMilanoB}

\author{Geoffrey Comp\`ere\orcidlink{0000-0002-1977-3295},}
\affiliation{\ULB}

\author{Carlo Contaldi\orcidlink{0000-0001-7285-0707},}
\affiliation{\ICL}

\author{Maxence Corman\orcidlink{0000-0003-2855-1149},}
\affiliation{\AEI}

\author{Francesco Crescimbeni\orcidlink{0009-0001-4088-5443},} 
\affiliation{\sapienza} 
\affiliation{\INFNRoma}

\author{Sergio Cristallo\orcidlink{0000-0001-9683-9406},}
\affiliation{\INAFTeramo}

\author{Elena Cuoco\orcidlink{0000-0001-6046-1237},}
\affiliation{\UniBo}
\affiliation{\INFNBologna}

\author{Giulia Cusin \orcidlink{0000-0001-6046-1237},}
\affiliation{\IAP}
\affiliation{\DPTunige}

\author{Tito Dal Canton \orcidlink{0000-0001-5078-9044},}
\affiliation{\UniParisSaclay}

\author{Gergely Dálya\orcidlink{0000-0003-3258-5763},}
\affiliation{\LtwoIT}

\author{Paolo D'Avanzo\orcidlink{0000-0001-7164-1508},}
\affiliation{\INAFMilano}

\author{Nazanin Davari\orcidlink{0000-0003-0708-0261},}
\affiliation{\INAFRoma}

\author{Valerio De Luca\orcidlink{0000-0002-1444-5372},}
\affiliation{\UPenn}

\author{Viola De Renzis\orcidlink{0000-0001-7038-735X},} 
\affiliation{\mrs}

\author{Massimo Della Valle\orcidlink{0000-0003-3142-5020},} 
\affiliation{\INAFNapoli}

\author{Walter Del Pozzo\orcidlink{0000-0003-3978-2030},}
\affiliation{\UniPisa}
\affiliation{\InfnPisa}

\author{Federico De~Santi\orcidlink{0009-0000-2445-5729},}
\affiliation{\UniPisa}
\affiliation{\InfnPisa} 

\author{Alessio Ludovico De~Santis \orcidlink{ 0009-0005-4288-3758 }, }
\affiliation{\gssi}
\affiliation{\infnlngs}

\author{Tim Dietrich \orcidlink{0000-0003-2374-307X},}
\affiliation{\UniPotsdam}
\affiliation{\AEI}

\author{Ema Dimastrogiovanni\orcidlink{0000-0001-9576-0608},}
\affiliation{\VSI}

\author{Guillem Domenech \orcidlink{0000-0003-2788-884X},}
\affiliation{\LUH}
\affiliation{\aeihannover}

\author{Daniela Doneva\orcidlink{0000-0001-6519-000X},}
\affiliation{\UniTuebingen}

\author{Marco Drago\orcidlink{0000-0002-3738-2431},}
\affiliation{\sapienza}
\affiliation{\INFNRoma}

\author{Ulyana Dupletsa\orcidlink{0000-0003-2766-247X},}
\affiliation{\gssi}
\affiliation{\infnlngs}

\author{Hannah Duval\orcidlink{0000-0002-2475-1728},}
\affiliation{\VUB}

\author{Irina Dvorkin\orcidlink{0000-0002-2353-9194},}
\affiliation{\IAP}
\affiliation{\IUF}

\author{Nancy Elias-Rosa\orcidlink{0000-0002-1381-9125},}
\affiliation{\INAFPadova}

\author{Stephen Fairhurst\orcidlink{0000-0001-8480-1961},}
\affiliation{\Cardiff}

\author{Anthea F. Fantina\orcidlink{0000-0003-2225-4100},}
\affiliation{\GANIL}

\author{Matteo Fasiello\orcidlink{0000-0002-2532-5202},}
\affiliation{\IFTUAM}

\author{Maxime Fays\orcidlink{0000-0002-4390-9746},}
\affiliation{\STARInstitute}

\author{Rob Fender\orcidlink{0000-0002-5654-2744},}
\affiliation{\Oxford}

\author{Tobias Fischer \orcidlink{0000-0003-2479-344X},}
\affiliation{\UniWroclaw}
\affiliation{\WroclawUniScienceTech}
\affiliation{\Opava}

\author{François Foucart \orcidlink{0000-0003-4617-4738},}
\affiliation{\UNewHampshire}

\author{Tassos Fragos\orcidlink{0000-0003-1474-1523},}
\affiliation{\AstroGeneve}
\affiliation{\gwsc}

\author{Stefano Foffa \orcidlink{0000-0002-4530-3051},}
\affiliation{\DPTunige}
\affiliation{\gwsc}

\author{Gabriele Franciolini \orcidlink{0000-0002-6892-9145},}
\affiliation{\CERN}

\author{Jacopo Fumagalli\orcidlink{0000-0001-8081-7640},}
\affiliation{\ICBarcelona}

\author{Jonathan Gair\orcidlink{0000-0002-1671-3668},}
\affiliation{\AEI}

\author{Rossella Gamba \orcidlink{0000-0001-7239-0659},}
\affiliation{\Berkeley}
\affiliation{\PennState}

\author{Juan~Garcia-Bellido\orcidlink{0000-0002-9370-8360},}
\affiliation{\IFTUAM}

\author{Cecilio García-Quirós \orcidlink{0000-0002-8059-2477},}
\affiliation{\UZurich}

\author{László Árpád Gergely \orcidlink{0000-0003-3146-6201},}
\affiliation{\SzegedU}
\affiliation{\Wigner}

\author{Giancarlo Ghirlanda\orcidlink{0000-0001-5876-9259},}
\affiliation{\INAFMilano}
\affiliation{\INFNMilanoB}

\author{Archisman Ghosh\orcidlink{0000-0003-0423-3533},}
\emailAdd{archisman.ghosh@ugent.be}
\affiliation{\UGent}

\author{Bruno Giacomazzo\orcidlink{0000-0002-6947-4023},}
\affiliation{\UniMilanoB}
\affiliation{\INFNMilanoB}
\affiliation{\INAFMilano}

\author{Fabian Gittins\orcidlink{0000-0002-9439-7701},}
\affiliation{\GRASP}
\affiliation{\Nikhef}
\affiliation{\STAG}

\author{Ines Francesca Giudice\orcidlink{0009-0003-5816-653X},}
\affiliation{\UniNa}
\affiliation{\INAFNapoli}

\author{Boris Goncharov\orcidlink{0000-0003-3189-5807},}
\affiliation{\aeihannover}
\affiliation{\leibnitzuni}

\author{Alejandra Gonzalez \orcidlink{0000-0002-5034-9353},}
\affiliation{\UniJena}
\affiliation{\UIB}

\author{Stéphane Goriély\orcidlink{0000-0002-9110-941X},}
\affiliation{\ULB}

\author{Luca Graziani\orcidlink{0000-0002-9231-1505},}
\affiliation{\sapienza}

\author{Giuseppe Greco\orcidlink{0000-0002-2353-6604},}
\affiliation{\INFNPerugia }

\author{Leonardo Gualtieri\orcidlink{0000-0002-1097-3266},}
\affiliation{\UniPisa}
\affiliation{\InfnPisa}

\author{Gianluca Maria Guidi\orcidlink{0000-0002-3061-9870},}
\affiliation{\UniUrbino}
\affiliation{\INFNFirenze}

\author{Ish Gupta\orcidlink{0000-0001-6932-8715},}
\affiliation{\PennState}

\author{Maria Haney\orcidlink{0000-0001-7554-3665},}
\affiliation{\Nikhef}

\author{Mark Hannam \orcidlink{0000-0001-5571-325X},}
\affiliation{\Cardiff}

\author{Jan Harms\orcidlink{0000-0002-7332-9806},}
\affiliation{\gssi}
\affiliation{\infnlngs}

\author{Arus Harutyunyan\orcidlink{0000-0002-3006-8618},}
\affiliation{\Byurakan} 
\affiliation{\UYerevan}

\author{Brynmor Haskell\orcidlink{0000-0002-8255-3519},}
\affiliation{\UniMilanoS}
\affiliation{\INFNMilanoS}
\affiliation{\CAMK}

\author{Andreas Haungs\orcidlink{0000-0002-9638-7574},}
\affiliation{\KR}

\author{Nandini Hazra\orcidlink{0000-0002-3870-1537},}
\affiliation{\gssi}

\author{Gary Hemming\orcidlink{0000-0001-5268-4465},}
\affiliation{\EGO }

\author{Ik Siong Heng \orcidlink{0000-0002-1977-0019}}
\affiliation{\UniGlas}

\author{Tanja Hinderer \orcidlink{0000-0002-3394-6105},}
\affiliation{\ITPUtrecht}

\author{Alexander van der Horst\orcidlink{0000-0001-9149-6707},}
\affiliation{\GWU}

\author{Qian  Hu\orcidlink{0000-0002-3033-6491},}
\affiliation{\UniGlas}

\author{Sascha Husa\orcidlink{0000-0002-0445-1971},}
\affiliation{\ICE}
\affiliation{\UIB}

\author{Francesco Iacovelli\orcidlink{0000-0002-4875-5862},}
\affiliation{\DPTunige}
\affiliation{\gwsc}
\affiliation{\JohnsHopkins}

\author{Giulia Illuminati\orcidlink{0000-0002-4138-8027},}
\affiliation{\UniBo}
\affiliation{\INFNBologna}

\author{Gianluca Inguglia\orcidlink{0000-0003-0331-8279},}
\affiliation{\AustrianAcademy}

\author{David Izquierdo Villalba\orcidlink{0000-0002-6143-1491},}
\affiliation{\UniMilanoB}

\author{Justin Janquart\orcidlink{0000-0003-2888-7152},}
\affiliation{\UniLouvain}
\affiliation{\BelObs}

\author{Kamiel Janssens\orcidlink{0000-0001-8760-4429},}
\affiliation{\UAntwerpen}

\author{Alexander C. Jenkins\orcidlink{0000-0003-1785-5841},}
\affiliation{\KavliCambridge}
\affiliation{\damtp}

\author{Ian Jones\orcidlink{0000-0002-0117-7567},}
\affiliation{\STAG}

\author{Balázs Kacskovics \orcidlink{0000-0001-9216-8713},}
\affiliation{\Wigner}

\author{Ralf S. Klessen\orcidlink{0000-0002-0560-3172},}
\affiliation{\zah} 
\affiliation{\IWRHeidelberg}

\author{Kostas Kokkotas\orcidlink{0000-0001-6048-2919},}
\affiliation{\UniTuebingen}

\author{Hao-Jui Kuan\orcidlink{0000-0002-3480-3448},}
\affiliation{\AEI}

\author{Sumit Kumar\orcidlink{0000-0002-6404-0517},}
\affiliation{\aeihannover}
\affiliation{\leibnitzuni}

\author{Sachiko Kuroyanagi\orcidlink{0000-0001-6538-1447},}
\affiliation{\IFTUAM}

\author{Danny Laghi\orcidlink{0000-0001-7462-3794},}
\affiliation{\LtwoIT}
\affiliation{\UZurich}

\author{Astrid Lamberts\orcidlink{0000-0001-8740-0127},}
\affiliation{\OCA}
\affiliation{\Artemis}

\author{Gaetano Lambiase\orcidlink{0000-0001-7574-2330},}
\affiliation{\unisa}
\affiliation{\INFNSa}

\author{François Larrouturou\orcidlink{0000-0003-3173-7227},}
\affiliation{\DESY}

\author{Paola Leaci\orcidlink{0000-0002-3997-5046},}
\affiliation{\sapienza}
\affiliation{\INFNRoma}

\author{Michele Lenzi\orcidlink{0000-0002-0131-2829},}
\affiliation{\ICE}
\affiliation{\IEEC}

\author{Andrew Levan \orcidlink{0000-0001-7821-9369}}
\affiliation{\Radboud}
\affiliation{\Warwick}

\author{T.~G.~F.~Li\orcidlink{0000-0003-4297-7365},}
\affiliation{\KULeuvenI}
\affiliation{\KULeuvenD}
\affiliation{\KULeuvenDD}

\author{Yufeng Li\orcidlink{0000-0002-4135-074X},} 
\affiliation{\BNU}

\author{Dicong Liang\orcidlink{0000-0001-5021-235X},}
\affiliation{\SMU}
\affiliation{\KIAA}

\author{Marco Limongi \orcidlink{0000-0003-0636-7834},}
\affiliation{\INAFRoma}
\affiliation{\IPMU}
\affiliation{\INFNPerugia}

\author{Boyuan Liu\orcidlink{0000-0002-4966-7450},}
\affiliation{\zah}
\affiliation{\Cambridge}

\author{Felipe J.\ Llanes-Estrada\orcidlink{ORCID 0000-0002-2565-4516},}
\affiliation{\UCM}

\author{Eleonora Loffredo \orcidlink{0000-0001-7888-9733},}
\affiliation{\INAFTeramo}
\affiliation{\gssi}

\author{Oliver Long \orcidlink{0000-0002-3897-9272},}
\affiliation{\AEI}

\author{Eva Lope-Oter\orcidlink{0000-0002-2567-1009},}
\affiliation{\UCM}

\author{Georgios Lukes-Gerakopoulos \orcidlink{0000-0002-6333-3094},}
\affiliation{\ASU}

\author{Elisa Maggio\orcidlink{0000-0002-1960-8185},} 
\affiliation{\AEI} 

\author{Michele Maggiore\orcidlink{0000-0001-7348-047X},}
\emailAdd{michele.maggiore@unige.ch}
\affiliation{\DPTunige}
\affiliation{\gwsc}

\author{Michele Mancarella\orcidlink{0000-0002-0675-508X},} 
\affiliation{\mrs} 

\author{Michela Mapelli\orcidlink{https://orcid.org/0000-0001-8799-2548},}
\affiliation{\zah}

\author{Pablo Marchant\orcidlink{0000-0002-0338-8181},}
\affiliation{\KULeuvenA}
\affiliation{\UGent}

\author{Annarita Margiotta\orcidlink{0000-0001-6929-5386},}
\affiliation{\UniBo}

\author{Alberto Mariotti\orcidlink{0000-0002-8320-4169},}
\affiliation{\VUB}

\author{Alisha Marriott-Best\orcidlink{0009-0008-9155-7889},}
\affiliation{\Swansea}

\author{Sylvain Marsat \orcidlink{0000-0001-9449-1071},}
\affiliation{\LtwoIT}

\author{Gabriel Mart{\'i}nez-Pinedo\orcidlink{0000-0002-3825-0131},}
\affiliation{\IKPDarmstadt}
\affiliation{\HFHF}
\affiliation{\GSI}

\author{Andrea Maselli\orcidlink{0000-0001-8515-8525},}
\affiliation{\gssi}
\affiliation{\infnlngs}

\author{Simone Mastrogiovanni\orcidlink{0000-0003-1606-4183},}
\affiliation{\INFNRoma}

% ...

\scanData

\abstract{Einstein Telescope  (ET) is the European project for a  gravitational-wave (GW) observatory of third-generation. In this paper we present a comprehensive discussion of its science objectives, providing state-of-the-art predictions for the capabilities of ET in both geometries currently under consideration, a single-site triangular configuration or two L-shaped detectors. We discuss the impact that ET will have on domains as broad and diverse as  fundamental physics, cosmology, early Universe, astrophysics of compact objects, physics of matter in extreme conditions, and dynamics of stellar collapse. 
We discuss how the study of extreme astrophysical events will be enhanced by multi-messenger observations. We highlight the ET synergies with ground-based and space-borne GW observatories, including  multi-band investigations of the same sources, improved parameter estimation, and complementary information on astrophysical or cosmological mechanisms obtained combining observations from different frequency bands. We  present advancements in waveform modeling dedicated to  third-generation observatories, along with open tools developed within the ET Collaboration for assessing the scientific potentials of different detector configurations. We finally discuss the data analysis challenges posed by third-generation observatories, which will enable access to large populations of sources and provide unprecedented precision.} 

\begin{document}

\maketitle

\flushbottom

%

\section*{General Introduction} \label{section:GeneralIntro}

\addcontentsline{toc}{section}{General Introduction}

Thanks to the extraordinary discoveries of the 
LIGO, Virgo and KAGRA (LVK) Collaboration in the last few years
(e.g.~\cite{LIGOScientific:2016aoc,LIGOScientific:2017vwq,LIGOScientific:2017zic,LIGOScientific:2017ync,LIGOScientific:2020ibl,KAGRA:2021vkt,KAGRA:2021duu,LIGOScientific:2021sio,LIGOScientific:2021aug}), the field of gravitational-wave (GW) science is blossoming. However,
the current second-generation detectors are nearing the limits of their capabilities, constrained by their infrastructures. 
As a result, the GW community is preparing the jump toward third-generation (3G) GW detectors, new observatories that are designed to detect GW sources along the cosmic history up to the early Universe. The 3G European observatory project is Einstein Telescope (ET)  \cite{Punturo:2010zz,Hild:2010id},   while the United States community effort is represented by the Cosmic Explorer (CE) project~\cite{Reitze:2019iox,Evans:2021gyd,Evans:2023euw}.

In recent years the ET project has undergone a significant acceleration, with the successful proposal in 2020 for including ET in the European Strategy Forum on Research Infrastructures (ESFRI) Roadmap (the roadmap which identify priority European research infrastructures), and the formal creation of the ET Collaboration  at the
XII Einstein Telescope Symposium in 2022.\footnote{See \url{https://indico.ego-gw.it/event/411/}.}
The ET Collaboration (ETC) currently counts over 1800 members, representing 264 Institutions in 31 different countries.
The ETC  activities have been structured in four boards: the Instrument Science Board (ISB), for the design of the instrument; the Site Characterization Board (SCB) for the geological/geographic characterization of the site(s); the e-Infrastructure Board (eIB) for computing and services; and the Observational Science Board (OSB), whose goal is to develop the science case of ET, along  
with the associated data analysis that will enable the full exploitation of its observations. In addition, the OSB aims at developing the synergies with other GW observatories and with multi-messenger observations. In parallel, the ET Organization (ETO) has the mandate to  coordinate the ET Project’s preparatory phase, which will conclude with the selection of the site,  the decision on the configuration of the detector and the governance for the future organization, and to  lay the foundations of the future legal entity that will be responsible for the operation of the ET on technical and administrative level.

The present paper summarizes  work performed within the OSB (which currently counts more than 700 members) over the last few years, and provides an updated and comprehensive study of the science that can be achieved with ET.
In the context of the recent developments of  ET, a first systematic discussion of the science case (building on many previous works) was presented in
\cite{Maggiore:2019uih}, which was part  of the ESFRI submission. In a subsequent work~\cite{Branchesi:2023mws} the ET science case was further developed, also in relation to the two possible  geometries that are currently considered for ET, i.e.  a single-site triangular geometry made of  3 nested detectors, 
and a network made of two identical  L-shaped detectors in two distant sites  (``2L'' in the following). The present paper  significantly extends  these studies, thanks to the large community work performed within the OSB. While the objective of ref.~\cite{Branchesi:2023mws} was to compare different ET configurations and designs by establishing a common set of modeling and astrophysical assumptions to build comparison metrics, the objective of the present work is to present a comprehensive overview of the current knowledge of GW sources detectable by ground-based observatories, identify the major open questions and the predicted impact of ET on the widest range of science cases.

The paper is organized as follows. Section~\ref{section:div1} discusses how ET can be used to probe fundamental physics, such as testing fundamental aspects of gravitational interactions, compact objects and near-horizon physics,  and dark matter candidates. Section~\ref{section:div2} is devoted to cosmology; ET can probe early-Universe cosmology through various mechanisms that can produce stochastic GW backgrounds, and late-Universe cosmology through ``standard sirens". Section~\ref{section:div3} discusses the impact that ET will have on our understanding of astrophysical populations; with its vast distance reach and large detection rates,  ${\cal O}(10^5)$
binary black hole  (BBHs) and binary neutron stars (BNSs) per year, ET will characterize these populations across the whole cosmic history, including the possibility of identifying black holes from the first stars (Population~III stars) and primordial black holes, and the access to low frequencies will make it possible to observe population of intermediate-mass black holes up to $10^4 M_{\odot}$. Section~\ref{section:div4} is dedicated  to multi-messenger astronomy, where ET will force a change of paradigm on electromagnetic searches and characterization, making the rare multi-messenger observations of today a routine with possibly hundreds of detections per year.
The synergies of ET with other GW detectors, such as Cosmic Explorer, LISA, deciHz detectors, or PTA are investigated in section~\ref{section:div5}.
Section~\ref{section:div6} is devoted to what can be learned on nuclear physics and matter under extreme conditions from ET observations of neutron star mergers, while 
section~\ref{section:div7} discusses the perspective for detection for core-collapse supernovae, burst sources, continuous and long-transient sources. 
Sections~\ref{section:div8}, \ref{section:div9} and \ref{section:div10} address different aspects related to 
the development of forecasts and of data analysis for 3G detectors.
In particular, section~\ref{section:div8} discusses the requirements and advances in waveform modeling at the level of accuracy required by ET; section~\ref{section:div9} presents public tools developed within the OSB to assess the scientific potential of ET in different configuration or networks, such as tools for
performing relatively fast forecasts on large populations of compact binaries, ringdown studies or stochastic searches. Section~\ref{section:div10} presents the challenges and current progresses toward  setting up the data analysis for ET. Each section ends with an executive summary of its key science takeaways.

Forecasts will  in general be presented using the same ET sensitivity curve already used in \cite{Branchesi:2023mws},\footnote{These curves  have been computed in the context of  the activities of ISB and are publicly  available at \url{ https://apps.et-gw.eu/tds/ql/?c=16492}, see in particular the Annex files; see also section~\ref{section:div9} of this paper for more details.  Some of the results presented have rather  been obtained with the somewhat older ET-D sensitivity curve, which however will not affect significantly the results obtained.} which are shown in figure~\ref{fig:asd}.
For parameter estimation forecasts that depend on astrophysical populations we  use in general 
the same BBH  and BNS catalogs  as in \cite{Branchesi:2023mws}.\footnote{These catalogs  are publicly available at  \url{https://apps.et-gw.eu/tds/?content=3&r=18321}.}
The comparison between the triangle and 2L geometry will be performed whenever significant.

\section{Fundamental Physics with ET} \label{section:div1}

In parallel with the anticipated diverse astrophysical applications of gravitational-wave astronomy in the forthcoming years, next-generation detectors hold the promise to explore foundational problems in gravity, high-energy, and particle physics. These explorations encompass scrutinizing the fundamental principles of the gravitational interaction, unraveling the nature of compact objects and horizon-scale physics, and identifying novel gravitational-wave signatures of dark matter surrounding or constituting compact sources. In this context, owing to its superior  sensitivity and larger bandwidth, ET will be a unique discovery machine that will open the precision-gravity frontier, while intersecting various scientific domains and establishing profound synergies with waveform modeling, data analysis, population studies, nuclear and high-energy physics, as well as other branches of astrophysics and cosmology. A~summary of the main---potentially game-changer---deliverables is discussed in this section.

\subsection{Introduction}  \label{div1_introduction}


Gravity is the weakest and arguably the most enigmatic among the fundamental interactions. Its understanding is paramount to comprehend the evolution of the Universe and is intertwined with some of the deepest longstanding puzzles in physics, from the tiniest to the largest scales. 
Thanks to the advent of GW astronomy and observational BH physics,
the last decade has witnessed the rapid increase in our understanding of the gravitational interactions of extreme sources. 
This momentum is far from being exhausted, with the forthcoming 
decades promising an even more dramatic improvement.

One of the fundamental science objectives of ET is to test General Relativity~(GR), Einstein's celebrated theory of gravity, the nature of GW sources, such as compact objects, and our understanding of matter and particle physics under the most extreme conditions. The output of ET observations will therefore have profound implications for diverse fields ranging from astrophysics, cosmology, nuclear and particle physics, 
to perhaps even the realms of quantum gravity. In parallel, the discovery potential in GW science has reinvigorated the ongoing efforts to construct high-accuracy waveform models, either through numerical simulations or using sophisticated analytic methods from particle physics that have been instrumental for “new physics” searches at particle colliders through various precision-data programs. The science of ET will then usher us into the precision-gravity frontier, leading the path forward to several tests which are inaccessible with present-day GW detectors, greatly impacting our understanding of the Universe. Devising these tests and describing the theoretical repercussions of GW observations is the main motivation of this section, developed in the context of the ET OSB ``Fundamental Physics" Division. 

This section will be dealing with several topics carrying a unique discovery potential, thus complementing the numerous deliverables for the science of ET discussed throughout other sections. It is divided into three broad areas: \emph{Testing the fundamental principles of the gravitational interactions} (section~\ref{div1_fundgravity}); \emph{Testing the nature of compact objects and horizon-scale physics} (section~\ref{div1_natureofcompact}); and \emph{Searches of dark-matter candidates and new fields} (section~\ref{div1_DMcandidates}). For each of them we will discuss ET's ability to detect (or constrain) fundamental physics beyond our present knowledge of the laws of nature, in the advent of a new era of explorations of the cosmos. Synergies with other OSB Divisions are briefly covered in section~\ref{div1_synergies}.

\subsection{Testing  the  fundamental  principles  of  the  gravitational  interaction}
\label{div1_fundgravity}

GR is an extremely successful theory describing the gravitational interaction in geometric terms.
It is rooted in Einstein's strong equivalence principle and based on a certain number of assumptions.
Lovelock's theorem states that GR is the only theory in four spacetime dimensions with second-order field equations that are diffeomorphism-invariant and built solely from the metric tensor~\cite{Lovelock:1972vz}.
Relaxing any of these hypotheses (e.g., considering extra dimensions, higher-order field equations, a massive graviton, or including new gravitational degrees of freedom) gives rise to multiple pathways toward alternative theories of gravity~\cite{Sotiriou:2014yhm,Berti:2015itd}.
The minimal and most explored deviation from GR is the inclusion of extra fields which, depending on the specific theory, can violate one of more of the pillars upon which GR is built.
Therefore, searching for the effects of extra fields in gravitational systems provide a way to test GR in various regimes.

The emission of GWs from binary systems offers a unique possibility to test fundamental aspects of the gravitational interactions which are also responsible for the binding of compact objects, such as BHs and NSs, in Nature. This can be achieved from an accurate reconstruction of the early stages of an inspiral phase, or from the strongly-coupled regime of the merger, or by an accurate characterization of the coalescence's remnant. In all these phases, GR deviations may manifest themselves in additional propagating degrees of freedom and dissipative channels, as well as other subtle nonlinear effects. The quest towards high-precision waveform models also unlocks new challenges in our understanding of the two-body dynamics that deserve further study. This is a fundamental problem on its own, independently of the phenomenological implications associated with GW science. For instance, both precision tests of the inspiral regime as well as numerical simulations for beyond-GR theories are crucial if one wishes to be agnostic about the origin of the GW signals. These topics are naturally framed within the context of fundamental aspects of gravitational interactions and covered in this section. 

\subsubsection{Tests of the inspiral dynamics}
\label{inspiral}

A model-independent approach to quantifying GR violations in the GW phase evolution is
to allow for modifications in a limited number of post-Newtonian~(PN) coefficients in the perturbative expansion 
\cite{Blanchet:1994ez,Arun:2006yw,Arun:2006hn,Yunes:2009ke,Mishra:2010tp,Li:2011cg,Agathos:2013upa,Meidam:2017dgf}. Deviations in the amplitude can also be searched for, which is especially interesting in the presence of detectable higher harmonics 
of the basic signal~\cite{Dhanpal:2018ufk,Islam:2019dmk,Capano:2020dix,Puecher:2022sfm}.
More concretely, one considers the frequency-domain phase in the stationary phase approximation:
\be
\Psi(f) = 2\pi f t_c - \varphi_c -\frac{\pi}{4}  + \frac{3}{128\eta}\left(\frac{v}{c}\right)^{-5}
\sum_{n=0}^7 \left[ \varphi_n + \varphi^{(l)}_n \log\left(\frac{v}{c}\right) \right]
\,\left(\frac{v}{c}\right)^n, 
\label{PNphase}
\ee
where $v/c = (\pi M f)^{1/3}$, with $M$ the total mass of the 
binary in the detector frame, $t_c$ and $\varphi_c$ are 
respectively the time and phase at coalescence, and 
$\eta = m_1 m_2/M^2$, with $m_1$, $m_2$ the component masses.
Within GR, the coefficients $\varphi_n$ and $\varphi^{(l)}_n$ 
depend on the masses and spins of the component objects~\cite{Blanchet:2013haa}, and 
a series of tests is done on them~\cite{LIGOScientific:2021sio}. For example, each of the 
$\varphi_n$ are replaced by $(1 + \delta\hat{\varphi}_n)\,\varphi_n$, and the fractional deviation $\delta\hat{\varphi}_n$
is measured as a variable together with all other parameters 
entering the waveform; similarly for the $\varphi^{(l)}_n$. 
Very broad classes of modified theories of gravity can be mapped, to leading PN order, to specific PN deviations~\cite{Yunes:2009ke,Berti:2018cxi}. For example, 
theories causing dipole radiation manifest themselves
to leading order at $n = -2$ in the notation of 
\eq{PNphase} (i.e., a $-1{\rm PN}$ correction). Since this term is absent in GR, one does a test  
with a coefficient $\delta\hat{\varphi}_{-2}$ multiplying the 
corresponding power of $v/c$; in this case the testing 
coefficient represents an absolute rather than a relative deviation. The same holds at 0.5PN, 
since $\varphi_1 = 0$ in GR.

\begin{figure}
    \centering
    \includegraphics[width=0.80\textwidth]{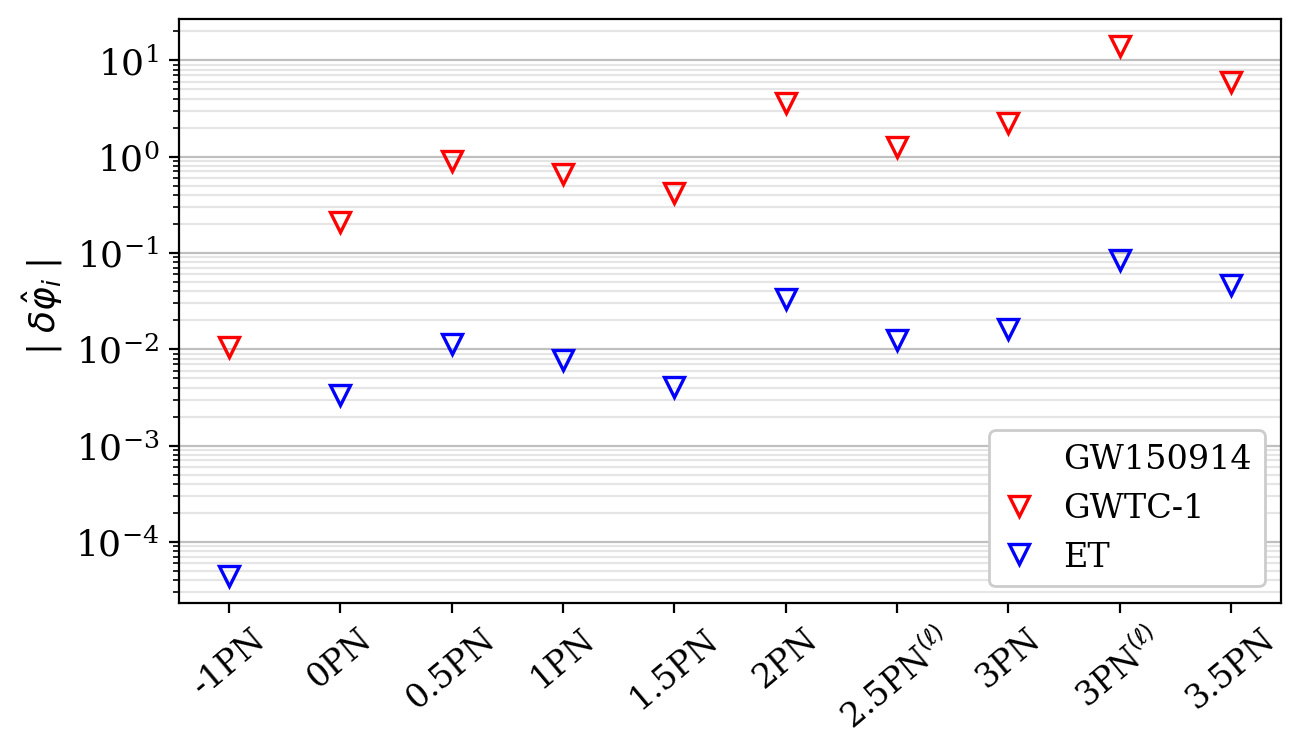}
    \caption{90\% upper bounds on the beyond-GR parameters $\delta\hat{\varphi}_n$ and $\delta\hat{\varphi}^{(l)}_n$ in the 
    inspiral phase of GW150914, from $-1{\rm PN}$ to 3.5PN order. 
    Red triangles  represent bounds from Advanced LIGO observations~\cite{LIGOScientific:2019fpa}, while blue triangles show bounds from simulated GW150914-like signals injected into synthetic noise for a single triangular ET.}
    \label{fig:parametrized:bound}
\end{figure}

To get an idea of what to expect with ET compared to existing detectors for signals from binary BH coalescences, it is convenient to use phenomenological 
inspiral-merger-ringdown waveforms such as IMRPhenomPv2
\cite{Husa:2015iqa,Khan:2015jqa,Hannam:2013oca}, which
includes the phase of \eq{PNphase} as part of its 
description of the inspiral. 
Figure~\ref{fig:parametrized:bound}
shows 90\% confidence bounds that were obtained on the 
$|\delta\hat{\varphi}_n|$ and $|\delta\hat{\varphi}^{(l)}_n|$ from Bayesian analyses of the first 
detected GW signal, GW150914, which indeed happened to be from 
a binary BH. These analyses also yielded maximum-likelihood
values for the testing parameters along with 
the other parameters (masses, spins, sky position, distance, ...) that characterize
the signal~\cite{LIGOScientific:2019fpa}. Subsequently, simulated signals for those same parameters were ``injected'' into stationary, Gaussian noise following the noise power spectral density of ET, and the analyses were repeated to predict 
90\% confidence bounds one would obtain with the latter when 
observing a signal such as GW150914. As one can see, 
for the given signal, ET improves over Advanced LIGO in its
first observing run by about two orders of magnitude for 
all the testing coefficients. However, it should be noted 
that it is possible to combine results from \emph{all} the binary 
BH detections made (as is already being done 
with current detectors 
\cite{LIGOScientific:2019fpa,LIGOScientific:2020tif,LIGOScientific:2021sio}). ET may observe $\mathcal{O}(10^5)$
such signals and, assuming the GR deviations are the same for all sources, measurement accuracies on the $|\delta\hat{\varphi}_n|$ will scale roughly as $1/\sqrt{N_{\rm det}}$, where 
$N_{\rm det}$ is the number of detections. Therefore, ET will 
enable upper bounds that are well below 0.01 -- or
reveal a GR violation of such size, if present.

\subsubsection{Extra polarizations}

Metric theories of gravity allow for up to six different polarization modes, 
which can be categorized into tensor modes (the two polarizations present in 
standard GR), two vector modes, and two scalar modes~\cite{Eardley:1973zuo,Eardley:1973br}, which
below will be denoted by T, V, and S, respectively. 

Extra polarizations occur in a wide variety of alternative theories of gravity \cite{Capozziello:2006ra,Corda:2008si,Hou:2017bqj,Liang:2017ahj,Moretti:2019yhs,Jacobson:2004ts,Gong:2018cgj,Hinterbichler:2011tt,Liang:2022hxd,Frolov:2002qm,Will:1977wq,Foster:2006az,Zhang:2019iim}, and are thus an important science target for 
fundamental physics with ET.
In the notation of Newman and Penrose~\cite{Newman:1961qr}, the scalar, vector and tensor polarizations can be written in terms of the Riemann tensor as
\be
\begin{array}{rll}
\Phi_{22} &= -R_{0x0x} - R_{0y0y}                  \qquad\qquad &(s=0) \, ,\\ [1.5mm]
\Psi_2 &= -\frac16 R_{0z0z}                        \qquad\qquad &(s=0) \, ,\\[1.5mm]
\Psi_3 &= -\frac12 R_{0x0z} + \frac{i}{2} R_{0y0z} \qquad\qquad &(s=\pm 1) \, ,\\ [1.5mm]
\Psi_4 &= -R_{0x0x}+R_{0y0y} + 2i R_{0x0y}         \qquad\qquad &(s=\pm 2)\, .
\end{array}
\ee
However, they are not  necessarily equivalent to dynamical degrees of freedom, and as pointed out in~\cite{Eardley:1973zuo}, the latter can be fewer or more than the number of polarization modes. 
For example, in various scalar-tensor theories of gravity, the number of dynamical degrees of freedom (three) is smaller than the number of polarization modes (four)~\cite{Liang:2017ahj}.
On the contrary, in other models, where different fields produce the same polarization mode,  the number of dynamical degrees of freedom is larger than the number of 
polarizations~\cite{Bahamonde:2021dqn}. In general, the metric perturbation $h_{\mu\nu}$ can be decomposed into irreducible pieces according to their properties under spatial rotations, which can then be recombined into six quantities invariant under infinitesimal coordinate transformations (i.e., gauge-invariant at linear order)  and which, under spatial rotations, transform as  two scalars, two components of a transverse vector and two components of a transverse traceless tensor~\cite{Bardeen:1980kt,Jackiw:2003pm,Flanagan:2005yc,Jaccard:2013gla,Maggiore:2018sht}. 
Examples of the polarization contents of some modified gravity theories considered in the literature are the following:

\vspace{1mm}
(i) Scalar-tensor theories (in the metric formalism) add an extra mode associated to the scalar (propagating) field~\cite{Capozziello:2006ra,Corda:2008si,Hou:2017bqj}, an example of this being $f(R)$ gravity~\cite{Liang:2017ahj,Moretti:2019yhs}. \vspace{1mm}

(ii) Einstein-aether theory predicts two tensor modes, two vector modes and one scalar mode~\cite{Jacobson:2004ts,Gong:2018cgj}.\vspace{1mm}

(iii) The ghost-free massive gravity proposed by de Rham-Gabadadze-Tolley (dRGT) ~\cite{deRham:2010kj}, in which gravity is mediated by a massive field $m_s$ of spin $s=2$,  propagates $2s+1=5$ polarizations~\cite{Hinterbichler:2011tt}.\vspace{1mm}

(iv) In Bumblebee gravity the polarization content is anisotropic, and depends on the angle between the background vector field and the propagation direction of GWs~\cite{Liang:2022hxd}.\vspace{1mm}

(v) Randall-Sundrum braneworld models yield five bulk graviton polarizations~\cite{Frolov:2002qm}.\vspace{1mm}

(vi)~In the version of non-local massive gravity free of ghosts on Minkowski spacetime~\cite{Modesto:2013jea}, one has five perturbative degrees of freedom. However, this model still has a cosmologically unstable mode~\cite{Foffa:2013vma}.\vspace{1mm}

(vii) In the non-local infrared  modification of gravity proposed in \cite{Maggiore:2013mea} (see \cite{Maggiore:2016gpx,Belgacem:2020pdz} for reviews) the non-local term can be localized using  auxiliary  fields that do not propagate, so only the two tensor modes propagate, while the non-propagating conformal mode becomes massive.\vspace{1mm}

(viii)~A class of UV-complete (unitary and renormalizable) non-local theories of quantum gravity features eight non-perturbative degrees of freedom~\cite{Calcagni:2018gke}, just like non-unitary local Stelle gravity~\cite{Stelle:1976gc}. Out of these degrees of freedom, only two modes corresponding to the massless graviton propagate in Minkowski spacetime but the others can become non-trivial on curved backgrounds.


Given the fact that the existence of extra polarizations means extra energy emission channels, this can accelerate the orbital decay of binary systems. There are many studies on constraining modified gravity with observations of such decays, e.g.~\cite{Will:1989sk,Yagi:2013qpa,Yagi:2013ava}, as well as the phase evolution in the tensor modes, 
e.g.~\cite{Berti:2015itd,Chatziioannou:2012rf,Barausse:2016eii}. In what follows we discuss some general features of using measurements of the GW polarization to test deviations from GR.

\paragraph{Detector response to extra polarizations.}

Polarization tests typically rely on the fact that interferometers have different 
antenna responses to the different polarizations. When multiple, differently oriented detectors are available then the presence of non-standard polarizations can in principle be established. Currently, there are three major types of beam detectors: ground-based interferometric detectors, space-based interferometric detectors, and pulsar timing arrays. They have different response to different polarizations, and we refer to~\cite{Nishizawa:2009bf,Amalberti:2021kzh,Liang:2019pry,Zhang:2019oet,2008ApJ...685.1304L,Chamberlin:2011ev,Romano:2016dpx} for details.

For the case of stochastic GW backgrounds, cross correlation of the outputs of multiple detectors is needed~\cite{Christensen:1992wi,Allen:1997ad}. The sensitivity to the stochastic background of the detector pair depends on the antenna response of each detector, and it gets reduced due to the separation and non-optimal orientations of the detector pair. The reduction of the sensitivity is characterized by the overlap reduction function~\cite{Flanagan:1993ix,Romano:2016dpx} (see section~\ref{sect:CharStocBackdiv9}).
For different polarizations, the overlap reduction functions of various of ground-based detector network configuration  are discussed in refs.~\cite{Nishizawa:2009bf,Amalberti:2021kzh}. Different polarizations are projected on the strain via the antenna pattern function. Thus, for a network of multiple non-coaligned detectors, we get different linear combination of polarizations. 
Theoretically, for a transient signal, the number of non-coaligned detectors should be at least the same as the number of the polarization modes in order to distinguish each of them. 
It should be observed, however,  that the detector pattern functions of the two scalar modes are degenerate in the limit $f\ll f_*$, where $f_*=c/(2 \pi L)$ (with $L$ the interferometer arm-length) \cite{Nishizawa:2009bf}; still, this degeneracy is lifted and the signals from the two scalar modes can in principle be disentangled at $f> f_*$ (see Fig.~24 in~\cite{Amalberti:2021kzh}). For  $L=10$~km (as in the 10~km triangle), $f_*\simeq 4.7$~kHz; for $L=15$~km, as in the  2L configuration of ET, $f_*\simeq 3.2$~kHz; for  a 40~km CE, $f_*\simeq 1.2$~kHz. Thus, for a stochastic background, only the part of the spectrum above such frequencies can be used to disentangle the two scalar modes.

The capability of the second generation detector network to separate the polarization modes for the inspiraling compact binary was studied systematically in ref.~\cite{Takeda:2018uai}.
ET has great sensitivity at lower frequency, and thus it can observe the early inspiral stage of the binary neutron star for a long time. In this sense, a single ET detector can be regarded as a virtual detector network due to the Earth's rotation~\cite{Takeda:2019gwk}, which helps distinguishing the scalar, vector and tensor  polarizations among them. Note, however, that for CBCs disentangling further the two scalar modes among them is  more challenging  since, particularly for a 10~km arm-length, basically all  of the observable signal is at $f\ll f_*$ 
(and the Earth's rotation in this case does not help, since the degeneracy between the pattern functions of the two scalar modes at $f\ll f_*$  is not affected by the  Earth's rotation).


\paragraph{Current and future constraints}

Preliminary tests were carried out with GW events seen in both LIGO detectors and Virgo, e.g.~GW170814~\cite{LIGOScientific:2017ycc} and GW170817~\cite{LIGOScientific:2017vwq}.
For the first detected GW event, GW150914~\cite{LIGOScientific:2016aoc}, there was no conclusive constraint on the polarization because there were only two detectors and they have similar orientation~\cite{LIGOScientific:2016lio}. 
In contrast, for the first two GW events detected by three detectors, GW170814~\cite{LIGOScientific:2017ycc} and GW170817~\cite{LIGOScientific:2017vwq}, it is found that the purely tensor polarization is strongly favored over purely scalar or vector polarizations~\cite{LIGOScientific:2017ycc,LIGOScientific:2018dkp}. 
The same test was also performed to GW170818 in GWTC-1~\cite{LIGOScientific:2019fpa}.
The test was based on the method proposed in~\cite{Isi:2017fbj}, which applied coherent Bayesian analysis using GR waveform template but with different antenna response.  
For  future studies, the precise waveform template in modified gravity is needed; for more details we refer to section~\ref{inspiral} for inspiral, section~\ref{sect:NRdiv1} for merger and section~\ref{ringdown} for ringdown.
The Bayesian model selection method is also applicable for  persistent sources, such as continuous waves from isolated neutron star~\cite{Isi:2017equ} and stochastic gravitational wave background~\cite{Callister:2017ocg}.

If the sky location of the GW source is known through an electromagnetic counterpart, one can linearly combine the multiple outputs from the detector network to construct a so-called \emph{null stream}, in which the two tensor modes are eliminated completely~\cite{Guersel:1989th,Chatziioannou:2012rf}.\footnote{For a given signal with known localization, the null stream can be constructed also in the 2L geometry of ET. In the triangular geometry, one can construct a null stream simultaneously for all (tensor) signals, independently of the knowledge of the sky position; see sections~\ref{sec:div9_null_stream} and \ref{subsec:null_stream} for extended discussion.}
The analysis of polarizations for signals like GW170817 with null stream was studied in refs.~\cite{Hagihara:2018azu,Hagihara:2019ihn}, and then developed in refs.~\cite{Pang:2020pfz,Wong:2021cmp}.
The null stream method does not require precise waveform templates for the extra polarizations; one only needs to check whether or not the null stream is statistically consistent with the presence of noise only. 
Such analyses have been applied to the GW events in the GWTC-2 catalog of LIGO and Virgo~\cite{LIGOScientific:2020tif}, and in the GWTC-3 catalog~\cite{LIGOScientific:2021sio},  though the resulting constraints are not yet very informative. 
For ET, combined with other next-generation detectors, it has been demonstrated that 
one can probe the relative ratio of the amplitude of the scalar mode to the tensor mode at the level of $A^S_T \simeq 0.045$, and $A^V_T \simeq 0.014$ for the vector modes, from a single BNS merger located at $100$ Mpc with optimal sky location~\cite{Hu:2023soi}. 
Considering multiple events uniformly distributed in volume within a sphere with 
a large radius corresponding to a luminosity distance of 
$\sim 1$~Gpc, the detection limit can reach $A^S_T \simeq 0.05$ and $A^V_T \simeq 0.02$ on average; see Fig. 4 in~\cite{Hu:2023soi}.


Though stochastic GW backgrounds have not yet been found in the LIGO-Virgo-KAGRA band~\cite{LIGOScientific:2018czr,LIGOScientific:2019vic,KAGRA:2021kbb}, they too offer opportunities for studying polarization content.
The latest upper limits on the energy density spectrum (normalized by the critical 
density) are $\Omega^{S}_{\text GW}(25 \,\text{Hz}) \leq 2.1\times 10^{-8} $ for scalar backgrounds, $\Omega^{V}_{\text GW}(25 \,\text{Hz}) \leq 7.9\times 10^{-9} $ for vector backgrounds, and $\Omega^{T}_{\text GW}(25\, \text{Hz}) \leq 6.4 \times 10^{-9} $ for tensorial ones~\cite{KAGRA:2021kbb}.
For ET together with CE as a network of next generation detectors, the optimal 
geometric configuration (location and orientation) to GW backgrounds of different polarizations is discussed in detail in~\cite{Amalberti:2021kzh}. In general, the sensitivity can reach $\Omega^A_{\text GW} \simeq 10^{-12} - 10^{-11} $ (A= T, V, S), for five years of observations~\cite{Amalberti:2021kzh}, constituting  two to three orders of magnitude improvement; see Table~\ref{tab:polarization} for details.

\begin{table}
\begin{tabular}{cccc}
\hline
\hline              
Tensor+Vector+Scalar     & ET $\Delta$ + CE            & ET 2L $\alpha = 0^{\circ}$ + CE & ET 2L $\alpha = 45^{\circ}$ + CE              \\ \hline
$h_0^2\Omega^T_{\rm GW}$ & $2.80 \times 10^{-10}$      & $2.63 \times 10^{-11}$          &  $1.24 \times 10^{-10}$           \\  
$h_0^2\Omega^V_{\rm GW}$ & $1.82 \times 10^{-9}$       & $3.24 \times 10^{-11}$          &  $1.18 \times 10^{-10}$           \\
$h_0^2\Omega^S_{\rm GW}$ & $1.13 \times 10^{-9}$       & $2.37 \times 10^{-11}$          &  $8.17 \times 10^{-11}$           \\
\hline
\end{tabular}
\caption{Detectable energy density (SNR$=5$ and $T=5\,{\rm yr}$, integrated in the sense of the PLS), for a SGWB made of combination of tensor, scalar and vector modes with ET and CE in different configurations. In particular, for ET we consider the triangle configuration, 2L parallel, and 2L at $45^{\circ}$; for CE we consider a single 40-km detector located at the Hanford site. The parameter $\xi_s$, which is used to characterize the contribution from breathing mode and longitudinal mode, is fixed to $\xi_s=1/3$.
}
\label{tab:polarization}
\end{table}

\paragraph{Polarization studies with lensed GW events.} Returning to transient sources of 
GWs, even if an EM counterpart is available to localize a source on the sky, the  current network of ground-based detectors is limited in its ability to disentangle the six polarizations, since this requires the number of detectors in the network to be at least as large as the number of polarizations to probe (and, even in this case, we have seen above that the two scalar modes can only be disentangled at $f>f_*$). While in the ET/CE era we will not have that many detectors, Nature offers us a way to artificially extend our detector network through \emph{strong lensing}~\cite{Takahashi:2003ix}. GW lensing is similar to that of light, where a massive object on the GW's travel path affects its propagation. A lensed GW signal is observed as multiple `images' that arrive on Earth at different times and have different (de)magnifications, but the same frequency evolution~\cite{Takahashi:2003ix, Ezquiaga:2020gdt, Janquart:2021qov, Lo:2021nae}. In between the arrival of two images, the Earth rotates and the detector network gets re-oriented, effectively becoming a different network. Thus, $N$ detectors observing $M$ images is equivalent to a network of $N \times M$ detectors observing a single signal. Note that strongly lensed images are separated on the arcsecond scale, while GW detectors have a precision of several degrees. So, the separation between the images is completely swamped in the relatively low sky location accuracy from the interferometers, and the $M$ images are effectively associated to the same direction in space.

If a BBH signal is strongly lensed, leading to e.g.~4 images, even a single triangular ET observatory becomes equivalent to a 12-detector network observing a single signal, with potentially excellent sky localizability. Since in the ET era one expects 
to observe tens to hundreds of strongly lensed events per year~\cite{Piorkowska:2013eww, Wierda:2021upe, Gupta:2023lga}, strong lensing indeed offers the possibility of 
probing the polarization contents of GWs without the need for an EM counterpart.

\begin{figure}
    \centering
    \includegraphics[keepaspectratio, width=0.8\textwidth]{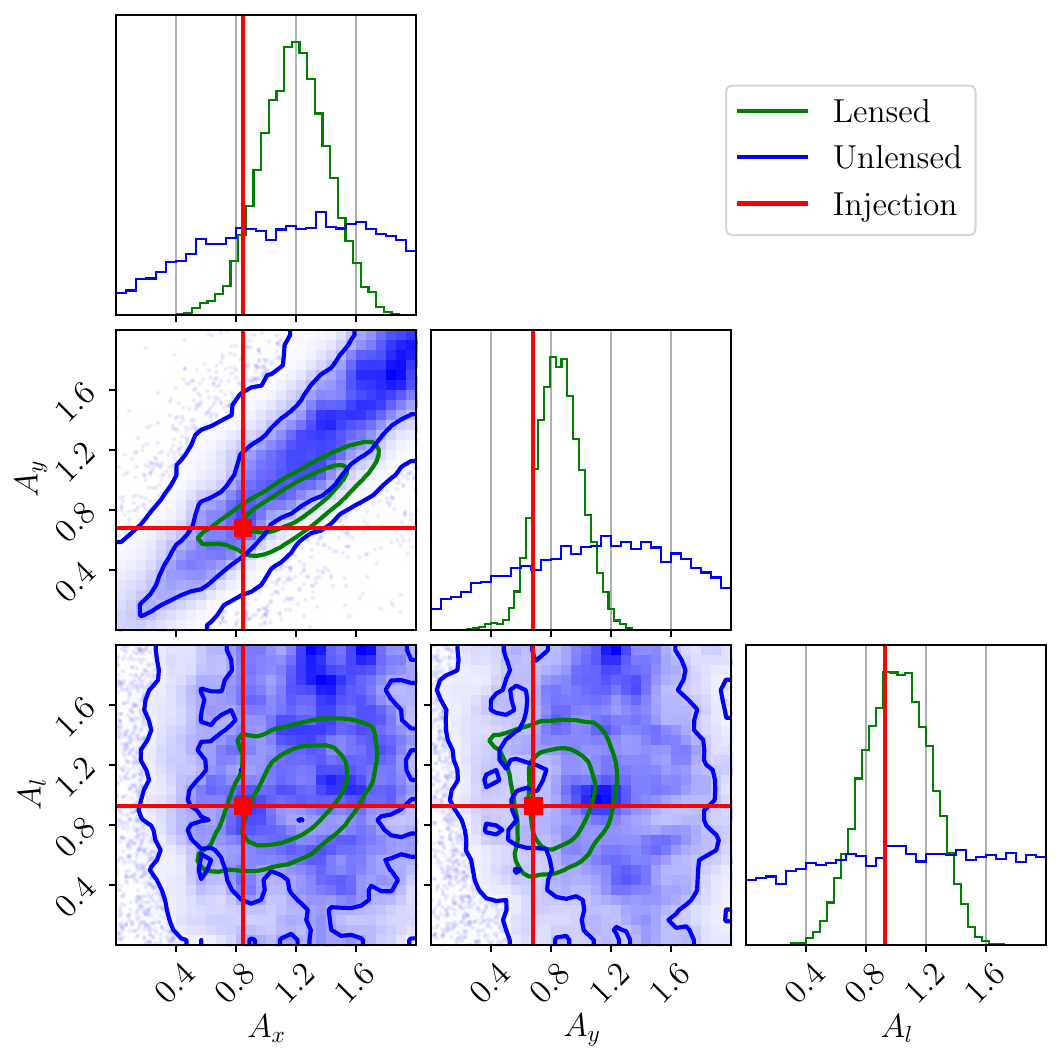}
    \caption{An illustration of strongly lensed GWs used to disentangle polarization states with a single triangular ET. Apart from the tensor polarizations we consider 
    the two vector polarizations, $x$ and $y$, and the longitudinal scalar polarization $l$. Posteriors are shown for the amplitudes of the vector polarizations relative to the $+$ and $\times$ amplitudes, which are denoted $A_x$ and $A_y$, and similarly for the scalar polarization, where the relative amplitude is denoted $A_l$. 
    Green indicates lensing, blue unlensed; solid lines show true values. Without lensing, posteriors are uninformative, while lensing enables individual polarization studies.   
    }    \label{fig:corner_amplitude_strongly_lensed}
\end{figure}

To demonstrate the capacity of lensing to identify extra polarizations with future detectors, we perform Bayesian parameter estimation runs in which we inject signals with all six polarizations present.  In figure~\ref{fig:corner_amplitude_strongly_lensed}, we show example posterior probability distributions for the relative amplitudes of the non-GR polarizations with a triangular ET, for the lensed and unlensed cases. For this event, the detector frame component masses are $73.6\, M_{\odot}$ and $71.7\, M_{\odot}$, and the signal-to-noise ratio is 100 (150) for the unlensed (lensed) case, the difference being indicative of what one can expect from the multi-image nature of lensed events. The relative amplitudes of the polarizations were  picked randomly between 0 and 1. It is evident that in the lensed case, these relative amplitudes are well measurable, and zero is excluded for the given parameter choices, while in the unlensed case, the posterior distributions are essentially uninformative. 

\subsubsection{Anomalous gravitational-wave propagation, tests of Lorentz violation and minimal length}

\begin{figure}
    \centering
    \includegraphics[keepaspectratio, width=0.65\textwidth]{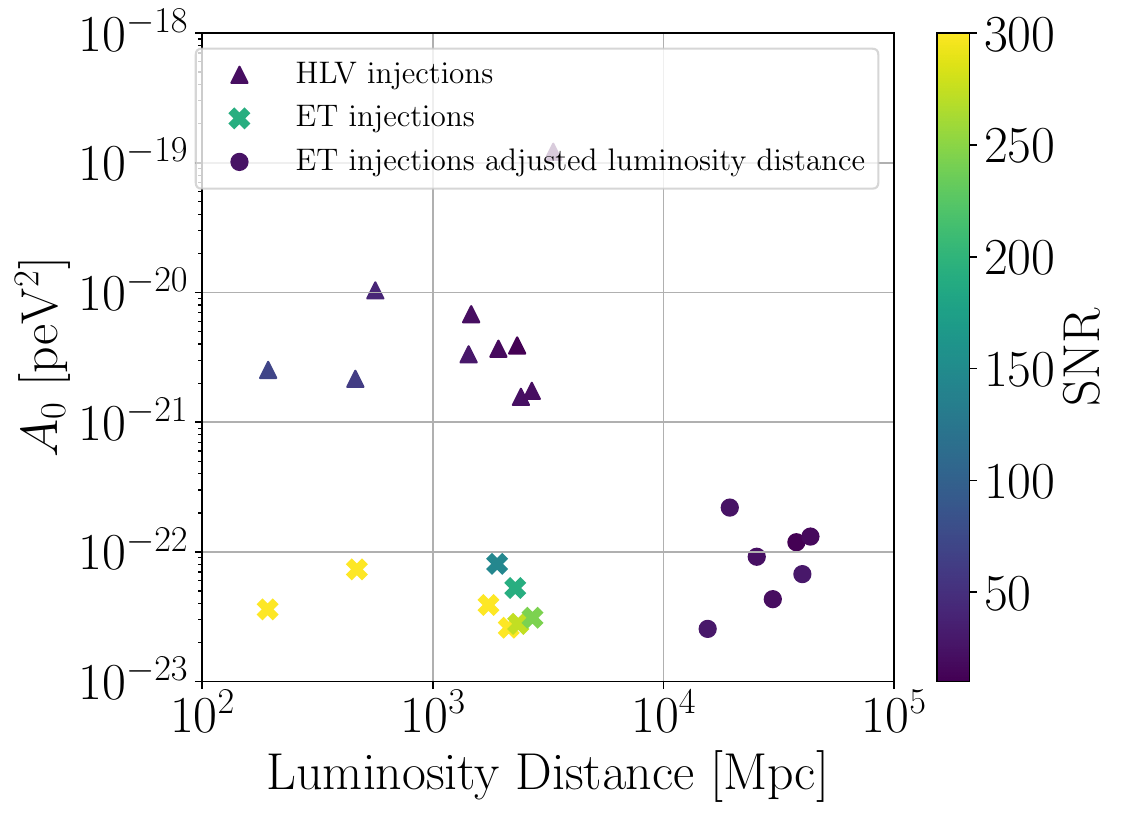}
    
    \caption{90\% confidence interval for the graviton mass, $A_0 = m_g^2$, as a function of luminosity distance for three GW injection sets. Triangles represent analyses with HLV, crosses and circles with a triangular ET; for the circles,  luminosity distance was increased to match SNRs of LVK.
    The results demonstrate that ET improves bounds on $A_0$ both due to increased SNR for close-by events, and its ability to access events at larger distances despite their lower SNR.   
    }
    \label{fig:mdr_comparison}
\end{figure}


\par\noindent
A feature of several extensions of the Standard Model (SM) of particle physics is that the dispersion relations of massive and massless particles gets modified. Furthermore, several arguments suggest that quantum gravity features a minimal length~\cite{Veneziano:1986zf,Gross:1987ar,Maggiore:1993rv,Maggiore:1993kv,Maggiore:1993zu,Doplicher:1994tu} and dispersion relations can be modified  also by models of quantum gravity that incorporate a minimal length~\cite{Majhi:2013koa,Das:2008kaa, Das:2010sj}.

A possibility to test physics beyond the SM in the gravitational sector utilizes the GW signal from the coalescence of compact binary systems~\cite{Kostelecky:2016kfm,Kostelecky:2016kkn,Shao:2020shv,Das:2021lrb,Calcagni:2019ngc,Ong:2018nzk,Ong:2018zqn,Prasetyo:2022uaa}. Another option is the induced anomalous GW propagation and its imprint on the waveform~\cite{Tso:2016mvv}. 
Violations of fundamental symmetries, such as Lorentz invariance, is also predicted in beyond-GR theories of gravity, which may also induce a modified dispersion relations~\cite{Mirshekari:2011yq, Samajdar:2017mka,LIGOScientific:2017bnn}. 
In figure~\ref{fig:mdr_comparison}, we show the results of a simulation where BBH signals are injected into stationary, Gaussian noise, and analyzed in a Bayesian way. We show upper bounds on $A_0 = m_g^2$, with $m_g$ the graviton's mass, for (a) a network consisting of LIGO Livingston, LIGO Hanford, and Virgo (LVK), and (b) a triangular ET. The ET bounds improve on LVK by nearly two orders of magnitude, with only weak dependence on distance to the source. At relatively low luminosity distance this results from the higher SNRs of ET signals, whereas at larger distances we see the advantage of the longer path traveled by GWs. Other possibilities to perform this test are offered by the signal from deformed rotating NSs, and the search for continuous GW sources is ongoing~\cite{LIGOScientific:2021nrg,Dergachev:2020fli}.  
Overall, it is expected that the next-generation of GW detectors, in particular ET, 
will allow us to set stringent bounds on parameters characterizing violations of fundamental symmetries with an unprecedented level of accuracy.


\subsubsection{Gravitational-wave memory} 

\begin{figure*}[t]
    \centering
    \begin{subfigure}[b]{0.5\textwidth}
        \includegraphics[width=\textwidth]{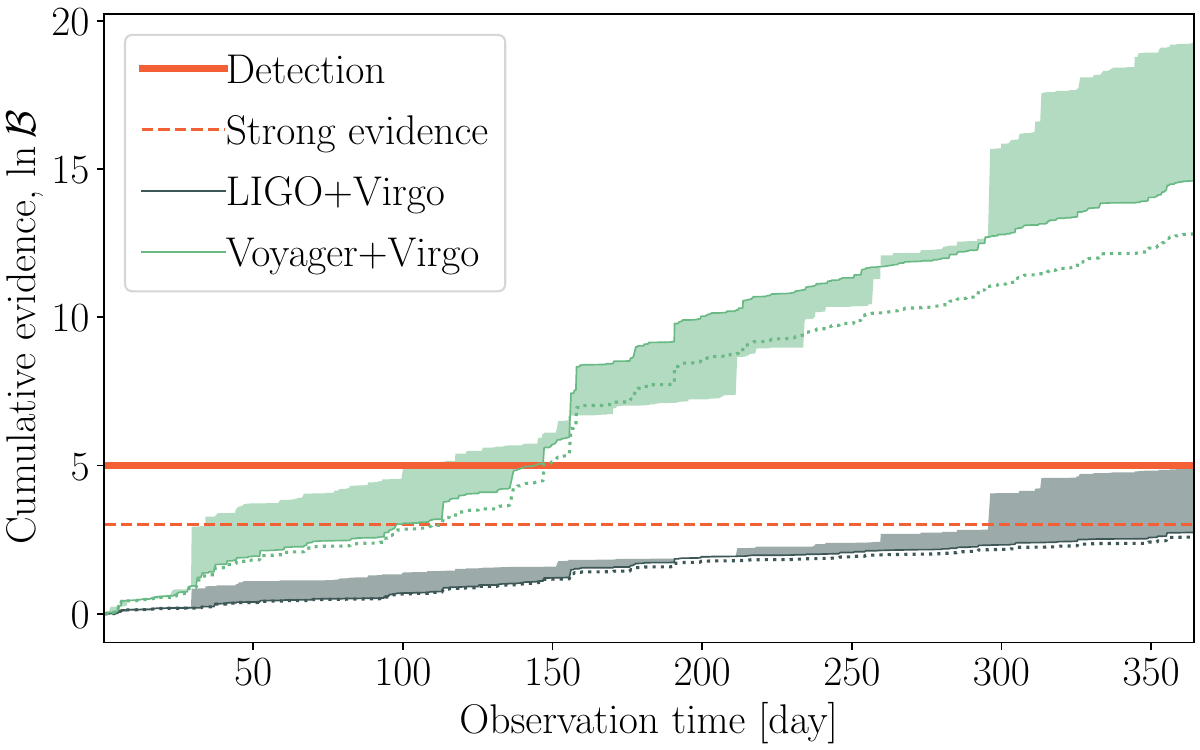}
        \caption{Displacement memory detection}
        \label{fig:combine:lvk_pop}
    \end{subfigure}
    \begin{subfigure}[b]{0.5\textwidth}
        \includegraphics[width=\textwidth]{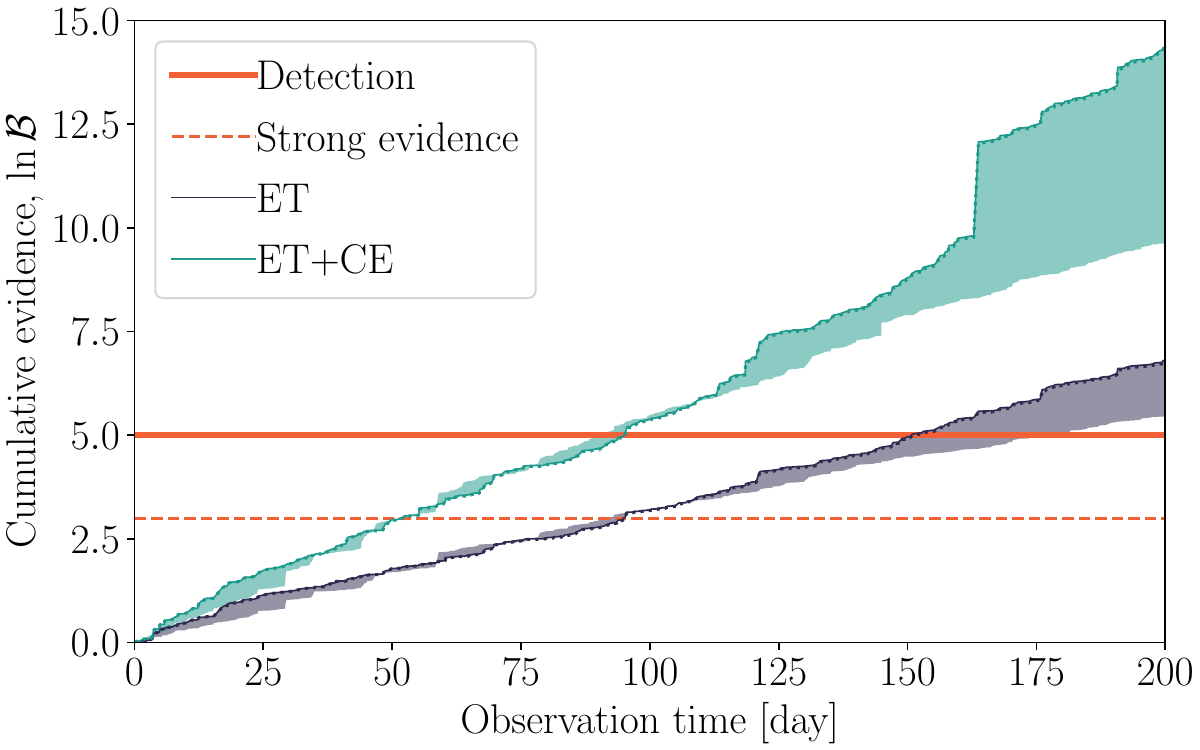}
        \caption{Spin memory detection}
        \label{fig:combine:etce_pop}
    \end{subfigure}
    \begin{subfigure}[b]{0.4\textwidth}
        \includegraphics[width=\textwidth]{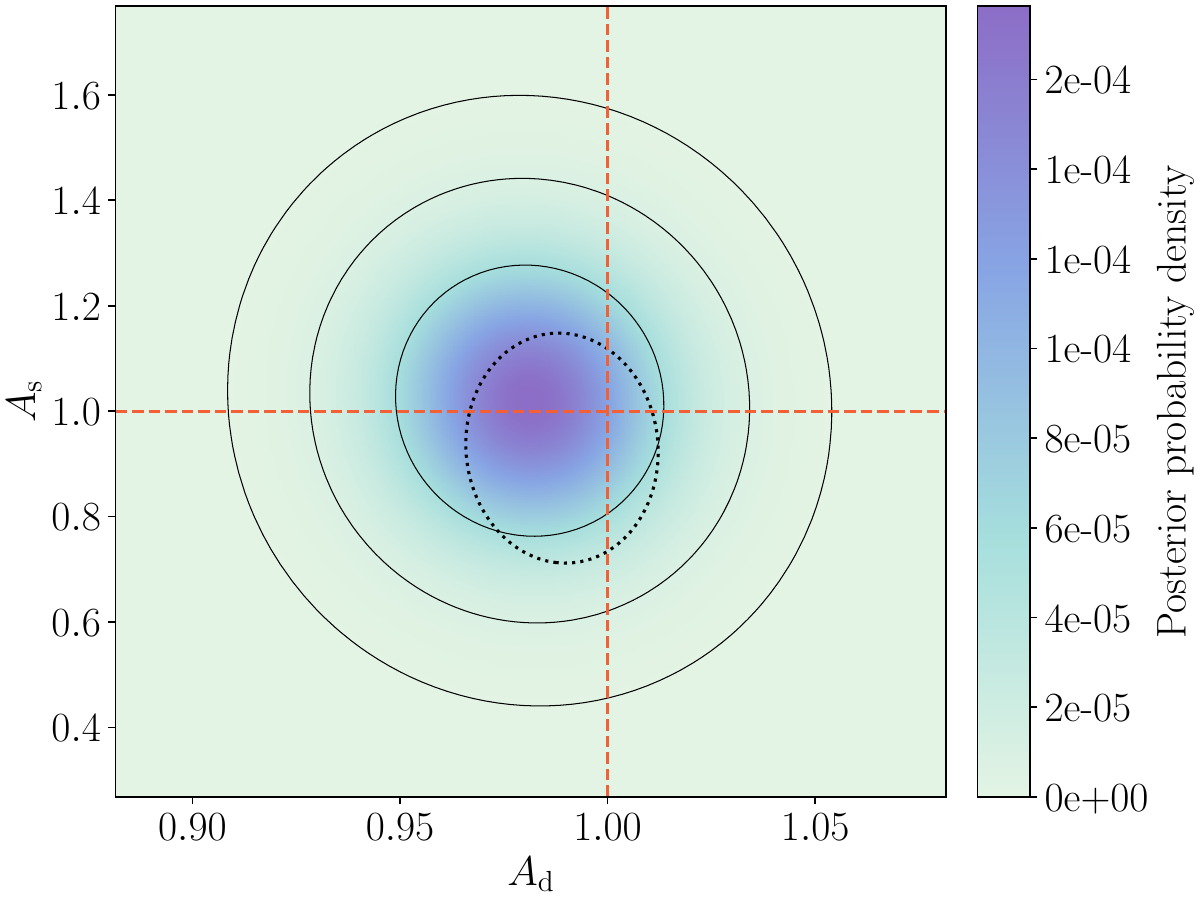}
        \caption{Parameter estimation}
        \label{fig:combine:etce_amp}
    \end{subfigure}
    \caption{Panel~(a): evidence (solid lines) or absence (dotted lines) for displacement memory based on multiple BH GW observations with the LIGO-Virgo network, as well as in a LIGO-Voyager network (where Voyager  is a proposed upgraded of LIGO making use of the same LIGO facility~\cite{Adhikari:2019zpy}). 
    Panel~{(b)}: spin memory evidence using ET. 
    Filled regions in both panels show differences between expected and pessimistic BH merger rates. Evidence is plotted against observation time at design sensitivity.
    %
    Panel~{(c)}: simulated measurements of displacement and spin memory strains $A_\text{d}$ and $A_\text{s}$, 
    from 1000 optimally-chosen BH mergers observed by ET. Solid curve denote the $1,2,3$-$\sigma$ credible levels, whereas the dotted curve denotes the
 $1$-$\sigma$ credible level up to redshift $z \approx 1$. We assume a $10\,{\rm km}$ triangular configuration for ET and a $40\,{\rm km}$ configuration for CE. See~\cite{Goncharov:2023woe} for more details.
    } 
    \label{fig:memory_gw}
\end{figure*}

A key aspect of GR is the expectation that GWs contribute also through non-oscillatory terms, known as the memory effect~\cite{Zeldovich:1974gvh,Christodoulou:1991cr,Blanchet:1992br,Thorne:1992sdb,Grant:2022bla}, often described as a 
 displacement that manifests as a permanent offset between test particles after the passage of GWs.  The gravitational memory represents a pivotal aspect of the so-called ``infrared triangle'', connecting it with Weinberg's soft graviton theorem~\cite{Weinberg:1965nx} and the Bondi-Metzner-Sachs (BMS) symmetry group of asymptotically-flat spacetimes~\cite{Strominger:2014pwa,Bondi:1962px,Sachs:1962wk}. 
Later studies of BMS symmetries proposed more relaxed fall-off conditions for the spacetime metric near null infinity and have been shown  to lead to new symmetries and new GW memory terms. In particular, one such null memory term is known as the spin memory~\cite{Pasterski:2015tva}. These studies reveal the profound interrelations among disparate aspects of long-distance physics, offering new insights into gravity's universal characteristics~\cite{Strominger:2017zoo}. The GW memory may be then used to experimentally unveil Nature's true asymptotic symmetries~\cite{Goncharov:2023woe}.

The detection of the displacement memory may be within reach of currently operating ground-based detectors~\cite{Lasky:2016knh,Hubner:2019sly,Grant:2022bla,Goncharov:2023woe}, and efforts on this direction are underway~\cite{Hubner:2021amk,Cheung:2024zow}. 
The displacement memory may further assist in breaking degeneracies in parameter estimation~\cite{Gasparotto:2023fcg,Goncharov:2023woe}, identifying low-mass BBHs~\cite{Ebersold:2020zah}, and help distinguish between BHs and NSs~\cite{Yang:2018ceq,Tiwari:2021gfl,Lopez:2023aja}.
Next-generation ground-based detectors, such as ET, may also be able to detect spin memory effects~\cite{Grant:2022bla,Goncharov:2023woe}. 
In figure~\ref{fig:memory_gw}, we show how evidence for the displacement and spin memory accumulates as a function of the observation time for various ground-based detectors and ET~\cite{Goncharov:2023woe}. 
In short, operating at design-sensitivity, ET may be able to detect displacement memory with only a handful of BBHs, and it may be able to detect spin memory in less than a year of uninterrupted observations.

\subsubsection{Fundamental aspects of the two-body problem}

\paragraph{Numerical simulations beyond General Relativity.}
\label{sect:NRdiv1}

Numerical simulations are needed for the purpose of performing model selection and comparing the beyond-GR with the standard GR hypothesis in the nonlinear regimes probed by merging binaries or by the gravitational collapse. This topic is discussed in detail in section~\ref{section:div8}. However, even before attempting to model the signal to an accuracy level matching the ET sensitivity, just performing  simulations beyond GR requires significant theoretical advances, notably in terms of viable theories and of a more rigorous understanding of the numerical evolution beyong GR. We discuss some paradigmatic examples below. 

\subparagraph{BBH mergers.}

Tracking the merger of two BHs requires fully nonlinear numerical simulations. While these are viable in GR (see e.g.~\cite{alcubierre2008introduction} and references therein), performing such simulations beyond GR can be challenging, not only because of the technical difficulties but also because of the need of a well-posed formulation. In fact, for many conjectured extensions, even the existence of such formulation remains an open issue. Nonetheless, in some cases, for instance Bergmann-Wagoner scalar-tensor theories~\cite{fujii2003scalar}, a well-posed formulation has been found~\cite{Salgado:2008xh}  and BBH simulations have been carried out, e.g., in~\cite{Healy:2011ef}. In this class of theories, however, stationary BHs are still described by the Kerr metric~\cite{Sotiriou:2011dz}, yielding deviations from GR which are difficult to probe with BBH mergers. 
This limitation may be overcome if the scalar field is supported by time-dependent boundary conditions or spatial gradients~\cite{Berti:2013gfa}. 

Other type of scalar-tensor theories, like Einstein-Maxwell-dilaton gravity~\cite{Hirschmann:2017psw} and cubic Horndesky gravity~\cite{Kovacs:2019jqj}, have also been proven to allow for a well-posed formulation, although hyperbolic evolution is guaranteed only in the so-called weak-coupling regime for the latter. Numerical simulations have been performed~\cite{Kovacs:2020pns,Kovacs:2020ywu,AresteSalo:2022hua}, and in some cases, if the scalar field is massive, a long-lived cloud may survive long enough to significantly affect the waveforms~\cite{Hirschmann:2017psw,Figueras:2021abd}. 
Furthermore, numerical simulations of coalescing BBHs have also been performed for theories for which stationary BH solutions are different from those of GR~\cite{East:2020hgw,Corman:2022xqg,AresteSalo:2022hua,AresteSalo:2023mmd}. However, such simulations have shown the formation of elliptic regions, signaling the lack of predictability, when curvature invariants become large~\cite{R:2022hlf,Doneva:2023oww}. 

An alternative approach consists of expanding the field equations in a small-coupling constant parameterizing deviations from GR. In principle, this can potentially circumvent the well-posedness issues, provided other constraints (such as, e.g., causality~\cite{Endlich:2017tqa}) are enforced. Along these lines, an initial value formulation has been studied in various theories~\cite{Delsate:2014hba,Figueras:2024bba}, and simulations have also been performed through an expansion of the coupling constant~\cite{Okounkova:2018abo,Okounkova:2019dfo,Okounkova:2021xjv,Okounkova:2022grv,Witek:2018dmd,Okounkova:2020rqw}.
However, these simulations show a secular growth in the inspiral phase, which signals the breakdown of the perturbative expansion at late times. These developments clearly illustrate the need to further strengthen the theoretical backbone of modifications of GR, and in particular a consistent embedding into a complete framework.  

Another interesting aspect of scalar-tensor theories is the presence of so-called {\it spontaneous scalarization}~\cite{Doneva:2017bvd,Silva:2017uqg,Dima:2020yac,Herdeiro:2020wei}, where BHs with a non-trivial scalar field configuration are also solutions of the theory. For certain values of the mass and/or spin, the Kerr solution becomes unstable, and spontaneously grows a scalar field. The dynamics of this mechanism was studied in the context of BBH merger, both in the limit of small coupling constant~\cite{Silva:2020omi,Elley:2022ept,Doneva:2022byd}, and for finite couplings, within the modified harmonic formulation~\cite{East:2021bqk,AresteSalo:2022hua,AresteSalo:2023mmd,Corman:2022xqg}. 

In conclusion, despite recent progress, theoretical control over modifications of GR, and the  the accuracy of corresponding waveforms for the merger phase of BBHs, is not yet at the required level, and further development is required, to be ready when ET will become operational.

\subparagraph{BNS mergers.}

\begin{figure}
    \centering
    \includegraphics[width=0.7\linewidth]{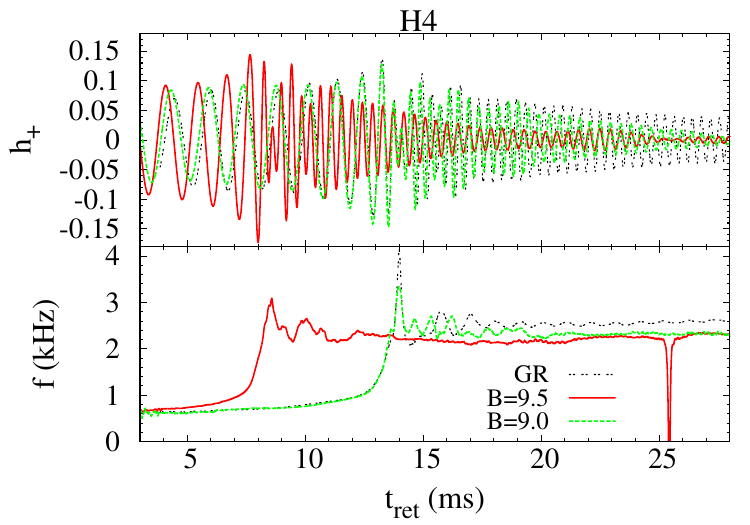}
    \caption{GWs from the coalescence of equal-mass NSs in a scalar-tensor theory compared to GR. Upper panel: Plus polarization along the axis perpendicular to the orbital plane. Lower panel:
    GW frequency vs.\ retarded time. 
    Simulations assume a binary total mass $m = 2.7M_\odot$ and the AH4 EoS, with $B$ denoting the scalar-Ricci coupling strength. Credits~\cite{Shibata:2013pra}.}
    \label{fig:Shibata2013_fig4b}
\end{figure}

Numerical simulations beyond GR have also been explored for BNS mergers, within the context of parameterized tests of GR (e.g.~\cite{Tahura:2018zuq}) or theory-specific effective models for waveform building (e.g.~\cite{Khalil:2022sii}). 
It is worth mentioning that probes of modifications of GR with NSs almost inevitably suffer from uncertainties in the nuclear EoS, that are often degenerate with modified-gravity effects, see section~\ref{subsec:fundphysandEoS} and section~\ref{section:div6}. While BBHs are cleaner from the point of view of matter ``contamination'', BNS may display very different phenomenology~\cite{Berti:2015itd}. An example of a strong-field effect that has been unveiled by numerical simulations is dynamical scalarization~\cite{Barausse:2012da, Palenzuela:2013hsa, Shibata:2013pra,Taniguchi:2014fqa}, originally discovered by Damour and Esposito-Farèse~\cite{Damour:1993hw}, who demonstrated that NSs may develop a nontrivial scalar charge. Observations of radio pulsars strongly constrain the scalar charge, and the parameter space for spontaneous scalarization~\cite{Damour:1996ke,Antoniadis:2012vy,Anderson:2019eay}. Yet,
BNS simulations have shown that large scalar charges are possible during the late inspiral and plunge phases, which can significantly impact the waveforms~\cite{Barausse:2012da, Palenzuela:2013hsa, Shibata:2013pra}. This is evident, for example, in figure~\ref{fig:Shibata2013_fig4b} where the scalarized NS merge faster compared to GR. Theory-agnostic models~\cite{Khalil:2019wyy} suggest that this effect is expected whenever the theory predicts scalarization of isolated compact objects, which has been confirmed, for instance, by numerical simulations of NS mergers in a class of theories that admit scalarization~\cite{Kuan:2023trn}.
The high sensitivity of ET in the kHz region makes it ideal for capturing (or constraining) such effects.



Numerical simulations in these theories have also revealed that the
post-merger BNS signal may be appreciably different from GR both in
scalar-tensor theory~\cite{Shibata:2013pra} and in theories of gravity
involving higher curvature corrections~\cite{East:2022rqi}. For
instance, the post-merger dynamics affects the lifetime of the remnant
hypermassive NS, as well as the amount of ejected material which can
source electromagnetic counterparts. The oscillation spectrum of the
newly formed hypermassive remnant can be significantly altered by the
presence of a scalar field and new types of scalar-field-induced modes
can be excited. Since the post-merger signal is known to depend
significantly on the nuclear EoS, many of the effects observed in
these simulations may be degenerate with the unknown EOS in the NS
core. Systematic investigations are therefore needed in order to
identify distinct signatures of modifications to GR, such that the
observation of the post-merger spectrum by ET will impact not only the
understanding of nuclear matter but also putative new degrees of
freedom.


\subparagraph{Stellar core collapse.}

Stellar core collapse beyond GR was first studied in the Damour--Esposito-Farèse scalar-tensor model~\cite{Novak:1999jg,Gerosa:2016fri}. The final outcome of the collapse strongly depends on the theory parameters, and it has been shown that three possibilities for the remnant are possible~\cite{Gerosa:2016fri}:  a GR (i.e non-scalarized) NS, a scalarized NS, or a short-lived proto-NS followed by the formation of a nonscalarized BH. An important advance was the extension to the case of massive scalar-tensor theories. The presence of a scalar mass reconciles the theory with binary pulsar observations for a larger range of parameters~\cite{Ramazanoglu:2016kul,Yazadjiev:2016pcb}, while providing interesting consequences for the GW signal~\cite{Sperhake:2017itk,Cheong:2018gzn,Rosca-Mead:2019seq,Rosca-Mead:2020ehn,Kuroda:2023zbz,Geng:2020slq}. The most prominent feature is the possible existence of an inverse-chirp signal that can last for years with a near monochromatic signature on time scales of the order of one month. The results indicate that for a source located at $D=10\,{\rm kpc}$, the resulting GW signal, observed for 2 months, can reach an SNR of over 1000 for ET~\cite{Rosca-Mead:2020ehn}. 
Stellar core collapse has also been considered in other modified theories, e.g.~\cite{Kuan:2021lol}, with rich phenomenology. For instance, the proto-NS oscillation modes, potentially detectable with ET, can be affected significantly by the presence of a scalar field~\cite{Blazquez-Salcedo:2021exm,Kruger:2021yay}.

\paragraph{Analytical approaches.}



Time and again, the rich history of the two-body problem in gravity has shown that with any major advance, beyond just accessing  the next level in the perturbative scheme of choice, post-Newtonian (PN)~\cite{Blanchet:2013haa,Porto:2016pyg,Goldberger:2022ebt},  post-Minkowskian (PM)~\cite{Dlapa:2023hsl}, or self-force (SF)~\cite{Barack:2018yvs} approximations, the resolution of new fundamental aspects of the problem was needed. For instance, the complete understanding of the 4PN dynamics in GR was only possible after the interplay between near- and
far-zone fields was clarified~\cite{Damour:2014jta,Galley:2015kus, Bernard:2017bvn, Foffa:2019yfl}, allowing for a systematic resolution of the IR/UV divergences which (separately) plagued the potential and radiation regions~\cite{Porto:2017dgs}. Furthermore, recent developments in the study of scattering processes using techniques inspired by collider physics, e.g.~\cite{Bern:2019nnu,Kalin:2020mvi,Kalin:2020fhe,Kalin:2020lmz,Mogull:2020sak,Bern:2021yeh,Dlapa:2021npj,Dlapa:2021vgp,Dlapa:2022lmu,Kalin:2022hph,Jakobsen:2023ndj,Jakobsen:2023hig,Driesse:2024xad,Dlapa:2024cje,Driesse:2024feo}, have also opened new venues towards simultaneously tackling near- and far-zone dynamics, while at the same time incorporating an infinite tower of velocity corrections at each order in Newton's constant. To quote an example, the study of hereditary and radiation-reaction terms~\cite{Damour:2020tta,Bini:2021gat,Bini:2022enm,Bern:2021yeh,Dlapa:2021npj,Dlapa:2021vgp,Dlapa:2022lmu} were instrumental in the determination of the imprint of nonlinear gravitational effects in the dynamics at 5PN order~\cite{Porto:2024cwd}, as well as in the resolution of apparent discrepancies with the previous literature.

One of the key aspects in the computations of gravitational observables is the role of time nonlocality, already present at 4PN, due to tail effects~\cite{Damour:2014jta,Galley:2015kus, Bernard:2017bvn, Foffa:2019yfl}. These are nonlinear contributions associated with the emitted GWs interacting with the background geometry, yielding orbit-dependent results. At the same time, nonlocality may also be present for memory terms at 5PN order, due to the interaction with the earlier radiation instead, as well as in corrections stemming from second-order effects in the leading radiation-reaction force~\cite{Porto:2024cwd}. All of these subtleties illustrate the intricate connection between several physical effects, as we move forward in the perturbative expansion, in a nonlinear theory such as GR. Fortunately, many of these issues are naturally handled by importing ideas from particle physics---such as the effective field theory approach~\cite{Goldberger:2004jt,Porto:2005ac,Goldberger:2005cd,Porto:2007qi,Goldberger:2022ebt,Porto:2016pyg}---which provide a systematic framework to tackle the two-body problem in gravity~\cite{Foffa:2019yfl,Blumlein:2022qjy,Almeida:2021xwn,Cho:2022syn,Amalberti:2024jaa,Porto:2024cwd}.

\begin{figure}[t]
	\centering
    \includegraphics[width=.65\textwidth]{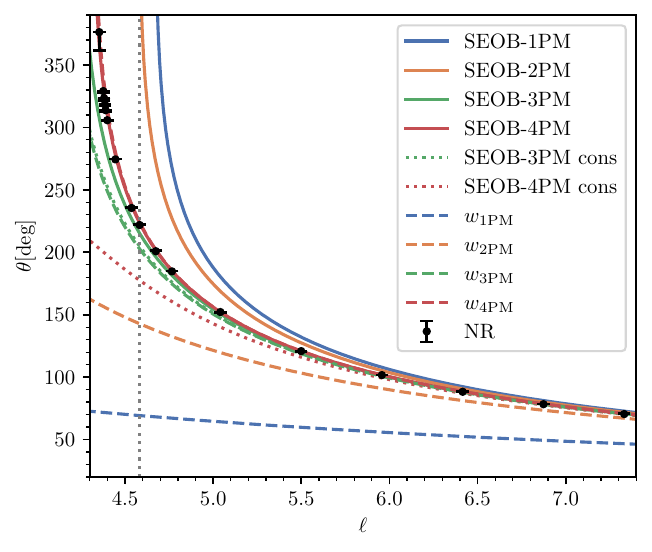}
	\caption{Comparison of the scattering angle in various PM approximations vs numerical simulations. See~\cite{Buonanno:2024vkx} for details.}
	\label{fig:chi4PM}
\end{figure}

Moving onto the case of scattering events, the state-of-the-art results in the PM expansion~\cite{Dlapa:2022lmu,Dlapa:2023hsl,Jakobsen:2023hig}, endowed with a resummation scheme~\cite{Kalin:2019rwq,Kalin:2019inp}, have proven to be very successful in matching numerical simulations with exquisite level of precision~\cite{Damour:2022ybd,Buonanno:2024vkx,Swain:2024ngs}, see figure~\ref{fig:chi4PM}. For the case of bound states, on the other hand, fundamental challenges arise connecting between observables for unbound/bound motion. The relevance of scattering dynamics for the understanding of the bound problem was originally triggered by the observation that data from asymptotic states can also provide information about an effective potential~\cite{Damour:2016gwp,Damour:2017zjx}, as well as by the ``boundary-to-bound'' dictionary relating observables for hyperbolic (deflection angle) and elliptic (periastron advance) orbits through analytic continuation of the radial  action~\cite{Kalin:2019rwq,Kalin:2019inp,Cho:2021arx}. The challenge, however, is to properly separate the universal  part of the dynamics (local-in-time) from the orbit-dependent part (nonlocal-in-time)  ~\cite{Galley:2015kus,Bini:2020nsb,Bini:2020hmy}. While the latter is amenable to the boundary-to-bound connection~\cite{Cho:2021arx}, the former depends on the type of trajectory, and therefore cannot be directly extrapolated from hyperbolic to elliptic motion. This observation also directly impacts the expected connection between the mass-dependence of the scattering result and bound observables~\cite{Kalin:2019inp,Kalin:2019rwq,Damour:2019lcq}, which becomes even more subtle at higher PN/PM orders~\cite{Dlapa:2024cje,Porto:2024cwd}. 

Progress in the direction of capturing the universal part of scattering data has already appeared in~\cite{Dlapa:2024cje}, where the local-in-time contribution to the scattering angle was obtained at 4PM order, and a total bound Hamiltonian derived using a hybrid PN/PM scheme for the remaining nonlocal-in-time effects. However,  due notably  to the challenging ``integration problem"~\cite{Dlapa:2023hsl,Driesse:2024xad}, a scalable approach, likely merging several perturbative (PN/PM/SF) schemes and novel ideas, will be sorely needed to achieve the level of precision required by  ET. As expected, the hunt for accuracy will inevitably lead to a deeper understanding of the gravitational interactions, and perhaps to novel (non-perturbative) descriptions of Einstein's theory, which will be put to the test in the new area of GW science. See section~\ref{section:div8} for more details. 

\subsection{Testing the nature of compact objects \& horizon-scale physics}
\label{div1_natureofcompact}


Current electromagnetic and gravitational observations are compatible with BHs as predicted by GR~\cite{EventHorizonTelescope:2019ggy,EventHorizonTelescope:2022xqj,GRAVITY:2020gka,LIGOScientific:2016lio,LIGOScientific:2019fpaxf,LIGOScientific:2020tif,LIGOScientific:2021sio}. 
An important open issue is to which level these observations confirm that dark compact objects have an event horizon concealing a curvature singularity, where it is also known that Einstein's classical field equations must break down. 
From the theoretical standpoint, the possibility that quantum effects might not only resolve the singularity but also modify the physics at horizon-size scales is attracting more
interest. 
Several recent investigations also suggest that quantum gravity effects
may show up at distances larger than the Planck scale~\cite{Bekenstein:1995ju,Dvali:2011nh,Dvali:2011aa,Casadio:2021eio,Cadoni:2022chn,Casadio:2023ymt,Bena:2022rna},
and make predictions that may be tested using GW
astronomy~\cite{Cardoso:2019apo,Laghi:2020rgl,Agullo:2020hxe,Brustein:2020tpg,Mayerson:2020tpn,Brustein:2021bnw,Chakravarti:2021clm,Coates:2021dlg,Nair:2022xfm,Chakraborty:2023zed} (see~\cite{Cardoso:2019rvt} for an overview).
Another possibility is the presence of exotic compact objects (ECOs), which are regular (horizon-less) objects predicted in various gravitational models (e.g., fuzzballs~\cite{Mathur:2005zp}, gravastars~\cite{Mazur:2004fk}), as well as produced by dark matter and/or exotic fields (e.g., boson stars~\cite{Liebling:2012fv}, wormholes~\cite{Morris:1988cz}).
In the following, we will discuss the phenomenology of horizon-scale physics and ECOs, emphasizing their detectability with ET in a model-agnostic way, but also occasionally referring to specific models.

\subsubsection{Inspiral tests}

\paragraph{Model-agnostic tests of the multipolar structure.}


The multipolar structure of a Kerr BH is uniquely determined by its mass and spin~\cite{Hansen:1974zz},
$M_{\ell}^{\rm BH}+ i S_\ell^{\rm BH}=M^{\ell+1}\left(i \chi\right)^\ell$, where $M_\ell$ and $S_\ell$ are the mass and current multipole moments, respectively, ${\chi:=J/M^2}$ is the dimensionless spin and $M$ the mass.
One can then parametrize deviations from the Kerr multipole moments for exotic compact objects in the following way:
\begin{equation}
\label{eq:momentsECOs}
M^{\rm ECO}_{\ell}=M^{\rm BH}_{\ell} + \delta M_{\ell}\quad\,,\quad
S^{\rm ECO}_{\ell}=S^{\rm BH}_{\ell} + \delta S_{\ell}\,,
\end{equation}
with $\delta M_{\ell}$ and $\delta S_{\ell}$ model-dependent corrections. This parametrization can also be extended beyond axisymmetry~\cite{Bena:2020see,Bianchi:2020bxa,Bena:2020uup,Bianchi:2020miz,Loutrel:2022ant}. Generically, the dominant multipolar effect comes from the quadrupole moment, $M_2$.
\begin{figure}
    \centering
    \includegraphics[width=0.7\linewidth]{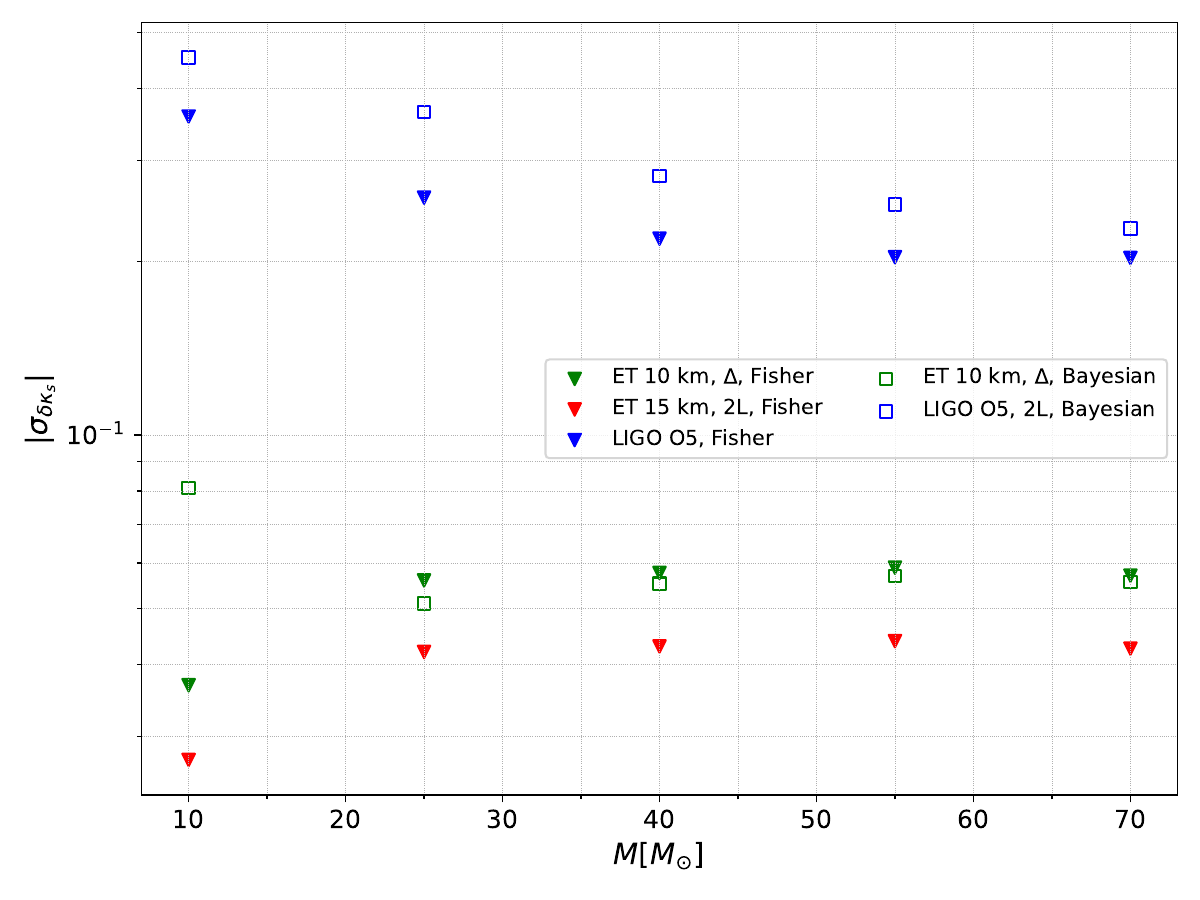}
    \caption{68\% constraints on $\delta k_s$, quantifying deviations from the Kerr quadrupole moment, as a function of binary total mass ($d_L=400\,{\rm Mpc}$). We assume mass ratio $q=3$, aligned spins $\chi_1=0.9$ and $\chi_2=0.8$, and inclination $\iota=\pi/3$. Blue/green triangles represent LIGO O5 and ET constraints (Fisher analysis); empty squares show Bayesian inference results. We show results for both a $10-{\rm km}$ triangular configuration and a $15-{\rm km}$ 2L configuration of the ET observatory.
    }
    \label{fig:quadrupole_ETvsO5}
\end{figure}
Figure~\ref{fig:quadrupole_ETvsO5} shows the projected constraints on the dimensionless parameter $\delta\kappa_s$ (defined as $\delta\kappa_s=\kappa_s-\kappa_s^{\rm Kerr}$, where $\kappa_s=(\kappa_1+\kappa_2)/2$ and $\kappa_i:= -M_2/(M^3\chi^2)$ for each i-th binary component) from a typical ET observation and compared with LVK bounds. The constraints were obtained assuming aligned binaries, using an augmented TaylorF2 waveform, and both using the Fisher-matrix code GWFISH~\cite{Dupletsa:2022scg} and through a Bayesian inference with BILBY~\cite{Ashton:2018jfp}. Interestingly, the Fisher-matrix and Bayesian results are in fairly good agreement between each other, especially for ET, as expected given the higher SNR.
Overall, ET bounds can improve by more than an order of magnitude compared to future LVK observations, and the constraints with the $15-{\rm km}$ 2L configuration are better than those with a $10-{\rm km}$ triangular configuration by a factor of a few, in agreement with the analysis in Ref.~\cite{Branchesi:2023mws}.


\paragraph{Model-agnostic tests of tidal effects.}\label{sec:ECO_tides}

\begin{figure}
    \centering
    \includegraphics[width=0.7\linewidth]{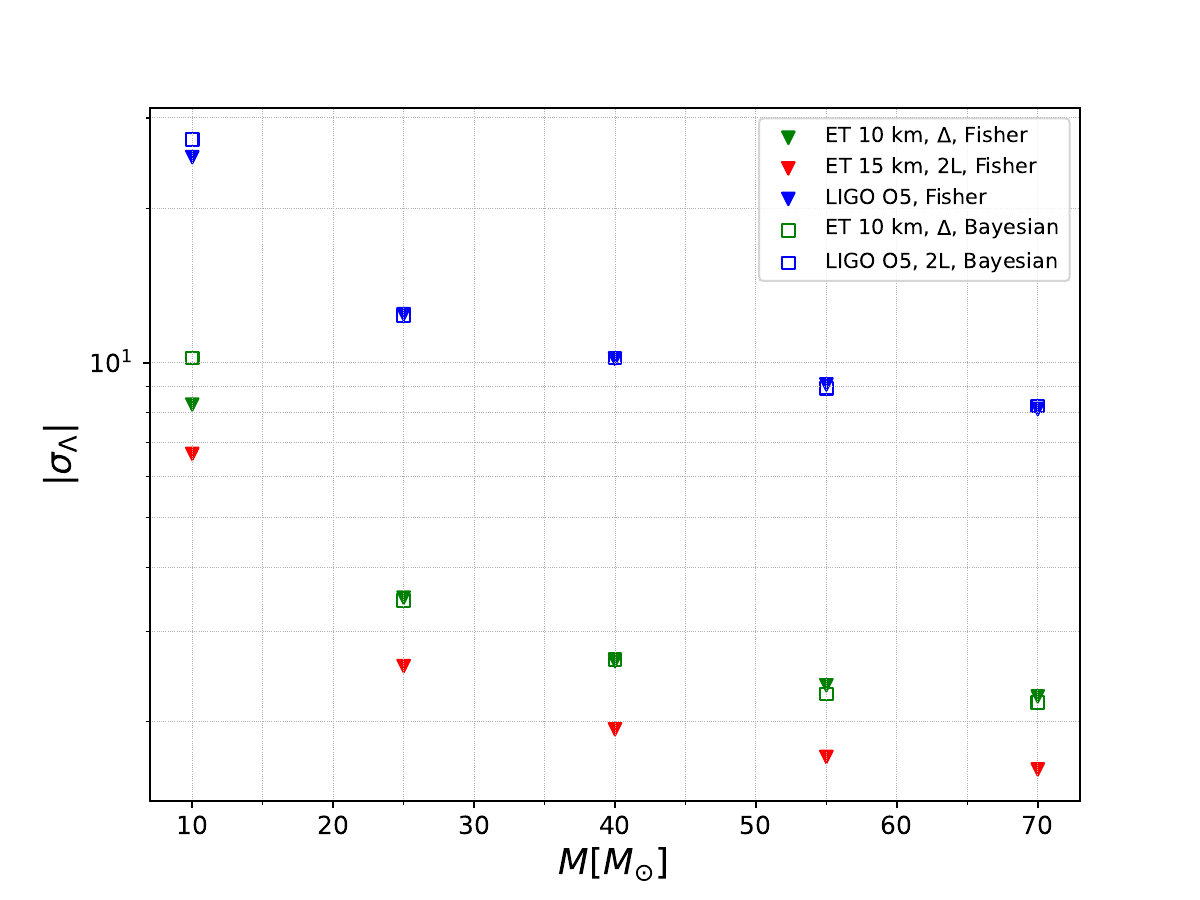}
    \caption{Same as in figure~\ref{fig:quadrupole_ETvsO5} but for  the tidal deformability parameter $\Lambda$ as a function of the 
    total mass of the binary ($d_L=400\,{\rm Mpc}$).
    In this case we assume mass ratio $q=3$, zero spins, and inclination $\iota=\pi/3$.
    We show results for both a $10-{\rm km}$ triangular configuration and a $15-{\rm km}$ 2L configuration of the ET observatory.
    }
    \label{fig:tidal_ETvsO5}
\end{figure}

Since the BH's of GR have been shown to have tidal deformability that vanishes in the static limit~\cite{Binnington:2009bb, Damour:2009vw,Chia:2020yla,Iteanu:2024dvx}, the detection of a nonzero value would constitute a smoking-gun signature of new physics~\cite{Porto:2016zng}, either in the form of deviations from GR~\cite{Cardoso:2017cfl,Cardoso:2018ptl}, ECOs~\cite{Cardoso:2017cfl}, or condensates of ultralight dark matter particles, e.g.~\cite{Baumann:2018vus}. 
Alternatively, it might signal dense environments around merging BHs~\cite{Cannizzaro:2024fpz,Katagiri:2024fpn}.
The largest effect is described by the quadrupolar, electric-type, Love number that parameterizes the proportionality between the gravitational field and the objects' deformation. The Love numbers of various ECOs, including boson stars, were computed in~\cite{Pani:2015tga, Uchikata:2016qku,Mendes:2016vdr,Giudice:2016zpa,Cardoso:2017cfl, Sennett:2017etc,Raposo:2018rjn,Berti:2024moe}.
%
General prescriptions for computing quantum mechanical contribution to the Love numbers of astrophysical BHs were suggested in~\cite{Brustein:2020tpg}, and the classical limit was discussed in~\cite{Brustein:2021bnw}.
For instance, the area 
quantization~\cite{Bekenstein:1973ur} may induce nonzero Love numbers~\cite{Nair:2022xfm}, and a model for dynamical Love numbers of ECOs was recently developed in~\cite{Chakraborty:2023zed}.
Despite the differences among models, an overall common feature of ECOs,  as well as regular (devoid of singularities)~\cite{Carballo-Rubio:2023mvr} or quantum~\cite{Casadio:2023iqt} 
BHs is the presence of internal structures. As a result, their tidal response is expected to scale  qualitatively similarly to those of NSs. It is thus straightforward to develop model-agnostic tests of the tidal deformability in compact binaries, by including tidal corrections to the waveform (starting at 5PN order)~\cite{Cardoso:2017cfl,Sennett:2017etc}. Figure~\ref{fig:tidal_ETvsO5} shows the prospects for measurability of the Love numbers with ET and compares it with future LVK bounds (see~\cite{Chia:2023tle} for a recent analysis with current LVK data). Also in this case we adopted an augmented TaylorF2 waveform, assuming aligned spins, and performed the analysis both with a Fisher matrix and with a Bayesian inference, finding good agreement and better constraints  with the $15-{\rm km}$ 2L configuration~\cite{Branchesi:2023mws}.
As we see in the plot, the bounds with ET improve by roughly an order of magnitude relatively to LVK O5. For the case of stellar-mass objects (which can be confused with ordinary NSs) approximately universal relations between different tidal Love numbers~\cite{Yagi:2013sva} can be exploited to break the degeneracy among different models~\cite{Berti:2024moe}. This requires measuring both the leading quadrupolar electric Love number, and a subdominant one (either the magnetic quadrupolar or the electric octupolar, entering at 6.5PN and 7PN, respectively), a challenging task that is possibly achievable only with next-generation GW detectors~\cite{JimenezForteza:2018rwr}.

Another peculiar property of classical BHs is the so-called tidal heating~\cite{Hartle:1973zz,Hughes:2001jr,Goldberger:2005cd,Porto:2007qi}, which may be also modified for quantum BHs, or suppressed for horizon-less compact objects. This effect is particularly significant for highly spinning BHs, and in the latest stages of the inspiral, since it enters the GW phase at $2.5{\rm PN}$  ($4{\rm PN}$) order for spinning (nonspinning) binaries. At the same time, it is also mildly correlated with the time and phase of coalescence in the waveform, and therefore hard to measure. Constraints of the amount of dissipation introduced by tidal heating are stronger for eccentric~\cite{Datta:2023wsn} or highly spinning binaries with large mass ratios~\cite{Maselli:2017cmm,Datta:2019epe} and are particularly stringent for EMRIs/IMRIs~\cite{Hughes:2001jr,Bernuzzi:2012ku,Taracchini:2013wfa,Harms:2014dqa,Datta:2019euh,Datta:2019epe,Maggio:2021uge,Zi:2023geb,Datta:2024vll}. For comparable mass binaries, generic PN deviations are searched for with test-of-GR pipelines~\cite{LIGOScientific:2021sio}, whereas a specific analysis with LIGO data was performed in~\cite{Datta:2020gem,Chia:2024bwc}. EMRI/IMRIs are among the main sources of space missions such as LISA~\cite{Colpi:2024xhw}, but they may also be detectable by ET if subsolar-mass compact objects exist~\cite{Barsanti:2021ydd}.
%
%

\paragraph{Constraining couplings to scalar fields.}

In addition to model-agnostic inspiral tests discussed above, for concrete models one can attempt to construct inspiral waveforms from first principles. In this case the deviations from the standard BH waveforms are not independent, but eventually dictated by the fundamental coupling constants of the model. For instance, in the case of a boson star,
the tidal deformability of a spherically symmetric configuration can be shown to depend directly on the couplings of the theory~\cite{Sennett:2017etc}, which in turn can be used to discern between fundamental interactions from measurements of finite-size effects in the waveforms~\cite{Pacilio:2020jza}.  

\begin{figure}[t!]
    \centering
    \includegraphics[width=0.48\textwidth]{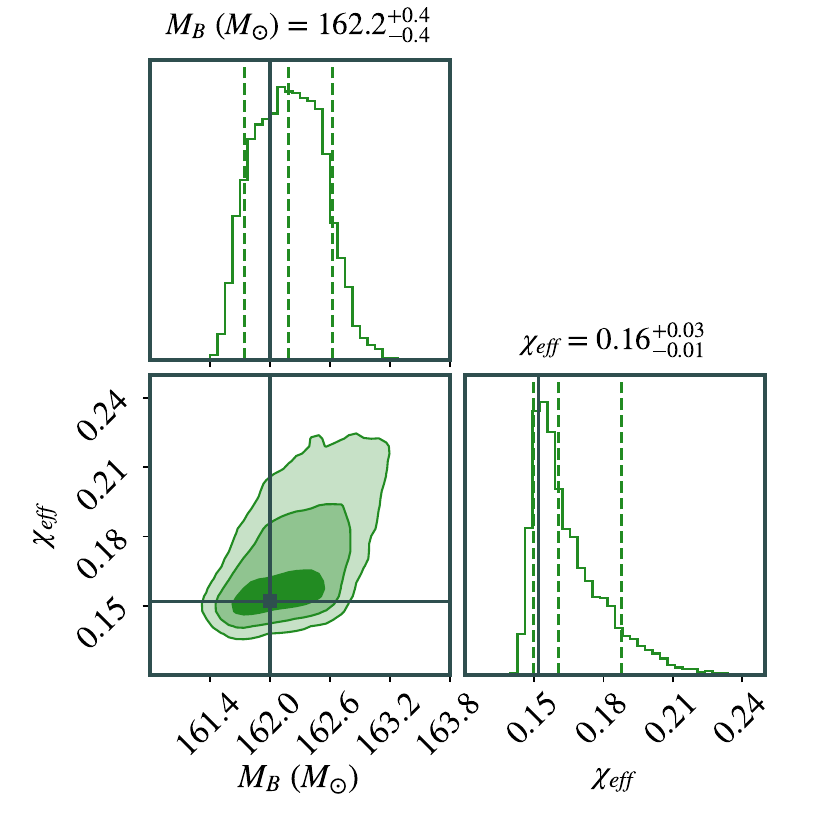}
    \includegraphics[width=0.48\textwidth]{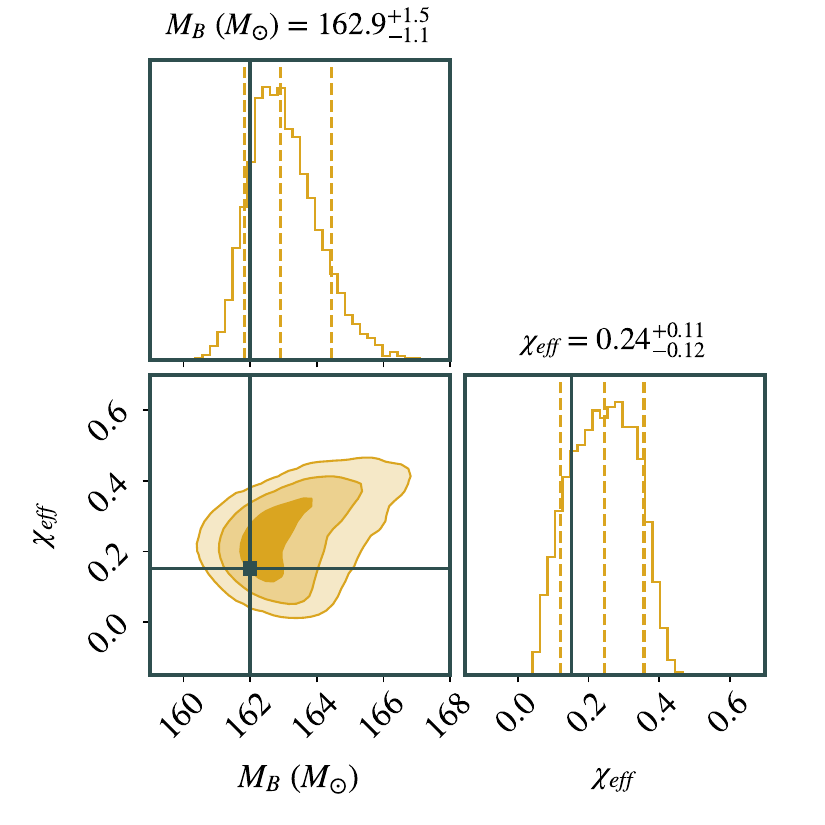}
    \caption{Posterior distributions for the effective spin parameter $\chi_{\rm eff} = (m_1\chi_1 + m_2\chi_2)/(m_1 + m_2)$ and $M_B\equiv \sqrt{\lambda}/\mu^2$, with $(\mu,\lambda)$ the boson's mass and self-interaction coupling, from a mock boson-star binary signal with ET (left) and LVK O4 (right) at $d_L=400\,{\rm Mpc}$. The binary has chirp mass $\mathcal{M}=8M_\odot$, mass ratio $q=0.6$, and component (aligned) spins $(\chi_1,\chi_2)=(0.2,0.1)$. The inspiral cutoff frequency is fixed at the Roche frequency $f_{\text{RO}}=120\,{\rm Hz}$, corresponding to the binary tidal disruption. 
 }
    \label{fig:comparison_corner}
\end{figure}
Figure~\ref{fig:comparison_corner} presents a subset of posterior distributions illustrating the analysis of a boson-star binary signal injected in both ET and O4 noises. 
Since the fundamental coupling constant $M_B$ is related to the tidal deformability only logarithmically~\cite{Pacilio:2020jza}, the constraints improve only by a factor three. Nevertheless, ET capability of constraining the fundamental couplings of a scalar field forming a boson star at the subpercent level will help discriminate between various classes of compact objects and ultimately facilitate model selection. 

\subsubsection{Ringdown and post-merger tests}


\label{ringdown}

The ringdown waveform, corresponding to the phase when the binary coalescence settles down and approaches stationarity, can be modeled as a superposition of quasinormal modes (QNM), writing $h(t)=h_+(t)+ih_\times(t)$, with
\begin{equation}
\label{eq:rdmodel}
   h_{+,\times}(t)=\sum_{lmn}{\cal A}_{lmn}\cos\left(2\pi f_{lmn}t+\phi_{lmn}^{+,\times}\right)e^{-t/\tau_{lmn}}\,\mathcal{Y}^{lm}_{+,\times}(\iota)\,,
\end{equation}
where $\phi_{lmn}^{\times}=\phi_{lmn}^{+}-\pi/2$, $\iota$ is the spin inclination angle  of the remnant with respect to the line of sight, and
$ \{f_{lmn}, \tau_{lmn}, \mathcal{A}_{lmn}, \phi_{lmn}\equiv\phi_{lmn}^+ \}$ are the frequency, damping time, amplitude, and phase of the $(lmn)$ quasinormal mode, respectively, and the 
$\{+,\times\}$ polarizations should be defined using a spin-weighted spheroidal harmonics basis.
The integers $(lmn)$ refer to the multipolar, azimuthal, and overtone index, respectively, where $n=0$ corresponds to the fundamental tone and $(lmn)=(220)$ is the dominant mode in a non-precessing quasicircular coalescence.
The amplitude scales as ${\cal A}_{lmn}\sim M_f/d_L$, where $M_f$ is the mass of the remnant BH and $d_L$ is the luminosity distance from the source~\cite{Gossan:2011ha}.

The (complex) QNM frequencies, $f_{lmn}-i/\tau_{lmn}$, 
depend only on the remnant's properties. If the latter is a Kerr BH, the entire spectrum depends only on the mass and spin. Hence, observing multiple ringdown modes from a single source provides a null-hypothesis test of the nature of compact objects.
Furthermore, most modified gravity theories and models of ECOs predict extra modes not present in GR~\cite{Crescimbeni:2024sam} and different amplitudes and phases compared to the standard BH binary case in GR~\cite{Forteza:2022tgq}, so both effects can be studied in a model agnostic way. 

The prospects for accurately observing ringdown signals with ET are extremely promising~\cite{Berti:2016lat,Bhagwat:2023jwv}.  ET will allow us to transition from observing such events occasionally, and with moderate SNR, to becoming almost daily occurrences extracted with great accuracy. For this reason, testing GR and the nature of the remnant with ringdown signals is a key science objective of ET, as we detail below.

\paragraph{Black-hole spectroscopy.}

\begin{figure*}
    \centering
    \includegraphics[width=0.495\textwidth]{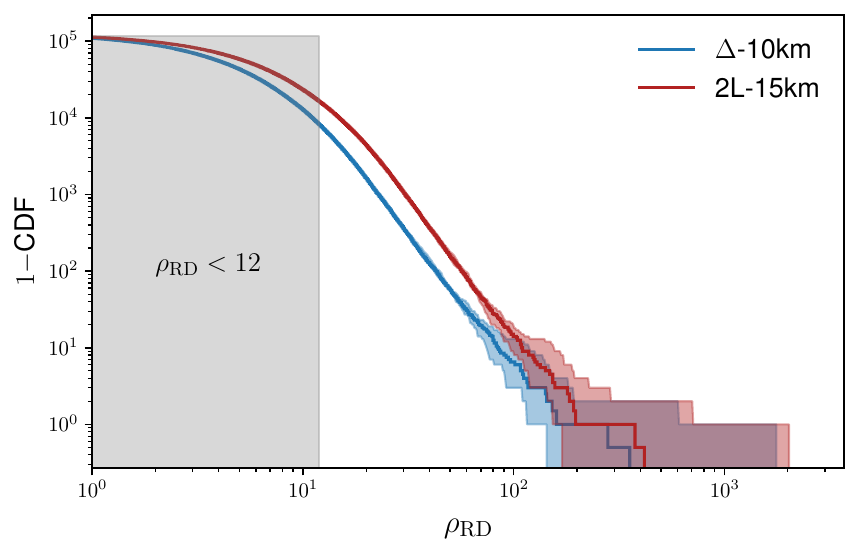}
    \includegraphics[width=0.44\textwidth]{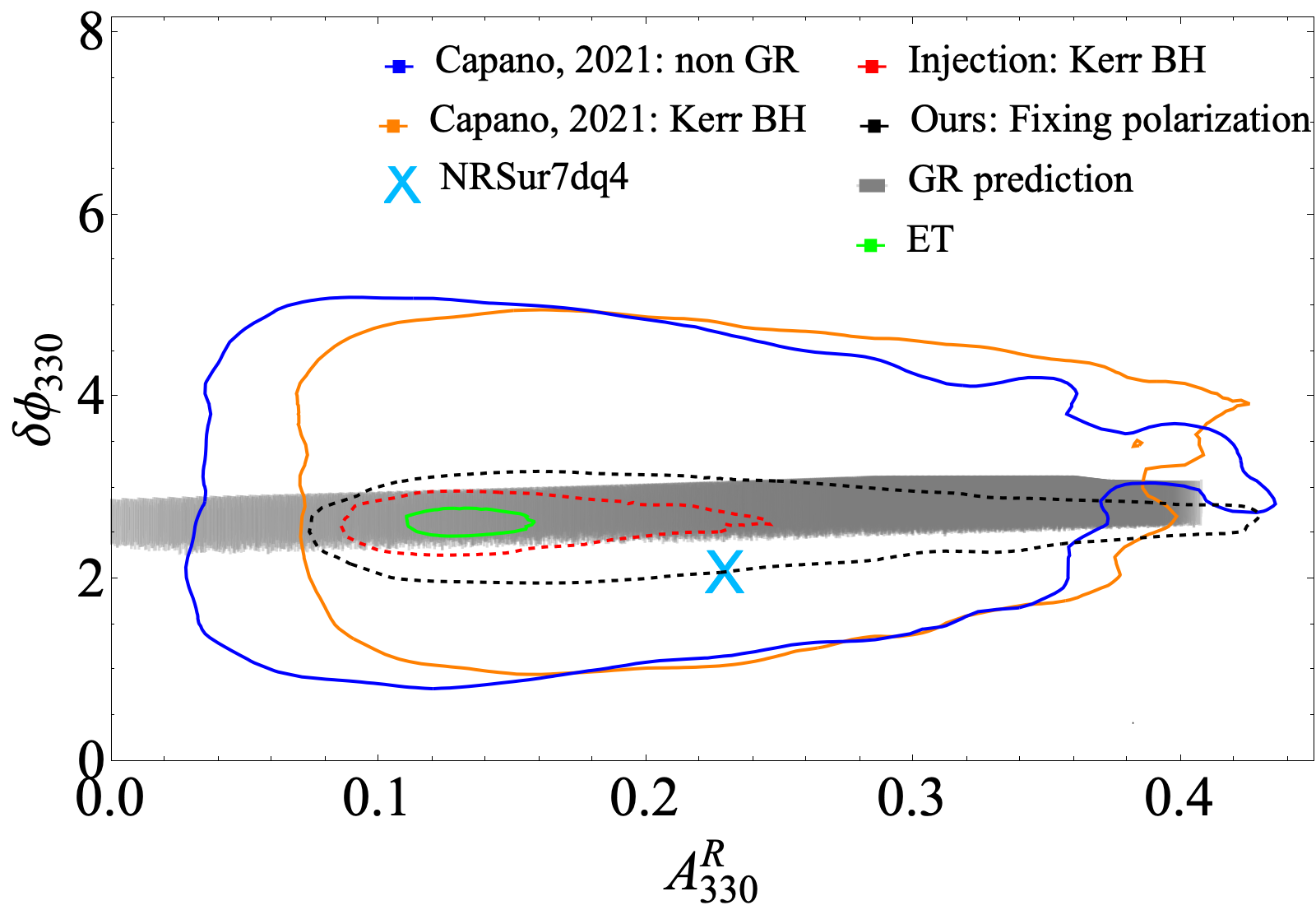}
    \caption{Left: Inverse cumulative distribution of ringdown SNRs $\rho_{\rm RD}$ for different detector configurations. The shaded band corresponds to signals with $\rho_{\rm RD}<12$. From ref.~\cite{Bhagwat:2023jwv}. 
    Right: Forecasts for measuring the amplitude and phase of a secondary ($lmn=330$) ringdown mode in a GW190521-like event detected by ET, compared with current uncertainties and with the GR prediction. Extended from~\cite{Forteza:2022tgq}.  
    }
    \label{fig:ringdown:snr}
\end{figure*}

Using stellar mass BBH population estimates calibrated to LVK catalogues up to O3~\cite{Mapelli:2021gyv}, ET is expected to observe $\sim 10^4$ events with a ringdown-only SNR ($\rho_{\rm RD}$) bigger than 10 of which $\sim 5/{\rm yr}$ will have $\rho_{\rm RD}>100$~\cite{Bhagwat:2023jwv, Branchesi:2023mws}. Furthermore, with ET we expect a ringdown-horizon up to $z\sim 10$~\cite{Baibhav:2018rfk}. Numerical simulations show that, in non-precessing quasi-circular BBH ringdowns, the dominant mode is the $(2,2,0)$ mode, while the most excited subdominant modes are the $(3,3,0)$, $(2,1,0)$ and $(4,4,0)$ modes~\cite{Gossan:2011ha,Kamaretsos:2012bs,JimenezForteza:2020cve,Forteza:2022tgq}.
The frequency and damping time of the dominant mode can be inverted to infer an independent measurement of the remnant's mass and spin, while any subdominant mode can be used as a null-hypothesis test of GR, since they are fully determined once the fundamental mode is fixed.
Ref.~\cite{Bhagwat:2023jwv} quantifies ET's measurability of subdominant modes and prospects of BH spectroscopy. ET is expected to measure $\mathcal{O} (10^3)$ ringdowns/yr with $\mathcal{O}(10)\%$ precision for deviations of the subdominant frequencies from the GR predictions, whereas $\mathcal{O}(10)$ ringdowns/yr are expected to have $\mathcal{O}(1)\%$ precision, see left panel of figure~\ref{fig:ringdown:snr}. Constraints on deviations of the damping times are less precise and therefore will produce weaker tests of the BH spectrum.
Apart from stellar mass BH binaries, ET will also be able to perform a reasonably good BH spectroscopy with a measurability $\sim 10 \%$ for $\sim 5-20$ events/yr from light seed supermassive BH population~\cite{Bhagwat:2021kwv}, thus providing complementary information to LISA~\cite{Colpi:2024xhw}. The possibility of using pre-merger information for ringdown based tests has also been explored in~\cite{Brito:2018rfr} which shows promising prospects for the SNRs attainable by ET. Furthermore, ET will be able to put strong constraints using consistency tests based on ringdown mode amplitudes and phases~\cite{Forteza:2022tgq}, see right panel of figure~\ref{fig:ringdown:snr}.

\begin{figure*}[t]
    \centering
    \includegraphics[width=0.495\textwidth]{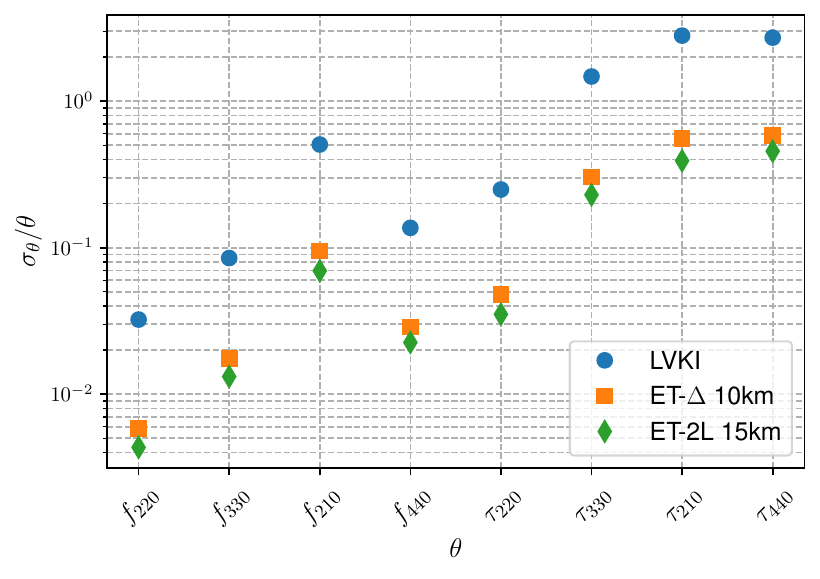}
    \includegraphics[width=0.495\textwidth]{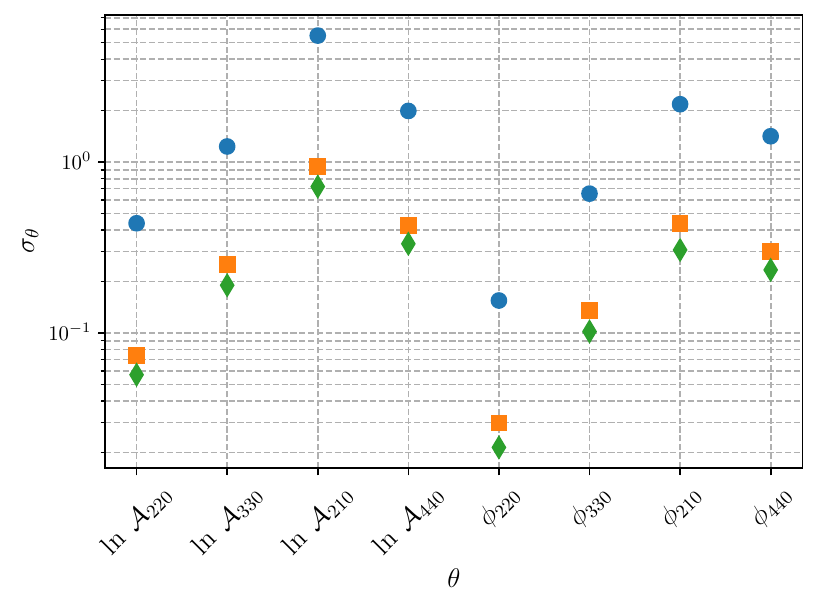}
    \caption{BH spectroscopy with a GW150915-like event using LVKI or ET. \textit{Left:} bounds on the frequencies $f_{lmn}$ and damping times $\tau_{lmn}$. \textit{Right:} bounds on the excitation amplitudes $\mathcal{A}_{lmn}$~\cite{Gossan:2011ha,Kamaretsos:2012bs} and phases $\phi_{lmn}$. }
    \label{fig:ringdown:tests}
\end{figure*}

\begin{figure*}[t]
    \centering
    \includegraphics[width=1\textwidth]{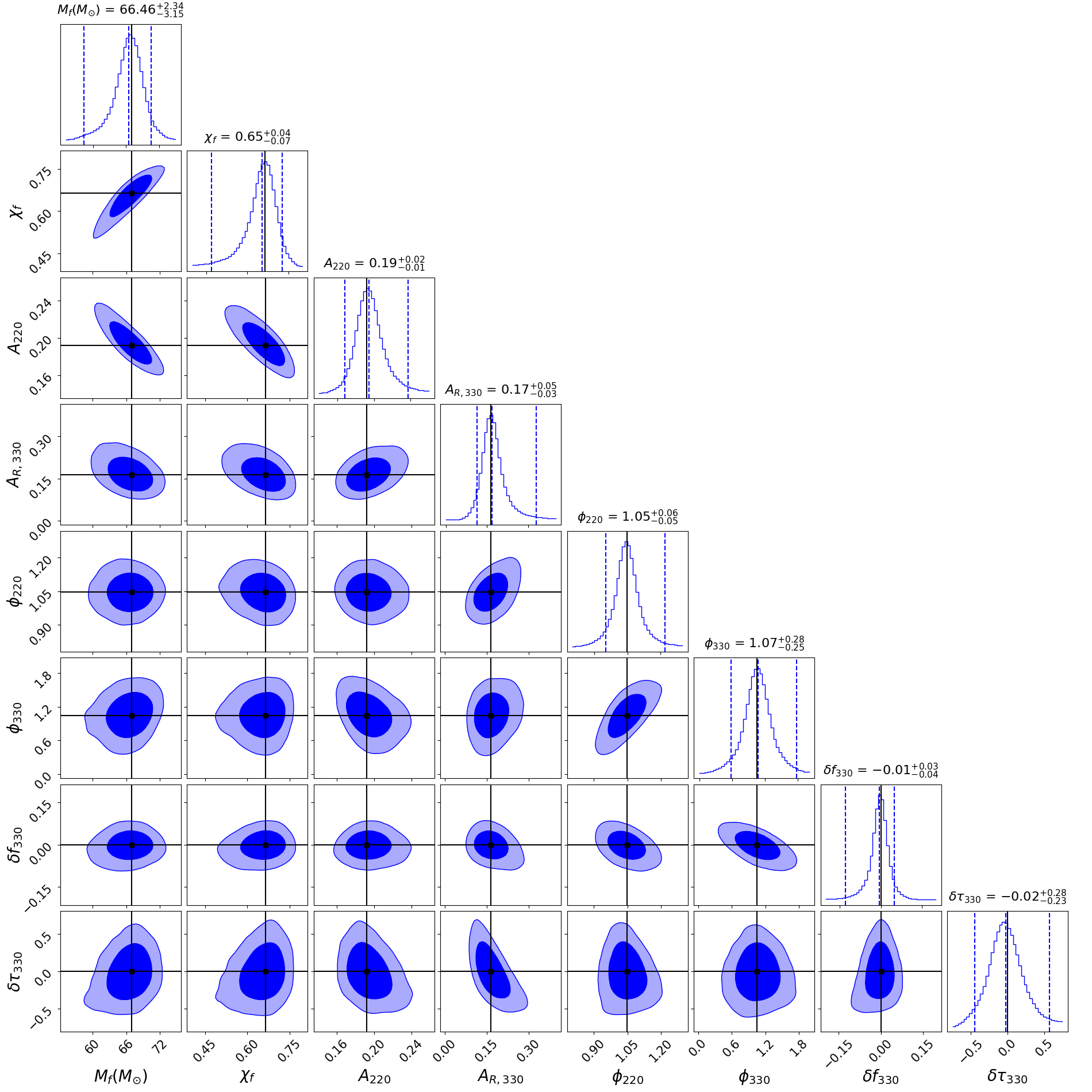}
    \caption{Posteriors of the final mass, spin, and amplitudes/phases of the 220 and 330 modes, and of the 330 GR deviations ($\delta f_{330}$ and $\delta \tau_{330}$) for a GW150914-like remnant, detected with ET in the standard 10-km triangular configuration. 
    Here, $A_R$ is the relative mode amplitude. The 2D distributions show 68\% and 90\% credible levels, while the dashed lines in the 1D distributions mark the 90\% credible intervals. Black lines indicate the injected values.
    }
    \label{fig:ringdown:testsBayesian}
\end{figure*}

In order to demonstrate the impact of ET in performing BH spectroscopy and the large improvement with respect to present GW detectors, 
figure~\ref{fig:ringdown:tests} (left) shows that ET will allow measuring multiple ringdown frequencies with $\mathcal{O}(1\%)$ precision from individual events, and (right) that ${\cal O}(10\%)$ constraints will be placed on the magnitudes of the ringdown amplitudes and phases, thus allowing for novel tests of the BH nature~\cite{Bhagwat:2021kfa,Forteza:2022tgq}.
These results are in agreement with a Bayesian analysis on the detectability of a secondary ($330$) mode 
(see figure~\ref{fig:ringdown:testsBayesian}), which is very challenging with current detectors, mostly due to the limited SNR~\cite{Capano:2020dix} (see also figure~\ref{fig:ringdown:tests}).

Exceptionally loud events with $\rho_{\rm RD}>100$ are expected in ET on a yearly basis and they will deliver the tightest constraints as individual events. Regarding the ability of distinguishing a ringdown waveform in GR from that in an alternative theories of gravity, a study in ref.~\cite{Pacilio:2023mvk} shows that percent-level deviations from GR can be identified with SNR $\rho_{\rm RD}\sim\mathcal{O}(150-500)$, which is well within the ET capabilities. These results are also corroborated by specific case studies in modified theories of gravity~\cite{Tattersall:2019pvx,Carullo:2021oxn,Pierini:2022eim,Cano:2023jbk,Antoniou:2024gdf}. They rely on the QNM spectra in modified theory which are currently available for Kerr-Newman BHs in Einstein-Maxwell theories~\cite{Dias:2015wqa}, slowly-spinning BHs in theories with quadratic curvature corrections~\cite{Pierini:2022eim,Cano:2020cao,Cano:2021myl,Wagle:2021tam,Cano:2023jbk}, Horndeski gravity~\cite{Tattersall:2018nve}, and higher-order effective-field-theory gravity~\cite{Cardoso:2018ptl} to different orders in spin and coupling constants.


Finally, at large SNR the ringdown analysis might be affected by next-to-leading order effects in perturbation theory~\cite{Cheung:2022rbm,Mitman:2022qdl,Lagos:2022otp,Kehagias:2023ctr,Baibhav:2023clw}. 
A recent analysis showed that the dominant nonlinear 
mode may be observable by next generation detectors for 
few tens of events per year, although such results depend on the 
event rates~\cite{Yi:2024elj}. 
The loudest signals can provide constraints on the quadratic mode frequency and 
damping times of the order of few and ten percent, respectively.

\paragraph{Tests of exotic compact objects with quasinormal modes and echoes.}  \label{div1_nearHorizon}




If~the remnant is an exotic compact object (ECO), the amplitude of the QNM will be different, and new modes associated with matter fields may be excited. For a nonspinning object the perturbations of the remnant can be classified as axial or polar, depending on their parity. For BHs, axial and polar modes are isospectral, whereas for a horizon-less compact object, axial and polar QNMs form a doublet~\cite{Maggio:2020jml}. In the limit in which the ECO is almost as compact as a BH, the deviations from the BH QNM spectrum are increasingly large and the QNMs are low-frequencies and long-lived~\cite{Cardoso:2016rao,Cardoso:2016oxy}. These modes are trapped in the cavity of the effective potential formed between the light ring and the compact object interior. Furthermore, the internal structures of regular~\cite{Carballo-Rubio:2023mvr} or (integrable) quantum~\cite{Casadio:2023iqt} 
BHs can induce modifications in the outer geometry predicted by GR~\cite{Calmet:2021stu}.
These modifications will affect the geodesic motion and ringdown phase after the merger.
For example, by modeling the external geometry with a quantum coherent state, one expects deviations as large as $5\%$ from the QNMs of the pure Schwarzschild geometry~\cite{Casadio:2021eio}. 
Measuring QNMs may also enable tests of black hole area quantization~\cite{Cadoni:2021jer,Agullo:2020hxe}. This is being explored through analytical methods incorporating a minimum-length metric to study the impact of area quantization on the ringdown~\cite{Kothawala:2013maa,Kothawala:2014fva,Pesci:2018syy}.

\begin{figure}[t]
\centering
\includegraphics[width=0.7\textwidth]{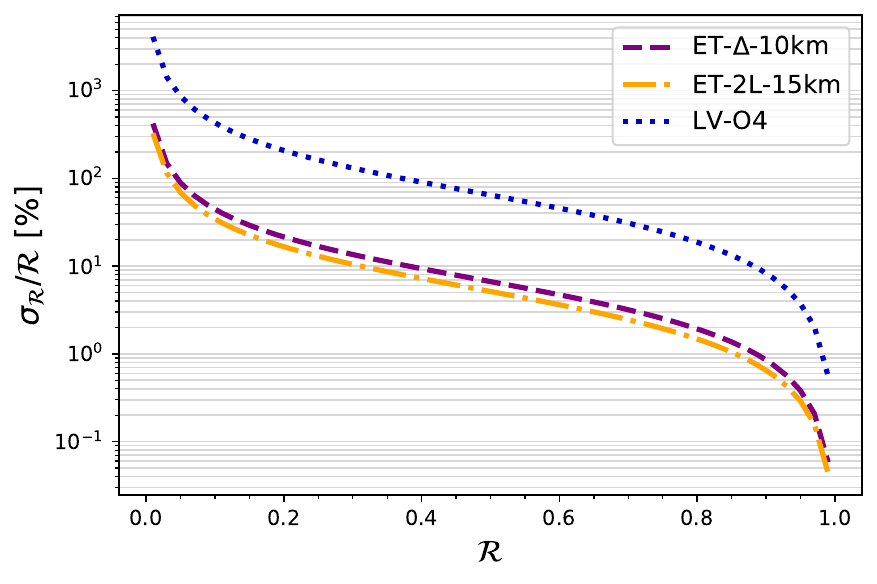}
\caption{Fractional percentage errors on the reflectivity of compact objects for a GW150914-like event~\cite{Branchesi:2023mws}.}
\label{fig:echo}
\end{figure}

In the time domain, the prompt ringdown signal of ultracompact objects would be nearly indistinguishable from that of a BH since it is due to the excitation of the light ring~\cite{Cardoso:2016rao}. If the object does not absorb GW radiation, a train of modulated pulses~--~known as GW echoes~--~would be emitted in the late post-merger stage in addition to the ringdown expected from BHs~\cite{Cardoso:2016rao,Cardoso:2016oxy,Cardoso:2017cqb,Abedi:2020ujo}.
The amplitude of GW echoes depends on the boundary conditions describing the compact object and the initial conditions of the perturbation~\cite{LongoMicchi:2020cwm,Annulli:2021ccn,Xin:2021zir,Srivastava:2021uku,Ma:2022xmp,Vellucci:2022hpl}. 
In particular, the echo amplitude depends on the reflectivity of the compact object to ingoing perturbations, whereas a BH is described by null reflectivity at the horizon.
%
Several searches for GW echoes have been performed based on the superposition of generalized wavelets adapted from burst searches~\cite{Tsang:2018uie,Tsang:2019zra}, Fourier windows using the fact that GW echoes should pile up at specific frequencies~\cite{Conklin:2017lwb,Conklin:2019fcs}, and the coherent WaveBurst pipeline~\cite{Miani:2023mgl}. Following some debated claim~\cite{Abedi:2016hgu,Conklin:2017lwb,Conklin:2019fcs,Abedi:2018npz,Westerweck:2017hus,Nielsen:2018lkf,Lo:2018sep,Uchikata:2019frs,Tsang:2019zra}, the LVK Collaboration found no evidence of echoes in GW data~\cite{LIGOScientific:2020tif,LIGOScientific:2021sio}, both using a template-based search~\cite{Lo:2018sep,Abedi:2016hgu} in the second GW transient catalog and a morphology-independent search~\cite{Tsang:2018uie,Tsang:2019zra} in the third catalog.
%
%
The ET observatory has the potential to observe (or rule out) GW echoes emitted by horizon-less compact objects in a much larger region of the parameter space relative to the one currently available~\cite{Maggio:2019zyv,Testa:2018bzd,LISA:2022kgy,Ma:2022xmp}. 
It is estimated that models with perfect reflectivities can already be detected or ruled out by Advanced LIGO and Virgo; whereas excluding or detecting echoes for models with generic reflectivities as small as a few percent requires ${\rm SNR}\gtrsim 100$ in the ringdown, which will be achieved by ET~\cite{Testa:2018bzd,Maggio:2019zyv}. Ref.~\cite{Branchesi:2023mws} assessed the measurability of a GW150914-like event with different ET configurations, particularly the triangular-10 km and 2L-15 km designs.
The parameters of the system are final mass $M=70M_\odot$ and final spin $\chi=0.68$ placed at a fiducial distance $d_L = 1 \ \text{Gpc}$.
If measured by ET, the first GW event would have had a ringdown SNR between 141 and 196 for the triangular-10~km and 2L-15~km configurations, respectively.
Figure~\ref{fig:echo} shows the fractional percentage errors on the reflectivity of compact objects expected for a GW150914-like event with ET compared to the Advanced LIGO and Virgo noise curve used for O4 simulations~\cite{KAGRA:2013rdx}.
The accuracy on the reflectivity is affected by an order of magnitude between the O4 LIGO-Virgo and ET configurations (see table~\ref{table:echo}).

\begin{table}[t]
\begin{center}
\begin{tabular}{ |p{4.5cm}|p{2cm}|p{2cm}|p{2cm}| }
\hline
\hfil {Detector configuration} & \multicolumn{3}{c|}{$\sigma_{\mathcal{R}}/\mathcal{R} [\%]$}\\ & \hfil {\bf $\mathcal{R}=0.01$} & \hfil {\bf $\mathcal{R}=0.5$} & \hfil {\bf $\mathcal{R}=0.99$} \\
\hline
\hfil LIGO-Virgo O4 & \hfil 4104 & \hfil 67 & \hfil 0.56 \\
\hfil ET $\Delta$-10 km & \hfil 422 & \hfil 6.9 & \hfil 0.058 \\
\hfil ET 2L-15 km & \hfil 327 & \hfil 5.3 & \hfil 0.044 \\
\hline
\end{tabular}
\end{center}
\caption{Fractional percentage errors on the measurability of the reflectivity of compact objects for small ($\mathcal{R} = 0.01$), medium ($\mathcal{R} = 0.5$) and large ($\mathcal{R} = 0.99$) reflectivity of a GW150914-like event~\cite{Branchesi:2023mws}. 
}
\label{table:echo}
\end{table}


\paragraph{Nonlinear dynamics of exotic compact objects.}


Previous tests focused on specific waveform parts (inspiral or ringdown) as these regimes are conveniently modeled perturbatively, forming a useful basis for model-independent parametrizations. More information can be extracted from the full coalescence signal (inspiral, merger, ringdown) by specifying an ECO model. This requires NR simulations which are primarily available for consistent models like boson stars~\cite{Liebling:2012fv}. For instance, for the case of boson stars, 
simulations of binary mergers have shown that collisions may result in a more massive star or a BH~\cite{Palenzuela:2006wp}. The stars' intrinsic properties affect the emitted GW energy and waveform~\cite{Palenzuela:2006wp,Sanchis-Gual:2022mkk}. For example, phase opposition between stars could lead to a repulsive force forming a dipole configuration~\cite{Bezares:2017mzk}. Orbital mergers of boson stars have confirmed that more compact stars may collapse into a BH, emitting a similar signal to a BH binary merger~\cite{Palenzuela:2007dm, Helfer:2021brt}. GW templates from Proca star collisions have been compared to events like GW190521, suggesting that ECO hypotheses can sometimes fit observations slightly better than traditional binary BH models, though more studies are needed to extend these models~\cite{Bustillo:2020syj, CalderonBustillo:2022cja}.
While overall it is challenging to distinguish ECOs from BBH mergers with current observations, the ET improved sensitivity, especially at low frequencies, could ultimately reveal smoking guns of ECOs during the coalescence. Furthermore, if ECOs are present in Nature, they could form hybrid binaries with BHs and NSs, possibly producing an electromagnetic counterpart~\cite{Dietrich:2018bvi}. Highly compact boson stars are also expected to eventually decay into a spherically symmetric state, with gravitational emission serving as a distinguishing signal~\cite{Croft:2022bxq}.

\subsubsection{Neutron stars and fundamental physics}
\label{subsec:fundphysandEoS}


\subparagraph{Modified gravity versus EoS.} NSs provide the densest accessible medium to carry out tests of the coupling of gravitational degrees of freedom to matter fields. Testing gravity with NSs, however, faces the problem of the degeneracies between the EOS of matter in the NS core (still unknown, although increasingly constrained, see section~\ref{section:div6}), and modified-gravity theories~\cite{Yazadjiev:2018xxk,Nobleson:2021itp,Moreno:2023xez}.
In particular, measurements of the tidal deformability is not sufficient to constrain/discard a modified theory of gravity, since varying the EOS can mimic beyond-GR effects, as  will be discussed in more detail  in section~\ref{subsec:newphysics}. In principle, the large number of BNSs detectable with ET, and the accuracy in the measurement of tidal parameters~\cite{Pacilio:2021jmq}, can help reduce the degeneracy. 
In addition, reconstructing the dependence of the tidal deformation with the mass from future ET data one can also constrain some of the parameter space of modified-gravity theories~\cite{Lope-Oter:2023urz,Biswas:2023ceq} .
Another promising (model-agnostic) approach to test gravity with BNSs is by exploiting quasi-universal relations, for example between the tidal deformability and the spin-induced quadrupole moment~\cite{Yagi:2013bca} or between different tidal Love numbers~\cite{Yagi:2013sva} (see~\cite{Yagi:2016bkt} for a review). ET will be crucial in this context, because measuring the quadrupole moment or subleading tidal terms requires a level of precision which is not achievable with current detectors. The same approach can be used to distinguish NSs from ECOs~\cite{Berti:2024moe}. 

Finally, it is relatively common in the numerical simulation community to employ simple polytropic EOS because of their smooth behaviour that avoids propagating errors in an evolution code. This common practice however 
disregards progress in nuclear and particle physics~\cite{Llanes-Estrada:2019wmz}, and leaves out the possibility of first-order phase transitions (non-analyticities) in NS matter. Although this further complicates tests of gravity with NSs, it would be important to systematically include such effects in future investigations (see section~\ref{section:div6}). New physics parameters whose effect is a strengthening of gravity of order 10\% are allowed and degenerate with variations of the EoS. ET data should supply ample opportunity to test best practices in the EOS inference, and modified gravity could also be used to benchmark the uncertainties in nuclear matter.

\begin{figure}
\includegraphics[width=0.47\textwidth]{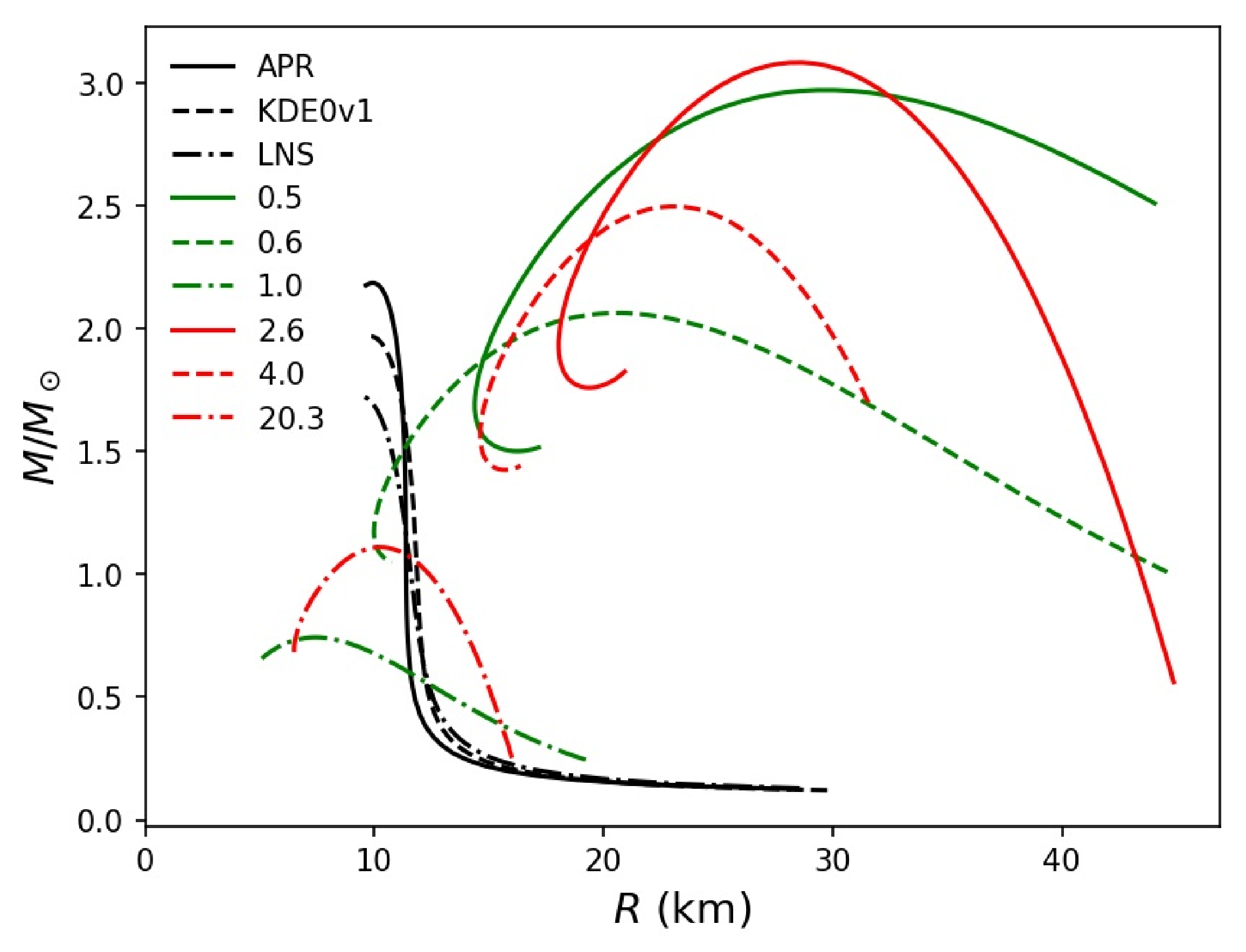}\ \ \ \ 
\includegraphics[width=0.47\textwidth]{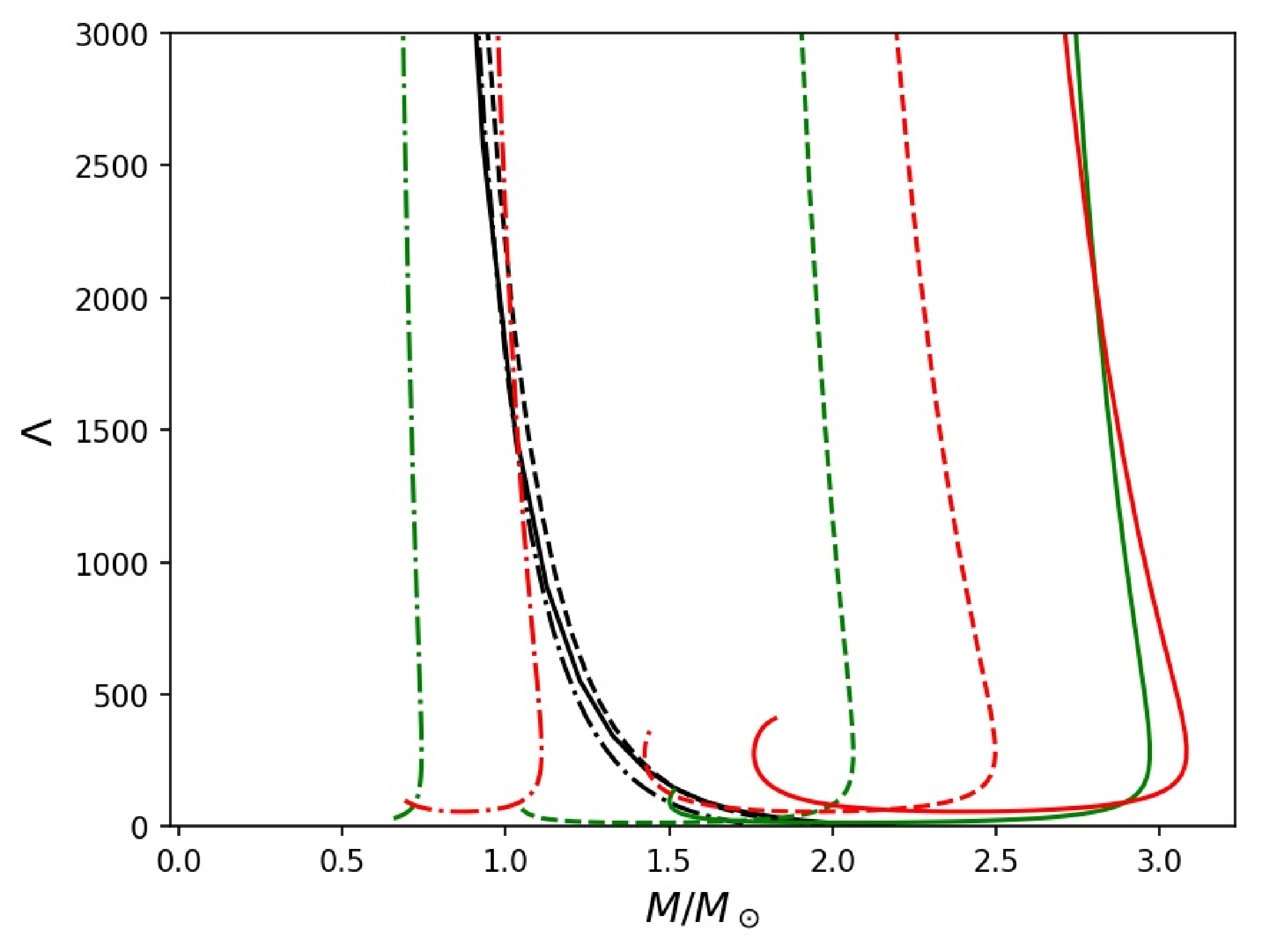}
\caption{ 
\label{fig:DMvsEoS} 
Mass-radius diagram (left) and dimensionless tidal deformability (${\Lambda}$) and mass diagram (right) of a NS with a DM core.
Stars containing fermionic DM (green lines) are labeled by the particle mass $\mu$ (in GeV). 
The ones with bosonic DM (red lines) are labeled by $\rho_0 \hbar^3$ (in $10^{-4}\,{\rm GeV}^4$).
Reproduced from~\cite{Leung:2022wcf} with license from the American Physical Society.}
\end{figure}

\subparagraph{Dark Matter versus EoS.}
%
The possibility of DM detection in NSs has been already emphasized in the literature, e.g.~\cite{Guver:2012ba}. Due to the large uncertainties, the impact of DM on the NS's properties are model dependent. 
Figure~\ref{fig:DMvsEoS} shows the mass-radius and the tidal-deformation/mass diagrams for three EOS (black lines) and for EOS with added DM (see also figure~\ref{fig:degeneracy_FR} for a comparison with the effect of some specific modified gravity theories). Since ET will substantially increase the population of observable NS-participated binaries, potentially detecting events near galactic centers where the DM density may be larger, one possible strategy is thus to employ BNSs to constrain the parameter space of DM models.

Another test of DM-admixed NSs is directly from the merger phase. These can be simulated using standard codes, for instance by adding a new fluid component~\cite{Emma:2022xjs,Bezares:2019jcb}. The presence of DM can impact the properties of the remnant. For example, for a sufficiently large boson-to-fermion ratio ($\sim 10\%$), the quadrupole symmetry of the massive remnant may be broken by the bosonic cores. This would lead to a suppression of the $l=m=2$ emission, and give rise to a new peak, at $f\sim 1$ kHz, related to the $m=1$ mode. Compact dark cores in NSs have also been modeled as point particles in~\cite{Bauswein:2020kor}. In this case, the compact dark objects undergo orbital motion within the merger remnant, generating a GW signal with a frequency of a few kHz. The signal is expected to be weak, but long lasting, and may be detectable if integrated during the remnant lifetime (assumed to be sufficiently long)~\cite{Bauswein:2020kor}.

\subsection{Searches for dark-matter candidates \& new fields}
\label{div1_DMcandidates}



There is overwhelming evidence for the existence of DM, coming from various observations, such as measurements of galaxy rotation curves, the CMB, gravitational lensing, and galaxy cluster dynamics (see e.g.~\cite{Bertone:2004pz, Bertone:2016nfn,Cirelli:2024ssz} for reviews). However, although the existence of DM is by now a nearly irrefutable fact, its nature remains one of the biggest unresolved puzzles in fundamental physics. At the same time, GW observations provide a unique and novel way to search for DM signatures, potentially unveiling its ultimate character~\cite{Bertone:2019irm}. In this section we summarize some of the most promising prospects to search and constrain DM candidates and new fundamental fields with the ET observatory. Notably, the possibility to detect new {\it ultralight} bosons (much below the eV scale), either directly interacting with the GW detectors or surrounding BHs in binaries, as well as other type of ``environmental" effects due to DM such as dynamical friction. As we shall see, these are complementary to searches at particle colliders~\cite{Bertone:2019irm,Adhikari:2022sve}, or with lower-frequency GW observatories such as LISA~\cite{LISA:2022kgy}, or pulsar timing arrays~\cite{NANOGrav:2023hvm}.


\subsubsection{Direct detection of ultralight dark matter with interferometers}

Searches for the imprint of ultralight DM fields in interferometric GW detectors have mainly focused, so far, on vector particles in the form of dark photons. The latter would cause time-dependent oscillations in the mirrors, leading to a differential strain on the detector~\cite{Pierce:2018xmy, Carney:2019cio} (a similar behavior has been predicted for tensor particles~\cite{Armaleo:2020efr}.) 
Dark photons cause a small differential change in the interferometer’s arm length in two ways. First,
the mirrors are subject to slightly different phases of the dark photon field~\cite{Pierce:2018xmy}. Second, an equivalent differential strain is produced by the finite travel time of the laser light in the arms~\cite{Morisaki:2020gui}. The overall signal at the detector's output is stochastic and narrow-band, with an amplitude proportional to the strength of the dark photon/baryon (or baryon-lepton) coupling, and a frequency spread of the order of $10^{-7}~f_0$, with $f_0=m_bc^2/h$, and $m_b$ the dark photon's mass which, in range $10^{-14}$~--~$10^{-11}$ eV, corresponds to frequencies in the ET band. 

\begin{figure}[t!]
	\centering
    \includegraphics[width=0.7\textwidth]{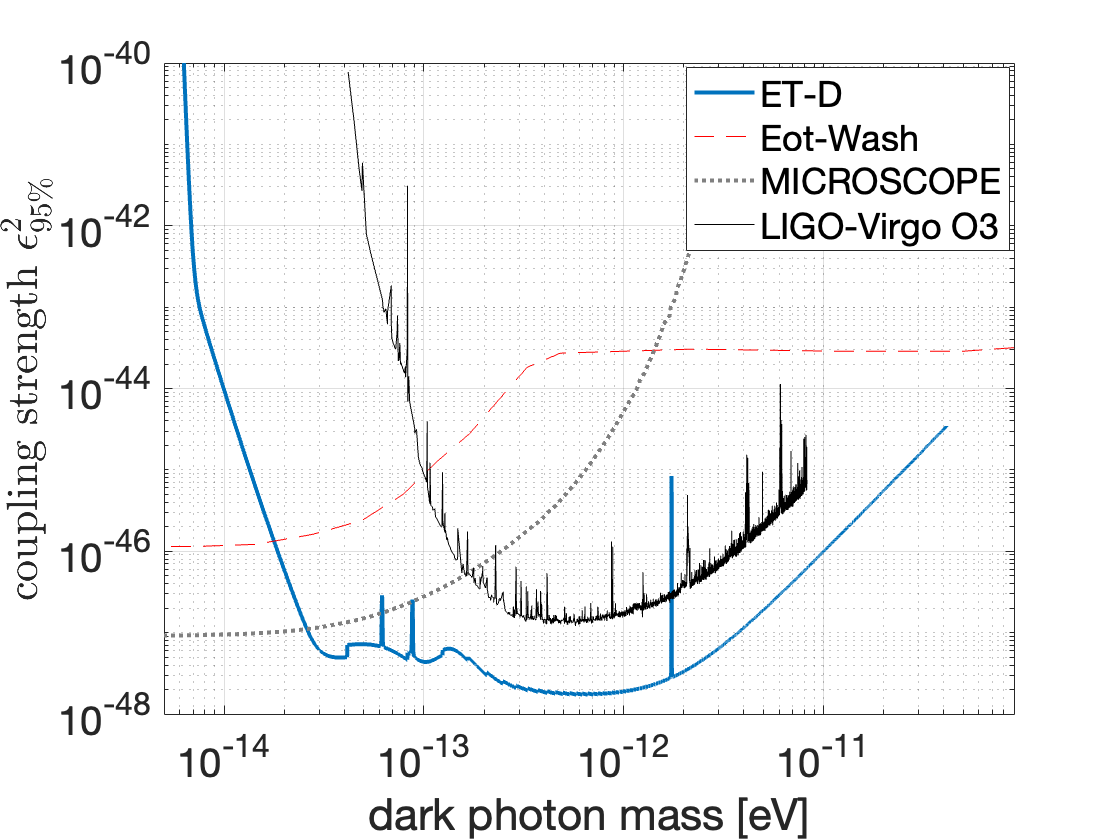}
	\caption{Predicted constraint on dark photon/baryon $U(1)_B$ coupling from ET-D (blue continuous), compared to LIGO O3 upper limit (black continuous) and current MICROSCOPE (dotted)~\cite{Berge:2017ovy} and E\"ot-Wash torsion balance experiment~\cite{Schlamminger:2007ht} (dashed) limits. Adapted from~\cite{LIGOScientific:2021ffg}.}
 \label{fig:dpdm_constraints}
\end{figure}

The dark photon signal requires the same type of methodology typically used for stochastic (or continuous) GW  searches~\cite{Pierce:2018xmy,Miller:2020vsl,Morisaki:2020gui}. In particular, cross-correlation and excess power methods have been used to establish competitive upper limits in the frequency range 25-2000 Hz~\cite{LIGOScientific:2021ffg}. Figure~\ref{fig:dpdm_constraints} is adapted from~\cite{LIGOScientific:2021ffg} and shows upper limits on the dark photon to baryon coupling $U(1)_B$ obtained analyzing O3 LIGO data, compared to MICROSCOPE constraints on scalar dilaton~\cite{Berge:2017ovy} and E\"ot-Wash torsion balance experiment on scalar particles~\cite{Schlamminger:2007ht}, and to the predicted constraint from ET-D.
As shown in~\cite{Michimura:2021hwr,Nakatsuka:2022gaf,LIGOScientific:2024dzy} differences in mirror materials (as it happens in KAGRA detector) can significantly enhance the sensitivity to $U(1)_{B-L}$ gauge symmetry. As a consequence of the variation of fermion masses and electromagnetic fine-structure constant induced by the presence of new interactions, extra scalar fields can also change the size and refractive index of objects. Constraints on scalar couplings, based on the variation of the refraction index and thickness of the beam-splitter of GEO600 detector, have been described in~\cite{Vermeulen:2021epa,Gottel:2024cfj}.

\subsubsection{Environmental effects}\label{sec:environ}

\paragraph{Boson clouds from superradiance.}\label{sect:bosoncloudsdiv1}

Among the most suitable DM candidates for which GW detectors are particularly relevant are ultralight bosons. The existence of such particles has been proposed in a multitude of Beyond Standard Model theories, e.g.~\cite{Peccei:1977hh,Peccei:1977ur,Wilczek:1977pj,Weinberg:1977ma,Svrcek:2006yi,Abel:2008ai,Arvanitaki:2009fg,Goodsell:2009xc,Freitas:2021cfi}, with the earliest candidate perhaps associated with the Peccei and Quinn's solution to the strong CP problem~\cite{Peccei:1977hh,Peccei:1977ur}, implying the existence of a new light (pseudo)-scalar particle: the ``axion''~\cite{Wilczek:1977pj,Weinberg:1977ma}. Bosonic fields with masses in the range $m_b \simeq 10^{-21}$~--~$10^{-11}$ eV have Compton wavelengths that are of the order of the size of known astrophysical BHs, with masses ranging from  stellar-mass, $M_{\rm BH} \simeq 10 M_{\odot}$, to supermassive BHs, $M_{\rm BH} \simeq 10^{10} M_{\odot}$, respectively. These ultralight fields can then render spinning BHs superradiantly unstable -- a process in which the BH's rotational energy is transferred to the boson field, which in turn condensates in a macroscopic ``boson cloud'' around the BH (see ref.~\cite{Brito:2015oca} for a review). 

The possible formation of such clouds has been shown to produce a number of GW signatures. For example, given that rotational energy is extracted from the BHs, a robust observable for this effect is the lack of BHs spinning above a certain threshold, which depends on $M_{\rm BH}$ and $m_b$~\cite{Arvanitaki:2010sy,Arvanitaki:2014wva,Cardoso:2018tly}. Therefore BH spin measurements, either from GW observations or other probes, can be used to constrain the mass of ultralight bosons~\cite{Arvanitaki:2014wva,Arvanitaki:2016qwi,Baryakhtar:2017ngi}. 
In this respect 3G detectors will be particularly useful, since they will both provide accurate measurements of the binary BH spins and will detected a large number of events, allowing for an accurate reconstruction of the ``gaps'' in the BH mass-spin diagram~\cite{Arvanitaki:2009fg,Arvanitaki:2010sy} and, in turn, for a measurement of the putative field mass~\cite{Brito:2017zvb,Cardoso:2018tly,Ng:2019jsx,Ng:2020ruv}. Besides these indirect signatures, the formation of boson clouds also leads to the emission of GWs from the cloud itself~\cite{Yoshino:2013ofa,Arvanitaki:2014wva,Brito:2017zvb,Siemonsen:2019ebd}. Moreover, various modifications with respect to the evolution in vacuum occur when the BH surrounded by a boson cloud is a member of  a binary system~\cite{Baumann:2018vus,Baumann:2019ztm,Boskovic:2024fga,Tomaselli:2024bdd}. We discuss these  possibilities in more detail below. 

\subparagraph{Continuous waves.} Once formed, boson clouds will in general dissipate via annihilation into gravitons, emitting a quasi-monochromatic signal (over a time scale much longer than the scale of superradiant instability) with frequency~\cite{LIGOScientific:2021rnv} 
\be\label{fgwbosoncloudsdiv1}
    f_\mathrm{gw}  \simeq 483\, \text{Hz} \left( \frac{m_\mathrm{b}}{10^{-12}\,\text{eV}} \right) \left[ 1-7\times 10^{-4}\left( \frac{M_\mathrm{bh}}{10M_\odot}\frac{m_\mathrm{b}}{10^{-12} \text{eV}} \right)^2 \right] \,,
\ee
where $m_\mathrm{b}$ is the boson mass  and $M_\mathrm{bh}$ is the black hole mass. The second term in parenthesis is a small correction to the leading term.
In general, scalar boson clouds~\cite{Yoshino:2013ofa,Arvanitaki:2014wva,Brito:2017zvb} emit longer but weaker signals, with respect to vector bosons~\cite{Baryakhtar:2017ngi,Siemonsen:2019ebd} or light tensor bosons~\cite{Brito:2020lup}. Moreover, the signal is characterized by a spin-up, due to the cloud contraction over time~\cite{Baryakhtar:2017ngi,Siemonsen:2022yyf} and, possibly, to the boson self-interactions~\cite{Baryakhtar:2020gao}.  

Searches for continuous gravitational waves from both scalar and vector boson clouds have been proposed~\cite{DAntonio:2018sff,Isi:2018pzk,Jones:2023fzz} and carried on LIGO-Virgo data, and they allowed  setting constraints on the existence and properties of ultralight bosons with mass in the range $10^{-13}$~--~$10^{-11}$ eV. Specifically, scalar clouds have been subject to all-sky searches (namely, without making any assumption about source position and signal parameters). This allows for the exclusion of a portion of the $m_\text{b}$~--~$M_\text{BH}$ plane, given assumptions on the distance, spin, and age of the BHs~\cite{Palomba:2019vxe,LIGOScientific:2021rnv}. A directed search toward a specific potential source, i.e. Cygnus X-1, has also been performed~\cite{Sun:2019mqb}.  

\begin{figure}[t!]
	\centering
    \includegraphics[width=0.7\textwidth]{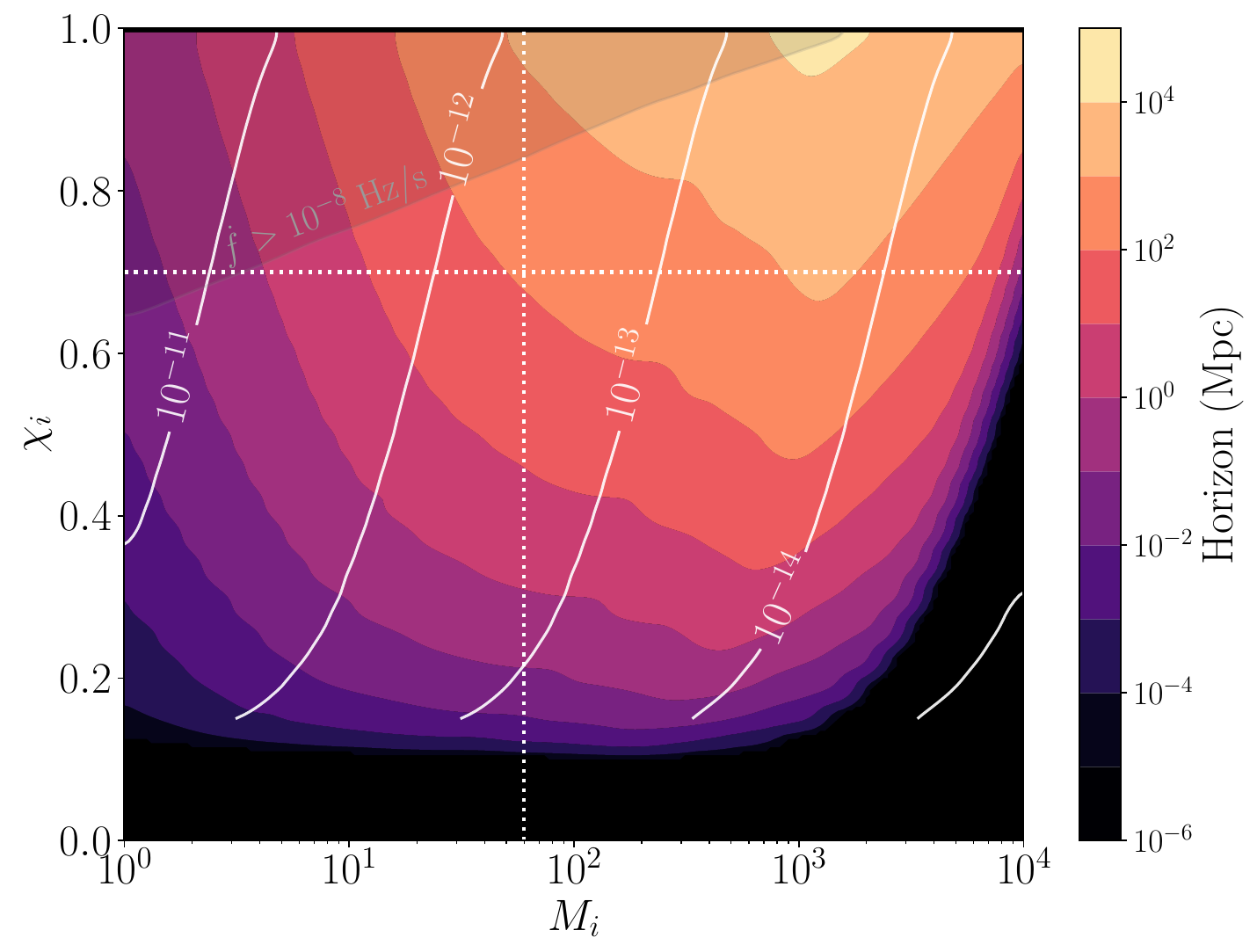}
	\caption{Maximum detectable luminosity distances for optimal scalar clouds around BHs with initial mass $M_i$ and dimensionless spin parameter $\chi_i$ for ET, assuming one year of observation and ET in the triangle configuration (results for the 2L configuration are almost indistinguishable). White contour lines indicate the values of the corresponding boson rest-energy in units of ${\rm eV}$. The dotted white lines highlight the mass and spin of a GW150914-like merger remnant. Adapted from ref.~\cite{Isi:2018pzk} and obtained using the python package gwaxion~\cite{gwaxion}. We are indebted to Max Isi for providing the code.}
	\label{fig:horizon_et}
\end{figure}

\begin{figure}[htb!]
	\centering
    \includegraphics[width=0.7\textwidth]{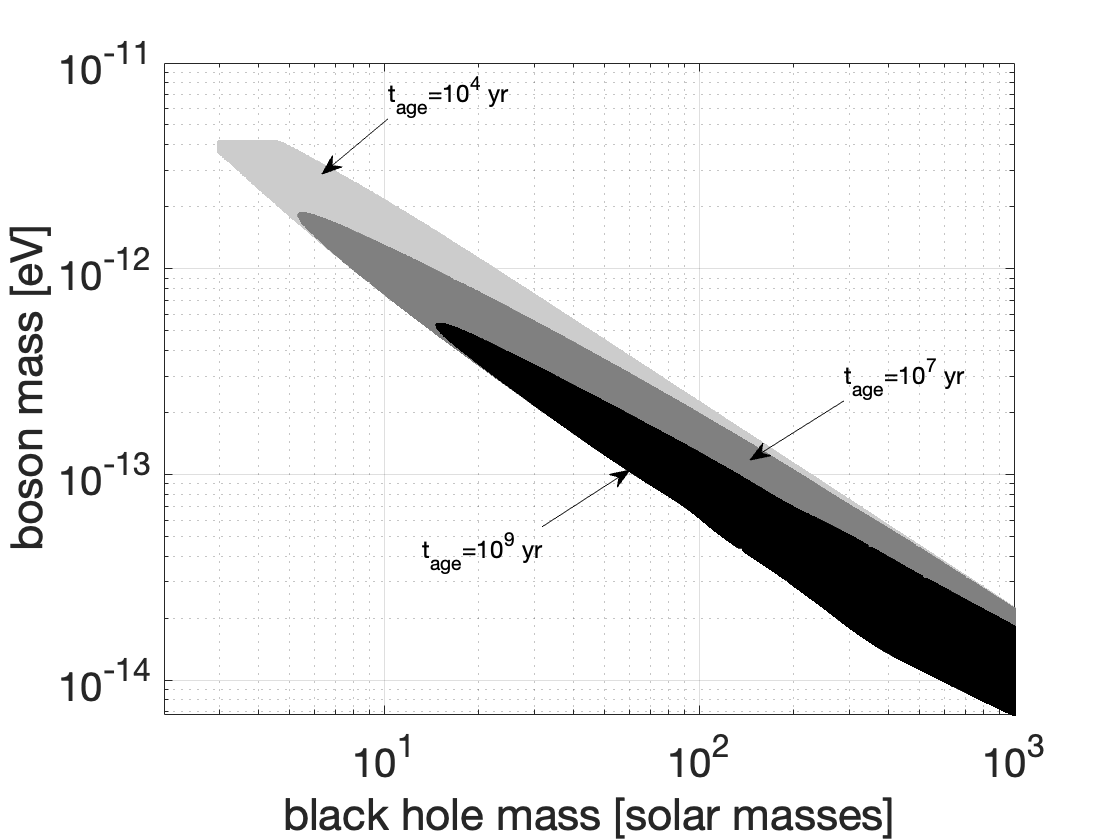}
	\caption{ET exclusion regions for scalar boson clouds assuming a BH distance of 15 kpc, an initial spin $\chi_i=0.6$ and three different BH ages of $10^4,~10^7$ and $10^9$ years.}
 \label{fig:cloud_mass_exclusion}
\end{figure}

The ET sensitivity will allow us to either make a direct detection, or set much more stringent constraints on the boson cloud mechanism.  As an example, figure~\ref{fig:horizon_et} shows the maximum detectable luminosity distance for direct searches over one year of observation, assuming the triangle configuration for ET.
The distance reach is well beyond the Galaxy for a portion of the parameter space, and this makes remnants of BH-BH mergers, for which the final BH mass and spins can be measured to good accuracy, potentially interesting sources. 
All-sky searches will be able to set constraints for galactic scalar boson clouds, as shown in figure~\ref{fig:cloud_mass_exclusion}, where the ET exclusion regions in the $m_\text{b}$~--~$M_\text{BH}$ plane are plotted assuming a BH distance of 15 kpc, an initial spin $\chi_i=0.6$ and three different BH ages of $10^4,~10^7$ and $10^9$ years.

It is worth stressing that strong bosonic self-interactions, couplings of the ultralight boson with Standard Model particles, or other environmental effects such as nearby objects, could also affect the GW signal, since such effects could slow down the superradiance extraction process or even destroy the cloud~\cite{Ikeda:2018nhb,Baumann:2018vus,Baryakhtar:2020gao,Baumann:2021fkf,Omiya:2022gwu,Tong:2022bbl,Spieksma:2023vwl}. One therefore expects that any constraints on ultralight bosons based on continuous GW searches to hold provided that ultralight bosons have weak self-interactions and/or interact weakly with other particles. 

\subparagraph{Stochastic background.}

Besides the continuous GW searches mentioned above, one also expects that the incoherent superposition of all non-resolved GW signals emitted by boson clouds produces a stochastic GW background. This background has been computed in refs.~\cite{Brito:2017zvb,Zhu:2020tht,Tsukada:2020lgt,Yuan:2021ebu} both for scalar and vector clouds and considering both galactic and extra-galactic sources. Constraints based on the lack of observation of a stochastic background in LIGO and Virgo have been set~\cite{Tsukada:2018mbp,Tsukada:2020lgt,Yuan:2022bem} showing that, within reasonable astrophysical assumptions, current observations rule out boson clouds formed by ultralight bosons with masses in a range around $m_b \sim 3 \times 10^{-13}$ eV. Such constraints, or potential detections, will be significantly improved with ET~\cite{Yuan:2021ebu}. In particular, the improved sensitivity at low-frequencies with respect to current detectors will allow us to probe stochastic backgrounds in a wider boson mass range compared to LIGO-Virgo, potentially probing the entire range $m_b \sim 9\times [10^{-13},  10^{-12}]\,{\rm eV}$~\cite{Yuan:2021ebu,Yuan:2022bem}. 

\begin{figure}
	\centering
    \includegraphics[width=0.7\textwidth]{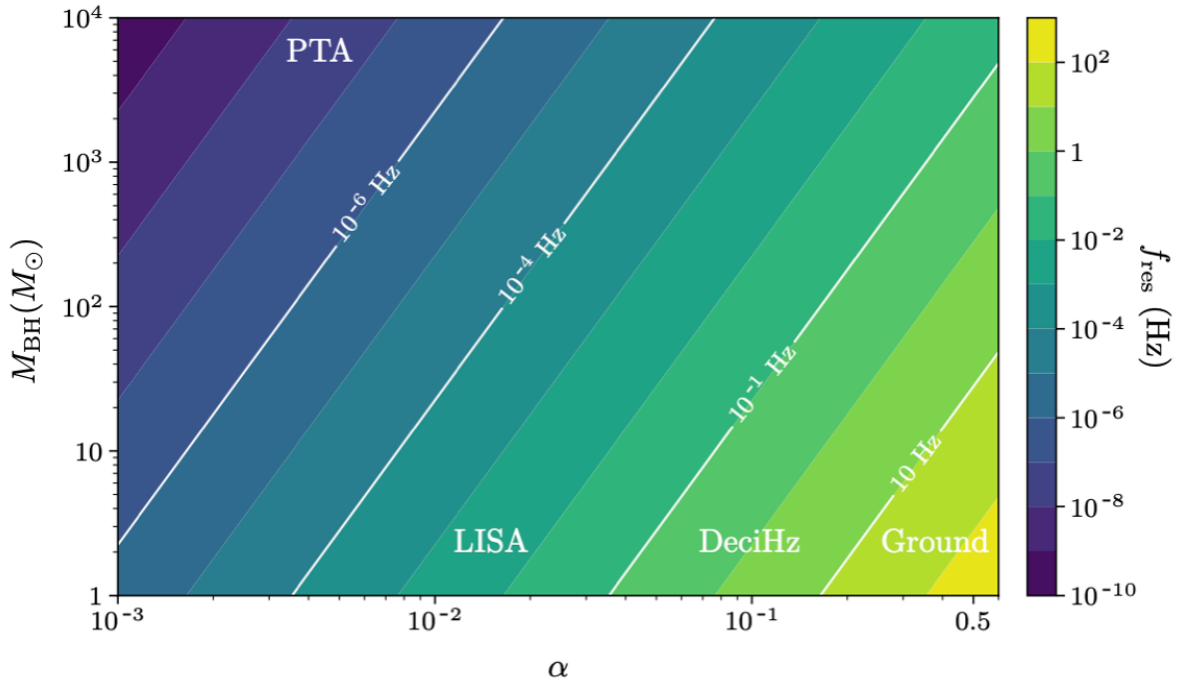}
	\caption{Resonance frequency of the (main) Bohr transition [see \eq{bohr_resonances}], as a function of the BH mass, $M_{\rm BH}$, and~$\alpha$. See ref.~\cite{Baumann:2019ztm}.}
 \label{fig:bohr_trans}
\end{figure}

\subparagraph{Binary systems.} 
Boson clouds can reach a mass density much higher than other astrophysical environments. As a result, they are able to dramatically affect the gravitational dynamics during the inspiral regime of a binary system containing a BH surrounded by it. There are two prominent types of cloud-binary interactions, orbital resonances and dynamical friction, alongside a number of associated effects. Orbital resonances~\cite{Baumann:2018vus,Baumann:2019ztm,Zhang:2018kib,Ding:2020bnl,Tong:2021whq,Takahashi:2021eso,Du:2022trq,Takahashi:2023flk} are excited whenever 
the binary's orbital frequency is equal to (or is an integer multiple of) the energy difference between the bound state initially occupied by the cloud and other eigenstates in the spectrum, often described (due to similarities with the hydrogen atom) in terms of the $(n,\ell,m)$ {\it quantum} numbers~\cite{Baumann:2018vus}. Resonances may be classified as {\it hyperfine} (only $\Delta m \neq 0$), {\it fine} ($\Delta \ell \neq 0,\Delta n=0$), and {\it Bohr} $(\Delta n \neq 0)$, respectively~\cite{Baumann:2018vus,Baumann:2019ztm}. Overtones of a resonance can also be excited for eccentric orbits~\cite{Berti:2019wnn,Boskovic:2024fga, Tomaselli:2024bdd}. For equatorial orbits, the main resonant Bohr frequency for GW emission, is given by 
\begin{equation}\label{bohr_resonances}
f_{\text{GW}}^{\text{res}}\simeq \frac{\SI{26}{Hz}}{|m_2-m_1|}\biggl(\frac{10M_\odot}{M_{\text{BH}}}\biggr)\biggl(\frac\alpha{0.2}\biggr)^3\bigg|\frac1{n_1^2}-\frac1{n_2^2}\bigg|\,,
\end{equation}
where $\alpha=M_{\text{BH}}m_b$ (in $G=\hbar=c=1$ units) is the gravitational coupling, $n_{1,2}$ and $m_{1,2}$ are the principal and magnetic quantum numbers of the two resonating states of the cloud, see figure~\ref{fig:bohr_trans}.

During the orbital evolution, the cloud undergoes a {\it Landau-Zener} type of transition between resonant states~\cite{Baumann:2019ztm}, resembling similar phenomena in atomic physics. The energy required for the process comes at the expenses of the binary systems, which in turn experiences  \emph{sinking} or \emph{floating} orbits, depending on the type of transition~\cite{Baumann:2019ztm,Boskovic:2024fga,Tomaselli:2024bdd}. During floating, the orbital frequency and associated GWs emitted remain at an approximately constant value (with a pre-factor which may vary for different resonances). The floating time scales as
\begin{equation}
\Delta t_{\text{float}}\sim\SI{0.2}{s}\,\biggl(\frac{M_{\text{BH}}}{10M_\odot}\biggr)\frac{(1+q)^{2/3}}{q^2}\biggl(\frac\alpha{0.2}\biggr)^{-3}\,,
\end{equation}
where $q$ is the binary mass ratio, which significantly alters the evolution of the GW phase~\cite{Baumann:2019ztm}. On the other hand, sinking orbits introduce  a temporary speed-up of the frequency chirp.

The low-frequency sensitivity of ET will be crucial to directly observe resonant transitions through the changes in the orbital evolution. Detecting \emph{multiple} resonances would provide strong evidence for the existence of the cloud, as the ratio of successive resonant frequencies is predicted and constrained by the cloud's energy spectrum. Another prominent effect, relevant for Bohr resonances, is the dynamical friction exerted by the cloud on the secondary object moving through it~\cite{Zhang:2019eid,Baumann:2021fkf,Tomaselli:2023ysb}. This is the backreaction of the excitation of the cloud to gravitationally unbound states, which carry mass, energy and angular momentum away from the system; by analogy with atomic physics, the process is also known as ``ionization''. Energy can be dissipated through ionization much more efficiently than through GW radiation. As a result, when the separation of the binary is of order $\mathcal O(M_{\text{BH}}\alpha^{-2})$, the dynamics is driven by ionization rather than GW emission. During this phase, the frequency chirps according to a universal shape~\cite{Baumann:2022pkl}, which includes, remarkably, quasi-discontinuous features at frequencies
\begin{equation}
    f_{\text{GW}}^{(g)} \simeq \frac{\SI{6.45}{Hz}}{g}\biggl(\frac{10M_\odot}{ M_{\rm BH}}\biggr)\biggl(\frac{\alpha}{0.2 }\biggr)^3\biggl(\frac{2}{n}\biggr)^2,\qquad g=1,2,\ldots
\end{equation}
The distinctive frequency evolution due to ionization makes the boson cloud distinguishable from other environments~\cite{Cole:2022yzw}. For high enough mass ratios $q$, however, ionization becomes efficient enough to disrupt the cloud significantly. Numerical investigations suggest that the total mass lost by the cloud throughout the inspiral is roughly $\sim qM_{\text{BH}}$, independent of $\alpha$. Only clouds more massive than the smaller compact object will then survive and still undergo ionization in the late stages of the inspiral.

    
Due to its relatively large extent (of the order $1/\alpha^2$ of the BH's size), the presence of a boson cloud can also be inferred through the measurement of non-vanishing finite size effects, such as larger-than-expected spin-induced quadrupole moments and nonzero tidal Love numbers~\cite{Baumann:2018vus,Baumann:2019ztm,DeLuca:2021ite,DeLuca:2022xlz}. This possibility is further enhanced by the intriguing result that, within GR in four spacetime dimension, the Love numbers of vacuum, isolated, BHs vanish~\cite{Binnington:2009bb,Damour:2009vw,Damour:2009va,Pani:2015hfa,Porto:2016zng,LeTiec:2020spy, Chia:2020yla,LeTiec:2020bos}. The existence of a boson cloud then provides an effective dressing of astrophysical BHs, yielding large tidal effects. That is because Love numbers of BHs surrounded by a bosonic scalar-field condensate scale as $\frac{1}{\alpha^n} \left( \frac{M_\text{\tiny cloud}}{M_\text{\tiny BH}} \right)$, where $M_\text{cloud}$ is the mass of the cloud (typically $M_\text{cloud} \simeq \alpha M_{\rm BH}$), whereas $n=8$ for scalar, vector and spin-2 odd tidal perturbations~\cite{Baumann:2018vus, DeLuca:2021ite,DeLuca:2022xlz}, and $n=10$ for the dominant spin-two even perturbations~\cite{Arana:2024kaz}. 

\begin{figure}[t!]
	\centering
    \includegraphics[width=0.5\textwidth]{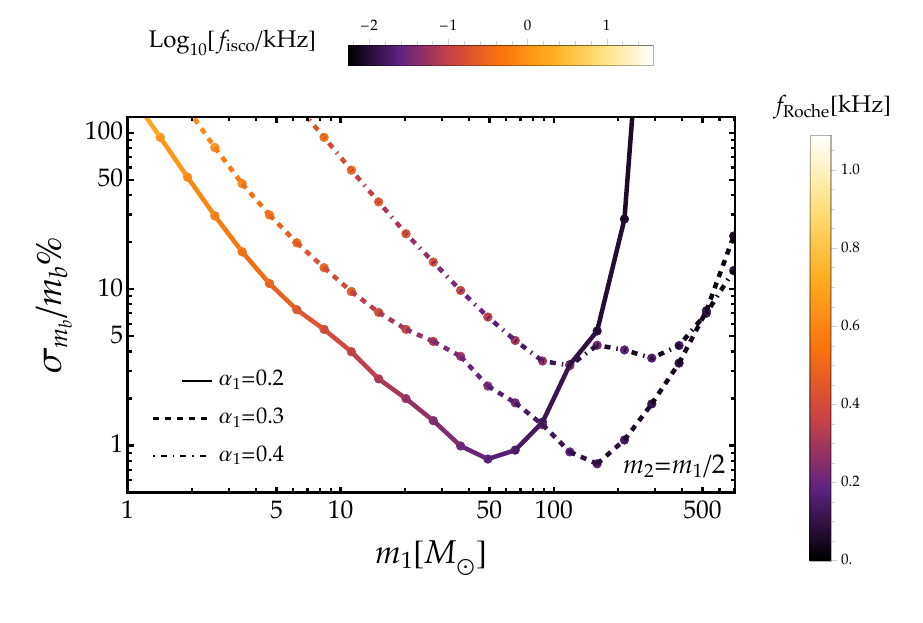}
    \includegraphics[width=0.45\textwidth]{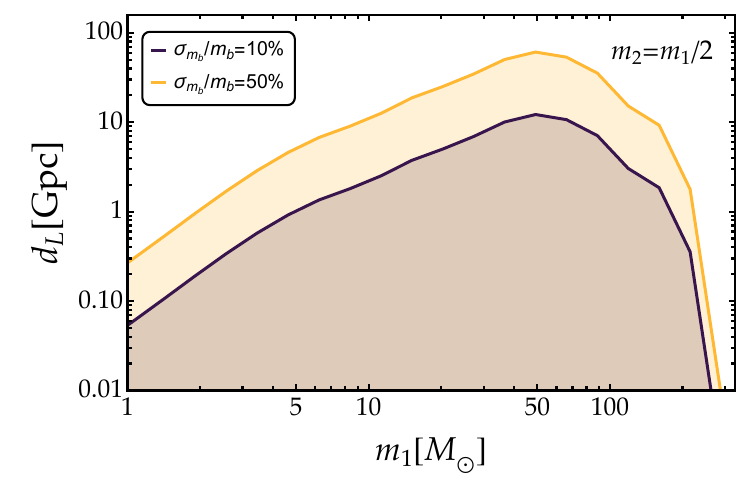}
	\caption{Left panel: relative percentage error on the 
scalar field mass $m_b$ for a BH
binary system with mass ratio $m_2/m_1 = 1/2$, 
detected by ET at 
$d_L = 1 \, {\rm Gpc}$. Bars 
show the values of the innermost stable compact orbit (ISCO) and Roche frequencies. 
 We fix the scalar cloud mass to the upper bound 
$M_\text{\tiny cloud} = 0.1 m_i$ for both objects ($i=1,2$).
%
Right panel: maximum luminosity distance for which the 
scalar field mass can be constrained by ET with a relative percentage accuracy of $10\%$ (magenta) and $50\%$ (yellow), assuming $\alpha_1=0.2$. 
See ref.~\cite{DeLuca:2022xlz}.}
	\label{fig:TLN_boson}
\end{figure}

Tidal effects may ultimately be unraveled by the gravitational interactions with a companion in binary systems towards the late stages of the inspiral phase. As the companion enters the cloud, finite size effects are suppressed (due to Gauss' law). This process can then be modeled by introducing a time-dependent tidal Love number, that smoothly approaches zero at higher frequency, leading to tidal effects that fade away at late times towards the merger phase~\cite{DeLuca:2021ite,DeLuca:2022xlz}. The time dependence of Love numbers is also manifest during resonance transitions, in which the cloud transitions into a {\it decaying} mode and is ultimately reabsorbed into the BH~\cite{Baumann:2018vus,Baumann:2019ztm}.
In~ref.~\cite{DeLuca:2021ite,DeLuca:2022xlz}, the detectability of Love numbers through GW observations was estimated (see also~\cite{Chia:2023tle}). The main results are plotted in figure~\ref{fig:TLN_boson}, which shows that ET could measure scalar field masses in the range $m_b \in [10^{-14}, 10^{-11}]$ eV with an accuracy of a few percent, for stellar mass binaries at distances of a few Gpc. Interestingly, these constraints extend within the $10^{-14}$ eV regime, which is harder to probe with other superradiance-driven observations. 
Finally, let us stress that an accurate measure of time-dependent Love numbers allows for the possibility to disentangle dressed BHs from neutron stars in the (sub)solar mass range and from exotic compact objects (see section~\ref{sec:ECO_tides}), since the latter are characterized by tidal effects which persist until the last stages of the merger~\cite{Crescimbeni:2024cwh,Crescimbeni:2024qrq}. 
    
    
Besides the effects discussed above, the cloud-binary system exhibits a number of other potentially observable phenomena, notably through the backreaction on the orbital parameters during a resonant transition. For instance, the presence of a cloud in the earlier cosmological history of the binary's evolution, even much prior to entering the detector's band, can induce larger-than-expected values for the orbital eccentricity that are later observed via GW emission~\cite{Boskovic:2024fga,Tomaselli:2024bdd}. This can have a big impact in the interpretation of future GW observations, in particular inferring the origin (e.g. dynamical or in isolation) of binary compact objects based on the eccentricity and mass distributions derived from pure vacuum evolution. Other potentially observable effects include accretion on the companion~\cite{Baumann:2021fkf}, backreaction of the cloud on the geometry~\cite{Ferreira:2017pth,Hannuksela:2018izj}, binary outspiral due to the cloud decay in GWs~\cite{Cao:2023fyv}, and the termination of superradiance due to level mixing~\cite{Fan:2023jjj,Tong:2022bbl}.


\paragraph{Dynamical friction, accretion, and overdensities.}


 
 Among the type of environmental effects that can play a significant role is dynamical friction, which emerges as a fundamental mechanism in dense environments~\cite{Chandrasekhar:1943ys}. The gravitational interactions between the binary and surrounding matter can lead to a gradual loss of orbital energy and angular momentum. This process, governed by the density and velocity distribution of the medium, plays a key role in altering the orbital parameters and phase evolution of compact binaries~\cite{Barausse:2014tra,Yue:2017iwc,Cardoso:2019rou, Kavanagh:2020cfn, Coogan:2021uqv, Becker:2021ivq, Speeney:2022ryg, Cole:2022yzw}. Accretion stands as another notable environmental effect, particularly prevalent in scenarios involving gas-rich environments or surrounding accretion disks~\cite{Barausse:2014tra, Becker:2022wlo, Cole:2022yzw}. As the binary interacts with these regions, matter accretes onto one or both binary components, inducing changes in their masses and influencing the orbital dynamics and phase evolution~\cite{Barausse:2014tra, Becker:2022wlo, Cole:2022yzw}. The intricacies of accretion processes, ranging from Bondi-Hoyle-Lyttleton accretion to collisionless accretion, dictate the degree of impact on the binary's evolution, further complicating the interplay between environmental factors and binary dynamics~\cite{Macedo:2013qea,Barausse:2014tra,Cardoso:2019rou, Becker:2022wlo, Cole:2022yzw}.

Most studies of environmental effects on GW signals have typically focused on sources relevant for LISA~\cite{LISA:2022yao,LISA:2022kgy}, mainly due to the fact that environmental effects tend to be more important in the early inspiral and more easily detectable in intermediate or extreme mass-ratio systems~\cite{Eda:2013gg, Eda:2014kra, Yue:2017iwc, Cardoso:2019rou, Kavanagh:2020cfn, Coogan:2021uqv, Becker:2021ivq, Becker:2022wlo, Speeney:2022ryg, Cole:2022yzw, Berezhiani:2023vlo}. In ref.~\cite{CanevaSantoro:2023aol}, it was estimated that ET will be sensitive to the effect of dynamical friction in a GW170817-like event evolving in mediums with densities $\gtrsim 10^{-3}$ g/cm$^{3}$. Such densities would be possible for compact binaries merging within dense thin accretion disks~\cite{Barausse:2014tra}, but it is unlikely that DM environments can reach densities $\gtrsim 10^{-3}$ g/cm$^{3}$ around such sources.
A more promising possibility, however, is that of primordial BHs \cite{Carr:2023tpt} (see section~\ref{section:div3}). If they do not make up all of the DM, which is likely on the mass range that ET is able to probe, it has been shown both analytically and numerically that DM particles should form dense spikes around them \cite{Bertschinger:1985pd,Mack:2006gz,Ricotti:2007jk,Boudaud:2021irr}. It is expected that DM spikes around equal-mass systems will be disrupted or destroyed before the objects are close enough  for the dynamical friction effect to be observable in the GW signal~\cite{Kavanagh:2018ggo}. Yet, in the regime where the mass ratio of the binaries is $q\lesssim10^{-2.5}$, the smaller BH will move through the DM, causing a dephasing with respect to an equivalent system inspiralling through vacuum~\cite{Kavanagh:2020cfn,Eda:2013gg}. This effect can then be observed by ET with week-long duration signals, and failing to include dynamical friction may lead to significant signal-to-noise ratio losses~\cite{Cole:2022ucw}. Furthermore, for BHs with sub-solar masses, ET will provide exquisite measurements of the binary parameters~\cite{Barsanti:2021ydd}, which are also sensitive to dynamical-friction feedback~\cite{Kavanagh:2020cfn, Coogan:2021uqv, Becker:2022wlo}, and secondary accretion~\cite{Nichols:2023ufs}. Finally, let us emphasize that the detection of primordial BHs, through their imprint in GW observables, would in turn also put among the strongest constraints one could set on dark matter candidates and extensions of the standard model \cite{Lacki:2010zf,Adamek:2019gns,Bertone:2019vsk}, see section~\ref{section:div3}.

\subsection{Synergies with other aspects of the ET Science Case}
\label{div1_synergies}

Here we discuss a list of outstanding ET science cases related to fundamental physics which require an interdisciplinary approach with topics discussed in other chapters. Such an approach is ubiquitous in most aspects of ET science, but  topics related to fundamental physics  are, by their own nature, particularly cross-cutting.
We envisage strong synergies in the following aspects:

\begin{itemize}
    \item {\bf Improve (beyond) GR waveforms to match ET requirements.}
    %
    The unprecedentedly loud signals to be detected by 3G detectors will require an accuracy improvement of the waveforms to be used to extract signals from data. A necessary condition for two waveforms to be indistinguishable is that their mismatch must be smaller than $ 1/(2\,{\rm SNR}^2)$, so that louder events exacerbate any possible modeling systematics (see~\cite{Gupta:2024gun} for a recent overview). 
    When available, coalescence waveforms in theories beyond GR or for ECO models are far less accurate than the prototypical binary BH waveforms in GR, both in absolute terms and in terms of coverage of the multidimensional parameter space. An outstanding task is to select the best motivated beyond-GR models and develop them at the level of accuracy required in the ET era.
    This will be achieved by extending various perturbative and numerical techniques discussed in section~\ref{section:div8} also for alternative theories and models, most likely on a case-by-case basis. 
\item {\bf Improve data analysis techniques beyond GR.} 
    As discussed in section~\ref{section:div10}, ET poses new urgent challenges for data analysis, mostly due to the long duration of certain signals, the possible overlap of several sources, and the huge number of expected detections. 
    While overlapping signals should not be a significant concern both in terms of detection~\cite{Regimbau:2012ir, Meacher:2015rex} and parameter estimation~\cite{Samajdar:2021egv, Pizzati:2021apa, Relton:2021cax, Himemoto:2021ukb}, it would be advisable to optimize current data-analysis pipelines for this scenario. Any fundamental physics effect that one wishes to search for in the datastream should be properly included in the pipeline. In the near future, synergy among data analysis (section~\ref{section:div10}),  waveform modeling (section~\ref{section:div8}), and the fundamental physics topics disucssed in this section will be needed, for example, to introduce beyond-GR effects in the ET mock data challenge (discussed in section~\ref{section:div10}).
\item{\bf Fundamental physics implications of primordial relics, cosmic strings, and primordial BHs.}
Early Universe cosmology is strongly tied with fundamental physics, GR extensions, and Beyond Standard Model physics. In particular, a putative GW detection of primordial BHs, cosmic strings, or any primordial relics (see section~\ref{section:div2}), or any direct signature of inflation, would have a huge impact on foundational physics.
For example, if microscopic horizon-less relics form in the early Universe either directly through gravitational collapse or as stable remnants of the Hawking evaporation of primordial BHs, they could evade cosmological constraints arising from evaporation~\cite{Carr:2020gox} and could explain the entirety of the DM. The stochastic GW background associated with the formation of microscopic DM relics in various scenarios is at most marginally detectable with current interferometers but could be measured by ET~\cite{Franciolini:2023osw}. This strengthens the connection between fundamental compact-object physics and cosmological GW sources, and calls for a synergy between the  topics developed in this section and those in section~\ref{section:div2}.

\item{\bf Identifying the origin of merging binaries across cosmic history.}
Likewise, detecting a new population of compact objects other than BHs or NSs, or BH binaries of primordial origin, would have a strong impact on fundamental physics. Indeed, it would imply the existence of Beyond Standard Model and/or beyond-GR effects, which should then be further developed and could be possibly explored by other complementary means.
Except for the smoking gun signatures of deviations from standard theory discussed in this section, evidence for new families of compact object could come from population studies (see section~\ref{section:div3}), underscoring a strong connection between the latter and fundamental physics.

\item{\bf Fundamental physics implications of NS mergers.}
As the densest known objects in the Universe, NSs are a unique laboratory for all fundamental interactions. As previously explained, NS-based tests of gravity are tantalizing but severely limited by the degeneracy with the unknown EOS in the NS core (see section~\ref{section:div6}). Finding ways to break this degeneracy calls for a synergy among  nuclear matter and fundamental physics studies. As discussed, a promising possibility is to make use of approximately universal relations among the NS parameters that are almost EOS insensitive. ET will be instrumental for these tests, since they require very high SNR to measure subleading parameters such as quadrupole moments or non-standard Love numbers.
\end{itemize}

\subsection{Executive summary}
We conclude this section by summarizing the advancements that ET will bring in Fundamental Physics, in comparison to the current state of the art, specifically emphasizing the distinct features that  ET will allow us to explore. The following are the major areas where ET's capabilities surpass the current generation of GWs detectors:

\begin{highlightbox}{Fundamental principles of the gravitational interaction}
ET will probe the spacetime dynamics, the generation, and the propagation of GWs in Einstein’s theory of gravity like never before, including:

\begin{itemize}
\item {\it Tests in the inspiral phase}, with up to two orders of magnitude improvements compared to current detectors;
\item {\it Searches for extra GW polarizations},  which are inaccessible with current detectors;
\item {\it Tests of the mass of the graviton and dispersion relations}, with up to two orders of magnitude improvements compared to current detectors;
\item {\it Tests of GW (displacement) memory effects}, for the first time;
\item {\it High-precisions tests of the fully nonlinear (perturbative/non-perturbative) structure of GR.}
\end{itemize}
Consequently, ET will be able to tell whether the gravitational interactions driving the coalescence of compact objects are described by GR, or if the data will point instead to alternative theories which manifest themselves in regions of parameter space that are not reachable by any other experiment.
\end{highlightbox}

\begin{highlightbox}{The nature of compact objects and horizon scale physics}

ET will provide high accurate measurements of the nature of GW sources, including:

\begin{itemize}
\item {\it Ringdown tests and BH spectroscopy} at unprecedented level, with ${\cal O}(10)$ event/yr expected to have ${\cal O}(1)\%$ precision in the measurement of multiple ringdown modes;
\item {\it Tests of intrinsic multipole moments and no-hair theorems}, with almost an order of magnitude improvement compared to LVK at design sensitivity;
\item {\it Tests of tidal deformability effects and absorption}, with almost an order of magnitude improvement compared to LVK at design sensitivity;
\item {\it Tests of the post-merger phase and of horizon-scale physics}, with more than an order of magnitude improvement compared to current detectors.
\end{itemize}

Consequently, ET will be able to tell whether the dark massive compact objects in the sky producing ripples of spacetime are indeed the BHs of GR, or whether ET will allow us to detect novel types of compact objects for the first time.  
\end{highlightbox}

\begin{highlightbox}{Dark-matter particles and new fields}

ET has a unique potential to discover the existence of new fields/particles in nature, including:

\begin{itemize}
    \item {\it Searches of ultralight particles} using BH mergers and the stochastic GW background at extra galactic distance, exploring the region between $10^{-14}\,{\rm eV}$ and $10^{-11}\,{\rm eV}$ of putative new bosonic matter;
    \item {\it Tests of dark-matter environmental effects, boson clouds, and dynamical friction}, in regions currently inaccessible by GW detectors.
\end{itemize}

Consequently, even if Einstein’s theory withstands all the aforementioned scrutiny, ET will then turn into a precision machine for searches of new particles and fields through GW precision data, opening entirely new windows of exploration. These searches will be synergetic with complementary experimental efforts, including space-based detectors such as LISA.   

\end{highlightbox}


\section{Cosmology with ET}\label{section:div2}

With its unprecedented sensitivity, ET will detect GWs from events such as merging black holes, neutron star collisions, and potential signals from the early Universe. By probing these cosmic events, ET will provide new insights into the fundamental properties of gravity, the expansion of the Universe, and the nature of dark matter and dark energy, serving as a powerful tool to explore previously uncharted aspects of cosmology.

One of the main targets for ET is the stochastic gravitational wave background (GWB), which is a diffuse signal composed of many unresolved gravitational wave sources, encompassing both astrophysical (AGWB) and cosmological (CGWB) components. 
Section~\ref{sect:SGWBdiv2} discusses the general formalism for GWBs, including anisotropies, parity violation, the separability of different contributions, and the impact of correlated noise on detectability. As we will discuss in section~\ref{sec:early_universe}, stochastic background of cosmological origin are probes of the early Universe, and potential sources in the ET band include inflationary models beyond single-field slow-roll scenarios, first-order phase transitions  occurring at energy scales of up to $10^8$ GeV, corresponding to timescales of $10^{-12}$ to $10^{-6}$ seconds after the Big Bang; topological defects, such as cosmic strings and domain walls, as well as individual GW bursts from cosmic string cusps and kinks; or signals related to the formation mechanism of primordial BHs.

ET will also offer unique opportunities for probing the late Universe and contributing to cosmography, by
detecting BBH mergers at high redshifts, up to $z \sim 20$ and higher. This will allow ET to observe black hole formation in the early Universe, tracing back to the population of primordial black holes.
A key feature of GWs from compact binary coalescences  is that they are `standard sirens', i.e.  they serve as self-calibrating distance indicators, enabling reconstruction of the distance-redshift relation and providing insights into the expansion history of the Universe. However, GWs alone generally cannot determine redshift of the binary. For bright standard sirens, which have an electromagnetic counterpart, the redshift can be directly measured from the electromagnetic observations, enabling constraints on cosmological parameters. Most GW events, however, are `dark sirens', i.e. do not have an observed  electromagnetic counterpart, and necessitate the use of correlation with galaxy catalogs or population inference techniques to perform statistical inference on the relation between luminosity distance and redshift. The potential of bright and of dark sirens for cosmology and for probing the late Universe with ET will be discussed in section~\ref{sec:lateuniversediv2}.

Finally, ET observations, particularly when combined with future galaxy surveys, are expected to help address various cosmic tensions, such as discrepancies in the Hubble constant $H_0$ and potential differences in cosmic dipole measurements from electromagnetic and large-scale structure (LSS) observations. The potential of the ET observations for probing LSSs of the Universe is discussed in section~\ref{sect:LSSdiv2}.

\subsection{Stochastic gravitational-wave backgrounds}\label{sect:SGWBdiv2}

Gravitational-wave backgrounds are of great interest in cosmology. The detection of a CGWB would be a smoking-gun probe of the physics of the early Universe, as it would allow for unique constraints on cosmic inflation, phase transitions, or cosmic strings. The anisotropies of the CGWB are a probe of the cosmological perturbations in the early Universe. Moreover, the AGWB, resulting from the superposition of numerous unresolved CBC signals, can also be used as a cosmological probe, as its frequency spectrum is sensitive to cosmological parameters, and its anisotropies are tracers of the large scale structures (LSS) of the Universe. Finally, the detection of a polarized CGWB could serve as key evidence for parity-violating mechanisms in the early Universe (even though the AGWB is also expected to be polarized  and constitutes an unavoidable foreground for a polarized CGWB). Indeed, the frequency spectrum, anisotropies, and polarization of the GWB are crucial observables for distinguishing between the CGWB and AGWB, as various cosmological and astrophysical sources are expected to produce GWBs with distinct and unique signatures.

\subsubsection{Definition and characterisation} 
\label{sec:stochback}


A GWB is a diffuse signal formed by the incoherent superposition of numerous unresolved GW sources. Some GWB components exhibit stochastic properties due to their generation process, while others result from the limited sensitivity of the specific detectors used for their observation. 
A variety of sources are expected to produce a gravitational-wave background, with different properties and frequency profiles (see~\cite{Maggiore:1999vm,Maggiore:2018sht,Caprini:2018mtu,Christensen:2018iqi,Renzini:2022alw} for reviews). Traditionally, we distinguish between an astrophysical GW background (AGWB) and a cosmological GW background (CGWB). In a first approximation, the GWB is assumed to be homogeneous and isotropic across the sky. This contribution (i.e., the monopole) is the dominant contribution and it is characterized through its frequency dependence. However, the GWB exhibits also some level of anisotropy due to inhomogeneous distribution of sources, line-of-sight effects accumulated during the propagation across the large-scale structure (LSS) \cite{Contaldi:2016koz,Cusin:2017fwz,Cusin:2017mjm,Jenkins:2018nty, Cusin:2018avf,Cusin:2018rsq, Bertacca:2019fnt,Pitrou:2019rjz, Cusin:2019jpv,Cusin:2019jhg, Capurri:2021zli,Bellomo:2021mer, Bartolo:2019oiq,Bartolo:2019yeu,ValbusaDallArmi:2020ifo}, and kinematic anisotropies arising from the observer's velocity relative to the rest frame of the GWB \cite{Cusin:2022cbb,ValbusaDallArmi:2022htu,Chung:2022xhv}.

Let us recall here the basic definitions for the characterization of stochastic backgrounds (we follow the conventions in~\cite{Maggiore:2007ulw}).
A continuous superposition of GWs coming from all directions, with propagation directions labeled by a unit vector $\hatn$,  can be written as 
\be\label{snrhabdiv2}
h_{ij}(t,{\bf x})
=\int_{-\infty}^{\infty} df \int_{S^2} d^2\hatn\,\sum_{A=+,\times}  \tilde{h}_A(f,\hatn) e^A_{ij}(\hatn) \, \,   e^{-2\pi i f(t-\hatn\cdot{\bf x} /c )}
\, ,
\ee
where the spatial indices $i,j$ take values $1,2,3$ and $e_{ij}^{A}$ are the polarization tensors, normalized as (the sum over the spatial indices $i,j$ is understood)
\be\label{eijnormalizationdiv2}
e_{ij}^{A}(\hatn)e_{ij}^{A'}(\hatn)=2\delta^{AA'}\, .
\ee
In a given detector,  located at the position ${\bf x}_a$, the  observed signal $h_a(t)$  is obtained convolving each polarization with the pattern functions $F_a^A(\hatn)$ that encode the detector response,\footnote{We assume here that the size of the detector is
much smaller than the GW wavelength, so we can neglect the spatial variation
of the GW over the  extension of the detector.  This condition is very well satisfied by ground-based interferometers: as the GW frequency $f$ ranges between 10~Hz and 1~kHz, the wavelength $c/f$ ranges between $3\times 10^4$~km and 300~km, much larger than the 10~km or 15~km length of the arms of ET.}
\be\label{hFhFdiv2}
h_a(t)= \int_{-\infty}^{\infty} df \int_{S^2} d^2\hatn\,\sum_{A=+,\times}\tilde{h}_A(f,\hatn) F_a^A(\hatn) \, \,   e^{-2\pi i f(t-\hatn\cdot{\bf x}_a /c )}
\, ,
\ee
and therefore the Fourier transform of the signal is given by
\be\label{tildehaFAdiv2}
\tilde{h}_a(f)=\int_{S^2} d^2\hatn \sum_{A=+,\times}\tilde{h}_A(f,\hatn) F_a^A(\hatn) \, \,   
e^{2\pi i f \hatn\cdot{\bf x}_a /c}\, .
\ee
If the signal is given by 
a  stochastic background of GWs  that is stationary,  isotropic and unpolarized, its two-point correlator  is given by
\be\label{avediv2}
\langle \tilde{h}_A^*(f,\hatn)
\tilde{h}_{A'}(f',\hatn')\rangle =
\delta (f-f')\, \frac{\delta(\hatn,\hatn')}{4\pi}\,
 \delta_{AA'}\, \frac{1}{2}S_h(f)\, ,
\ee
where  the brackets $\langle \cdot \rangle$ denote the ensemble average and 
$\delta(\hatn,\hatn')$ is a Dirac delta over the two-sphere,  
$\delta(\hatn,\hatn')=\delta (\phi -\phi ') \delta (\cos\theta -\cos\theta ')$, where $(\theta ,\phi)$ are the polar angles that define $\hatn$.
\Eq{avediv2}  defines the (single-sided) spectral density of the signal, $S_h(f)$. 
The factor $1/(4\pi)$ in \eq{avediv2} is a standard choice of normalization, chosen so  that
\be\label{normShdiv2}
\int_{S^2} d^2\hatn\, d^2\hatn'
\langle \tilde{h}_A^*(f,\hatn)
\tilde{h}_{A'}(f',\hatn')\rangle=
\delta (f-f')\delta_{AA'}
\frac{1}{2}S_h(f)\, ,  
\ee
so that, for each polarization,  the right-hand side has   the  normalization used also for the (single-sided) noise spectra density, see \eq{eq:PSD_definition} below.\footnote{Here we follow the conventions used in refs.~\cite{Maggiore:1999vm,Maggiore:2007ulw}, and in most of the literature. Observe, however, that these differ by a factor of 2 from those used in \cite{Romano:2016dpx}, see their eq.~(2.14).}

However, more generally, a stochastic GW background can be polarized and anisotropic. Assuming again stationarity (which, in Fourier space, implies that the correlator is proportional to $\delta(f-f')$),
the most general form of the two-point correlator is then~\cite{Seto:2008sr,Romano:2016dpx,Renzini:2022alw,Belgacem:2024ohp}  
\be\label{Stokes1div1}
\langle\tilde{h}_A^*(f,\hat{\bf{n}})\tilde{h}_{A'}(f',\hat{\bf{n}}')\rangle=\delta(f-f')\,
\frac{\delta(\hatn,\hatn')}{4\pi}
\frac12 H_{A A'}(f,\hat{\bf{n}})\, .
\ee
Note that we still assume that signals coming from different directions are uncorrelated, so that the correlator is still proportional to $\delta(\hatn,\hatn')$; however, anisotropy is encoded in the fact that the function $H_{A A'}(f,\hat{\bf{n}})$ could depend on $\hatn$, and  the possibility of a polarization   in the fact that $H_{A A'}(f,\hat{\bf{n}})$ is  a  $2\times 2$  matrix in the polarization indices, not necessarily proportional to $\delta_{AA'}$.
The reality of $h_a(t)$ implies that $\tilde{h}_A^*(f,\hat{\bf{n}})=\tilde{h}_A(-f,\hat{\bf{n}})$ so, taking the complex conjugate of \eq{Stokes1div1}, one sees that $H_{A A'}(f,\hat{\bf{n}})$ is a  Hermitian matrix in the polarization indices. As any $2\times 2$ Hermitian matrix, it can then be decomposed (with real coefficients)
into the basis of the identity matrix  and the three Pauli matrices,
\be\label{Stokes2div2}
H_{A A'}(f,\hat{\bf{n}})=I(f,\hat{\bf{n}})\,\delta_{A A'}+U(f,\hat{\bf{n}})\,\sigma^{1}_{A A'}+V(f,\hat{\bf{n}})\,\sigma^{2}_{A A'}+Q(f,\hat{\bf{n}})\,\sigma^{3}_{A A'}\, ,
\ee
where the  Pauli matrices are given by
\begin{equation}\label{Pauli}
\sigma^{1}=\,\nonumber
\begin{pmatrix}
0       & 1 \\
1       & 0
\end{pmatrix}\,,\quad \sigma^{2}=\,\nonumber
\begin{pmatrix}
0       & -i \\
i       & 0
\end{pmatrix}\,,\quad \sigma^{3}=\,\nonumber
\begin{pmatrix}
1       & 0 \\
0       & -1
\end{pmatrix}\, .
\end{equation}
In matrix form 
\be\label{Stokes2matrix}
H(f,\hat{\bf{n}})=
\begin{pmatrix}
I (f,\hat{\bf{n}}) +Q (f,\hat{\bf{n}})    & U (f,\hat{\bf{n}}) -iV (f,\hat{\bf{n}})\\
U (f,\hat{\bf{n}})  +iV (f,\hat{\bf{n}})     & I (f,\hat{\bf{n}}) -Q (f,\hat{\bf{n}})\,
\end{pmatrix}.
\ee
The coefficients of this decompositions define the Stokes parameters of the GW stochastic background, and are real functions of $f$ and $\hat{\bf{n}}$ describing intensity ($I$), linear polarization ($U$ and $Q$) and circular polarization ($V$).
Therefore,\footnote{Here, again, one must be aware of the existence of different conventions in the literature, in which a factor $4\pi$ or $8\pi$ in \eq{Stokes1div1} is reabsorbed into the definition of the Stokes parameters. Our conventions are the same as in \cite{Smith:2019wny} (see, e.g., also \cite{Cruz:2024esk,Belgacem:2024ohp}).}
\begin{equation} \label{eq:gwb_stokes}
\begin{aligned}
    &\hspace*{-10mm}\begin{pmatrix}
        \langle \tilde{h}^*_{+}(f,\hat{\textbf{n}}) \tilde{h}_{+}(f',\hat{\textbf{n}}')\rangle & \langle \tilde{h}^*_{+}(f,\hat{\textbf{n}}) \tilde{h}_{\times}(f',\hat{\textbf{n}}')\rangle \\
        \langle \tilde{h}^*_{\times}(f,\hat{\textbf{n}}) \tilde{h}_{+}(f',\hat{\textbf{n}}')\rangle & \langle \tilde{h}^*_{\times}(f,\hat{\textbf{n}}) \tilde{h}_{\times}(f',\hat{\textbf{n}}')\rangle
    \end{pmatrix} = \\ 
    & = \frac{1}{8\pi}\, \delta(\hatn,\hatn')\delta(f-f')
    \begin{pmatrix}
        I(f,\hat{\textbf{n}})+Q(f,\hat{\textbf{n}}) & U(f,\hat{\textbf{n}})-iV(f,\hat{\textbf{n}}) \\
        U(f,\hat{\textbf{n}})+iV(f,\hat{\textbf{n}}) & I(f,\hat{\textbf{n}})-Q(f,\hat{\textbf{n}})
    \end{pmatrix}\, .
\end{aligned}   
\end{equation}
If the GWB is unpolarized, $U=V=Q=0$ and \eq{eq:gwb_stokes} reduces to 
\be\label{eq: iso_unpo}
\langle \tilde{h}_A^*(f,\hatn)
\tilde{h}_{A'}(f',\hatn')\rangle =
\delta (f-f')\, \frac{\delta(\hatn,\hatn')}{4\pi}\,
 \delta_{AA'}\, \frac{1}{2}S_h(f,\hatn)\, ,
\ee
with $S_h(f,\hatn)=I(f,\hatn)$. If, furthermore, the background is isotropic, we get back \eq{avediv2} with a spectral density $S_h(f)=I(f)$ independent of $\hatn$.
If we  assume that the GWB is Gaussian, then the 2-point correlation function, \eq{eq:gwb_stokes} in the most general case or \eq{eq: iso_unpo}  in the unpolarized case, is sufficient to fully describe its statistical properties.

An important property of a GWB is its energy density per unit logarithmic frequency interval, which is conventionally normalized to the critical energy density for closing the Universe $\rho_c$,
\begin{equation}
\label{eq: OmegaGW}
\Omega_{\rm GW}(f)\equiv\frac{1}{\rho_c}\frac{{\mathrm d}\rho_{\rm GW}}{{\mathrm d}\ln f}\, ,
\end{equation}
where $\rho_c = 3 c^2 H_0^2/\left(8\pi G\right)$ and $H_0$ is the present-day value of the Hubble parameter. Note that $H_0$, in the denominator of \eq{eq:Sh},  entered through the normalization factor $\rho_c$, so the observational uncertainty in this factor  $H_0$ is not related to an intrinsic uncertainty of the energy density of the stochastic background considered, but rather of the quantity that we have chosen for normalizing it. In order to get rid of this uncertainty, it is often convenient to work in terms of the quantity $h^2\Omega_{\rm GW}(f)$, where
the reduced Hubble constant $h$ is defined from $H_0=100 h~{\rm km} \, {\rm s}^{-1}\, {\rm Mpc}^{-1}$.

For a stationary, isotropic  and unpolarized background, described by \eq{normShdiv2}, the energy density can be written in terms of the spectral density as
(see eq.~(7.202) of~\cite{Maggiore:2007ulw})
\begin{equation}
\label{eq:Sh}
\Omega_{\rm GW}(f) =  \frac{4\pi^2} {3 H_0^2} f^3S_h(f) \, .
\end{equation}
The same expression holds for a polarized background (as long as it is stationary and anisotropic), since the Stokes parameters related to linear and circular polarization do not contribute to the energy density, and only the intensity $I(f)\equiv S_h(f)$ contributes~\cite{Belgacem:2024ohp}. 
For a generic anisotropic and polarized GW background (but still assuming  stationarity), one can show that the energy density is given by~\cite{Belgacem:2024ohp}
\be\label{rhogwIfhatn}
\rho_{\rm gw}=\frac{\pi c^2}{2 G}\,
\int_{0}^{\infty} d\log f\, f^3 \int_{S^2} \frac{d^2\hatn}{4\pi}\,  I(f,\hatn)\, ,
\ee
so that
\be\label{drhodlofdn}
\frac{d\rho_{\rm gw}}{d\log fd^2\hatn}=\frac{c^2}{8G}\, 
f^3 I(f,\hatn)
\, .
\ee
Note that this expression is obtained using the full correlator given in \eqs{Stokes1div1}{Stokes2div2}. However, in the computation the traceless terms associated to polarizations give zero~\cite{Belgacem:2024ohp}, and only the intensity Stokes parameter contributes to the energy density. 
Then, we have 
\bees\label{eq: OmegaGW_f_n}
\Omega_{\rm GW}(f, \hatn) &\equiv & 
\frac{1}{\rho_c} \frac{d\rho_{\rm GW}(f, \hatn)}{d\log f \, d^2 \hatn}\nn\\
&=& \frac{\pi}{3H_0^2}\, f^3\, I(f,\hatn)
\, .
\ees    
In the typical situation in which the anisotropic part is a small perturbation of the isotropic part, we can write
\be\label{Omegaanisotdiv2}
\Omega_{\rm GW}(f, \hatn)= \bar{\Omega}_{\rm GW}(f) + \delta \Omega_{\rm GW}(f, \hatn)\,,
\ee
i.e.,
the energy density is  split into a homogeneous component  $\bar{\Omega}_{\rm GW}(f)$, and an anisotropic perturbation, $\delta \Omega_{\rm GW}(f, \hatn)$, which is generally small compared to the monopole.
Under the assumption of statistical isotropy, the sky distribution of the GWB energy density can be characterised in terms of a frequency-dependent angular power spectrum, $C_{\ell}(f)$, defined through
\begin{equation}\label{OmegaLegendrediv2}
    \langle  \delta \Omega_{\rm GW}(f, \hatn) \,  \delta \Omega_{\rm GW}(f, \hatn') \rangle = \sum_{\ell=1}^{\infty} C_{\ell}(f) P_{\ell}(\hatn \cdot \hatn'), 
\end{equation}
where $P_{\ell}(\hatn \cdot \hatn')$ are the  Legendre polynomials and the multipoles $\ell$ are inversely proportional to the angular scale of the perturbation.

A useful tool in the study of the sensitivity of a network of detectors to an unpolarised and isotropic  GWB is the power-law sensitivity (PLS) curve~\cite{Thrane:2013oya}. We refer the reader to section~\ref{sec:div9_PLSdefinition} for details on its construction, its physical meaning and some properties, and here we limit ourselves to a brief reminder of the main ingredients needed. When constructing the PLS curve, one considers the family of power-law GW backgrounds
\begin{equation}
\label{eq: pow law GW div2}
\Omega_{\rm GW}(f;\beta)=\Omega_\beta\left(\frac{f}{f_{\rm ref}}\right)^\beta \,,
\end{equation}
where the exponent $\beta$ is the family parameter, which can take any real value, and $f_{\rm ref}$ is an arbitrary fixed reference frequency (whose change simply amounts to a redefinition of $\Omega_\beta$). The dependence of the amplitude coefficient $\Omega_\beta$ on $\beta$ is determined by requiring that, in a given coincident observation time $T$, the integrated signal-to-noise ratio (SNR) in the frequency range $[f_{\rm min}, f_{\rm max}]$ probed by a given detector network takes an assigned value $\rho$. This leads to
\begin{equation}
\label{eq: Omega_beta div2}
\Omega_\beta=\rho\left[2T\int_{f_{\rm min}}^{f_{\rm max}} \dd{f}~\Omega^{-2}_{\rm eff}(f)\left(\frac{f}{f_{\rm ref}}\right)^{2\beta}\right]^{-1/2}\,,
\end{equation}
where the function $\Omega_{\rm eff}(f)$ depends on the noise properties of the detectors (their PSD or noise correlations, if present) and by the geometry of the detector configuration via the overlap reduction functions for each pair of detectors, see \eq{defOmeffdiv9} for the explicit expression.

The PLS curve is then defined as the envelope of the family of 
curves~(\ref{eq: pow law GW div2}), with $\Omega_\beta$ as in eq.~(\ref{eq: Omega_beta div2}),  and it is given by
\begin{equation}
\label{eq: PLS def}
\Omega_{\rm PLS}(f)=\underset{\beta}{\max}~\Omega_{\rm GW}(f;\beta)\,.
\end{equation} 
By construction, in a log-log plot any line tangent to the PLS curve represents a power-law GW background with integrated SNR equal to the value of $\rho$ used in the PLS definition.

\subsubsection{Anisotropies of the GWB} 
\label{sect:anisoSGWBdiv2}

\paragraph{Anisotropies of the Cosmological Background.}
\label{sub:anisotropies cosmo}

The anisotropies of the CGWB can be described by introducing a distribution function of the primordial gravitons (in the high frequency approximation, where anisotropies are typically studied) and solving the Boltzmann equation in cosmological perturbation theory~\cite{Contaldi:2016koz,Bartolo:2019oiq,Bartolo:2019yeu,Pizzuti:2022nnj}. The solution of the Boltzmann equation shows that the anisotropies are generated by any inhomogeneity and anisotropy in the production mechanism that sourced the background (i.e., initial conditions) and by the metric perturbations around a  Friedmann-Lemaître-Robertson-Walker background crossed by gravitons along their geodesics. In addition, kinematic anisotropies, in particular a dipole, are induced by the peculiar motion of the observer with respect to  the rest frame of the CGWB~\cite{Jenkins:2018nty,Cusin:2022cbb}. 

The anisotropies due to the propagation of gravitons and kinematic effects are proportional to $4-n_{\rm gwb}$, where $n_{\rm gwb}$ is the spectral tilt of the CGWB monopole. So, the spectral tilt may have an impact on the amplitude of the anisotropies. Such a possibility has been discussed in~\cite{Dimastrogiovanni:2022eir,LISACosmologyWorkingGroup:2022kbp} for the case of sharply peaked CGWB spectra generated at second order by curvature perturbations, making these anisotropies a potential probe of primordial black-holes physics.

The contribution due to the initial conditions largely depends on the model considered and could dominate the spectra in a number of different scenarios. For instance, if the CGWB is produced by the decay of a field which gives the dominant contribution to the curvature perturbation of the Universe, the initial conditions are expected to be adiabatic, in analogy with the CMB. For the CGWB generated by quantum fluctuations of the metric during inflation a large non-adiabatic contribution in the initial conditions is expected, which can increase the anisotropies by an order of magnitude~\cite{ValbusaDallArmi:2023nqn}. In the case of scalar-induced gravitational waves (SIGWs), the initial conditions are sensitive to any amount of local primordial non-Gaussianity, parametrized by $f_{\rm NL}$~\cite{Bartolo:2019zvb}, while for GWs from phase transitions large isocurvature perturbations due to a decoupled sector at the production time could enhance the spectrum~\cite{Bodas:2022urf}. The angular power spectrum of the CGWB for all these mechanisms and the corresponding predictions for ET, can be computed with the code \texttt{GW\textunderscore CLASS}\footnote{The code is publicly accessible as a new branch (\texttt{GW\textunderscore CLASS}) on the \texttt{CLASS} repository \url{www.github.com/lesgourg/class_public}.}~\cite{Schulze:2023ich}.

Another important source of anisotropies in the form of a contribution to the initial condition term arises in inflation models predicting a sizable tensor or mixed (scalar-tensor-tensor) bispectrum in the squeezed configuration. A non-trivial correlation between two short-wavelength modes and one long-wavelength mode of the primordial fluctuations would indeed result in a modulation of the gravitational wave two-point function, giving rise to anisotropies in the energy density of the GW background \cite{Dimastrogiovanni:2019bfl}. If observed, these signatures would be a smoking gun for inflation models beyond the basic single-field slow-roll paradigm \cite{Dimastrogiovanni:2021mfs}. Another interesting aspect of this class of anisotropies is that they may be tested through cross-correlations with the cosmic microwave background: the same large-scale fluctuations modulating the GW energy density at interferometer frequencies, would contribute to large-scale CMB anisotropies, making GW-CMB cross-correlations a new probe of squeezed primordial non-Gaussianity, complementary to CMB bispectrum measurements \cite{Adshead:2020bji,Malhotra:2020ket,Perna:2023dgg}. 

Since gravitational interactions decouple at the Planck scales, the CGWB could be used to test the geometry and the particle content of the Universe at early times. In~\cite{ValbusaDallArmi:2020ifo} it was shown that the presence of exotic (beyond Standard Model) relativistic and decoupled particles could modify the amplitude of the anisotropies on large scales. In~\cite{Malhotra:2022ply} the dependence of the anisotropies on the equation of state parameter of the Universe was analysed, finding a universal behavior for adiabatic primordial fluctuations. It was shown that deviations from this behavior (which could be in the form of an amplification of the anisotropies) would point to the presence of isocurvature fields during inflation. In~\cite{Schulze:2023ich} a joint Bayesian analysis of the CMB and CGWB anisotropies for ET+CE has been performed, exploiting the large correlation expected between the CMB and the CGWB~\cite{Ricciardone:2021kel}. It has been shown how the CGWB could be used to estimate cosmological parameters, breaking some degeneracies which are present in CMB measurements, such as the one between the amplitude of the primordial scalar power spectrum and the reionization epoch of the CMB photons.  The parameter in this case is the angular power spectrum of the CGWB. A Fisher matrix analysis of the detectability of the angular power spectrum from the case of phase transitions has been done in~\cite{Cui:2023dlo}.

\begin{figure}[t]
    \centering
    \includegraphics[scale=0.5]{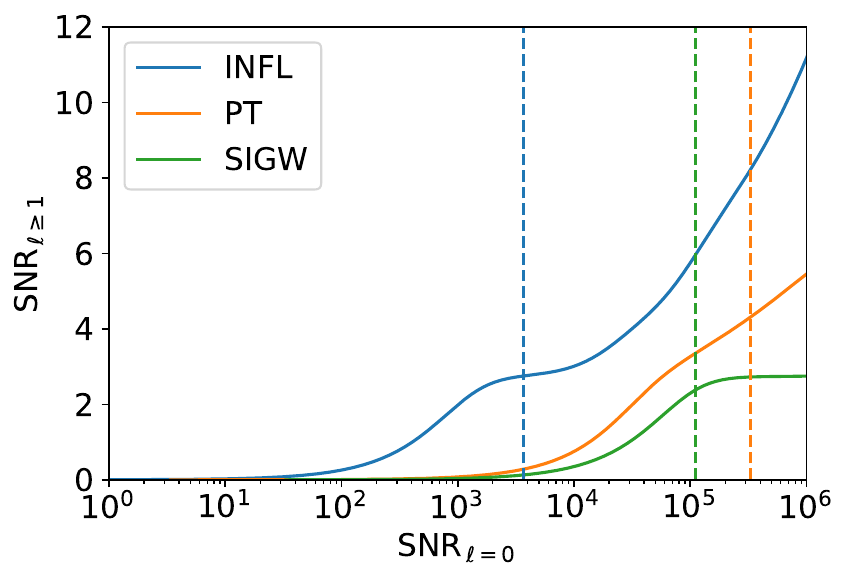}
    \includegraphics[scale=0.5]{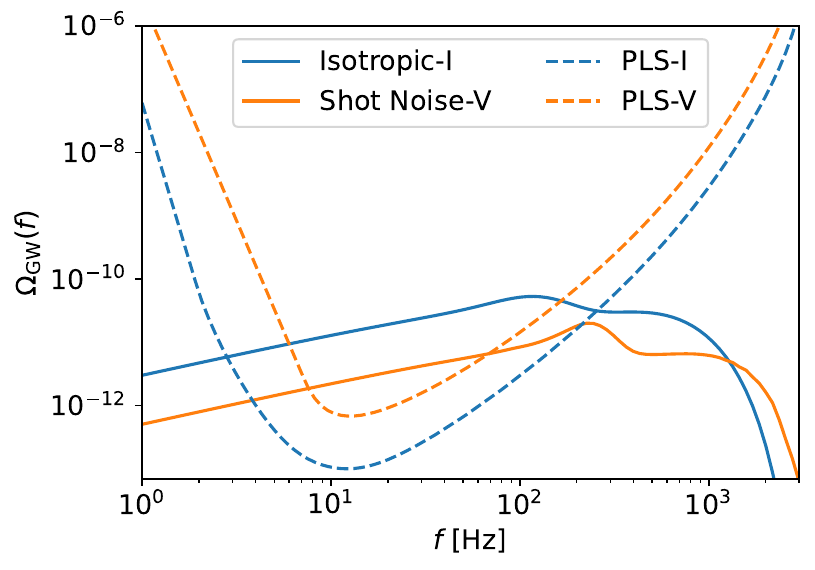}
    \caption{Left: plot of the cumulative SNR of the anisotropies as a function of the SNR of the monopole for ET (triangular) + CE ($40\, \rm km$) for five years of observation. The dashed lines represent the LVK upper bounds on the amplitude of the monopole for the three mechanisms considered (inflation, PT and SIGW). Right: plot of the monopole of the intensity (solid blue), circular polarization (solid orange) of the AGWB, and of the PLS of the intensity and circular polarization (dashed blue and orange respectively) for ET+CE in one year of observation.}
    \label{fig:SNRs_SGWB_anis}
\end{figure}

In the left panel of figure~\ref{fig:SNRs_SGWB_anis} we plot the cumulative SNR of the angular power spectrum of the auto-correlation of the CGWB and cross-correlation between the CGWB and CMB as a function of the SNR of the monopole. We consider as detector network ET in the triangular configuration in combination with one CE of $40\, \rm km$, for 5 years of observation. Because of the limited angular resolution of the network, the cumulative SNR of the anisotropies is sensitive up to $\ell_{\rm max}\approx 10$ (for the largest values of the monopole in the figure). We show the result for the CGWB from inflation with $n_T = 0.35$, phase transitions with $f_*=1\, \rm Hz$ and a Gaussian SIGW spectrum peaked at $f_*=100\, \rm Hz$. The dotted lines identify the maximum values of the SNR of the monopole consistent with the upper bounds of LVK~\cite{KAGRA:2021kbb}, which are distinct for the three signals, because of the different dependence on the frequency of the monopoles. For low values of the SNR of the monopole, the dominant contribution to the SNR of the anisotropies is given by the kinetic dipole, which is orders of magnitude larger than the cosmological anisotropies, while for larger monopoles the cosmological anisotropies become comparable with the instrumental noise, giving rise to detectable signals.

\paragraph{Anisotropies of the Astrophysical Background.}
\label{sec:Anisotropies of Astrophysical Backgrounds}

The distribution of the number of sources, and thus the AGWB, is expected to exhibit a uniform and dominant contribution across all directions in the sky (i.e., a monopole component). Superimposed on this dominant term, small anisotropies of the order of $10^{-2}-10^{-5}$ are anticipated. These deviations from a homogeneous and isotropic spectrum arise due to cosmological perturbations (i.e., intrinsic anisotropies), which also account for the overdensity and peculiar velocity of galaxies, as well as the anisotropies in the CMB. Additionally, Poisson fluctuations result from the finite number of sources contributing to the background within a limited timeframe and volume (i.e., shot noise), and the observer's peculiar motion relative to the sources' rest frame introduces kinetic effects, such as Doppler boosting and aberration in the AGWB.

Following~\cite{Phinney:2001di,Regimbau:2011rp}, the monopole energy density of the AGWB generated by the superposition of many, unresolved, astrophysical sources can be computed by integrating the product of the energy spectrum and the source rate over the entire parameter space describing the population of GW events (see also section~\ref{sect:Populationbackdiv3}),
\begin{equation}
    \bar{\Omega}_{\rm GW}(f) 
    = \frac{f}{\rho_c c^2}\int \frac{dz}{(1+z)H(z)}R(z)\int d\vec{\theta}\, p(\vec{\theta})\frac{dE}{dfd\Omega}(f,z,\vec{\theta}) \,.
    \label{eq:omega_mon}
\end{equation}
Here, $z$ is the redshift of the source, $\vec{\theta}$ represents the intrinsic parameters of the binary, e.g., in the case of BBH, the masses, spins and inclination angle of the orbit (i.e. the angle between the angular momentum of the binary and the direction of observation); $dE/dfd\Omega$ denotes the energy spectrum per solid angle and frequency emitted by individual binaries~\cite{Phinney:2001di,Regimbau:2011rp}. The merger rate per comoving volume $R(z)$ has been assumed here independent of the distribution of the intrinsic parameters of the binaries $p(\vec{\theta})$; see section~\ref{div3:bh-masses}
for a more detailed discussion of the evolution of the mass distribution with the redshift. 

The other Stokes parameters of the AGWB, introduced in eq.~\eqref{eq:gwb_stokes}, can be computed by replacing the energy density of the individual binaries in eq.~\eqref{eq:omega_mon} with an equivalent ‘‘polarized energy spectrum'', which has the same amplitude as the intensity, but varies differently with respect to  the inclination angle; see, e.g.,~\cite{ValbusaDallArmi:2023ydl}.\footnote{Note, however, that only the Stokes parameter associated to the intensity, $I$,  corresponds to  an  energy density, and no actual energy density is associated to the Stokes parameters describing linear and circular polarizations~\cite{Belgacem:2024ohp}.}
The circular polarization of the AGWB has been discussed in the previous subsection, while cosmological models which produces polarized signals will be discussed in section~\ref{sec:parity_violation}. 

The intrinsic anisotropies have been computed in~\cite{Cusin:2017fwz,Cusin:2017mjm,Cusin:2018rsq,Jenkins:2018uac,Jenkins:2018kxc,Bertacca:2019fnt, Pitrou:2019rjz} and can be classified into density anisotropies, which are related to the distribution of GW sources in the sky; Kaiser and Doppler anisotropies, which are related to the peculiar velocities of the hosts; and anisotropies imprinted by gravitational potentials along the graviton's path. A fully relativistic, gauge-invariant computation using the \textit{cosmic rulers} formalism~\cite{Schmidt:2012ne} has been performed in~\cite{Bertacca:2019fnt}. The monopole and anisotropies of the AGWB generated by BBH, BNS and BHNS have been fully characterized in~\cite{Capurri:2021zli,Bellomo:2021mer}. The public available code \texttt{CLASS$\_$GWB}~\cite{Bellomo:2021mer} allows  computing the angular power spectrum of the intrinsic and shot noise anisotropies for different populations of CBC. The relative perturbation in the energy density of the AGWB at the observer can be written as
\begin{equation}
\label{Eq::GW_Anisotropies}
    \frac{\delta\Omega^{\rm intr}_{\rm GW}}{\bar{\Omega}_{\rm GW}}(f) = \int \frac{d^3k}{(2\pi)^3}\zeta\left(\vec{k}\right)\int_0^{\eta_0}d\eta\left[\Delta^{\rm den}(\eta,k,f)+\Delta^{\rm rsd}(\eta,k,f)+\Delta^{\rm gr}(\eta,k,f)\right]\, , 
\end{equation}
where $\zeta$ is the primordial curvature perturbation and $\eta_0$ is the conformal time today. The expression for the source functions $\Delta^i$ can be found e.g.  in~\cite{Bertacca:2019fnt,Bellomo:2021mer}.  The cross-correlation of the intrinsic anisotropies with other LSS tracers will be discussed in  section~\ref{subsec:agwb_lss}, focusing in particular on the dependence of the intrinsic anisotropies on the bias of the GWs. The intrinsic anisotropies are typically considered as Gaussian random variables of zero mean and covariance proportional to the power spectrum of the primordial curvature perturbation, although the current Planck bounds admit the possibility of the presence of a non-negligible amount of primordial non-Gaussianity in the cosmological perturbations~\cite{Planck:2019kim}. In section~\ref{sec:cross-corrAGWB-CMB} we will discuss  the effect of primordial non-Gaussianity on the anisotropies of the AGWB in cross-correlation with the CMB.

The number of events that contribute to the AGWB fluctuate according to a Poisson distribution, inducing a shot noise term in the AGWB~\cite{Jenkins:2019nks,Jenkins:2019uzp,Alonso:2020mva,Belgacem:2024ohp}. In the formalism adopted here, consistently with~\cite{Bellomo:2021mer,ValbusaDallArmi:2022htu,ValbusaDallArmi:2023ydl}, the shot noise of the energy density is a Gaussian random variable $\delta\Omega_{\rm GW}^{\rm SN}(\hat{n},f)$, with zero mean and covariance given by
\begin{equation}
    \begin{split}
        \mathcal{C}^{\rm SN}(\hat{n},f;\hat{n}^\prime,f^\prime)=& \int dz\, d\vec{\theta} \frac{R(z)p(\vec{\theta})\, \delta(\hat{n}-\hat{n}^\prime)}{(1+z)^2H^2(z)\frac{dV}{dzd\Omega}(z)T_{\rm obs}}\prod_{\tilde{f}=f,f^\prime} \frac{\tilde{f}}{\rho_c c^2}\frac{dE}{dfd\Omega}(\tilde{f},z,\vec{\theta}) \, ,
        \label{eq:sn_covariance}
    \end{split}
\end{equation}
where we have neglected Poisson fluctuations proportional to the square of the number of GW events, suppressed by the low number of GWs per host galaxy. The term $\delta(\hat{n}-\hat{n}^\prime)$ indicates that Poisson fluctuations are uncorrelated for objects in different regions in the sky, thus the angular power spectrum is constant in the multipole space.

\subparagraph{Kinematic anisotropies.}

The motion of the observer relative to the rest frame of the sources induces kinematic anisotropies in the AGWB on the order of $10^{-3} c$, consistent with measurements of Doppler boosting in the CMB~\cite{Planck:2013kqc,Planck:2018vyg}. The velocity of the Local Group (LG) relative to the large-scale structure (LSS) generates a dipole in the anisotropies~\cite{ValbusaDallArmi:2022htu}, while the motion of the Earth relative to the LG induces a Doppler boosting of the signal, resulting in a dipole of order one in the velocity and in a mode coupling between different multipoles, linear in both the dipole and the intrinsic anisotropies~\cite{Cusin:2022cbb,Chowdhury:2022pnv}. In~\cite{ValbusaDallArmi:2022htu}, a comprehensive discussion of all contributions to the dipole of the AGWB is provided, demonstrating how the Einstein Telescope (ET) in combination with the Cosmic Explorer (CE) could be used to accurately measure the velocity of the Local Group (LG), yielding results competitive with large-scale structure (LSS) observations.
Ref.~\cite{Chung:2022xhv} showed that prior knowledge of the direction of the LG velocity (e.g., from CMB observations) can enhance the sensitivity to the AGWB kinematic dipole. Additionally, it showed that the Earth's peculiar motion must be accounted for to avoid biases of up to $\sim 10\%$ in the inferred dipole amplitude.

\subparagraph{Component separation of the anisotropies.}

\begin{figure}[t]
    \centering
    \includegraphics[scale=0.5]{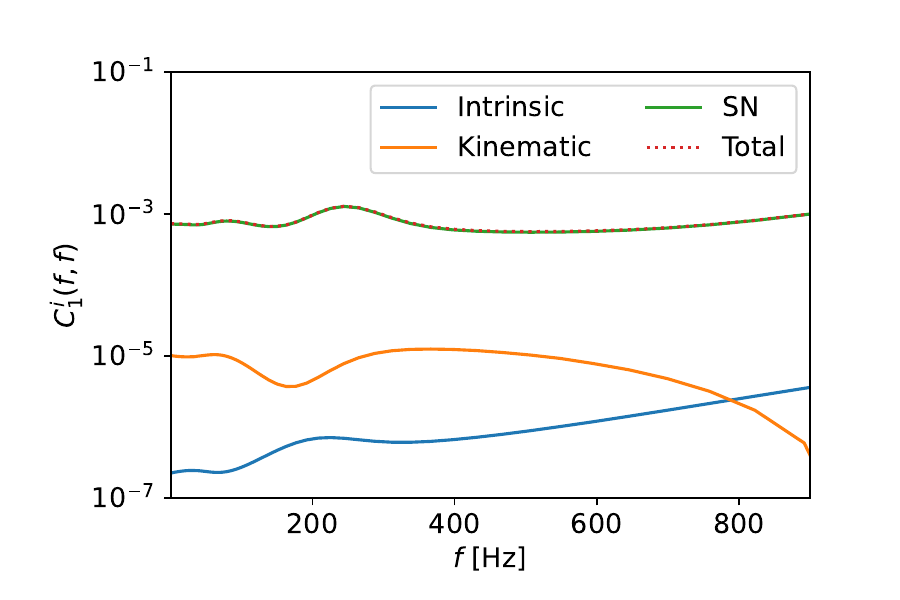}
    \includegraphics[scale=0.5]{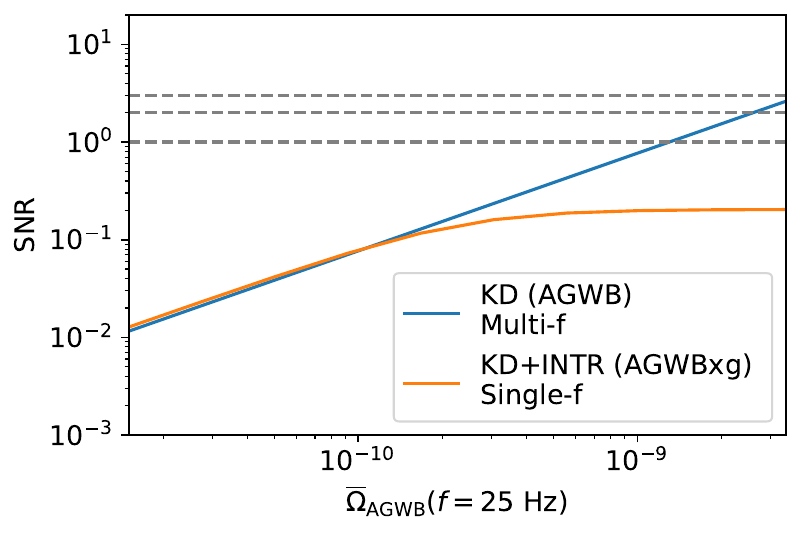}
    \caption{Left: plot of the intrinsic, shot noise and kinetic contributions to the dipole of the AGWB as a function of the frequency. Right: plot of the SNR of the kinetic dipole of the AGWB, obtained with the multi-frequency analysis of the anisotropies, and of the intrinsic and kinetic anisotropies of the AGWB, in cross-correlation with the galaxy survey SKAO2, as a function of the amplitude of the monopole of the AGWB. Here a network with the triangular ET configuration plus CE with the two interferometers (of 20km and 40km) placed in Hanford and Livingston for 5 yrs has been considered.  }
    \label{fig:AGWB_dipole}
\end{figure}

The anisotropies of the AGWB are expected to be dominated by the shot noise fluctuations, which are of the order of $10^{-1}-10^{-3}$, because of the low number of CBC mergers in the Universe per year. Therefore, the detection of the kinetic and intrinsic anisotropies could be very challenging. Several attempts have been made in order to reduce the impact of the shot noise, for instance by cross-correlating the AGWB with galaxies~\cite{Alonso:2020mva} or lensing~\cite{Capurri:2021zli}, but, because of the low angular resolution of ET to the anisotropies of the GWB, it is challenging to get a SNR larger than one for the intrinsic and kinetic components using standard techniques. In~\cite{ValbusaDallArmi:2022htu}, it has been realized that the three contributions to the angular power spectrum exhibit different frequency dependencies for $f \gtrsim 100\, \rm Hz$. Therefore, the distinct frequency scaling of these anisotropies could potentially be used for component separation. In the left panel of figure~\ref{fig:AGWB_dipole} we plot the three dipoles at different frequencies, while in the right one we show the SNR of the anisotropies with ET+CE in 5 years of observations. In the right plot, the blue line is the SNR of the kinetic dipole of the AGWB with the multi-frequency analysis. When the instrumental noise is taken into account, it becomes evident that a kinetic dipole in an AGWB with an amplitude consistent with the current LVK bound could be detected with an SNR $\simeq 3$. In order to show the power of the multi-frequency analysis, we also plot the SNR of the intrinsic and kinetic anisotropies of the auto- and cross-correlation of the AGWB with the SKAO2 galaxy catalogue~\cite{Maartens:2021dqy}, up to $\ell_{\rm max}=100$ (orange line), without considering the frequency dependence of the anisotropies. When instrumental noise does not dominate the AGWB signal, it become evident that multi-frequency analysis seems to be the most efficient technique for addressing the shot noise issue, improving by orders of magnitude the SNR.

\subsubsection{Polarization of the GWB and parity violation}
\label{sec:parity_violation}

\paragraph{Polarization of the AGWB.}

In GR, the AGWB is expected to be unpolarized because the average over polarization  angles set to zero linear polarizations and the average over inclination angles set to zero the circular polarization  (see \cite{Belgacem:2024ohp} for the explicit computation).
However,  because of the  effect of sampling a realization, shot noise  can generate a net amount of both linear and circular polarization (even for isotropic source distributions~\cite{Belgacem:2024ohp}). In particular, circular polarization   could be observed by ET+CE with an SNR greater than 2 in one year of observations~\cite{ValbusaDallArmi:2023ydl}. A computation analogous to eq.~\eqref{eq:sn_covariance} for the $V$ Stokes parameter indicates that the amplitude of the circular polarization is indeed comparable to the shot noise in the intensity and exhibits an angular power spectrum that is constant in multipole space. In the right panel of figure~\ref{fig:SNRs_SGWB_anis} we plot the monopole (intensity) and amplitude of the shot noise as a function of frequency, along with the PLS of the intensity (monopole) and circular polarization (monopole plus anisotropies). The circular polarization of the AGWB represents a foreground for the detection of polarized cosmological signals, and component separation techniques based on the frequency and angular dependence of the signal should be adopted~\cite{ValbusaDallArmi:2023ydl}.

Various mechanisms in the early Universe can generate parity violation, leading to an uneven production of left- and right-handed circularly-polarized GWs. Since parity-violating GWs are expected to be cosmologically sourced, polarization could be used together with the intensity to identify the primordial origin of the signal, although the circular polarization of the AGWB discussed  is an unavoidable foreground for polarized cosmological backgrounds.  Searches for polarized GWBs can allow one to place constraints on parity-violating theories, as we discuss next.

\paragraph{Chiral GW Sources.}
\label{sec: PV_Sources}


A number of mechanisms can result in parity violation in the early Universe~\cite{Alexander:2004us}. Two compelling, well-studied, classes of parity-violating inflationary models that could lead to a circularly polarized GW background include Chern-Simons gravity ~\cite{Satoh:2007gn,Bartolo:2020gsh, Takahashi:2009wc,Bartolo:2018elp}, and axion-gauge field inflation \cite{Garretson:1992vt,Anber:2006xt,Barnaby:2010vf,Cook:2011hg,Sorbo:2011rz,Anber:2012du,Adshead:2012kp,Dimastrogiovanni:2012ew,Adshead:2016omu,Dimastrogiovanni:2016fuu,Agrawal:2017awz,Caldwell:2017chz,Thorne:2017jft,Dimastrogiovanni:2018xnn,Fujita:2018vmv,Domcke:2018rvv,Lozanov:2018kpk,Watanabe:2020ctz,Holland:2020jdh,Domcke:2020zez,Iarygina:2023mtj,Dimastrogiovanni:2023oid,Ishiwata:2021yne,Durrer:2024ibi,Dimastrogiovanni:2024xvc}. More details on axion inflation mechanism can be found in section~\ref{subsubsec: inflation_multifieldScenarios}.
Other sources such as GWB from turbulence in the primordial plasma induced either from cosmological first-order (electroweak or QCD phase transition beyond the Standard Model) phase transitions~\cite{Kamionkowski:1993fg,Witten:1984rs, Breitbach:2018kma} and from the primordial magnetic fields that are coupled to the cosmological plasma~\cite{Brandenburg:1996fc, Christensson:2000sp, Kahniashvili:2010gp, Brandenburg:2017rnt, Brandenburg:2021aln} can be chiral. Although an AGWB is typically expected to be unpolarized,  polarization can occur when considering shot noise fluctuations, as we discussed in section~\ref{sec:Anisotropies of Astrophysical Backgrounds}.

\paragraph{Detection Formalism for Chiral Gravitational Waves.}
\label{sec: PV_formalism}

\begin{figure}[t]
    \centering
    \includegraphics[width=\textwidth]{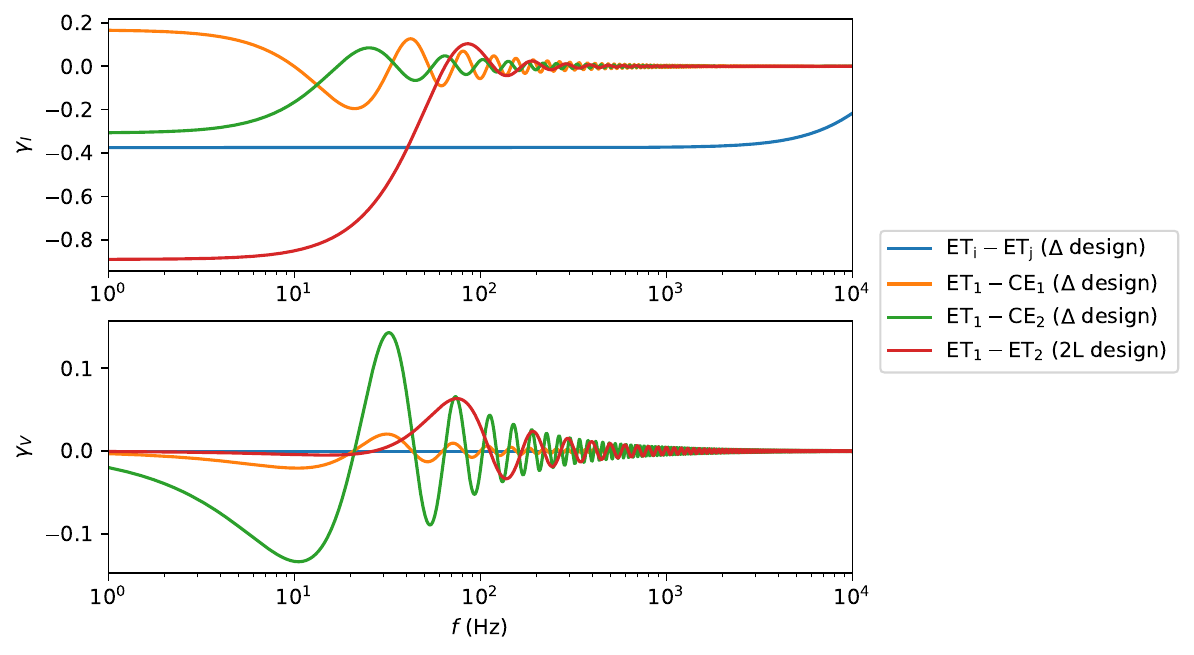}
    \caption{Standard and parity-violating overlap reduction functions plotted over ground-based detector frequency range. When assuming a triangular configuration for ET,  we plot the result for a pair of ET detectors (out of the three composing the triangle), and for  one ET detector  (out of the three composing the triangle)  and a Cosmic Explorer (CE), with CE taken  either in the current Hanford site or in the current Livingston sites. When assuming the 2L configuration for ET,  we plot the result for this pair of L-shaped detectors. The triangular ET design is assumed in Virgo location. We denote the triangular pairing with $i$, $j$ as all pairings produce the same overlap reduction functions.}
    \label{fig:ORF_I_V}
\end{figure}

We now discuss the formalism used in detecting parity-violating backgrounds, following~\cite{Seto:2007tn, Seto:2008sr}. This formalism was used in practice in the LVK data in~\cite{Crowder:2012ik, Martinovic:2021hzy}, and in simulated third-generation networks in~\cite{Crowder:2012ik, Badger:2021enh}. We use the circularly polarized bases $e^R = (e^{+} + ie^{\cross})/\sqrt{2}$ and $e^L = (e^{+} - ie^{\cross})/\sqrt{2}$ (with + and $\cross$ the plus and cross polarizations, respectively)  to obtain the right- and left-handed modes $h_R = (h_{+} - ih_{\cross})/\sqrt{2}$ and $h_L = (h_{+} + ih_{\cross})/\sqrt{2}$, respectively. 
Using \eq{eq:gwb_stokes}, the right-right and left-left correlators can then be written as
\be\label{eq:RL_correlators}
    \begin{pmatrix}
        \langle h^*_{R}(f,\hat{\bf{n}}) h_{R}(f',\hat{\bf{n}}') \rangle \\
        \langle h^*_{L}(f,\hat{\bf{n}}) h_{L}(f',\hat{\bf{n}}') \rangle
    \end{pmatrix}
    = \frac{1}{8\pi} \delta(f-f')\delta(\hatn,\hatn')
    \begin{pmatrix}
        I(f,\hat{\bf{n}}) - V(f,\hat{\bf{n}}) \\
        I(f,\hat{\bf{n}}) + V(f,\hat{\bf{n}})
    \end{pmatrix},
\ee
We use the standard cross-correlation 
estimator~\cite{Romano:2016dpx,Cornish:2013nma,Allen:2002jw,Drasco:2002yd}
\begin{eqnarray}
    \langle \hat{C}_{d_1 d_2} \rangle &=&\int_{-\infty}^{\infty}df\int_{-\infty}^{\infty}df'\delta_T(f-f')\langle s_{d_1}^{*}(f)s_{d_2}(f') \rangle \tilde{Q}(f')\nonumber \\
    &=& \frac{3H_0^2 T}{10\pi^2}\int_0^{\infty}df\frac{\Omega'_{\rm GW}(f)\gamma_I^{d_1 d_2}(f)\tilde{Q}(f)}{f^3},
    \label{eq:Y_estimator}
\end{eqnarray}
where
\begin{eqnarray}
    \Omega'_{\rm GW} &=& \Omega_{\rm GW}\bigg[1+\Pi(f)\frac{\gamma_V^{d_1 d_2}(f)}{\gamma_I^{d_1 d_2}(f)}\bigg],\nonumber \\
    \gamma_I^{d_1 d_2}(f) &=& \frac{5}{8\pi}\int d\hat{\Omega}(F_{d_1}^{+}F_{d_2}^{+*} + F_{d_1}^{\cross}F_{d_2}^{\cross*})e^{2\pi if\hat{\Omega}\cdot\Delta\Vec{x}}, \nonumber\\
    \gamma_V^{d_1 d_2}(f) &=& -\frac{5}{8\pi}\int d\hat{\Omega}(F_{d_1}^{+}F_{d_2}^{\cross*} - F_{d_1}^{\cross}F_{d_2}^{+*})e^{2\pi if\hat{\Omega}\cdot\Delta\Vec{x}},
    \label{eq:PVOmeg&ORF}
\end{eqnarray}
with $H_0$ the Hubble parameter, $T$ the observing time, $\delta_T(f) = \sin(\pi fT)/(\pi f)$, $\tilde{s}_{d_1}(f)$ and $\tilde{s}_{d_2}(f)$ the Fourier transforms of the strain time series of two GW detectors (denoted by $d_1, d_2$). We apply the usual $\tilde{Q}(f)$ as the optimal filter taking into account detectors' strain power spectral densities, and $F_n^A = e_{ab}^A d_n^{ab}$ stands for the contraction of the tensor modes of polarization $A$ to the \textit{n}th detector's geometry. We denote by 
$\gamma_I^{d_1 d_2}$ 
the standard overlap reduction function of two detectors $d_1, d_2$, and by $\gamma_V^{d_1 d_2}$ 
the overlap function associated with the parity violation term. The polarization degree, $\Pi(f) = V(f)/I(f)$, takes on values between -1 (fully left polarization) and 1 (fully right polarization), with $\Pi = 0$ being an unpolarized isotropic GWB.
We build the Gaussian log-likelihood for a network of $N$ detectors
\begin{align}
    \log p(\hat C(f) | \boldsymbol{\theta})
    \propto \sum_{i, j > i}^N\sum_{f}\frac{\left[\hat{C}_{d_i d_j}(f) - \Omega'_{\rm GW}(f, \boldsymbol{\theta}) \right]^2}{\sigma_{d_i d_j}^2(f)},
    \label{eq:pe_ms:likelihood}
\end{align}
where 
$\hat{C}_{d_i d_j}(f)$
is the frequency-dependent cross-correlation estimator of the GWB  calculated using data from detectors $d_i, d_j$, and 
$\sigma^2_{d_i d_j}(f)$ 
is its variance, and $\boldsymbol{\theta}$ are the associated parameters to the tested GWB model.
\paragraph{Detection Prospects for Chiral Gravitational Waves.}
\label{sec: PV_prosepcts}
In the context of ET, this formalism has a number of ramifications on parity 
violation detection. We plot the standard and parity-violating associated overlap reduction functions in figure~\ref{fig:ORF_I_V}. One can see that the parity violating associated overlap 
reduction function between two detectors of an assumed triangular design\footnote{We follow the  terminology of the 2020 ET Design Report Update,
\url{https://apps.et-gw.eu/tds/?r=18715}: the high-frequency (HF) and low-frequency  (LF) interferometers that make the so-called ``xylophone'' configuration are indeed referred to as ``interferometers''. The combination of a HF interferometer  and a LF interferometer (whether in a L-shaped geometry, or with arms at $60^{\circ}$ as in the triangle configuration)
is called a ``detector''.  The whole ensemble of detectors is called an ``observatory''. So, ET in the triangle configuration is made of three detectors for a total of six interferometers, while in the 2L configuration it is made of two detectors, for a total of four interferometers.\label{foot:nomenclature}} would 
be approximately 0 due to their co-planarity to one another.
From \eq{eq:PVOmeg&ORF} this would imply that $\Omega_{\rm{GW}} \approx\Omega'_{\rm{GW}}$ for even entirely chiral GWs. Thus, parity- violation inferences on GW data with this formalism is difficult with a triangular ET configuration alone. However, ET in combination with other detectors (such as CE) should have $\gamma_V^{d_1 d_2}(f) \neq 0$ due to their non-coplanarity - thus allowing for parity violating inferences in extended networks whilst taking advantage of ET’s sensitivity. A 2L design for ET, and any additional detectors in the network, would not face such problems for similar reasons. 
Reconsidering the previously described formalism while taking into account the full response functions and noise curves across the entire frequency band (for planar detectors), it is found that $\mathcal{O}(1)$ net polarization can be detected in the GWB, with a magnitude of $h^2\Omega_{\rm{GW}}  \simeq 10^{-11}$ and an SNR of order one~\cite{Domcke:2019zls}. Fully analytical and covariant details of these parity-sensitive response functions, which are critical for parity violation detection, can be found in~\cite{Domcke:2019zls}. Extensions to search for GWBs with anisotropies and circular polarization have been developed and studied in the context of ground-based detectors~\cite{Mentasti:2023gmg}. Many results are in agreement with~\cite{Martinovic:2021hzy} in an LVK context; however, an ET+CE network can improve this by about three orders of magnitude.
As discussed in section~\ref{sec:Anisotropies of Astrophysical Backgrounds}, the presence of a circularly polarized AGWB may provide an additional challenge for CGWB parity violation detection~\cite{ValbusaDallArmi:2023ydl}. A simplified example showed that, by taking advantage of different background features, such as their frequency dependence and features in the sky maps, it is possible to reconstruct the primordial background with good precision using ET+CE, primarily limited by instrumental noise, as in the case of the intensity. That said, separation corrections of the order of $20\%$ to the total SNR must be carefully taken into account.

\subsubsection{Source separability}
\label{subsec:source_separability}


Once a GWB is detected, the next important question will be to relate the signal to the sources that contribute to it. In particular, one will be interested in separating compact binary coalescences (namely a pure astrophysical contribution) from possible sources of a cosmological origin (for instance inflation, cosmic strings or first-order phase transitions). A Bayesian analysis has shown that current terrestrial interferometers (i.e., Advanced LIGO and Advanced Virgo network), even once operating at design sensitivity will not allow for separation of the sources~\cite{Martinovic:2020hru}. Assuming one is able to do an individual source subtraction~\cite{Zhou:2022nmt}, the residual CBC contribution to the GWB will be dominated by unresolved binary neutron star mergers at the level of $\sim 10^{-11}$ at 10 Hz.\footnote{A more detailed analysis shows that the result actually depends on the shape of the cosmological spectrum that is being searched, see \cite{Belgacem:2024ntv} and section~\ref{sect:subtractastrobkgdi9}.} Third generation (3G) detectors, like Einstein Telescope, and in particular a network of 3G detectors will allow simultaneous separation of astrophysical and cosmological GWBs, provided reasonable levels of individual source subtraction can be achieved~\cite{Martinovic:2020hru}.

The separability of sources in the ET frequency band has several challenges given that multiple components contribute to the overall signal. This makes the parameter space large and complex. This complexity, along with degenerate characteristics, complicates such a task making it also computationally challenging. A computationally less demanding approach could be performed in a two-step process: first, reconstruct the GWB spectrum agnostically (as in LISA data analysis techniques~\cite{Caprini:2019pxz,Karnesis:2019mph,Pieroni:2020rob,Flauger:2020qyi}); second, convert the spectrum constraints into population and cosmological parameter constraints (see e.g.,~\cite{Braglia:2024kpo} for a recent application). This method allows to balance accuracy and feasibility and can help identify relevant parameter spaces for full Bayesian techniques.

Another limitation in the detection of the GWB signal  is the impact of significant residuals, primarily from uncertainties in coalescence phase and luminosity distance, when resolved sources are subtracted~\cite{Zhou:2022nmt,Belgacem:2024ntv}. Enhanced signal subtraction methods are necessary to improve the observation of other astrophysical and cosmological GWBs.

\subsubsection{Impact of correlated noise on GWB}\label{sect:correlated_noise_div2} 


To be able to resolve the weak signal from the GWB, one needs to rely on cross-correlation methods to suppress independent noise sources at the different detectors. In the absence of correlated noise, this implies that any remaining correlated signal is from GWs. However, past research has shown that such a correlation analysis is prone to noise contamination from correlated noise sources, both for co-located as well as widely separated detectors. 
As a first order approach one should consider to reduce the noise couplings, so to lower or entirely remove such correlated noise disturbances. If this is deemed impossible one could consider  performing a joint parameter estimation taking both the correlated noise and the GWB signal into account (see section~\ref{sect:RecCorrNoise_div2}). Finally, if the correlated noise has a much larger amplitude and/or is not known with high accuracy, one has to consider whether it is more efficient to remove all the contaminated frequency bins/time segments and only use the part of the data with the highest data quality.

In the next sections we will discuss two noise sources, which have been considered to potentially contaminate the search for an isotropic GWB with the ET. Here we only focus on fundamental, environmental noise sources and their impact. Earlier work has indicated that co-located detectors could have additional correlated noise from detector infrastructure, which could potentially have significantly higher effects \cite{Janssens:2021cta}. However, more work is needed to understand the levels of correlation of infrastructural noise in a realistic ET environment.

\paragraph{Noise correlations from electromagnetic phenomena.}

On the Earth, every second about hundred lightning strikes occur. This vast amount of electromagnetic discharges creates standing waves in the cavity formed by the Earth's surface and the ionosphere, known as the Schumann resonances \cite{1952ZNatA...7..149S,1952ZNatA...7..250S}. With the fundamental mode and subsequent harmonics occurring at $\sim$7.8~Hz, $\sim$14~Hz, $\sim$~21Hz, etc., this broadband noise source falls within ET's low frequency sensitivity range. Due to the global nature of this noise source, it has the potential to couple coherently to GW interferometric detectors, no matter their separation. In past work the Schumann resonances were coherently observed in the magnetometers at both the LIGO sites as well as at Virgo \cite{Thrane:2013npa,Thrane:2014yza,Coughlin:2016vor,Coughlin:2018tjc,Janssens:2022tdj}.

Earlier work has highlighted how correlations due to electromagnetic fluctuations can impact the search for an isotropic SGWB up to $\sim$ 30~Hz if one assumes similar levels of magnetic coupling at the ET compared to 2G detectors such as LIGO and Virgo \cite{Janssens:2021cta}. The coupling should be reduced by three to four orders of magnitude per detector around 5~Hz to remove any noise contamination \cite{Janssens:2021cta}. Furthermore, above 30~Hz there is potential of additional contamination in narrow frequency bands around resonant features in the magnetic coupling.
More work is needed to understand whether the magnetic coupling to ET can be reduced compered to 2G detectors to (partially) deal with this noise contamination. Also the effectiveness of noise subtraction, as described in \cite{Thrane:2014yza,Coughlin:2016vor,Coughlin:2018tjc}, or joint estimation, as discussed in \cite{Meyers:2020qrb}, in case of such a loud noise sources should be further investigated. However, it is to be noted the correlated signal from magnetic origin has a steeply declining behavior as a function of frequency, whereas many of the expected GW backgrounds have a flat or slowly increasing $\Omega_{\rm GW}$ signature. This characteristic behavior originates from the steeply declining magnetic coupling function observed at 2G detectors.

\paragraph{Noise correlations from seismic phenomena.}\label{noisecorrseismicdiv2}

More recent work has also investigated the effect of coherent seismic noise and the subsequent Newtonian Noise (NN) \cite{Janssens:2022xmo,Janssens:2024jln}.\footnote{NN is a force directly exerted on GW test-masses caused by density fluctuations in the surrounding medium.} This noise source will only appear correlated between different ET detectors if they are located close to each other. In this context, the detectors are considered to be located close to each other  if their input and/or end towers are separated by a distance of several hundreds of meters up to several kilometers. This will be the case  for ET in the triangular configuration. The exact distance at which the coherence of seismic noise decreases enough in the relevant frequency band for the ET is likely to be somewhere between 2~km and 10~km; however, additional research is needed to confirm this.

\begin{figure}[t]
    \centering
    \includegraphics[width=0.7\linewidth]{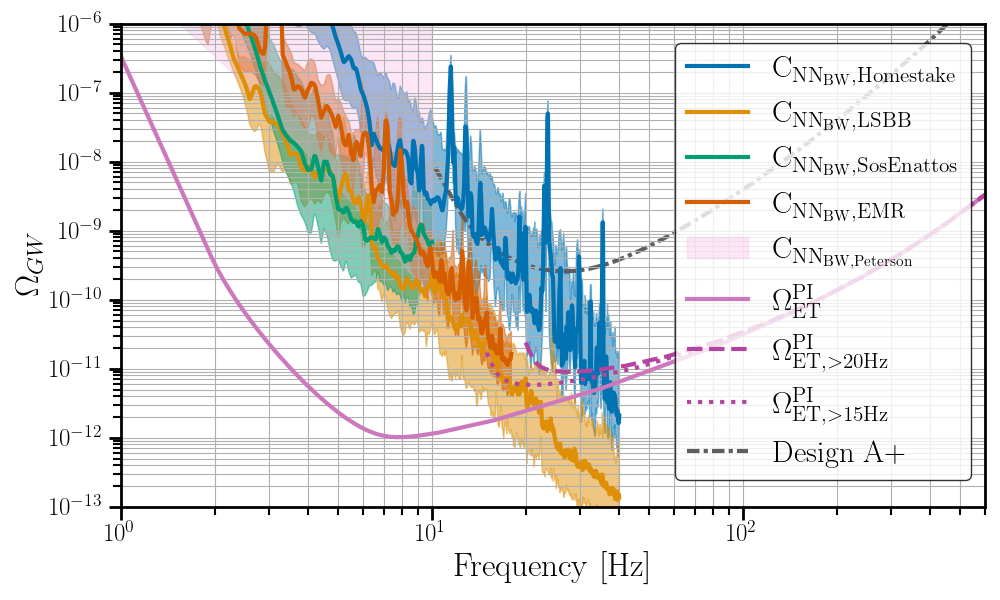}
    \caption{The projected impact from correlated NN from body-waves on the search for an isotropic GWB. We present budget based on observed seimic correlations from four different seismic sites. Note that many parameters, such as depth, horizontal separation, etc., vary for the different sites; see~\cite{Janssens:2024jln} for an in-depth discussion. Figure  from \cite{Janssens:2024jln}.}
    \label{fig:NNBudget_GWB}
\end{figure}

Depending on the specific seismic environment of ET's site, the coherent NN noise originating from seismic body-waves will contaminate the sensitivity to an isotropic SGWB up to 20~Hz--40~Hz~\cite{Janssens:2022xmo,Janssens:2024jln}, as shown in figure~\ref{fig:NNBudget_GWB}. Similar to the magnetic noise scenario, the lowest frequencies are affected more seriously and the NN is predicted to have a declining behavior as a function of frequency (when expressed in terms of $\Omega_{\rm GW}$). This might facilitate spectral separation if one is looking for a GWB which has a flat or increasing $\Omega_{\rm GW}$ spectrum as function of frequency. 
In section~\ref{sect:RecCorrNoise_div2} we will discuss the recovery of a cosmological signal in the presence of such correlated noise.

\subsubsection{Reconstruction of GWBs in presence of correlated noise}\label{sect:RecCorrNoise_div2}

As discussed in section~\ref{sect:correlated_noise_div2}, a critical challenge in detecting the GWB is the precise characterization of correlated detector noise, which is essential for distinguishing the signal from spurious instrumental and environmental fluctuations. We have seen that the triangular geometry suffers from correlated noise  due to the relative proximity of interferometer components, leading to potential contamination in GWB searches. 

The effect of correlated noise on the detection and characterization of the GWB with ET has been recently studied in~\cite{Caporali:2025mum}, where a Bayesian framework has been introduced to simultaneously reconstruct the parameters of the GWB and the correlated noise, both modeled as power laws in the frequency domain. 
For detectors $I$ and $J$, the covariance of the noise is defined as the two-point correlation function,
$N_{IJ}(f)\equiv \langle\tilde{n}_I(f)\tilde{n}^*_J(f')\rangle$.
For the triangle configuration, in ref.~\cite{Caporali:2025mum}  is made the assumption that all diagonal terms in the PSD are given by the same function $N_d(f)$, and that all  off-diagonal terms are also all given by the same function $N_o(f)$,  so that
\be\label{NIJfdiv2}
N_{IJ}(f)=
\left(
\begin{array}{ccc}
N_d(f) & N_o(f)  & N_o(f) \\
N_o(f) & N_d(f)  & N_o(f) \\
N_o(f) & N_o(f)  & N_d(f) 
\end{array}
\right)\, ,
\ee
and it is assumed that the off-diagonal terms  can be parametrized by a simple power-law
\be\label{Nopowerlawdiv2}
N_o(f) =N_o(2.75\, {\rm Hz})\,    \(\frac{f}{\rm 2.75\, Hz}\)^{n_{\rm noise}}\, .
\ee
Defining $r= N_o(2.75\, {\rm Hz})/N_d(2.75\, {\rm Hz})$, so that 
$r$ is the correlation coefficient at the reference frequency 2.75~Hz, we then write
\be\label{NoNdpowerlawdiv2}
N_o(f) =N_d(2.75\, {\rm Hz})\,  r  \(\frac{f}{\rm 2.75\, Hz}\)^{n_{\rm noise}}\, ,
\ee
The value of the tilt expected from Newtonian noise is  
$n_{\rm noise}=-8$~\cite{Badaracco:2019vjq,Janssens:2022xmo,Janssens:2024jln}, while different values of the correlation coefficients, in the physical range
$r\in [-0.5, 1]$, have been tested.

The stochastic GW signal is also modeled as a simple power-law,\footnote{For the GW signal, it is customary to take a reference frequency closer to the minimum of the PLS, such as $25$~Hz here, while for the correlated noise in \eq{NoNdpowerlawdiv2} is used  a reference frequency 2.75~Hz closer to the region where it is more relevant. Of course, different choices of the reference frequency can be reabsorbed into different definitions of the corresponding amplitudes.} 
\be
\Omega_{\rm GW} (f)= A_{\rm GW}\, \(\frac{f}{\rm 25\, Hz}\)^{n_{\rm GW}}\, 
\ee
Under these assumptions,
ref.~\cite{Caporali:2025mum} derives
a general likelihood for the GWB  that accounts for correlated noise and performs parameter estimation for 
the whole set of parameters $\{A_{\rm GW},n_{\rm GW},r,n_{\rm noise} \}$ that characterize the signal and the correlated noise.

\begin{figure}[t!]
    \centering
    \includegraphics[width =.8 \textwidth]{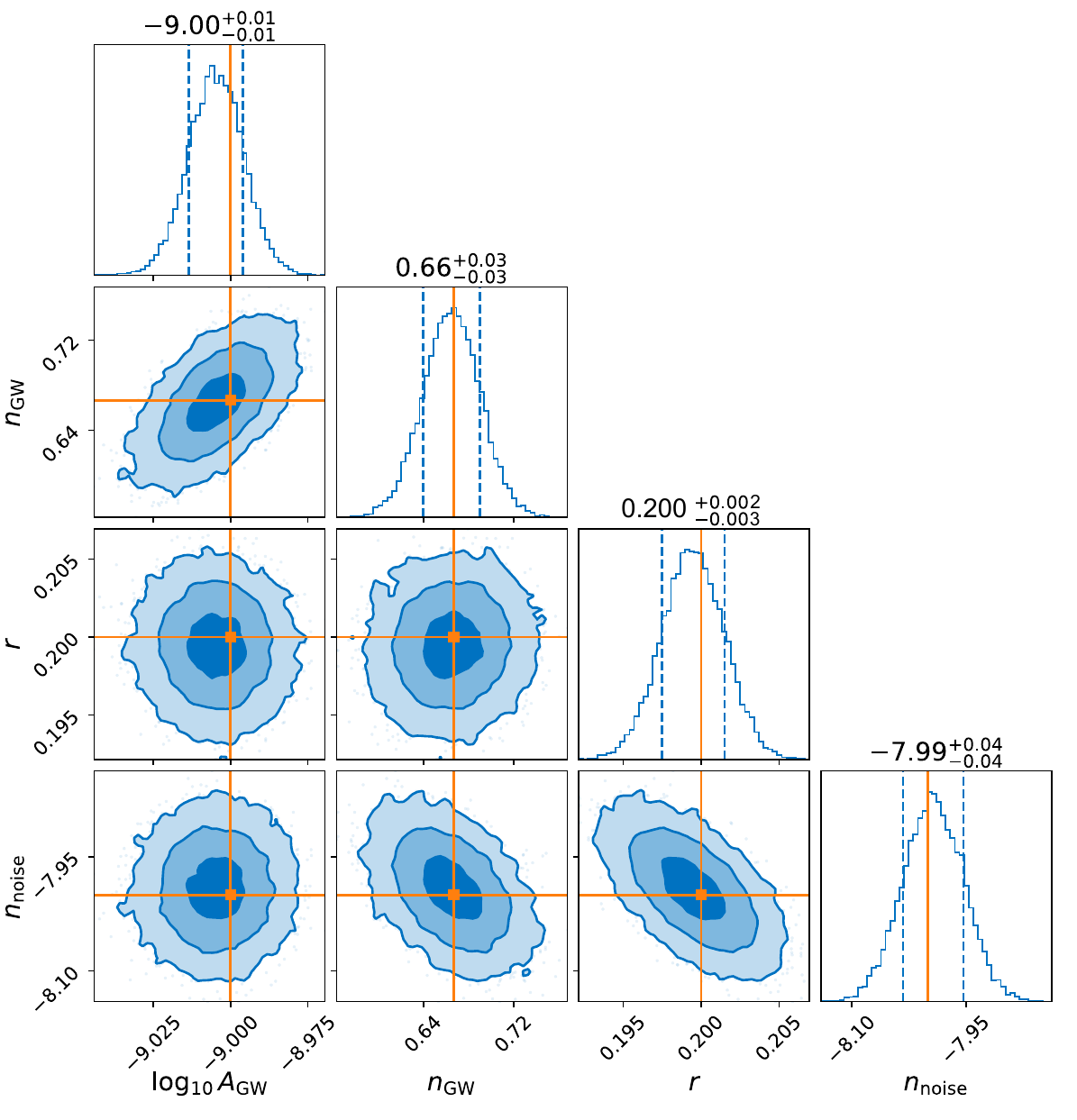}
    \caption{Corner plot of the posterior distributions for the log-amplitude and tilt of the GWB, $\{ \log_{10}A_{\rm GW}, n_{\rm GW}\}$ for the correlation coefficient and tilt of the noise,  $\{r, n_{\rm noise} \}$,
    when the injected signal has $\log_{10}A_{\rm GW}=-9$ and  $n_{\rm GW}=2/3$, while the injected noise has 
    $r^{\rm inj}=0.2$, $n_{\rm noise}=-8$.
    The shaded areas represent the 1-, 2-, and 3-$\sigma$ credible regions. The orange lines indicate the injected values for the four parameters. Figure from \cite{Caporali:2025mum}.} 
    \label{fig:corner_triangular}
\end{figure}

\begin{figure}[t!]
    \centering
    \includegraphics[width =.8 \textwidth]{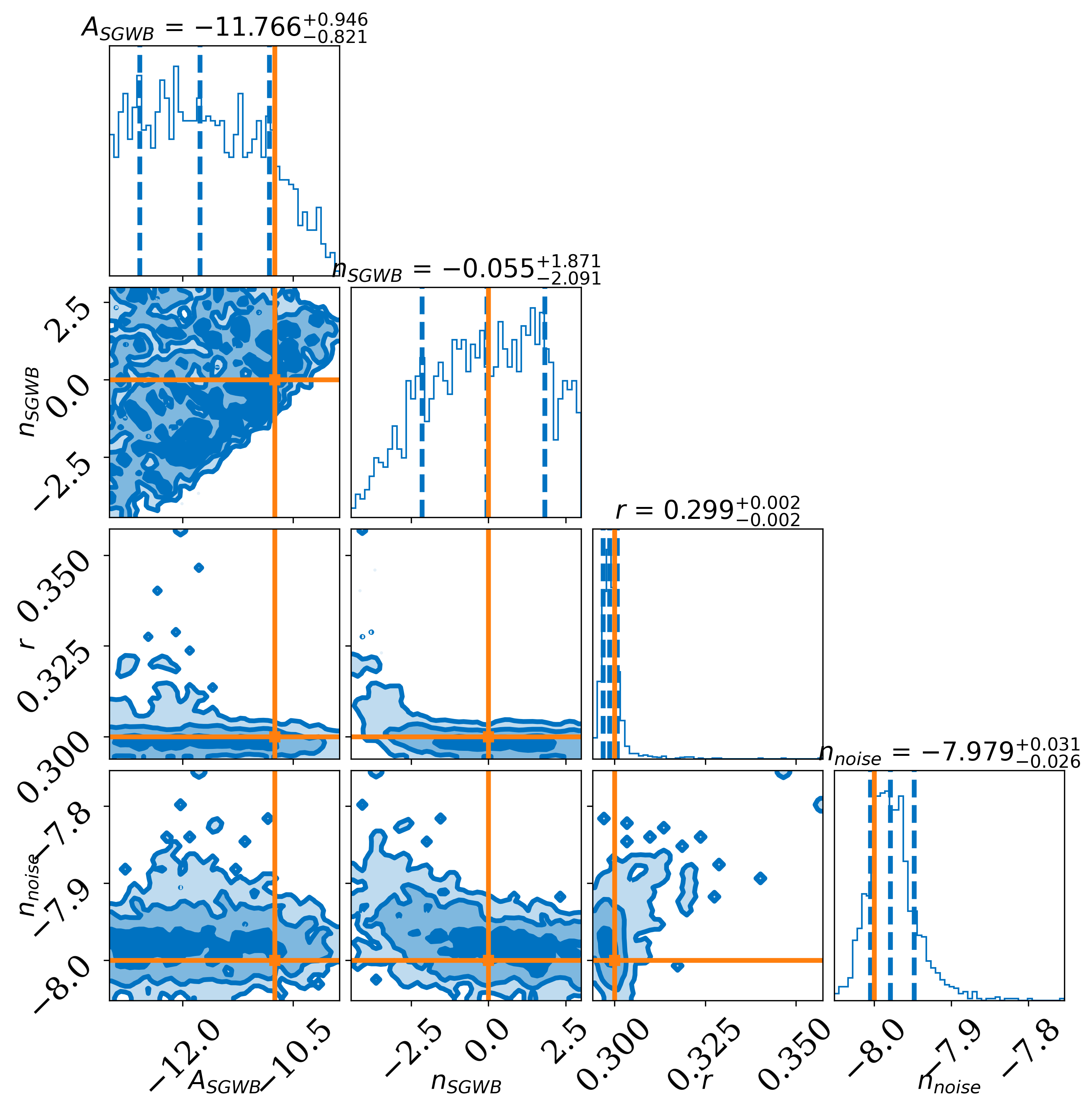}
    \caption{As in figure~\ref{fig:corner_triangular} when
    the injected signal has $\log_{10}A_{\rm GW}=-11$ and  $n_{\rm GW}=0$, while the injected noise has 
    $r^{\rm inj}=0.2$, $n_{\rm noise}=-8$.
    } 
    \label{fig:Tobs1day_snr3}
\end{figure}

The quality of the reconstruction depends, of course, on the amplitude of signal with respect to the noise. Figure~\ref{fig:corner_triangular} shows the result of the parameter reconstruction when the injected correlated noise has the correlation coefficient $r^{\rm inj}=0.2$ and a tilt
$n^{\rm inj}_{\rm noise}=-8$, while  the  injected GW signal  is taken to have $A^{\rm inj}_{\rm GW}=10^{-9}$, $n^{\rm inj}_{\rm GW}=2/3$, which are the amplitude and tilt expected for the stochastic background generated by the superposition of coalescing binaries in the inspiral phase (see figure~\ref{fig:{monopole_total}}).
This  injected stochastic GW background accumulates  an SNR  of about  67 in just 1~day of observation (restricting the analysis to frequencies below 200~Hz, where the CBC background can be modeled as a power law with $n^{\rm inj}_{\rm GW}=2/3$),  while the correlated noise, if interpreted as a signal, has ${\rm SNR}\sim 135$
(again for 1~day of observation).\footnote{These are the signal-to-noise ratio  computed using only the diagonal term $N_d$ for the noise, i.e. we are considering the off-diagonal term as a competing correlated signal.} Since the SNR for a two-detector (or multiple-detector) correlation grows with the integration time as $T^{1/2}$, see \eq{eq:SNRcross}, these spectra correspond to an 
SNR of about 1280 and 2580, respectively, for 1~yr of integration.

In this case, where the GW signal has a very large SNR and a mild blue slope, quite  different from the steep red slope of the noise, we see from the figure that, already 
with one day of observation, both the GWB and noise parameters could be reconstructed with percent-level accuracy. 

We have repeated the analysis for a GW background with 
$A^{\rm inj}_{\rm GW}=10^{-11}$, $n^{\rm inj}_{\rm GW}=0$. Such a background would produce a signal-to-noise ratio ${\rm SNR}=3$ in 1~day of observation (i.e., ${\rm SNR}\simeq 57$ in 1~yr). For the noise we still take $r^{\rm inj}=0.2$ and a tilt
$n^{\rm inj}_{\rm noise}=-8$. The results of the parameter reconstruction with 1~day of data are shown in
figure~\ref{fig:Tobs1day_snr3}. We see that in this case the quality of reconstruction of the signal is less good (and in particular the tilt cannot be  reconstructed), as expected from the fact that it has a smaller SNR, while the parameters of the noise, which is the same as before and therefore has a very large SNR (again, about 135 in just 1~day), can still be reconstructed well.
Note that this  reconstruction is obtained from just 1 day of data, but this is due to the fact that  also the injected GW signal reaches an SNR equal to 3 in just 1~day. 
The effect of the same   correlated noise will be significantly  more important in the reconstruction of  GW backgrounds with even lower amplitude,  such as a background that would accumulate, say, ${\rm SNR}=3$ in 1~yr of data, i.e. 
near the nominal sensitivity limit determined by the diagonal terms in \eq{NIJfdiv2}.

This  shows that the presence of correlated noise  in the triangle configuration is a relevant limitation, although it could still possible to 
reconstruct the GWB parameters, provided that the SNR of the signal is sufficiently large and that a reliable noise model is available. 
This conclusion, however, holds when the injected  data are  generated using the same power-law noise model that is then employed in the recovery of the signal. When dealing with real data, more sophisticated noise models, possibly informed by seismographic data taken on-site, will be necessary to achieve the required level of precision in noise reconstruction. 
The 2L configuration, where correlated noise is (to a first approximation) negligible, is immune to this problem.

\subsection{Probing the early Universe}
\label{sec:early_universe}

\subsubsection{GWs from inflation}
\label{subsec: inflation}


\paragraph{The single-field slow-roll paradigm: vacuum fluctuations.}

In its simplest realization, inflation is driven by a single scalar field slowly rolling down its potential \cite{Guth:1980zm,Starobinsky:1980te,Linde:1981mu,Albrecht:1982wi}. There exist several such ``minimal'' inflationary models that fare well compared to  observations, such as the Starobinsky model \cite{Starobinsky:1980te} or Higgs inflation \cite{Bezrukov:2007ep}. The general prediction associated with single-field slow-roll models is that of a slightly red-tilted GW spectrum sourced by vacuum fluctuations~\cite{Grishchuk:1974ny,Starobinsky:1979ty}. Typically, CMB probes are then expected to provide the most stringent constraints on this class of inflationary models.

Let us recall that, in a generic inflationary model, the primordial spectrum of  tensor perturbations ${\cal P}_{T,\rm in}(k)$ is defined by\footnote{We follow the definitions and notation of Chapter~19 of \cite{Maggiore:2018sht}.} 
\be\label{4corrhTTijP}
\langle h^{\rm TT}_{ij}(\eta_{\rm in},{\bf x}) h^{\rm TT}_{ij}(\eta_{\rm in},{\bf x})\rangle=\int_0^{\infty} \frac{dk}{k}\, {\cal P}_{T,\rm in}(k)\, ,
\ee
where $\eta_{\rm in}$ is chosen deep into radiation dominance, and such that all modes of interested are super-horizon, and is taken as the initial value of conformal time  for the subsequent evolution until the present epoch.
A useful parametrization for the primordial GW spectrum in single-field slow-roll (SFSR) models is a power law
\be\label{4primtensP}
{\cal P}_{T,\rm in}(k)=A_T(k_*)\, \(\frac{k}{k_*}\)^{n_T}\, ,
\ee
where $k_*$ is a suitably chosen pivot scale (for CMB studies typical values are   $k_*=0.05\, {\rm Mpc}^{-1}$, or 
$k_*=0.002\, {\rm Mpc}^{-1}$), while $n_T$ is the tensor spectral index, which in general is a function of $k$. In similar way, can be  parameterized the primordial scalar power spectrum, with an amplitude $A_{\cal R}(k)$, where ${\cal R}$ is the primordial curvature perturbation. Using the equivalent notation $A_S(k)\equiv  A_{\cal R}(k)$ for the amplitude of scalar perturbations, the ratio of the amplitude of the tensor and scalar power spectra allows to define the 
tensor-to-scalar ratio $r(k)=A_T(k)/A_S(k)$.

\begin{figure}[t!]
    \centering
    \includegraphics[width=0.6\textwidth]{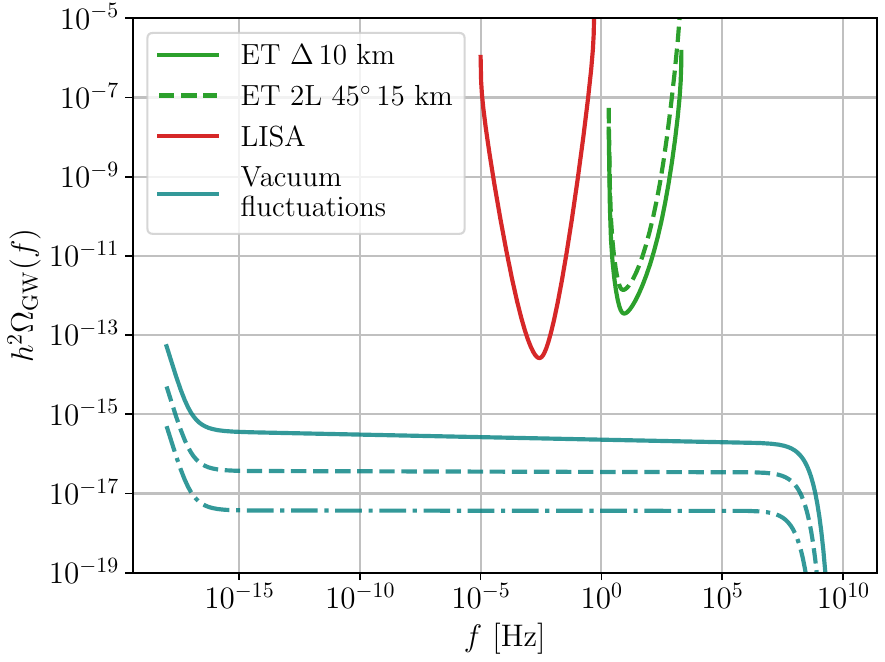}
    \caption{Spectrum of the single-field slow-roll inflation as a function of frequency, together with the PLS of LISA and ET in two different configurations. The solid, dashed, dot-dashed standard inflation curves are computed by adopting $r(k_{\star}) = 0.1$, $r(k_{\star}) = 0.01$ and $r(k_{\star}) = 0.001$, respectively.}
    \label{fig:standard_inflation}
\end{figure}

In the SFSR paradigm, 
for $f\,\gsim\, 10^{-4}$~Hz (and therefore at the frequencies relevant for ground-based detectors, as well as for LISA and even PTA) the GWB produced by the amplification of quantum vacuum fluctuations can then be written as
\be\label{4OgwcompactT}
h_0^2\Omega_{\rm gw} (f)\simeq 1.44 \times 10^{-16}\, \(\frac{r(k_*)}{0.1}\)\, 
\(\frac{A_S(k_*)}{2.14\times 10^{-9}}\) \(\frac{f}{f_*}\)^{n_T}\,
  \(\frac{106.75}{g_*(T_k)}\)^{1/3}
\, ,
\ee
(see, e.g., eq.~(21.357) of \cite{Maggiore:2018sht}).\footnote{The expression of $h_0^2\Omega_{\rm gw} (f)$ for 
$f_{\rm eq}\lsim f \lsim 10^{-4}\, {\rm Hz}$, where $f_{\rm eq}\simeq 3\times 10^{-17}\, {\rm Hz}$,  is also quite similar,  see 
eq.~(19.291) of \cite{Maggiore:2018sht}.}
CMB observations  constrain the amplitude of scalar perturbations to $A_S(k_*)\simeq 2.14\times 10^{-9}$ 
at a pivot scale $k_*=0.05\, {\rm Mpc}^{-1}$, and put an upper bound on $r(k_*)$ of about 0.03, when the consistency relation $r=-8n_t$ is assumed~\cite{Tristram:2021tvh}(see also~\cite{Galloni:2022mok} for a constrain at $k_*=0.01$). If the consistency relation is not assumed the bounds are less stringent with $r_{0.01}< 0.066$ and $-0.76 < n_{T} < 0.52$ at $95\%$CL~\cite{Planck:2018jri}. The frequency $f_*$ corresponding to the pivot scale $k_*$ is extremely low, compared to the frequencies explored by ground-based detectors. For instance, a pivot scale $k_*=0.002\, {\rm Mpc}^{-1}$ corresponds to 
$f_*\simeq 3\times 10^{-18}\, {\rm Hz}$. Therefore, in \eq{4OgwcompactT},  at the frequencies of ground-based detectors (or of LISA), in the term $(f/f_*)^{n_T}$,
the factor $f/f_*$ is very large, and 
the prediction for $h_0^2\Omega_{\rm gw} (f)$ at frequencies relevant for ET depends crucially on the sign of $n_T$.  For single-field slow-roll models of inflation, $n_T=-r(k_*)/8$, so the tilt is negative (and small in absolute value, given the observational bound $r(k_*)<0.1$). Then,  at the frequencies relevant for ET, $h_0^2\Omega_{\rm gw} (f)$ is below $10^{-16}$, and is not detectable by 3G detectors, see figure~\ref{fig:standard_inflation}.

\paragraph{Multi-field scenarios.} 
\label{subsubsec: inflation_multifieldScenarios}

Remarkably, there exist a non-trivial number of multi-field inflationary scenarios supporting a detectable signal at interferometer scales. Amongst these, a promising class of models that has gained considerable attention is axion inflation (see \cite{Pajer:2013fsa} for a review). In these models the axion potential has an (approximate) shift symmetry protecting its light mass from large quantum fluctuations. When the axion-like particle plays the role of the inflaton the symmetries of the theory addresses directly the $\eta$-problem. Well studied single-field realisation of axion inflation (such as natural inflation~\cite{Freese:1990rb}) have been recently ruled out \cite{BICEP:2021xfz}. Remarkably, multi-field realizations exist~\cite{Barnaby:2010vf, Sorbo:2011rz,Anber:2012du,Garretson:1992vt, Turner:1987bw,Anber:2006xt,Cook:2011hg,Adshead:2012kp,Dimastrogiovanni:2012ew,Bartolo:2012sd,Adshead:2016omu,Dimastrogiovanni:2016fuu,Agrawal:2017awz,Caldwell:2017chz,Thorne:2017jft,Dimastrogiovanni:2018xnn,Fujita:2018vmv,Domcke:2018rvv,Lozanov:2018kpk,Watanabe:2020ctz,Holland:2020jdh,Domcke:2020zez,Iarygina:2023mtj,Dimastrogiovanni:2023oid,Ishiwata:2021yne,Durrer:2024ibi,Dimastrogiovanni:2024xvc} with intriguing gravitational wave phenomenology and the added feature of a sub-Planckian axion decay constant, a very desirable feature from the top-down perspective~\cite{Baumann:2014nda}. In multi-field constructions the axion is typically coupled to a (hidden) gauge sector via a Chern-Simons  term. Note that models with multiple axion-like particles (ALPs) have also been explored \cite{Kim:2004rp,Dimastrogiovanni:2023juq}, with ALPs populating also spectator sectors of the inflationary Lagrangian. The GW signatures of theories rich in spectator fields are very similar to those of their more minimal counterparts.

Let us consider the example of a pseudo-scalar inflaton field $\phi$ and $\mathcal{N}$ gauge fields $A^I_{\mu}$ (see e.g.~\cite{Barnaby:2010vf,Sorbo:2011rz,Anber:2012du,Garretson:1992vt,Turner:1987bw,Anber:2006xt,Maleknejad:2011jw,Maleknejad:2012fw}):
\begin{equation}
    \mathcal{L} = -\frac{1}{2}\partial_{\mu}\phi\partial^{\mu}\phi - \frac{1}{4}F_{\mu\nu}^I F^{\mu\nu}_I - V(\phi) - \frac{\lambda^I}{4 f}\phi F_{\mu\nu}^I \tilde{F}_I^{\mu\nu}\,,
    \label{mainaxion}
\end{equation}
with scalar potential $V$ typically of the natural inflation type $V(\phi)= \Lambda^4 [1+\cos\left(\phi/f \right)]$,
and where $F^{I\,a}_{\mu \nu} = \partial_{\mu}A_{\nu}^{I\,a} - \partial_{\nu}A_{\mu}^{I\,a}- g\,\epsilon^{abc} A_{\mu}^{I\,b} A_{\nu}^{I\,c}$ is the field strength tensor. Notice that the gauge group index ``a'' is suppressed in eq.~(\ref{mainaxion}). The field strength reduces to the Abelian case for vanishing $g$. The dual field strength tensor is $\tilde{F}_I^{\mu \nu} = \epsilon^{\mu\nu \alpha\beta} F^I_{\alpha\beta}/(2\sqrt{-g})$ and $f$ is the mass scale suppressing higher dimensional operators, whilst $\lambda$ parameterizes the strength of the Chern-Simons coupling. This setup leads to exponential gauge boson production, with potentially detectable GWs and possibly amplified scalar perturbation sourced by the gauge field. 

Common to Abelian and non-Abelian scenarios is the following dynamics: the axion-inflaton rolls down its potential and dissipates some of its kinetic energy into the gauge sector. The parity-violating nature of the Chern-Simons term results in only one polarisation of the gauge fields being amplified when sourcing GWs (and scalar fluctuations). Such a sourcing mechanism may give rise to a blue or bump-like features in the GW and scalar spectra depending also on the inflaton/spectator role played by the axion(s). 

The corresponding GWB can be characterised as~\cite{Barnaby:2010vf, Linde:2012bt, Domcke:2018rvv}

\begin{equation}
    \Omega_{\rm GW} \simeq \begin{cases}
        \frac{1}{12} \Omega_{R,0} \left( \frac{H^2}{\pi^2 M_{\rm Pl}^2} \right) 
        \left( 1 + 4.3 \times 10^{-7} \frac{H^2}{M_{\rm Pl}^2 \xi^6} e^{4\pi\xi} \right), & \text{for } U(1) \\[10pt]
        \frac{1}{12} \Omega_{R,0} \left( \frac{H^2}{\pi^2 M_{\rm Pl}^2} \right)_{\xi = \xi_{\text{cr}}}
        \left[ 1 + \frac{1}{2} \xi_{\text{cr}}^6 \left( \frac{2^{7/4} H}{g\sqrt{\xi}} e^{(2-\sqrt{2})\pi\xi} \right)^2_{\xi = \xi_{\text{ref}}} \right], & \text{for } SU(2)
    \end{cases}
    \label{eq: GW_Eq_AxInf}
\end{equation}
where $M_{\rm Pl}$ is the reduced Planck mass, the radiation energy density today $\Omega_{R,0} = 8.4\times 10^{-5}$ and the parameter $\xi$ is proportional to the inflaton velocity
\begin{equation}
    \xi \equiv \frac{\lambda \dot{\phi}}{2f H}~.
    \label{eq: xi_def}
\end{equation}
where $\xi_{\text{cr}} = \xi(x=1)$ and $\xi_{\text{ref}} = \xi\left(x=(2+\sqrt{2})\xi_{\text{cr}}\right)$, with $x = -k \tau$ for conformal time $\eta$. The result for the $SU(2)$ case is valid when $0.008 e^{2.8 \xi_{\text{cr}}} \gtrsim 
1/g$.

The exponential sensitivity of the gravitational wave spectrum is a key feature that allows for promising detection prospects. The GWB in eq.~(\ref{eq: GW_Eq_AxInf}) is expected to be also non-Gaussian and fully chiral for large $\xi$~\cite{Sorbo:2011rz, Domcke:2018rvv,Maleknejad:2016qjz}, two features that can assist in the separation of sources in the GW data.

  \begin{figure}[t]
    \centering
    \includegraphics[width=0.7\textwidth]{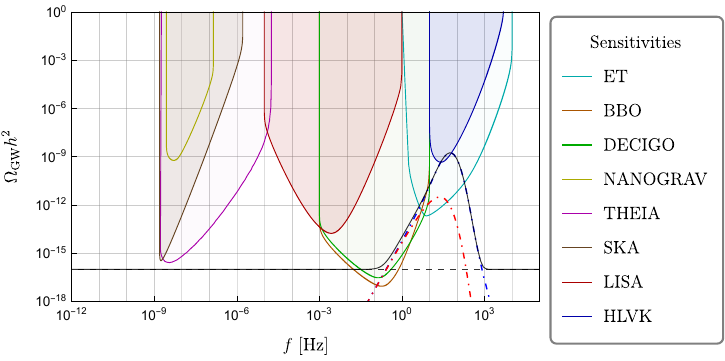}
    \caption{The GW density $\Omega_{\rm GW}$ obtained in the case of a spectator axion superimposed with the sensitivity curves of GW detectors, including ET. The blue (red) dashed line corresponds to the strongly (weakly) amplified polarization. The continuous black line gives the total signal. }
    \label{fig:GW_spectator_Abelian}
\end{figure}

\begin{figure}[t]
    \centering
    \includegraphics[width=0.75\textwidth]{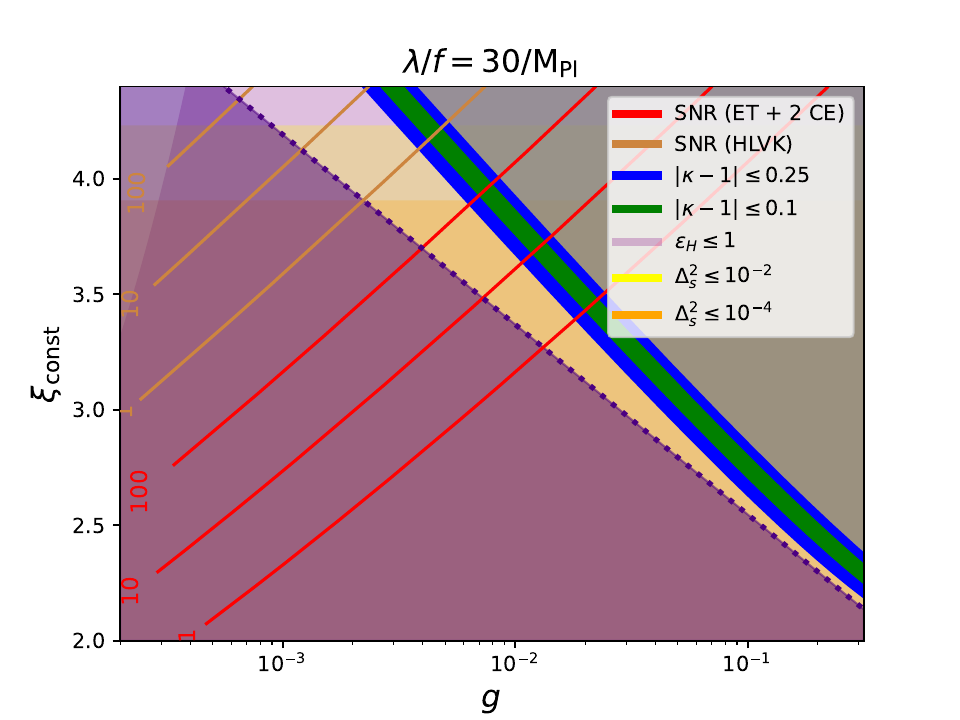}
    \caption{Backreaction and SNR prospects as a function of the gauge coupling constant $g$ and the velocity parameter $\xi(f) = \xi_{\rm{const}}$ for two values of the coupling constant, $\lambda /f = 30/M_{\rm Pl}$. The Hubble rate on the CMB scale is assumed to be $H_{\rm{CMB}} = 2.1 \times 10^{-5}$ $M_{\rm Pl}$. The brown and red lines are the SNR contours calculated assuming 1 year of observation 
    with the design LVK and ET + 2 CE network  sensitivities, respectively. The blue and green bands represent $|\kappa -1|\leq 0.25$ and $0.1$, where we expect that the PBH overproduction bound could be relaxed in the strong backreaction regime. The light purple region represents the slow-roll parameter satisfying $\epsilon_H \leq 1$, where inflation is still ongoing. Yellow and orange regions correspond to the parameter space where the primordial curvature spectrum stays below the PBH upper limit assuming Gaussian ($\mathcal{P}_{\mathcal{R}} \leq 10^{-2}$) and $\chi^2$  ($\mathcal{P}_{\mathcal{R}} \leq 10^{-4}$) statistics. Additionally, we shade the Abelian regime in magenta (below the black dotted line) and the strong backreaction regime in gray (upper region of the $|\kappa -1|$ band), in both of which the estimation of the SNR is not reliable.}
    \label{fig:SU2_Gen_Study}
\end{figure}

\begin{figure}[t!]
    \centering
    \includegraphics[width=\linewidth]{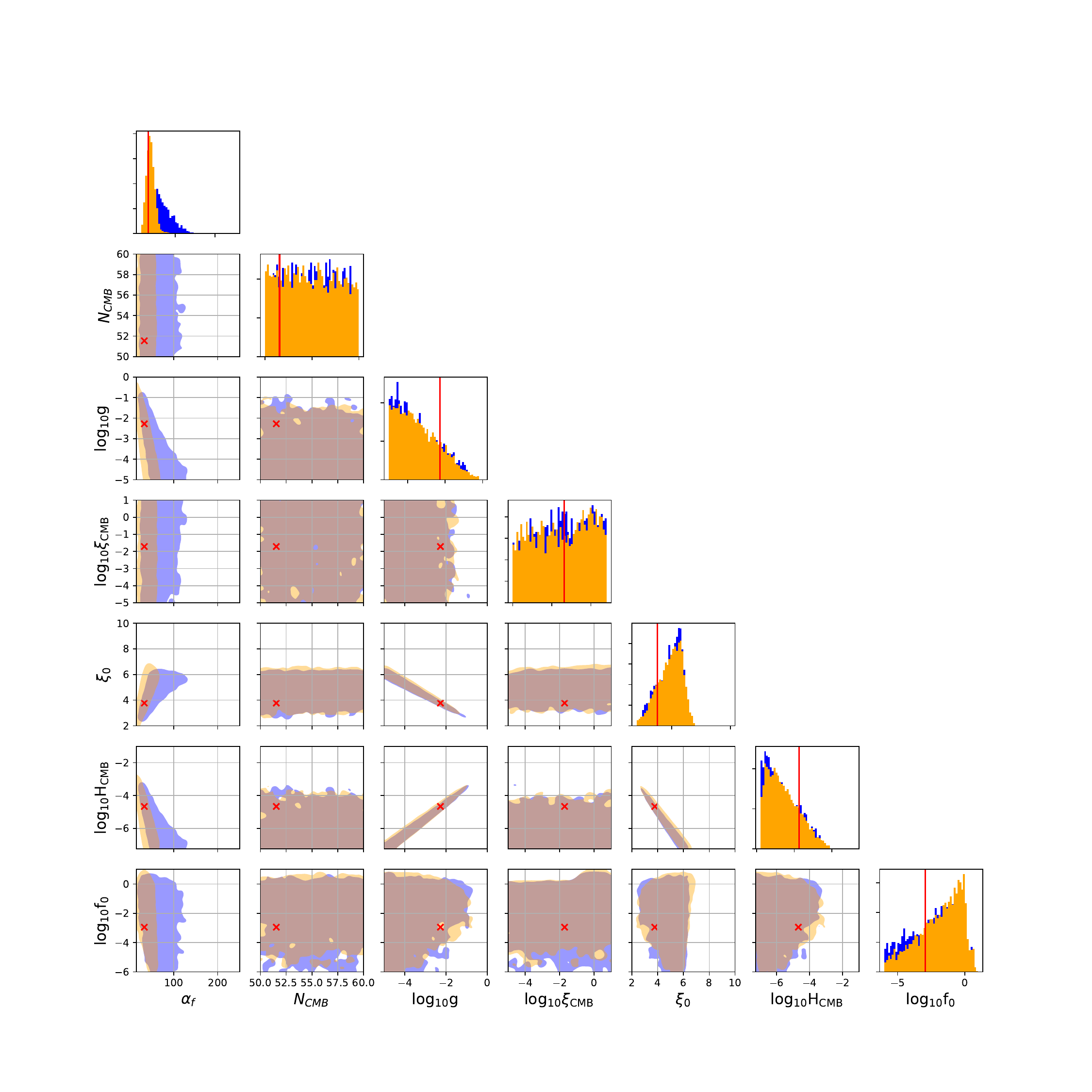}
    \vspace{-16mm}
    \caption{Posterior distributions from Bayesian inference assuming a piecewise, toy model potential, with inflaton velocity $\xi_{\rm CMB}$ before pivot frequency $f_0$, and velocity $\xi_0$ after. This search is done assuming $\chi^2$ statistics. The shaded regions correspond to the $95 \%$ CL.
    The different colors used are: blue for ET and orange for a network consisting of ET+2CE. The red cross and line correspond to the injected values.}
    \label{fig:SU2_PE_prospects_ChiSq}
\end{figure}


The ability of gauge fields to significantly amplify GW notwithstanding, the GW observational prospects can be limited, for Abelian models, by the risk of overproducing scalar modes~\cite{Garcia-Bellido:2016dkw, Linde:2012bt}. The non-observation of primordial black holes (PBH) places upper bounds on the scalar power spectrum. PBHs are produced when large curvature fluctuations re-enter the horizon during the radiation epoch. These PBHs must satisfy constraints on their abundance, derived from cosmological and astrophysical data~\cite{Carr:2009jm, Carr:2020gox}. The statistics of the comoving curvature perturbation plays a key role in this upper bound~\cite{Garcia-Bellido:2016dkw, Linde:2012bt}: a Gaussian assumption places an approximate upper bound of $\mathcal{P}_{\mathcal{R}} \leq 10^{-2}$, whereas an assumed $\chi^2$-statistics places a tighter upper bound of $\mathcal{P}_{\mathcal{R}} \leq 10^{-4}$.

Enforcing PBH constraints may result in a GWB sourced by axion inflation that is undetectable by ET. There are however several models that can generate a detectable GW signal all the while satisfying PBH bounds. For example, $\mathcal{N} > 1$ coupled $U(1)$ gauge fields can suppress the scalar power spectrum by $1/\mathcal{N}$ in the $\xi \gg 1$ regime. Such an extension was shown to create a working linear potential for $\mathcal{N} = 6$ couplings and a working quadratic model for $\mathcal{N} = 10$ couplings~\cite{Garcia-Bellido:2016dkw,Badger:2021enh}. Recent lattice simulations of axion $U(1)$ inflation assuming a quadratic potential showed that strong backreaction close to the end of inflation can suppress non-Gaussian behavior in comoving curvature perturbation, softening the PBH bound in ground detector frequency regime~\cite{Caravano:2022epk} (see also \cite{Figueroa:2023oxc} for lattice studies on Abelian models). This relaxed bound would allow this model to be detectable by GW detectors, including ET, LISA and even advanced LIGO. Extending axion inflation models with an additional pseudoscalar field different from the inflaton can provide a localised bump in the tensor and scalar spectrum. It can also create detectable GW signal that conform to PBH and CMB related bounds~\cite{Namba:2015gja,Garcia-Bellido:2016dkw}, see figure~\ref{fig:GW_spectator_Abelian}.

The GW sourcing mechanism for non-Abelian models is rather different:  GWs are sourced already at the linear level by tensor degrees of freedom in the gauge sector \cite{Adshead:2012kp}. The sourcing mechanism for scalars being less tied to the fate of tensors amplification, PBH bounds are typically less stringent in these scenarios. 

A PBH constraint in non-Abelian field contexts can limit, but does not exclude GWB detection prospects~\cite{Dimastrogiovanni:2024xvc,Badger:2024ekb}. A general parameter space study performed in ~\cite{Badger:2024ekb} found that assuming,  conservatively, a constant inflaton velocity,\footnote{Inflaton velocity $\xi$ generally increases exponentially during inflation.} a GWB from non-Abelian fields can be 3G detectable - even with more constraining power in the $\chi^2$-statistical assumptions, as reported in figure~\ref{fig:SU2_Gen_Study}. An injection of a GWB sourced by a piece-wise scalar potential analysed using a triangular ET detector and a ET+2CE network yielded a logBayes factor for an GWB versus noise greater than 79, a confident detection.  The posterior distribution of the non-Abelian and of the cosmological parameters is shown in figure~\ref{fig:SU2_PE_prospects_ChiSq} for different detector configurations.

\paragraph{Pre-big-bang cosmology.}


\begin{figure}[t]
    \centering
    \includegraphics[width=.8\textwidth]{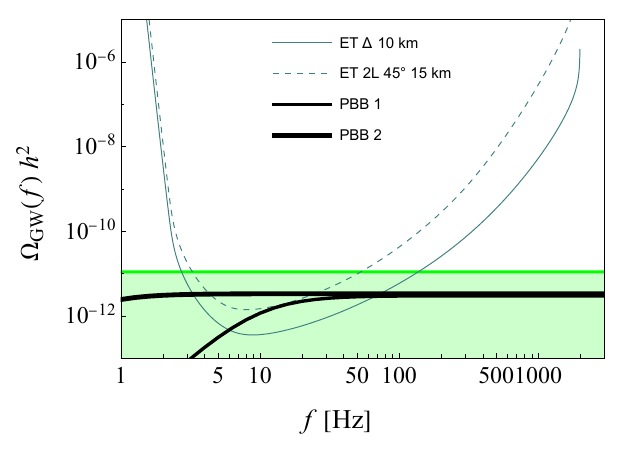}
    \caption{GWB predicted by pre-big-bang cosmology, compared with the PLS of the considered ET configurations. Here we assume an observation time of one year and $\text{SNR} = 1$. The shaded green area is the range of the GWB allowed by a set of consistency conditions on the parameter space of the model \cite{Ben-Dayan:2024aec}. The GWB spectral shape is computed assuming $z_s=5\times 10^6$, $z_d=10^3$, $z_\sigma=2.2$ (``PBB 1'' curve, broken power law within the ET sensitivity window) and $z_s=10^8$, $z_d=10^3$, $z_\sigma=2.2$ (``PBB 2'' curve, almost flat plateau within the ET sensitivity window). Here $z_s=\eta_s/\eta_1$, $z_d=\eta_d/\eta_1$ and $z_\sigma=\eta_\sigma/\eta_1$ are ratios of conformal-time scales marking four different regimes: de Sitter evolution for $-\eta_s<\eta<-\eta_1$, radiation domination for $-\eta_1<\eta<\eta_\sigma$, matter domination for $\eta_\sigma<\eta<\eta_d$ and another radiation-domination phase for $\eta>\eta_d$. \label{fig:relic_gravitons}}
\end{figure}

As we have seen in the previous subsection, the inflationary prediction from single-field slow-roll inflation is in general too small to be detectable by third-generation detectors. From \eq{4OgwcompactT} we see that the key to having a detectable spectrum is to have $n_T>0$ (i.e., a ``blue" spectrum), at least up to the ET frequencies. Eventually, the slope will have to flatten until it reaches a cutoff, in order not to violate upper bounds on $h_0^2\Omega_{\rm gw} (f)$, such as the Big-Bang Nucleosynthesis (BBN) bound, that requires that the integral of $h_0^2\Omega_{\rm gw} (f)$ over $d\log f$ will be smaller than $\mathcal{O} \left( 10^{-6} \right)$  \cite{Pagano:2015hma}.

A model that displays such a behavior is the so-called ``pre-big-bang" cosmology~\cite{Gasperini:1992em,Gasperini:2002bn}, which emerges quite naturally from the low-energy effective action of string theory, and predicts a blue tensor spectrum~\cite{Brustein:1995ah,Buonanno:1996xc}. 
The GWB of this model roughly has a bell-shaped form characterized by a maximal flat plateau whose length in frequency can vary from zero (maximum peak) to a wide interval, depending on the value of the theoretical parameters \cite{Ben-Dayan:2024aec}. The portion of this shape falling into the ET frequency range is either a flat signal (the plateau at maximal amplitude) or a broken power law; see figure~\ref{fig:relic_gravitons}. In the latter case, the GWB raises as $f^3$ at low frequencies and then, at some frequency $f_s$, it changes slope to $f^0$; $f_s$ must be taken as a free parameter, since it depends on the details of the cosmological evolution in the high-curvature phase of the model~\cite{Gasperini:1996fu,Foffa:1999dv}. The spectrum then drops down at higher frequencies not shown in the plot.

\subsubsection{GWs from phase transitions} 
\label{sec:phasetransitions}


One of the possibly strongest sources of GWs from the early Universe is a cosmological first-order phase transition \cite{Kamionkowski:1993fg} that could lead to large observable signals at next-generation GW observatories.  The peak frequency of the signal is fixed by the temperature of the Universe at the time of the phase transition.
Particularly interesting is the case of the electroweak phase transition that would give rise, if it is first-order, to a GW signal in the milliHertz range observable at LISA~\cite{Maggiore:1999vm,Apreda:2001us,Grojean:2006bp,Caprini:2015zlo,Caprini:2019egz,Caprini:2024hue}. ET would be a window on phase transitions happening much earlier, and would therefore probe particle physics that is well beyond the reach of particle colliders.

While the Standard Model of high energy physics does not exhibit any first order phase transitions, numerous Beyond Standard Model (BSM) scenarios do. 
Quantum field theory and string theory typically feature a complicated vacuum manifold with several coexisting minima. The unknown evolution of the early Universe might have populated, possibly via the mechanism of symmetry restoration, false vacua and lead to a series of phase transitions (PTs) between the minima of the manifold. 
The larger the latent heat released during the transition, the stronger is the resulting GW signal. 
In this context, first-order PTs (FOPTs) proceed via the stochastic nucleation of spherical bubbles of the new phase in the old phase, with the probability of nucleation per unit of time and volume controlled by stochastic processes 
\begin{equation} 
\mathcal{P}_{\rm nucleation} \propto A(t)e^{-\mathcal{S}_c(t)}\,,
\end{equation}
where $\mathcal{S}_c(t)$ is the euclidean action of the critical bubble. The bubbles subsequently grow and fill the space until the total conversion of the plasma to the true minimum. In this process they produce GWs that can be observed at ET. 
There are two crucial steps on mapping the BSM physics parameters (couplings, mass scales) to the induced GW signal. The first step consists in mapping the  BSM parameters to the thermal description in terms of energy budget parameters.
The energy released during the PT can be  quantified by the ratio between the energy stored in the vacuum and in the plasma at the moment of the transition~\cite{Caprini:2015zlo,Caprini:2019egz,Ellis:2018mja}
 \begin{equation}
\alpha(T) \equiv \frac{\Delta V(T)-T \frac{\partial \Delta V}{\partial T}}{\rho_{\text{rad}}}\qquad \text{(Strength parameter)},
\label{eq:alpha}
\end{equation} 
which has to be evaluated at the percolation temperature.
A second important parameter is the duration of the phase transition. The duration from the nucleation to the percolation can be approximated by expanding the rate of nucleation at first order and leads to the definition
\begin{equation}
\tilde{\beta} \equiv \frac{\beta}{H}= -\frac{d(\mathcal{S}_c/T)}{H dt}  \qquad \text{(Duration of the transition)} \,. 
\label{eq:duration_phase_t}
\end{equation}
The normalised speed of nucleation $\tilde{\beta} \propto t_{\text{expansion}}/t_{\text{transition}}$, with $t_{\text{transition}}$ the typical time the transition takes to complete and $t_{\text{expansion}}$ the Hubble time, measures how fast a bubble nucleates with respect to the expansion of the Universe, giving an estimate of the speed of completion of the transition. This value can be related to the size of the bubbles at collision via \cite{Enqvist1992NucleationAB,Hindmarsh:2017gnf,Caprini:2019egz}
\begin{equation}
R_{\rm coll} = \frac{(8\pi)^{1/3}}{H_{\rm reh}\beta} \, .
\end{equation}
Note that in the case in which the phase transition is seeded by topological defects~\cite{Witten:1984rs,Steinhardt:1981mm,Steinhardt:1981ec}, the duration of the transition and the bubble size are instead related to the number of defects per Hubble volume, affecting also the GW signal~\cite{Blasi:2022woz,Agrawal:2022hnf,Blasi:2023rqi,Agrawal:2023cgp,Blasi:2024mtc}.

 Assuming that the transition occurs in less than an Hubble time, we can apply conservation of energy arguments to obtain the \emph{reheating temperature} $T_{\rm reh}$:
\begin{equation} 
(1 - \Omega_{\rm GW}) \big(\Delta V + \rho_{\rm rad}\big|_{T= T_{\rm per}}\big) = \rho_{\rm rad}\big|_{T= T_{\rm reh}} \qquad \Rightarrow T_{\rm reh} \approx (1+\alpha)^{1/4} T_{\rm nuc}\,,
\end{equation} 
where we neglected the energy lost in the GW signal. 

The terminal velocity of the bubble wall $v_w$ of the bubble has been the subject of intense investigation, as it requires to solve a very non-trivial set of coupled equations. Estimates of the terminal wall velocity are usually based on kinetic theory~\cite{Dine:1992wr,Liu:1992tn,Moore:1995ua,Moore:1995si}, improvement of it~\cite{Cline:2020jre, Cline:2021iff,Laurent:2022jrs}, or holographic methods for strongly coupled theories~\cite{Bea:2021zsu, Baldes:2020kam,Bigazzi:2020avc,Bigazzi:2021ucw,Bea:2022mfb} (see ref.~\cite{Kang:2024xqk} for a quasiparticle (quasigluons) method). Typically, calculating the terminal wall velocity is a very challenging task, and has only been performed for a  limited number of models~\cite{Liu:1992tn,Moore:1995si,Moore:1995ua,Dorsch:2018pat,Laurent:2022jrs,Jiang:2022btc}. In such computations, one needs to solve the scalar field equation of motion (EoM) coupled to the Boltzmann equations describing the particles in the plasma. Simplifications can be obtained by assuming local thermal equilibrium (LTE)~\cite{Konstandin:2010dm, BarrosoMancha:2020fay,Balaji:2020yrx, Ai:2021kak,Wang:2022txy,Ai:2023see,krajewski2023hydrodynamical, Sanchez-Garitaonandia:2023zqz} or in the ultrarelativistic regime~\cite{Bodeker:2009qy, Bodeker:2017cim,Hoche:2020ysm, Azatov:2020ufh,Gouttenoire:2021kjv, Ai:2023suz,Azatov:2023xem}.
 Schematically, this mapping can be written  
\begin{equation}
 \boxed{\text{BSM parameters} ({\tt couplings}, {\tt masses} ... )\Rightarrow (\alpha, \beta, T_{\rm reh}, v_w) \,. } 
\end{equation}
The second matching consists of the prediction of the features of the GW signal, namely, the shape of GW spectrum, the peak amplitude $\Omega_{\rm GW,peak}$, and the peak frequency $f_{\rm peak}$, from a set of $(\alpha, \beta, T_{\rm reh}, v_w)$. Schematically 

\begin{table}[t]
\begin{center}
\begin{tabular}{lll}
\hline
\multicolumn{1}{|l|}{Reference and group} & \multicolumn{1}{l|}{IR tail} & \multicolumn{1}{l|}{UV tail}  \\ \hline
\multicolumn{1}{|l|}{Envelope simulation\cite{Huber:2008hg} $\Omega_\phi$} & \multicolumn{1}{l|}{$ 3$} & \multicolumn{1}{l|}{$ 1$}\\
\multicolumn{1}{|l|}{Envelope simulation\cite{Konstandin:2017sat} $\Omega_\phi$} & \multicolumn{1}{l|}{$ 2.9^r-2.95$} & \multicolumn{1}{l|}{$ 0.9-1$}\\
\multicolumn{1}{|l|}{Envelope analytic\cite{Jinno:2016vai, Jinno:2017fby} $\Omega_\phi$} & \multicolumn{1}{l|}{$ 3$} & \multicolumn{1}{l|}{$ 1$}\\
\multicolumn{1}{|l|}{Bulk flow\cite{Konstandin:2017sat} $\Omega_{\phi}$} & \multicolumn{1}{l|}{$ 0.95-0.9^r$} & \multicolumn{1}{l|}{$ 2.9-2.1^r$}\\
\multicolumn{1}{|l|}{Lattice Simulation \cite{Cutting:2018tjt} $\Omega_{\phi}$} & \multicolumn{1}{l|}{$ 3$} & \multicolumn{1}{l|}{$\sim  1.5$}
\\
\multicolumn{1}{|l|}{Lattice Simulation\cite{Cutting:2020nla} $\Omega_{\phi}$} & \multicolumn{1}{l|}{$ 0.68-1.56$} & \multicolumn{1}{l|}{$ 1.35-2.25$}\\
 \hline
\multicolumn{1}{|l|}{Sound Shell model\cite{Hindmarsh:2016lnk,Hindmarsh:2019phv} $\Omega_{sw}$} & \multicolumn{1}{l|}{$ 9$} & \multicolumn{1}{l|}{$4$}
\\
\multicolumn{1}{|l|}{Analytic\cite{Jinno:2017fby} $\Omega_{sw}$} & \multicolumn{1}{l|}{$ 3$} & \multicolumn{1}{l|}{$ 1$}\\
\multicolumn{1}{|l|}{Hybrid simulation\cite{Jinno:2020eqg} $\Omega_{sw}$} & \multicolumn{1}{l|}{$ 3$} & \multicolumn{1}{l|}{$ 3$}\\
\multicolumn{1}{|l|}{Lattice Simulation\cite{Hindmarsh:2017gnf} $\Omega_{sw}$} & \multicolumn{1}{l|}{$ 3$} & \multicolumn{1}{l|}{$ 4$}
\\ \hline
\multicolumn{1}{|l|}{Analytic and simulations\cite{Jinno:2022fom} $\Omega_{\rm FS}$} & \multicolumn{1}{l|}{$ 1$} & \multicolumn{1}{l|}{$ 2$}\\
\hline
\end{tabular}
\end{center}
\caption{Shape of the GW signals from  some of the different groups working on GW from FOPT in the UV and the IR. Results are shown under the form of $(f/f_{p,0})^{n}$ for the IR and $(f/f_{p,0})^{-n}$ for the UV tail. The suffix $r$ means that the result is specific to relativistic walls. The mismatch between numerical simulations and the current models available (Sound shell and Bulk flow) motivates us to investigate further the analytic understanding of GW from bubble collision.
\label{Table:shapespectrum}}
\end{table}

\begin{equation}
    \boxed{
 (\alpha, \beta, T_{\rm reh}, v_w) \Rightarrow (f_{\rm peak}, \Omega_{ \text{peak}}, \text{GW spectrum})\,.} 
\end{equation} 
 The spectrum of the GW signal depends strongly on the source and requires to be simulated.  The process of expansion and collision of those bubbles of true phase will generate GWs via three distinct mechanisms: i) from the \textit{bubble collision}, where the energy collected in the scalar shell on the bubble wall is released at collision~\cite{Huber:2008hg, Konstandin:2017sat, Jinno:2016vai, Jinno:2017fby, Cutting:2018tjt, Cutting:2020nla}, ii) from the \textit{sound waves} where the energy injected in the plasma propagates under the form of sound waves~\cite{Hindmarsh:2016lnk,Hindmarsh:2019phv,Jinno:2019jhi, Cutting:2019zws, Hindmarsh:2017gnf,Cutting:2019zws,Jinno:2020eqg,Jinno:2022mie}, iii) from the \textit{turbulence} occurring when the sound waves decay~\cite{RoperPol:2019wvy}
(see also \cite{Jinno:2022fom} for another source of GWs).
It is typically found to be well fitted by a \textit{broken power-law} of the type 
\begin{equation}\label{OmegaBPLdiv2}
\Omega_{\rm GW}(f) \propto \frac{(a+b)^c f_{\rm peak}^b f^a}{\big(b f_{\rm peak}^{\frac{a+b}{c}}+a f^{\frac{a+b}{c}}\big)^c}\,,
\end{equation}
with parameters $a, b,c$ given by simulations. 
In table \ref{Table:shapespectrum} we report the slope in the UV and in the IR for different simulations for the $\Omega_{\phi}$ (bubble collision source), the $\Omega_{sw}$ sound wave sources and from the free-streaming source $\Omega_{\rm FS}$. Even if it seems that the spectra from the sound waves and the bubble collisions are very different for small $\alpha$, it was recently argued that the spectra from the sound-waves in the regime of large $\alpha$ resembles the spectrum of the bubble collision~\cite{Lewicki:2022pdb}.

\begin{figure}[t!]
    \centering
    \includegraphics[width=0.6\textwidth]{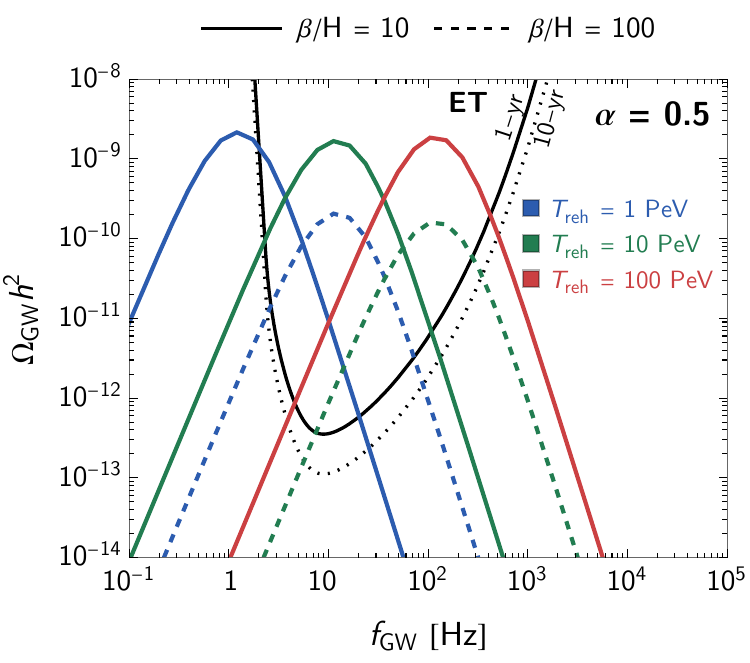}
    \caption{GW spectra from sound waves produced by first-order FOPTs at temperature $T_{\rm reh}$ \cite{Caprini:2019egz}, with the wall velocity that gives the maximal energy transfer to sound waves~\cite{Espinosa:2010hh, Gouttenoire:2022gwi}. The black curves are the power-law integrated sensitivities of ET assuming  a triangular xylophone
configuration with 10km arms and the 1 and 10 years observation time with ${\rm SNR}=1$.}
    \label{fig:pt_spectrum}
\end{figure}

\begin{figure}[th]
    \centering
    \includegraphics[width=0.495\textwidth]{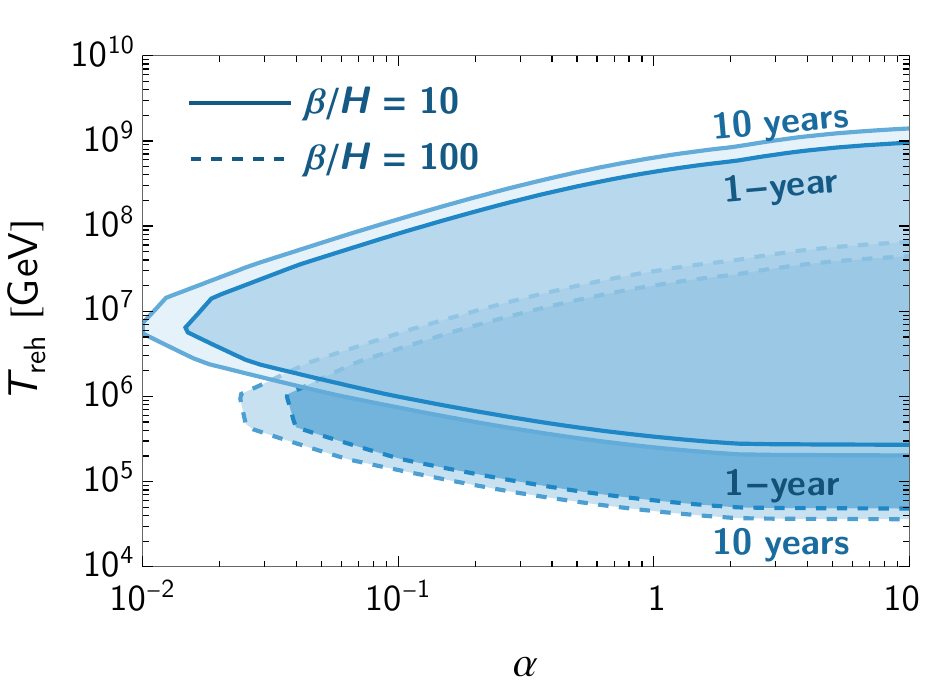}\hfill
    \includegraphics[width=0.495\textwidth]{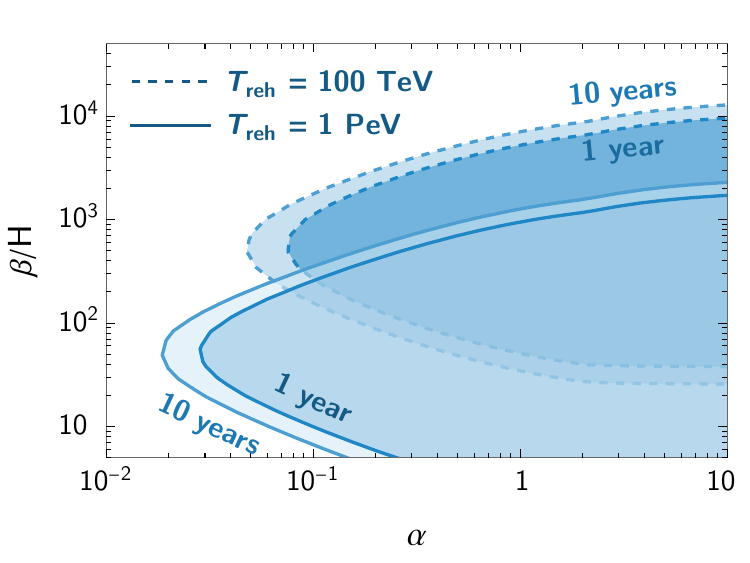}\\[-1em]
    \caption{Regions of FOPT parameter spaces that can be probed by ET assuming triangular xylophone
configuration with 10km arms for two values of $\beta/H$ (left) and $T_{\rm reh}$ (right), assuming GW from sound waves \cite{Caprini:2019egz} shown in figure~\ref{fig:pt_spectrum}. The darker regions can be probed with ET at SNR $\geq 1$ and 1 year of observation time, while the lighter regions require 10 years.}
    \label{fig:pt_contour}
\end{figure}

The typical frequency peak $f_{\rm peak}$ is controlled by the temperature $T_{\rm reh}$ after the completion of the PT and the size of bubbles at collision $R_{\rm coll}$:
\begin{equation}
\label{eq:f_peak}
f_{\rm peak} \approx \mathcal{O}(1) \times 10^{-5} \frac{1}{H_{\rm reh} R_{\rm coll}} \frac{T_{\rm reh}}{100 \text{GeV}} \text{ Hz}\,,
\end{equation}
where $H_{\rm reh}$ is the Hubble constant after the completion of the PT and a $\mathcal{O}(1)$ factor depending on the details of the source. 
Figure~\ref{fig:pt_spectrum} shows the expected spectra from sound waves, while figure~\ref{fig:pt_contour} shows the parameter spaces of FOPT that can be probed by ET.

Testing particles physics model would consist in inverting the steps that we have made so far. Assuming an observed stochastic signal, we have to perform the inverse problem
\begin{equation}
  \boxed{(\text{GW spectrum}, f_{\rm peak}, \Omega_{\rm GW, \text{peak}})  \Rightarrow (\alpha, \beta, T_{\rm reh}, v_w) \Rightarrow \text{BSM parameters} \, .}  
\end{equation} 
Supercooled phase transitions in particular lead to very strong signals (large $\alpha$, small $\beta/H$). This arises in particular naturally in models featuring nearly-conformal dynamics \cite{Iso:2017uuu,vonHarling:2017yew,Baratella:2018pxi}. Such supercooled PTs and their associated signals were studied extensively in the context of new TeV physics \cite{Randall:2006py,Konstandin:2011dr,Bruggisser:2022rdm} but the same physics can be readily applied to strong dynamics occurring at higher energies that could be connected to dark matter and that would lead to signals in the ET window \cite{Baldes:2021aph}.

Going back to eq.~(\ref{eq:f_peak}), as can be seen from figure~\ref{fig:pt_spectrum}, this GW signal would roughly lie in the detection range of the ET detector, $f$ between a few Hz and a few $10^3$ Hz,    if the ratio $(H_{\rm reh} R_{\rm coll})^{-1} ({T_{\rm reh}}/{100 \text{GeV}})$ is between a few times  $10^5$ and a few times $10^8$, i.e. for PTs largely above the EW scale. In this detection range, the LVK collaboration could also constrain large regions of parameter space, specially in the case of models featuring large supercooling, such as the conformal models~\cite{Badger:2022nwo,Romero:2021kby,
Romero-Rodriguez:2021aws},  Pati-Salam models~\cite{Athron:2023aqe},  high energy $B-L$ breaking PT~\cite{Brdar:2018num}, SUSY breaking \cite{Craig:2020jfv,Fornal:2021ovz}, Peccei-Quinn symmetry breaking \cite{DelleRose:2019pgi,VonHarling:2019rgb}, flavour models \cite{Greljo:2019xan,Ringe:2022rjx}, models with extended gauge groups \cite{Huang:2020bbe,Fornal:2022qim,Fornal:2023hri,Bosch:2023spa}, and the remnant of the confinement of a hypothetical composite Higgs \cite{Bruggisser:2018mrt,Agashe:2019lhy, Azatov:2020nbe,Bruggisser:2022rdm} (see also ~\cite{Ellis:2020awk, Ellis:2019oqb, Azatov:2019png} for more general studies). Other physical phenomena related to signals in ET have also been envisioned, such as bubble wall enhanced phenomenon like baryogenesis \cite{Hall:2019ank,Azatov:2021irb, Baldes:2021vyz}, leptogenesis~\cite{Huang:2022vkf,Chun:2023ezg}, Dark Matter production~\cite{Baker:2019ndr,Baldes:2021aph,Azatov:2021ifm,Azatov:2022tii, Fornal:2022qim}, and could be in the reach of detection of ET. This exciting possibility motivates us to strengthen the relation between model building parameters and observables in ET telescope.

\subsubsection{GWs from cosmic strings} \label{subsec:cosmic_strings}

\paragraph{Local and global strings.}
Topological defects of various kinds -- including cosmic strings and domain walls (see section~\ref{sec:domain_walls}) -- may form in symmetry-breaking phase transitions in the early Universe~\cite{Kibble:1976sj}.
In particular, if the vacuum manifold contains non-contractible loops then a network of strings can be produced, and (if they are dynamically or energetically stable) they will still be present in the Universe until today.
Cosmic strings are referred to as \textit{local/global} depending on the nature of the underlying symmetry groups (see~\cite{Vilenkin:2000jqa,Hindmarsh:1994re,Sakellariadou:2006qs,Vachaspati:2015cma} for reviews on cosmic strings).
Local cosmic strings are ubiquitous in supersymmetric grand unified theories: indeed all models which solve the monopole problem, lead to baryogenesis after inflation and are consistent with proton lifetime measurements form local strings~\cite{Jeannerot:2003qv}. 
Global strings on the other hand are characteristic of axion models, and are formed for instance when the Peccei-Quinn symmetry is broken after inflation~\cite{Vilenkin:1982ks, Sikivie:1982qv,Vilenkin:1984ib}.
The global strings in these models can become unstable at a later stage when a subsequent symmetry breaking phase transition leads to the formation of domain walls~\cite{Vilenkin:1982ks,Sikivie:1982qv, Vilenkin:1984ib, Davis:1986xc, Hiramatsu:2010yn, Hiramatsu:2012gg} (see also section \ref{sec:domain_walls}).

Both local and global string networks (and indeed domain-wall networks) can source gravitational waves (GWs), and this production occurs throughout the lifetime of the network, from the initial phase transition to the present day.
As a result, strings produce a GWB  spectrum which generally covers a very broad range of frequencies (see e.g.~\cite{Auclair:2019wcv}), and hence can be probed by experiments in different frequency bands---from the nHz and pulsar timing array (PTA) experiments~\cite{NANOGrav:2023hvm,EPTA:2023xxk,EuropeanPulsarTimingArray:2023lqe}, to the mHz and LISA~\cite{Auclair:2019wcv}, up to the $1\text{--}10^4\,\mathrm{Hz}$ band of ET and other ground-based interferometers.
Furthermore, if a string is sufficiently close to us, the characteristic bursts of GWs it emits~\cite{Damour:2001bk,Siemens:2006vk} can be searched for individually~\cite{LIGOScientific:2021nrg}.
Both type of observations (bursts or GWB) can put constraints on the energy scale of the phase transition in which strings are formed, $\eta$, and can thus reveal the properties of the phase transition and particle physics at that energy scale.

Local strings have a characteristic width $w\sim 1/\eta$ and finite energy per unit length which (in  units $\hbar=c=1$), is given by 
\be
G\mu \sim 10^{-6} \left(\frac{\eta}{10^{16} ~\mathrm{GeV}}\right)^2\, .
\ee
Local strings predominantly lose energy into GWs. In contrast, global strings dominantly radiate Nambu-Goldstone bosons, and 
the existence of this Nambu-Goldstone radiation leads to a time-dependent, logarithmically-divergent energy per unit length
\be
G\mu(t) \sim 10^{-6} \left(\frac{\eta}{10^{16} ~\mathrm{GeV}}\right)^2\,  \log\(\frac{H^{-1}(t)}{w}\)\, ,
\ee
where $H(t)$ is the Hubble parameter.

In the remainder of this section, focusing on the ET frequency band, we summarise the burst and GWB signatures of both local and global strings, taking into account existing observational constraints.
Beyond the minimal cosmic-string models -- depending only on $G\mu$ (or equivalently $\eta$) -- discussed here, there are several other well-motivated cosmic-string models that depend on additional parameters, modifying the string-network dynamics and leading to interesting features on GWB. These include the superstrings~\cite{Sarangi:2002yt,Dvali:2003zj,Copeland:2003bj,Sakellariadou:2004wq,Rajantie:2007hp,Sakellariadou:2008ay}, the metastable cosmic strings \cite{Vilenkin:1982hm, Copeland:1987ht, Preskill:1992ck, Vilenkin:2000jqa, Monin:2008mp, Leblond:2009fq,Dror:2019syi, Buchmuller:2019gfy, Buchmuller:2021mbb,Buchmuller:2023aus,Servant:2023tua,Chitose:2023dam}, Hubble-induced strings~\cite{Bettoni:2018pbl} and superconducting strings~\cite{Witten:1984eb,Nielsen:1987fy,Auclair:2022ylu,Rybak:2022sbo}. Although their GWB and features are relevant for ET, these models are subjected to some uncertainties and will need more numerical studies in order to draw robust predictions. 

\paragraph{Gravitational waves from local strings.}
Due to the vast separation of scales between the length and the width of a cosmic string, it is often convenient to describe their dynamics in the zero-width limit. For local strings, this results in a Nambu-Goto action characterised purely in terms of $G\mu$. Cosmic string networks can also be studied using lattice field theory simulations (see, e.g.,~\cite{Vincent:1996rb,Vincent:1997cx, Hindmarsh:2017qff}). The results from these simulations have sparked a long-standing debate on how to reconcile these two types of simulations. Recent studies however suggest that indeed field theory string loops coming from cosmic string network simulations evolve as Nambu-Goto dictates \cite{Blanco-Pillado:2023sap}. On the other hand, it is clear that more work in this direction is needed to conclusively show the agreement between these two approaches. Here we will concentrate of the predictions from Nambu-Goto strings.

This Nambu-Goto approximation has been used extensively in the literature to investigate the dynamics of local string networks (see, e.g., refs.~\cite{Hindmarsh:1994re,Sakellariadou:1987kc,Vilenkin:2000jqa,Auclair:2019wcv} for reviews).
The picture that emerges is that each Hubble patch contains $\order{1}$ `long' superhorizon strings, which continually produce closed string loops by self-intersecting.
These loops evolve periodically over time, expanding and contracting at relativistic velocities due to their tension.
This relativistic dynamics results in strong GW production from loops, particularly via pronounced features such as cusps (string segments which move instantaneously at the speed of light), kinks (discontinuities in the tangent vector along the loop), and kink-kink collisions~\cite{Vilenkin:1981bx,Vachaspati:1984gt,Damour:2001bk}.
As a result, loops  decay in size and energy via GWs long kinky strings~\cite{Sakellariadou:1990ne} at a rate $\dv*{E}{t}\approx-\Gamma_{\rm GW}G\mu^2$ (with $\Gamma_{\rm GW}\approx50$ a dimensionless coefficient \cite{Vachaspati:1984gt,Burden:1985md,Blanco-Pillado:2017oxo}), radiating at increasingly high frequencies until they vanish from the network.
This decay of individual loops, combined with the continuous production of new loops from the self-intersection of `long' strings, leads to an attractor solution in which the distribution of loop sizes and the total energy contained in the network scale with the expansion of the Universe~\cite{Kibble:1984hp}. This is the so-called {\it scaling solution}. Evidence of this attractor solution has been found in many of the numerical simulations of string networks (see, e.g., \cite{Albrecht:1984xv,Bennett:1987vf,Sakellariadou:1987kc,Allen:1990tv}).

Early attempts to detect local strings focused mostly on their imprints on the cosmic microwave background (CMB) due to the deflection of CMB photons, resulting in constraints of $G\mu\lesssim10^{-7}$~\cite{Planck:2013mgr,Lizarraga:2016onn,Charnock:2016nzm,Torki:2021fvi}.
However, the advent of GW astronomy has enabled much more sensitive searches, and therefore much more stringent constraints on the string tension, as we discuss below.
The main theoretical uncertainty in these searches is the loop size distribution reached by the network in the scaling regime.
The two most well-established models for this distribution -- which we refer to here as model A~\cite{Blanco-Pillado:2013qja} and model B~\cite{Lorenz:2010sm} (based on~\cite{Ringeval:2005kr}) -- predict different abundances of small loops, and therefore different GW signatures, particularly at the higher frequencies relevant to ground-based interferometers such as ET and LIGO/Virgo/KAGRA. For a discussion of the origin of this discrepancy on the loop number density of non-self-intersecting loops computed from different simulations see \cite{Blanco-Pillado:2019vcs,Blanco-Pillado:2019tbi}.

Another important theoretical uncertainty on the computation of the GW signatures of cosmic strings comes from the role played by gravitational back-reaction on the loop's evolution. The emission of gravitational radiation from loops modifies their
shape altering their power spectrum. This effect has been recently studied analytically as well as numerically from a small number of loops \cite{Quashnock:1990wv,Wachter:2016hgi,Blanco-Pillado:2018ael,Chernoff:2018evo,Blanco-Pillado:2019nto}. However, its impact on the final gravitational wave signals coming from strings has only been discussed on simple toy model description in \cite{Blanco-Pillado:2017oxo}.

\paragraph{GWs from global strings.}
The dynamics of global strings can be described effectively as Nambu-Goto strings coupled to the Goldstone-boson field using the Kalb-Ramond action~\cite{Kalb:1974yc,Vilenkin:1986ku,Davis:1988rw}. 
This additional coupling affects GW production in two ways.
First, the evolution of the global-string network is modified by the back-reaction from Goldstone emission~\cite{Klaer:2017qhr,Martins:2018dqg,Klaer:2019fxc} in competition with the loop-chopping; this affects the loop number density.
Second, a global-string loop only lives for a short time after its production because it decays quickly into Goldstone bosons at the rate $\dv*{E}{t}\approx-\Gamma_{\rm gold} \mu/ \log(H^{-1}/w)$ with $\Gamma_{\rm gold} \approx 65$~\cite{Vilenkin:1986ku}, which is $\mathcal{O}(1/(G\mu))$ more efficient than GW emission. Similarly to local strings, long global strings emit Goldstone bosons~\cite{Sakellariadou:1991sd}.
As a result, the GW production from loops is roughly the same as the time of loop production, and is therefore diluted more by Hubble expansion than in the local-string case.
The resulting GWs are highly suppressed compared to those from local networks formed at the same symmetry-breaking scale~\cite{Chang:2019mza,Gouttenoire:2019kij,Chang:2021afa,Gorghetto:2021fsn}.
In the case of axion strings, the spectrum is cut off at frequencies below a threshold set by the axion mass due to the strings decaying.
However, for axion mass below $10^{-22}$ eV and axion decay constant $\sim 10^{15}$ GeV that do not cause dark matter overproduction~\cite{Dabholkar:1989ju,Davis:1986xc,Hiramatsu:2010yn,Hiramatsu:2012gg,Buschmann:2019icd, Gorghetto:2018myk,Gorghetto:2020qws,Gorghetto:2021fsn, Buschmann:2021sdq}, this cutoff is well below the ET frequency band, so can be neglected here.
(This cutoff is relevant however for low-frequency GW observations, e.g. with PTAs~\cite{Gorghetto:2021fsn,Servant:2023mwt}.)

GW predictions for global strings must be treated with care, as there are several unresolved theoretical uncertainties.
Many lattice simulations have attempted to verify whether the global string network actually reaches the scaling regime or has its string correlation length corrected with $\log(H^{-1}/w)$~\cite{Gorghetto:2018myk,Gorghetto:2020qws, Kawasaki:2018bzv, Vaquero:2018tib, Klaer:2017qhr, Klaer:2019fxc, Hindmarsh:2019csc, Figueroa:2020lvo, Buschmann:2019icd, Buschmann:2021sdq, Gorghetto:2021fsn, Hindmarsh:2021vih}. 
Moreover, the recent debate on the powers of emissions into GW, Goldstone bosons, and massive radiation in different numerical simulations is open~\cite{Gorghetto:2021fsn, Drew:2019mzc, Drew:2022iqz, Drew:2023ptp, Baeza-Ballesteros:2023say}, e.g., the amplitude of the GW signal can be smaller by $\log^2(H^{-1}/w)$ than the previous expectation~\cite{Baeza-Ballesteros:2023say}.

In addition to GWs, the massless Goldstone production from global strings is subject to strong constraints on the number of relativistic degrees of freedom present at BBN and CMB scales, $N_\mathrm{eff}$.
Accounting for uncertainties in the global-string network evolution, this constraint bounds the energy scale of the string network formation to be $\eta \lesssim (1 -3) \times 10^{15} $ GeV~\cite{Chang:2021afa,Gorghetto:2021fsn}.
The bound from deflection of CMB photons ($G\mu(t_{\rm CMB}) \lesssim 10^{-7}$) also gives similar constraints~\cite{Chang:2019mza,Gouttenoire:2019kij,Chang:2021afa,Gorghetto:2021fsn}.

\begin{figure}[t!]
    \centering
    \includegraphics[width=\textwidth]{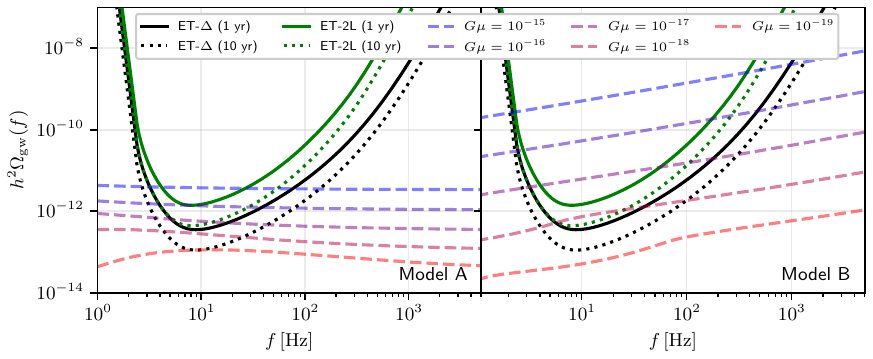}
    \caption{Forecast power-law-integrated GWB sensitivity of ET, compared with signals from local Nambu-Goto strings of various tensions $G\mu$.
    We consider both the triangular  configuration for ET with 10~km arms, and the 2L configuration with 15~km arms, misaligned as in \cite{Branchesi:2023mws}; the solid black curves corresponds to one year of observations, while the dotted black curve corresponds to 10 years.
    The left and right panels show the predictions for models A~\cite{Blanco-Pillado:2013qja} and B~\cite{Lorenz:2010sm} of the loop network, respectively.
    Both models predict that, in the triangle configuration, ET will be sensitive to $G\mu\gtrsim10^{-18}$ after one year of observations with SNR $\geq 1$.}
    \label{fig:cs_stochastic_forecasts}
\end{figure}

\begin{figure}[t!]
    \centering
    \includegraphics[width=0.55\textwidth]{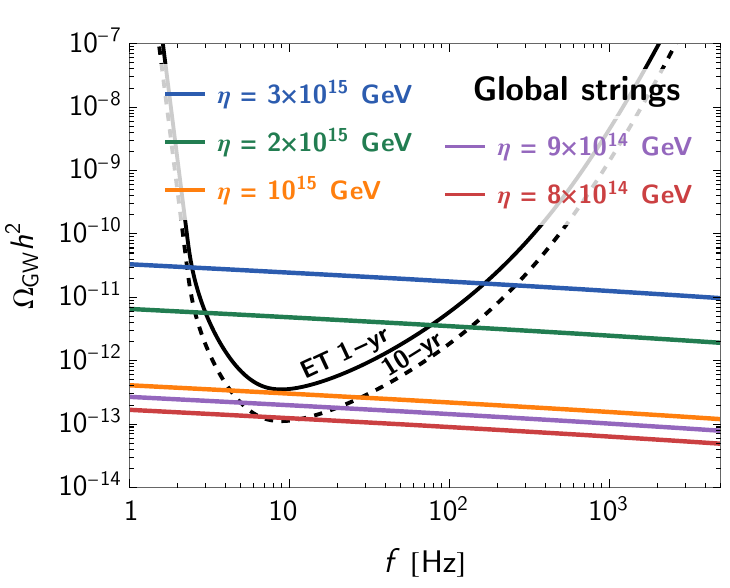}
    \caption{Spectra of GWB from global strings formed at energy scale $\eta$ are shown in solid colored lines. The black lines are the same ET-triangle sensitivity curves as shown in figure~\ref{fig:cs_stochastic_forecasts}. We used the  semi-analytic calculation in~\cite{Gouttenoire:2019kij}; see also~\cite{Chang:2019mza,Chang:2021afa}.}
    \label{fig:cs_global}
\end{figure}

\paragraph{GW background from strings.}

The incoherent superposition of GWs from many string loops gives rise to a GWB signal, whose amplitude and spectral shape are determined by the symmetry-breaking scale $\eta$ (which in the case of local strings corresponds one-to-one with the string tension $G\mu$), as well as by the loop network modelling as discussed above.
As mentioned above, the signal spans many decades in frequency, due to the very broad range of loop sizes and emission epochs involved in the network.
As a result, it is possible to set complementary constraints on string scenarios with multi-band GW observations.
For local strings, results from the LVK O3 stochastic search rule out $G\mu\gtrsim10^{-8}$ for model A and $G\mu\gtrsim4\times10^{-15}$ for model B~\cite{LIGOScientific:2021nrg}, while the most recent results from NANOGrav and EPTA each rule out $G\mu\gtrsim10^{-10}$ for both models A and B~\cite{NANOGrav:2023hvm,EPTA:2023xxk,EuropeanPulsarTimingArray:2023lqe}.
(This assumes that the GWB signal detected by these PTAs is due to inspiralling supermassive binary black holes~\cite{NANOGrav:2023hfp}; several more exotic explanations for this signal have been proposed, including cosmic strings.)

These constraints will improve drastically with next-generation GW observations in the 2030s.
The space-based interferometer mission LISA, which has recently undergone ESA adoption~\cite{Colpi:2024xhw}, is expected to be sensitive to local string networks with $G\mu\gtrsim10^{-17}$ (for both models A and B)~\cite{Auclair:2019wcv}.
As shown in figure~\ref{fig:cs_stochastic_forecasts}, we can expect even stronger constraints from ET. In the triangular 10km configuration we can probe string tensions $G\mu\gtrsim10^{-18}$ with just one year of data both for models A and B, while with the 2L~15km aligned configuration we can probe $G\mu\gtrsim10^{-19}$, for model A, and slightly smaller for model B~\cite{Branchesi:2023mws}.

For global strings, as shown figure~\ref{fig:cs_global}, we can expect ET to probe $\eta\gtrsim8\times10^{14}\,\mathrm{GeV}$.
This sensitivity is comparable to that of LISA \cite{Chang:2021afa,Gouttenoire:2019kij,Gorghetto:2021fsn}, while for the most recent PTA data from e.g., NANOGrav, one obtains $\eta\gtrsim3\times10^{15}\,\mathrm{GeV}$~\cite{Servant:2023mwt}, which is comparable to the current constraints from Nambu-Goldstone radiation.
ET will therefore improve upon these current constraints; however, these other constraints are also likely to improve by the time ET is operational, e.g., due to stage-4 CMB surveys~\cite{CMB-S4:2016ple}.

\paragraph{GW bursts from strings.}
\label{sect:GWburststringsdiv2}

\begin{figure}[t]
    \centering
    \includegraphics{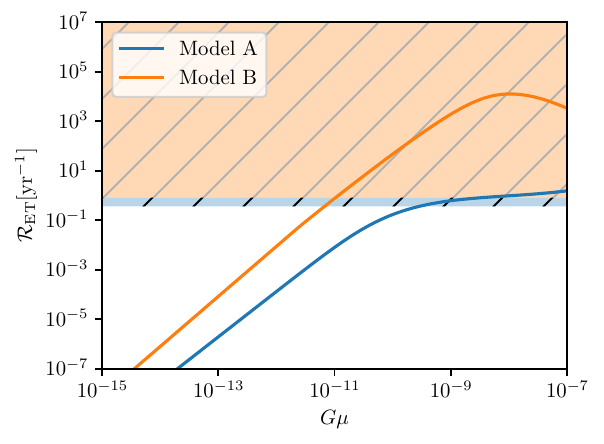}
    \caption{Expected rate of detected bursts in Einstein Telescope as a function of the string tension for models A and B. In case ET does not detect bursts from cosmic string cusps, the orange hatched region is excluded after 4 years of observations and the blue hatched region is excluded after 8 years.}
    \label{fig:cs_burst_rate}
\end{figure}

Strong bursts from nearby cosmic string loops may be resolved in GW detectors and searches are currently undertaken with the hope to detect individual cusps or kinks in LVK datatsets ~\cite{LIGOScientific:2017ikf,LIGOScientific:2021nrg} as well as PTA data \cite{Yonemaru:2020bmr}.
Similar searches will also be carried out in the future for LISA~\cite{Babak:2008aa,Auclair:2023brk,Colpi:2024xhw}.
Here we forecast the capabilities of ET to detect individual cusps assuming a single detector with a triangular xylophone configuration with $10$km arms (ET LF+HF)~\cite{Branchesi:2023mws}. 

As a first approximation, we compute the combined SNR of individual bursts using the response function for a triangular detector~\cite{Regimbau:2012ir} and claim a detection if the SNR is above $20$.
Accurate data analysis would rely on match-filtering methods or new methods such as machine learning~\cite{Meijer:2023yhn}.
Similarly to the procedure already described for LISA~\cite{Auclair:2023brk}, we simulate a set of sources with random polarization and sky-localization, and obtain the average response function of ET to cosmic string bursts as a function of their amplitude.
We then combine the response function of ET with the rate of bursts predicted for different loop population models to obtain the expected rate of bursts detected in ET.

The expected burst rates for both models A and B are presented in figure~\ref{fig:cs_burst_rate}.
For $G\mu \approx 10^{-11}$ compatible with the PTA signal~\cite{EuropeanPulsarTimingArray:2023lqe,EPTA:2023xxk,NANOGrav:2023hvm}, one expects to detect about one burst per year for B and $10^{-2}$ for A.
If ET does not detect bursts from cosmic string cusps, it would place constraints on the abundance of loops in our cosmic neighbourhood, and constrain the string tension to $G\mu \lesssim 10^{-11}$ for B and $G\mu \lesssim 10^{-10}$ for A.

Note however that the computation of these rates have not taken into account the effect of gravitational back-reaction. This could not only affect the typical number of cusps per loop but also their persistence in time.

\subsubsection{GWs from domain walls}\label{sec:domain_walls} 


Domain walls (DWs) are two--dimensional defects (surfaces) that separate regions of space belonging to physically different vacua of the theory. These vacua are typically related by a global discrete symmetry, so that they are exactly degenerate in energy and the DWs are topologically stable. As the DW configuration interpolates between discrete minima, the corresponding field is forced to leave the vacuum manifold at the center of the wall, so that its energy density can in fact be very high. 
Domain walls can possibly lead to a cosmological problem if they are absolutely stable and eventually dominate the energy density of the Universe. Solutions to this problem come with
interesting phenomenological consequences for the domain wall network, as we will discuss.

Domain walls arise in many beyond Standard Model theories where the symmetry of the SM is extended and a discrete group (or subgroup) of this symmetry is spontaneously broken during the cooling of the Universe.
The list of motivated models is large and includes:
discrete flavor symmetries \cite{Riva:2010jm,Gelmini:2020bqg}, two-Higgs \cite{Battye:2020jeu, Arkani-Hamed:2020yna}, twin-Higgs \cite{Chacko:2005pe, Batell:2019ptb}, composite Higgs \cite{DiLuzio:2019wsw,Panico:2015jxa} models, non-Abelian flavor symmetries \cite{Riva:2010jm, Gelmini:2020bqg},  axion monodromy  \cite{Hebecker:2016vbl, Krajewski:2021jje}, supersymmetry \cite{Witten:1982df,Ellis:1986mq,Abel:1995wk,Dvali:1996xe,Kovner:1997ca}, grand unification \cite{Lazarides:1981fv, Everett:1982nm, Lazarides:1982tw}, Hubble-induced phase transitions \cite{Bettoni:2019dcw,Bettoni:2021zhq} and discrete spacetime symmetries \cite{Dashen:1970et,Rai:1992xw,Dvali:1994wv,Smilga:1998dh,Gaiotto:2017tne,Craig:2020bnv}.

QCD axion models \cite{Peccei:1977hh,Peccei:1977ur,Weinberg:1977ma,Wilczek:1977pj} are of particular interest. While solving the strong CP problem of the Standard Model, they lead naturally to DWs \cite{Sikivie:1982qv,Vilenkin:1982ks,Huang:1985tt}. The underlying Peccei-Quinn (PQ) mechanism introduces a new  global $U(1)_{\rm PQ}$ symmetry that is spontaneously broken at a high scale $F_a$ and it is anomalous under QCD ~\cite{Peccei:1977hh, Peccei:1977ur}. At low energies, the model consists of the (pseudo-)\-Goldstone boson, the axion, coupled to topological defects: global strings, and DWs. See also \cite{Preskill:1982cy,Abbott:1982af,Dine:1981rt,Dine:1982ah,Arias:2012az} and \cite{Svrcek:2006yi,Arvanitaki:2009fg,Ringwald:2012hr,Marsh:2015xka,Bauer:2017ris,Brivio:2017ije,CidVidal:2018blh,Bauer:2020jbp,Goh:2008xz,Bellazzini:2017neg,Belyaev:2015hgo,DiLuzio:2020wdo,Choi:2020rgn,OHare:2024nmr} for recent reviews on several aspects related to axion models and phenomenology.

In post-inflationary scenarios, the cosmological PQ phase transition leads to the formation of a network of $U(1)_{\rm PQ}$ global strings. At much lower temperature, $T\sim \Lambda_{QCD}$,  QCD further breaks $U(1)_{\rm PQ}$ to a discrete subgroup $Z_{N_\text{DW}}$, and DWs appear. 
The integer $N_\text{DW}$ is the number of DWs attached to each string and is a model dependent parameter (it relates to the PQ charges of SM and/or heavy quarks). If $N_\text{DW}=1$, the string network is quickly annihilated at $T\sim \Lambda_{QCD}$ and negligible GWs are produced. Instead, if $N_\text{DW}>1$ a string-wall network is formed and some additional $U(1)_{\rm PQ}$ breaking sector enforces DW annihilation. 

As a result, QCD axion models with $N_\text{DW}>1$ naturally lead to a DW epoch and so to potentially a sizeable GW signal. Numerical simulations in this scenario were performed in \cite{Hiramatsu:2012sc}. The signal is strongest in heavy QCD axion, which are themselves of particular interest because they address the axion `quality’ problem \cite{Holdom:1982ex,Holdom:1985vx,Flynn:1987rs,Barr:1992qq,Kamionkowski:1992ax,Kamionkowski:1992mf,Holman:1992us,Berezhiani:1992pq,Ghigna:1992iv,Senjanovic:1993uz,Dobrescu:1996jp,Rubakov:1997vp,Choi:1998ep,Berezhiani:2000gh,Banks:2010zn,Hook:2014cda,Fukuda:2015ana,Dimopoulos:2016lvn,Agrawal:2017ksf,Gaillard:2018xgk}, see also \cite{ZambujalFerreira:2021cte,Ferreira:2022zzo,Jiang:2022svq, Blasi:2022ayo,Blasi:2023sej,Caprini:2024ofd}. Importantly, in these models the amplitude of GW observables is correlated with the proton and neutron electric dipole moments \cite{ZambujalFerreira:2021cte,Ferreira:2022zzo}, which is a welcome feature towards possible degeneracies in interpretations. 

The relevance of DWs in cosmology has been pointed out a long time ago\,\cite{Zeldovich:1974uw,Kibble:1976sj}. The typical setup involves the spontaneous breakdown of a discrete symmetry during the thermal history of the Universe. Regions of space separated by distances that are larger than the correlation length of the field around the temperature of the phase transition will randomly select one among the possible vacua of the theory. DWs will then form at the boundaries between these causally--disconnected regions where the field remains trapped at the maximum of the potential. 

The resulting DW network is characterized by the tension $\sigma$, which measures the DW energy (or mass) per unit surface, and the correlation length, $R$, indicating the typical distance between the walls as well as their average curvature radius. The evolution of the network in a cosmological setting is determined by the tendency of the walls to reduce their energy by reducing their surface, as well as by the Hubble expansion and the effect of friction with the surrounding plasma. When particle friction can be neglected, the network is known to approach an attractor solution independently of the initial conditions in which the correlation length is set by the Hubble horizon, $R = \xi H^{-1}$, with constant $\xi = \mathcal{O}(1)$ indicating the average number of DWs per Hubble patch. In this so--called scaling regime, the energy density of the network is
\be
\label{eq:rhoDW}
\rho_{\rm DW} = 2 \mathcal{A}  \, \sigma H,
\ee
with $\mathcal{A} = \mathcal{O}(1)$. It appears from the scaling in \eqref{eq:rhoDW} that the DW energy density will eventually take over the energy in matter and radiation leading to an inconsistent cosmology\,\cite{Zeldovich:1974uw,Sikivie:1982qv,Vilenkin:1981zs}. This conclusion can be avoided if the DWs still remain a subdominant component of the energy budget today. This can occur if the DW tension is small enough, or in case a period of inflation has provided sufficient dilution of the network. 
These two options inevitably suppress most of the signals coming from the DWs.\footnote{Another possibility is that the domain wall tension decreases with time, eventually leading to the melting of the DW network \cite{Ramazanov:2021eya,Babichev:2021uvl,Dankovsky:2024ipq}.} We will instead consider a third possibility in which the DW network is actually metastable and it undergoes annihilation at a temperature $T_\text{ann}$ when its energy density is still far from domination.

As mentioned before, DWs related to exact discrete symmetries are topologically stable and cannot annihilate. To allow for annihilation one then needs  to introduce an explicit breaking of the underlying discrete symmetry so that the degeneracy of the vacua is lifted by a small bias $\Delta V$. 
This introduces an additional force acting on the network, namely a vacuum pressure towards the true vacuum of the theory. At the early stage of the evolution, the bias has no effect and the network can still reach the scaling regime. However, at the temperature $T_\text{ann}$ when the vacuum pressure balances the tension force $\sigma/R$, the friction with the plasma as well as the Hubble expansion, the network begins to collapse. One estimates this temperature to be related to $\Delta V$ as\,
\cite{
Hiramatsu:2010yz,Hiramatsu:2012sc,Hiramatsu:2013qaa,Saikawa:2017hiv}
\begin{equation}
\label{Tannsimple}
T_\text{ann} \simeq 270 ~\text{MeV} 
\left(\frac{\sigma^{1/3}}{100~ \text{TeV}} \right)^{-3/2}
\left(\frac{\Delta V^{1/4}}{100~\text{MeV}} \right)^{2}
\, \bigg(  \frac{g_*}{10} \bigg)^{-1/4}.
\end{equation}
Annihilation at $T_\text{ann}$ results in the copious production of gravitational waves, particle excitations and (as discussed in section~\ref{sec:DW_pbh}) possibly additional relics such as primordial black holes.
The energy budget available from the DW annihilation can be parameterized by $\alpha_\text{ann}$ given by
\begin{equation}
\label{eq:alphastar}
    \alpha_\text{ann} \equiv \frac{\rho_{\text{DW}}}{\rho_{\rm tot}} \Big|_\text{ann}  \simeq 0.02
    \left(\frac{\sigma_{}^{1/3}}{100~ \text{TeV}} \right)^{3}
    \left(\frac{T_\text{ann}}{100~\text{MeV}} \right)^{-2}
    \, \bigg( \frac{g_*}{10} \bigg)^{-1/2},
\end{equation}
with $\rho_{\rm tot} = 3 H^2 M_{\rm Pl}^2$ the total energy density, and $\alpha_\text{ann} < 1$ to avoid DW domination. The corresponding spectrum of GWs is characterized by a peak amplitude
\be
\Omega_{\text{peak}} 
\simeq 
 1.64\times 10^{-6} \left( \frac{\tilde{\epsilon}_\text{gw}}{0.7} \right)
 \left(\frac{\mathcal{A}}{0.8}
 \right)^{2}
 \left(\frac{g_*(T)}{10}\right)\left(\frac{g_{*s}(T)}{10}\right)^{-4/3}\; \alpha_\text{ann}^2\, ,
\ee
with $\tilde{\epsilon}_\text{gw}$ an $O(1)$ efficiency factor, and the peak frequency is
\be
f_{\text{peak}}  \simeq 
1.15\times 10^{-9} ~ \text{Hz}\left(\frac{g_*(T)}{10}\right)^{1/2}\left(\frac{g_{*s}(T)}{10}\right)^{-1/3}\left(\frac{T_{\text{ann}}}{10\ \text{MeV}}\right)\, .
\ee
As expected, the GW amplitude scales like $\alpha_\text{ann}^2$ while the peak frequency is set by the correlation length of the network just before annihilation begins, which for the scaling solution corresponds to the Hubble horizon at $T_\text{ann}$.

\begin{figure}[t]
    \centering
    \includegraphics[scale=0.52]{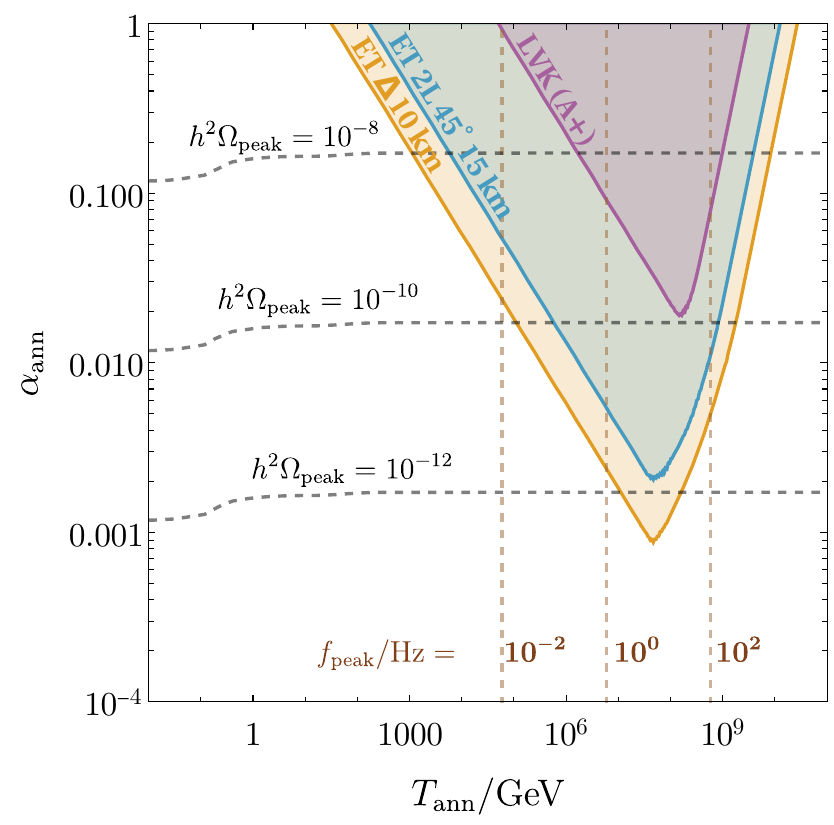}\hspace{3mm}
    \includegraphics[scale=0.48]{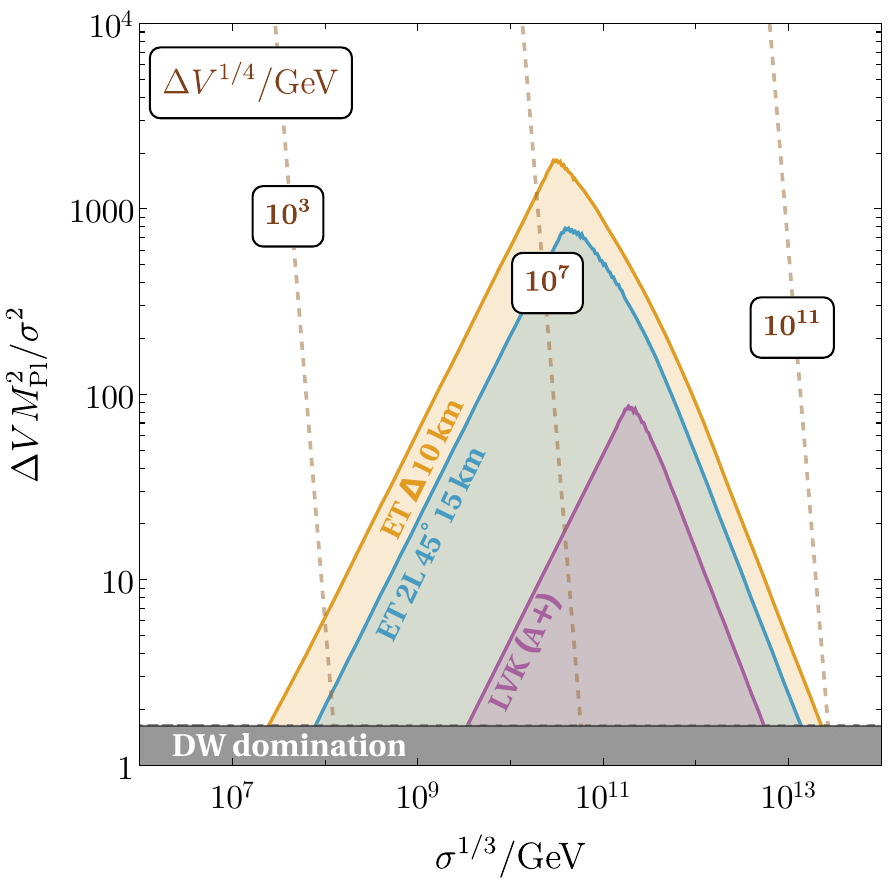}
    \caption{Left: $(\alpha_\text{ann}, T_\text{ann})$ parameter space which can be probed by ET, assuming a triangular xylophone configuration with 10km arms (orange), a 2L misaligned configuration with 15km arms (blue), both for an observation time of $T = 1$ year and $\text{SNR} = 1$. For comparison, we show the sensitivity region of LVK A+ (purple) with $T = 1$ year and $\text{SNR} = 1$. Contours of $h^2 \Omega_\text{peak}$ and $f_\text{peak}$ are shown by black and brown dashed lines respectively. Right: bias vs. tension parameter space with similar sensitivity regions as on the left plot. Brown dashed lines indicate the size of the bias. The gray region corresponds to a forbidden DW dominated Universe, for which $\alpha_\text{ann} \geq 1$.}
    \label{fig:DWplot}
\end{figure}

Figure~\ref{fig:DWplot} shows the reach of ET to these signals, in terms of the model independent parameters, $\alpha_\text{ann},T_\text{ann}$, and of the parameters in the minimal $Z_2$ model. We choose to show the energy scales associated to the DW tension and the bias, $\sigma^{1/3}$ and $(\Delta V)^{1/4}$, which in the simplest model mark the scales of spontaneous and explicit breaking of the $Z_2$ symmetry. The breadth of the ranges of scales is remarkable.

It is worth mentioning that GWs are actually continuously radiated by the DW network from the moment of its formation. However, since the DW energy density grows with time with respect to the critical density, the contribution to the GW spectrum is maximal at $T=T_\text{ann}$ and it outshadows the previous emission. Let us also mention that our expressions for the amplitude and peak frequency correspond to the spectrum coming from the last moment of scaling, and does not include the contribution from the actual collapse of the network. This has been recently investigated in ref.\,\cite{Kitajima:2023cek,Ferreira:2024eru}, see also \cite{Cyr:2025nzf,Notari:2025kqq,Babichev:2025stm}. Additional effects may arise if the friction from the thermal plasma on the DW network is not negligible around the time of annihilation, \cite{Blasi:2022ayo, Blasi:2023sej}.

Finally, note that collapsing DW networks also can lead to substantial PBH formation, see section~\ref{sec:DW_pbh}. This adds signatures to the DW models that are correlated with GW observables, which should be useful to discriminate among different  candidate sources.

\subsubsection{Primordial black holes} 
\label{sect:PBHdiv2}

Primordial Black Holes (PBHs) can form in the very early Universe through a variety of mechanisms and have been studied as a possible candidate to explain the dark matter (see e.g.,~\cite{LISACosmologyWorkingGroup:2023njw} for a recent review). 
Nevertheless, even if they accounted for a subdominant contribution to the current energy density in the Universe, they could have played a crucial role in the cosmological history, leading to various GW signatures.
As the mass of these objects is linked by model-dependent ${\cal O}(1)$ factors to the mass of the Hubble scale at formation, 
they could have formed within a very large range of masses, potentially even lighter than  $m_{\rm PBH}\lesssim 10^{-18} M_\odot$ corresponding to unstable objects through Hawking evaporation within a timescale comparable to the age of the Universe \cite{Hawking:1974rv}.
The Einstein Telescope has the opportunity to constrain this putative population of BHs by searching for the associated GW signatures. 

A population of PBHs is expected to form binaries which could merge in the relatively late-time Universe and lead to GWs signatures.\footnote{Note, however, that for PBHs formed at the early Universe, the expansion of the Universe cannot be neglected, questioning the choice of metric that most adequately describes PBHs. The choice of the appropriate metric will have implications for the conditions for two PBHs to form a decoupled binary, and therefore for calculating the merger rate~\cite{Boehm:2020jwd}. 
} These signatures of PBH binaries falling within the ET frequency band would come from PBH of masses around the stellar mass and thus need to be distinguished from the expected signatures of astrophysical origins, i.e. coming from stars and astrophysical black holes. This is discussed in detail in Chapter \ref{section:div3}, dedicated to GW population studies. 

In this section, instead, we will focus on the truly cosmological signatures of PBH formation scenarios, i.e., the cosmological GWB associated with the physical mechanisms that lead to the collapse of large overdensities in the early Universe. As the formation of PBHs requires extreme overdensities, it is generically associated with the emission of sizeable GWB. To gain an intuition on the typical mass range probed by ET, we can consider the relation between the present-day GW frequency and the mass contained in the Hubble sphere at Hubble crossing time (taken to be a proxy for the PBH mass)
\begin{equation}\label{GW_peak_frequency2}
    f_{\rm GW} = 
    4.1 \,  {\rm Hz}
    \left(\frac{\kappa}{2.51}\right) 
    \left( \frac{g_*}{106.75} \right)^{-1/12} 
    \left(\frac{M_H}{10^{-17} M_\odot} \right)^{-1/2},
\end{equation}
where $\kappa = k r_m$ relates the perturbation size $r_m$ (see, e.g., \cite{Musco:2018rwt}) to the wavenumber $k$ and $g_*$ is the number of  effective degrees of freedom in the energy density (see, e.g., \cite{Maggiore:1999vm,Maggiore:2018sht}), that in \eq{GW_peak_frequency2} has been  normalized to the value that it has in the Standard Model at high energy. 
From this relation, we immediately see that ET will probe the formation of ultra-light PBHs, potentially narrowing the open window where no bounds prevent PBHs from being the dark matter (see, e.g., \cite{Carr:2020gox} for a recent review).

\paragraph{PBHs from primordial perturbations.}
\label{sect:PBHprimpertdiv2}

The simplest formation scenario assumes the collapse of extreme inhomogeneities during the radiation-dominated era~\cite{Zeldovich:1967lct,Hawking:1971ei,Chapline:1975ojl,Carr:1975qj}.
The PBH mass resulting from the collapse is related to $M_H$ 
 by an order-unity factor controlled by the critical collapse parameters~\cite{Musco:2008hv}
 (see~\cite{Escriva:2021aeh} for a recent review).
Adopting threshold statistics, we compute the mass fraction $\beta$ assuming Gaussian primordial curvature perturbations 
(see e.g.,~\cite{Atal:2019cdz,Yoo:2019pma,Young:2022phe,Ferrante:2022mui,Gow:2022jfb,Pi:2022ysn} for extensions alleviating this assumptions) 
and accounting for the nonlinear relationship between curvature and density perturbations~\cite{DeLuca:2019qsy,Young:2019yug,Germani:2019zez}.
One obtains
\begin{align}
\beta & 
= 
\mathcal{K}
\int_{\delta_l^{\rm min}}
{\rm d} \delta_l
\left(\delta_l - \frac{3}{8}\delta_l^2 - \delta_c\right)^{\gamma}
P_\text{\tiny G}(\delta_l)\,,
\label{eq:GaussianTerm1}
\end{align}
where $P_\text{\tiny G}$ is the Gaussian distribution for the linear component of the density contrast $\delta_l$, related to the primordial curvature perturbations by $\delta_l = - 4 r \partial_r \zeta(r) / 3 $. 
We indicate with $\sigma(r_m)$ the variance of the linear density field computed at horizon crossing time and smoothed on a scale $r_m$ (see e.g.,~\cite{Franciolini:2022tfm} for more details), while $\delta_c$ is the threshold for collapse derived by dedicated numerical relativity simulations \cite{Musco:2004ak,Escriva:2019phb,Yoo:2020lmg,Musco:2020jjb}.
We also introduced the parameters ${\cal K}$ and $\gamma$ to include the effect of critical collapse \cite{Choptuik:1992jv,Evans:1994pj,Niemeyer:1997mt,Green:1999xm,Musco:2008hv}.

The same large adiabatic perturbations required in this scenario of PBH formation are also inevitably responsible for the GW emission at second order in perturbation theory~\cite{Tomita:1975kj,Matarrese:1993zf,Acquaviva:2002ud,Mollerach:2003nq,Carbone:2004iv,Ananda:2006af,Baumann:2007zm,Wang:2019kaf}.
The current energy density of these 
scalar-induced GWs (SIGW, whose effect on anisotropies of the CGWB we already discussed in section~\ref{sub:anisotropies cosmo}) is given, in terms of the curvature power spectrum ${\cal P}_\zeta$, by (see e.g.,~\cite{Domenech:2021ztg} for a recent review)
\begin{equation} \label{eq:SIGW}
 \Omega_{\text{\tiny GW},0} = c_g
 \, \Omega_{R,0}
 \int_{0}^{\infty}\!\!{\rm d} v
 \int_{|1-v|}^{1+v}\!\!{\rm d} u
 \mathcal{T}(u,v)
 \mathcal{P}_{\zeta}(ku)\mathcal{P}_{\zeta}(kv).
\end{equation}
We introduced the parameter $c_g$, which  accounts for the suppression of SIGW amplitude induced by the evolution of the degrees of freedom in thermal equilibrium, with $c_g \simeq 0.4$ assuming that the Standard Model remains valid up to the relevant energy, while
$\mathcal{T}(u,v)$ is the so-called transfer function 
\cite{Espinosa:2018eve,Kohri:2018awv} which accounts for the time evolution of the source.
For a sufficiently narrow spectrum of curvature perturbations, the SIGW spectrum features a low-frequency tail that scales as $\Omega_\text{\tiny GW} \sim f^3$, due to the causality limited efficiency of the super-Hubble emission~\cite{Caprini:2009fx,Cai:2019cdl,Hook:2020phx,Brzeminski:2022haa,Loverde:2022wih}, while the high-frequency tail that depends on the drop-off of the curvature spectrum.
The peak amplitude is expected to be $\Omega_{\rm GW} \simeq  10^{-5} A^2$, where $A$ parametrizes the enhanced amplitude of ${\cal P}_\zeta$ at small scales.

\begin{figure}[t!]
    \centering
    \includegraphics[width=0.6\textwidth]{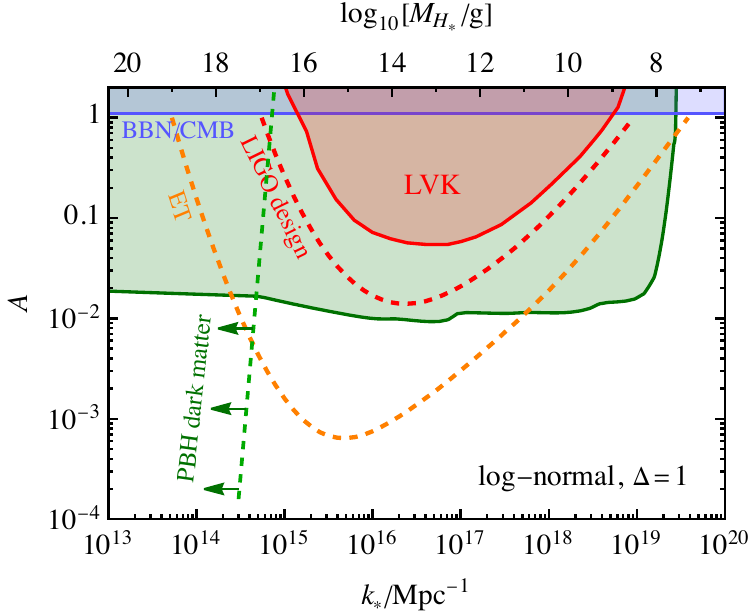}
    \caption{Upper bound on the amplitude of the curvature power spectrum at small scales $A$ as a function of the wavenumber $k_*$ in case of null detection of a SIGW. The green and blue regions at the top show, respectively, the exclusions from PBH constraints and from BBN/CMB constraint on the number of relativistic degrees of freedom. In red the bound from LVK is shown for comparison.}
    \label{fig:SIGW_PBH}
\end{figure}

In figure~\ref{fig:SIGW_PBH}, we show the region of parameter space that ET will be able to constrain by searching for SIGW signals (orange dashed line).
This plot assumes the curvature power spectrum to feature a lognormal enhancement at around $k_*$ with a width $\Delta = 1$ (see \cite{Romero-Rodriguez:2021aws} for more details).
The green curve indicates the amplitude required to generate PBH abundance that saturates the current constraints in the mass range shown in the upper edge of the plot. 

\begin{figure}[t!]
    \centering
    \includegraphics[width=0.6\textwidth]{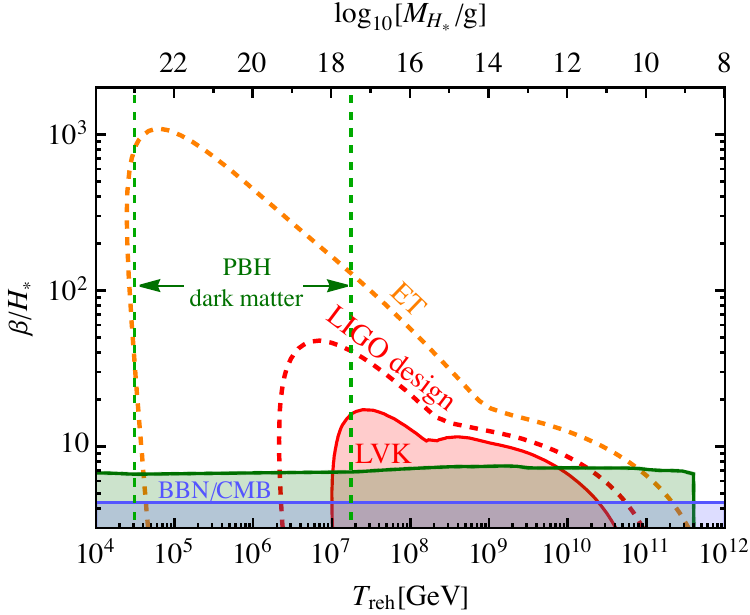}
    \caption{Lower bound on the inverse time duration of the phase transition $\beta/H_*$ as a function of the temperature $T_{\rm reh}$ right after the phase transition in case of null detection of a SIGW. The green and blue regions at the bottom show the exclusions from PBH constraints and from BBN/CMB constraint on the number of relativistic degrees of freedom.}
    \label{fig:PT_PBHs}
\end{figure}

\begin{figure}[t!]
    \centering
    \includegraphics[width=0.6\textwidth]{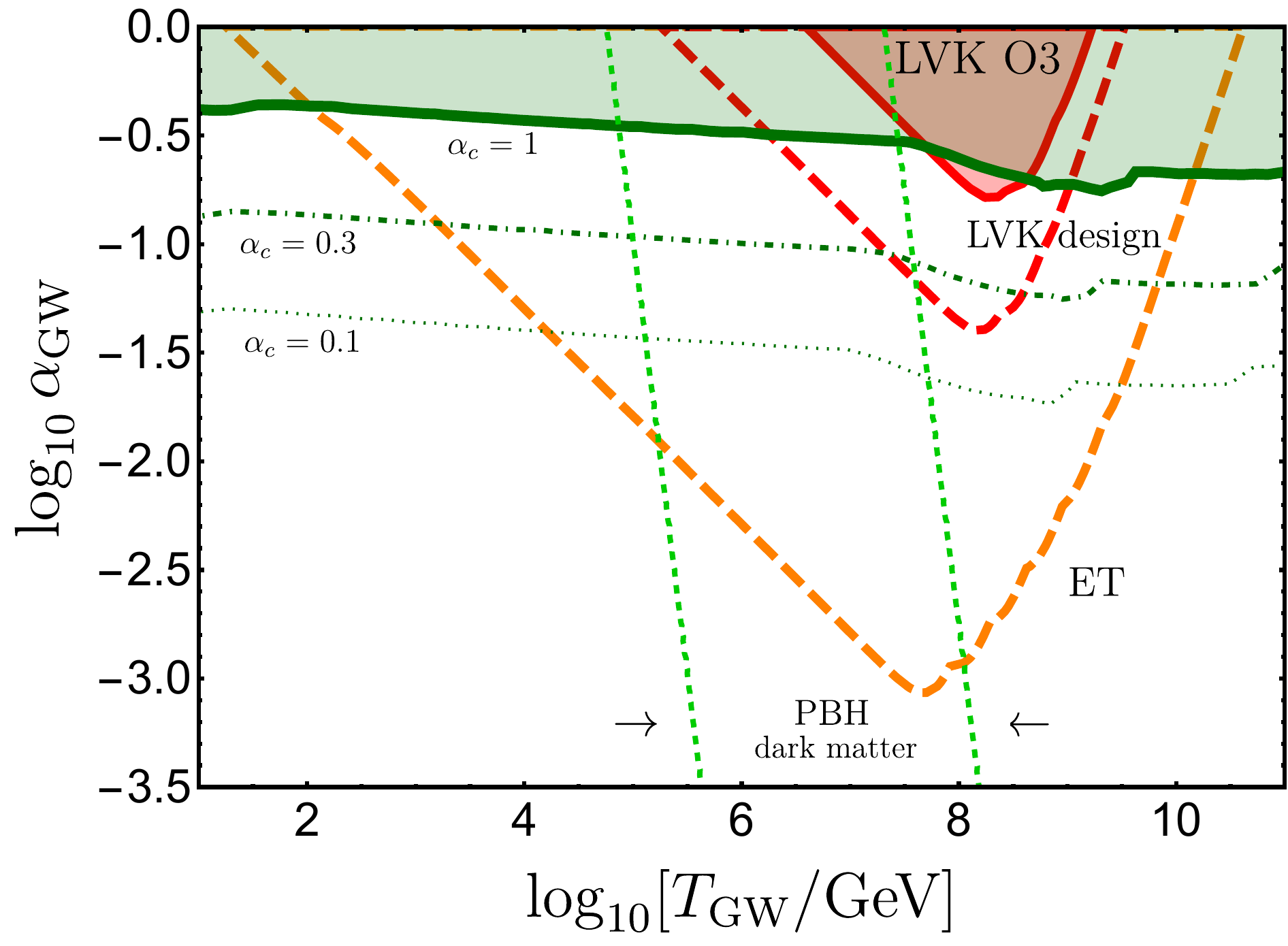}
    \caption{Constraints from PBH overproduction as a function of temperature $T_\text{GW}$ and DW abundance at the peak of GW emission (adapted from \cite{Ferreira:2024eru}). Superimposed, the detectable  regions with ET and LKV design as well as the exclusion from LVK O3 data. The shaded region above the solid line is excluded from the PBHs formed at $t_\text{PBH}$ or later. The dash-dotted and dotted lines refer to sub-Hubble PBHs, for representative values of $\alpha_c$, giving a reasonable sense of current systematic uncertainties. The vertical dashed lines indicate the asteroid mass range (assuming $\alpha_c=1$).}
    \label{fig:DW_PBHs}
\end{figure}

The parameter space on the left of the dashed diagonal line in figure~\ref{fig:SIGW_PBH} (see also the range between the dashed lines in figures~\ref{fig:PT_PBHs} and~\ref{fig:DW_PBHs}) indicates the asteroid mass window where the PBHs can constitute all of dark matter. For lower masses, i.e., to the right side of the dashed line, Hawking evaporation leads to the emission of particles that could be detected with different probes (mainly BBN, CMB anisotropies and distortions and cosmic rays, among others \cite{Carr:2020gox}).
However, these bounds may be avoided in scenarios where black hole evaporation deviates from the semi-classical result \cite{deFreitasPacheco:2023hpb,Alexandre:2024nuo,Thoss:2024hsr,Anchordoqui:2024dxu}.
Furthermore, microscopic horizonless relics could form in the early Universe either directly through gravitational collapse or as stable remnants of the Hawking evaporation of primordial black holes. In both cases they completely or partially evade cosmological constraints arising from Hawking evaporation and they could still explain the entirety of the dark matter in certain mass ranges, a scenario that will be tested by ET 
\cite{Domenech:2021wkk,Domenech:2023mqk,Franciolini:2023osw}.

\paragraph{PBHs from phase transitions.}

Cosmological phase transitions are typically associated with the breaking of global or local symmetries in a process where the vacuum expectation value of a scalar field changes. They can be either continuous transitions where the scalar field rolls down to the new minimum or first-order transitions that proceed by nucleation and expansion of bubbles inside which the field is in the true vacuum, see  the discussion in section~\ref{sec:phasetransitions}. Either way, the transition can lead to the formation of large inhomogeneities, GWs and even PBHs. For the latter, the transition typically needs to be so strongly supercooled that the vacuum energy of the scalar field in the false vacuum causes a period of thermal inflation. The formation of PBHs can occur during the transition in collisions of several bubbles~\cite{Hawking:1982ga,Kodama:1982sf,Lewicki:2023ioy} or after it when the scales that exited the cosmological horizon during the thermal inflation reenter the horizon~\cite{Dimopoulos:2019wew,Liu:2021svg,Kawana:2022olo,Gouttenoire:2023naa,Lewicki:2024ghw}. In terms of the phase transition parameters defined in section~\ref{sec:phasetransitions}, the first-order phase transitions where PBH formation is efficient correspond to $\alpha\gg 1$ and $\beta/H_* \lesssim 8$.

In case of a first-order phase transition, the distribution of the density contrast $\delta$ generated for the modes that exited horizon during the thermal inflation has a negative non-Gaussianity~\cite{Lewicki:2024ghw}. For comparison, models of primordial inflation including an ultra-slow-roll period during which the fluctuations can became very large predict a positive non-Gaussianity~\cite{Tomberg:2023kli}. This implies that the SIGW signature~\eqref{eq:SIGW} of the PBH formation from fluctuations generated during a first-order phase transition is stronger because the PBH abundance is sensitive to the large $\delta$ tail of the distribution that is suppressed by the negative non-Gaussianity, whereas the SIGW abundance is sensitive to the typical values of $\delta$, i.e. the width of the distribution. Moreover, as discussed in section~\ref{sec:phasetransitions}, also the bubble collisions source GWs. Consequently, the GW spectrum related to PBH formation in a strongly supercooled first-order phase transition has two peaks~\cite{Lewicki:2024ghw}: the SIGW peak at the scale corresponding to the horizon size at the end of the thermal inflation, and the bubble collisions GW peak at the scale corresponding to the typical bubble size.

In figure~\ref{fig:PT_PBHs} we show the parameter space ET will be able to probe by searching the GWB from a strongly supercooled ($\alpha\gg1$) first-order phase transition. For $\beta/H_*\lesssim 20$ the double-peak structure of the spectrum can be detectable, whereas for $\beta/H_*\gtrsim 20$ the component sourced by bubble collisions fully dominates the spectrum. The range of ET extends  up to PBH masses $\simeq 10^{-10}\,M_\odot$, covering the asteroid mass window where PBHs can constitute all DM. At these masses ET will probe the high-frequency tail of the GW spectrum sourced by the bubble collisions.

\paragraph{PBHs from collapsing domain wall networks.}
\label{sec:DW_pbh}

We have summarized the motivation and observational consequences of Domain Walls  in section~\ref{sec:domain_walls}. In short, DW networks have a strong impact on cosmology and so they are an excellent means to probe models with spontaneously broken discrete symmetries. The strong impact translates into abundant production of cosmological relics, like a GW background. 

The ability of DW networks to form PBHs has been re-examined recently \cite{Garriga:2015fdk,Deng:2016vzb,Vachaspati:2017hjw,Ferrer:2018uiu,Gelmini:2022nim,Pujolas:2022qvs,Gelmini:2023ngs,Ferreira:2024eru,Dunsky:2024zdo}. The PBH formation process is tied to the annihilation regime, the end of the DW epoch. As in section~\ref{sec:domain_walls}, we specialize the discussion to the simplest annihilation mechanism, a small explicit breaking of the discrete symmetry. This uplifts the degeneracy in the vacuum energies of the vacua, $\Delta V$, which acts as a pressure that pushes the walls to increase the true vacuum volume. During annihilation most DWs collapse due to tension $\sigma_\text{DW}$ and vacuum difference $\Delta V$. The annihilation time $t_\text{ann}$ is estimated from $\sigma_\text{DW} H_\text{ann} \simeq \Delta V$. At that time the fraction in DWs is given by $\alpha_\text{ann}$, see  \eqref{eq:alphastar}, which must satisfy $\alpha_\text{ann} < 1$ in order not to run into a DW problem. 

The PBH abundance is mainly governed by two factors \cite{Ferrer:2018uiu}: the likelihood that Hubble-sized false vacuum  pockets shrink within their own Schwarzschild volume $R_S^3$; and the number of Hubble-sized pockets at each time. The first factor is small at $t_\text{ann}$ (for $\alpha_\text{ann}\ll1$), but it increases in time \cite{Ferrer:2018uiu}. 
This leads to a rather sharp notion of `PBH formation time’, $t_\text{PBH}$, defined as when the Schwarzschild radius of a Hubble-sized pocket coincides with the Hubble length. Surviving Hubble-sized pockets at $t\gtrsim t_\text{PBH}$ convert into BHs efficiently, independently of shape and angular momentum (and have a baby-Universe wormhole topology, see \cite{Garriga:2015fdk,Deng:2016vzb}). 
Notice that there is a parametric delay between between the the time of GW emission (set by $t_\text{ann}$ up to a numerical factor \cite{Ferreira:2024eru}) and PBH formation: $t_\text{PBH}\sim t_\text{GW} / \alpha_\text{ann}^{1/2}$ \cite{Ferrer:2018uiu,Ferreira:2024eru}. 

The second factor (the abundance of Hubble sized pockets) requires a dedicated computation of how fast the network decays in time, which is challenging from the point of view of numerical simulations. Recent work combining numerical simulations and semi-analytical methods \cite{Ferreira:2024eru} concluded that this abundance decays like $\exp[ - (t/\tau)^{3/2} ]$, see also \cite{Chang:2023rll}. For Hubble-sized pockets the decay time $\tau$ is shorter than $t_\text{ann}$ by a factor $2-3$ \cite{Ferreira:2024eru}. This leads to a very strong dependence of the abundance, of the form
$\exp\big[ -c\, \alpha_{\rm ann}^{-3/4}\big]$ with $c$ a constant.

False vacuum  pockets entering before $t_\text{PBH}$ can also form BHs, though they require some amount of sub-Hubble contraction. Their collapse probability depends significantly on asphericities, angular momentum, etc, but on the other hand these pockets are more abundant. A proper computation of this population is challenging and, at present, lacking. 
A pragmatic way to account for these sub-Hubble BHs is to introduce a lower collapse criterion, $\alpha_c$ (the minimum value of $R_S\,H$ when the pocket enters the Horizon).

The resulting constraints are shown in figure~\ref{fig:DW_PBHs} as a function of temperature $T_\text{GW}$ and DW abundance at the peak of GW emission, together with the detectable regions in ground based observatories. Remarkably, ET will be able to test whether DM is composed of asteroid mass PBHs produced during the collapse of a DW network.

\subsubsection{GWs as probes of the early Universe expansion history} 


When studying modifications to the standard cosmological model based on GR and $\Lambda$CDM, it is convenient to distinguish between the early Universe evolution, defined as the evolution from the earliest instants of the Universe until, say, Big Bang Nucleosynthesis (BBN), and the late Universe evolution, from BBN until the present epoch. The late Universe evolution is strongly constrained by observations, and the role of GWs in probing it will be discussed in details in section~\ref{sec:lateuniversediv2}. Early Universe cosmology is much less constrained. Deviations from the predictions of  GR and $\Lambda$CDM could appear as a consequence of modifications of gravity relevant at high energies, which indeed emerge in different contexts \cite{Clifton:2011jh}, such as scalar-tensor theories \cite{Perivolaropoulos:2009ak, Hohmann:2013rba, Jarv:2014hma}, braneworld theories \cite{Nojiri:2002wn, Kaminski:2009dh}, compactified extra dimensions and Kaluza-Klein models 
\cite{Arkani-Hamed:1998sfv,Perivolaropoulos:2002pn},
$f(R)$  theories and theories with higher-order terms in the curvature invariants \cite{Berry:2011pb, Capozziello:2014mea}, string loop effects~\cite{Damour:1994ya} or string cosmology \cite{Gasperini:1994xg,Buonanno:1996xc}. The evolution predicted by the standard $\Lambda$CDM cosmological model can also be affected, in the early Universe, by physics beyond the Standard Model, phase transitions, etc. In the previous subsections we have seen for instance how such physics beyond the Standard Model could lead to GW signals produced by cosmic strings, domain walls, or related to primordial BH formation. 

Another way in which GWs can be used to probe the early Universe, and in particular modifications of its expansion history, is through the imprint that the expansion history leaves on  primordial GWs (PGWs)  produced in the early Universe, as they  propagate until the present epoch. This information is contained in the transfer function, that connects the primordial power spectrum to the power spectrum observed today. To fix the notation, let us recall that the primordial power spectrum ${\cal P}_{T,\rm in}(k)$ of the tensor perturbations was defined  in
\eq{4corrhTTijP}.
The simplest form of the spectrum, at least in a range of comoving wavenumbers $k$ close to a ``pivot scale'' $k_*$, is just a power-law, see \eq{4primtensP},
characterized by an amplitude $A_T$ (which depends on the value chosen for $k_*$) and a tilt $n_T$ (which to a first approximation can be taken as constant, or else can also be made to ``run'' with $k$). A typical choice for $k_*$ is $0.05\, {\rm Mpc}^{-1}$, corresponding to the wavenumbers probed by the CMB. The tensor amplitude $A_T(k_*)$ can  be traded for the tensor-to-scalar ratio $r(k_*)=A_T(k_*)/A_S(k_*)$, where $A_S(k_*)$ is the analogous amplitude for scalar perturbations, whose value is fixed by CMB observations: e.g.,  $A_S(k_*=0.05\, {\rm Mpc}^{-1})\simeq 2.14\times 10^{-9}$.

Given a cosmological model, such as $\Lambda$CDM, the evolution in time of the Fourier modes of the GW perturbations over a FRW background can be computed, and the result is encoded in the transfer function $T_{\rm GW}(k)$,
\be\label{4defTGW12}
\tilde{h}_A(\eta_0,k)=T_{\rm GW}(k)\tilde{h}_A(\eta_{\rm in},k)\, ,
\ee
(where $A=+,\times$ labels the two polarizations and, at least in GR as well as in any modifications of it that preserves parity, the transfer function is the same for both polarizations). Then,
the power spectrum observed today,  ${\cal P}_{T,0}(f)$,  is related to the primordial tensor power spectrum by 
\be\label{4PT0TPtin}
{\cal P}_{T,0}(k)=|T_{\rm GW}(k)|^2\, {\cal P}_{T,\rm in}(k)\, .
\ee
Instead of the comoving wavenumber $k$, we
can equivalently express all results in terms of the GW frequency observed today, $f$, which (in units $c=1$) is related to $k$ by $f=k/(2\pi)$.
The relation between $\Omega_{\rm GW}(f)$ and ${\cal P}_{T,0}(k)$ is given by [see eq.~(19.288) of ref.~\cite{Maggiore:2018sht}]
\be\label{4OgwandcalPT}
\Omega_{\rm GW}(f)=\frac{\pi^2}{3H_0^2}\, f^2{\cal P}_{T,0}(f)\, ,
\ee
and therefore
\be\label{4OgwandcalPTin}
\Omega_{\rm GW}(f)=\frac{\pi^2}{3H_0^2}\, f^2|T_{\rm GW}(f)|^2\, {\cal P}_{T,\rm in}(f)\, ,
\ee
where we have used $f=k/(2\pi)$ as the argument of the functions.
In particular, for modes that re-entered the horizon during radiation dominance, i.e. modes  with $f\gg f_{\rm eq}\sim 1.7\times 10^{-17}\, {\rm Hz}$ (and therefore for modes relevant for pulsar timing arrays, space-borne and ground based  GW interferometers), we have
\be\label{4hogwperfgsimfeq}
h_0^2\Omega_{\rm GW} (f)\simeq \frac{1}{24}\, h_0^2\Omega_{\rm R} \,
\(\frac{g_*(T_k) }{3.363}\)\(\frac{3.909}{g^S_{*}(T_k)}\)^{4/3}
\, {\cal P}_{T,\rm in}(f)\, ,
\ee
where $\Omega_{\rm R}$ is the energy fraction of radiation (numerically  $h_0^2\Omega_{\rm R}\simeq 4.184\times 10^{-5}$); $T_k$ is the temperature at which a mode with comoving wavenumber $k$ re-enters the horizon; and $g_*(T)$ and  $g^S_{*}(T)$ are  the effective number of species in the energy density and in the entropy, respectively, at temperature $T$. 
For the frequencies relevant for 
space-borne and ground-based interferometers we have $T_k\gg 100$~GeV and, in the Standard Model,
$g_*(T_k)=g^S_{*}(T_k)\simeq 106.75$, and therefore
\be\label{4hogwperfgrande}
h_0^2\Omega_{\rm GW}(f)\simeq 6.73\times 10^{-7}\,
{\cal P}_{T,\rm in}(f)\, ,
\ee
where we have used the numerical value $h_0^2\Omega_{\rm R}\simeq 4.184\times 10^{-5}$. Equivalently,  using \eq{4primtensP} for the primordial power spectrum,
and writing 
$A_T(k_*)=r(k_*)A_S(k_*)$,
$h_0^2\Omega_{\rm GW}(f)$ can be written as in \eq{4OgwcompactT}.

If the cosmological evolution is modified, with respect to that obtained assuming the validity of GR and $\Lambda$CDM, the transfer function
in \eq{4defTGW12} will be different and, given a mechanism that produces a known primordial spectrum 
${\cal P}_{T,\rm in}(f)$, the stochastic GW background observed today, given by \eq{4OgwandcalPTin}, will  be different from that expected in  
GR and $\Lambda$CDM. Stochastic GW backgrounds can therefore be a probe of the expansion history of the Universe, and in particular of the early Universe, for which it is in general more difficult to obtain observational constraints.

\subparagraph{Power-like phenomenological modifications of the expansion rate.}
At the  level of background evolution,   modification of the expansion rate can be described writing $H(t)$ as   $H_{\rm MC}(T) = A(T)H_{\rm GR}(T)$ \cite{Schelke:2006eg, Catena:2009tm, Allahverdi:2020bys}, where $H_{\rm MC}(T)$ refers to a given modified cosmology while $H_{\rm GR}(T)$ refers to the  standard cosmological model (GR+$\Lambda$CDM).  The factor $A(T)$, which depends on the temperature and  accounts for the underlying modified cosmological model, can be parameterized phenomenologically in different ways \cite{Schelke:2006eg, Catena:2009tm}. In all cases, 
to avoid conflict with the successful predictions of BBN \cite{Kawasaki:2000en, Hannestad:2004px, deSalas:2015glj},  the parametrization must satisfy the conditions $A(T)\to 1$ at the onset of the BBN period. At early times, however, one can have $A(T)$ significantly different from one. A number of models, in this regime, can be described by a power-law function
$A(T)\simeq A_* (T/T_*)^\nu$ \cite{Schelke:2006eg, Catena:2009tm},  with $\{A_*, \nu\}$ dimensionless constants and $T_*$ a reference temperature.  
In particular, $\nu=2$ corresponds to a  Randall-Sundrum type II brane cosmology; $\nu=1$ to kination models in GR; $\nu=0$ gives a model  with a constant rescaling of the Hubble expansion rate; $\nu <0$ in scalar-tensor cosmology; and $\nu = 2/n -2$ in $f(R)$ cosmology, with $f(R) = R+\alpha R^n$.   This power-law behavior  is then smoothly 
interpolated toward $A(T)=1$ at some  temperature  $T_*$  \cite{Schelke:2006eg, Catena:2009tm}.
For the analysis of the PGWs spectrum, it is convenient to refer to  the frequency $f_*$, instead of the corresponding temperature $T_*$. For $f \ll f_*$ and $f \gg f_*$, the following general cases can be analyzed \cite{Bernal:2020ywq,Bernal:2019lpc}:   

\begin{figure}[t]
\begin{center}
\includegraphics[height=0.7\textwidth]
{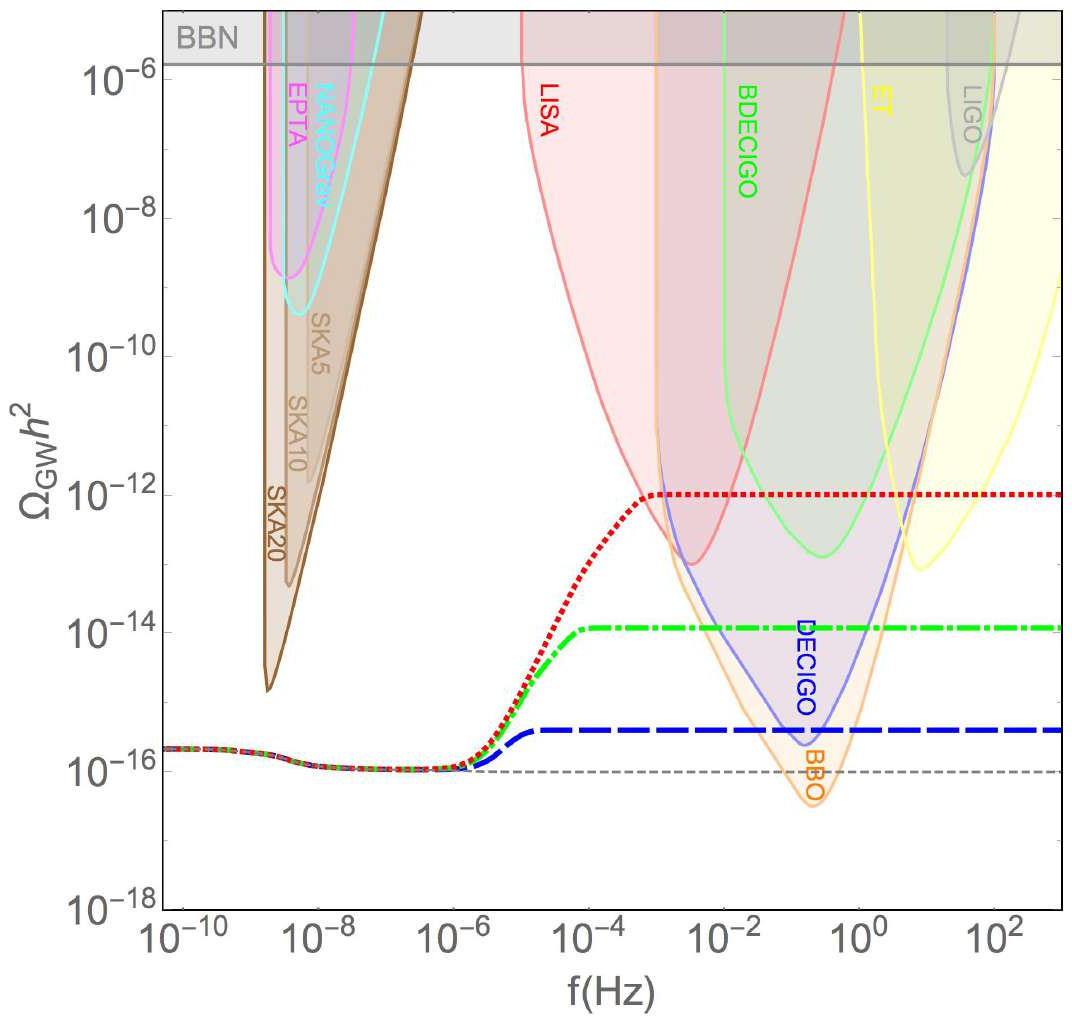}
\includegraphics[height=0.7\textwidth]
{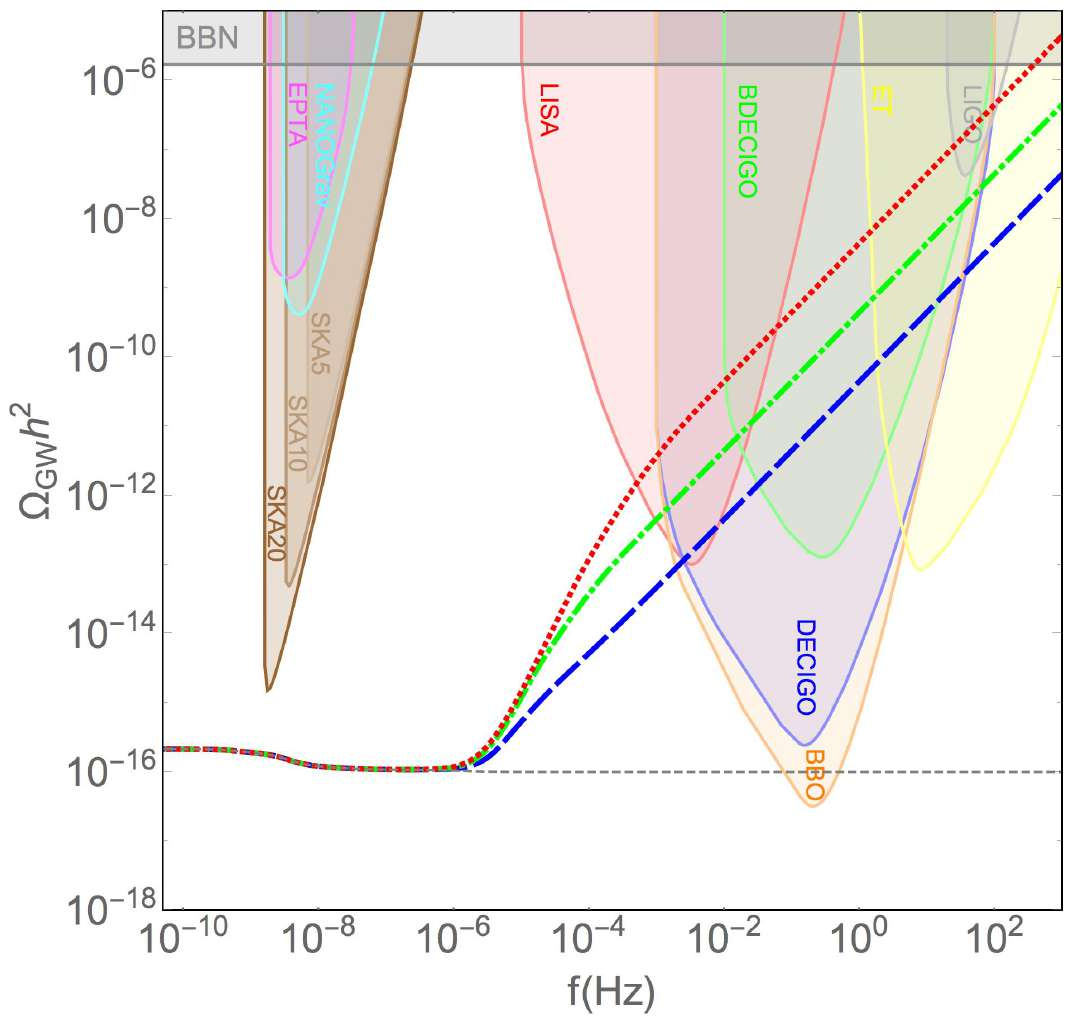}\\[-3em]
\caption{$h^2\Omega_\text{GW}(f)$ 
for the amplification of vacuum fluctuations in standard single-field slow-roll inflation. The gray dashed lines corresponds to the result in GR+$\Lambda$CDM, assuming a tensor-to-scalar ratio $r=0.07$ and a scale-invariant primordial tensor spectrum, $n_T=0$. The colored lines show the result obtained with the modified cosmological  discussed in the text,
for fixed $f_{*}\simeq 2.5\times10^{-6}$~Hz, and $A_*=1$ (blue dashed lines), $A_*=10$ (green dot-dashed lines), $A_*=100$ (red dotted lines), and $\nu=0$ (left panels), $\nu=1$ (right panels). The colored regions refer to the BBN constraint and the projected sensitivities of various GW observatories.}\label{fig:OmegaMCdiv2}
\end{center}
\end{figure}

\begin{figure}[th]
\centering
\includegraphics[width=0.485\textwidth]{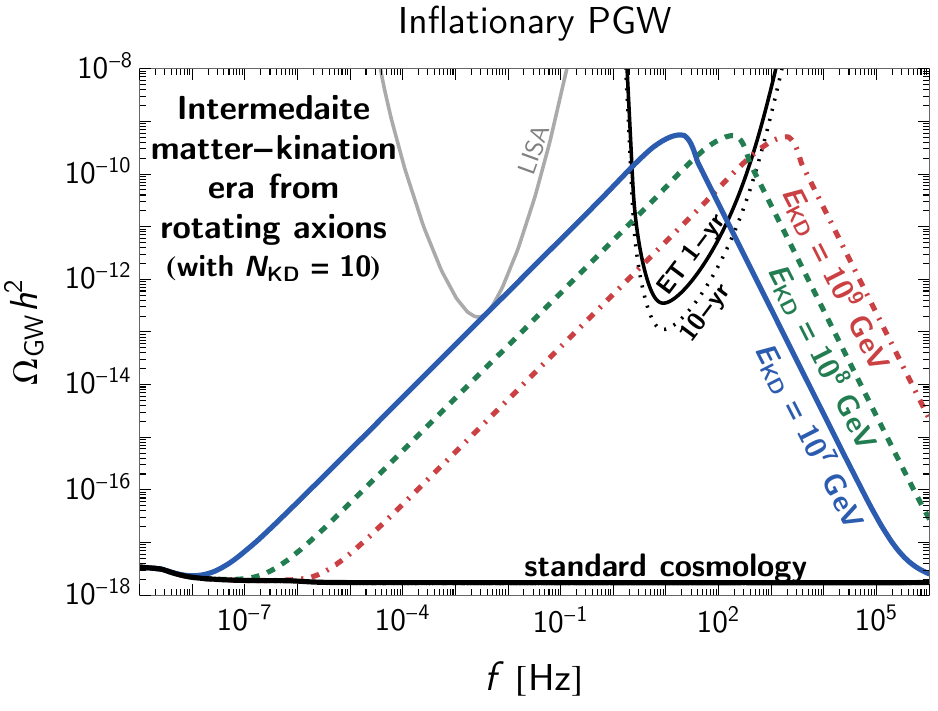}\hfill
\includegraphics[width=0.485\textwidth]{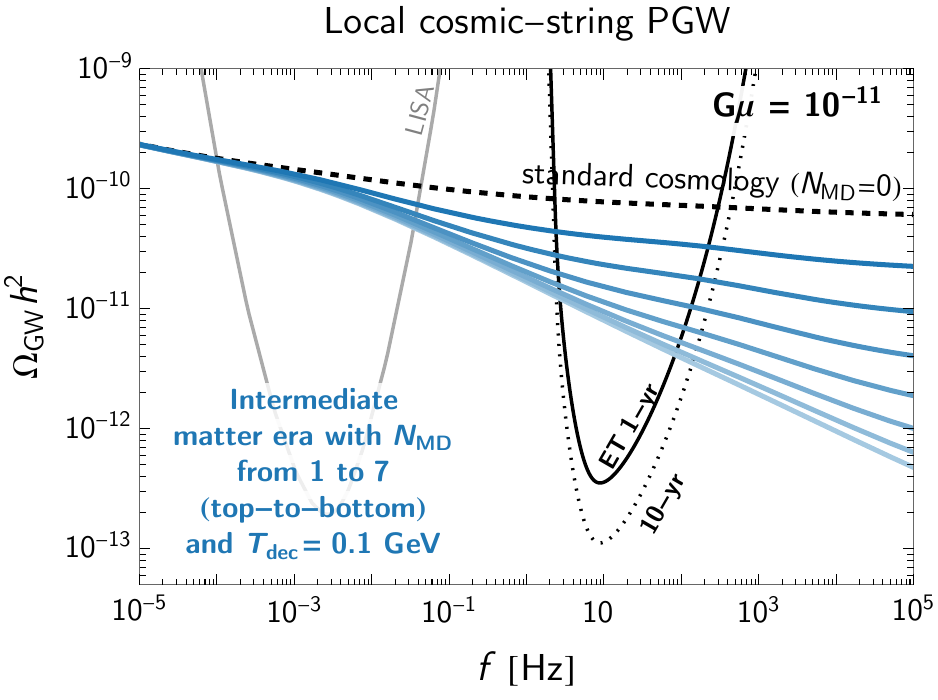}\\[-1em]
	\caption{(left) Inflationary spectra 
 induced by the intermediate matter-kination era are shown as colored lines, while the prediction from standard $\Lambda$CDM cosmology is the black line. Following the period of matter domination, the intermediate kination era starts at energy scale $E_{\rm KD}$ and lasts for $N_{\rm KD}$ e-folds of the scale-factor expansion. We assume the scale invariant tensor perturbation with inflationary energy scale $E_{\rm inf} \simeq 1.6 \times 10^{16}$ GeV. The sensitivity curves of ET assume
 respectively 1 and 10 years of observation with SNR $=1$, and the LISA sensitivity comes from \cite{Flauger:2020qyi}. (right) Effect of an intermediate matter era---lasting for $N_{\rm MD}$ efolds and ending at temperature $T_{\rm dec}$---on the prediction for $h^2\Omega_{\rm GW}(f) $ from local cosmic strings, shown by the blue lines  \cite{Gouttenoire:2019kij,Gouttenoire:2019rtn,Ghoshal:2023sfa}.}
 \label{fig:axion_kination}
\end{figure}

\begin{itemize}
    \item $f\ll f_*$. In this range of frequencies, the cosmology converges to GR, $H(a)=H_\text{GR}(a)$, so the transfer function is the same and, for a given primordial spectrum ${\cal P}_{T,\rm in}(f)$, the spectrum ${\cal P}_{T,0}(f)$ observed today is the same. Therefore $h_0^2\Omega_{\rm GW} (f)\propto {\cal P}_{T,\rm in}(f)$, see \eqs{4hogwperfgsimfeq}{4hogwperfgrande}.

\item $f\gg f_*$. In this range of frequency, the factor $A(f)$ plays a major role. The regimes where $\nu>0$, $\nu=0$, and $\nu<0$ are separately discussed: 
   
 -- $\nu>0$: if $\nu$ takes positive values, the PGW relic density reads $\Omega_\text{GW}(f) \propto {\cal P}_{T,\rm in}(f)\,f^\frac{2\nu}{1+\nu}$,
in which appears an extra factor $f^\frac{2\nu}{1+\nu}$, that generates a blue-tilted modification to the original tensor power spectrum. This enhancement follows from the reduction of the friction term entering the equation of GWs.

-- $\nu=0$:  In this case, the Hubble rate is enhanced by a constant factor $A=1+A_*$, and the PGW spectrum is therefore not distorted and is modified by an overall shift $\Omega_{\rm GW}(f) \propto (1+A_*)^2 {\cal P}_{T,\rm in}(f)$.

--  $\nu<0$: for negative values of $\nu$, the factor $A \to 1$ both for low ($f\ll f_*$) and high frequencies ($f\gg f_*$), with a maximum at $f=\bar f\gtrsim f_*$ given by $A(\bar f)=A_*$. In these limits the PGW spectrum  presents the same tilt as the original tensor power spectrum, with a characteristic bump at $f=\bar f$.

\end{itemize}

\noindent
Figure~\ref{fig:OmegaMCdiv2} shows $h^2\Omega_{\rm GW}(f)$ for the amplification of vacuum fluctuations in standard single-field slow-roll inflation. The result in GR+$\Lambda$CDM is given by the gray dashed line, while the colored lines shows the results for $\nu=0$ (left panel) and $\nu=1$ (right panel) 
for different values of $A_*$. The colored regions refer to projected sensitivities for different GW experiments.

\subparagraph{Modified equation of state in the early Universe.}
As another example of modified expansion history,  if there were some pre-BBN epochs in the cosmological history with a modified equation of state associated for instance with either a matter domination or a stiff era, this would lead to
a tilt in $\Omega_{\rm GW}(f)$ with spectral index  $\beta=-2(1-3w)/(1+3w)$, with $w$ the equation of state of the Universe, at the frequencies corresponding to the modes which re-entered the horizon in this non-standard era~\cite{Figueroa:2019paj,Allahverdi:2020bys,Gouttenoire:2021jhk}. 
For example, the kination era right after inflation can result in the enhanced GW spectrum shown in the right panel of figure~\ref{fig:OmegaMCdiv2}, although any observable signal at ET would be strongly constrained by $\Delta N_{\rm eff}$ bound, as shown in \cite{Figueroa:2019paj,Gouttenoire:2021jhk}.
The most striking case comes from an intermediate matter-kination era, 
well-motivated by rotating axions \cite{Gouttenoire:2021wzu,Gouttenoire:2021jhk,Co:2021lkc},
that largely amplifies  the inflationary signal and induces the ``triangular signature" [with $\beta = 1$ during kination era ($w = 1$) and $\beta = -2$ slope during matter era ($w = 0$)] shown in the left panel of figure~\ref{fig:axion_kination}. This amplification can make the signal from slow-roll inflation  observable at ET,  for kination energy scales of order $(10^6-10^9)$ GeV, that connect to axion mass $m_a$ and axion decay constants $f_a$ in the range $m_a \lesssim 6.5 ~ \mu {\rm eV}(10^9 ~ {\rm GeV}/f_a)(\Omega_a/\Omega_{\rm DM})$, where the last bracket is the energy fraction of axions as dark matter \cite{Gouttenoire:2021jhk}.

\begin{figure}[t]
    \centering
    \includegraphics[width=0.7\textwidth]{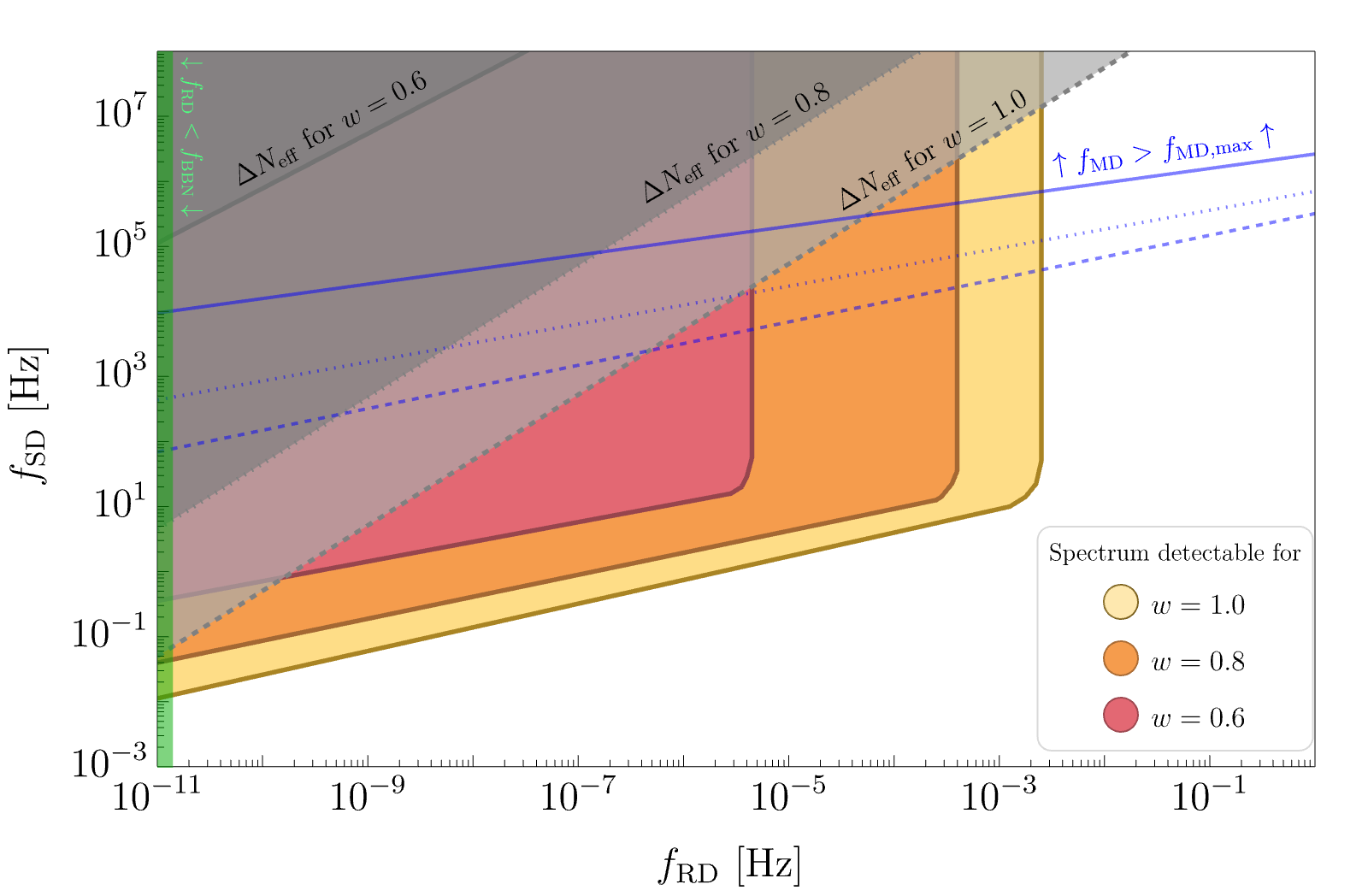}
    \caption{The ($f_{\rm RD}, f_{\rm SD}$) parameter regions accessible by ET are shown. The colored regions represent the detectable parameter spaces for three different values of $w$: red for $w=0.6$, orange for $w=0.8$, and yellow for $w=1.0$ (corresponding to kination). The gray regions indicate the parameter space excluded by indirect limits from BBN and CMB, with the darkest gray for $w=0.6$ and the lightest gray for $w=1$.  We again assume the scale invariant tensor perturbation with inflationary energy scale $E_{\rm inf} \simeq 1.6 \times 10^{16}$ GeV. The parameter $f_{\rm MD}$ is chosen as a function of $f_{\rm RD}$,  $f_{\rm SD}$ and $w$, as described in the text, and we denote the parameter region where $f_{\rm MD} > f_{\rm MD,max}$ with blue lines (dashed for $w=1$, dotted for $w=0.8$ and thick for $w=0.6$).
    The sensitivity curve of ET assumes 1 year of observation with SNR $=1$. The green area on the left is excluded, since there $f_{\rm RD}< f_{\rm BBN}$, which is not allowed.}
    \label{fig:stiffeq-paramspace}
\end{figure}

Furthermore, one can study scenarios with a stiff epoch weaker than kination ($1/3 <w < 1$) \cite{Figueroa:2019paj, Duval:2024jsg}. As a case study, we focus on a cosmological history similar to the one studied above,
where after inflation there is an intermediate radiation-matter-stiff era  (denoted with RD1, MD and SD respectively), followed by the conventional radiation dominated Universe. We assume instantaneous transitions between the different epochs.   
This scenario also leads to a peak-like spectrum that can be testable at Advanced LIGO A+ sensitivity and by third-generation detectors \cite{Duval:2024jsg}.
The frequency of the stiff-to-radiation transition is denoted by $f_{\rm RD}$, and the one of matter-to-stiff transition by $f_{\rm SD}$. 
The parameter space of the model that is testable at ET is illustrated in figure~\ref{fig:stiffeq-paramspace},
for three different values of the equation of state $w$ during the stiff period. 
Assuming no entropy variation, the frequency at the RD1-to-MD transition can be expressed  
as $f_{\rm MD} = f_{\rm RD}^{\frac{1-3w}{1+3w}} f_{\rm SD}^{\frac{6w}{1+3w}}$.
This transition cannot occur earlier than the end of inflation, implying a maximum value of $f_{\rm MD, max}=1.8 \times 10^8$ Hz. 
This limit is indicated by the blue lines in   figure~\ref{fig:stiffeq-paramspace}.

The non-standard era beyond the radiation era at high temperatures not only impact the inflationary PGW, but it can also lead to significant spectral distortions in GW from other cosmological sources, in particular, from cosmic strings \cite{Cui:2017ufi,Cui:2018rwi,Gouttenoire:2019kij,Gouttenoire:2019rtn,Cui:2019kkd,Gouttenoire:2021jhk,Co:2021lkc,Ghoshal:2023sfa,Chang:2019mza,Chang:2021afa}, and causality tails \cite{Hook:2020phx,Racco:2022bwj,Franciolini:2023wjm,Gouttenoire:2021jhk,Servant:2023tua}. The frequencies at which the spectral slope gets modified correspond to the scales at which the GW is being produced; the slope depends on the equation of state of the Universe and can differ from the case of inflationary PGW; see e.g., \cite{Allahverdi:2020bys,Simakachorn:2022yjy} for reviews. This is illustrated for the case of local cosmic strings in the right panel of figure~\ref{fig:axion_kination}.

\subparagraph{Oscillation features in the GWB.}


Interferometers such as ET have the capability not only to observe the general coarse-grained trend of a primordial GWB, but also its fine-grained features. In particular, periodic oscillatory patterns in the GWB have been identified as distinctive signatures of primordial features~\cite{Fumagalli:2020nvq,Braglia:2020taf,Fumagalli:2021cel}. These imprints reflect early-universe dynamics that deviate significantly from scale invariance and arise across a broad range of theoretical frameworks, including both inflationary models and their alternatives (see \cite{Slosar:2019gvt,Chluba:2015bqa,Achucarro:2022qrl} for reviews on the subject). The GWB provides a powerful observational probe of primordial features at scales well beyond the reach of the CMB and galaxy surveys, where most searches have traditionally been concentrated. Detecting such signatures would unveil new physics beyond the standard cosmological model and opening an unprecedented window into the dynamics of the primordial universe.

Primordial features often manifest as oscillations in the scalar power spectrum. These have been found to be of two types.
  \textit{Sharp features} arise from brief deviations in background parameters from their attractor behavior, producing transient, linear oscillations in the scalar power spectrum. This linear oscillatory shape typically appears in inflationary models with a step-like feature in the potential
~\cite{Starobinsky:1992ts, Adams:2001vc,Chen:2006xjb, Bean:2008na,Achucarro:2010da,Bartolo:2013exa,Palma:2014hra,Hazra:2014jka,Hazra:2014goa,Hazra:2016fkm}, turns in the multi-field trajectory \cite{Ashoorioon:2008qr, Achucarro:2010da,Cespedes:2012hu,Palma:2020ejf, Fumagalli:2020adf} or sudden transition in the speed of sound \cite{Park:2012rh,Achucarro:2012fd,Achucarro:2013cva,Ballesteros:2018wlw}. \textit{Resonant features} result from periodic background oscillations that resonate with quantum modes of the primordial curvature fluctuations, leading to sustained oscillations at constant frequency across a wide log-scale range. These log-oscillations can be produced in inflationary scenarios where the inflaton potential has oscillatory bumps~\cite{Chen:2008wn}, multi-field dynamics~\cite{Achucarro:2010da,Chen:2011zf,Chen:2011tu,Shiu:2011qw,Gao:2012uq,Gao:2013ota,Chen:2014cwa}, other settings such as axion monodromy inflation~\cite{Flauger:2009ab,Flauger:2014ana,Behbahani:2011it,Bhattacharya:2022fze} and axion Chern--Simons gravity~\cite{Mavromatos:2022yql}. 
A hybrid of both types appears in models known as classical primordial standard clocks \cite{Chen:2011zf, Chen:2011tu, Chen:2012ja, Battefeld:2013xka, Gao:2013ota, Noumi:2013cfa, Saito:2012pd, Saito:2013aqa, Chen:2014joa, Chen:2014cwa, Huang:2016quc, Domenech:2018bnf, Braglia:2021ckn, Braglia:2021sun, Braglia:2021rej}.

These features in the scalar sector are imprinted in the GWB through nonlinear interactions \cite{Fumagalli:2020nvq,Braglia:2020taf,Fumagalli:2020adf,Witkowski:2021raz,Fumagalli:2021cel,Fumagalli:2021dtd}, where scalar perturbations source tensor modes, if their amplitude is sufficiently large. A broad class of oscillatory patterns in the GWB can then be captured by the following generic templates, either of linear oscillations,
\begin{equation}
	 \Omega_{\textsc{gw}}(f) = \bar\Omega_{\textsc{gw}}(f)\left\{1+\sum_{l=1}^{l_{\rm max}} \left[A^{\rm c}_l\cos\left(l\omega\frac{f}{f_{*}}\right)+A^{\rm s}_l\sin\left(l\omega\frac{f}{f_{*}}\right)\right]\right\},\label{masterlin}
\end{equation}
or of log-periodic ones,
\begin{equation}
\Omega_{\textsc{gw}}(f) = \bar\Omega_{\textsc{gw}}(f)\left\{1+\sum_{l=1}^{l_{\rm max}} \left[A^{\rm c}_l\cos\left(l\omega\ln\frac{f}{f_{*}}\right)+A^{\rm s}_l\sin\left(l\omega\ln\frac{f}{f_{*}}\right)\right]\right\},\label{master2}
\end{equation}
where $\bar\Omega_{\textsc{gw}}(f)$ is the envelope or main spectral shape, $f$ is the observed frequency of the GW signal, $f_*$ is the pivot frequency scale of the observatory, $l_{\rm max}$ denotes the maximum harmonic determined by the sourcing mechanism. The coefficients $A^{\rm c,s}_l$ represent (not necessarily constant) amplitudes, and $\omega$ is the angular frequency of the oscillations—not to be confused with the frequency $f$.

Sharp features during inflation lead to the template in Eq. \eqref{masterlin} in its simplest form, i.e. $l_{\mathrm{max}} = 1$ ~\cite{Fumagalli:2020nvq,Braglia:2020taf}, for GWs produced when scalar fluctuations—affected by the feature—re-enter the horizon during the post-inflationary era.\footnote{For the template of primordial GWs sourced directly at the time of the feature during inflation, see Ref.~\cite{Fumagalli:2021mpc}.
} In this context, the characteristic frequency is $\omega \sim 2/f_\mathrm{f}$, where $f_\mathrm{f}$ corresponds to the scale that exited the horizon at the time of the feature during inflation.  Typically, the resulting GW spectrum exhibits $O(10\% - 20\%)$ modulations, arising from the resonant amplification of $O(1)$-level oscillations in the sourcing scalar power spectrum. These oscillations arise from moderate to copious particle production induced by the sharp feature \cite{Fumagalli:2020nvq}. This also leads to a characteristic relation between the frequency at which $ \Omega_{\mathrm{GW}} $ peaks, denoted $ f_* $, and the feature scale $ f_f $, typically given by $ f_* \sim \mathcal{O}(10\text{--}100) \, f_f $. Consequently, for a signal within the ET band, linear oscillations can be probed in the range $ 0.1/\mathrm{Hz} < \omega < 10/\mathrm{Hz} $. Additionally, both the amplitude of the oscillations and the shape of the envelope are influenced by the background equation of state at the time of GW production~\cite{Witkowski:2021raz}.

\begin{figure}[t!]
\centering
\includegraphics[keepaspectratio, scale=0.65]{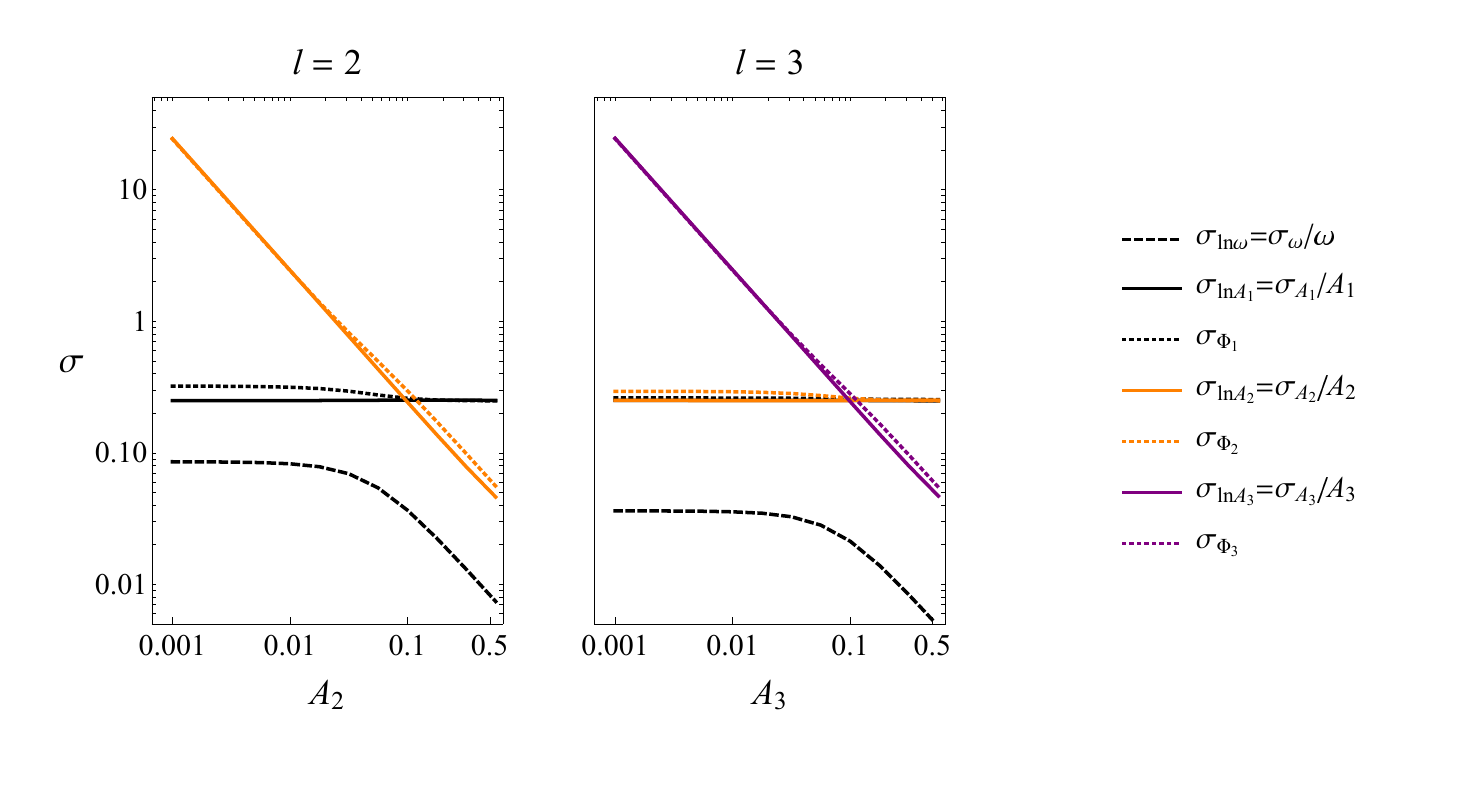} 
\caption{Dependence of the expected relative errors $\sigma$ for $l_{\rm max}=2$ (left) and $l_{\rm max}=3$ (right) harmonics. The oscillation amplitude $A_2$ and $A_3$ vary, while the other parameters are fixed as follows. For $l_{\rm max}=2$: $\Omega_0=10^{-11}$, $n_{\rm t}=0$, $\omega=10$, $A_1=0.1$, $\Phi_1=0$. For $l_{\rm max}=3$: $A_2=0.1$, $\Phi_2=0$ and $\Phi_3=0$. The black, orange and purple curves correspond to the oscillation parameters (oscillation amplitude $A_l$ and phase $\Phi_l$) for $l=1$, $l=2$, and $l=3$, respectively. From~\cite{Calcagni:2023vxg}.}
\label{figgwb}
\end{figure}

Resonant features, on the other hand, lead to a template as given in Eq. \eqref{master2}, with $l_{\rm max} = 2$. In this case, the GW spectrum exhibits a superposition of two oscillatory components, with frequencies $\omega_\mathrm{log}$ and $2\omega_\mathrm{log}$ \cite{ Fumagalli:2021cel}. Here, $\omega_\mathrm{log}$ corresponds to the logarithmic oscillations in the primordial power spectrum, often linked to a mass parameter in units of the Hubble scale. The relative amplitudes of these two harmonics depend on $\omega_\mathrm{log}$, the shape of the peak in the scalar power spectrum, and the background equation of state at the time of GW production. The templates that have been derived in~\cite{Fumagalli:2020nvq, Braglia:2020taf, Fumagalli:2021cel, Witkowski:2021raz} for the post-inflationary contribution to the~GWB encapsulate these dependencies.

The log-oscillatory profile in Eq. \eqref{master2} can also be produced directly in the tensor sector as a manifestation of a broken discrete scaling symmetry in scenarios of quantum gravity~\cite{Calcagni:2023vxg,Calcagni:2017via}. In this case, the energy spectrum in \eqref{master2} is given by \eq{4hogwperfgsimfeq} with
\begin{equation}\label{cpt2}
{\cal P}_{T, \rm in}(k)  = \bar{\cal P}_{T, \rm in}(k)\left[1+A^{\rm c}_1 \cos\left(\omega\ln\frac{k}{k_{*}}\right)+A^{\rm s}_1 \sin\left(\omega\ln\frac{k}{k_{*}}\right)\right],
\end{equation}
where the prefactor $\bar{\cal P}_{T, \rm in}(k)$ depends on the model. A function $F$ is invariant under a discrete scale symmetry if $F(\lambda_\omega^m f)=F(f)$, where $\lambda_\omega:= e^{\frac{2\pi}{\omega}}$ and $m$ is an integer. The oscillatory modulation of (\ref{master2}) enjoys this symmetry but the envelope $\bar\Omega_{\textsc{gw}}(f)$ in general does not. A discrete scale invariance in frequency as the one above corresponds to a symmetry in momentum space, mirrored by an invariance under discrete dilations $x^\mu\to \lambda_\omega x^\mu$ in position space. Such a micro-structure of spacetime geometry would correspond to the one of a deterministic (multi-)fractal~\cite{1983JSP....33..559D,1984CMaPh..94..115D,1997PhRvL..78.3245E,1984JSP....34...75B,NIGMATULLIN20052888} replicated at all scales. This everpresence of the same structure transcending a UV/IR divide is responsible for the appearance of log oscillations at all frequencies in (\ref{master2}).

Expressions (\ref{masterlin}) and (\ref{master2}) can serve as templates to fit any signal displaying linear or logarithmic oscillations, regardless of its physical origin~\cite{Calcagni:2023vxg}. To check the detectability of the log oscillations in ET, in~\cite{Calcagni:2023vxg} the Fisher matrix for the cross-correlation analysis was calculated to estimate the expected error on the parameters in (\ref{master2}) for the ET-D sensitivity curve assuming a three-year observation run. It has been shown that, for oscillation amplitudes $A_l\sim 0.1$ and a frequency $\omega\sim 10$, we need a detection with signal-to-noise ratio $\gtrsim 100$ to determine all the oscillation parameters with an $O(10\%)$ precision. 
If the GWB is primordial and the envelope is parameterized by $\bar\Omega_{\rm GW}=\Omega_0 (f/f_*)^{n_t}$, the minimum GWB amplitude $\Omega_0$ and the minimum tensor tilt $n_{\rm t}$ needed to achieve such SNR at the ET frequencies satisfying the current upper bound $r<0.036$ of the tensor-to-scalar ratio at the CMB scale $k_0=0.05\,{\rm Mpc}^{-1}$~\cite{BICEP:2021xfz} are roughly
\begin{equation}
\Omega_0\gtrsim 10^{-11}\,,\qquad n_{\rm t}\gtrsim 0.28\,.
\end{equation}

\noindent
Relative errors on oscillation frequency, oscillation amplitude, and the oscillation phase scale as $\sigma_{\ln \omega}\propto 1/(A_l\omega)$, $\sigma_{\ln A_l}\propto 1/A_l$ and $\sigma_{\Phi_l}\propto 1/A_l$, respectively, and they all decrease as $\propto 1/{\rm SNR}$. Notably, the error is insensitive to the inclusion of higher-order harmonics, for the reason that oscillation frequencies are ordered hierarchically as integer multiples of $\omega$ and each term in the harmonic expansion can be constrained independently. The $l_{\rm max}=2$ and $l_{\rm max}=3$ cases are shown in figure~\ref{figgwb}. The range of $\omega$ accessible by ET is roughly 
\begin{equation}
10 < \omega < 10^3\,,
\end{equation}
where the lower bound is fixed by requiring to have enough oscillation cycles inside the sensitivity band of ET and the upper bound is determined by $\omega< f/\Delta f$, where $\Delta f$ is the frequency resolution of the instrument. Our estimation is obtained by $f\sim 10$Hz and $\Delta f\sim 0.01$Hz, which is a similar value used in the LVK stochastic analysis as a result of coarse graining.

\subsection{Probing the late Universe with Einstein Telescope}\label{sec:lateuniversediv2}

In the previous section we have discussed how ET can probe the early Universe. The relevant observable, in that case, is the stochastic GW background of cosmological origin that, as we have seen,  can be generated by many different early Universe mechanisms. For probing the late Universe with GWs, in contrast, the most important  signals are those due to the coalescence of compact binaries. A crucial feature of GWs for late time Universe studies is the fact that CBC sources are self-calibrating distance indicators. When the redshift of the source can also be somehow inferred, one can reconstruct the distance-redshift relation, tracking the expansion history of the Universe and constraining cosmological parameters such as the Hubble constant \cite{Schutz:1986gp,Holz:2005df}, similarly to the use of standard candles. In this context, CBC are then called ``standard sirens".
This reconstruction is, however, non-trivial, since from a GW alone one can not in general infer the redshift, which is degenerate with the chirp mass. While breaking this degeneracy is possible in special cases, like for {\it Love sirens} \cite{Messenger:2011gi}, or for matter structure/cosmological effects considered in \cite{Bonvin:2016qxr}, in general direct redshift measurements are not possible with GW detectors.  On the other hand, for {\it bright standard sirens}, which are those events followed by electromagnetic counterparts, direct redshift measurements are enabled by host galaxy identification. However, most GW events seen so far by LVK have not been followed by electromagnetic counterparts, hence the name of {\it dark (standard) sirens}. With dark sirens, the most common techniques for cosmological parameter estimation rely on combinations with galaxy catalogs from which redshift information can be extracted \cite{Schutz:1986gp,DelPozzo:2011vcw}, allowing to track the cosmic expansion history \cite{Finke:2021aom,Gray:2021sew,Gair:2022zsa,Mastrogiovanni:2023zbw,Borghi:2023opd}.
Additionally to these methods, one can make use of known merger redshift distribution \cite{Chernoff:1993th,Ding:2018zrk,Leandro:2021qlc} or
features in the mass distribution of BHs and NS to obtain redshift information~\cite{Taylor:2011fs,Taylor:2012db,Farr:2019twy,Mancarella:2021ecn,Finke:2021eio,Ezquiaga:2022zkx,Chen:2024gdn}, a technique called {\it spectral sirens}. We highlight, however, that 
cosmological and astrophysical population inference are deeply intertwined
\cite{Mastrogiovanni:2021wsd, Gair:2022zsa, Moresco:2022phi, Mastrogiovanni:2023emh,Gray:2023wgj, Borghi:2023opd}. In this section we discuss various types of methods for doing cosmography with ET and their prospects.

\subsubsection{Cosmography with coalescing binaries}\label{sec:Cosmography}

\paragraph{Bright sirens.}
\label{sec:brightSirens}

Bright sirens are expected to have two main types of prompt electromagnetic counterparts: kilonova (KN) explosions, and short GRBs, as we  know from the celebrated example of   GW170817 \cite{LIGOScientific:2017ync,LIGOScientific:2017zic}; see section~\ref{section:div4} for a detailed discussion of the perspective for multi-messenger observations with ET.

Estimates of the accuracy on cosmological parameters that can be obtained from bright sirens depend on  the BNS and BHNS rates, which are currently less constrained than the BBH rate, and also on  uncertainties on the network of electromagnetic observatories that will operate at the time of ET, as well as on the amount of telescope time that they will devote to the EM follow-up of GW events. 
Here we assume the two different scenarios for the joint GW+EM detection,  already considered in \cite{Branchesi:2023mws}. In the first one, the merger of a BNS is followed by a Kilonova (KN) emission that is potentially observed electromagnetically by the Vera Rubin Observatory (VRO) \cite{LSST:2008ijt}. In the second scenario, under proper circumstances, a short Gamma Ray Burst (GRB) detection by THESEUS \cite{THESEUS:2017wvz,Rosati:2021yjd} is associated to the GW event from a BNS. For the ET+VRO study we consider the multimessenger sources detected in one year of observations, while for ET+THESEUS, in order to have a statistically similar sample, we considers a $5$ years catalog;
the resulting dataset of  bright sirens is the same as in tables $6$ and $10$ of \cite{Branchesi:2023mws}. 
We also use the same BNS population  as in \cite{Branchesi:2023mws}.\footnote{This catalog is obtained from  state-of-the art population models tuned to the O3 LVK data. The absence, to date,  of BNS  candidates in the currently ongoing O4 run  implies that the rate used for BNS is now toward the upper range, rather than the mean value.\label{foot:BNSrate_div2} } 

As shown in  \cite{Branchesi:2023mws} (see tables 28 and 29), the resulting marginalized error on $H_0$  in the GW+GRB case (and 5~yr of observations) is $5.7\%$ for  the 10-km triangle and $4\%$ for 2L-15km-$45^{\circ}$  configuration, while for GW+KN (and 1~yr of observations) it is $0.9\%$ for  the 10-km triangle and $0.6\%$ for 2L-15km-$45^{\circ}$.

\begin{figure}[t!]
    \centering
\makebox[\textwidth][c]{
    \includegraphics[width=.5\textwidth]{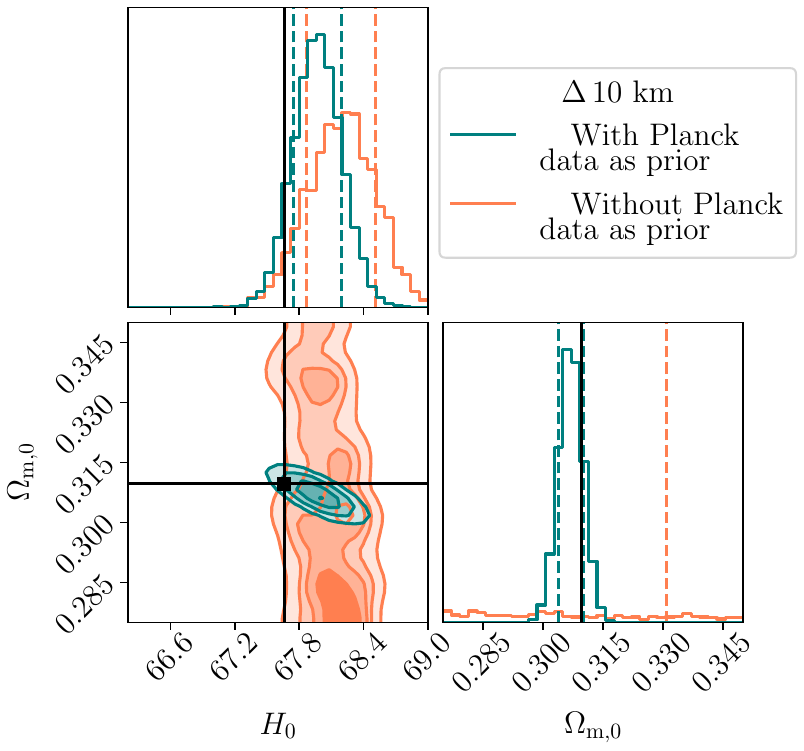}
    \includegraphics[width=.5\textwidth]{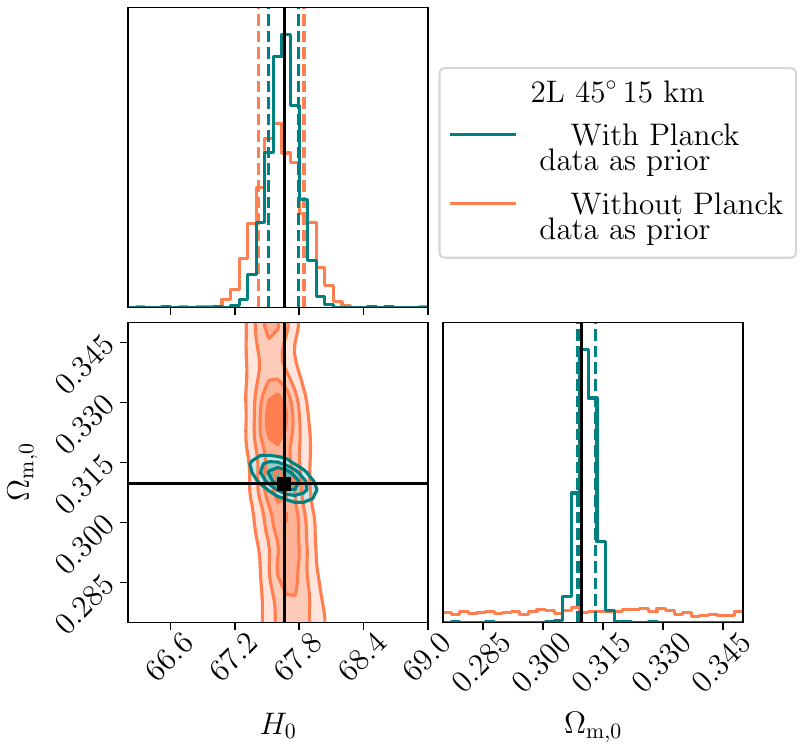}
    }
    \caption{Results of the joint inference on the cosmological parameters $H_0$, $\Omega_{\rm m,0}$, employing GW+KN events detected in one year of observations by the $10$ km triangular (left panel) or the   2L-15km-$45^{\circ}$ (right panel) ET configurations,  with the EM counterpart detected by the Vera Rubin Observatory. Vertical dashed lines represent the $68\%$ CI of each distribution, while the black solid lines label the fiducial values.}
    \label{fig:multimessenger_KN_H0Om0_only}
\end{figure}

\begin{figure}[th]
    \centering
\makebox[\textwidth][c]{
    \includegraphics[width=.5\textwidth]{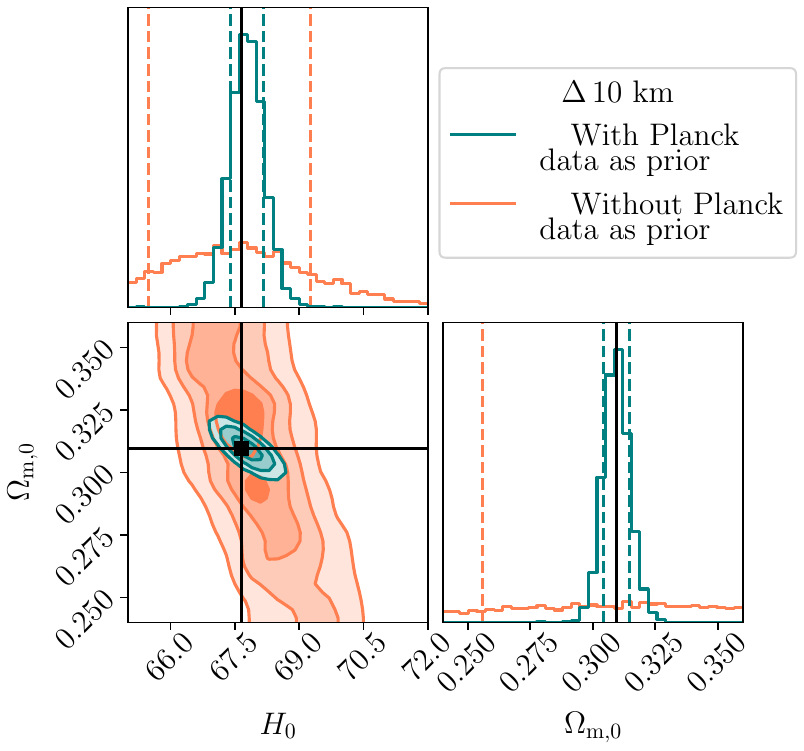}
    \includegraphics[width=.5\textwidth]{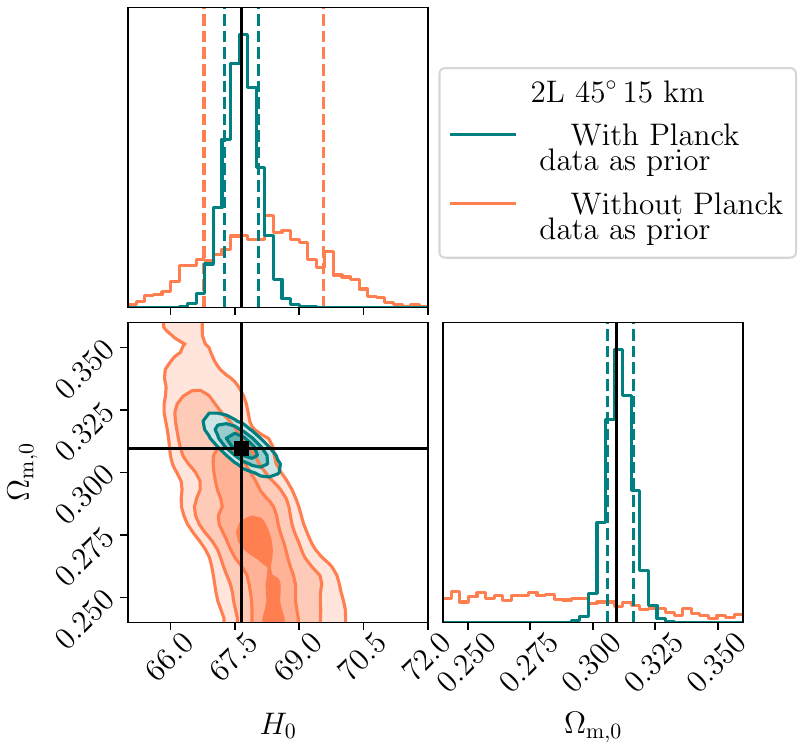}
    }
    \caption{Results of the joint inference on the cosmological parameters $H_0$, $\Omega_{\rm m,0}$ employing GW+GRB events detected in $5$ years of observations by the $10$ km triangular (left panel) or the   2L-15km-$45^{\circ}$ (right panel) ET configurations,  with the EM counterpart detected by THESEUS. Vertical dashed lines represent the $68\%$ CI of each distribution, while the black solid lines label the fiducial values.}
    \label{fig:multimessenger_GRB_H0Om0_only}
\end{figure}

We have repeated the analysis showing also the impact of using the Planck 2018 \cite{Planck:2018vyg} posterior distributions as a prior.
The corresponding forecasts for the accuracy on $H_0$ and $\Omega_M$ are  shown in figure~\ref{fig:multimessenger_KN_H0Om0_only} for GW+KN,  and  in \ref{fig:multimessenger_GRB_H0Om0_only} for GW+GRB.
In particular, from these results we find that, when adding the Planck prior, the relative error on $H_0$ in the GW+GRB case (and 5~yr of observations) becomes $1.1\%$ for both the 10-km triangle and for 2L-15km misaligned (so in this case the result is clearly dominated by the Planck prior), while for GW+KN (and 1~yr of observations) it becomes $0.7\%$ for  the 10-km triangle and $0.4\%$ for 2L-15km misaligned.

We stress, however, that on top of the uncertainties related to the BNS rate and to the facilities and telescope time that will be available for the electromagnetic follow-up  of  GW events,
the analysis performed in the GW+GRB case has been conducted under the  assumption that the relative error on the luminosity distance can be approximated as $2/$SNR, as assumed also in previous studies \cite{Belgacem:2019tbw,Branchesi:2023mws}. While the $\text{SNR}^{-1}$ scaling theoretically represents a good proxy for most of the parameter space, it reaches important limitations in some specific corners \cite{Vallisneri:2007ev}.
In particular, 
the GRB counterpart associated to a GW event can be detected only for small inclination angles, since the emission mechanism typically ejects the material in the same direction of the angular momentum of the binary. This circumstance introduces two non-trivial challenges in the determination of the luminosity distance uncertainty: in the limit of face-on/face-off binaries, the Fisher Matrix becomes ill-conditioned, leading to an unfaithful reconstruction of the infamous $d_L$--$\iota$ degeneracy \cite{Iacovelli:2022bbs,Dupletsa:2024gfl} and to a regime where the $2/$SNR approximation is not appropriate anymore; additionally, this intrinsic correlation gets stronger as the inclination angle approaches smaller values.
The two above-mentioned effects, which inevitably appear in the framework of GW+GRB observations, represent non-negligible sources of bias that can jeopardize the whole inference if not taken into account for. A comprehensive description of how to properly deal with such issues has been recently proposed in \cite{Mancarella:2024qle}, where the Fisher Matrix limitations are overcome and the $d_L$--$\iota$ correlation is mitigated by the underlying fact that the EM detection implicitly select a specific corner of the parameter space, see also the discussion in section~\ref{sec:div9_fim_improvements}.

Besides model-based cosmological parameter estimation with bright sirens, one can also employ model-independent methods to reconstruct the distance-redshift relation and test for deviations of the standard cosmological model or some of its basic assumptions. That can be done for example with the use of Gaussian Process as in \cite{Belgacem:2019zzu, Hogg:2020ktc,Cozzumbo:2024vxw}, or by combining GW distances with other distance indicators, such as standard candles and angular diameter distances from the large-scale structure \cite{Matos:2023jkn,Cozzumbo:2024vxw}. In ref~\cite{Cozzumbo:2024vxw}, it was shown that fewer than 40 GW-GRB detections, in combination with forthcoming cosmological surveys, will enable unprecedented precision on $H_0$ and 
$\Omega_m$, and accurately reconstruct the DE density evolution. Taking the ratio of different distances is an interesting approach as it avoids the intrinsic degeneracies present in the distance-redshift relation between standard cosmological parameters, notably $\Omega_m$, and effects that can modify the GW luminosity distance or other cosmic distances, such as violations of the transparency of the Universe (cosmic opacity) or modified gravity (see 
section~\ref{sect:ModGWpropdiv2}). In \cite{Matos:2023jkn}, it was shown that, in combination with other distance measurements expected for the near future with surveys like Euclid, LSST, DESI and Roman, bright sirens from ET would allow tests of cosmic opacity together with deviations in the GW propagation due to modified gravity models that can be achieved with percent level precision, in various redshift bins up to $z = 0.9$, with no assumption on the redshift evolution of the mechanisms causing deviations from standard physics.

As shown in~\cite{Alfradique:2022tox}, bright sirens will be useful tracers of both the density and velocity fields in cosmology.  In particular, a combination of ET sirens with EM counterparts with LSST allows for a 30\% improvement on the measurements of $H_0$ with ET alone, but also percent-level measurements of $\Omega_{k0}$ (an improvement of over one order of magnitude). We can also get good constraints on the growth of structure in particular on $\sigma_8$ and the growth-rate index $\gamma$.
In~\cite{Alfradique:2022tox} it was also proposed the use of preliminary GW distances, obtained for the alerts for follow-ups, to inform the optimal telescope exposure time. It was shown that this technique reduces the necessary telescope time to obtain the same kilonova completeness by around 25\%.  This could be useful since in the ET era the bottleneck for bright sirens catalogs will possibly be due to available telescope time.

\paragraph{Dark sirens.}\label{sect:darksirensdiv2}
While bright sirens provide the most direct method  of using  resolved sources for late--time cosmology, the other side of the coin is the difficulty in the observation of the EM counterpart.
Events with direct EM counterparts will constitute only a tiny fraction of the total detected events, with a strong selection from the EM side limiting these sources to redshifts generally of order $z\lesssim \mathcal{O}(1)$~\cite{Belgacem:2019tbw, Ronchini:2022gwk}. On the other hand, an unprecedented opportunity offered by ET will be the detection of compact binary mergers across all cosmic history, in particular BBH mergers up to redshift $\mathcal{O}(20)$~\cite{Iacovelli:2022bbs,Branchesi:2023mws}. At redshifts below the star formation peak, ET can in particular detect almost $100\%$ of the population~\cite{Iacovelli:2022bbs,Branchesi:2023mws}. 
The vast majority of these events will not have EM counterparts but contain precious information about the evolution of the Universe, especially at redshifts that are not accessible to the current generation of GW detectors, not even at their most advanced sensitivity. This will open a window on the evolution of the Universe well before the Dark Energy dominated era and beyond the star formation peak, possibly even in region where compact objects formed by standard stellar--evolution scenarios are expected to be absent \cite{Haemmerle:2020iqg,Klessen:2023qmc}. 
 Even if their actual high--redshift nature will have to be carefully assessed on an event-by-event basis~\cite{Ng:2021sqn,Ng:2022vbz,Mancarella:2023ehn,Fairhurst:2023beb}, inferring their statistical distribution might lead to clearer signatures~\cite{Ng:2022agi}. 
Therefore, methods that target the use of the full sample of detected events will be crucial for cosmography with ET. Several of those are actively being developed by the community, and are generally referred to as ``dark siren'' methods.
The general principle behind dark sirens is the use of prior information on the properties of the underlying distribution of sources, of which the observed catalog is a biased sample, to break the degeneracy between source--frame masses and redshift in the GW waveform at the statistical level~\cite{Moresco:2022phi}.

\subparagraph{The line-of-sight redshift prior.}
The GW observations provide a direct measurement of the luminosity distance. For a dark siren measurement of $H_0$ or the other cosmological parameters, the GW distance information needs to be convolved with some ``prior'' information on the redshift. This prior comes from the underlying cosmology and binary merger rate evolution (uniform part), as well as potential host galaxies (non-uniform part). Cosmology inference algorithms allow us to construct this prior along each line-of-sight. This line-of-sight redshift prior essentially consists of an in-catalogue part and an out-of-catalogue part. The in-catalogue part is constructed using redshifts and luminosities of the galaxies in the region. The luminosity (in a certain observation band) is used to effectively weight the galaxy in proportion to the probability of sourcing a merger. Astrophysical assumptions go into this construction, given that the luminosity in a certain band may trace a certain aspect of the astrophysics. The out-of-catalogue is used to model the galaxies which are beyond the catalogue, due to their distance or due to low intrinsic brightness. This part usually depends on galaxy luminosity parameters (or Schechter function parameters) in addition to the assumed cosmological model. A detailed prescription of the construction of the line-of-sight redshift prior can be found in~\cite{Gray:2023wgj}. In the following paragraphs we describe the galaxy catalogs and surveys which can be used to construct this prior in the 3G era.

\subparagraph{Galaxy catalogs.}
The lack of an EM counterpart poses a tough challenge in GW cosmology. However, as first proposed by Schutz in 1986 \cite{Schutz:1986gp}, statistical approaches that rely on available galaxy catalogs can counterbalance the missing source's redshift information. Specifically, the potential host galaxies that fall in the localization error volume of a detected GW event display an inhomogeneous distribution that can be employed as redshift prior in the Bayesian framework.
Contrarily to bright sirens, a single or few of such events are likely unable to deliver tight constraints on the cosmological parameters. As a statistical tool, the efficiency of this approach strongly relies on the size of the sample of dark sirens used, and we expect the precision to shrink with the square root of the number of events. 
In the past decade, this methodology has successfully been applied both to simulated \cite{DelPozzo:2011vcw,Chen:2017rfc,Gray:2019ksv,Gair:2022zsa,Borghi:2023opd,Muttoni:2023prw,Mastrogiovanni:2023emh} and real data 
with different galaxy catalogs, including GLADE/GLADE+~\cite{Dalya:2018cnd,Dalya:2021ewn,LIGOScientific:2018gmd,LIGOScientific:2019zcs,Finke:2021aom,LIGOScientific:2021aug,Gray:2023wgj,Naveed:2025kgk}, the Dark Energy Survey~\cite{DES:2019ccw,DES:2020nay}, the DESI Legacy Survey~\cite{Palmese:2021mjm}, the DECam Local Volume Exploration Survey~\cite{Alfradique:2023giv}, DESI~\cite{DESI:2023fij}, and to the PSZ2 and eRASS galaxy cluster catalogs~\cite{Beirnaert:2025wcx}.

\subparagraph{Spectral Sirens.}
A complementary possibility  (also called the ``spectral sirens'' method) is the use of features in the source--frame mass distribution~\cite{Taylor:2011fs,Taylor:2012db,Farr:2019twy,Mancarella:2021ecn,Finke:2021eio,Ezquiaga:2022zkx}. The population properties, in particular, the mass and redshift distributions, are given in source--frame; for example, one can predict the probability that a black hole has a certain mass, or forms at a certain redshift, from population synthesis codes~\cite{Breivik:2019lmt,Giacobbo:2017qhh,Stevenson:2017tfq}.
On the other hand, GW detections measure detector--frame quantities (redshifted masses and luminosity distance). The population properties reconstructed in source--frame from detector--frame observations will thus be reshaped by a change of the cosmology. In case source--frame distributions are informative, the degeneracy between source--frame masses and redshift in the GW waveform can be broken by combining an ensemble of sources. 
This can drive constraints on cosmology, as the presence of preferential scales breaks the mass--redshift degeneracy at the statistical level.

\subparagraph{Potential systematic effects.}
Recent advances have led to the development of pipelines incorporating information from both the mass distribution and galaxy catalogs~\cite{Mastrogiovanni:2023emh,Gray:2023wgj,Borghi:2023opd}. Thus one no longer needs to assume a fixed mass distribution for an inference using galaxy galaxy catalogs, which may otherwise have biased the measurement. Taking care of this leading order systematics exposes us to smaller sources of potential systematic uncertainties. These can come from the GW data, the EM data, or the underlying assumptions. It is crucial to understand and address even the smallest of the potential systematic effects as the statistical power of the data increases, in order to converge towards an unbiased measurement. On the GW side, mis-modelling the source population or its redshift evolution could pose a problem~\cite{Pierra:2023deu}. On the EM side, it is important to understand redshift measurement errors~\cite{Turski:2023lxq}. Uncertainties in the measurement of galaxy luminosities and the measurement of the Schechter luminosity function parameters (and their evolution) come next~\cite{Turski:2025flk}. It has recently been shown that the results could be biased due to incorrect assumptions on the luminosity weighting of galaxies~\cite{Perna:2024lod,Hanselman:2024hqy}. A large body of work is underway largely in the context of upcoming LVK observations~\cite{Agarwal:2024hld}, to understand and address small systematic effects and ensure an accurate measurement as we move towards precision. Targeted studies to determine the potential of these techniques for ET will be necessary in the future.

\begin{figure}[t]
    \centering
    \includegraphics[width=.45\textwidth]{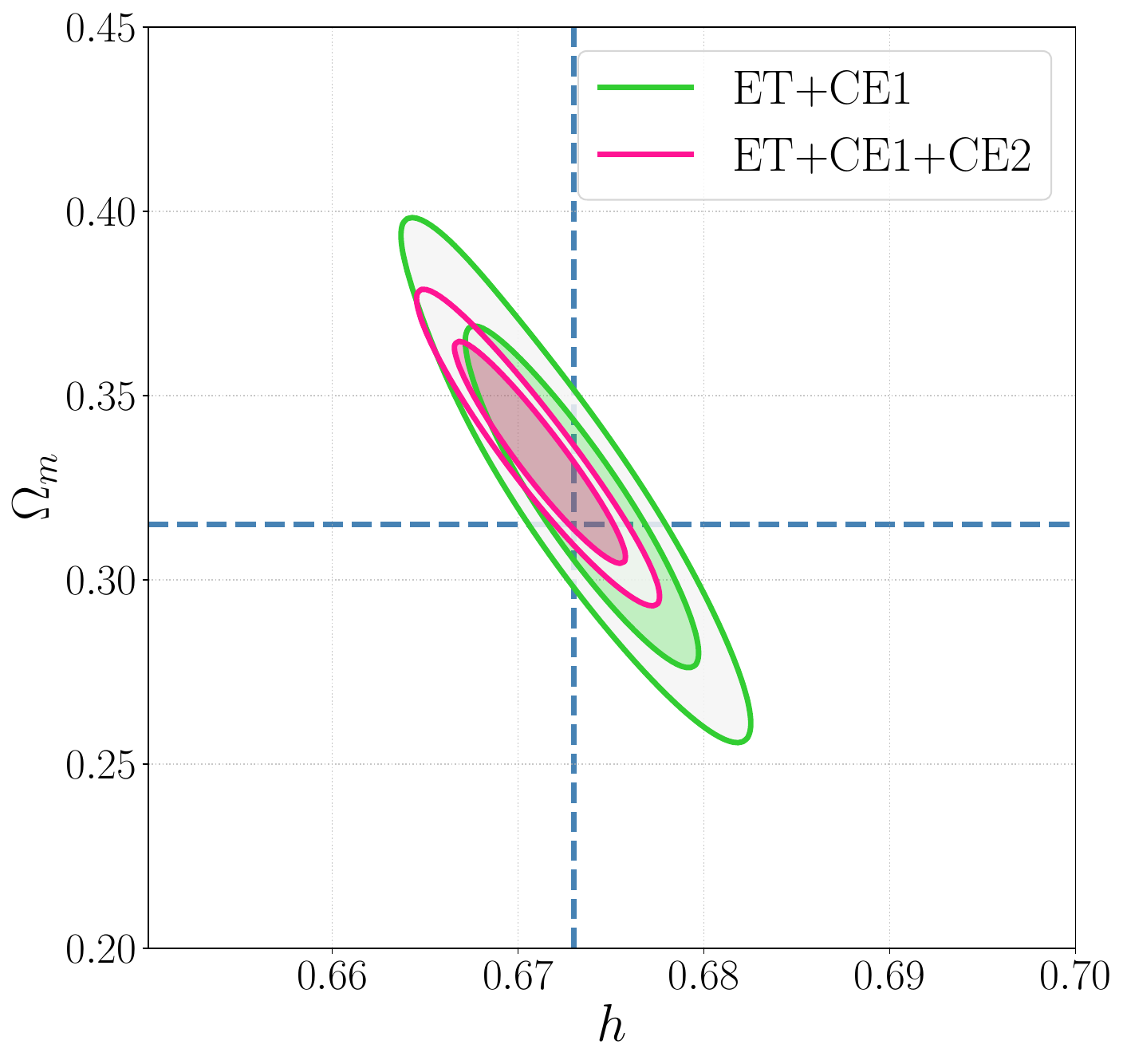}
    \includegraphics[width=.44\textwidth]{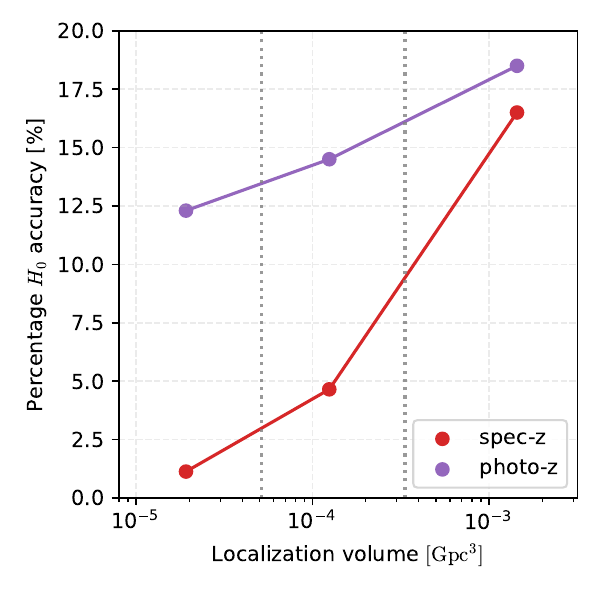}
    \caption{Left panel: Posterior joint distribution on $H_0$ and $\Omega_{\rm m,0}$ obtained by employing BBHs as dark sirens and cross correlating the loudest GW event localization volumes with a  galaxy catalog. The contours, which refer the $68\%$ and $90\%$ confidence level, are shown for two networks of 3G detectors assuming one year of continuous observations. Here ET is in a $10$ km-triangular configuration and CE1 and CE2 correspond to $40$ km  and $20$ km Cosmic Explorer, respectively. Figure  from \cite{Muttoni:2023prw}. Right panel: Dependence of the accuracy on $H_0$ on the GW event localization volume with a photometric (\texttt{photo-z}) or spectroscopic (\texttt{spec-z}) galaxy catalog.}
    \label{fig:H0_dark_sirens_galaxy_catalog_3G}
\end{figure}

The 3G era is expected to bring $\mathcal{O}(10^{4-5})$ BBH detections per year. Albeit computationally challenging, this abundance naturally meets the conditions that make the galaxy catalog approach a powerful statistical tool. Indeed, current forecasts predict that a sub-percent measurement of  $H_0$  and a $\sim10\%$ measurement of $\Omega_{\rm m,0}$  can confidently be attained by 3G detectors by employing only the loudest dark sirens observed in one year \cite{Muttoni:2023prw,Zhu:2023jti}. Yet, as shown in figure~\ref{fig:H0_dark_sirens_galaxy_catalog_3G}, reaching this precision requires a mandatory joint effort between more than a single 3G detector, as the localization power plays a crucial role in the statistical machinery. 

We stress, however, that these predictions are subject to variability, which depends on the properties of the galaxy catalog adopted in these studies. A promising candidate will be the database created from the observations of the Vera Rubin Observatory. It is predicted to contain $10^{10}$ galaxies up to a redshift  $z \leq 6$, covering 20,000 square degrees. For about $4\times 10^9$ galaxies, six-band photometric measurements and shape measurements will be available, making the estimation of photometric redshifts much more precise \cite{LSSTScience:2009jmu}.  Early estimations show that a faint sample of all types of galaxies having photometric information in 6 bands would have a statistical photo-$z$ error of $\Delta z\sim 0.05\,(1+z)$ per galaxy \cite{LSST_photo_z}. Even assuming improved photometric redshift estimating techniques, the errors from these measurements would be an important source of uncertainty for the dark siren results. As demonstrated in \cite{Borghi:2023opd}, the accuracy in redshift determination plays a crucial role in the estimate of cosmological and astrophysical parameters, since already at O5 sensitivities the accuracy on the determination of $H_0$ can improve down to the percent level when a spectroscopic catalog is considered (see figure~\ref{fig:H0_dark_sirens_galaxy_catalog_3G}, right panel). Hence, a spectroscopic survey would be crucial to maximising the potential of this method.

The galaxy database from the Vera Rubin Observatory, therefore,  should be complemented with other galaxy catalogues to provide spectroscopic full-sky coverage as deeply as possible. From this point of view, several ongoing and planned galaxy surveys can be exploited. Currently, the largest spectroscopic galaxy surveys that are taking data are the Dark Energy Spectroscopic Instrument survey (DESI, from the ground, \cite{DESI:2023dwi}) and the ESA mission Euclid (from space, \cite{EUCLID:2011zbd}). DESI is planned to observe order of $10^7$ galaxies at $0<z<3.5$ over $\sim14000$ square degrees, at a depth of 24, 23.4, and 23 for AB magnitudes g, r, and z, respectively; on the other hand, covering $\sim14000$ square degrees, Euclid will provide 1.5 billions photometric redshifts in the range $0<z<2$ and order of 30 millions spectroscopic redshifts at $0.84<z<1.88$, down to a Euclid H$_E$ magnitude of 24. In the meanwhile, future spectroscopic surveys are already being planned and, pushing significantly deeper in redshift and completeness, they could constitute a fundamental resource to exploit the dark sirens provided by ET; in this respect, the most promising ones are the Wide-Field Spectroscopic Telescope (WST \cite{Mainieri:2024kkb}) and Spec-S5\footnote{See \url{https://www.spec-s5.org/pubs/}.} facilities.

Besides utilising existing data, another possibility is performing dedicated galaxy surveys for the best localised GW events without a counterpart to maximise their impact on the dark siren method. In parallel to that, dedicated spectroscopic follow-up surveys could be conducted for the best-localized events.

Formally, dark sirens analyses are conducted with a hierarchical Bayesian approach~\cite{Loredo:2004nn,Mandel:2018mve,Vitale:2020aaz}, which allows the propagation of uncertainties on the individual events to the population properties. Crucially, when using the population properties to infer cosmology, one must take into account the fact that those are actually unknown, but determine the selection bias of the experiment, which needs to be carefully deconvolved to obtain unbiased estimates of the cosmological parameters. This is usually done with large injections campaigns to determine the sensitivity of the instrument to the targeted population~\cite{Tiwari:2017ndi,Farr:2019rap,Essick:2022ojx}. The scaling of this technique to ET needs to be assessed, both in terms of accuracy and computational cost. Furthermore, the likelihood evaluation, involving the computation of multi--dimensional Monte Carlo integrals at each iteration to marginalize over the individual source properties and/or the construction of redshift priors from the interpolation of large galaxy catalogs, is computationally demanding. Alternative approaches such as simulation--based~\cite{Leyde:2023iof} and/or machine--learning leveraged pipelines~\cite{Stachurski:2023ntw}  and/or GPU-accelerated codes \cite{Tagliazucchi:2025ofb}, might be beneficial for tackling these challenges. 

\begin{figure}[t]
  \includegraphics[width=75mm]{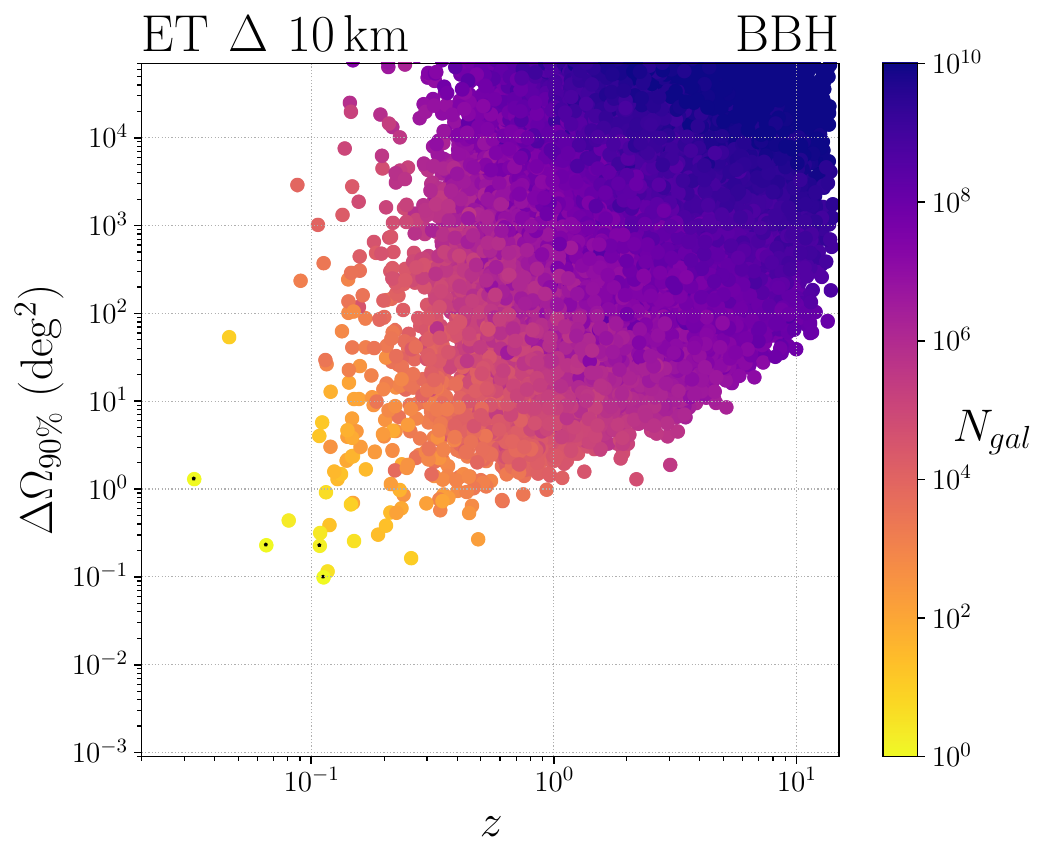}  
  \includegraphics[width=75mm]{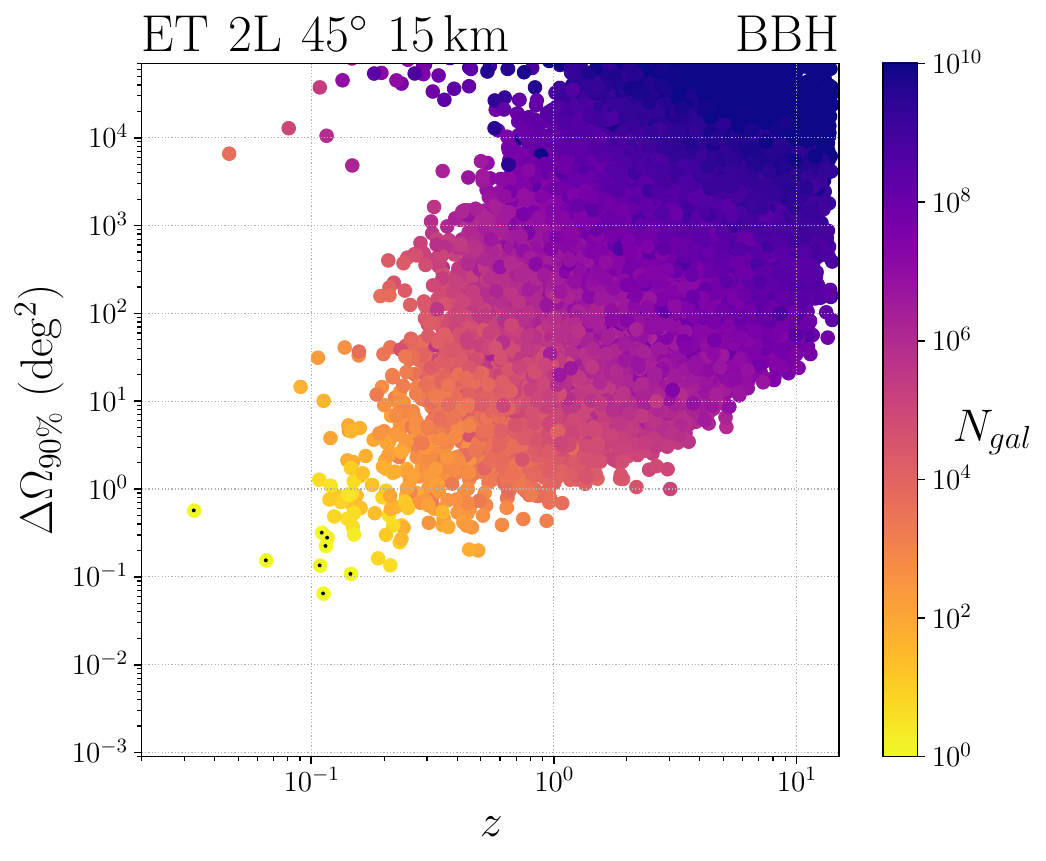} 
    \caption{Localization capabilities for BBHs of ET  in its triangular (left panel) and 2L-15km-$45^{\circ}$  (right panel) configurations. The color scale denotes the number of galaxies expected in the $90\%$ localization volume. Events marked with black dots are localized to one galaxy only.}
    \label{fig:ET_loc_BBH}
\end{figure}

\begin{figure}[t]
    \centering
    \includegraphics[width=\textwidth]{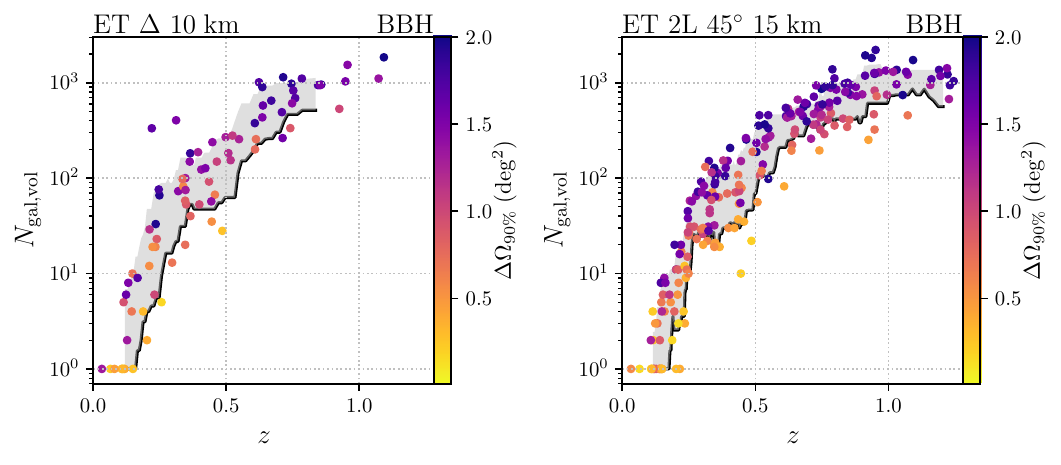}
    \vspace*{-0.5cm}
    \caption{Number of galaxies inside the localization volume of ET golden BBH events ($\Delta\Omega<2\,\mathrm{deg^2}$) for the triangular (left) and 2L (right) configuration. The color bar represents the 90\% sky localization area. The black line and shaded region indicate the 16th and 84th percentiles, respectively. }
    \label{fig:ET_NgalVol}
\end{figure}

We can identify several milestones that ET can bring in this context, regardless of the technical details. The construction of a redshift prior from galaxy catalogs benefits in particular from the accurate sky localization.\footnote{Similar arguments hold for the cross--correlation of resolved sources with other tracers of the LSS, see section~\ref{sec:GWLSSCC}, as well as for the identification of a direct counterpart, see section~\ref{sec:brightSirens}.}
Figure~\ref{fig:ET_loc_BBH} shows the localization capabilities of ET in the triangular and 2L configurations.\footnote{We refer to ref.~\cite{Branchesi:2023mws} for the details on the catalogs, software, and detector shape and orientations used.} In particular, we show the $90\%$ sky localization area as a function of redshift, while the color refers to the number of galaxies contained in the localization volume, assuming a constant galaxy density of $10^{-2}\, \rm galaxies/Mpc^{3}$. In particular, ET has the potential of localizing $\mathcal{O}(10^3)$ sources per year to better than $10\, {\rm deg}^2$~\cite{Branchesi:2023mws}.
Remarkably, figure~\ref{fig:ET_loc_BBH} also shows that ET alone could allow pinpointing the host galaxy of the merger at a rate of a few per year without the need of a direct EM counterpart. Events with a single host galaxy are further marked with a black dot in figure~\ref{fig:ET_loc_BBH}. This scenario alone would lead to a  determination of the Hubble constant, and is a unique possibility of ET. Figure~\ref{fig:ET_NgalVol} shows the number of galaxies enclosed in the GW localization volume of ET  golden BBH events (i.e.,  with $\Delta\Omega<2\,\mathrm{deg^2}$) in the triangular and 2L configurations using the catalog of massive MICEcat galaxies, as in \cite{Borghi:2023opd}. In particular, ET has the potential of localizing $\mathcal{O}(10^2)$ sources per year to better than $2\, {\rm deg}^2$~\cite{Branchesi:2023mws}. We note that, besides cosmology, this could bring invaluable information about the relation between galaxy properties and GW emission \cite{Perna:2024lod}.
Indeed, assigning a weight to galaxies according to their properties such as mass or star formation rate can help increasing the accuracy on $H_0$ when using also GW sources with a large number of galaxies in their localization volume~\cite{Finke:2021aom,LIGOScientific:2021aug}. However, the dependence on the galaxy properties will have to be correctly modeled and marginalized over, in order to avoid systematics~\cite{Perna:2024lod}.
%

Besides subsamples of ``golden'' events, as already remarked, it is important to stress that ET will detect almost the entire population of BBHs below the star formation peak, and more than $90\%$ of BBHs of stellar origin across all redshifts~\cite{Iacovelli:2022bbs,Branchesi:2023mws}. In this large--statistics regime, information from sources with poorer sky localization will still contribute to reconstruct precisely the source--frame mass distribution, hence to determine cosmology~\cite{Chen:2024gdn}. In this context, the possible systematics related to the correct modeling of source--frame features~\cite{Pierra:2023deu}, and their correlation to redshift evolution~\cite{Rinaldi:2023bbd}, will need to be addressed. However, we stress that the presence of features in the mass spectrum can in principle be inferred directly from data, without the need of a specific modeling, using non--parametric approaches~\cite{Rinaldi:2021bhm,Ray:2023upk,Rinaldi:2023bbd}. This will allow a robust measurement of the cosmic expansion history.

\subparagraph{Love Sirens.}\label{sect:Lovesirensdiv2}
Neutron stars taking part in compact binary coalescences, whether as BNS or NSBH systems, undergo tidal deformation during the later stages of the inspiral process. The tidal deformability parameter $(\Lambda)$ is related to the compactness $(\mathcal{C})$ of the neutron star via the second tidal Love number $(k_2)$ by,
\begin{equation} \label{eq_lam_C}
    \Lambda = \frac{2}{3} k_2 \mathcal{C}^{-5},
\end{equation}
where $\mathcal{C} = m/R$, where $R$ is the radius of the NS and $m$ its source-frame mass \cite{Hinderer:2009ca}.  The parameters $k_2$ and $\mathcal{C}$ depends on the equation of state governing the NS (see section~\ref{section:div6} for details). These deformations leave an imprint on the GW amplitude and phase;  in particular, the dominant effect of tidal deformation appears in the phase at the fifth post-Newtonian order, characterized by a combination of the individual tidal deformability parameters (note that $\Lambda=0$ for BHs) and the mass ratio of the binary \cite{Wade:2014vqa}. This combination, called the reduced tidal deformability $\tilde{\Lambda}$ (see \eq{deftildeLambdadiv6} or \eq{eq:tildeLam_def} for the explicit expression), is the best-measured tidal parameter using GW. The next-to-leading order contribution to the GW phase is characterized by another linear combination of individual $\Lambda$s and the mass ratio, denoted by $\delta \tilde{\Lambda}$, see \eq{eq:tildeLam_def}. The contribution of $\delta \tilde{\Lambda}$ to the phase is subdominant compared to that of $\tilde{\Lambda}$, even at the sixth post-Newtonian order, which is why it is not expected to be constrained very well with GW observations. Furthermore, if the neutron stars are spinning, the individual $\Lambda$s also contribute to the phase from spin-induced quadrupolar deformations which appear at the 2.5 post-Newtonian order and beyond \cite{Marsat:2014xea,Bohe:2015ana,Nagar:2018zoe,Dietrich:2019kaq}. Given these effects, one can constrain the individual tidal deformability of the two compact objects using GW detections, as we detail in section~\ref{section:div6}. 

With an established NS equation of state, inferred $\Lambda$ can be converted to constraints on source-frame mass using \eq{eq_lam_C}. As GW observations also provide precise constraints on detector-frame mass, the measured detector-frame mass can be compared to the inferred source-frame mass, giving constraints on the redshift associated with the source \cite{Messenger:2011gi}. Together with  the measurement  of the luminosity distance $d_L$ obtained from the GW signal, one can then obtain measurements of the cosmological parameters. Furthermore, exploiting the post-merger signals of BNS mergers can enable the resolution of redshift-mass degeneracy by exploiting spectral features dependent on the source-frame masses of the binary \cite{Messenger:2013fya}. This method, which leverages the constraints on tidal deformability parameters, knowledge of the neutron star equation of state, and, possibly, the spectral features from neutron star post-merger, to break the mass-redshift degeneracy and obtain a redshift estimate, is referred to as the Love siren approach.

If using only the inspiral part of the signal, the redshift constraints obtained using the Love siren method depend on the precision with which $\tilde{\Lambda}$ and the detector-frame masses are inferred. The sensitivity improvement in ET at both low and high frequencies greatly improves these two measurements. Specifically, for $\mathcal{O}(100)$ BNS events detected with the triangular ET configuration every year, the fractional error in $\tilde{\Lambda}$ is expected to be less than $10\%$ and that in chirp mass is expected to be less than $\mathcal{O}(10^{-5})$ \cite{Iacovelli:2022bbs}. With the 2L-$45^{\circ}$ geometry, this number can increase to $\mathcal{O}(10^3)$ every year \cite{Iacovelli:2023nbv}. With the triangular ET geometry, the redshift can be measured with a precision of $\mathcal{O}(10\%)$ for systems up to $z\sim1$ \cite{Messenger:2011gi,Li:2013via,DelPozzo:2015bna}. With the post-merger spectra, this can be achieved for systems up to $z\sim0.04$ \cite{Messenger:2013fya}. In fact, the post-merger signal for a BNS system can be detected with an SNR of 10 for systems up to $\sim 90$ Mpc with ET alone, resulting in $\mathcal{O}(1\%)$ measurement of the post-merger peak frequency $f_2$ \cite{Prakash:2023afe}. Together with CE, the redshift can be resolved to $\sim1\%$ for $\mathcal{O}(10^4)$ yearly detections, which translates to $0.1\%$ error in $H_0$ and $0.61\%$ error in $\Omega_m$ in a year of observation \cite{Dhani:2022ulg} (see also~\cite{Jin:2022qnj,Li:2023gtu}).

The Love siren method represents an approach to cosmological parameter estimation that only relies on the GW signal, requiring  a robust understanding of the neutron stars equation of state, expected to be significantly enhanced by forthcoming observatories \cite{Finstad:2022oni,Huxford:2023qne,Iacovelli:2023nbv,Gupta:2023lga}. An equation of state agnostic approach can also be used,  that jointly estimates the equation of state and the cosmological parameters \cite{Ghosh:2022muc}. However, caution is warranted in post-merger spectrum-based forecasts, which require improved modeling of gravitational waves from hypermassive neutron star remnants. 

\paragraph{Cosmic dipole tension.}
The Earth's motion relative to the Hubble flow generates a dipole anisotropy in both the temperature of the CMB \cite{Planck:2013kqc,Planck:2018vyg,Ferreira:2020aqa} and in the angular distribution of electromagnetic sources observed at cosmological distances \cite{Colin:2017juj,Siewert:2020krp,Secrest:2022uvx}. This type of angular anisotropy is typically called the ``cosmic dipole''. There is currently a $\sim 5 \sigma$ tension between the estimation of the cosmic dipole from the early Universe (CMB) and the one inferred from the angular distribution of active galactic nuclei (AGN) \cite{Planck:2018vyg,Colin:2017juj}. 
Whereas this tension is robust to effects like the redshift evolution of sources \cite{Dalang:2021ruy,vonHausegger:2024jan},  it has been shown that systematic uncertainties related to the spatial distribution of sources and the sky coverage may have been under-estimated, and that accounting for those may remove the tension \cite{Abghari:2024eja}. Hence nowadays this tension is still open for debate
(see also the discussion in~\cite{Boubel:2024cmh,Secrest:2025wyu} and the references therein). 

GW sources observed with 3G detectors offer a new avenue to solve this cosmic dipole tension. The cosmic dipole could be estimated from the stochastic GW backgrounds generated by cosmological and astrophysical sources (see section~\ref{sec:stochback} for more details), but can also be estimated from the angular distribution of GW sources \cite{Mastrogiovanni:2022nya, Grimm:2023tfl}. ET embedded in a detector network with CE will provide $\sim 3 \cdot 10^4$ BBH sources per year with a 90\% localization area lower than 10 deg$^2$. The angular distribution of these well-localized sources can be used to detect and estimate the cosmic dipole.

By considering the angular number density of GW sources, with just  2-3 years of observations, 3G detectors have a 75\% probability of detecting the cosmic dipole if its value is consistent with the one inferred from AGN \cite{Mastrogiovanni:2022nya}. The probability of detecting the dipole increases to 100\% if we consider 10 years of joint observations. Therefore, with just few years of observations, there is the concrete possibility of ruling out the AGN value of the cosmic dipole and solving this tension. The possibility of detecting and measuring the cosmic dipole are enhanced if also binary neutron star (BNSs) sources are considered. For this class of sources, it is also possible to inspect the presence of a cosmic dipole by studying the angular dependence of the detector-frame mass spectrum of BNSs, which is also expected to be modified by the cosmic dipole. Including BNSs and the angular distribution of the detector-frame mass spectrum, it will be possible to enhance the probability of detecting the cosmic dipole by 30-50\% \cite{Grimm:2023tfl} for a given observing time, thus making the detection of an AGN cosmic dipole in just 2-3 years of observation almost certain.

{Furthermore, as discussed in detail in section~\ref{sec:Anisotropies of Astrophysical Backgrounds}, in~\cite{ValbusaDallArmi:2020ifo} it has been shown that a detection of the dipole of the AGWB at different frequencies with ET and CE would allow to measure the dipole of the AGWB emitted by BBH and to distinguish between the kinematic, intrinsic and shot noise contributions, see figure~\ref{fig:AGWB_dipole} for more details. The component separation of the dipole of the AGWB which could be done by ET and CE is crucial to understand if the cosmic dipole tension is due to a relative velocity between the CMB and LSS frame (by comparing the CMB and AGWB kinetic dipoles) or to an anomalous intrinsic dipole of the CMB or LSS.}

\paragraph{The astrophysical background as a cosmological probe.} \label{subsec:astro_sgwb_cosmo_probe}

\begin{figure}[t]
    \centering
    \includegraphics[width=.9\textwidth]{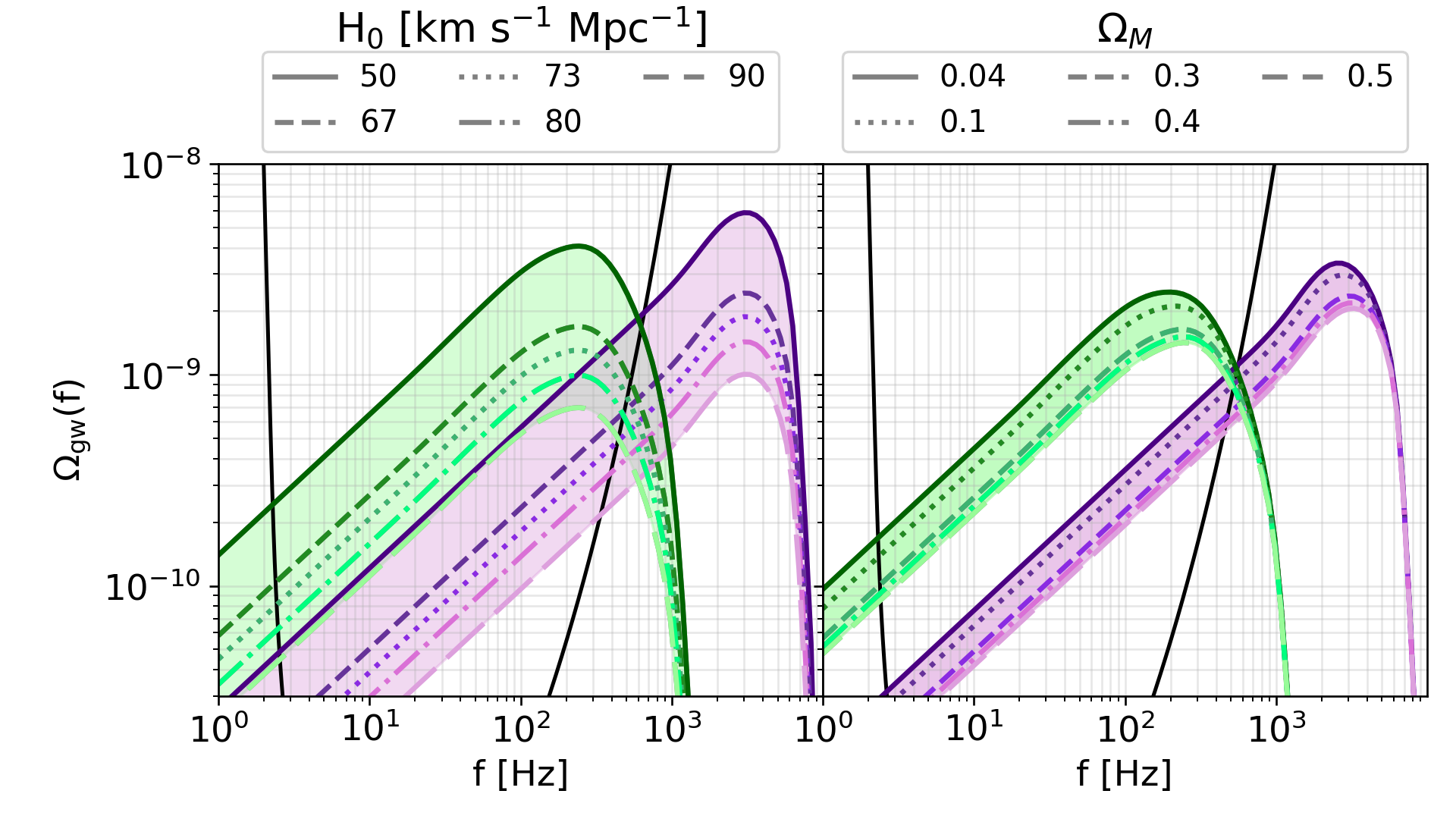}
    \caption{Effect of varying the value of $H_{0}$ and $\Omega_{M}$ on the energy density of the stochastic background from BBHs (green) and BNSs (purple). The astrophysics is fixed to the models described in \cite{Capurri:2021zli,Boco:2020pgp}. The black curve is the PLS of ET for the triangular 10 km configuration \cite{Branchesi:2023mws}.}
    \label{fig:sgwb_cosmo_params}
\end{figure}

Another observable that can be used to constrain cosmological parameters with ET is the astrophysical GW background (AGWB) produced by coalescing binaries. The frequency spectrum of the AGWB is determined by the complex interplay of various astrophysical factors, as discussed in section~\ref{section:div3}, but also by the underlying cosmology. The region of the spectrum which contains most of the information is the high-frequency peak, whose shape and position are very sensitive to all the various astrophysical and cosmological factors at play. For instance, in figure~\ref{fig:sgwb_cosmo_params} we show the effect of varying the value of the Hubble parameter and the matter density parameter on the peak of the AGWB produced by BBHs and BNSs. With the sensitivity of ET in the 1-100 Hz range, the frequency spectrum of the astrophysical GWB will be measured precisely. This, in turn, will open the possibility for a Bayesian reconstruction of both astrophysical and cosmological parameters, in line with the methodology outlined in~\cite{Capurri:2023jsy} for the specific case of the GWB from BNSs. This approach complements existing methods such as GW sirens, discussed in the previous sections, which constitute the main strategy to constrain cosmological parameters with GWs. By employing these techniques concurrently, it becomes feasible to mitigate degeneracies and potentially refine parameter constraints further, particularly if posterior contours are orthogonal.

\subsubsection{Modified GW propagation}  \label{sect:ModGWpropdiv2}

\paragraph{Observables for the dark-energy sector.}
In $\Lambda$CDM, on sufficiently large scales where cosmological perturbation theory applies, one  performs a separation between a homogeneous  Friedmann-Lemaitre-Robertson-Walker (FLRW)   background, and  scalar, vector and tensor perturbations  over it.  Whenever gravity is modified on cosmological scales, both the background evolution and  scalar and tensor perturbations are modified (while, in $\Lambda$CDM and in most modified gravity models, vector perturbations usually only have decaying modes and are irrelevant).
At the level of the background evolution, the effect of modified gravity
is encoded in a  dark energy  (DE)   density $\rho_{\rm DE}(z)$ and a DE pressure 
$p_{\rm DE}(z)$. One can then introduce the DE 
equation of state $w_{\rm DE}(z)$, defined by $p_{\rm DE}(z)=w_{\rm DE}(z)\rho_{\rm DE}(z)$; $\Lambda$CDM  is recovered for $w_{\rm DE}(z)=-1$. Using the conservation of the DE energy-momentum tensor, one finds that
the DE density is given as a function of redshift by 
\begin{equation}\label{4rdewdeproofs}
\rho_{\rm DE}(z)  =\rho_0 \Omega_{\rm DE}\, \exp\left\{ 3\int_{0}^z\, \frac{d\tilde{z}}{1+\tilde{z}}\, [1+w_{\rm DE}(\tilde{z})]\right\}\, ,
\end{equation}
where $\Omega_{\rm DE}=\rho_{\rm DE}(0)/\rho_0$ is the DE density fraction and $\rho_0=3H_0^2/(8\pi G)$ is the critical density today.
At the background level, the properties of DE are therefore encoded just in one function $w_{\rm DE}(z)$.
The corresponding expression for the luminosity distance   is
\begin{equation}\label{dLemmod}
d_L(z)=\frac{c}{H_0}\, (1+z) \,\int_0^z\, 
\frac{d\tilde{z}}{ \sqrt{\Omega_{\rm M} (1+\tilde{z})^3+ 
\Omega_{\rm R} (1+\tilde{z})^4+ \rho_{\rm DE}(\tilde{z})/\rho_0} }\, ,
\end{equation}
where $\Omega_{\rm M}$  is the present matter  fraction  and $\Omega_{\rm R}$ that of radiation (which we have written for completeness, but is completely negligible at the redshifts where astrophysical compact binaries merge; we also write for simplicity our equations for a flat cosmology, so $\Omega_{\rm M}+\Omega_{\rm R}+\Omega_{\rm DE}=1$). When $w_{\rm DE}=-1$,
$\rho_{\rm DE}$ in \eq{4rdewdeproofs} becomes a constant, that we denote  $\rho_{\Lambda}$, and 
\eq{dLemmod} reduces to the $\Lambda$CDM expression
\begin{equation}\label{dLemLCDM}
d_L(z)=\frac{c}{H_0}\, (1+z) \, \int_0^z\, 
\frac{d\tilde{z}}{\sqrt{\Omega_{\rm M} (1+\tilde{z})^3 +\Omega_{\rm R} (1+\tilde{z})^4+\Omega_{\Lambda} }}\, ,
\end{equation}
where   $\Omega_{\Lambda}=\rho_{\Lambda}/\rho_0$.  At low redshift, both \eq{dLemmod} and \eq{dLemLCDM} reduce to Hubble's law $d_{L}(z)\simeq (c/H_0)z$, so standard sirens at low redshift  allows us to measure  $H_0$. At the redshifts accessible to ET, however, we can also test the DE equation of state, and  any deviation   from the $\Lambda$CDM value $w_{\rm DE}=-1$ would  provide evidence for a dynamical dark energy.
In general,  however, it is difficult to extract from the data a full function of redshift, such as $w_{\rm DE}(z)$, and a parametrization in terms of a small number of parameters is useful. For the DE equation of state a  standard choice is the $(w_0,w_a)$ parametrization \cite{Chevallier:2000qy,Linder:2002et},
\begin{equation}\label{w0wa}
w_{\rm DE}(z)= w_0+\frac{z}{1+z} w_a\, .
\end{equation} 
The evolution of the cosmological background can be tested both with standard electromagnetic cosmological probes, such as CMB, Supernovae, Baryon Acoustic Oscillations (BAO) or structure formation, 
and by GW observations, combining the GW measurement of the luminosity distance of coalescing binaries with information on the redshift, obtained either with an electromagnetic counterpart, if it is observed (``standard sirens") or with statistical methods (``dark sirens"), as discussed in section~\ref{sec:Cosmography}.
A first observable for GW observations, in the context of dark energy and modified gravity, is therefore the DE equation of state $w_{\rm DE}(z)$ or, more simply, the two parameters $(w_0,w_a)$. 
Modifications of the cosmological model in the sector of scalar perturbations are less directly relevant for GW experiments, and rather manifest themselves in cosmological structure formation; these modifications are among the targets of current galaxy surveys.
However, for GW observation, there are further quantities that are accessible, to test modified gravity: GW observation can in fact access
modifications in the tensor perturbation sector (i.e., GWs propagating over a FLRW background). 
In GR the equation that governs the  propagation of tensor perturbations over FLRW is given by 
\begin{equation}\label{4eqtensorsect}
\tilde{h}''_A+2{\cal H}\tilde{h}'_A+c^2k^2\tilde{h}_A=0\, ,
\end{equation}
where $\tilde{h}_A(\eta, {\bf k})$ is the Fourier-transformed GW amplitude,  $A=+,\times$ labels the two GW polarizations, the prime denotes the derivative with respect to cosmic time $\eta$ [defined by $d\eta=dt/a(t)$], $a(\eta)$ is the scale factor, and 
${\cal H}=a'/a$. In modified gravity theories, 
this propagation equation is  modified, as has been found on many explicit 
modified gravity models~\cite{Saltas:2014dha,Lombriser:2015sxa,Nishizawa:2017nef,Arai:2017hxj,Belgacem:2017ihm,Amendola:2017ovw,Belgacem:2018lbp,Nishizawa:2019rra,Belgacem:2019lwx,Belgacem:2020pdz,LISACosmologyWorkingGroup:2019mwx}.
A modification of  the coefficient of the $k^2\tilde{h}_A$ term induces a speed of GWs, $c_{\rm gw}$, different from that of light. After the observation of GW170817, this is  now excluded  at a level  $|c_{\rm gw}-c|/c< {\cal O}(10^{-15})$ \cite{LIGOScientific:2017zic} (at least for scale-independent modifications, see \cite{deRham:2018red}). However, the modified gravity models that pass this constraint still, in general, induce a change in the above equation, that  becomes
\begin{equation}\label{prophmodgrav}
\tilde{h}''_A  +2 {\cal H}[1-\delta(\eta)] \tilde{h}'_A+c^2k^2\tilde{h}_A=0\, ,
\end{equation}
for some function  $\delta(\eta)$ that encodes the modifications from GR. As a result, the amplitude of a GW propagating across cosmological distances, from the source to the observer,  is modified, because of this different friction term;
one can  then show that the amplitude of the GW from a coalescing binaries
no longer allows us to obtain the standard luminosity distance $d_L(z)$ of the source [that, in this context, we will denote by $\dem(z)$, since this is the quantity that would be measured, for instance, using the  electromagnetic signal from a counterpart]. Rather, the quantity extracted from GW observation is a  ``GW luminosity distance''  $\dgw(z)$
\cite{Belgacem:2017ihm}, related to $\dem(z)$ by~\cite{Belgacem:2017ihm,Belgacem:2018lbp}
\begin{equation}\label{dLgwdLem}
\dgw(z)=\dem(z)\exp\left\{-\int_0^z \,\frac{dz'}{1+z'}\,\delta(z')\right\}\, ,
\end{equation}
where the function $\delta$ that appears in \eq{prophmodgrav} has now been written as a function of redshift. 

Of course, to perform a full inference on the  function $\delta(z)$ from future GW data is difficult (although it can be performed, to some extent, with the technique of Gaussian process reconstruction~\cite{Belgacem:2019zzu}) and, similarly to what is done for the  DE equation of state,  it is convenient to introduce a parametrization. 
A very convenient choice,  proposed in \cite{Belgacem:2018lbp}, takes the form
\be\label{eq:fit}
\frac{d_L^{\,\rm gw}(z)}{d_L^{\,\rm em}(z)}=\Xi_0 +\frac{1-\Xi_0}{(1+z)^n}\, ,
\ee
in terms of two parameters $(\Xi_0,n)$.
This parametrization interpolates, with a power-law in the scale factor $a=1/(1+z)$,  between   $d_L^{\,\rm gw}/d_L^{\,\rm em}=1$ at $z= 0$ (i.e. when there is no propagation), and a constant value large redshifts; the latter limit is motivated by the fact that,
in  typical  modified gravity  models, the deviations from GR only appear in the recent cosmological epoch, so $\delta(z)$ goes to zero at large redshift, and  the integral in \eq{dLgwdLem}
saturates to a constant; GR corresponds to $\Xi_0=1$ (for any $n$).
This simple parametrization has been shown to work remarkably well for practically all best-studied modified gravity models~\cite{LISACosmologyWorkingGroup:2019mwx} (except bigravity, where $d_L^{\,\rm gw}/d_L^{\,\rm em}$ displays some oscillations). 

Another parametrization  of modified GW propagation which  has been used in the literature is given by 
\be\label{alphaM}
\alpha_M(z)\equiv -2\delta(z)=
c_M\frac{\ode(z)}{\ode}\, ,
\ee 
where $c_M$ is a constant and $\ode(z)=\rde(z)/\rho_0$.
This parametrization  was  introduced  in \cite{Bellini:2014fua} to study the scalar sector of Horndeski theories and, in these theories, it turns out that  the same function  also describes the modification of the tensor sector. A drawback of the parametrization 
(\ref{alphaM}) is that it requires to postulate an expression for  $\ode(z)$, i.e. for $\rde(z)$, so it is less suitable for model-independent studies. 
In practice, when applied to modified GW propagation, this parametrization has been used assuming that the dark energy density is constant, $\rde(z)=\rde(0)$~\cite{Lagos:2019kds,Ezquiaga:2021ayr} (which, however, is not expected to  be the case, since if the tensor sector  of the theory is modified with respect to $\Lambda$CDM, also the background evolution will be different). In any case, 
as long as $\ode(z)$ is such that the integral in \eq{dLgwdLem} saturates to a finite value (which is also the case of a constant $\ode$), this parametrization is still qualitatively of the same form as \eq{eq:fit}, since the ratio $\dgw(z)/\dem(z)$ goes from unity at $z=0$ to a constant $\Xi_0$ at large $z$. However,
this parametrization is less general than the $(\Xi_0,n)$ parametrization, since it only has one parameter, $c_M$, so it  implicit assumes a relation between $\Xi_0$ and $n$ (see App.~F of \cite{Mancarella:2021ecn} for detailed comparison). { Other proposed parametrizations include
\begin{equation}
    \Xi(z) = \sqrt{\frac{1 + \Omega_0}{1 + \Omega_0a^{\beta}}}\,, \label{eq:Omega_0}
\end{equation}
from \cite{Matos:2022uew}, which has the same behavior as that of eq. (\ref{eq:fit}) when the deviations from GR are small ($|\Omega_0| \ll 1$), with the correspondence $\Xi_0 = 1 + \Omega_0/2$ and $n = \beta$; also
\begin{equation}
    \alpha_M(z) = \alpha_{M0}a^{\beta} \quad (\beta > 0)\,,
\end{equation}
used in the context of the EFT of dark energy \cite{Planck:2015bue, Planck:2018vyg}; and the one of \cite{Matos:2021qne}.}

\begin{figure}[t]
\centering
\includegraphics[width=0.48\textwidth]{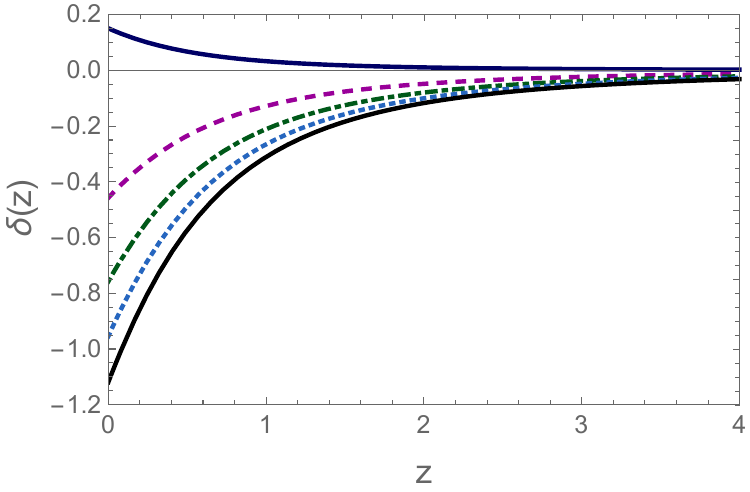}
\includegraphics[width=0.48\textwidth]{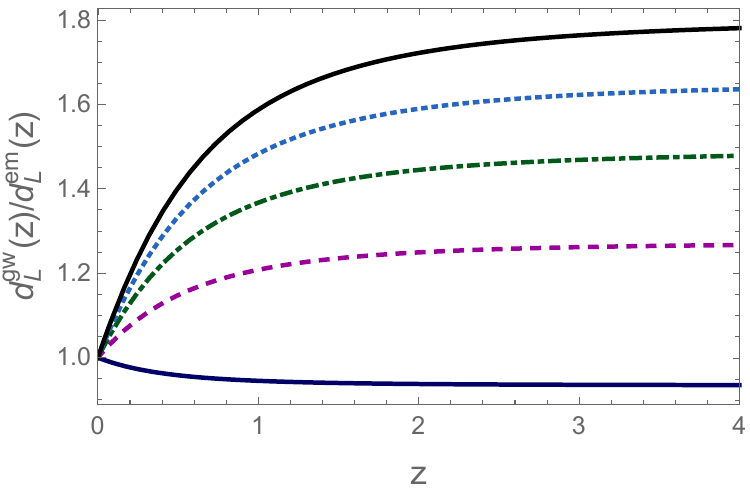}
\caption{The functions $\delta(z)$ (left panel) and $\dgw(z)/\dem(z)$ (right panel),
for the non-local gravity model proposed in \cite{Maggiore:2013mea},  for different values of a free parameter of the theory. From 
ref.~\cite{Belgacem:2020pdz}.
}
\label{fig:deltadgw}
\end{figure}

It is interesting to note that, with electromagnetic observations,  the background evolution and the scalar perturbations have been tested to a few percent level. For instance, combining many different cosmological datasets (and setting $w_a=0$), one finds $w_0=-1.031^{+0.030}_{-0.027}$ \cite{DES:2021wwk}, consistent with $\Lambda$CDM at the $3\%$ level (although  the recent DESI results~\cite{DESI:2025zgx}, even if  consistent with a flat $\Lambda$CDM model, when combined with other datasets prefer an evolving dark energy equation of state with phantom crossing below $w=-1$ at redshift $z\simeq 0.4$), while for the parameters $\Sigma_0$ and $\mu_0$ that parametrize deviations in the scalar perturbation sector (with $\mu_0=\Sigma_0=0$ in GR) one finds
$\mu_0=-0.04\pm 0.25$ and $\Sigma_0=-0.024\pm 0.054$
\cite{eBOSS:2020yzd}, so the latter corresponds to a test of GR and $\Lambda$CDM at the $5\%$  level.
Naively one would then imagine that, if a modified gravity model complies with existing bounds in the background evolution and in the scalar perturbation sector, even in the tensor perturbations its deviation from GR will not exceed a few percent.  Indeed, this is true for some important classes of models, like the $f(R)$ or several scalar-tensor theories, where the GW friction is the time variation of the effective Planck mass,
\begin{equation}
\alpha_M(z) = \frac{d\log M_{\ast}^2}{d\log a}\,. \label{eq:Planck_mass}
\end{equation}
There, $\alpha_M$ is also sourcing gravitational slip $\eta \equiv \Phi/\Psi = 2\Sigma/\mu -1$ \cite{Saltas:2014dha, Bellini:2014fua, Matos:2022uew}, therefore linking tests of gravity based on standard sirens with those from large-scale structure. 
 However, this link is not generally true; as an explicit example, the model originally proposed in \cite{Maggiore:2013mea} complies with all observational limits in the background evolution and in the scalar sector but, in the tensor sector, predicts a value of $\Xi_0$ that (depending on a free parameter related to initial conditions) can be as large as $1.8$~\cite{Belgacem:2020pdz}, corresponding to a $80\%$ deviation from GR. 
Figure~\ref{fig:deltadgw}, from ref.~\cite{Belgacem:2020pdz}, shows the prediction for $\delta(z)$ (left panel) and 
$d_L^{\,\rm gw}/d_L^{\,\rm em}$ in this non-local model, for different values of a free parameter of the theory (see \cite{Belgacem:2020pdz} for details).

This shows that the window of cosmological tensor perturbations, that has been opened for the first time by the LVC and LVK observations, and that will be deeply explored by ET, is extremely interesting and could reserve us remarkable surprises.

\paragraph{Current limits on $\Xi_0$.}\label{sect:limitsXi0}

Using the various methods for bright and for dark sirens discussed previously in this chapter, several limits on modified GW propagation have already been obtained from the current dataset of GW detections.
 
A first limit was obtained in~\cite{Belgacem:2018lbp}   using GW170817 as a standard siren with counterpart
(see also \cite{Lagos:2019kds}).  In this case  the electromagnetic luminosity distance of the host galaxy can be  obtained   from surface brightness fluctuations. Because of the very small redshift 
of GW170817, what is obtained is really a limit on $\delta(z=0)$, independently of the parametrization used, and the result from~\cite{Belgacem:2018lbp} is
$\delta(0)=-7.8^{+9.7}_{-18.4}$.
In terms of $\Xi_0$, this corresponds to a rather broad limit $\Xi_0\,\lsim\, {\cal O}(10)$.
This limit is not stringent because GW170817 has a very small redshift,
$z\simeq 0.01$, and the effect of modified GW propagation disappears as $z\ra 0$. 

A stronger bound  was obtained
in \cite{Finke:2021aom} using  BBH dark sirens from the O1, O2 and O3a LIGO/Virgo run, and performing a correlation  with the GLADE~\cite{Dalya:2018cnd} galaxy catalog. This gives (at $68\%$ c.l.)
\be\label{Xi0limit1}
\Xi_0=2.1^{+3.2}_{-1.2} \, .
\ee
A  more stringent bound can obtained if one accepts the  tentative  identification of the flare  ZTF19abanrhr as the electromagnetic counterpart of the BBH coalescence GW190521, as proposed in \cite{Graham:2020gwr}.\footnote{In \cite{Ashton:2020kyr} it was observed that the volume localization of GW190521 is not statistically sufficient for a confident identification of ZTF19abanrhr as the electromagnetic counterpart. However, this assumes no modified GW propagation. Introducing the possibility of modified GW propagation has precisely the effect of restoring the consistency between the volume localization of GW190521 and the flare ZTF19abanrhr.}
Then, the analysis in \cite{Finke:2021aom} (keeping fixed $H_0$ to the Planck value) gives  
\be\label{Xi0limit2}
\Xi_0=1.8^{+0.9}_{-0.6}\, ,
\ee 
($68\%$ C.L.). A similar results is obtained in \cite{Mastrogiovanni:2020mvm}, again assuming the identification of   ZTF19abanrhr with GW190521, and combining this with GW170817. Using again a narrow prior on $H_0$ the most stringent limit obtained in 
\cite{Mastrogiovanni:2020mvm} (for different choices of the waveform) is $\Xi_0<2.9$.

Another approach is to constrain
modified GW propagation  using  the BBH mass distribution,   following the strategy proposed in \cite{Farr:2019twy}. 
Using the BBH mass function and the GWTC-3 catalog, ref.~\cite{Mancarella:2021ecn} finds
\be\label{Xi0limit3}
\Xi_0 = 1.2^{+0.7}_{-0.7}\, ,
\ee
with a flat prior on $\Xi_0$, and $\Xi_0 = 1.0^{+0.4}_{-0.8}$ with a prior uniform in $\log\Xi_0$, while using the GWTC-2 catalog of GW events and the parametrization (\ref{alphaM}) with a constant $\rde$,  ref.~\cite{Ezquiaga:2021ayr},
finds $c_M=-3.2^{+3.4}_{-2.0}$ (see App.~G of \cite{Mancarella:2021ecn} for a discussion of the technical differences between  the analyses of refs.~\cite{Ezquiaga:2021ayr} and \cite{Mancarella:2021ecn}).

\paragraph{Forecasts for ET.} Constraints on deviations from General Relativity and $\Lambda$CDM, as those parametrized by the dark energy equation of state and by modified GW propagation, are among the primary targets of the ET science case. Information on these parameters  can come from both bright sirens and dark sirens, as we discuss next.

\subparagraph{Constraints from bright sirens.}

Forecasts on the accuracy of the cosmological parameter $H_0,\Omega_M$ that can be obtained from bright sirens in the framework of $\Lambda$CDM have been presented in section~\ref{sec:brightSirens}. Here, using the same methodology, we enlarge the analysis to modified gravity, 
extending the analysis performed in \cite{Branchesi:2023mws} -- where the inference was carried out on two cosmological parameters at a time -- to a more comprehensive investigation. Specifically, this study is conducted on the same dataset of bright sirens (see tables $6$ and $10$ of \cite{Branchesi:2023mws}), but aims to extract a measurement for all the $6$ considered $\Lambda \rm CDM$+DE+modified GW propagation parameters $H_0$, $\Omega_{\rm m,0}$, $w_0$, $w_a$, $\Xi_0$ and $n$ simultaneously.
While this choice inevitably leads to broader constraints, we stress that this wider approach is necessary if no other complementary survey's results are readily available to fix one or more parameters in the inference. Furthermore, it represents a standalone benchmark for the additional science that GW astronomy can deliver regardless of the available previous measurements.
\begin{figure}[t!]
    \centering
\makebox[\textwidth][c]{
    \includegraphics[width=.55\textwidth]{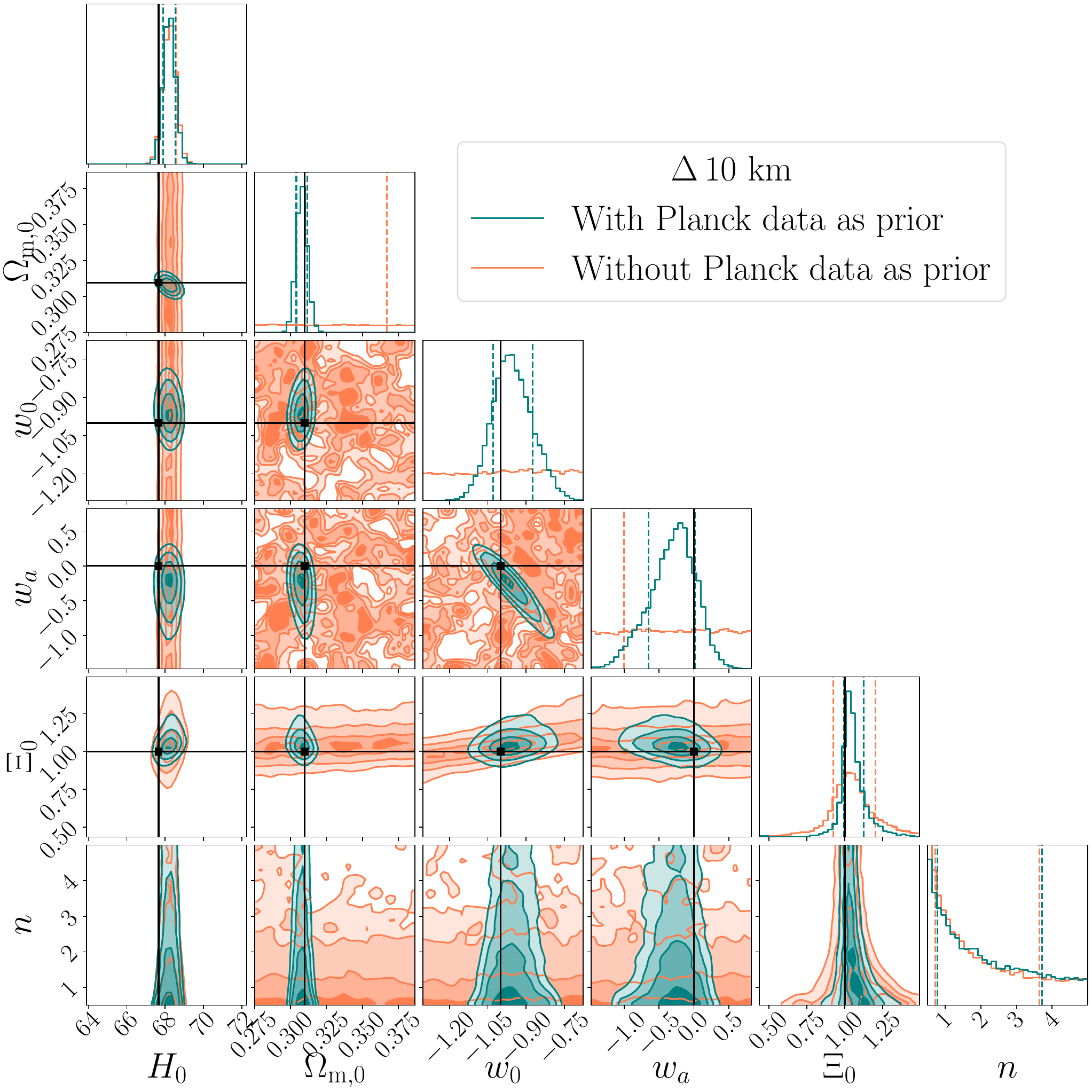}
    \includegraphics[width=.55\textwidth]{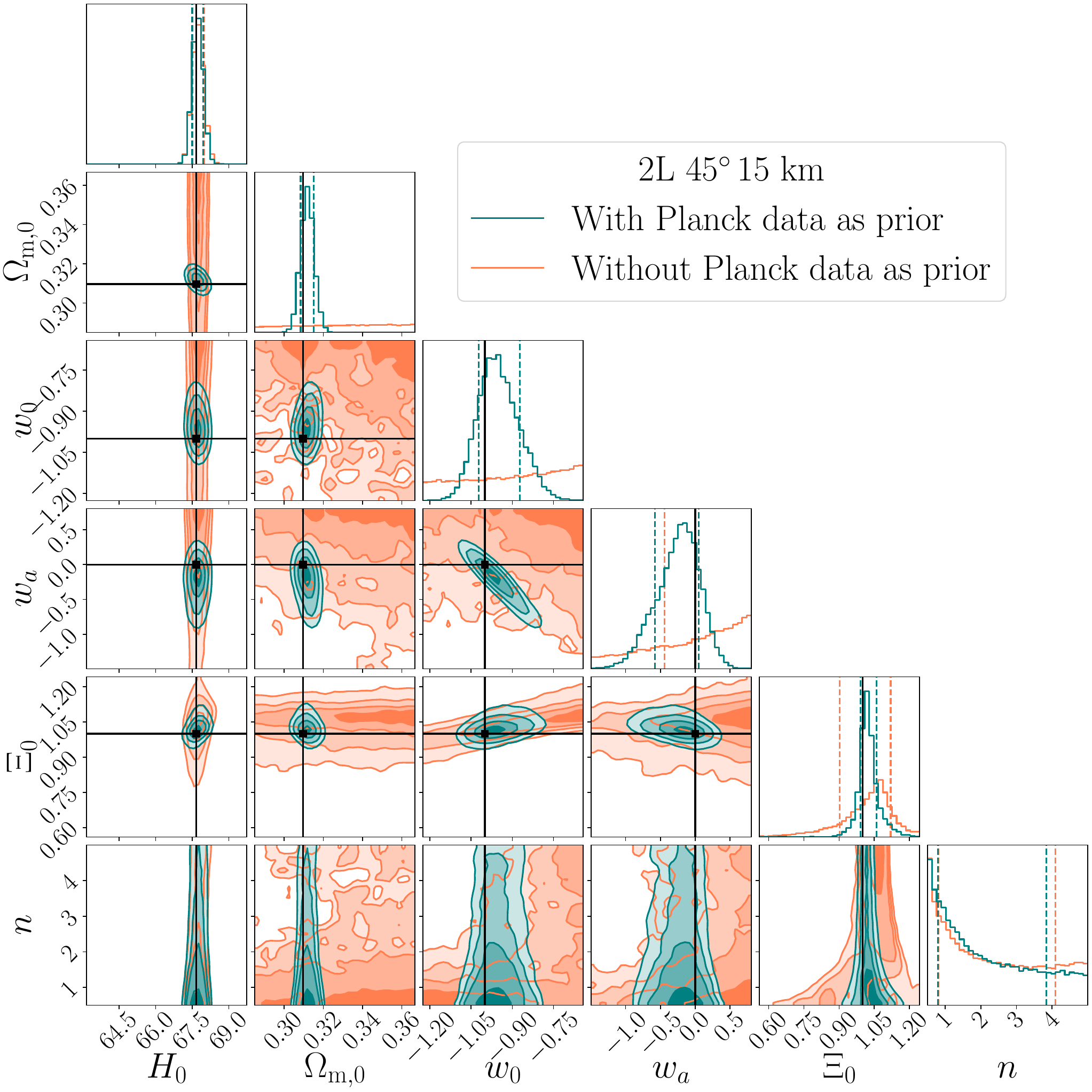}
    }
    \caption{Results of the joint inference on the cosmological parameters $H_0$, $\Omega_{\rm m,0}$, $w_0$, $w_a$, $\Xi_0$ and $n$ employing GW+KN events detected in one year of observations by the $10$ km triangular (left panel) and the 2L-15km-$45^{\circ}$  (right panel) ET configurations, together with the Vera Rubin Observatory. Vertical dashed lines represent the $68\%$ CI of each distribution, while the black solid lines label the fiducial values.}
    \label{fig:multimessenger_KN}
\end{figure}

\begin{figure}[th]
    \centering
\makebox[\textwidth][c]{
    \includegraphics[width=.55\textwidth]{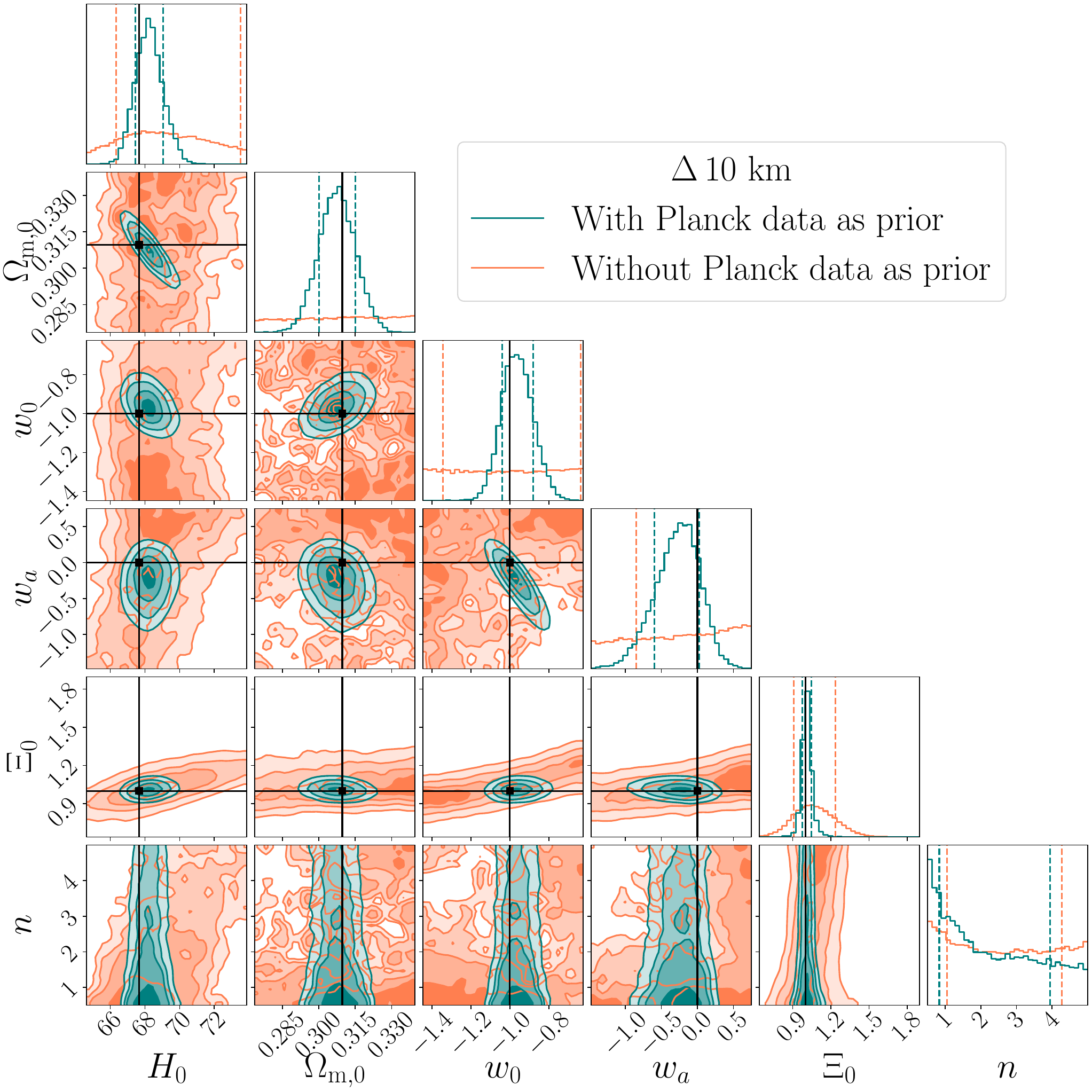}
    \includegraphics[width=.55\textwidth]{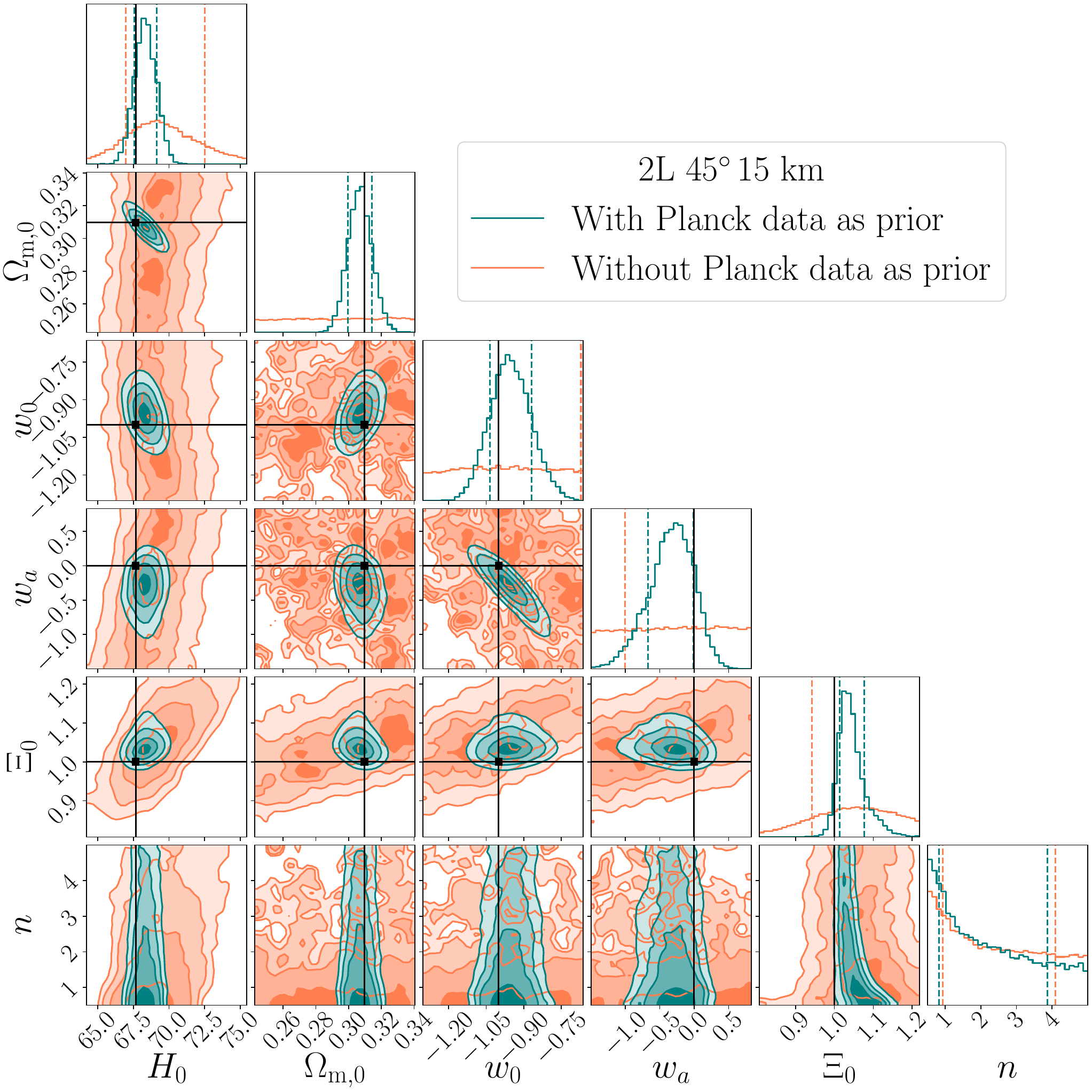}
    }
    \caption{Results of the joint inference on the cosmological parameters $H_0$, $\Omega_{\rm m,0}$, $w_0$, $w_a$, $\Xi_0$ and $n$ employing GW+GRB events detected in $5$ years of observations by the $10$ km triangular (left panel) and the 2L-15km-$45^{\circ}$  (right panel) ET configurations, together with  THESEUS. Vertical dashed lines represent the $68\%$ CI of each distribution, while the black solid lines label the fiducial values.}
    \label{fig:multimessenger_GRB}
\end{figure}

We report these results in figures~\ref{fig:multimessenger_KN}, \ref{fig:multimessenger_GRB} and in table~\ref{tab:cosmo_results}, where we also show the impact of using the Planck 2018 \cite{Planck:2018vyg} posterior distributions as a prior. As expected, extending the analysis to all $6$ cosmological parameters worsen the precision on $\Xi_0$ by a factor of order $\sim 5$ with respect to the results reported in \cite{Branchesi:2023mws}, while $n$ remains undetermined in both cases due to the strong degeneracy with $\Xi_0$. On the other hand, however, using  the early Universe from Planck as priors strongly increase the precision on $\Xi_0$, which ranges from $\sim 12\%$ in the worst scenario, to $\sim 6\%$ in the best case.
The error on $H_0$, on the other hand, increases only marginally compared to the case when we restrict to $\Lambda$CDM. For instance, as we found in section~\ref{sec:brightSirens}, 
for GW+KN (and 1~yr of observations) in $\Lambda$CDM, without using the Planck prior,  the relative error on $H_0$ was $0.9\%$ for  the 10-km triangle and $0.6\%$ for 2L-15km misaligned, and these numbers become, respectively,  $1.0\%$ and $0.7\%$ when enlarging the parameter space to $\{H_0, \Omega_{\rm m,0}, w_0, w_a, \Xi_0, n\}$.
It is also interesting to observe that, as we   saw in section~\ref{sec:brightSirens},  in $\Lambda$CDM  the constraint  on $H_0$ from GW+GRB are much less stringent than from GW+KN. However, we see that this is no longer the case for the $\Xi_0$ parameter. This is due to the fact that  KN observation are restricted to smaller redshifts compared to GRB, while modified GW propagation is a cumulative effect that increases with the distance to the source, and therefore is better probed by high-redshift sources.\footnote{We also stress again that, as in the discussion in section~\ref{sec:brightSirens}, in this study  the relative uncertainty on the luminosity distance has been  approximated as $2/$SNR, since the Fisher matrix approach breaks down for face on/face off sources.  This approximation could be improved in future studies along the lines 
recently proposed in \cite{Mancarella:2024qle} and discussed  in section~\ref{sec:div9_fim_improvements}.}

\begin{table*}[t]
\centering
\setlength{\tabcolsep}{0.3em} 
\makebox[\textwidth][c]{%
\begin{tabular}{c | c || c | c || c | c }
\hline
\hline
\multirow{2}{*}{\textbf{EM Source}} & \multirow{2}{*}{\textbf{ET configuration}} & \multicolumn{2}{c||}{Without Planck data priors} & \multicolumn{2}{c}{With Planck data priors} \\ \cline{3-6}
       &                                 & $\Delta \Xi_0 / \Xi_0$ & $\Delta n$ & $\Delta \Xi_0 / \Xi_0$ & $\Delta n$ \\
\hline
\multirow{2}{*}{KN} & $\Delta \, 10$ km                       &  $0.27$ & $> 1$               &  $0.12$  & $> 1$\\
& 2L $45^{\circ} \, 15$ km                &  $0.21$ & $> 1$               &  $0.07$  & $> 1$\\
\hline
\multirow{2}{*}{GRB} & $\Delta \, 10$ km                       &  $0.31$ & $> 1$               &  $0.07$  & $> 1$\\
& 2L $45^{\circ} \, 15$ km                &  $0.27$ & $> 1$               &  $0.06$  & $> 1$\\
\hline
\hline
\end{tabular}
}
\caption{$68\%$ CI relative uncertainties on the modified GW propagation parameters $\Xi_0$ and $n$ obtained from multimessenger observations and adopting the two shown ET configurations.}
\label{tab:cosmo_results}
\end{table*}

Further improvement might come when combining these observations with other distance measurements expected from future standard candles and large-scale structure surveys, in which case  the precision on $\Xi_0$ could reach the $\sim 1\%$ level, while keeping model-independence \cite{Matos:2023jkn}.
In the context of EFT of dark energy, when identifying $\Xi$ as the effective Planck mass ratio of \eq{eq:Planck_mass}, also LSS and CMB data will help constraining $\Xi_0$ when combined with bright sirens, since in this case the modification in the tensor perturbation sector are related to those in the scalar perturbation sector, and can therfore be constrained also by structure formation data. The price one has to pay is of course to be more model-dependent in order to describe the modifications of gravity in the scalar sector. In the analysis of \cite{Matos:2022uew}, it was shown that the addition of 500 bright sirens with the triangular 10 km configuration of ET would improve current Planck constraints on $\Xi_0$ by $\sim 40\%$, for typical late time modifications of gravity where the same $\Xi$ is generating gravitational slip. The further combination with galaxy clustering and weak lensing from an Euclid-like survey would allow reaching an error of $0.3\%$ (more precisely, $\sigma(\Omega_0) = 0.6\%$ with the equivalent parametrization of \eq{eq:Omega_0}).

\begin{figure}[t]
\centering
\includegraphics[width=0.48\textwidth]{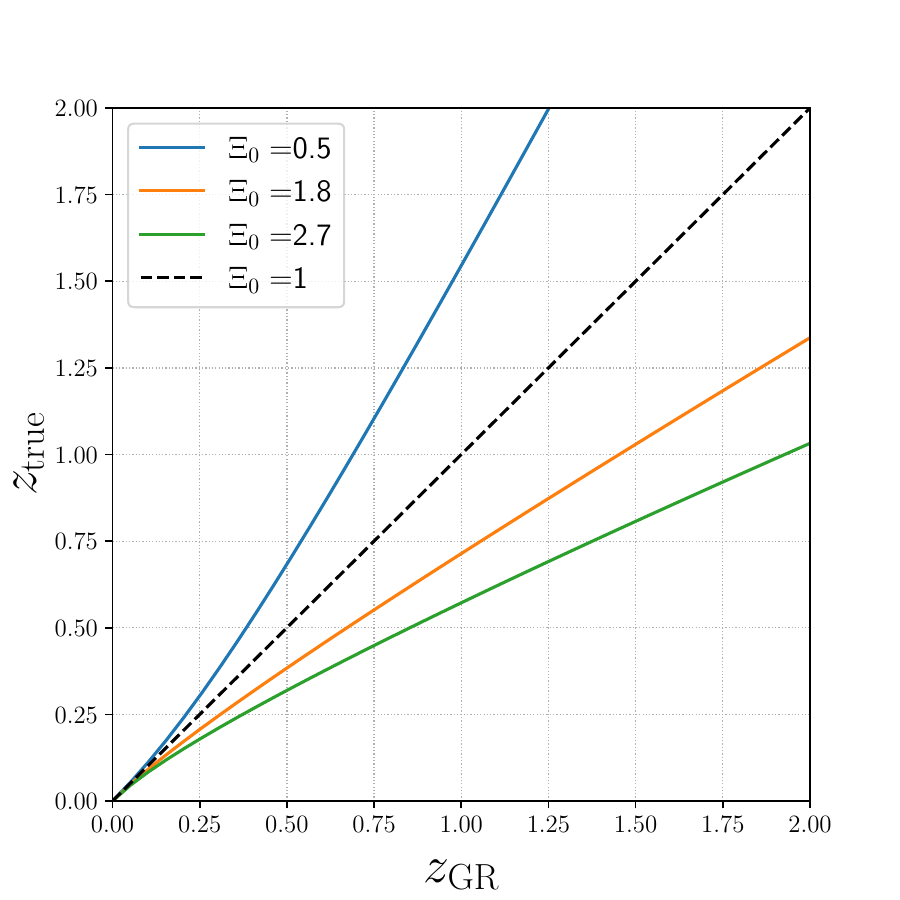}
\includegraphics[width=0.5\textwidth]{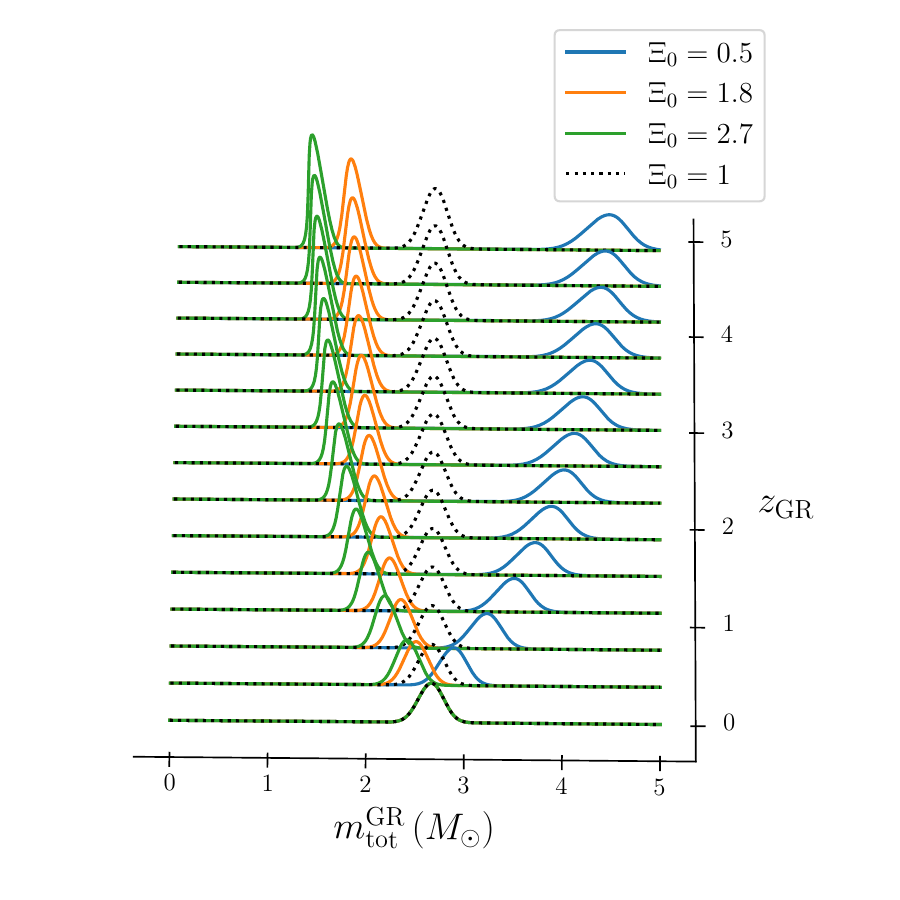}
\caption{Left: the  redshift $z_{\rm true}$ of a source, as a function of the value $z_{\rm GR}$ that would be incorrectly inferred using GR
if Nature is  described by a modified  gravity  theory with $\Xi_0\neq 1$, for different values of $\Xi_0$. Right: the effect on the distribution of the total mass of the binary from a `wrong' reconstruction using GR, $m_{\rm tot}^{\rm GR}$, 
assuming that the true 
distribution of the   source-frame total mass of the binary is a  (redshift-independent) Gaussian, with mean $2.66 \Msun$ and standard deviation $0.13 \Msun$. 
Adapted from \cite{Finke:2021eio}.}
\label{fig:zgw_from_zem}
\end{figure}

\subparagraph{Constraints from dark sirens.} As we have seen in section~\ref{sect:limitsXi0}, the dark siren method based on features on the mass function can provide stringent limits on modified GW propagation. Currently, the best limits are obtained from the BBH mass function, given that the current BNS sample is very small and limited to small redshifts (where the effect of modified GW propagation is suppressed). However, ET will detect a large number of BNS, of order $10^4-10^5$ per year, and up to large redshifts of order 2--3. This will allow to use this statistical method also on BNSs, which has the  advantage that the BNS mass function is quite narrow,  compared to the BBH mass function. The logic behind the possibility of bounding (or observing)  modified GW propagation from  the BNS mass function can be described as follows~\cite{Finke:2021eio}. From a GW detection of a BNS  we obtain  its GW luminosity distance, $\dgw$. If Nature is described by a modified gravity theory with 
$\Xi_0\neq 1$, $\dgw$ is different from 
the electromagnetic luminosity distance $\dem$; therefore, the source redshift  $z_{\rm GR}$ that would be inferred from the measured $\dgw$ assuming (incorrectly) the validity of GR and $\Lambda$CDM, would  differ from the true value $z_{\rm true}$. The effect is shown in the left panel of figure~\ref{fig:zgw_from_zem}, for some values of $\Xi_0$ consistent with the limits 
given in \eqst{Xi0limit1}{Xi0limit3} at about the (1-2)$\sigma$ level.\footnote{In this plot, we use for illustration the same values of $H_0$ and $\oma$ in the modified gravity theory and in GR, setting them to $H_0=67.9\, {\rm km}/({\rm s\, Mpc})$ and $\oma=0.30$. In a full analysis, all cosmological parameters will have to be determined self-consistently in GR and in modified gravity, by fitting the prediction of each theory to the cosmological observations.} We see that the effect can become very significant at large redshifts: for instance, if Nature were not described by  GR but rather by a modified gravity theory with  $\Xi_0=1.8$, a source whose true redshift is $z_{\rm true}=1$,  would be assigned a wrong redshift $z_{\rm GR}\simeq 1.45$, if the redshift is inferred assuming $\Xi_0=1$.

In turn, this would produce a systematic (and redshift-dependent)  bias in the inference of the actual (``source-frame'') masses $m_i$ ($i=1,2$) of the component neutron stars, since these are obtained from the measured ``detector-frame" masses  $m_{{\rm (\rm det)}, i}$ through the relation $m_i=m_{{\rm (\rm det)}, i}/(1+z)$. Getting  $z$ wrong would then result in a wrong estimate of $m_i$. More precisely, 
if Nature is described by a modified gravity theory with $\Xi_0\neq 1$, the true values  of the source-frame masses, $m_{{\rm true}, i}$ are obtained from the observed detector-frame masses by 
$m_{{\rm true}, i}= m_{{\rm (\rm det)}, i} /(1+z_{\rm true})$, while the values incorrectly inferred assuming GR would be
$m_{{\rm GR},i}  = m_{{\rm (\rm det)}, i}/(1+z_{\rm GR})$, and therefore
\be\label{msource_true}
m_{{\rm true}, i}=
\(\frac{1+z_{\rm GR} }{ 1+z_{\rm true} }\)\,  m_{{\rm GR},i}\, .
\ee
For instance (assuming again, as an example,  $\Xi_0=1.8$), for a  NS with $m_{\rm true}=1.35\Msun$ at $z_{\rm true}=1$,  the   mass incorrectly inferred from GR would be  $m_{\rm GR}\simeq 1.10\Msun $, while for $z_{\rm true}=2$, in GR one would infer $m_{\rm GR}\simeq 0.99\Msun$. The effect is quite striking because it would bias, in the same way, all NS masses at a given $z$ (so, in particular, both masses of a BNS, contrary to what could be expected from evolutionary paths leading to accretion of matter on one of the two neutron stars), shifting the whole distribution of NS masses in a characteristic redshift-dependent manner. This is  shown (for the distribution of the total mass of the binary) in the right panel of figure~\ref{fig:zgw_from_zem}.

Given the large number of BNS that ET will detect, this method can in principle reach a very good accuracy on $\Xi_0$. The effect can be tested  directly on the chirp mass of the system, which is very accurately measured, so in the end, with this technique, the main error on $\dgw/\dem$ from a single BNS is expected to be of order of the relative width of the BNS mass function. With $N$ detections at sufficiently large redshift, where the effect of modified GW propagation can be sizable, the error scale approximately as $1/\sqrt{N}$, so sub-percent errors on $\Xi_0$ could be obtained, as long as other competing sources of biases, such as lensing or a possible redshift evolution of the BNS mass function~\cite{Roy:2024oxh}, can be adequately accounted for.




\subsubsection{GW lensing}

According to GR, if a GW encounters a massive object on its travel path from source to observer, it can undergo gravitational lensing. Depending on the lens mass and the lens-source alignment, one can get different effects; for the lowest masses and more misaligned systems, one has microlensing, where the GW undergoes frequency-dependent modulations~\cite{Takahashi:2003ix, Wright:2021cbn}. On the other hand, for more massive lenses and better-aligned systems, one is in the geometric optics regime, and the wave is split into multiple copies of the original system, with an overall magnification, and overall phase shift and a time delay~\cite{Takahashi:2003ix, Dai:2017huk}. If the system is less aligned and/or the lens lighter, the various images have a small time delay and overlap, leading to a single non-trivial signal in the detector, called millilensing~\cite{Liu:2023ikc}. For larger masses, the images are individually resolvable, generally referred to as strong lensing~\cite{Dai:2017huk}. While rates for micro and millilensing are uncertain---due to their strong dependency on the nature of dark matter--- strong lensing is expected to be already observable for BBH signals in current detectors~\cite{Ng:2017yiu, Wierda:2021upe}.  Hundreds of such lensed signals~\cite{Gupta:2023lga} and tens of lensed BNSs should be observed in next-generation detectors every year~\cite{Magare:2023hgs}. A fraction of the latter could also have an observable lensed EM counterpart, making their correlation easier to do~\cite{Smith:2022vbp}.

\paragraph{GW lensing as probe of cosmological parameters.}

In the geometric optics, according to GR, GW and light are bend the same way, which can be leveraged to study our Universe, with unique possibilities at high redshift. Indeed, if a GW coming from a BBH is lensed, then the EM emission from its host galaxy is also lensed. One can then match both to use their complementary information to probe various physical phenomena~\cite{Hannuksela:2020xor, Wempe:2022zlk}. Matching a lensed BBH and its host galaxy is also easier than for an unlensed signal because lensing offers a much more precise sky localization~\cite{Janquart:2021qov}. In addition, GW measurements lead to millisecond precision measurements of the arrival times, and one can use time delay cosmography to constrain cosmological parameters~\cite{Liao:2017ioi}, provided the EM observations complete the observations by giving access to the lens and source redshifts. Moreover, one can also have extra constraints via the reconstructed magnifications of the images. The total measure obtained for the Hubble constant is relatively accurate, even for a single observation~\cite{Hannuksela:2020xor}. Additionally, in next-generation detectors, several observations can be stacked to improve the final constraints, similarly to what is already done nowadays for dark sirens~\cite{LIGOScientific:2021aug}. When the BBH host galaxy cannot be identified,
one can still develop a dark sirens-like approach for lensing, which accounts for a list of possible counterparts and the probability to miss the genuine system in the EM. While less precise than with a direct access to the counterpart in the EM, the list of candidates will be more restrained than for the usual dark siren method thanks to the selection done using the observed lens characteristics~\cite{Hannuksela:2020xor}. Moreover, these events would also allow for measurements at higher redshifts than unlensed systems, giving extra insights on the early Universe. 

\begin{figure}[t!]
    \centering
    \includegraphics[width=0.46\textwidth]{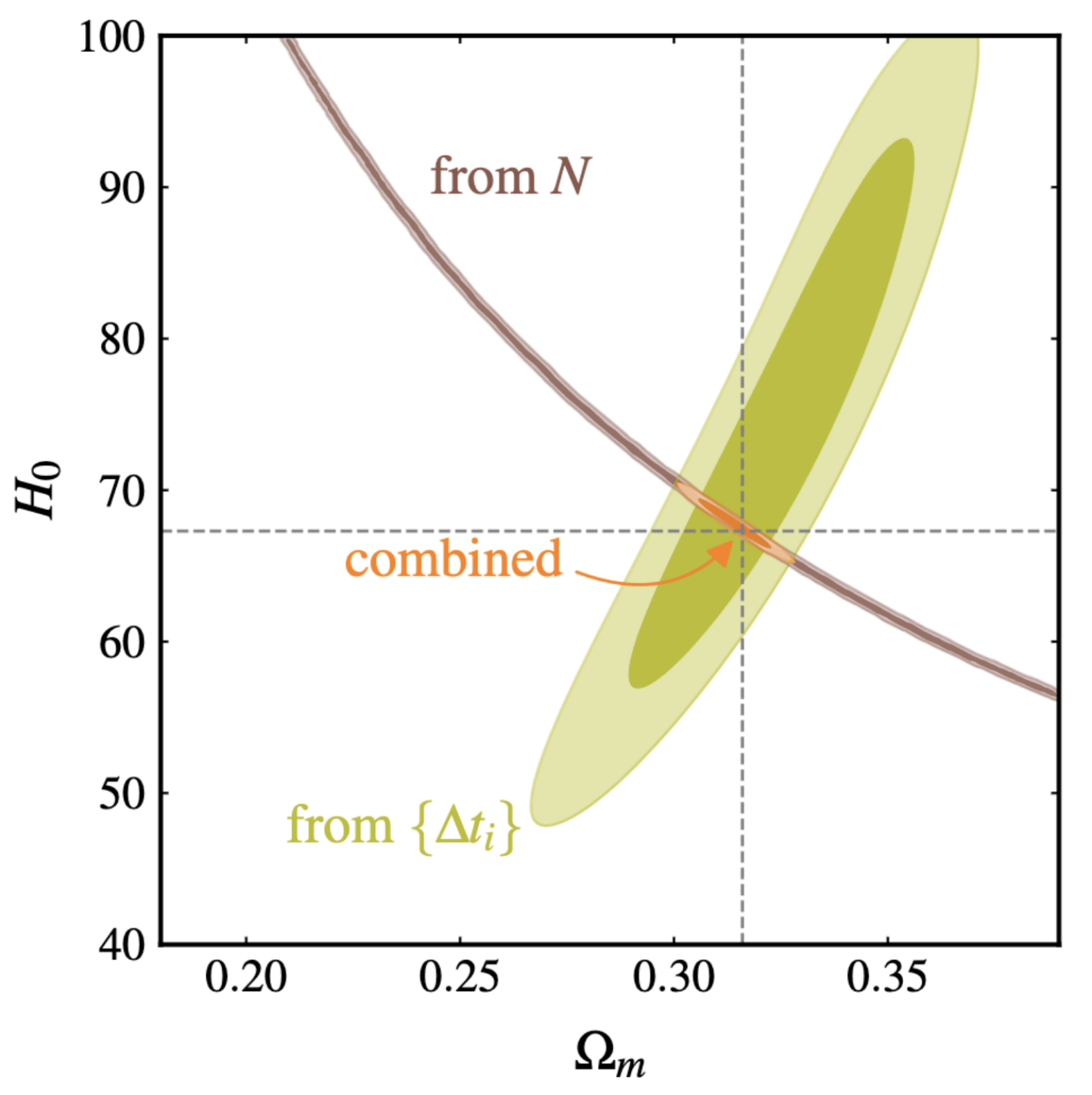}
    \includegraphics[width=0.5\textwidth]{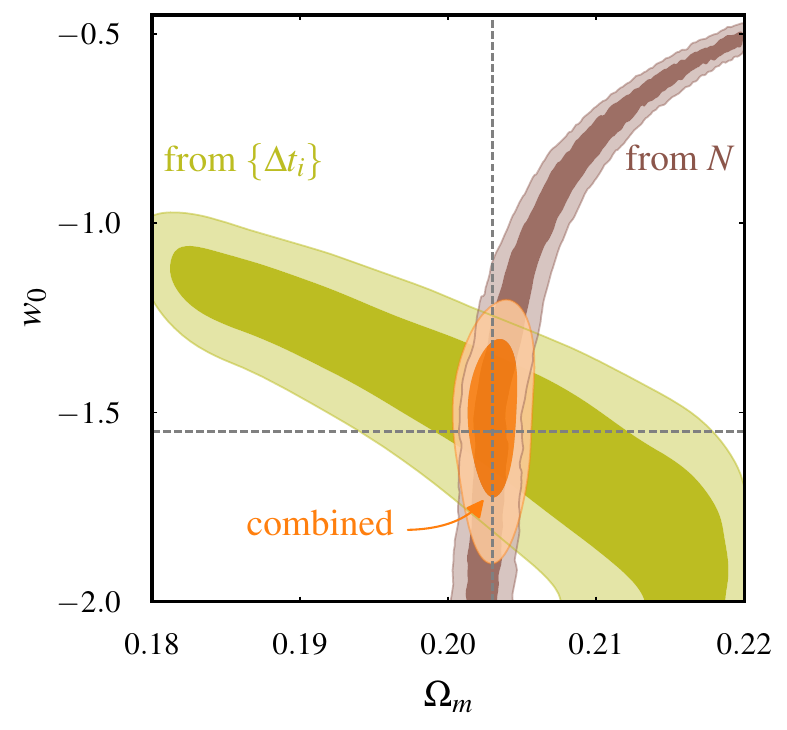}
    \caption{Constraints obtained from the statistical method relying on lensed BBH events to constrain the cosmological parameters presented in~\cite{Jana:2022shb}. Left: $H_0$ vs. $\Omega_m$. Right: $w_0$ vs. $\Omega_m$. These results are obtained for a ten year observation period, assuming that a generic network of 3G detectors will have a nominal BBH detection rate of $5\times 10^5\, \mathrm{yr}^{-1}$ upto a typical redshift of $z_\text{max}\sim15$ (exact value dependent on the cosmology and population models), and that their time of arrivals will be measured precisely. The bounds obtained are competitive with other studies, and the BBH signals originate from further away, potentially carrying information from earlier periods in the Universe. Figures adapted from~\cite{Jana:2022shb}.}
    \label{fig:bounds_cosmology_lensing}
\end{figure}

Furthermore,  lensed events without EM counterpart can be used to constraint cosmology based on their statistical properties, using the fact that  the expected number of lensed
events and the distribution of the time delay between lensed images depend on the cosmology. In particular, in ref.~\cite{Jana:2022shb} is shown how the expected distributions of lensed GW events can be used to constrain the cosmological parameters.  Denoting  by $N$ the number of  lensed BBHs events observed over an extended period of time  $T_{\rm obs}$ (each one assumed to produce two copies of the observed GW signal), and by $\{ \Delta t_{i}\} \, , i = 1, \dots, N$ their corresponding time delays, one can construct a probability distribution for the cosmological parameters $\vec{\Omega}$ as 
\begin{equation}\label{eq:proba_params_lensing}
p(\vec{\Omega} | N, \dtens, T_{\rm obs}) = \frac{p(\vec{\Omega}) p(N | \vec{\Omega}, T_{\rm obs}) p(\dtens | \vec{\Omega}, T_{\rm obs})}{Z} \, ,
\end{equation}
where one uses the conditional independency between the time delays and the number of lensed BBH detections given the parameters and the observation time $T_{\rm obs}$; $p(\vec{\Omega})$ is the prior and $Z$ is the normalization constant. By performing various simulations on a coarse grid of values for the cosmological parameters, it is then possible to construct the two probability functions. Then, one can evaluate those with a given observation time and find a probability distribution for the cosmological parameters. With this approach, using 10 years of observation and a BBH merger rate of $5\times 10^5 \, \mathrm{yr}^{-1}$ with 3G detectors (made of CE and ET), one obtains $H_0 = 67.8\pm 1.1 \, \mathrm{km} \,\mathrm{s}^{-1}\mathrm{Mpc}^{-1}$ and $\Omega_m = 0.3142 \pm 0.0056$ at 1-$\sigma$ C.I. for the standard $\Lambda$CDM cosmology. Figure~\ref{fig:bounds_cosmology_lensing} shows the posterior distribution for the cosmological parameters found with this methods. The bounds obtained depend on the detection rate, with broader constraints for lower rates. Besides, the approach also works for other cosmological models, such as $w$CDM, where the same observational conditions lead to competitive bounds compared to other traditional approaches (see right plot in Figure~\ref{fig:bounds_cosmology_lensing})~\cite{Jana:2022shb}. An interesting feature of these observations is that (a fraction of) such BBHs can come from high redshift, and, therefore, such studies could also probe cosmological parameters in a regime that is not explored by other methods.

As discussed in section~\ref{sect:ModGWpropdiv2}, modified GW propagation can be a smoking gun for modifications of gravity at cosmological scales, and 
lensing can also be used to study it~\cite{Finke:2021znb, Narola:2023viz}. Basically, if one observes a strongly lensed GW and its EM counterpart, and performs the lens reconstruction, one can access two different measures of the luminosity distance: one via GWs---by correcting the measured luminosity distance by the magnification, and one by the EM redshift measurements. If GR is the correct theory of gravity, the two values should match. However, in the presence of  modified GW propagation,  the two quantities will be different, and the difference can be used to constrain the parameter(s) linking the two quantities. This can be done for the $\Xi_{0}$ parametrization mentioned above~\cite{Finke:2021znb} but also for other theories such as large extra spatial dimensions or a time-varying Planck mass~\cite{Narola:2023viz}. This method can be used for lensed BNS, NSBH, and BBH events. The former have the advantage of offering a better sky location thanks to the use of Earth rotation in addition to the multiple images~\cite{Baral:2023xst}, while the latter would lead to better constraints  thanks to  the fact that, for a given SNR, they can be detected to larger distance. 

Another way to look into possible modifications in long-distance propagation is to test the speed of gravity by comparing the arrival time of photons and GWs~\cite{LIGOScientific:2017zic}. For unlensed events, the intrinsic difference in emission time between photons and GWs is one of the main sources of uncertainty, and varies from one model to the other~\cite{LIGOScientific:2017zic}. As mentioned above, if a GW is lensed, the light emitted by its source is also lensed. Therefore, we would now have multiple measures of the difference in arrival time between GW and EM. Moreover, by comparing the differences between EM and GW measured for the various images, one can cancel out several nuisance terms in the measure, and in particular the intrinsic difference in emission time~\cite{Baker:2016reh, Fan:2016swi}. Therefore, one may expect a more precise measurement of the speed of gravity, depending on the source of EM emission. Also, this method suffers from different sources of uncertainties (related to the lens itself and not the time of emission), so it can be used to cross-check the measurements made with other observations.

\paragraph{GW lensing and dark matter.}

When a GW is lensed, the lensing effect encodes the mass and density profile of the lens. Therefore, it can be used to probe the lens itself. In particular, millilensing and microlensing originate from smaller-scale objects, which can potentially match some dark matter candidates such as dark matter halos or primordial black holes. A particularly interesting case is when one searches for micro- or millilensing effects on top of strong lensing; if a GW is deflected by a galaxy or a galaxy cluster, one image or more can interact with smaller objects in the macrolens. These can be stars or dark matter~\cite{Diego:2019lcd, Mishra:2021xzz, Meena:2022unp}. In particular, relying on different mass models, and looking how the mass is distributed in the lens, it is possible to probe various dark matter models~\cite{Fairbairn:2022xln}. The same is true for millilensing, where objects are heavier than for microlensing, corresponding to the expected mass regime of dark matter halos. However, the number of images, their magnifications and time delays inform us about the objects within the galaxy that lead to the extra effect~\cite{Liu:2023ikc}. Knowing more about the scale helps constraining dark matter models since, from one model to the other, different scales are favored or prohibited~\cite{Kravtsov:2009gi, Oguri:2020ldf}. Alternatively, isolated objects can also lead to such lensing effects, and they can be used to study the abundance of intermediate-mass and primordial black holes~\cite{Lai:2018rto, Cai:2022kbp}. For these studies, micro- and milli-lensing effects from baryonic substructures need to be carefully accounted for. Effects similar to microlensing can stem from other sources, such as orbital precession, missing waveform features or noise artefacts. Recent and ongoing work attempt to break the degeneracies in order to distinguish true microlensing from such false positives~\cite{Liu:2023emk,Mishra:2023ddt,Keitel:2024brp}. For micro- and millilensing searches, ET offers the interesting advantage of enabling the detection of more subtle features thanks to its enhanced sensitivity. 

The above assumes the detection of micro or millilensing. However, non-detection can also be particularly useful to constrain the fraction of dark matter made of compact objects such as MACHO's~\cite{Basak:2021ten}. Indeed, depending on the model for dark matter, the fraction of it made of compact objects, and the BBH distribution, one can expect a different number of detected microlensed events. So, in the absence of  detection, one can constrain the dark matter fraction made of compact objects. This is done by simulating many realisations with different fractions and checking which ones are compatible with the non-observation, leading to an upper bound on this fraction~\cite{Basak:2021ten, LIGOScientific:2023bwz}. If a microlensed GW is detected, it is also possible to use this to put a lower bound on the dark matter fraction. For this, ET offers extra opportunities thanks to its additional sensitivity within the volume already probed, but also thanks to its larger horizon distance, increasing the probability for lensing.

\paragraph{Impact of lensing on standard sirens measurements.} 
In the previous paragraphs we have discussed  the prospects of lensing as an extra source of information. On the other hand, it is important to highlight that lensing can also act as an additional source of nuisance or systematic uncertainty in standard siren measurements. Besides unrecognized strongly lensed events, the phenomenon of weak lensing due to the gravitational potential of large-scale structures in the Universe is ubiquitous, as it accumulates along the propagating paths of all GWs  \cite{Holz:2005df,Laguna:2009re,Bertacca:2017vod,Fonseca:2023uay}. Within the geometric optics approximation, lensing induces frequency-independent (de)magnification of the GW amplitude, which directly translates to the inference of a biased luminosity distance of the GW source:
\begin{equation}
    d_{\rm L}^{\rm obs}(z,\mu)=\frac{d_{\rm L}^{\rm true}(z)}{\sqrt{\mu}}\; ,
\end{equation}
where $d_{\rm L}^{\rm true}$ is the true luminosity distance of the GW source and $\mu$ is the lensing magnification. The probability distribution function of the lensing magnification depends on the redshift of the source; it is sharply peaked at $\mu\approx1$~\cite{Canevarolo:2023dkh}, such that the majority of events will suffer very small modification, but it has a tail towards high magnifications that becomes more relevant for increasing redshifts. Moreover, lensing selection effects, i.e. the fact that magnified events are more likely to be detected than demagnified ones, tend to worsen its impact \cite{Cusin:2019eyv,Cusin:2020ezb,Shan:2020esq}. This effect is usually accounted as an additional source of noise in the luminosity distance uncertainty \cite{Hirata:2010ba}, worsening the estimation of the cosmological parameters, particularly those relying on distant events. Focusing on the case of bright standard sirens, lensing can also act as a systematic uncertainty in the estimation of such cosmological parameters. Using idealized mock catalogs of bright standard sirens events, relevant for ET, \cite{Canevarolo:2023dkh} shows that in some scenarios the lensing bias on the cosmological parameters can be comparable or greater than the expected statistical uncertainty. In this regard, \cite{Mpetha:2024xiu} highlights the importance of properly modeling the lensing magnification probability distribution function when evaluating the expected lensing uncertainty.
Furthermore, in the absence of an EM counterpart, and when the mass-redshift degeneracy cannot be broken by any other external information, lensing can also impact the inference of the source-frame masses of the GW sources \cite{Dai:2016igl,He:2022ett,Smith:2022vbp,Canevarolo:2024muf}. In particular, in the case of BNS mergers, the masses can be magnified above the intrinsic maximum mass of the distribution, thus falling into the lower mass gap. In the era of ET, it would be important to account for this effect, to properly infer the maximum mass of NSs and to avoid misinterpretation of NS events as BH ones, spoiling for example the multi-messenger searches. In the coming years, the development of mitigation strategies in the data analysis will be necessary to properly account for lensing in ET events.

\subsection{Probing the large scale structure of the Universe}\label{sect:LSSdiv2}

\subsubsection{Cross-correlation GWxLSS}
\label{sec:GWLSSCC}

As a cosmic tracer of the underlying matter distribution, resolved GWs and the GWB offer the unique possibility to map the large scale structure (LSS) of the Universe though gravitational radiation~\cite{Raccanelli:2016fmc,Bertacca:2019fnt, Libanore:2020fim}.
In the ET era, the anisotropic distribution of GWs will provide an independent cosmological probe and effectively become a useful LSS tracers in a complementary way to other LSS surveys (that trace cosmic structures through redshift measurements), as well as to emerging techniques such as Line Intensity Mapping or HI intensity mapping. 
Since all those map the same underlying density field via complementary observables, their cross-correlation offers a new opportunity for Cosmology and probing modified gravity (see also~\cite{Oguri:2016dgk, Raccanelli:2016fmc, Scelfo:2018sny, Mukherjee:2019oma, Mukherjee:2020mha, Scelfo:2022lsx, Bosi:2023amu, Libanore:2023ovr,Afroz:2024joi,Afroz:2024lou}). 

\paragraph{Cross-correlation of GW resolved events with LSS.}
\label{Cross-correlation of GW Resolved Evenets with LSS}


Besides information that can be extracted from individual detections (such as in the case of bright sirens) or from ensemble of GW detections (such as from dark sirens), large GW catalogs of resolved events open more general possibilities. 
They naturally trace structure in a luminosity distance space~\cite{Libanore:2020fim}, whereas galaxy surveys probe the redshift space~\cite{Bonvin:2011bg}. The expected number count exhibits different effects in both spaces, and the mapping of one into the other depend on both cosmology and gravity theory. The expression for the relativistic number count in luminosity distance space was initially considered in~\cite{Zhang:2018nea} (see~\cite{Libanore:2020fim,Alfradique:2022tox} for subsequent studies), and then calculated in full generality in~\cite{Namikawa:2020twf,Fonseca:2023uay}. Importantly, these works showed that the result is different with respect to its analogue in redshift space, with contributions such as lensing including additional terms. Those differences will become relevant for ET, which could allow using tomographic redshift bins up to $z\sim 2$~\cite{Fonseca:2023uay}. Ref.~\cite{Zazzera:2023kjg} further models the evolution and magnification biases for 3G detectors which affect those contributions. 

\begin{figure}[t!]
    \centering
    \includegraphics[width=0.9\textwidth]{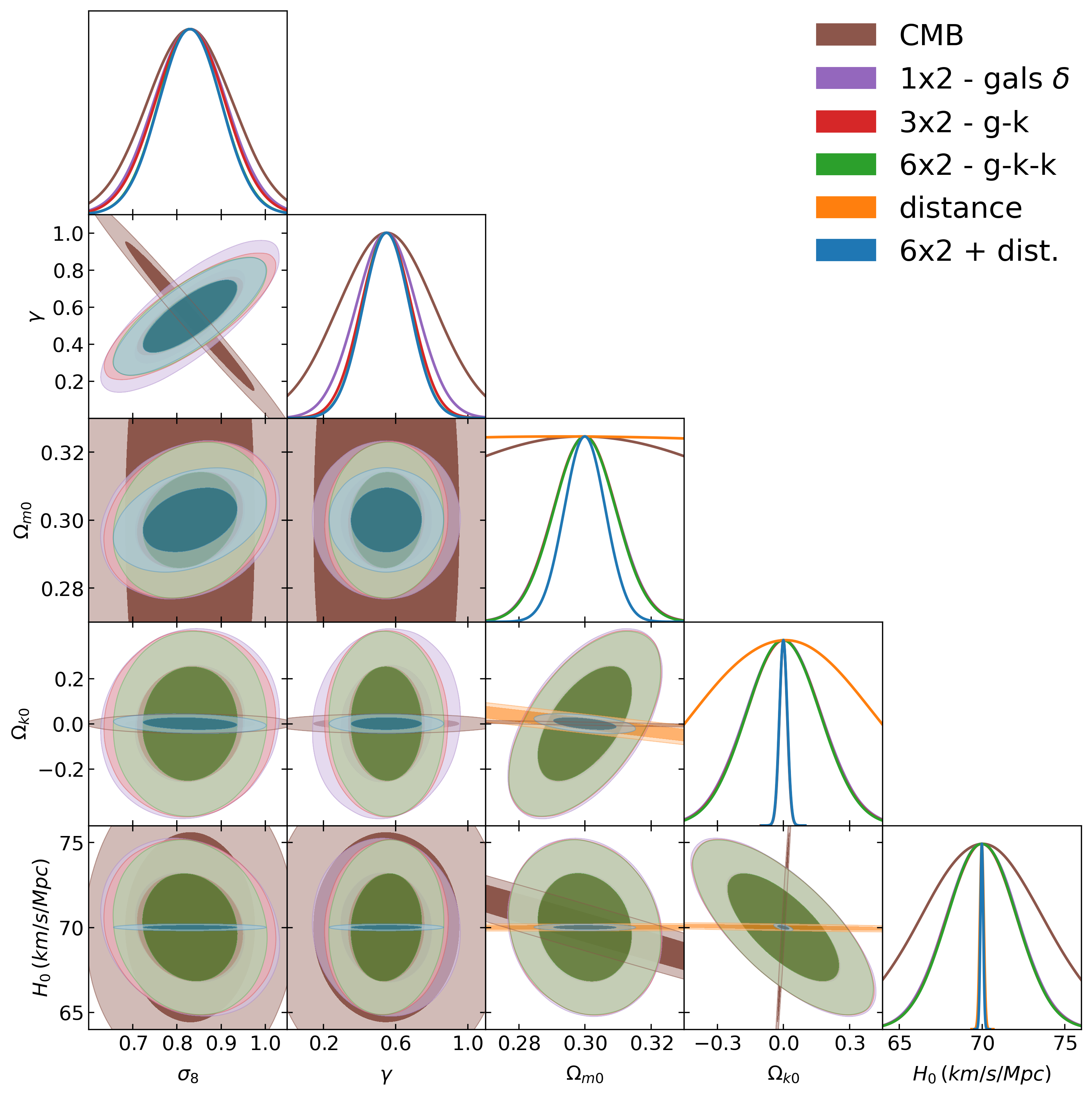}
    \caption{Marginalised constraints 
($1\sigma$ and $2\sigma$ confidence regions)
    combining ET KN and galaxies for $z\le0.5$, separated according to the method used. Purple: only DESI BGS galaxies; Red: DESI+KN velocities ($3\times 2$pt); Green: DESI+KN in the full $6\times 2$pt; Orange: KN distances only; Blue: $6\times 2$pt + distances; Brown: CMB. Figure from~\cite{Alfradique:2022tox}.}
    \label{fig:cornerplot-6x2pt}
\end{figure}

Interestingly, similarly to supernovae, bright sirens are useful tracers not only of the density but also of the velocity field~\cite{Wang:2017jum,Palmese:2020kxn}, because the latter introduces correlations in the distances which are well modelled by linear theory~\cite{Gordon:2007zw}. Combining both measurements, one can improve precision substantially by breaking the degeneracy between fundamental parameters such as $\sigma_8$ and the tracer bias. Moreover, combining with a galaxy catalog covering the same area of the sky, one gains access to a total of one velocity and two density tracers. One can thus probe six power spectra with the different combinations, and perform a $6\times2$pt analysis~\cite{Quartin:2021dmr,Alfradique:2022tox}. In particular, it was shown in~\cite{Alfradique:2022tox} that, using GW signals from BNS with ET with kilonova follow-ups from the Vera Rubin Observatory,\footnote{\url{https://www.lsst.org/}} together with DESI galaxies\footnote{\url{https://www.desi.lbl.gov/}} in the range $0<z<0.5$, such an analysis could, when compared to a standard BNS Hubble diagram one, improve $H_0$ precision by $30\%$, improve precision in $\Omega_{k0}$ by a factor of more than 10 (reaching $\sigma(\Omega_{k0}) < 0.02$ in 5 years), and constrain well both $\sigma_8$ and the growth-rate index simultaneously. Therefore, a joint analysis of GWs and LSS will help address both $H_0$ and $\sigma_8$ tensions. Figure~\ref{fig:cornerplot-6x2pt} shows  forecasts obtained in~\cite{Alfradique:2022tox} for BNS with KN obtained with a Vera Rubin Observatory follow-up, together with DESI BGS galaxies in an area of 14000 deg${}^2$, assuming a maximum wavenumber $k_{\rm max}=0.1\, h\rm{/Mpc}$ and $z_{\rm max}=0.5$.  
The forecast is for $\{\sigma_{8}, \gamma, \Omega_{m0}, \Omega_{k0}, H_{0}\}$, where $\gamma$ is the growth-rate index, which is assumed constant as a simple parametrization of beyond-GR gravity, marginalized over two free bias parameters in each of 5 redshift bins and three global non-linear redshift-space distortion parameters.

\subparagraph{Basic formalism.}
The anisotropic distribution of resolved GW events is described by the relative GW number count fluctuation in a direction~$\mathbf{n}$ and at redshift~$z$ defined as
\begin{equation}
    \Delta(\mathbf{n},z) = \frac{N-\langle N\rangle}{\langle N\rangle}.
\end{equation}
Given that GW interferometes are effectively all-sky experiments, it is convenient to study the $n$-point functions of this stochastic field in a base that reflects the intrinsic symmetry of the sky.
Therefore, following the treatment presented in~\cite{Scelfo:2018sny}, we decompose the number count fluctuation in a spherical harmonics base and we study the angular power spectrum given by
\begin{equation} \label{angular_ps}
    C_\ell^{XY}(z_i,z_j)=\frac{2}{\pi}\int \frac{dk}{k}\mathcal P(k)\Delta^{X,z_i}(k)\Delta^{Y,z_j}(k)\,,
\end{equation}
where we allow for the possibility of including different tracers~$X,Y$ (GWs and/or galaxies) in different redshift bins, and~$\mathcal P(k)$ is the (almost-scale invariant) primordial scalar power spectrum.
The number count fluctuation~$\Delta^{X,z_i}(k)$ reads as
\begin{equation}\label{eq:source_num_den}
    \Delta^{X,z_i}(k) = \int_{0}^{\infty} dz \frac{dN_X}{dz}w(z,z_i)\Delta^X(k,z),
\end{equation}
where~$dN_X/dz$ is the {\it observed} source number density per redshift interval, $w(z,z_i)$ a normalized window function centered at redshift $z_i$, and~$\Delta^X(k,z)$ encodes contributions from density, velocity, lensing and gravity effects~\cite{Bonvin:2011bg, Bellomo:2021mer}. 


\subparagraph{Noise, dependence on the population model and detector sensitivity. }
Since GWs are a discrete tracer of the LSS, their angular power spectrum measurement has a shot noise contribution analogous, for instance, to that of galaxies.
Assuming the noise of different experiments and in different redshift bins to be uncorrelated, the shot noise reads as~$\mathcal{N}_\ell^{XY} = \delta_{XY}\delta_{ij}(dN_X/ d\Omega)^{-1}$, where~$dN_X(z_i) / d\Omega$ is the {\it{total}} number of {\it{observed}} sources per steradian in the $i$-th redshift bin.
This quantity depends on the population properties, in particular on the total merger rate and on detector sensitivity.
Conversely, we note that the signal in eq.~\eqref{eq:source_num_den} depends on a normalized window function, hence the dependence on the population model is through its shape rather than through its overall amplitude~\cite{Scelfo:2018sny}.

For GW sources, an important difference with LSS studies is the limited angular resolution of GW detectors. The best angular resolution determines a limiting scale in the angular power spectrum~\cite{Oguri:2016dgk,Libanore:2020fim,Scelfo:2022lsx,Bosi:2023amu}, $\ell_{\rm max} \sim  180^{\circ}/\Delta \Omega_{\rm max}^{1/2}$ with $\Delta \Omega_{\rm max}$ being the best $1 \sigma$ sky localization of the GW catalog. 
From figure~\ref{fig:ET_loc_BBH}, we see that the best angular resolution of  ET alone corresponds to roughly $\Delta \Omega_{\rm max}\sim 0.1\, \rm deg^2$, hence a maximum multipole of at most $\sim300$ in a local redshift shell (up to $z\sim 0.2$), while for higher redshifts the range will be further limited. 
The effect of limited spatial resolution can be effectively modeled as a smoothing of the underlying overdensity field by an observational ``beam'', similarly to CMB experiments. 
A common choice in the literature is to parametrize this effect with a Gaussian beam, inducing an exponential damping in the GW-GW correlation signal~\cite{Calore:2020bpd,Libanore:2020fim,Scelfo:2022lsx,Bosi:2023amu}.
Moreover, error in distance/redshift localization can be accounted as showed in ref.~\cite{Oguri:2016dgk,Calore:2020bpd,Libanore:2020fim,Bosi:2023amu}, which suggested to implement a probabilistic treatment similar to those employed in weak lensing analysis, where photometric redshifts carry a non-negligible uncertainty.

\subparagraph{Cosmography.} In presence of tracers of the same density fields in distance and redshift spaces, the cross-correlation signal depends on the parameters of the distance-redshift relation.  
Here we consider in particular GW sources correlated with a galaxy catalog, as well as the respective auto--correlations.
The window function in eq.~\eqref{eq:source_num_den} takes the schematic form $w_i(z|\mathrm{bin\;in}\;z)$ for the galaxy survey and $w_i(d_L(z,H_0, ...)|\mathrm{bin\;in}\;d_L)$ for the GW catalog. That is, the binning is performed in the spaces where sources are observed - redshift for galaxies and luminosity distance for GWs, but GW sources distributed according to the source--frame distribution $dN_{\rm GW}(z_i) / d z$ are observed in different distance bins depending on the value of the parameters determining the distance--redshift relation, with the likelihood of the observed signal being maximised for the correct cosmology~\cite{Namikawa:2015prh,Oguri:2016dgk,Pedrotti:2025tfg}. 
\begin{figure}[t]
    \centering \hspace*{-10mm}
  \begin{minipage}[b]{0.60\linewidth}
  \vspace{0pt}
    \includegraphics[keepaspectratio, width=0.8\textwidth]{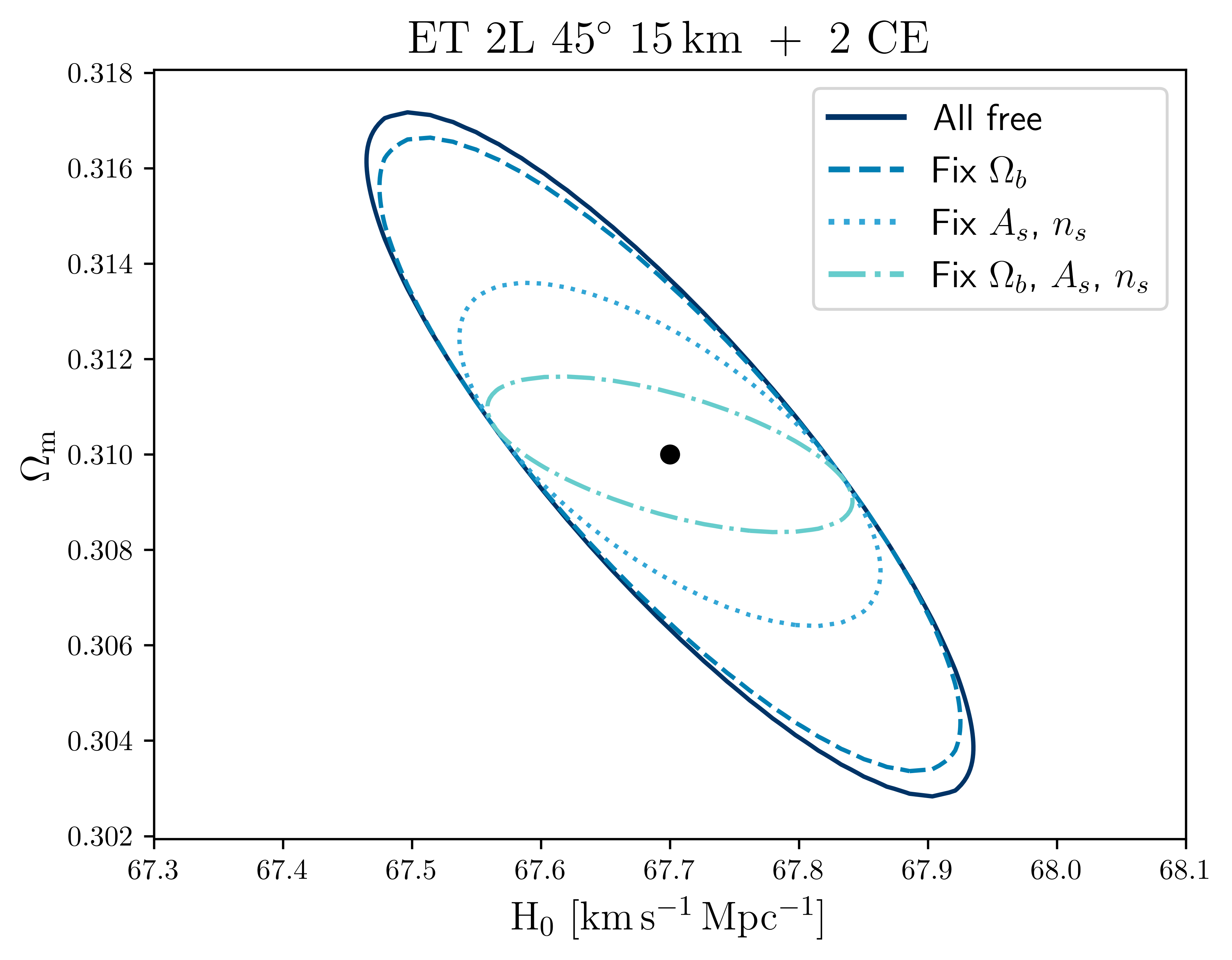}
  \par\vspace{0pt}
  \end{minipage} \hspace*{-15mm}
  \begin{minipage}[b]{0.50\linewidth}
  \vspace{-1pt}
    \resizebox{\textwidth}{!}{%
\begin{tabular}{|c||c|c|c|c|}
    \hline
    Parameter & \thead{All nuisance \\ parameters free} & Fix $\Omega_b$ & Fix $A_s, n_s$   & Fix $\Omega_b,\ A_s, n_s$ \\
    \hline
    \hline
     \multicolumn{5}{|c|}{$\rm{ET\ 2L\ 45^{\circ}\ 15\ km\ +\ 2CE}$} \\
    \hline
    $H_0$ & 0.8 & 0.75 & 0.55 & 0.47 \\
    \hline
    $\Omega_m$ & 5.3 & 4.9 & 2.6 & 1.2\\
    \hline
    \hline \multicolumn{5}{|c|}{$\rm{ET\ \Delta\ 10\ km\ +\ 2CE}$} \\
    \hline
    $H_0$ & 0.8 & 0.79 & 0.6 & 0.53 \\
    \hline
    $\Omega_m$ & 5.3 & 4.9 & 2.6 & 1.2\\
    \hline
\end{tabular}%
}
\par\vspace{60pt}
\end{minipage}
\caption{Left: $1\sigma$ constraints in the $H_0 - \Omega_{\rm m}$ plane from the  tomographic angular auto- and cross-correlation of resolved BBHs with a Euclid--like photometric galaxy catalog, for ET in the 2L configuration in combination with two CE detectors. 
Different lines correspond to fixing other cosmological parameters (baryon density $\Omega_b$, amplitude $A_s$ and spectral index $n_s$ of the primordial power spectrum), representing the case where prior knowledge is assumed on those, while still marginalising over the tracers' bias.
The case of a triangular ET is not displayed since almost indistinguishable. 
Right: relative errors (in $\%$) on $H_0,\ \Omega_{\rm m}$ for the 2L and triangular configurations of ET. Results from~\cite{Pedrotti:2025tfg}. }
\label{fig:contour_H0Om_crosscor}
\end{figure}

Figure~\ref{fig:contour_H0Om_crosscor} shows the forecasts in the $H_0-\Omega_{\rm m}$ plane obtained in~\cite{Pedrotti:2025tfg} for the auto-- and cross--angular power spectrum in tomographic redshift/distance bins [see 
\eq{angular_ps}] of GW sources with a galaxy catalog with the specifics of the Euclid photometric survey~\cite{EUCLID:2011zbd,Amendola:2016saw}. 
While the application of this technique to ET alone is possible, the angular resolution highly benefits from the presence of a network of detectors. So, we focus on 10 years of observations of ET in combination with two  CE detectors. We refer to~\cite{Branchesi:2023mws} for a description of the BBH population used.
We employ a Fisher matrix analysis with 20 equally populated redshift bins for the galaxy catalog, converted into luminosity distance bins for GW events assuming a fiducial cosmology. 
We account for the dominant contributions to the number count fluctuation, namely density and lensing~\cite{Fonseca:2023uay}. 
We also include the effect of the limited angular resolution with an exponential damping in the signal, with a redshift dependence computed in each redshift bin from the median localization in the simulations shown in figure~\ref{fig:ET_loc_BBH}. Furthermore, we restrict to a maximum multipole $\ell_{\rm max}=300$ in any case, and for GW sources we impose a stricter hard cut on the maximum multipole in each bin (denoted with the index $i$) at $\ell_{\rm max, i}=\pi/\sqrt{\Delta \Omega_{\rm min, i}}$ being $\Delta \Omega_{\rm min, i}$ the best sky localization in the bin.
The result is marginalised over the amplitude and spectral index of the primordial power spectrum, the baryon density, and the amplitude of the GW and galaxy bias in each bin, for a total of 43 nuisance parameters.
Therefore, importantly in light of the Hubble tension, this constraints are independent on any other external probe or prior (e.g. BBN, CMB, SNe). 
We also show the result obtained fixing the baryon density $\Omega_b$ and primordial power spectrum amplitude $A_s$ and spectral index $n_s$ to their fiducial values, corresponding to the case where external priors are added.
The relative error is reported in the table in the right panel of figure~\ref{fig:contour_H0Om_crosscor}. 
In the best case, we forecast a $0.8\%$ measurement of $H_0$, which can be reduced to $\sim 0.5\%$ when assuming prior knowledge of $\Omega_b,\ A_s,\ n_s$, in which case a $\sim 1\%$ measurement of $\Omega_m$ is also obtained.
In conclusion, the cross--correlation technique for ET in combination with CE can lead to a promising sub--percent determination of $H_0$, with percent--level precision already reached after one year of operation. 

\begin{figure}[t!]
    \centering
\includegraphics[width=0.54\columnwidth]{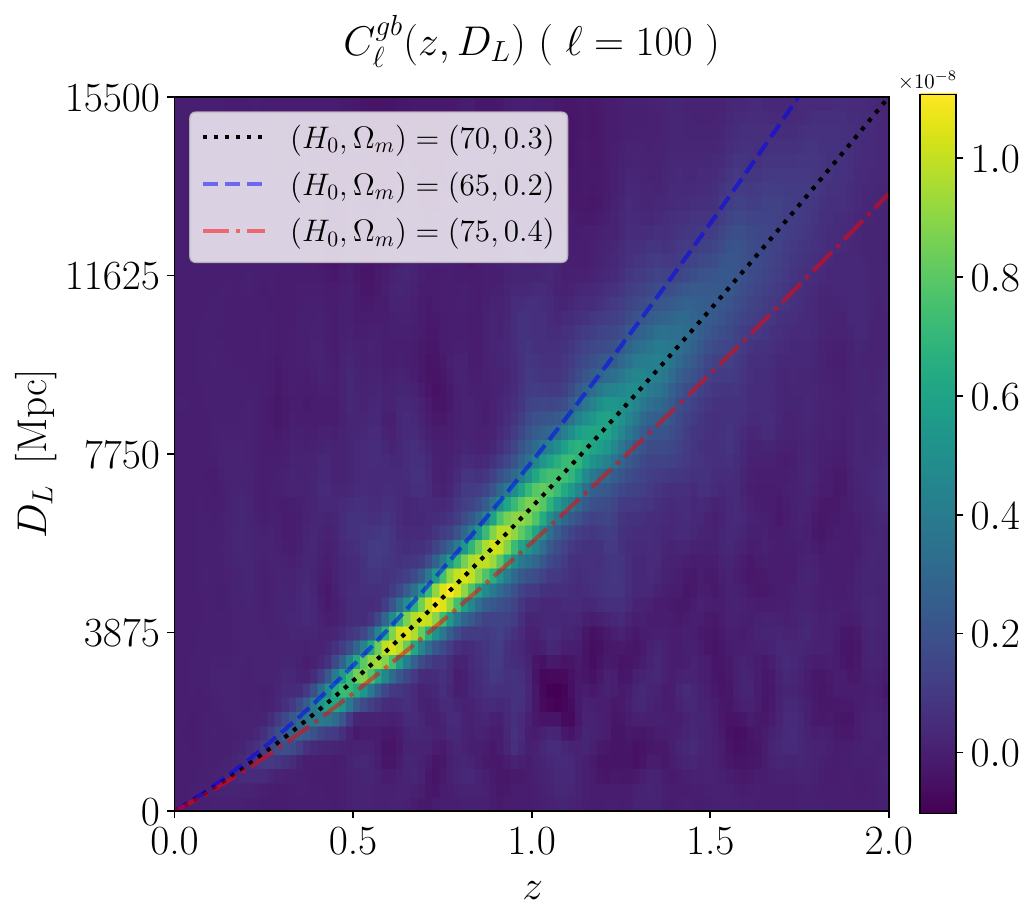}
    \caption{The cross-correlation between the harmonic modes of galaxies ($g$) at $z$ and BBHs ($b$) at $D_L$ leads to a cross-angular power spectrum $C^{gb}_\ell(z,D_L)$.
    We show the average cross-spectrum for 1000 light-cone simulations of galaxies and BBHs, for $\ell=100$ ($\sim 2^\circ$). When a shell in $z$ coincides with a shell in $D_L$, the correlation between the galaxy and BBH maps is maximal.
    The red, blue and black lines correspond to the Hubble diagrams $D_L(z)$ of three different cosmologies -- in the fiducial model $H_0=70.0\,\, {\rm km}\, {\rm s}^{-1}\, {\rm Mpc}^{-1}$ and $\Omega_m=0.3$. Figure taken from \cite{Ferri:2024amc}.      \label{fig:Intro12}}
\end{figure}

Reconstructing the Hubble diagram from cross-correlations of GW sources and tracers in redshift space can also be done directly through full-sky cosmological galaxy mock simulations. In \cite{Ferri:2024amc}, the authors employ thousands of relativistic galaxy simulations together with binary black-hole sky maps populating the same large-scale structures of the galaxy maps, taking into account various systematic sources of anisotropies: the direction dependent sensitivities of both 2G and 3G GW networks, galaxy mask, the lensing magnification of luminosity distances, and modeling BBHs as extended sources due to the much larger errors in the determination of their 3D positions (RA, Dec, $D_L$) as compared to the ones for galaxies.
Ref.~\cite{Ferri:2024amc} shows that  the method is robust to all these effects and to the choice of fiducial cosmology behind the simulations. This means that the position of the crest in the $(z, D_L)$ plane illustrated in fig. \ref{fig:Intro12} is a cosmic standard ruler, not depending strongly on the details of the individual large-scale structures in the two spaces (distance or redshift) and rather on their matching. With such technique, one can get sub-percent measurements of the Hubble constant with 5 years of ET+2CE, while measuring also other cosmological parameters, which is one order of magnitude more precise than what is expected from the next observing run of LVK (O5). Moreover, if LVK is still active at the time of ET, LVK would help ET on localizing binary sources, even when those are below the detection thresholds for the 2G network, also potentially leading to sub-percent level measurements of $H_0$.

\subparagraph{Modified gravity.}
One of the goals of all next generation cosmological experiments is to test General Relativity in the late Universe.
In particular, General Relativity has not been fully tested on scales comparable to today's cosmological horizon.
At those scales projection effects have the largest impact on the clustering of LSS tracers, while, at the same time, being very sensitive to deviations from General Relativity.
Therefore GW clustering offer the unique possibility to fulfill this task in a complementary fashion to galaxy surveys or line intensity mapping experiments.

Two possible extensions that can be tested are: an equation of state for dark energy~$w_\mathrm{DE}\neq -1$, and a modified growth of structure, often parametrised via the (possibly) time- and scale-dependent~$\mu,\eta$ functions defined as~\cite{Amendola:2007rr}
\begin{equation}
    k^2\Psi(k,z) = -4\pi G a^2 \mu(k,z) \bar{\rho} D, \qquad \Phi(k,z) = \eta(k,z) \Psi(k,z), 
\end{equation}
where~$\Psi,\Phi$ are the Bardeen potentials, $a$ is the scale factor, $\bar{\rho}$ is the background energy density, and~$D$ is the gauge-invariant energy density fluctuation.
This parametrization allows to test a broad class of models, from scalar-tensor theories~\cite{Horndeski:1974wa, Deffayet:2009mn, Gleyzes:2014dya, Kobayashi:2019hrl} to models that invoke extra dimensions~\cite{Dvali:2000hr}, by choosing an appropriate time- and scale-dependence of the~$\mu,\eta$ variables.

Many different authors employed different parametrisations to forecast sensitivity of next generation GW observatory, such as ET, on deviations from General Relativity, both as unique probe or by cross-correlating this datasets with galaxy clustering, weak lensing and HI line intensity mapping~\cite{Scelfo:2021fqe, Scelfo:2022lsx, Bosi:2023amu}.
Constraints on these parameters are heavily dependent on the chosen parametrization, however the general consensus is that in all the cases ET alone, or in synergy with Cosmic Explorer, will provide unparalleled constraints on extensions of General Relativity.


\subparagraph{Primordial black holes.}
The clustering properties of PBH binaries can be substantially different from those of traditional stellar-origin BH binaries~\cite{Bird:2016dcv, Raccanelli:2016cud, Libanore:2023ovr}.
While astrophysical BH binaries are the result of intense stellar formation history in galaxies, and thus they trace the galaxy (anisotropic) spatial distribution, PBH binaries trace a different kind of environment, depending on their formation channel (see section~\ref{section:div3} for detailed discussion).
In particular, there are two well-understood formation channels that are worth mentioning: early-time~\cite{Sasaki:2016jop} and late-time binaries~\cite{Bird:2016dcv, Clesse:2016vqa}.
In the former channel, binaries start forming in the early Universe, during radiation domination, and therefore they are thought to evenly trace all sort of dark matter structure that form in the late Universe.
In the latter formation channel, binaries are expected to form in the late Universe in low-mass dark matter halos ($M_\mathrm{halo} \simeq 10^2-10^3\ M_\odot$) via a direct capture process because of the low relative velocity of halo particles.

Clustering properties of GWs depend mainly on the bias parameter~$b_{\rm GW}$.
Since astrophysical binaries trace luminous galaxy distributions, we expect them to have a bias close to that of galaxies, i.e.,~$b_{\rm GW} \sim b_\mathrm{gal} \sim 1.2-3.0$.
On the other hand, early-time binaries, since they effectively trace the entire dark matter distribution, are expected to have~$b_{\rm GW}\sim 1$.
Finally, late-time binaries, because of the peculiarity of their formation mechanism, trace only low mass halos, mostly in filaments, thus they have~$b_{\rm GW} \sim 0.5$.
Therefore, it is possible to disentangle PBHs from astrophysical binaries by measuring the GW bias with sufficient precision.
Even in this case, the exact numbers depend on the modeling of PBHs; however the consensus reached by different studies is that by measuring the bias with a precision of order few percent it is possible to constraint PBH abundance at the ten percent level~\cite{Raccanelli:2016cud, Scelfo:2018sny, Scelfo:2021fqe, Bosi:2023amu, Libanore:2023ovr}.
Interestingly, reaching that level of precision will be well within the reach of ET in combination with many current and future LSS surveys, making it potentially capable of resolving the long-standing issue of what the nature of dark matter is~\cite{Libanore:2023ovr}.


\paragraph{Cross-correlation of AGWB with LSS.} \label{subsec:agwb_lss}

In section~\ref{sec:Anisotropies of Astrophysical Backgrounds} 
it has been discussed that the inhomogeneous distribution of sources and the line-of-sight effects accumulated during propagation contribute to the anisotropies of the AGWB.
These intrinsic anisotropies are generated by the primordial perturbations in the Universe on large angular scales, and therefore  constitute a tracer of the LSS. This implies that the cross-correlation of the anisotropies of the AGWB with other tracers of the LSS, such as galaxy number counts~\cite{Cusin:2018rsq,Cusin:2019jpv,Canas-Herrera:2019npr,Alonso:2020mva,ValbusaDallArmi:2022htu,Yang:2023eqi}, weak gravitational lensing \cite{Cusin:2018rsq,Cusin:2019jpv}, CMB \cite{Ricciardone:2021kel}, and CMB lensing \cite{Capurri:2021prz}, could be very large in light of the common seeds of the perturbations which generate the anisotropies of these signals.

It is possible indeed to characterize the anisotropies of the AGWB as a tracer of the LSS by computing the redshift distribution, the bias, and the magnification bias of the AGWB energy density, as shown in \cite{Bertacca:2019fnt,Capurri:2021zli,Bellomo:2021mer}. Among these three functions, the bias is of particular interest, since it contains information about GW clustering, quantifying how effectively the AGWB traces the overdensity of CDM. The analysis of the bias of the AGWB could be complementary to the one of the anisotropies of the resolved GW sources discussed in section~\ref{Cross-correlation of GW Resolved Evenets with LSS}. The bias of the AGWB energy density can be theoretically predicted starting from the bias of the host galaxies as 

\begin{equation}
    b_{\textrm{GW}}(z) = \frac{\int d\theta_{\textrm{gal}} \frac{d\bar{\Omega}_{\rm GW}}{dz d\theta_{\textrm{gal}}} b_{\textrm{gal}}(z, \theta_{\textrm{gal}}) }{\int d\theta_{\textrm{gal}} \frac{d\bar{\Omega}_{\rm GW}}{dz d\theta_{\textrm{gal}}}},
\end{equation}
where $\theta_{\textrm{gal}}$ is the key parameter describing the properties of the host galaxies. It could be the star-formation rate $\psi$ as in \cite{Capurri:2021prz}, or the total stellar mass $M_{\star}$ as in \cite{Bellomo:2021mer}. In~\cite{Bellomo:2021mer} an analog expression for the bias has been given in terms of the mass function and the bias of the host halos. Since the dominant contribution to AGWB anisotropies is given by the clustering term \cite{Bertacca:2019fnt,Cusin:2019jpv}, even a measurement of the first few multipoles of the angular power spectrum can yield informative constraints on $b_{\textrm{GW}}$. The astrophysical implications of the cross-correlation of the AGWB with other probes are discussed in section~\ref{subsect:crosscorr_EM_conterparts}.

In addition, the cross-correlation of the AGWB with other LSS probes could play a crucial role in the detection of the anisotropies. As discussed in section~\ref{sec:Anisotropies of Astrophysical Backgrounds}, the shot noise and instrumental noise make the direct measurement of the anisotropies very challenging with current standard techniques. However, cross-correlating the GWB with other LSS tracers~\cite{Cusin:2018rsq,Cusin:2019jpv,Canas-Herrera:2019npr,Alonso:2020mva,ValbusaDallArmi:2022htu,Yang:2023eqi,Ricciardone:2021kel,Capurri:2021prz}, could help increasing the SNR by a significative factor. Thus, the cross-correlation with electromagnetic probes represents one of the best tools to detect GWB anisotropies with ET, if used together with the multi-frequency analysis of the anisotropies discussed in section~\ref{sec:Anisotropies of Astrophysical Backgrounds} or other data analysis techniques. Indeed, even detecting the first few multipoles of the cross-correlation angular power spectrum would provide insights into the aforementioned underlying astrophysical and cosmological processes.


\subsubsection{Cross-correlation of AGWB with CMB}
\label{sec:cross-corrAGWB-CMB}

Cross-correlations of GW and CMB anisotropies are an important probe for characterising the GW background and possibly disentangling its various components (e.g., the cosmological versus astrophysical contributions). The two backgrounds are indeed affected by the same large-scale (scalar and tensor) perturbations, both at production and during propagation through the Universe towards the Earth \cite{Contaldi:2016koz,Bartolo:2019oiq,Bartolo:2019yeu,Bartolo:2019zvb,Dimastrogiovanni:2019bfl,Adshead:2020bji,ValbusaDallArmi:2020ifo,Malhotra:2020ket,Ricciardone:2021kel,Dimastrogiovanni:2021mfs,Dimastrogiovanni:2022eir,Malhotra:2022ply,Bodas:2022urf,Cui:2023dlo,Schulze:2023ich,ValbusaDallArmi:2023nqn}. An overview of GW background anisotropies and their properties in the context of different early Universe scenarios can be found in section~\ref{sub:anisotropies cosmo}. 

As discussed before, a possible source of anisotropies for the GW signal comes from the distribution of GW sources in the sky. Astrophysical GWs events, in fact, are expected to be hosted in galaxies, whose distribution follows the underlying dark matter one. Hence such sources reside in the large-scale gravitational potentials. Among the different anisotropies imprinted on the CMB signal, instead, the relevant one for the cross-correlation with GW sources is the late-times integrated Sachs-Wolfe effect~\cite{Sachs:1967er}. This effect arises from the evolution of gravitational potential in times: the energy of a photon going trough the potential is changed leaving an imprint on the CMB signal. The cumulative change of energy during the photon path in the dark-energy era gives rise to the late-Integrated Sachs-Wolfe (ISW) effect. This is not only a late-time effect but happens to be also relevant on the largest scales. The physical explanation for the cross-correlation, hence, becomes clear: the large-scale gravitational potential giving rise to the late-ISW are the same potentials hosting GW sources, hence tracing their distribution in the sky. The dominant contribution to the GWxCMB cross-correlation is thus expected from the density anisotropies of the former with the ISW of the latter.

The cross-correlation between GW and the CMB could also provide a complementary way to constrain primordial non-Gaussianity. As first shown 
by~\cite{Dalal:2007cu,Matarrese:2008nc,Slosar:2008hx}, the presence of local non-Gaussianities induces an additional scale-dependent term in the bias of the tracer considered. This correction affects the density term in \eqref{Eq::GW_Anisotropies}, being the only one proportional to the GW bias~\cite{Bertacca:2019fnt,Bellomo:2021mer}. The scale dependence scales as $1/k^2$, hence it dominates on the largest scales. As explained above, on those scales the cross-correlation of GWs and the CMB is expected to be important and hence to be affected by the presence of non-Gaussianities, suffering of an enhancement or a suppression which can be probed by 3G detectors~\cite{Perna:2023dgg}.

\subsubsection{Probing LSS with GWs alone}

The study of the LSS of the Universe is performed by constructing a two-point correlation function of galaxies, which describes their clustering properties at large scales $(\gtrsim 10~ \text{Mpc})$. These studies are usually performed using galaxy catalogues constructed from large EM surveys. One result of these studies is the identification of the Baryon Acoustic Oscillation (BAO) peak at $\approx 100\ h^{-1}\text{Mpc}$, which puts constraints on the density parameter $\Omega_{m0}$ of the cosmological model and on the equation of state of dark energy models. We expect that the 3D localization of coalescing compact binaries could serve as another probe providing complementary information, as these compact objects trace the underlying DM distribution in a possibly different way than the luminous matter of galaxies does, due to the astrophysical processes leading up to the formation of CBCs. This discrepancy is quantified by the clustering bias. Compact binary source localizations thus probe the {\em GW bias} $b_\text{GW}$, which is inherently different from the galaxy bias $b_\text{gal}$ probed by luminous matter distribution. One should note that $b_\text{GW}$ is potentially dependent on redshift as well as the astrophysical environment for compact binary formation and merger. It goes without saying that GW observations are needed to probe the GW bias. Earlier in this section, we have discussed the prospects of measuring $b_\text{GW}$ using a combination of GW and EM tracers. It is worth exploring the possibility of measurement of $b_\text{GW}$ using GW probes alone.

The proposed paradigm comes with certain challenges. Firstly, instead of point-like objects, the new method has to deal with extended GW detection volumes, which are currently of the same size or larger than the clustering scale itself. Secondly, the current GW detection rates are too low to perform this kind of statistical analysis. Hence, this endeavor requires a new detector network with enhanced sensitivity. Einstein Telescope and the 3G detector network in general open up the unique prospect of probing the LSS of the universe using GW alone. 

\begin{figure}
\centering
\includegraphics[width=0.42\textwidth,trim={2cm 2cm 0 0},clip]{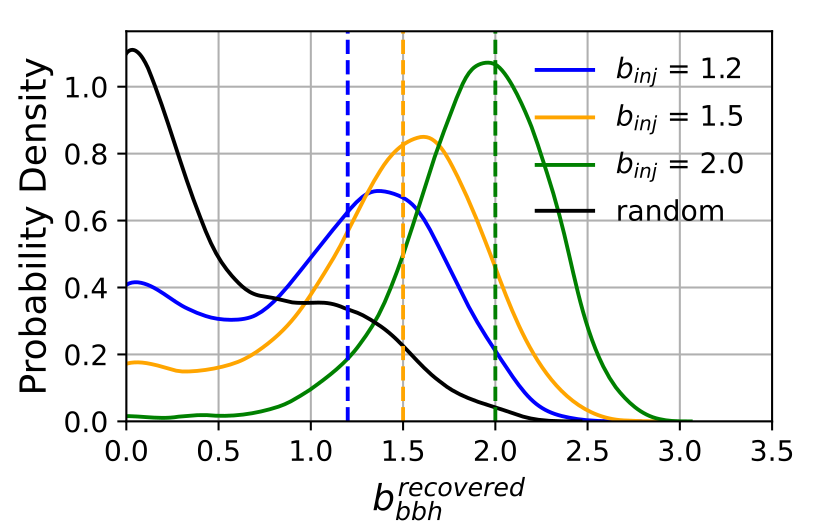} \hspace{0.08\textwidth}
\includegraphics[width=0.4\textwidth,trim={1.6cm 1.2cm 0 0},clip]{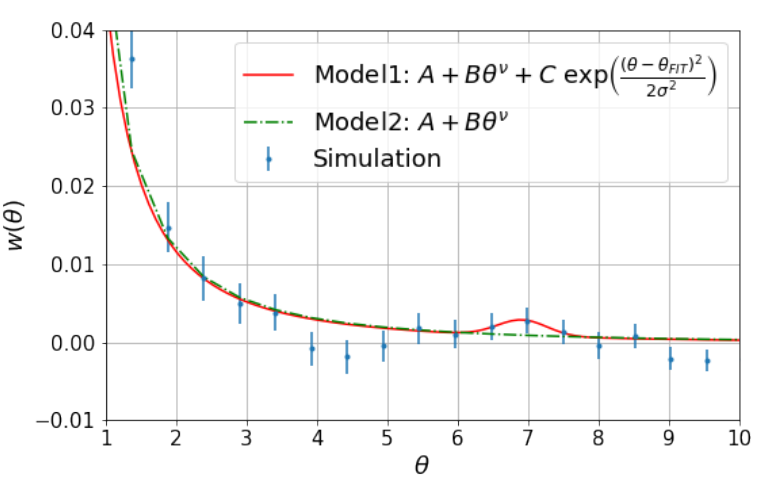}
\vspace{1.5em}
\put(-375,-10){\small Gravitational-wave bias $b_\text{GW}$}
\put(-420,13){\rotatebox{90}{\small Probability density}}
\put(-145,-10){\small Angular scale $\theta$ [degrees]}
\put(-195,10){\rotatebox{90}{\small Ang.~correlation $w(\theta)$}}
\caption{{Left:} Recovery distributions of the GW bias $b_\text{GW}$ correspond to the injected values~\cite{Vijaykumar:2020pzn}. These results are for simulated BBH observations at a redshift $z=0.5$ (shell thickness $350\,h^{-1}\,\text{Mpc}$) and with 10 years of observation by a worldwide ET+2CE network (ET 10\,km $\Delta$ + CE 40\,km US + CE 40\,km Australia). {Right:} Recovery of the angular two-point correlation function $w(\theta)$ reveals the BAO peak at the injected angular scale, $\theta=6.9^\circ$~\cite{Kumar:2021aog}. These results are for simulated BNS observations at redshift $z=0.3$ (shell thickness $150\,h^{-1}\,\text{Mpc}$) observed by the same detector network and over similar observation times.}
\label{fig:lss-gw-alone}
\end{figure}

Using third-generation ground based detectors consisting of ET and CE, we expect to detect thousands of mergers per year with sky localization areas within a square degree. It has been shown that with 5 to 10 years of observation time with such a network, we can extract the clustering bias of the BBH population~\cite{Vijaykumar:2020pzn} (figure~\ref{fig:lss-gw-alone}, left panel). In that same observation time, it is possible to detect the BAO peak solely from GW observations~\cite{Kumar:2021aog} (figure~\ref{fig:lss-gw-alone}, right panel). This reconstruction  constrains the value of $H_0$. Therefore, the separate galaxy and GW measurements of BAO peak are complementary to each other, and combining their data will enable us to constrain cosmological parameters using a standard ruler, i.e.~the BAO scale~\cite{Kumar:2022wvh}. The results of these preliminary studies are the first indications that there are indeed prospects in probing the LSS with GW alone and more-realistic follow-up studies are in order. 

Another method for probing LSS with third-generation GW observatories is by leveraging its impact on the event rate of BBHs and other CBCs~\cite{Mosbech:2022nkk}.
In particular, the rate of high-redshift BBHs is a sensitive tracer of the abundance of low-mass dark matter halos that form at early times.
Measuring this rate therefore allows us to test for suppression in the matter power spectrum on small scales compared to $\Lambda$CDM, as predicted by various alternatives, such as warm dark matter, fuzzy dark matter, and interacting dark matter~\cite{Boehm:2000gq}.
Ref.~\cite{Mosbech:2022nkk} found that a 3G network composed by ET+2CE could improve current constraints on these dark matter alternatives by as much as two orders of magnitude.

\subsection{Executive summary}
Here we summarize the advancements that ET will bring in Cosmology, in comparison to the current state of the art, specifically emphasizing the distinct features that the ET will allow us to explore. The following are the major areas where ET's capabilities surpass the current generation of GWs detectors:


\begin{highlightbox}{Early Universe Cosmology with Einstein Telescope}
\begin{itemize}
    \item 	\textbf{Gravitational Wave Background (GWB)}
    
    \begin{itemize}
        \item 	\textbf{Sensitivity:}  Einstein Telescope will provide a strain sensitivity improvement of up to a factor of $10$ compared to Advanced LIGO-Virgo, achieving a targeted sensitivity of $S_h^{1/2}(f)\simeq 10^{-24}\, {\rm Hz}^{-1/2}$ at $10$ Hz  and  a broader frequency range extending down to $3-5$~Hz, compared to the $\sim 10-20$ Hz lower limit of current detectors (see figure~\ref{fig:asd} for a plot of the strain sensitivity in different configurations). This expanded range will allow ET to capture more sources and trace features of the  GWB that are inaccessible today.
        \item 	\textbf{Detailed Study of GWB Components:} With its enhanced sensitivity and wider frequency range, ET will be able to:
        
        \begin{itemize}
            \item 	\textbf{Cosmological GWB:} Detect the CGWB with a sensitivity reaching energy density values of $\Omega_{\rm GW} \sim 10^{-12}$, allowing us to probe early Universe events like inflation, phase transitions, and cosmic strings.
            \item 	\textbf{Astrophysical GWB:} Extract astrophysical population properties more precisely, tracing large-scale structures with better sensitivity.
        \end{itemize}
          \item 	\textbf{Anisotropies and Parity violation:} with its improved sensitivity and new layout ET will have better resolution to probe anisotropies of the GWB, and will be also sensitive to chiral GWBs, which are indicators of new early Universe physics.
    \end{itemize}

    \item 	\textbf{Inflation, Phase Transitions and Cosmic Strings}
    
    \begin{itemize}
        \item 	\textbf{Inflation:} ET is expected to be sensitive to GWs originating from inflationary models beyond standard single-field slow roll and from multifield models, providing an opportunity to test fundamental physics and the presence of extra field in the early Universe.
        \item 	\textbf{First-Order Phase Transitions (PTs):} ET is expected to be sensitive to gravitational waves originating from PTs, with an observable energy density $\Omega_{\rm GW}$ potentially as low as $10^{-11}$ for strong PTs at temperatures of $\sim 10^3$ GeV, corresponding to  earlier epochs of the primordial Universe with respect to those probed by  LISA.
        \item 	\textbf{Cosmic Strings:}  ET will probe local string networks with tensions as low as $G\mu \sim 10^{-18}$, which is an improvement of several orders of magnitude compared to existing bounds ($G\mu \gtrsim 10^{-10}$ from current PTA experiments).
    \end{itemize}

\end{itemize}

\end{highlightbox}


\begin{highlightbox}{Late Universe Cosmology with Einstein Telescope}

\begin{itemize}
    \item 	\textbf{Cosmography and Cosmological Parameters}
    
    \begin{itemize}
        \item 	\textbf{Bright Sirens:} ET will detect hundreds of bright standard siren events (neutron star mergers with kilonovae/short GRB counterparts) per year, allowing direct measurements of the Hubble constant $H_0$ with sub-percent precision.
        \item 	\textbf{Dark Sirens:} For dark sirens (BBHs without EM counterparts), ET will utilize galaxy catalog information, or correlation with the mass function,  to statistically infer redshifts, enabling a $\sim 10\%$ determination of $H_0$ and other cosmological parameters.
        \item 	\textbf{Distance-Redshift Relation:} The precision in measuring the distance-redshift relation will enable model-independent tests of the expansion history of the Universe, addressing current tensions in $H_0$ measurements.
    \end{itemize}

    \item 	\textbf{Modified Gravity and New Physics}
    
    \begin{itemize}
        \item 	\textbf{Gravitational Wave Propagation:} The ability to compare GW and EM luminosity distances will allow ET to test deviations in GW propagation that might arise from modifications to General Relativity, potentially detecting differences  of order $10^{-2}$.
        \item 	\textbf{Cosmic Dipole and $H_0$ Tensions:} ET will also contribute to resolving cosmic dipole discrepancies by comparing dipoles in the GW and EM sectors with a precision that is unachievable with current second-generation detectors.
    \end{itemize}

    \item 	\textbf{Detection of Individual Cosmological Events}
    
    \begin{itemize}
        \item 	\textbf{Gravitational Wave Bursts from Cosmic Strings:} ET will be capable of detecting GW bursts from cosmic string cusps if $G\mu \sim 10^{-11}$, while non-detection could constrain $G\mu$ below this level for different string loop models.
        \item 	\textbf{Primordial Black Holes:} With its improved sensitivity, ET could detect signatures from PBH mergers, originating not only from inflationary mechanisms but also from domain walls or phase transition. In this way, it will allow to shed light on Dark Matter.
        \item \textbf{Ultralight dark matter:} ET will probe the existence of scalar or vector ultralight dark matter, that can form clouds around spinning black holes.
    
    \end{itemize}

\end{itemize}

Einstein Telescope will provide a significant step forward in our ability to study our early and late Universe. It will offer deeper insight into processes occurring at high redshifts and across a broader range of frequencies than ever before, making it an essential tool in the pursuit of our understanding of cosmology.

\end{highlightbox}

\section{Population studies and astrophysical background}\label{section:div3}

The first detection of gravitational waves (GWs) in 2015 is often compared to Galileo's first astronomical observation with an optical telescope. Moving from LIGO-Virgo-KAGRA to the Einstein Telescope (ET) and Cosmic Explorer marks an equally transformative leap, comparable to advancing from Galileo's crude instrument to the James Webb Space Telescope, which now reveals the cosmos in unparalleled detail. ET will observe approximately $10^5$ binary black hole (BBH) and $10^4-10^5$ binary neutron star (BNS) mergers every year, up to a redshift of 100, and for a few thousand BBH events per year and several tens to ${\cal O}(10^2)$ BNS per year, the signal-to-noise ratio will exceed 100~\cite{Iacovelli:2022bbs,Branchesi:2023mws}. With such capabilities, ET will unveil the mass function and spin distribution of BHs and neutron stars (NSs) in exquisite detail, allowing us to probe their evolution with redshift. While LIGO and Virgo have already paved the way for studying intermediate-mass BHs ($10^2-10^5 \Msun$) with GWs, ET will observe systems with masses of $\approx{10^3} \Msun$ up to a redshift of $\approx{}8$, possibly via joint multi-band observations with LISA, the first space-borne GW detector. This will be pivotal  for understanding the formation of supermassive black holes in the Universe.

 At high redshift, Population III stars, the very first stars formed in the Universe, continue to elude our observations, even with the James Webb Space Telescope. The Einstein Telescope will routinely observe mergers of black holes born from Population~III stars,  with the main challenge being distinguishing such high-redshift messengers from other events. The individual observation of a binary BH merger at redshift $\gtrsim{}30$, or of a sub-solar mass non-tidally deformable object would be a smoking gun for the existence of primordial black holes, with profound implications for our understanding of dark matter. This will ultimately pave the way for interpreting the formation  channels of binary compact object mergers.

\subsection{Introduction: formation channels of binary compact objects}
\label{div3:formationchannel}

After three observation runs, the LVK detected $\sim 100$ compact binary coalescences (CBCs, \citep{KAGRA:2021vkt}), a number that may double or triple by the end of the ongoing fourth observation run. The number of detected CBCs enabled the scientific community to start placing constraints on the properties of the underlying population of compact objects. Nonetheless, the origin of CBCs is still unknown and widely debated. Broadly speaking, we can distinguish two main formation channels: the isolated channel, according to which compact binaries are the byproduct of binary stellar evolution; and the dynamical channel, in which compact binaries --- or their progenitors --- pair up through gravitational interactions in dense stellar environments.
ET will contribute key improvements to reveal the formation channels and distinguish between them, for example through golden (high SNR) events that will pin down properties currently difficult to measure  like spin, as well as through multidimensional constraints, such as the redshift dependence of masses and spins. Here below, we discuss this in detail. 

\subsubsection{Isolated channel: compact binary mergers from pairs and multiples in galactic fields}\label{sect:isolatedchannel}

\paragraph{Binary evolution with mass transfer.}\label{sect:binarywithmasstransfer}

For most close binaries, it can become unavoidable for the two stars to interact at some point during their life to eventually form a GW source. A circular binary with two compact objects can only experience a merger due to GW emission within a Hubble time if the orbital separation  is less than a few solar radii (for NS+NS) to less than several tens of solar radii (for BH+BH) \cite{Peters:1964zz}. 
However, the radius of their progenitor stars is already larger than a few solar radii at birth, and easily reaches tens to hundreds of solar radii during the giant phases \cite{Maeder:2000wv}. For that reason, the story of the formation of GW sources in stellar binaries is one of stellar interactions and mass transfer. 

A mass transfer phase strongly affects the stellar components (i.e., their masses and mass ratio, their spin properties), as well as the orbital separation, and consequently the delay time (i.e. the time elapsed from binary formation down to the merger). Progenitors of CBCs typically undergo two mass transfer interactions: a first one in which the initially most massive (primary) star is stripped of its hydrogen envelope, and the second one where the companion star is stripped. If the process occurs in a self-regulating manner, it is referred to as ``stable mass transfer'', while ``unstable mass transfer'' is often assumed to lead to a common-envelope (CE) phase \cite{1976IAUS...73...75P,Webbink:1984ti, Ivanova:2012vx, Ivanova20}. 
Interestingly, the process of stable and unstable mass transfer affects the system in different ways, leaving different imprints on the system. These imprints provide us with a wonderful opportunity to distinguish between (or potentially determine the) GW progenitors from the source properties. 
 Here we mention the two main channels in binary evolution with mass transfer and discuss the corresponding source properties. Other channels likely make up only a few per cent of the total yield (e.g., \cite{Broekgaarden:2021iew, Broekgaarden:2021efa}).

\begin{figure}
\centering
	\includegraphics[trim= 40 0 50 0, clip, width=0.5\columnwidth]{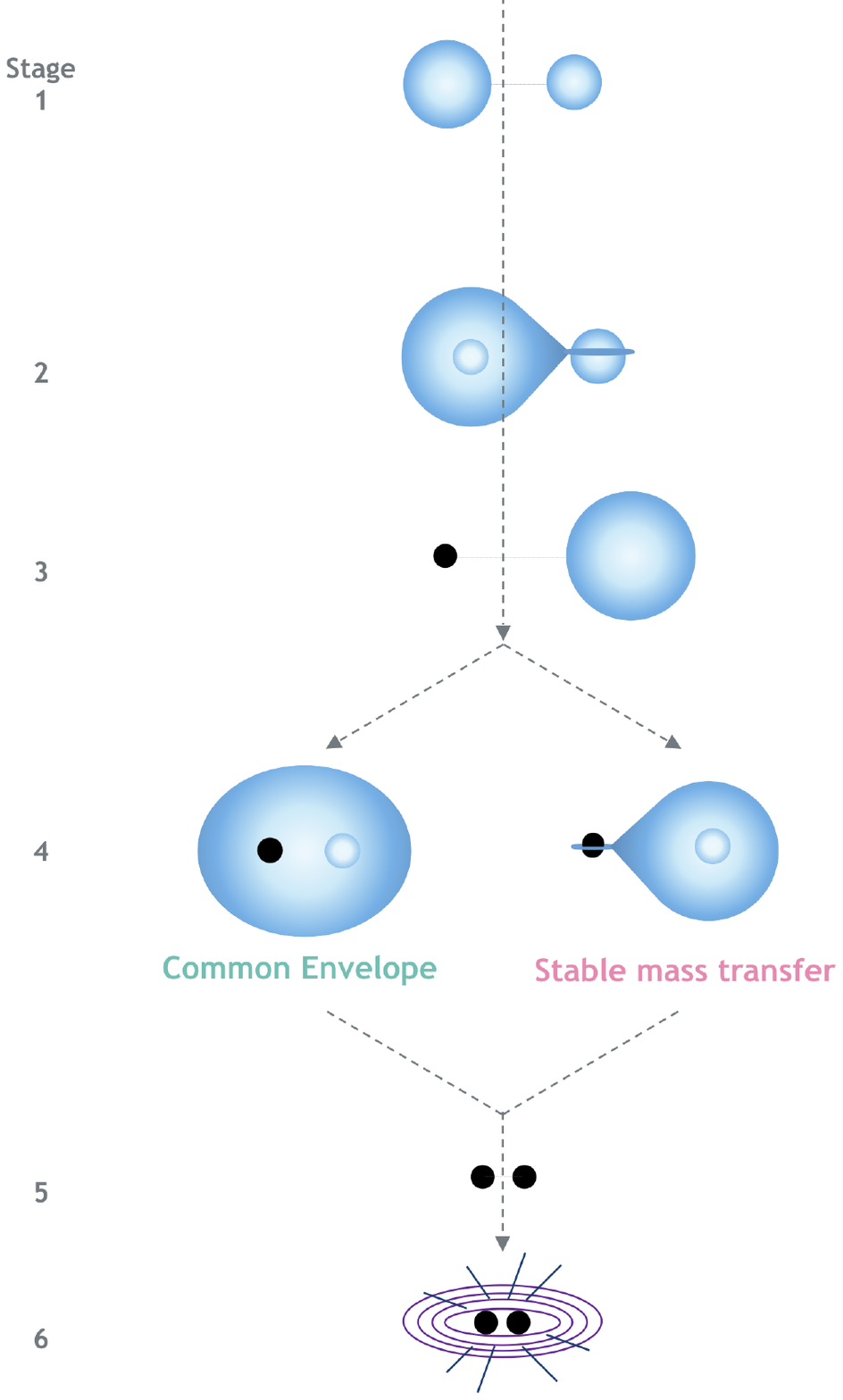}
        \includegraphics[trim= 280 50 320 100, clip, width=0.4\columnwidth]{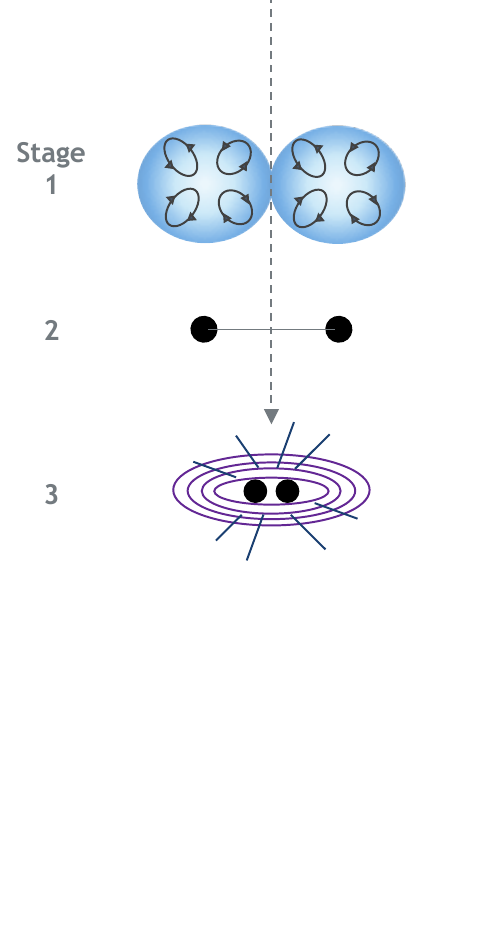}
    \caption{Evolutionary pathways to form a GW source for isolated binary evolution. Left: the common-envelope channel and the stable mass transfer channel. Stage 1) Zero-Age-Main-Sequence (ZAMS) 2) First phase of mass transfer 3) Formation of BH or NS 4) Second phase of mass transfer 5) Formation of double compact object 6) GW merger. Right: chemically homogeneous evolution. Stage 1) ZAMS 2) Formation of binary BH 3) GW merger.} 
   \label{fig:classical_bin_channel}
\end{figure}

\subparagraph{Common envelope channel.}
A well-known evolutionary pathway is the common-envelope channel (hereafter CE channel: an example is given in figure~\ref{fig:classical_bin_channel}, left). This channel leads to plenty of BBH mergers \cite{Belczynski:2006zi, Postnov:2014tza}, BHNS mergers \cite{Giacobbo:2018etu, Neijssel:2019irh,Broekgaarden:2021iew}, as well as BNS mergers \cite{Tauris:2017omb, Chruslinska:2017odi, Vigna-Gomez:2018dza, Giacobbo:2018etu}. Initially on the zero-age main-sequence (ZAMS), the stars are not interacting (stage~1).
After expanding, the primary star fills its Roche lobe and the first phase of mass transfer commences (stage~2).  It proceeds in a stable manner, typically leading to widening of the orbit. The donor star is slowly stripped of its envelope, and at the end of its nuclear evolution, it will collapse to a compact object (i.e. BH or NS) at stage~3 (see also section \ref{sec:sn-direct-collapse}). 
The next stage in the channel is driven by the evolution of the secondary star, leading to a second phase of mass transfer at stage 4 (see also section \ref{sec:ns-mass}).
 For this channel, the second phase of mass transfer occurs in an unstable manner due to the extreme mass ratio and the response of the donor to mass loss, and a common-envelope phase is expected to develop. The name of the channel is derived from this second phase of mass transfer. During the common-envelope phase, the companion plunges into the envelope of the secondary star, and continues spiraling in due to friction. A merger is avoided if and when the envelope becomes unbound and is ejected from the system. The remaining binary has a much reduced orbital separation. 
 In case of BHNS and NSNS formation, the secondary star may fill its Roche lobe a second time as a hydrogen-poor helium-star (i.e. case BB Roche Lobe Overflow), before collapsing to a NS. 
At stage~5, the system consists of two compact objects in a close orbit. Due to GW emission, the orbit shrinks further, eventually leading to a double compact object merger (stage~6).

\subparagraph{Stable mass transfer channel.}
The stable mass transfer channel differs from the CE channel in only one key aspect \cite{vandenHeuvel:2017pwp, Neijssel:2019irh, vanSon:2022myr, Briel:2021bpb}; the second phase of mass transfer (stage~4) does not lead to a CE-phase, but in fact the mass transfer is self-regulating and therefore proceeds in a stable manner (stage~4 in figure~\ref{fig:classical_bin_channel}). 

It is important to realise that in order to form a GW merger, this second phase of mass transfer should lead to a shrinkage of the orbit. This occurs for those systems that have relatively high mass ratios (i.e. the secondary star is more massive than the compact object) and that experience relatively large angular momentum loss compared to their mass loss, as expected for a BH accretor. As the compact object binaries tend to be wider at formation (stage~4 in figure~\ref{fig:classical_bin_channel}) than for the classical CE-channel, the stable channel typically leads to GW events with longer time delays \cite{Bavera:2020uch, vanSon:2021zpk}.

\subparagraph{GW signatures.}

As the GW sources from isolated binary evolution can be formed through multiple sub-channels (e.g., CE channel and stable mass transfer channel), one can ask: what is the relative contribution of each channel? Historically, binary evolution calculations have found that the CE channel dominates the total merger rate \cite{Tutukov:1993bse, Broekgaarden:2021efa}. However, recently there are indications that its contribution  to the BBH merger rate may have been overestimated \cite{Pavlovskii:2016edh, Klencki:2021hxe,  Marchant:2021hiv, Gallegos-Garcia:2021hti, Olejak:2021fti}. These works suggest that the common envelope may not be successfully ejected (stage~4 in figure~\ref{fig:classical_bin_channel}) in many binaries, but instead leads to a merger or to stable mass transfer. The level of mixture between the channels therefore depends on the assumptions made when modeling binary evolution \cite{Broekgaarden:2021efa, Briel:2022cfl, Dorozsmai:2022wff}.

ET can play an important role in distinguishing formation channels, since the intrinsic level of mixture is very likely redshift-dependent. As the stable channel typically leads to binary compact objects with relatively wide orbits (see above), its GW sources have longer delay times ($t_{\rm delay}\gtrsim 1$Gyr) merging predominantly in the local Universe. Correspondingly,  binary compact objects from the CE channel are born with more compact orbits, and merge with shorter delay times ($t_{\rm delay}\lesssim 1$Gyr) and higher redshifts on average. 
It is found that the BBH merger rate from isolated binary evolution is dominated by sources from the CE channel at large redshifts, while in the local Universe the contributions from the CE and stable channel are similar \cite{vanSon:2021zpk, Bavera:2022mef}. 

In addition ET can put constraints on the physics of the CE phase that is poorly understood \cite{Ivanova:2012vx} by inferring from the merger rate the  size of the post-CE orbits that depends on the efficiency of the deposition of energy into the envelope, in particular for BNS\footnote{It is advantageous to use BNS over BBH or NSBH, as the BNS merger yield is virtually independent of metallicity \cite{Chruslinska:2017odi, Giacobbo:2018etu, Neijssel:2019irh}.}
mergers \cite{Safarzadeh:2019pis}. This is not possible with current instruments, given the limited redshift coverage, even with the help of precise redshift determinations from electromagnetic counterparts.

There are also correlations between the formation channel and the compact object masses that ET can uncover. 
For example, the typical mass ratio of BBHs formed through the stable mass transfer channel range between 0.6--0.8, whereas the CE channel provides sources with a relative flat mass ratio distribution between\footnote{The shape of the mass distribution depends on the CE efficiency \cite{Giacobbo:2018etu,Bavera:2020uch}.} 0.2--1.0 \cite{Giacobbo:2018etu, Neijssel:2019irh, vanSon:2021zpk, Bavera:2020uch,Olejak:2024qxr}. Similarly, the mass ratio of BHNS ($M_{\rm NS}/M_{\rm BH}$) peaks at $\sim0.3$ for the stable channel, and spans from 0.1 to 0.5 for the CE channel \cite{Broekgaarden:2021iew, Giacobbo:2018etu}. 
In addition, there is a link between the BH mass, the delay time and the formation channel of BBH mergers \cite{Spera:2018wnw, vanSon:2021zpk, Briel:2022cfl, Belczynski:2022wky}. 
The CE channel preferentially produces BHs with masses below about $30\Msun$, while the stable channel primarily forms systems with BH masses above $30\Msun$. 
As BBH mergers from the stable mass transfer channel typically merge with long delay times, mergers of massive BHs are expected to predominantly occur at low redshifts and be relatively less common at higher redshifts. This counter-intuitive result holds despite the fact 
that massive BHs are easier to form at low metallicities which are more common at higher redshifts \cite{Belczynski:2008nh, Spera:2018wnw, vanSon:2021zpk}.

\paragraph{Chemically homogeneous evolution (CHE).}\label{subsec:CHE}

An example of non-mass-transfer binary evolution is chemically homogeneous evolution \cite{Maeder87,Heger:2004qp,Yoon:2006fr}. In the context of GW progenitors, the CHE scenario considers compact binaries that are near contact or even in contact at the start of the hydrogen burning phase (see stage~1 in figure~\ref{fig:classical_bin_channel}, right panel \cite{deMink:2016vkw, Mandel:2015qlu, Marchant:2016wow, duBuisson:2020asn,Riley:2020btf}). 
In such binaries, tidal interaction will enforce rapid stellar rotation, which in turn induces meridional currents in the radiative layers of the rotating star \cite{1924MNRAS..84..665V, Eddington25, Sweet50}. 
The mixing currents prevent the build-up of a chemical gradient inside the star and keep the star chemically homogeneous during most of the core hydrogen burning phase. Contrary to classically evolving (i.e. slowly rotating) stars, chemically homogeneously evolving stars 
do not experience significant 
expansion during the post-main-sequence phase. 
After the core hydrogen burning phase ends, the stars will become helium-burning stars, before collapsing to a black hole (stage~2). The binary black hole then merges due to GW emission (stage~3). 

\subparagraph{GW signatures.}
For the CHE scenario to produce mergers in a Hubble time, stellar winds need to be weak, as is expected at low metallicity \cite{Vink:2001cg, Vink:2005zf}. The reason for this is threefold: 1) stellar winds cause a spin-down of the star which halts rotational mixing, 2) stellar winds cause a widening of the orbit in which a tidally locked star has to spin down, 3) the widened orbit may be too wide to merge within a Hubble time due to GW emission alone. Therefore we expect the CHE channel to take place only at low metallicities. In addition, as the wide orbits give rise to long delay times, mergers at high redshifts are less likely to be observed for this channel in comparison with the classical channels (e.g. \cite{Bavera:2022mef}).
Lastly, naively one may expect large effective spins from this channel due to the high rotational velocities required for the stars. However, Marchant et al. \cite{Marchant:2016wow} argue for moderate spins ($\sim$0.4) due to stellar winds during the Wolf-Rayet phase. 

Other signatures that may be used to distinguish events from the CHE channel from other channels are related to the expected masses. In the CHE channel, there is a strong preference for high mass black holes ($\gtrsim 25M_{\odot}$) as CHE is not effective for lower mass progenitors \citep{deMink:2008iu}. Therefore, it only produces binary black hole mergers, and not binary neutron star or BH-NS mergers. Secondly, there is a preference for detecting events with nearly equal-mass components \citep{Mandel:2015qlu, Marchant:2016wow}. ET will thus be able to determine if a population of BBH with these properties exists or not.

The CHE channel is competitive with the classical isolated binary evolutionary channel with respect to rates \citep{Mandel:2015qlu, Marchant:2016wow, duBuisson:2020asn,Riley:2020btf} (see also section \ref{div3:rates} for a discussion on rates and their large uncertainties). In particular, Riley et al. ~\cite{Riley:2020btf} showed that the CHE channel may contribute 50-75\% of all aLIGO BBH detections from isolated binary evolution. For the highest chirp masses ($M_{\rm chirp}>30M_{\odot}$), this percentage may be as high as 80\%. 

ET can also measure the effective inspiral spin parameter, $\chi_{\rm eff} = (m_1 \vec{s_1} + m_2 \vec{s_2})/(m_1 + m_2)\,{}\cdot{}\hat{\textbf{L}}$, where $\vec{s_1}$ and $\vec{s_2}$ are the dimensionless spins of the two components, while $\hat{\textbf{L}}$ is the orbital angular momentum versor. 
For isolated binary evolution, at first sight it does not strongly constrain the formation channels in isolated binary evolution. This is because the effective spin is predominantly determined by the natal BH spin and the orientation of the BH spins \cite{Gerosa:2013laa,Gerosa:2018wbw}. If angular momentum is transported efficiently through the progenitor star, the resulting BH natal spin is low, as preferred by current-day GW observations (e.g., \cite{Fuller:2019sxi, Belczynski:2017gds}). Low effective spin is also possible if the BH spins are misaligned. It is possible to form GW sources with high spins in isolated binaries if  the stellar angular momentum transport is inefficient, or by increasing BH spins through accretion \cite{Bavera:2020uch,vanSon:2020zbk,Steinle:2022rhj, Shao:2022jzo} or through tidal interactions \cite{Kushnir:2016zee, Zaldarriaga:2017qkw,Hotokezaka:2017esv, Bavera:2022mef}. 
Interestingly, the latter implies a relation between the effective spin and the redshift of the merger that should become visible with third-generation instruments \cite{Bavera:2022mef}. 
The fraction of highly spinning BBHs from isolated binary evolution is expected to increase with redshift, as low metallicity stars experience less wind mass loss and less radial expansion, enabling the formation of more compact binaries with stronger tidal interaction and prompt GW mergers. At low redshift ($z\sim{0.5}$ i.e., in the currently observable population) highly spinning BHs are expected to be dominated by BHs formed through chemically homogeneous evolution. In addition, for low CE efficiencies that translate to tighter post-CE orbits and more strongly spun up BHs, the highly spinning BHs in the CE channel merge at high redshifts ($z\sim{2}$) close to the peak of the cosmic star formation rate, but outside current detector horizons \cite{Bavera:2020uch}.
The advent of third-generation instruments would finally enable us to detect the redshift dependence of the intrinsically highly spinning BBH population from isolated binaries. 
This would enable us to clearly identify the population of BBHs that are of isolated origin.

\paragraph{Evolution in higher order systems.}
\label{sub:three}
Observations of massive stars in the Milky Way and the Magellanic Clouds suggest that their companion fraction is high; typically massive stars not only have a stellar companion, but two or three companions \citep{Evans:2010qb, 2017ApJS..230...15M, Sana:2012px, Kobulnicky:2014ita, bordier24, offner23}. It is therefore interesting to ask if and how this may change the GW signatures compared to isolated binary evolution. In the following section, we consider the evolution of small and isolated stellar systems. 

\begin{figure}
\includegraphics[trim= 200 250 355 50, clip, width=\columnwidth]{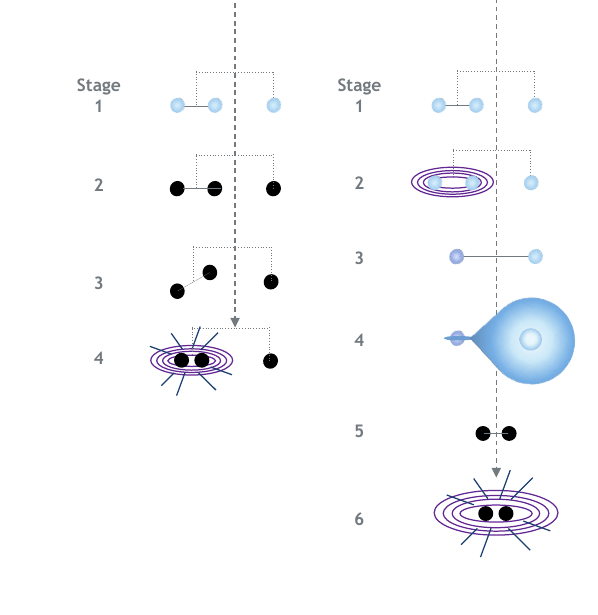}

\caption{Evolutionary pathway to form GW sources  a) according to the non-interacting triple channel, b) in an interacting triple with a stellar merger in the inner binary. On the left: Stage 1) ZAMS 2) Formation of triple BH 3) ZLK cycles and orbital dissipation 4) GW merger. On the right: Stage 1) ZAMS 2) Stellar merger in the inner binary 3) Formation of binary with a rejuvenated star 4) Various phases of mass transfer 5) Formation of binary BH 6) GW merger. }
\label{fig:triple_channels}
\end{figure}

\subparagraph{Non-interacting triples.}

If both the inner and outer orbit are sufficiently wide that mass transfer will not take place, the stars evolve as if they were single stars. After a triple compact object has been formed, the system experiences strong three-body dynamical effects such as von Zeipel-Lidov-Kozai cycles \cite{Zeipel10, Lidov:1962wjn,Kozai:1962zz}. During these cycles, the eccentricity of the inner binary can reach high values, which enhances the GW emission and can lead to a merger  \cite{Thompson:2010dp,Silsbee:2016djf, Antonini:2017ash, Fragione:2019poq,Su:2020vda, Martinez:2021tmr, Rodriguez:2018jqu,Fragione:2019zhm,Fragione:2019mpq}. The predicted BBH merger rate for this channel events is estimated at $\sim 0.5-5$ yr$^{-1}$ Gpc$^{-3}$ for solar metallicity and 25/yr/Gpc$^3$ at low metallicity. Due to the strong impact of supernova kicks, the rate for NSBH mergers is more uncertain, $10^{-4}-25$ yr$^{-1}$ Gpc$^{-3}$ \cite{Fragione:2019zhm, Fragione:2019mpq}.

\subparagraph{Interacting triples.}

When the orbits are closer,  triples do experience mass transfer at some point in their evolution (right column in figure~\ref{fig:triple_channels}, see \cite{Toonen20, Kummer23, Stegmann:2021jen, Hamers:2019oeq}). The existence of a tertiary star can lead to an earlier interaction in the inner binary, a merger of the inner binary turning the triple into a binary system, or even multiple mergers in case of higher-order systems. In a third of the triple systems, the tertiary plays a fundamental role in the evolutionary pathway, leading to configurations that are not predicted for isolated binaries. The predicted merger rate in this channel can reach $25\, {\rm yr}^{-1} {\rm Gpc}^{-3}$ (e.g. \cite{Stegmann:2022ruy}).

\subparagraph{GW signatures.} The GW signatures of these mergers are high eccentricity  \cite{Antonini:2017ash, Liu:2019gdc} and a mass ratio $q$-distribution that is significantly flatter than for isolated binary evolution or for dense stellar clusters \cite{Martinez:2021tmr,Su:2021ogc,Stegmann:2022ruy,Dorozsmai:2023iij}. Multiple mergers can produce masses in the upper and lower mass-gap \cite{Vigna-Gomez:2020fvw, Fragione:2020aki, Lu:2020gfh, Safarzadeh:2019qkk, Liu:2020gif}, see also section \ref{div3:bh-masses}. Moreover, in triples with chemically homogeneously evolving inner binary stars, commonly the third star will transfer mass to the inner binary black hole \cite{Dorozsmai:2023iij}, possibly leading to ``wet'' mergers in which the orbital decay is driven by gas drag instead of GW emission \cite{Kummer:2024kyi}. Of particular interest is the effect of triple stars on the effective spin of BBH mergers. Although complicated by GR effects, many studies have found that, independently of their magnitude, the coupling of triple dynamics with GW energy loss leads to misaligned spins and $\chi_{\rm eff}\sim 0$  \citep{Liu:2017yhr,Antonini:2017tgo,Liu:2018nrf,Rodriguez:2018jqu,Liu:2019gdc,Fragione:2019mpq,Yu:2020iqj, Su:2020vda}. ET will be able to uncover a significant population of mergers originating in triple systems (e.g., \cite{Chen:2017gfm}).

\subsubsection{Dynamical channel: stellar systems as factories of merging compact objects}\label{sec: formation channels dynamical}

With densities much larger than galactic fields, star clusters represent ideal factories for the production of dynamical CBCs. Theoretical models and observations generally gather clusters into three main categories: open and young massive clusters, with typical masses in the range $10^2-10^5 \Msun$ \citep{PortegiesZwart:2010cly}, globular clusters ($\sim 10^5-10^7 \Msun$) \citep{2010arXiv1012.3224H}, and nuclear clusters ($10^6-10^8 \Msun$) \citep{Neumayer:2020gno}. 
A primary difference among different cluster types is the typical density, which varies from $<10^4 \Msun$ pc$^{-3}$ in 
open clusters \citep{PortegiesZwart:2010cly,2021MNRAS.507.3312G} and can reach extreme densities of $10^8 \Msun$ pc$^{-3}$ in the densest nuclear clusters \citep{2016MNRAS.457.2122G}. Additional differences pertain to the possible presence of gas in young clusters, the development of multiple populations in globular clusters, and the peculiar star formation history or the possible presence of a central supermassive black hole in nuclear clusters. 

Within the dynamical scenario, there is a large variety of processes that can lead to the formation of CBCs, as we briefly describe in the following. The branching ratio of such processes intrinsically depends on the binary properties and the cluster type and structure.


Dynamical formation in a star cluster is expected to be more effective for BBHs than for both BNSs and NSBHs, because 
BHs dominate the dynamics in the cluster central regions, preventing the segregation of NSs and their progenitors and suppressing their interaction rate. 

As the star cluster evolution proceeds, though, BHs are slowly evacuated from the cluster in what is called the BH burning process \citep{Breen:2013vla}. As the number of BHs diminishes, the probability for NSs to interact with other BHs and among each other increases. Therefore, theoretical models predict that the number of NSBH and BNS mergers with a dynamical origin is much smaller than that of BBHs, and their occurrence requires that the host cluster lost a significant fraction of its original BH population \citep{Clausen:2012zu, Ye:2019xvf, Rastello:2020sru, Sedda:2020wzl, Fragione:2020wac}. 

Despite their rarity, a subpopulation of dynamical NSBH mergers may exhibit characteristic features that differentiate them from the isolated ones. These include BHs heavier than $10 \Msun$ \citep{Sedda:2020vwo,Rastello:2020sru}, a chirp mass $>4 \Msun$ \citep{Sedda:2020wzl}, the absence of an electromagnetic counterpart \cite{Foucart:2012nc,Ackley:2020qkz,Sedda:2020wzl}, and longer delay times, with most of them likely forming over a cluster relaxation time \citep{Sedda:2020vwo,Ye:2019xvf}. 

BNS mergers are even rarer than NSBH and their occurrence in star clusters is expected to be several orders of magnitudes smaller than in the field \citep{Belczynski:2017mqx,Ye:2019xvf}. Given their rarity, it is also unclear whether dynamical BNS mergers can exhibit distinctive features compared to those forming in galactic fields \citep{Gupta:2019nwj,Rastello:2020sru}. Most likely, the only way to infer a dynamical origin is via the association of the resulting EM counterpart with a dense cluster. 

Moreover, the low detection rate of both BNS and NSBH by current ground-based detectors, alongside with the limited number of BNS detected through other means of observation, make it hard to assess the existence of a ``lower mass-gap'' separating NSs and BHs, in the mass range $2-5 \Msun$. In this sense, the recent discovery of a CBC by the LVK Collaboration involving a NS and an object with a mass in the range $2.5-4.5 \Msun$ \citep{LIGOScientific:2024elc} could make the case for a dynamical origin \citep{Gupta:2019nwj,Rastello:2020sru,Ye:2024wqj} or push new studies to explore more some poorly understood mechanism of stellar evolution, like the common envelope physics \citep{Zevin:2020gma}.
\\
\\
\paragraph{How can CBCs form in star clusters and merge within a Hubble time?}

\subparagraph{Primordial binaries.} 
The simplest way to form CBCs in a star cluster is from the evolution of binary stars already paired at birth, the so-called ``primordial binaries''. Observations of stellar nurseries and young star clusters suggest that up to $50-70\%$ (and even $100\%$ in some extreme cases) of stars are born in binary systems and higher-order multiples \citep{Sana:2012px,2017ApJS..230...15M}. The binary fraction seems to increase with the primary mass \citep{2017ApJS..230...15M}, thus suggesting that many, if not all, compact object progenitors may have been part of a ``primordial" binary. 

At the simplest level, the binary star can evolve into a CBC via stellar evolution  making it indistinguishable from the same binary evolving in isolation. If, however, the timescale for binary--single stellar interactions is shorter than the binary delay time,  the perturbations induced by stellar flybys and close encounters on the orbit can significantly alter the binary evolution, either shortening its merging time or delaying it, leading to a swap between an interloper and one of the binary components, or causing its ionization. 
 
Even in the extreme case that no stars are born in a binary system, there are several ways to form CBCs via dynamics.

\subparagraph{Gravitational wave captures.}
If the two CBC components, initially unbound, undergo a close hyperbolic encounter, GWs emitted during the passage at pericentre (GW bremsstrahlung) can carry away enough energy to bind the two objects. This process, called gravitational wave capture, requires an environment with densities typical of galactic nuclei \citep{OLeary:2008myb,Hoang:2020gsi} and dense globular clusters \citep{Samsing:2017xmd,Rodriguez:2019huv}. In this scenario, the pericentre distance between the two objects must be so small that CBCs forming through this channel merge minutes to hours after the interaction \citep{OLeary:2008myb}. Given their short formation time, their merger rate will directly follow the binary formation rate, which is expected to be in the range $0.2 -150$ Gpc$^{-3}$ yr$^{-1}$ \cite{Sedda:2023big}. Up to $90\%$ of mergers formed this way can retain an eccentricity $e>0.9$ at 10 Hz \citep{OLeary:2008myb,Gondan:2017wzd,Gondan:2018khr,Rasskazov:2019gjw}.

\subparagraph{Three-body binary formation.}
Another way to form tight binaries is via three-body binary formation. In this case, three initially unbound objects find each other and closely interact up to the point that two objects pair up and the third carries away the extra energy. The timescale of this process is extremely sensitive to the cluster velocity dispersion, being $\propto \sigma^9$ \citep{Lee:1994nq}, and generally exceeds a Hubble time for the densest globular clusters and nuclear clusters. Nonetheless, three-body binary formation represents a primary channel to build-up hard binaries, i.e. binaries with a binding energy larger than the average kinetic energy of the environment \cite{Heggie:1975rcz}, containing at least one compact remnant \citep{OLeary:2005vqo,Morscher:2014doa}, but see also \cite{2024ApJ...970..112A}. Such binaries can either merge outside the cluster or harden via binary--single and binary--binary interactions.

\subparagraph{Binary--single interactions.}
If the cluster harbours a population of binaries, either primordial or formed via three-body interactions (see paragraph above), CBCs can form through interactions between the binary and a third compact object (binary--single interactions) depending on the binary dynamical status (i.e. if it is hard or soft) and the star cluster structure and evolution \cite[]{Miller:2001ez,Samsing:2013kua}. This type of interaction can account for up to $20-70\%$ of all mergers occurring in Galactic nuclei 
\citep{Leigh:2017wff}, and is a primary mechanism for forming CBCs
with large eccentricity (e.g., \cite{Tagawa:2020jnc,Samsing:2013kua,Rodriguez:2018pss,DallAmico:2023neb}) or with at least one component in the lower or upper mass gap by triggering the merger of the inner binary system \citep{Samsing:2018isx,Tagawa:2020qll,Kremer:2020wtp}.  

When the distance between a binary and an incoming third body falls below a few times the binary semimajor axis, the three bodies can undergo a so-called {\it resonant} interaction, a chaotic phase during which the three bodies pair up and exchange continuously (e.g., \cite{Hut:1982di,Hut:1992wz,Sigurdsson:1993tui}). This process proceeds up to the point in which one of the objects, generally the least massive, is expelled and the other two form a hard binary \citep{Sigurdsson:1993tui,Clausen:2012zu,Sedda:2020wzl,DallAmico:2023neb,2024arXiv240212429A}. 

\subparagraph{Gas-driven dynamics.}
Star clusters rich in gas, such as embedded open clusters or young massive clusters, can nurture the development of CBCs as the gaseous medium can alter the orbit of a binary, for example by inducing a dynamical friction term on the components that shrinks the binary and can enhance the probability for the binary to merge within a Hubble time \citep{2024arXiv240401384R}.

Gas-driven CBC formation can also occur in the dense disks that surround AGNs \citep{Mckernan:2017ssq,Tagawa:2019osr,Yang:2020lhq}. In such extreme environments, the disk can dissipate the orbital energy of compact objects passing through its medium and capture them, thus their orbits settle in the plane of the disk and undergo inward migration that steepens the radial density profile around the SMBH \citep{Artymowicz:1993xz}. The aforementioned process acts faster than relaxation in AGN disks, and can nurture the formation of CBCs in the plane of the disk aided by a combination of gaseous accretion, gas dynamical friction, and stellar scattering \citep{Tagawa:2019osr, Sedda:2023big}.

In the case of BBHs, it has been suggested that mergers occurring in an AGN can power an EM counterpart \citep{McKernan:2019hqs}. GW190521, the first GW source to produce a final mass in the intermediate-mass black hole (IMBH) range ($\sim 150 \Msun$) \citep{LIGOScientific:2020iuh}, has been associated with a high-energy transient detected by the Zwicky Transient Facility \citep{Graham:2020gwr}, although it is rather hard to assess whether this association can be robustly confirmed \citep{Ashton:2020kyr}.

\subparagraph{Secular mechanisms.}

Secular effects generally develop when the binary is perturbed by a third object or a series of close passages by different objects. 
 Depending on the perturber orbital properties, the binary can undergo Kozai-Lidov oscillations \citep{Kozai:1962zz,Lidov:1962wjn}, a mechanism that could boost the merger probability of CBCs in dense clusters \citep{Antonini:2015zsa,Arca-Sedda:2018qgq,Martinez:2020lzt} and galactic nuclei \citep{Wen:2002km,Seto:2013wwa,Antonini:2012ad,Hoang:2017fvh,Fragione:2018yrb,Sedda:2020jvg}. 
Weak perturbations acting on soft binaries in loose star clusters, for example, can induce a steady increase of the binary eccentricity on a timescale shorter than the binary evaporation time, up to the point that the binary coalescence occurs within a Hubble time \citep{Hamilton:2019yij,Michaely:2019aet}.

\paragraph{Properties of dynamical CBCs}
\label{sect:propdynCBCdiv3}
Dynamical CBCs exhibit peculiar properties, some of which may be unique marks of this formation channel, as we briefly discuss in the following.

\subparagraph{Eccentricity of dynamical CBCs.}
Several works have pointed out that orbital eccentricity may represent a key parameter to identify dynamical CBCs, as according to most stellar evolution theories, isolated CBCs are expected to circularise before reaching the Hz frequency band \cite{Nishizawa:2016eza}. Moreover, eccentricity could be used to untangle the population of mergers occurring after their ejection or inside a cluster, either triggered by repeated interactions or owing to hierarchical secular dynamics \citep{DOrazio:2018jnv,Zevin:2018kzq,Rodriguez:2018pss,Arca-Sedda:2018qgq,Sedda:2023qlx}.
In general, we expect that in-cluster mergers are more eccentric than mergers occurring after the ejection of the CBC from the parent cluster \citep{DOrazio:2018jnv,Zevin:2018kzq}. 

The different eccentricity distribution for in-cluster and ejected mergers may be particularly evident in the 1--10 Hz frequency band. In-cluster mergers forming from three-body and binary--single interactions have typical eccentricity in the $10^{-4}-10^{-2}$, whilst those forming from GW captures have very high eccentricities, $>0.9$, and merge on timescales of days or years. In the case of ejected binaries, instead, their eccentricity may be too small $<10^{-5}$ to be accurately measured in the $1-10$ Hz band, and a detector sensitive at lower frequencies could be more suited to detect them.  In these regards, the possibility to access the $0.001-0.1$~Hz frequency band, for example with LISA, DECIGO, or the LGWA, could enable us to observe the eccentricity evolution of the binary weeks to months prior to the merger, helping us to place constrain on the origin of stellar mass CBCs. However, particular attention is needed when modeling the signal, owing to the spin precession-eccentricity degeneracy \cite[and reference therein]{Romero-Shaw:2022fbf}, which can significantly hinder eccentricity measurements (e.g., \cite{Gayathri:2020coq,Romero-Shaw:2020thy, Romero-Shaw:2022fbf}).

\subparagraph{Stellar interactions, hierarchical mergers, and spins.}

A binary traveling in a stellar environment is subjected to  binary--single interactions that can impart on its motion a Newtonian recoil and remove part of the binary orbital energy. As a consequence, the binary progressively shrinks ({\it hardening} process), making the new interactions rarer but more violent \cite{Heggie:1975rcz}. At some point, GW emission takes over and drives the binary to coalescence inside the cluster, provided that the binary shrinking rate due to GWs is larger than the binary--single interaction rate. Otherwise, a new interaction can, on the one hand, increase the binary eccentricity and possibly trigger a rapid GW coalescence, and, on the other hand, transfer enough kinetic energy to expel the binary, which then will merge outside the cluster. 

\begin{figure}
\centering
\includegraphics[width=0.47\columnwidth]{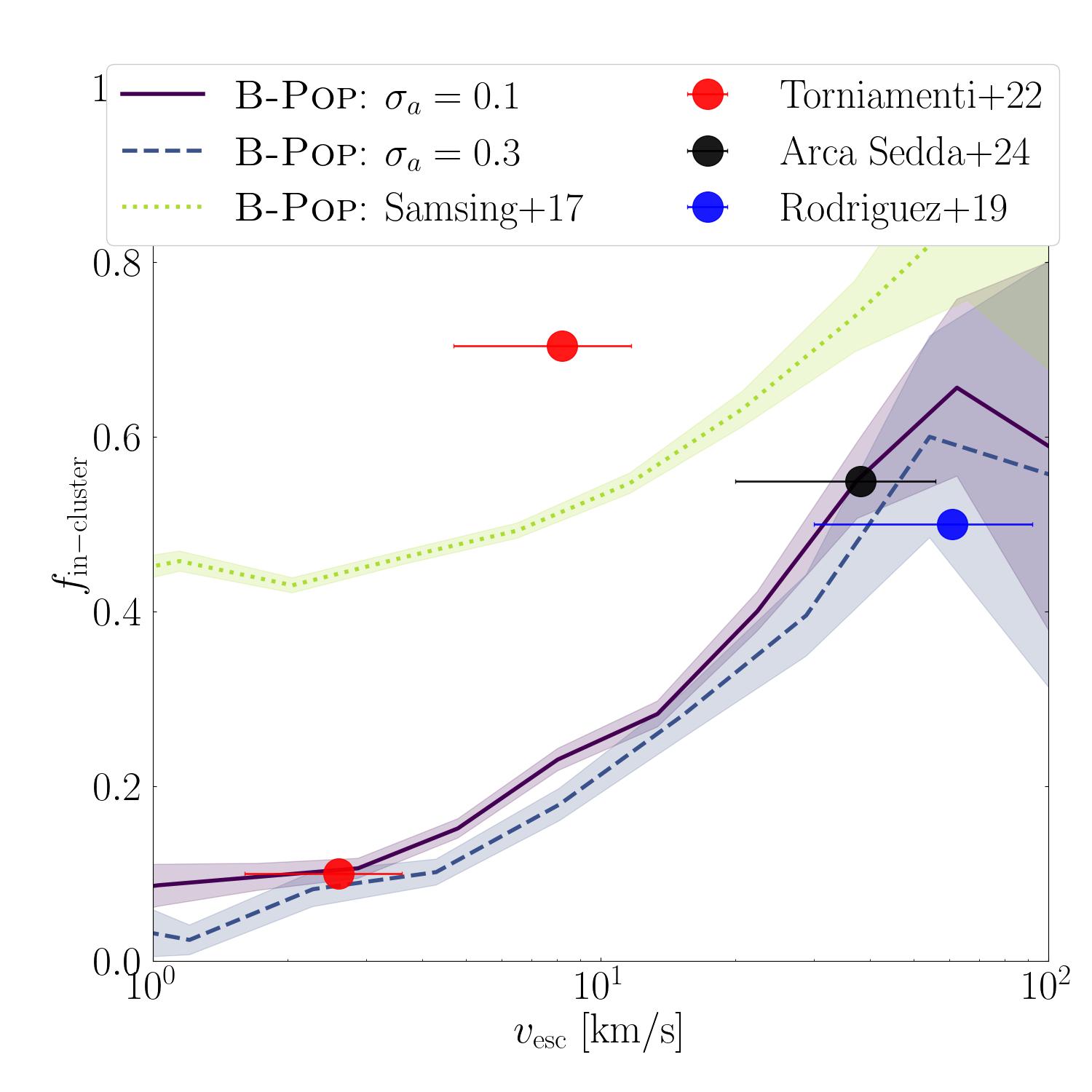}
\includegraphics[width=0.47\columnwidth]{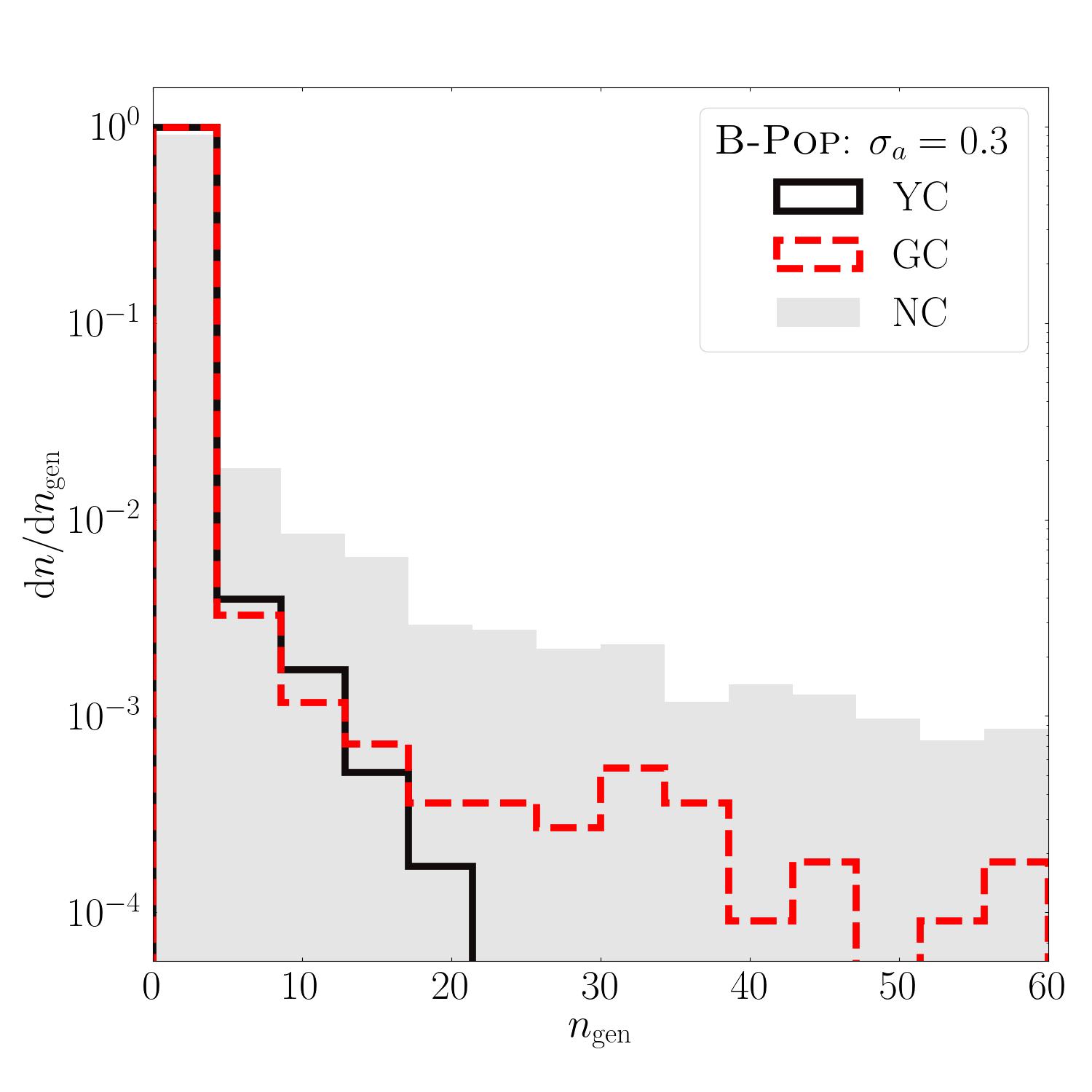}
\caption{Left panel: Fraction of BBHs that merge inside the cluster for different assumptions about the binary semimajor axis sampling: assuming a Gaussian distribution peaked around the hard binary separation with $\sigma_a=0.1$ (purple straight line), or with $\sigma_a=0.3$ (blue dashed line), or following the assumption of \cite{Samsing:2017xmd}, i.e. that $a$ is distributed between $0.1-0.2$ AU according to a flat distribution. Shaded areas encompass the Poissonian error associated with the samples. The points represent data from self-consistent $N$-body and Monte Carlo simulations of star clusters. Right panel: distribution of merger generation for different cluster types. All models are performed with the \textsc{B-Pop} population synthesis tool \cite{Sedda:2021vjh}.}
\label{fig1:ejected}
\end{figure}

Using the population synthesis code \textsc{B-Pop} \citep{Sedda:2021vjh}, we show in figure~\ref{fig1:ejected} how the fraction of in-cluster mergers varies with the cluster escape velocity, calculated at the time of the binary merger for different assumptions about the initial binary semimajor axis distribution. Specifically, we either assume a Gaussian distribution peaked around the hard-binary separation \citep{Sigurdsson:1993tui} with dispersion $\sigma_a=0.1,~0.3$, or a flat distribution between 0.1-0.2 AU \citep{Samsing:2017xmd}. \textsc{B-Pop} models are compared against direct $N$-body \citep{Torniamenti:2022txt, Sedda:2023qlx} and Monte Carlo simulations \citep{Rodriguez:2018pss}.  Note that for all the simulation database considered, the fraction of in-situ merger is calculated taking into account only dynamically assembled binaries, and the range of simulated velocity dispersion is shown, generally limited to within a factor of a few from the average value. Despite \textsc{B-Pop} relies on an approximated method to simulate BBH mergers, their results agree surprisingly well with both few-body and self-consistent simulation of star clusters. In general, it is evident that the fraction of in-cluster mergers sensibly increases with the cluster escape velocity, especially in the case of tight BBHs \citep{Gerosa:2019zmo, Rodriguez:2019huv, Yang:2019, Zevin:2022bfa, Kimball:2020qyd}. This aspect can have crucial implications for the development of multiple mergers and the formation of higher-generation BHs, as we discuss in the next section.

Asymmetries in the emitted GW signal impart a relativistic kick on the merger remnant \citep{Campanelli:2007cga,Lousto:2010xk,Lousto:2012su}. If the amplitude of such recoil is smaller than the cluster escape velocity, the remnant can remain in the cluster and merge again, leaving behind a second-generation remnant. The development of repeated mergers is usually referred to as the hierarchical merger mechanism \citep{Miller:2001ez}.
Relativistic kicks can be as large as $10^4$ km s$^{-1}$, therefore hierarchical mergers are expected to develop only in the most massive and dense star clusters \citep{Miller:2001ez,2017PhRvD..95l4046G,Rodriguez:2019huv,Arca-Sedda:2020lso,Sedda:2021vjh,Antonini:2018auk,Antonini:2016gqe,Vaccaro:2023cwr,Torniamenti:2024uxl,Mapelli:2021syv,Chattopadhyay:2023pil,Fragione:2023kqv,Fragione:2021nhb}, especially because the escape velocity of star clusters decreases over time owing to relaxation processes, which cause mass-loss and expansion \citep{Antonini:2020xnd, Sedda:2021vjh, Torniamenti:2024uxl}.

Higher-generation BHs are expected to be characterised by clearly different spins compared to first-generation mergers. For example, second-generation mergers are expected to have dimensionless spin distribution tightly peaked around $\chi \sim 0.7$ \citep{Pretorius:2005gq,Scheel:2008rj,Berti:2008af}. The ability to identify a second-generation merger thus depends on the natal spin distribution of their progenitors \citep{Gerosa:2021mno,Sedda:2021abh,Sedda:2021vjh}. For higher generation remnants, instead, the older the generation the smaller the spin \citep{Arca-Sedda:2020lso,Kritos:2022non}. Higher generation merger remnants also feature typical precessing spin values $\chi_p = 0.5-0.8$, which are hardly achievable with isolated and 1st generation mergers \citep{Mapelli:2021gyv,Torniamenti:2024uxl,Baibhav:2020xdf,Baibhav:2021qzw}.

The mass, spin, and generation of a hierarchical merger product encode information about the host cluster and the properties of the previous chain of mergers. For example, the GW recoil is generally smaller if the merging components have small spins, therefore the development of a long merger chain requires a natal spin distribution that favors low spins \citep{Rodriguez:2019huv,Zevin:2022bfa}. The amplitude of the kick requires a host environment with a large escape velocity, generally $>100$ km s$^{-1}$, thus favoring dense and massive globular and nuclear clusters as preferred environments for higher generation merger products \citep{Fragione:2020nib, Gerosa:2019zmo,Kritos:2022non,Mapelli:2021syv,Sedda:2021vjh}.

The hierarchical merger mechanism represents a viable way to form IMBHs (see section \ref{sec:IMBH}  and references therein) and BHs in the upper mass gap \citep{1987ApJ...321..199Q,1993ApJ...418..147L,Miller:2001ez,OLeary:2005vqo,Antonini:2016gqe,Antonini:2018auk,Kremer:2020wtp,DiCarlo:2020lfa,ArcaSedda:2023mlv,Chattopadhyay:2023pil,Sedda:2023qlx}. Upper mass-gap BHs can also form via stellar evolution under particular conditions \citep{Mapelli:2019ipt,Costa:2020xbc,2020A&A...640L..20B,2021MNRAS.504..146V} or via stellar collisions \citep{Spera:2018wnw,
DiCarlo:2019fcq, Kremer:2020wtp, Ballone:2022ugp,Costa:2022aka, Sedda:2023qlx}. Mergers involving upper mass-gap BHs can thus carry information about their formation origin. Precise measurements of masses and spins could help untangle different formation channels for mergers involving one or two BHs with masses in the gap.
In this sense, at masses below O($500 \Msun$)  ET will enable to place stringent constraints on the primary spin, chirp mass, location and distance \citep{Huerta:2010tp,Iacovelli:2022bbs,Branchesi:2023mws,Gupta:2023lga,Fairhurst:2023beb,Reali:2024hqf}, offering a unique opportunity to infer the origin of mergers with one or both components in the upper mass gap.

\subsubsection{Fingerprints of different formation channels} \label{sec: fingerprints}

As we have seen in the previous sections, CBCs forming from the isolated and dynamical channels exhibit some peculiar features that could be used to infer the contribution of different channels to the cosmic population of CBCs \citep{Zevin:2020gbd, Sedda:2021vjh, Mapelli:2021gyv}. With the prospective detection of $10^5$ BBH and BNS events per year \citep{Iacovelli:2022bbs,Branchesi:2023mws}, it will be possible to characterise in  detail the impact of the isolated and dynamical channels on the formation of CBCs.
This can be achieved by constraining crucial quantities like merger rates, redshift dependencies, masses and spins of merging CBCs.

Merger rates are highly uncertain, as they depend on the cosmic star formation history, the metallicity evolution, and the star cluster formation process. Isolated binary models predict a BBH merger-rate density $\mathcal{R}\sim (0.5-5,000)\,  {\rm yr}^{-1} {\rm Gpc}^{-3}$ at redshift $z \lesssim 1$,  while dynamical models suggest $\mathcal{R}\sim (10^{-3}-10^2)\,  {\rm yr}^{-1} {\rm Gpc}^{-3}$ (for a comprehensive discussion about rates from different channels, \cite{Mandel:2021smh,Sedda:2023big}). A similar difference in the expected merger rates is also estimated for BHNS mergers, while for BNS most models suggest that isolated mergers outnumber dynamical mergers by one order of magnitude, despite the large uncertainties \citep{Mandel:2021smh}. Note that the upper limits to theoretical predictions are already in tension with the local merger-rate densities inferred from LVK data \citep{KAGRA:2021duu}. See also section \ref{div3:rates} for further details about CBC merger rates. 
With ET, we will be able to verify the possible evolution with redshift of the merger rate \citep{KAGRA:2021duu} and other quantities, like the median mass or the primary mass of merging BBHs \citep{Sedda:2021vjh,Ye:2024ypm,Torniamenti:2024uxl}, and their mass ratio \citep{KAGRA:2021duu}.

The binary mass, primary mass, and mass ratio distributions can be diagnostic quantities to unveil features from different channels. For example, mergers with large primary mass and low-mass ratios are most likely byproducts of dynamical processes \citep{Miller:2001ez,2017PhRvD..95l4046G,Rodriguez:2019huv,Arca-Sedda:2020lso,Sedda:2021vjh,Antonini:2018auk,Antonini:2016gqe,Vaccaro:2023cwr,Torniamenti:2024uxl,Mapelli:2021syv,Chattopadhyay:2023pil,Antonini:2018auk,Fragione:2023kqv,Fragione:2021nhb,Ye:2019xvf,Sedda:2020wzl,Rastello:2020sru}, while isolated mergers predominantly favor the formation of lower mass mergers \citep{2020A&A...636A.104B,Ghodla:2023ymi,Iorio:2022sgz,Belczynski:2010tb,Dominik:2012kk,Klencki:2018zrz,Spera:2018wnw,Zevin:2020gma,Broekgaarden:2021iew}, although uncertainties in binary stellar evolution and pair-instability  physics leave room for massive BH formation also through this channel \cite{Belczynski:2020bca,Olejak:2024qxr}. 

Spin magnitude and alignment are other quantities that ET can accurately measure, at least for the closest BBH mergers, to provide crucial insights about the unknown distribution of BH natal spins and the origin of the merging binary,  despite the large uncertainties in stellar dynamics and evolution, and in the physics of stellar explosions  \cite{Kalogera:1999tq,Gerosa:2015hba,Hotokezaka:2017esv,Hotekezaka:2017uxl}. Isolated BBHs are predicted to have mildly aligned spins, depending on the impact of SN explosions on the binary orbit \citep{Kalogera:1999tq,Wysocki:2017isg,Baibhav:2022qxm}, and the establishment of precessional instabilities that can completely change the initial spin configuration (e.g., \cite{Gerosa:2018wbw,Gerosa:2015hba}), whilst dynamical BBHs should have randomly oriented spins, unless some effect efficiently aligns them, e.g. like disk torques in AGN disks  \cite{Tagawa:2020dxe}. 

Eccentricity probably is the most representative parameter of the dynamical channel. The isolated channel largely produces nearly circular mergers at $f > 1$~Hz. 
The eccentricity of dynamically assembled CBCs, gravitational-wave captures, and hierarchical triple systems, instead, can attain high values ($e>0.9$) depending on the processes that affected the binary evolution. ET can measure eccentricities above $e>0.01$ with accuracy at $1\%$ level at redshift $z \lesssim 0.1$ \citep{Branchesi:2023mws,Saini:2023wdk}, making it possible to identify mergers coming from high-energy dynamical interactions in dense star clusters, e.g. GW captures or collisions from resonant interactions in dense star clusters,  or from dynamics in isolated triples.

\subsection{Merger rate density of CBC across cosmic time}\label{div3:rates}

A common metric of the capabilities of GW detectors for population studies, which demonstrates the promise of next-generation detectors, is the fraction of the population that can be detected above a certain signal-to-noise (SNR) threshold as a function of redshift.
The upgraded network of current detectors (Advanced LIGO, Advanced Virgo, and KAGRA with sensitivity curves as shown in \cite{Borhanian:2022czq})  will be able to detect complete population of stellar BBH mergers with ${\rm SNR} \geq 10$ up to $z\sim0.1$ and only up to $z\sim0.01$ for the lower mass BNS mergers \cite{Borhanian:2022czq,Gupta:2023lga}. 
Therefore, even the upgraded network of current GW detectors will provide limited insight into the evolution of CBC properties through cosmic history. As we discuss in this section, this dependence encodes valuable information about the formation and evolution of stars and stellar multiples in environments very different from our cosmic neighborhood.
It also probes the cosmic history of star clusters and galaxies. 
In addition, constraints on the evolution of the merger rate at redshifts $z>2$ (above the peak of the cosmic star formation history \cite{Madau:2014bja}) may prove key to overcoming current degeneracies in the astrophysical interpretation of the population properties of BBH mergers.

Although efforts are made to constrain the evolution of the CBC merger rate with current telescopes through sources such as gamma-ray bursts, many challenges (such as understanding selection effects, source classes, and environments)  limit our ability to constrain the CBC merger rate even at small redshifts, let alone at $z\gtrsim2$ \cite{Mandel:2021smh}. 
Such constraints can be obtained with next generation GW detectors. Iacovelli et al. ~\cite{Iacovelli:2022bbs} estimate that ET alone can detect 100\% of the BBH (BNS) mergers with ${\rm SNR}\geq 12$ up to $z=1$ ($z=0.2$). 
For the network of at least two next generation GW detectors, the complete BBH (BNS) population can be detected up to $z\approx{}2$ ($z\approx0.3$) \cite{Iacovelli:2022bbs,Gupta:2023lga}. Moreover, more than $90\%$ of the BBH population can still be detected with ${\rm SNR} \geq 12$ out to $z = 20$. This redshift is more than a factor of two higher than that of the most distant known electromagnetic signal associated with the formation of a stellar compact object \citep{Tanvir:2009zz,Salvaterra:2009ey,Cucchiara:2011pj} and further back in the history of our Universe than any galaxy has ever been observed \cite{2024arXiv240521054A}. ET will allow us to map the BBH merger rate beyond the current limits on the onset of the cosmic star formation and to access the unexplored epoch of the formation of the first stars and the first galaxies with BBH mergers.

\subsubsection{The key ingredients of compact binary coalescence rates}
With the exception of primordial BH mergers (see section \ref{sec:PBHsdiv3}), CBCs detectable by ground-based GW detectors originate from compact objects left over from the evolution of massive stars ($\gtrsim 8 \Msun$).
Given the short evolutionary timescales of such stars (a few $10^{6}-10^{7}$~yr), the formation rate of stellar BHs and NSs follows the overall cosmic star formation history. 
However, only a small fraction of those compact objects evolve into observable CBC.
Furthermore, depending on the efficiency of the processes that bring the two compact objects to merger (see section \ref{div3:formationchannel}), these events can occur with a wide range of time delays, $t_{\rm delay}$, with respect to the formation of their progenitor stars. These delays can range from a few Myrs to the age of the Universe $T_{\rm Hubble}\sim14$~Gyr. 
Consequently, the observable CBC population contains a mixture of systems formed throughout the Universe: at different times, with different chemical compositions, and in very different environments.
 Both the birth chemical composition and the environment in which stars evolve strongly influence the formation of CBCs and their population properties.

\subparagraph{Birth chemical composition leaves imprint on stellar afterlives (CBC properties).}
 Metallicity (i.e. the abundance of elements heavier than helium, $Z$) is a necessary initial condition for modeling the evolution of massive stars \cite{Woosley:2002zz} and stellar binaries/multiples, and influences the properties of stellar-origin compact objects such as masses and spins.
Especially important is the abundance of iron, which determines the mass loss in stellar winds \cite{Vink:2001cg,Vink:2005zf,Mokiem:2007ji,Graefener:2008pq,2020MNRAS.499..873S}. Iron-poor stars can end their lives as heavier black holes \cite{Belczynski:2009xy}, whose mergers lead to stronger GW signals. Weaker wind mass loss reduces the amount of angular momentum removed from the stars, possibly allowing the formation of compact objects with higher spin magnitudes.
Stellar winds also remove mass and angular momentum from the system's orbit, typically leading to wider binaries at higher metallicities. This can change the type of interaction the binaries undergo later in their evolution, which has a non-trivial effect on the formation of double compact objects and CBC rates.
Metallicity appears to be particularly important for the formation of mergers involving stellar BHs, whose formation efficiency in isolated channels may be greatly enhanced at low metallicity \cite{Belczynski:2010tb,Dominik:2012kk,Stevenson:2017tfq,Klencki:2018zrz, Chruslinska:2018hrb,Giacobbo:2018etu,Broekgaarden:2021efa}.
 In addition, certain CBC formation scenarios (e.g., chemically homogeneous evolution; see section \ref{subsec:CHE}) may operate exclusively for metal-poor progenitors. 
Modeling the CBC population requires specifying the birth metallicity distribution of the progenitor stars forming at different times. 

\subparagraph{Environment plays a crucial role in the formation of CBCs.}
Other differences in the environments in which stars evolve allow for unique merger formation channels  \cite{Barack:2018yly}) (see section \ref{div3:formationchannel}). 
In sparse stellar environments (galactic field), CBC formation involves exchanges of mass and angular momentum between stellar companions in isolated binary/multiple systems. 
In regions with a sufficiently high density of stars and compact objects, dynamical effects can further influence the formation of CBCs.
In the densest environments (e.g., globular/nuclear clusters), compact object binaries can efficiently assemble dynamically and possibly undergo multiple mergers.
Further differences in the CBC formation are expected in the presence of gas (e.g., in AGN disks), or in the case of extremely metal-poor/metal-free progenitors (see section \ref{div3:pop3}).
Therefore, modeling the CBC population requires specifying the cosmic formation history of the progenitors in the considered type(s) of environment.

\begin{figure}[t]
\centering
\includegraphics[width=.65\textwidth]{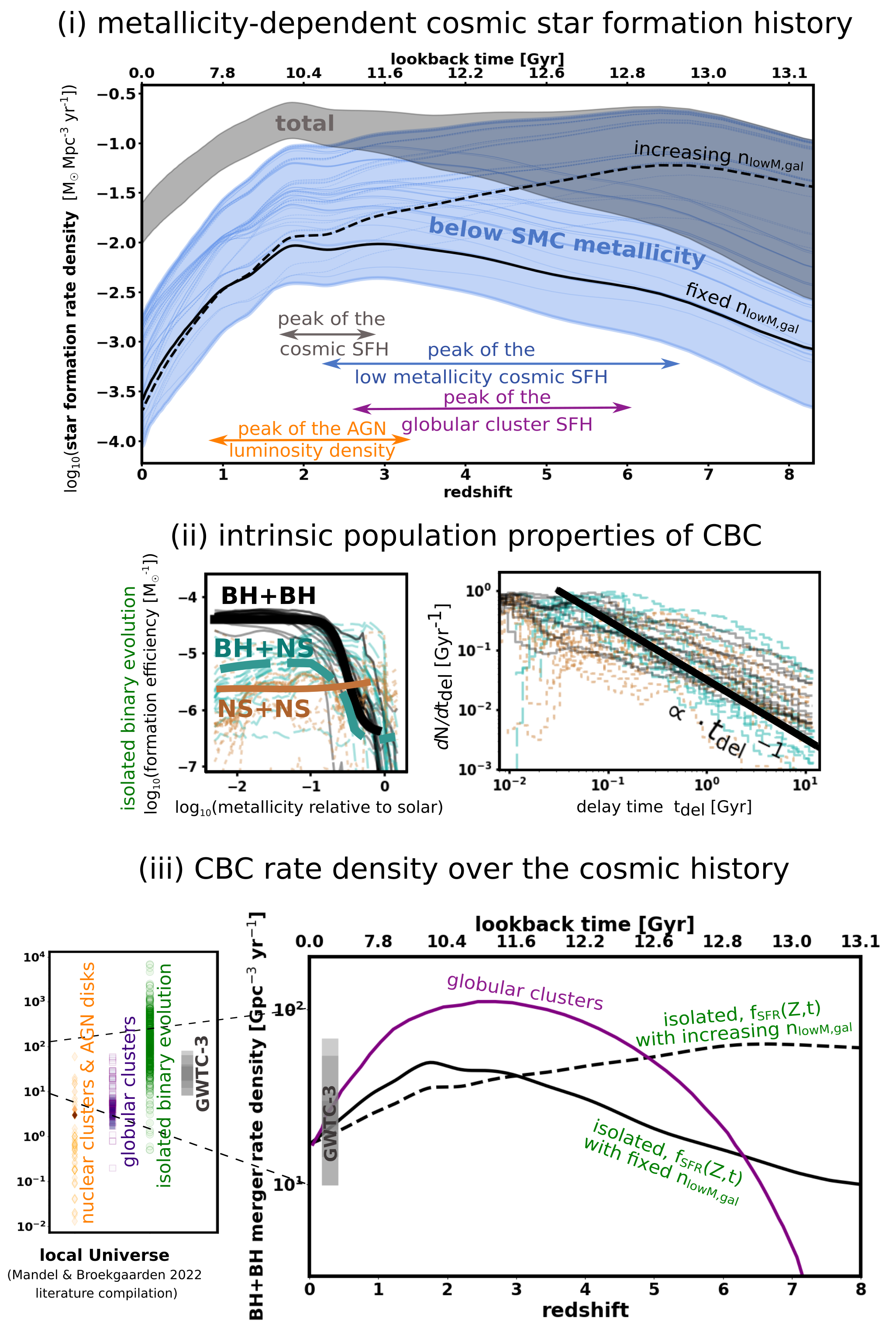}
\caption{\footnotesize{
The key ingredients of compact binary coalescence rates.
\underline{Top:} star formation rate density (gray - total, blue - at metallicity lower than in the Small Magellanic Cloud) as a function of redshift/lookback time, spanned by observation-based $f_{\rm SFR}(Z,t)$ models from \cite{10.1093/mnras/stz2057,10.1093/mnras/stab2690}. 
Solid and dashed black lines show variation in the evolution of the low-metallicity Star Formation History (SFH) due to the uncertain high redshift evolution in the number density of low-mass galaxies (n$_{\rm lowM, gal}$). The arrowheads indicate the predicted range of redshifts for the peak of the star formation history in various types of environments and peak of the AGN luminosity density.
\underline{Middle:}  $\eta_{\rm form}$ and delay-time distribution (DTD) illustrative of merging BBH (black), BHNS (turquoise) and BNS (brown) formed in isolated binary evolution channel. Thick lines: simplified main trends found in the literature. Thin lines: $\eta_{\rm form}$ and DTD  for example binary population synthesis model variations from \cite{Broekgaarden:2021efa} (left) and \cite{Boesky:2024msm} (right, normalised to the same value at peak) showing a diversity of shapes.
\underline{Bottom left:} literature compilation of the local BBH $R_{\rm{merger}}$ from \cite{Mandel:2021smh}.
\underline{Bottom right:} BBH $R_{\rm{merger}}$ redshift evolution examples normalised to the same local rate. Black dashed/solid lines - calculated using $f_{\rm SFR}(Z,t)$ variations corresponding to those shown as black dashed/solid lines in the top panel and the  $\eta_{\rm form}$ and DTD shown as black lines in the middle panel. Purple line - example BBH $R_{\rm{merger}}$ for globular cluster channel from \cite{Ng:2020qpk}.
}} 
\label{fig:CBC rate example}
\end{figure}

\subparagraph{CBC population – a mixed bag of everything.}
The overall observable CBC population is likely a superposition of subpopulations formed in multiple channels, whose relative contributions are expected to vary over cosmic time. The resulting cosmic merger rate density for a certain type of double compact objects (either BBH, BHNS or BNS binaries) is the sum over the contributions from different environments:
\begin{align}
&R_{\rm{merger,\ DCO}} (t_{\rm merger}) = \frac{d^2 N_{\rm merger,\ DCO} }{d t_{s}d V_{c}} (t_{\rm merger})  \notag =\\
		&= \sum_{\rm env} \int d Z  \int_0^{t_{\rm merger}} d t_{\rm delay} \, \left( \frac{dM_{\rm SFR}}{dt_{\rm form}dV_{c}dZ} (t_{\rm form}, Z) \right)^{(i)}_{\rm env} \times \left(\frac{d^{2} N_{\rm form,\rm DCO}}{dM_{\rm SFR} dt_{\rm delay}} (Z) \right)^{(ii)}_{\rm env} 
%
\label{eq:MSSFR-merger-rate}
\end{align}
where $t_{s}$ is the time in the source frame of the merger, $V_{c}$ is the
comoving volume, $t_{\rm merger} = t_{\rm form} + t_{\rm delay}$ and $Z$ stands for metallicity.
The term (ii) in the second parenthesis describes the \textit{intrinsic} properties of the CBC populations.
Specifically, ${d^{2} N_{\rm form,\rm DCO}}/{dM_{\rm SFR} dt_{\rm delay}}$ is the number of binary compact objects that form per unit stellar mass and merge per unit delay time with $t_{\rm delay}<T_{\rm Hubble}$. In general, term (ii) depends on the formation channel/environment (e.g. field, globular cluster with certain properties). We can call merger efficiency the term $\eta_{\rm form,\ DCO}= dN_{\rm form,\ DCO}/dM_{\rm SFR}$ and delay time distribution (DTD) the term
 ${\rm DTD} =dN_{\rm form,\ DCO}/dt_{\rm delay}$.

For a population of field stars formed with the same chemical composition, $\eta_{\rm form}$ and  DTD depend on factors such as the fraction and initial parameters of stellar binaries/multiples, their evolution and interactions (mass transfer, tides), core collapse physics \cite{Chruslinska:2018hrb, Santoliquido:2020axb,Broekgaarden:2021efa,vanSon:2021zpk,Chruslinska:2022ovf}.
For the CBC population formed in dense environments, the initial cluster properties (density profile, mass, binary fraction) can override the effects of stellar evolution on  $\eta_{\rm form}$ and DTD \cite{Antonini:2020xnd}.
The term (i) in the first parenthesis of \eq{eq:MSSFR-merger-rate} describes the cosmic volume-averaged star formation rate density at a given time and metallicity that ends in the considered type of environment (section \ref{sec: step (i)}).  Its role is to weight the intrinsic population properties according to the chemical and star formation history representative of that environment. Term (i) is sensitive to factors such as the chemical evolution of galaxies, the physics of star formation and cluster formation history \cite{Chruslinska:2022ovf}.

\subparagraph{The interpretation challenge.}
Many of the factors that go into the $R_{\rm{merger,\ DCO}}$ calculation are subject to large uncertainties. Taken together, this leads to a wide range of model CBC rate predictions in the literature (see \cite{Mandel:2021smh} and references therein).
Strikingly, these predictions span a range that is orders of magnitude larger than the current GW-based constraints on the local $R_{\rm{merger}}$.
This is the case for all CBC types, but is particularly striking for the BBH merger rate $R_{\rm{merger}, BBH}$, which is the best measured.
The lower left panel of figure~\ref{fig:CBC rate example} shows that the GW constraints already rule out most current models, which tend to overestimate $R_{\rm{merger}, BBH}$ (especially when considering the possible contribution from multiple formation channels). However, the astrophysical interpretation of this fact, and hence the lesson to be learned from it, is unclear.
The interpretation of these constraints is challenging because, even when considering a single CBC formation channel, $f_{\rm SFR}(Z,t)$ (term i),  $\eta_{\rm form}$ and DTD (term ii) degenerate into $R_{\rm{merger}}$. In particular, different combinations of modeling assumptions about terms (i) and (ii) and the relative contributions of merger formation channels operating in different environments can lead to similar $R_{\rm{merger}}$ in the local Universe.
The accuracy of local $R_{\rm{merger}}$ measurements for each type of CBC will continue to improve with ongoing and future GW observations, narrowing the overall parameter space of viable models. However, information on the redshift dimension is likely to be required to resolve the degeneracies in the interpretation.
The role of ET constraints on the CBC rate in addressing this challenge is twofold:
\begin{itemize}
    \item First, ET will provide joint constraints on the merger rate of all BBH, NSBH and BNS out to at least $z\sim$1, mapping the rate evolution for all CBCs over the bulk of the cosmic history.
This joint measurement is key: although BBH, BNS and BHNS are expected to be sensitive to different modeling uncertainties (e.g. due to the mass dependence of stellar evolution or the efficiency of dynamical merger formation), one must be able to reproduce  $R_{\rm{merger}}(z)$ of all three from  the same metallicity dependent cosmic star formation history.
\item Second, ET will constrain the BBH merger rate well beyond the peak of the cosmic star formation history, where the uncertain factors associated with the environment (term (i)) are likely to dominate and lead to divergent $R_{\rm{merger, BBH}}(z)$ predictions, and they may affect the mass-specific rate even more strongly, see section \ref{div3:bh-masses} on mass distribution. These factors, discussed further in the following subsection \ref{sec: step (i)}, could potentially be constrained with the Einstein Telescope.
\end{itemize}

\subsubsection{Part (i): cosmic star formation history as a function of metallicity and environment} \label{sec: step (i)}
Each subsequent generation of stars forms with broadly different metallicities. 
Their exact chemical composition depends on the complex interplay between the timing and metal yields of enrichment events, the mixing of metals in the interstellar medium, metal-poor inflows and metal-rich outflows.
These processes shape the chemical evolution of individual galaxies. Their combined evolution determines the metallicity dependent cosmic star formation history $f_{\rm SFR}(Z,t)$, which characterises term (i) in \eq{eq:MSSFR-merger-rate} for stars in the field and young/open clusters.
The fraction of $f_{\rm SFR}(Z,t)$ forming in massive bound clusters (the progenitors of today's globular clusters), and the properties of these clusters, are expected to vary with the physical conditions in the local star-forming environment. Such factors further affect term (i) in \eq{eq:MSSFR-merger-rate} characterising CBC progenitor formation in dense environments.
The possibility of enhanced dynamical BBH merger formation in galactic nuclei  (nuclear star clusters, gaseous AGN disks) introduces yet another class of environments with uncertain cosmic formation history.
In principle, all environments characterised by different cosmic star formation histories can leave a unique imprint on the redshift evolution of the BBH rate, which can potentially be unravelled with future GW data.

\paragraph{Metallicity-dependent cosmic star formation history.}

Metallicity-dependent cosmic star formation history can be derived by combining statistical galaxy properties constrained by galaxy surveys \cite{Dominik:2013tma,10.1093/mnras/stz2057,Boco:2019teq,Boco:2020pgp,10.1093/mnras/stab2690} or extracted from cosmological simulations of galaxy evolution \cite{Schaye:2014tpa,Crain:2015poa,Pillepich:2017fcc,Nelson:2018uso,Pakmor:2022vnb}. Mixed approaches and simplified analytical prescriptions for the birth metallicity distribution of stars normalised to the chosen cosmic SFH are often used in the context of GW source population modeling (see \cite{Chruslinska:2022ovf} for a recent discussion and comparison of literature assumptions).

The limitations of current methods result in $f_{\rm SFR}(Z,t)$ being particularly poorly constrained at low metallicity and high redshift (figure~\ref{fig:CBC rate example}, top panel).
Due to limited resolution, low-mass and starburst galaxies, which are crucial for low-metallicity star formation, are currently not accurately described in cosmological simulations. Many processes that are key to chemical evolution (e.g., enrichment by different types of supernovae, feedback processes) rely on prescriptions with poorly constrained parameters. 
Studies of the effects of such parameter choices in the full cosmological volume are limited by the high computational cost.
Empirical bounds on $f_{\rm SFR}(Z,t)$ can be derived by combining observational distributions of galaxy properties (star formation rate, gas phase metallicity, stellar mass) and number statistics \cite{Dominik:2013tma,10.1093/mnras/stz2057,Boco:2019teq,Boco:2020pgp,10.1093/mnras/stab2690}. These are increasingly incomplete at high redshift and low mass, i.e. for typically metal-poor galaxies. Constraints on the gas-phase metallicity are typically inferred from the oxygen abundances, which have large systematic uncertainties \cite{Kewley:2008mx,2019A&ARv..27....3M} that translate directly into the uncertainty in $f_{\rm SFR}(Z,t)$ (see blue range in figure~\ref{fig:CBC rate example}, which is based on oxygen abundance measurements). These systematics are one of the dominant factors driving the $f_{\rm SFR}(Z,t)$ uncertainty at low redshifts.
Furthermore, the metallicity dependence of stellar evolution (and CBC formation) is thought to be primarily driven by iron. While constraints on gas-phase oxygen abundances are rapidly becoming available for much larger samples and redshifts, constraints on iron abundances in the star-forming material are scarce and likely to remain a challenge. Consequently, oxygen is often used as a proxy for iron abundance.  However,  the two elements  are produced by sources operating on different timescales and the link between them is not straightforward  - failure to account for that can lead to significant errors \cite{2023arXiv230800023C}.

\subparagraph{The case for next generation GW detectors.}
Combined, the issues discussed above lead to a wide range of observationally allowed $f_{\rm SFR}(Z,t)$. This uncertainty cannot be ignored when interpreting the GW-based CBC population constraints, as 
the assumed $f_{\rm SFR}(Z,t)$ can strongly affect the model CBC properties, especially $R_{\rm{merger}}$ \cite{Chruslinska:2018hrb,Neijssel:2019irh,Tang:2019qhn,Santoliquido:2020axb,Boco:2020pgp,Broekgaarden:2021efa,Briel:2022cfl,Chruslinska:2022ovf, vanSon:2022ylf}.  This is particularly important for BBH mergers, where $f_{\rm SFR}(Z,t)$ alone can affect the local merger rate by more than an order of magnitude.
If the strong low-metallicity preference of the BBH  $\eta_{\rm form}$ seen in isolated channels is correct (black line, middle left panel in figure~\ref{fig:CBC rate example}), the low-metallicity cosmic star formation history becomes the determining factor  for the high redshift $R_{\rm{merger, BBH}}$ of such formed BBH mergers.
This part of the $f_{\rm SFR}(Z,t)$ is shaped by the uncertain properties of low-mass and faint galaxies \cite{10.1093/mnras/stab2690}, which map to a distinct $R_{\rm{merger, BBH}}$ evolution at redshifts $>$2 (compare the upper and lower panels of the figure~\ref{fig:CBC rate example}). 
Constraints in this regime will provide a potential new way, complementary to electromagnetic observations, to study cosmic star formation, the early (iron) enrichment history and the properties of low-mass galaxies in the reionisation epoch \cite{Vitale:2018yhm,Chruslinska:2022ovf,Turbang:2023tjk,Fishbach:2023pqs}.

Different types of mergers are expected to have a different  $\eta_{\rm form}$ dependence on metallicity (figure~\ref{fig:CBC rate example}, middle left panel), and hence their $R_{\rm{merger}}(z)$ may probe different parts of $f_{\rm SFR}(Z,t)$.
Given the exciting prospect that the properties of BBH mergers will be mapped to $z>10$ by the next generation GW detectors, these systems received the most attention in the literature as probes of cosmic history.
However, the interpretation of the redshift evolution of BBH mergers is complicated by the potential competing contributions of different formation channels.
This is less likely to be the case for mergers involving NS (see 
section~\ref{div3:formationchannel}), where isolated binary evolution is thought to dominate \cite{Barack:2018yly, Mandel:2021smh}. 
While the GW constraints will not be available for the lower-mass BNS and BHNS at such high redshifts as for BBH mergers, their $R_{\rm{merger}}$ will be measured for most of cosmic history.
Due to the expected weak metallicity dependence of BNS formation, their $R_{\rm{merger}}(z)$, modulo the effect of the DTD, is likely to track the total cosmic SFH (i.e. $f_{\rm SFR}(Z,t)$ integrated over all metallicities).
BHNS formation is found to be enhanced at low metallicity, although less so than BBH.
This makes the relative rate of BNS and BHNS mergers another promising probe of $f_{\rm SFR}(Z,t)$ \cite{Broekgaarden:2021efa,Chruslinska:2022ovf}.

\paragraph{Cluster formation history and properties.}

Both observational \cite{2017A&A...606A..85L,Adamo:2020jkb,2023MNRAS.525.4456B} and theoretical studies \cite{Kruijssen:2012ia,2019MNRAS.482.4528E,Grudic:2023jkb} indicate that the efficiency of cluster formation and their properties depend on the star forming conditions.
The emerging consensus is that denser and more actively star-forming regions (such as those found in local starbursts \cite{2020MNRAS.499.3267A} and in the early Universe \cite{2023MNRAS.520.2180C,2023ApJ...945...53V}) favor the formation of massive and dense star clusters that can remain bound and evolve into today's globular clusters. Conversely, 
unclustered star formation leads to stellar systems that dissolve on short timescales and/or evolve into less dense open clusters.
However, the full picture, from star formation through young clusters (where isolated CBC formation dominates) to the potential globular cluster stage (where dynamical interactions drive the CBC), remains fragmentary.  This evolution is challenging to model because of the variety of spatial and temporal scales that need to be considered as clusters co-evolve with their host galaxies (but see \cite{2023MNRAS.521..124R,2024A&A...686A.106B}, for the recent efforts in the GW context).
Related, the observational link between globular clusters and their progenitors is unclear because of their long history during which clusters can lose mass or dissolve completely. 

Nevertheless, it is clear that only a fraction of $f_{\rm SFR}(Z,t)$ will eventually end up in environments where dynamical CBC formation becomes efficient.
Given the properties of local globular clusters \cite{2013ApJ...775..134V, 2019MNRAS.490..491U} and the conditions that favor the formation of their progenitors, this fraction is expected to be higher at higher redshifts and lower metallicities.
Furthermore, the efficiency with which CBCs can be dynamically produced depends on the uncertain distributions of initial cluster properties \cite{Askar:2016jwt, Rodriguez:2016kxx,Hong:2018bqs, Choksi:2019, Antonini:2020xnd}. In other words, term (ii) in \eq{eq:MSSFR-merger-rate} is sensitive to cluster characteristics, such as their masses and densities/half-mass radii. Calculating the total $R_{\rm{merger}}$ from cluster environments requires summing over the contributions from clusters with different characteristics.

Finally, we note that while dynamical formation in globular clusters can make a substantial contribution to the overall BBH merger rate, this is less likely to be the case for BNS and BHNS mergers. Their lower masses make them less likely to reside in cluster central regions where dynamical interactions are frequent \cite{PortegiesZwart:2010cly} and natal kicks associated with NS formation cause many of them to escape the cluster \cite{Pfahl:2001df} (see section \ref{sec: formation channels dynamical}). 
The potential imprint of the dynamical formation on other population properties of BBH mergers (mass, effective spin and eccentricity distributions) further motivates the focus on this class of CBC in the literature in the context of this channel (see section \ref{sec: fingerprints}). 

\subparagraph{The case for next generation GW detectors.}
Both the possibility that the majority of today's globular cluster progenitors formed within the first Gyr of the Universe, and the possibility that the majority formed near the peak of cosmic star formation are considered in the literature.
Such scenarios could potentially be discriminated with the $R_{\rm{merger, BBH}}$ constraints provided by the next generation GW detectors \cite{Romero-Shaw:2020siz,Fishbach:2023xws}.

Several authors discuss the promise of future GW data to jointly constrain the formation history of globular clusters and some of their initial properties \cite{Antonini:2020xnd,Fishbach:2023xws}. 
It is important to note that GW data alone cannot constrain all these uncertain properties, since many of them are degenerate in BBH $R_{\rm{merger}}$ (e.g. the distribution of virial radii and the cosmic formation rate of globular clusters). However, it can provide complementary information to improve constraints from other studies, at least as long as the subpopulation of cluster-formed mergers among the detected CBCs can be identified  (e.g. by using GW-based information on masses, eccentricities and spins, see section \ref{sec: fingerprints}).

\paragraph{Nuclear star clusters.}
A somewhat separate class of cluster environments are nuclear star clusters. 
These are the densest stellar systems in the Universe, and as such were proposed as promising sites for efficient dynamical CBC formation \cite{Sedda:2023big}.
They are found in the centres of different types of galaxies, are more common in massive star-forming objects, and can make up $\sim (0.1 - 1)\%$ of the stellar mass of the galaxy \cite{Neumayer:2020gno}. They are thought to form from a combination of in-situ star formation  \cite{Neumayer:2020gno,Fahrion:2021wmz} and migration of  globular clusters that spiral to the galactic center due to dynamical friction. 
These two scenarios imply different cosmic star formation histories for nuclear clusters, which could range from following the total $f_{\rm SFR}(Z,t)$ in the second case, to resembling the globular cluster formation history in the first case.
They would lead to different properties of the stellar and compact object populations in the cluster \cite{Sedda:2023big}.
Recent observational studies support an intermediate scenario, in which globular cluster accretion dominates the mass build-up of the nuclear cluster in dwarf galaxies, while in situ star formation dominates the formation of the most massive nuclear clusters in massive galaxies \cite{2021A&A...650A.137F,Fahrion:2021wmz}. It is currently unclear whether the history of nuclear cluster formation can lead to a distinguishable imprint on CBC $R_{\rm{merger}}$ evolution \cite{Mapelli:2021gyv}.

\paragraph{AGN disks.}

The absolute contribution to $R_{\rm{merger}}$ from the AGN channel is particularly uncertain and depends strongly on factors such as the number density of galaxies hosting an AGN, the lifetime and the characteristics of the AGN disk \cite{Bartos:2016dgn,Tagawa:2019osr,2020ApJ...898...25T,Sedda:2023big,Vaccaro:2023cwr}. 
Current population predictions for the AGN channel focus mostly on the local Universe, and the discussion of trends in the high redshift regime that will be probed by ET is lacking.
Nonetheless, because of the short BBH merger delay times expected from this channel, the resulting $R_{\rm{merger}}(z)$ is expected to be proportional to the number density of AGNs with the appropriate properties.
The dominant AGN property influencing the BH merger rate appears to be the mass accretion rate onto the central supermassive BH \cite{Yang:2020lhq}, also powering the AGN luminosity. However, the exact dependence is different for several processes that can drive BBH mergers in AGN disks \cite{Tagawa:2019osr}. 
If the dominant processes are strongly enhanced at high accretion rates, then the BBH merger rate will scale with the abundance of the most luminous AGNs. Conversely, if such a dependence is weak, then low luminosity AGNs will contribute a substantial fraction of the total BBH merger rate, and the rate will scale better with the total AGN number density. Since the number density of faint AGNs appears to peak at lower redshifts than that of bright ones \cite{Shen:2020obl}, in the latter case the AGN-driven BBH merger rate would peak at lower redshifts. 
This has been found by Yang et al. \cite{Yang:2020lhq}, who use the AGN luminosity function \cite{Shen:2020obl} to derive AGN-driven BBH merger rate at $z<2$. Their $R_{\rm{merger}}(z)$ shows a milder evolution with a peak at lower redshifts ($z<1$) than expected from other BBH merger formation channels and when compared to the peak of the cosmic SFH and of the cosmic luminosity density of AGNs \cite{2023MNRAS.524.1448D}. However, there are considerable uncertainties in the number density and luminosity distribution of (particularly faint) AGNs as a function of redshift \cite{2023MNRAS.524.1448D} which need to be explored further in future studies.
Such a distinct $R_{\rm{merger}}(z)$ evolution would help to identify the CBC subpopulation formed in AGN disks.
Next-generation detectors will allow to select such subpopulations, even for subdominant channels, by dramatically increasing the number of detections and the accuracy of their estimated astrophysical parameters.
Information from the spatial correlation between observed GW events and AGN already constrains this formation channel at low redshifts and high AGN luminosities \cite{Bartos:2017ggb,Veronesi:2023ugk,Veronesi:2024tef}. Future GW data could help to constrain it in the faint and high redshift regimes, where the 
approach based on the AGN luminosity function described above is challenged by the incompleteness of galaxy catalogs.

\subsubsection{Part (ii): intrinsic properties of the CBC populations set by binary compact object formation efficiencies and delay times}\label{sec: step (ii)}
The binary compact object formation efficiency and delay time distributions play a key role in setting the properties and rates of CBC populations across cosmic time. Binary compact objects are a rare outcome of massive star evolution -- to form a binary compact object that can merge in a Hubble time the progenitor system typically has to survive many phases of massive star evolution and/or dynamical interactions.  
The physics underlying these processes impact the fraction of progenitor stars that can form binary compact objects: the so-called formation efficiency. 
This can be written down as a `Drake equation': a multiplication of factors that a population of stars needs to satisfy to form a CBC. For the formation of binary compact objects through the isolated massive binary star channel, the key factors can be written as something like:
\begin{align}\label{DrakeIsolated}
 \eta_{\rm form} &= f_\textrm{init}  \times f_\textrm{survive SNe} \times f_\textrm{mass transfers}\times f_\textrm{merge},  
\end{align}
where each factor $f$ is a factor selecting only a subset of a given initial population of stars that has the properties to become a CBC source, as we will describe in more detail for the channels below.

The delay time distribution (DTD) spans a broad range of values (from many Myr to Gyr) and as a result, for binary compact objects formed at a given redshift in our Universe, can significantly impact at which redshift these sources can contribute to the merger population and thus what sources GW detections probe as they reveal the CBC populations across cosmic time \cite{Chruslinska:2022ovf}.  
This is because the orbital angular momentum loss by {GW} radiation is typically very small. 
For a binary consisting of two compact objects with masses $M_1$, $M_2$, eccentricity $e$, and semi-major axis $a$ the change in separation due to {GW} radiation is given by \citep{Peters:1964zz}  
\be
\frac{da}{dt}
=  -\frac{64G^3}{5c^5} \frac{{M_{\rm{1}} M_{\rm{2}}} (M_{\rm{1}} + M_{\rm{2}})}{a^3} 
\frac{1}{(1-e^2)^{7/2}}\, 
\(1 + \frac{73}{24}e^2 + \frac{37}{96}e^4\)\, \, ,
\ee
which, for a circular orbit and a system with masses $M=M_{\rm{1}}=M_{\rm{2}}$, simplifies to
$ t_{\rm insp} \approx 0.1  \, {\rm Gyr} \,  (a/\Rsun)^4 (M/\Msun)^{-3}$.
This means that the delay times depend strongly on the  separation, as $t \propto a^{4}$.
If the distribution of orbital separations at the moment of binary compact object formation is given by some power law distribution $d n / d a \propto a^{-\gamma}$, we can use this to derive that the DTD is approximately described as:
\begin{equation}
    \frac{d n}{d t} \propto  \frac{d n}{d a}    \frac{d a}{d t} \propto   a^{-\gamma} a^{-3} \propto t^{-(\gamma+3)/4}  
\end{equation}
If we take that, based on observational evidence, binaries are born with a separation distribution described by $\gamma \approx 1$ \citep{Abt:1983tr, Sana:2012px, Duchene:2013cba, Almeida:2017}, the equation reduces to $ {d n}/{d t} \propto t^{-1}$, which is a commonly used distribution and also found by observations (e.g., \cite{Berger:2014, Fong:2015, Nugent:2022paq, Zevin:2022dbo}).

The formation efficiency and DTD are still uncertain and can vary for different formation channels based on different underlying physical processes leading to binary compact object formation. The details of these ingredients play a major role in shaping the CBC properties as a function of redshift.  Observations with Einstein Telescope of the CBC merger rate as a function of redshift will probe the formation efficiency and DTD in enough detail to enable inferring the underlying physics \citep{Punturo:2010zz, Sathyaprakash:2012jk,Reitze:2019iox,Maggiore:2019uih,Singh:2021zah,Gupta:2023lga}. We discuss this in more detail below.

\paragraph{Isolated massive stars (the field).}
In \eq{DrakeIsolated}, 
the key factors for determining the formation efficiency of CBCs in the field and their DTDs are the initial conditions, and those relating to mass transfer episodes, stellar winds, and supernovae -- fundamental phases in the lives of massive stars that are poorly understood and can be uniquely probed with the Einstein Telescope. %
The key factors entering the CBC efficiency depend on the specific formation channel within the field (only stable mass transfer, classic common-envelope, CHE, triples, see the overview in section~\ref{div3:formationchannel}). 
Here below, we list some of the most important among such factors.

First, the initial conditions of the stars need to be sufficient for binary compact object formation, $(f_{\rm{init}})$. The system needs to have stars that are massive enough to form NSs or BHs, the system typically needs to be born with a small enough separation to initiate a mass transfer phase during its lifetime (as mass transfer interactions are expected to be required to bring the binary compact object to a tight enough orbit to merge within a Hubble time) but not too small for it to undergo a stellar merger at ZAMS, and the system needs to be born as a binary or a triple star system (typically described by the ``multiplicity factor''). The initial conditions and binary fractions of massive stars are uncertain and can be informed by observations \cite{Opik:1924, Ulrich:1976, KippenhahnMeyerHofmeister:1977, Neo:1977, Wellstein:2001, deMink:2007af, Schneider:2015, Kroupa:2000iv, Sana:2012px}.
Second, the binary (or hierarchical) system typically  disrupts during the supernova by the significant natal kicks of hundreds of $\rm{km}/\rm{s}$ that the system can receive when the {NS} (or BH) forms (e.g., \cite{Hobbs:2005, Verbunt:2017zqi}). This likely disrupts the majority of the systems during the supernovae (especially the first supernova when the system typically has a wide orbit), though uncertainties remain  \cite{Nelemans:1999ux, Vigna-Gomez:2018dza, Renzo:2019}.  This probability is expressed by the factor $f_\textrm{survive SNe}$.
Third, the system is typically thought to undergo several mass transfer phases, either stable or an unstable common-envelope phase. The stability of mass transfer and the mass and angular momentum loss during these phases might well be one of the most uncertain aspects of massive star evolution. What stars and what fraction of stars can successfully eject the envelope and form a tight system is uncertain and adds an important factor in the formation efficiency equation $f_{\rm{mass\, transfer}}$  \cite{Giacobbo:2018etu, Vigna-Gomez:2018dza, Klencki:2018zrz, Neijssel:2019irh, Dorozsmai:2022wff, Broekgaarden:2021efa, Santoliquido:2020axb, Roman-Garza:2020uou}. 
Last, even once the  binary compact object has formed it needs to have a tight enough orbit to merge within the age of the Universe so that it can merge and be observable today,  likely removing  another half of the systems described by the factor $f_{\rm{merge}}$ \citep{Peters:1964zz}.  
Other field-based channels, including CHE, triples and hierarchical systems have similar ingredients impacting the DTD and formation efficiency, but often include extra Drake factors, corresponding to questions such as: What is the metallicity at which CHE formation channel operates \citep{duBuisson:2020asn, deMink:2016vkw, Mandel:2015qlu}? How do interactions with a triple or hierarchical companion contribute to binary compact object formation efficiency and DTD \citep{2023A&A...678A..60K, Toonen:2020,  Stegmann:2021jen, Hamers:2019oeq, Hamers:2021tyz}? How do Population III stars form, evolve, and produce CBCs \citep{Kinugawa:2014zha, Kinugawa:2020ego, Belczynski:2017gds, Hijikawa:2021hrf, Liu:2021svg,  Santoliquido:2023wzn, Costa:2023xsz}?
These factors combined with the DTD decide the merger rate of CBCs. Observing the CBC merger rate across redshift with the Einstein Telescope will be crucial to infer the formation efficiency and DTD and study the underlying physical processes in massive star evolution \citep{Vitale:2018yhm, Fishbach:2023pqs, Singh:2023cxn}.

\paragraph{Cluster formation (dynamical).}
The formation efficiency and DTD distributions for binary compact objects formed in dense star clusters is still under debate and depends on the dynamical formation channel at play and the expected binary compact object properties, as described in detail in section~\ref{div3:formationchannel}. Important open questions that ET will probe by inferring the CBC merger rate include the following. 
First, what are the CBC host cluster properties? The formation rate and history of globular cluster populations across cosmic time is still under debate as well as the (initial) mass distribution of globular clusters, which is important as the formation efficiency and DTD are found to be significantly impacted by the cluster mass \citep{2019MNRAS.482.4528E, Askar:2016jwt, Rodriguez:2016kxx,Hong:2018bqs, Choksi:2019, Antonini:2020xnd, 2019MNRAS.486..331C, Kritos:2022ggc, Sedda:2021vjh,Antonini:2018auk,Antonini:2016gqe, Fragione:2021nhb}. 
What fraction of binary compact objects can be retained in the cluster and undergo multiple interactions thought to be required to harden the binary enough to form a CBC \citep{Quinlan:1996vp, 2003gmbp.book.....H}? What fraction of merger products receives low enough recoil kicks such that it can be retained in the cluster center and form hierarchical mergers \citep{Herrmann:2007ex, Pollney:2007ss, Rodriguez:2018jqu,Gerosa:2021mno}? 
How does the presence of a supermassive black hole impact the formation and merger rate of CBCs \citep{MillerLauburg:2008, Antonini:2016gqe, 2018MNRAS.477.4423A}? What is the role of accretion disks around supermassive black holes and potential migration traps in forming CBC sources across cosmic time \citep{Bellovary:2016, Bartos:2016dgn, Stone:2016, Mckernan:2017ssq, Yang:2019, Secunda:2019, PengChen:2021}?  Einstein Telescope is poised to detect the CBC merger rate with sub-percent precision; this, together with measuring the properties of CBC across cosmic time, is expected to unveil the formation channels and uniquely inform the key ingredients contributing to the CBC rate: the star formation histories,  formation efficiencies, and DTDs  \citep{Fishbach:2023xws, Gerosa:2021mno, Punturo:2010zz, Sathyaprakash:2012jk, Reitze:2019iox, Maggiore:2019uih, Singh:2021zah, Gupta:2023lga}.

\subsubsection{Future outlook}
\begin{itemize}

\item {\em CBC merger rate is expected to evolve as a function of redshift.} Its evolution is not expected to follow directly the cosmic SFH.
The simple link to cosmic SFH is modified by the delay time distribution (relevant for all types of mergers), further affected by the metallicity dependence of the CBC formation efficiency (BH-NS and BBH mergers), and possibly by contributions from multiple formation channels (expected primarily for BBH mergers) whose relative importance may vary with redshift.
$R_{\rm{merger}}(z)$ encodes a wealth of astrophysical information, especially at high redshifts that will be accessible to ET, where much of the uncertainty in current models lies.

\item {\em There is a large parameter space to explore with ET, and a challenge for the astrophysical interpretation.}
There are substantial uncertainties due to factors related to both the formation environment (metallicity dependent cosmic star formation history, cluster formation history and initial properties) and intrinsic properties of the CBC merger population (delay time distribution and formation efficiency for different formation channels). All of these factors affect the predicted evolution of the CBC merger rate with redshift. However, many of them are degenerate in $R_{\rm{merger}}$. Future constraints on the joint redshift dependence of all CBC and on the high redshift evolution of  BBH merger rate will be key to break such degeneracies in the interpretation.
These constraints will be most informative if subpopulations of mergers formed in specific channels/environments can be identified, and if improving electromagnetic constraints are used to place bounds on the population models. From this perspective, the greatest promise of ET lies in the use of future GW constraints to complement the electromagnetic studies.

\item {\em It is unclear whether the redshift-dependent merger rate of the entire BBH population will show multiple features} (peaks) or not.
The former case can be expected if different formation channels dominate or give similar contributions to the merger rate at different redshifts.
The latter case could occur if one of the formation channels dominates the BBH formation, or if multiple channels lead to similar merger rate evolution over cosmic history. 
Both scenarios seem possible within the uncertainties of current models and ET will provide constraints to distinguish between them.
More detailed astrophysical inference requires: (a) on the theoretical side, reliable models for $R_{\rm{merger, BBH}}(z)$ from various formation channels/environments, or the ability to confidently identify the corresponding subpopulations in the data based on other properties, and (b) on the empirical side, accurate $R_{\rm{merger, BBH}}(z)$ measurements up to high redshifts and the ability to accurately infer the redshift of sources and associated features in population properties \cite{Ng:2020qpk,Singh:2023cxn,Mancarella:2023ehn}. For the latter, a network of next-generation GW detectors is advantageous. In general, a network of GW detectors leads to better redshift measurements for individual sources and thus to tighter constraints on the evolution of CBC population properties compared to a single detector (which requires longer observing times to reduce statistical uncertainties). 
The fraction of GW detections at $z>1$ with accurate luminosity distance measurements increases by almost an order of magnitude for a network of 3 next-generation detectors compared to a single ET \citep{Iacovelli:2022bbs, Gupta:2023lga}. 

\item {\em Full population inference for CBC catalogs observable by a detector with ET capabilities is needed.} 
As discussed in this section, some of the poorly-constrained astrophysical assumptions lead to divergent predictions of the evolution of the BBH merger rate only at high redshift $z>3$. These are of particular interest in the context of next-generation detectors, which are going to provide unique constraints on the rate in this regime.
To better quantify which astrophysical assumptions we can expect to constrain with ET, it is important to quantify
(i) what is the minimum factor to which we can constrain the CBC merger rate, especially at high redshifts and
(ii) what is the minimum amplitude/width of a feature in the BBH merger rate that can be confidently recovered from the population analysis of ET-like data (as a function of redshift).

The scalability of currently employed population inference schemes to the numbers of detections and accuracy on the reconstruction of the single events' parameters achievable at ET is limited by the high computational cost. Work in this direction is ongoing (see also the discussion in section~\ref{sect:inferencepopdiv9}), extending methodologies to more flexible models, improving their efficiency, as well as exploring new tools, such as machine learning techniques.

\item {\em ET is essential to make significant progress in observing the CBC merger rate.}
In the coming years updating current ground-based detectors (O4 and O5) will likely bring the number of BBH mergers to around $\sim 500 $ with detections out to $z\sim 1$ \citep{KAGRA:2013rdx}, but this is not enough to reveal important parts of the CBC merger rate such as the peak of the merger rate (expected around $z\sim2$), multiple features or peaks in the merger rate especially expected at $z\gtrsim2$ as different formation channels might be dominating the rate at different redshifts. ET will  detect stellar-mass black hole mergers out to (and beyond) redshifts $z\gtrsim 10$, and BNS and BHNS beyond $z\gtrsim 2$, which will drastically change the field by increasing our understanding of the redshift behavior of CBCs \cite{Punturo:2010zz, Sathyaprakash:2012jk, Maggiore:2019uih, Singh:2021zah, Gupta:2023lga}.

\end{itemize}

\subsection{Mass function of BHs and its evolution with redshift}\label{div3:bh-masses}

\subsubsection{The scientific potential of the mass function}\label{sec:BH-mass-measure}

GW astronomy is well-suited to measure the masses of binary components. 
From the physics encoded in the observed gravitational waveform, we can infer the source properties~\cite{LIGOScientific:2016vlm,LIGOScientific:2016dsl}. 
The inspiral portion of a binary coalescence is determined, to leading order in the post-Newtonian expansion, by the chirp mass of the binary, with the mass ratio contributing at the next order~\cite{Cutler:1994ys,Blanchet:1995ez,Poisson:1995ef,Baird:2012cu}. The transition to merger and the subsequent ringdown are principally determined by the binary's total mass~\cite{Echeverria:1989hg,Graff:2015bba,Haster:2015cnn}. 
The sensitivity of next-generation detectors means that for the closest (and hence highest SNR) sources, these mass measurements will have unprecedented precision~\cite{Vitale:2016icu,Kalogera:2021bya,Borhanian:2022czq,Iacovelli:2022bbs,Branchesi:2023mws}.

The masses of BHs are a key tracer of their evolutionary history. 
Considering a star in isolation, its remnant mass is primarily set by its initial mass~\cite{Woosley:2002zz}. 
Given the initial mass function of stars, we therefore expect more low-mass BHs \cite{Salpeter:1955it,Kroupa:2000iv}. 
However, the mapping between initial and final mass is not monotonic: different evolutionary histories can lead to different amounts of mass loss, e.g., through stellar winds and supernova explosions. 
The amount of mass loss also varies with metallicity~\cite{Puls:2008mk,Smith:2014txa,Vink:2021dxo}. 
Stellar winds are stronger for higher metallicity stars, meaning that lower mass remnants are typically formed. 

With GWs, we observe BHs in merging binaries. 
Therefore, we must consider the processes that form binaries, and drive them to be close enough to merge. 
As discussed in section~\ref{sect:binarywithmasstransfer}, 
periods of mass transfer between binary stars will alter the orbital configuration and remnant masses. 
Dynamical exchanges may preferentially select the most massive objects in a cluster to form a merging binary~\cite{Sigurdsson:1993tui,Rodriguez:2019huv,Bouffanais:2021wcr}. 
Accretion for binaries embedded within active galactic nucleus disks will grow BHs~\cite{McKernan:2012rf,Tagawa:2019osr}. 
This wealth of diverse astrophysical processes leaves imprints on the masses of observed BHs ---both on the individual component masses and their mass ratio--- and through reconstructing the mass distribution we may uncover the underlying physics that shaped these binaries. 

The component masses are one piece of information about the origins of BHs. 
Further information is encoded in other measurements. 
The overall merger rate (section~\ref{div3:rates}), the spins (section~\ref{div3:spins}), the distribution in redshift, and the orbital eccentricity all contain information about the origins of binaries. 
Crucially, looking at the correlations between parameters (such as whether spin alignment is correlated to mass ratio) can reveal information that is overlooked when considering each parameter in isolation~\cite{Callister:2021fpo,Adamcewicz:2022hce,Heinzel:2023hlb}.
Only by combining all this information can we untangle the complicated physics of how BHs form (section~\ref{div3:formationchannel}), distinguishing  primordial from stellar BHs (section~\ref{sec:PBHsdiv3}), and BHs that formed from Population III or from later stars (section~\ref{div3:pop3}). 

Understanding the distribution of masses also has implications beyond understanding the sources themselves. 
As we discussed in section~\ref{sec:lateuniversediv2}, 
GW observations can be used to measure the expansion of the Universe~\cite{LIGOScientific:2017adf,LIGOScientific:2021aug,Chen:2024gdn}. 
While the GW measurement provides the distance~\cite{Schutz:1986gp,DelPozzo:2011vcw}, the redshift has to be obtained through other means. 
If no electromagnetic counterpart is detected (which is expected to be the case for the majority of GW sources)
we can link the GW source to its host galaxy (and thus merger redshift) through a statistical association with galaxy catalogs, as has been done in recent years for both simulated data
\cite{DelPozzo:2011vcw,Chen:2017rfc,Gray:2019ksv,Gair:2022zsa,Borghi:2023opd,Muttoni:2023prw,Mastrogiovanni:2023emh} and actual data \cite{Dalya:2018cnd,Dalya:2021ewn,LIGOScientific:2018gmd,LIGOScientific:2019zcs,Finke:2021aom,LIGOScientific:2021aug,DES:2019ccw,DES:2020nay,Palmese:2021mjm,Alfradique:2023giv,DESI:2023fij};
see section~\ref{sect:darksirensdiv2} for extended discussion.
Alternatively, features in the mass distribution, such as mass gaps or peaks, can be used to determine the redshift \cite{Taylor:2011fs,Taylor:2012db,Farr:2019twy,Mancarella:2021ecn,Finke:2021eio,Ezquiaga:2022zkx}, see again section~\ref{sect:darksirensdiv2}; however, this requires that we understand how these features of the mass distribution evolve with redshift~\cite{Mukherjee:2021rtw,Pierra:2023deu}.
In either case, calculation of the distance is influenced by mass measurement: a signal can potentially be from a louder source (higher mass) at greater distance or a quieter source (lower mass) at small distance. 
Therefore, when performing an inference to measure cosmological parameters, we must simultaneously infer the underlying astrophysical source population~\cite{Mastrogiovanni:2021wsd,LIGOScientific:2021aug,Gray:2023wgj}.  

 \begin{figure*}[t!]
\centering
\includegraphics[width=0.9\linewidth]{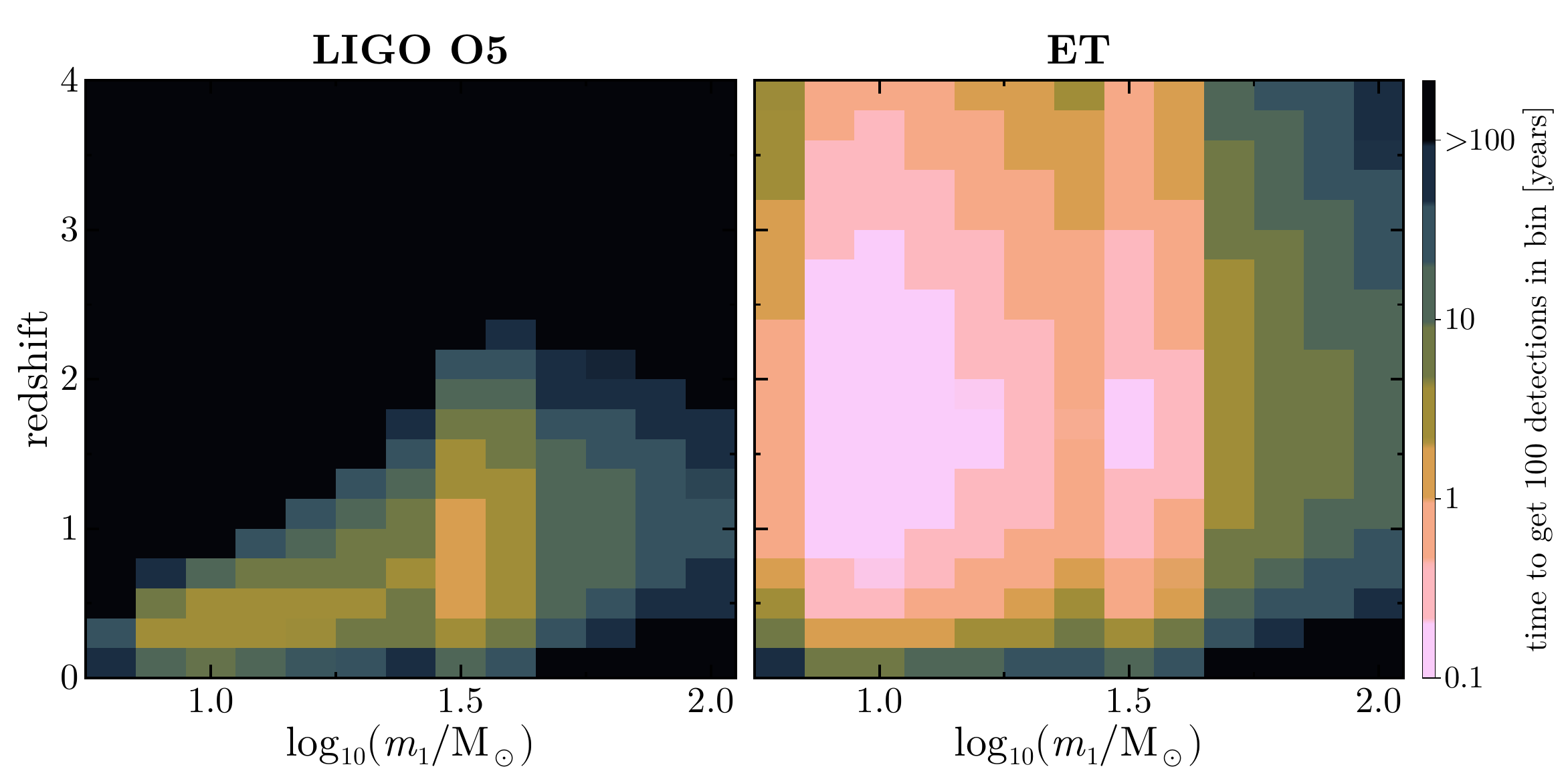}
\caption{Time to detect $100$ merging binary BHs in bins of size $\Delta z=0.2$ in redshift and $\Delta \log m_1=0.1$ in primary mass, indicative of the number of measurements required to provide a constraint on the rate with an error of $10\%$ in each bin. The distribution of masses is taken to be the \textsc{Power Law+Peak} parametric model derived from GWTC-3~\cite{KAGRA:2021duu} using the median values for all parameters (which include a localized Gaussian peak at $34\Msun$), and increasing the upper cut on BH mass to $m_\mathrm{max}=250 \Msun$. This figure assumes the  sensitivity curves adopted in \cite{Branchesi:2023mws} for a 2L configuration with 15 km arm-length.} 
\label{fig:GW_mass_dist}
\end{figure*}

Next-generation ground-based GW observatories such as ET are well-designed to uncover the details of the mass distribution of BHs. 
Their sensitivity means that we can detect a large number of BH binaries, and make detections out to large distances~\cite{Kalogera:2021bya,Iacovelli:2022bbs,Branchesi:2023mws,Fairhurst:2023beb}. 
This sensitivity compared to current-generation GW detectors is illustrated in figure~\ref{fig:GW_mass_dist}, which shows the time necessary to obtain a significant number of detections to constrain the merger rate as a function of redshift and primary mass. This is computed based on the \textsc{Power Law+Peak} parametric model determined from GWTC-3 \cite{KAGRA:2021duu}, although the upper limit on BH masses of the model has been increased to $m_\mathrm{max}=250\Msun$ to highlight the expected sensitivity of ET at high masses.\footnote{The median value derived for $m_\mathrm{max}$ from GWTC-3 is poorly constrained, with large support up to $m_\mathrm{max}=100\Msun$ which is the edge of the uniform prior used \cite{KAGRA:2021duu}.} The total merger rate has been taken to follow the cosmic star formation history as given by \cite{Madau:2014bja}, with the rate at $z=0.2$ fixed to the value inferred from GWTC-3 \cite{KAGRA:2021duu}.
The precision of population measurements typically scales with the square root of the number of detections~\cite{Stevenson:2017dlk,Barrett:2017fcw}, so the increased sample size will enable a revolutionary advancement compared to existing GW observatories. 
The long-distance reach enables reconstruction of the mass distribution over cosmic time---although figure~\ref{fig:GW_mass_dist} was constructed assuming it is redshift independent, we expect this distribution to change with time due to the evolution of the source population, e.g., through changes in the metallicity distribution~\cite{Giacobbo:2017qhh,Neijssel:2019irh,vanSon:2020zbk}, in the balance of star formation in the field compared to clusters~\cite{Zevin:2020gbd,Singh:2021zah}, and due to the prevalence of primordial compared to stellar-origin BHs~\cite{Ng:2022agi,Franciolini:2023opt}. 
 
\begin{figure}[t]
\centering
\includegraphics[width=0.8\linewidth]{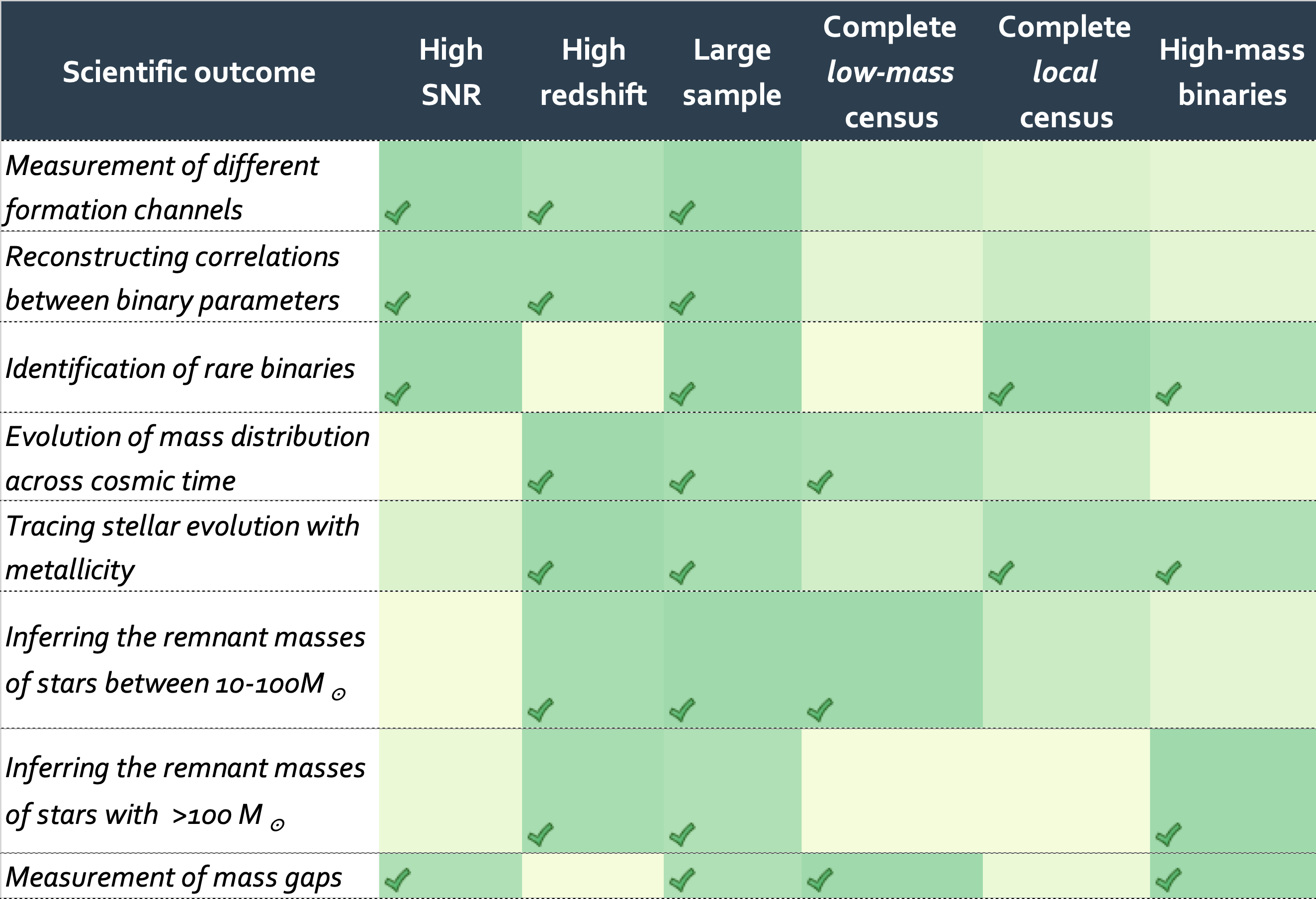}
\captionof{table}{The interdependence of observational capabilities of next-generation GW detectors and how they may uniquely answer key scientific questions about the masses of BHs. Shades of green indicate how relevant an observational capability is for the envisioned science goal, while check marks designate essential capabilities. 
\label{tab:mass-checklist}} 
\end{figure}

The combination of the large number of detections with the increased detection horizon means that with these observations we can precisely reconstruct not only the overall mass distribution, but the distribution as a function of redshift, and explore correlations at different points in the Universe's history~\cite{Kalogera:2019sui}. 
As summarised in table~\ref{tab:mass-checklist}, this is precisely the information needed to unravel the intricate mysteries of how (the progenitors of) BHs form and evolve. 
Throughout the rest of this section, we link the scientific outcomes listed in table~\ref{tab:mass-checklist} to the science enabled by ET.

\subsubsection{Reconstructing the mass distribution}\label{sec:BH-mass-inference}

To uncover the astrophysics of the population of GW sources, we must perform a sequence of analyses. 
First, signals must be identified in the data stream, and their source properties inferred~\cite{LIGOScientific:2019hgc,KAGRA:2021vkt}. 
The sensitivity of next-generation detectors like ET means that we can detect sources at large distances, and that the closest sources (as currently detected by the LVK~\cite{KAGRA:2021vkt}) will be measured with high SNR, meaning that the source properties will be precisely measured~\cite{Cutler:1994ys,Vitale:2016icu,Kalogera:2021bya,Borhanian:2022czq,Iacovelli:2022bbs,Branchesi:2023mws}. 
To map from the observed sample to the underlying astrophysical population, we must correct for selection effects~\cite{Mandel:2018mve}. 
Some signals are easier to detect than others, hence the observed distribution is biased compared to the true source distribution. 
This is the case for current GW observations~\cite{KAGRA:2021vkt}. 
However, the sensitivity of next-generation detectors means that there will be a range of masses and redshifts where the probability of detection is effectively equal to $1$; hence all these sources will be detected, and the observed population is a complete, unbiased census of all the merging compact-object binaries. 
To consider the full ranges of masses and redshifts, selection effects will still need to be accounted for. 
Finally, we can combine the inferred properties of the observed sources with the calculated selection effects to constrain the astrophysical population: this requires an assumption about the underlying distribution.

There are multiple approaches to modeling the underlying distribution when performing a population inference. 
Broadly, these may be categorised into three approaches:
\begin{itemize}
  \item \emph{Using predictions driven by population models}. 
  We can use predictions from theoretical models of different formation channels to match observations~\cite{Stevenson:2015bqa,Zevin:2017evb,Bouffanais:2019nrw}. 
  The observed population is a mix of different channels, and so it is necessary to consider a combination of channels in this inference (omitting channels leads to biased results~\citep{Zevin:2020gbd}). 
  The observations can be used to infer the branching ratio between different channels, and any uncertain model inputs (section~\ref{div3:formationchannel}). 
  Hence, this approach provides the most direct probe of the underlying physics. 
  However, this approach also requires accurate model predictions~\citep{Belczynski:2021zaz}, and the calculation of (potentially computationally expensive) models across a range of input parameters.
  \item \emph{Using parametric models}. 
  We can alternatively use a flexible parametric model, such as a power-law, to fit the population~\cite{Kovetz:2016kpi,Fishbach:2017zga,Wysocki:2018mpo}. 
  This does not rely upon theoretical predictions for the inference, making it easier to identify unexpected features in the data. 
  However, it is still necessary to define a model, and if the model is not a good fit to the underlying data, the inferred fit will not be a good representation of the truth. 
  The parameters of the model need not have direct physical meanings, making their interpretation more difficult than if using theoretical predictions. 
  However, the model features, such as the position of a peak in the mass distribution, may drive development of theoretical models to explain them.  
  \item \emph{Using non-parametric models}. 
  All models require parameters to define them, but some approaches allow the number of parameters to be adapted during the fitting (potentially growing to an infinite number). 
  These are referred to as non-parametric models. 
  Examples include using a Dirichlet process Gaussian mixture model~\cite{Rinaldi:2021bhm,Rinaldi:2023bbd} or an autoregressive process~\cite{Callister:2023tgi}.
  The flexibility means that they can recover unexpected features in the observations. 
  The result of the inference is a reconstruction of the underlying population, but this does not directly encode information about the source physics.  
\end{itemize}
In each case, the expected measurement precision on the population properties scales with the square root of the number of detections. 
The high detection rate of next-generation detectors will mark a significant advancement in our understanding of these populations.

In practice, we expect a combination of different approaches to be used to understand results. 
This may be illustrated in considering the evolution of LVK results. 
The first inference of the astrophysical population used a simple power-law for the masses---given the small number of detections at the time, that was the most complicated model that was considered justified~\cite{LIGOScientific:2016dsl}. 
As the population has grown, more complicated parametric models have been introduced, adding peaks and dips to the model~\cite{Talbot:2018cva,LIGOScientific:2018jsj,LIGOScientific:2020kqk,Farah:2021qom}. 
Flexible features have been added to the parametric models (such as using a variable number of Gaussian kernels to reconstruct the mass distribution~\cite{Tiwari:2020vym}, using a binned distribution~\cite{Mandel:2016prl,Heinzel:2024jlc,Ray:2024hos}, or adding a spline to model additional features in the mass distribution~\cite{Edelman:2021zkw}), starting to bridge the gap to non-parametric models~\cite{KAGRA:2021duu}. 
These allow for checks of whether the parametric models are capturing all of the features in the astrophysical population. 
With the current population of order $\sim100$ observations, it is still difficult to understand if a lack of detections is the sign of a physical dip in the population, or the result of a Poisson fluctuation~\cite{Farah:2023vsc}. 
Despite the relatively small sample size, non-parametric approaches have also been applied, showing that some features found by the parametric models are robust (such as a peak at low masses)~\cite{Tiwari:2021yvr,Sadiq:2021fin,Edelman:2022ydv,Callister:2023tgi,Rinaldi:2023bbd}. 
Direct comparison with population-synthesis predictions have also been made~\cite{Wong:2020ise,Bouffanais:2021wcr,Franciolini:2021tla,Iwaya:2023mse}, demonstrating that there is probably a mix of different formation channels~\cite{Zevin:2020gbd,Cheng:2023ddt,Stevenson:2022djs}. 
As the number of detections has grown, we have gone from identifying general trends, to starting to identify features and pointing out limitations in our current theoretical models.

Over the coming years, we expect the current generation of GW detectors will yield hundreds of new detections~\cite{KAGRA:2013rdx}. 
Our understanding of the astrophysics of these sources will therefore advance greatly. 
This knowledge, however, will be eclipsed by that resulting from next-generation detectors, as demonstrated in figure~\ref{fig:GW_mass_dist}. 
Detectors like ET will detect large numbers of sources ($\sim10^5~\mathrm{yr}^{-1}$), allowing us to perform the same types of analyses as we do currently for individual redshift bins---we will not only be able to map out the mass spectrum in high detail, but also see how it varies across cosmic time~\cite{Kalogera:2019sui}.

\subsubsection{BH mass spectrum and outstanding questions}\label{sec:BH-mass-physics}

The BH mass spectrum holds crucial information about the evolution and final fate of stars. 
Here, we address BHs of stellar origin only (primordial BHs are discussed in section~\ref{sec:PBHsdiv3}) and the richness of the physics that could be inferred from the mass distribution. 
The masses of such BHs depend on the final moments of the life of their stellar progenitors and whether they explode in a supernova, collapse directly into a BH, or experience partial explosion and fallback of mass onto the forming compact remnant, possibly giving rise to a BH. 
The question of the explodability of stars relates to the supernova mechanism (e.g., electron-capture supernovae,
core-collapse supernovae and pulsational pair-instability supernovae) and the final structure of stars, which is set by their evolutionary history.
This evolution depends strongly on the metallicity of the stars and the formation channel of the binaries (i.e., their interactions with their environment or stellar companion). 
For example, for single stars, there is a critical metallicity beyond which stars lose so much mass by stellar winds that virtually no star can form a BH anymore; see  \cite{Heger:2023jjg} for a recent review of many of such aspects related to the formation of BHs in stars. 

As shown in figure~\ref{fig:GW_mass_dist},  ET will reconstruct the BH mass spectrum at various redshifts, and hence at different average metallicities. 
Therefore, it will be possible to directly probe theoretical models and a wide range of basic physics, as highlighted in table~\ref{tab:mass-checklist}. 
In the following, we describe possible features in the BH mass spectrum that will be observable with  ET, and explain the essential physics that governs them. 
In essence,  ET can help solve the complex question ``Which stars form BHs?''.
This question encompasses aspects ranging from stellar explodability to the physics of pair-instability pulsations, but also concerns the formation channels (i.e., stellar collisions and hierarchical mergers).
We start the discussion with the explodability of stars (section~\ref{sec:sn-direct-collapse}) and then dive into various features of the BH mass spectrum. 
We consider the transition from the neutron star to the BH mass spectrum and the possibility of the  neutron star--BH mass gap (section~\ref{sec:ns-bh-mass-gap}); we explain theoretical ideas and observations of global peaks in the BH mass spectrum (section~\ref{sec:global-peaks-BH-mass-spectrum}), and we finally move to the upper mass region of the BH mass spectrum, where is expected  an upper mass gap with a pile-up due to pulsational pair-instability supernovae (section~\ref{sec:upper-bh-mass-gap}).


\paragraph{Supernova or direct collapse?}\label{sec:sn-direct-collapse}

Stars with low iron-core masses and those undergoing electron-capture supernovae are thought to leave behind neutron star remnants, discussed in section~\ref{sec:ns-mass}. 
We now focus on iron-core collapse situations within the framework of neutrino-driven supernovae (pulsational pair-instability supernovae are discussed in section~\ref{sec:upper-bh-mass-gap}). 
The pre-supernova stellar structure is often summarized by the so-called compactness parameter $\xi_M = (M/\Msun) / (R(M) / 1000\,\mathrm{km})$, which is the dimensionless mass-to-radius ratio at a given mass coordinate $M$, usually taken to be $2.5\,\Msun$ to measure the compactness just outside the iron core \cite{OConnor:2010moj}. 
High values of compactness ($\xi_{2.5}\gtrsim0.45$) signal BH formation, while neutron star formation may occur for lower values. 
Over the last decade, various groups reported a qualitatively similar landscape of compactness showing clear peaks and troughs as a function of the core mass of stars. 
The peaks could indicate islands of BH formation, but given the uncertainties in the supernova explosion mechanism and core-collapse dynamics, it remains an open question which pre-supernova stellar structures lead to BH formation. 
If there was a direct link between high compactness and BH formation, an almost universal picture would emerge with probable BH formation at carbon--oxygen core masses of $7\text{--}8\,\Msun$ and ${>}\,13\,\Msun$: a qualitatively similar landscape is found for different stellar-evolution codes \cite{Woosley:2002zz, Sukhbold:2013yca, Chieffi:2020gxh, Patton:2020tiy, Schneider:2020vvh, Takahashi:2023jfk}, largely independent of  evolutionary histories of stars (i.e., single stars, binary-stripped stars, accretors of binary mass transfer and even stellar mergers) \cite{Schneider:2020vvh, Schneider:2024mue},  metallicities \cite{Woosley:2019nnp, Aguilera-Dena:2022ypv, Schneider:2023mxe}, and uncertain physics such as convective boundary mixing \cite{Temaj:2023nuo}. 
Since pre-supernova red supergiants follow a tight carbon--oxygen core mass-to-luminosity relation, the high compactness at carbon--oxygen core masses of $7\text{--}8\,\Msun$ predicts a luminosity range of red supergiants that shall collapse into BHs and not explode in a supernova. 
The best candidate red supergiant to have collapsed into a BH, N6946-BH \cite{Gerke:2014ooa, Basinger:2020iir}, had an inferred luminosity in general agreement with this prediction \cite{Temaj:2023nuo, Schneider:2024mue}.

On top of the general compactness landscape, there is considerable variability, and the connection of the pre-supernova stellar structure to BH formation is still uncertain. 
While three-dimensional simulations of iron core collapse supernovae seem to converge on the explosion mechanism \cite{OConnor:2018sti, Burrows:2020qrp}, the exact outcome remains uncertain because it is computationally challenging to reach a conclusive statement regarding compact-object formation and possible fallback of envelope material onto the proto-neutron star. 
Starting from one realisation of one-dimensional simulations of core collapse, a complex picture emerges that reproduces the general picture described above, but additionally shows sub-structure in the form of smaller islands of BH formation \cite{Ertl:2015rga, Sukhbold:2015wba}. 
Another set of such computations gives opposite results in terms of explodability and hence compact remnant masses \cite{Boccioli:2024kvw}. 
With ET, there will finally be a direct observation of the BH mass spectrum that will allow us to put constraints on the question of BH formation in stars and the explosion mechanism of core-collapse supernovae (section~\ref{sec:ns-bh-mass-gap}, section~\ref{sec:global-peaks-BH-mass-spectrum} and section~\ref{sec:upper-bh-mass-gap}).


\paragraph{The neutron star--BH mass gap.}\label{sec:ns-bh-mass-gap}
More than two decades ago, studies on the mass distribution of BHs based on X-ray binary observations suggested an absence of BHs in the $(3-5)\,\Msun$ range \cite{Bailyn:1997xt,Fryer:1999ht,Ozel:2010su,Farr:2010tu}. 
This idea of a gap between the most massive neutron stars and the least massive BHs, also known as the neutron star--BH mass gap (or sometimes, the ``lower-mass gap''), hinges on the mass measurement in one of the X-ray binaries (GRO\,J0422\,+32),  and has been a topic of active debate ever since it was first suggested, due to expected observational and evolutionary selection biases \cite{Kreidberg:2012ud,Wyrzykowski:2019jyg,Jonker:2021rkb,Siegel:2022gwc}. 
Nonetheless, multiple theoretical models have been proposed to explain this gap as the result of a discontinuous remnant-mass distribution, driven by either the supernova engine \citep{Fryer:2011cx,Fryer:2022lla} or the progenitors' density profile, which determines if a supernova fails \citep{Kochanek:2013yca,Kochanek:2014mwa}.

The first catalogs of GW observations showed tentative support for a relative dearth of merging BBHs with component masses between $2.5\text{--}5 \Msun$ \citep{Farah:2021qom,Ye:2022qoe,Biscoveanu:2022iue}. 
Studies have linked this  dearth of low-mass BHs observed with GWs to the underlying supernova engine (e.g., \cite{Zevin:2020gma,Olejak:2022zee}). 
However, the population of GW sources is a heavily biased population in terms of detected events and should receive equal, if not more scrutiny over its evolutionary selection effects as X-ray bright sources.
Binary evolution alone can also produce a dearth of low-mass BHs without invoking a discontinuous remnant-mass distribution \cite{vanSon:2022myr}.

Moreover, at the time of writing, there has been at least one confident detection of a GW source with a primary component falling into this neutron star–-BH mass gap (GW230529\_181500)~\cite{LIGOScientific:2024elc}, while there have been two detections with tentative secondary component masses detected in this mass range (GW190814, GW200210\_092254) \cite{KAGRA:2021duu}.

%

How sparsely populated is this mass range? Is this the result of the remnant mass function or of evolutionary selection effects? Is this dearth related to the metallicity and/or redshift of star formation? ET will be perfectly positioned to answer these questions, since the improvement of ET over current LVK detections will be most significant in the relevant mass regime of this purported mass gap (figure~\ref{fig:GW_mass_dist}).

\paragraph{Global peaks of the BH mass spectrum.}
\label{sec:global-peaks-BH-mass-spectrum} 

There are several lines of evidence that point towards an abundance of $\sim9\,\Msun$ BHs. 
First, there seems to be a peak at about this mass in BH mass distribution of X-ray binaries \cite{Kreidberg:2012ud}; second, a similar peak is found in BBHs observed with GWs \cite{KAGRA:2021duu}, and third, BHs found in field binaries often have a similar mass (e.g., Gaia~BH1 and~BH2 \cite{El-Badry:2022zih, Chakrabarti:2022eyq, El-Badry:2023pah}). 
The current GW data further reveals a significant peak in the primary BH mass distribution at $\sim35\,\Msun$ and a tentative peak at $\sim16\,\Msun$ \cite{KAGRA:2021duu, Tiwari:2020otp}.
While the significance of these features still needs to be assessed \cite{Talbot:2018cva, Edelman:2022ydv, Farah:2023vsc}, we discuss the possible origins of these and other features, and what information they may hold for our understanding of BH formation in light of the capabilities of ET.

The isolated binary evolution channel for BBH mergers makes the prediction of peaks in the BH mass spectrum at $\sim9\,\Msun$ and $\sim16\,\Msun$ \cite{vanSon:2022myr, Schneider:2023mxe}, and the $\sim9\,\Msun$ peak is even predicted to exist because of two different mechanisms. 
Both mechanisms probe different physics, and revealing such features in observed populations of BHs would allow us to directly constrain this physics.

First, the stable mass transfer channel has been proposed to naturally lead to a dearth of low-mass compact objects \cite{vanSon:2022myr}.
The stability of mass transfer depends on the mass ratio of the systems involved in the mass transfer. 
This places strong constraints on the progenitor population masses that can lead to merging BBH systems. 
This should lead to a minimum primary compact-object mass that can follow from the stable mass transfer channel. 
If this is the case, then the global peak at $9\Msun$, and a decrease towards lower masses could be the transition from the stable mass transfer channel as the dominant formation channel, to the common-envelope channel (cf. \cite{Klencki:2020kxd,Pavlovskii:2016edh,Marchant:2021hiv,Gallegos-Garcia:2021hti,Olejak:2021fti}). 
This is in line with recent work that suggest that the dominant formation channel for binary neutron star mergers is the common-envelope channel, while BBH formation has been suggested to be dominated by the stable mass transfer channel \cite{Wagg:2021cst,Iorio:2022sgz}

Second, both BHs in BBH mergers from isolated binary evolution are from progenitor stars that have been stripped of their hydrogen-rich envelopes. 
Envelope removal is known to affect the core structures at the pre-supernova stage: almost independently of metallicity, binary-stripped stars with carbon--oxygen core masses of $\sim7\text{--}8\,\Msun$ achieve stellar structures at core collapse that are difficult to explode within the framework of neutrino-driven supernovae, and may thus give rise to direct collapse BHs. 
Since binary-stripped stars at a large range of metallicities follow a carbon--oxygen core mass--final stellar mass relation, they are predicted to form BHs of similar mass around $\sim9\,\Msun$ and largely independent of metallicity \cite{Schneider:2023mxe}. 
A similar reasoning holds for higher carbon--oxygen core masses of $\gtrsim13\,\Msun$ and binary-stripped stars with such cores are found to exhibit core structures that appear likely to directly collapse into BHs \cite{Woosley:2019nnp, Aguilera-Dena:2022ypv, Schneider:2023mxe}. 
Indeed, one-dimensional, core-collapse simulations show that these stars likely produce BHs of $\sim9\,\Msun$ and $\gtrsim12\,\Msun$ \cite{Ertl:2019zks}. 
The models clearly predict the existence of such BH masses largely independent of metallicity and this can be directly probed by ET. The physics governing these BH masses is a combination of stellar, nuclear and SN physics \cite{Schneider:2023mxe}. 
On the stellar physics side, one can, e.g., test binary evolutionary channels leading to the formation of BBH mergers and convective boundary mixing during core helium burning. 
In terms of nuclear physics, the characteristic carbon--oxygen core masses are set by the complex interplay between nuclear burning processes during the late stages of massive star evolution and thermal neutrino losses, and particularly the generally important $^{12}\mathrm{C}(\alpha,\gamma)^{16}\mathrm{O}$ reaction \cite{Farmer:2020xne}. 
Regarding supernovae, a broad range of physics is relevant to establish the connection of pre-supernovae core structures that appear difficult to explode and the actual explodability of stars. 
Another important question is how much of the final stellar mass may form the resulting BH and how much is still ejected in the collapse of the star.

From data of the third GW Transient Catalog \cite{KAGRA:2021duu}, a clear peak at $\approx35\Msun$ has been inferred. Notably, also the third BH discovered by Gaia, Gaia~BH3, with a mass of about $33\,\Msun$ falls into this peak \cite{Gaia:2024ggk}. As of today, there is no consensus on the origin of this feature. Among other scenarios, mergers in globular clusters \cite{Antonini:2022vib} and repeated mergers of initially lower-mass BHs from other peaks in the BH mass spectrum have been suggested \cite{Tiwari:2020otp}, as well as a pile-up of BHs from pulsational pair-instability supernovae \cite{Talbot:2018cva, Stevenson:2019rcw, Woosley:2020mze} (see also section~\ref{sec:upper-bh-mass-gap} for more details).  

Peaks in the BH mass spectrum should persist across a wide range of metallicities and redshifts. These features are expected to evolve with redshift due to a combination of cosmological redshift, and the redshift-dependence of their formation channels. As long as these two effects can be disentangled, the redshift evolution of features in the mass distribution can serve as ``spectral sirens'', see the discussion in section~\ref{sect:darksirensdiv2}. This offers an independent measurement of the Hubble constant, and will help resolve the Hubble constant tension \cite{Planck:2020, Riess:2021jrx}, as well as the possibility of constraining or revealing deviations from the standard cosmological model, 
see section~\ref{sect:ModGWpropdiv2}.

\paragraph{Upper mass gap and (pulsational) pair-instability supernovae.}
\label{sec:upper-bh-mass-gap}

Stellar theory predicts a gap in the BH mass function between approximately $45\text{--}80 \Msun$ and $120\text{--}160 \Msun$, known as the pair-instability supernova mass gap \cite{Woosley:2002zz,Heger:2001cd,Woosley:2016hmi,Woosley:2019nnp}.
The progenitor stars of these BHs reach temperatures and densities in their centers that allow for electron-pair production, triggering a premature collapse of the carbon-oxygen core that results in an explosion that completely disintegrates the star \cite{Fowler:1964zz,Barkat:1967zz}. 
For helium cores of $M_{\mathrm{He}} \gtrsim 130 \Msun$, photo-disintegration prevents the disruption of the star and we expect BHs to form again, see e.g. \cite{Heger:2001cd}. 
Pair-instability theory thus predicts a mass gap in the distribution of BH masses. 
The expected location of this gap is robust against most uncertainties in stellar evolution, but is only sensitive to the $\rm ^{12}C(\alpha,\gamma)^{16}O$ reaction rate \cite{Farmer:2019jed,Marchant:2020haw,Renzo:2020rzx}.
This has renewed interest in constraining this rate \citep{Mehta:2021fgz,Farag:2022jcc,Shen:2023rco}, and it has conversely been suggested that an observed mass gap in GW sources can be used to assess its value \citep{Farmer:2020xne,Franciolini:2024vis}. 

ET is particularly promising in this regard because its lower frequency range will increase our sensitivity from $\sim100\mathrm{\Msun}$ to up to $\sim500\mathrm{\Msun}$, allowing us to observe both the lower, \textit{and the upper} edge of the PISN mass gap.
Initial GW detections supported a dearth of BHs with masses above $45\Msun$ \citep{Fishbach:2017zga, LIGOScientific:2018jsj}, but more recent observations have shown that this proposed gap is not empty \citep{KAGRA:2021duu}.
Since pair-instability theory predicts that isolated binaries cannot form BHs in this mass range, other formation channels have been suggested to explain the low but non-zero rate of mergers in the gap, including hierarchical mergers of BHs. 
%
Confirming the location of the PISN mass gap is thus linked to fundamental questions, such as ``What is the most massive BH we can form from stars?” (also closely related to IMBH formation, see section~\ref{sec:IMBH}), ``Which merging BBHs are formed from isolated binaries vs.  dynamical environments?”.

\subparagraph{Have we detected the lower edge of the PISN gap?}
Just below the lower edge of the PISN mass gap, stars exhibit pulsations that are not strong enough to completely disrupt the star, but instead map a range of core masses to similar BH masses. 
We thus expect a pile-up of BH masses just below the lower bound of the PISN mass gap \cite{Woosley:2021xba}. 
Over the past six years, the feature observed in the primary BH-mass distribution at $35\Msun$ has consistently been attributed to this pile-up resulting from pulsational-PISN \cite{Talbot:2018cva, Stevenson:2019rcw, Belczynski:2017gds, Woosley:2020mze, Karathanasis:2022rtr}.
However, recent studies seem to pile on evidence that the location of this feature is in tension with the expected location of PISN \cite{Farag:2022jcc,Golomb:2023vxm,Hendriks:2023yrw}, and whether the lower edge of the PISN mass gap has been observed remains an open question \cite{Ulrich:2024nez, Antonini:2024het}.

Addressing this requires detecting \textit{both} edges of the PISN mass gap. 
The lower frequency range of ET provides a unique chance to measure the \textit{full} range of the PISN gap, thereby enabling us to answer this question and distinguish between different BH formation channels.


\subsection{Constraining the mass function of neutron stars}

    
\subsubsection{Observations of radio pulsars and accreting NSs} \label{sec:ns-mass}

Neutron star (NS) properties \cite{Lattimer:2004pg} are determined by a complex interplay between stellar evolution, 
supernova physics, binary interactions (mass transfer), plasma physics, and ultra-dense matter behavior. 
Therefore, powerful insights into these processes can be obtained from macroscopic observables
such as NS masses, radii, moments of inertia, spins, and magnetic fields. 
NS masses in particular have been proven to be an exceptionally informative probe of NS physics. 

\begin{figure*}[t]
\centering
\includegraphics[width=0.9\textwidth]{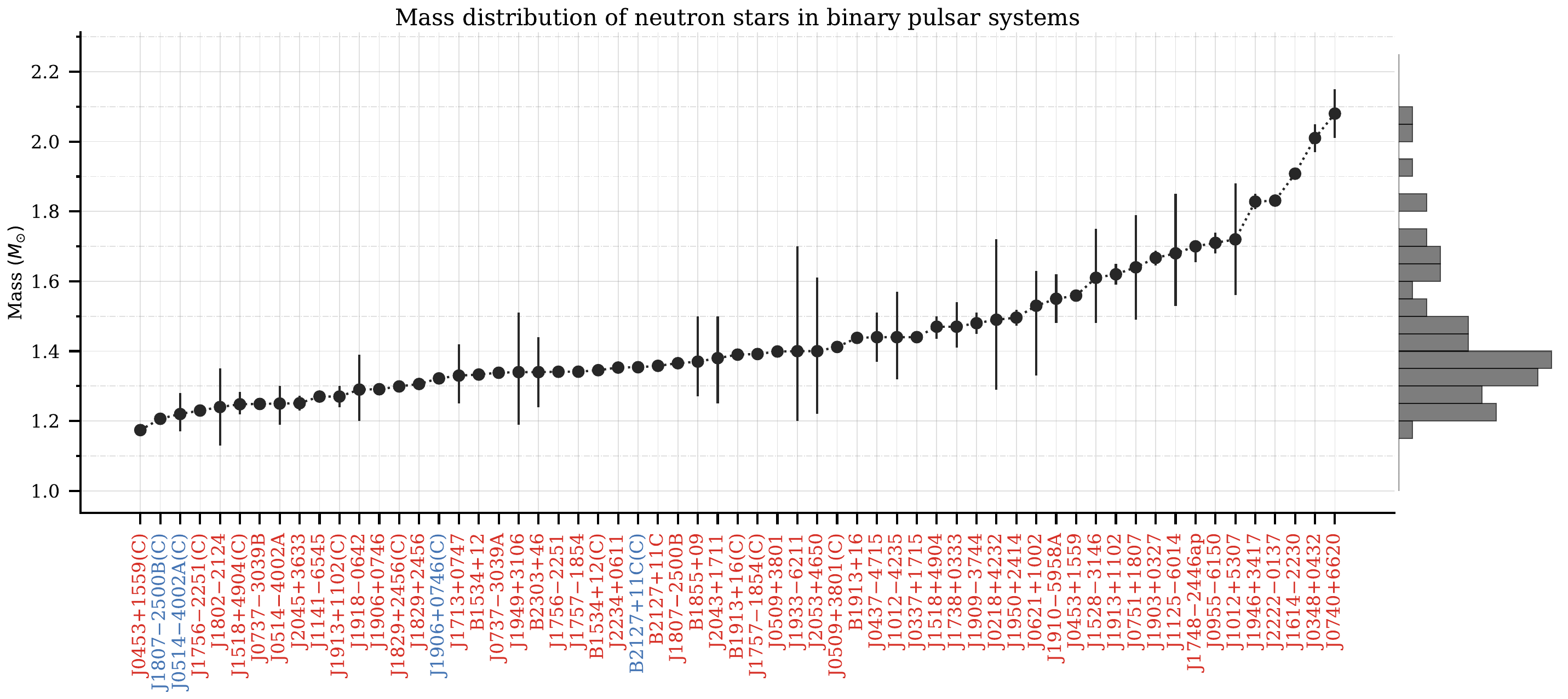}
\caption{NS mass measurements in binary pulsar systems, sorted by increasing mass. Names followed by (C) indicate companions to radio pulsars that are consistent with being NSs. The names in blue show systems in which the companion could also be a massive WD.
The error bars indicate the 68\% confidence intervals. The histogram on the right is based on the median of the mass PDFs. Figure reproduced from V. V. Krishnan and P. Freire; for up-to-date versions, see \url{https://www3.mpifr-bonn.mpg.de/staff/pfreire/NS_masses.html}.  
} 
\label{fig:MSPmasses}
\end{figure*}

Because BNS mergers are intrinsically much fainter than BBHs, second-generation detectors do not provide a detailed view of the NS mass spectrum. Instead, currently the only detailed insights are provided by binary radio pulsars within our Galaxy, which provide the most precise mass measurements \cite{Kramer:2021jcw}. 
Most of these are derived from long-term  monitoring of pulse arrival times, which are modulated by binary motion, thereby giving access to the orbital dynamics. In tight binaries and/or favorably oriented systems,
constraints on leading-order post-Keplerian (pK) parameters such as the Shapiro delay, 
orbital precession, and orbital decay often become possible \cite{Lorimer:2012abc}. 
Component masses can be measured in systems for which at least two pK parameters can be
determined \cite{Wex:2014nva}. For systems in which the pulsar companion is optically bright, 
spectroscopy and photometry can also be used to constrain the mass ratio and the companion mass \cite{Antoniadis:2013pzd,Linares:2018ppq,Kandel:2022qor}. 
Despite significant progress in recent years, acquiring precise pulsar
mass measurements still presents a formidable challenge. 
Although more than 600 binary radio pulsars are now known, only 56 masses in 46 binaries have been constrained with a precision better than 15\% (see figure~\ref{fig:MSPmasses}). 
Most of these objects are millisecond pulsars (MSPs) with low-mass companions, 
with only 22 being double NSs (DNSs). It is therefore unclear whether the general conclusions
drawn from these systems are directly applicable to merging DNS binaries. 

Determination of NS masses in X-ray binaries is also possible due to
the eclipsing and accreting, or even pulsating, nature of such binaries where a NS orbits 
a visible companion star. In such double-lined spectroscopic binaries, for which the
orbital inclination angle can be determined, one can determine the mass of the 
NS \cite{Tauris:2023sxf}. Unlike the situation for binary radio pulsars, 
the empirically 
determined masses of NSs in X-ray systems are typically subject to 
uncertainties of order $\sim 10$\%, or more  \cite{Falanga:2015mra,Ozel:2016oaf}. 
For this reason, individual NSs in X-ray binaries can rarely be used to constrain NS structure and formation theories. The mass distribution of NSs in X-ray 
binaries appears to roughly follow the spectrum of NS masses obtained from radio pulsar binaries.

\subsubsection{Overall shape of empirical NS mass distribution}
As a first guess, the masses of NSs are expected to spread between the Chandrasekhar limit and the Tolman-Oppenheimer-Volkoff limit. 
In principle, from an EOS point of view, the NS mass distribution could
be wider: for instance, if part of the iron core is somehow ejected during the supernova (SN), or if very fast differential rotation prevents a massive (e.g. post-merger) NS from collapsing \cite{Rezzolla:2017aly}. 

The empirical NS mass distribution for pulsars in our Galaxy is overall consistent with this picture. The lightest known in a DNS system, J0453+1559, has a mass of $m=1.174(4)\, \Msun$, 
while the most massive known NS is PSR J0740+6620 with a 
mass of $m=2.08(7)\, \Msun$ \cite{NANOGrav:2019jur, 
Fonseca:2021wxt}, although there are candidates such as 
PSR\,J0952$-$0607, that may have even higher NS masses 
\cite{Linares:2018ppq,Romani:2022jhd}. More recently, a 
binary NS was discovered \cite{Barr:2024wwl} with a total 
mass of $3.887(4)\,\Msun$ in NGC 1851. The compact 
companion has a mass between 2.09 and 2.71\,$\Msun$ (95\% confidence), but it is not yet clear whether it is a massive NS or a low mass BH. 

The observed masses are not uniformly distributed between the two extremes, but rather
seem to cluster around a small number of characteristic values. Approximately one third
of objects appear to have masses greater than $1.6\,\Msun$ (figure~\ref{fig:MSPmasses}). Several studies
suggest that the mass spectrum is multimodal, with at least two groups clustered 
around $\sim 1.4\,\Msun$ and $\sim 1.8\,\Msun$ \cite{Ozel:2016oaf,Antoniadis:2016abc,Alsing:2017bbc}. 
There is also marginal evidence for a high-mass cutoff around $2.1\Msun$, 
that could signify the dividing line between NSs and BHs \cite{Antoniadis:2016abc,Alsing:2017bbc,Rocha:2023xwp}. 
However, the presence of such a cutoff appears to be sensitive to
modeling choices and requires more measurements to be constrained with confidence.  

Differences between the mass spectra of different NS populations also appear to be present. 
For example, DNS systems seem to cluster with symmetric NS masses around $(1.35-1.40)\,\Msun$, thus generally having smaller masses than MSPs in binaries that host the most massive NSs. However, recent discoveries of
asymmetric DNS systems challenge this picture. Progress on this
front requires robust quantitative and qualitative comparisons with the
DNS merger population, which will become possible in ET era.

\subsubsection{Theoretical expectations}
Several theoretical studies suggest that the complexity of the NS mass spectrum can be largely explained by the nonlinear nature of late stellar evolution and an element of stochasticity in SN dynamics \cite{Wongwathanarat:2012zp}. For instance, subtle differences in the way late shell burning phases proceed may give rise to large explodability fluctuations between stars that had very similar properties at the beginning of their evolution \cite{Ertl:2019zks,Laplace:2020hum, Antoniadis:2021dhe,Aguilera-Dena:2022ypv}. In turn, these fluctuations seem to significantly influence not only the nature of the final remnant (NS or BH), but also the amount of matter ejected and falling back during the SN. 

Additional complexity may be introduced by more elusive core-collapse events such as electron-capture supernovae and accretion-induced collapse. 
For example, for stars which evolve to develop low-mass metal cores, the exact boundary in outcome between an iron core-collapse SN and an electron-capture SN depends on the location of ignition of the off-center neon and oxygen burning and the propagation of the neon--oxygen flame \citep{2014ApJ...797...83J}. The location and narrow range ($\sim 0.2-0.6\,\Msun$) of initial ZAMS masses that produce electron capture SNe, and thus the total number of low-mass NSs produced, are a function of metallicity \cite{Ibeling:2013bm,2017PASA...34...56D}. However, it should be noted that the lowest mass NSs ($<1.25\,\Msun$) are not expected from electron-capture SNe but from low-mass iron core-collapse SNe \cite{Timmes:1995kp}.
The reasons for this is that the cores of stars with ZAMS masses $\ge 12\,\Msun$ during and beyond carbon burning are not degenerate, and therefore their collapse is not directly related to the Chandrasekhar mass, as is the case for electron-capture collapse. As a consequence of this, NSs with the very lowest masses are expected to be below even the Chandrasekhar limit.
In electron-capture SNe, the core will collapse rapidly and thus produce a NS accompanied by a dim SN explosion. 
The related NS formation channel of an accretion-induced collapse (AIC) of a massive WD may also leave behind a NS.
Because of the promptness of these core collapses, caused by the steep density gradient at the edge of the core in
the progenitor stars, there is therefore little time for asymmetries to develop and it is therefore believed that both electron-capture SNe and AIC events produce small NS kicks 
and NS masses very close to the Chandrasekhar limit.

Mass transfer from the binary companion after the NS formation may also affect the mass spectrum towards higher masses. 
Interestingly enough, the mass difference between non-recycled and recycled NSs need not be dominated by accretion of matter after the formation of the NS. As argued by \cite{Tauris:2017omb}, the first-born NSs in NS+NS systems probably only accrete $\lesssim 0.02\,\Msun$, based on an analysis of the various accretion phases and observational data of both mass and spin periods. While there is no doubt that NSs in low-mass X-ray binary (LMXB) systems (producing MSPs with He~WD companions or H-rich ``spiders'') are able to accrete significantly more material than the first-born NSs in DNS systems, because the timescale of evolution in the former systems is much longer, there are still many examples of fully recycled MSPs which have small masses of $(1.2-1.4)\,\Msun$. Since the original birth masses of these NSs could not be much smaller than $1.2\,\Msun$, these systems show clear evidence of very inefficient accretion. (For a reference, it only takes $\sim 0.1\,\Msun$ of accreted material to spin up a NS to a rotation period of 2~ms \cite{Tauris:2012jp}.)
Therefore, the general differences in masses between non-recycled and recycled NSs is likely caused by a difference in the birth masses of NSs born first or second in a binary system. Thus, these mass differences are likely to be the result of the progenitor star evolution. 
At the upper end of the mass scale, it is possible that heavy accreting NSs may reach a critical mass and collapse to produce low-mass BHs \cite{Chen:2023etu}, even with masses below this critical threshold due to loss of gravitational binding energy when the BH is formed. 

\subsubsection{Anticipated impact from ET data} 
A high-resolution NS mass spectrum (and ultimately its dependence on redshift) is a key ET objective.
ET BNS detection rate is expected to be up to two orders of magnitude higher compared to second-generation  detectors \cite{Branchesi:2023mws}. For a BNS population following the Galactic DNS mass
distribution, this could translate to several thousands of events per year, detected up to redshifts
$z\sim 5$. A considerable fraction of these events (few hundred per year) will be recorded with
sufficient precision during both the inspiral and post-merger phases to accurately separate the
component masses. Hence, these data will provide a completely new probe of the NS mass distribution
and its evolution with redshift. In turn, this will allow for the first time a direct link to
physical processes before and after the NS formation. For example, the different nature of core collapse SNe (iron core collapse vs. electron capture) is guaranteed to produce a distinct signature in a high-resolution NS mass spectrum obtained at different metallicities \cite{Chanlaridis:2022egz}. 

Synergies with existing (e.g. the Five-hundred-metre Aperture Spherical radio; FAST) or upcoming (e.g. the Square Kilometre Array; SKA) facilities that will provide an order of magnitude more mass measurements for galactic binary pulsar systems \cite{Watts:2014tja}, will also help resolve degeneracies between the effects of the EoS, core-collapse physics, and binary evolution \cite{Antoniadis:2016abc}.  
Likewise, nearby golden events that will be observable for several days to weeks will also allow for joint EM monitoring and follow-up. Such events will further help to identify non-equilibrium effects (see section~\ref{section:div6} for details) during the inspiral and use them to obtain more accurate estimates of NS macroscopic properties. 
It is important to note that the LF instrument will be critical for achieving the sensitivity required to meet these objectives.

\subsection{Spins of stellar-origin BHs and NSs}\label{div3:spins}

\subsubsection{BH spins -- theoretical expectations}

The origin of spin in astrophysical BHs remains an active research topic. However, from first principles, BH spin is related to four factors \cite{Mandel:2020lhv}: \textit{(i)} the angular momentum content of the BH's immediate progenitor star, \textit{(ii)} the angular momentum of material accreted onto the BH after it formation, \textit{(iii)} an instability during the core collapse that transfers significant amount of angular momentum to the core of the collapsing star, while a small amount of material carrying the opposite angular momentum is ejected from the system, and \textit{(iv)} the conversion of orbital angular momentum into spin angular momentum in a binary BH (BBH) system. In the next paragraphs we examine each of these four aspects in the process of forming coalescing BBHs and NSBHs.

In the last several years, it has been argued that a BH formed from a single star or the first-born\footnote{One should note  that the terms ``first-born'' and ``second-born'' here refer to the order in evolution of the BBH progenitor system that the two BHs were formed, and should not  be used interchangeably with the terms ``primary'' and ``secondary'', often used in observational GW astronomy, referring to the more and the less massive BH, respectively, at the time of merger.} BH in an isolated BH binary systems is likely to have very slow spin as long as there is efficient angular momentum transport within the star \cite{Fragos:2014cva,Fuller:2019sxi}. As the BH progenitor star is expanding in its post-main-sequence evolution, angular momentum is transported to its outer layers, depleting the core from angular momentum. The outer layers of the BH progenitor star are subsequently removed via stellar winds and binary interactions, typically leaving behind a slowly spinning stripped star which eventualy forms a slow-spinning BH. For single stars, the exact BH spin depends on the initial rotation of the star and the details of the assumed angular momentum transport mechanism, but in any case it is expected to be $\lesssim 0.1$ \cite{Qin:2018vaa,Fuller:2019sxi,Belczynski:2017gds}. However, when member of a binary system, the progenitor star of the first-born BH may maintain a slightly higher angular momentum content, due to tidal interactions during it evolution, resulting in BH spins up to 0.2 \citep{Bavera:2023cyt}. One caveat to the argument above is the possibility for chemically homogeneous evolution (CHE) \cite{Maeder:2000wv}.  The rapid rotation of a massive star, which can be either its initial rotation or  induced by a binary interaction, can induce strong internal mixing, maintaining a uniform chemical composition throughout the star. CHE stars do not expand to become supergiants, and thus can avoid angular momentum  losses, especially at very low metallicity where stellar winds are significantly reduced, and can form BHs with higher spins. It is believed that CHE occurs for low-metallicity, fast rotating, massive stars, but the exact conditions that lead to CHE remain uncertain (e.g., \cite{deMink:2009jq,2016A&A...585A.120S,2015A&A...581A..15S,duBuisson:2020asn})


In all formation channels emerging from the evolution of isolated binaries, the progenitor of the second-born BH is a stripped He star. The spin of the second-born BH is determined by the net effect of the stellar wind and the tidal interaction of the BH–He-star binary system. Because of the efficiency of the angular momentum transport, as explained above, the He-star emerges from the second mass-transfer event with a negligible spin. If the orbital separation is small enough, and stellar winds do not widen the system significantly, the He-star, however, can be spun up by tides. These conditions are typically met at low metallicities for BBHs formed through the common-envelope formation channel. In contrast, in the case of stable mass-transfer, the orbits shrink less efficiently leading to less tidally spun-up second-born BHs (e.g. \cite{Kushnir:2016zee,Zaldarriaga:2017qkw,Hotokezaka:2017esv,Qin:2018vaa,Belczynski:2017gds,Fuller:2019sxi,Bavera:2020inc}).


Accretion of material onto a BH can alter its spin. While it is relatively easy to spin up a non-degenerate star (only $\sim 1\%$ of the star's initial mass needs to be accreted in order to significantly alter its spin \cite{deMink:2012xx}), a BH mass must be roughly doubled due to accretion to produce a rapidly spinning object. A more accurate calculation (e.g., \cite{1974ApJ...191..507T,King:1999aq,2002ApJ...565.1107P,Fragos:2014cva}) shows that 20\% of the BH's mass must be added to bring the spin of a non-spinning BH to 0.5, and the initial BH mass must be more than doubled to bring the spin to 0.99.  The mass doubling timescale for a BH accreting at the Eddington limit is of order 100 million years -- far longer than the $\lesssim 10^5$ year lifetime of high-mass X-ray binaries composed of a BH and its mass-shedding companion.

Stable Roche-lobe overflow accretion onto a BH can result in  mass-transfer rates that can exceed the Eddington rate by many orders of magnitude. In the context of binary evolution and binary population synthesis modeling, a common assumption is to limit the accretion onto the BH to the Eddington, or a few times the Eddington, rate. However, hypercritical accretion has been claimed as a way to allow for a significant amount of matter to be accreted at super-Eddington rates \cite{MorenoMendez:2008jkj}, and radiation GR-MHD simulations of super-critical accretion disk (e.g., \cite{Sadowski:2015hia}) partially support this argument. Binary population synthesis models that consider super-Eddington accretion show that BHs can indeed increase significantly their masses during a stable mass-transfer episode (e.g., \cite{Briel:2022cfl}). 

The question of angular momentum gained during such a super-Eddington mass-transfer episode is a more delicate one. The common assumption that the accreted material carries the angular momentum of the ISCO may not be valid at highly super-Eddington rates. Radiation GR-MHD simulation of super-Eddington accretion disk around BHs show that, if significant structured magnetic fields are present in those disks, then the disks may enter in a magnetically arrested disk  state, where the specific angular momentum of the material accreted by the BH is much less that that of the ISCO \cite{Lowell:2023kyu}. Furthermore, magnetically arrested disks can lunch relativistic jets, which extract rotational energy from the BH \cite{Tchekhovskoy:2012up}. In that case accretion can lead to spin down of the BH instead of spin up. In the absence of significant structured magnetic fields, the assumption that the accreted material carries the specific angular momentum of the ISCO is closer to what numerical simulations predict \cite{Gammie:2003qi}.

It is plausible that core collapse itself is a process that spins up a BH. If during a supernova explosion the ejecta somehow were to carry away a significant amount of angular momentum, while the collapsing part of the star acquired the conjugate angular momentum, then a spinning BH could be formed. 
However, black holes more massive than $\gtrsim 15 \Msun$ are likely to form through complete fallback \cite{Fryer:2011cx}. In this context, it has being suggested that  black holes could spin up  if part of the ejecta were torqued by the binary companion before faling back, effectively converting part of the orbital angular momentum to BH spin. For this process to be efficient, however, a fine-tuning of the ejecta velocity distribution is required,  making it unlikely to lead to significant spin-up in most cases \cite{Batta:2017yag,Schroder:2018hxk}. Finally, the standing accretion shock instability during core collapse and gravity waves during the late evolutionary stages of massive stars have been suggested as potential mechanisms that can stochastically transfer angular momentum to the stellar core right before or during its collapse, potentially explaining the spin distribution of young NSs. However, it is not expected that either of these two mechanisms can significantly affect the spin of a BH \cite{2015ApJ...810..101F,MorenoMendez:2015sxk}.

While all the above processes were discussed in the context of the evolution of isolated binaries, they are also relevant for most other formation channels, e.g. dynamical formation in triple and higher multiplicity systems (e.g., \cite{Silsbee:2016djf,Antonini:2017ash}),  globular clusters (e.g., \cite{Sigurdsson:1993zrm,PortegiesZwart:1999nm}), nuclear clusters (e.g., \cite{Antonini:2016gqe}), young star cluster (e.g., \cite{DiCarlo:2020lfa}) or AGN disks (e.g., \cite{Tagawa:2019osr}). In such environments,  dynamics and stellar/binary evolution act together to form coalescing double compact objects. Dynamical encounters are expected to lead to a perfectly isotropic orientation of BH spins in the densest environments, where each binary system experiences multiple close gravitational interactions and exchanges \citep{Rodriguez:2016vmx}. Only in open clusters and low-mass young clusters, spin orientation can retain some memory of the initial alignment with the  orbital angular momentum of the binary system \citep{Trani:2021tan}.

In the densest stellar systems, globular and nuclear clusters, the merger product of a coalescing BBH can be retained. This second generation BH retains part of the orbital angular momentum of the progenitor BBH, such that even if the two progenitor BHs where non-rotating, the resulting BH would have a spin of $\sim 0.7$ \cite{Barausse:2009uz}. Although a kick is imparted onto the BH merger product, second generation BHs can sometimes be retained in the clusters and be involved in the formation of a new BBH. Merging BBH that contain at least one second generation BH, although account only for a small fraction of the total BBH population, can leave a distinct imprint in the BH spin distribution inferred from BBH mergers (e.g., \cite{Rodriguez:2019huv,Mapelli:2021syv}).

\subsubsection{BH spins -- empirical evidence from X-ray binaries and GWs} 

\begin{figure*}[t]
\centering
\includegraphics[width=.8\textwidth]{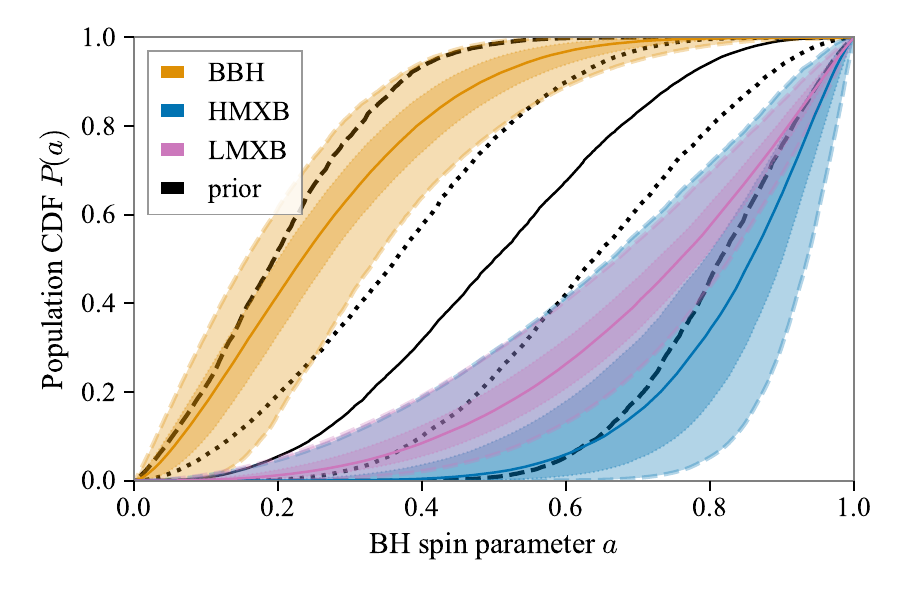}
\caption{Cumulative distributions of BH measured from BBHs (yellow), low-mass X-ray binaries (blue) and high-mass X-ray binaries (magenta). The distribution of prior CDFs is shown by the unfilled black bands~\cite{Fishbach:2021xqi}. The choice of priors is discussed in \cite{LIGOScientific:2020kqk}.
} 
\label{fig:BHspins}
\end{figure*}

The first efforts to constrain the spins of stellar-mass BHs were done using electromagnetic observations of BH X-ray binaries, and specifically by studying either the spectrum of the emitted X-ray radiation or the X-ray timing variability of the systems. There are three widely used  methods for estimating the spins of stellar-mass BHs observed in X-ray binaries \cite{Remillard:2006fc}: \textit{(i)} fitting the thermal continuum spectrum of the accretion disk, \textit{(ii)} modeling the reflection spectrum of the disk and specifically the distortion of the Fe K line due to GR effects, and \textit{(iii)} modeling high-frequency ($\sim{100-450}$ Hz) quasi-periodic oscillations (HFQPOs). While for the former two methods there are well-understood physical models underpinning them, there is no established model for the origin of HFQPOs. Over the last decade,  the results from  disk continuum fitting and the reflection line fitting methods have broadly converged.  
The picture that emerges is that BHs in low-mass X-ray binaries have spins that cover the entire range, from 0 to 1, while BHs in high-mass X-ray binaries tend to be fast, almost maximally, spinning (figure~\ref{fig:BHspins}). 

Gravitational-wave observations have provided insights into the spins of black holes. 
Spins leave an imprint of the gravitational waveform, enabling measurement of spin magnitudes (between $0$ and $1$) and orientations (typically defined relative to the orbital angular momentum)~\cite{LIGOScientific:2016vlm,LIGOScientific:2016dsl}. 
These imprints are more subtle that the effects of the masses, making them more difficult to precisely measure~\cite{Poisson:1995ef,Purrer:2015nkh,Vitale:2016avz,Pratten:2020igi}. 
The low-frequency sensitivity of ET may be especially important for measuring spin precession, that takes place when the spins are misaligned with the orbital angular momentum~\cite{Apostolatos:1994mx}.

In contrast to X-ray observations, the majority of the detected gravitational-wave sources are consistent with having zero spin~\cite{KAGRA:2021vkt}. 
The more massive component of the source GW190814 has the most precise spin-magnitude measurement, $\chi \leq 0.08$~\cite{LIGOScientific:2020zkf,LIGOScientific:2021usb}.
Several sources, such as for GW151216~\cite{LIGOScientific:2016sjg,LIGOScientific:2021usb}, have been observed to have non-zero spin~\cite{KAGRA:2021vkt}. 
Considering the observed population, half of spin magnitudes are below $\chi = 0.25$~\cite{KAGRA:2021duu}. The statistical analysis by ref.~\cite{Fishbach:2021xqi} shows that the spins of BBHs from gravitational waves and those of BHs in X-ray binaries cannot be drawn from the same underlying distribution (figure~\ref{fig:BHspins}).

Spins are often discussed using effective spin parameters: the effective spin parameter $\chi_{\rm eff} = (m_1\chi_1 + m_2\chi_2)/(m_1 + m_2)$ is  a mass-weighted combination of the spins aligned with the orbital angular momentum~\cite{Ajith:2009bn,Santamaria:2010yb}, and the effective precession spin $\chi_\mathrm{p}$ characterises a mass-weighted in-plane that contributes to spin precession~\cite{Schmidt:2014iyl}.
To date, most detected sources are consistent with $\chi_\mathrm{eff} = 0$~\cite{KAGRA:2021vkt}, which could indicate either spin magnitudes are close to zero, spins are anti-aligned with each other, or spins are primarily in the orbital plane. 
The systems with  non-zero $\chi_\mathrm{eff}$ currently all have $\chi_\mathrm{eff} > 0$ indicating a preference for spins aligned, rather than antialigned, with the orbital angular momentum~\cite{KAGRA:2021vkt,KAGRA:2021duu}. 
GW190814's source has the best measured value of $\chi_\mathrm{p}$, with $\chi_\mathrm{p} \leq 0.07$ consistent with having a low-spin primary component. 
Identifying an individual source showing spin precession has been more difficult: GW200129\_065458 is potentially the best candidate~\cite{KAGRA:2021vkt,Hannam:2021pit,Islam:2023zzj}, but this result is sensitive to the noise subtraction procedure used~\cite{Payne:2022spz,Macas:2023wiw} and it has been suggested that the signs of precession could be a misdiagnosis of the influence of orbital eccentricity on the signal~\cite{Gupte:2024jfe}. 
Considering the entire population, however, there is evidence that some sources have non-zero $\chi_\mathrm{p}$~\cite{LIGOScientific:2020kqk,KAGRA:2021duu}. 
Therefore, black hole spins can be misaligned with their orbital angular momentum.

\subsubsection{NS spin periods -- theoretical expectations and limitations}

\begin{figure*}[t]
\centering
\includegraphics[width=.8\textwidth]{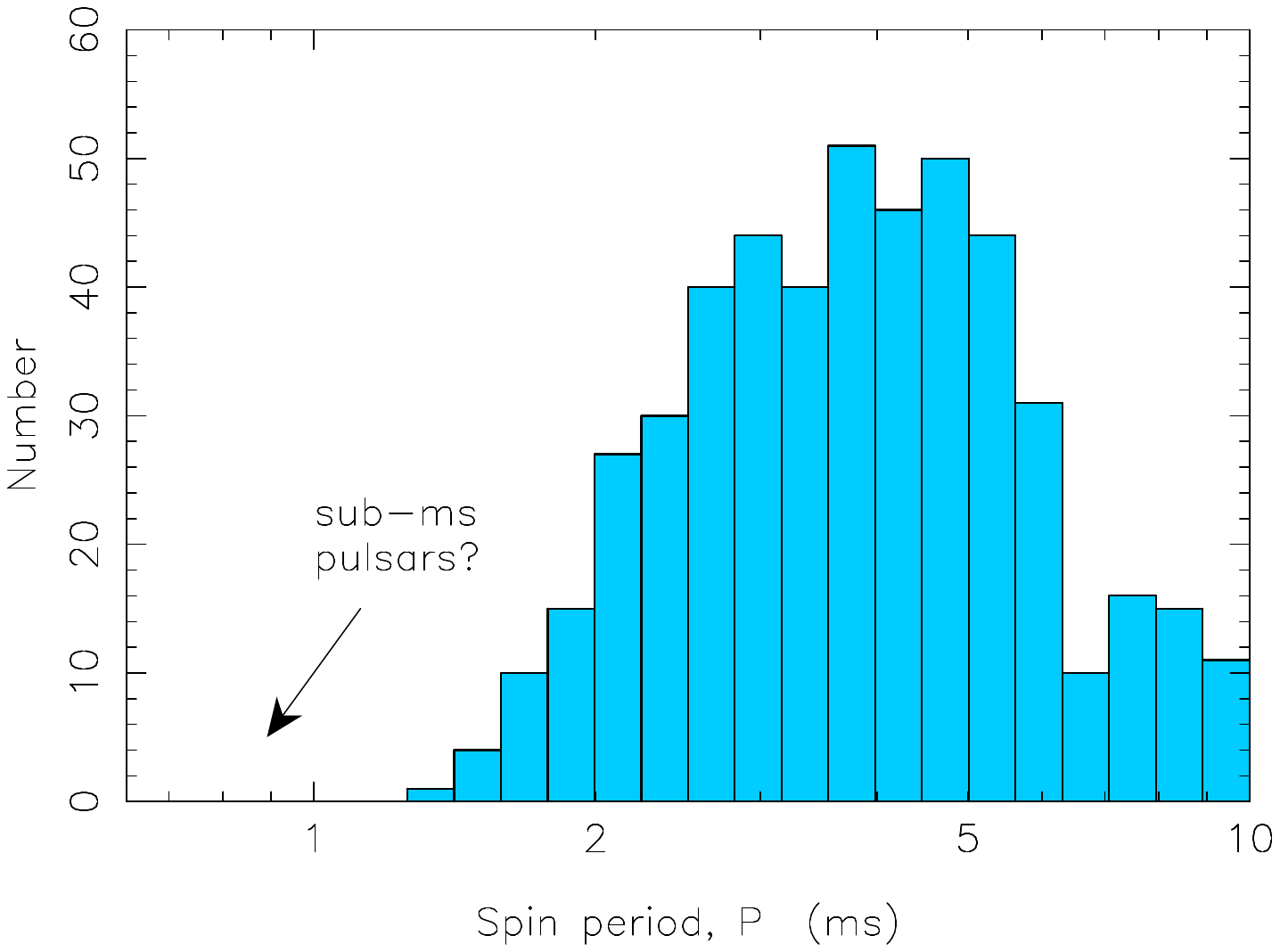}
\caption{Distribution of spin periods of 485 radio MSPs with $P<10\;{\rm ms}$. Despite intense searches, no sub-ms MSP has been discovered thus far. ET may reveal such a sub-ms MSP in a GW merger -- see text. After~\cite{Tauris:2023sxf}.
} 
\label{fig:MSPspins}
\end{figure*}

NS spin periods measured so far span over 7~orders of magnitude \cite{Tauris:2023sxf}: from $\sim$1.4~ms for the radio millisecond pulsar  J1748$-$2446ad \cite{Hessels:2006ze} to more than 5~hr for the symbiotic X-ray binary 4U~1954+31 \cite{Corbet:1998hr}.  
The possible existence of sub-ms pulsars ($P<1\;{\rm ms}$) is a long standing and intriguing question. 
The distribution of MSP spins is shown in figure~\ref{fig:MSPspins}.
The discovery of a sub-ms MSP is very important for constraining the equation-of-state of NSs \cite{Haensel:1989mvc}. Given the physical mass shedding frequency limit of about 1500~Hz ($P~\simeq 0.67\;{\rm ms}$), it is perhaps puzzling why no sub-ms pulsars have been found so far in light of the intense efforts to detect these objects in both radio and X-rays. Nowadays, there seems to be no instrumental or computational selection effects that would prohibit such a discovery. 
From a theoretical point of view, it is an open question whether the lack of detected sub-ms pulsars might be due to magnetospheric conditions \cite{Tauris:2023sxf} or GW radiation during the accretion phase that halts the spin period above a certain level \cite{Bildsten:1998ey,Chakrabarty:2003kt,Patruno:2017oum,Haskell:2018nlh}.
A GW discovery of an accreting sub-ms pulsar would thus be a major discovery. The emitted GW signal would have a frequency above 2000~Hz, and although this is not optimal for ET where the sensitivity peaks at 200--300~Hz, any GW strain above $10^{-24}\;{\rm Hz}^{-1/2}$ @ 2000~Hz is potentially detectable.

\subsubsection{Spins of double NS systems: evidence from Galactic sources and expectations for mergers}
Radio pulsar data on recycled and young pulsars in Galactic field NS+NS systems reveal that the spin values of the second-born (i.e. non-recycled) pulsars are equivalent to those of single young pulsars ($\sim 0.1-3\;{\rm s}$) and that the recycling process of the first-born NS is rather inefficient: the most rapidly spinning NS measured in a Galactic NS+NS system has a spin period of $P\simeq 21\;{\rm ms}$. This is expected \cite{Tauris:2023sxf} because NS+NS systems descend from high-mass X-ray binaries (HMXBs), which means that the relatively short lifetime of the massive (ultra-stripped) progenitor of the second-born NS prevented sufficient supply of material to spin up the first-born accreting NS \cite{Tauris:2015xra}.
However, the situation is different for NS+NS produced in dense stellar environments, like globular clusters \cite{Ye:2021lnk}, because binary exchange encounters may pair up an MSP (formed in a low-mass X-ray binary systems) with another NS --- a process which is evident from exactly such MSP+NS pairs detected in globular clusters \cite{Lynch:2011aa}. Thus we do expect ET to eventually detect mergers of NS+NS systems in which one NS component (or even both) may be an MSP, although their rate will likely only be a small fraction of the total detected NS+NS merger rate \cite{Ye:2019xvf,Rosswog:2023rqa}.

A natural question is what the spin rate will then be at the moment of the merger since NS+NS system may have long delay times of several Gyr, during which magnetodipole radiation and magnetospheric plasma currents give rise to loss of rotational energy. It turns out that many MSPs will lose only a small amount of rotational energy \cite{Tauris:2012jp} because their surface B-fields are very low ($10^7-10^8\;{\rm G}$) as evident from their $\dot{P}$ values down to $\sim 10^{-21}$. Thus we do expect ET to detect some NS+NS mergers with NS component spin periods of order ms, although the rate is likely low and the resulting post-merger GW signal from such NS+NS mergers with one high-spin component is slightly weaker than for slower spinning NSs \cite{Rosswog:2023rqa}.

\subsubsection{BH/NS spin-axis direction and core collapse}
 Whereas the spin magnitudes of BHs and NSs are expected to be determined by progenitor star evolution and binary interactions (mass accretion and tides) after their formation, the directions of BH and NS spins are of uttermost importance for probing core collapse and understanding SN explosion physics. Measurements of individual spin components will still remain challenging with ET. However, the distribution of measured effective spins ($\chi_{\rm eff}$, i.e. spins projected along the orbital angular momentum vector during in-spiral) can teach us not only about the formation channel of GW mergers \cite{Mandel:2009nx,Rodriguez:2016vmx,Kalogera:1999tq,Farr:2017uvj} but also about SN explosion physics. Evidence has recently been suggested for spin-axis tossing (flipping the rotation axis in a random direction) at birth based on empirical data from LVK observing runs O1--O3 \cite{Tauris:2022ggv} and theoretical investigations of this phenomenon (at least for NSs and low-mass BHs) has been put forward \cite{Janka:2021deg}. A large increase in GW merger data from ET will significantly boost the empirical statistics to firmly infer both progenitor formation channels and the question of tossing the spin axis at birth. 

\subsubsection{BH/NS spins in light of ET}
ET offers a significant increase in detector sensitivity compared to LVK. This improvement enhances the possibility of detecting continuous GW signals at high frequencies from Galactic spinning NS with sufficiently large ellipticities \cite{Andersson:2010ufc,Andersson:2021qdq,Hua:2023aff,Pagliaro:2023bvi}, see section~\ref{sec:nscw} for extended discussion. Detection of continuous GWs from a spinning NS would be a major breakthrough and open up a new path to study the structure and evolution of NSs. If GWs from a spinning NS are detected in a tight binary, from which LISA will be able to detect the orbital motion in low-frequency GWs, i.e. a dual-line GW source \cite{Tauris:2018kzq,Suvorov:2021mhr}, the prospects for investigating the interior of these NSs are even brighter.

\subsection{Primordial versus stellar-origin BHs}\label{sec:PBHsdiv3}

As we already discussed in section~\ref{sect:PBHdiv2},
primordial black holes (PBHs) are a putative population of BHs that could have formed in the early Universe, much before the formation of structures and stars. The simplest formation scenario assumes the collapse of extreme inhomogeneities during the radiation--dominated era~\cite{Zeldovich:1967lct,Hawking:1971ei,Chapline:1975ojl,Carr:1975qj} (although various alternative mechanisms were suggested, e.g. based on early matter dominated eras~\cite{Khlopov:2008qy,Harada:2016mhb}, cosmological phase transitions~\cite{Baker:2021sno,Flores:2024lng,Lewicki:2023ioy,Liu:2021svg,Gouttenoire:2023naa,Flores:2024lng}, the collapse of domain walls~\cite{Liu:2019lul,Ferreira:2024eru,Dunsky:2024zdo} and others \cite{Dvali:2021byy,Cotner:2019ykd,Flores:2020drq,Alonso-Monsalve:2023brx,Flores:2024eyy}).
While remaining one of the most studied candidates to explain the dark matter, even a subdominant population of PBHs could serve as the seed of supermassive black holes at high redshift~\cite{2010A&ARv..18..279V,Clesse:2015wea,Serpico:2020ehh,DeLuca:2022bjs}, explain small scale structure features \cite{Carr:2018rid,Boldrini:2019isx,Carr:2023tpt}, and lead to boosted signatures of other dark matter candidates (such as WIMPs) \cite{Lacki:2010zf,Adamek:2019gns,Bertone:2019vsk,Carr:2020mqm}.

The characteristic natal mass of these objects is linked by model-dependent factors to the mass within the Hubble sphere $m_H$ at formation. 
For example, if the formation takes place in a radiation dominated Universe, it was shown to follow a critical scaling as a function of the collapsing overdensity above the threshold. For typical perturbations, $m_{\rm PBH}/m_{\rm H}$ takes values of order unity, although different hierarchies can be attained. 
Overall, PBHs can generically be formed in a range of masses which spans several orders of magnitude~\cite{Ivanov:1994pa,GarciaBellido:1996qt,Ivanov:1997ia,Blinnikov:2016bxu}, from the ultra-light masses $m_{\rm PBH}\lesssim 10^{-18} \Msun$ that are unstable through Hawking evaporation, up to supermassive scales.  
There are many constraints on the  PBH abundance $f_{\rm PBH} \equiv \Omega_{\rm PBH}/ \Omega_{\rm DM}$ (see~\cite{Carr:2020gox} for a comprehensive review).
In this landscape, ground-based detectors will be able to provide strong bounds in the mass range that goes from planet size masses, 
$m_{\rm PBH} \sim {\cal O} (10^{-6}) \Msun$, up to intermediate masses $m_{\rm PBH} \sim {\cal O} (10^4) \Msun$, by searching for the GW signatures of PBH mergers,  exceeding in most of this mass range the existing bounds derived from microlensing surveys and various phenomena related to particle emission from accreting PBHs.
The complementary constraints that will be set by searching for GWs associated with the formation mechanisms are discussed in section~\ref{sect:PBHdiv2}.

While current LVK observations are still compatible with a fraction of the events to have primordial origin ~\cite{Franciolini:2021tla} (see also refs.~\cite{Franciolini:2022tfm,Escriva:2022bwe} for the most recent analyses, and \cite{Sasaki:2018dmp,LISACosmologyWorkingGroup:2023njw}  for reviews on PBH mergers as GW sources),
it is exciting that next-generation GW detectors,  such as ET, would greatly enhance our capabilities to search for GW signatures of PBHs \cite{Chen:2019irf,DeLuca:2021wjr,Pujolas:2021yaw}, leading to unprecedented tight bounds on their abundance, as low as $f_{\rm PBH}\lesssim 10^{-10}$ in some extreme scenarios \cite{DeLuca:2021hde}. 

PBHs form in the early Universe. Their distribution is random unless specific correlations are induced in the primordial perturbations \cite{Young:2015kda,Tada:2015noa,Ali-Haimoud:2018dau,Desjacques:2018wuu,MoradinezhadDizgah:2019wjf,Auclair:2024jwj,Animali:2024jiz},
 and they can pair to form binaries through different processes \cite{Raidal:2024bmm}:
 \begin{itemize}
     \item[{\it i)}]  via gravitational decoupling from the Hubble expansion, which typically happens before the matter--radiation equality~\cite{Nakamura:1997sm,Ioka:1998nz}. Such binaries are initially highly eccentric, with a semi-major axis distribution allowing for a large fraction to merge in the late-time Universe;
\item[{\it ii)}] through gravitational capture within present-day small scale structures and galaxies \cite{Bird:2016dcv,Clesse:2016vqa}; 
\item[{\it iii)}] through three-body interactions in Poisson-induced PBH small scale structures~\cite{Vaskonen:2019jpv,Kritos:2022ggc}.
It is also possible that GW bursts are produced via PBH hyperbolic encounters \cite{Garcia-Bellido:2017qal}, although with an expected subdominant rate within most of the ground-based observable band \cite{Garcia-Bellido:2021jlq}.
 \end{itemize}
Out of those channels, it was shown that the dominant contribution comes from binaries formed in the early Universe~\cite{Ali-Haimoud:2017rtz,Raidal:2017mfl,Vaskonen:2019jpv,DeLuca:2020jug}.\footnote{
This conclusion was also tested with dedicated N-body simulations for relatively narrow mass functions \cite{Raidal:2017mfl} (up to 2 orders of magnitude in width). Whether disruption of binaries is more or less efficient for broader mass distributions is still subject to debate (see also \cite{Carr:2023tpt}). }

It is important to stress that understanding clustering properties, and in particular clustering evolution at low redshift, is crucial in this context, as non-trivial PBH correlation beyond what is induced by Poisson initial conditions can enhance binary formation in both the early- and late-time Universe, while at the same time inducing stronger suppression factors due to disruption of binaries through more frequent binary-PBH interactions. 
More broadly, a precise description of the clustering evolution of PBHs after their formation is challenging, especially at low redshift. Numerical simulations of PBH cosmologies were able to describe only high redshift ($z\gtrsim 100$) \cite{Inman:2019wvr} or individual clusters of PBHs \cite{Raidal:2018bbj,Jedamzik:2020ypm,Trashorras:2020mwn}, while the impact of clustering on the merger rate was investigated with advanced analytical studies \cite{Vaskonen:2019jpv,DeLuca:2020jug}. The detection of PBH mergers would provide valuable insights and allow to narrow down theoretical uncertainties.
Furthermore, due to the range of scales involved with the formation of PBHs of masses that can be probed by  ET, synergies with other observables are possible. On the one hand, scenarios induced by large stellar-mass PBH clustering require non-trivial correlations at the scales probed by CMB distortions \cite{DeLuca:2021hcf} with future experiments (e.g. PIXIE and Voyage 50 \cite{Chluba:2019kpb}), while the enhanced density perturbations required to form PBHs would fall in the range probed by PTA experiments.

Searching for evidence on the existence of PBH binaries is very challenging due to large uncertainties on their properties, such as their mass distribution, which is inherently model-dependent and linked to the early Universe initial conditions. An additional challenge is posed by the expected astrophysical BBHs that act as foreground, and trying to distinguish the two contributions is mostly limited by existing theoretical uncertainties and degeneracies between the astrophysical and primordial channels. 
However, it is important to stress here that some distinctive features exist, and in the following, we will present what are considered the smoking-gun signatures of PBH binaries.  

\begin{figure}[t]
\centering
\includegraphics[width=.8\textwidth]{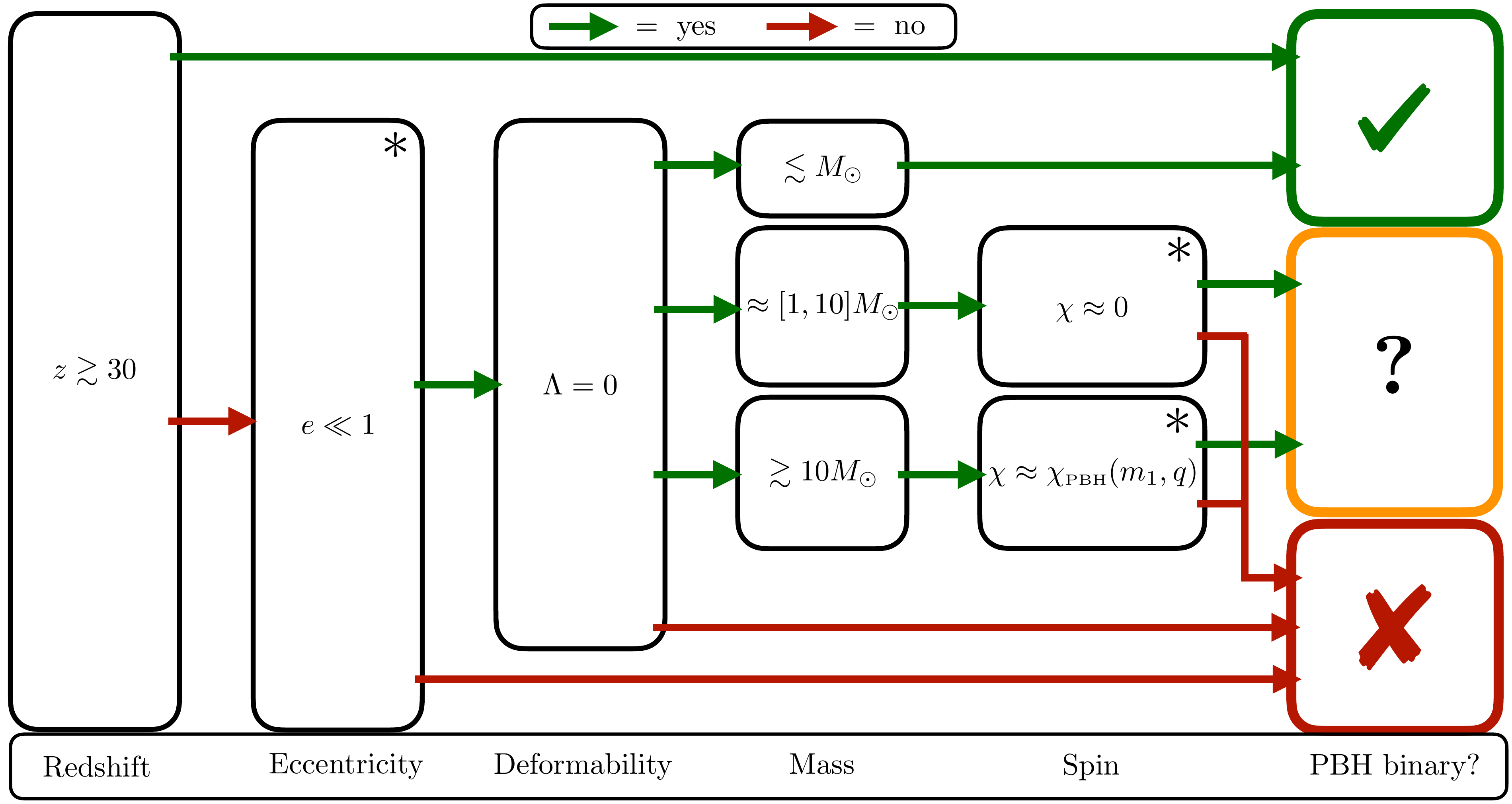}
\caption{ 
A flowchart to test the primordial nature of a binary. Green and red refer to the condition in the box being met or violated, respectively. Adapted from ref.~\cite{Franciolini:2021xbq}.
The asterisks indicate that the condition is evaluated in the scenario where PBHs are formed in a radiation-dominated Universe, and a model-dependent alternative is possible.} 
\label{fig:flowchart}
\end{figure}

In figure~\ref{fig:flowchart} we summarize the main properties of PBH binaries formed in the radiation-dominated early Universe.
As we will see, detecting high redshift mergers and/or subsolar binaries would allow for a confident discovery of a PBH population. In most scenarios, one also expects specific mass-spin correlations that could be searched for in the GW data. In this endeavor, ET will be a game-changer compared to current GW detectors~\cite{Franciolini:2021xbq}. 
These smoking guns could be exploited on an event-by-event basis or through population studies. We will discuss both approaches below. In section~\ref{subsec:pbh} we will discuss the potential contribution from PBH mergers to the stochastic GW background. 

Current LVK observations set the most stringent bound on the merger rate of primordial binaries. However, next-generation detectors would boost our capabilities to constrain this scenario. In table~\ref{tab:maxnumberPBH} we report the intrinsic and detectable number of mergers in the interesting mass and redshift range (assuming the PBH abundance saturates current PBH upper bounds, see ref.~\cite{Franciolini:2021xbq} for more details). 

{\footnotesize
\renewcommand{\arraystretch}{1.2}
\setlength{\tabcolsep}{10pt}
\begin{table}[t!]
\begin{tabularx}{1.\columnwidth}{||X||c|c|c||c|c||}
\hline\hline
        $N^\text{\tiny tot}$ &
        $N^\text{\tiny SS}$ &
        $N^{z>10}$ &
        $N^{z>30}$ &
        $N^\text{\tiny LMG}$ &
        $N^\text{\tiny UMG}$ 
\\
\hline\hline
        1920000 &
        708487 &
        1400384 &
        795904 &
        300220 &
        7774 
\\
       \hline\hline
        13347 & 1650 & 336 & 17 & 2638 & 235 \\
        \hline\hline
\end{tabularx}
\caption{
The total number of PBH ($N^\text{\tiny tot}$),  subsolar events ($N^\text{\tiny SS}$), high redshift ($z>10$ and $z>30$) events, and ``mass gap" events (LMG and UMG for the lower and upper mass gap, respectively), for the intrinsic numbers of  
PBH mergers per year predicted by maximizing current LVK bounds~\cite{Franciolini:2023opt}  (second row) and for the 
events per year detected  with ET at ${\rm SNR}\geq12$ (third row).
}
\label{tab:maxnumberPBH}
\end{table}
}

\subsubsection{Constraints on high-redshift merger rate evolution}\label{sec:PBH_rate evolution}

The unique feature of the PBH merger rate density is that it is monotonically increasing with redshift as
$
{\cal R}_\text{\tiny PBH} (z) \propto t^{{-34}/{37}} (z) 
$~\citep{Ali-Haimoud:2017rtz,Raidal:2018bbj,DeLuca:2020qqa},
extending up to early epochs at $z\gtrsim{\cal O} (10^3)$. Interestingly, this property is dictated by how pairs of PBHs decouple from the Hubble flow
before the matter-radiation equality. Therefore, it is a robust feature of the PBH channel.\footnote{
In the case of initially strongly clustered PBHs, the merger rate evolution with cosmic time is only slightly modified to
${\cal R}_\text{\tiny PBH} \sim t^{-1}(z)$ \cite{Raidal:2017mfl,DeLuca:2021hde,Stasenko:2024pzd}.}

As high-redshift mergers are difficult to produce by astrophysical channels, one can set a conservative threshold above which no astrophysical mergers are expected to exist within standard cosmologies~\cite{Koushiappas:2017kqm}. One can adopt the conservative redshift threshold  $z\gtrsim30$ to be a smoking gun for primordial binaries~\cite{Nakamura:2016hna,Koushiappas:2017kqm,DeLuca:2021wjr,Ng:2021sqn,Ng:2022agi,Martinelli:2022elq,Ng:2022vbz}.

\begin{figure*}[t]
\centering
\includegraphics[width=.6\textwidth, trim=0 -1.1cm 0 0]{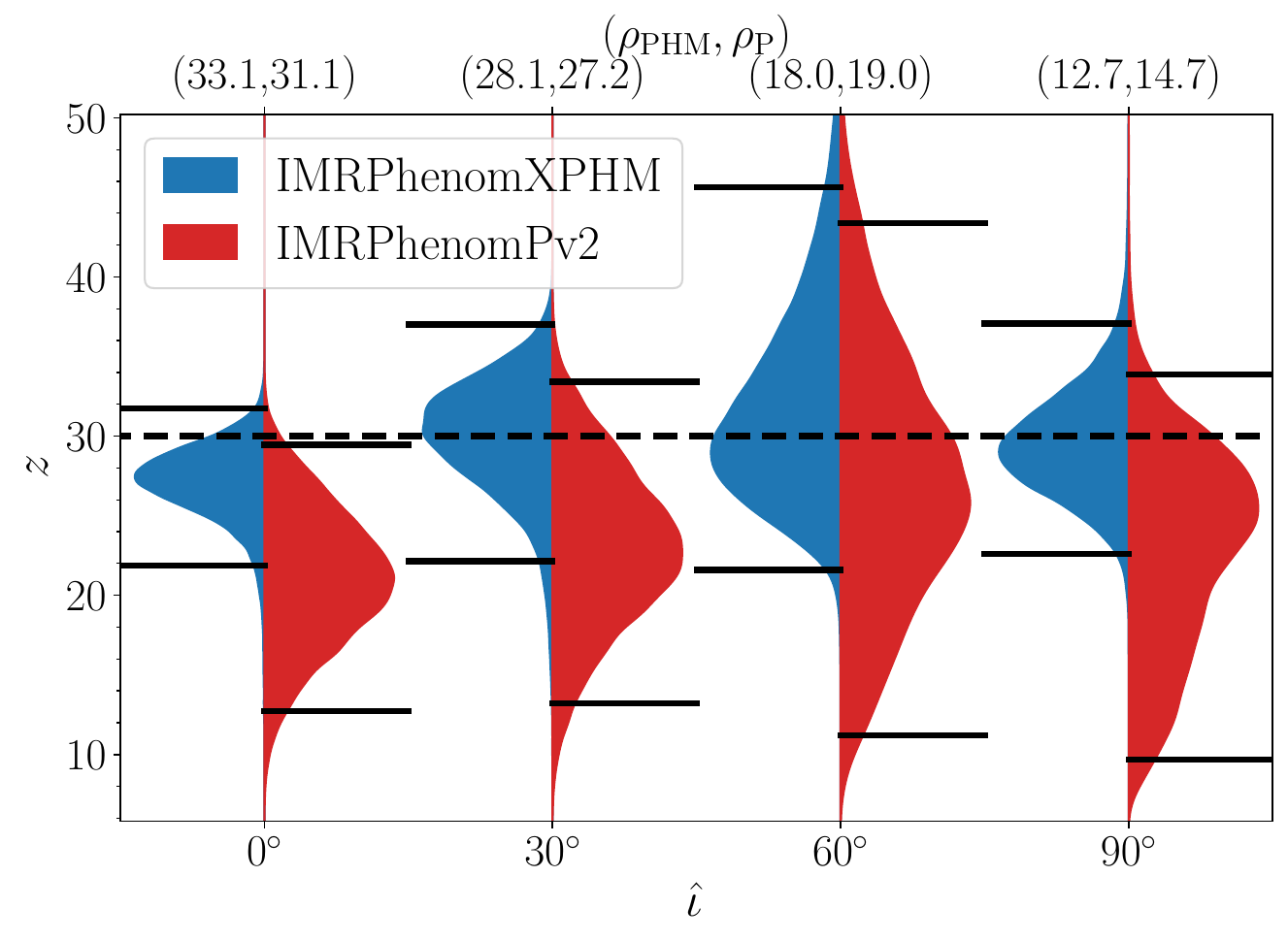}
\caption{ 
Redshift posteriors for sources with $(M_{\rm total}, z, q)=(40 \Msun,30,1)$ at $\hat{\iota}=0^{\circ}, 30^{\circ}, 60^{\circ}$ and $90^{\circ}$, obtained with a waveform with (blue) and without (red) higher-order modes and for a detector network composed of two CE and one ET (see ref.~\cite{Ng:2022vbz} for more details on the assumed detector network). The solid horizontal lines show the 95\% credible intervals, whereas the dashed lines mark the injected value $z=30$. The top axis shows the optimal SNR of the two waveforms. 
From ref.~\cite{Ng:2022vbz}.
}
\label{fig:PBHredshift}
\end{figure*}

The superior sensitivity and wider bandwidth of ET at low frequencies are crucial to detect putative mergers at $z\gtrsim30$, and up to hundreds of them per year are expected when  the PBH abundance saturates current bounds~\cite{DeLuca:2021wjr,Branchesi:2023mws}. However, an important aspect is the measurement uncertainty on the source redshift $z$ associated with a putative detection.
In simple terms, this is because an individual detection that is compatible with $z\approx 10$ within uncertainties could be confused with black holes originating from Pop~III stars, see section~\ref{sec:pop3}
~\cite{Ng:2021sqn,Franciolini:2021xbq,Ng:2022vbz,Mancarella:2023ehn,Fairhurst:2023beb}.
Figure~\ref{fig:PBHredshift} shows the posterior of the redshift in various cases~\cite{Ng:2021sqn,Ng:2022vbz} for an example network composed of ET+2CE. Note that analogous conclusions are reached with similar detector configurations. 
For a correct recovery of the true redshift, it is important to include higher-order modes in the waveform~\cite{Ng:2021sqn,Ng:2022vbz}, which alleviate the distance-inclination degeneracy and allow for a sufficiently accurate measurement of $z$. 
Still, uncertainties could remain as large as $\Delta z/z \sim 30\%$ for a source at $z=30$, depending on the source properties, and would quickly degrade at larger distances.

\begin{figure*}[t]
\centering
\includegraphics[width=.492\textwidth, trim=0 -1.cm 0 0]{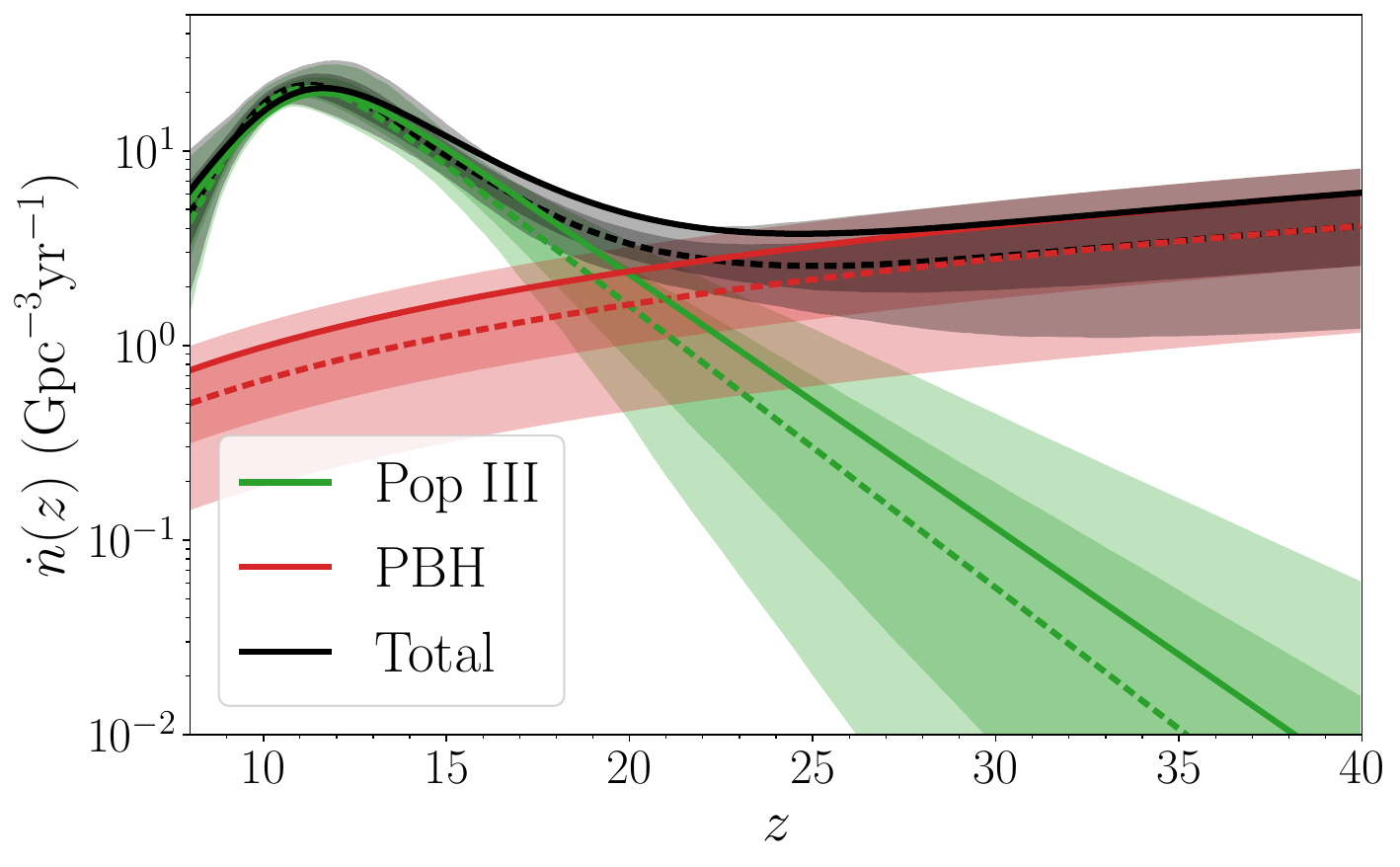}
\includegraphics[width=.488\textwidth]{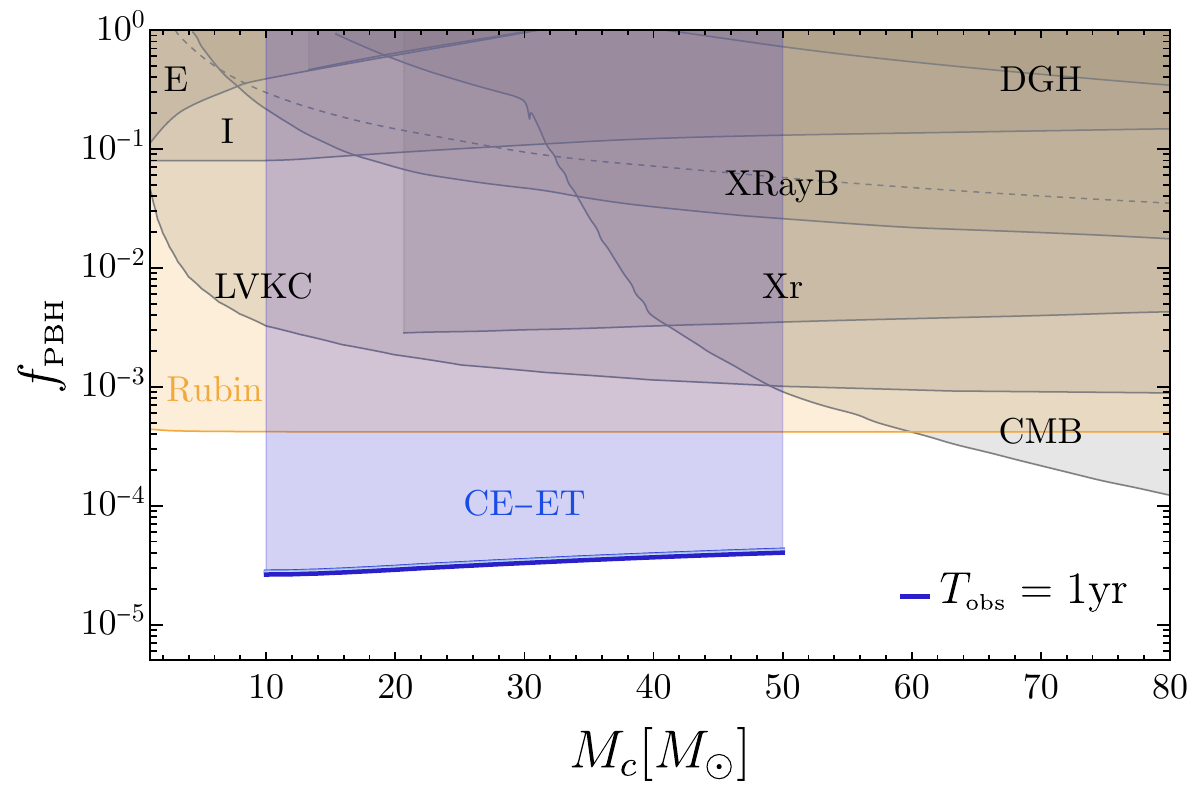}
\caption{ 
Left panel: Reconstructed merger rate at $z>10$ assuming ET+CE and two populations of events: astrophysical (Pop~III) and primordial. We consider the most conservative scenario in which both populations have the same mass function, we neglect PBH accretion, and we adopt an optimistic Pop~III merger rate~\cite{Belczynski:2016ieo}.
Right panel: a null observation for a primordial population with ET+CE can be translated into an upper bound on the PBH abundance.
Taken from ref.~\cite{Ng:2022agi}.
} 
\label{fig:PBHrateevohighz}
\end{figure*}

Further information can be gathered by reconstructing the merger rate as a function of redshift for a population of events. Figure~\ref{fig:PBHrateevohighz} (left panel) shows the posterior predictive distributions at $z>10$ reconstructed from a hierarchical Bayesian inference of 4 months of synthetic data, assuming two populations of mergers: a representative astrophysical merger channel from Pop~III stars and a primordial one with abundance saturating current upper bounds and a mass function centered around $M_c\sim 30 \Msun$. 
In the right panel, we report the corresponding upper bound that will be set on $f_{\rm PBH}$
from a non-detection of PBHs in 1 year of data taking.

\subsubsection{Subsolar PBH binary searches}\label{sec:PBH_subsolar}

\begin{figure*}[t]
\centering
\includegraphics[width=.49\textwidth, trim=0 -0.2cm 0 0]{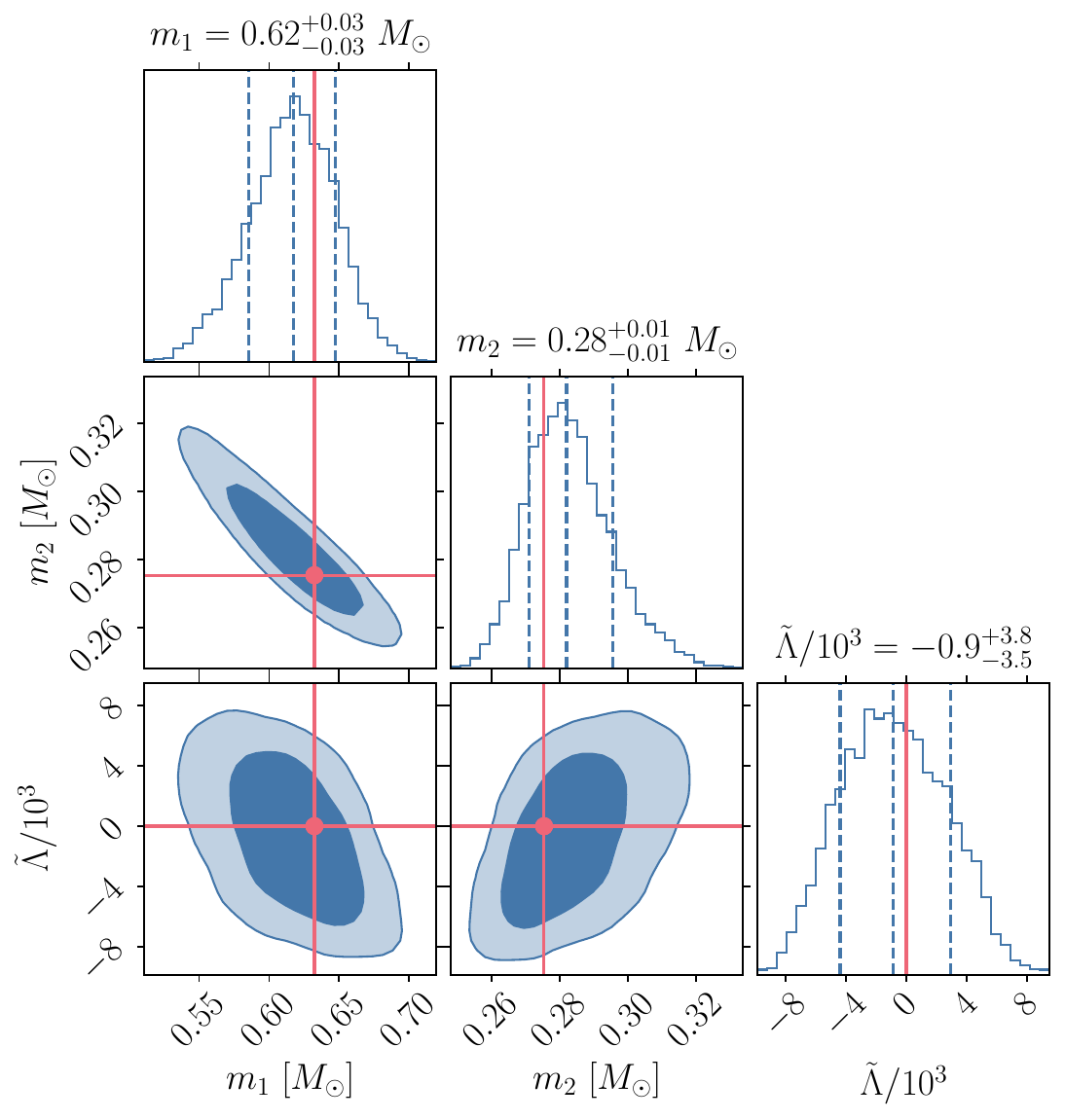}
\includegraphics[width=.49\textwidth]{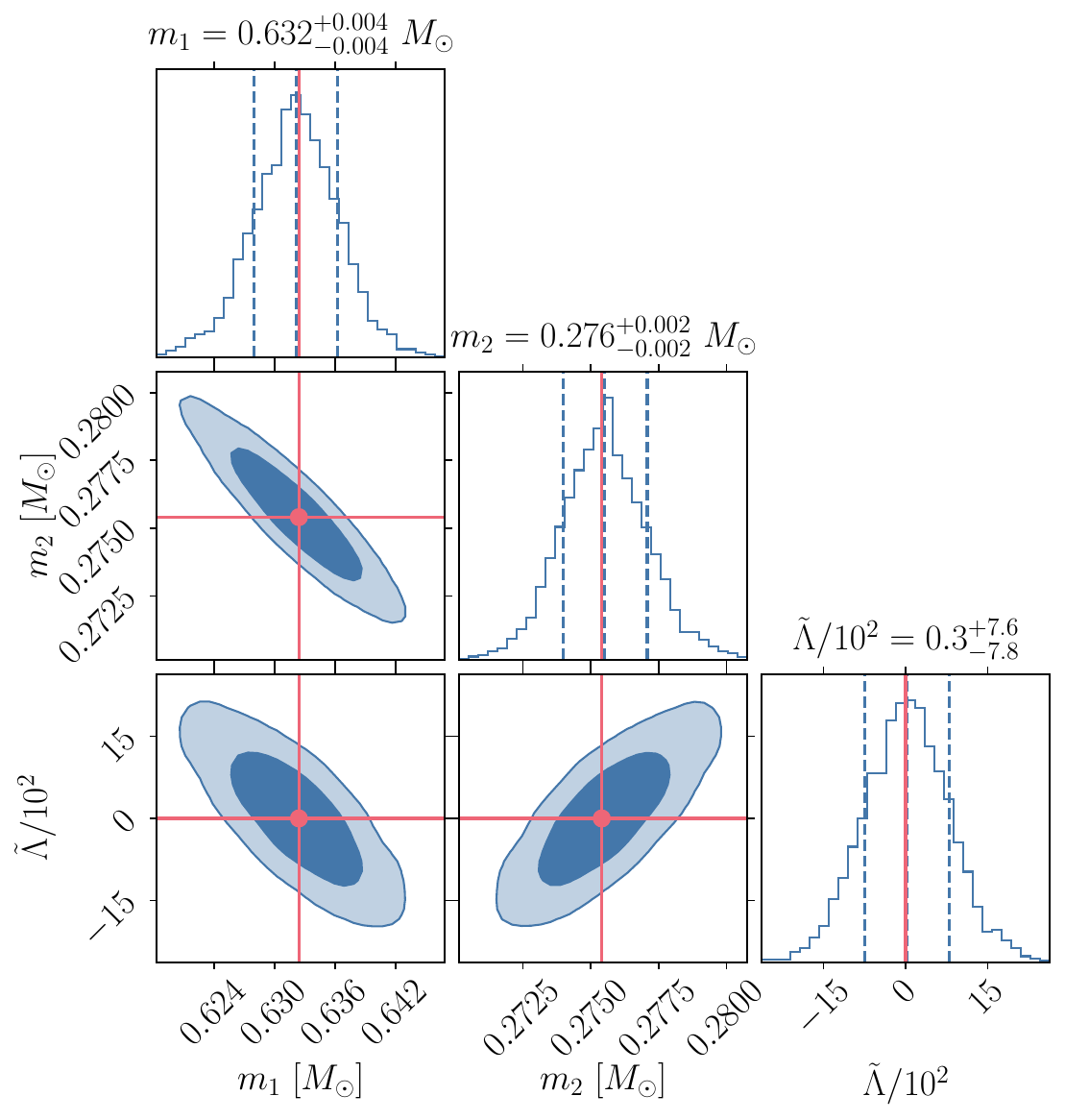}
\caption{ 
Posterior distribution of the masses $(m_1,m_2)$ and the tidal deformability $\tilde{\Lambda}$ parameter for LVK O5 (left) and ET+2CE (right)
(see ref.~\cite{Crescimbeni:2024cwh} for more details on the assumed detector network).  The red lines represent the injected values of a typical subsolar event. In ET+2CE case, the upper bounds on $\tilde{\Lambda}$ would exclude neutron stars or more exotic alternatives \cite{Cardoso:2019rvt}.
Taken from ref.~\cite{Crescimbeni:2024cwh}.
} 
\label{fig:PBHsubsolarCorner}
\end{figure*}

The detection of a subsolar object in a compact binary merger is regarded as one of the smoking gun signatures of a population of PBHs.\footnote{
Alternative new physics scenarios could also lead to the formation of subsolar BHs. For example, dissipative dark matter models allow for the generation of dark black holes, with similar properties compared to the formation from Population-III stars with, however, potentially different masses \cite{Shandera:2018xkn,Singh:2020wiq}.
Additionally, nearly solar-mass BHs can form out of NS transmutation in certain scenarios~\cite{Bramante:2017ulk,Takhistov:2017bpt,Takhistov:2020vxs,Dasgupta:2020mqg,Giffin:2021kgb,Abramowicz:2022mwb,Bhattacharya:2023stq,Bhattacharya:2024pmp}.
If one observed both the NS and the transmuted BH populations, one could infer implosion timescale, and eventually the properties of the dark matter particle that induced the transmutation, such as mass and interaction cross section \cite{Singh:2022wvw}.} 
Current bounds from LVK observations are already forcing the PBH abundance to be smaller than unity in the subsolar mass range \cite{Nitz:2021vqh,LIGOScientific:2022hai,Nitz:2022ltl}. Also, microlensing constraints are setting comparable bounds in the nearly-stellar mass range \cite{EROS-2:2006ryy,Gorton:2022fyb,Petac:2022rio,DeLuca:2022uvz,Mroz:2024mse}.
Even complying with current bounds, however, the number of potential subsolar mergers in 1 yr of observation of ET could be large, as indicated in table~\ref{tab:maxnumberPBH}, given the much higher sensitivity of the future detector. 
In case one candidate detection is found, it would be crucial to:
\begin{itemize}
   \item[{\it i)}] check that the supersolar regime is excluded within uncertainties; 
   \item[{\it ii)}] exclude possible confusion with other subsolar objects, such as light NSs or more exotic compact objects.
\end{itemize}
PBH binaries could be distinguished from stellar binaries based on the measurability of tidal effects (i.e. tidal Love number as well as tidal disruption of the compact objects)~\cite{Crescimbeni:2024cwh,Golomb:2024mmt}. At variance with PBHs, the GW signal from stellar binaries is affected by tidal effects (see also sections~\ref{section:div1} and \ref{section:div6}), which dramatically grow for moderately compact stars, as those expected in the subsolar range.
The prospects of measuring tidal effects in subsolar binaries are good already with the LVK network, but only the next generation of GW detectors such as ET will allow to claim a putative subsolar detection at ($>5\sigma$) level~\cite{Crescimbeni:2024cwh}.
This is shown in figure~\ref{fig:PBHsubsolarCorner}, summarizing the results of a model-agnostic Bayesian analysis~\cite{Crescimbeni:2024cwh} of a simulated GW event from a PBH binary with masses $m_1=0.62\,\Msun$ and $m_2=0.27\,\Msun$ at $z=0.02$, and in table~\ref{tab:SSM}. This example is motivated by the low significance trigger found by LVK \cite{LIGOScientific:2022hai}, see also \cite{Prunier:2023cyv}.
{\footnotesize
\renewcommand{\arraystretch}{1.2}
\setlength{\tabcolsep}{4pt}
\begin{table}[t!]
\begin{tabularx}{\columnwidth}{|X|c|c|c|c|}
\hline
\hline
Network  & LVK O3 & LVK O4 & LVK O5 & ET+2CE \\
\hline
\hline
\multicolumn{5} {|c|} {BNS
($\tilde \Lambda  = 1.5 \cdot 10^5,  
\delta \tilde \Lambda =  4.9 \cdot 10^4, 
\tilde \lambda_f = 0.075 $)} 
\\
\hline
SNR & 7.90 & 12.8 & 22.4 & 398 \\
\hline
$\Delta m_1/ m_1$   &
0.47 & 0.22 & 0.082&  0.0017\\
\hline
$\Delta m_2/ m_2$   & 
0.39  & 0.19  & 0.070& 0.0015 \\
\hline
$\Delta \tilde \Lambda / \tilde \Lambda$    & 
0.86  & 0.66  & 0.55 & 0.047 \\
\hline
$\Delta \tilde \lambda_f / \tilde \lambda_f $    & 
 0.38 & 0.24  & 0.13  & 0.015  \\
\hline
\hline
\multicolumn{5} {|c|} {BPBH 
($\tilde \Lambda = \delta \tilde \Lambda = 0, 
\tilde \lambda_f = 1$)} 
\\
\hline
SNR & 8.38 & 13.4 & 23.9 & 403 \\
\hline
$\Delta m_1/ m_1$   &
 0.20 &  0.13 & 0.044 & $6.6\cdot 10^{-3}$   \\
 \hline
$\Delta m_2/ m_2$   &  
0.17 & 0.11 & 0.037 & $5.6\cdot 10^{-3}$  \\
\hline
$ \Delta \tilde \Lambda $    & 
 $9.1 \cdot 10^3$ & $5.8 \cdot 10^3$  & $3.0 \cdot 10^3$ & $7.5 \cdot 10^2$  \\
\hline
\hline
\end{tabularx}
\caption{Fisher parameter estimation uncertainties with current and future GW experiments. We inject a system with $m_1=0.62\Msun$ and $m_2=0.27 \Msun$, assuming the binary was either a BNS (top rows) or a BPBH (bottom rows).  Adapted from ref.~\cite{Crescimbeni:2024cwh}.
}
\label{tab:SSM}
\end{table}
}

In particular, table~\ref{tab:SSM} shows two types of analyses, assuming the event was either a BNS (including tidal deformability in the waveform and tidal disruption frequency $f_{\rm max}$ beyond which the signal shuts off, parametrized by the effective tidal deformability $\tilde\Lambda$ and $\tilde \lambda_f \equiv f_{\rm max}/f_{\rm ISCO}$, respectively, where $f_{\rm ISCO}$ is the frequency corresponding to the innermost-stable circular orbit) or a BBH (assuming an injection with zero tidal effects but including $\tilde\Lambda$ in the recovery).
While the errors on the masses are sufficiently small to be confidently ascribed to the subsolar range already for LVK O4, in the BNS case the errors on the tidal deformability are relatively large, so that $\tilde\Lambda=0$ cannot be excluded with a high confidence level. Only next-generation detectors such as ET will allow to exclude $\tilde\Lambda=0$ at least at $3\sigma$ level (at $>5\sigma$ level, in fact). Nonetheless, the cut-off frequency for tidal disruption (parametrized by $\tilde\lambda_f$) would be constrained away from unity at $4\sigma$ level already by O4, which would still support the interpretation of the observed signal as coming from a BNS.
Likewise, if the subsolar event is a binary PBH, model-agnostic measurements of $\tilde\Lambda$ with ET+2CE (or similarly with ET alone) will exclude the NS hypothesis and also certain classes of exotic compact objects, such as compact boson stars~\cite{Crescimbeni:2024cwh}.

Even lighter binaries could be searched for using ad-hoc techniques \cite{Miller:2024fpo}. As light PBH binaries would only appear in ET frequency band during the inspiral phase, the signal evolution is very slow, and appears as a nearly monochromatic signal. While the current LVK sensitivity is still not sufficient to derive constraints which are competitive with microlensing upper bounds on $f_{\rm PBH}$ \cite{Mroz:2024mse} for $m_{\rm PBH} \lesssim 10^{-2} \Msun$ \cite{Miller:2024fpo}, ET will drastically improve the sensitivity leading to much tighter bounds \cite{Miller:2020kmv,Miller:2024rca}, tightening the ones currently derived from microlensing surveys.

\subsubsection{Spins of primordial black holes}\label{sec:PBH_masses}

A broad class of models predicts the Kerr dimensionless spin parameter 
at PBH formation to be very small, of the order $a\lesssim 10^{-2}$ \cite{DeLuca:2019buf}. This includes the most studied formation mechanism where the collapse of extreme inhomogeneities takes place during the radiation-dominated Universe.
However, alternative formation scenarios \cite{Flores:2024eyy}, such as collapse during an early matter-dominated phase,
or through the subhorizon assembly of massive objects, can lead to larger natal spins \cite{Flores:2021tmc}.
This model dependence is inevitably linked to the primordial scenario due to the currently poor knowledge of the initial phases of cosmological evolution. 
After formation, PBHs may spin up due to the accretion of baryonic matter throughout the cosmological evolution, in a process that is expected to quench at the onset of the reionization era, due to both enhanced PBH peculiar velocities within structures as well as gas heating up \cite{DeLuca:2020bjf,Hasinger:2020ptw}.
The dominant contribution to the PBH merger rate comes from binaries formed in the early Universe, through decoupling from the Hubble expansion, and thus no spin alignment is expected at binary formation. 
Spin induced by accretion, which is only expected to be efficient for masses larger than ${\cal O}(10)\Msun$, leads to a correlation between large total binary masses and large spins which can be used to distinguish this scenario from others.

\subsubsection{Tests based on mass and spin distributions}\label{sec:PBH_tests}

PBHs can be formed in a wide range of masses and, at odds with the astrophysical scenarios, there are no forbidden mass gaps where they cannot form. In particular, as the formation mechanisms rely on the collapse of large density perturbations in the early Universe, they are not affected, for example, by the physics dictating pair-instability supernova \cite{Fowler:1964zz,Barkat:1967zz,Woosley:2021xba}.
While large uncertainties on the exact location of the astrophysical mass gaps remain (see section~\ref{sec:ns-bh-mass-gap} and \ref{sec:upper-bh-mass-gap}), the presence of events in such regions would potentially require different astrophysical mechanisms (e.g. astrophysical hierarchical mergers) or point towards the primordial nature of the binary. 

\begin{figure}[t]
\centering
\includegraphics[width=0.49\textwidth]{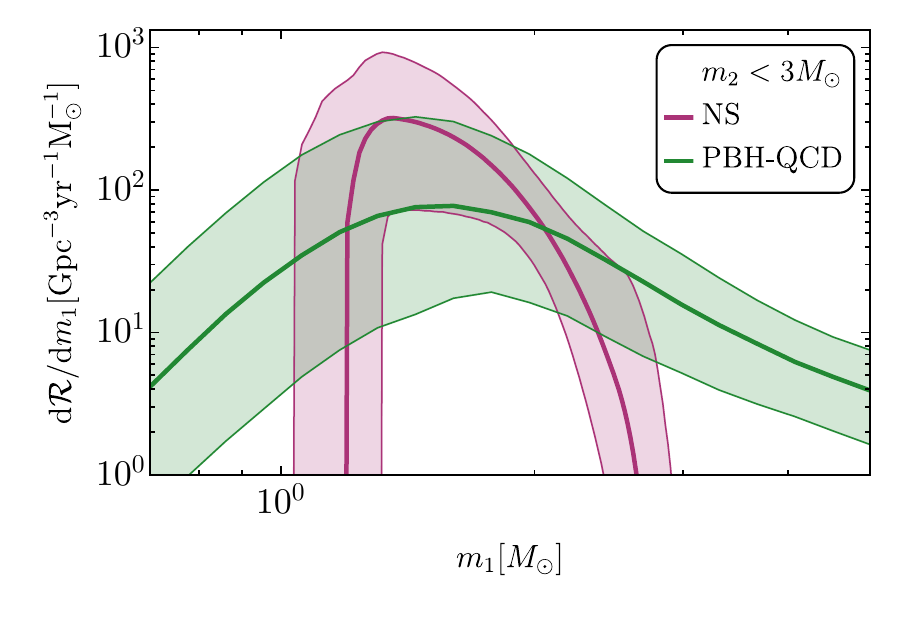}
\includegraphics[width=0.49\textwidth]{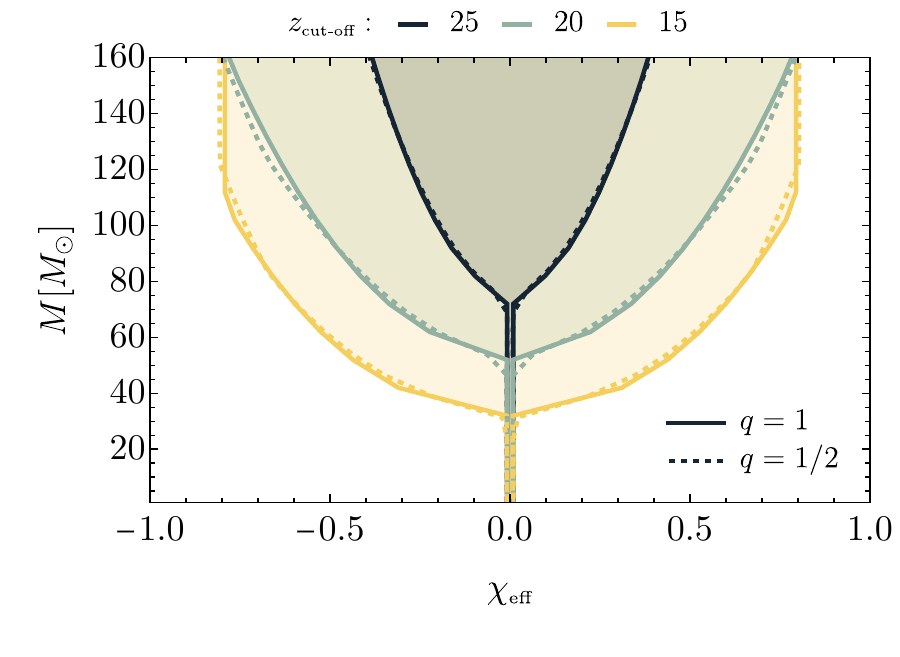}
\caption{
Left panel: posterior predictive distribution for events within the NS range based on the LVK GWTC-3 catalog \cite{Franciolini:2022tfm}.
Right panel:
Distribution of $\chi_\text{\rm eff}$ expected for PBH binaries as a function of the total mass $M$ depending on the accretion efficiency (see \cite{DeLuca:2020bjf,DeLuca:2023bcr} for more details). We stress that this prediction applies to PBHs born in a radiation-dominated Universe or to alternative formation scenarios for which the natal spin is negligible, and therefore represents a weaker test of the primordial nature of the binary. 
}\label{fig: PBH mass_spin}
\end{figure}

Let us focus first on the so-called lower mass gap, the putative gap in the distribution of masses separating astrophysical NS and BH binaries in the range $\approx [3,5] \Msun$. If PBH binaries were contaminating the NS range, the mass distribution would also contaminate  the lower mass gap, due to the difficulty of generating a sharply peaked PBH mass distribution in the early Universe. 
In figure~\ref{fig: PBH mass_spin} (left panel) we show an example of this, following ref.~\cite{Franciolini:2022tfm}.  
The green band indicates the mass distribution assuming the primordial scenario to be responsible for the light events in the GWTC-3 catalog \cite{KAGRA:2021vkt}, to be compared with the NS predictions in purple. It shows the larger width necessarily induced by the QCD effects and critical collapse \cite{Byrnes:2018clq,Clesse:2020ghq,Escriva:2022bwe,Musco:2023dak}, 
which would then correspondingly predict events both in the adjacent subsolar and lower mass gap range. 
Similar considerations apply to the upper mass gap, in the range approximately between $50-130\,M_{\odot}$ where, as discussed in section~\ref{sec:upper-bh-mass-gap}, a dearth of BHs is theoretically expected due to the pair-instability supernova mechanism \cite{Fowler:1964zz,Barkat:1967zz,Woosley:2021xba}. As contamination of the gap by astrophysical binaries is possible (see section~\ref{div3:formationchannel} for more details), it is important to devise robust tests for the primordial nature of the binary \cite{DeLuca:2020sae,Clesse:2020ghq}.

One possibility to distinguish between these two populations is based on searching for features in the mass distribution, or searching for correlations between masses and spins. 
In a general sense, remnants of second-generation BHs, which are the product of a prior merger, are identified by a final spin of around $\chi_f\simeq0.68$~\cite{Hofmann:2016yih}, assuming that the spin of first-generation BHs is negligible or isotropic distribution of BH spin directions \cite{Berti:2008af}. Additionally, it is expected that the mass ratio will be less than unity. Conversely, the relationship between masses and spins in the PBH scenario exhibits a different pattern. By concentrating on the formation in a Universe dominated by radiation (which is generally applicable even in extended scenarios \cite{Yoo:2024lhp}), the main predictions of the primordial scenario for the distribution of the effective spin parameter $\chi_{\rm eff}$ are as follows \cite{Franciolini:2021xbq} (see also figure~\ref{fig: PBH mass_spin}): 
\begin{itemize}
\item[{\it i)}] Binaries possess negligible $\chi_{\rm eff}$ at low masses within the observable range of ground-based detectors; 
\item[{\it ii)}] At larger masses, one expects a correlation between large binary total masses and wide distributions of $\chi_{\rm eff}$ induced by accretion effects;
\item[{\it iii)}] The distribution of $\chi_{\rm eff}$ is only dependent on the total mass of the binary, with negligible dependence on the mass ratio (see also figure~\ref{fig: PBH mass_spin}). 
Therefore, for fixed primary mass $m_1$, the $\chi_{\rm eff}$ distribution widens for larger $q$. In other words, there is a positive correlation between larger spin magnitude and mass ratios close to unity.
This trend with $q$ is {\it opposite} to the one expected for dynamically formed astrophysical BBH, where more massive, higher-generation mergers tend to have smaller mass ratios, i.e. negative correlation between large spin magnitude and $q$.
\end{itemize}
While current GWTC-3 observations allow for a positive correlation between binary mass and width of the $\chi_{\rm eff}$ distribution \cite{Franciolini:2022iaa}, poorly measured spins and the low statistics do not allow to confirm this trend and to distinguish between the PBH and dynamical astrophysical channels. 
ET is expected to gather a much larger number of detections with more precise spin measurements, potentially reaching sub-percent precision on individual spin magnitudes for a large fraction of detections (see e.g. \cite{Branchesi:2023mws,Franciolini:2023opt}), and allowing for tests of  the presence of mass-spin correlation consistent with the PBH scenario.

\subsection{Revealing Population III stars with the first BHs}\label{div3:pop3}

\subsubsection{The nature of Pop III stars}
\label{sec:pop3}

The formation of the first (Pop III) stars marks a primary transition phase of cosmic evolution. Their light ends the so-called ‘dark ages’, and they play a key role in cosmic metal enrichment and reionization, thereby shaping the Universe at large and the present-day galaxy population. 
The field of Pop III star formation and how Pop III stars impact cosmic evolution has been rapidly growing in the past 25 years, and it is therefore a relatively young field of research. 
As a consequence, there is still considerable debate about the physical processes that govern stellar birth at high redshift, and the overall properties of the first stars. Here we provide a very brief summary of the latest findings and highlight if and how ET observations of the GW emission from their BH remnants may help us improve our understanding of the first stars.
For the interested readers, we suggest some of the latest excellent reviews which have appeared on this subject \cite{Klessen2023, Haemmerle:2020iqg, Greif2015, Bromm:2013iya, Yoshida:2012me}.

\paragraph{Cosmological context of Pop~III star formation.}
Initially, primordial star formation was proposed to be very simple, governed by well defined initial conditions provided by Gaussian fluctuations of the cosmic density field. In this context, the formation of the first stars occurs in regions where the cosmic fluid fulfils two conditions. First, it needs to decouple from the global expansion and begin to contract due to the self-gravity of dark matter, in a so-called dark matter halo. Second, the gas within the dark matter halo needs to be able to cool and go into runaway collapse to eventually reach stellar densities. For a gas of primordial composition, the only available coolant are atomic and molecular hydrogen. The combination of $\Lambda$CDM and primordial gas cooling leads to an initial collapse phase driven by H$_2$ cooling in dark matter (DM) mini-halos with masses $M_{\rm h} \sim 10^5 - 10^6\ \Msun$ which form at the intersect of cosmic web filaments, at $20 \lesssim z \lesssim 30$ \citep{Tegmark:1996yt,Trenti:2009cj,Nebrin:2023yzm}.   

At these high redshifts, electromagnetic observations of individual Pop III forming regions are very challenging even for the James Webb Space Telescope (JWST) \cite{Windhorst:2018wft,Rhodes:2020xwp,Schauer:2020nzl,2024MNRAS.533.2727Z}. However, Pop III stars forming in pockets of metal-free gas at $z \lesssim 10$ have been predicted by simulations \cite{Liu:2020krj,Venditti:2023ddx} and may have already been tentatively detected by JWST \cite{Schauer:2022ucs,Maiolino:2023wwm}. In addition, their cumulative emission leaves an important signature on the 21cm signal from cosmic dawn and the epoch of reionization \cite{Barkana:2016nyr}, which will provide important constraints on the onset and efficiency of Pop III star formation \cite{Fialkov:2014wka,Magg:2021jyc,Schauer:2019ihk,Gessey-Jones:2022njt,Pochinda:2023uom}.

We shall also note that other important physical processes affect the onset of Pop III star formation and the minimum dark matter halo mass where this can occur. 
One example is the existence of streaming velocities, a relative motion between dark matter and baryons which stems from second-order cosmic perturbation theory and decreases linearly as the Universe expands \citep{Tseliakhovich:2010bj}. Simulations that include streaming velocities suggest that their presence delays the onset of cooling, and leads to a larger critical mass for collapse to set in \cite{Klessen2023}. 
Another important aspect that may affect the minimum halo mass to host star formation is the nature of dark matter (DM). Warm DM or fuzzy DM models suppress structure formation on small scales that depends sensitively on the mass of DM particles, which therefore affects the predicted minimum scale for structure formation and the onset of Pop III star formation \cite{Sullivan:2018szg, Mocz:2019uyd,Maio:2014qwa, Dayal:2014nva, Magg:2016rkb}. 
This may result in a delay of Pop III star formation into more massive (atomic-cooling) halos ($M_{\rm h} \gtrsim 10^{7}\,\Msun$), where the primordial gas can cool by the collisionally excited Lyman-$\alpha$ line.

    \begin{figure}[t]
        \centering
        \includegraphics[width=0.8\linewidth]{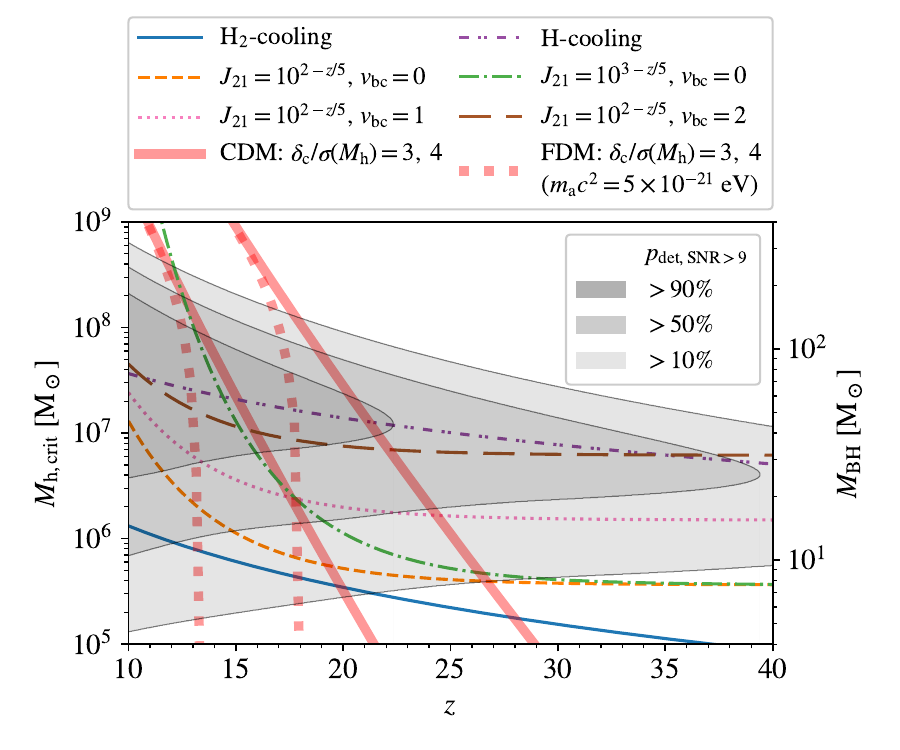}
        \vspace{-16pt}
        \caption{Redshift evolution of the critical halo mass threshold above which Pop~III star formation can happen (left axis) and the BH mass range in BBH mergers detectable by ET (right axis). We show $M_{\rm h, crit}$ for efficient $\rm H_{2}$- and H-cooling from \cite{Trenti:2009cj} with the thin solid and dash-dot-dotted curves. We plot $M_{\rm h, crit}$ regulated by Lyman-Werner (LW) radiation and baryon-DM streaming motion in four cases based on \cite{Schauer:2020gvx} (see also \cite{Nebrin:2023yzm}), adopting different models for the LW background intensity $J_{21}$ (in the unit of $10^{-21}\ \rm erg\ s^{-1}\ cm^{-2}\ sr^{-1}$) and streaming velocities $v_{\rm bc}$ (in the unit of the cosmic root-mean-squared velocity). The thin dashed, dotted, and long-dashed curves show the results for $v_{\rm bc}=0$, 1, and 2, respectively, under the model of $J_{21}$ from \cite{Greif:2006nr,Hartwig:2022lon}, while $J_{21}$ is boosted by a factor 10 under $v_{\rm bc}=0$ for the dash-dotted curve. We also show the halo masses corresponding to 3- and 4-$\sigma$ density peaks for the standard $\Lambda$CDM model and a fuzzy DM (FDM) model given a boson particle mass of $m_{\rm a}c^{2}=5\times 10^{-21}\ \rm eV$ with the thick solid and dotted curves. 
        For the right axis, the inner, intermediate, and outer shaded regions show the BH mass ranges in which more than 90\%, 50\%, and 10\% of \textit{equal-mass} BBH mergers can be detected at $\rm SNR > 9$ by ET. }
        \label{fig:mhcrit_pop3}
    \end{figure}    

Finally, once the first Pop III stars form, there are other additional effects which control Pop III star formation, such as the presence of radiation fields in the LW UV band, which photo-dissociate H$_2$, and hence suppress cooling, or in the X-ray band, which instead promote H$_2$ formation by increasing the free electron fraction and hence favor H$_2$ cooling. These additional physical processes however, depend on the properties of the radiation sources, hence on their mass spectrum and multiplicity, which we will discuss below.

In figure~\ref{fig:mhcrit_pop3} we illustrate how some of the above effects may be constrained by ET by detecting the high-$z$ BBH mergers formed as remnants of Pop III stars. The left axis shows the redshift evolution of the halo mass threshold, $M_{\rm h, crit}$ above which Pop~III star formation can happen. In particular, we report $M_{\rm h, crit}$ for efficient $\rm H_{2}$- and H-cooling from \cite{Trenti:2009cj}, and $M_{\rm h, crit}$ regulated by LW radiation and baryon-DM streaming motion in four cases based on \cite{Schauer:2020gvx}. 
We also show the halo masses corresponding to 3- and 4-$\sigma$ density peaks for the standard $\Lambda$CDM model and a fuzzy DM model with a boson particle mass of $m_{\rm a}c^{2}=5\times 10^{-21}\ \rm eV$. The onset of Pop~III star formation can be identified by the intersects of these halo masses with $M_{\rm h, crit}$, which span a broad range of redshifts ($z\sim 12 - 30$).     
To appreciate these results in the context of GW observations, we show the BH mass ranges in which more than 90\%, 50\%, and 10\% of \textit{equal-mass} BBH mergers can be detected at $\rm SNR > 9$ by ET with the right axis of figure~\ref{fig:mhcrit_pop3}. 
It is clear that ET has the potential to probe the formation history of Pop~III stars up to $z\sim 30$ via mergers of Pop~III BH remnants (at least in the mass range of $\sim 20-40\ \Msun$) (see e.g. \cite{Santoliquido:2023wzn} for detailed discussions on the relation between the detection rate of Pop~III BBH mergers by ET and the Pop III star formation history), which will provide interesting constraints on the factors that regulate the onset of Pop~III star formation, such as the nature of DM, baryon-DM streaming motion, and UV radiation fields.

\paragraph{Initial mass function of Pop~III stars.}
\label{sec:pop3_imf}
    
Starting from a pristine dark-matter mini-halo, the sequence of events which characterize the initial collapse phase is dictated by the physics of H$_2$. The formation of the dense cores, which are hotter and more massive (typically a few 1000 $\Msun$) then those in present-day conditions, corresponds to the point where H$_2$ cooling becomes inefficient.\footnote{This occurs when the gas temperature is $\simeq 200$K and the gas density is $\simeq 10^4 {\rm cm}^{-3}$, and the H$_2$ levels are no longer populated according to non-local thermodynamic equilibrium (NLTE). As a result, the cooling becomes less efficient \citep{Bromm:2013iya}.} From this point on, the cloud continues to collapse without further fragmentation until it becomes optically thick and forms the central protostar. Early simulations were stopped at this point, when the object was surrounded by only a small disk-like structure which was more strongly supported by pressure than by rotation \cite{Bromm:1999du, Bromm:2001bi, Abel:2000tu, Abel:2001pr, Nakamura:2000ez, Yoshida:2003rw, Yoshida:2006bz, Tan:2003bs, OShea:2006eik, Yoshida:2008gn}.
The authors of these studies argued that all the inflowing mass would end up in one single high-mass star.

In addition, due to the larger gas temperatures in the star-forming cores, the accretion rates onto Pop III protostars is expected to be several orders of magnitude higher than for present-day star formation and at these high rates the evolution was predicted to continue up to the arrival of the growing protostar onto the Zero-Age-Main-Sequence, where its mass was expected to be already of $M_{\rm ZAMS}\sim 100\ \Msun$ \citep{Hosokawa:2012pp}. 

Because the luminosity increases rapidly with protostellar mass, radiative feedback is expected to regulate the mass accretion and ultimately shut off the accretion flow, setting the final mass to a few tens of $\Msun$ \cite{Hosokawa:2011qa, Hosokawa:2012pp}, although this value is likely to depend on environmental conditions, with values ranging from just under ten to more than one thousand solar masses \cite{Hirano:2013lba, Hirano:2015wxa}. 
It is important to stress that these results where based on 2D radiation-hydrodynamic simulations that only follow the evolution of the central collapsing core. Hence, by construction, they describe a mass spectrum of solitary Pop III stars, one for each halo. 

    \begin{figure}[t]
        \centering
        \includegraphics[width=0.8\linewidth]{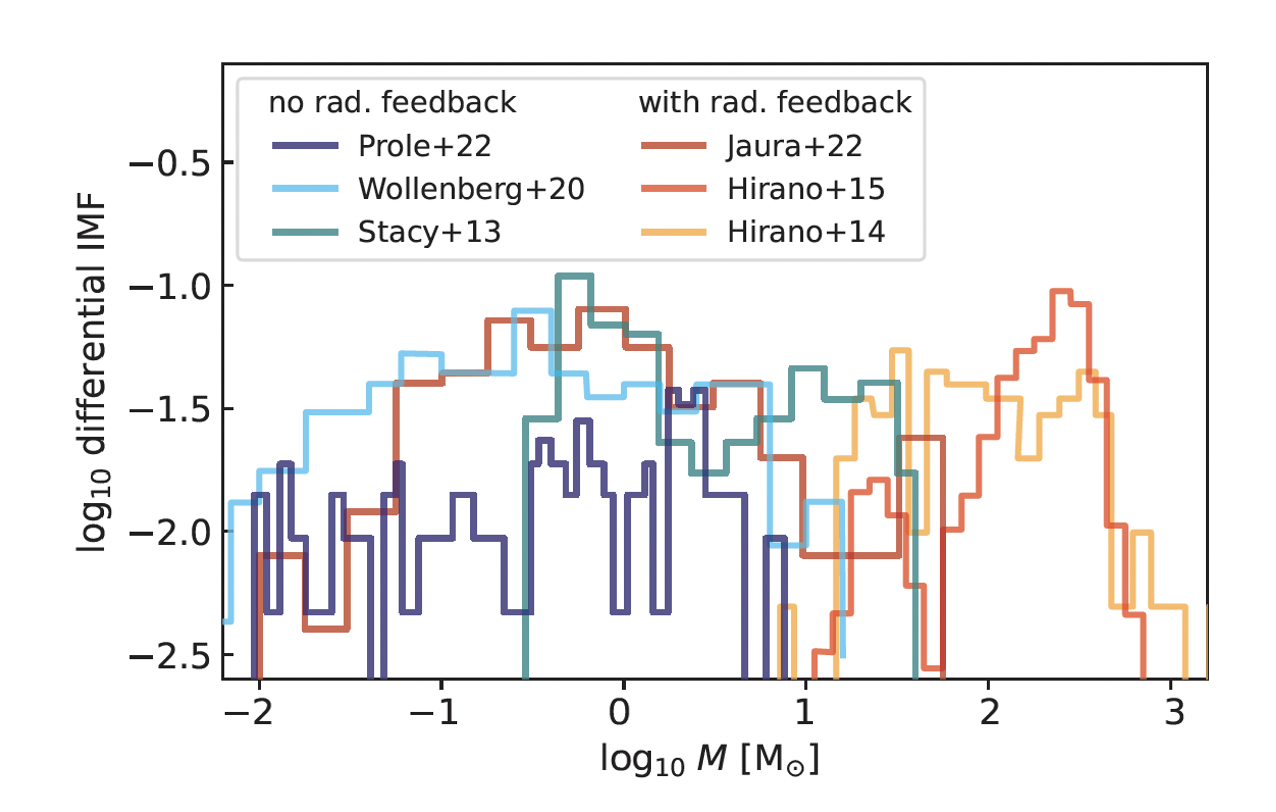}
        \vspace{-5pt}
        \caption{Illustrative examples of the Pop III stellar IMF predicted 
        by three high-resolution models without stellar feedback \cite{Prole:2021nym, Wollenberg2020, Stacy:2012iz} and three simulations with radiative feedback included \cite{Jaura:2022sny, Hirano:2013lba, Hirano:2015wxa}. This figure is adapted from figure~6 in \cite{Klessen2023}.}
        \label{fig:imf_pop3}
    \end{figure}    

In the last two decades, the simple scenario of solitary formation of one massive Pop~III star per minihalo has significantly evolved into a more complex picture that primordial star-forming gas is highly susceptible to fragmentation and Pop~III stars typically form with broad mass distributions in (small) clusters where binary and multiple systems are common. 
In figure~\ref{fig:imf_pop3} we show some illustrative examples of the Pop III stellar initial mass function (IMF) predicted by numerical studies with and without the inclusion of radiative feedback \cite{Klessen2023}. Despite the large variation in range (due to numerical limitations in resolution and time coverage), the predicted distributions are rather flat (in log space) compared to the highly bottom-heavy present-day IMF, and can extend up to $\sim   1000\ \Msun$. 
Pop III stars may therefore leave behind black holes with a variety of masses, ranging from remnants of core-collapse SN explosions, to black holes formed by failed explosions, up to black holes with masses above the PISN mass gap, reaching the mass range expected for intermediate-mass black holes (IMBHs, see section~\ref{sec:BH-mass-measure} and \ref{sec:IMBH}). In this new picture, fragmentation has significant implications not only for the resulting Pop III stellar mass spectrum but also for the formation of binary/multiple systems of stars and BHs, as we discuss below.

\paragraph{Pop~III star clusters and binaries.}
\label{sec:pop3_sc}
    
In the standard pathway of Pop~III star formation, a star-forming disk forms at the center following the monolithic collapse of pristine gas by runaway molecular cooling in a minihalo ($M_{\rm h}\sim 10^{5}-10^{6}\ \Msun$). The disk fragments to produce multiple protostars, which eventually become a small cluster of a few to few tens stars and a total mass of $\sim 10-1000\ \Msun$ when the disk is evaporated by stellar feedback. 
The initial configurations of Pop~III star clusters and statistics of binary/multiple systems are shaped by poorly understood processes of protostar formation and evolution 
\cite{Belczynski:2016ieo,Liu2021}. 
Suffering from uncertainties due to limited resolution and imperfect modeling of stellar feedback, some simulations predict that Pop~III binary stars typically have large separations $a\sim (2\times 10^2-4\times 10^{4})$~AU due to outward migrations of nearly equal-mass protostars by rapid accretion of gas with high angular momentum \cite{Park:2022usr,Park2024,Sugimura:2020hug,Sugimura:2023knx}, while others favor the formation of closer binaries ($a\lesssim 10^{3}$~AU down to the resolution limit of a few AU) with inward migrations by, e.g., dynamical friction in dense gas structures 
\cite{Susa:2014moa,Riaz2023}. 
    
The situation becomes even more complex in non-standard pathways where Pop~III star formation is delayed to occur in more massive (atomic-cooling) halos ($M_{\rm h}\gtrsim 10^{7}\ \Msun$) by, e.g., baryon-DM streaming motion and LW radiation (see figure~\ref{fig:mhcrit_pop3} and section~3.3 in \cite{Klessen2023}).  
The rapid gas inflows ($\dot{M}\sim 0.1-10^{3}\ \Msun\ {\rm yr}^{-1}$) in such massive halos are expected to produce massive ($\gtrsim 10^{4}\, \Msun$), dense star clusters, where competitive accretion and runaway stellar collisions lead to the formation of a small number of very/super massive stars with $M_{\rm ZAMS}\gtrsim 300\ \Msun$ up to a few $10^{5}\ \Msun$ \cite{Sakurai:2017opi,Boekholt:2018gbw,Chon2020,Regan:2020drm,Wang:2022unj,Liu:2023zea,Prole:2023toz,Reinoso:2023nfx,Liu2024}, which are promising progenitors of IMBHs (see section~\ref{sec:IMBH}). 
However, the abundance and properties of such massive Pop~III clusters (e.g., mass, size, virial factor, internal structure, and IMF of stars) are also uncertain in current theoretical models.  

In summary, our knowledge of Pop~III star clusters and binaries is still incomplete due to computational limitations of simulations (e.g., resolution, time coverage, volume, treatment of feedback) 
and the lack of direct observations 
(which will remain challenging in the near future). 
Nevertheless, as discussed below, the properties of Pop~III star clusters and binaries, as well as Pop~III stellar evolution, are encoded in the properties of compact object mergers from Pop~III stars. Therefore, GW observations by ET will bring us new insights into the physics of Pop~III star formation/evolution. 
    
\subsubsection{The evolution of Pop~III stars and binary black holes} 
\label{sec:pop3_evo}

Compared with metal-enriched Population~I/II (Pop~I/II) stars, Pop~III stars are more compact and experience negligible mass loss from radiation-driven winds because radiation escapes from them more easily in the lack of metal-line opacity (see figure~13 of \cite{Klessen2023}). The special metal-free stellar evolution, massive nature (section~\ref{sec:pop3_imf}), and common multiplicity (section~\ref{sec:pop3_sc}) of Pop~III stars indicate  that they are efficient progenitors of BBHs. 
Similar to BBH mergers from Pop~I/II stars, the evolution of Pop~III BBHs to mergers can also be classified into the isolated (section~\ref{sec:pop3_bbh_ibse}) and dynamical (section~\ref{sec:pop3_bbh_sc}) channels. As summarized in figure~\ref{fig:pop3_flow_chart} and discussed in detail below, these two channels are closely connected to the special cosmological context and pathways of Pop~III star formation, and they can potentially produce BBH mergers\footnote{Pop~III stars also produce BH-NS and NS-NS mergers but at much lower efficiencies compared with BBH mergers (see, e.g., \cite{Kinugawa:2016skw,Hartwig:2016nde,Belczynski:2016ieo,Liu:2021jdz}). Such less massive mergers are also more difficult to detect at high $z$. Therefore, we focus on Pop~III BBH mergers.} with unique signatures identifiable in GW observations by ET, which makes ET an indispensable probe of Pop~III stars (section~\ref{sec:pop3_gw}).

\begin{figure}[t]
    \centering
    \scriptsize
\begin{tikzpicture}[node distance=1.3cm]
\node (cosmic) [startstop] {Cosmic structure formation + (metal-free) primordial gas};
\node (standard) [process, below of=cosmic,xshift=-3.2cm] {Molecular cooling};
\node (small) [process, below of=standard, align=left,yshift=0.2cm] {Standard pathway ($M_{\rm h}\sim 10^{6}\ \Msun$):\\ Small star clusters ($\sim 10-1000\ \Msun$)};
\node (non-standard) [process, below of=cosmic,xshift=3.2cm] {Delay of star formation, atomic-cooling};
\node (massive) [process, below of=non-standard, align=left,yshift=0.2cm] {Non-standard pathways ($M_{\rm h}\gtrsim 10^{7}\ \Msun$):\\ Massive star clusters ($\gtrsim 10^{4}\ \Msun$)};
\node (ibse) [decision, below of=small, align=left,yshift=-0.2cm] {Isolated binary stellar evolution \\ $\rightarrow$ Initially isolated BBHs};
\node (dyn) [decision, below of=massive, align=left,yshift=-0.2cm] {Dynamical interactions\\ (capture, hardening, triples, ...)};
\node (iso_merger) [decision, below of=ibse, xshift=-0.7cm]{Isolated BBH mergers}; 
\node (nsc) [decision, below of=ibse, xshift=3.5cm]{Nuclear star clusters (Pop~I/II)}; 
\node (dyn_merger) [decision,right of=nsc, xshift=3cm]{Dynamical BBH mergers};
\node (gap) [io, below of=iso_merger, xshift=0cm] {Mass gap at $\sim 100-130\ \Msun$};
\node (ecc) [io, below of=dyn_merger, xshift=0cm, align=left] {Massive, eccentric, high-$z$};
\node (sgwb) [io, below of=nsc, xshift=0.2cm] {GW background at $f\lesssim 10\ \rm Hz$}; 
\draw [arrow] (ibse) -- node[xshift=1.3cm,yshift=0.1cm]{Dynamical friction} (nsc);
\draw [arrow] (nsc) -- node[xshift=0.6cm,yshift=-0.1cm]{Ex-situ} (dyn);
\draw [arrow] (cosmic) -- (standard);
\draw [arrow] (cosmic) -- node[xshift=2.25cm, align=left]{Baryon-DM streaming,\\ LW radiation, ...} (non-standard);
\draw [arrow] (standard) -- (small);
\draw [arrow] (non-standard) -- (massive);
\draw [arrow] (small) -- node[xshift=1.25cm]{Cluster dispersion} (ibse);
\draw [arrow] (massive) -- node[xshift=0.5cm]{In-situ} (dyn);
\draw [arrow] (ibse) -- node[xshift=0.9cm,yshift=-0.1cm]{GW emission} (iso_merger);
\draw [arrow] (dyn) -- (dyn_merger);
\draw [arrow] (iso_merger) -- (gap);
\draw [arrow] (dyn_merger) -- (ecc);
\draw [arrow] (iso_merger) -- (sgwb);
\draw [arrow] (dyn_merger) -- (sgwb);
\end{tikzpicture}
    \caption{Connections between the cosmological context (red) and pathways (orange) of Pop~III star formation (section~\ref{sec:pop3}), evolution of Pop~III BBHs (green, see section~\ref{sec:pop3_evo}), and identifiable signatures of Pop~III BBH mergers in ET observations (blue, see section~\ref{sec:pop3_evo} and \ref{sec:pop3_gw}).}
    \label{fig:pop3_flow_chart}
\end{figure}

\paragraph{Pop~III binary black hole mergers from isolated binary stellar evolution.}
\label{sec:pop3_bbh_ibse}

BBH mergers from Pop~III stars have been studied most intensively for the isolated binary stellar evolution  channel (see section~\ref{sect:isolatedchannel}) \cite{Hartwig:2016nde,Belczynski:2016ieo,Inayoshi:2017mrs,Kinugawa:2014zha,Kinugawa:2015nla,Kinugawa:2020ego,Kinugawa:2020xws,Kinugawa:2021qee,Tanikawa:2020abs,Tanikawa:2020cca,Tanikawa:2021qqi,Hijikawa:2021hrf,Costa:2023xsz,Santoliquido:2023wzn,Tanikawa:2024mpj}. 
The isolated binary stellar evolution channel is expected to be suitable for the standard pathway of Pop~III star formation that produces small Pop~III clusters with short dynamical and relaxation timescales, where most dynamical interactions happen in the early stage (within $\lesssim 1\ \rm Myr$), and Pop~III binaries\footnote{Triples and quadruples also form in small Pop~III star clusters \cite{Ryu2016,Sharda:2020aio,Liu2021}. The BBH mergers produced by such higher-order systems can be different from those from standard isolated binary stellar evolution (see section 1.1.3). However, this aspect has not been explored for Pop~III stars.} evolve effectively in isolation thereafter. 
Although well-motivated, the merger efficiency and properties of Pop~III BBH mergers from isolated binary stellar evolution are highly sensitive to the obscure initial statistics of Pop~III binary stars, binary interaction processes, and Pop~III stellar evolution. 

For instance, if the initial binary statistics of Pop~III stars is dominated by wide binaries with $a\sim 10^2-10^{4}\ \rm AU$ as found in the simulations from \cite{Stacy:2012iz}, 
rather than close binaries with $a\sim 0.1-10\ \rm AU$ like in observations of present-day stars \cite{Sana:2012px}, the merger efficiency of Pop~III BBHs will be reduced by up to a factor of $\sim 20$, and the dominant evolutionary pathway changes from stable mass transfer to common envelope evolution \cite{Kinugawa:2020ego,Costa:2023xsz,Santoliquido:2023wzn}.     
The evolution of individual stars has strong effects on the BH mass distribution in Pop~III BBH mergers, in particular for massive BHs inside or above the PISN mass gap\footnote{Pop~III single stars with initial masses $\sim (50-100)\ \Msun$ can avoid pair instability and retain most of their hydrogen envelopes before collapse to produce BHs of similar masses in the PISN mass gap \cite{Farrell:2020zju,Volpato:2022nwe}. If their spins remain small ($\lesssim 0.2$) after birth \cite{Fuller:2019sxi,Ghodla:2023ymi,Marchant:2023ncp}, Pop~III BHs in the mass gap can be distinguished from the merger products of smaller BHs \cite{Gerosa:2021hsc}, although it is challenging to tell them from PBHs with similar masses and spins (see section~\ref{sec:PBHsdiv3}). How such Pop~III BHs show up in BBH mergers is still in debate \cite{Tanikawa:2020abs}.} $\sim (50-130)\ \Msun$. 
In general, models with stronger stellar expansion disfavor the formation of massive Pop~III BBH mergers via isolated binary stellar evolution, since larger stars are more likely to suffer strong mass loss and undergo stellar mergers during binary interactions. 
If massive Pop~III stars with $M_{\rm ZASM}\gtrsim 100\ \Msun$ become red super giants with radii $R\gtrsim 10^{3} R_\odot$ when they approach the end of main sequence (MS) or during post-MS evolution, Pop~III BH mergers mostly contain BHs below $50\ \Msun$ \citep{Costa:2023xsz,Santoliquido:2023wzn}. 
In contrast, if stellar expansion is suppressed, for example, by inefficient convective overshooting, close ($a\lesssim 1\ \rm AU$) binaries of Pop~III stars with $M_{\rm ZAMS}\sim (50-100)\ \Msun$ can produce 
BBH mergers involving BHs of similar masses inside the PISN mass gap 
via stable mass transfer with little mass loss, and binaries of more massive ($M_{\rm ZAMS}\gtrsim 240\ \Msun$) Pop~III stars can produce BBH mergers with BHs above the PISN mass gap ($\gtrsim 130\ \Msun$) avoiding stellar mergers \cite{Tanikawa:2020abs,Tanikawa:2020cca,Tanikawa:2021qqi,Hijikawa:2021hrf,Tanikawa:2024mpj}. 
    
Moreover, these results point out the intriguing possibility that the BH mass distribution in Pop~III BBH mergers from isolated binary stellar evolution shows a unique\footnote{BBH mergers from extremely metal-poor ($Z\lesssim 10^{-4}$) stars have similar properties as Pop~III mergers if given the same conditions \cite{Costa:2023xsz,Tanikawa:2021qqi}. However, it is unknown whether such stars can be as abundant and massive as Pop~III stars to have noticeable signatures in GW observations. } narrow mass gap around $\sim (100-130)\ \Msun$, under the peculiar condition that (1) close ($a\lesssim 10\ \rm AU$) binaries of massive Pop~III stars are common, 
and (2) massive Pop~III stars do not expand significantly before collapse. 
The identification or non-detection of this feature in GW observations of BBH mergers (up to $z\sim 20$ by ET, see figure~\ref{fig:mhcrit_pop3}) will place novel constraints on the binary statistics of Pop~III stars and Pop~III (binary) stellar evolution.  

\paragraph{Dynamical Pop~III binary black hole mergers in dense star clusters.}
\label{sec:pop3_bbh_sc}

Other than isolated binary stellar evolution, dynamical Pop~III BBH mergers can also occur 
in dense star clusters in two scenarios. (1) In-situ: Pop~III BBHs form in massive ($\gtrsim 10^{4}\ \Msun$), dense clusters of Pop~III stars/remnants themselves \cite{Wang:2022unj,Liu:2023zea,Mestichelli:2024djn} hosted by atomic-cooling halos ($M_{\rm h}\gtrsim 10^{7}\ \Msun$). (2) Ex-situ: Pop~III BBHs (initially born in isolation) fall into nuclear star clusters made of Pop~I/II stars by dynamical friction when their natal minihalos grow/merge into larger halos \cite{Liu:2020lmi,Liu:2021jdz,Liu:2024mkh}. 
The main difference between the two scenarios is that Pop~III mergers in nuclear star clusters mostly happen at relatively low redshifts $z\lesssim 6$ because only massive ($M_{\rm h}\gtrsim 10^{9}\ \Msun$) halos can possibly host nuclear star clusters, while a large fraction ($\sim 0.4-0.8$) of in-situ dynamical Pop~III mergers can happen at $z\sim 6-20$ providing that massive Pop~III clusters are common. 
Both scenarios produce mergers of more massive BHs with lower mass ratios compared with the case of isolated evolution, 
especially for BHs above $\sim 50\ \Msun$ including IMBHs 
that can be extremely rare in isolated mergers \cite{Costa:2023xsz,Santoliquido:2023wzn,Mestichelli:2024djn,Liu:2024mkh}. 
Besides, a significant fraction (up to $\sim30\%$) of dynamical Pop~III BBH mergers will have large enough eccentricities $e\gtrsim 10^{-3}$ detectable by ET at the observer-frame frequency $f=10\ \rm Hz$ \cite{Saini:2023wdk}.
    
These results are consistent with the common properties of dynamical BBH mergers (see sections \ref{sec: formation channels dynamical} and \ref{sect:propdynCBCdiv3}), although the predicted merger rates are subject to large uncertainties in the abundance, initial conditions, and long-term environments of massive Pop~III clusters, as well as the assumptions on high-$z$ nuclear star clusters and dynamics of Pop~III BBHs in high-$z$ galaxies. 
It will be challenging to identify Pop~III BBH mergers in nuclear star clusters where they are likely overwhelmed by dynamical mergers of Pop~I/II BBHs with similar masses \cite{Mapelli:2021syv}. However, observations of massive, eccentric mergers (with low mass ratios) at very high redshifts ($z\gtrsim 10$) by ET will strongly constrain the in-situ Pop~III scenario, since BBH mergers from other origins either occur at lower redshifts (Pop~I/II BBHs) or on quasi-circular ($e\lesssim 10^{-5}$) orbits (isolated Pop~III BBHs and PBHs). 
    
\subsubsection{Gravitational waves from Pop III remnants}
\label{sec:pop3_gw}

\begin{figure}[t]
        \centering
        \includegraphics[width=0.7\linewidth]{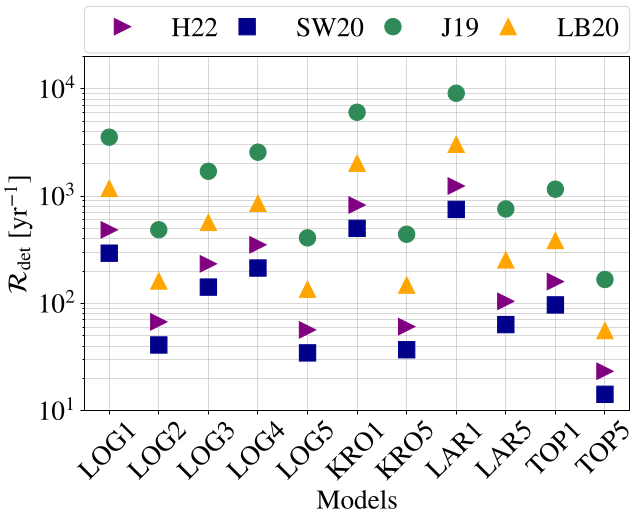}
        \vspace{-5pt}
        \caption{Detection rates of Pop~III BBH mergers from isolated binary stellar evolution by ET with $\rm SNR>9$ for different models of the initial statistics of Pop~III stellar binaries ($x$ axis) and cosmic star formation histories of Pop~III stars (denoted by different markers) adapted from Fig.~4 in \cite{Santoliquido:2023wzn}. The rates predicted by other studies for isolated mergers and dynamical mergers in dense star clusters are within the range shown here.}
        \label{fig:pop3_et_rate}
    \end{figure}

Due to large uncertainties in the cosmic star formation history of Pop~III stars (see figure~2 of \cite{Klessen2023}), properties of Pop~III clusters and binaries, as well as Pop~III (binary) stellar evolution, theoretical predictions 
Pop~III BBH mergers 
are highly uncertain. Their merger rate (density) is expected to peak around $z\sim 6-15$ with $\mathcal{R}_{\rm PopIII}^{\max}\sim (0.1-200)\ \rm yr^{-1}\ Gpc^{-3}$, and the local rate is $\mathcal{R}_{\rm PopIII}(z=0)\sim (10^{-3}-100)\ \rm yr^{-1}\ Gpc^{-3}$ \cite{Hartwig:2016nde,Belczynski:2016ieo,Kinugawa:2014zha,Kinugawa:2015nla,Liu:2020ufc,Kinugawa:2020ego,Kinugawa:2021qee,Tanikawa:2020cca,Tanikawa:2021qqi,Hijikawa:2021hrf,Santoliquido:2023wzn,Wang:2022unj,Liu:2023zea,Liu:2021jdz,Tanikawa:2024mpj,Mestichelli:2024djn,Liu:2024mkh}. 
In general, Pop~III BBH mergers will dominate the merger rate at least for $z\gtrsim 20$ (unless PBH mergers are abundant), and they typically contain BHs above $20\ \Msun$, more massive than the low-mass ($\sim 10\ \Msun$) BHs that dominate the observed local merger rate. 
Depending on the Pop~III IMF and (binary) stellar evolution parameters, Pop~III BBH mergers in isolation may contain massive ($\gtrsim 50\ \Msun$) BHs inside and above the PISN mass gap, while such BHs will be common in dynamical Pop~III BBH mergers in dense star clusters. 
The high sensitivity of ET, especially at low-frequencies, makes it an ideal instrument for observing Pop~III BBH mergers at very high redshifts ($z\sim 30$), reaching the smallest structures hosting the first generation of stars in the Universe (figure~\ref{fig:mhcrit_pop3}). A promising detection rate of $\sim 10-10^{4}$ events per year is estimated from current theoretical models, as illustrated in figure~\ref{fig:pop3_et_rate}.

More importantly, even if Pop~III BBH mergers remain a sub-dominant component in the detectable events, they can be identified with their potentially unique features, such as a narrow mass gap $\sim 100-130\ \Msun$ in the BH mass distribution and high eccentricities ($e\gtrsim 10^{-3}$ at $f=10$~Hz) at $z\gtrsim 10$, and with the aid of machine learning \cite{Iwaya:2023mse,Franciolini:2024vis,Saini:2023wdk,Santoliquido:2024oqs}. Beyond resolved merger events, under certain conditions, Pop~III BBH mergers can possibly make significant 
contributions to the GW background, producing a spectral feature at $f\lesssim 10$~Hz accessible to ET observations, in particular for the residual background where individual detected sources are properly subtracted, see also section~\ref{subsec:popIII} \cite{Perigois:2021ovr,Martinovic:2021fzj,Kouvatsos:2024eok,Liu:2024mkh}. Since it is incredibly difficult to observe individual Pop~III stars directly via electromagnetic waves \cite{Rhodes:2020xwp,Schauer:2020nzl}, GW observations by 3rd-generation detectors like ET, combined with other indirect probes of Pop~III stars \cite{Dvorkin:2016wac,Inayoshi:2016hco,Tanikawa:2022qms}, will be extremely valuable for revealing the elusive nature of Pop~III stars and early structure formation.

\subsection{Intermediate-mass BHs (IMBHs): formation channels and merger rate}
\label{sec:IMBH}
 

Intermediate-mass black holes (IMBHs) are an elusive class of objects that may bridge stellar-mass black holes  and supermassive black holes (SMBHs). IMBHs are expected to densely populate the $10^2-10^5 \Msun$ mass range, although there is little knowledge about their properties due to scarce observational evidence.

During the first three observing runs, the LVK collaboration identified 8 IMBH candidates (3 of them at $90\%$ credibility level) formed from stellar BH mergers \cite{LIGOScientific:2020iuh,LIGOScientific:2020ufj,LIGOScientific:2021tfm}. Several IMBHs, generally heavier than $50,000\, \Msun$, have been observed in the nucleus of dwarf galaxies \cite[]{Reines:2013pia,2015ApJ...809L..14B,2017ApJ...836..237N,2019ApJ...872..104N,2019NatAs...3..755W,2021ApJ...921...98C,Mezcua:2017npy,Greene:2019vlv}, and in the center of massive systems in the outskirts of massive galaxies, suggesting they are the nucleus of a dwarf galaxy that was stripped during a minor merger (e.g., \cite{Farrell:2009uxm,Mezcua:2013mua,Mezcua:2013goa,Mezcua:2015pra,2018MNRAS.480L..74M}).

IMBHs are one possible engine powering the so-called ultra-luminous X-ray sources (ULXs) (e.g., \cite{Kaaret:2000sh, Miller:2004bu}), off-centered sources with an X-ray luminosity above the Eddington limit for a stellar BH, $L_X\sim 10^{39}$ erg s$^{-1}$ (see the review  \cite{Kaaret:2017tcn}). Most of the known population of ULXs are associated with emitting NSs, but the most energetic ones -- also known as hyper-luminous X-ray sources (HLXs) -- with $L_X>10^{41}$ erg s$^{-1}$, can be powered only by an object with mass $>10^3 \Msun$ \citep{Webb:2012cii,Cseh:2014qsa}.

In the $(10^3-10^4)\, \Msun$ mass range, putative IMBHs have been also observed in a few galactic and extragalactic star clusters (e.g., \cite{Gebhardt:2005cy,Lutzgendorf:2012xc,Lutzgendorf:2012sz,2022ApJ...924...48P,2024arXiv240506015H}). 
However, most observations of these mid-range IMBHs are controversial (e.g., \cite{2017Natur.542..203K,Abbate:2019qoc,Perera:2017jrk,2018MNRAS.473.4832G}). The absence of compelling evidence for IMBHs with masses $(10^3-10^4)\, \Msun$ leads to a natural, fundamental question about IMBHs: do
IMBHs constitute a unique class of astrophysical BHs? Or, rather, do they populate the high end of the stellar BH mass distribution and the low end of the SMBH mass distribution?

Several scenarios have been proposed to explain IMBH formation. We can broadly divide them into three main ensembles: direct collapse, Pop III stellar evolution, and dynamical formation.

\subsubsection{Formation Scenarios}
\paragraph{Direct collapse.}

Direct collapse BHs can have masses of $\sim (10^4-10^5)\, \Msun$ and result from the collapse of a supermassive star of $\sim 10^5 \Msun$ \cite{Loeb:1994wv,Eisenstein:1994nh,Bromm:2002hb,Lodato:2006hw,Ferrara:2014wua,Woods:2021lgj,2023MNRAS.521..463H}. Such supermassive stars form via a rapid inflow of dense gas in metal-free, or very metal-poor, protogalaxies, in which an intense Lyman-Werner radiation field prevents gas fragmentation and star formation. This Lyman-Werner radiation can be provided by close star-forming galaxies \citep{Dijkstra:2016jwm,Agarwal:2014hla,Habouzit:2016nyf}. The high supersonic turbulence in the primordial gas reservoirs could however also yield the formation of direct collapse BHs without the need for a strong Lyman-Werner field \cite{Latif:2022vwc}.
Because of all these constraints, direct collapse BHs are thought to be less abundant than IMBHs formed via other mechanisms (e.g. Pop III stars) and their mass range is out of ET operating band. They could however be detected at near-infrared wavelengths at z $\gtrsim$ 20 \cite{Natarajan:2016rii,Whalen:2020ypj} and in the radio regime at  $z \sim 8-10$  \cite{Whalen:2020goy,2023ApJ...956..133W}.

\paragraph{Pop III stars.}
One possible pathway to form IMBH seeds with masses in the range $(10^2-10^3)\, \Msun$ is via single and binary stellar evolution of Pop III stars. This channel has been  discussed in detail in section~\ref{sec:pop3_sc}. In this scenario, IMBHs form in the early Universe, at a redshift $z \sim 8-20$ (for a discussion about the various uncertainties in Pop III star formation history and their impact on BH formation, see \cite{Santoliquido:2023wzn})  in the center of protogalaxies. Given their mass, IMBH from Pop III stars are usually referred to as ``light seeds", in contrast to the ``heavy seeds" formed from direct collapse discussed above. In fact, light seeds can, in principle, further grow via gaseous accretion and reach final masses that cross the IMBH-SMBH boundary. However, theoretical and numerical models suggest that the IMBH mass growth is not sufficiently rapid to explain the observation of SMBHs at redshift $z>10$, as recently discovered by JWST \cite{Bogdan:2023ilu,Maiolino:2023zdu}, favoring the direct collapse scenario.

A viable way to increase the IMBH mass is via repeated mergers with smaller BHs and accretion of stellar material. This ``dynamical" process is poorly explored for Pop III stars, owing to our poor knowledge of the physics of clustering in the first stages of life of the Universe, but it is quite robustly explored in the case of Pop I and Pop II star clusters, as we discuss in the next section.

\paragraph{Dynamical formation of IMBHs.}
Several galactic and extragalactic GCs exhibit peculiar features that may be ascribed to a central IMBH with masses from a few $10^2 \Msun$ to $\sim 10^4 \Msun$ (e.g., \cite{2024arXiv240506015H}). Three main mechanisms can support the seeding and growth of an IMBH in a star cluster: stellar collisions, stellar feeding, and hierarchical BH mergers (for a recent review, see \cite{Askar:2023pmd}).

\subparagraph{Stellar collisions.}
The combination of mass segregation, by which the most massive objects in a stellar ensemble lose orbital energy to the lighter ones, and relaxation, which is driven by close dynamical interactions, causes over time a continuous increase of the central density of star clusters up to a critical point, referred to as ``core-collapse", at which the density reaches a peak and interactions become so violent to favor the formation of tight binaries. The binaries act as a heating source, scattering off stars and causing the expansion and cooling of the cluster core. During this complex phase, binaries can get so tight that their components merge, and densities become high enough to favor close hyperbolic interactions. 

\begin{figure}
\centering
\includegraphics[width=0.7\columnwidth]{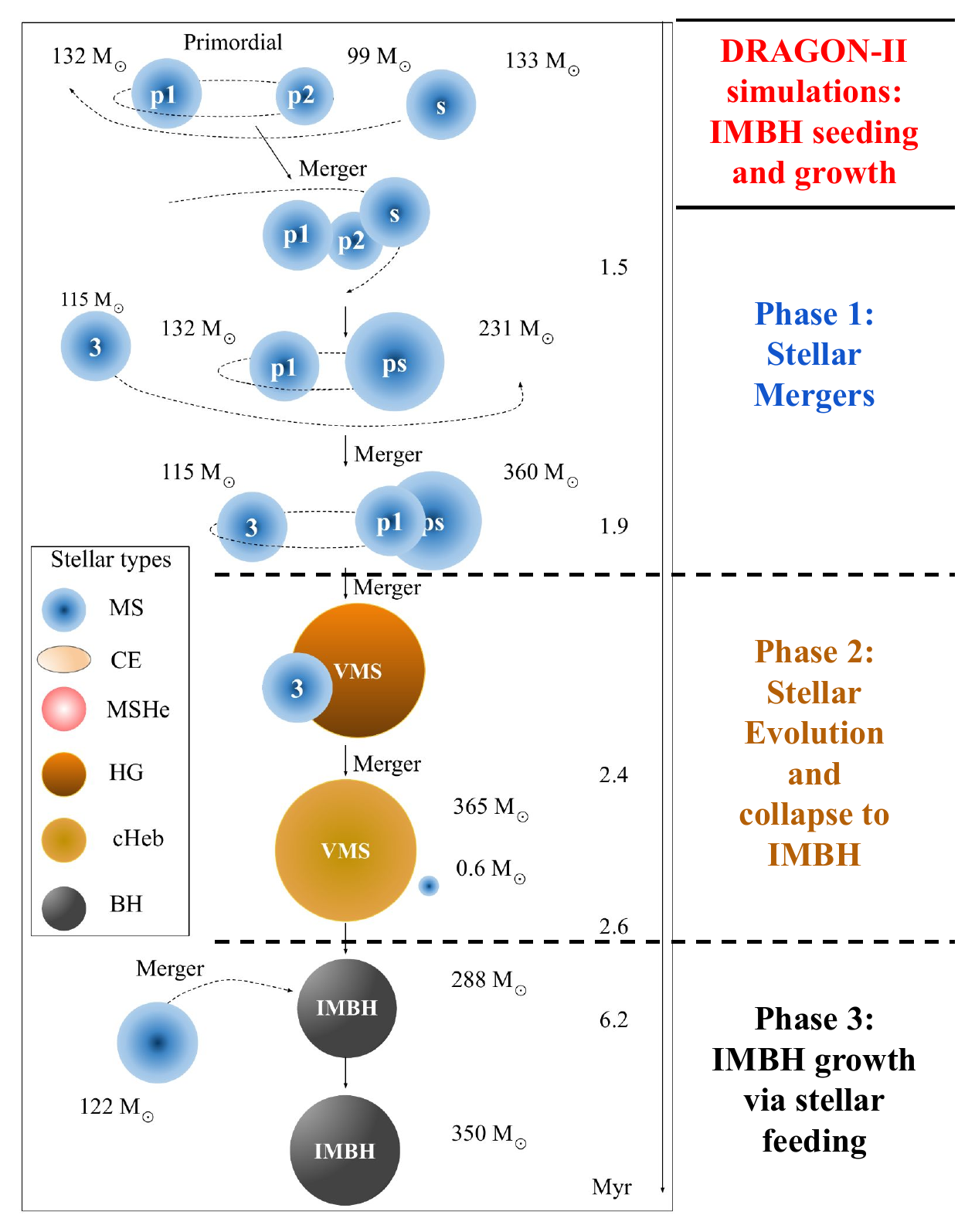}
\caption{Stellar mergers leading to the formation of an IMBH in one of the \textsc{Dragon-II} numerical simulations \citep{ArcaSedda:2023mlv}.}
\label{fig:dragonII}
\end{figure}

 If the timescale of core-collapse is shorter than the typical stellar evolution time for massive stars, this process can favor stellar mergers and collisions. This generally happens in clusters with a mass $\lesssim 10^5 \Msun$ and half-mass radii in the range $\sim (0.2-1)$~pc \citep{PortegiesZwart:2002iks, 2008gady.book.....B}. In such low-mass and dense clusters stellar collisions can proceed in a runaway fashion, rapidly building up a very massive star (VMS) with a mass $\sim (10^2-10^3)\, \Msun$ \citep{PortegiesZwart:2002iks,Giersz:2015mlk,Mapelli:2016vca,2021MNRAS.501.5257R,Gonzalez:2020xah,ArcaSedda:2023mlv}. The VMS ultimately collapse to an IMBH, which possibly grows later, at a slower pace, by capturing passing by stars or merging with smaller BHs, see figure~\ref{fig:dragonII}.

There may be a threshold value of the cluster core density above which the stellar collision process is favored over others to build-up an IMBH, likely in the range $\rho_{\rm thre} \sim (3\times 10^5-10^7)\,  \Msun$ pc$^{-3}$ \citep{PortegiesZwart:2002iks,Giersz:2015mlk,ArcaSedda:2023mlv}. The time over which this runaway process onset depends critically on several quantities that characterise the cluster structure, like possible primordial mass segregation, primordial binaries, or the stellar mass function. 

The possibility of forming an IMBH seed through this mechanism depends on the mass of the VMS, which must overcome the mass range affected by PISN and pulsational pair-instability supernova (PPISN). Typically, the onset of PISN, which completely disrupts the star, becomes inefficient at $m_{\rm VMS} > 240 \Msun$, although the minimum star mass above which PISN significantly increases at increasing the metallicity (e.g., \cite{Spera:2017fyx}). Unfortunately at metallicities $Z \geq 0.008$, the VMS collapse would result in a BH of $\sim (20-30)\, \Msun$, regardless its mass \citep{Spera:2017fyx}, thus suggesting that metal-rich clusters are unlike site of formation of IMBHs via stellar collisions, unless the internal structure of the VMS significantly differs from that of a single massive star owing to the material reshuffling promoted by the stellar collision process. Accretion of gaseous material during the phase in which clusters are still embedded in their progenitor clouds can significantly contribute to the VMS growth, possibly favoring the formation of supermassive stars as massive as $(10^3-10^4)\, \Msun$ (e.g., \cite{2018MNRAS.478.2461G,Reinoso:2023nfx}), that would likely undergo direct-collapse.

\subparagraph{Stellar feeding.}

In dense stellar environments, close encounters between stars and BHs can trigger star-BH collisions, with consequent fall of stellar material onto the BH. This can sustain the IMBH growth, especially for massive ($>50\, \Msun$) stars. Whether, and in what amount, such material can be fed to the BH is unclear.

Little accretion is expected in the case of a normal star colliding with a stellar BH, as in such a case most of the energy is expected to be radiated away via jets \citep{Guillochon:2012uc,MacLeod:2015bpa,Cruz-Osorio:2020dja,Kremer:2022xgm}. Nonetheless, some models suggest that during the common envelope phase, a BH can accrete the star core and expel the envelope \citep{Schroder:2019xqq}, a process that can significantly spin up the BH. If the accreting star is a byproduct of previous stellar collisions, the merger product should be characterised by a tenuous envelope and dense core \citep{Glebbeek:2009jr}. In such a case, the BH can accrete most of the star mass. 

IMBHs forming through this mechanism may be preceded by SN-like \citep{Schroder:2019xqq} and other high-energy transients \citep{Kremer:2022xgm}.  These IMBHs may be characterised by peculiar spins, depending on their formation history. If most of its mass comes from the accretion of one massive star, the accretion process can significantly spin-up the IMBH up to extremal values. If, instead, the IMBH mass growth proceeds through episodic accretion from random directions, the cumulative effect should lead the IMBH to spin-down.

\subparagraph{Hierarchical BH mergers.}

IMBHs can also form via repeated mergers among BHs, or {\it hierarchical} mergers   \citep{Miller:2001ez,Antonini:2018auk,Arca-Sedda:2020lso,Fragione:2020nib,ArcaSedda:2023mlv}. The growth of an IMBH through repeated mergers with smaller BHs can be hindered by both Newtonian and Relativistic recoil, which can expel the IMBH from the host cluster (see also section \ref{div3:formationchannel}). 
Newtonian kicks on IMBHs with $m<10^3 \Msun$ can be as large as $10^2$ km s$^{-1}$ \citep{Goodman:1992cy,ArcaSedda:2023mlv}. Relativistic kicks typically exceed $10^2$ km s$^{-1}$ for binary mass ratio $>0.1$\footnote{Although depending on the spin alignment the kick can remain in the range $10-100$ km s$^{-1}$ for a binary mass ratio value $>0.7$.} \cite[]{Sedda:2021abh}. 
Owing to the typical mass and density of young massive and globular clusters, IMBHs in these clusters are expected to reach maximum masses in the range $(100-1000)\, \Msun$ (e.g., \cite{Fragione:2018vty,Arca-Sedda:2020lso,Sedda:2021vjh,Torniamenti:2024uxl}). Heavier IMBHs could form in nuclear clusters, where escape velocities can reach values $(10^2-10^3)\, {\rm km}\,  {\rm s}^{-1}$ \citep{Antonini:2016gqe,Fragione:2020nib,Mapelli:2021syv,Sedda:2023big}.

Long merger chains can leave observable imprints on the properties of the IMBH population. The increase in mass through successive mergers and the reduction in spin driven by the accretion of smaller BHs from isotropic directions can lead to a peculiar mass-spin-redshift relation (e.g., \cite{Arca-Sedda:2020lso}), although  such complex relation is severely affected by the uncertainties in theoretical models. Accurate measurements of spin and masses can thus possibly help to discriminate between IMBHs that grew prevalently via BH mergers and those that accreted a significant amount of stellar matter.

\subparagraph{Dynamical formation in AGN disk.}

An alternative formation scenario for IMBHs is via repeated mergers and gaseous accretion in the disk of AGN \citep{McKernan:2012rf,Bellovary:2011jq}. In this framework, orbits of stars and BHs crossing the disk are damped by dissipation forces and settle in the disk plane.  Additionally, massive stars can form directly in-situ from fragmentation in marginally stable discs \citep{1987Natur.329..810S,Nayakshin:2005bm}, although the slope and extent of the mass function may depend on the SMBH mass and environment \citep{Derdzinski:2022ltb}.

The reduced relative velocities in the disk plane, the impact of gaseous dynamical friction, and the possible emergence of regions of the disk where inward and outward forces cancel out (so-called migration traps) favor scatterings and collisions that can rapidly build up an IMBH \citep{Tagawa:2020qll}. This channel can leave peculiar imprints in the distribution of  merger masses and spins, especially if the IMBH accretes gas from the disk after the formation (see the recent review by \cite{Sedda:2023big}).

This type of process could, in principle, favor the formation of an IMBH binary, although it is hard to assess the likelihood for this process to occur. 

\subparagraph{Multiple scenarios.}
The aforementioned scenarios do not work necessarily exclusively. For example, stellar mergers can lead to the formation of an IMBH seed that subsequently grows via repeated BH mergers. If the formation of an IMBH depends on the cluster structure, the cosmic distribution of IMBH mass should encode information on the structure and cosmic formation history of star clusters. 
The possibility to observe $\sim (10-10^3)$  mergers/yr involving an IMBH with ET \citep{Arca-Sedda:2020lso} will give us insights on the global IMBH mass distribution, at least at masses $< 10^3 \Msun$ \citep{Fragione:2022ams}, and their nursing environments \citep{Fishbach:2023xws}.

\subsubsection{IMBH binaries}

\paragraph{Massive star clusters.}
In dense star clusters, forming an IMBH-IMBH binary requires that multiple merger chains separately seed the two BHs (e.g., \cite{Gurkan:2005xz,Rasskazov:2019tgb}). This process is unlikely in young massive and globular clusters, where typically one massive star or BH grows and soon dominates the dynamics cannibalizing the population of BHs or their progenitors and preventing the onset of a new merger chain (e.g., \cite{2015MNRAS.454.3150G,2021MNRAS.501.5257R,Rantala:2024crf}). Two chains may develop if the cluster is sufficiently massive, dense, and rich in binaries \citep{2006ApJ...640L..39G}.

The merger of massive, Pop III, binaries can represent a viable way to form IMBHs with mass $<10^3 \Msun$ in high-redshift star clusters (see also section \ref{div3:pop3} for further details on this scenario). 

Alternatively, star cluster collisions can aid IMBH binary formation (e.g., \cite{Amaro-Seoane:2006zlf,Sedda:2019btz,Askar:2021rll}). 
In a similar scenario, the collision of molecular clouds and protoclusters in starburst environments can lead to the rapid build-up and pairing of IMBHs with nearly equal masses \citep{Rantala:2024crf,Shi:2024skj}.

The detection of a merging IMBH binary with mass $<10^3 \Msun$ by ET can help place constraints on the properties of the star clusters that can host such systems.

\paragraph{Dwarf galaxies.}
According to the $\Lambda$CDM scenario for structure formation, dwarf galaxy mergers are expected to be relatively frequent, as also supported by cosmological simulations \cite{Fakhouri:2010st,Deason:2014eea}. 
This process can aid IMBH binary formation (e.g., \cite{Bellovary:2018gbb,Tamfal:2018dfh,Khan:2021jqf}), although it is unclear whether dynamical friction is efficient enough as to prevent the IMBHs from staling and not reaching coalescence \cite{Tamfal:2018dfh}. 

Observing IMBH mergers in the mass range $\gtrsim 10^3 \Msun$ with ET could thus provide us with crucial insights about dwarf galaxy formation processes.

\subsubsection{Can ET decipher the origin of IMBHs?}

An instrument like ET has the potential to place stringent constraints on the mass distribution of double IMBH binaries with total masses $m_{\rm IMBH}\lesssim 3\times 10^3 \Msun$ at redshift $z>1$ \citep{Branchesi:2023mws,Fairhurst:2023beb}. Compared to the equal-mass binary case, the horizon redshift slightly changes if an IMBH paired with a smaller compact object is considered. 

\begin{figure}
\centering
\includegraphics[width=0.485\columnwidth]{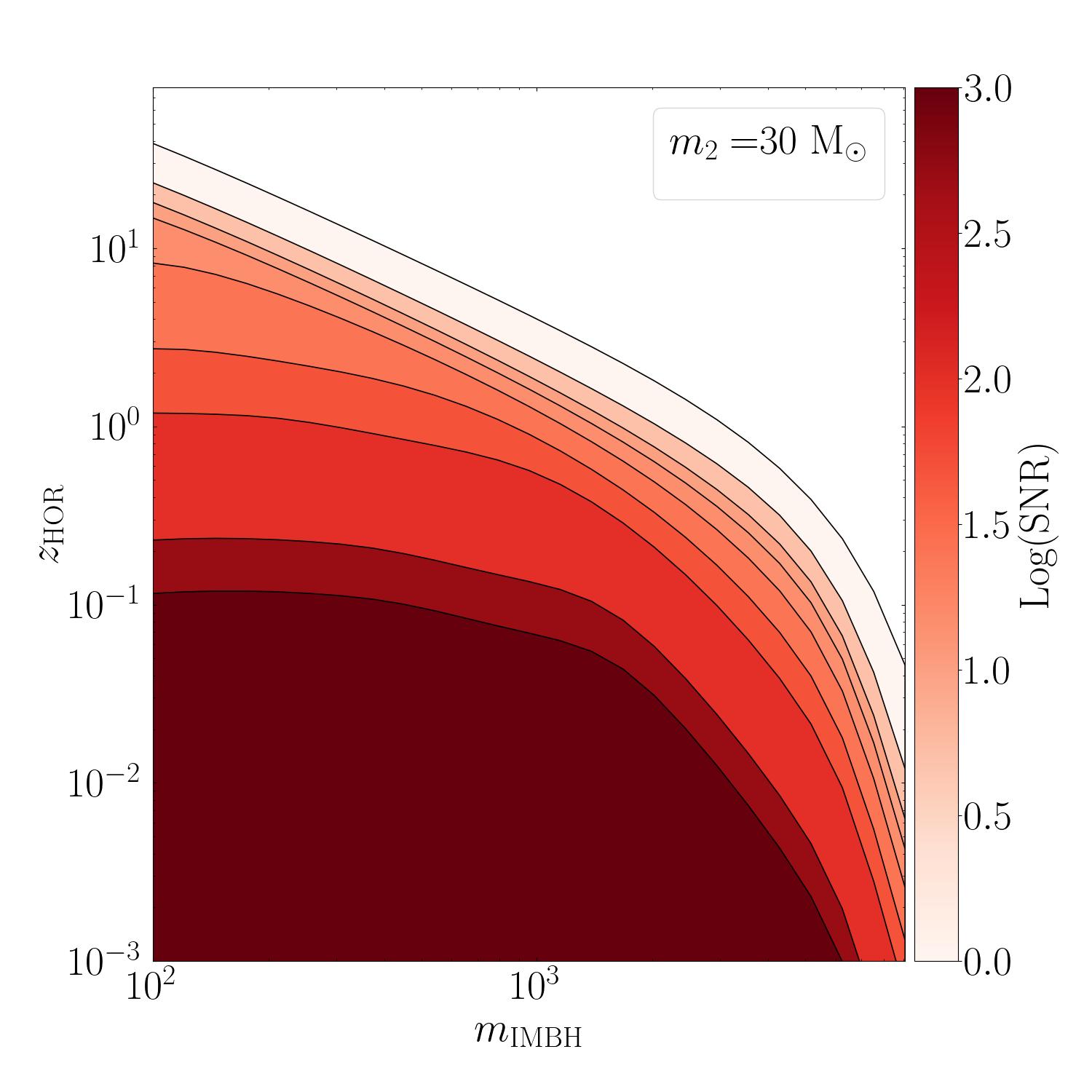}
\includegraphics[width=0.485\columnwidth]{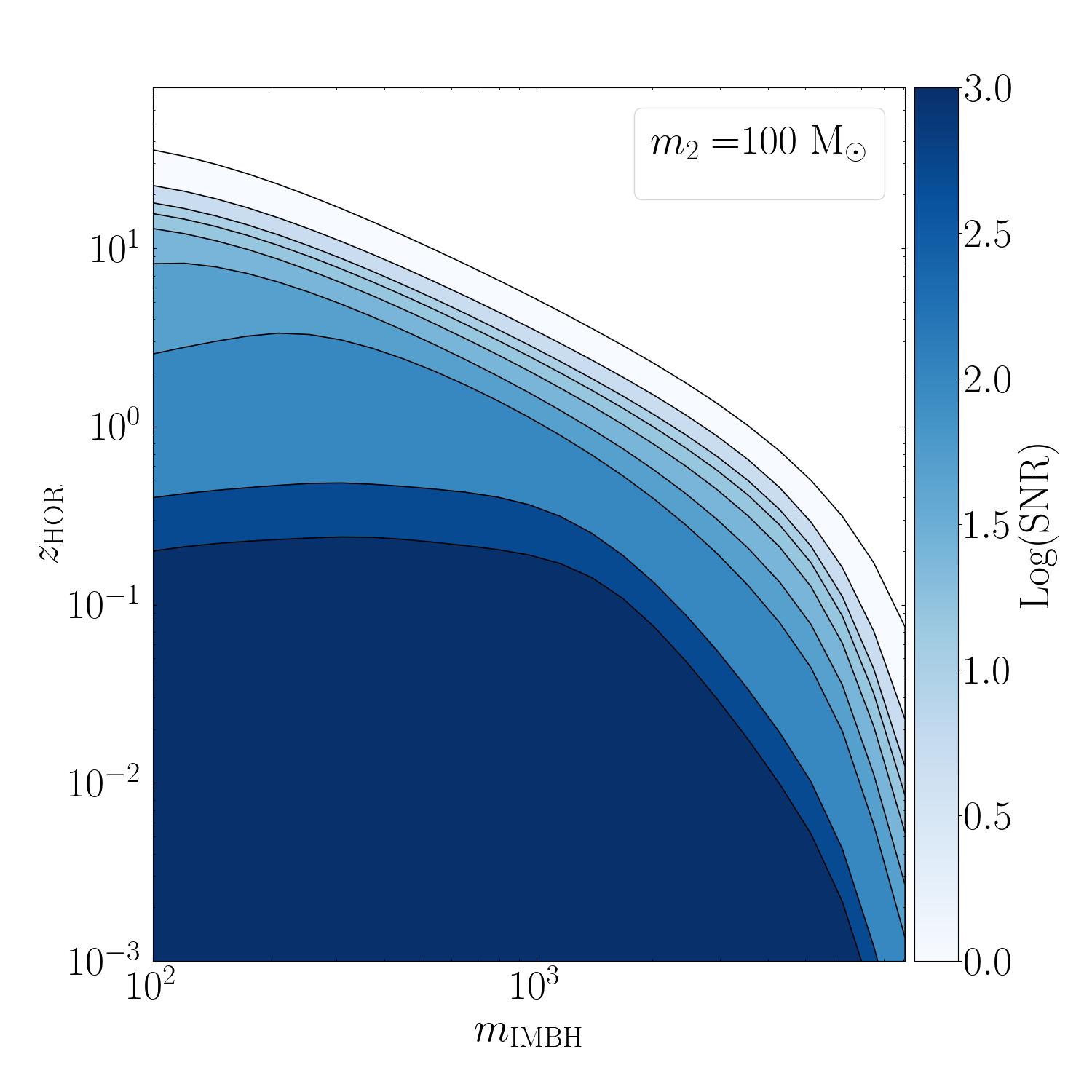}
\caption{Horizon redshift for IMBH-BH mergers assuming a companion mass of $30 \Msun$ (left panel) and $100 \Msun$ (right panel). From brighter to darker colours, contours encompass regions with an SNR $> 1,~5,~10,~15,~25,~50,~100,~500,~1,000$, respectively. Calculations are performed through the \textsc{GWFish} package \citep{Dupletsa:2022scg}, assuming a PhenomD waveform approximant.}
\label{fig:horimri}
\end{figure}

\begin{figure}
    \centering
    \includegraphics[width=0.75\linewidth]{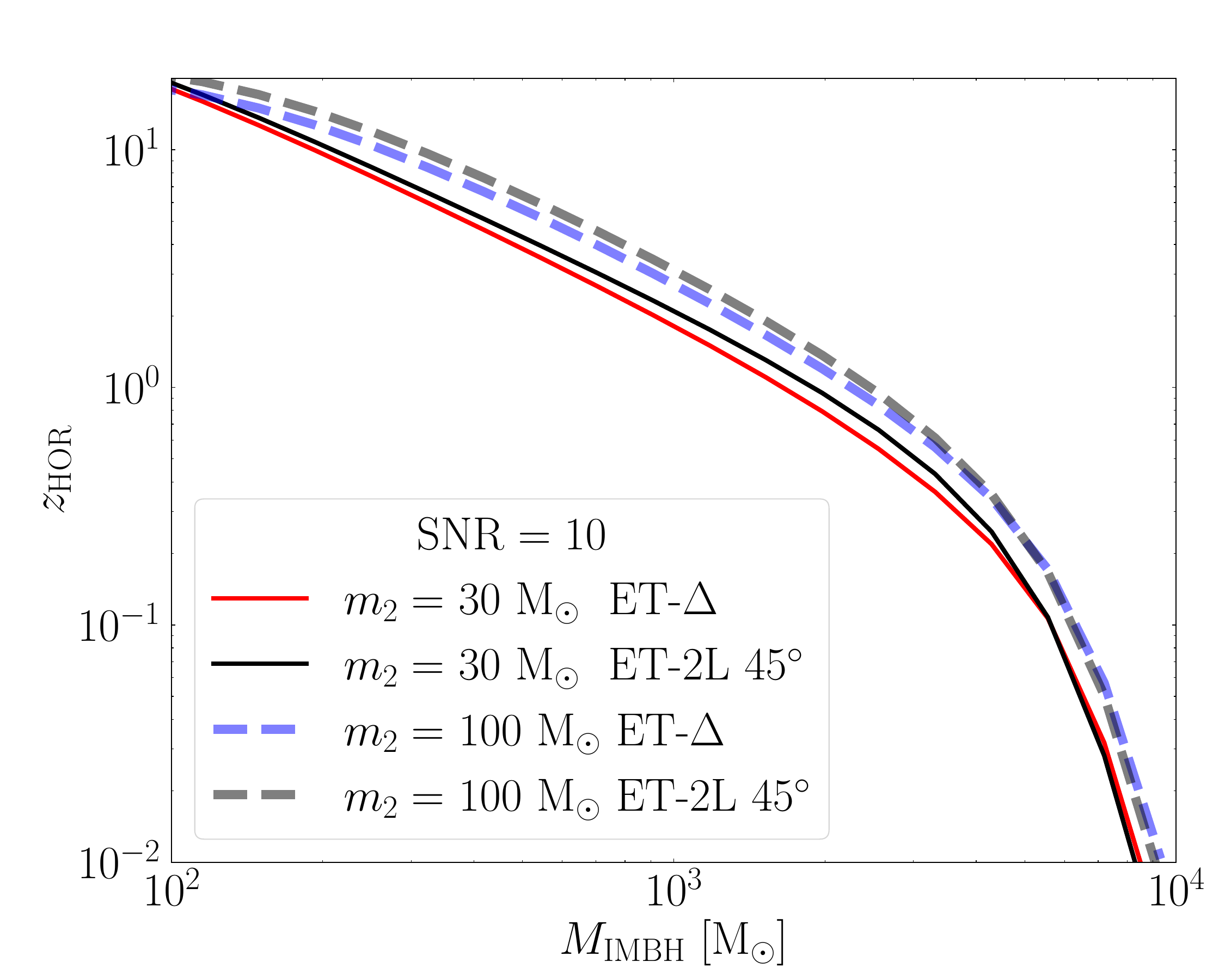}
    \caption{Horizon redshift for IMBH-BH mergers assuming a companion mass of $30 \Msun$ (solid line) and $100 \Msun$ (dashed line) for a fixed SNR$=15$ and comparing a 15km 2L ET configuration with $45\deg$ inclination (black) and a 10km triangle ET configuration (red and blue for the 30 and 100 M$_\odot$ companion, respectively). The sensitivity curves come from \cite{Branchesi:2023mws}. We used the \textsc{GWFish} package \citep{Dupletsa:2022scg}, assuming a PhenomD waveform approximant, i.e. the same as Fig.~\ref{fig:horimri}.}
    \label{fig:imri2lt}
\end{figure}

Figure~\ref{fig:horimri} shows ET horizon redshift for different values of the SNR and assuming a companion BH with a mass $m_2$ equal to  $30\, \Msun$ or to $100\,  \Msun$. ET can potentially detect merging IMBHs with masses in the mass range $(10^2-10^3)\, \Msun$ with a stellar (upper-mass gap) BH out to a redshift $\lesssim 3 (6)$ with an SNR $\gtrsim ~15$. Golden events at $z<1$ (with an SNR $>~ 100$) could help place stringent constraints on the mass of both the spin and mass of the merging components. ET can thus significantly boost our understanding of IMBHs and their cosmic evolution. We also compare a 2L ET configuration with 15 km arm length and $45\deg$ inclination and a triangle configuration with 10 km arm length and found minimal differences, see figure~\ref{fig:imri2lt}.

Essentially, ET will be able to detect IMBH binaries and intermediate-mass ratio inspirals (IMRIs). However, while the theory about SMBH-BH mergers, i.e. extreme mass-ratio inspirals (EMRIs), is  robustly defined, despite the main unsolved uncertainties, the same theoretical framework is hard to apply to IMRIs. SMBHs remain anchored at the host galaxy center, whilst IMBHs wander in the cluster core. Therefore, the EMRI formation is driven by relaxation of stellar motion in the cusp, and most EMRIs are expected to appear highly eccentric in LISA band. For IMBHs, the wandering motion can have a non-trivial effect on the global distribution of the eccentricity at the moment of IMRI's formation \citep{Amaro-Seoane:2007osp}. Depending on the measurement accuracy of IMRI eccentricities, ET could potentially help to place constraints on the IMRIs eccentricity distribution. For example, an excess of high-eccentricity IMRIs in ET band would support the idea that IMRIs form in massive and dense star clusters via close single-single and single-binary encounters (e.g., \cite{Konstantinidis:2011ie}).

Whether IMBHs are spinning depends on many uncertain quantities, like the natal spin distribution of stellar BHs, or the impact of mass transfer on the spin evolution. Recent numerical simulations based on both stellar dynamics and evolution suggest that young massive clusters can nurture the formation of IMBHs with masses in the range $(100-500)\, \Msun$ \citep{Maliszewski:2021jci, Gonzalez:2020xah, 2021MNRAS.501.5257R, ArcaSedda:2023mlv} and expected spins $\chi > 0.6$ \citep{ArcaSedda:2023mlv}. More massive IMBHs, $m_{\rm IMBH} > 10^3 \Msun$, forming through long chains of repeated BH mergers, instead, are expected to develop a final spin relatively small $<0.1$, leading to an anti-correlation between the IMBH spin and its mass \citep{Arca-Sedda:2020lso,Kritos:2022non}.
Stellar feeding can also have a significant effect: coherent accretion, for example, represents a viable way to spin up the IMBH while feeding on stellar material, similar to what is expected for SMBHs (for a recent review on the subject, see \cite{Reynolds:2020jwt}).  This would be the case in which a single, very massive star, collides with a stellar-mass BH, leading the resulting IMBH to gain most of its mass in a single event. Differently, when the mass growth occurs through slow accretion of small stars (e.g., \cite{Giersz:2015mlk, Stone:2016ryd}), the IMBH should spin-down. This may lead to a variety of IMBH sub-populations characterised by peculiar spin distributions at different cosmic times -- as the timescale regulating the IMBH accretion can vary by several orders of magnitudes -- depending on the process.  
Repeated mergers with BHs can damp the IMBH spin once the IMBH mass reaches a given threshold, i.e. the binary mass ratio falls below a certain value \citep{Arca-Sedda:2020lso}. If the IMBH forms in a gaseous medium, like an AGN disk, the accretion of gaseous material can also spin up the IMBH \citep{Tagawa:2020dxe}. 

Depending on the assumptions, models predict that ET could observe $10-10^3$ IMBH-BH mergers per year involving IMBHs with masses up to $\lesssim 10^3 \Msun$ mass range \citep{Arca-Sedda:2020lso}. This is somehow supported by the population of mergers observed during the O1-O3 runs by LVK, which lead to a predicted rate for equal-mass binaries with mass $200\, \Msun$ of 0.056\, ${\rm Gpc}^{-3}\,  {\rm yr}^{-1}$\citep{LIGOScientific:2021tfm}. Assuming that ET can detect all IMBH mergers and a constant merger rate up to redshift 8 leads to $\sim 160$ detections per year. 
Since ET can measure the IMBH spin down to $\sigma_\chi \simeq 10^{-3}$ \citep{Huerta:2010tp}, the large number of possibly detectable IMBHs can provide crucial clues on the spin distribution of merging IMBHs.  

 For the sake of comparison, we show in table~\ref{tab:IMRI} the SNR and the relative error in measuring the IMBH mass and spin for different assumptions of the IMBH-BH binary properties. From our analysis, performed with the \textsc{GWFish} package, we find that ET could observe IMBH-BH binaries with mass $<10^3 \Msun$ out to redshift $z<4$ with an ${\rm SNR}\, \gg\,  1$.

\begin{table}
\centering
\begin{tabular}{ccccccc}
\hline\hline
$z$ & $m_1$ & $m_2$ & $\chi_{1,2}$& SNR & $\sigma_{m_1}/m_1$ &$\sigma_{\chi_1}/\chi_1$ \\
\hline
\multicolumn{7}{c}{IMRI}\\
\hline
$1$ & $100$ & 30 & $0.5$ & $117$& $0.006$ &$0.005$\\
$1$ & $500$ & 30 & $0.5$ & $209$& $0.0006$&$0.002$\\
$1$ & $1000$& 30 & $0.5$ & $67$ & $0.006$ &$0.01$\\
$1$ & $3000$& 30 & $0.5$ & $6$  & $0.04$  &$0.07$\\
$4$ & $100$ & 30 & $0.5$ & $39$ & $0.02$  &$0.04$\\
$4$ & $500$ & 30 & $0.5$ & $33$ & $0.01$  &$0.02$\\
$4$ & $1000$& 30 & $0.5$ & $4$  & $0.1$   &$0.3$\\
$4$ & $3000$& 30 & $0.5$ & $0.1$& $3.2$   &$10.1$\\
\hline
\multicolumn{7}{c}{IMBH binaries}\\
\hline
$1$ & $100$ & $100$ & $0.5$ & $90$ & $0.04087$ & $0.08624$\\
$1$ & $500$ & $500$ & $0.5$ & $564$ & $0.03564$ & $7.186e-07$\\
$1$ & $1000$ & $1000$ & $0.5$ & $242$ & $0.01833$ & $2.578e-07$\\
$1$ & $3000$ & $3000$ & $0.5$ & $16$ & $0.07471$ & $3.146e-06$\\
$4$ & $100$ & $100$ & $0.5$ & $46$ & $0.1018$ & $0.001492$\\
$4$ & $500$ & $500$ & $0.5$ & $44$ & $0.03316$ & $7.034e-07$\\
$4$ & $1000$ & $1000$ & $0.5$ & $4$ & $0.1561$ & $3.331e-06$\\
\hline
\end{tabular}
\caption{Col. 1: redshift. Col. 2-3: primary and companion mass. Col. 4: component spins. Col. 5: SNR. Col. 6-7: relative error on the IMBH mass and spin. Simulations are carried out with \textsc{GWFish} \citep{Dupletsa:2022scg}, assuming a PhenomD waveform approximant}.
\label{tab:IMRI}
\end{table}

\begin{figure}
    \centering
    \includegraphics[width=0.75\linewidth]{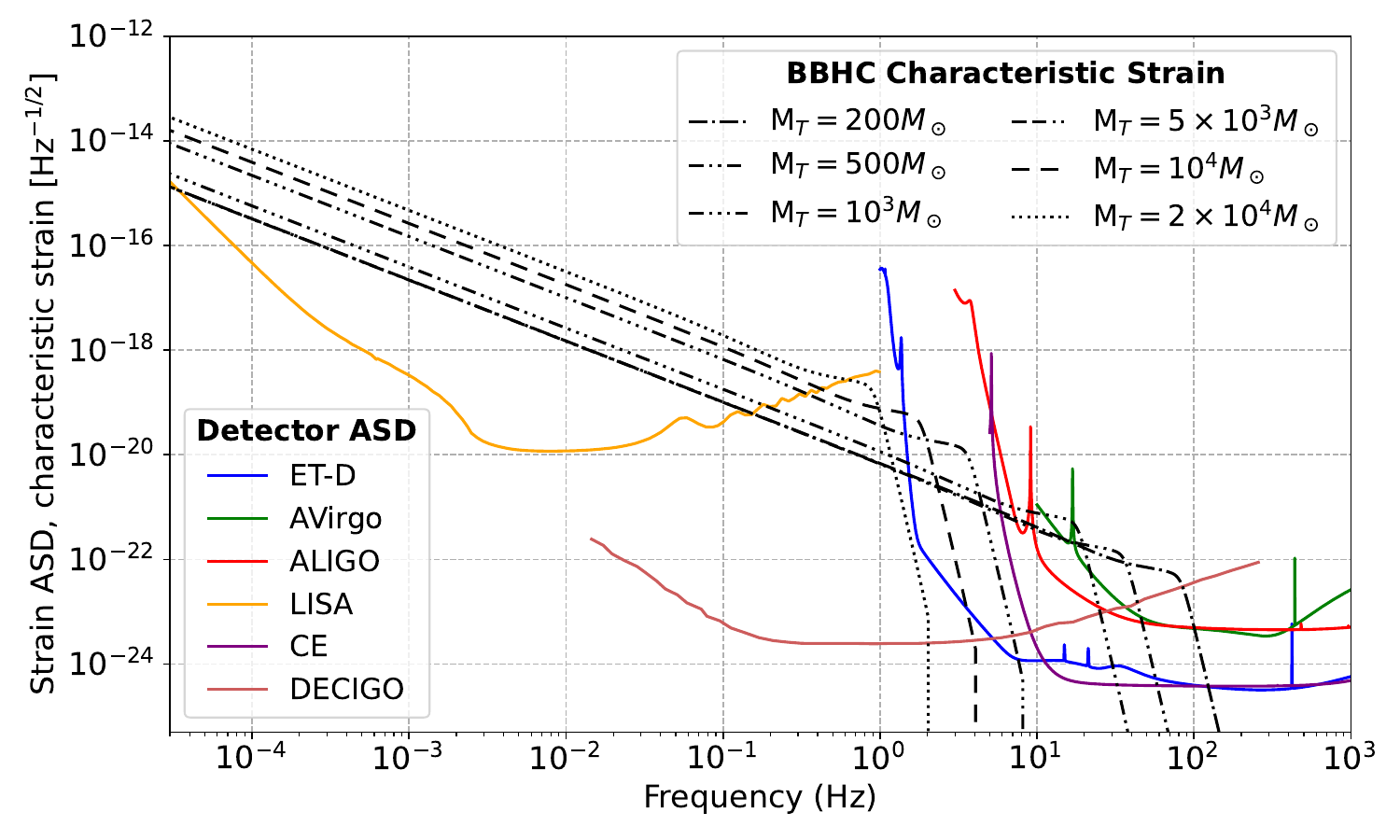}
    \caption{Characteristic strain amplitude and frequency for IMBH-IMBH binaries with different masses, assuming a luminosity distance $d_L = 1$ Gpc. The signal is overlaid to the sensitivity curves of various detectors at low-frequency (LISA), mid-range (DECIGO), and high-frequency (ET, CE, AVirgo, and ALIGO).}
    \label{figIMBHASD}
\end{figure}

Therefore, ET has the unique possibility to deeply probe the IMBH mass range within $m_{\rm IMBH} < 10^3 \Msun$, a region of the phase space that cannot be accessed with detectors operating at lower frequencies, like LISA.  In some cases, the emitted signal could also be observed in different bands, depending on the binary location, distance, and orbital parameters. An example of multiband sources is shown in figure~\ref{figIMBHASD}, where the signal coming from different IMBH-IMBH binaries is compared against the SNR of several detectors. A network of terrestrial detectors could also represent a viable way to place strong constraints on IMBH properties, allowing to precisely measure masses, spins, and redshift within O$(0.1-10)\%$ error at redshift $z<10$ (see for example figures~6 and 7 in \cite{Reali:2024hqf}). Therefore, exploiting the synergies among ET and lower-frequency detectors can enable us to obtain a full picture of the IMBH mass range (see section~\ref{section:div5} for a more detailed discussion on these topics).

\subsection{The host galaxies of binary compact object mergers}

The properties of host galaxies where binary compact objects form and merge provide complementary information to understand how these binary systems evolved. Identifying the host galaxies can be 
challenging and, in principle, requires an electromagnetic counterpart emission after the GW event. 
While some of the GW events from the coalescence of binary neutron stars (BNS) and black hole - neutron star binaries (BHNS) would be accompanied by an electromagnetic counterpart detection, this is possibly not the case for binary black hole (BBH) mergers, which are likely to be EM-dark (although BBH mergers in dusty regions such an AGN disk is a possible scenario for an EM counterpart event, see \cite{Graham:2020gwr}).

To date, the sole unequivocally identified host is NGC 4993, an early-type galaxy with a stellar mass of $(0.3 - 1.2)\, \times 10^{11} ~{\rm M_{\odot}}$, a mean stellar age of $3$~Gyr and a metallicity of $\sim (0.2 - 1.0)\, Z_\odot$ \cite{Levan:2017ubn,Im:2017scv,Ebrova:2018gtz,Kilpatrick:2021aav}. The merging binary neutron star, observed as the GW event GW170817, occurred in this galaxy \cite{LIGOScientific:2017bnn,LIGOScientific:2017zic, Goldstein:2017mmi,Savchenko:2017ffs, Margutti:2017cjl, Coulter:2017wya, DES:2017kbs, Chornock:2017sdf, Cowperthwaite:2017dyu, Nicholl:2017ahq, Pian:2017gtc, Alexander:2017aly}. This event coupled with its electromagnetic counterpart shed light on a wide range of physical phenomena. It confirmed that GWs travel at the speed of light and validated binary neutron star mergers as both progenitors of short gamma-ray bursts \cite{LIGOScientific:2017vwq} and prolific sites for the formation of heavy elements \cite{Kasen:2017sxr}.

Studying the host galaxies of binary compact object mergers is crucial for several reasons. First, the identification of the most likely host galaxy can help to improve the electromagnetic counterpart detection of GW events (section \ref{sec:multimessenger}) which, as discussed in detail in section~\ref{sec:Cosmography}, provides an alternative measurement of current methods to constrain the cosmological parameters such as the Hubble constant. Second, galaxy formation and evolution throughout the history of the Universe affects stellar populations, thereby shaping the likely pathways for compact object mergers (section \ref{sec:HGformationchannel}). Here, we explore these aspects in anticipation of future detections with ET.

\subsubsection{Multi-messenger detections and sky localization of the host galaxies}    
\label{sec:multimessenger}

\begin{figure}
    \centering
    \includegraphics[width=0.9\textwidth]{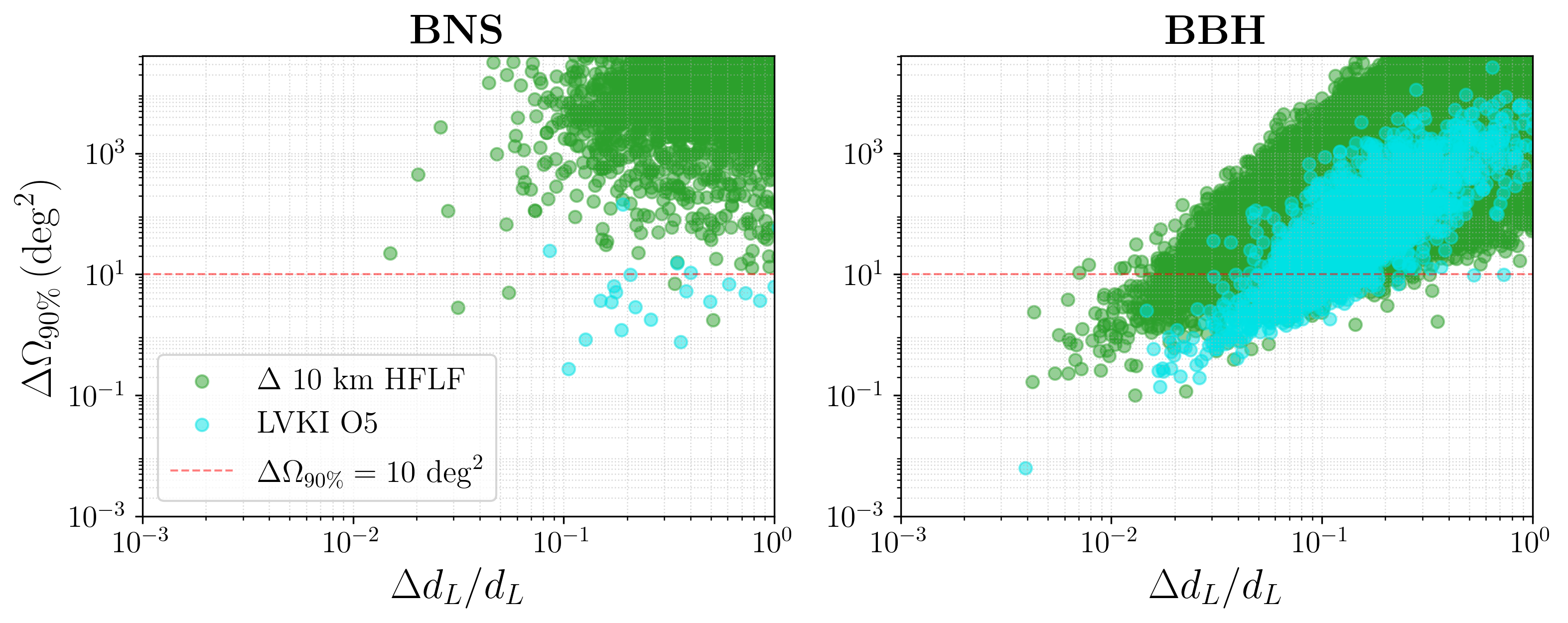}
    \caption{The joint accuracy on luminosity distance $(\Delta d_L/d_L)$ and angular resolution $(\Delta \Omega_{90\%})$ for BNS (left) and BBH (right). Green shows results from the triangular ET detector (10-km arms) with both high- and low-frequency instruments (HFLF). Light blue indicates forecasts for LVKI O5, the most advanced 2G detector network (LIGO Hanford, LIGO Livingston, Virgo, KAGRA, and LIGO India) \citep{KAGRA:2013rdx}. The dashed red line represents a 10 deg$^2$ sky localization error. Adapted from \citep{Branchesi:2023mws}.
}
    \label{fig:skyloc}
\end{figure}

The large sky localization inferred with current GW detectors, spanning hundreds of square degrees \cite{KAGRA:2013rdx}, represents a challenge for host galaxy identification. 
The impact of this uncertainty on the efficiency of host galaxy identification is substantial, as it determines the number of pointings required by electromagnetic instruments to cover the GW signal location and the corresponding observational time to be used. This challenge, however, holds promising prospects with the deployment of ET. Its low-frequency sensitivity promises to observe binary compact objects for a longer period of time before the merger occurs \cite{Branchesi:2023mws}. For instance, a BNS signal analogous to GW170817 can be visible by ET $> 20$ hours before merger \cite{Iacovelli:2022bbs} (see also  figure~\ref{fig:GW170817atET} in section~\ref{section:div10}).  This aspect will significantly increase the number of BHNS and BNS mergers with a sky localization of less than 10 square degrees \cite{Iacovelli:2022bbs,Branchesi:2023mws,Mo:2024frl}. Figure~\ref{fig:skyloc} shows that the angular resolution expected for the most advanced 2G network at full O5 sensitivity (LVKI O5: LIGO Hanford, LIGO Livingston, Virgo, KAGRA, and LIGO India) is similar to that of a single 10-km ET triangular observatory. However, ET detects over 10 times more events, including those with high SNR, resulting in a nearly tenfold reduction in the error on luminosity distance \citep{Branchesi:2023mws}. 

Despite the improved sky localization achieved with ET, the electromagnetic follow-up process still tipically entails scanning through multiple galaxies (although in some case the angular resolution could be sufficiently good  that there will be only one galaxy in the localization volume, see figure~\ref{fig:ET_loc_BBH} in section~\ref{section:div2}). Therefore, optimized follow-up strategies can be accomplished through simulation-informed ranking of galaxies within a given sky localization. We can derive the probability that a galaxy hosts a compact object coalescence based on various properties, such as its stellar mass and star formation rate (see e.g. \cite{Toffano:2019ekp,Artale:2020swx,Rinaldi:2022kyg,Ducoin:2022ftb}. The refinement of simulation-informed ranking, where multiple formation scenarios of compact object mergers are taken into account, is particularly crucial for identifying the host galaxies of compact objects where electromagnetic emission is impossible or improbable, such as BBH mergers and BHNS mergers \cite{Colombo:2023une}. 

As we saw in section~\ref{sec:Cosmography},
host galaxy identification is also crucial in the use of compact binary mergers as ``bright'' sirens, in order to infer cosmological parameters.
As discussed in particular in section~\ref{sect:darksirensdiv2}, even with the lack of an EM counterpart detection, galaxies with known redshift within the sky localization can be ranked and used to infer the cosmological parameters with statistical methods, particularly  if the localization volume contains a limited number of possible host galaxies.
The thousands of  GW events that  ET will detect with its improved sky localization will provide a significant step forward in determining cosmological parameters.

\subsubsection{Formation channels and their link with galaxy properties}
\label{sec:HGformationchannel}

As we have discussed, the improvements expected for third-generation detectors will provide a number of simultaneous identification of GW events and their host galaxies at different cosmic times \cite{Branchesi:2023mws}, either through the detection of an electromagnetic counterpart (at least for BNSs) or, for BBHs, through the statistical association with a limited number, or possibly just a single host galaxy in the localization volume. These concurrent detections (or statistical associations)  offer a unique opportunity to gain additional insights into the underlying physics governing the formation channels and evolution of compact objects (see e.g. \cite{OShaughnessy:2016nny,Lamberts:2018cge,Artale:2019doq,Adhikari:2020wpn,Perna:2021rzq,Rose:2021ykg,Chu:2021evh,Vijaykumar:2023bgs}).

As mentioned previously (see section \ref{div3:formationchannel} and reference therein), the two main proposed pathways are the isolated formation channel, which includes processes like  common envelope evolution, chemically homogeneous evolution, and stable mass transfer; and the  dynamical formation channel taking place in dense stellar clusters (e.g. galactic center, open star clusters, globular clusters and young star clusters). Current detections suggest that multiple formation channels are needed to explain the origin of BBHs (see e.g. \cite{Zevin:2020gbd}). Furthermore, most binary compact object systems are formed in galaxies. Therefore, the host galaxy properties are an excellent complement to understand the role of the different formation channels of gravitational-wave events across cosmic time.

Galaxy formation and evolution is regulated by a large diversity of baryonic processes shaping their star formation history, chemical enrichment, stellar mass, gas abundance, among others \citep{Somerville:2015}. These factors impact the stellar populations' properties and, in turn, the most favorable formation channels of binary compact object mergers.

For instance, galaxies in high-dense regions often undergo major mergers increasing the probability of hosting a large number of stellar clusters \cite{2018RSPSA.47470616F,2019ApJ...879L..18L}. These galaxies, in turn, would produce more BBHs through dynamical interactions \cite{Bruel:2023bxl}. Notably, observations from the local Universe indicate a strong dependence of the number and mass of globular clusters on the total stellar mass of the galaxies, and a tendency to be more numerous in early-type galaxies \cite[]{Harris:2013,Brodie:2006,Beasley:2020}. Furthermore, stellar clusters and active galactic nuclei are considered potential factories of BBH mergers in the pair-instability mass gap and above (see e.g. \cite{Tagawa:2020qll,Bruel:2023bxl,Vaccaro:2023cwr}).

The delay time, defined as the time for compact objects to merge, is important for connecting galaxy properties with binary compact object formation channels. It
depends on the outcome of specific physical processes experienced by stars in binary systems such as the common envelope phase in the isolated scenario (see section \ref{div3:formationchannel}). In particular, the number of mergers that occur in massive and passive galaxies tends to increase when the delay time is longer \cite{Artale:2020swx,Santoliquido:2022kyu}. The wide redshift range expected for ET detections \cite{Punturo:2010zz,Reitze:2019iox,Maggiore:2019uih}, together with their electromagnetic counterparts will play a fundamental role in deciphering the contribution of the different formation channels across cosmic time.

\begin{figure}
    \centering
    \includegraphics[width=0.8\textwidth]{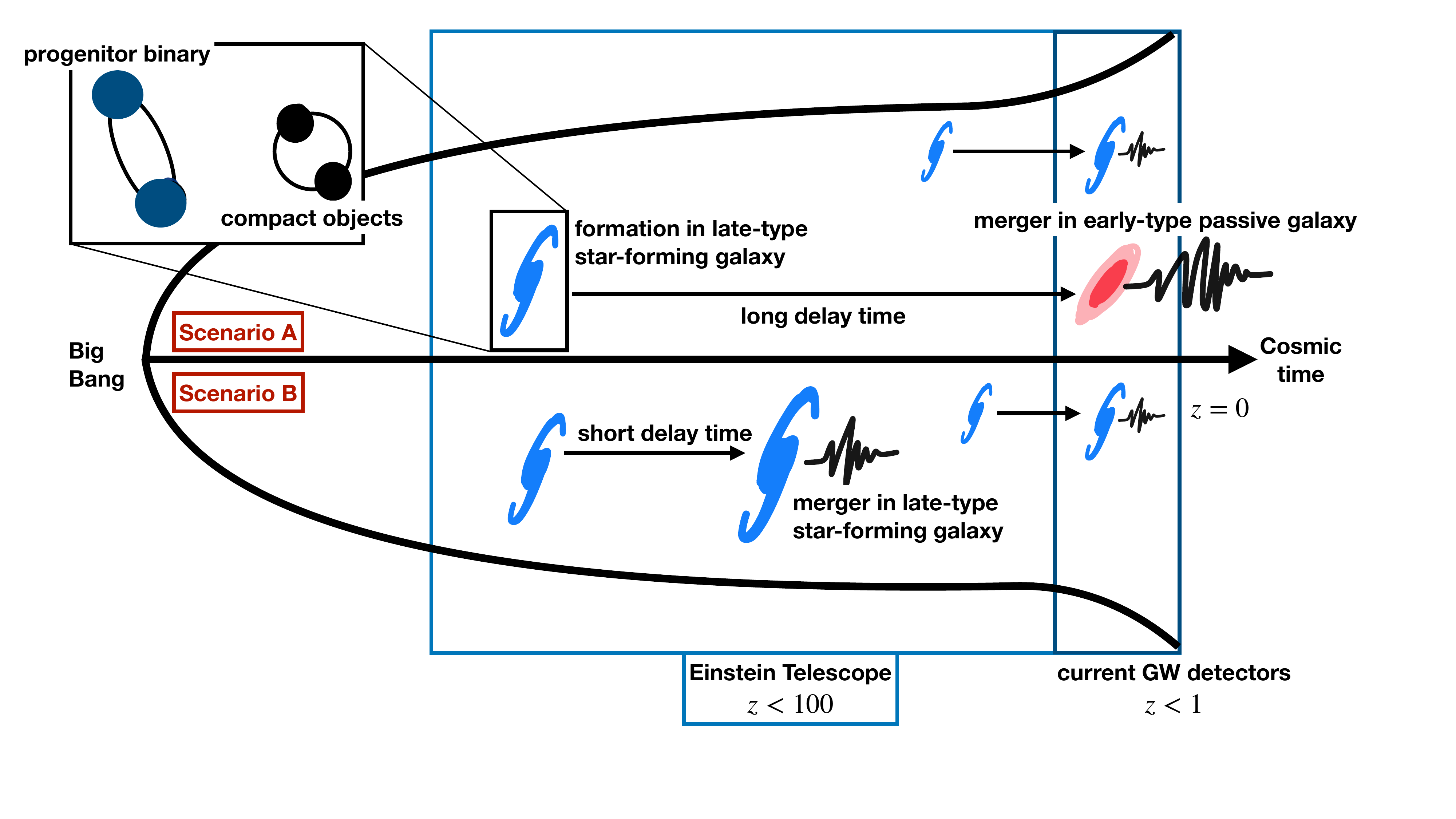}
    \caption{ET will explore the interplay between the physics of compact objects, originating from massive stars, and the galaxies hosting their mergers. Presently, with current GW detectors, exploration is confined to the local Universe ($z < 1$), where mergers can occur in early-type passive galaxies, like NGC 4993 which hosted GW170817. In contrast, ET will be able to observe BBH mergers with sky localization error of less than $10$ deg$^2$ up to $z \sim 6$ (see figure~10 in \cite{Branchesi:2023mws}), opening to completely new perspectives in host galaxy identification. 
    }
    \label{fig:scheme}
\end{figure}

The chemical abundance of the stellar populations within galaxies is another key piece of information. Theoretical inferences from binary population synthesis simulations suggest the efficient formation of BBHs at sub-solar metallicities \cite{Giacobbo:2017qhh, Kruckow:2018slo, Chruslinska:2018hrb, Santoliquido:2020axb, Broekgaarden:2021efa}. Consequently, low-mass galaxies ($< 10^8 \Msun$) emerge as promising sites for detecting these mergers \cite{Boco:2019teq, Boco:2020pgp}. However, their faintness and high-redshift nature ($z > 4$) make them challenging to constrain through EM observations. As a result, ET emerges as a promising tool for delving into the chemical evolution of galaxies \cite{Vitale:2018yhm, Graziani:2020ekn}. As host galaxies leave distinctive imprints on compact object mergers throughout cosmic time, the identification capability of ET becomes crucial. If these galaxies have features like low mass and low metallicity, it implies a relatively short delay time for BBHs on average. This breakthrough will effectively resolve the degeneracy between the impacts of formation metallicity and delay time distribution \cite{Chruslinska:2022ovf}, addressing a significant challenge in gravitational-wave astrophysics. This is also shown in figure~\ref{fig:scheme} schematically.

\subsection{Populations backgrounds}\label{sect:Populationbackdiv3}

The incoherent superposition of the GW signal from all astrophysical sources in our Universe creates a stochastic background whose amplitude and spectral shape reflect the properties of each source population contributing to it. The stochastic GW background (SGWB) is thus a powerful astrophysical probe, complementary to the detection of individual astrophysical sources.

As discussed in section~\ref{sec:stochback} a SGWB, either of astrophysical or cosmological origin, is characterised in terms of the dimensionless energy density parameter, defined in \eq{eq: OmegaGW}, that we recall here, 
\begin{equation} \label{eq:omega_gwdiv3}
    \Omega_{\mathrm{GW}}(f) = \frac{1}{\rho_c} \frac{\mathrm{d}\rho_{\mathrm{GW}}(f)}{\mathrm{d} \ln f}\,,
\end{equation}
where $\rho_c$ is the critical energy of the Universe. The astrophysical component can be schematically parametrised as:
\begin{equation}
    \Omega_{\mathrm{GW}}(f) = \frac{f}{\rho_c c^2}\, \sum_{\mathrm{i}} \int^{z_\mathrm{max}}_0 \int_{\lambda} \,\,\frac{R_\mathrm{i}(z,\lambda)\,  \, P_i(\lambda) }{ (1+z)\, H(z)}\frac{\mathrm{d}E_\mathrm{GW,i}(f_s,\lambda)}{\mathrm{d}f_s}\,\mathrm{d}\lambda \,\mathrm{d}z\,,
    \label{eq:sgwb-astro}
\end{equation}
where $H(z)$ is the Hubble parameter, the index $i$ runs over different source classes (such as BBH, BNS, supernovae, neutron stars) and the parameter $\lambda$ encodes all the source properties of each class (such as masses and spins for CBC, ellipticity for rotating neutron stars etc.). Finally, $R_\mathrm{i}(z,\lambda)$ is the rate of each source and $\mathrm{d}E_\mathrm{GW,i}(f_s,\lambda)/\mathrm{d}f_s$ is the energy spectrum it emits.

We will focus our discussion here on CBCs and the unique information that can be obtained with ET. The main challenges are studying BBH formation channels, searching for Population III stars and detecting the residual BNS background. We will also mention other sources and their contribution to the SGWB and discuss synergies with electromagnetic observations.

We note that the background defined in \eq{eq:sgwb-astro} is the total background. In the case of CBCs, it can be relevant to study the residual background obtained after subtracting individually detected sources. On the one hand this background is complementary from individual detection and can provide statistical information about the faintest sources at the largest distances, and  on the other hand subtracting sources can help observing backgrounds that are masked by the total background. 
For LIGO-Virgo-Kagra, the number of sources detectable individually is sufficiently small so that the distinction between total background and residual one (obtained after resolvable sources are filtered out), is very mild. However, with its highly improved sensitivity, ET will detect a very large number of individual sources:  their accurate subtraction is crucial and will pose itself an interesting challenge. Indeed, as discussed in section~\ref{sect:subtractastrobkgdi9}, suboptimal individual sources subtraction will manifest itself as an additional stochastic component. In the rest of this section we will assume that source subtraction is completely noiseless and discuss the scientific questions that SGWB observations can address. We will discuss both the total background (without subtraction) and the residual background (obtained after noiseless subtraction).

We note that, even if most of the BBHs are expected to be detectable individually by ET, this will not be the case for a population of BNSs: since they are fainter  than BBH and with a longer time in band, they are expected to produce superposed signals, and the bulk of the population is expected to be subthreshold (see figure~14 of \cite{Iacovelli:2022bbs} or
figure~\ref{fig:cumul_SNR-Mc-Om-dL_BNS_allconf_ETCE} in section~\ref{section:div9}). A background approach will provide us with information on this population that cannot be achieved with a catalog (individual detections) approach alone. However, even for BBHs, using a background approach to reconstruct parameters such as merger rate and mass distribution is useful to perform a cross-check with the same quantities extracted from a catalog. A discrepancy might be the indication of the presence of a faint subpopulation, that is not captured when looking at individual events. A population of PBHs extends to redshift higher than the instrument horizon, and the properties of a high-$z$ population can be extracted only with a background approach. Finally, since astrophysical backgrounds represent a foreground for sources of primordial origin, understanding their properties is crucial for accurately subtracting their contribution from an observed background map and looking at cosmological components.

\subsubsection{Study of BBH formation channels } 
\label{subsec:BBHchannel}


Detection and characterization of the SGWB will provide additional information on source population properties. One of the main contributions to the signal in ET is expected to be from BBH mergers occurring at relatively low redshifts ($z<5$), with masses in the range $\sim (10-10^3)\, \Msun$ \cite{Branchesi:2023mws}. The formation channels of these BBH, as discussed in section~\ref{div3:formationchannel}, can be broadly divided into the isolated channel (including formation from isolated field pairs and multiples), and the dynamical channel. The latter including formation in globular (GC), nuclear (NSC) or young star clusters (YSC), and gas-driven formation in AGN disks. These formation channels leave an imprint on the masses, spins and eccentricities of the resulting BBH, as well as the merger rate, the distribution of time delays and the evolution of the mass spectrum with redshift. These differences will manifest themselves in the SGWB detectable with ET.

    \begin{figure}[t]
    \centering
    \includegraphics[width=0.8\textwidth]{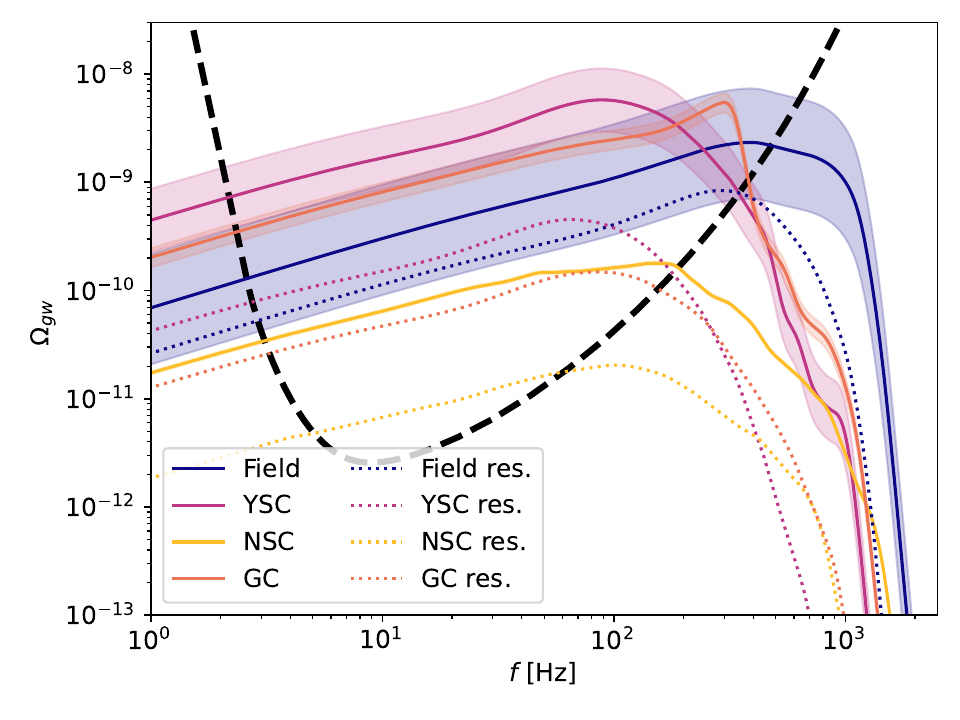}
    \caption{Background energy density spectrum for BBHs assuming four evolution channels, isolated (Field), Young Star Cluster (YSC), Globular Cluster (GC) and Nuclear Star Cluster (NSC). Uncertainties, depicted by the shaded areas, are derived from the merger rates given in table~2 of \cite{Mapelli:2021gyv}. The black dashed curve shows the sensitivity of ET for the triangular configuration.}
    \label{fig:BBH_channels}
    \end{figure}

As an example, in figure~\ref{fig:BBH_channels} we show the contributions to the SGWB from BBH formed via the isolated channel (Field) and three variants of the dynamical channel: YSC, NSC and GC. All the computations were performed with the public code \texttt{PRINCESS}\footnote{See \url{https://gitlab.com/Cperigois/Princess} .} using population models from \cite{Mapelli:2021gyv}. We show both the total and residual backgrounds, where the latter are  computed by removing all sources having an individual SNR $\rho_{\rm thrs.}>20$ (neglecting the error in the subtraction;
see section~\ref{sect:subtractastrobkgdi9} for the treatment of the subtraction error)
and assuming a random inclination and position in the sky. The shaded areas shown for the Field, YSC and GC channels reflect the uncertainty in the merger rates for each population model, and are derived from table~2 in \cite{Mapelli:2021gyv}. The model uncertainty of the NSC channel is relatively small (up to $30\%$).
We also show the power-law integrated sensitivity curve of ET to stochastic backgrounds, computed for the triangular configuration and assuming SNR of $2$ and $1$ year of observing time with duty cycle of $50\%$. As discussed in \cite{Branchesi:2023mws}, this sensitivity depends on the  configuration: choosing two L-shaped detectors decreases the sensitivity above $\sim 100$ Hz because the overlap reduction function drops significantly. However, L-shaped configurations where the detectors are co-aligned perform better at low frequencies than the triangular ones.

As can be seen in figure~\ref{fig:BBH_channels}, the spectral shape at low frequencies is dominated by the inspiral phase of each BBH and therefore has a constant slope close to $f^{2/3}$. At higher frequencies, however, we start observing features specific to each formation channel that reflect the distribution of masses and redshifts of each population \cite{Bavera:2021wmw, Perigois:2021ovr}. In particular, the frequency at which the spectrum deviates from a power-law corresponds to the minimal mass in the BBH population (the peak is relatively wide due to redshifting effects). In figure~\ref{fig:BBH_channels}, this deviation starts at a few hundred Hz for the Field formation channel, and at $\sim 100$ Hz for the YSC and GC channels. Thus, observing the shape of the SGWB can provide important clues as to the formation scenario of the sources.

\subsubsection{Study of Population III stars}
\label{subsec:popIII}

    \begin{figure}[t]
    \centering
    \includegraphics[width=0.8\textwidth]{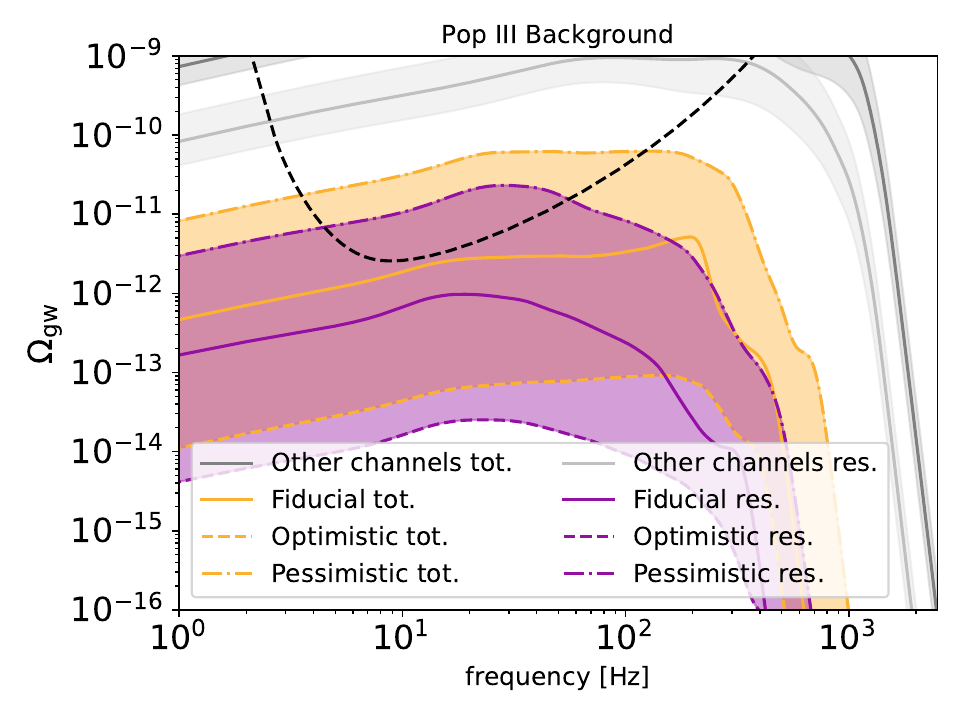}
    \caption{Background energy density spectra for BBHs born from Pop.~III stars. Fiducial, Optimistic and Pessimistic models are the LOG\_H22, LAR\_H22 and LAR\_J19 respectively populations models from \cite{Santoliquido:2023wzn}. The residuals correspond to $\rho_{\rm thrs.} > 20$, and the grey lines correspond to the background from Pop.~I/II channels (see section~\ref{subsec:BBHchannel}).}
    \label{fig:popIII_bkg}
    \end{figure}

As discussed in section~\ref{sec:pop3}, Pop III stars are the first stars that formed in the Universe. Similarly to Pop I/II stars, they form BBHs via the isolated or dynamical channels. However, Pop III remnants are expected to have unique properties that may help to distinguish them from Pop I/II remnants, in particular merger rate evolution with redshift and mass spectrum. These properties are also imprinted on the SGWB they produce \cite{Inayoshi:2016hco,Perigois:2020ymr,Martinovic:2021fzj,Kouvatsos:2024eok}.

As an example, in figure~\ref{fig:popIII_bkg} we show the total and residual SGWB, where the source populations are taken from \cite{Santoliquido:2023wzn}. The residual signal was calculated assuming all sources with an individual signal to noise ratio $\rho_{\rm thrs.} > 20$ are subtracted noiselessly. Note that the SGWB from Pop III remnants starts deviating from a power-law at lower frequencies compared to Pop I/II remnants, as a consequence of their higher masses and higher merger redshifts. 

We note that the residual signal form Pop I/II remnants in the model shown in figure~\ref{fig:popIII_bkg} is more than two orders of magnitude above that of Pop III remnants due to their low merger rate. Therefore a better understanding of Pop III remnants is needed in order to efficiently extract information from the SGWB. Nevertheless, even if the residual Pop III signal is indeed much below the one from Pop I/II, SGWB observations will contribute to putting upper limits of high-$z$ merger rates. In addition, the predictions for Pop III remnant merger rates and masses are much more uncertain that those for Pop I/II. Finally, we note that the ability to accurately subtract the signal from individually detected Pop III remnants  highly depends on ET performance at low frequencies.

\subsubsection{Primordial black hole contribution}\label{subsec:pbh}

Primordial black hole mergers may lead to a sizeable SGWB  \cite{Clesse:2016ajp,Chen:2018rzo,Chen:2019irf,Mukherjee:2021ags,Bagui:2021dqi,Bavera:2021wmw}.\footnote{The SGWB associated with the PBH formation mechanisms is discussed in section~\ref{sect:PBHprimpertdiv2}.}
As indicated in \eq{eq:omega_gwdiv3}, the SGWB results from the incoherent superposition of stellar and BH mergers, and it gets contributions from the entire merger history. In other words, it is sensitive to the rate evolution with redshift of the channels responsible for GW emission.
Crucially, the PBH scenario is characterized by a merger rate that grows monotonically with redshift~\citep{Ali-Haimoud:2017rtz,Raidal:2018bbj,DeLuca:2020qqa} 
$
{\cal R}_\text{\tiny PBH} (z) \propto t^{{-34}/{37}} (z)\,, 
$
extending up to early epochs at $z\gtrsim{\cal O} (10^3)$, 
see also discussion in section~\ref{sec:PBH_rate evolution}.\footnote{We do not consider the contribution to the SGWB from PBH hyperbolic encounters, since it is typically subdominant \cite{Garcia-Bellido:2021jlq}.}

\begin{figure*}[t]
\centering
\includegraphics[width=.53\textwidth]{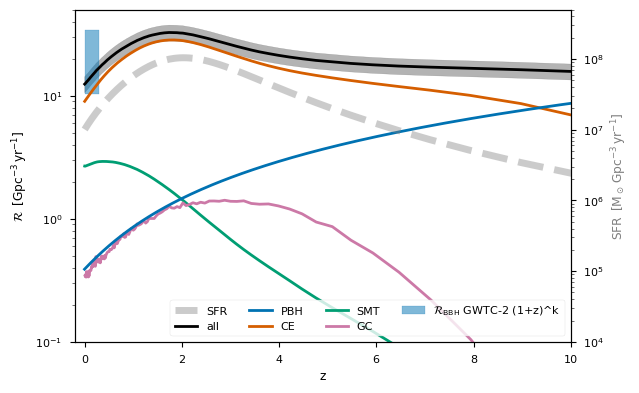}
\includegraphics[width=.46\textwidth]{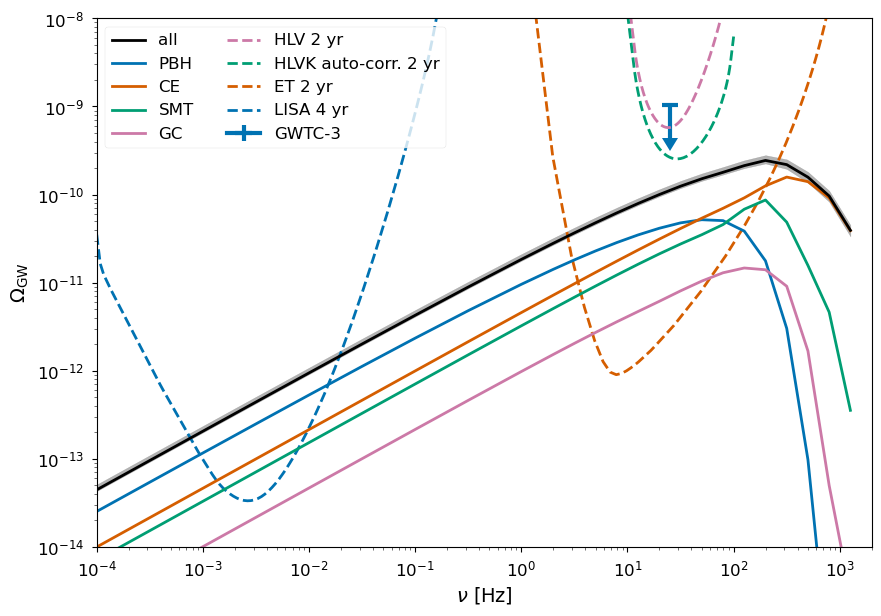}
\caption{ 
{\it Left panel:}
Merger rate evolution for representative astrophysical models and the PBH channel. As for the astrophysical models, CE, SMT, and GC refer to common envelope mass transfer, stable mass transfer, and globular clusters, respectively. The normalization comes from the Bayesian inference on the GWTC-2 catalog in ref.~\cite{Bavera:2020uch}. The blue band reports the local merger rate bound from LVK while the dashed gray band the star formation rate (SFR). 
{\it Right panel:}
Total SGWB (black line) coming from the various channels shown in the left panel, while we report with colored lines their contributions. We do not show the residual SGWB.
Figures adapted from ref.~\cite{Bavera:2021wmw}. } 
\label{fig:SGWB_PBHs}
\end{figure*}

In figure~\ref{fig:SGWB_PBHs} (left panel), we show the merger rate distribution of PBHs as a function of redshift (blue line), compared to other representative astrophysical models both from isolated (common envelope, stable-mass transfer) and dynamical (globular cluster) formation scenarios from ref.~\cite{Zevin:2020gbd} (see \cite{Bavera:2020uch} for more details). The chosen normalization reflects the result found in ref.~\cite{Franciolini:2021tla} with a multi-channel Bayesian inference of the GWTC-2 catalog. 
While we do not discuss the mass distribution of the populations assumed in these plots, we highlight the large contribution from PBHs at low frequencies even if they only contribute to a subdominant fraction of the local merger rate density in this specific example.
This property allows testing the presence of PBH binaries, adding an independent constraint to the model
selection, as a cross-correlation between local rate and SGWB amplitude 
would provide valuable information on the merger rate distribution of the various channels at a relatively high redshift.
The degeneracy between the PBH models and other astrophysical scenarios would then be broken by the fact that the former predicts a monochromatic growth of the merger rate density with redshift, thus the integrated SGWB would be larger for the same local (i.e. low redshift) observed rate.

\subsubsection{Astrophysical Uncertainties in background description}
    \label{subsect:astro_uncertainties}


While the recent detections of CBC by ground-based interferometers have provided a wealth of information on the BBH, BNS and NSBH populations, many uncertainties remain.  These uncertainties can have an important impact on our predictions of the SGWB and can be divided into two categories: CBC source properties, which can be modelled with stellar models and constrained with current observations; and galaxy evolution, including SFH,  Initial Mass Function (IMF) and metallicity evolution, which are taken from observations. On the other hand, observations with ET offer an opportunity to lift these uncertainties and thus obtain indirect constraints on these physical processes.

As was shown in \cite{Lehoucq:2023zlt}, uncertainties in the mass distribution of stellar-mass BBH affect not only the overall amplitude of the SGWB (see also \cite{Babak:2023lro}), but also the spectral shape around the peak and the location of the peak. Moreover, current instruments are sensitive only to relatively low redshifts ($z\sim 0 $ for BNS and $z\lesssim 1$ for BBH). Therefore in order to compute the SGWB we need to extrapolate our models up to $z \sim 10$. However, not only the merger rate may evolve with redshift, but also source parameters, such as mass distribution, which would strongly influence the SGWB, as shown in \cite{Lehoucq:2023zlt}.

Another major source of uncertainty is the SFH, which serves as the basis for SGWB modeling since it determines the total number of stars formed at each epoch \cite{Santoliquido:2022kyu}. The SFH can be measured from the UV and IR luminosity functions of galaxies; however, in addition to various observational biases, such as dust attenuation, surveys often suffer from incompleteness at higher redshifts, beyond $z\sim 3$ (see e.g. \cite{Boco:2020pgp}). A closely related parameter is the IMF, which describes the number of \emph{massive} stars formed in a given population. The IMF is typically assumed to be universal across environments and redshifts. However, some recent studies show that this may not be the case, which could strongly influence the number of compact objects formed and therefore the SGWB. Finally, the metallicity evolution is highly important for modeling the SGWB, since, as discussed in section~\ref{div3:rates}, the efficiency of forming CBCs strongly depends on metallicity (see e.g. \cite{Spera:2015vkd,Chruslinska:2018hrb,Neijssel:2019irh,Santoliquido:2022kyu}). The reason for this dependence is two-fold: first of all, stellar winds become stronger at high metallicity, and can remove a large fraction of the initial stellar mass before the star reaches core-collapse. This process leads to reduced masses of the resulting BHs. Second, metallicity can impact the efficiency of mass transfer phases during binary star evolution (like the common envelope phase), which can result in less efficient mergers and a lower probability of forming BBHs. In addition, metallicity influences the properties of supernovae that produce BHs through fallback. Measuring the metallicity evolution remains challenging, especially for $z\gtrsim 3$ (e.g., \cite{Boco:2020pgp}) and these observational uncertainties lead to large variations in the predicted SGWB \cite{Perigois:2021ovr,Lehoucq:2023zlt}. 



We expect several of these uncertainties to be lifted in the next few years: the number of detected CBCs is expected to increase, leading to a better determination of the mass spectrum; the reach of electromagnetic observations, in particular with JWST, is also extending, and a better measurement of the SFH at high redshift is foreseeable. Nevertheless, some of the open questions will remain until the deployment of ET, such as the mass distribution and merger rate of CBCs at high redshifts. The metallicity evolution and IMF universality also remain challenging targets to study. Thus, future observations of the SGWB with ET will help constraining not only the properties of CBCs but also several crucial aspects of galactic evolution. In this regard, studying the SGWB is complementary to the direct catalog approach: combining both resolved and unresolved populations will give the most complete picture of the compact binary evolution across cosmic times.

\subsubsection{Sources other than CBCs}
\label{subsec:astro_sources}


While CBC are the most studied class of GW sources, and the only one detected so far, other sources are expected to be detectable with ET, including stellar collapses and isolated neutron stars that are described in detail in section~\ref{section:div7}. Recent works have studied the SGWB produced by core-collapse SNe \cite{Abdikamalov:2020jzn,Finkel:2021zgf} and rotating neutron stars \cite{Regimbau:2011rp,Sieniawska:2019hmd}. In the case of core-collapse supernovae (CCSN), individual detections with ET would be possible only for the local Universe, whereas the SGWB includes also the contribution from more distant sources. 
Ref.~\cite{Finkel:2021zgf} studied a large range of CCSN explosion models and calculated the stochastic GW signal. They showed that for most models the expected SGWB is below the sensitivity of ET, with $\Omega_{\mathrm{GW}}\simeq 10^{-16}$ at $f\sim 10$ Hz and $\Omega_{\mathrm{GW}}\simeq 10^{-14}$ at $f\sim 100$ Hz. The only exception are extreme models that feature rapidly rotating progenitors with a low ratio of rotational to gravitational potential energy that produces a rotational instability (low-$T/|W|$ instability). This instability produces strong GW emission between $200$ and $300$ Hz. Therefore, if a large fraction of CCSN progenitors share this property, the resulting stochastic background would be much stronger, with $\Omega_{\mathrm{GW}}\simeq 10^{-10}$ at $f\sim 200$ Hz and possibly detectable with ET. It should be noted that such extreme models are considered unrealistic, since only a small fraction of massive stars are rapidly rotating. On the other hand, there are currently large uncertainties in the predictions of the GW signal from CCSN, and the resulting GW background differs by several orders of magnitude between models.

Rotating non-axisymmetric neutron stars also emit GW due to time varying mass quadrupole. This signal is expected to be much weaker than that of merging binary compact objects, but to last for much longer times. The loudest sources will thus be possibly detectable individually, but the contribution from unresolved sources will constitute a confusion background. This signal depends on the highly uncertain equation of state of neutron stars, their deformability, the distribution of rotation periods and magnetic fields (that contribute to the deformation of the star) and possible accretion from a companion. In ref.~\cite{Regimbau:2011rp}, the SGWB from rotating neutron stars was estimated as $\Omega_{GW}\simeq 2\cdot 10^{-8}$ at $f=760$ Hz, which would be potentially detectable with ET.

In both cases, CCSN and isolated neutron stars, the modeling uncertainties are still very large and accurate predictions are challenging. More theoretical work is needed, in particular in the areas of hydrodynamical simulations of core-collapse SNe and the nuclear equations of state of neutron stars.

\subsubsection{Anisotropies and cross-correlation with electromagnetic observables}
\label{subsect:crosscorr_EM_conterparts} 

The astrophysical SGWB is not perfectly homogeneous across the sky, but  exhibits some level of anisotropy. As we already saw in 
\eqs{Omegaanisotdiv2}{OmegaLegendrediv2}, these  can be addressed introducing a directional dependence in \eq{eq:omega_gwdiv3}, writing
\begin{equation}
    \Omega_{\mathrm{GW}}(f, \boldsymbol{\hat{n}}) = \frac{1}{\rho_c} \frac{\mathrm{d}\rho_{\mathrm{GW}}(f, \boldsymbol{\hat{n}})}{\mathrm{d} \ln f \ \mathrm{d} \boldsymbol{\hat{n}}} = \bar{\Omega}_{\mathrm{GW}}(f) + \delta \Omega_{\mathrm{GW}}(f, \boldsymbol{\hat{n}})\, ,
\end{equation}
where $\boldsymbol{\hat{n}}$ is the direction of observation, $\bar{\Omega}_{\mathrm{GW}}(f)$ is the homogeneous component of the energy density, and $\delta \Omega_{\mathrm{GW}}(f, \boldsymbol{\hat{n}})$ is the anisotropic perturbation. The statistical properties of the anisotropies can be characterised in terms of the frequency-dependent angular power spectrum:

\begin{equation}
    \langle  \delta \Omega_{\mathrm{GW}}(f, \boldsymbol{\hat{n}}) \,  \delta \Omega_{\mathrm{GW}}(f, \boldsymbol{\hat{n}'}) \rangle = \sum_{\ell} C_{\ell}(f) P_{\ell}(\boldsymbol{\hat{n}} \cdot \boldsymbol{\hat{n}'}), 
\end{equation} 
where $P_{\ell}(\boldsymbol{\hat{n}} \cdot \boldsymbol{\hat{n}'})$ are Legendre polynomials.
The SGWB anisotropies are generated by the intrinsic clustering of sources and the line-of-sight effects accumulated during the propagation through the large-scale structure \cite{Contaldi:2016koz,Cusin:2017fwz,Cusin:2017mjm,Jenkins:2018nty, Cusin:2018avf, Bertacca:2019fnt,Pitrou:2019rjz}. Moreover, kinematic anisotropies arise from the motion of the observer with respect to the rest frame of the SGWB \citep{Cusin:2022cbb,ValbusaDallArmi:2022htu,Chung:2022xhv}. The theoretical expression of the SGWB anisotropies 
and their detection prospects have been subject of intense study in the last years;  see section~\ref{sect:anisoSGWBdiv2}
for a review of this topic, for both the cosmological and astrophysical background components.

The astrophysical SGWB is given by the superposition of a large yet finite number of unresolved signals. In the ET band, the GW events from BBHs are expected to be separated in time and with a limited time overlap (while some overlap is expected for BNS mergers). Hence, the sources of the SGWB are discrete in space and time, and their number fluctuates according to a Poisson distribution, resulting in a significant shot noise. The total SGWB angular power spectrum is therefore $C_{\ell}^{\textrm{tot}} = C_{\ell} + N^{\textrm{shot}}$, where the shot noise contribution is flat in the $\ell$-space and dominates over the intrinsic anisotropies. Indeed, the intrinsic anisotropies are typically at the level of $\delta \Omega_{\textrm{GW}}/\Omega_{\textrm{GW}}\sim 10^{-3}-10^{-2}$ \cite{Cusin:2018rsq,Cusin:2019jpv,Cusin:2019jhg, Capurri:2021zli,Bellomo:2021mer}, while the shot noise is around one order of magnitude larger \cite{Capurri:2021prz,Jenkins:2019uzp,Jenkins:2019nks,Alonso:2020mva}. Given the sensitivity of ET to various multipoles (see e.g. the analysis of \cite{Alonso:2020rar}) 
and the presence of shot noise, directly detecting the anisotropies due to clustering and line-of-sight effects will be extremely challenging. However, shot noise is expected to be measured with good precision, at least within the range of amplitudes predicted by most astrophysical scenarios. Despite being insensitive to the sky distribution of the sources, the shot noise still carries astrophysical information as its amplitude depends on the overall number of sources contributing to the SGWB and thus reflects the underlying astrophysics.

An effective way to overcome the shot noise issue is to cross-correlate the SGWB with electromagnetic observables. Indeed, the SGWB anisotropies are correlated with other cosmic fields that trace the large-scale structure, such as galaxy number counts \cite{Cusin:2018rsq,Cusin:2019jpv,Canas-Herrera:2019npr,Alonso:2020mva,Yang:2023eqi}, weak gravitational lensing \cite{Cusin:2018rsq,Cusin:2019jpv}, CMB \cite{Ricciardone:2021kel}, and CMB lensing \cite{Capurri:2021prz}. All of these studies have shown that the SNR of the cross-correlation outperforms that of the auto-correlation by several orders of magnitude.
Measuring the angular power spectrum of the cross-correlation between the SGWB and an electromagnetic observable can provide constraints on various astrophysical and cosmological quantities. Moreover, cross-correlating with a redshift-dependent observable (such as the number counts of galaxy, at different redshifts),  allows one to have a tomographic reconstruction of redshift information, i.e. to extract the contribution to the total background from sources in a specific redshift band.  

In \cite{Yang:2023eqi}, the cross-correlation between LVK data and the Sloan Digital Sky Survey catalog was exploited to derive upper limits on a set of effective astrophysical parameters describing the local process of GW emission at galactic scales. As the sensitivity of the GW detector network improves, the approach presented in \cite{Yang:2023eqi} will become more effective, enabling more stringent constraints on the astrophysical kernel and its effective parameters. 
ET will improve the sensitivity to the angular power spectrum of the cross-correlation by more than a factor $\sim 1000$: ET is expected to reach and explore the astrophysically interesting region of the parameter space. Yet another approach could be to use the hierarchical Bayesian approach to estimate the BBH SGWB anisotropy~\cite{Banagiri:2020kqd}, which could also lead to $\sim 1000$ times sensitivity improvements relative to the approach presented in \cite{Yang:2023eqi}, hence resulting in a very significant improvement when applied to ET data. This is a promising approach which might help to set stringent constraints on the parameter space of the underlying astrophysical models. 
The cosmological significance of cross-correlating the astrophysical SGWB with electromagnetic observables, such as measuring the bias of the SGWB and constraining non-gaussianities in the large-scale structure, is discussed in section~\ref{subsec:agwb_lss}.
 


Given a generic anisotropic signal, an important  challenge is mapping its angular distribution \cite{LISACosmologyWorkingGroup:2022kbp}. For an unpolarized  stochastic signal, the stochastic search with ET aims to reconstruct the SGWB power spectrum $S_h(f,\hatn)$ defined in 
\eq{eq: iso_unpo}, while in the more general case of a polarized signal we want to reconstruct the quantity $H_{A A'}(f,\hat{\bf{n}})$ defined in
\eq{Stokes2div2}, which also includes the Stokes parameters describing circular and linear polarization.
For most of the astrophysical backgrounds, the frequency and the angular dependence of the power spectrum of the SGWB in eq. \eqref{eq: iso_unpo} are factorized.\footnote{ Although Doppler anisotropies induced by the motion of the detector with respect to
the rest frame of the SGWB induce a mixing between frequency and angular dependence, see~\cite{Cusin:2022cbb}. For a more general treatment we refer to the formalism developed in \cite{Mentasti:2023uyi}.} Then, restricting to the unpolarized case
\begin{equation}
\label{eq:anisotropic_spectrum}
S_h(f,\hat n)= S_h(f) E \left(  \hatn \right)\,.
\end{equation}
It is then convenient to decompose the angular part of the power spectrum in spherical harmonics\footnote{The so-called pixel domain is an alternative basis that can be employed to decompose the angular dependence of the power spectrum \cite{KAGRA:2021rmt,Agarwal:2023lzz}. However, for the astrophysical and most of the cosmological models of a SGWB the spherical harmonics choice is the most natural one.}
\begin{align}\label{eq:spher_harm}
E(\hatn)=\sum_{\ell=0}^\infty\sum_{m=-\ell}^\ell \delta_{\ell m}Y_{\ell m}(\hatn)\,,
\end{align}
and then reconstruct the value of the anisotropic (map) coefficients $\delta_{\ell m}$ from the cross-correlation of the instruments' data streams that are available \cite{Allen:1996gp,Mentasti:2023gmg}. The reconstruction is performed in the framework of Bayesian analysis, by writing the likelihood function of the cross-correlators, expressing it as a function of the coefficients $\delta_{\ell m}$ and performing a maximum likelihood estimation, which can be done by a Markov-chain Monte Carlo (MCMC) or by a Fisher analysis. 

Several factors may affect the quality of the reconstruction of the anisotropic map.
Firstly, the instrumental noise puts a threshold on the detectability of the  anisotropy coefficients $\delta_{lm}$.
In particular, the higher values of $\ell$ in the expansion of eq. \eqref{eq:spher_harm} produce a signal in the high-frequency side of the power spectrum. This is due to the fact that the anisotropic overlap reduction functions of ground-based interferometers (which quantify the sensibility of a cross-correlator to a specific anisotropy) show a peak at a frequency $f\propto\ell b/c$, where $b$ is the baseline between the pair of instruments taken into account.

Furthermore, up to today most of the map-making analyses have been performed for a Gaussian SGWB \cite{KAGRA:2021kbb,Mentasti:2023gmg}, with stationary statistics in its reference frame. Under this assumption, the frequency domain is a natural choice to perform the data analysis.\footnote{The rotation of the Earth results in an observed GW background which is cyclo-stationary in the reference frame of the detectors. In \cite{Allen:1996gp,Mentasti:2023gmg} a folding method is employed in order to take into account this effect in the data analysis process.}
The signal-to-noise ratio (SNR) grows with the square root of the time and it is dominated by the frequencies at which the overlap reduction functions, for each anisotropic coefficient of the map, display a peak.
In fact, the astrophysical GW background is expected to be  nonstationary due to the finite number of sources generating it \cite{Jenkins:2019uzp,Jenkins:2019nks,Alonso:2020mva,Capurri:2021zli}. More sophisticated analyses that aim to work beyond the frequency domain must be employed if we want to tackle the challenge of extracting a weak anisotropic signal, as expected by both the astrophysical and cosmological models. 
A joint analysis of the SGWB along with the information given by the angular distribution of the resolved GW sources and the electromagnetic surveys will certainly help in that direction.

Another aspect that must be considered when studying the SGWB with ET (and in general in next generation detectors) is the nuisance effect due to the intrinsic variance of the signal \cite{Mentasti:2023icu}. This effect impacts the measurement of the anisotropy coefficients in the limit where the instrumental noise is lower in absolute value with respect to the total effect of the SGWB on the interferometers. In fact, the current analyses work in the opposite regime, i.e. the so-called noise-dominated assumption. It has been shown \cite{Mentasti:2023gmg} that current detectors live in the noise-dominated regime, while for ET and the new generation interferometers, the intrinsic variance of the signal will  no longer be a negligible effect when compared to the low instrumental noise power spectrum.

As a last important remark, the mapping of the high $\ell$ values for the SGWB angular power spectrum and all the odd $\ell$ values is suppressed when the baseline between the instruments is small. This results in enhanced sensitivity for the proposed configuration \cite{Branchesi:2023mws} of 2 L-shaped Einstein Telescopes in two different sites, improving the detectability for those anisotropies of several orders of magnitude with respect to the original triangular-shaped design. In a similar way, the analysis obtained by cross-correlating ET with other new-generation instruments (such as Cosmic Explorer) will result in a significantly greater sensitivity to the SGWB anisotropies.

\subsubsection{Spectral shape reconstruction}

As already mentioned in section~\ref{subsec:popIII}, the SGWB spectral shape contains plenty of information on the underlying astrophysical populations from which the sources responsible for the SGWB generation are drawn. Techniques for SGWB spectral shape reconstruction are thus crucial for extracting this information from the data. While it is known that the astrophysical background is not perfectly isotropic, stationary, or unpolarized (see, e.g.,~\cite{Jenkins:2018uac,Jenkins:2018kxc,Cusin:2018rsq,Jenkins:2019nks,Alonso:2020mva,Capurri:2021zli,ValbusaDallArmi:2023ydl}), in the remainder of this section, we neglect these features and focus on the spectral shape only. 
In this limit, all the information on the SGWB is contained in the function $S_h(f)$ in \eq{eq:anisotropic_spectrum}, which fully characterizes the spectrum. This quantity is uniquely determined, through \eqs{eq:Sh}{eq:omega_gwdiv3}, by the population properties, and inherits the dependence on the population (hyper)parameters, say $\Lambda_i$. These (hyper)parameters characterize, among others, the mass function and the merger rates for all the subpopulations contributing to the SGWB, and impact the overall amplitude and the position and sharpness of the high-frequency cutoff in the frequency spectrum~\cite{Bavera:2021wmw,Braglia:2021wwa}. 

In a  
consistent Bayesian approach, all the $\Lambda_i$ would have to be included in the analysis, and their parameter space would have to be sampled simultaneously. 

There are several challenges to be faced while following the above procedure in the context of ET. For example, since several components contribute simultaneously to the overall signal, the dimensionality of the parameter space may be significant (if not prohibitive), and all signal components will effectively act as an additional nuisance to each other. Sampling such a parameter space, which might be highly degenerate and non-Gaussian, can be quite a non-trivial problem from a numerical point of view. For this reason, an alternative (and computationally less demanding) avenue could be to separate the data analysis problem into two steps. In an initial step, one would attempt an agnostic reconstruction of the SGWB spectrum (for some techniques developed in the context of LISA data analysis, see, e.g.,~\cite{Caprini:2019pxz,Karnesis:2019mph,Pieroni:2020rob,Flauger:2020qyi,Alvey:2023npw,Dimitriou:2023knw}) to capture the overall frequency shape. In a second step, the constraints on the spectrum are converted into constraints on the population (hyper)parameters. Such an approach has the merit of compromising between accuracy and feasibility, and it could be used, e.g., to identify the relevant part of the parameter space to be sampled with the full Bayesian techniques described in the previous paragraph.

\subsection{Executive summary}

Here we summarize the advancements that ET will bring in  population studies and astrophysical backgrounds, in comparison to the current state of the art, specifically emphasizing the distinct features that the ET will allow us to explore. The following are the major areas where ET's capabilities surpass the current generation of GWs detectors:

\begin{highlightbox}{Astrophysical CBCs across cosmic time}
  \begin{itemize}
  \item{\bf{Formation channels.}} Current models describe a remarkable zoo of CBC formation channels, including (stable and unstable) mass transfer in massive binaries, chemically homogeneous evolution, dynamics in star clusters and gas rich AGNs. With the prospective detection of $10^5$ BBH and BNS events per year, it will be possible to characterise in  detail the impact of the isolated and dynamical channels on the formation of CBCs. This can be achieved by constraining crucial quantities like merger rates, redshift dependencies, masses and spins of merging CBCs.
  \item{\bf{Merger rate density evolution.}} Current detectors enable us to infer the merger rate density of BNSs and BHNSs at redshift zero, and to reconstruct the BBH merger rate density up to a redshift of $z\approx{1.5}$, under several assumptions. ET will be a paradigm shift for measuring the merger rate evolution of astrophysical CBCs: we will estimate the intrinsic BBH merger rate up to a redshift of $z\approx{}100$ and the BNS merger rate at least up to cosmic noon. This will yield unique opportunities to constrain the formation of CBCs, their possible dependence on stellar metallicity, their delay time, the evolution of their stellar progenitors, and the link with their host galaxies.
  \item{\bf{CBC masses.}} From current detections, several peaks have already emerged in the mass function of primary black holes. ET will not only constrain these features to an exquisite level of accuracy, but will also probe their evolution with cosmic time and the existence of the claimed pair-instability mass gap.
  \item{\bf{CBC spins.}} Current gravitational-wave data suggest a tension between the spin of black holes in gravitational-wave detections and X-ray binaries. Our ability to interpret this claimed tension in terms of formation history of X-ray and gravitational-wave black holes is hampered by large observational uncertainties. Moreover, hierarchical merger models predict a peak of the precessing spin at $\chi_{\mathrm{p}}\sim{}0.7$ for second-generation black holes. ET will yield an accurate reconstruction of the spin distribution of black holes, providing a conclusive answer about the claimed tension and the existence of moderate to high-spin peaks.
  \item{\bf{Population III stars.}} ET will yearly observe between 10 and $10^4$ binary black hole mergers from the first generation of stars that appeared in the Universe (Population III stars). The main challenge will be to disentangle such population from primordial black holes and black holes born from other stellar progenitors.
   \item{\bf{Intermediate-mass black holes.}} Intermediate-mass black holes are crucial to explain the formation of supermassive black holes observed at the center of most galaxies, but have largely eluded our efforts to characterize them with electromagnetic surveys. GW190521 demonstrated the potential of gravitational-wave detectors to probe the intermediate-mass black hole regime. ET will enable us to study a rich population of intermediate-mass black holes, up to a redshift $z\approx{}10$. This will offer unique opportunities for multi-band gravitational-wave observations with space-based detectors.
    \end{itemize}

\end{highlightbox}

\begin{highlightbox}{Primordial Black Holes}
PBHs are BHs which could have formed in the early universe and might  compose a non negligible fraction of the dark matter.  ET will allow  to pin down the presence of PBHs to   unprecedented precision through its capability  to search their GW signatures. On one side, thanks to its  superior sensitivity and wider bandwidth  at low frequencies,   ET will test  the unique feature of the PBH merger rate density, its   monotonically increase
with redshift, up to redshifts larger than 30 where 
astrophysical sources are not present. Secondly,  ET will drastically improve the sensitivity leading to a possible detection of subsolar PBHs, a smoking gun for their existence  given the fact that subsolar astrophysical BHs may not exist.

\end{highlightbox}

\begin{highlightbox}{Population backgrounds}

\begin{itemize}
\item {\bf{Sources.}} Most BBHs can be detected individually by ET. However, using a background approach for parameter reconstruction is valuable for cross-checking results against catalog data, as any discrepancies may indicate faint subpopulations that individual analyses might miss. For BNSs, the bulk of the population will be sub-threshold, and a background study provides insights that catalog approaches cannot offer. ET is also expected to detect sources other than CBCs, such as stellar collapses and isolated neutron stars (see section~\ref{section:div7}). Finally, a population of PBHs extends at redshifts beyond ET's detection horizon, making a background approach crucial for studying them.

\item {\bf{Astrophysical information.}} Different formation channels for BBHs will influence their masses, spins, and eccentricities, as well as the merger rate, distribution of time delays, and the evolution of the mass spectrum with redshift. These variations are expected to be reflected in the stochastic gravitational wave background (SGWB) observable by ET. Pop III remnants are expected to possess unique characteristics that could be detected in the SGWB they generate. 

\item {\bf{Detection prospects.}} The angular power spectrum of the cross-correlation between the SGWB and electromagnetic observables is a promising observable for a first detection of anisotropies. ET is expected to enhance sensitivity to this angular power spectrum by over a factor of approximately 1000, allowing access to astrophysically significant regions of parameter space. Similar improvements are expected from the hierarchical Bayesian approach.

\item {\bf{Mapping.}} The  configuration of two L-shaped detectors at different sites greatly enhances the detectability of anisotropies compared to the  triangular design. Moreover, cross-correlating ET with new-generation instruments like Cosmic Explorer will further improve angular sensitivity to the SGWB.

\item {\bf{Spectral shape.}} An agnostic reconstruction of the frequency shape allows converting constraints on the spectrum into limits on population (hyper)parameters. 

\end{itemize}

\end{highlightbox}
\section{Multi-messenger observations in the ET era}\label{section:div4}

The cosmological reach of the Einstein Telescope is such that it will detect extremely large numbers of compact object mergers which should also create luminous electromagnetic counterparts. If successfully paired, this can lead to substantial samples of multi-messenger sources that can be used to understand heavy element nucleosynthesis, probe physics at high density and extreme velocity, measure cosmological parameters, and constrain alternative theories of gravity. These objects will typically lie in the mid-to high-redshift Universe, and so while their detection in gravitational waves may be straightforward, we must enhance and optimize our electromagnetic capabilities if we are to exploit the treasure trove of scientific results that can flow from multi-messenger observations. Here we review the likely signatures occurring in coincidence with ET-detected mergers. We construct both population models and up-to-date implementations of the emission physics for electromagnetic counterparts to determine their recovery with both electromagnetic and neutrino facilities that may be available in the 2030s. We highlight the requirements of localisation capability, and that EM capability should be co-developed along with ET to ensure that the critical science questions enabled by multi-messenger observations can be answered.

\subsection{State of the art}

\subsubsection{Multi-messenger Astronomy}

Traditionally, we have studied the Universe with light, from millennia of observations with the naked eye to the insights enabled by the discovery of the telescope and its continued refinement. In the 20th century, astronomers made spectacular progress with the development of multiwavelength astronomy. Multiwavelength astronomy led to the realisation that physical conditions in the Universe enabled some astrophysical sources to emit light across the electromagnetic spectrum. A myriad of astrophysical insights stem from multiwavelength observations, from the cosmic microwave background \cite{Penzias:1965wn} and its many applications 
\cite{Planck:2018vyg} to new exotic source classes, including the radio-emitting pulsars \cite{Hewish:1968bj} and the high-energy dominated gamma-ray bursts \cite{Klebesadel:1973iq}. 

Throughout the twentieth century, there was also a realisation that distant astrophysical sources may also create signals that are not electromagnetic in origin. The journey to understanding these alternative carriers of information (messengers) began with the discovery of cosmic rays in 1912 
and flows through solar neutrinos in the 1960s 
\cite{Cowan:1956rrn} to the increasingly complex and sophisticated detectors for gravitational waves, which discovered the first sources in 2015 
\cite{LIGOScientific:2016aoc}.

Multi-messenger astronomy is the combination of two or more of these astrophysical messengers, although one of these is typically electromagnetic light. The value of such detections is immense; the non-photonic messengers often carry information from regions that are essentially invisible to light. Gravitational waves cannot be shielded and so provide information about the behaviour of mass in kilometre-sized regions where compact objects such as neutron stars and black holes merge. Neutrinos can escape from the cores of collapsing stars, hidden from electromagnetic view by the star itself. Therefore, multi-messenger astronomy provides a direct probe of matter in conditions far exceeding those available in any Earth-bound laboratory. Strikingly, many objects that emit panchromatic electromagnetic light are also likely to be multi-messenger emitters, and this combination of messengers is critical both in studying the full details of the events in question and in using them as probes of extreme astrophysics and cosmology (see figure~\ref{fig:mm_info}). 

The first multi-messenger detection was of thermal neutrinos, coincident with SN~1987A in the Large Magellanic Cloud \cite{Bionta:1987qt}. However, supernovae neutrinos have not been seen in the 35 years since, despite the substantial improvements in detectors. The gravitational wave multi-messenger era began in 2017 with the discovery of gravitational waves from the binary neutron star merger GW170817 \cite{LIGOScientific:2017vwq} and its subsequent identification from the gamma-ray to radio regime \cite{LIGOScientific:2017ync}. This detection enabled a raft of new scientific insights, which we describe in detail in section~\ref{gw170817}. To date, there has yet to be a further gravitational wave - electromagnetic source, despite intensive efforts. These endeavours will likely unveil further electromagnetic counterparts as the current generation of interferometers reaches their design sensitivities in the next few years. However, the rates of such detections mean that building samples beyond a handful of events will be challenging, and these events will all be at low redshift, ideal for detailed study of extreme physics in compact objects but of less utility in understanding source evolution or in deriving cosmological parameters. 

Large samples of electromagnetic bright sources will require next-generation gravitational wave detectors such as the Einstein Telescope (ET), whose astrophysical reach is sufficient to uncover populations of thousands of likely electromagnetically bright events. This dramatic rate increase is possible because of the increased horizon of ET. It comes at the cost that the events will be more distant and will require novel approaches for their identification and characterisation, combined with the next generation of electromagnetic observatories that will be active in the 2030s. The realisation of the multi-messenger promise of ET will rely on developing electromagnetic capability in tandem with the gravitational wave detectors. 

This section outlines the science enabled by multi-messenger observations. It reviews the progress made to date via studies of both GW170817 and similar systems, perhaps seen so far only in electromagnetic light. We subsequently construct detailed source population models to make predictions for both the rates of events and, critically, the expected electromagnetic signals. We then use these, along with a census of likely telescopes available to the community in the mid-2030, to make robust, if broad, predictions for the recovery of electromagnetic counterparts. 

\subsubsection{GW170817}
\label{gw170817}

To date, there is only a single unambiguous event in which gravitational waves and electromagnetic light have been observed in tandem, GW170817 \cite{LIGOScientific:2017vwq,LIGOScientific:2017ync}. This binary neutron star merger provided an exceptionally rich set of data that continues to provide new scientific insight. Observations of GW170817 hence offer a template for what future observations may enable and also an opportunity to learn lessons and improve follow-up. 

Both the LIGO detectors detected GW 170817, while an upper limit from VIRGO was decisive in breaking location degeneracies and providing an error box of only $\sim 30$ sq. degrees \cite{LIGOScientific:2017ync}. Only 1.7 seconds after the merger of the two neutron stars, a short-duration gamma-ray burst, GRB 170817A, was also detected by both the Fermi \cite{Goldstein:2017mmi} and INTEGRAL \cite{Savchenko:2017ffs} satellites. The rate of short GRB detection by these satellites is approximately one per week, so the time coincidence alone made identifying the GRB a highly significant event. 

The source location was non-optimal for electromagnetic observations, lying only $\sim$ 50 degrees from the Sun, with very limited possibilities for follow-up. This location was also not promptly visible to any observatories at the point at which it was disseminated. Targeted electromagnetic observations did not begin immediately, but several hours later, when observatories in South America could observe the sky localisation, and the {\em Swift} satellite started its search \cite{Evans:2017mmy}. These searches were rapidly successful, with \cite{Coulter:2017wya} reporting a new transient, subsequently named AT2017gfo, in the galaxy NGC 4993 in observations taken 11 hours after the merger. Indeed, before this first announcement, five more observatories had also observed the source \cite{Arcavi:2017xiz,DES:2017kbs,Lipunov:2017dwd,Tanvir:2017pws,Valenti:2017ngx}. Initially, it was unclear if this source was related to GW170817 or was an unrelated supernova explosion. However, imaging and spectroscopy over the next 48 hours revealed a spectral evolution unlike any previously observed transient and cemented the association {\cite{Pian:2017gtc,Smartt:2017fuw,Chornock:2017sdf,Kasliwal:2017ngb,Shappee:2017zly}. This spectral evolution was extremely rapid, with the source spectrum moving from a spectral peak at $\sim 5000$\AA\, to $\sim 15000$\AA\, on a timescale of only a few days. Strikingly, this signature was qualitatively extremely similar to what had been predicted for kilonova emission \cite{Barnes:2013wka,Metzger:2016pju}, in which the high opacity of heavy elements synthesised in the explosion effectively blocks the optical light. IR and optical observations of the source continued for as long as allowed by the encroaching Sun, and on a timescale of ~10-20 days, the counterpart was also detected in both X-ray \cite{Margutti:2017cjl,Troja:2017nqp} and radio \cite{Hallinan:2017woc} observations. Indeed, while a faint source was detectable by the Hubble Space Telescope ~100 days after the merger \cite{Lyman:2018qjg}, the X-ray and radio afterglows continued to brighten, reaching their peaks on timescales of hundreds of days before decaying \cite{Alexander:2018dcl,Mooley:2017enz,Margutti:2018xqd,Troja:2018uns} . 

The comprehensive observational campaign for GW170817 led to a myriad of scientific advances, which we highlight below.

\begin{figure*}
    \centering
    \includegraphics[width=0.9\textwidth]{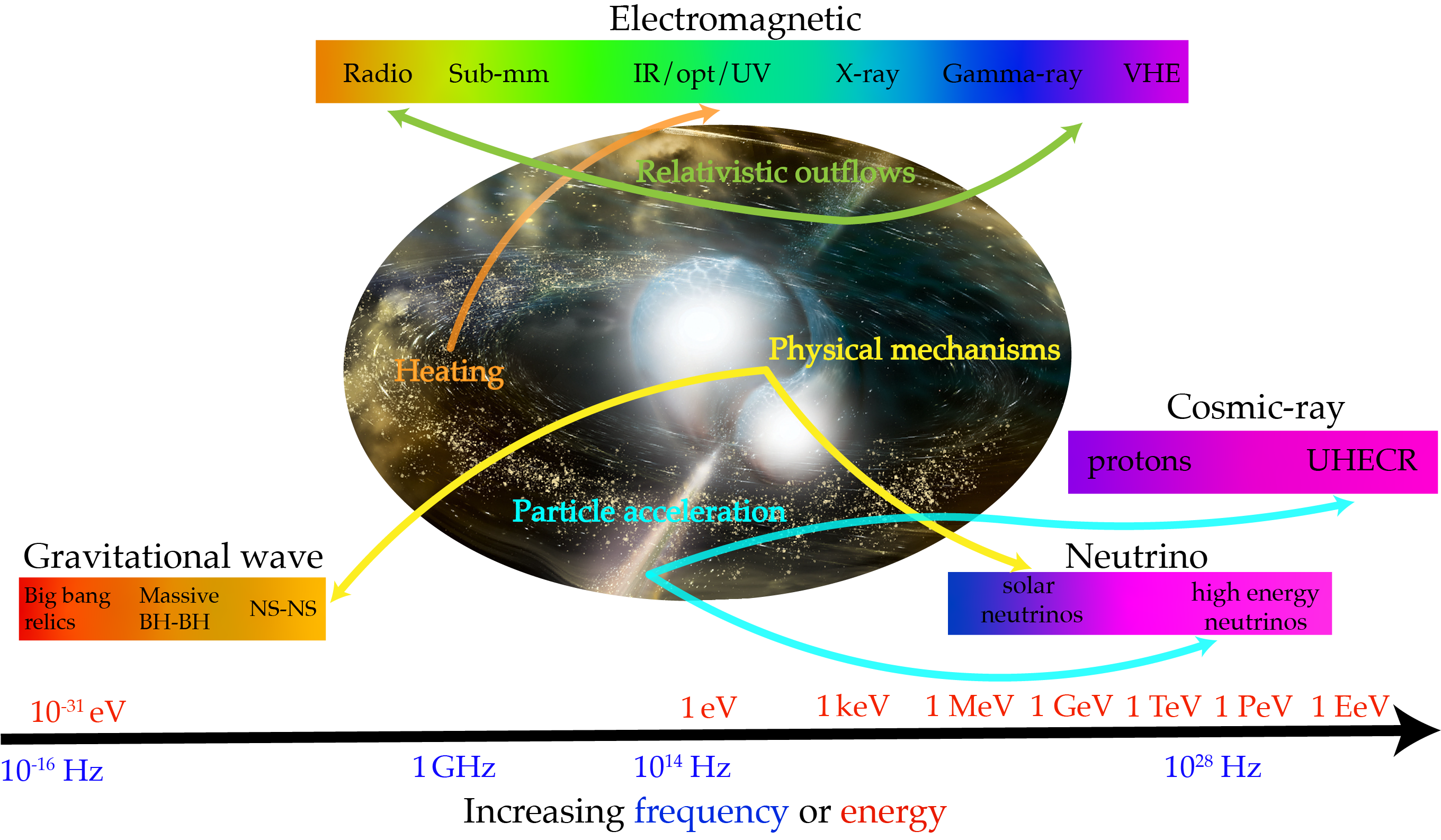}
    \caption{The promise of multi-messenger astrophysics. While much astronomy to date has been undertaken in the electromagnetic bands, in particular in optical light, many extreme systems in the Universe emit light across the electromagnetic spectrum and also produce additional messengers in the form of cosmic rays, neutrinos, or gravitational waves. Understanding the details of these systems and the insight they offer into central questions in astronomy, cosmology, and fundamental physics is something that can be done via multi-messenger observations.  }
    \label{fig:mm_info}
\end{figure*}

\begin{figure*}
    \centering
    \includegraphics[width=0.9\textwidth]{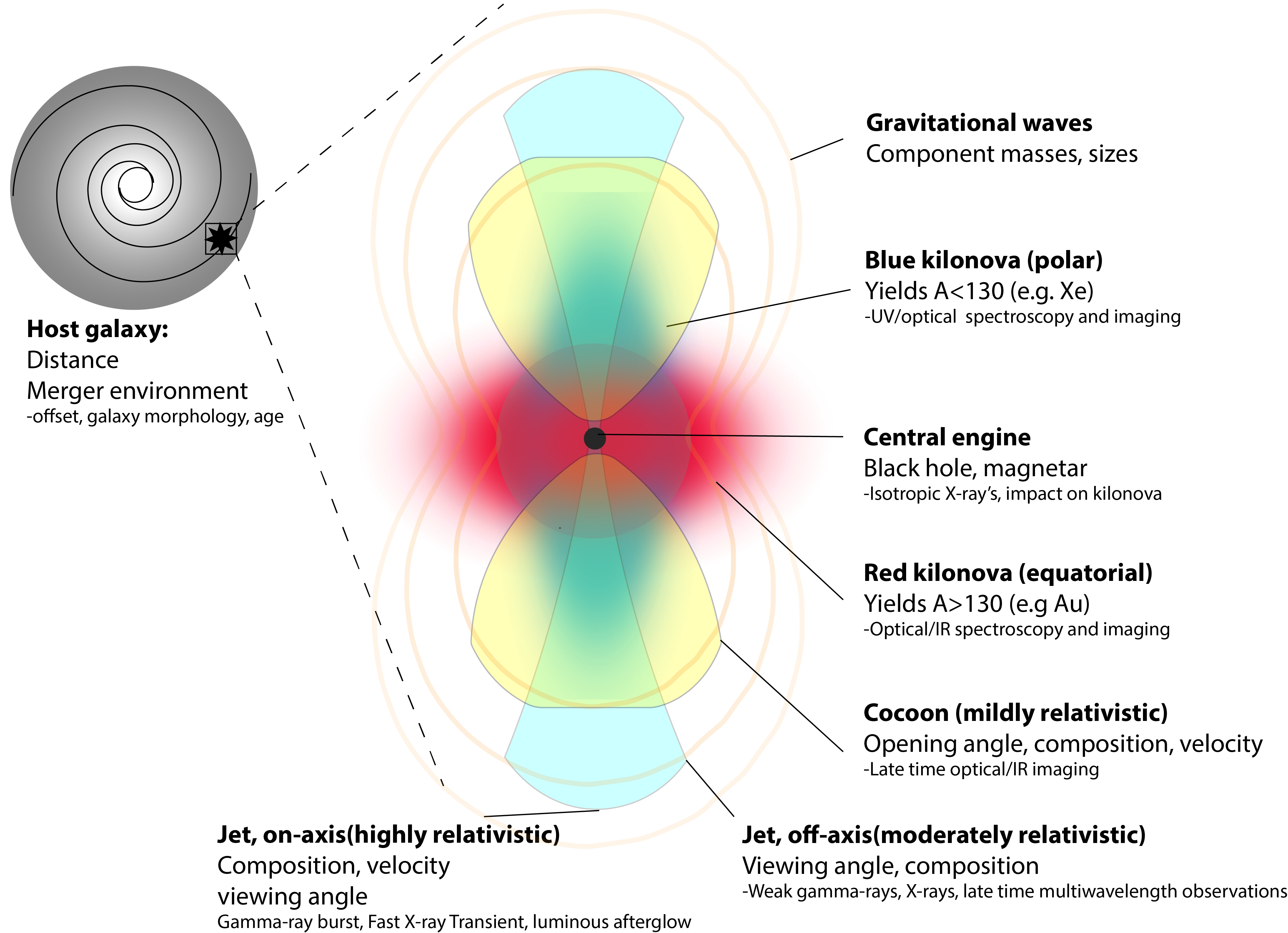}
    \label{fig:kn_info}
    \caption{The physical process ongoing in the merger of a binary system of neutron stars (or a neutron star and a black hole), their multi-messenger observational signals and the scientific insight enabled from them (updated from an original concept by \cite{Metzger:2016pju}). Gravitational wave observations provide robust measurement of the component masses, and constraints on their sizes and spins. Many possible electromagnetic signals are also possible, including the detection of associated GRBs, cocoon emission, and kilonova signatures.  }
\end{figure*}

\paragraph{The kilonova and the origin of the heavy elements.} 

The optical and IR light created in GW170817 was caused by the decay of newly synthesized elements in the merger. During the merger, neutrons are ejected both in tidal tails from the merger and a range of additional processes, such as the wind from a newly formed accretion disc. These neutrons then undergo neutron capture reactions via the so-called $r$-process, which builds extremely neutron-rich nuclei far from the valley of stability \cite{Burbidge:1957vc}. The energy released as these unstable isotopes decay then powers the faint and fast transient that we know as a kilonova \cite{Metzger:2016pju} while also populating the high-mass regions of the periodic table. Indeed, it is thought that the $r$-process is responsible for the creation of around half the elements heavier than iron and is the dominant source of elements with significant social, geological, and even biological importance, such as gold, thorium, and iodine. 

Before the discovery of GW170817 and its associated kilonova (AT2017gfo) several candidate kilonovave had been uncovered in the afterglows of (short) GRBs \cite{Tanvir:2013pia,Berger:2013wna}. However, AT2017gfo is by far the best-studied event, with a comprehensive set of both imaging and spectroscopic observations that are outlined in figure~\ref{fig:specseq}. These observations provide strong evidence for the production of $r-$process elements. 

\begin{figure}
    \centering
    \includegraphics[width=1.1\textwidth]{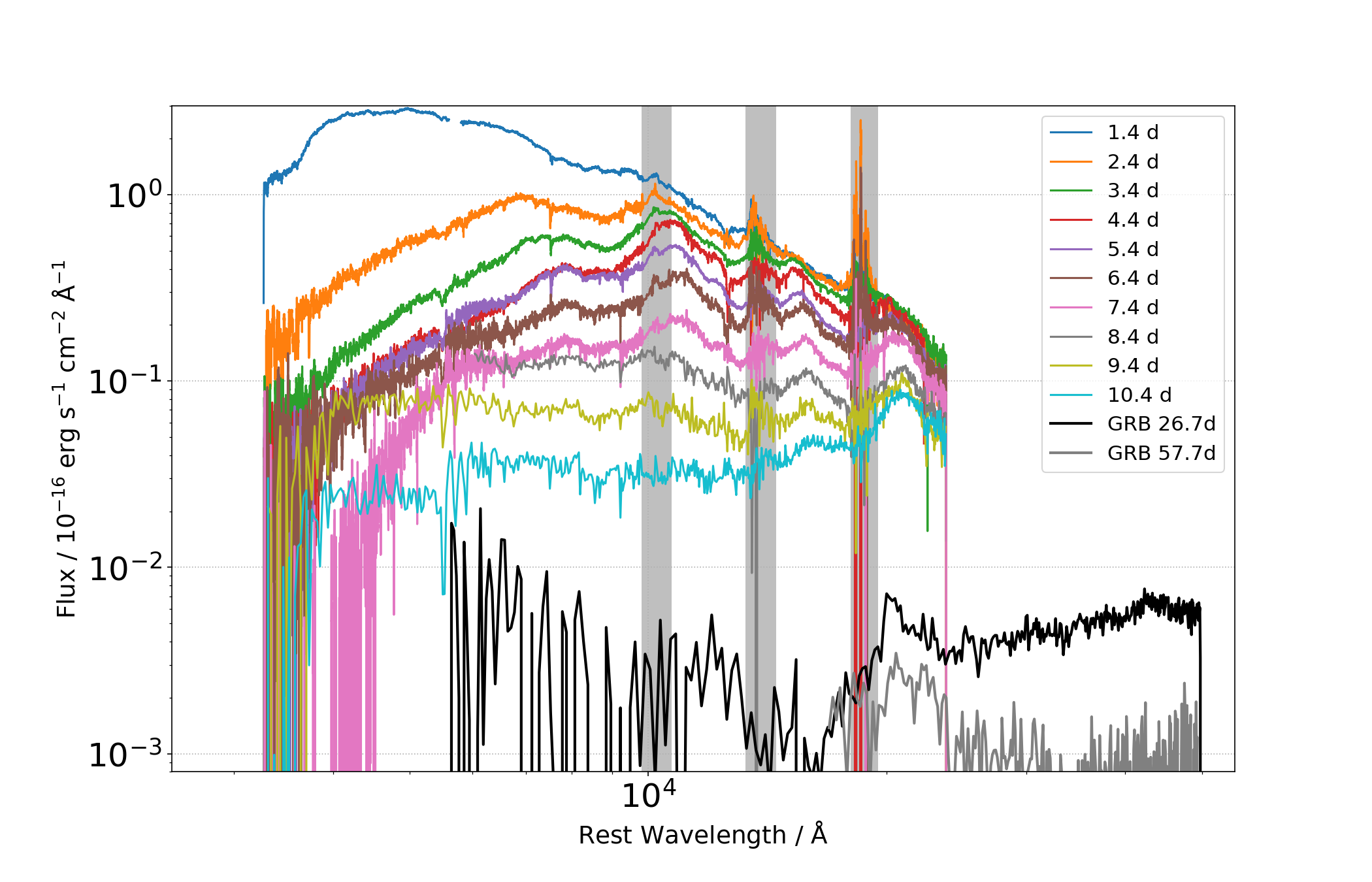}
    \caption{The spectroscopic evolution of kilonovae. The upper (colored) lines show the X-shooter spectral sequence for AT2017gfo \cite{Pian:2017gtc,Smartt:2017fuw}, demonstrating the early blue emission with the subsequent transition to a much redder spectrum. The lower (grey/black) lines show much later time spectra of the kilonova in GRB 230307A (AT 2023vfi) obtained with {\em JWST} \cite{JWST:2023jqa}. The spectra contain numerous spectra features both early in absorption and later in emission, and these have been linked to several different $r$-process elements. }
    \label{fig:specseq}
\end{figure}

The abundance patterns of heavy elements show three apparent ``peaks", often referred to as the first, second, and third peak elements. These are broad regions of higher elemental abundance centered at atomic numbers around krypton (atomic number 36, first peak), xenon (atomic number 54, second peak), and platinum (atomic number 78, third). The production of elements around each peak depends on the nuclear physics ongoing within the material forming the heavy elements (in the case of a compact object merger, this is the ejecta). In particular, different nucleosynthetic pathways are available depending on how neutron-rich the material is. Perhaps counterintuitively, the neutron-richness of the ejecta is defined by its so-called electron fraction ($Y_e$), which is equivalent to the proton-fraction for material that is neutral on a bulk scale. In broad terms, access to the more massive elements becomes increasingly straightforward for decreasing $Y_e$. In general, the production of material in the second and third $r$-process peaks is thought to require $Y_e < 0.15$ \cite{Metzger:2016pju}.  

Within the merger of a binary system of neutron stars, or a neutron star with a black hole, material can be ejected via various mechanisms. Firstly, the initially less massive object is larger and will be tidally stretched and disrupted as the gravitational wave-driven inspiral reaches its final phases. Material from these (or plausibly both) stars is ejected in the plane of the orbit in so-called tidal dynamical ejecta. Having been stripped directly from a neutron star, this material is extremely neutron-rich and should be able to undergo $r-$process reactions to the most massive elements. As the merger proceeds, the vast majority of the total mass remains with the stars and, ultimately, the central merged object. However, some also form a disc around this object. This disc may be a source of $r-$process synthesis itself \cite{Siegel:2018zxq}, and also drives a wind which creates further ejecta in the polar directions. Because of strong radiation from this disc (in particular in the form of neutrinos which trigger inverse beta-decays), the material in this polar ejecta is less neutron-rich than in the tidal ejecta and likely yields intermediate mass elements around the first $r-$process peak.

While the elemental abundances within the ejecta are determined by nuclear physics, the observed light in the kilonova is determined by the atomic physics of the newly synthesized material. While this atomic physics, and in particular the energy levels of a given isotope, are well known for low-mass elements, there are substantial gaps and uncertainties for the most massive nuclei due to the immensely complex electron structures. Of particular importance for the appearance of kilonovae are the so-called lanthanides and actinides that have open electron $f$-shells, and extremely high numbers of electron transitions throughout the UV and optical region. The effect of this is to provide extremely high opacity to outgoing light rendering kilonovae which produce these elements extremely red \cite{Barnes:2013wka}.

This work, much of which was undertaken prior to the discovery of AT2107gfo (but has been refined since), provided a broad expectation for what a kilonova should look like. It will have a low luminosity at peak (compared to most transients), evolve quickly because of the low ejecta mass, and either be very red or transition from blue to red on a timescale of a few days. 

In bulk terms, this is exactly what was observed in AT2017gfo, and it was the similarity of the observed properties to the pre-existing theoretical predictions that gave strong evidence that AT2017gfo was indeed the counterpart of GW170817. However, it is also true that when investigated in detail there were many aspects of the observed event that did not match the models, and demonstrated the requirement of more sophisticated approaches to fully exploit the potential of kilonovae in probing heavy element enrichment. In part, this explains why continuing analysis of the AT2017gfo observations reveals new insight 7 years after its discovery. 
Indeed, the kilonova, AT2017gfo has been the most widely studied component of GW170817, with several hundred articles dedicated to its study. While many of these studies disagree on the details of the source, often because of the use of disparate sets of models, the broad picture appears to be generally agreed upon. The source emitted in the range 0.02-0.1 M$_{\odot}$ of $r-$process material \cite{Arcavi:2017xiz,Chornock:2017sdf,Smartt:2017fuw,Pian:2017gtc,Tanvir:2017pws,Villar:2017wcc,Kasen:2017sxr}. The composition of this material is a mixture of lighter $r$-process elements which are responsible for the early bluer emission, and heavier elements, responsible for the redder, later emission \cite{Villar:2017wcc,Margutti:2020xbo}.

Most the these conclusions are drawn from the study of the multi-wavelength lightcurves of the kilonova. However, very substantial effort has also gone into the evolving spectral series, which shows significant variations, even on the sub-hour timescale at early times \cite{Sneppen:2024gnt}. These observations reveal the presence of strontium at early times \cite{Watson:2019xjv}, an identification now generally agreed upon in the community \cite{Gillanders:2022opm} and supported by theoretical expectations~\cite{Perego:2020evn}. These early observations also potentially contain several further spectral features, such as yttrium \cite{Sneppen:2023lgo}, although the challenge in such identifications is substantial given the uncertainties in the atomic physics of heavy elements. 

There has also been substantial interest in the later phases of the spectrum, which because of the low ejecta mass and high velocities rapidly become optically thin. At this point, it is plausible that emission lines may be observed. In the later IR spectra of AT2017gfo a prominent line at 2.2 microns has been interpreted as tellurium \cite{Hotokezaka:2023aiq,Gillanders:2023zys}, a line also seen in the JWST spectrum of a further KN in a GRB \cite{JWST:2023jqa,Gillanders:2023zys}.

The rich dataset for AT2017gfo continues to be exploited as models improve. The complex spectral and temporal evolution of the kilonova provides the opportunity to dissect the details of its composition and ejecta mass, although are currently limited substantially by model uncertainties. As these improve it should be possible to robustly infer the contribution of kilonova to the heavy element budget of the Universe. However, doing so concretely is likely to require samples of tens of kilonovae to understand source diversity and pinpoint rates. The current paucity of detections means that it is unclear if such observational samples can be built in a reasonable time with the current generation of gravitational wave detectors.

\paragraph{The gamma-ray burst and its afterglow.}

In addition to diagnostics from the core of the merging binary, electromagnetic observations also probe the relativistic outflow that created the GRB. In particular, GW observations (in particular when combined with the EM distance to the galaxy) provide a route to measuring both the energetics of the outflow, the inclination angle, and the structure of the relativistic jet. These provide substantial insight into the physics of GRB jets, as well as potentially into larger physics questions relating to the acceleration of material to ultra-relativistic velocity. 

GRB 170817A was observationally a rather typical, if faint, short duration GRB, with a duration of $T_{90}=2.0 \pm 0.5$ \cite{Goldstein:2017mmi} and a spectral shape consistent with that seen for other GRBs. In the absence of a distance measurement, it would not have stood out as exceptional. However, at a distance of only 40 Mpc the inferred energetics of the GRB are at the extremes of the distribution. In particular, the so-called isotropic equivalent energy release (the energy from the burst on the (false) assumption that it emits equally in all directions) is $E_{iso} = 10^{46}$ erg \cite{Goldstein:2017mmi,Savchenko:2017ffs}. This is four orders of magnitude lower than typical for short-GRBs \cite{Fong:2015oha,Nugent:2022paq}, which are typically found at much larger distances (these closest short-GRB after GRB 170817A is GRB 080905A at $z=0.125$, or $\sim$ 600 Mpc \cite{Rowlinson:2010jb}, a factor of $>15$ more distant than GRB 170817A, although there are some merger-related GRBs likely at lower redshift).

This likely explains why GRB 170817A was so underluminous compared to the bulk of the short GRB population -- it was viewed off-axis. In turn, studies of the GRB afterglow provide a route to studying the jet structure and imply that the jet in GRB 170817A was highly structured, with a lower Lorentz factor in the wings, but a luminous, highly collimated event to an on-axis observer \cite{Mooley:2017enz,Lazzati:2017zsj,Lyman:2018qjg,Lamb:2018qfn,Margutti:2018xqd,Troja:2018ruz,Ryan:2023pzk}. Indeed, in the years following the merger, deep observations utilizing radio interferometry \cite{Ghirlanda:2018uyx,Mooley:2018qfh} and the {\em Hubble Space Telescope} \cite{Mooley:2022uqa} could directly resolve the motion of the relativistic GRB jet. The implication is that GRB 170817A was viewed $\sim 25$ degrees from its jet-axis. It was an extremely faint afterglow, only just detectable to the premier facilities in the X-ray ({\em Chandra}/{\em XMM-Newton}), optical ({\em HST}) and radio (JVLA). Indeed, it may be that multi-messenger observations are one of the most effective routes of identifying GRB jets off-axis, and in doing so to study the structure of the GRB jets, itself of great importance in understanding the details of ultra-relativistic motion and particle acceleration.

\paragraph{The host galaxy and environment.}
In addition to the light from the transient itself, we can also glean much information about a given event from the environment in which it is born. Indeed, early in the history of studies of supernovae, it was apparent that type II events occurred exclusively in young, star-forming galaxies, while SN Ia also occurred in older populations, implying different progenitor systems. There is now a long history of using the environments of supernovae, gamma-ray bursts \cite{Bloom:2000pq,Fruchter:2006py,Savaglio:2008fj,Svensson:2010tj,Kelly:2011vi,Hakobyan:2012sf,Irani:2021oxf,Nugent:2022paq} and other transients \cite{Kasliwal:2011se,Lyman:2014qza,Lyman:2020ppm} as a route to understanding the stars that form them.

The host galaxy of GW170817/AT2017gfo is NGC 4993, a massive lenticular galaxy. This has also been studied in great detail on both the scale of the galaxy \cite{Fong:2017ekk,Levan:2017ubn,Im:2017scv,Kilpatrick:2021aav}, and the local environment \cite{Levan:2017ubn,Fong:2019vgn}.  These studies generally concur that the host galaxy is dominated by an old stellar population ($>1$ Gyr, \cite{Fong:2017ekk,Levan:2017ubn,Pan:2017jem}), with evidence for minor merger tens of millions of years ago \cite{Palmese:2017yhz,Kilpatrick:2021aav}. The vast majority of the stellar mass resides in the older populations, suggesting it is most likely the binary was formed in this population \cite{Stevance:2023byv}. There is no evidence for any underlying globular cluster to limits that probe the majority of the globular cluster luminosity function \cite{Lyman:2018qjg,Kilpatrick:2021aav}. In principle, observations of these locations are highly diagnostic of progenitor production, natal kicks, and delay times, since the host of GW 170817 is much closer than those of short GRBs for which such studies are typically undertaken \cite{Church:2011gk,Fong:2022mkv,Nugent:2022paq}

\paragraph{Tests of cosmology and fundamental physics.} 

Observations of GW170817, GRB170817A, and the kilonova AT2017gfo enabled a unique route of probing the extreme physics at play in compact object mergers. For example, gravitational wave observations measure (or place constraints) the so-called tidal deformability of the neutron stars during the merger. The degree of deformability places restrictions on plausible equations of state (or mass-radius relations) for the matter at nuclear densities within neutron stars \cite{LIGOScientific:2018hze}. Consistency checks on the results from these analyses can be run in tandem with electromagnetic observations; for example, are the allowed equations of state consistent with the mass ejected in the kilonova, or even how the equations of state rule in (or out) via tidal deformability studies compare with those inferred for different neutron stars via high-resolution X-ray spectroscopy of millisecond pulsars \cite{LIGOScientific:2018hze,Raaijmakers:2019dks}. 

A critical use of combined EM and GW observations is a novel one-stop route (distance ladder independent) to measure cosmological parameters. In particular, the waveforms of merging compact objects encode their masses and are so-called standard sirens \cite{Schutz:1986gp}, where the measurement of the gravitational wave amplitude directly provides the distance to the source. The Hubble constant can be determined if paired with a measure of the velocity, for example, in the form of a redshift of either the electromagnetic counterpart or its host galaxy. For GW 170817, initial estimates were of $H_0 = 70^{+12}_{-8}$ km s$^{-1}$ Mpc$^{-1}$ \cite{LIGOScientific:2017adf}, but improved to $H_0 =71.9 \pm 7.1$ km s$^{-1}$ Mpc$^{-1}$  \cite{Cantiello:2018ffy} using improved estimates of the distance to NGC 4993 or to $H_0 = 70.3^{+5.3}_{-5.0}$ via the determination of the inclination angle due to superluminal motion measurements of the GRB afterglow \cite{Hotokezaka:2018dfi}. 

Finally, the joint observations can constrain core questions in fundamental physics. In particular, the time delay between the gravitational waves and the first electromagnetic emission provides robust tests on any deviation between the speed of gravitational waves and the speed of light, which are several orders of magnitude better than previously available, suggesting any variations are smaller than one part in $\sim 10^{15}$ \cite{LIGOScientific:2017zic}. In turn, this rules out numerous alternatives to General Relativity which predict a difference  between the two speeds. While more complex scenarios remain plausible (e.g., frequency-dependent speed of GWs), the observations are a strong demonstration of the ability to multi-messenger observations to pose new tests for the validity of General Relativity.

\paragraph{Limits on other messengers associated with GW170817.}

GW170817 was the first source to be identified in both gravitational waves and light, but there is also significant interest in the prospect that it also created further messengers, such as high-energy neutrinos accelerated in the relativistic jet. Several such searches were undertaken using time windows around GW170817 including both close to the merger time \cite{Baikal-GVD:2018cya} and over a much longer $\sim 2$ week time frame \cite{ANTARES:2017bia}. The detection of such neutrinos would be a major step forward in our understanding of the details of particle acceleration in compact object mergers, although despite the relatively close distance to GW170817 the limits were generally significantly shallower than expected for the majority of models \cite{Biehl:2017qen,ANTARES:2017bia}.

\subsubsection{GRBs and KN as counterparts of compact binary mergers} 

While we have, to date, only uncovered one electromagnetic counterpart of a binary neutron star merger detected in gravitational waves, we likely studied hundreds of such events in electromagnetic light only. In particular, the detection of a GRB in GW170817 strongly links at least some short-duration gamma-ray bursts to the mergers of compact objects. Indeed, this joint detection cemented a growing body of evidence that already linked short-GRBs to such mergers. For example, the host galaxies of the bursts appear to span a broad range of ages, from young, star-forming dwarfs to giant ellipticals \cite{Fong:2013eqa}. Such a broad range of ages is inconsistent with any origin in young, massive stars. Furthermore, the locations of the short-GRBs in and around their host galaxies were very different from that seen in long-duration gamma-ray bursts \cite{Fruchter:2006py} or in supernovae \cite{Kelly:2011vi}. In particular, a substantial fraction of short-GRBs arise at large offsets from their host galaxies \cite{Berger:2010ag,Fong:2013iia,Tunnicliffe:2014ioa}, making it difficult to explain their locations without invoking progenitors that have received some kind of ``kick" (though see \cite{Perets:2021pfl,Gaspari:2023euw}). Such kicks are naturally expected to occur in compact object systems in which the combination of mass loss at the time of supernova (the so-called Blauw-kick) and natal kicks to neutron stars on formation \cite{Hobbs:2005yx,Verbunt:2017zqi} can propel the binary from its birth site at tens to hundreds of km s$^{-1}$. Finally, the first suggestions of 
kilonova emission were in the short-GRB 130603B \cite{Tanvir:2013pia,Berger:2013wna} and several possible examples had been uncovered before 2017 \cite{Yang:2015pha,Jin:2016pnm}. Combined, this evidence suggests that many, if not all, short-duration GRBs arise from compact object mergers and provides us with a much larger sample of events from which we can hone our expectations for future gravitational wave sources. 

A remarkable realisation in the past two years has been the identification of kilonova signatures in long-duration gamma-ray bursts. In particular, two events, GRBs 211211A \cite{Rastinejad:2022zbg,Troja:2022yya,Yang:2022qmy,Mei:2022ncd}, and GRB 230307A \cite{JWST:2023jqa,Yang:2023mqt,Gillanders:2023zys}, both have durations which place them in the bulk of the long duration GRB population, but also appear to contain kilonovae remarkably similar to AT2017gfo. The implication is that mergers powering long-GRBs may be comparably common in the local Universe to mergers powering short-GRBs, and motivates using both categories for GRB-GW coincidence searches. 

A complexity in using GRBs to predict the properties of gravitational wave counterparts is that they are (mostly) selected as events in which we view down the relativistic jet. The GRB and its afterglow are bright in this scenario,  and we cannot expect off-axis events to be such luminous high-energy emitters. Indeed, even the kilonovae themselves are expected to have a varying appearance depending on the viewing angle and if we observe primarily the polar (light $r-$process) or equatorial (heavy $r-$process) material.

In principle, the kilonovae associated with mergers can be identified without any pre-selection from either gravitational wave or GRB identification. A suite of small telescopes
now routinely scan essentially all the visible sky every few nights (these telescopes include the Zwicky Transient Factory (ZTF, \cite{Graham:2019qsw}), the ATLAS telescopes \cite{Rest:2018amw}, the MASTER telescopes \cite{Lipunov:2009ck} and the Gravitational Wave Optical Transient Explorer (GOTO) \cite{Steeghs:2021wcr}), while other, slightly narrower but deeper telescopes also conduct surveys of much of the sky (e.g. DECCAM at CTIO, PanSTARRS, BlackGEM). These predominantly optical surveys can uncover kilonovae and potentially have unrecognised events within their large catalogues of transients. However, there are substantial challenges in moving from detection to discovery for these events. In particular, the rate of very nearby events is likely low, with even the current complete range of uncertainties from the GWTC-3 predicting between 0.05 and  7 events per year at 100 Mpc \cite{KAGRA:2021vkt}. Given some loss of events is to be expected (direction too close to the Sun or behind the Galactic plane, poor weather, technical problems, some events too far from host galaxies to be associated at $<100$ Mpc), it is perhaps unsurprising that these searches have, to date, not yielded a robust kilonova detection. Indeed, this problem is compounded because the number of sources that may mimic the peak luminosities and timescales of kilonovae likely substantially outnumber the kilonovae. Indeed, intensive searches for faint fast transients have been run for the past several years and have identified many such sources. While none show the combined hallmarks of rapid evolution and fast transition from blue to red, substantial follow-up resources have been necessary to demonstrate this. Indeed, most of these faint, fast sources appear to be some combination of outbursts from Luminous Blue Variables, unusual white dwarf explosions (e.g., SN Iax), or ultra-stripped supernovae \cite{ENGRAVE:2022kkg}. These are all of substantial interest in their own right but complicate the identification of kilonovae. Indeed, with the imminent arrival of the Vera Rubin Observatory (VRO) and its Legacy Survey of Space and Time (LSST), the ability to isolate candidate kilonovae, either found blindly or in the error box of gravitational wave searches, becomes an urgent challenge to overcome.

\subsubsection{Alternative signatures of mergers}

The vast majority of effort in the detection of multi-messenger signals has been directed at the mergers of CBs that contain a neutron star, or more specifically that contain a neutron star and a mass ratio such that material is expected to remain outside the innermost stable orbit of the merged compact object \cite{Foucart:2012nc,Foucart:2018rjc}. This is natural because it is in these scenarios where baryonic matter is left outside the merger remnant producing luminous electromagnetic emission. The merger of a BBH binary in a vacuum should yield no detectable EM emission. Despite this, it is striking to note that the first BBH detection, GW150914 \cite{LIGOScientific:2016aoc}, was associated with a weak detection of a short GRB \cite{Connaughton:2016umz} that remains controversial \cite{Greiner:2016dsk}. 

Partly motivated by this coincidence, several works explored the possibility of deriving light from BBH mergers. These critically consider that BBH mergers do not happen in a true vacuum, but contain material in the form of the surrounding medium which may take a variety of forms. For example, it may be an ionized plasma \cite{Kelly:2020vpv}, a remnant disc around one of the black holes \cite{Martin:2018iov,Perna:2019pzr} or a BBH embedded in the dense disc around a supermassive black hole \cite{Bartos:2016dgn}. 

It is the latter of these suggestions that has received the most interest to date, with the suggestion from \cite{Graham:2020gwr} of an AGN outburst consistent in time and space with the extremely high mass BBH mergers GW190521 \cite{LIGOScientific:2020iuh}. This was suggested to be an unusual AGN outburst triggered by the merger of the two components within the disc of the AGN. Subsequent analysis of a large sample of BBH mergers that received follow-up from the Zwicky Transient Factory revealed evidence of a larger number of unusual AGN outbursts in the error boxes than expected randomly, perhaps implying a substantial fraction of BBH mergers that could be formed in the discs of AGN \cite{Graham:2022xxu}. Although the rates of such mergers are expected to be much lower than for those formed in the field, the higher masses result in a larger horizon. Through studies of GRBs, there is now some limited evidence for neutron star mergers in AGN discs \cite{Levan:2023qdh,Lazzati:2023ayh}, and so BBH mergers may naturally be expected as well. 

However, this result has also proved to be controversial. In particular, the challenge of associating a given AGN outburst with a BBH merger in a large error box is non-trivial \cite{Veronesi:2022hql,Veronesi:2023ugk} and the details of the wide range of AGN variability remain poorly understood. Nonetheless, these claims suggest that we should remain alert to the possibility of multi-messenger signals from BBH mergers, and the science they may enable.

\subsection{Modeling the EM counterparts of ET detected CB mergers}
\label{simu}

This section presents predictions of the observational properties of the electromagnetic counterparts of compact binary mergers, namely BNS and BHNS binaries, detectable by the Einstein Telescope.  

We consider the kilonova and the jet-related emission, the latter including both the prompt emission (appearing as a short GRB when the jet is observed close to on-axis) and the broadband afterglow. Four state-of-the-art population synthesis models are considered, with predicted gravitational and electromagnetic signals for each binary. Despite the considerable knowledge advancement made possible by GW170817, these model predictions are still based on a variety of astrophysical assumptions which need to be tested with new detections in the years to come until ET operations. Therefore, any description of the expected EM properties of the GW sources detected by ET is subject to a considerable amount of uncertainty. To partly account for this modeling uncertainty, we adopted four models published in the literature, each of which provides an end-to-end simulation of the GW and EM signals. We highlight that the kilonova properties could be impacted if some of the compact objects are quark stars instead of neutron stars \cite{Kluzniak:2002dm,Drago:2013fsa, Drago:2015cea,Drago:2017bnf, DePietri:2019khb,Bucciantini:2019ivq,DiClemente:2021dmz, Mathias:2023dyn,Miao:2024qik}.  An overview of the main assumptions of our models is provided in sections~\ref{sec:popmod} and \ref{sec:progmod}. Specifically, the parameter estimation analysis of the GW signal is presented in section~\ref{sec:gwmod}, while the models adopted for the prediction of the electromagnetic signals are described in section~\ref{sec:emmod}. 

For each binary class and for each emission component, we consider the subsample of events detectable by ET. For these we provide in section~\ref{sec:emcomp} two kind of information:
\begin{itemize}
    \item the properties (e.g., light curves, energetics, characteristic timescales) of the KN and jet-related GRB emission components; 
    \item the annual joint detection rates as a function of the individual EM component flux at specific frequencies or within selected energy bands. 
\end{itemize}

While there are a number of present or planned major facilities for the search and follow-up of GW-detected sources that are expected to operate in synergy with ET, the specific evolution of the observational landscape is hard to predict. This section, therefore, aims to provide a few general tools that can be used to estimate the expected joint detection rates by scaling for basic facility/instrument configurations. 

Figure~\ref{fig:donut} shows a possible 1 year of multi-messenger detections by ET and compares BNS (left) with BHNS (right). The jet-related emission (cyan double cones) and the kilonova (red butterflies) show the EM horizons reachable by a typical GRB satellite (with a sensitivity comparable to that of Fermi/GBM) and an optical search facility (reaching a 26 magnitude limit), respectively. Jets of BNS are brighter than BHNS systems and can be detected out to and above redshift $z=1$. For BNS systems an EM horizon at $z\sim0.5$ is identified. Within this horizon most of the multi-messenger events will be KN-GW systems and a fraction will have GRB-KN-GW signals. At $z>0.5$ most of the BNS joint detections will be GRB-GW signals. This selection effect is mainly due to the orientation of the jet, with off-axis jets being detectable only within relatively smaller distances.  Compared to the BNS case, a much smaller number of joint detections is expected for the BHNS channel and the relative EM horizons are much closer to the Earth.  


\begin{figure}
    \centering
    \includegraphics[width=\textwidth]{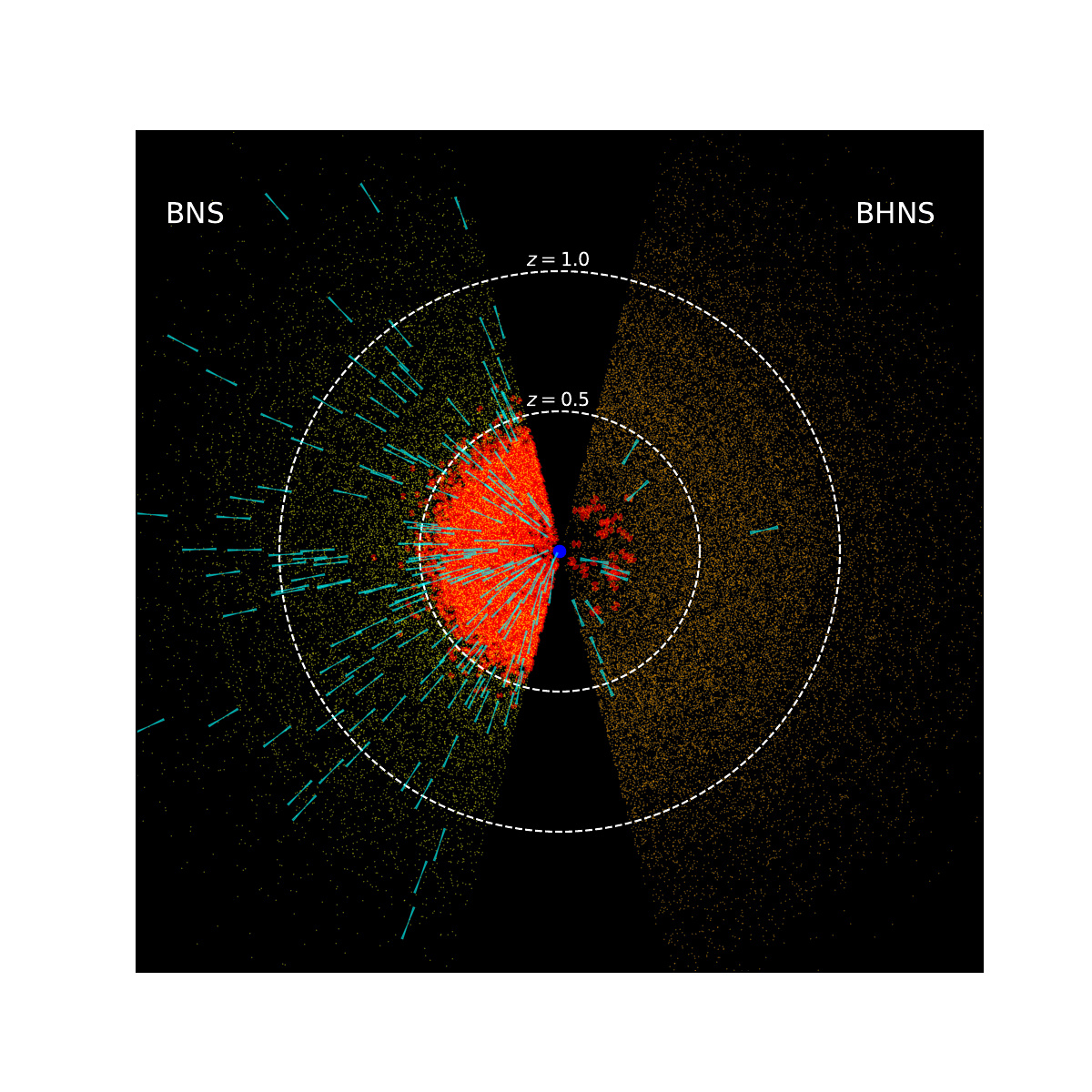}
    \footnotesize
    \caption{Representation of two of our reference multi-messenger synthetic populations in a geocentric Universe. Compact binaries in the synthetic populations of C22 and C23 that are detected by ET-$\Delta$ are represented by yellow (BNS -- left semicircle) and orange (BHNS -- right semicircle) dots. The distance of each dot from the Earth (blue circle) is proportional to the redshift of the corresponding compact binary. If the simulated merger produces a relativistic jet whose prompt ($>4$ ph cm$^{-2}$ s$^{-1}$) or afterglow emission ($>0.01$ mJy, radio band) is detectable, a cyan jet is plotted centered on the dot, with its axis inclined by the actual viewing angle with respect to the line of sight to the Earth. If a kilonova is also produced, and if it is detectable ($<26$ mag, $g$ band), then a red butterfly shape is also plotted. The total number of binaries is representative of one year of ET operation.}
    \label{fig:donut}
\end{figure}

\subsubsection{Dynamics and emission components of compact binary mergers}

\begin{figure}
    \centering
    \includegraphics[width=0.73\textwidth]{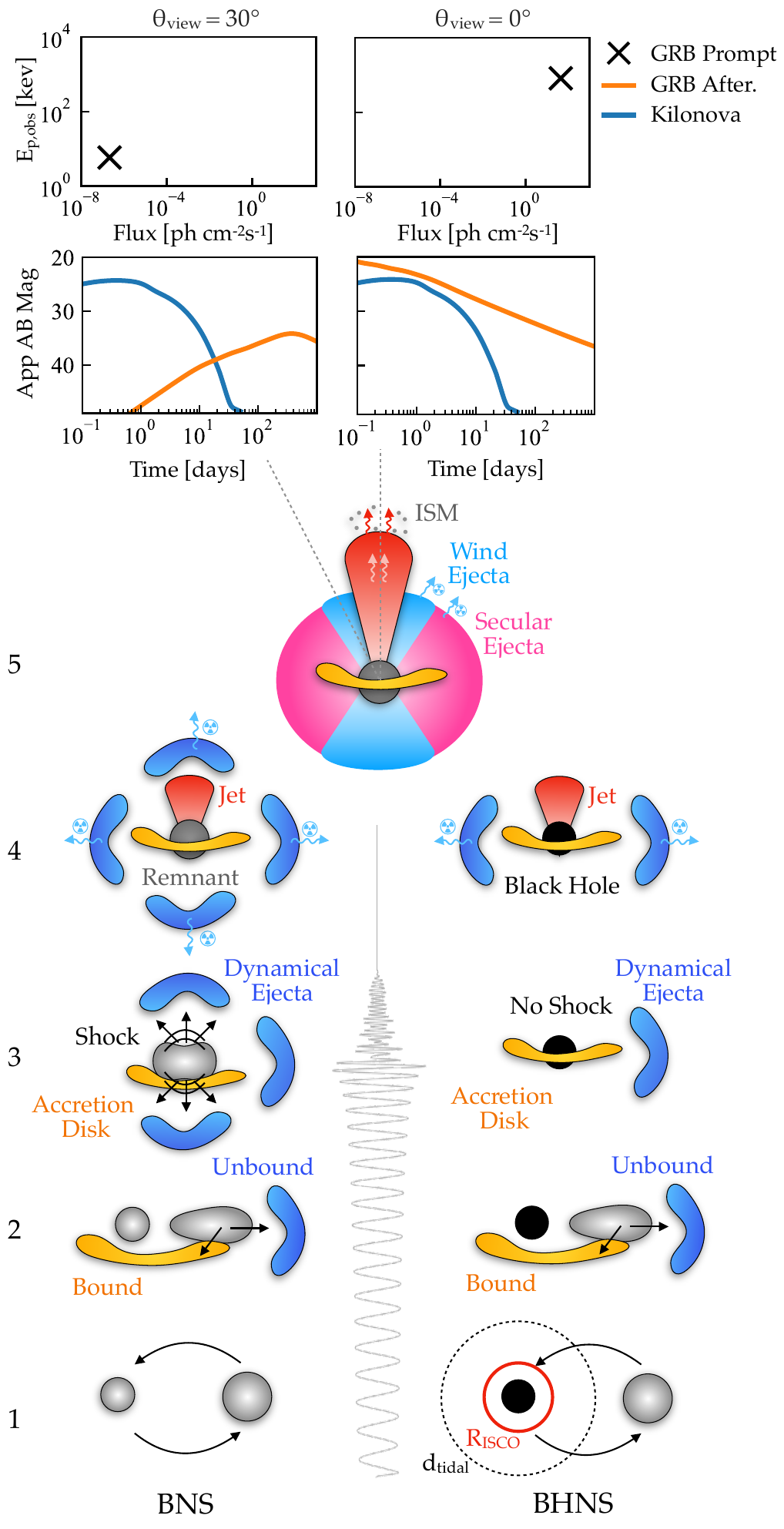}
    \caption{Schematic representation of BNS and BHNS mergers and their multi-messenger emissions.}
    \label{fig:dynamic_sketch}
\end{figure}

In figure~\ref{fig:dynamic_sketch} we present a schematic representation of the dynamics of BNS and BHNS mergers and their respective multi-messenger emissions. For BNS systems, during the final phase of the inspiral, the development of tidal forces leads to the partial disruption of the stars, producing an emission of neutron-rich material. This material can be divided into two main components: a gravitationally unbound component, called dynamical ejecta, which leaves the merger region, and a bound component, which forms an accretion disk around the remnant. Compression and shock heating lead to additional emissions when the two stars' surfaces collide. The emitted neutron-rich material is an ideal site for r-process nucleosynthesis, which produces the heaviest elements in the universe. The synthesized nuclei are unstable, and their radioactive decay can power a KN emission. On longer timescales, other emissions originate from the disk: some ejecta are produced by magnetic pressure and neutrino-matter interactions, called wind ejecta, and some ejecta are produced by viscous processes inside the disk, called secular ejecta.

When the NS is disrupted and some of the released material forms a disk, its accretion onto the remnant can induce the launch of a relativistic jet. Internal shocks inside the jet and/or magnetic reconnection phenomena lead to the dissipation of a fraction of the jet's kinetic energy, which is radiated, giving rise to the prompt emission of the GRB. The jet then continues its expansion into the interstellar medium, where it begins to decelerate. This causes the formation of a shock near which electrons are accelerated, producing synchrotron emission: the afterglow emission of the GRB.

BHNS mergers can also be accompanied by an EM counterpart, as in the BNS case. The condition for ejecta production is that the tidal disruption distance $d_\mathrm{tidal}$, at which tidal disruption occurs, must be larger than the BH's innermost stable circular orbit $R_\mathrm{ISCO}$. This condition is favored if the BH is of low mass and/or rapidly spinning. Otherwise, the NS directly plunges into the BH, and no ejecta are produced (and, consequently, the event has no EM counterpart). Unlike BNS mergers, where the impact of the two surfaces provokes additional emissions making the distribution of the dynamical ejecta much more isotropic, BHs do not have a crust, so no shock occurs. Therefore, these systems produce dynamical ejecta close to the equatorial plane and mainly in about half of the azimuthal angle. Moreover, the absence of a hypermassive or supramassive NS phase can cause differences in the properties of the wind and secular ejecta.

In figure~\ref{fig:dynamic_sketch} we present (in the middle of the stages labeled 1-4) a representation of the GW waveform of a BNS merger.\footnote{The time domain GW signal corresponds to the BNS simulation BAM:0095~\cite{Dietrich:2017aum} in the \textsc{CoRe} database~\cite{Gonzalez:2022mgo}.} At the top, we show a representation of the expected EM emission. Specifically, in the first-row panels we show the observed peak energy and photon flux for the GRB prompt emission, assuming a viewing angle of 0 and 30 degrees. In the second-row panels we present the optical light curve of the KN (blue) and the GRB afterglow 
(orange), again assuming viewing angles of 0 and 30 degrees. The EM emission was calculated using the C22 model for a BNS of 1.4 and 1.4 $M_\odot$ at the median redshift of the population, $z \sim 0.22$. It can be noted that the KN emission is much more isotropic, not showing substantial differences in the light curve when moving to larger viewing angles. In contrast, the GRB emission, being collimated, is much more sensitive to variations in this parameter. When shifting from 0 to 30 degrees, the GRB prompt flux decreases by several orders of magnitude, and the afterglow light curve peaks several months after the merger at a much lower apparent magnitude compared to the on-axis case.

\subsubsection{Multi-messenger BNS \& BHNS population models}
\label{sec:popmod}

We considered three population models for BNS: Colombo et al. 2022 (C22), Ronchini et al. 2022 (R22) \cite{Ronchini:2022gwk} and Loffredo et al. 2025 (L25) \cite{Loffredo:2024gmx}. For the BHNS population we consider the model of Colombo et al. 2023 (C23) \cite{Colombo:2023une}. These models are based on different assumptions of the hyperparameters describing the population of binaries, their evolution and their EM emission components. Table \ref{tab:compare_models} lists the main assumptions of the models providing the key references. In this section, we provide a brief description of these models and compare their different assumptions. The reader is referred to the relevant publication for a more in-depth description of each model. For C22 and C23 updated populations are given in  \citep{Colombo:2025sdm}.

\begin{table}[!h]
  \caption{List of the main characteristics of the four different models used to compute EM emission from BNS and BH-NS mergers detectable by ET. 
  }
  \label{tab:compare_models}
  \scriptsize
  \scalebox{0.93}{
  \begin{tabular}{lllll}
  \hline \\ [-1.5ex]
  & \textbf{C22} \cite{Colombo:2022zzp} & \textbf{R22} \cite{Ronchini:2022gwk} & \textbf{L25} \cite{Loffredo:2024gmx} & \textbf{C23} \cite{Colombo:2023une}\\ 
  \hline \\ [-1.5ex]
  CBC System & BNS & BNS & BNS & BH-NS \\
  CBC Pop. & R0+delayed SFR & Santoliquido+21 & Iorio+23 & Broekgaarden+21 \\
  $\mathcal{R}_0$ [Gpc$^{-3}$ yr$^{-1}$] & $347^{+536}_{-256}$ & $365$ & $107(23)$\tablefootnote{The value for the local merger rate corresponds to the fiducial population in L25, while the value for the pessimistic population is provided in parenthesis. These populations are consistent with the BNS mergers observed by LVK, as detailed in \citep{Loffredo:2024gmx}.} & $149^{+153}_{-87}$\\
  Mass distr. & Analytic & Uniform  & Gaussian & Broekgaarden+21 \\
  EOS & SFHo & $-$ & BLh & SFHo \\
  Fisher analysis & \texttt{GWFast} & \texttt{GWFish} & \texttt{GWFish} & \texttt{GWFast}\\
  SNR cut & 12 & 8 & 8 & 12 \\
  EM Transients & KN, GRB & GRB & KN & KN, GRB \\
  KN ejecta model & \cite{Radice:2018pdn, Barbieri:2019kli, Kruger:2020gig}  & $-$ & \cite{Loffredo:2024gmx, Radice:2018pdn, Kruger:2020gig} & \cite{Kawaguchi:2016ana, Kruger:2020gig} \\
  KN emission model & \cite{Perego:2017wtu, Breschi:2021tbm} & $-$ & \cite{Ricigliano:2023svx} & \cite{Barbieri:2019kli, Perego:2017wtu} \\
  Jet launch \& breakout & Yes & No & $-$ & Yes\\
  Cocoon Emission & Yes & No & $-$ & No\\
  \hline \\ [-1.5ex]
  \end{tabular}
  }
\end{table}

\paragraph{Progenitor compact binary population properties.}
\label{sec:progmod}

The population models considered here predict different local merger rates, $\mathcal{R}_0$, compatible with current constraints provided by the Advanced LIGO and Virgo detectors in GWTC3 \cite{KAGRA:2021vkt} ($\mathcal{R}_{0} \in [10,~1700]$ Gpc$^{-3}$ yr$^{-1}$ for BNS and $\mathcal{R}_{0} \in [7.8,~140]$ Gpc$^{-3}$ yr$^{-1}$ for BHNS). Given the lack of BNS merger detections in the ongoing run (O4) of LVK, we tune our populations on a fiducial local merger rate closer to the lower bound of the GWTC3 constraint. In particular, for BNS the local rates considered correspond to $\mathcal{R}_0 = 347,~365,~107$ Gpc$^{-3}$ yr$^{-1}$ for C22, R22, and L25, respectively
We adopt $\mathcal{R}_0 = 149$ Gpc$^{-3}$ yr$^{-1}$ for the BHNS population of C23.

The mass distribution of NSs (isolated and in binaries) is still uncertain, therefore we explore three possibilities for BNS systems. C22 models the mass distribution of BNSs via an analytic fit to both Galactic 
BNS and the two BNS mergers detected through GWs (GW170817 and GW190425). R22 adopts a uniform NS mass distribution between $1.0~M_\odot$ and $2.5~M_\odot$, which is consistent with the GW detections, while L25 considers a Gaussian distribution centred at $1.33 M_\odot$ with a standard deviation of $0.09 M_\odot$, which fits the mass of Galactic NSs in binaries. The NS and BH mass distributions in C23 are derived from the population synthesis model of \cite{Broekgaarden:2021iew} ensuring that the mass distribution of the population of BHNS is compatible with the constraints imposed by GW observations \cite{Biscoveanu:2022iue}. NSs in binaries are predicted to have a negligible spin, therefore we assume spinless NSs in all our populations. BHs might have a non-negligible spin before the merger. However, since larger BH spins in BHNS binaries favour a stronger EM emission, we choose a conservative approach imposing a BH spin parameter $\chi_{\rm BH} = 0$. R22 and L25 compute the redshift distribution of BNS mergers using the semi-analytic code {\sc{cosmo$\mathcal{R}$ate}} \cite{Santoliquido:2020bry}, which models the star formation rate density and the average metallicity evolution of the Universe as in \cite{Madau:2016jbv}. Conversely, C22 models the BNS redshift distribution analytically, using a delay time distribution compatible with the results of population synthesis models \cite{Dominik:2012kk}. For BHNS C23 assumes the fiducial metallicity specific star formation rate density from \cite{Broekgaarden:2021iew}, based on the phenomenological model described in ref. \cite{2019MNRAS.490.3740N}.

The EOS of NSs determines the tidal deformability of NSs in binaries and affects the properties of the EM emission. Given the present uncertainties on the NS EoS, we assume two different nuclear EOSs, namely SFHo \cite{Steiner:2012xt} (for C22 and C23) and BLh \cite{Bombaci:2018ksa, Logoteta:2020yxf} (for L25). These EOSs are built within very different theoretical frameworks, in particular, SFHo is based on relativistic mean field theory while BLh is a microscopic EOS based on the Brueckner-Bethe-Goldstone many-body theory. However, these EOSs predict similar values for the NS maximum mass and corresponding radius, specifically $M_{\rm max} = 2.06 M_\odot$ and $R_{\rm max} = 10.3$ km for SFHo, while $M_{\rm max} = 2.10 M_\odot$ and $R_{\rm max} = 10.4$ km for BLh. Both the EOSs are compatible with current experimental constraints on nuclear matter properties and astrophysical constraints on NS maximum mass, radius, and tidal deformability. They both produce quite compact NSs, with SFHo predicting slightly more compact NSs compared to BLh. By selecting the SFHo and BLh EOSs we made a conservative choice for the modeling of the KN emission since more compact NSs in binaries are associated with fainter KNae. A different approach is followed in R22, which does not make any assumption on the NS EoS. Accordingly, C22, L25, and C23 compute the NS dimensionless tidal deformability, $\Lambda$, from the mass and the EOS of each NS, while R22 samples $\Lambda$ uniformly in the interval $[0, 2000]$ and independently for each NS. In all the considered models, the BNS and NS-BH mergers are distributed isotropically in the sky with a random inclination of the orbital plane to the line of sight.


\paragraph{Gravitational wave detection and parameter estimation simulation.}
\label{sec:gwmod}

The parameter estimation analysis of the BNS population for L25 and for R22 was carried out using the Fisher matrix code \texttt{GWFish} \cite{DupletsaHarms2023}. For the BNS and BHNS populations of C22 and C23, respectively, \texttt{GWFast} \cite{Iacovelli:2022bbs, Iacovelli:2022mbg} is adopted. These codes have been cross-checked within the activities of the \textit{Common Tools} OSB division and yield comparable results. For a comprehensive description of the codes, we refer to section~\ref{section:div9}.

The settings of the gravitational analysis are common across the three BNS population models and the two different codes. We used the \texttt{IMRPhenomD\_NRTidalv2} waveform approximant \cite{Khan:2015jqa, Dietrich:2019kaq}, which allows us to include tidal effects. We used the full cryogenic (HFLF-cryo) sensitivity curve for ET, including both low and high-frequency\footnote{All the available curves, including HFLF-cryo, can be found at \href{https://apps.et-gw.eu/tds/?content=3&r=18213}{https://apps.et-gw.eu/tds/?content=3\&r=18213}.}, with the starting frequency corresponding to $2$\,Hz. 

We analyzed the two following detector configurations~\cite{Branchesi:2023mws}:
\begin{enumerate}
    \item the reference triangular ET design of 10\,km located in Sardinia (Sos Enattos) with the HFLF-cryo sensitivity curves;
    \item two L-shaped interferometers of 15\,km arms misaligned at 45 deg located, respectively, in Sardinia and in the Meuse-Rhine region.
\end{enumerate}

For BHNS systems of C23, the GW analysis,  
we consider the same configurations, sensitivities, and starting frequency of 2~Hz as for the BNS case, but we adopt the waveform approximant \texttt{IMRPhenomNSBH}, which is more suitable for this kind of systems \cite{Pannarale:2015jka,Dietrich:2019kaq}. In particular, it is tuned up to mass ratios of $q\sim100$, it includes tidal effects and the signatures on the signal due to different possible outcomes of the merger (plunge or disruption).

\paragraph{Electromagnetic emission models.}
\label{sec:emmod}

\subparagraph{Kilonova.}

C22 computes the BNS KN light curves from 0.1--50 days, using the anisotropic three-component model of \cite{Perego:2017wtu,Breschi:2021tbm}. The model accounts for the emission from three different outflow components: 
\begin{itemize}
\item dynamical ejecta, produced by tidal interactions on a timescale of few milliseconds; 
\item winds ejecta, produced by the accretion disk, due to neutrino matter-interactions and magnetic pressure on a time scale of tens of milliseconds; 
\item secular ejecta, also produced by the accretion disk, due to viscous processes of magnetic origin throughout the duration of the accretion.
\end{itemize}
The ejecta are assumed to be in homologous expansion and thermal emission is computed following \cite{Grossman:2014,Martin:2015}.
The expected ejecta mass, ejecta average velocity and accretion disk mass are computed using numerical relativity-informed fitting formulae \cite{Radice:2018pdn,Barbieri:2021,Kruger:2020gig}, assuming the SFHo EOS \cite{Steiner:2012xt} and the masses of the wind and secular ejecta are computed as fractions of the disk mass: $\xi_\mathrm{w} = 0.05$ and $\xi_\mathrm{s} = 0.2$, respectively. The KN model parameters, such as the ejecta geometry, opacity, and the velocity of the wind and secular ejecta, are assumed as the best-fit parameters from \cite{Barbieri:2019kli}.

For the BHNS population in C23, the KN light curves are computed in the same way as for the BNS in C22, accounting for the expected differences in the KN model parameters \cite{Barbieri:2019kli}. In this case, they compute the expected mass $m_\mathrm{dyn}$ and average velocity $v_\mathrm{dyn}$ for the dynamical ejecta and the mass $m_\mathrm{out}$ remaining outside the remnant BH, using the numerical relativity-informed fitting formulae from \cite{Kruger:2020gig,Kawaguchi:2016ana,Foucart:2018}. For the binaries where $m_\mathrm{dyn}>0.5 m_\mathrm{out}$, they assume $m_\mathrm{dyn} = m_\mathrm{dyn}^\mathrm{max}(m_\mathrm{out}) = 0.5 m_\mathrm{out}$ \cite{Foucart:2019}. Finally, the mass of the accretion disk is computed as $M_\mathrm{disk} = m_\mathrm{out} - m_\mathrm{dyn}$.

L25 models the BNS KN emission from 1 hour to 10 days assuming a two-component anisotropic ejecta. The first ejecta component (C1) includes matter ejected on a dynamical timescale (dynamical ejecta) and matter unbound from the disk through spiral-wave winds. The second component (C2) contains matter unbound from the disk through viscous processes on a secular timescale (secular ejecta). The mass and velocity of the dynamical ejecta and the amount of mass in the disk are computed employing numerical relativity-informed fitting formulas reported in \cite{Loffredo:2024gmx}, assuming the BLh EOS \cite{Bombaci:2018ksa, Logoteta:2020yxf}. These formulas model the possibility of the prompt collapse of the remnant to a BH, which is favoured for more massive and asymmetric binaries. A prompt collapse to BH affects the ejecta properties, reducing the amount of dynamically ejected matter and preventing the formation of spiral wave winds. Therefore, if the remnant undergoes prompt collapse to BH, the amount of matter unbound through spiral-wave winds is set to zero, while it corresponds to a fraction $\xi_{sw} = 0.15$ of the disk mass, otherwise. The total amount of mass in the first component, $m_{\rm C1} = m_{\rm dyn} + m_{\rm sw}$, is distributed as $\sin\theta$, where $\theta$ is the polar angle. The corresponding velocity, $v_{\rm C1}$, is obtained as a weighted average of the velocity of the dynamical ejecta, $v_{\rm dyn}$, and the velocity of the spiral-wave wind $v_{\rm sw} = 0.17 c$ \cite{Nedora:2019jhl}. The mass in the second component, $m_{\rm C2}$, is modelled as a fraction $\xi_{\rm s} = 0.3$ of the disk mass and the velocity is set to $v_{\rm C2} = 0.06c$. The ejecta properties such as opacity, entropy, expansion timescale, and floor temperatures are chosen as in \cite{Loffredo:2024gmx}, ensuring that the KN light curves of the BNS in the population with chirp mass compatible with GW170817 reproduce AT2017gfo data. 
The KN light curves are computed using the semi-analytic model \texttt{xkn-diff} \cite{Ricigliano:2023svx}, based on a semi-analytic solution of the radiative transfer equation for homologously expanding matter.

\subparagraph{Jet-related emission.}

C22 and C23 adopt, for both the BNS and the BHNS population, a jet angular structure inspired by GRB170817A \cite{Ghirlanda:2019} with an half-opening angle of 3.4 deg. They compute the GRB afterglow light curves from $0.1$ to $1000$ days in the radio ($1.4$ GHz), optical ($g$ band) and X-rays (1 keV) bands, fixing the median interstellar medium density $n=5\times10^{-3}\,\mathrm{cm^{-3}}$ \cite{Fong:2015} and assuming the microphysical parameters characteristic of GW170817, $\epsilon_\mathrm{e}=0.1$, $\epsilon_\mathrm{B}=10^{-3.9}$ and $p=2.15$. 

In C22 and C23, the methodology for GRB prompt emissions is based on models similar to those in \cite{Barbieri:2019kli} and \cite{Salafia:2019}. They assume that a constant fraction $\eta_\gamma=0.15$ of the jet energy density, confined to regions with a bulk Lorentz factor $\Gamma\geq 10$, is emitted as photons. The resulting observed spectrum is the integration of this radiation over the jet's solid angle, incorporating relativistic beaming effects. Finally, photon fluxes in specific energy bands are calculated, assuming a standard burst duration of 2 seconds.

In C22 they assume the launch of a relativistic jet just for BNS mergers where the remnant mass $M_\mathrm{rem}\geq 1.2 M_\mathrm{TOV}$, which suggests either an immediate or delayed collapse into a black hole. For BHNS mergers in C23, they consider the formation of a relativistic jet with energy $E_\mathrm{c}$ if the disk mass $M_\mathrm{disk}>0$. In both works, if the jet energy exceeds a specific threshold (as defined by \cite{Duffell:2018}), it successfully breaks through the ejecta, resulting in GRB prompt and afterglow emissions.

For BNS systems observed within a viewing angle $\theta_\mathrm{v}\leq 60^\circ$, C22 also incorporates a cocoon shock breakout component, modeled in line with the properties of GRB170817A, namely a luminosity $L_\mathrm{SB} = 10^{47}$ erg s$^{-1}$, a cutoff power-law spectrum with $\nu F_\nu$ peak photon energy $E\mathrm{p,SB} = 185$ keV and low- energy photon index $\alpha = -0.62$. In BHNS systems, C23 does not include this additional emission due to the absence of observational constraints for this type of source.

For the prompt emission, R22 considers a universal jet structure, corresponding to the profile derived for GW170817. The jet is assumed to have a core with an opening angle of 3.4 deg. The profile outside the jet core follows a power-law with index $s=4$. The values of jet energy, on-axis rest-frame peak energies and pulse duration are extracted from Eqs. (6), (7) and (8) of R22, respectively. The parameters of the distributions, including also the probability that a BNS produces a successful jet, are calibrated performing a comparison with existing samples of short GRBs observed by Fermi. 

For the computation of the afterglow emission, the \texttt{afterglowpy} model is adopted~\cite{Ryan:2019fhz}. The isotropic-equivalent kinetic energy of the jet is derived starting from the posterior distribution of the isotropic-equivalent prompt energy taken from R22. The micro-physical parameters are $\epsilon_e=0.1$, $p=2.2$, $\log_{10}{\epsilon_B}$ uniformly distributed in the range $[-4,-2]$ and $n \in [0.25,15] \times 10^{-3}$ cm$^{-3}$.

\subsubsection{EM properties of the GW-detectable sub-population}
\label{sec:emcomp}

In this section we describe the observed properties of the EM emission of sources detected in GW by ET. We also provide the rates of joint detectable sources as a function of a variable detection limiting flux in different energy bands for each emission component. It is important to distinguish between detectable and detected joint EM-GW events. The former are those sources, detected as GW signals by ET, which are above a given sensitivity limit, i.e., they are in principle bright enough to be detectable by any facility reaching that sensitivity. We will consider this condition in defining detectable joint events. However, the rate of joint detected sources may be smaller than the rate of detectable ones. Indeed, the latter should be scaled by any EM facility property other than its sensitivity, such as the field of view, the observability constraints, the duty cycle etc. In this section, we will only report the joint detectable event rates. We will provide more realistic detected rates in section~\ref{sect:Obsfac_div4} by considering specific mission/facilities.\footnote{We adopt the convention to distinguish between ``missions" and``facilities" referring to space-based detectors and ground based telescopes, respectively. } 

\paragraph{Kilonova.}
\label{KNGG}

\begin{figure*}
    \centering
    \includegraphics[width=\textwidth]{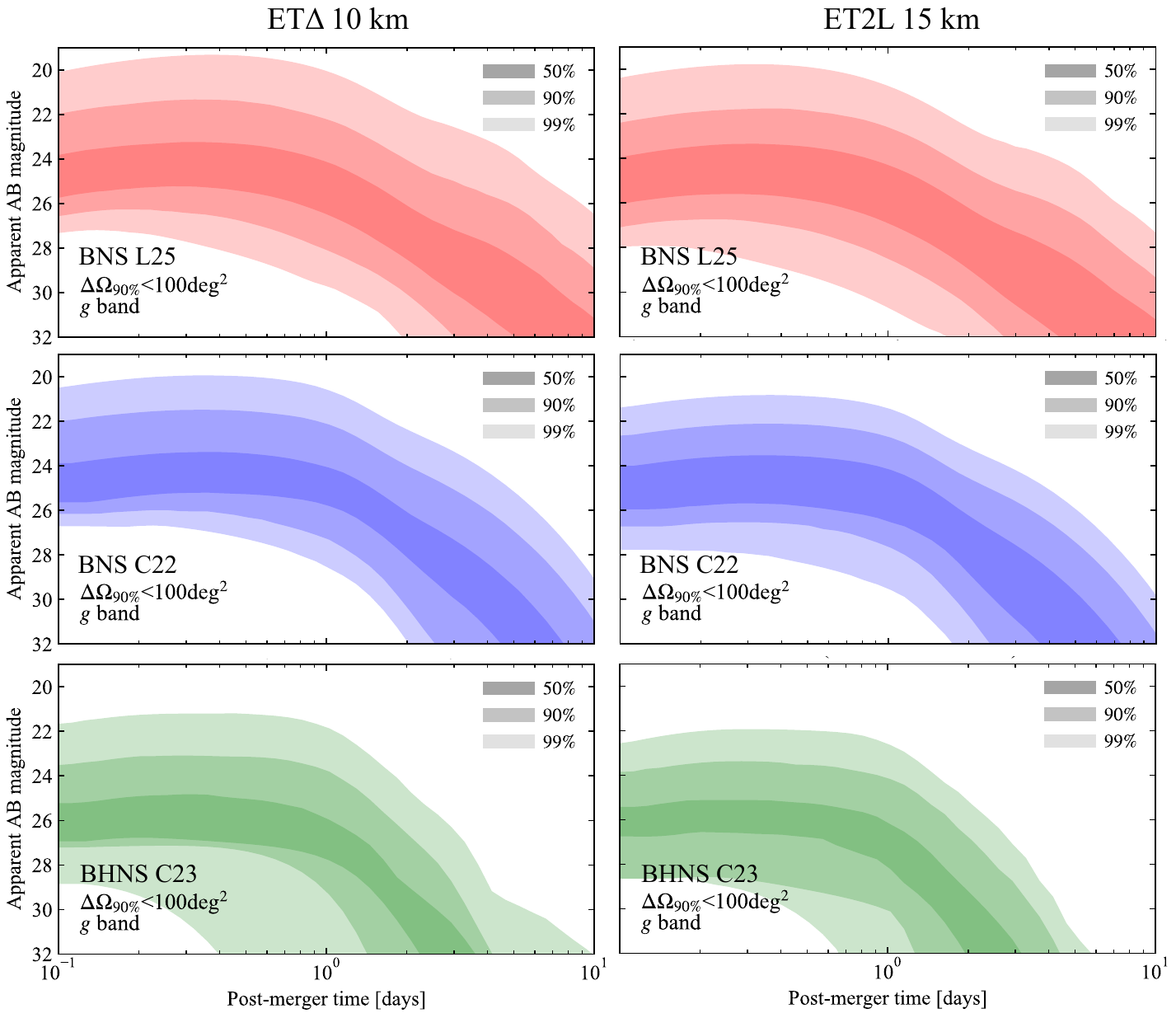}
    \caption{Distribution of ET optical KNe as a function of time assuming the BNS population from L25 (red, first row) and C22 (blue, central row), and the BHNS population from C23 (green, last row). The left column assumes the ET-$\Delta$ configuration, the right column is the 2L configuration. The plots represent the apparent AB magnitude versus post-merger time (days) for our simulated KN light curves in the $g$ (484 nm) band, restricting to ET GW-detectable sources with $\Delta\Omega_{{\rm 90}\%}<100\,\mathrm{deg}^2$. The shaded regions contain $50\%$, $90\%$, and $99\%$ of the KN light curves.}
    \label{fig:kn}
\end{figure*}

\begin{figure*}
    \centering
    \includegraphics[width=\textwidth]{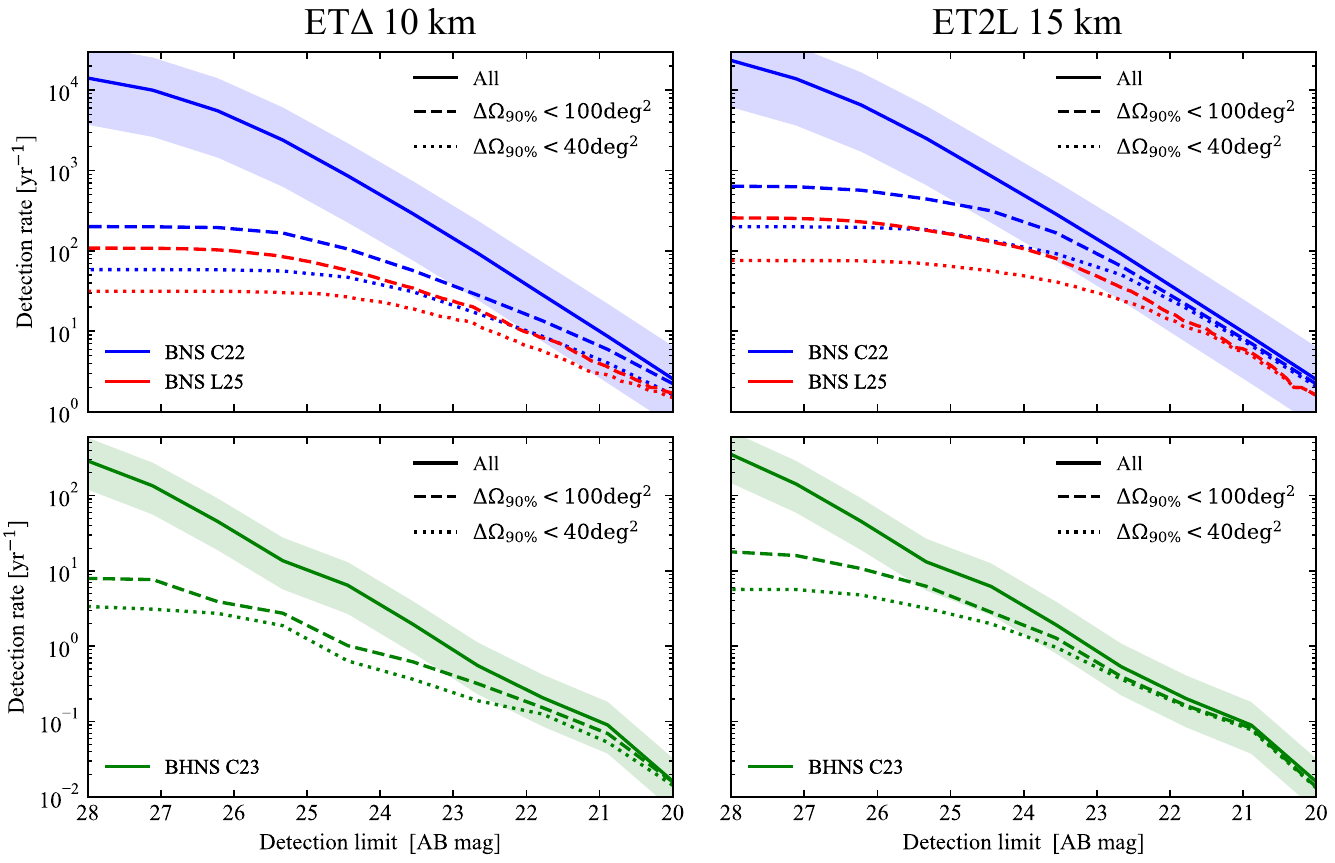}
    \caption{Detection rate as a function of detection limit threshold for the KN BNS population presented in L25 and C22 (upper panels, red and blue, respectively) and for the BHNS population from C23 (lower panels, green). The left column assumes the ET-$\Delta$ configuration, and the right column is the 2L configuration.
    The lines indicate the KN+GW ($g$ band) detectable binaries, assuming all the BNSs (solid line from C22 upper panels and C23 lower panels), the ones with $\Delta\Omega_{{\rm 90}\%}<100\,\mathrm{deg}^2$ (dashed lines) and the ones with $\Delta\Omega_{{\rm 90}\%}<40\,\mathrm{deg}^2$ (dotted lines). The shaded region indicates the uncertainty due to the local merger rate.}
    \label{fig:kn_rate}
\end{figure*}


Figure~\ref{fig:kn} shows the distributions of the KN lightcurves in the \textit{g} band of the subsample of sources (BNS and BHNS) detected as GW events and localized with a sky accuracy of up to 100 deg$^{2}$ by ET in two different configurations (10 km triangle and 15 km 2L - left and right panels, respectively). There are no appreciable differences in the light curves detected by the two configurations for either the BNS or BHNS populations. BNS kilonovae typically exhibit AB magnitudes in the range of 20-27 (99\% confidence region) at approximately 2.4 hours post-merger, and their brightness remains almost constant (or slightly increases) until about 1 day, and then decreases rapidly over the 10-day timescale. Despite still being perfectly consistent within their dispersion, the BNS-KNae lightcurves of L25 show a shallower decay with respect to those of C22 owing to the different KN model parameters, as well as the different ejecta distribution in the two populations. 

BHNS systems (bottom panels of figure~\ref{fig:kn}) produce KN light curves, which are typically fainter (AB magnitudes in the range 22-29 at 2.4 hours post-merger) than BNS and evolve towards a faster decline within 1 day from the merger. This may be explained, respectively, by the larger GW horizon of BHNS systems with respect to BNS and the different ejecta properties resulting from the assumed mass and spin distributions and EoS. 

Figure~\ref{fig:kn_rate} depicts the joint detection rate as a function of different AB magnitude detection limits for the BNS populations in L25 (red) and C22 (blue), and for the BHNS population in C23 (green). The panels on the right assume the ET triangle configuration, while those on the left assume ET-2L. Solid lines indicate all detectable KN+GW systems, while dashed and dotted lines represent only events with a sky localization $\Delta\Omega_{{\rm 90}\%}<100\,\mathrm{deg}^2$ and $\Delta\Omega_{{\rm 90}\%}<40\,\mathrm{deg}^2$, respectively. 

The colored band represents the uncertainty due to the local merger rate, which propagates as a constant relative fraction to all other rates. Despite different assumptions for the initial BNS population and the KN model, L25 and C22 predict rates that differ by only a factor of about 2-3, and when considering the uncertainty due to the local merger rate, they are entirely compatible.

Using this plot, given an instrument with specific sensitivity, one can estimate the rate of GW events with a KN brighter than this limit, thus estimating the rate of potentially detectable KNe. For instance, the Vera Rubin Observatory can reach an AB magnitude of 26 in the $g$ band with an exposure time of 180s. The majority of KNe lightcurves are brighter than this limiting magnitude up to 1-2 days post-merger, see figure~\ref{fig:kn}). 
Considering this representative limit and events with $\Delta\Omega_{{\rm 90}\%}<100\,\mathrm{deg}^2$, L25 yields a rate of detectable  KNe  $100(21)$ yr$^{-1}$,\footnote{The rate of detectable into parenthesis refers to the pessimistic population of L25~\citep{Loffredo:2024gmx}} while C22 yields a rate of $191^{+295}_{-141}$ yr$^{-1}$. 

For BHNS events, the rate is $5.1^{+5.2}_{-3.0}$ yr$^{-1}$. In general, the BHNS rates are systematically lower than those of BNS events. The reasons, as explained in C23, are due to the fact that only a small fraction (about 2\%) of events can emit ejecta, and the KN population is generally less luminous.

\paragraph{Gamma-Ray Burst prompt emission.}

\begin{figure*}
    \centering
    \includegraphics[width=\textwidth]{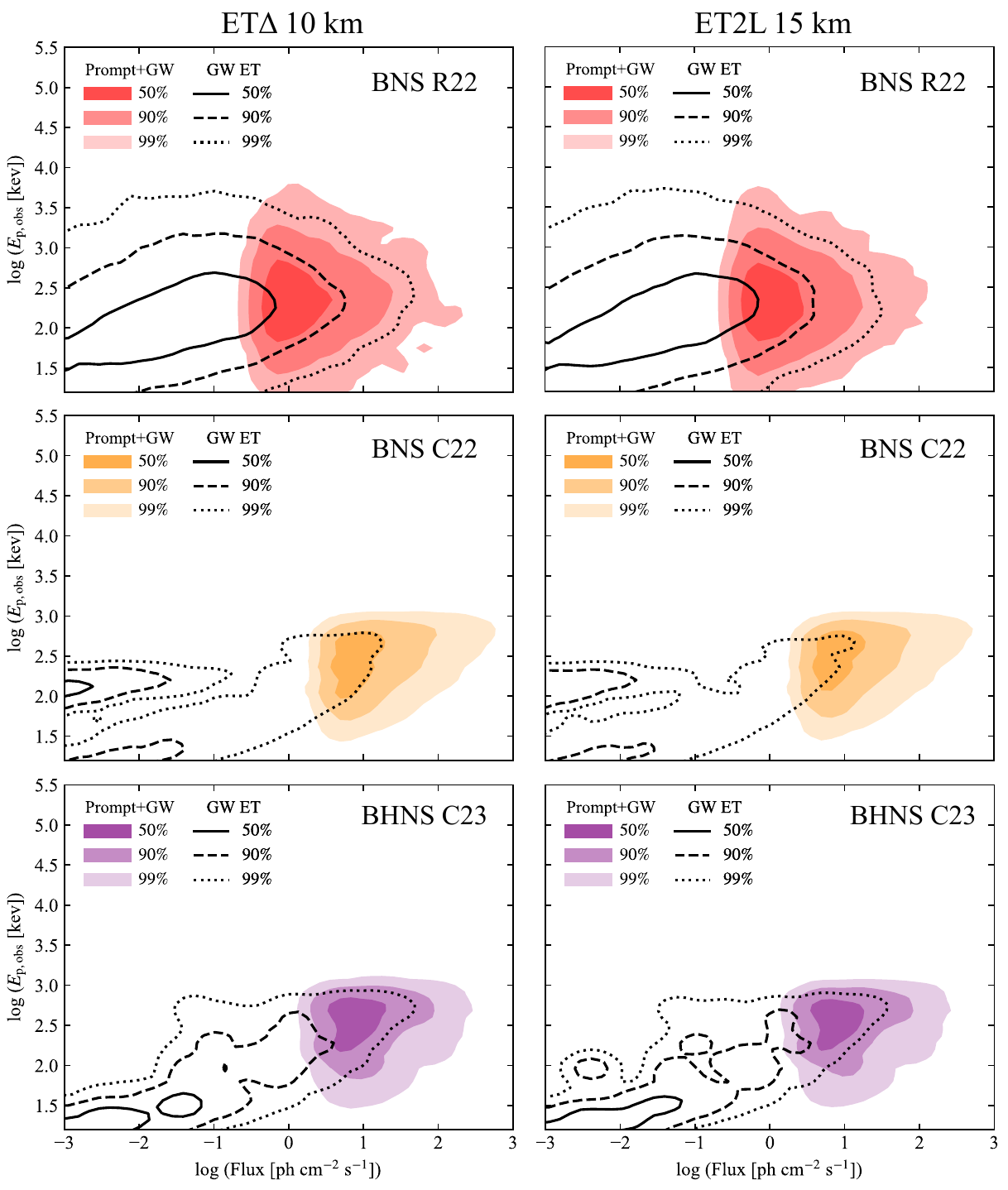}
    \caption{Observed spectral energy distribution peak photon energy $E_\mathrm{p,obs}$ vs photon flux for the GRB prompt BNS population presented in R22 (first row) and C22 (central row), and for the BHNS population from C23 (last row). The colored-filled regions contain 50\%, 90\%, and 99\% of the binaries, both GRB Prompt and ET detectable. The black lines contain 50\%, 90\%, and 99\% (solid, dashed, and dotted, respectively) of the ET-detectable binaries.}
    \label{fig:prompt_ep_pf}
\end{figure*}

Figure~\ref{fig:prompt_ep_pf} presents the observed peak energy $E_{\mathrm{p,obs}}$ as a function of the photon peak flux for the BNS populations of R22 (upper panels) and C22 (mid panels), and for the BHNS population of C23 (lower panels). The colored regions include events that are both GW and GRB prompt-detectable. For prompt detection, Fermi/GBM was used as a reference instrument, with threshold values chosen to replicate the short GRB observations of Fermi/GBM: 0.5 ph cm$^{-2}$ s$^{-1}$ for R22 and 4 ph cm$^{-2}$ s$^{-1}$ for C22. Note that R22 considers the 50-300 keV band, while C22 assumes the 10-1000 keV band.

Generally, GW+GRB prompt events extend from an $E_{\mathrm{p,obs}} \sim 30$ keV to $E_{\mathrm{p,obs}} 
10$ MeV, concentrating around $E_{\mathrm{p,obs}} \sim 300$ keV. For C22, the events do not extend beyond $1$ MeV due to different assumptions regarding the jet peak energy. Concerning the photon flux, all models reach up to $100$ ph cm$^{-2}$ s$^{-1}$, while the 50\% of events concentrates between 0.5-10 ph cm$^{-2}$ s$^{-1}$, depending on the flux cut assumption.

The black lines enclose 50\% (solid), 90\% (dashed), and 99\% (dotted) of events detectable by ET, without the requirement of prompt detection. In this case, the population extends to much lower fluxes and peak energies and becomes dominated by off-axis events.  While these events are out of the reach of Fermi/GBM mainly due to its sensitivity and low energy threshold, next-generation GRB detectors may extend their energy band below 10 keV with an improved sensitivity, thus capturing the bulk of the prompt emission. 

The BNS C22 population shows two clusters when considering only the events detected by ET. This effect is due to the assumption that for events with a viewing angle less than 60 degrees, there is also a shock breakout component whose properties mimic those of GRB170817A. Events dominated by the cocoon shock breakout component form the upper cluster and become insignificant when also considering the detection by Fermi/GBM, which is dominated by events with a higher photon flux.

Figure~\ref{fig:prompt_cumul} shows the detection rate as a function of different photon flux detection limits for the BNS populations of R22 (red), C22 (orange), and the BHNS population of C23 (purple). The solid curve refers to all BNS events, while the dashed and dotted lines consider events with sky localization smaller than 100 deg$^2$ and 10 deg$^2$, respectively. The colored bands represent the uncertainty in the local merger rate for C22 and C23, and the Poisson uncertainty due to variations in hyperparameters for R22. Considering the flux cuts assumed by the respective studies using Fermi/GBM, the rates obtained are $57^{+56}_{-35}$ yr$^{-1}$ and $59^{+91}_{-44}$ yr$^{-1}$. Additionally, these rates can be scaled by a factor accounting for the duty cycle and field of view of the instrument. We consider the latter two parameters yielding, for Fermi/GBM, a scaling factor 0.6 so that ET(triangle)-Fermi/GBM joint detection rate for the prompt emission would be $34^{+34}_{-21}$ yr$^{-1}$ and $35^{+55}_{-26}$ yr$^{-1}$. This means that the majority of short GRBs currently detected by Fermi/GBM would have a GW counterpart detected by ET.

\begin{figure*}
    \centering
    \includegraphics[width=\textwidth]{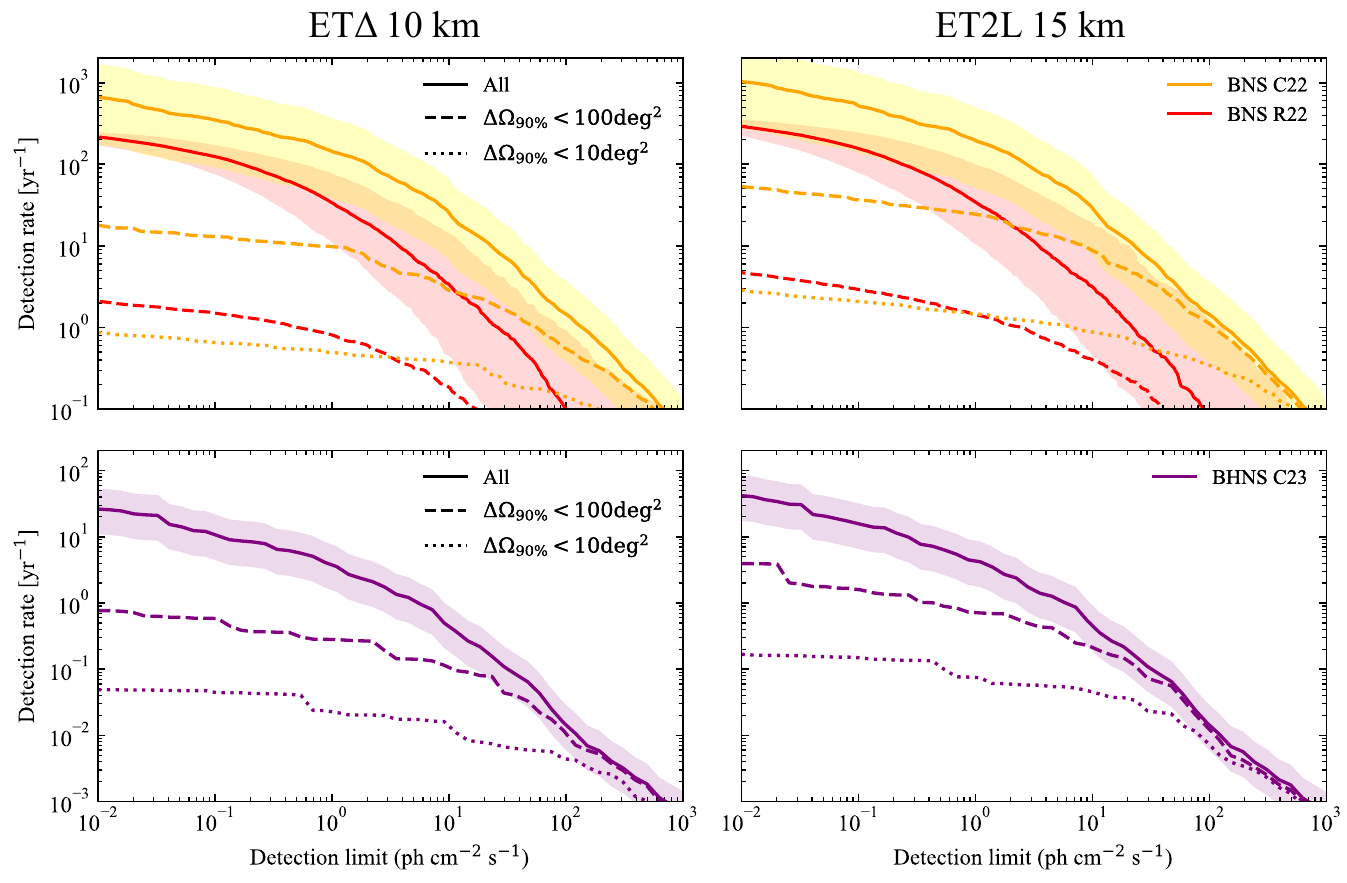}   
    \caption{Detection rate as a function of detection limit threshold for the GRB prompt BNS population presented in R22 and C22 (upper panels, red and orange, respectively) and for the BHNS population from C23 (lower panel). The left column assumes the ET delta configuration, and the right column assumes the 2L configuration.
    The lines indicate the GRB prompt+GW detectable binaries, assuming all the BNSs (solid lines), the ones with $\Delta\Omega_{{\rm 90}\%}<100\,\mathrm{deg}^2$ (dashed lines) and the ones with $\Delta\Omega_{{\rm 90}\%}<10\,\mathrm{deg}^2$ (dotted lines). The shaded region indicates the uncertainty at $90\%$ credible level. For C22 and C23, it indicates the uncertainty due to the local merger rate.}
    \label{fig:prompt_cumul}
\end{figure*}

\paragraph{Gamma-Ray Burst afterglow emission.}
\label{afterglow}

\begin{figure*}
    \centering
    \includegraphics[width=\textwidth]{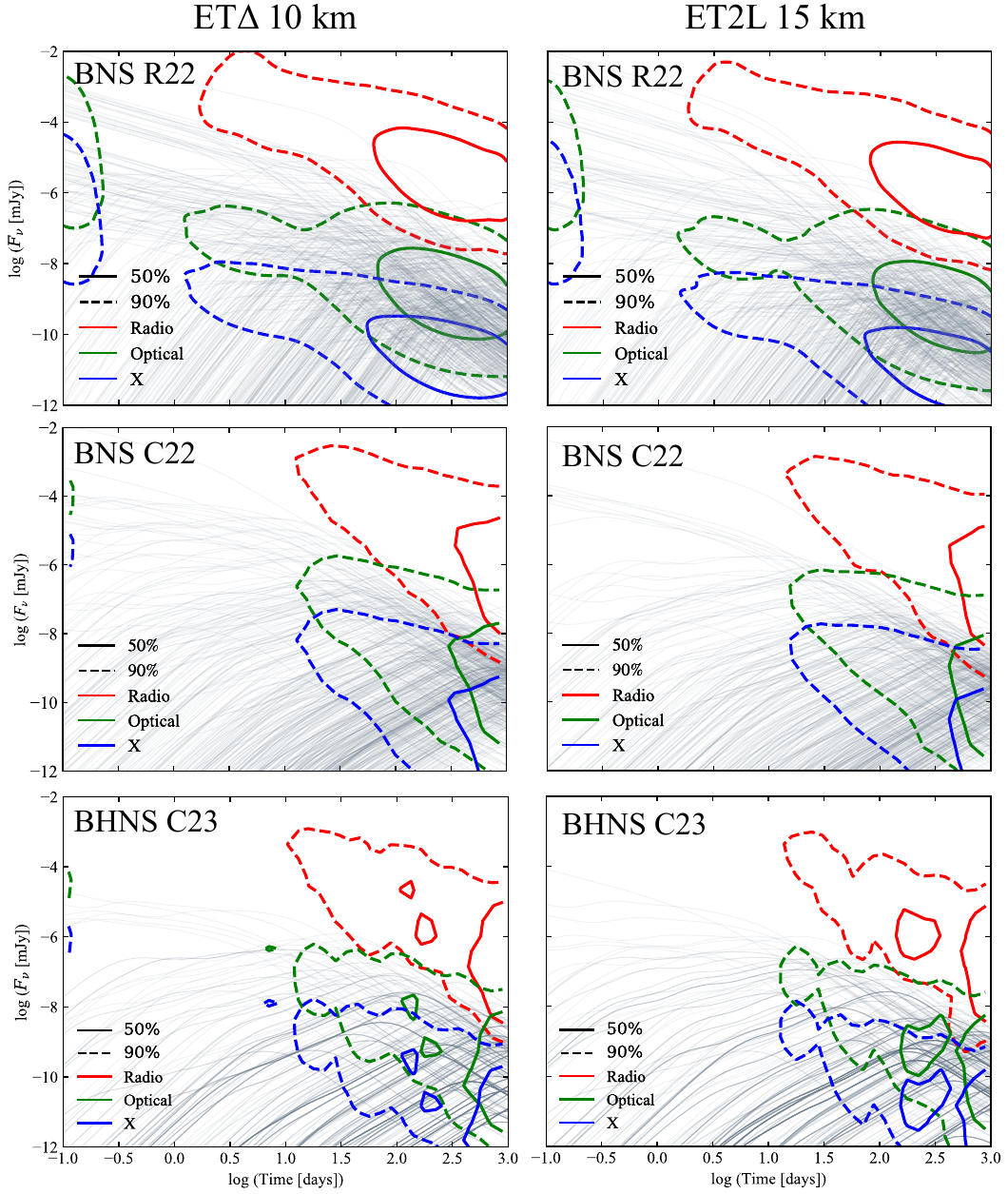}
    \caption{Afterglow lightcurves in Radio (red), Optical (green) and X-ray (blue), for the delta (left) and 2L (right) configurations, assuming the BNS population from R22 (first row) and from C22 (central row), and the BHNS population from C23 (last row). The dashed and solid lines are the 50$\%$ and 90$\%$ containment regions of the afterglow peaks, in the respective bands. The grey lines in the background are randomly sampled optical light curves.}
    \label{fig:aft}
\end{figure*}

Figure~\ref{fig:aft} shows the average properties of the afterglow lightcurves of all the BNS and NSBH mergers detected by ET, considering the delta (left panel) and 2L (right panel) configurations. The figure reports the 50\% (solid line) and 90\% (dashed line) contours containing the peak fluxes of afterglow light curves in the radio (red), optical (green), and X-ray (blue) bands for the BNS populations of R22 (upper panels) and C22 (mid panels), and for the BHNS population of C23 (lower panels). The gray curves in the background are a sample of optical light curves for the respective populations.

\begin{figure*}
    \centering
    \includegraphics[width=\textwidth]{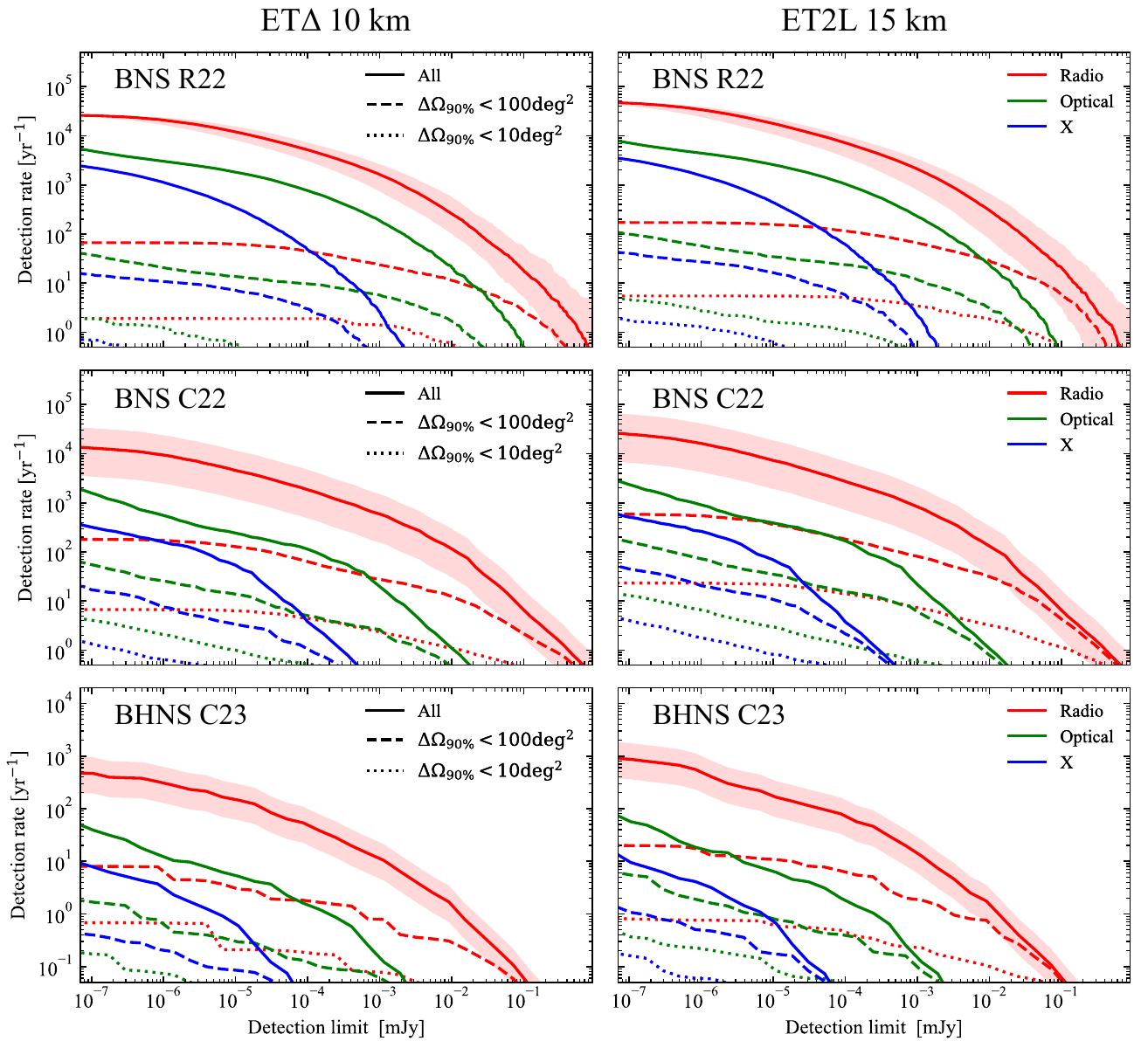}
    \caption{Detection rate as a function of detection limit threshold for the GRB afterglow BNS population presented in R22 (first row) and C22 (central row), and for the BHNS population from C23 (last row). The left column assumes the ET delta configuration, the right column the 2L configuration.
    The lines indicate the GRB afterglow+GW  detectable binaries in the radio (red), optical (green) and X (blue) bands, assuming all the BNSs (solid line), the ones with $\Delta\Omega_{{\rm 90}\%}<100\,\mathrm{deg}^2$ (dashed lines) and the ones with $\Delta\Omega_{{\rm 90}\%}<10\,\mathrm{deg}^2$ (dotted lines). The shaded region indicates the uncertainty at $90\%$ credible level. For C22 and C23, it indicates the uncertainty due to the local merger rate.
    }
    \label{fig:aft_cumul}
\end{figure*}

Generally, all populations have peaks concentrated after $10^2$ days, with a tail extending to shorter timescales. Specifically, the R22 population is concentrated at around $10^{2.5}$ days, while the C22 and C23 populations show a slower growth, with peaks concentrated around $10^3$ days.

Most populations exhibit a subset of bright optical and X-ray events with peaks at very short timescales (hours). These are the on-axis events of the population, as can be seen from the monotonous gray light curves in the background, representing light curves observed at viewing angles comparable to the jet opening angle. This subset is always present, even when the population is dominated by off-axis events, which do not show such curves. Overall, there are no noticeable differences between the delta and 2L configurations, in terms of peak brightness of the afterglow and distribution of the peak times.

Figure~\ref{fig:aft_cumul} represents the detection rates as a function of different flux limits in mJy for the BNS populations of R22 (upper panels) and C22 (mid panels), and for the BHNS population of C23 (lower panels). The lines represent the detectable GW+GRB afterglow events, assuming all binaries (solid line), binaries with sky localization within 100 deg$^2$ (dashed line), and within 10 deg$^2$ (dotted line). In the X-rays, for a representative limiting flux of 10$^{-14}$ erg cm$^2$ s$^{-1}$, R22 predicts around $10^3$ detectable BNS per year,
while C22 predicts around 200 detectable BNS per year. For both R22 and C22, the detectable sources reduce to $\sim$10 and less then 1, considering the ones localized better than 100 deg$^2$ and 10 deg$^2$. The difference can be attributed to the different GW SNR detection threshold and modelling between R22 and C22. 
Overall, the 2L configuration gives approximately a factor two more BNS and NSBH mergers with detectable afterglow.

\subsubsection{GW sky localization of multi-messenger events}

\begin{figure*}
    \centering
    \includegraphics[width=\textwidth]{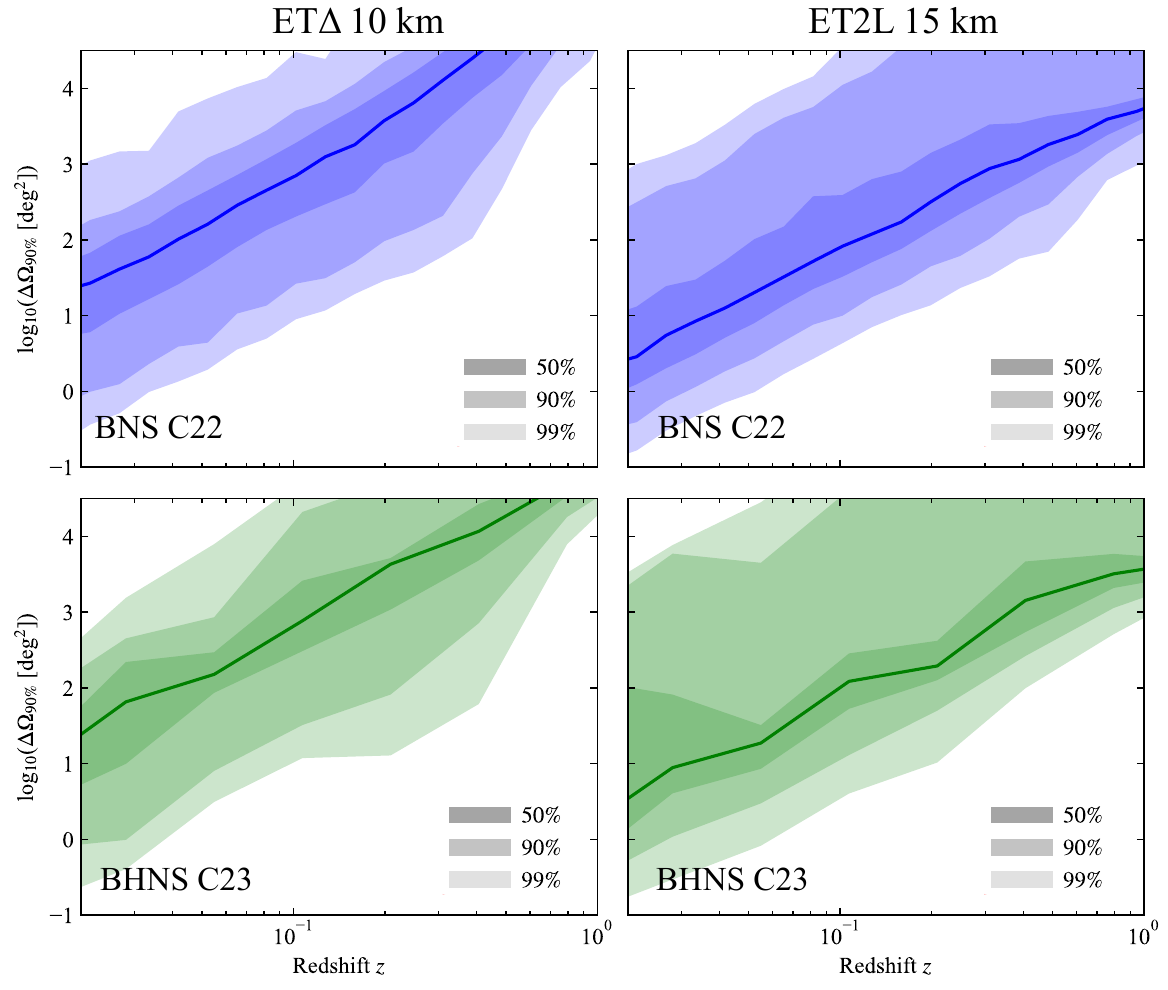}
   
    \caption{Distribution in redshift of the sky localizations of BNS events (upper panel) and BHNS events (lower panel) that can power and EM counterpart, for the delta (left) and 2L (right) configurations. The line indicates the median value, the colored bands are the 50$\%$, 90$\%$ and 99$\%$ credible regions.}
    \label{fig:skyloc_redshift}
\end{figure*}

Figure~\ref{fig:skyloc_redshift}  shows the $\Delta\Omega_{{\rm 90}\%}$ sky localization distribution as a function of redshift for the BNS population from C22 (upper panel, blue) and the BHNS population from C23 (lower panel, green), considering only events capable of emitting an EM counterpart, thus having non-zero ejecta and/or accretion disk. The colored bands represent the 50\%, 90\%, and 99\% credibility regions, while the solid line indicates the median value. The left column assumes the ET configuration with a triangle, while the right column assumes the 2L configuration.

The figure illustrates the gradual broadening of sky localization with increasing redshift. For the triangle configuration, within a redshift of approximately $z \lesssim 0.05$, 50\% of the events have a sky localization between 10 and 100 deg$^2$. Beyond a redshift of $z \gtrsim 0.2$, 50\% have a sky localization exceeding 1000 deg$^2$. In contrast, for the 2L configuration, there is a general decrease in the median value by about an order of magnitude. Specifically, within a redshift of $z \lesssim 0.05$, most events have a sky localization between 1 and 10 deg$^2$, while 50\% have a sky localization exceeding 1000 square degrees for redshifts greater than $z \gtrsim 0.5$. The greater broadening of the distribution for the 2L configuration as compared to the triangle can be traced mainly to events for which only one of the detectors is operational due to our choice of the duty cycle. For a discussion regarding the comparison of the sky localization capabilities of the triangular and 2L ET configurations within the context of Fisher forecasts we refer the reader to section~3 of~\cite{Branchesi:2023mws}.

\subsubsection{ET in a network with Cosmic Explorer}

\begin{figure*}
    \centering
    \includegraphics[width=\textwidth]{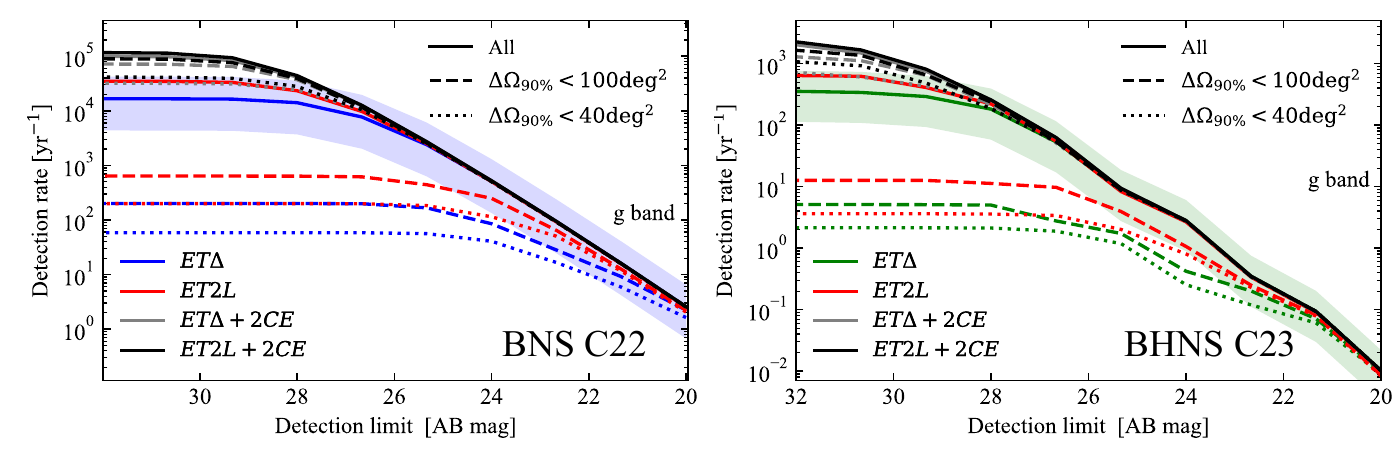}
   
    \caption{Detection rates as a function of the detection limit threshold for the KN BNS population from C22 (left panel) and the BHNS population from C23 (right panel). The different colors represent various ET configurations and detector networks: blue (or green) for ET-$\Delta$, red for ET2L, gray for ET-$\Delta$+2CE, and black for ET2L+2CE. The lines correspond to KN+GW ($g$-band) detectable binaries: solid lines indicate all BNSs (or BHNSs), dashed lines represent events with $\Delta\Omega_{{\rm 90}\%}<100\mathrm{deg}^2$, and dotted lines represent events with $\Delta\Omega_{{\rm 90}\%}<40\mathrm{deg}^2$. The shaded region illustrates the uncertainty arising from the local merger rate.}
    \label{fig:ET_config}
\end{figure*}

ET  may operate together with another 3G detectors such as Cosmic Explorer. Predicted performances of an ET-CE network have been explored in a number of works (e.g. \citep{Iacovelli:2022bbs,Ronchini:2022gwk,Branchesi:2023mws, Loffredo:2024gmx,Corsi2024}), and the synergies between ET and CE will be further discussed in section~\ref{section:div5}.

To compare ET as a standalone observatory with ET as part of a network of other 3G detectors, figure~\ref{fig:ET_config} shows the KN detection rates as a function of detection limits for the BNS population from C22 (left plot) and the BHNS population from C23 for different GW detector networks. The different line colors represent various ET configurations and network setups. Specifically, the blue (or green) lines indicate ET-$\Delta$, while the red lines correspond to ET2L. The gray and black lines represent the networks ET-$\Delta$+2CE and ET2L+2CE, respectively. In this context, we assume the 2 CE network as the most optimistic scenario, consisting of two L-shaped detectors with arm lengths of 40 km and 20 km located in the USA. Solid, dashed, and dotted lines represent, respectively, all BNS (or BHNS) events, events with $\Delta\Omega_{{\rm 90}\%}<100\mathrm{deg}^2$, and events with $\Delta\Omega_{{\rm 90}\%}<40\mathrm{deg}^2$, respectively.

The comparison between the two ET configurations has been previously discussed; however, it is noteworthy that in a network including 2 CE detectors, the number of joint detections significantly increases due to the enhanced precision in sky localization. Specifically, considering all events with $\Delta\Omega_{{\rm 90}\%}<100\mathrm{deg}^2$, the number of detectable events (assuming a detection limit of AB magnitude 26) increases by more than an order of magnitude. In a complete network, the differences between the two ET configurations become negligible up to a magnitude of 26. For deeper limits, the increased EM horizon allows for the observation of more distant events, which results in a small but noticeable difference in detection rates between the two configurations.

\subsection{Observational facilities and strategies}\label{sect:Obsfac_div4}

\subsubsection{The ET change of paradigm}

ET will revolutionize the field of multi-messenger astrophysics, particularly the study of BNS and BHNS mergers. It will transform the field from being observationally constrained to the local universe and to a poor number of events into a domain exploring more than 10 billion years of cosmic history, encompassing hundreds of thousands of objects.

This paradigm shift necessitates the development of new observational strategies to detect, identify, and characterize the EM counterparts of GW events detected by ET.
Different strategies should be developed depending on the scientific goals, and taking into account four intrinsic factors well illustrated in the previous sections: (i) the large localization uncertainty for the majority of the GW events; (ii) the large number of ET detections; (iii) the faintness of the KN and GRB off-axis afterglow counterparts; (iv) the brightness of on-axis GRBs and afterglows at any distances explored by ET. 
Indeed, while studies involving cosmic evolution should take advantage of the entire volume covered by the ET sensitivity, the study of KNe or off-axis afterglows may rely on the numerous nearby events detected by ET with a relatively good localization. 

The precision of GW localization, although introducing biases favoring nearby events, 
will be a key selection criterion for the EM observations, because of the extremely large observing time required to scan large sky region at the necessary depths, and of the extremely high number of ET events. About hundred ET-detected BNS events per year at $z<0.5$ (a few tens at $z<0.2$) will have a GW localization to within a few tens of square degrees (see figure~\ref{fig:Scumul_extrinsicpars_BNS_allconf_cryoHF} in section~\ref{sec:div9_skyloc} and ref. \cite{Loffredo:2024gmx}). However, if a scenario with ET + CE observations is considered, more than ten thousand events will have such a sky-localization, several thousands of which will have localization uncertainty smaller than $10\rm ~deg^2$, decreasing to less than $1\rm ~deg ^2$ for a few hundreds of them (see figure~\ref{fig:cumul_SNR-Mc-Om-dL_BNS_allconf_ETCE} in section~\ref{sec:CBC-ground-based}). 
In such a scenario, additional selection criteria will therefore be crucial to optimize the use of the limited observational time. Prioritization of the alerts and observational strategies will have to be adapted for each specific scientific case.

\subsubsection{Kilonovae}
The best spectral range to look for KNe  is the UV-optical-(near)infrared one. Through photometry it is possible to obtain images of the sky where to look for unidentified transients, and to follow their lightcurves in search of the time and color evolution signatures expected for KNe. However, for a certain KN identification, spectroscopic observations provide more precise (r-process element) features and are often required. Considering the intrinsic faintness of such EM counterparts (see section~\ref{KNGG}) instruments with a high sensitivity are required. Unfortunately, to date, he best instruments to obtain deep images have a small Field of View (FoV), i.e. they are able to observe only a very limited area of the sky, much smaller than the typical GW sky-localization regions. Spectroscopic observations are even more limited because the fibers or slit cover only few arcsec, unless Integral Field Spectroscopy (IFS) is used. This technique enables the possibility to obtain a spectrum for each pixel of the image of a field, but nowadays sensitive IFS are limited to FoV of the order of the arcmin$^2$.  The ideal facility for the most efficient multi-messenger observations of KN would be a highly sensitive integral field spectrograph with a very large FoV (several square degrees) and (sub-)arcsecond spatial resolution.

The simulations presented in Section \ref{simu} on the predicted brightness of the KN (and GW error regions) imply that, with the most advanced facilities expected to be available by the late 2030s (see the table in figure~\ref{fig:table_div4_EM}}), KN detections will remain limited to redshifts $z < 0.5$. 
Among the planned instruments, photometric and spectroscopic facilities with unprecedented sensitivity and coverage, such as the Vera Rubin Observatory and the Wide Survey Telescope (WST), will be particularly well-suited for detecting KNe up to  $z < 0.4-0.5$ from the events localized to within a few tens of square degrees \citep{Loffredo:2024gmx,Mainieri:2024}. 
Target of Opportunity observations by the Vera Rubin Observatory will provide the photometric depth and spectral constraints needed to detect  candidate KN counterparts. Realistic simulations including ToO multi-filter and multi-epoch observations (more effective to identify the counterparts among large number of contaminants), mosaic of the sky-localization region, Rubin sky coverage, and Galactic absorption predict several tens of KN per year associated with BNS detected by ET \citep{Loffredo:2024gmx}.
This numbers may improve by a factor of $\sim 10$, providing to dispose of enough observing time, if ET will operate in the network with current detectors LIGO and Virgo at the design sensitivity and including a detector in India. A further increase of detections by Rubin is obtainable when ET operates with CE  but it is limited to a factor of a few because of the Rubin detection efficiency drop already at z larger than 0.3 \citep{Loffredo:2024gmx}.

\begin{figure*}
    \centering
    \includegraphics[width=0.48\textwidth]{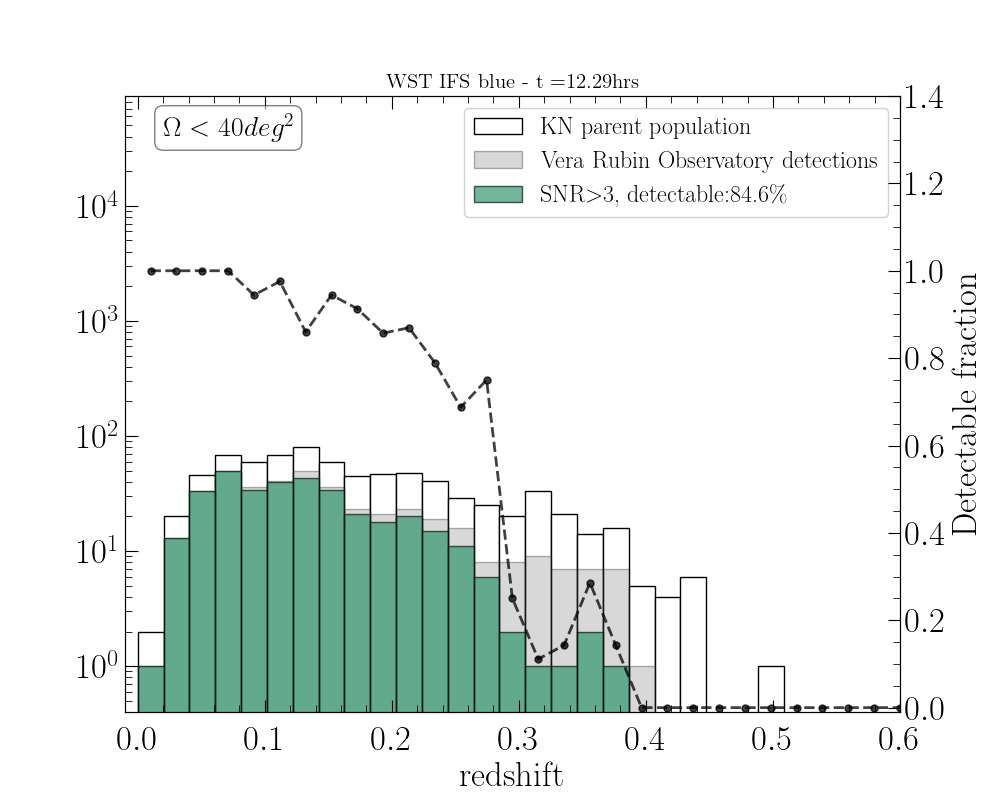}
\includegraphics[width=0.48\textwidth]{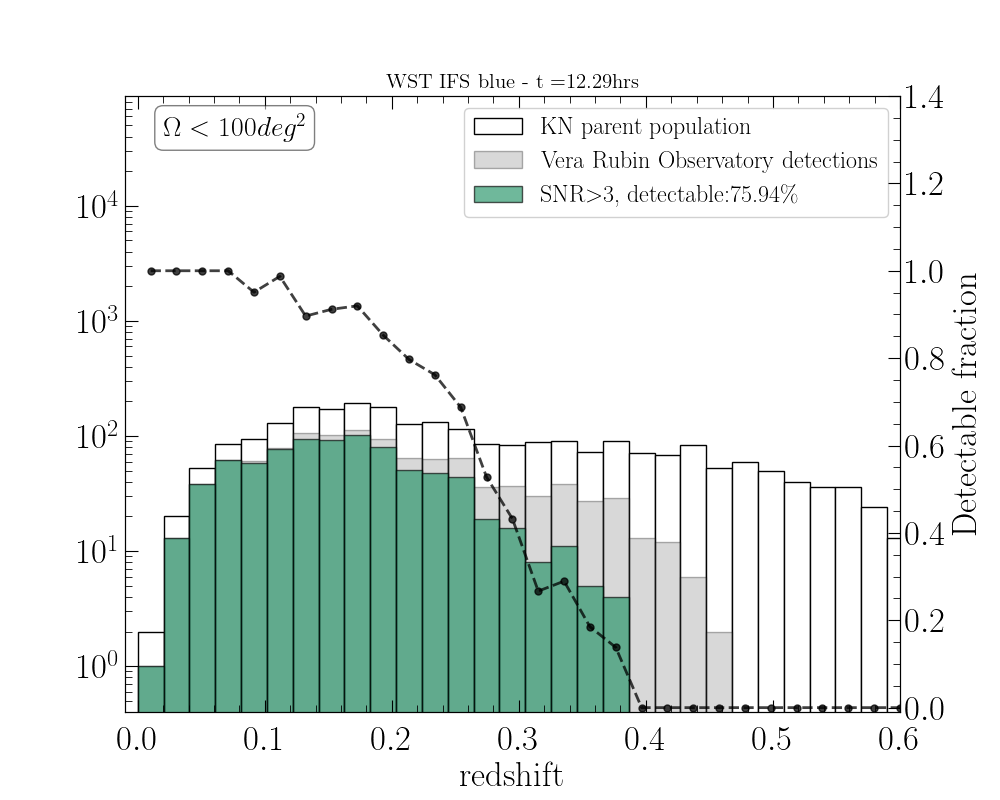}
   \includegraphics[width=0.48\textwidth]{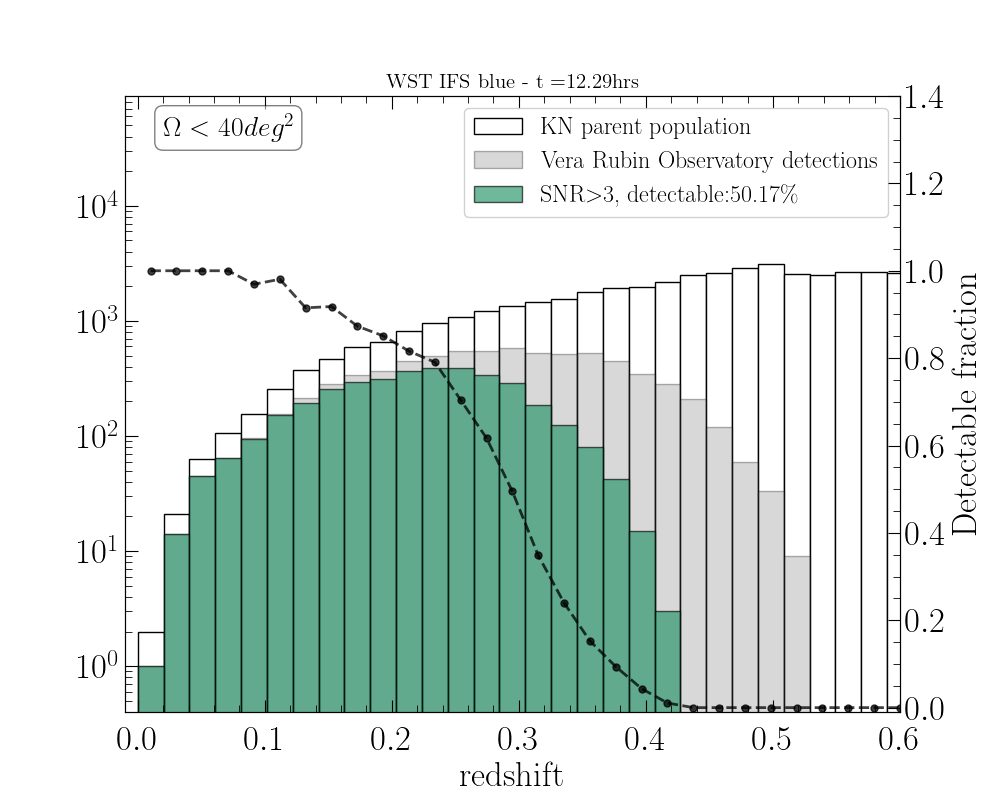}
    \caption{Example of redshift distribution of simulated KNe detected by Vera Rubin and WST, associated with BNS detected by ET in the 2L (top panels) and 2L + CE(40 km) (bottom panel) configurations, and having error regions of $<40$\,deg$^2$ (top left and bottom panles), and $<100$\,deg$^2$ (top right panel). 
The background distribution in white corresponds to the parent BNS+KN population. The green colored distribution corresponds to the KNe that are detectable with WST IFS (SNR $> 3$). KNe detectable with Vera Rubin are shown in grey. Black points refer to the y-axis scale on the right-hand side and show the fraction of detectable  KN with WST with respect to those that are detectable with Vera Rubin. From Bisero et al., in preparation.}
    \label{fig:WSTET}
\end{figure*}

Facilities such as WST could work in synergy with Rubin or independently and detect about a hundred KN counterparts per year at a signal-to-noise ratio
(S/N) $> 10$ (thousands at lower S/N), if that observations are collected within $\sim 12-24$ hours after the merger (see figure~\ref{fig:WSTET}). It is evident that, in both cases, the observing time available will be the main limitations to the number of events for which KN can be detected. Therefore, in addition to the distance and position indications provided by ET, other parameters (such as the inclination angle) will have to be considered to select the events triggering the search for a KN counterparts. For more details on the synergies between ET and WST, see \cite{Bisero:2025tkw}.

Current and forthcoming sky surveys providing galaxy redshift catalogs could enable photometric and/or spectroscopic targeted observations, leveraging distance information from ET detections. However, this would not be an efficient strategy to find KNe beyond $z \sim 0.1$. In fact, in this case, the number of galaxies requiring observation quickly becomes vast---thousands, within the expected sky localizations of ten - few tens square degrees. Targeted spectroscopic searches at higher redshifts could be useful in a scenario where ET operates with two next-generation observatories distributed around the globe, for example a CE(40 km) in USA and one in Australia (or also with the addition of an interferometer in China, e.g. \citep{Li:2024}), as, for $\sim$ a thousand events per year, the sky localizations would have error regions less than $1\rm ~deg ^2$ also for events at $z \sim 0.5$, and the errors on the distances will be reduced by a factor of ten, allowing targeted searches (with IFS / multi-object spectrographs) up to $z \sim 0.3$. 
We stress, however, that the Rubin Observatory can easily cover such sky regions in just one exposure. In this case, redshift catalogues will also be useful to discard contaminant transients from KN candidates.

Detecting EM candidate counterparts is only the first step in multi-messenger studies. Proper characterization of KN counterparts will require the largest ground-based observatories and space telescopes, such as the \textit{James Webb Space Telescope} (JWST), which offers unprecedented sensitivity and spatial resolution. From ground, 30-meter-class telescopes as the \textit{Extremely Large Telescope} (ELT) will be fundamental to track spectroscopically the evolution of KNe and study their role in the chemical enrichment of the universe, because in most cases 10-meter-class telescopes will lack the sensitivity for spectral studies.
Given the high-demand of such kind of telescopes, it is likely that such deep studies will be performed only for a limited number of detected KN. In any case, it is of the highest importance for  science that multi-messenger studies are well integrated in the highest priority observations of such facilities.

\subsubsection{Off-axis afterglows}
Another EM counterpart of BNS and BHNS mergers  whose detection will also be limited to the low-redshift regime (for viewing angles larger than $\sim 15^{\circ}$) is the off-axis afterglow of GRBs associated with GW events. Observing strategies for these counterparts partially align with those for KNe but, in this case, the search at lower (radio) and higher (X-rays) frequencies will play a significant role. Radio frequencies result in being very appropriate for off-axis afterglow detections. SKAO-Mid, with its unprecedented sensitivity and survey speed (see the Table in figure~\ref{fig:table_div4_EM}), will be the ideal facility to detect (and follow, in synergy with the ngVLA) off-axis afterglows (see Section \ref{afterglow}). It is of great importance that specific SKAO GW follow-up programs are established. X-rays can also be an efficient way to look for off-axis afterglows, provided that suitable  facilities joining sensitivity and FoV will be operational at the ET epoch. Missions projects such as THESEUS and \textit{NewAthena} at ESA can be key to the identification and characterization of these EM counterparts, detecting $\sim$ dozens of them per year (e.g. \citep{Cruise:2025,Ciolfi:2021,Ronchini:2022gwk}).

\subsubsection{On-axis GRBs}
As stated, while KN and off-axis afterglows will be likely too faint to be detected and studied at $z > 0.5$, ET will be able to detect BNS and BHNS mergers up to redshifts beyond the cosmic star formation peak. To exploit  the entire volume probed by ET for multi-messenger studies, the detection of on-axis (or slightly off-axis, i.e., viewing angles $ <10^\circ $) GRBs and afterglows will be  indispensable. Realizing this opportunity will depend  on the availability of facilities capable of detecting and localizing GRBs. Missions like the current \textit{Neil Gehrels Swift Observatory} or the SVOM satellite have the required capabilities, but ensuring that similar missions are operational beyond 2035 is critical. As of now, no approved missions are slated to fly in the late 2030s or beyond. The proposed THESEUS  satellite, currently competing for selection as ESA's M7 mission, could fulfill these requirements and offers the additional advantage of onboard redshift determination for a significant fraction of GRBs.

\subsubsection{ET pre-merger detection and ET and early warnings alerts}
\begin{figure*}
\centering
\includegraphics[width=0.32 \linewidth]{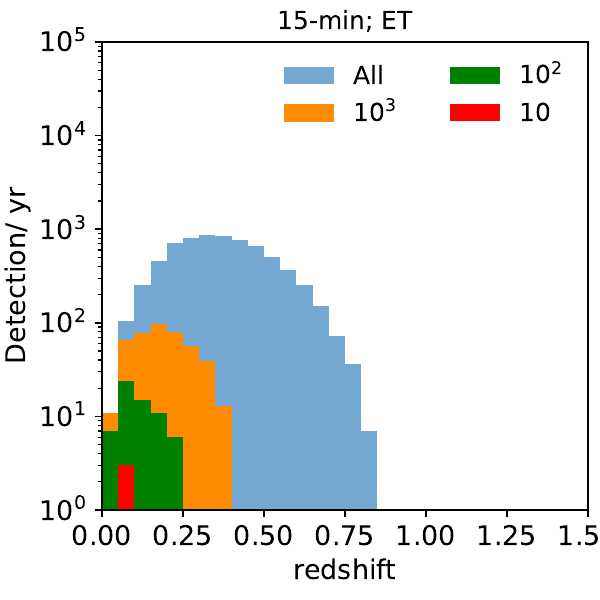}
\includegraphics[width=0.32 \linewidth]{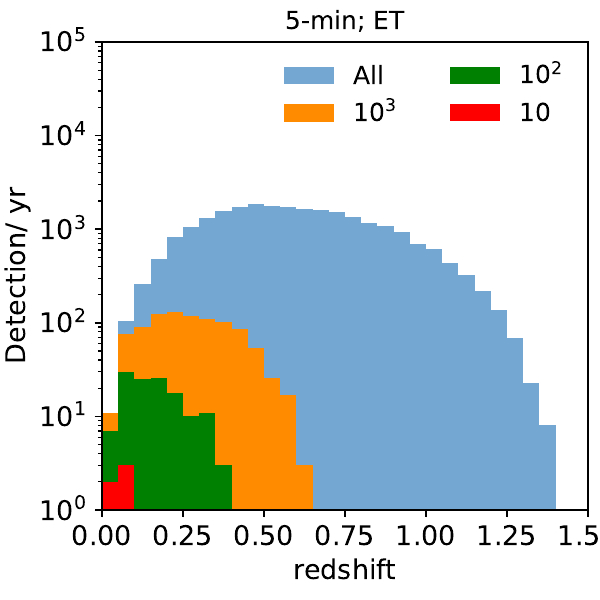}
\includegraphics[width=0.32 \linewidth]{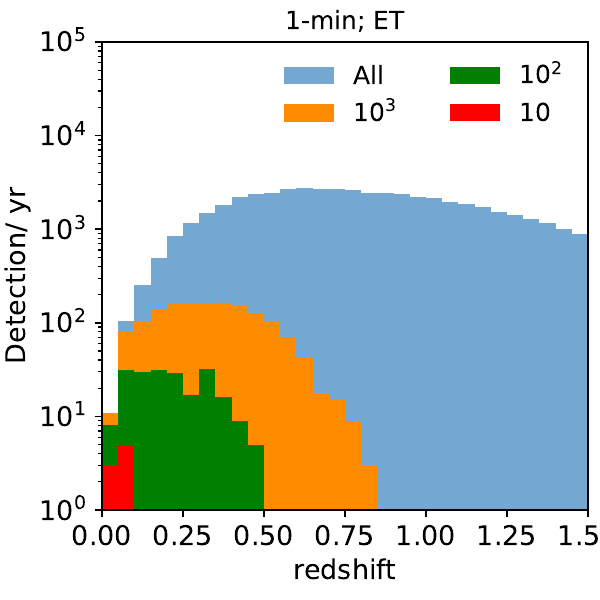}
\includegraphics[width=0.32 \linewidth]{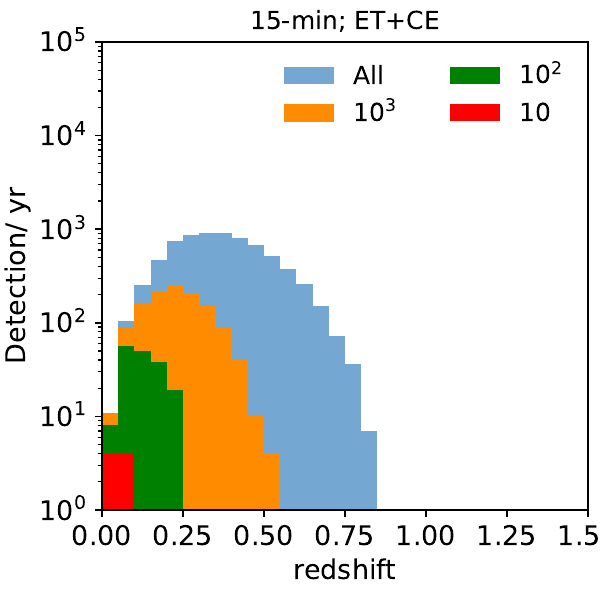}
\includegraphics[width=0.32 \linewidth]{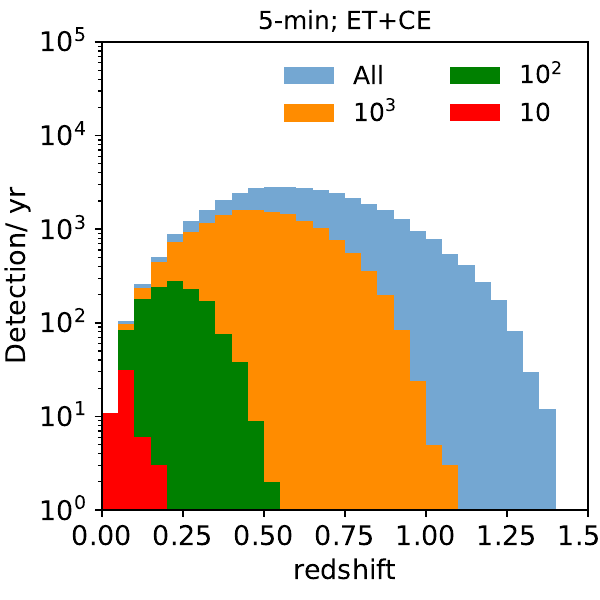}
\includegraphics[width=0.32 \linewidth]{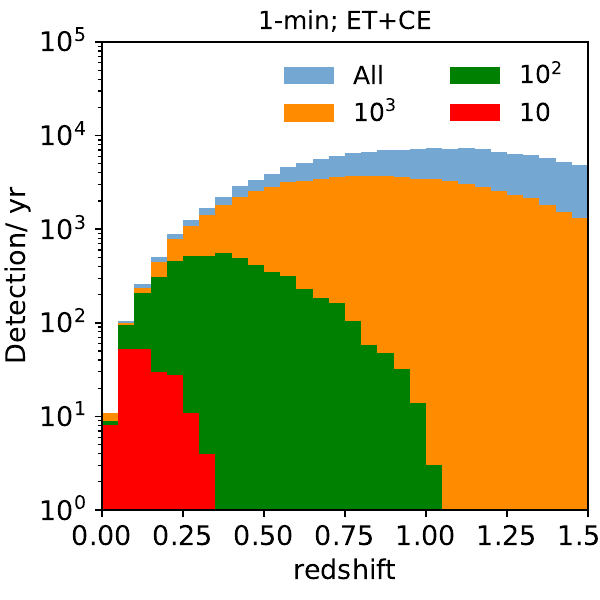}
    \caption{Redshift distribution of the sky-localization uncertainty (given as 90\% credible region) for ET and ET+CE configurations. 
The panels show the BNS detections and the corresponding sky-localizations as a function of the
redshift 15, 5, and 1 minute(s) before the merger. The blue histogram represents all detected sources, while the other colors indicate sources with sky localizations more precise than 10$^3$\,deg$^2$ (orange), 10$^2$\,deg$^2$ (green), and 10\,deg$^2$ (red). Adapted from \cite{Banerjee:2023}.}
    \label{fig:early}
        \end{figure*}

The Einstein Telescope's access to low frequencies makes it possible to observe signals, such as the inspirals of BNS systems, for a long time (hours), and thus to use the Earth rotation imprint in the GW signal itself to detect and localise BNS systems before the merger \citep{Branchesi:2023mws}. This allows early warnings to be sent to the astronomical community hours to minutes before mergers, with an increasingly precise sky-localization as the event approaches merger. Several studies have been conducted to assess the pre-merger capabilities of ET. ET is capable of localizing several tens to hundreds of BNS mergers per year with a sky localization of less than $\sim100$\,deg$^2$ minutes before the merger\citep{Nitz:2021pbr,Banerjee:2023,Miller:2024109d3021M}. This number increases to a hundred of BNS mergers per month for the network of ET operating with CE \citep{Hu:2023hos}. Figure~\ref{fig:early} shows the improvement in sky localization capabilities as a function of the redshift and how increasingly distant events can be localized as they get closer to the time of merger. 

Detecting localized BNS pre-merger opens the possibility of following the BNS during the inspiral phase with the EM observatories and being on the source before and during the merger. Indeed, the rapid ($\sim$ minute scale) pointing capabilities of optical transient survey facilities will allow for the exploration of the very early emission associated with KNe, which is fundamental for probing the first ejecta and processes at play in KN emission. Furthermore, GW early warnings will allow the search for associated very high energy (VHE) emission by VHE observatories such as CTAO \citep{Banerjee:2023} that will have suitable rapid pointing, FoV and sensitivity. 
As recently demonstrated by {\it Swift}, high-energy satellites can react to triggers by ground-based observatories on $\sim$minute scale. Wide FoV X-ray instruments with such rapid pointing could detect the early X-ray emission of off-axis GRBs in absence of the prompt gamma-ray detection \citep{Ronchini:2022gwk}.

\subsubsection{ET as an alert receiver}
Finally, we should not forget that ET will also operates as alert {\it receiver}. Due to the huge amounts of events, it will be computationally prohibitive to perform a complete parameter estimation for all the GW detections, especially in low-latency. However, EM facilities will regularly detect short GRBs (space-based missions), afterglows (space-based missions and ground-based facilities, scanning the sky at different wavelengths covering the entire EM spectrum, from VHE, as CTAO, to radio, as SKAO) or KN (especially optical and near-infrared sky surveys). ET will {\it listen} to these alerts from the EM community, and use them to perform an informed search and parameter estimation for the corresponding GW signal. It will be crucial to prioritize the GW analysis of these events and send in low-latency refined parameter estimations for them to inform the astronomical community and optimize the use of the observational resources.

\subsubsection{Available and expected EM facilities characteristics}

\begin{figure*}
    \centering
    \vspace*{-10mm}
    \hspace*{-5mm}\includegraphics[width=1.009\textwidth]{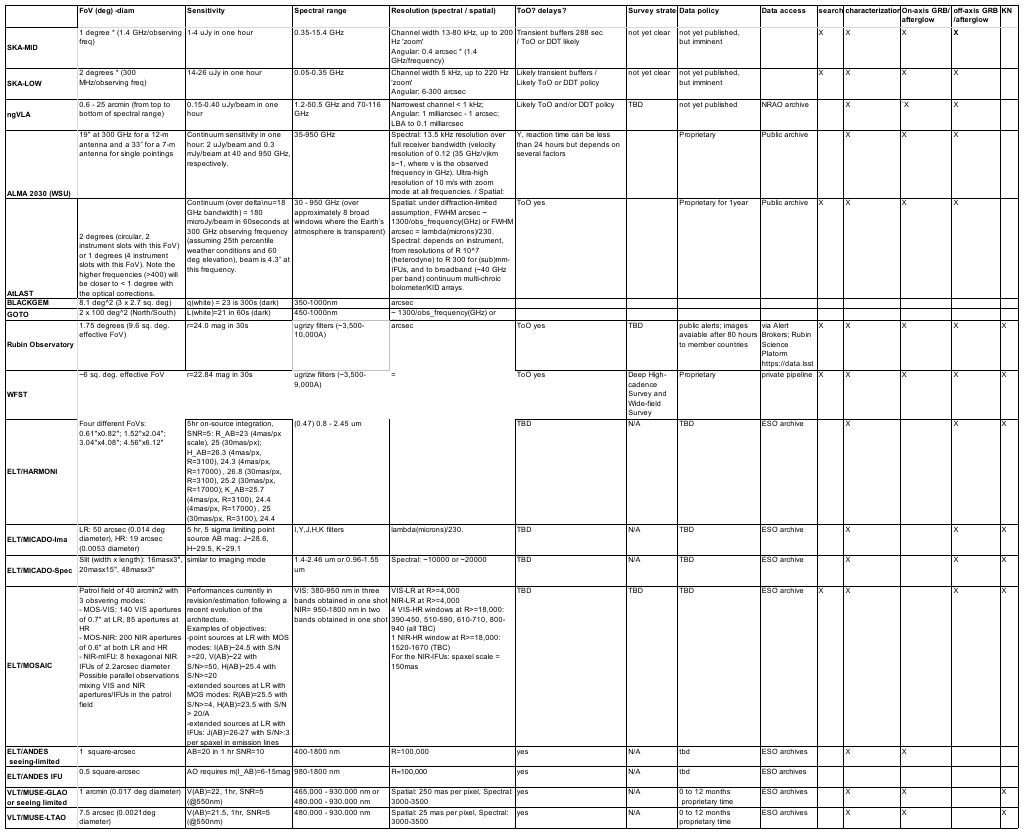}
    
    \vspace*{-1mm}
    \hspace*{-5.5mm}\includegraphics[width=1.0\textwidth]{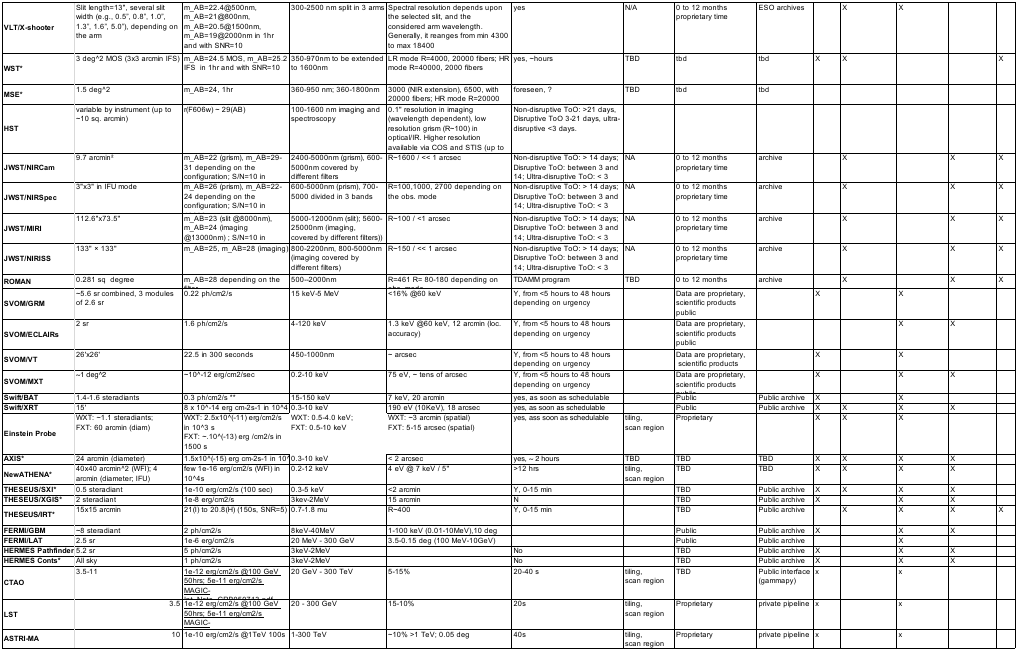}

    \vspace*{0mm}
    \caption{Characteristics of  most advanced EM facilities expected to be available by the late 2030s.  Also available as an Excel table at \url{https://apps.et-gw.eu/tds/ql/?c=17760}. Asterisk indicates facilities and missions that are still in the project phase and have not yet been approved or adopted.}
    \label{fig:table_div4_EM}
\end{figure*}

In figure~\ref{fig:table_div4_EM} we report the results of a survey within the ET observational science board on the best EM  current facilities and future projects for multi-messenger observations (column 1), as well as their corresponding characteristics, in terms of Field of View (FoV; column 2), sensitivity (column 3), spectral range (column 4), resolution (column 5), capability and time delay to perform  Target of Opportunity (ToO) (column 6), survey strategy (column 7), data policy and data access (columns 8 and 9). Each facility may be of good use to look and/or to follow-up and characterize one or more type of EM counterparts (GRBs, off-axis afterglows, KNe). This is indicated from column 10 to 15. Projects that are still at the proposal stage and not yet adopted are indicated with an asterisk.

\subsubsection{Requirements for ET BNS and BHNS multi-messenger observations based on simulations.}

\begin{figure*}
    \centering
    \hspace*{-18mm}\includegraphics[width=1.2\textwidth]{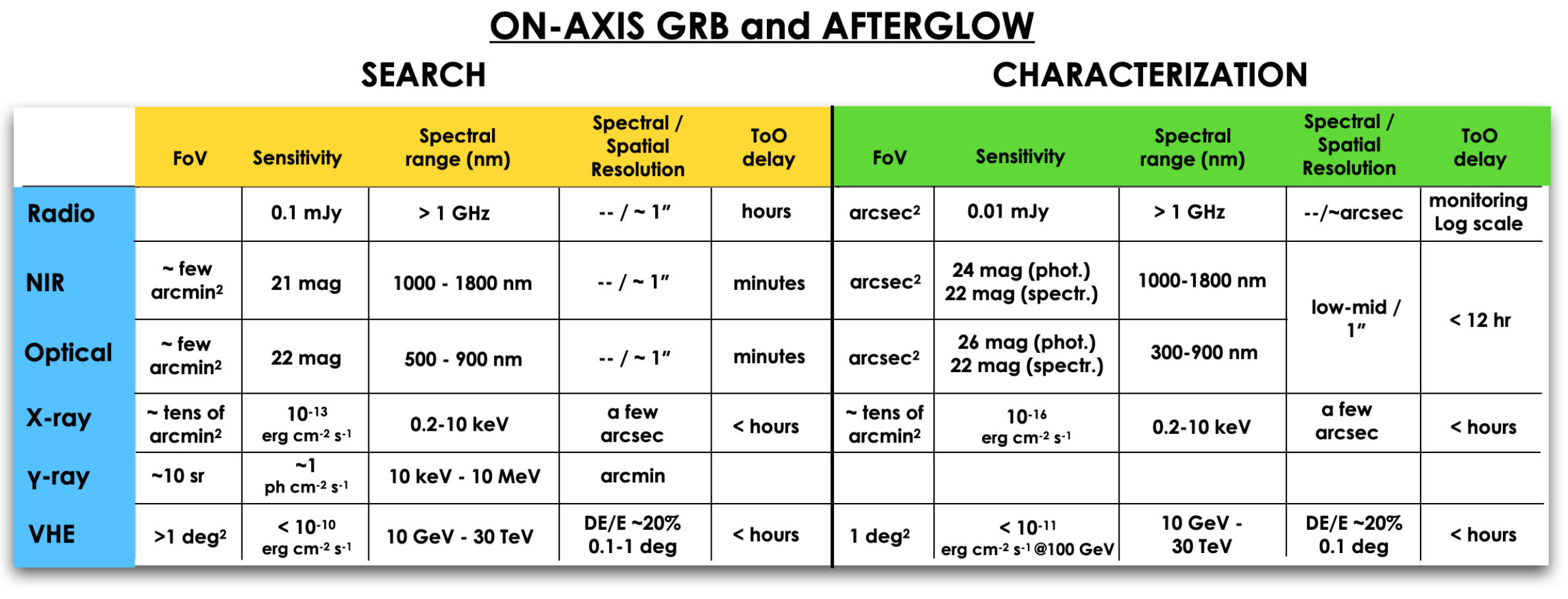}
    \caption{Minimal requirements for detecting and characterizing EM counterparts of BNS and BHNS mergers detected by ET, considering on-axis GRB and afterglow emission and EM facilities operating across different wavelength ranges. The gamma range in the table extends to the X-ray band considering the characteristics of the gamma-rays instruments for GRB detection.}
    \label{fig:Requirements_div4_On-axis}
\end{figure*}

\begin{figure*}
    \centering
    \hspace*{-18mm}\includegraphics[width=1.2\textwidth]{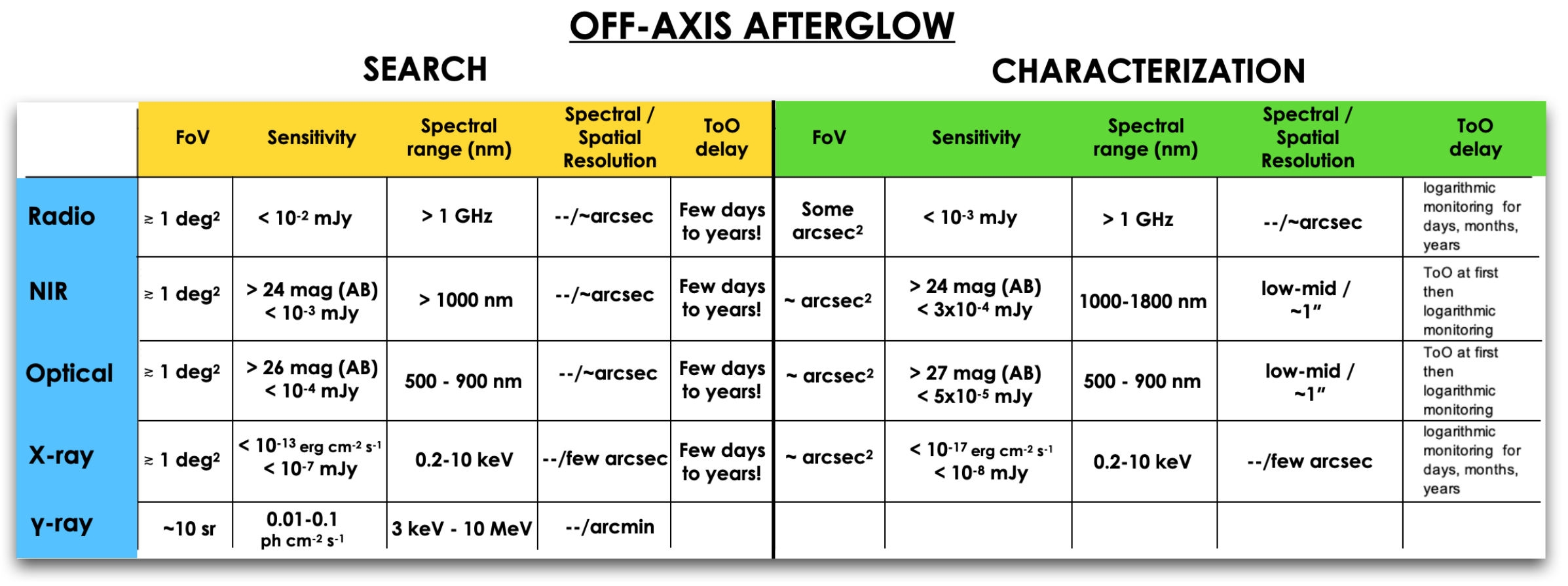}
    \caption{Minimal requirements for detecting and characterizing EM counterparts of BNS and BHNS mergers detected by ET, considering off-axis afterglow emission and EM facilities operating across different wavelength ranges. In this case, we consider a field of view equal to $1,\mathrm{deg}^2$ (for radio, NIR, optical, and X-ray), in contrast to figure \ref{fig:Requirements_div4_KN}, where a larger field of view is required. This choice accounts for the possibility of detecting gamma-ray or X-ray emission from slightly off-axis systems, which could enable source localization. The gamma range in the table extends to the X-ray band considering the characteristics of the gamma-rays instruments for GRB detection. 
    }
    \label{fig:Requirements_div4_Off-axis}
\end{figure*}

\begin{figure*}
    \centering
    \hspace*{-18mm}\includegraphics[width=1.2\textwidth]{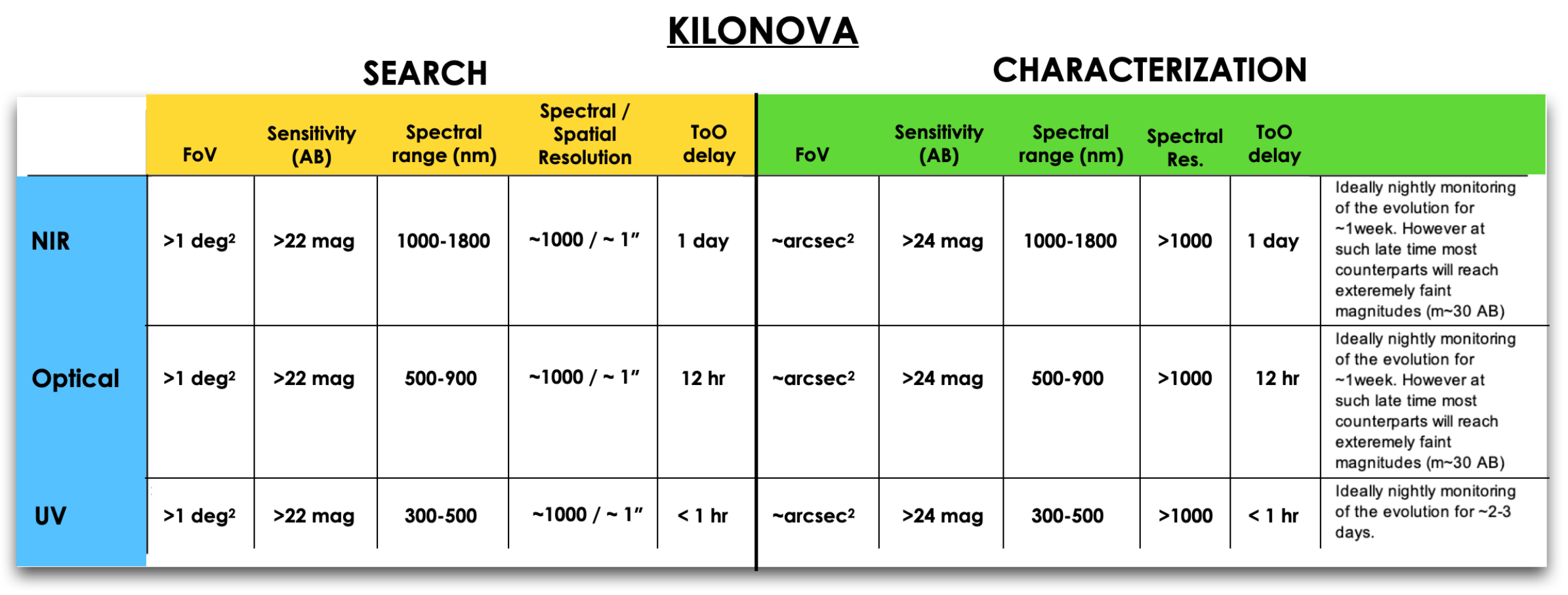}
    \caption{Minimal requirements for detecting and characterizing EM counterparts of BNS and BHNS mergers detected by ET, considering KN emission and EM facilities operating across different wavelength ranges.}
    \label{fig:Requirements_div4_KN}
\end{figure*}

Based on the simulations presented in this work, in figures~\ref{fig:Requirements_div4_On-axis}--\ref{fig:Requirements_div4_KN}
we report the minimal requirements to detect and study the EM counterparts (on-axis GRB, off-axis afterglow, KN) of BNS and BHNS detected by ET, for EM facilities working in different ranges of the EM spectrum. The evaluation of the requirements takes into account a balanced amount of time to be required to the EM observatories.  We consider the FoV, sensitivity, spectral range, resolution, and capability and time delay to perform ToOs. We separate the {\it Search} for the EM candidate counterparts (orange columns) from their in depth study for the {\it Characterization} of their properties (green columns). For on-axis GRBs, the search requirements are based on the assumption of a GRB prompt detection by a satellite with arcmin localization capability. For the VHE the requirements take into account the possibility to point pre-merger to detect the emission during the prompt/early phase. We stress that, except for the scenario of a well-localized GRB detection in temporal coincidence with a GW signal which can drive the multi-wavelength search and characterization, also in the ET era sensitive wide FoV instruments ($> \rm 1\, deg^2$) will be necessary for the search. In addition, to increase the chance to detect off-axis afterglows also observations at late time (months to years) might be required. Except for on-axis GRBs, very sensitive instruments are required at any wavelength for the characterization.

\subsection{Neutrinos from Compact Binary Coalescence}
\subsubsection{Common environments for neutrinos and GWs}
The existence of a common astrophysical source of neutrinos and GWs is theoretically well-motivated. One such example is compact binary mergers. The most promising ones are binary systems with at least one neutron star component, such as binary neutron stars (BNS), and neutron star-black hole (NSBH) systems \cite{GRB_Kimura_Review, Kimura:2017kan,Waxman:1997ti}. Otherwise, binary black holes (BBH) embedded in Active Galactic Nuclei (AGN) disks could also have a favourable environment for neutrino production \cite{Zhou:2023rtr,Zhu:2023hsq}.

Based on theoretical predictions, following the merger, a central compact object is formed. Consequently, powerful relativistic jets efficiently eject the remaining accretion disc matter into the surrounding environment after the merging phase. There, non-thermal charged particles are accelerated in the presence of strong magnetic field. The accelerated protons in and around the jet actively participate in hadronic interactions. They interact with the strong ambient photon fields, leading to photo-pion interactions. As one of the secondary particles, high-energy neutrinos (HENs) are produced via charged pion decay. It could happen during the prompt phase \cite{Rees:1994nw}, and also later, coincident with the longer afterglow radiation \cite{Meszaros:1996sv}. It is worth mentioning that HENs can also be produced via hadronuclear (pp) interactions. Such processes also result in the production of charged pions and kaons, which subsequently decay into charged leptons and neutrinos. Furthermore, neutral pions, produced from the same primary hadrons, decay into high-energy photons. These photons, if within the observable range from Earth, strengthen the connection between high-energy neutrinos and gravitational waves, reinforcing the multi-messenger approach.

\subsubsection{Current and future neutrino telescopes}

In view of promising theoretical predictions, since the last decade, multiple neutrino detectors have been looking for the correlation between the detection of gravitational waves and any possible detection of astrophysical neutrinos.
At the time of writing, three major installations, IceCube, KM3NeT, and GVD-Baikal, are active and taking data although at different stages of realisation. All of them come with a target instrumented volume of the size of 1 km$^3$. 

The IceCube Neutrino Observatory \cite{Gaisser:2014foa,IceCube:2016zyt} completed its installation in 2013. It is a cubic kilometer ice-Cherenkov detector located under the Antarctic ice in the geographic South Pole. Motivated by its rich science output, further enhancements offering better sensitivity across a vast energy range are envisioned. The sensitivity of IceCube in the sub-TeV range will improve with the installation of IceCube Upgrade \cite{IceCube-PINGU:2014okk,Ishihara:2019aao}, i.e.\, seven new densely packed strings, planned for 2025-26. 
Another extension of the existing IceCube infrastructure in planning is IceCube-Gen2~\cite{IceCube-Gen2:2023vtj}. 
It will be an 8 km$^3$ and multi-component detector with better sensitivity to HENs with energies above 1 PeV. 

Although already operational, KM3NeT and GVD-Baikal are still in construction stages in the abyssal site of the Mediterranean Sea and at the bottom of Baikal Lake in Siberia, respectively. In the Northern Hemisphere, the  KM3NeT neutrino telescope  
\cite{KM3Net:2016zxf} is composed of two undersea detectors, namely ORCA and ARCA. KM3NeT/ARCA, with a target instrumented volume of 1 km$^3$, is optimised for the detection of astrophysical neutrinos from 10 TeV up to hundreds of PeV. KM3NeT/ORCA 
is designed to study the physical properties of neutrinos, such as the mass hierarchy and the oscillation parameters, by detecting the atmospheric neutrinos in the sub-GeV up to the TeV energy range.  Early in 2025, the KM3NeT Collaboration published the discovery of a record-breaking neutrino detected by KM3NeT/ARCA with an estimated energy of about 220 PeV \cite{KM3NeT:2025npi}, demonstrating the excellent capabilities of underwater neutrino telescopes.

Baikal-GVD \cite{Baikal-GVD-GENERAL} is located at a depth of $\sim$ 1400 m in the Baikal Lake. Promising results of the measurement of the astrophysical neutrino diffuse flux above 100 TeV were recently published \cite{Baikal-GVD:2022fis}. 

A group of next-generation TeV-PeV neutrino telescopes are under design. In some cases, the prototypes of the new technologies are already being tested. They are concentrated in the Northern Hemisphere in different sites around the World. 
P-ONE \cite{P-ONE:2020ljt}, currently at a prototypal stage, aims at the realisation of a 1 km$^3$ array of multi-PMT optical modules off Canada in the Pacific Ocean. The Chinese project TRIDENT \cite{TRIDENT} is studying the feasibility of a neutrino telescope in the South China Sea, aiming at an instrumented volume of 8 km$^3$. In China, other ambitious projects are under discussion (e.g. NEON \cite{Zhang:2024slv} or HUNT \cite{Huang:2023mzt}) to realise very large telescopes of 10 km$^3$ or 30 km$^3$.

Binary neutron star mergers are potential sources of ultrahigh-energy neutrinos (above 1 PeV), and next-generation radio detectors such as GRAND~\cite{GRAND:2018iaj}, IceCube-Gen2 Radio array, and RNO-G~\cite{RNO-G:2020rmc} are designed to detect these elusive particles.

\subsubsection{Multi-messenger neutrino and GW frameworks}
Despite all the decade-long observational efforts, coincident detection of HEN and GWs is still missing \cite{ANTARES:2018bmu,ANTARES:2017bia, IceCube:2020xks, IceCube:2022mma,IceCube:2023atb}. To date only upper limits on the HEN content of GRBs \cite{IceCube:2017amx} and observed GW sources have been obtained, owing, most likely to the fact that the current neutrino experiments are not sensitive enough to the neutrino flux of individual sources.

At present, both IceCube and KM3NeT are following up on the existing individual GW events to identify their neutrino counterparts. The current GW follow-up strategy consists of both real-time and archival searches for individual GW events, using different event topologies, ranging from MeV to PeV energies \cite{Valtonen-Mattila:2023pgk,IceCube:2023yfi,KM3NeT:2023cdr, thwaites_ICRC2023_GWsearch,KM3NeT_ONLINE}.
Sub-threshold GW events from the Ligo-Virgo-KAGRA third observation run (O3) are also separately followed up by IceCube in an archival search to identify potentially interesting multi-messenger sources \cite{mukherjee_ICRC23}. Similar studies would play a crucial role in the era of next-generation GW detectors.
For GeV-TeV neutrinos, the typical observation time window is selected to be $\pm$ 500 s concerning the GW event detection time. We expect the neutrino signal from the source to dominate over the background within this time window \cite{Baret:2011tk,Kimura:2018vvz}. However, in IceCube, for BNS mergers, an extended time window of two weeks is also considered, to account for certain model predictions \cite{Fang:2017tla}. Additionally, real-time analysis to identify MeV neutrino bursts with IceCube within four different time windows: 0.5\,s, 1.5\,s, 4\,s, and 10\,s associated with individual GW events is under development \cite{Valtonen-Mattila:2023pgk}.
On the other hand, in KM3NeT, an additional [-500 s, +6 h] time window is considered for real-time searches of GeV-TeV neutrinos, while a 2\,s window is used for MeV neutrinos. So far, no significant correlation has been observed. An alternative archival search strategy could also be explored in the future, aiming to identify neutrino counterparts from a population of GW candidates via stacking analysis. This approach enhances the detection potential by accumulating signal neutrinos within an optimal time window, allowing them to stand out against the background neutrino distribution.

Such observational prospects will surely increase with next-generation GW like Einstein Telescope (ET) and HEN detectors, specifically with KM3NeT, and Baikal-GVD. Once they reach their nominal size and with the active participation of P-ONE and IceCube-Gen2, the combined performance of all these advanced neutrino detectors can offer maximum sensitivity to detect neutrino sources all over the sky, which would benefit GW follow-up efforts \cite{Schumacher:2021hhm}. According to the current plan, the Einstein Telescope \cite{Maggiore:2019uih,Branchesi:2023mws} and the US-lead Cosmic Explorer (CE) \cite{Evans:2021gyd,Evans:2023euw} will be able to detect BNS mergers out to O(10) Gpc. During their active operation time, IceCube and KM3NeT will continue actively participating in the GW follow-up efforts. As neutrinos do not get absorbed or deflected while propagating toward us from the respective astrophysical sources, the prospect of coincident detections from the furthest of sources in the future remains a key objective. With the expected boosts in the sensitivities of future neutrino detectors, extending the current searches for coincident detection of HEN and GWs for next-generation detectors is crucial for the future of multi-messenger astronomy. Ref.~\cite{Mukhopadhyay:2023niv} investigates the prospects of performing triggered-stacking searches for coincident high-energy neutrinos from BNS mergers, using next-generation GW detectors in synergy with IceCube-Gen2. The study quantifies optimal distance thresholds and time windows to minimize background, and highlights that precise sky localization, achieved by the combined network of ET and CE, is essential for enabling either neutrino detections or placing meaningful constraints on emission models within approximately 10 years. A fraction of BNS mergers is expected to produce long-lived magnetars as remnants, which, according to some models, can lead to enhanced electromagnetic and neutrino emission. Ref.~\cite{Mukhopadhyay:2024ehs} shows that GW-triggered stacking searches, using the ET and CE network together with IceCube-Gen2, offer a realistic chance of detecting neutrinos from these events within an operational timescale of about 20 years. Considering ultrahigh-energy neutrinos, instruments such as GRAND and IceCube Gen2 Radio, using triggers by the ET+CE network, have a significant probability of detecting at least one coincident  GW/neutrino within 15 years~\cite{Mukhopadhyay:2024lwq}.

While such a joint detection can be very informative about the physical processes happening in extreme conditions, it will also improve the sky localisation of the event by orders of magnitude compared to the localisation with GWs only. Considering the huge data we will receive daily with ET, identifying neutrino counterparts for GW events before triggering EM follow-up searches will allow us to improve GW event localisation by identifying joint best-fit GW + $\nu$ source location. This will enable us to schedule EM follow-up search initiatives more efficiently, using telescopes with narrow fields of view. All these prospects motivate us to question further the probability of finding a common source of GW and HEN in the era of ET.

\subsection{Multi-messenger infrastructure challenges}
The large number of alerts generated by ET during both the inspiral phase and the merger of compact objects in binary systems poses significant computational, organizational and logistical challenges. There is a need for dynamic databases capable of efficiently managing a high number of candidate events and the flow of updated information for each of them to send to the astronomical community. These systems must also be able to automatically prioritize events to be characterized with precise parameter estimation. Additionally, ET will need to operate as an EM trigger listener, requiring infrastructure also capable of receiving and processing a large number of external alerts detected by the EM observatories, in order to search for potential temporal and spatial coincidences with GW signals.

The search for gravitational wave signals will not rely solely on the match filtering technique but will also incorporate innovative methodologies aimed also at identifying new classes of signals (see section~\ref{section:div7}). In this context, advanced multi-messenger analyses that simultaneously integrate spatial and temporal information will be extremely valuable to optimize the identification, characterization, and interpretation of gravitational wave signals. 

Multi-messenger astronomy presents critical challenges in terms of interoperability and the application of FAIR (Findable, Accessible, Interoperable, and Reusable) principles. The rapid exchange of information is enabled by the development and adoption of shared standards, facilitating seamless integration and data sharing across heterogeneous research infrastructures. The standardization of processes will be fundamental to allowing scientists to efficiently work with more and more complex datasets from diverse observational sources, including gravitational waves, neutrinos, and traditional electromagnetic signals. 

Founded in 2002, the International Virtual Observatory Alliance (IVOA) is a freely accessible consortium that brings together scientific communities, research institutions, and software developers. It functions as an active community and a collaborative forum where members define and discuss protocols for the management, exchange, and analysis of astronomical data. The IVOA protocols currently used by the scientific community for transient events provide a valuable basis for advancing the handling of data and computational demands on a much larger scale. LSST, with its live stream of millions of transients detected every night, will define the protocols and standards for decades to come. 
A key element of the ET integration in mainstream of multi-messenger astronomy 
will be the implementation and adoption of open and shared standards, essential for ensuring interoperability across different observational platforms and facilitating data integration and analysis. The use of common protocols will help overcome technological barriers, optimizing the exchange of information among the various components of the multi-messenger network and enabling an effective response to fundamental scientific questions that only a truly integrated approach can address. The ET GW community is collaborating with the EM communities to include the peculiarities of gravitational transients in the upcoming standards and infrastructures. 

This section of the paper shows how the realization of ET’s full multi-messenger potential will depend on the coordinated development of electromagnetic observational capabilities. Here, we highlight that to maximize scientific return of next-generations multi-messenger observatories, it will also be critical to develop infrastructures, protocol and standards in close coordination.

\subsection{Executive Summary}

Einstein Telescope will, each year, identify thousands of gravitational wave signals from sources that also emit electromagnetic light. This rich stream of data will provide the opportunity to build the first substantial samples of multi-messenger events. In turn, these objects will enable transformative science across many of the core areas of ET science. 

\begin{highlightbox}{Discovery of sources with multiple messengers}
\begin{itemize}
\item ET-alone is capable of localizing a few  hundreds of BNS and NSBH mergers with a localization of tens to hundred of square degrees in size. Although these regions may contain thousands of galaxies, they are sufficiently informative for the electromagnetic observatories to discover an electromagnetic counterpart able to pinpoint the GW source on the sky, identify host galaxies and provide measurements of the distance (redshift) to the source.

\item
There are many plausible electromagnetic signals that a merging compact object binary may create during or after the merger, including high-energy transients such as {\bf gamma-ray bursts (GRBs)}, {\bf multi-wavelength afterglows} seen from the TeV range through X-ray/Optical/NIR  to the radio band, and radioactively powered {\bf kilonovae} emitting dominantly in the optical to mid-infrared. In addition to these EM signals which are dealt with in this chapter, ET may also open new science cases with the joint association of GW to 
high energy neutrino. 

\item
The identification of these sources in the large gravitational wave sky-localizations is not straightforward, and in this chapter we utilize two state of the art approaches to provide self-consistent estimates of event rates and detectability to the current and next generation of electromagnetic observatories for two proposed configurations of the Einstein Telescope (triangle and 2L) and for ET in a network of 3G detectors. 
These simulations show that, while ET may identify $>10,000$ compact object mergers that create electromagnetic light, the majority will remain undetectable to near-term electromagnetic facilities. However, tens to hundreds of events per year will be accessible, providing the first opportunity to build substantial {\bf samples of object detected both in gravitational waves and electromagnetic light}.
\end{itemize}
\end{highlightbox}

\begin{highlightbox}{Enabling new science}
Knowledge of the precise source locations offers a route to addressing a wealth of astrophysical questions, many of which are described in other chapters of this book. In particular, these observations will enable ET data, in concert with electromagnetic observations to:
\begin{itemize}
\item Determine the contribution of compact object mergers to {\bf heavy element enrichment} via studies of their kilonovae.
\
\item Study the launching, collimation and emission processes of {\bf relativistic jets} from the post-merger remnant. 

\item Use the properties of the associated electromagnetic signals to probe the {\bf equation of state of matter at nuclear densities} (see section~\ref{section:div6}). 
\item Combine electromagnetically determined redshifts with gravitational wave measured distances to measure key {\bf cosmological parameters} (see section~\ref{section:div2}).
\item Use source locations and the demographics of the galaxies which host them to inform the {\bf evolution of the stellar populations} that create compact object binaries (see section~\ref{section:div3}). 

\end{itemize}
\end{highlightbox}

\begin{highlightbox}{A new paradigm for multi-messenger astronomy}
Focus in the current era has been on identifying multi-messenger signals in the relatively nearby Universe. In contrast, the cosmological reach of ET will identify them across cosmic time. This dramatically changes the most likely electromagnetic counterparts and the associated search techniques.
\begin{itemize}
\item 
Our simulations suggest that the best route to growing the multi-messenger samples in the ET era will arise via the detection of coincident gamma-ray burst (or associated 
high-energy transient) signals. {\bf Realising the scientific return of ET will critically rely on the availability of 
high-energy instrumentation with good localisation capabilities.}

\item 
{\bf Future observational facilities} will play a crucial role in identifying and characterizing EM counterparts. Photometric (e.g Vera Rubin) and spectroscopic (e.g. WST) instruments 
will be pivotal for detecting KNe up to $z\sim 0.5$ while large ground and space based telescopes (e.g. ELT, JWST) will be essential for spectroscopic characterization. 

\item 
{\bf Radio and X--ray} follow-up will be critical for detecting off--axis afterglows with facilities such as SKA, ngVLA, NewAthena and THESEUS. GW with on-axis GRB counterparts, detected up $z>2$, will provide valuable insights into the cosmic star formation history of double compact systems. 

\item 
The capability of ET to provide several minutes of advanced warning prior to a given merger raises the prospect of sensitive (but still wide field) electromagnetic observations observing  the merger in real-time. This opens new EM observational windows for studying {\bf precursor emission, kilonova shock physics and VHE counterparts}. The latter could be efficiently detected by CTAO and other next-generation gamma-ray observatories. 

\item 
The identification of EM-signals (e.g. gamma-ray bursts) may be crucial as a route of identifying the GW signals  of events of particular interest within the ET data flow.

\item 
{\bf High-energy neutrino observatories} such as IceCube-Gen2, KM3NeT, and Baikal-GVD will play a key role in the multi-messenger network, enhancing the probability of detecting coincident GW–neutrino–EM events.

\end{itemize}
\end{highlightbox}

\section{Synergies of ET with other gravitational-wave observatories}\label{section:div5}

This section aims to explore the scientific synergies that arise from combining gravitational waves  detected by ET with those from other GW observatories.
Synergies with other ground based detectors operating between a few Hz and a few kHz, such as the Cosmic Explorer, are crucial for consolidating the science case of ET~\cite{Maggiore:2019uih,Kalogera:2019sui,Kalogera:2021bya,Branchesi:2023mws}.
Synergies with GW observatories at frequencies that are not accessible from Earth-based detectors ($\lesssim$ Hz) broaden instead the scientific reach of ET, allowing for connections to be drawn with different GW sources.
This chapter investigates key themes that highlight the scientific advances that are most impactful for the science of ET. The consequences of these synergetic observations span astrophysics, fundamental physics, nuclear physics, late-Universe cosmography, and early-Universe cosmology, offering immense potential for breakthroughs across these fields.

\subsection{Introduction}
\label{sec:SOs}

The synergies of ET with other GW  observatories
enable a {\it deeper, sharper, and wider} exploration of our Universe, enhancing our understanding of the physical laws that govern it. In certain areas, ET can benefit from complementary observations by other GW observatories, while in others it can uniquely contribute to them, thanks to its high sensitivity and extended  observational horizon, especially in the lower range of the high-frequency spectrum around a few Hz.

With  ET operating as part of a network of other third-generation (3G) ground-based observatories, such as the Cosmic Explorer (CE) \cite{Evans:2023euw}, the accuracy of measurements for physical parameters, luminosity distance, and sky localization of GW sources will improve significantly \cite{Iacovelli:2022bbs,Branchesi:2023mws,Gupta:2023lga}. The scientific advantages include enhanced cumulative sensitivity, longer observation periods and duty cycles, an extended reach to distant sources, and larger sample sizes of detected events. The baseline configuration considered here includes ET (in either a triangular or 2L configuration) alongside two CE detectors, covering a frequency range from approximately 3 Hz to the kHz region. We will also consider a 3G network made by ET (either in the triangular or in the 2L configuration) together with a single 40~km CE detectors in the US and a LIGO-India detector.
In the ``high-frequency'' range, we also highlight synergies with the proposed Neutron Star Extreme Matter Observatory (NEMO) \cite{Ackley:2020atn}, a GW interferometer optimized for studying nuclear physics at frequencies above a kHz.

At lower frequencies, we consider different GW observatories in space: the Laser Interferometer Space Antenna (LISA) \cite{Colpi:2024xhw}, recently adopted as a flagship mission by ESA in collaboration with NASA, and the concept designs Tian~Qin \cite{TianQin:2015yph} and Taiji \cite{Ruan:2018tsw}, all of which are sensitive to GWs from below $10^{-4}$ Hz to about 0.1 Hz. In the decihertz band, we consider the Lunar Gravitational Wave Antenna (LGWA) \cite{LGWA:2020mma, Ajith:2024mie} as an example of Moon-based projects (see also LILA; \cite{Jani:2020gnz}) and the proposed space mission DECIGO \cite{Seto:2001qf, Kawamura:2011zz, Nakamura:2016hna, Kawamura:2020pcg}, with sensitivities extending from about 0.01 Hz up to Hz frequencies.
Finally, at even lower frequencies, around the nHz band, we will discuss possible synergies with Pulsar Timing Array (PTA) experiments, using expected radio observations from the Square Kilometre Array (SKA) \cite{Dewdney:2009tmd} as a primary example of a radio telescope infrastructure of interest for our considerations.

\begin{figure}
\centering
\includegraphics[width=0.95\textwidth]{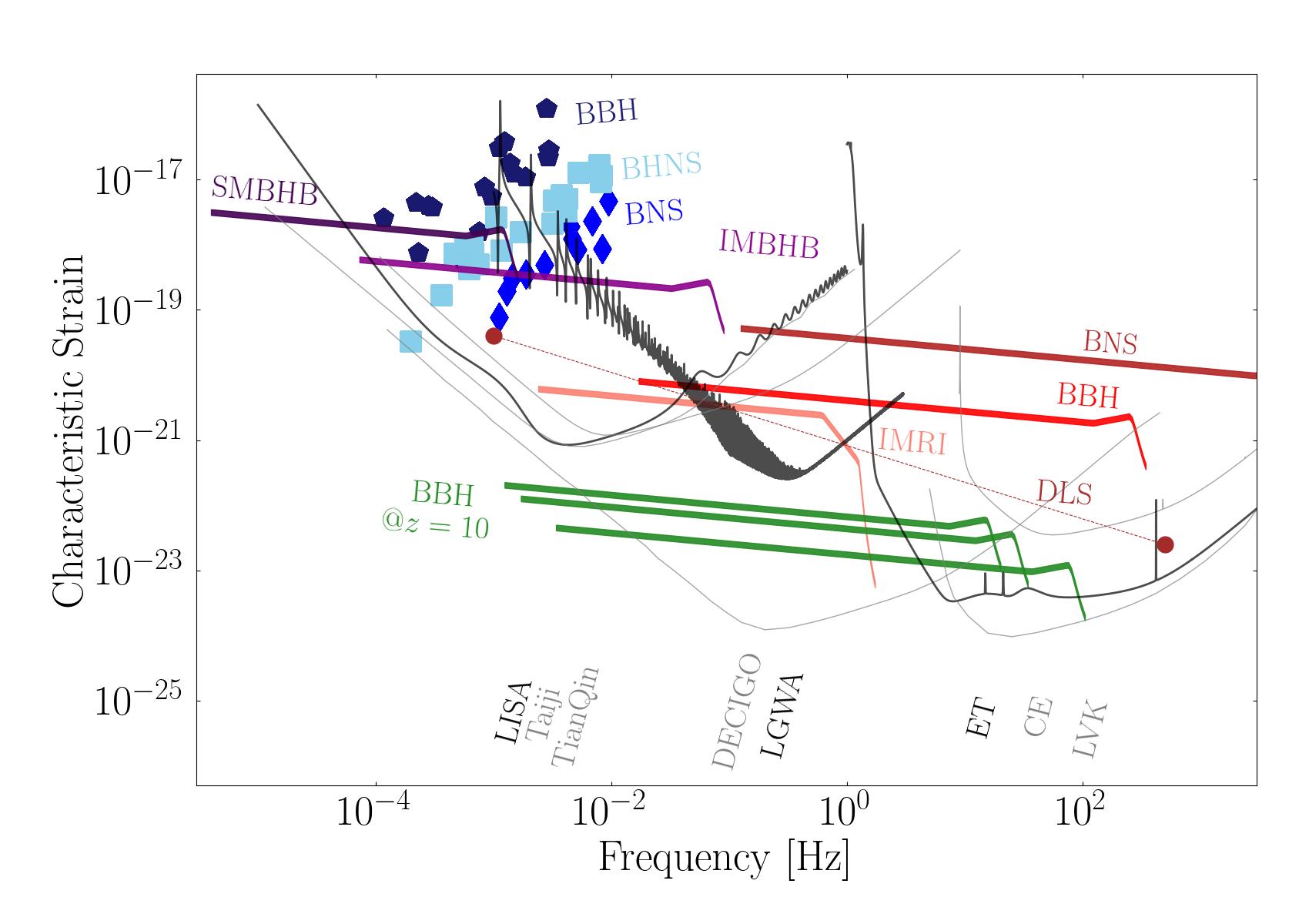}
\caption{
The GW landscape --- a schematic representation of synergetic sources in the GW frequency-amplitude plane. The figure shows the sensitivities for ET and CE (including the current LVK detectors) at higher frequencies, and for LISA, Taiji, Tian Qin, DECIGO and LGWA  at lower frequencies (sensitivities for the latter are from \cite{Colpi:2024xhw, Ajith:2024mie, Wang:2020vkg, Seto:2001qf, Kawamura:2020pcg, Li:2023umb}).
Tracks (left to right) represent: an equal mass black hole binary merger (labelled SMBH) of $10^6\,\Msun$ at $z\sim 6;$ an equal-mass intermediate mass black hole merger (IMBHB) of $ 10^5 \Msun$ at $z=1.5$; a multi-band source hosting an inspiralling equal-mass stellar black hole binary (BBH) of $ 60\Msun$ at $z\sim 0.1$ merging in the ET bandwidth; an intermediate mass ratio inspiral (IMRI) hosting a stellar black of $30\,\Msun$ orbiting around a black hole of $5\times 10^3\,\Msun$ at $z=1$. The ``BBH@z=10" track denote mergers at $z=10$ of equal-mass black hole binaries forming in metal poor early galaxies, with a total mass, from bottom to top, of $20,~60,~100\, \Msun$. Squares, pentagons and diamonds refer to Galactic compact binaries detected by LISA, millions to billions of years away from merging.  DLS indicates a dual line system. Each section of this chapter presents each of these sources and their scientific opportunities in detail. }
\label{fig:GW_Landscape}
\end{figure}

Although we may mention other GW observatories that have been proposed in the literature (e.g.~\cite{Adhikari:2019zpy,Sesana:2019vho, Baibhav:2019rsa, Kuns:2019upi, Sedda:2019uro, Baker:2019ync, Jaraba:2023djs, Moore:2017ity, Crosta2024, 2021QS&T....6d4003A, Canuel:2017rrp, Canuel:2020cxb, Zhan:2019quq, Badurina:2019hst,Proceedings:2024foy}), at the present stage of developments the ones mentioned above represent the current standard for expected detection capabilities across different frequency ranges during the observational epoch of ET. We will therefore consider these as primary examples providing synergetic observations for ET. In figure~\ref{fig:GW_Landscape}, we show the sensitivities of some of the GW observatories mentioned above, along with the frequency tracks of few example GW sources.

In section~\ref{sec:CBC} we quantify the scientific gains from combining observations of merging binary neutron star (BNS) systems and stellar-mass binary black holes (BBHs) from a network of 3G Earth-based observatories, including ET. These joint observations will deepen our understanding of the properties and evolution of BNSs and stellar-mass BBHs across cosmic epochs and will provide insights into the equation of state of nuclear matter, the dynamics of the Universe on large scales, and the possible existence of exotic compact objects. While these scientific topics are covered in later sections, section~\ref{sec:CBC} will focus on the improvements in detection and parameter estimation for these sources.
The synergy of combining GW observations of compact binary mergers, such as BNSs and BBHs, can extend to observatories operating at frequencies below a few Hz, which are inaccessible to ground-based observatories. Section~\ref{sec:CBC} also discusses nearby extragalactic stellar-mass BBHs, which serve as intrinsic {\it multi-band GW sources}. When far from coalescence, they emit GWs at frequencies twice their Keplerian orbital frequency, appearing as slowly chirping signals. These low-frequency signals contain unique information about their formation pathways, which is otherwise inaccessible. Detectors such as LISA, Tian~Qin, and Taiji can capture these signals at frequencies as low as mHz, long before they enter the Deci-Hz band and eventually merge within ET’s frequency range (above a few Hz).
Deci-Hz detectors like LGWA and DECIGO will not only track the inspiral of stellar BBHs closely but also observe BNS inspirals, providing early estimates of their masses, distances, and sky locations. These early estimates will allow electromagnetic (EM) telescopes to target these sources long before merger, greatly improving the potential for EM follow-up observations and hence, significantly advancing multi-messenger science.

In section~\ref{sec:nuclphys} we review the expected insights into nuclear physics.
The combined measurements of a 3G network of Earth-based observatories will allow for unprecedented constraints on the tidal deformability parameters of NSs and in turn on the state and properties of nuclear matter at extreme densities.
These scientific breakthroughs 
will complement those from nuclear physics experiments at lower energy as well as those from particle physics experiments at higher energy, yielding a complete panorama of the behavior of nuclear matter at all energy scales.

In section~\ref{sec:fundphys} we present the implications for fundamental physics obtained by the combination of ET with other GW observations.
BHs are scale-invariant, vacuum solutions of Einstein's field equations. Probing the spacetime geometry of BHs, i.e.~whether they are described by the Kerr metric or not, from the stellar-mass scale to the supermassive scale, is of paramount importance. The benefit of
combining ET with CE or with LISA has far reaching consequences.  Tests of the nature of black hole horizons,  of the orbital evolution during the inspiral phase and  of the propagation properties of GWs can be carried out at better precision, given the wider dynamic range that becomes accessible.

The synergy among different GW observatories has the potential to revolutionize late-time cosmology, as discussed in section~\ref{sec:cosmo}. With ET operating as part of a 3G network, the larger dataset, and significantly improved sky localization, will enable better identification of the host galaxies of sources through the detection of EM counterparts of BNS and NSBH systems. These events will serve as unique cosmic tracers, allowing precise measurements of the Universe's expansion, delivering stringent constraints on the Hubble constant, and enabling new tests of the nature of dark energy. Even in the absence of an EM counterpart, the ET+CE network will provide accurate measurements of source masses, sky locations, and distances to  enable powerful cosmological analyses. These include measuring the Hubble constant with dark sirens, cross-correlating localization data with galaxy catalogues and exploiting features in the mass distribution of black holes (BHs) and neutron stars (NSs). The large number and extended distance reach of detected GW sources will facilitate precise measurements of the Universe on large scales, offering new insights into current cosmological enigmas and serving as valuable probes at high redshifts.

Section~\ref{sec:BHseeds} discusses an additional, albeit indirect, intersection of ET with LISA. This connection addresses an unresolved problem in astrophysics: the origin of supermassive black holes (SMBHs), which range from millions to billions of solar masses and reside at the centres of bright galaxies in the present-day Universe. The discovery of quasars near and beyond the epoch of cosmic reionization raises significant questions about the rapid growth of these massive, accreting black holes.
One hypothesis is that these SMBHs form from {\it seed black holes} of about $ 10^2 \,\Msun$ to $10^5\,\Msun$, and grow by accretion and mergers. Among these, light seeds -- those lighter than about $ 10^3\Msun$ -- are expected to be the relics of the first forming, metal-free massive stars present in the earliest dark matter halos assembling around $z\sim 20,$ when the Universe was only a few hundreds of millions of years old. Observing the merger signals of black holes from this first pristine population of stars with ET would establish a connection with the population of merging massive black holes that LISA will detect at comparable and lower redshifts in the mass range around $(10^{4} - 10^7) \,\Msun$. Matching these two independent observations will provide invaluable clues into the physical origin of the seeds, their masses and the processes through which they grew to become supermassive during galaxy evolution.
This connection is unique and is made possible since both ET and LISA have comparable cosmic horizon-reaches.

Section~\ref{sec:earlycosmo} explores potential synergies between ET, LISA and PTA in the detection of a stochastic GW background (SGWB) from the early Universe (see in particular section~\ref{sec:early_universe} for discussion of cosmological backgrounds and section~\ref{sect:Populationbackdiv3} for discussion of astrophysical backgrounds). By combining these independent observations we can cover a much broader range of GW frequencies and achieve a precise characterization of the expected cosmological background(s).  The detailed morphology of the SGWB encapsulates information about the mechanisms that originated the emission of GWs in the early Universe. Complementary observations at different frequencies will thus greatly enhance our ability to precisely characterise these early-Universe phenomena that have produced the detected SGWB. Additionally, the synergy with other Earth-based detectors is crucial for cross-correlating the data of ET and differentiating between noise and stochastic signal. A single interferometer alone will struggle to extract more than an upper bound on  SGWBs from the observed data.

In each of the following sections, readers will find two boxes highlighting important text.
Cyan boxes outline the main scientific questions associated to the section, while teal boxes summarize key take-home messages of the section.

\subsection{Compact object binaries}
\label{sec:CBC}

\bigskip{}{}
\begin{tcolorbox}[standard jigsaw, colframe=cyan1, colback=black!10!white, opacityback=0.6, coltext=black,title=\sc The stellar-mass compact object binaries in the Universe]

{\sc 
  \begin{itemize}
           
            \item What scientific questions on the astrophysics of neutron stars and black hole binary systems can only be answered with ET in synergy with a network of  next generation ground-based observatories?
            \item What is the promise of multi-band GW observations for understanding compact object binaries?
            \item Are there dual-frequency GW sources and how will they inform us about the physics of compact objects?
        \end{itemize}
}
\end{tcolorbox}


The majority of ET sources will consist of two stellar-mass compact objects with masses in the range $\sim1$--$1000\,M_{\odot}$. ET is expected to detect hundreds of thousands of these sources throughout the Universe each year   \cite{Borhanian:2022czq, Gupta:2023lga, Iacovelli:2022bbs}.   
The properties
and origin of the population of compact object binaries have only started
to be explored, with many formation pathways suggested (see section~\ref{section:div3}). There is no doubt that ET on its own will enable very significant advances in our understanding of the population and origin of compact object binaries. However,  in synergy with other GW detectors, several additional science results are expected, which are highlighted here. We discuss both synergy with high-frequency detectors and with mid-frequency detectors.

\begin{figure}
\centering
\includegraphics[scale=0.6]
{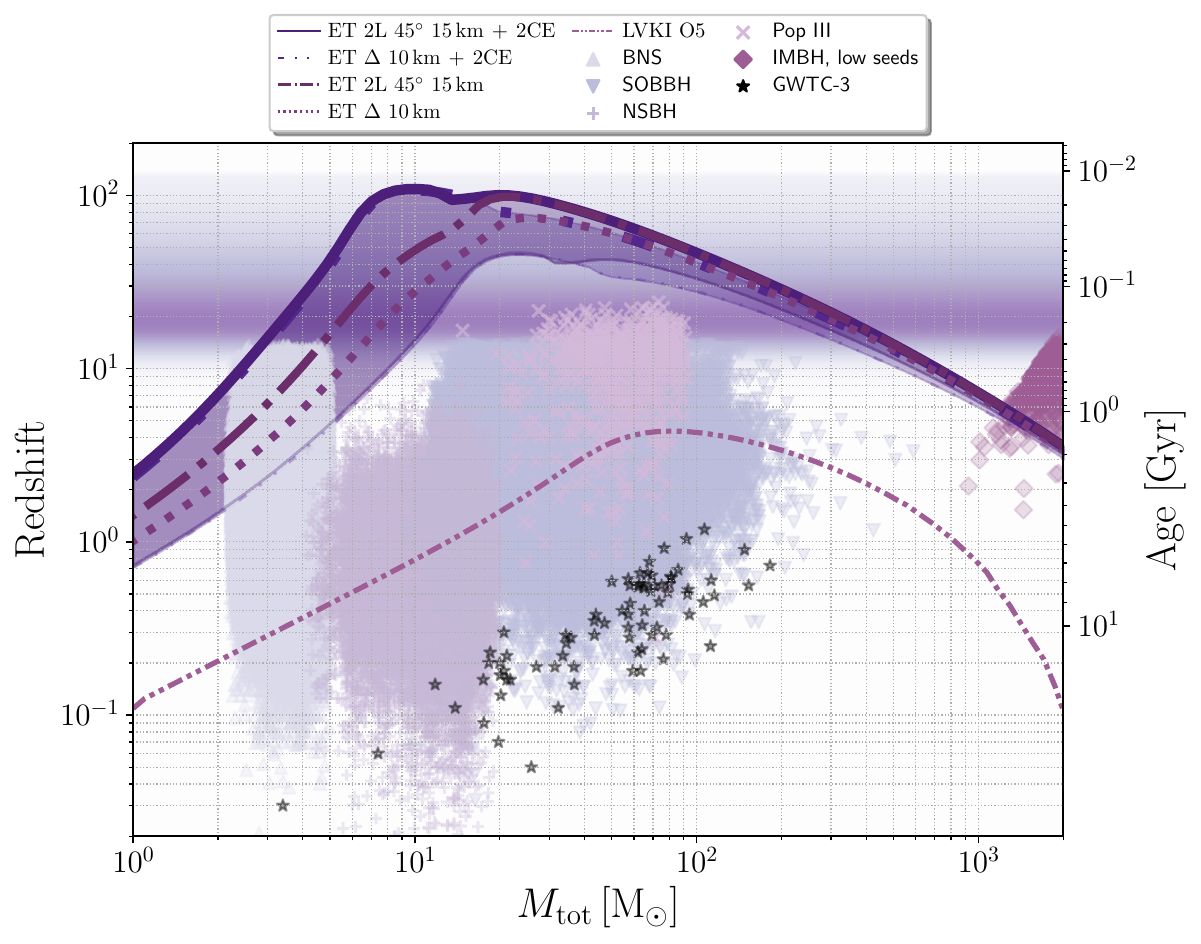}
    \caption{Detector horizons for the two baseline configurations of ET, alone and in conjunction with a network of two CE as described in the legend. The shaded bands, reported only for the combinations with CE, denote the range within which $50\%$ of the sources are detected, when averaging over sky location.
    Known classes of target sources are reported for illustration: stellar-mass black holes (SOBBH) and double neutron star systems   (catalogs from ~\cite{Branchesi:2023mws}), neutron star-black hole binaries (catalogs from \cite{Iacovelli:2022bbs} based on~\cite{Broekgaarden:2021iew,Giacobbo:2018etu,Zhu:2020ffa}), Population III stars~\cite{Costa:2023xsz,Santoliquido:2023wzn}, and intermediate mass black holes (IMBH) \cite{Colpi:2024xhw}.  The orange band is representative of a population of high--redshift primordial black holes (PBHs) ~\cite{Franciolini:2021tla,Ng:2022agi}.
    Triangles show the events detected by the LVK collaboration~\cite{KAGRA:2021vkt}.}
    \label{fig:det_horizons}
\end{figure}

\subsubsection{Synergy with ground-based  detectors} 
\label{sec:CBC-ground-based}

There are several other detectors in the frequency range of ET, from a few Hz to a few kHz, that could provide synergy with ET. The most important one is Cosmic Explorer (CE) \cite{Evans:2023euw}, a US project for a third generation detector. 
Einstein Telescope and Cosmic Explorer strengthen and complement each other in important ways. An important difference is that  CE is expected to be slightly more sensitive in the 10--100 Hz range which will complement the ET sensitivity curve \cite{Evans:2023euw}, whereas ET is expected to have better sensitivity compared to CE at lower frequencies as a result from ET’s underground location that suppresses seismic disturbances, reducing the Newtonian noise that limits CE  at low frequencies.  To highlight this synergy,  figure~\ref{fig:det_horizons} shows the detection horizons for equal--mass, non--spinning binaries, with a network $\rm SNR \geq 8$, for ET in its triangular and 2L configurations, both alone and in a network of two CE detectors.

\begin{figure}[t]
    \centering
    \includegraphics[width=1.\textwidth]{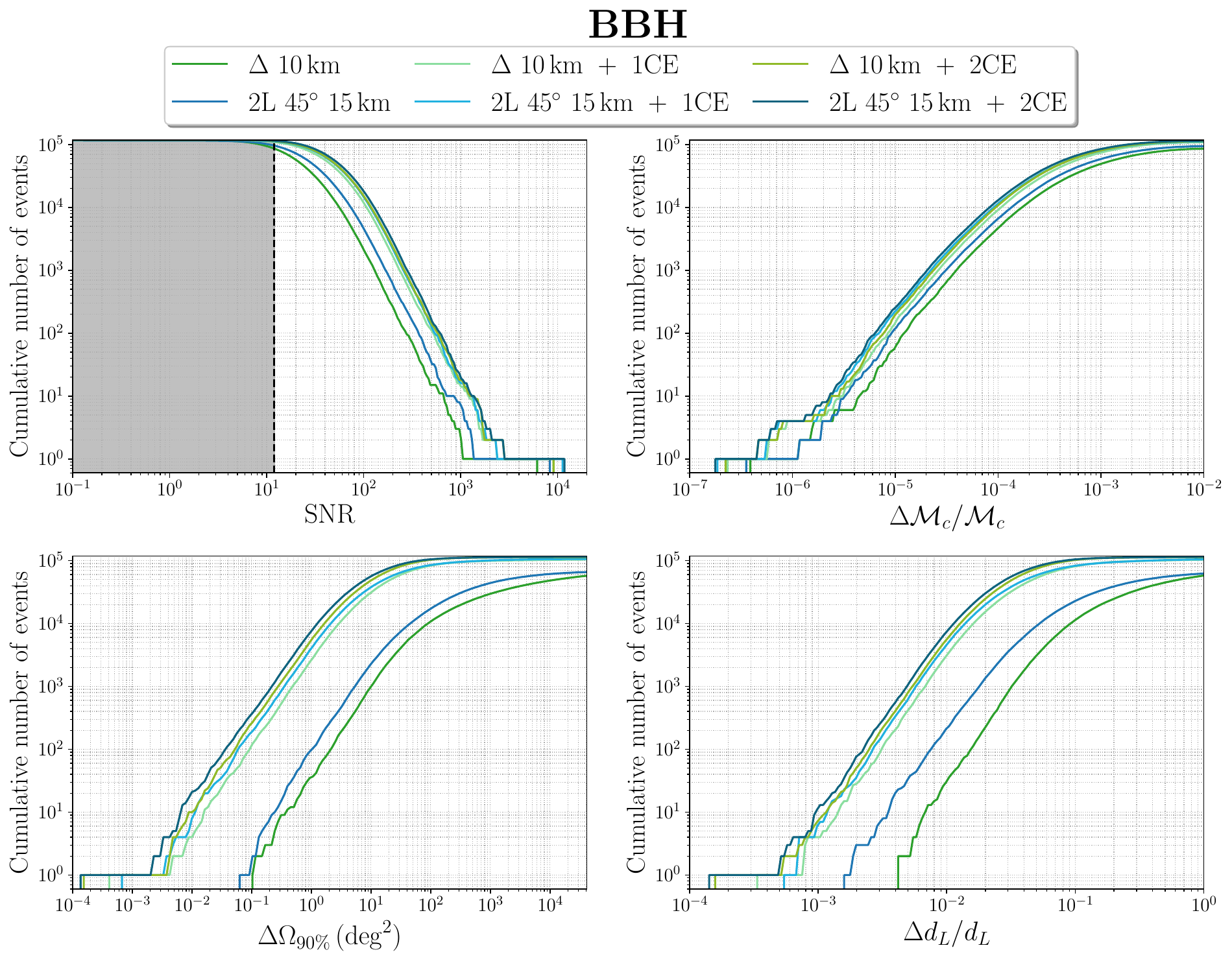}
    \caption{\small Comparison of SNR and parameter estimation error for the chirp mass, angular localization and luminosity distance for BBHs. We show the results for ET alone, for ET +1CE and for ET+2CE and, in all cases, for ET we consider both the triangle 10~km and the  2L-$45^{\circ}$ configurations. The results are obtained with the \texttt{GWfast} code~\cite{Iacovelli:2022mbg}; technical details as in \cite{Branchesi:2023mws}.}
    \label{fig:cumul_SNR-Mc-Om-dL_BBH_allconf_ETCE}
\end{figure}

\begin{figure}[t]
    \centering
    \includegraphics[width=1.\textwidth]{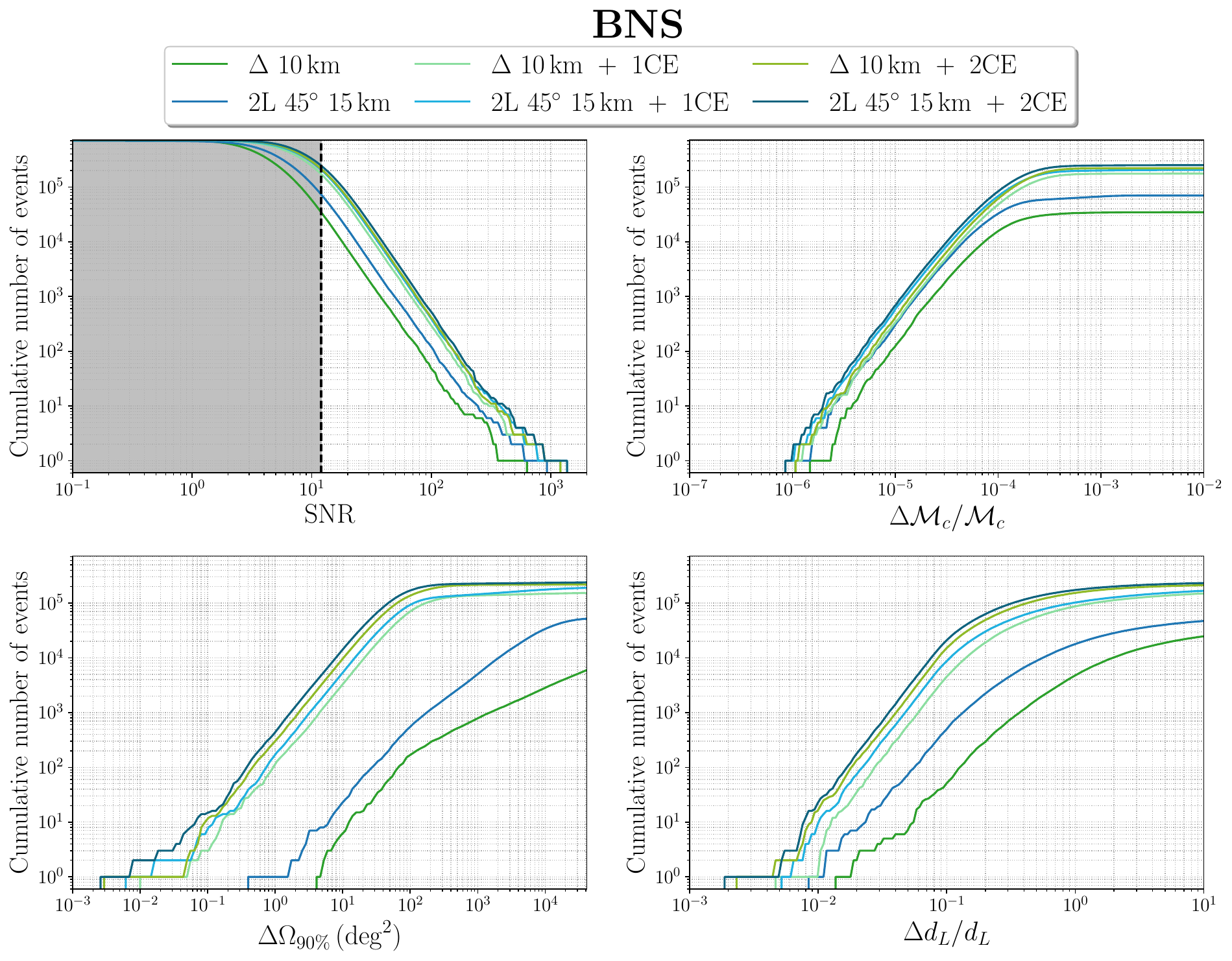}
    \caption{\small As in figure~\ref{fig:cumul_SNR-Mc-Om-dL_BBH_allconf_ETCE}, for BNSs.}
    \label{fig:cumul_SNR-Mc-Om-dL_BNS_allconf_ETCE}
\end{figure}

From just a point of view of detection, ET alone is expected to detect more than $90\%$ of BBHs of stellar origin (SOBBHs) and about $26\%$($15\%$)  of the binary neutron stars in the 2L (triangular) configuration~\cite{Branchesi:2023mws}. This corresponds to $\mathcal{O}(10^5)$ BBH and BNSs detected per year. ET and CE operating in network will allow us to significantly improve range and number of BNS detections. In the ET2L+2CE configuration, the number of BNS sources detectable per year will be $>4\times 10^5$  (in the population model considered). 
Moreover, a network of next generation GW observatories with both ET and CE is crucial to provide orders-of-magnitude improvement in parameter estimation for coalescing binaries, with significant impact on most of the science goals outlined in the ET and CE studies.
As an example, figure~\ref{fig:cumul_SNR-Mc-Om-dL_BBH_allconf_ETCE} shows the SNR distribution as well as the parameter estimation error for the chirp mass, the angular localization and the luminosity distance for the BBHs mergers corresponding to one year of data (using the same population as in \cite{Branchesi:2023mws}), while figure~\ref{fig:cumul_SNR-Mc-Om-dL_BBH_allconf_ETCE} shows the same results for BNS. In both plots we show the results for
ET alone, for ET  together with a 40~km CE (labeled ET+1CE)  and  for ET  together with a 20~km and a 40~km CE 
(ET+2CE). In all cases, for ET we consider both the 10~km triangle  and the  configuration with 2L of 15km at a relative angle of $45^{\circ}$. We see that the improvement is particularly remarkable for angular localization and for luminosity distance (while the improvement in the accuracy of  the mass measurement is less dramatic).

Examples of scientific results that can only be found by the combination of ET and CE are: (i) detections of BBHs (BNSs) with  90$\%$-credible sky area less than $0.01 \,(0.1)\, {\rm deg}^2$, important for multi-messenger observations \cite{Iacovelli:2022bbs,Branchesi:2023mws,Gupta:2023lga}; (ii) regular detection of BNS (BBH) systems at $z>1\, (z>10)$ with mass uncertainty $\delta m_1/m_1 < 0.3 (0.2)$ \cite{Gupta:2023lga}, important for understanding BH and NS properties across cosmic time, probing fundamental physics and performing precision cosmology; (iii) detection (or strong upper limits on the rate of) primordial black hole mergers at $z\gtrsim 25$ with $\delta z/z <0.2$ and population III binary black holes at $z \gtrsim 10 $ with $\delta z/z <0.1$, which will be crucial to inform BH formation channels \cite{Ng:2021sqn,Gupta:2023lga}. 
Importantly, by being located in the USA, CE will provide a longer baseline that will improve sky localizations and parameter estimations \cite{Iacovelli:2022bbs, Borhanian:2022czq, Gupta:2023lga}, as we see from figures~\ref{fig:cumul_SNR-Mc-Om-dL_BBH_allconf_ETCE} and \ref{fig:cumul_SNR-Mc-Om-dL_BNS_allconf_ETCE}. Having both CE and ET will also be beneficial for increasing the completeness of the observed population: for example, a network with both ET and CE  will achieve $50\%$ completeness of BNS sources up to $z\sim 2$,  as can be seen in Figure~\ref{fig:det_horizons}, whereas an (A+)-like network will only achieve that for $z\sim0.2$ \cite{Borhanian:2022czq}.  Furthermore,  having both ET and CE will increase the time that at least one detector is always online to assure measurements are made during exceptional events such as a galactic supernova \cite{Borhanian:2022czq, Gupta:2023lga, Iacovelli:2022bbs}.  

Overall, the presence of a next-generation detector network, including ET and CE, will enhance the science obtainable from ET observing alone. Fields such as those related to the measurement of cosmological parameters, unraveling the nuclear equation of state, and making new discoveries such as the presence of dark matter around NSs or BHs, multi-messenger merger observations can largely benefit from the improved capabilities of a network of observatories~\cite{Evans:2023euw, Gupta:2023lga, Borhanian:2022czq, Iacovelli:2022bbs}.
Similar synergies can be outlined between ET and the Neutron Star Extreme Matter Observatory (NEMO) \cite{Ackley:2020atn}, a proposed GW interferometer optimized to study nuclear physics with merging neutron stars which,  together with ET, could contribute  to inform the physics of BHs and NSs and their progenitors.

\begin{figure}[t]
    \centering
    \includegraphics[width=1.\textwidth]{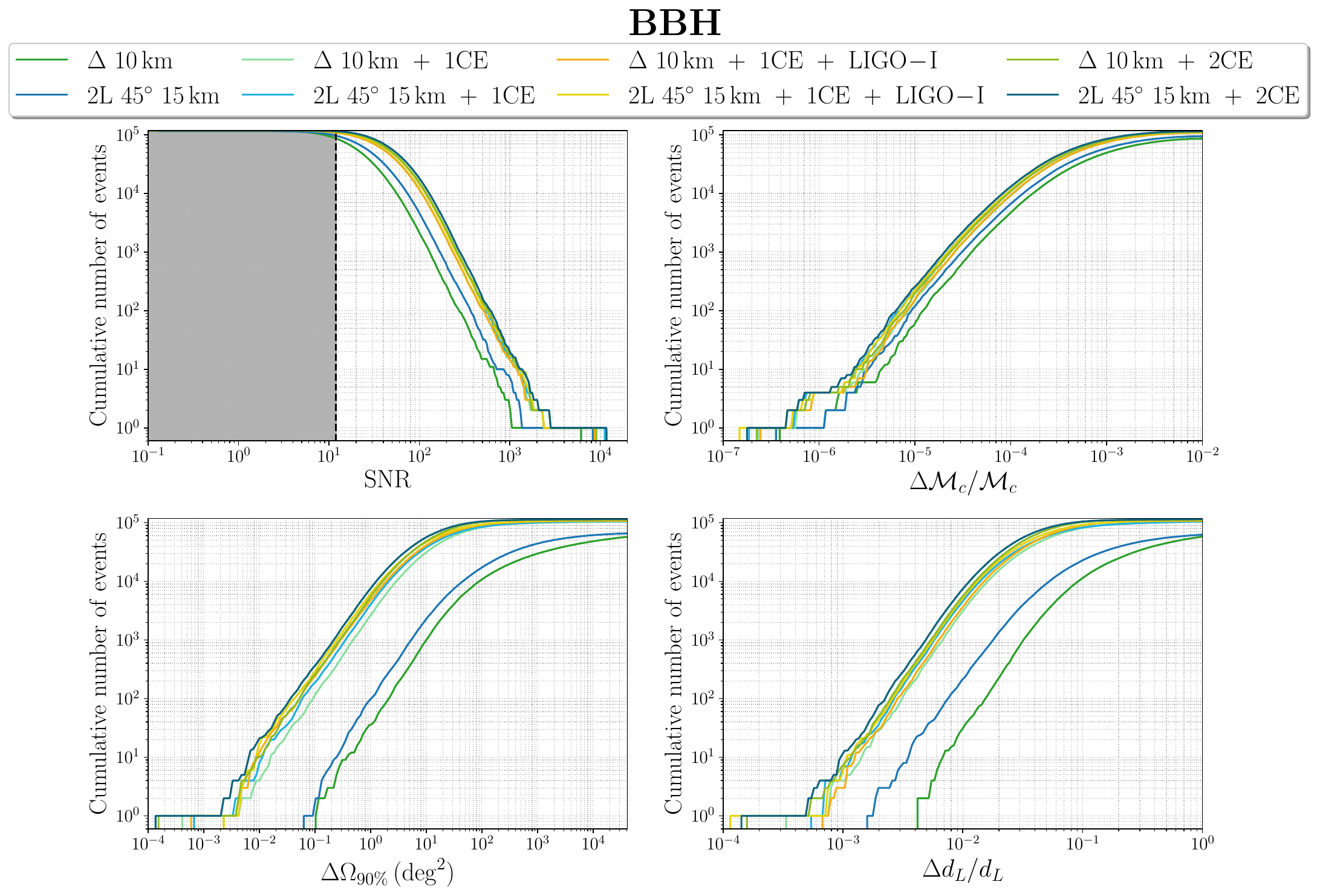}
    \caption{\small Comparison of SNR and parameter estimation error for the chirp mass, angular localization and luminosity distance for BBHs for different networks (as in the legend), including networks involving LIGO-India. The results are obtained with the \texttt{GWfast} code~\cite{Iacovelli:2022mbg}; technical details as in \cite{Branchesi:2023mws}.}
    \label{fig:cumul_SNR-Mc-Om-dL_BBH_allconf_ETCEIndia}
\end{figure}

\begin{figure}[t]
    \centering
    \includegraphics[width=1.\textwidth]{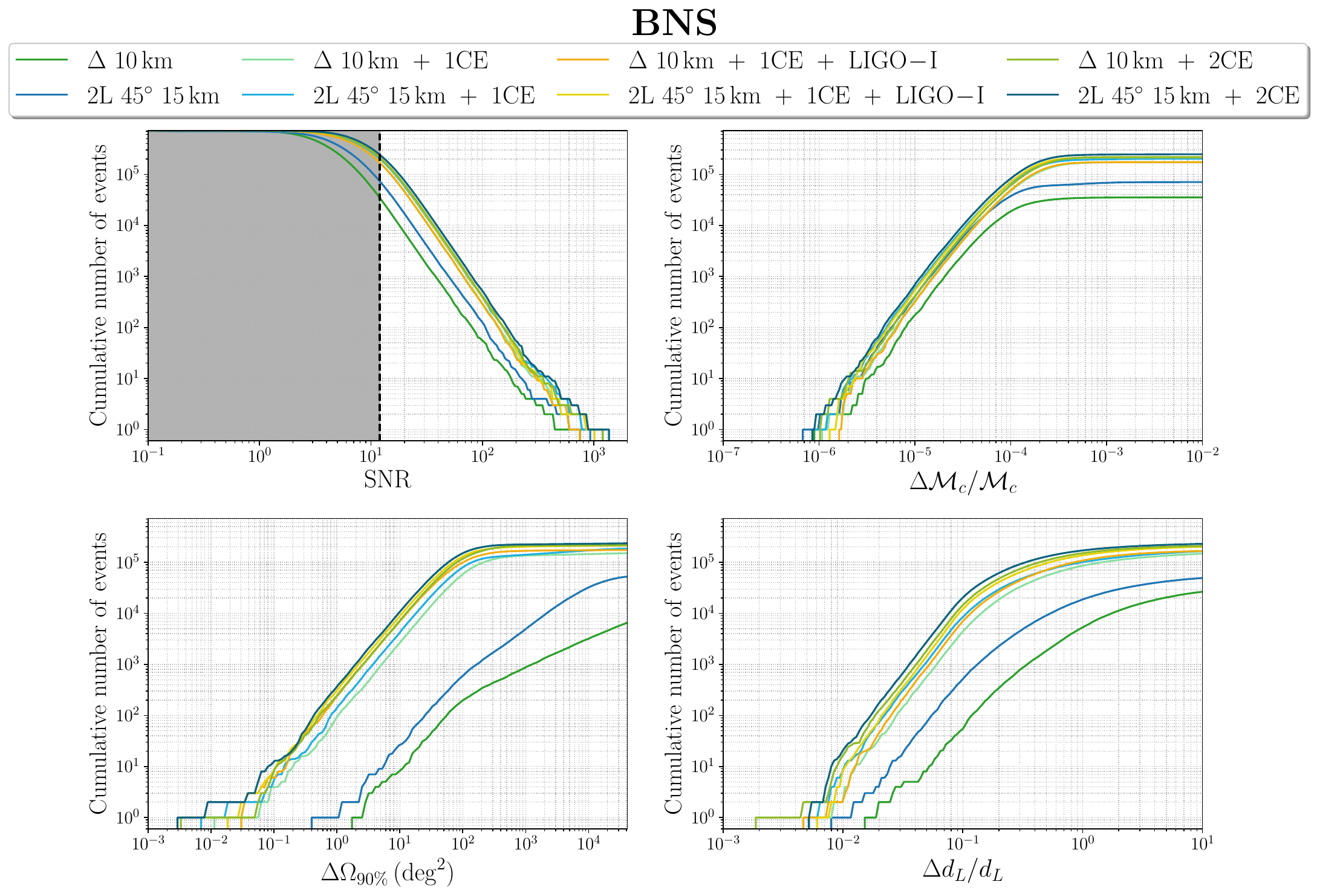}
    \caption{\small As in figure \ref{fig:cumul_SNR-Mc-Om-dL_BBH_allconf_ETCEIndia}, for BNSs.}
    \label{fig:cumul_SNR-Mc-Om-dL_BNS_allconf_ETCEIndia}
\end{figure}

Another network configuration which could be possible in the next-generation era is made of  ET (either in its triangular or in its 2L configuration) together with a single 40~km CE detector in the US  and LIGO-India. This configuration has been recently studied in \cite{Gupta:2023lga,Pandey:2024mlo} (restricting however to ET in its triangular configuration).
LIGO-India, currently under construction in Aundha, India, will feature 4~km arms as its US counterparts
and is expected to be operational in the 2030s at A+ sensitivity \cite{KAGRA:2013rdx}, with upgrades
at A$^{\#}$ sensitivity levels \cite{Fritschel:2022}.
As shown in \cite{Gupta:2023lga,Pandey:2024mlo}, the results  obtained with this network are quite interesting. Indeed, with respect to 
a network made by ET together with a 40~km  CE and a 20~km CE one finds that, for  angular localization and accuracy on luminosity distance, 
replacing the 20~km CE
detector with LIGO-India operating at A$^{\#}$  sensitivity gives nearly identical performances, since the factor of five in arm length is  offset by the increase in baseline relative to a second CE (although other aspects of the science case are  limited by the shorter arm length and narrower sensitivity band).

Here we perform again the analysis for a network made by ET together with a single 40~km CE detector and LIGO-India, extending the analysis of 
\cite{Pandey:2024mlo} by considering  both the triangular and the 2L configuration of  ET, and assuming A$^{\#}$ sensitivity for LIGO-India. For uniformity with the results shown in figures~\ref{fig:cumul_SNR-Mc-Om-dL_BBH_allconf_ETCE} and \ref{fig:cumul_SNR-Mc-Om-dL_BNS_allconf_ETCE}, as well as with similar analysis presented elsewhere in this paper (see in particular section~\ref{section:div9}) we use the same population model, sensitivity curve and technical details as in \cite{Branchesi:2023mws}.\footnote{To compare with the results presented for BNS in \cite{Pandey:2024mlo} one must  take into account some technical differences. Our population model, restricting (as in \cite{Pandey:2024mlo}) to $z<0.5$, contains $1.1\times 10^4$ BNS (and corresponds to a local merger rate  $R_0= 250\, {\rm Gpc}^{-3} {\rm yr}^{-1}$), to be compared with $1.6\times 10^4$  BNS (and 
$R_0= 320\, {\rm Gpc}^{-3} {\rm yr}^{-1}$)
in 
\cite{Pandey:2024mlo}. Furthermore, we  assume an uncorrelated 85\% duty cycle in each L-shaped detector, and in each of the three instruments composing the triangle, while \cite{Pandey:2024mlo} uses $100\%$ for the duty cycle; this affects in particular the number of rare events with the very best accuracy of reconstruction. We have checked that, if we use $100\%$ duty cycle and take into account the different BNS rate, the results  are consistent, within a factor of 2, with those in  \cite{Pandey:2024mlo}.} The results are shown in figures~\ref{fig:cumul_SNR-Mc-Om-dL_BBH_allconf_ETCEIndia} for BBHs and in figure~\ref{fig:cumul_SNR-Mc-Om-dL_BNS_allconf_ETCEIndia} for BNSs, and show that a network made by ET together with a single 40km CE detector and LIGO-India can be quite competitive.

\subsubsection{Synergies with space-borne detectors}

The synergies with space-borne detectors, in particular LISA, the space mission recently adopted by ESA  in collaboration with NASA \cite{Colpi:2024xhw}, and operating in the ``mid-frequency''  band $\sim 10^{-4}-10^{-1}$ Hz,
come in three distinct ways: i) observations of \emph{different instances} of a class of compact object binaries at lower frequency, that provide complementary insight in their population properties and their origin; ii) observations of \emph{multi-band} sources, that are first observed at mid-frequency and then later in the ET band; and iii) \emph{dual-line sources} that emit simultaneously in both mid-frequency band and the ET band.


\subparagraph{Observations of the same source class.} Different origins of compact object binaries (from isolated binaries,  from dynamical processes, or from primordial black holes) give rise to different population properties that can be used to disentangle their origin, e.g. \cite{Breivik:2016ddj}. Orbital eccentricity is thought to be a significant indicator of formation via dynamical processes and can be detected by LISA, but cannot easily be measured by high-frequency detectors, since the eccentricity is radiated away by GW and it very small by the time the signals enter the detectable band, even for large initial eccentricity \cite{Breivik:2016ddj,Nishizawa:2016eza}. 
Observing the same type of sources at much lower frequency is an excellent way to constrain the importance of dynamical formation, albeit only in the nearby Universe, since LISA and similar detectors typically observe stellar-mass binary black holes in the Milky Way and nearby galaxies \cite{He:2023ywk}, unless they are at very high frequency when they can be observed 
to several tens or at most hundreds of Mpc, see \cite{LISA:2022yao}.
Deci-Hz observatories such as DECIGO would in fact be even better at detecting eccentricity given the large number of sources they can observe up to cosmological distances \cite{Chen:2017gfm}.

\subparagraph{Multi-band sources.} These are binary black holes that are observed in the LISA band and then later, on human time scales (years), merge in the high-frequency band \cite{Sesana:2016ljz}. In order for this to happen the binaries must have heavy black holes of around 30 $M_\odot$ or higher. Such heavy sources have been observed already with the current detectors. However, the expected number of multi-band systems that will be observed with LISA is likely small \cite{Gerosa:2019dbe,Buscicchio:2024asl}. The scientific value of these measurements would be high: the LISA measurements can make a very accurate prediction of the time and sky localisation of the merger in the ET band and the properties of these systems would be determined to very high accuracy, see  e.g. \cite{Klein:2022rbf}. Another proposed special class of multi-band sources is a low-frequency binary black hole orbiting around a super-massive black hole, where high induced eccentricity could move the system temporarily into the ET band \cite{Zhang:2024ibf}.
Multi-band sources would as well be ideal targets for Deci-Hz detectors, as these observatories will be able to closely follow their inspiral almost until the merger, providing a strong complementarity with ET measurements and excellent pre-localisation accuracy for multi-messenger studies \cite{Nair:2018bxj,Nair:2015bga,Nakano:2021bbw,Liu:2021dcr,Ajith:2024mie}.

\subparagraph{Dual-line sources.} These are proposed sources emitting simultaneously in two frequency bands. An example are ultra-compact X-ray binaries consisting of a rapidly rotating NS (with spin frequencies in the hundreds of Hz) orbiting a companion star \cite{Tauris:2018kzq}. GW from the orbital motion can be detected by LISA while continuous GWs can be detected in the ET band  due to the presence of accretion-built mountains, toroidal magnetic fields, and/or r-mode oscillations \cite{Suvorov:2021mhr}. A dual-line detection could provide percent-level constraints on the mass, radius, and internal magnetic field strength of the neutron star. With ET,  at least four of the known ultra-compact binaries become potentially dual-line detectable \cite{Suvorov:2021mhr}.

\subsubsection{Stochastic background from compact binaries}

Finally, there is added value in correlating GW observations between different instruments for the detection of broadband stochastic signals that make up the astrophysical GW background, e.g. \cite{Regimbau2022,KAGRA:2021kbb} to both characterize the population of astrophysical GW sources and distinguish these from cosmological backgrounds, as discussed in section~\ref{sect:subtractastrobkgdi9}. In the ET frequency band, the astrophysical GW background is the collective signal produced by unresolved black hole and neutron star binaries, and can be detected by ET with significant confidence (see section~\ref{sect:Populationbackdiv3}). However,  its broadband features imply it may be difficult to distinguish from instrumental noise, particularly when the latter is not very well known, see e.g.~\cite{Renzini:2022alw}. Cross-correlating measurements of different, independent instruments, such as ET and CE will thus increase detection confidence. %

In addition, there is synergy with detection efforts in other frequency bands: the same binary background extends to lower frequencies and can in principle be detected by detectors such as LISA and Tian Qin. A multi-band detection and characterization effort would provide additional constraints on compact binary population features, including information on compact binary progenitors, see e.g. \cite{Babak:2023lro,Lehoucq:2023zlt,Bavera:2021wmw,Dvorkin:2016okx}, complementary to those coming from resolved source analyses. Furthermore, low frequencies such as those monitored by LISA will be dominated by a broadband stochastic foreground due to unresolved white dwarf binaries (that, due to the relatively large size of the white dwarfs, does not contribute to the ground-based detector GW band). Recent studies suggest that the white dwarf background may well dominate LISA data, hindering detection prospects for the stellar-mass compact object background~\cite{Staelens:2023xjn,Hofman:2024xar}. In this scenario, detection of this background signal with ET will be all the more valuable, and will be used to inform analyses at lower frequencies.

A notable data analysis challenge in this context will be component separation. Differentiating between resolved events and background and between background contributions of various nature requires novel analysis techniques, and in particular  the residual subtraction error could dominate stochastic measurements~\cite{Biscoveanu:2020gds,Zhong:2024dss}. However, 
recent work that performed explicitly the
correlation between data streams of different detectors, after 
reconstruction and subtraction of CBCs  
from the individual detector data streams, 
show that 
the enhanced resolving power of a network made of ET+CE can reduce significantly the errors in the reconstruction of resolved sources, to the extent that the sensitivity to cosmological stochastic background will only be marginally affected, see \cite{Belgacem:2024ntv} and section~\ref{sect:subtractastrobkgdi9}.

\begin{tcolorbox}[standard jigsaw, colframe=teal, colback=black!10!white, opacityback=0.6, coltext=black,title=\sc Take Home Message]
\sc 
\begin{itemize}
   \item  
      ET in a 3G network alongside other ground-based detectors, such as Cosmic Explorer,  will extend the cosmic horizon reach, enable deeper and more accurate  surveys, and yield significantly more precise physical measurements of stellar-mass compact object binaries.

   \item Joint observations with a ground-based GW network (ET with CE or NEMO) will enable us to map neutron stars across cosmic time.

   \item  Synergies between ET and lower-frequency space-based detectors, such as LISA, will enable multi-band observations of the same source classes, offering unique insights into their formation pathways. 
   
   \item Multi-band observations of the same source are essential to discover dual-line.

   \item 
   Astrophysical, stochastic GW signals from compact binaries of diverse type and  in different evolutionary states are expected to be detected by ET and LISA.  These observations will provide unique statistical information into the underlying source populations.
   
\end{itemize}
\end{tcolorbox}


\subsection{Nuclear physics}
\label{sec:nuclphys}



\begin{tcolorbox}[standard jigsaw, colframe=cyan1, colback=black!10!white, opacityback=0.6, coltext=black,title=\sc Nuclear Physics in the Realm of GW Sources]
\sc 
\begin{itemize}
   \item How many more BNS and NSBH mergers we can observe with ET+CE? How does it translate to constraints on EoS?
   \item How much better can we observe the post-merger phase of a BNS merger? How does it translate to QCD phase transition constraint?
\end{itemize}
\end{tcolorbox}

A principal goal of heavy-ion collision experiments such as Relativistic Heavy Ion Collisions (RHIC)~\cite{Trainor:2013bma} and Facility for Rare Isotope Beams (FRIB)~\cite{Bollen:2010leu} is to understand the nature of hot and dense hadronic matter. These experiments have improved our understanding of the quark-gluon plasma, a state of matter thought to have existed in the early moments after the Big Bang. However, much about the properties of this matter remains unknown. At the other end of the spectrum, neutron star cores are believed to contain matter at even higher densities, though at much colder temperatures than those explored in heavy-ion collision experiments.

During the inspiral phase of a neutron star merger, the properties of the cold supranuclear matter are encoded in the tidal dephasing of the signal. The supranuclear matter can reach a temperature of 50 MeV~\cite{Hammond:2021vtv, Raithel:2021hye, Most:2022yhe} during the post-merger phase. Therefore, the observation of a binary neutron star merger provides a unique opportunity to explore the QCD phase diagram comprehensively, spanning both cold and hot regimes of dense matter, as we discuss in detail in section~\ref{section:div6}. 

In this section, we showcase how the synergy between ET and CE can: (1) constrain the cold supranuclear matter with the thousands of neutron star merger detections; (2) detect the currently undetectable post-merger signal; (3) gain insights about the supranuclear matter with the post-merger signal.

\subsubsection{Population approaches for nuclear physics}

Next-generation GW observatories will detect thousands of binary coalescences involving a neutron star (NS) \cite{Branchesi:2023mws,Gupta:2023lga}. These mergers are rich laboratories to study the behavior of matter under extreme conditions \cite{Baiotti:2016qnr}, as also demonstrated by the binary neutron star GW170817 \cite{LIGOScientific:2018hze,LIGOScientific:2017ync}.
When the NS approaches its companion, in particular a BH, it may undergo tidal deformation and be disrupted. This process imprints a distinctive signature on the GW waveform that is parametrized with a tidal deformation parameter $\Lambda$ for the NS (see \eqs{eq:biglambdadef}{deftildeLambdadiv6} below). From the knowledge of $\Lambda$ and the NS mass, it becomes possible to estimate the NS radius. Finally, a joint estimation of the NS mass and radius is crucial for determining the Equation of State (EoS) that governs the star mass-radius relation \cite{LIGOScientific:2018cki}. However, the EOS for neutron stars remains unknown and measuring NS radii is extremely challenging as the $\Lambda$ parameters are not often well constrained. The BNS merger GW170817 currently provides the best constraint on the NS radii, with a precision of a few kilometres \cite{LIGOScientific:2018cki}. This precision was only sufficient to exclude  the stiffest EoSs.

\begin{figure*}
    \centering
    \includegraphics[width=0.65\textwidth]{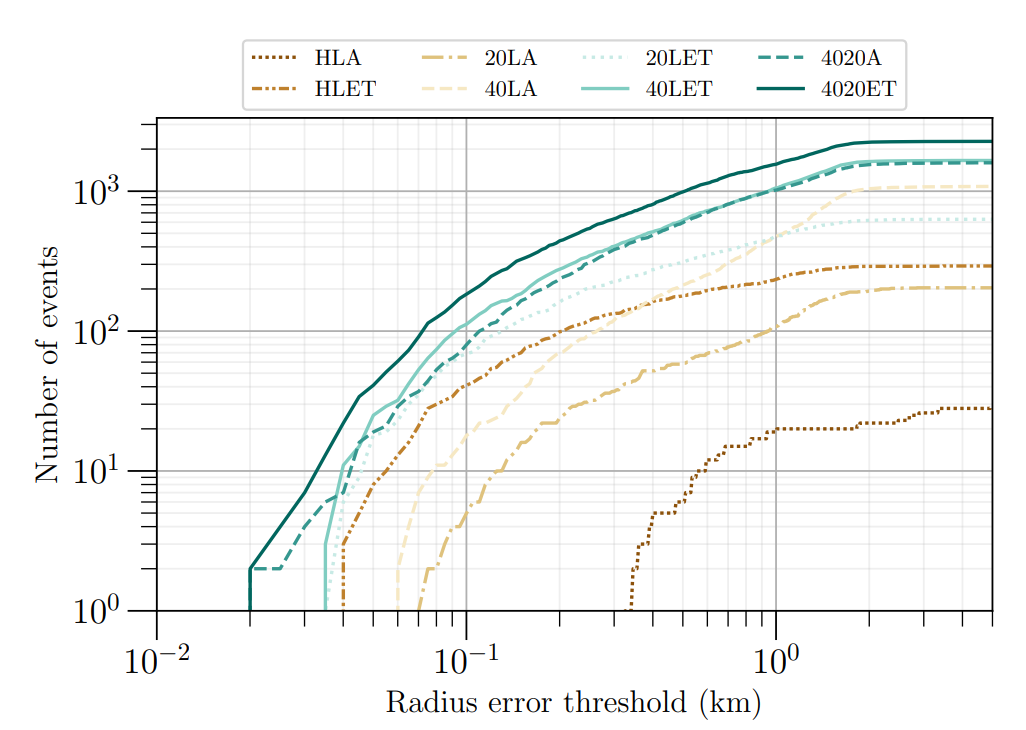}
    \caption{The cumulative number of GW detections (vertical axis) for which it will be possible to constrain NS radii below a specified threshold (horizontal axis) is shown. The lines represent different GW detector configurations: HLET, 20LA, and 40LA indicate networks with one next-generation detector, while 4020A, 20LET, and 40LET represent networks with two next-generation observatories. Figure reproduced from \cite{Gupta:2023lga}.}
    \label{fig:radii_constraints}
\end{figure*}

Next-generation observatories will detect thousands of GW events like GW170817, including hundreds of sources with higher signal-to-noise ratio \cite{Iacovelli:2022bbs,Branchesi:2023mws,Gupta:2023lga}. Figure~\ref{fig:radii_constraints} shows the forecasted constraints on the NS radii achievable in the next-generation GW era. For a GW detector network that includes at least one next-generation GW detector, it is anticipated that NS radii can be measured with a  precision of 1 km (approximately 10\% of the nominal NS radius) for about  200 detections per year \cite{Gupta:2023lga}. In a network with at least two next-generation GW observatories, this number increases to around  1,000 detections annually. These observations would provide the robust data necessary to precisely test and refine proposed EoSs for NSs \cite{Huxford:2023qne, Iacovelli:2023nbv}. The ability of 3G observatories to constrain the NS EOS from populations of compact binary coalescences is not significantly affected by the specific geometries of the detectors but rather depends on the number of 3G observatories available to capture these signals \cite{Iacovelli:2023nbv}.

\subsubsection{Measurability of the post-merger phase signal}

The post-merger phase explores a different regime in the QCD phase diagram, characterized by significantly higher densities and temperatures compared to the inspiral phase. During the inspiral, the neutron stars remain largely intact, and the densities reached are limited to their central densities, approximately 3–4 times the nuclear saturation density. On the other hand, the post-merger phase probes densities exceeding five times nuclear saturation density \cite{Pang:2022rzc}. At temperatures of approximately 50 MeV, the effects of different transport coefficients become prominent, leaving noticeable imprints on the data \cite{Hammond:2021vtv, Raithel:2021hye, Most:2022yhe}. However, the amplitude of the post-merger phase signal is weaker than that of the inspiral phase due to the reduced quadrupole component \cite{Kastaun:2014fna, Bauswein:2012ya, Takami:2014tva, Bernuzzi:2015rla, Bauswein:2015yca}. 
This challenge is compounded by the reduced sensitivity of current detectors at the higher frequencies where the post-merger signal lies.
However, a next-generation GW detector significantly improves access to the post-merger phase. As shown in table~\ref{tab:pmsnr}, for three sources taken at the same distance of 68~Mpc,\footnote{This specific value of the distance was chosen in \cite{Puecher:2022oiz}
because,  as we see from the row  labeled LHV in table~\ref{tab:pmsnr}, in a network
made by  Advanced LIGO+ and Advanced Virgo+, it corresponds
to a signal-to-noise ratio of 100 for Source~1 (which is defined by the masses and tidal deformability given in the table caption).}
but different values of the masses and of the tidal deformability, the post-merger phase SNR is five times higher  in the next-generation GW detector network as compared to current generation networks.

\begin{figure*}
	\centering
	\includegraphics[width=1\textwidth]{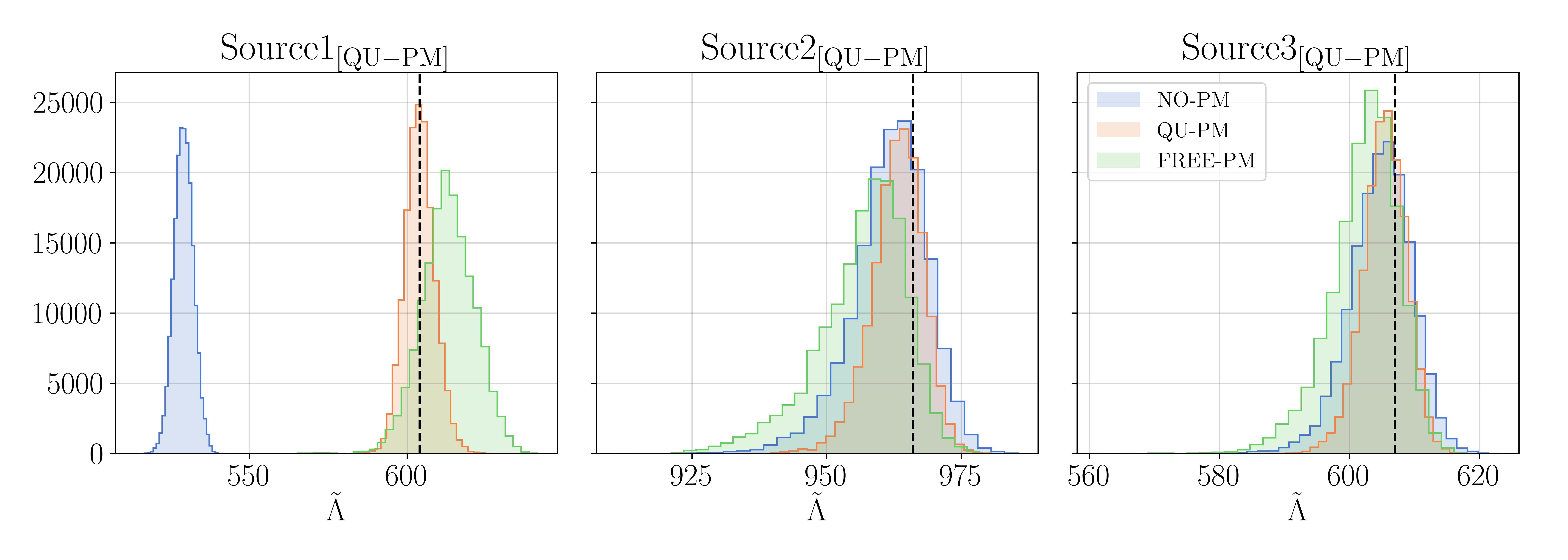} \\
	\caption{Posterior probability density for $\tilde{\Lambda}$ (see \eq{deftildeLambdadiv6} for definition) for sources  observed by the ET+CE network, recovered with three different models NO-PM (without post-merger phase), QU-PM (with post-merger phase and employing quasi-universal relation), and Free-PM (with post-merger phase and its parameters freed from that of the inspiral phase), in blue, orange, and green, respectively. 
    The black dashed lines correspond to the injected values. The parameters (distance, masses, tidal deformability $\tilde{\Lambda}$) of the three sources are the same as in table~\ref{tab:pmsnr}. Figure from \cite{Puecher:2022oiz}.}
	\label{fig:lambda_bcs}
\end{figure*}

\begin{figure}[t]
	\centering
	\includegraphics[width=0.65\columnwidth]{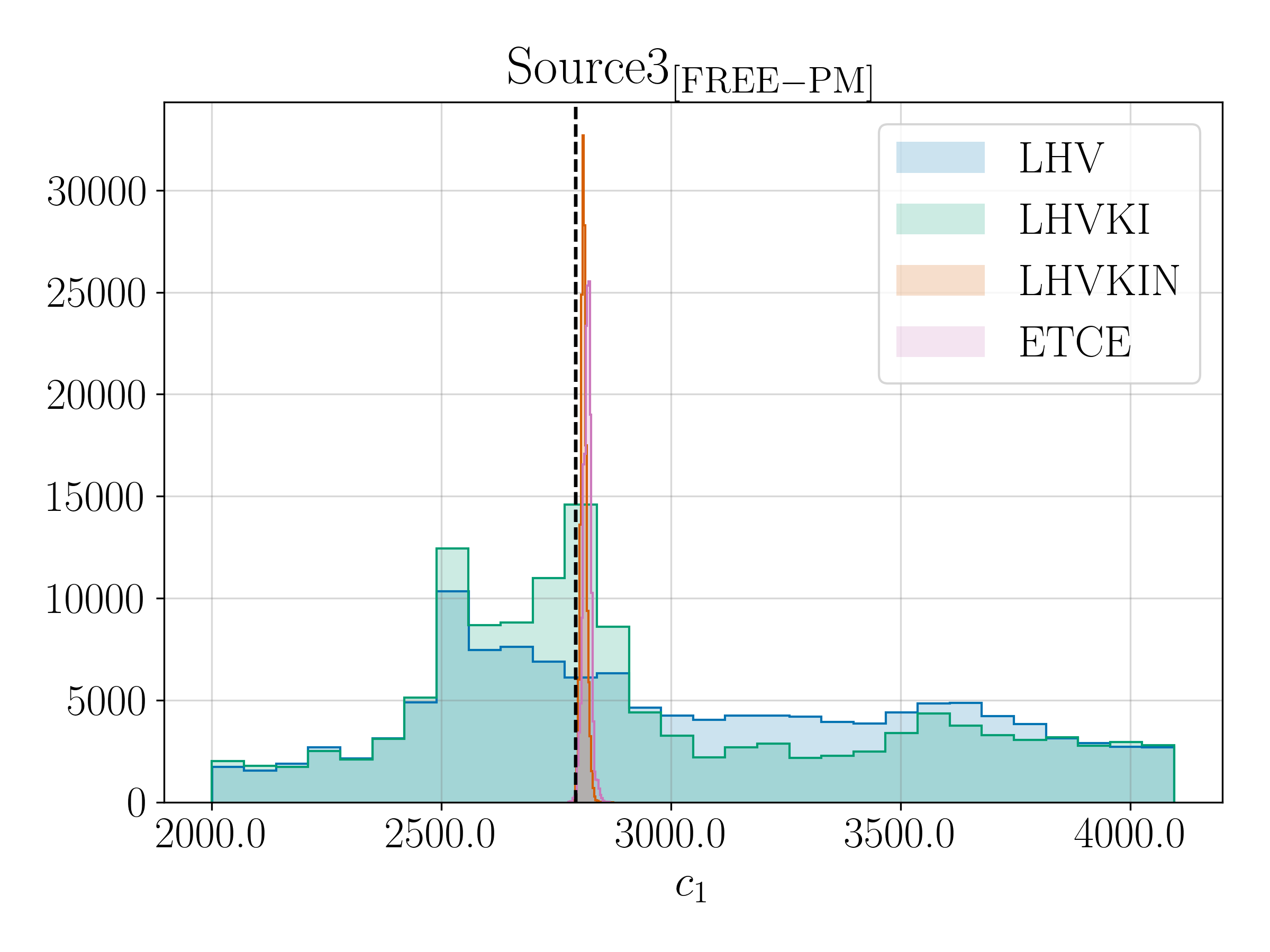}
	\caption{Posterior probability density for $c_1$, one of the post-merger phase Lorentzian parameter for different detector networks. The dashed vertical line indicates the injected value. Figure  from ~\cite{Puecher:2022oiz}.}
	\label{fig:free_det}
\end{figure}

To translate the SNR  into physically meaningful parameters, parameter inference must be conducted within a next-generation GW detector network. Figure~\ref{fig:lambda_bcs} illustrates that including the post-merger phase and linking its parameters to those of the inspiral phase using quasi-universal relations (QU-PM) results in significantly tighter posteriors for tidal deformability compared to analyses that exclude the post-merger phase (NO-PM). This demonstrates the potential of the post-merger phase to enhance studies of cold supranuclear matter.
Additionally, as shown in figure~\ref{fig:free_det}, the parameters of the post-merger phase are only measurable with a next-generation GW detector network. This highlights the critical role of the post-merger phase in studying hot supranuclear matter, contingent on the availability of a next-generation GW detector network.

A network of next generation GW observatories is essential for breaking degeneracies among waveform parameters. Adding interferometers such as NEMO \cite{Ackley:2020atn}, which are optimised for the higher-end of the high-frequency GW spectrum, would provide even stronger constraints on the EoS, thanks to a more precise characterisation of the post-merger phase.

\begin{table}[t]
\centering
	\setlength\extrarowheight{2pt}
	\renewcommand{\arraystretch}{1.3}
		\begin{tabular}{l|ll|ll|ll|}
			\cline{2-7}
			& \multicolumn{2}{c}{Source1}                           & \multicolumn{2}{c}{Source2}                           & \multicolumn{2}{c|}{Source3}                           \\ \cline{2-7} 
			\multicolumn{1}{l|}{}        & \multicolumn{1}{c|}{Total}   & \multicolumn{1}{c|}{PM}    & \multicolumn{1}{c|}{Total}   & \multicolumn{1}{c|}{PM}    & \multicolumn{1}{c|}{Total}   & \multicolumn{1}{c|}{PM}    \\ \hline
			\multicolumn{1}{|l|}{LHV}    & \multicolumn{1}{c}{100}  & \multicolumn{1}{c|}{2.0}  & \multicolumn{1}{c}{94}   & \multicolumn{1}{c|}{2.5}  & \multicolumn{1}{c}{100}  & \multicolumn{1}{c|}{2.7}  \\
			\multicolumn{1}{|l|}{ET-CE}   & \multicolumn{1}{c}{1267} & \multicolumn{1}{c|}{10.2} & \multicolumn{1}{c}{1190} & \multicolumn{1}{c|}{12.3} & \multicolumn{1}{c}{1268} & \multicolumn{1}{c|}{13.3} \\  \hline
		\end{tabular}

	\caption{The SNRs for three  source located at a distance of 68~Mpc and with chirp mass, mass ratio and tidal deformability  as follows. Source 1: $\{ {\cal M}_c=1.17524\,\Msun, q=0.8,\tilde{\Lambda}=604\}$; 
    Source 2: $\{ {\cal M}_c=1.08819\,\Msun, q=1.0,\tilde{\Lambda}=966\}$; Source 3: $\{ {\cal M}_c=1.17524\,\Msun, q=1.0,\tilde{\Lambda}=607\}$.  The table includes the SNR for the entire waveform (labeled as ``Total") and for the post-merger phase specifically (labeled as ``PM"). See ref.~\cite{Puecher:2022oiz} for details.}
	\label{tab:pmsnr}
\end{table}

\subsubsection{Inferring exotic nuclear phenomena with binary neutron star mergers}

The merger of BNS systems can result in a dense NS remnant that can resist gravitational collapse for timescales ranging from several milliseconds to a few minutes after merger \cite{Rosswog:2001fh,Shibata:2006nm}.  The post-merger evolution of these remnants results in GW emission \cite{Bernuzzi:2015opx}, making them promising candidates for understanding the behavior of matter at extreme densities. These remnants have been shown to potentially exhibit several exotic phenomena, such as phase transitions to deconfined quarks \cite{Most:2018eaw,Most:2019onn}, deconfined quarks {\it via} quark-hadron crossover \cite{Kedia:2022nns,Fujimoto:2022xhv}, 
formation of absolutely stable strange quark matter \cite{Bauswein:2015vxa,DePietri:2019khb},
the appearance of hyperons \cite{Sekiguchi:2011mc,Radice:2016rys}, and thermal effects \cite{Perego:2019adq, Fields:2023bhs}; see section~\ref{ssec:postmerger} for more detailed discussion.

Among other outcomes, these effects can alter the compactness of the remnant (compared to the cold hadronic equation of state), thereby affecting the value of the peak post-merger frequency, $f_2$. Hence, precise measurement of $f_2$ is important for inferring the nuclear processes occurring within the dense remnant. Ref.~\cite{Prakash:2023afe} finds that deviations in $f_2$ caused by phase transitions can be detected with $90\%$ confidence for a post-merger SNR of $10$. This corresponds to an inspiral SNR of approximately 600, achievable by ET alone at distances $\sim 90$ Mpc. The 20 km CE detector, when optimized towards post-merger frequencies, can accumulate the same post-merger SNR at $\sim120$ Mpc. For the crossover scenario, ref.~\cite{Harada:2023eyg} employs a Bayesian model selection to claim that a crossover to the quark phase will be detectable with ET and CE observatories for a few BNS mergers every year. Further, the deviation in $f_2$ due to thermal effects can be detected with $90\%$ credibility for a post-merger SNR of $15$, achievable for by ET for events at $50$ Mpc \cite{Fields:2023bhs}. Together with CE, such events can be detected with the same SNR up to higher distances, which will increase the rate of detection of these events.  

While phase transitions to quark matter are expected to affect post-merger signals due to their relevance at high densities, several studies suggest that signatures of strong phase transitions may also appear in the inspiral phase. Single or sequential phase transitions can lead to significant deviations in the mass-radius ($M-R$) relationship compared to their hadronic counterparts \cite{Sieniawska:2018zzj, Han:2018mtj}. EoSs with first-order phase transitions may exhibit small deviations in the tidal deformability parameter $\Lambda$ while causing variations in the radius on the order $\mathcal{O}(100)$ m \cite{Raithel:2022efm}. Detecting such changes due to phase transitions using GWs alone requires precise measurements of both radius and $\Lambda$, which can only be achieved with next-generation GW observatories. Using a non-parametric approach, ref.~\cite{Essick:2023fso} finds that a catalog of several hundred BNS mergers would be necessary to confidently detect phase transitions, a capability that would only be feasible in the next-generation GW detector era. However, 
ref.~\cite{Mondal:2023gbf} demonstrates that if the phase transition occurs at twice the nuclear saturation density, it could be confidently inferred from a single BNS event at a distance of $\mathcal{O}(100)$ Mpc with next-generation GW observatories.  

Hence, precise measurement of NS parameters in both pre- and post-merger GW signals from BNS mergers will allow the inference of exotic nuclear processes. Such detailed and numerous estimates will only be achievable with a network that includes both ET and CE observatories. 
\begin{tcolorbox}[standard jigsaw, colframe=teal, colback=black!10!white, opacityback=0.6, coltext=black,title=\sc Take home message]
{\sc 
  The synergies between ET and other Earth-based GW observatories are essential for achieving a comprehensive and robust understanding  matter at supranuclear densities.
}
\end{tcolorbox}


\subsection{Fundamental physics}
\label{sec:fundphys}


Stellar- and intermediate-mass BBH systems serve as exceptional laboratories for testing General Relativity (GR). Stellar-mass BBHs evolve over long timescales within the LISA frequency band, where their dynamics can be accurately described using the post-Newtonian (PN) approximation \cite{Blanchet:2013haa}. As they gradually inspiral out of LISA's range and transition into the frequency band of ET, their dynamics enter a highly non-linear regime that eventually necessitates numerical relativity (NR) simulations. Multi-band observations of these systems thus provide a unique avenue to probe both lower-order PN predictions and the strong-field effects captured by NR simulations, offering a comprehensive test of GR across a wide range of physical regimes.

In contrast, intermediate-mass BBH systems have shorter lifespans in both the LISA and ET bands, but their significantly larger signal strains and SNRs enable unique tests of GR. In the LISA band, these systems allow exploration of mildly non-linear dynamics during the inspiral phase, while in the ET band, their post-merger signals probe the strongly non-linear regime. Additionally, concurrent observations by ET and CE will provide an unprecedented opportunity to test the polarization properties of the waves, further advancing our understanding of gravitational phenomena in the strong-field regime. The synergetic fundamental physics exploration with ET are enumerated in the box below.

\begin{tcolorbox}[standard jigsaw, colframe=cyan1, colback=black!10!white, opacityback=0.6, coltext=black,title=\sc Synergetic Fundamental Physics]
{\sc 
\begin{itemize} 
  \item  Multi-band observations of the inspiral, merger and ringdown phases of compact binary coalescences offer a unique opportunity for precision tests of GR.
  \item  Joint observations by ET and CE can help break degeneracies.
\end{itemize}
}
\end{tcolorbox}

As we have discussed in detail in section~\ref{section:div1}, one  of the fundamental science objectives of ET is to test General Relativity~(GR).
ET alone will be able to test GR to incredible precision with the detection of black hole binaries like GW150914 with an SNR of more than 1000. With multi-band observations with space interferometers the precision to which the theory can be tested increases by a order of magnitude. Additionally, combining ET and CE data will help test the theory in new regimes such as polarization states and lensing of the observed gravitational waves as they would require accurate sky pointing and distance measurements. In the reminder of this subsection we discuss how ET can achieve greater sensitivity to test GR with data from other terrestrial and space-based GW observatories.

\subsubsection{Tests of the inspiral phase}

In GR, the phase evolution of a binary system is driven by a balance equation between the change in orbital energy and the energy flux emitted in GWs. During the inspiral, this evolution is well represented by the post-Newtonian approach~\cite{Blanchet:2013haa}, that permits to derive a perturbative series for the phase evolution written in terms of the small PN parameter $v = (G M \omega/c^{3})^{1/3}$ for a total mass $M$ and orbital frequency $\omega$, with $v^{2}$ denoting the PN order parameter. In GR, the dominant term in the phase evolution is driven by the quadrupole radiation and is  $\propto v^{-5}$; it is often called Newtonian, or 0PN, term. High and low (or negative) PN orders affect most the late or early inspiral, respectively.

In alternative theories of gravity, this structure can be modified due to either the existence of extra fields or a modification of the gravitational dynamics. One well-known example is dipolar radiation, present generically in theories featuring scalar or vector fields~\cite{Will:2018bme}, which appears at leading order at the $-1$PN order with respect to the quadrupole radiation. Because dipole radiation appears at a relative lower order of $-1$PN with respect to the quadrupole radiation, it affects most prominently the early inspiral of a binary and represents a valuable window to look for clues of new physics in the gravity sector~\cite{Shao:2016ezh}

Among proposed modifications from GR (see e.g. Table VII in~\cite{Perkins:2020tra}), time-dependent variations of the gravitational constant itself create a negative $-4\mathrm{PN}$ correction; modifications of black hole evaporation in large extra dimensions also yield a $-4\mathrm{PN}$ term; scalar-tensor theories such as Jordan-Fierz-Brans-Dicke and Einstein-Dilaton-Gauss-Bonnet theories, as well as Einstein-Aether theories will all feature dipolar radiation with a $-1\mathrm{PN}$ leading term; introducing a graviton mass yields a $1\mathrm{PN}$ term; parity-violating Dynamical Chern-Simmons theory and some models of non-commutative gravity will introduce a modification at the $2\mathrm{PN}$ order.

In this landscape, negative PN orders will affect most prominently the early inspiral, while positive PN orders will affect the late inspiral the most. This creates a strong trend across instruments: observations sensitive to the late inspiral and merger of a binary system, such as massive BH binaries (MBHBs) in LISA or stellar-mass BH binaries (SBHBs) in ET, are typically best to constrain high PN orders, while observations sensitive to the early inspiral, such as SBHBs for LISA or binary pulsar timing, are best to constrain negative PN orders. Notably, dipolar radiation might even dominate over the quadrupolar radiation at low enough frequencies. However, if multi-band GW observations are made, the high-frequency band data can also help a lot in inferring parameters that are otherwise degenerate if only low-frequency data are available.

This synergy has been explored in many recent works. Ref.~\cite{Barausse:2016eii} proposed a theory-agnostic test on black-hole dipolar radiation, combining the LISA band and the band of ground-based instruments. Ref.~\cite{Carson:2019rda} explored this multi-band synergy for parity-violating gravity, ref.~\cite{Gnocchi:2019jzp} for parity-violating gravity as well as scalar-tensor theories, ref.~\cite{Gupta:2020lxa} in the form of a theory-agnostic test at every PN order. These studies relied on the Fisher matrix approximation for parameter recovery. Ref.~\cite{Toubiana:2020vtf} used Bayesian inference for the LISA data with priors from ground-based instruments to simulate an analysis of dipolar radiation and graviton mass. The emerging view from these studies is clear: multi-band observations would significantly improve over the constraints that could be achieved with LISA or ET alone.

Beyond multi-band observations with LISA, a similarly synergy would exist with instruments in the Deci-Hz band. Ref.~\cite{Liu:2020nwz} demonstrated that by combining data from ground-based observatories such as ET and CE  with data from milli-Hertz observatory (e.g., LISA) or Deci-Hertz observatory (e.g. DECIGO)  the parameter precision on the dipolar-radiation parameter indeed improves by several orders of magnitude. Similar improvements were also found~\cite{Zhao:2021bjw} by performing joint parameter estimation using ET and atom-interferometer gravitational-wave observatories that are also investigated for operating in the Deci-Hertz band~\cite{Proceedings:2023mkp,Proceedings:2024foy}.

In cases where dipolar GW radiation happens, it is common to simultaneously have a varying gravitational constant, e.g. in the scalar-tensor theory~\cite{Damour:1992kf, Will:2018bme}. Ref.~\cite{Wang:2022yxb} showed that a synergy between ET and other deci-hertz gravitational-wave observatories will be able to simultaneously test the effects from dipolar radiation and a linear-in-time evolving gravitational constant, thus constraining the underlying gravity theory from multiple angles.

A somewhat different scenario has been proposed where multi-band observations would be valuable: in presence of ultralight boson clouds around black holes, a GW detection in the LISA band could direct a search in ET for a continuous wave generated by clouds orbiting the remnant of the merger~\cite{Ng:2020jqd}.

\subsubsection{Tests of the merger and ringdown}

Multi-band observations with LISA and ET would provide a different view for model-agnostic tests checking the consistency of the signal in its inspiral and merger-ringdown phases, by providing independent information on the inspiral and improving the overall parameter estimation~\cite{Vitale:2016rfr}. This would be particularly important in providing a view of the inspiral of massive systems only detectable at merger in the ET band~\cite{Carson:2019rda, Datta:2020vcj}. 

Multi-band observations also provide an important operational input for ringdown analyses: SBHBs observed by LISA before their merger will be among the loudest targets of ground-based instruments, and an advance warning would allow us to make sure that the ground network will be fully operational and at optimal sensitivity by the time the merger occurs. It is also possible, in principle at least, to tune ground-based interferometers to achieve an increased sensitivity in a narrow frequency band that can be chosen to maximize the precision of ringdown tests for a particular binary, achieving an improvement by a factor of 2~\cite{Tso:2018pdv}.

\subsubsection{Tests of frequency-dependent speed of GWs}

A particularly important consequence of the observation of the bright siren event GW170817 has been  the determination of the propagation speed of GWs to be the speed of light within $10^{-15}$~\cite{LIGOScientific:2017vwq,LIGOScientific:2017ync}. This bound applies to GWs in the frequency range probed by ground--based interferometers. 
However, in an effective field theory approach to
dark energy, the
GW frequencies probed by ground-based interferometers lie around the typical strong coupling scale of the EFT, leaving the possibility that an appropriate ultraviolet completion of dark energy scenarios could avoid present constraints on the GW speed, while still showing deviations at lower frequencies~\cite{deRham:2018red,Baker:2022eiz}. 
Refs.~\cite{Baker:2022eiz} and~\cite{Harry:2022zey} showed however that a single multi-band observation of a GW150914-like event by LISA and a ground--based interferometer would allow to constrain the speed of GWs to match the speed of light within $10^{-15}$ across a much wider frequency range.

\subsubsection{Tests of the GW polarization}

In modified theories of gravity, GW signals can typically feature additional modes of polarization beyond the usual transverse-traceless tensor modes $h_{+},h_{\times}$~\cite{Will:2018bme}. In the most general case, the GW can comprise up to 6 modes, two transverse tensorial modes, two vectorial modes, a breathing scalar mode and a longitudinal scalar mode~\cite{Will:2018bme, Chatziioannou:2012rf}. Although the dominant effect of modifications to GR tends to be the contribution to the orbital phasing, and thereby to the phase of tensorial modes, the GW  itself will also be modified by the addition of these extra modes of polarization, for instance due to the additional energy loss in dipolar radiation. Since this is in essence an amplitude correction rather than a phase correction, constraints on the parameters of specific modified gravity theories are typically weak. However, checking the tensorial nature of GWs can also be seen as a theory-independent verification of the consistency of GR with the data, and different polarization hypotheses can be compared using null-stream methods~\cite{Wong:2021cmp, LIGOScientific:2021sio}.

Because each detector (or TDI channel in the LISA case) is only sensitive at any given time to a projection of the gravitational wave, constraining additional modes of polarization requires at least as many detectors as targeted modes~\cite{Takeda:2018uai}, although, for long-lived signals, the detector motion can also help in breaking degeneracies by changing the detector's pattern function over time~\cite{Takeda:2019gwk}. Theory-independent null-stream methods benefit from combining multiple detectors as well~\cite{Pang:2020pfz, Hu:2023soi}. Overall, this test presents a strong synergy between detectors, where combining detectors in a network, and in particular combining ET with one or two CE observatories, would directly improve our ability to disentangle additional polarizations~\cite{Takeda:2019gwk}.

\begin{tcolorbox}[standard jigsaw, colframe=teal, colback=black!10!white, opacityback=0.6, coltext=black,title=\sc Take home message]
{\sc 
\begin{itemize}
\item Synergetic observation of stellar and intermediate BBH systems by ET and LISA can bound GR by an order of magnitude better than either observatories.
\item Joint observation by ET and CE is a unique opportunity to measure polarization modes of the waves and constrain alternative gravity theories.
 \item Dipolar-radiation test is boosted by combining ET with milli-Hertz and Deci-Hertz GW observatories.
\end{itemize}

}
\end{tcolorbox} 

\subsection{Hubble tension and cosmography}
\label{sec:cosmo}



\begin{tcolorbox}[standard jigsaw, colframe=cyan1, colback=black!10!white, opacityback=0.6, coltext=black,title=\sc Cosmography and Cosmology: Synergies with CE and LISA]
{\sc 
  \begin{itemize}
           
            \item What are the key improvements that a 3G network of Earth-based GW observatories can bring to cosmology?
            \item Can the synergy among GW experiments be also crucial for other observatories, particularly in the context of multi-messenger astronomy?
        \end{itemize}
}
\end{tcolorbox}

The determination of the expansion history of the Universe largely benefits from the combination of different datasets, as in the current ``concordance'' scenario. The use of GW sirens provides a new player in this landscape, which will be further bolstered by the synergy among future GW observatories. 

The discrepancy between local~\cite{Riess:2019cxk, Freedman:2019jwv, Wong:2019kwg,Freedman:2020dne,Riess:2021jrx} and CMB-based~\cite{Planck:2018vyg} measurements of the Hubble constant $H_0$ either points to unresolved systematics or new physics. As such, the field of cosmology greatly benefits from new classes of cosmological sources for which it is possible to measure the Universe's expansion. The presence of a network of next-generation GW observatories will give us access to a plethora of sources at cosmological scales, thus allowing us to corroborate the origin of the $H_0$ tension. In the case of new physics, the possibility of determining $H_0$ to sub-percent level with tracers complementary to EM and CMB observations would be of paramount importance.

Furthermore, the increased detection horizon of 3G observatories, as shown in Figure~\ref{fig:det_horizons}, will give us access to thousands of GW events from the dark matter-dominated era (beyond redshift 0.3). Contrary to current generation GW observatories at their designed sensitivity, 3G generation observatories will allow us to constrain cosmological parameters beyond $H_0$ such as the dark matter fraction $\Omega_m$.
GWs can further allow us to probe the presence of dark energy, through the imprint of its EOS  on rate of expansion of the Universe, and more importantly the phenomenon of ``modified GW propagation'' taking place whenever GR is modified at cosmological scales, as discussed in section~\ref{sect:ModGWpropdiv2}.
The coordination of ET with a network of next-generation GW observatories will significantly boost current techniques for GW cosmology and also allow for studies of GW sources as large-scale structure tracers that are currently impossible. In general, we can identify several common benefits for GW cosmology: \textit{(i)} the improved distance reach allowing to target high--redshift sources; \textit{(ii)} the improved localization capabilities of a network of GW observatories; \textit{(iii)} the overall boost in the number of detections per observation time and \textit{(iv)} the possibility of multi--band observations and constraints.
We discuss how these common benefits will impact future prospects for GW cosmology.

\begin{figure}
    \begin{tabular}
    {c@{\hskip -.1mm}c@{\hskip -4mm}c}
  \includegraphics[width=70mm]{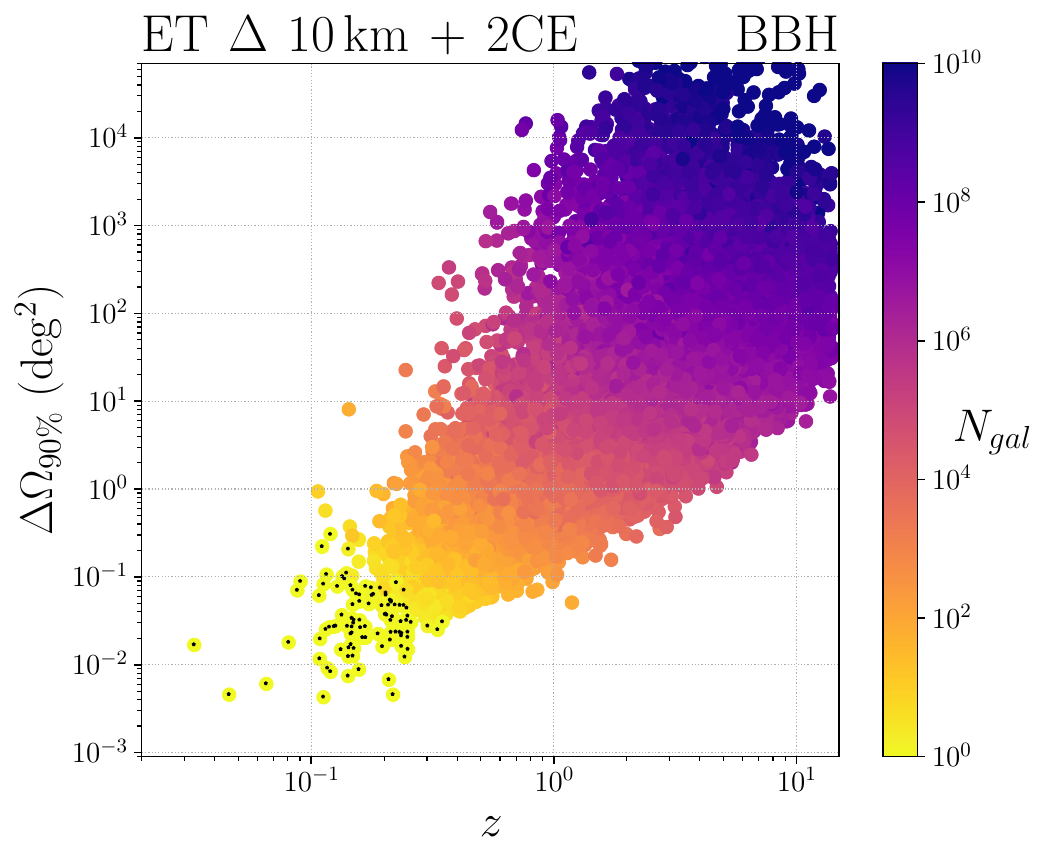} &   \includegraphics[width=70mm]{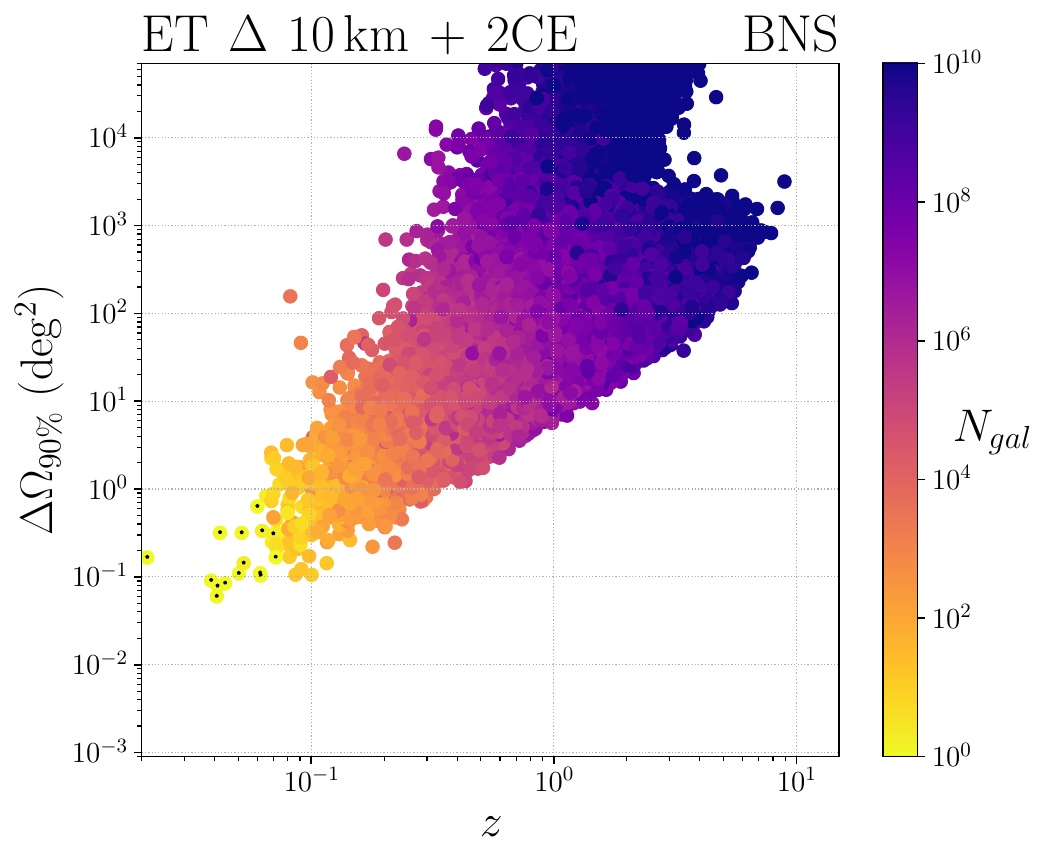} \\[-.5cm]
  \includegraphics[width=70mm]{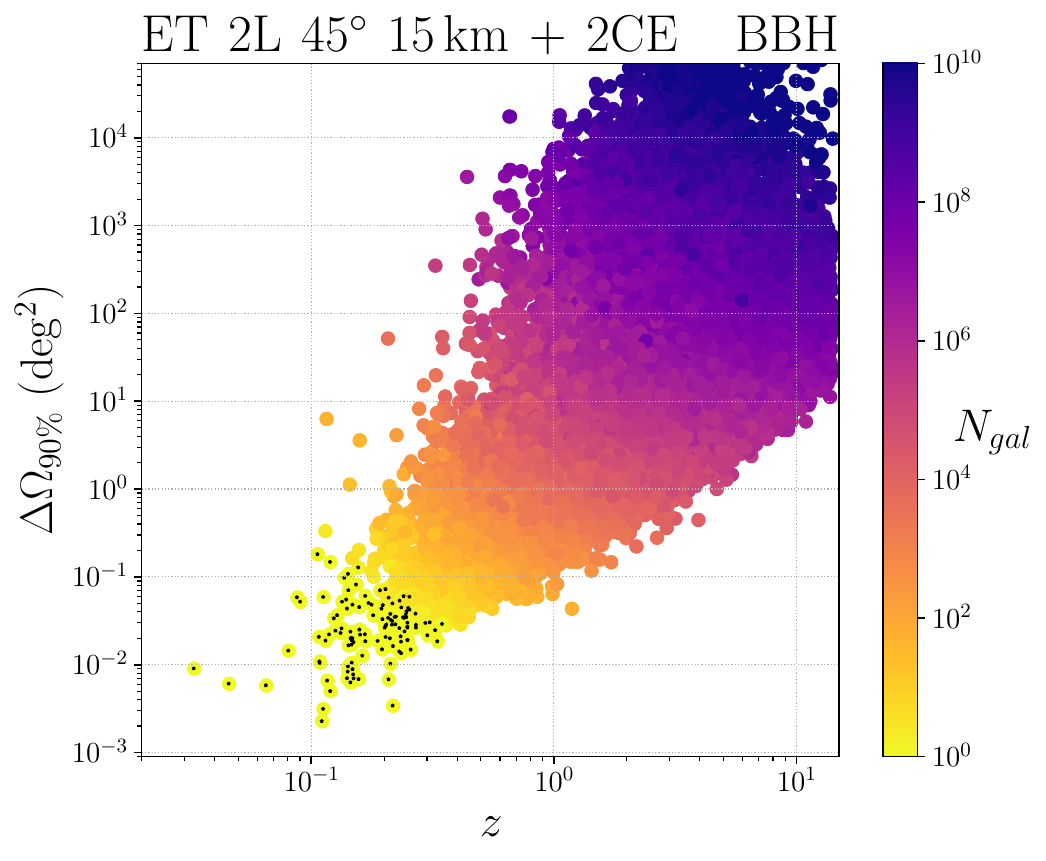} &   \includegraphics[width=70mm]{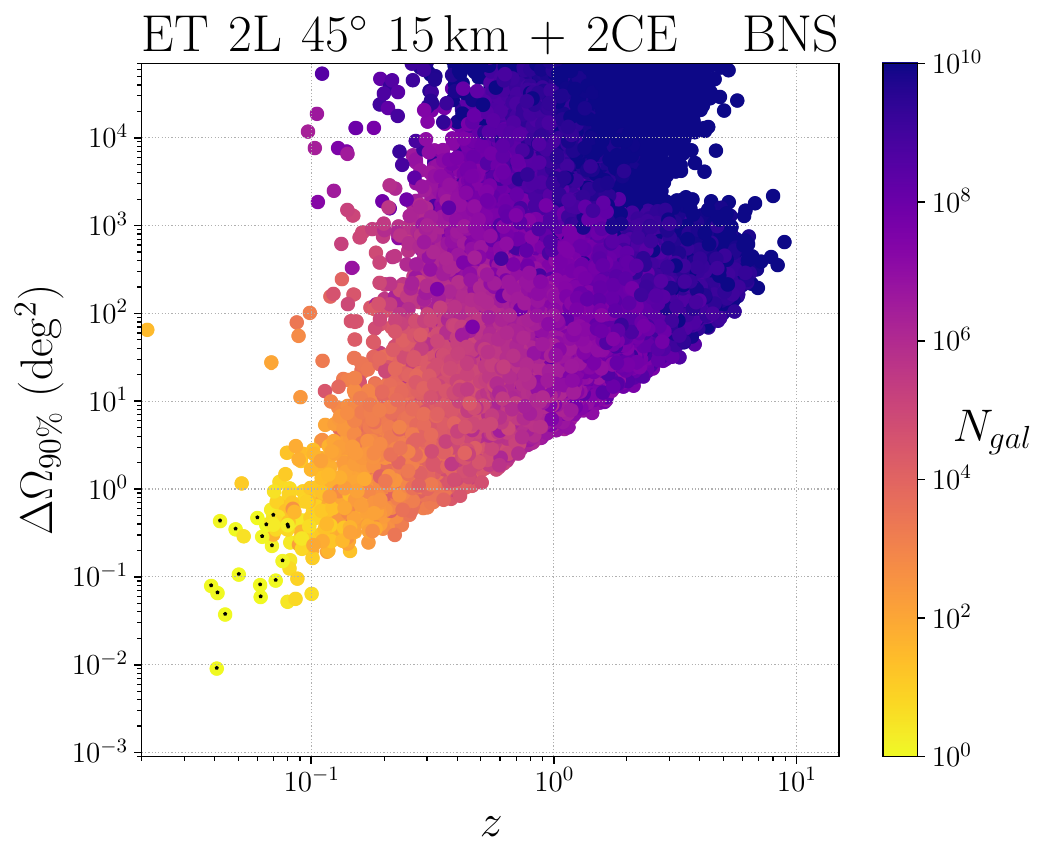} 
\end{tabular}
    \caption{Localization capabilities for BBH (left panels) and BNS (right panels) of ET in its triangular and 2L configurations (first and second rows respectively), in a network with two CE. The color scale denotes the number of galaxies expected in the $90\%$ localization volume. Events marked with black dots are localized to one galaxy only.  }
    \label{fig:ET2CE_loc}
\end{figure}

\subsubsection{Bright sirens cosmology}

Localization is of crucial importance for GW cosmology. A precise localization of GW events can allow us to promptly identify a possible EM counterpart but also statistically identify the source galaxy from galaxy surveys. 

Figure~\ref{fig:ET2CE_loc} 
shows the localization capabilities of ET in synergy with two CE observatories, for BBH and BNS respectively.\footnote{We refer to ref.~\cite{Branchesi:2023mws} for the details on the catalogs and software used.} In particular, we show the $90\%$ sky localization area as a function of redshift, while the color refers to the number of galaxies contained in the localization volume, assuming a constant galaxy density of $10^{-2}\, \rm galaxies/Mpc^{3}$, where the galaxy density is derived integrating the Schechter function of ref.~\cite{Conselice:2016zid} from $10^9$ to $10^{12}$ solar masses. Events with a single host galaxy in their localization volume are further marked with a black dot. Specifically, ET triangle (2L) operating with 2CE is able to localize within a volume containing a single galaxy (for galaxy in the above mass range), 47 (55) BBH mergers per year at redshifts below 0.4 and 17 (16) BNS mergers per year at redshifts below 0.1.\footnote{We underline again that these numbers are subject to uncertainties dominated by the limited knowledge of the local merger rate of compact objects. For BNS mergers, taking into account the absence, to date, of detection in the current O4 run, we can consider these numbers as an upper limit, see also footnote~\ref{foot:BNSrate_div2} on page~\pageref{foot:BNSrate_div2}.}

In particular, BNS systems are targets for follow-up multimessenger campaigns, where sky localization plays a major role. 
The synergy with CE brings a dramatic improvement in localization:
ref.~\cite{Branchesi:2023mws} finds that the number of sources localized to better than 10  $\rm deg^{2}$ increases from 8  to $\sim 6500$ for ET in the triangular configuration, and from 25  to $\sim 9800$  for ET in the 2L configuration;
for localizations better than 100 $\rm deg^{2}$  the number of sources increases from 184 to $\sim 10^{5}$
for ET in the triangular configuration, and from 559 to 
$\sim 1.5\times 10^5$ for ET in the 2L configuration. 
As shown in figure~\ref{fig:ET2CE_loc} there is subset of $\mathcal{O}(10)$ well-localized GW events for which it will be possible to identify the host galaxy solely from the GW localization volume, without the need of any transient EM counterpart. This is not currently possible with current-generation GW detectors. 

For the impact on the measurement of $H_0$, it would already be possible to obtain sub--percent accuracy with one year of observations of ET alone (see section~6.4 
of \cite{Branchesi:2023mws}), if a precise redshift measurement following the localization of the source through the kilonova emission produced by BNS mergers in the optical band can be obtained. In particular, this can be possible in synergy with the Vera Rubin Observatory~\cite{Branchesi:2023mws} for sources up to redshift $z\sim0.3-0.4$.
The addition of two CE detectors would boost this accuracy to an unprecedented level, as well as  decreasing  the time required to reach such a goal (even to just a few days of observations). 
Even more interestingly, we note that besides $H_0$, other parameters affecting the distance--redshift relation might not be constrained to better than a few percent with ET alone. In particular, ref.~\cite{Branchesi:2023mws} finds that the matter density $\Omega_{\rm M}$ would be constrained at $20-30 \%$,  the dark energy EOS parameter $w_0$ at $\geq 10 \%$ and the parameter $\Xi_0$ describing modified GW propagation~\cite{Belgacem:2017ihm,Belgacem:2018lbp} at  $\sim 3 \%$. Crucially, the addition of CE would allow to reach percent or sub--percent level accuracy for all these parameters, allowing to reconstruct the local expansion history in full generality (as well as to provide a stringent test of $\Lambda$CDM) rather than just the local value of the Hubble constant~\cite{Cozzumbo:2024vxw}.

\subsubsection{Dark sirens cosmology}

3G interferometers are expected to detect most of the stellar BBHs merging up to $z \sim 100$ \cite{Iacovelli:2022bbs,Branchesi:2023mws}. Albeit no associated EM counterpart is emitted, the abundance of such dark sirens allows for statistical methods, discussed in section~\ref{sect:darksirensdiv2}, to become efficient. Within this framework, the localisation power plays a crucial role in the precision that this approach can reach, as galaxy catalogues are employed to counterbalance the lack of the source's redshift \cite{Gupta:2022fwd,Muttoni:2023prw,Zhu:2023jti,Muttoni:2021veo,Borghi:2023opd,Borhanian:2020vyr,Finke:2021aom, Gray:2021sew, Mastrogiovanni:2023emh, Gray:2023wgj, Gair:2022zsa}. 
While ET alone might not be able to pin down the sky position of a large number GW source accurately enough, 
a network of ET and CE can restrict the potential hosts to a few candidates for a large number of events. A sub-percent $\hubble$ measurement can confidently be attained by employing only the loudest dark sirens observed in 1~yr  of observations~\cite{Muttoni:2023prw}.
Furthermore, we expect to localise a few tens of GW events per year accurately enough that only one host galaxy falls inside the credible region (left panels of figure~\ref{fig:ET2CE_loc}). These unique events, which contain most of the information, would constrain $\hubble$ down to $\mathcal{O}(1\%)$ precision alone \cite{Borhanian:2020vyr, Muttoni:2023prw} if provided with complete galaxy catalogues. 

Dark sirens detected by 3G detectors will play a crucial role in cosmology even in the absence of any galaxy survey information. As discussed in section~\ref{sect:darksirensdiv2}, by exploiting the relation $m_{\rm det}=(1+z)m$ between detector-frame mass $m_{\rm det}$ and the actual (``source-frame") mass $m$, it is possible to obtain a statistical redshift estimation of the source \cite{Taylor:2011fs,Taylor:2012db} and jointly fit for the source mass spectrum and cosmological expansion parameters \cite{Farr:2019twy, Ezquiaga:2022zkx,Mastrogiovanni:2021wsd,Borghi:2023opd, Mastrogiovanni:2023emh,Gray:2023wgj}. Dark sirens that are employed for this type of analysis have been recently named ``spectral sirens'' \cite{Ezquiaga:2022zkx}. To reach a percent-level estimate of $H_0$, we would require at least thousands of spectral sirens \cite{Farr:2019twy, Mastrogiovanni:2021wsd, Borghi:2023opd} that could be provided at the end of the planned observing runs in 2027 for current GW detectors. While current GW detectors could already provide in the next years a percent-level measure of $H_0$ with spectral sirens, they will not provide a high enough number of sources to consistently check for systematics in the $H_0$ measure that could arise from  poor calibration of mass distribution models \cite{Pierra:2023deu}.
In contrast, 3G observatories will make accurate and precise cosmology with spectral sirens possible;  3G observatories will provide $\mathcal{O}(10^5)$ GW sources at all the cosmic times in just one year of observation, as we see also from   figure~\ref{fig:det_horizons}. The large number of GW detections will allow us to perform spectral siren analyses with independent subsets of thousands of GW sources. The use of an independent subset of GW observations, divided accordingly to their mass and distance from us, will provide an in-depth understanding on the most critical assumptions that could possibly introduce systematics in the estimate of $H_0$. 

Another possibility is the use of GWs as tracers of the Large Scale Structures, and the extraction of cosmological information from their two--point cross-correlation with galaxy surveys~\cite{Namikawa:2015prh,Oguri:2016dgk,Nair:2018ign,Libanore:2020fim,Bera:2020jhx,You:2020wju,Mukherjee:2020hyn,Kumar:2022wvh,Scelfo:2022lsx,Bosi:2023amu,Pedrotti:2025tfg}, 
see the discussion in section~\ref{sec:GWLSSCC}. The main limitation of this method comes from the poor angular resolution of GW sources as compared to galaxies. In particular, the minimum resolution of the GW survey determines a limiting scale in the angular power spectrum~\cite{Oguri:2016dgk,Libanore:2020fim,Scelfo:2022lsx,Bosi:2023amu}, $\ell_{\rm max}\sim 180^{\circ}/\Delta \Omega_{\rm max}^{1/2}$ with $\Delta \Omega_{\rm max}$ being the maximum resolution of the instrument.  
While the angular resolution of ET alone corresponds to roughly $\Delta \Omega_{\rm max}\sim 0.1\, \rm deg^2$, hence a maximum multipole of at most $\sim 300$ only in a local redshift shell (up to $z\sim0.2$), from figure~\ref{fig:ET2CE_loc} we see that  the addition of CE could allow to reach $\Delta \Omega_{\rm max}\lesssim 0.01\, \rm deg^2$, corresponding to a maximum multipole of $\geq 1800$ at low redshift, and a maximum multipole of $\geq 300$ up to redshift $z\sim1$. In practice, going to high multipoles would require to model non-linear scales; however, refs.~\cite{Oguri:2016dgk,Pedrotti:2025tfg} find that sub-percent accuracy on $H_0$ could already be achieved with a cut $\ell_{\rm max}=300$ which is reachable with ET in combination with CE.
 
Finally another dark sirens method which can be applied to 3G observatories, consists in the simultaneous estimation of cosmological parameters and the EOS of NSs, as discussed in section~\ref{sect:Lovesirensdiv2}.
A network of 3G observatories will have the potential to determine the EOS of NSs with great accuracy, in turns providing an associated redshift measurements for all compact binary mergers including a NS.
Such a method does not require an EM counterpart and could deliver constraints on $H_0$ comparable to other siren methodologies \cite{Messenger:2011gi,Messenger:2013fya,DelPozzo:2015bna,Chatterjee:2021xrm}.

\subsubsection{Multi-band sources}

The presence of multiple GW observatories covering complementary frequency ranges will open further unique possibilities.
ET detection horizon is complementary, in terms of source masses, to LISA, as well as   to the Deci-Hz detectors discussed in section~\ref{sec:SOs}.  This complementarity is crucial for GW cosmology for two reasons. These low-frequency detectors could provide early-warning alerts to prepare ET for the detection of the merger of BNSs and SOBBHs~\cite{Muttoni:2021veo}. They will scan the Universe for a population of BBHs with masses $\gtrsim 10^3 M_\odot$, on which we currently lack knowledge. If this population of GW sources exists, we will be able to perform GW cosmology with an independent set of data that can be used to corroborate possible systematics present in ET \cite{Yang:2021qge}.

In addition, there are classes of compact objects for which detections of the same source in different experiments will be possible.
In particular, BBHs originating from massive stars, which exceed the pair-instability mass gap and with masses in the range $120 M_{\odot}-300 M_{\odot}$, are considered prime dark siren candidates for GW observations across multiple frequency bands~\cite{Muttoni:2021veo,Ezquiaga:2020tns,Zhu:2021bpp}. The prolonged inspiral signals emitted by these systems can be observed by LISA, or even better by Deci-Hz detectors, while the later stages of inspiral and merger will be detected by ET. The combination of the observations from LISA and ET will improve the sky localization of these sources by a factor $\mathcal{O}(10-100)$ compared to those obtained with the same sources using only ET.
Ref.~\cite{Muttoni:2021veo} finds that with 4 years of LISA observation this multiband approach may allow for a $\mathcal{O}(2\%)$ estimate of the Hubble constant $H_0$ with $\mathcal{O}(10)$ detected sources.
An even better precision is expected from Deci-Hz detectors, which can follow the inspiral up to minutes before merger \cite{Seymour:2022teq,Dong:2024bvw}.

\begin{tcolorbox}[standard jigsaw, colframe=teal, colback=black!10!white, opacityback=0.6, coltext=black,title=\sc Take home message]
{\sc 
 The synergy among different Earth--based GW observatories can be a game-changer for late-time cosmology with resolved sources, due to the improved localization capabilities which are crucial for both bright and dark siren cosmology. 
 %
        %
The interplay with GW observatories in other frequency bands will further allow to enlarge the spectrum of GW sources used for cosmology and to target well-localized multi--band sources.
}
\end{tcolorbox} 

\subsection{The origin of supermassive black holes}
\label{sec:BHseeds}



\begin{tcolorbox}[standard jigsaw, colframe=cyan1, colback=black!10!white, opacityback=0.6, coltext=black,title=\sc The Link Between Stellar and Supermassive Black Holes]
{\sc 
  \begin{itemize}
              \item What can joint observations by ET and LISA reveal about the nature of the seed black holes that eventually grow supermassive? Are stellar-mass black holes, formed at cosmic dawn, the sole building blocks of the supermassive black holes powering quasars observed as early as redshift $z\sim 11$, and now residing almost ubiquitously in the centers of quiescent galaxies?
              \item What can ET and LISA tell us about  intermediate-mass black holes in the range $10^2-10^5\,\Msun$, which bridges the gap  between the stellar and the supermassive black holes?         
        \end{itemize}
      
}
\end{tcolorbox}

The discovery of  rare quasars at redshifts as high as $z\sim 11$ \cite{Harikane:2022rqt,Fan:2022fhc, Maiolino:2023bpi,Bogdan:2023ilu, Goulding:2023gqa, CEERSTeam:2023qgy,
Kovacs:2024zfh,2024ApJ...966..176Y,Greene:2024phl}, of the large population of active galactic nuclei (AGN) emerging at $z\lesssim  6$ and extending to $z\sim 1$ \cite{Shen:2010aa}, and of dark massive objects in the nuclei of today’s quiescent galaxies \cite{2013ARA&A..51..511K} provides evidence of the ubiquity of supermassive black holes (SMBHs) in the near and far Universe.  Understanding the origin of SMBHs is important and goes hand in hand with understanding the formation and evolution of stars and galaxies, which is a process rooted in our current model of cosmic structure formation \cite{Volonteri:2021sfo}.

The most massive BH binary  currently detected by GW observations, GW231123, resulted in a final BH with a mass of about  $225\, \Msun$
\cite{LIGOScientific:2025rsn}, providing  evidence of hierarchical assembly of heavy BHs from
stellar-mass parents.
ET on Earth and LISA from space will probe an immense volume of the Universe as both observatories have sensitivities to detect GW signals from merging binaries out to $z\sim 15-20$. This capability  enables the measurement of  black hole masses across the largely unexplored segment of the spectrum, ranging from about $10\,\Msun$ up to $10^7\,\Msun$, and over cosmic time, offering crucial insights into when, where and how SMBHs form from their  seeds (see \cite{2017PASA...34...31V, Inayoshi:2019fun,Volonteri:2021sfo}, for detailed reviews). 
ET will have the capability of discovering the earliest black holes, relics of the first population III stars. Inside gas-rich regions of the host halos, some of these black holes  can rapidly grow via accretion and mergers  becoming viable seeds for SMBHs ~\cite{Dubois:2011aa, Lupi:2014vza, pezzulli2016, Trinca:2022txs}. This establishes a close connection between stellar-mass and supermassive black holes.   While ET can reveal the existence of the first seeds, LISA will trace their subsequent growth, uncovering the evolving population of accreting light seeds over cosmic time ~\cite{Valiante:2020zhj}.

Independent,  synergetic observations of the two different populations, the stellar and massive ones, will help breaking degeneracies among different evolution pathways. This is achieved by matching, in a multi-dimensional space and within a defined theoretical framework, the information resulting from  the measurements of the black hole  masses, spins (when possible) and redshift of individual sources, and by confronting the relative rates of coalescence events detected by ET and LISA, respectively. 

\subsubsection {Seed black holes}

In section~\ref{div3:pop3} we presented a detailed overview  of the latest advances in understanding
the properties of the first stars, known  as population  III stars, forming in the earliest collapsing dark matter halos of $10^{5-6}\,\Msun$ at $z\lesssim 30$.  Their initial mass spectrum is unknown and theoretical investigations indicate that it is in the range from 1 M$_\odot$  to several $10^2\,\Msun$. This population is expected to dominate at $z >13$ and to persist down to $z\sim 6$, albeit in subdominant fractions compared to the growing population II stars. 
Despite the increasing number of hints, recently provided by the JWST telescope, on possible indications of population III star's signatures in the spectra of high redshift galaxies or in lensed sources \cite{Maiolino:2023wwm, Schauer:2022ucs, 2023A&A...678A.173V, 2024ApJ...967L..42W}, direct EM  observations  
of single stars are beyond foreseen capabilities.

\begin{figure}[t]
\centering
\includegraphics[width=0.88\columnwidth]{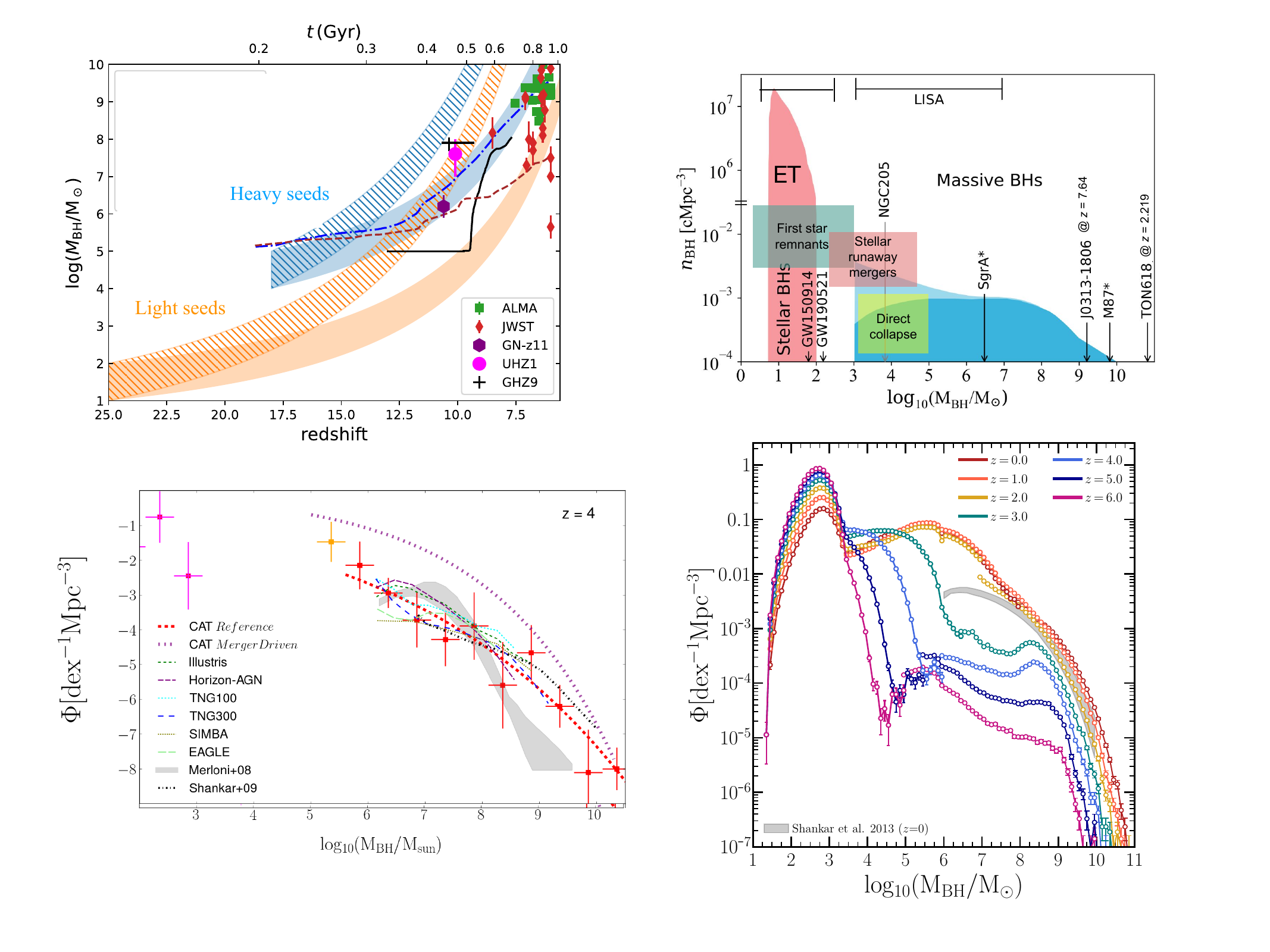}
\caption{\small {{\bf Upper left panel}: Tracks of BH growth, for different initial {\it seed} masses assuming uninterrupted accretion at the Eddington rate and at rates twice above this limit.
The different colored symbols labeled in the figure refer to $z>6$ observations of quasars by ALMA (green squares \cite{Neeleman:2019knu}) and JWST (red diamonds ~\cite{2024ApJ...966..176Y, 2024ApJ...964...90S, 2023ApJ...959...39H, Maiolino:2023bpi, 2023Natur.621...51D}), including the $z\sim 10$ massive black holes in UHZ$1$ (magenta circle \cite{Bogdan:2023ilu}), GHZ$9$ (black cross \cite{Kovacs:2024zfh}) and the candidate super-Eddington accreting black hole in GN$-$z$11$ (violet hexagon \cite{Maiolino:2023zdu}).
Light seeds can explain these quasars if they  
accrete at a super-Eddington rate. Heavy seeds, forming at lower redshifts, do not necessarily need sustained super-Eddington accretion.  There is a debate whether sustained accretion above the Eddington limit is likely in high-redshift galaxies~\cite{Pezzulli:2017ikf, 2019MNRAS.486.3892R, Trinca:2022txs, Massonneau:2022uwg, Lupi:2023oji, Greene:2024phl}.
The black, blue and red tracks refer to state-of-the-art high-resolution cosmological zoom-in simulations~\cite{Lupi:2023oji, Lupi:2019jgo} where a $10^5\Msun$ black hole was implanted at $z=10$ and let evolve including accretion and feedback.
{\bf Upper right panel}: Cartoon sketching the mass spectrum of black holes and the different  formation paths. 
Arrows and labels indicate selected quasars, the Milky Way black hole, Sgr A$^*,$~\cite{Genzel:2010zy, EventHorizonTelescope:2022wkp} M$87$~\cite{EventHorizonTelescope:2019pgp}, and the post-merger mass of GW$150914$ and GW$190521$, the first and the most massive coalescence events detected to date by LVK, respectively~\cite{LIGOScientific:2016wkq, LIGOScientific:2020iuh}. Shown are the windows of exploration of ET and LISA. 
{\bf Lower left panel}: BH mass spectrum at $z=4$ from \cite{Trinca:2022txs} as inferred from a suite of semi-analytical models and numerical simulations to show the still large theoretical uncertainties in predicting BH evolution. Data points show the distribution of light (magenta) and heavy (red, orange) BH seeds, with the best fit for the distribution of heavy seeds in red and violet dotted lines, and the results from large-scale cosmological simulations: Illustris (dark green, cyan, and blue \cite{Sijacki:2014yfa, weinberger2017}), SIMBA (olive \cite{Dave:2019yyq}), and EAGLE (light green \cite{McAlpine:2018dua}). The observational constraints by \cite{Merloni:2008hx} and \cite{Shankar:2007zg} are shown in the grey shaded area and black dashed-dotted lines.
{\bf Lower right panel}: evolution of the BH mass function from the semi-analytical model L-Galaxies \cite{Izquierdo-Villalba:2023ypb}. The figure shows how the intermediate-mass range is being progressively  filled due to the growth of light seeds from population III relics (Courtesy of Izquierdo Villalba). } }
\label{fig:BH mass spectrum}
\end{figure}

The properties of population III stars are encoded in those of  their relics, i.e. the black holes that ET can uncover, see section~\ref{div3:pop3}. The role of ET is therefore central
in inferring the properties of these first stars, but not only. 
These early black holes are expected to represent the channel of formation of the  light seeds of up to a few to several $10^2\Msun$ \cite{Volonteri:2021sfo}. 
These may not be the only seeds to grow SMBHs. 
Heavy seeds of $10^{3-5}\,\Msun$ have been also hypothesized, resulting either form direct collapse of massive gas clouds, via runaway stellar mergers or hierarchical black hole mergers \cite{Lupi:2014vza,2021NatAs...5..749G}. LISA can observe these heavy seeds, whose relative contribution in explaining  the formation of high-$z$ quasars ($z>4$) is under  debate~\cite{2017PASA...34...31V, Inayoshi:2019fun, Regan:2024wsu}.

The upper left panel of figure~\ref{fig:BH mass spectrum} illustrates the concept of black hole growth from light seeds and  heavy seeds in selected mass ranges.  In the same figure (upper right panel) we show the extension of the BH mass spectrum, known as of today, indicating the main formation pathways.  Note the presence of a desert around $10^{2-3}\,\Msun$, and the uncertainty in the rise or decline of mass distribution  around $10^{3-5}\Msun$, where intermediate mass black holes (IMBHs) sit \cite{Mezcua:2017npy,Greene:2019vlv}.
IMBHs  are in the midst of the high-end tail of the stellar BH mass distribution and the low-end tail of the SMBH mass distribution. At present, whether a discontinuity exists in the families of stellar and supermassive BHs, related to the existence of uncorrelated physical processes, is not known. In this regard, the lower right panel of figure~\ref{fig:BH mass spectrum} shows the evolution of the BH mass distribution as a function  of redshift from the semi-analytical cosmological model L-Galaxies \cite{Izquierdo-Villalba:2023ypb} that describes the formation and evolution of SMBHs  from light and heavy seeds. This model indicates that the IMBH segment is being progressively filled by growing light seeds which evolve to account for the vast population of AGN, with heavy seeds  partially explaining the rare high-$z$ quasars.

There are large theoretical uncertainties related to the way the seeding process and seed growth is treated in inferring the BH mass spectrum as a function of redshift. This is illustrated in the left lower panel of figure~\ref{fig:BH mass spectrum}, taken   from \cite{Trinca:2022txs}.
Different models give different predictions on the shape and normalization of the mass function \cite{2016MNRAS.457.3356V, Ricarte:2017ihq, Sassano:2021maj, Trinca:2022txs, Spinoso:2022ahp, Schneider:2023xxr}.  Therefore, future GW  observations will be key to help reducing these uncertainties, jointly with forthcoming deeper studies and observations of quasars and AGN in the EM sector.

\subsubsection{Synergies between ET and LISA}

\begin{figure}
\centering
\includegraphics[width=0.7\columnwidth]{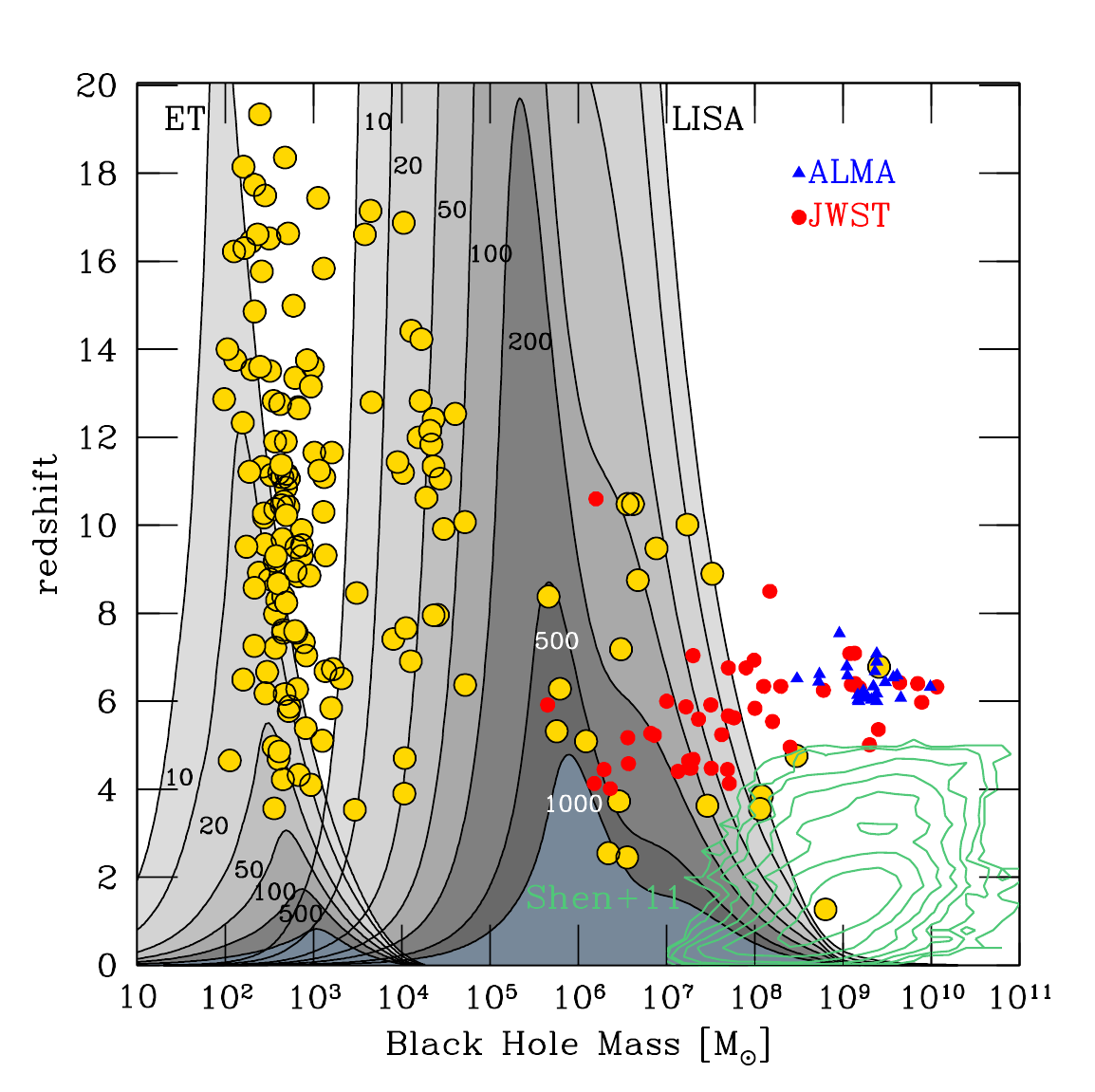}
\caption{
The ET and LISA cosmic horizons. The figure shows contour lines of constant signal-to-noise ratio (SNR) in the plane redshift 
versus total mass of the black hole binaries, as measured in the source frame. Binaries have all mass ratio 0.5. Yellow dots denote loci of BH mergers extracted from the semi-analytical model of \cite{Valiante:2020zhj}, which tracks the evolution of light and  heavy seeds forming in merging halos. Light seeds form at $z\sim 20-30$ as relics of population III stars in molecular cooling dark matter halos of $10^6\,\Msun$.   Heavy seeds 
form later ($z\sim 10-20$) in atomic cooling halos of $10^8\,\Msun$ under contrived conditions (see the review by \cite{Inayoshi:2019fun} and references therein). The bulk of the merging black holes in LISA come from the growing light seed population that ET can detect when seeds form in binaries. Red dots and blue triangles  denote the quasars observed at $4 < z < 11$, 
from JWST \cite{2024ApJ...966..176Y, 2024ApJ...964...90S, 2023ApJ...959...39H, Maiolino:2023bpi, 2023Natur.621...51D} and ALMA\cite{Neeleman:2019knu}, respectively. Green contour lines depicting the vast population of AGN are from \cite{Shen:2010aa}.
}
\label{fig:ET+LISA}
\end{figure}

Participating in the formation, growth, and assembly of cosmic structures, massive black hole in merging galaxies form binaries that coalesce, becoming sources of low-frequency GWs \cite{Colpi:2024xhw,LISA:2022yao}. LISA, operating from below  $0.1$ mHz to about $1$ Hz, aims at detecting  the signals from these sources in the  segment of the mass spectrum between about $10^4\,\Msun $ and $10^7\,\Msun$,  out to  redshift $z\sim 10-15$. The LISA's cosmic horizon together with ET's horizon are shown in figure~\ref{fig:ET+LISA}.

\subparagraph{Signals.} The GW signals detectable by LISA encode key information on the black hole source-frame masses, luminosity distance, and when sufficiently loud the spins \cite{Colpi:2024xhw}.  GW signals from binaries of $\sim 10^{4}-10^5\,\Msun$ feature an inspiral phase lasting several months before merger, with median signal-to-noise ratios (SNRs) typically reaching up to $\lesssim 100.$  The source-frame masses in this case can be determined with uncertainties from 10\% to 100\%. ET will measure with comparable accuracies the physical parameters of potential light seeds, whose signal display only few inspiral-cycles before final plunge  \cite{Fairhurst:2023beb}. 
Below $z\sim 8$, the inspiral, merger and ringdown signal of LISA's binaries contains all information to enable accurate measurements of the  source-frame masses  at the level of 10\% and effective spin with accuracy of 1\%. At $z\sim 2,$   SNRs are as large as a few thousands,  with black hole masses and  luminosity distances measured with accuracy at the level of 0.1\%, and the spin of the primary black hole, which provides clues on the way accretion proceeded \cite{Berti:2008af,LISA:2022yao}, with an absolute error of 1 part in  $10^{3}$~\cite{Colpi:2024xhw}.\footnote{It should be observed that many LISA massive black hole  mergers  are expected to occur in gas-rich (``dirty'') environments. Gas-induced effects, such as those from accretion disks, generally introduce waveform dephasing measurable during the end of the inspiral, the merger, and post-merger phases. This is known to be critical  for extreme mass ratio inspirals (EMRIs), which however are  sources unique to LISA.  For the high-redshift massive black holes observed by LISA during their long inspiral, and which could then later merge in the ET band, these environmental effects are smaller than the statistical uncertainties. For detailed discussions on environmental effects in GW astrophysics, see \cite{Barausse:2014pra}.} 

\subparagraph{Rates.}
The ET merger rate density associated to population  III stars (and in turn to potential seeds)  has been estimated to range from $0.1$ to about $200\,\rm yr^{-1} \,Gpc^{-3}$, extending from $z\sim 15$ to $6$, corresponding to a detection rate of about ten to hundreds of events per year in ET (e.g.~\cite{Hijikawa:2021hrf, Tanikawa:2021qqi,Santoliquido:2023wzn, Wang:2022unj, Liu:2021jdz, Liu:2024mkh, Mestichelli:2024djn, Davari:2024jhn}).
The LISA merger rates range between a few to several tens per year, with the uncertainties in these values rooted in the largely unknown process of pairing of massive black holes down to the milli-parsec scale below which GWs drive the late inspiral and final plunge \cite{volonteri2003, sesana2007, barausse2012, Barausse:2020mdt, ricarte2018a, Valiante:2020zhj,Izquierdo-Villalba:2023ypb,Spadaro:2024tve}. The physics of binary formation in LISA is completely different from that of ET stellar binaries, as it occurs via galaxy mergers: orbit's sinking in a galactic collision is controlled by the complex interaction black holes have with their surrounding dark matter, stars and gas, during a major or/and minor galaxy merger \cite{LISA:2022yao}.   
LISA rates at $z\gtrsim 8$ depends largely on how short  is the sinking time compared to the ``then" Hubble time, and depending on the model rates range again from a few to several per year  \cite{Barausse:2020mdt,Izquierdo-Villalba:2023ypb, Spadaro:2024tve}.

\subparagraph{ET and LISA together.}
In figure~\ref{fig:ET+LISA}
we show the synergy between ET and LISA. 
In \cite{Valiante:2020zhj}, a semi-analytical cosmological model was developed,  aimed at reproducing  typical quasars observed at redshift 8 and 2. The seeding mechanisms include both light and heavy seeds and merging binaries (yellow dots in the figure) form during halo-halo mergers, under the hypothesis that dynamical interactions lead to their formation and pairing.   The model shows that most of the LISA events are rooted in the process of formation of stellar black holes from population III stars, with only a few resulting from heavy seeds.

Comparing GW merger events  of $\sim 10^2 \, \Msun$  binaries at $z\sim 10-14$ detected by ET, with events of  $\sim 10^4-10^5 \, \Msun$ at comparable or lower redshifts with LISA (just on the edge of the left side of the LISA waterfall plot shown in figure~\ref{fig:ET+LISA}) paves the way for identifying light seeds as progenitors of SMBHs. 
Lack of ET high-$z$ events around a few $10^2\Msun$ could indicate that either they rarely form, or that they rarely pair in binaries which merge shortly after. From the LISA perspective this might indicate that heavy seeds only are the progenitor of the SMBHs or that light seeds grow at a very fast (super-Eddington) rate, following their formation without experiencing mergers. All of this needs to be quantified through careful modeling of black hole cosmic formation and evolution.

\subsubsection{IMBH populations across frequency bandwidths and detectors}

Intermediate mass black holes (IMBHs) in the  mass range between a few hundreds $\Msun$ and  ${\cal O}(10^5)\,\Msun$, bridging the known stellar and supermassive BHs, are ideal sources to exploit the advantages of having more than one detector in different frequency bands. As described in section~\ref{sec:IMBH}, IMBHs are expected to form as outcome of close and repeated dynamical encounters among stars or stellar black holes inside dense and massive clusters \cite{PortegiesZwart:2002iks,Miller:2001ez,Giersz:2015mlk}.  IMBHs can also from in the collapse of gaseous clouds in protogalaxies \cite{Begelman:2006db}, or from gaseous accretion onto massive stellar remnants originating from the first stars \cite{Madau:2001sc}. Despite the plethora of formation scenarios and the growing evidence of IMBH candidates in dwarf galaxies \cite{Mezcua:2017npy, Greene:2019vlv}, the existence of IMBHs and their role in galaxies, whether ubiquitous or rare, is still unclear and actively debated. In these regards, the detection of GWs from merging IMBHs is key to understanding where, when, and how IMBHs form along cosmic history.

ET can detect equal-mass  IMBH binaries  with source-frame masses of $\lesssim 10^3\,\rm  M_\odot$ at redshift $z\lesssim  7$~\cite{Branchesi:2023mws}, 
and  IMBH-BH  systems having a stellar BH and a primary of $\sim 10^3\, \Msun$,  at redshift $z \gtrsim  1$~\cite{Arca-Sedda:2020lso}.  
Detecting IMBH and IMBH-BH  binaries (the last referred also as  intermediate-mass ratio inspirals or IMRIs) will provide us information about the efficiency of dynamical processes in population I, II, and III star clusters in producing IMBHs.

As discussed above (see also figure~\ref{fig:ET+LISA}), the horizon reach of ET for IMBHs of mass  close to $(1-5) \times 10^2\, \Msun$ extends up to $z \sim 10$ \cite{Branchesi:2023mws,Fairhurst:2023beb,Reali:2024hqf}. However, binary IMBHs can also be present at lower redshifts. With ET operating with two CEs in network, we can measure  the mass of an equal-mass IMBH binary of $\sim 10^3\Msun$ with accuracy of  $\lesssim 0.1\%$ ($1\%$) at redshift $z=0.5$ ($2$) \cite{Reali:2024hqf}. Such a network can localise binaries at $z=0.1-0.5$  with  mass ratio $q \sim 0.1$ and primary mass of $\sim (5\times 10^2-10^3)\, \Msun$ within an angular localization $\Delta\Omega \sim (0.01-1)$ deg$^{2}$ \cite{Reali:2024hqf}. Good localisation and accurate measurements of the properties of merging components can help distinguishing both the environment nurturing the IMBH growth and the IMBH formation channel. 

\begin{figure}
    \centering
    \includegraphics[width=0.54\columnwidth]{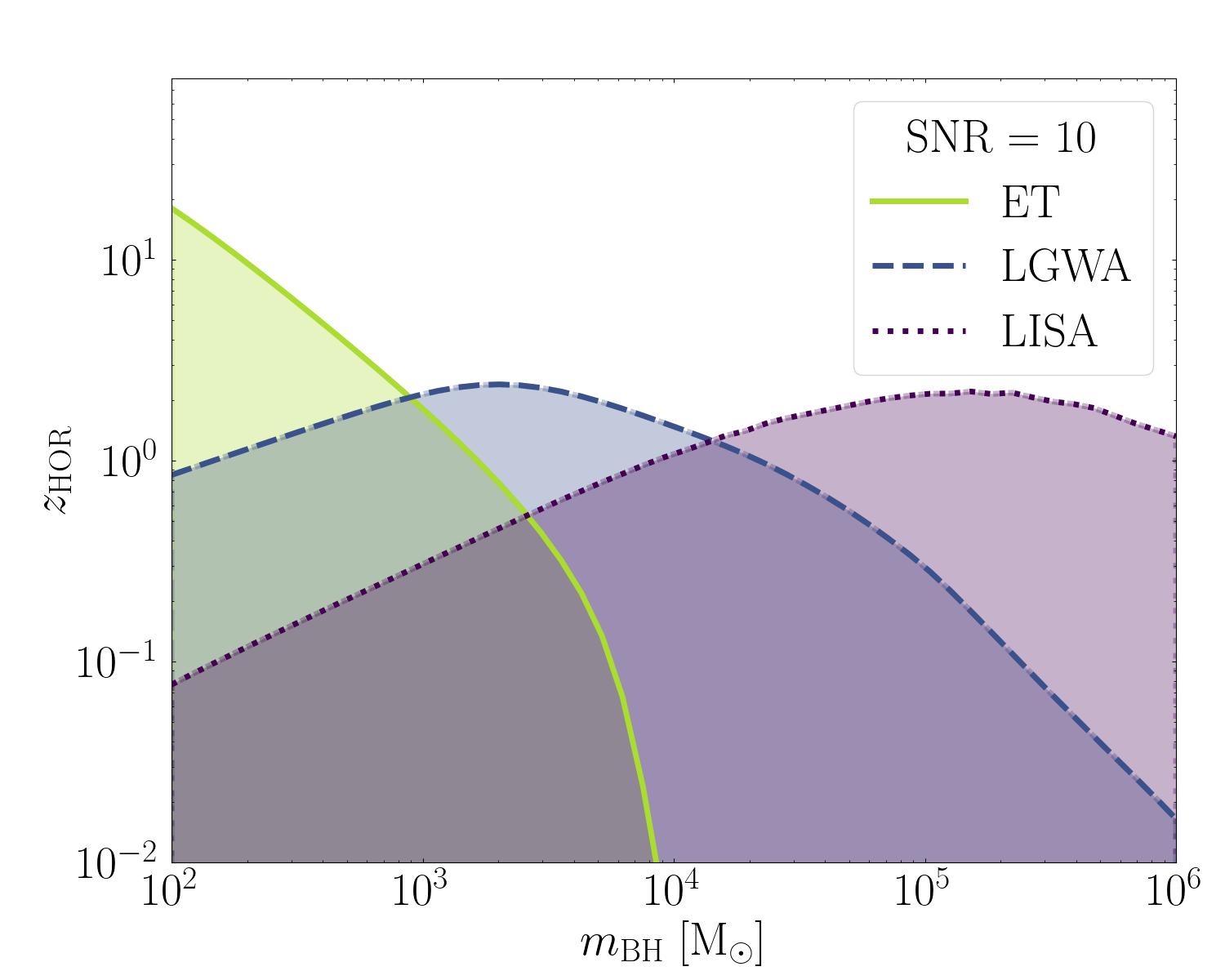}
    \includegraphics[width=0.44\columnwidth]{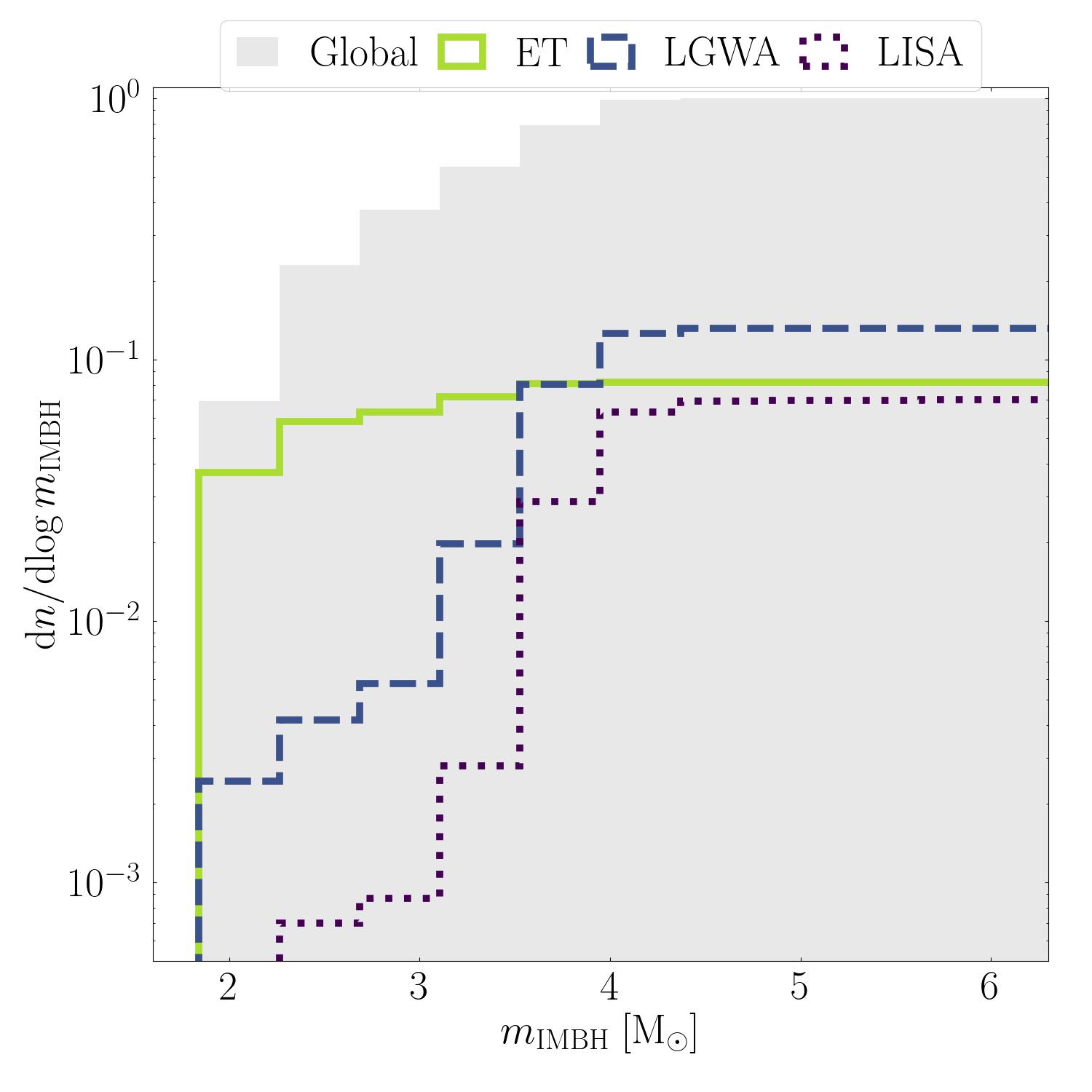}
    \caption{{\small {\bf Left panel}: horizon redshift for IMRIs involving a stellar BH of $30\Msun$ as a function of the mass of the more massive primary $m_{\rm IMBH}$. Detection is set at SNR=15 using the \textsc{GWFish} tool \cite{Dupletsa:2022scg} for ET (green straight curve), LGWA (blue dashed curve), and LISA (purple dotted curve). {\bf Right panel}: cumulative mass distribution of IMRI detection versus the IMBH mass. These systems are dynamically formed in young, globular, and nuclear clusters and their overall population is shown by the filled grey histogram.  The detectable  sub-population of IMRIs are shown for ET, LGWA and LISA (as in the legend). Models are obtained from combined population synthesis simulations performed with the \textsc{BPop} code \cite{Sedda:2021vjh} and parameter estimation performed with \textsc{GWFish}. Adapted from Arca Sedda (in prep). }}
    \label{fig: bpop}
\end{figure}

At $z\lesssim 1,$ ET can detect signals from binaries made of an IMBH and a stellar BH, which could in principle also be detected by low-frequency observatories such as LISA or Deci-Hz detectors. This is illustrated in figure~\ref{fig:GW_Landscape}. IMRI binaries are exquisite probes of the physics of formation of IMBH in dense environments as these systems form exclusively from the dynamical channel.
Combining different detectors operating independently at different frequencies makes it possible to cover the IMBH mass range fully and to place stringent constraints on the IMBH mass distribution, shedding light on the origin of IMBHs. As an example, the right panel of figure~\ref{fig: bpop} shows the mass distribution of a population of IMRIs resulting from dynamical interaction in dense young, globular, and nuclear clusters modeled with the \textsc{B-pop} population synthesis model \cite{Sedda:2021vjh} and the corresponding distribution of detectable mergers in ET and other detectors,
while  the left panel of the figure  shows the horizon redshift for ET, LISA, and LGWA in the case of an IMRI hosting a stellar BH of 
$30 \, \Msun$ inspiralling/merging  with a primary in the mass-range $(10^2-10^6) \, \Msun$. As shown 
in figure~\ref{fig: bpop}, an IMRI can be detected during the inspral phase in LISA and at merger in LGWA.  The cumulative  mass distribution for this selected IMRI (right panel of figure~\ref{fig: bpop}) shows the saturation mass in the three detectors. 
Finally, it is interesting to remark that with LGWA and Deci-Hz detectors we have the possibility to probe the region of the IMBH mass spectrum that is poorly covered by ET or LISA, i.e. roughly $(5\times 10^3-10^4)\,\Msun$~\cite{Sedda:2020vwo, Cozzumbo:2023gzs,Ajith:2024mie}.

Joint multi-band observations of the same event will be possible for $(10^2 - 10^4)\Msun$ binaries  out to redshift $z\sim 2-4$, enhancing the ability to carry on precise measurements of the source parameters at such high redshifts \cite{Jani2019}. Multi-band detection of distant lower (higher) mass binaries would be instead limited by the sensitivity for LISA at frequencies around and above $0.1$ Hz (for ET/CE at frequencies around and below $3$ Hz).

\begin{tcolorbox}[standard jigsaw, colframe=teal, colback=black!10!white, opacityback=0.6, coltext=black,title=\sc Take home message]
{\sc 
\begin{itemize}
    \item 
 ET will be uniquely capable of detecting black hole relics from the first generation of massive stars. Its sensitivity will allow good measurements of their masses and redshifts, offering valuable insights into the formation of pristine stars in binary systems. Some of these early black holes may serve as “light” seeds in the hierarchical growth of massive and supermassive black holes.
  
    \item By comparing  merger events observed by ET and LISA, each sensitive to different mass intervals -- around $100\,\Msun$ for ET and $10^5\Msun$ for LISA --it will be possible to statistically probe the mechanisms of black hole seeding and growth. 
    \item Intermediate-mass black holes of $10^3-10^4 \,\Msun$ paired with stellar black hole companions, are unique probes of the dynamical formation channel. ET, in synergy with lower-frequency detectors, offers a distinctive opportunity to detect these still elusive black holes. 
\end{itemize}
}
\end{tcolorbox}

\subsection{Early Universe cosmology}
\label{sec:earlycosmo}



\begin{tcolorbox}[standard jigsaw, colframe=cyan1, colback=black!10!white, opacityback=0.6, coltext=black,title=\sc Cosmological Backgrounds]
{\sc 
  \begin{itemize}
            \item What particle physics models predict early Universe phase transitions or the formation of cosmic strings?  Can we precisely pin down the properties of such models by measuring the GWs they produce, using ET in synergy with other GW observatories?  
            \item What is the source of cosmic inflation? Can
            it generate a broad spectrum of SGWB 
            with a sufficiently large amplitude  to be detected by ET, and by other GW experiments operating at lower frequencies?
            \item Can              
            we extract information on the early Universe cosmic history (before Big Bang nucleosynthesis) by precisely measuring the frequency dependence of 
             the stochastic GW background  produced by primordial sources? 
            Does a synergy between ET and other instruments offer any advantage in answering this  question? 
            \item What is the source of the GW background for which evidence was recently found by several PTA collaborations? If it has a primordial origin in terms of cosmic phase transitions, inflation, or cosmic strings, can a future detection
            by ET help in characterizing its origin?

        \end{itemize}
      
}
\end{tcolorbox}

\bigskip

\bigskip

Stochastic GW backgrounds can be detected across various frequency bands and have an astrophysical or cosmological origin \cite{Maggiore:1999vm, Maggiore:2018sht,Caprini:2018mtu,Christensen:2018iqi,Renzini:2022alw}.
In particular, as discussed in detail in section~\ref{sec:early_universe},
the early Universe immediately after the Big Bang could create a stochastic GW background (SGWB), through mechanisms such as the amplification of vacuum fluctuations in models beyond the single-field slow-roll paradigm, first-order phase transitions,   cosmic strings, or domain walls.  From the observed spectrum of the backgrounds, ET could learn about the physical processes that generated them and hence obtain a picture of the Universe at the earliest moments of its creation. The sensitivity to stochastic backgrounds of ET as a single observatory, both in the triangle and in the 2L configurations, are discussed in detail in section \ref{sec:div9_stoch_searches}, see in particular figure~\ref{fig:PLS_plots}, and in section~\ref{sec:pec_ET} where, for the triangle configuration,  is also discussed  the use of the null stream, see section~\ref{subsec:null_stream}, as well as the the potential caveat of the masking effect of correlated noise, see section~\ref{subsec:corr_noise}. 
The correlation of ET with other ground-based observatories would help rule out whether the observed backgrounds originate from instrumental or environmental disturbances. For example, ET in conjunction with Cosmic Explorer will be able to identify and characterize most binary coalescence events, subtract them from the data, and potentially measure the underlying primordial background (although the accumulation of the errors in the subtraction of resolved sources will also produce an important foreground~\cite{Zhou:2022nmt,Zhong:2024dss,Belgacem:2024ntv},
see also section~\ref{sect:subtractastrobkgdi9}).

Several mechanisms that generate stochastic GW background produce a spectrum that extends over many decades in frequency. The use  of observatories operating in very different frequency bands is then crucial for characterizing them (as well as for increasing the confidence in a detection).
Working with PTAs, the 10-year long GAIA database, LISA, as well as with  Deci-Hz detectors, ET could help understand stochastic backgrounds over six orders of magnitude in frequency range from nanohertz to kilohertz. In this section,
we explore the synergies between ET and LISA (section \ref{sec_cEL}), and ET and PTA (section \ref{sec_cEP})
towards the aim of a joint detection of the
cosmological SGWB.

\subsubsection{Synergies between ET  and LISA}
\label{sec_cEL}

Although  ET and LISA operate across different frequency ranges, both are expected to have comparable integrated sensitivities to the amplitude of a SGWB.
In this subsection we explore the potential synergies between these two observatories for analyzing GW sources from the early Universe that generate a SGWB with a high amplitude and a broad frequency spectrum.
In combining the results of LISA and ET, we are able to cover a much larger range of frequencies, of order 
\begin{equation}\label{eq:range}     {\cal O}(10^{-5})\, {\rm Hz}\leq f \, \leq {\cal O}(10^3)\, {\rm Hz}\,.
\end{equation}
As discussed in section~\ref{sec:early_universe},
SGWB sources that could  produce an observable stochastic  GW signal in this range include first order phase transitions, cosmic strings, and inflation (beyond single-field slow roll). For such broad signals, combining data from both LISA and ET offers a clearer signal characterisation and more precise extraction of physical properties than using either instrument alone, since we can study the signal across the full frequency band in \eq{eq:range}.

\begin{figure}[t]
    \centering
    \includegraphics[width=0.31\linewidth]{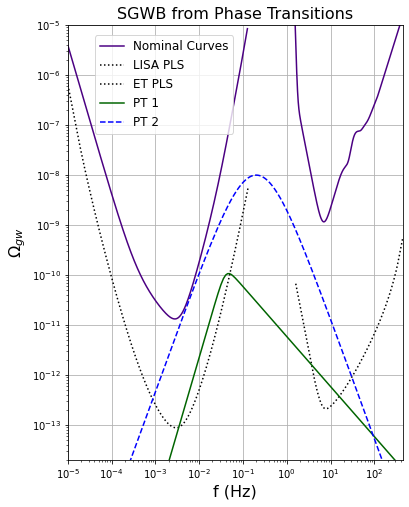}
\includegraphics[width=0.31\linewidth]{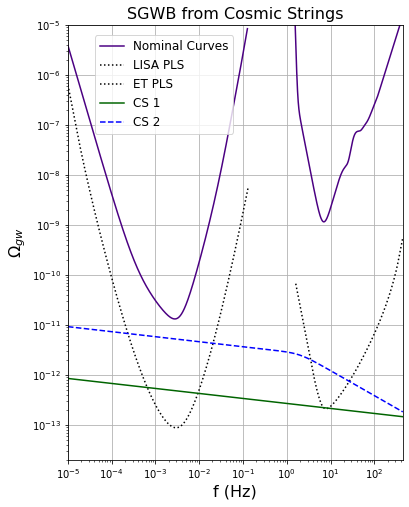}
    \includegraphics[width=0.31\linewidth]{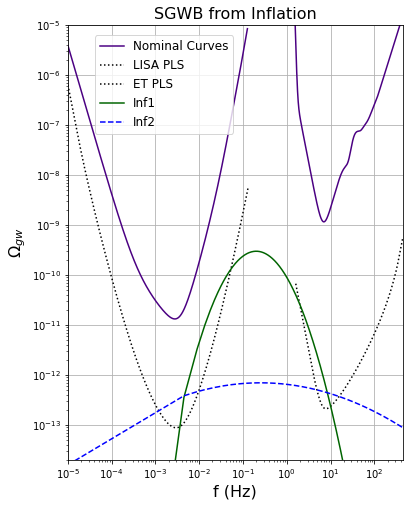}
    \caption{ 
\small    {\bf Left panel:} Examples of SGWB from phase transitions (PT).
 {\bf Middle panel:} Examples of SGWB from cosmic strings (CS). {\bf
 Right panel:} Examples of SGWB from cosmic inflation. Besides LISA and ET
 sensitivity curves (PSD) and the corresponding PLS (see \eq{eq: PLS def} and section~\ref{sec:div9_PLSdefinition} for definitions), in blue and green lines
 are represented the benchmark scenarios discussed in each of the aforementioned  sections.}
    \label{fig:benchmarks}
\end{figure}

We analyse this possibility using Fisher forecasts to assess the detectability of the SGWB shape in synergy among the two instruments. 
We consider a Gaussian likelihood on the 
GW energy density  ${\Omega}_{\rm GW}$ with a covariance matrix given by a combination
of the nominal sensitivity
curves for LISA \cite{Flauger:2020qyi} and ET in its triangle configuration.
The Fisher plots are produced using the {\texttt{GetDist}} package~\cite{Lewis:2019xzd}.
For the theoretical SGWB frequency  profile,
a useful SGWB template is the broken power law (BPL) function. This model provides a good first approximation for GW spectra generated by various early Universe phenomena \cite{Kuroyanagi:2018csn}. We follow ref.~\cite{Caprini:2024hue} for the form of the BPL function
\begin{equation}\label{eq:bpl}
    \Omega^{\rm th}_{\text{GW}}(f)=\Omega_\star\left(\frac{f}{f_\star} \right)^{n_1}\left[\frac{1}{2} +\frac{1}{2}\left(\frac{f}{f_\star} \right)^{\sigma} \right]^{\frac{n_2-n_1}{\sigma}}\, ,
\end{equation}
where $f_\star$ determines the break position, and $\Omega_\star$ is the amplitude of the spectrum at the break. The parameters $n_1$ and $n_2$ correspond to the spectral indices before and after the break, while $\sigma$ controls the break's smoothness. We use the BBN bound $\Omega_\text{GW}\leq1.7\times10^{-6}$ \cite{Pagano:2015hma} as the reference for the SGWB’s maximal amplitude, even when considering post-BBN sources. The previous profile 
will be applied to investigate GW from
phase transitions and cosmic strings; a different, log-normal 
Ansatz will be considered for the case of inflation. We show in figure~\ref{fig:benchmarks} the benchmark scenarios for
the examples we are going to discuss next, together with the nominal and power law
sensitivity curves for LISA, and for ET in the triangle configuration (using the ET-D sensitivity curve).\footnote{See figure~\ref{fig:PLS_plots} for the ET PLS with the updated sensitivity curve used in \cite{Branchesi:2023mws} as well as for the PLS of the 2L configurations, with different relative angles and arm lengths.}
For more details about our scenarios
and our methods of analysis, see ref.~\cite{Marriott-Best:2024anh}.

\begin{table}[t]
\begin{tabular}{| c | c | c | c | c | c | }
\hline
 {\rm }& \cellcolor[gray]{0.9} $\Omega_\star$  &\cellcolor[gray]{0.9}$n_1$&\cellcolor[gray]{0.9}$n_2$ &\cellcolor[gray]{0.9}$\sigma$&\cellcolor[gray]{0.9}$f_\star$ \\
\hline
\cellcolor[gray]{0.9} {\rm PT1} & $1\times10^{-10}$ & $3$ & $-1$ & $7.2$ & $0.04$ 
\\
\hline
\cellcolor[gray]{0.9} {\rm PT2} & $1\times10^{-8}$ & $2.4$ & $-2.4$ & $1.2$ & $0.2$
\\
\hline
\cellcolor[gray]{0.9} {\rm CS$1$} & $4\times10^{-13}$ & $-0.1$ & $-0.1$ & $-$ & $0.02$ 
\\
\hline
\cellcolor[gray]{0.9} {\rm CS$2$} & $2.5\times10^{-12}$ & $-0.1$ & $-\frac{1}{2}$ & $3$ & $2$
\\
\hline
\end{tabular}
\hskip1cm
\begin{tabular}{| c | c | c | c | c | c | }
\hline
 {\rm }& \cellcolor[gray]{0.9} $\Omega_\star$  &\cellcolor[gray]{0.9}$f_\star$&\cellcolor[gray]{0.9}$\rho$\\
\hline
\cellcolor[gray]{0.9} {\rm Inf1} & $3\times10^{-10}$ & $0.2$ & $0.45$
\\
\hline
\cellcolor[gray]{0.9} {\rm Inf2} & $7\times10^{-13}$ & $0.25$ & $1.6$
\\
\hline
\end{tabular}
\caption{ {\bf Left:} Benchmark values associated with Ansatz 
\eqref{eq:bpl} for the scenarios corresponding to cosmological  phase transitions  and cosmic string models.  {\bf Right:} Benchmark  values  for  cosmological inflation scenarios described
by Ansatz \eqref{ans_lnp}\label{tab:PT}. }
\end{table}
%

\subparagraph{Phase transitions.} The production of stochastic GW background from phase transitions has been discussed in section~\ref{sec:phasetransitions}. In particular, in the GWs produced by  bubble collisions, the SGWB follows the BPL profile~\cite{Caprini:2024hue}, as  in \eq{eq:bpl} or similar parametrizations such as  \eq{OmegaBPLdiv2}. Other mechanisms that produce  GWs from PTs, such as sound waves in primordial plasma \cite{Hindmarsh:2013xza, Hindmarsh:2015qta} and turbulent motion \cite{Kosowsky:2001xp, Dolgov:2002ra, Caprini:2009yp},  rather 
follow a double broken power-law profile, see \cite{Caprini:2024hue}. However, 
the focus here is on the overall structure across the broad frequency band in \eq{eq:range}, and so we use the BPL template \eq{eq:bpl} as a reasonable approximation.

We consider the scenarios in table~\ref{tab:PT}: scenario PT$1$ has spectral index $n_1 = 3$ in the infrared~\cite{Caprini:2009fx} and $n_2 = -1$ in the UV. This scenario matches cases where GWs are produced by sound waves of the bubble surrounding plasma, or by
effects of turbulent behavior in the fluid. 
For the second benchmark scenario, PT$2$, we consider the model of ref.~\cite{Lewicki:2022pdb} in the context of highly relativistic fluid shell dynamics, which finds $n_1 = -n_2 \simeq 2.4$ and $\sigma \simeq 1.2$.
These scenarios are shown in the left panel of figure~\ref{fig:benchmarks}. Notice that the SGWB amplitude is well below the nominal sensitivity curves (i.e. the Power Spectral Density, PSD): nevertheless, the SGWB can be detected with a sufficient signal-to-noise ratio by integrating over frequencies. In fact, in all panels
of figure~\ref{fig:benchmarks}, besides 
the  PSD , we also
include the power-law integrated
sensitivity curves  (PLS) introduced
in ref.~\cite{Thrane:2013oya} (see also section~\ref{sec:div9_PLSdefinition}), which  exploit the aforementioned integration over frequencies. In plotting the curves, we assume three years of observation, and we take a threshold of detection
corresponding to SNR$=5$.

\begin{figure}[t]

  \includegraphics[width=0.3\textwidth]{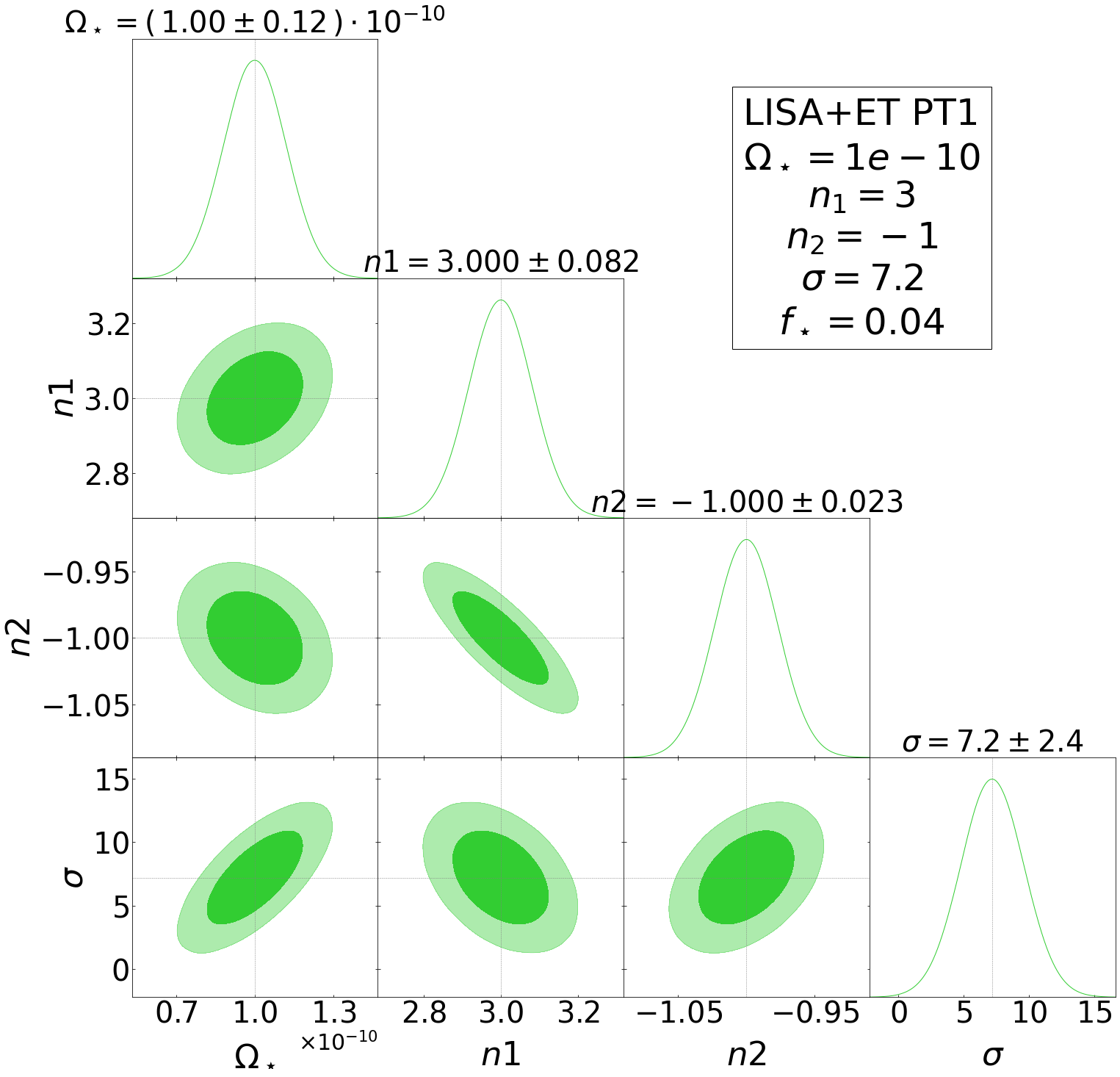}
  \includegraphics[width=0.3\textwidth]{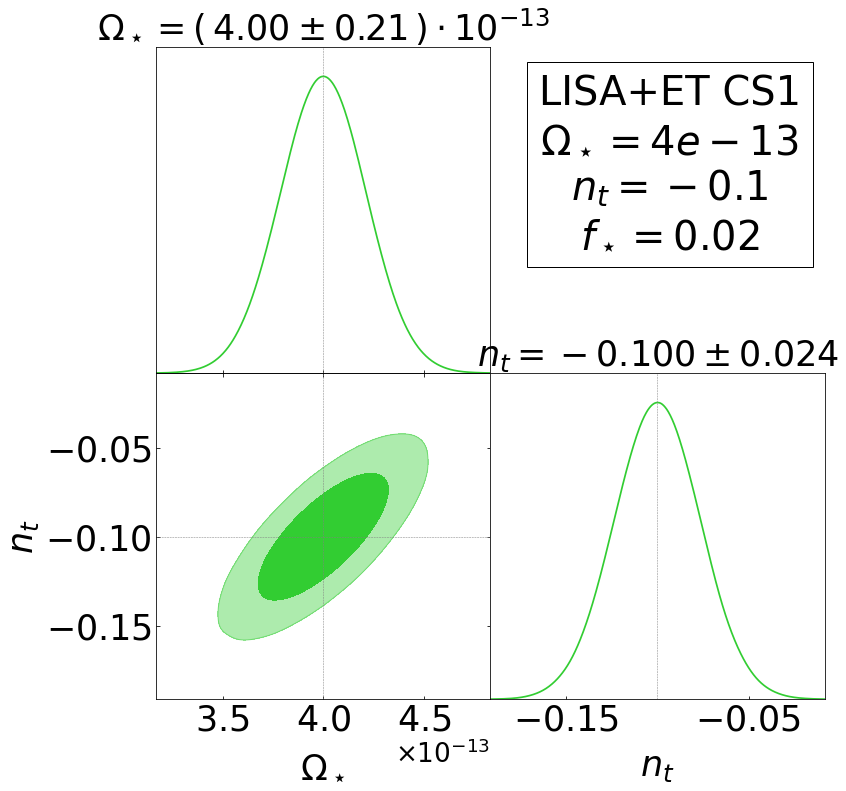}
  \includegraphics[width=0.29\textwidth]{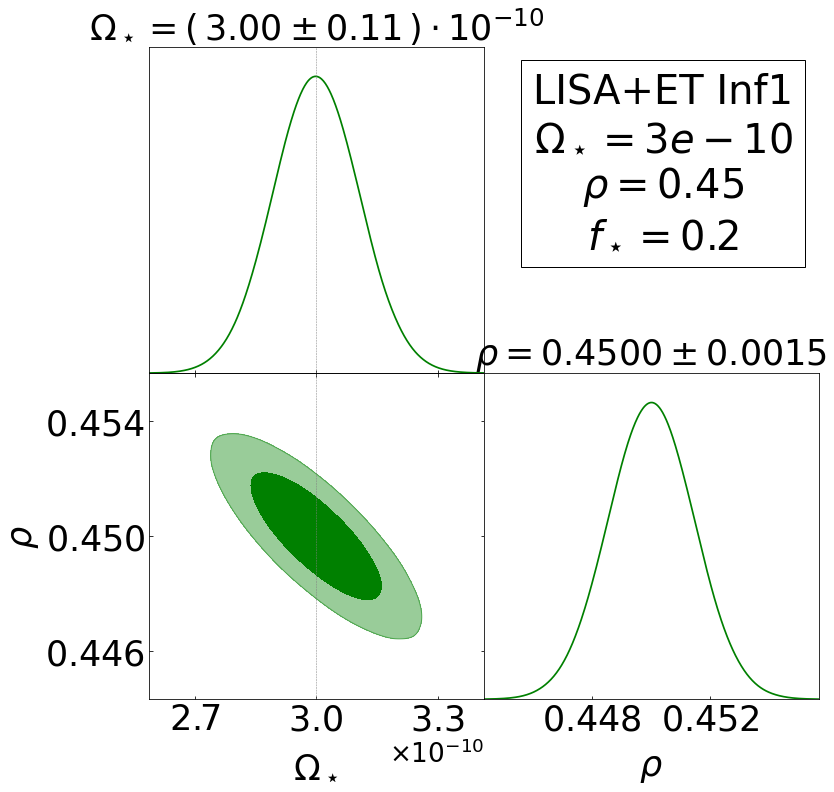}
  \includegraphics[width=0.35\textwidth
  ]{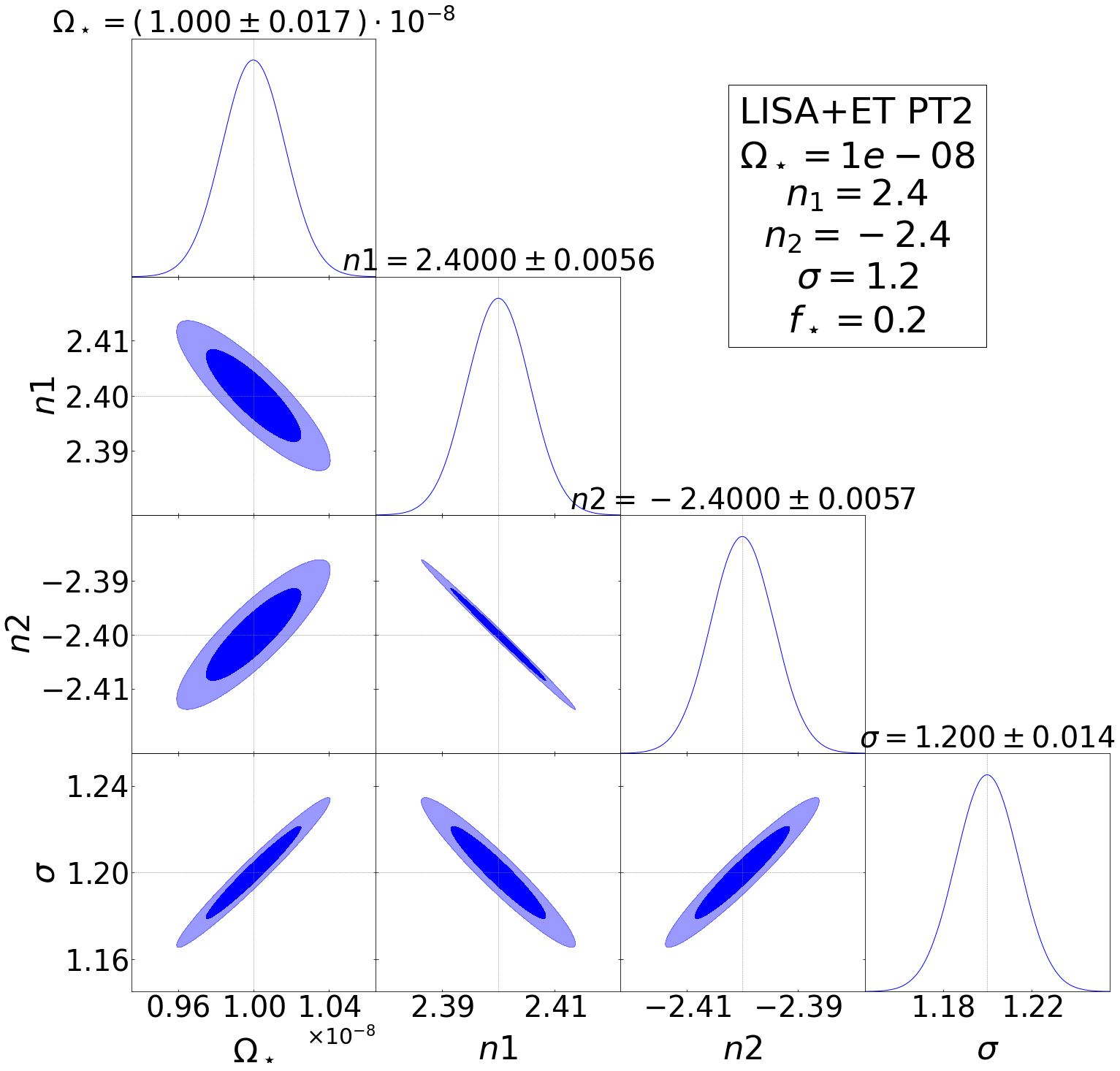}
  \includegraphics[width=0.31\textwidth]{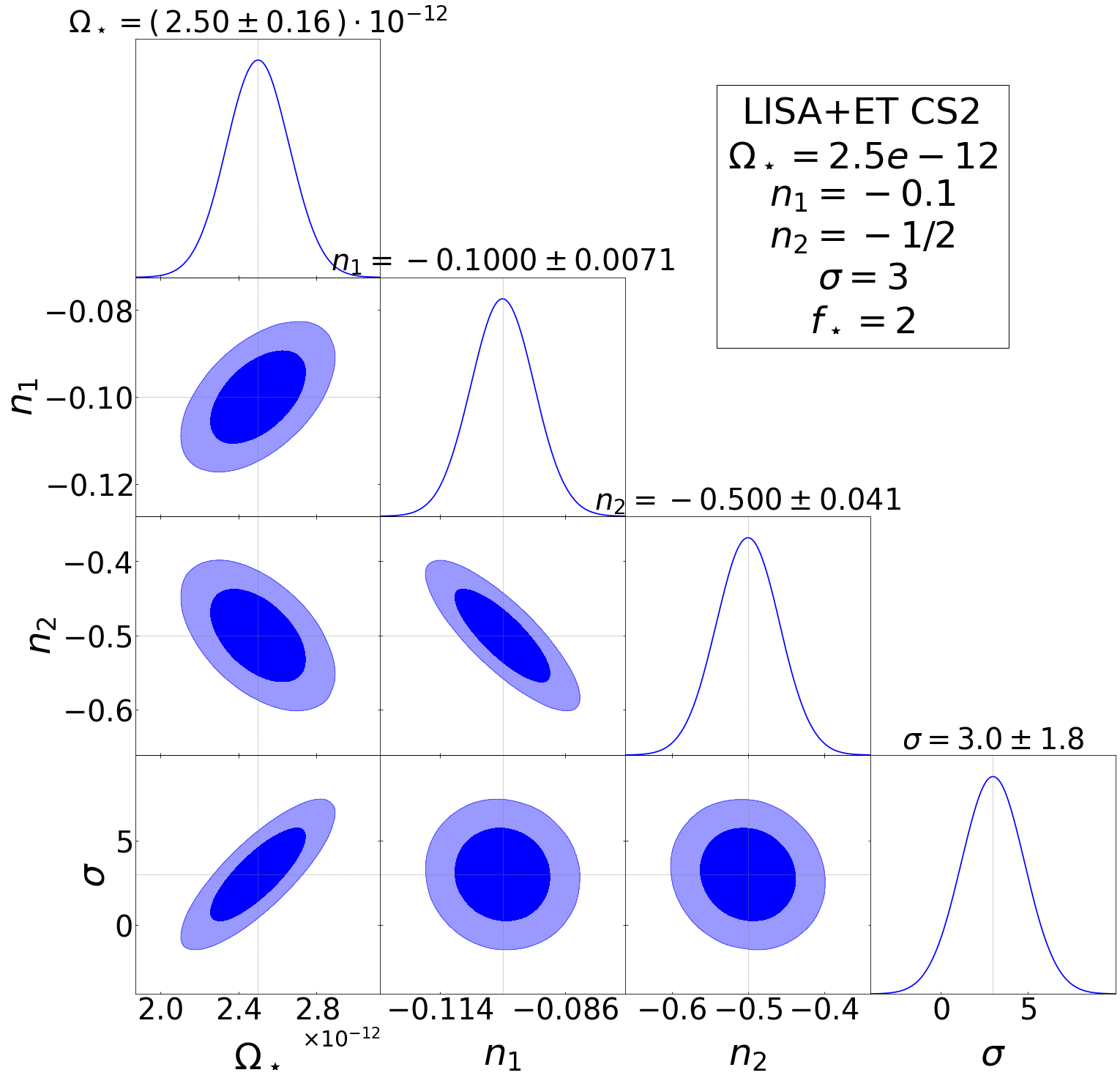}
  \includegraphics[width=0.29\textwidth]{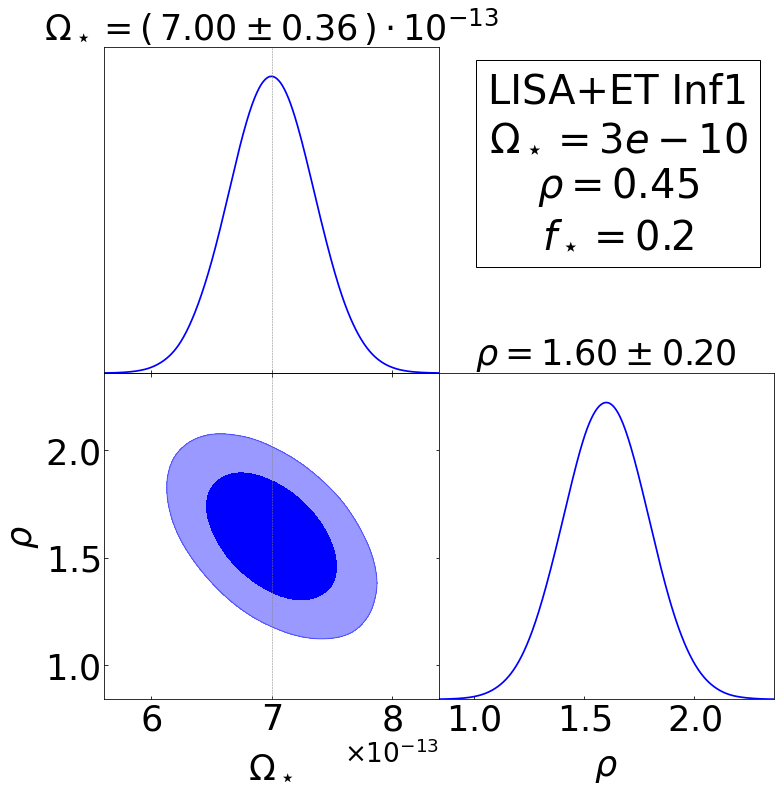}
    \caption{First and second column: Fisher forecasts for the phase transition and cosmic string scenarios.  Third column: Fisher forecasts for the inflation benchmark scenarios. See Table \ref{tab:PT}.
    } 
\label{fig:Fishphasel}
\end{figure}

For both scenarios PT1 and PT2 in figure~\ref{fig:Fishphasel}, LISA and ET individually can accurately measure only one of the spectral indices $n_1$ or $n_2$ (for more details and additional Fisher plots comparing different cases, see ref.~\cite{Marriott-Best:2024anh}). However, the two experiments  together  can measure both quantities with at least $10\%$ accuracy. The parameter $\sigma$, which controls the smoothness of the transition, can also be measured more precisely with combined data. This highlights that using LISA and ET together provides richer insights into high-temperature PT physics, also
reducing  parameter degeneracies.

\subparagraph{Cosmic strings.}  

As discussed in section~\ref{subsec:cosmic_strings}, cosmic strings form networks where interactions create loops of varying sizes, which decay and generate a SGWB spanning a broad frequency range.   Determining the spectral tilts $ n_1$, $n_2$ and the break position will provide insights into both the cosmic string properties and the Universe’s evolution before BBN. In fact, 
by analysing the spectrum’s frequency profile, we can also be able to constrain non-standard pre-BBN cosmic epochs. For normal
cosmological
expansions, we expect the SGWB from cosmic strings to
approximately follow a power law profile
in the band of \eq{eq:range}: this corresponds to our benchmark scenario CS1 of
table~\ref{tab:PT} (see e.g.~ref.~\cite{Kuroyanagi:2018csn}).
We obtain the values for the benchmark scenario CS$2$ from \cite{Blanco-Pillado:2024aca}, corresponding to a spectral tilt being dependent on the equation of state $w$ in a non-standard cosmological expansion, where $n_2\,=\,-2 {\left(3 w-1 \right)}/{\left(3 w+1\right)}$ for values where $w \geq 1/4$. The middle panel of figure~\ref{fig:benchmarks} illustrates the benchmark values used in our CS cases.
  Only with both instruments together can the full set of parameters be measured within $10\%$ accuracy.
See figure~\ref{fig:Fishphasel}, middle column,  and ref.~\cite{Marriott-Best:2024anh} for more details.
 

\subparagraph{Cosmic inflation.}  
%
As discussed in section~\ref{subsec: inflation}, see in particular figure~\ref{fig:standard_inflation}
while the most minimal inflation models  based on single-field slow-roll predict a SGWB amplitude too small for both ET and LISA to detect, several well-motivated scenarios can boost the amplitude to observable levels within the frequency band given by \eq{eq:range}. These scenarios include dynamics of vector and axion fields \cite{Barnaby:2010vf,Sorbo:2011rz}; spontaneous space-time symmetry breaking \cite{Endlich:2012pz, Cannone:2014uqa,Bartolo:2015qvr}; and effects associated with the production of primordial black holes \cite{Ananda:2006af, Baumann:2007zm}. 
Generally, a SGWB from inflation exhibits a more complex frequency profile than those from other sources, potentially featuring log-normal shapes, multiple peaks, or oscillatory patterns (see ref.~\cite{Braglia:2024kpo} for a discussion of various inflation templates.).
We consider the so-called log-normal peak Ansatz:
\begin{equation}
    \Omega_{\rm GW} = \Omega_{\star}  \exp{\left[{-\frac{\ln^2\left({f}/{f_\star}\right)}{2\rho^2}}\right]},
    \label{ans_lnp}
\end{equation}
where $\Omega_{\star}$, $f_\star$, and $\rho$ control the amplitude, position, and smoothness of the peak. This scenario can be the result of axion or axion spectator inflation models~\cite{Namba:2015gja, Dimastrogiovanni:2016fuu, Thorne:2017jft}. Table \ref{tab:PT} gives the values used in our benchmark scenarios, which are also shown in the right panel of figure~\ref{fig:benchmarks}. 
%
%
The corresponding Fisher forecasts are presented in figure~\ref{fig:Fishphasel}, third column. Particularly for scenario Inf$2$ we learn there are significant benefits from the combining the results of the two instruments for
reducing the errors in the measurements (again see ref.~\cite{Marriott-Best:2024anh} for a more detailed analysis).

\subsubsection{Synergies between ET and PTAs}
\label{sec_cEP}

A primordial SGWB can span a large range of  frequencies, from Cosmic Microwave Background frequencies of $~10^{-17}$ Hz, Pulsar Timing Array (PTA) frequencies of around $10^{-9}$ Hz, LISA frequencies in the milli-Hertz range as well as ground-based observatories above a few Hz and up to a few kHz. ET observations can thus be studied in synergy with CMB, PTA and LISA observations. Specific scenarios are however different for each type of cosmological background, e.g. inflation, first-order phase transitions, cosmic strings, domain walls or second-order scalar perturbations associated with primordial black holes: see for example  figure~17 of \cite{Ellis:2023oxs}.

Single field slow-roll inflation
is detectable by CMB instruments such as Planck at the lowest frequencies, but its red tilt implies that it cannot be detected by either PTAs, LISA or ET, see figure~\ref{fig:standard_inflation}. However, as we discussed in section~\ref{subsec: inflation} other, more complex inflationary scenarios  can produce a GW spectrum with a positive 
spectral index $n_t$ (blue tilt), making them potentially detectable by PTAs, LISA and ET. Ref.~\cite{Lasky:2015lej} used actual CMB, PTA and LIGO measurements to place constraints on a SGWB of inflationary origin, in particular on  $n_t$  and the tensor-to-scalar ratio $r$ (characterizing the amplitude of a primordial GW signal as compared to the amplitude of  scalar perturbations as detected in the CMB). This is an example of how different instruments across the GW spectrum can be used to place constraints on a SGWB.

More recently, in 2023, PTA results across the globe have shown evidence for a SGWB \cite{NANOGrav:2023gor,EPTA:2023fyk,Reardon:2023gzh,Xu:2023wog,InternationalPulsarTimingArray:2023mzf}. Work is in progress to confirm these results with a $5\sigma$ significance. If confirmed as a true SGWB, the most likely source is a background of supermassive black holes. In this case, because of the completely different masses,  there would be little synergy between PTAs and ET (but there would be synergy between PTAs and LISA in the sense that PTA sources can inform expected LISA findings on massive black hole binaries). However, another possible source is a cosmological background. In refs.~\cite{NANOGrav:2023hvm,EPTA:2023xxk}, the observed signal was interpreted as a cosmological background and the spectral index $n_t$ and tensor-to-scalar ratio $r$ were estimated. While, in principle, these estimates could be used to search for this same background in the ET data, the observed spectral index in PTA data is quite high and not compatible with the high frequencies of ET if the spectral index is assumed to be fixed; however, in the case of a running index across GW frequencies, a PTA background could be compatible with a background in ET. Another possibility is 
a non-conventional cosmological
history with a period characterised by a stiff
equation of state \cite{Duval:2024jsg}. Additionally, a CMB detection of a GW background would be groundbreaking; a combined detection of a primordial SGWB in both the CMB and PTA data could make it easier to find the same signal in ET. However, it is much more likely that in PTA data, the cosmological background is present as a secondary background with a foreground of supermassive black hole binaries. Work is ongoing to model a search for both backgrounds. Cosmic strings, domain walls, first-order phase transitions  and second-order perturbations or Primordial black holes (PBH) could also be detected by PTAs, LISA and ET. The recent PTA results \cite{NANOGrav:2023hvm,EPTA:2023xxk} are not compatible with a cosmic string signal, but are potentially compatible with a first-order phase transition or PHBs. Ref.~\cite{Figueroa:2023zhu} looked at the combined NANOGrav and EPTA results and their possible interpretation as a cosmological background. Many papers on the interpretation of PTA results as a cosmological background have in fact been published after the June 2023 release of results by the main PTA collaborations, and trying to summarize them all is a research avenue in and of itself.

In conclusion, PTA data can help inform an ET search for a cosmological background. On the one hand, PTA data can help constrain a SGWB in the nanohertz GW frequency range, as the sensitivity of PTAs improves, in particular in light of the upcoming Square Kilometre Array. Recent results (from 2023) provide evidence for a correlated common red noise that could be a SGWB that is compatible with a background generated in exotic inflationary scenarios, first-order phase transitions or second-order scalar perturbations, but requires in each case a surprisingly large effect. More probably, the cosmological background is only secondary to a foreground of supermassive black holes, in which the spectral index of the cosmological background would be lower and thus more compatible with a similar index at high ET frequencies. Studying scenarios of running spectral indices across the GW frequency range as well as differing equations of state are good avenues for linking the PTA frequencies with the ET frequencies and thus finding synergies.

\begin{tcolorbox}[standard jigsaw, colframe=teal, colback=black!10!white, opacityback=0.6, coltext=black,title=\sc Take home message]
{\sc 
\begin{itemize}
\item Possible early Universe GW sources can provide a SGWB profile spanning many decades in frequency: cosmic phase transitions, cosmic strings, and cosmic inflation.
\item SGWB from primordial sources have a rich frequency dependence, whose characterization improves up to a 10\% accuracy by means of a multiband detection by ET and by other GW detectors.

\item Synergies between ET and other GW instruments thus provide invaluable information on primordial sources of the SGWB, and on the early history of our Universe expansion.
\end{itemize}
}
\end{tcolorbox} 

\subsection{Executive summary}
\label{sec:ExecSumm_div5}

In this section we have explored how the science case of ET is enhanced by 
exploiting synergies with other GW observatories, operating either  in a similar  frequency band, as Cosmic Explorer, or in completely different bands, as LISA, deciHz detectors such as LGWA, or PTA.

Joint observations of ET with Cosmic Explorer will consolidate the science of this emerging and transformative field. The synergy with lower-frequency GW observatories, such as PTA at nHz, LISA in the mHz bandwidth, and design concepts like LGWA in the deciHz interval, will  broaden the scientific case for ET. These synergies allows for connections to be made with other GW sources also in different astrophysical environments, enabling the investigation of some of the most compelling questions in astrophysics, physics and cosmology. In particular:

\begin{highlightbox} {Synergies of ET with other GW observatories:}
 
\begin{itemize}

\item
Synergies between CE and ET enable a more comprehensive exploration of the Universe by extending  further the cosmic horizon (see figure~\ref{fig:det_horizons}),  enlarging the search for intermediate-mass black holes --an entirely uncharted territory-- and enhancing the accuracy of measurements of the source's physical parameters, resolving degeneracies and obtaining a comprehensive and robust understanding of the nature of neutron stars and black holes.

\item Joint observations of ET with CE offer a unique opportunity to measure the GW polarization modes and constrain alternative gravity theories. The dipolar radiation test is significantly enhanced by combining ET with milli-Hertz and Deci-hertz gravitational wave detectors.

\item 3rd-generation observatories like ET and CE will be the only observatories capable of discovering the relics of the first massive stars expected to form during cosmic dawn. These observatories will provide better measurements of the masses and redshifts of these black holes, shedding light on the formation of pristine stars in binaries. Some of these black holes could serve as “light” seeds for the growth of massive black holes.

\item The collaboration between ET and CE represents a transformative advancement for late-time cosmology with resolved sources, due to the improved localization capabilities that are critical for both bright and dark siren cosmology. Furthermore, the synergy with other gravitational wave observatories, such as LISA, will expand the range of GW sources available.

\item Multi-wavelength observations of merging black holes, captured by LISA in their early inspiral phase and by ET at the point of merger, are crucial for understanding the formation pathways of stellar-mass and intermediate-mass binary black holes. The combined observations of these  systems by ET and LISA can further constrain General Relativity (GR) with an order of magnitude better precision than either observatory could achieve alone.

\item BBHs originating from massive stars with masses in the range $120 M_{\odot}-300 M_{\odot}$, are prime candidates for GW observations across multiple frequency bands. The prolonged inspiral signals emitted by these systems can be observed by LISA and by Deci-Hz detectors, while the later stages of inspiral and merger will be detected by ET. The combination of the observations from LISA  or  from Deci-Hz detectors such as LGWA,  with  the ET observations will dramatically improve the sky localization of these sources, making them effective dark sirens for the determination of cosmological parameters.

\item By comparing merger events detected by ET and LISA, which are sensitive to different mass ranges (around  $100\, \Msun$  for ET and $10^5\, \Msun$  for LISA), we can gain statistical insight into the mechanisms behind black hole seeding and growth.

\item Intermediate mass ratio inspiral, where a stellar mass BHs merges with an intermediate mass black hole (IMBH), are exquisite probes of the physics of formation of IMBH in dense environments.
Combining observations from ET, LGWA and LISA  makes it possible to cover the IMBH mass range fully and to place stringent constraints on the IMBH mass distribution, shedding light on their origin, see figure~\ref{fig: bpop}.

\item GW sources from the early Universe, such as cosmic phase transitions, cosmic strings, and cosmic inflation, may provide a stochastic gravitational wave background (SGWB) profile across multiple frequency bands. The SGWB from primordial sources has a rich frequency dependence, which can be characterized with up to 10\% accuracy through multi-band detection by ET and other GW detectors such as LISA, detectors operating in the deciHz such as LGWA,  and PTA. Consequently, synergies between ET and other instruments offer invaluable information about primordial sources of the SGWB and the early history of the Universe's expansion.

\item  A confident detection of a stochastic GW background requires control of correlated noise among the detectors involved in the correlation (the two distant detectors in the ET-2L configuration, the three colocated detectors in the ET-triangle configuration, and the three independent channels of LISA).
Bringing in fully independent observatories, such as CE on the ground, or TianQin and Taiji in space, provides more robust detection prospects. Cross-correlating data from distant sites eliminates (or, on ground, further reduces) the possibility of correlated environmental or instrumental disturbances (see sections~\ref{sect:correlated_noise_div2}, \ref{sect:RecCorrNoise_div2} and \ref{subsec:corr_noise} for discussions), yielding a far more robust claim of correlated signal, rather than just an excess of correlated broadband noise. At the same time, each new baseline adds signal-to-noise, improving overall sensitivity.

\end{itemize}
\end{highlightbox}

\section{Subatomic Physics with ET}\label{section:div6}

The first detection of GWs from a merging neutron star binary system and the accompanying observations of electromagnetic counterparts in 2017 demonstrated the enormous potential of multi-messenger astronomy for nuclear physics. Neutron stars comprise the highest densities of matter that can stably exist in the Universe. The gravitational-wave signals emitted from events involving neutron stars encode detailed and clean information on the dense matter in their interiors and, for a subset of binary neutron star mergers, on the subatomic microphysics in regimes of yet higher density and temperature. Such invaluable insights into states of subatomic matter under extreme conditions are inaccessible to any other probes and, in conjunction with electromagnetic counterparts, enable a multitude of additional opportunities, such as understanding the nucleosynthesis of very heavy elements. This section emphasizes the advances in nuclear and neutron star physics poised to open up with future gravitational-wave detections by ET, which are synergistic yet go beyond what can be accomplished with other astrophysical and terrestrial experiments. We also advocate progress in microscopic nuclear and hadron theory to fully exploit ET and test gravity in the regime of largest achievable densities and pressures, that of neutron stars.

 
\subsection{Introduction}
\begin{figure}[t]
\begin{center}
\includegraphics[width=0.45\textwidth]{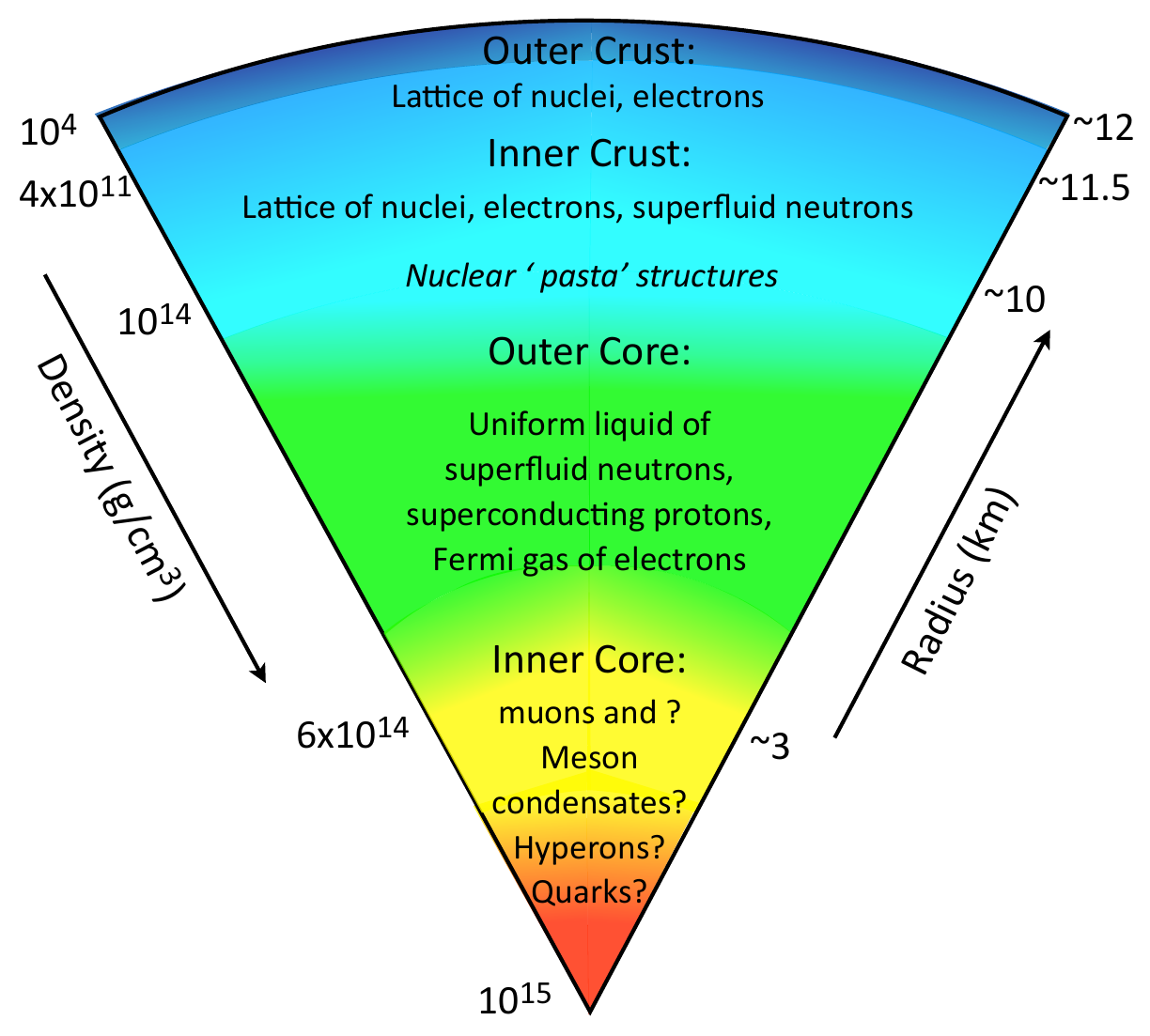}
\caption{Conjectured structure of a cold, mature neutron star. In the interior, matter ranges over nearly an order of magnitude in density and encompasses a variety of different phases and physical properties. Figure adapted from ref.~\cite{Maggiore:2019uih}.}
\label{fig:phases}
\end{center}
\end{figure}

Astrophysical environments involving neutron stars (NSs) -- objects
where gravity compresses matter to several times the normal nuclear
density -- are unique testbeds for a rich variety of subatomic
physics. A NS encompasses different phases of matter over a large
range in density: from neutron-rich nuclei and free electrons in the
outer parts to a uniform liquid of primarily neutrons, until the
densities become so high in the very cores that other particles
may be populated such as hyperons or mesons, which could
also form condensates, and/or a transition to degenerate quark matter. This is illustrated in figure~\ref{fig:phases}
and makes NSs exquisite probes for unexplored regimes of strongly interacting matter, as well
as for nuclear many-body phenomena and their variations with density
and isospin asymmetry.

Among the messengers probing NSs and related phenomena,
gravitational waves (GWs) are unique because they are purely based on gravitational interactions,
and thus travel nearly uncontaminated from the densest regions of
their sources to the detectors. The potential for nuclear physics with
GWs has already become evident with the spectacular first detection of
a binary NS (BNS) merger,
GW170817~\cite{LIGOScientific:2017vwq}. For this event, only the
inspiral epoch, when the NS and its companion are still orbiting and
matter is near its ground state, was discernible in the GW data. This
enabled inferring constraints on the cold equation of state (EOS) above
nuclear density from measurements of the NS's tidal deformability, a
characteristic parameter that depends on the properties of matter and
-- together with the electromagnetic counterpart-- provided the first
definite identification of the central engine of a kilonova and the association of a BNS merger with a GRB. Since
then, at least four more events likely involving one or two NSs in a binary
were discovered, but apart from the masses of the coalescing objects,
no new information relevant for nuclear physics could be extracted~\cite{LIGOScientific:2020aai,LIGOScientific:2021qlt,LIGOScientific:2024elc}.

Current detectors are mainly sensitive to the inspiral of NS binary coalescences. The outcomes beyond that vary depending on the type of binary system and its parameters. 
For mergers of NSs with a black hole (BH), the NS either becomes tidally disrupted, which leads to a strong suppression of the GW signal, or plunges into the BH, or is partially disrupted. In all such scenarios, the final remnant is a BH, and cases with tidal disruption are accompanied by electromagnetic counterparts.
The mergers of two NSs have different
outcomes depending on the initial systems: prompt collapse to a BH or the formation of a rotating massive NS
heated by shocks during the merger process, which either collapses to
a BH after shedding enough angular momentum or forms a stable NS
remnant. In the latter two cases, the post-merger behavior is highly
dynamical, involving differential rotation, oscillations, and
dissipation, along with various microphysical processes, small-scale
magnetohydrodynamic instabilities, and higher temperatures. This aftermath of a BNS merger provides a
window onto a completely unexplored swath of the parameter space of
subatomic physics that represents a complex nonperturbative regime
in-between our current understanding of matter at low baryon densities
from nuclear physics and heavy-ion collisions, and the regime of
unreachable high densities where first principle calculations within quantum chromondynamics (QCD), the theory of the strong interaction, are valid.

The coalescences of NS binaries are also among the key production sites
of elements heavier than iron in the cosmos. Heavy elements can be
synthesized from the neutron-rich material expelled during BNS mergers
or the tidal disruption of NSs by BHs and winds from the remnant
accretion disks. The subsequent radioactive decay of these elements
powers a kilonova electromagnetic transient. 

In addition to binary mergers, future multi-messenger observations
include the possible discoveries of signals from (i) a nearby
(galactic) core-collapse supernova (CCSN), where the GWs and neutrinos
contain imprints of the yet poorly understood explosion mechanism,
dense matter properties in regimes complementary to heavy-ion
collisions and BNS mergers, and the possible hot proto-NS,
and (ii) rotating isolated NSs with a deformation (e.g.\ elastically-
or magnetically-induced mountains on their surface or unstable
oscillation modes) encoding information about the structure of the
crust and the behavior of dense matter under the influence of a strong
magnetic field. 

Due to the projected increase in sensitivity of the existing detector
network of the LIGO, Virgo, and KAGRA (LVK) facilities in the coming years,
and based on the event rates inferred from the past detections, we
expect GW measurements of potentially dozens of binary merger events
involving NSs in the coming years~\cite{KAGRA:2013rdx,KAGRA:2021duu}. This will represent a larger sample of the diverse
population of NSs, with the nearby events measured with higher
accuracy and possible access to the tidal deformability characterizing the dominant EOS signatures in GW signals. Several cases
with electromagnetic counterparts are also expected~\cite{Colombo:2023une,Patricelli:2022hhr,Petrov:2021bqm,Bhattacharjee:2024wyz,Shah:2023ozh}.
Further upgrades, A\#~\cite{Fritschel2022} and
VirgoNext, are planned to fill the gap between the LVK run O5 and the onset of
next-generation observatories projected for the late 2030's~\cite{Bailes:2021tot}. These detections
will already enable significant advances in our understanding of dense
matter in NSs.

However, even after these planned upgrades, the capabilities of the
current detectors will remain limited in sensitivity and bandwidth,
especially at the high frequencies corresponding to the signals of the
post-merger phase. Next-generation observatories such as ET will have a significantly higher
sensitivity, e.g., GW170817 would have a signal-to-noise ratio of up to fifty times larger than current detectors due to the higher sensitivity and the larger frequency range that ET will cover. Moreover, ET is expected to detect several tens of thousands of BNS merger events per year~\cite{Maggiore:2019uih,Branchesi:2023mws} and thus probe the diverse population of NSs. Its increased low frequency sensitivity is important for sending early-warning alerts to electromagnetic telescopes before
the BNS mergers and for higher precision measurements during the inspiral phase. The increased high frequency sensitivity will enable higher precision measurements of matter effects in the late inspiral and  tracing the details of the behavior of matter during the
merger and its aftermath. For inspirals, this will not only enable precision measurements of tidal deformability along with the NS masses and spins, but also give access to subdominant effects, such as the tidal excitations of characteristic NS modes, among others, which will provide further details about NS matter.
The dominant GW observables from the
post-merger regime are the frequency associated with the rotation of
the dynamical remnant and its lifetime, with the frequency spectrum
also involving a number of additional features related to various
microphysics properties. For NS-BH binaries with tidal disruption, the disruption frequency provides additional information on the cold EOS. The possibility of performing a GW observation
from a CCSN event and the subsequent proto-NS (PNS) evolution, 
as well as of detecting GWs from isolated NSs also significantly increases with ET. 
Measurements with ET are thus poised to yield profoundly transformative 
advances for nuclear physics. 
Accompanying electromagnetic and neutrino signals to the
mergers will yield complementary information and, when combined, provide
copious insights into a variety of microphysics of dense matter and
the production of heavy elements. Conversely, with adequate nuclear theory support, NS binaries observed with ET can be used to test the theory of gravity and its matter couplings in the largest density and pressure regimes available in nature and probe for dark matter, as detailed in section~\ref{section:div1}.

This section is organized as follows.  In section~\ref{sec:microphysics} we   recapitulate the current
status of our knowledge on the properties of dense and hot matter in
NS-related environments. In section~\ref{sec:constraints} we 
discuss the potential of GW detections with ET to impact our
understanding of subatomic physics. The inspiral phase of a BNS merger, the tidal disruption of NSs in NS-BH systems, and the GWs
from an isolated NS thereby probe essentially cold matter, whereas in
the BNS post-merger phase and during a CCSN event or within a PNS, matter
is hot. 
Section~\ref{sec:degeneracies} is devoted to a discussion of the
different uncertainties in the theoretical modeling that underpins the interpretation of the data for nuclear physics and will be the main limiting factor for the science returns from measurements with ET. We also discuss  
degeneracies with other physical effects, such as those due to modified gravity or to dark matter, in 
the interpretation of the detected signal for subatomic physics. We conclude in section~\ref{sec:conclusion} with a summary of the key takeaway points and synergies with other facilities.

\subsection{Current status of microphysics properties}
\label{sec:microphysics}
To motivate the science potential of ET, we first review the current status of our knowledge and remaining open questions for the properties of dense and hot matter, which also enter the modeling of the various astrophysical systems considered in this section. This includes the EOS discussed in section~\ref{sec:eos}, as well as the different reactions rates from out-of-equilibrium processes which cannot be included in the EOS, and are discussed in section~\ref{sec:reactions}. 
\subsubsection{Equation of state modeling}
\label{sec:eos}

\begin{figure}[t]
\begin{center}
\includegraphics[width=0.6\textwidth]{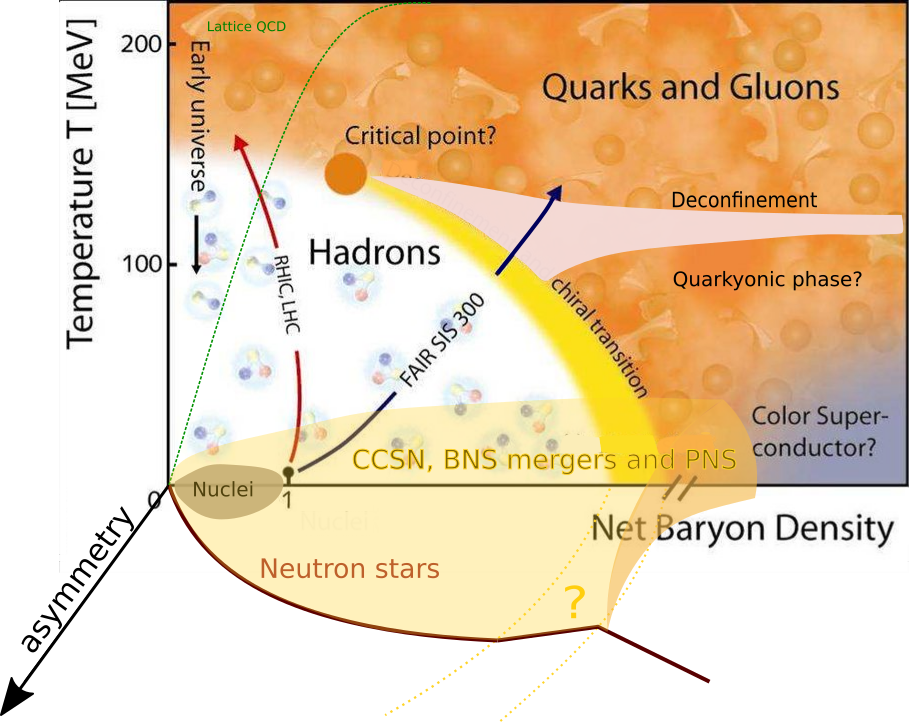}
\caption{
Depending on the thermodynamic conditions, matter can exist in different forms, like nuclei, gas of nuclear clusters, gas of hadrons, and strongly-coupled quark-gluon plasma states.
The temperature-density-isospin asymmetry diagram illustrates the different thermodynamic conditions spanned in NSs, BNS mergers, CCSN, PNSs, as well as in nuclei and the early universe. The regimes that can be explored in different terrestrial facilities are also indicated. 
Figure adapted from \cite{GSIphasediagram}.
}
\label{fig:phasediagram}
\end{center}
\end{figure}

The structure of cold, mature NSs can be described by a one-parameter EOS that relates the pressure to the energy density, as the temperature is well below the Fermi temperature and weak ($\beta$-)equilibrium has been reached. 
The typical density range extends from values encountered on Earth
up to several times the nuclear saturation density, 
$n_{\rm sat} \approx 0.16~\mathrm{fm}^{-3}$, which corresponds to $\rho_0 \approx  2 \times 10^{14} \mathrm{g/cm^3}$. 
In addition, the study of related astrophysical phenomena, e.g.,
the post-merger phase of a BNS merger, CCSN, PNSs and the formation of stellar BHs, 
requires EOSs, dubbed general purpose EOSs, that also include a wide range of temperatures 
$0 \lesssim T \leq 150~\mathrm{MeV}$ and charge fractions $0 \leq Y_Q \leq 0.6$, where $Y_{\mathrm Q}=n_{\mathrm Q}/n_B$ with $n_{\mathrm Q}$ defined as the electric charge density in hadrons and $n_B$ the baryon number density. By electric charge neutrality, $Y_Q$ equals the fraction of charged leptons, i.e., electrons and muons, which appear in the higher density regions of NSs. The regimes of baryon density, temperature, and isospin parameter space covered by NSs are illustrated in the phase diagram in figure~\ref{fig:phasediagram}. Describing the different properties of matter over this vast range of thermodynamic parameters is a challenge. Firstly, over these large ranges of density scales, matter characteristics change dramatically, from a mixture of different nuclei to uniform strongly
interacting matter, containing, in the simplest case, just nucleons but
potentially more exotic components such as hyperons or mesons. Moreover, a
transition to deconfined quark matter might occur at high densities. Secondly, the extreme densities and asymmetries involved in NSs are inaccessible to both terrestrial experiments and, so far, theoretical first-principles calculations, which are hindered by the non-perturbative character of this strongly interacting many-body system.

\paragraph{NS EOS.}
Over the last decades, numerous models for the NS EOS have been developed.
In what follows, we review some of the main classes of models, 
and expose the change of paradigm caused by large computing resources and 
availability of new astrophysical data, 
advantages and disadvantages of different strategies, and
the potential onset of non-nucleonic degrees of freedom.

\subparagraph{Agnostic EOS models.} Agnostic EOS models can either be non-parametric, e.g., based on Gaussian processes, or parametric, e.g., based on a piece-wise polytropic or spectral representation, or a parametrization of the speed of sound. The parameters of such a representation can be constrained based on a number of considerations that often include causality and thermodynamic stability, but also astrophysical measurements and ab initio calculations of pure neutron matter; see, e.g., refs.~\cite{Landry:2020vaw, Raithel:2022efm, Tews:2018iwm, Essick:2021ezp,Essick:2021kjb} and references therein. 
Some models also include constraints on the high density part of the EOS based on requiring compatibility between the behavior of the EOS at densities $n \lesssim 5 n_{\mathrm{sat}}$ and in the quark phase at ultrahigh densities of $n \sim 40 n_{\mathrm{sat}}$ accessible to perturbative QCD calculations, see, e.g., refs.~\cite{Annala:2017llu,Oter:2019rqp,Komoltsev:2021jzg}. 

Agnostic models enable exploring the relation between properties of the EOS and global NS properties. The latter include mass, radius, compactness, moment of inertia, tidal deformability, and Kepler frequency\footnote{The Kepler frequency is the maximum possible frequency of a rigidly rotating star.}.
While flexible, computationally cheap, and straightforward to use, these models lack a connection with the underlying microphysics. Hence, no information on the interactions and composition of the matter is available, and compliance with known properties of nuclear matter has not always been checked.

\subparagraph{Semi-agnostic models.} Semi-agnostic models have been designed to account for basic knowledge of nuclear matter while still allowing for significant flexibility. Some of these models rely on a meta-modeling approach, i.e., a flexible model where the parameters can be tuned to reproduce present information from nuclear physics and/or constraints from astrophysics. For instance, a useful basis~\cite{Margueron:2017eqc,Zhang:2018vrx,Zhang:2019fog,Tsang:2020lmb} is a Taylor expansion of the energy per particle in terms of particle number density of neutrons and protons $n=n_n+n_p$ and the isospin asymmetry $\delta=(n_n-n_p)/n$ (see refs.~\cite{FiorellaBurgio:2018dga, Gulminelli:2021pth} for a discussion)
\begin{eqnarray}
    \frac{E(n, \delta)}{A} &=& E_{\rm sat} + \frac{1}{2!} K_{\rm sat} x^2 + \frac{1}{3!} Q_{\rm sat} x^3 + \ldots \nonumber\\
    &&
    +\left[ E_{\rm sym} + L_{\rm sym} x  + \frac{1}{2!} K_{\rm sym} x^2 + \frac{1}{3!} Q_{\rm sym} x^3 + \dots \right] \delta^2 \ +\ldots,
    \label{eq:e_taylor}
\end{eqnarray}
with 
$x=(n - n_{\rm sat})/(3 n_{\rm sat})$ denoting the deviation from saturation. This enables directly incorporating constraints on the coefficients of the expansion, that is, the empirical nuclear matter parameters such as the energy per particle at saturation $E_{\rm sat}$, the nuclear compressibility $K_{\rm sat}$, the symmetry energy $E_{\rm sym}$, the slope $L_{\rm sym}$ and curvature $K_{\rm sym}$ of the symmetry energy; for a relativistic version see~\cite{Char:2023fue}. While low-order parameters, up to $K_{\rm sat}$ ($L_{\rm sym}$) for the isoscalar (isovector) channel are relatively well constrained by ab initio nuclear theory and low-energy nuclear experiments, large uncertainties still affect higher order parameters. Other models employ a chiral effective field theory ($\chi$EFT)-inspired expansion in terms of the neutron and proton Fermi momenta~\cite{Lim:2019som}. The increased flexibility of these models makes them suitable for Bayesian inferences. The related meta-model approach, which is based on an empirical parameterization inspired by~\eq{eq:e_taylor}, enables predicting observables with uncertainties controlled by our present theoretical and experimental knowledge of nuclear and NS physics. 

\subparagraph{Phenomenological models.} Phenomenological models employ effective interactions to describe the properties of atomic nuclei. The forms of the interactions are motivated by ab initio theory while the values of parameters are adjusted by extensive fits to different data from nuclear structure, particle production and flow in heavy-ion collisions, microscopic calculations of neutron matter and, most recently, multi-messenger observations of NSs. The non-relativistic models use the Hartree-Fock technique to solve for the nucleonic equations of motion based on a Hamiltonian. Most models of NS EOS rely on the zero-range Skyrme interaction~\cite{Dutra:2012mb} while others employ momentum dependent Gogny interactions~\cite{Gonzalez-Boquera:2017rzy}. For densities above a few times $n_{sat}$, a number of deficiencies have been identified, for instance, superluminal sound speeds and Fermi velocities~\cite{Duan:2023amg} or negative symmetry energies. These make the extrapolation to arbitrarily high densities questionable unless these deficiencies are fixed. The relativistic models, also known as covariant density functional models, are based on relativistic Lagrangian densities. Most models of NS EOS correspond to the Hartree approximation and are finite-range~\cite{Dutra:2014qga}. 
Relativistic Hartree-Fock models exist as well~\cite{Sun:2008zzk}; the Fock term is responsible for a strong density increase of the symmetry energy, which leads to large proton fractions in the core and low threshold densities for the onset of direct Urca processes, responsible for fast NS cooling. The ensemble of these models is evocative for how the limited information on nucleon-nucleon effective interactions away from saturation and isospin symmetry gets translated into large uncertainties in dense and/or isospin asymmetric matter~\cite{Fortin:2016hny}. Use of both non-relativistic and relativistic phenomenological models in Bayesian inferences of dense matter EOS and composition~\cite{Ghosh:2021bvw,Beznogov:2022rri,Beznogov:2023jqp,Imam:2024gfh} helps identify correlations between nuclear matter parameters and NS properties and test their robustness with respect to the energy density functional.
The Induced Surface Tension model, a multicomponent parametric model for the EOS~\cite{Sagun:2018sps,Sagun:2018cpi} enters this category, too. 
An advantage of phenomenological models is their physical underpinning and the possibility to apply the constrained effective interactions to other physical environments, such as nuclear structure calculations. Extensions to finite temperature are straightforward. Extensions to incorporate non-nucleonic degrees of freedom (d.o.f.) have also  been developed. 

\subparagraph{Ab-initio models.} Ab-initio models represent a theoretical approach in which the EOS is derived by solving the complicated nuclear many-body problem~\cite{Baldo99} starting from few-body interactions. For the nucleon-nucleon case, very high precision interactions exist from meson-exchange~\cite{Machleidt:1989tm,Maessen:1989sx,Haidenbauer:2005zh,Cottingham:1973wt} and potential models~\cite{Lagaris:1981mm,Wiringa:1994wb}, or, more recently, from $\chi$EFT~\cite{Weinberg:1990rz,Weinberg:1991um,Epelbaum:2008ga}. The latter has the additional advantage that three-body forces can be constructed in a consistent manner rather than adding them purely phenomenologically. For other particles, such as hyperon-nucleon interactions, the situation is much less favorable, in particular due to the scarcity of experimental data.  
The most commonly used microscopic many-body approaches include the (Dirac)-Brueckner-Bethe-Goldstone theories \cite{Baldo99}, the self-consistent Green's function technique \cite{Kadanoff,Dickhoff08}, the variational method \cite{Akmal:1998cf}, the correlated basis function formalism \cite{Fantoni98}, Quantum Monte Carlo techniques \cite{Wiringa:2000gb,Gandolfi:2009fj}, ladder resummations that reorganize scattering data without resorting to assumptions on microscopic dynamics~\cite{Alarcon:2022vtn}, and the use of renormalisation group improved interactions~\cite{Bogner:2009bt} or many-body perturbation theory (see e.g. ref.~\cite{Drischler:2021kxf}). Details of these approaches are reviewed in~\cite{Oertel:2016bki,Burgio:2021vgk}. 

\subparagraph{Non-nucleonic degrees of freedom.} An interesting feature at high densities is the potential appearance of new d.o.f.
Generic arguments based on the energetics suggest that extra baryonic d.o.f., e.g., $\Lambda$, $\Sigma^{-,0,+}$ and $\Xi^{-,0}$ hyperons or $\Delta^{-,0,+,++}$ resonances can be populated in dense matter~\cite{Glendenning:1984jr}. High densities can also favor the appearance of mesons, e.g., kaons and pions, as well as meson condensates~\cite{Glendenning:1984jr}.
This includes inhomogeneous condensates~\cite{Maedan:2009yi} which entail an anisotropy and require a description of the stress-energy tensor $T^{\mu\nu}$ beyond a simple perfect-fluid EOS, as further discussed in the paragraph on nonhomogeneous energy-momentum tensors below.
Generally, the onset of any extra d.o.f.\ entails a softening of the EOS, that is a reduction of the pressure, and consequently a reduced NS maximum mass.
A wealth of studies nevertheless testify that the appearance of any of the above-mentioned particles is compatible with the current observational lower bounds on the NS maximum mass. 
The vast majority of models with these ``exotic'' d.o.f.\ currently available rely on covariant energy density functional models, see e.g.~\cite{Oertel:2014qza,Chatterjee:2015pua,Tolos:2016hhl,Fortin:2017cvt,Sedrakian:2022ata,Raduta:2021xiz}. The parameters governing the interactions of the non-nucleonic d.o.f.\ are, however, very poorly constrained. Rare measured properties of hypernuclei set some constraints on the hyperonic interactions~\cite{Gal:2016boi,Drago:2014oja} (more are expected from femtoscopy studies of baryon-baryon correlations in heavy-ion collisions) and many models apply symmetry arguments from an underlying simple quark model to relate the different coupling parameters and thus reduce the parameter space, see e.g.~\cite{Weissenborn:2011kb,Weissenborn:2011ut}. Under certain conditions, the onset of these d.o.f.\ can occur via a density-driven phase transition~\cite{Oertel:2014qza,Raduta:2021xiz}. 

Another possibility for matter above a few times $n_{\rm sat}$ is that deconfined quark matter phases emerge. Hybrid compact stars, i.e., stars with a quark core surrounded by a nuclear envelope containing a minimal liquid inner shell and a solid crust, represent an interesting subclass of compact stars where a phase transition to quark matter is assumed to take place. The fundamental theory of the strong interactions, QCD, predicts an asymptotically-high density phase of color superconductivity, which shares features with a perturbative-QCD gas but in which color and flavor symmetries are partially broken, locked into a remaining subgroup ``CFL'' phase~\cite{Alford:1997zt,Manuel:2000wm}. This provides insights  into the EOS of strongly interacting matter at presumably unreachably high densities. At lower densities, additional phases such as the 2SC in which only two quark flavours ($u$,$d$) participate in the condensate may exist. However, the theory cannot currently be employed in the low-temperature, finite density and neutron-proton asymmetry regime typically encountered in cold NSs. 
Therefore, in most cases, effective models such as the Nambu-Jona-Lasinio model, the MIT bag model, or a constant sound-speed parametrization are often used to model quark phases in hybrid stars, see, e.g. refs.~\cite{Alford:2013aca, Zdunik:2012dj}, and ref.~\cite{Blaschke:2018mqw} for a review. An alternative possibility for the state of quark matter in NSs is that a quarkyonic phase emerges, where a Fermi sea of quarks defines the matter and confining forces are only realized at the surface. Such a model can allow quarks to exist at rather low densities \cite{McLerran:2018hbz, Margueron:2021dtx} and interestingly does not necessarily lead to smaller radii for hybrid stars compared with purely hadronic ones~\cite{Somasundaram:2021ljr}.   An even more radical possibility is that strange quark matter is absolutely stable \cite{Witten:1984rs}. In that case compact stellar objects entirely made of strange quark matter can exist and have large masses and radii, while stars made of nucleons, hyperons and resonances can also exist and have small radii but not too large masses \cite{Drago:2013fsa,Drago:2014oja}.
Due to the phenomenological  nature and the impossibility to explore dense quark matter in terrestrial experiments, the parameters of the different models are difficult to determine, or to relate to one another and to QCD. However, these models allow testing the consequences of a hadron to quark phase transition on the NS's structure, cooling, oscillation modes, etc. Moreover,  there have been recent advances on informing the NS EOS using holographic approaches for describing strongly-interacting QCD matter at high baryon density~\cite{Hoyos:2021uff,Kovensky:2021kzl,Fraga:2013qra,Jarvinen:2021jbd,Hoyos:2016zke}, or with functional renormalisation group techniques~\cite{Otto:2019zjy,Otto:2020hoz,Leonhardt:2019fua}.

\subparagraph{Crust models.}
Thus far, we discussed models for matter at high density, which is the most relevant for describing the NS core. At lower densities, nuclear clusters are formed under screened electromagnetic or frustrated nuclear interactions and arrange in crystalline structures. The properties of the crust are important, for instance, for (accreting) NS cooling \cite{Horowitz:2014xca}, for magnetic, thermal and rotational evolution of pulsars \cite{Pons:2013nea}, NS ellipticity that can be potentially measured by GW observations \cite{Gearheart:2011qt}, and for explaining pulsar glitches \cite{Haskell:2017lkl}. In these regimes, i.e. at low density and relatively low temperatures (though this also holds at slightly higher temperatures for the so-called general-purpose EOSs, see section~\ref{sec:general-purpose}), different approaches are employed to compute the EOS and the composition, either (i) using the so-called single-nucleus (SNA) or one-component plasma (OCP) approximation, where the composition of matter is assumed to be made of one representative heavy nucleus (the one that is energetically favored), and possibly unbound nucleons, or (ii) considering the distribution of an ensemble of nuclei at finite temperature or using a multi-component plasma approach (often called nuclear statistical equilibrium, NSE, in the context of general purpose EOS). 
While the internal constitution of the outermost layers of a non-accreted, cold NS crust is completely determined by atomic masses that are either experimentally measured or predicted by theoretical mass tables, the description of the inner crust, where unbound nucleons are also present, relies on theoretical models.
Methods to calculate the NS crust usually treat nucleons as non-uniformly distributed within a cell of a given shape and size, either considering them as localized particles, or assuming a nucleon density distribution, in a charge neutralizing electron background.
Despite remarkable progress in the development of many-body methods, ab initio calculations of the crust are still unattainable, thus phenomenological models are employed.
Most of the crust calculations rely on the Wigner-Seitz cell approximation, and while a spherical cell is usually assumed at lower density, since nuclear clusters are (quasi-)spherical, at higher densities in the inner crust, other geometries have been considered that take into account the possible appearance of non-spherical structures collectively known as `pasta phases'~\cite{Ravenhall:1983uh}.
Within this picture, different methods have been used in the literature to model the inhomogeneous crust.
Fully quantum mechanical approaches that incorporate shell and pairing effects, which are important in determining the ground-state composition of the crust, include Hartree-Fock, Hartree-Fock+BCS,
and Hartree-Fock-Bogoliubov  approaches (see e.g.\ the pioneering calculations of ref.~\cite{Negele:1971vb}, and the more recent works~\cite{Grill:2011dr, Pais:2012js, Fattoyev:2017zhb}).
These calculations are however computationally expensive.
For this reason, many studies have employed compressible liquid-drop models (e.g.~refs.~\cite{Baym:1971ax, Baym:1971pw, Ravenhall:1983uh} and the more recent ones of refs.~\cite{Douchin:2001sv, Lim:2017luh, Gulminelli:2015csa, Carreau:2019tiv, Balliet:2020nsh, DinhThi:2023hfp, Grams:2021lzx, Scurto:2022vqm}), that parameterize the energy of the system in terms of individual contributions, such as bulk, surface, and Coulomb energy; nucleons inside the neutron-proton clusters and free nucleons outside are assumed to be uniformly distributed and treated separately.
A more realistic effective 
treatment involves the use of semi-classical methods such as the (extended) Thomas-Fermi model, that determines the density distribution of nucleons in the cell using a given energy density functional, thus allowing for a consistent modeling of nucleons ‘inside’ and ‘outside’ clusters (see e.g.\ refs.~\cite{Sharma:2015bna, Lim:2017luh, Shelley:2021pnx} for recent works); although shell effects which play a role for the composition for the inner crust are usually lost, they can be restored perturbatively (see e.g.\ the recent works of refs.~\cite{Pearson:2018tkr, Pearson:2022vep, Shchechilin:2023hkv} and references therein).
Alternative approaches such as time-dependent Hartree-Fock methods based on a wavelet representation (e.g. \cite{Sebille:2011zz}), classical molecular dynamics (e.g. \cite{Horowitz:2014xca,Barba22,barba24}), or quantum molecular dynamics (e.g. \cite{Nandi:2017aqq,Maruyama:2012bi,Mehta:2021ube}), have also been employed, see e.g. refs.~\cite{Haensel:2007yy, Chamel:2008ca, Oertel:2016bki, FiorellaBurgio:2018dga, Blaschke:2018mqw} for a review. 

\begin{figure}[t]
	\centering
		\includegraphics[width=0.9\textwidth]{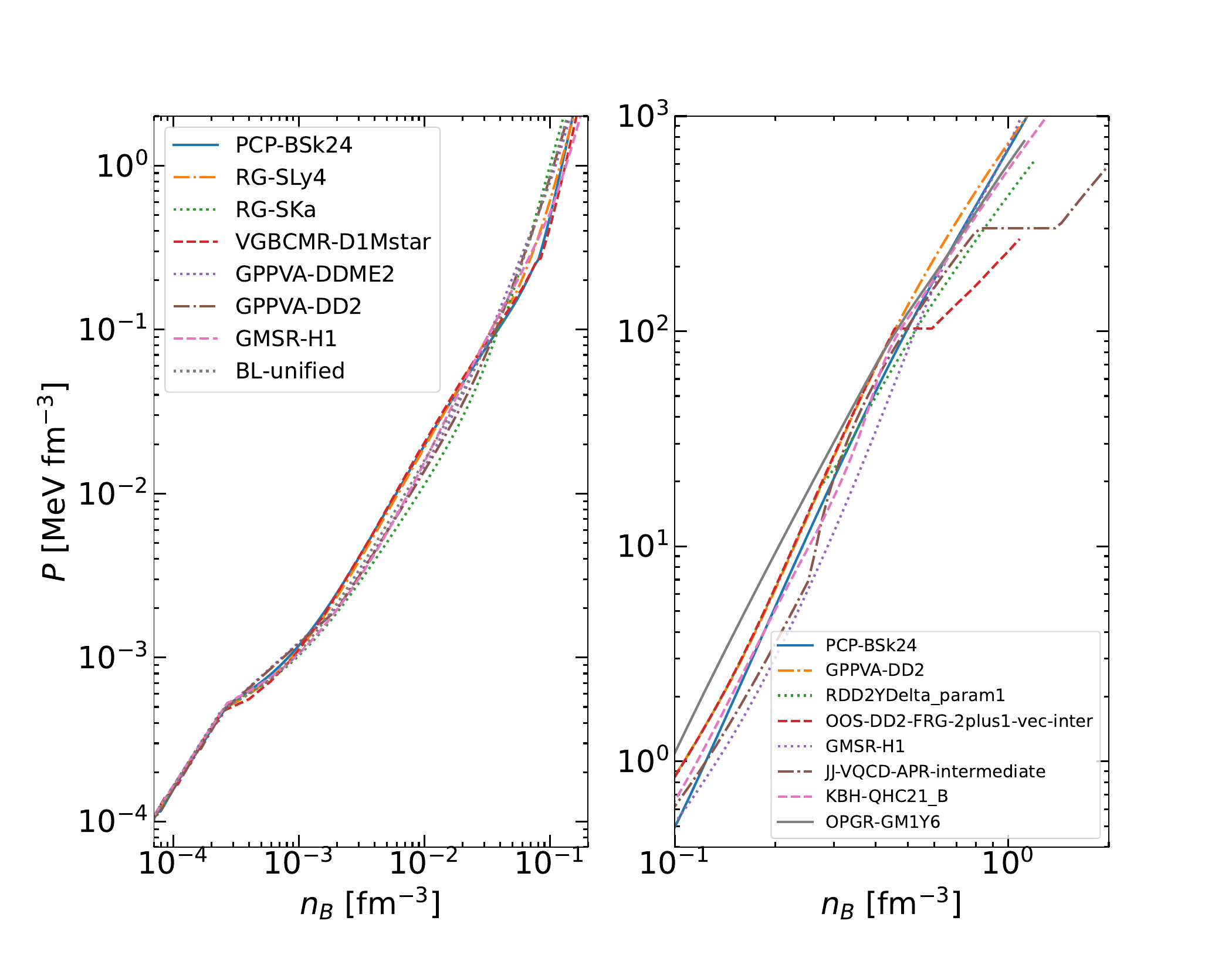}
		\caption{
    Pressure versus baryon number density for different EOS models taken from the CompOSE database~\cite{CompOSECoreTeam:2022ddl}. Left panel: different unified nucleonic EOSs in the crust-core density regime (changes of slope indicate the transition from the outer to the inner crust and from the crust to the core); right panel: different EOS models (nucleonic, with admixtures of hyperons and $\Delta$s, and with a phase transition to quark matter) in the high-density regime in the core.}
    \label{fig:EoS}
\end{figure}

We note that ensuring consistency between the description of the NS core and crust is important for EOS models, as has been thoroughly discussed in e.g.~\cite{Fortin:2016hny}. Most approaches to construct full NS EOSs have relied on an ad-hoc matching of three EOSs, corresponding to the outer and inner crusts and the core, computed using different models. Often, arbitrary values have been employed for the various transition densities. The rationale behind this procedure is that NS mass and radius are determined mainly by the core EOS. However, the crust-core matching procedures still alters the predicted values of NS tidal deformability and radius by a few to ten percent~\cite{Fortin:2016hny, Ferreira:2020wsf, Suleiman:2021hre}. 
Although these uncertainties seem relatively small compared with the current precision of observational data, they will become particularly relevant for next-generation GW observatories such as ET.
In order to minimize these uncertainties, the matching must be smooth, causal and thermodynamically consistent.
To avoid any arbitrariness in the matching procedure, so-called unified EOSs, where the same nuclear model is employed for the different regions of the NSs, must be used. Providing such unified EOS models has been the focus of much recent effort by the nuclear physics community, see e.g. refs.~\cite{Douchin:2001sv, Pearson:2018tkr, Gulminelli:2015csa, Fantina2013, Sharma:2015bna,Grill:2014aea,Scurto:2024ekq} and the \href{https:compose.obspm.fr}{\textsc{CompOSE} database}~\cite{Typel:2013rza,CompOSECoreTeam:2022ddl} for a collection. The \textsc{CUTER} tool~\cite{Davis:2024nda} enables constructing a consistent crust from a given core EOS.

Figure~\ref{fig:EoS} shows examples of a collection of beta-equilibrated cold NS EOSs from the \textsc{CompOSE} database: the left panel shows unified nucleonic EOSs (changes of slope indicate the transition between the outer and the inner crust, and between the crust and the core) in the crust-core density regime, while the right panel illustrates different EOSs (some of which include heavy baryons or a phase transition to quark matter) in the core. The largest uncertainties are indeed observed in the high-density regime in the NS core.

\subparagraph{Nonhomogeneous energy-momentum tensors.}
It is customary to approximate the energy-momentum tensor of NS matter by that of a perfect fluid in thermodynamic equilibrium, $T^{\alpha\beta}=(\epsilon+P)u^\alpha u^\beta+P g^{\alpha\beta}$ with a local fluid velocity $u$ satisfying $u^2=-1$ (in geometric units), energy density $\epsilon$ and pressure $P$.  Later on we will discuss dissipative effects, for example in subsection~\ref{subsec:reactionsandtransport}, which can be implemented by adding a non-ideal fluid piece $\tau_{ij}$ to the energy-momentum tensor. Further,  anisotropies in NS matter can  develop which require a non-rotationally invariant form for the tensor, affecting the formulation of the Einstein equations. This can happen for various reasons~\cite{Baiko:2017hnf}. One is the presence of external privileged directions such as magnetic fields~\cite{Bocquet:1995je,Pili:2013bwa,Chatterjee:2014qsa}, see \ref{sec:Magnetic} below, or rotation vortices which affect matter at smaller scales. But matter can also organize itself in crystalline structures to relax the electrostatic energy (Coulomb interactions in the crust) or to alleviate the pressure (because neutrons exclude a finite volume due to the repulsive $n-n$ short-range potential, arranging them in compact packings can relax the free energy)~\cite{Llanes-Estrada:2011nal} and this entails a breaking of rotational invariance. 
Anisotropies can also arise when two phases coexist~\cite{Glendenning:2001pe}.
Likewise, if quark phases emerge in NSs, it is possible that inhomogeneous chiral condensates may form, although this is still under active theoretical investigation~\cite{Motta:2023pks}.Thus, it is interesting to explore the capabilities of detectors such as ET to probe for anisotropic stress-energy tensors, which would have important implications for understanding the microscopic physics of NSs.

\paragraph{General purpose EOS.}
\label{sec:general-purpose}
Matter in CCSN, PNSs, or the post-merger phase of a BNS coalescence, is heated and not necessarily in $\beta$-equilibrium. Therefore, thermal and compositional effects have to be incorporated in the EOS. Matter is assumed to consist of hadrons or, alternatively, deconfined quarks, leptons and photons and taken to be in thermal equilibrium for hadrons, photons and charged leptons, as well as chemical equilibrium with respect to the strong interaction. Commonly, the corresponding EOSs, called ``general purpose'', are provided as functions of temperature $T$, baryon number density $n$, and the charge fraction $Y_Q$ (if electrons are the only considered charged leptons and charge neutrality holds, $Y_Q = Y_e$; if muons are included, then $Y_Q = Y_e + Y_\mu$, with $Y_\mu$ the muon fraction); see e.g. ref.~\cite{CompOSECoreTeam:2022ddl}. 
    
Charged leptons and photons are in general simply considered as ideal gases and the discussion below therefore focuses on the contribution of strongly interacting matter, hadrons and/or quarks. Due to the computational efforts needed to cover the entire range of thermodynamic parameters and the different matter compositions from a gas of (interacting) nuclei to potentially a color-flavor locked condensate phase, most of the existing general purpose EOS employ phenomenological models for the interactions with a few exceptions. EOSs with various particle blends have been built. Many of these general purpose EOSs are available for public use on \textsc{CompOSE}~\cite{Typel:2013rza,CompOSECoreTeam:2022ddl}. Information on thermodynamic properties and optionally on particle composition and microscopic quantities such as interaction potentials is provided in tabular form
with a mesh fine enough to allow usage in numerical astrophysical simulations or data interpolation.

In the absence of constraints from nuclear experiments, 
insights into the thermal behavior of matter can only be gained by confronting the predictions of available models against each other and by comparing with ab initio calculations. 
Such comparisons reveal that -- at least for purely nucleonic models -- the effective mass and its density derivative entering the statistical partition function play the dominant role in the thermal corrections to thermodynamic state variables~\cite{Constantinou:2014hha,Constantinou:2015mna,Raduta:2024awt}. However, due to the different ways the effective mass depends on the underlying interaction within the different types of models, large qualitative differences exist between the predictions (e.g.~\cite{Raduta:2021coc}), and it is difficult to attribute a generic thermal behavior to the EOS.  

Non-nucleonic degrees of freedom can also be thermally excited and are generally found to increase in abundances with temperature, see e.g.~\cite{Banik:2014qja,Marques:2017zju,Oertel:2016xsn,Dexheimer:2017nse,Raduta:2020fdn,Sedrakian:2022ata,Kochankovski:2022rid}.
The thermodynamic properties are very sensitive to the number of particle d.o.f.\ and less sensitive to the details of the underlying interaction~\cite{Raduta:2022elz}. A potential phase transition from hadronic to quark matter
along with the technical solution adopted to build the phase coexistence impacts all thermodynamic state variables, too, see~\cite{Raduta:2022elz} and references therein. In particular, a phase coexistence built according to the Maxwell construction, where the pressure in the mixed phase remains constant, leads to vanishing
values for the speed of sound, while a Gibbs construction leads to a smooth density dependence in most cases.  

Approximate methods for extending cold EOSs to finite temperatures exist as well, with flexibility and low computational cost being their most important assets.
The most commonly used example is the $\Gamma$-law~\cite{Hotokezaka:2013iia}, which
consists of supplementing the cold EOS with an ideal-gas-like contribution. 
As such, any density-, temperature-, composition- and interaction-dependence is disregarded, which represents a serious drawback. An improved solution is provided by attempts which use the finding that the effective mass dominates thermal effects in the EOS, see for instance~\cite{Huth:2020ozf,Raithel:2019gws}. The analytical expressions proposed in \cite{Raithel:2019gws} are able to reproduce the thermal pressure of a given class of models to within a few percent. However, as stated above, it cannot cover the model dependence and thus  uncertainties in thermal effects. 

\subsubsection{Current constraints on EOS and matter composition}
\label{ssec:current-constraints}
At present, the different approaches to the NS EOS yield generally similar results up to about twice the saturation density and significantly diverge at larger density~\cite{Burgio:2021vgk}. 
The main source of uncertainty is the extrapolation of the nucleonic interaction to large density. 
In fact, nuclear data from both experiments and theoretical ab initio calculations
provide information only up to less than two $n_{\rm sat}$. For the phenomenological EOS models, their spread at high density can be related to the use of parameters whose values are fitted on experimental data that is approximately available for symmetric matter around saturation density; for the ab-initio approaches, it is due to the different high-density behavior of nucleonic three-body forces, which are mainly constructed at saturation density and then extrapolated. NS matter is also poorly constrained due to it being neutron rich, i.e. very isospin asymmetric. 
Apart from first-principle constraints including causality, current constrains on the EOS come from nuclear physics (mainly from nuclear structure data, heavy-ion collision experiments, and theoretical ab initio calculations of pure neutron matter) and from astrophysical observations.

The theoretical ab initio calculations of pure neutron matter are particularly interesting since they impose constraints on the symmetry energy, i.e. the behavior of the EOS upon passing from symmetric nuclear matter to more neutron rich matter as in NSs. They have been implemented in several approaches to the EOS, see e.g.~\cite{Oter:2019rqp,Essick:2020flb,Tews:2018iwm,Thi:2021jhz}, but again the range of validity is limited to below $2 n_{\rm sat}$. At higher density, much progress has recently been achieved on completing higher-loop calculations in perturbative QCD to describe matter at densities $\gtrsim 40 n_{\rm sat}$, see e.g. refs.~\cite{Ghiglieri:2020dpq,Komoltsev:2023zor}. As discussed in refs.~\cite{Kurkela:2022elj,Komoltsev:2021jzg}, improving the theoretical knowledge in such high-density regimes well beyond what is encountered in NSs can have implications for NS EOS models. 

Laboratory experiments also give information up to or slightly above saturation density, and especially for isospin-symmetric matter (see e.g. refs.~\cite{Oertel:2016bki,FiorellaBurgio:2018dga,Burgio:2021vgk} for a review). Note that the recent PREX-II (CREX) measurement of neutron skins in $^{208}$Pb ($^{48}$Ca) (see ref.~\cite{Reed:2023cap} and refs. therein) is sensitive to the symmetry energy, too; see, however, e.g. refs.~\cite{Essick:2021kjb,Mondal:2022cva} for a critical discussion of the interpretation for parameters in~\eq{eq:e_taylor} in view of other constraints. 
(Semi-)agnostic approaches, too, show the impact of the lack of data beyond that range in density and for asymmetric matter. As discussed previously, the situation is even more complicated for hot and dense matter, since currently there are no experimental or observational data in the relevant parameter regimes, and only benchmark ab initio calculations can provide some theoretical information. 

Astrophysical observations have led to constraints on the NS EOS in various ways.
 For example, information on the EOS comes from NSs with high masses $\gtrsim 2M_{\odot}$~\cite{Demorest:2010bx,Antoniadis:2013pzd,NANOGrav:2017wvv,NANOGrav:2019jur,Fonseca:2021wxt}, which sets a lower limit on the maximum mass the EOS needs to support, joint estimates of NS mass and radius by NICER~\cite{Miller:2019cac,Riley:2019yda,Miller:2021qha,Riley:2021pdl}, and tidal deformability measurements from GW170817~\cite{LIGOScientific:2017vwq,LIGOScientific:2017ync}. To be more precise on the latter, during the inspiral of coalescing BNSs,  
the dominant EOS-dependent effects in the emitted GWs is characterized by the tidal deformability parameter $\Lambda$ that quantifies how easily one object in a binary system is deformed due to the gravitational field of the companion; see~\cite{Chatziioannou:2020pqz} and references therein for a review. 
Considering a static quadrupolar tidal field $\mathcal{E}_{ij}=C_{0i0j}$ in the rest frame of the NS, with $C_{0i0j}$ the time-space components of the Weyl curvature tensor due to the companion, the induced quadrupole moment $Q_{ij}$ characterizing the quadrupolar deviation of the NS exterior spacetime away from spherical symmetry at large distances, is given to linear order by \cite{Damour:1984rbx,Damour:2009vw,Hinderer:2007mb,Binnington:2009bb}
\begin{equation} \label{eq:tidal-deformability}
	Q_{ij} = - \lambda_2 \mathcal{E}_{ij} \ .
\end{equation}
The tidal deformability $\lambda_2$ is expressed in terms of the $l=2$ tidal Love number $k_2$ (in units $c=G=1$) as\footnote{In the Newtonian limit the perturbing tidal field $\mathcal{E}_{ij}$ is defined as the second spatial derivative of the external field, resulting in units of inverse length squared \cite{2014grav.book.....P} whereas the quadrupole moment has units of length cubed.}
\begin{equation}
	\lambda_2 = \frac{2}{3}k_{2}R^{5} \ ,
\end{equation}
with $R$ the star radius, and $k_{2}$ has typical values of order $0.1$ but strongly depends on the NS mass and EOS~\cite{Chatziioannou:2020pqz}.
A related quantity is the dimensionless tidal deformability defined as
\begin{equation}
\label{eq:biglambdadef}
	\Lambda \equiv \frac{\lambda_2}{M^{5}} = \frac{2}{3}k_{2}\frac{R^{5}}{M^{5}} = \frac{2}{3}k_{2}C^{5},
\end{equation}
where $M$ is the mass of the star and $C\equiv M/R$ is its compactness. Although $\lambda_2$ and $k_{2}$ are the most commonly considered tidal parameters in the context of NS binary systems, they are only the first in a series of tidal parameters of different types and orders that arise from the study of generic static tidal perturbations. However, higher multipolar contributions of $\lambda$ beyond the quadrupole $\lambda_2$ (e.g. the octupolar $\lambda_3$ or hexadecupolar $\lambda_4$ contributions) are much smaller. Their values are computed by solving the Einstein field equations coupled with the matter equations of motion for linear, static perturbations around an equilibrium configuration. Current models used for measurements also include effects of spin-induced multipole moments, which contribute GW imprints from EOS-dependent rotational Love numbers~\cite{Laarakkers:1997hb}. In addition, effects of dynamical tides due to the fundamental modes of the NS introduce a further dependence of the GW signal on the EOS-dependent fundamental mode frequency. Current data are not sensitive to all these parameters, and thus in the data analysis for GW170817 they were eliminated in favor of $\Lambda$ by using quasi-universal relations. At this point, it is worth highlighting that although EOS-independent relations between NS macroscopic parameters have often been employed when interpreting GW data, e.g.,~\cite{LIGOScientific:2018cki}. For next-generation observatories, the use of quasi-universal relations in parameter extraction leads to larger uncertainties on the extracted parameter than direct EOS inference \cite{Suleiman:2024ztn,Kashyap:2022wzr}. 

A variety of EOS models from the different categories discussed in section~\ref{sec:eos} have been employed with different levels of input from nuclear experiments and theory to study the impact of the measurement of $\Lambda$ from the event GW170817 (see e.g. ref.~\cite{Chatziioannou:2020pqz} for a review) and other astrophysical data on the NS EOS, the composition of dense matter, empirical nuclear matter parameters or parameters of a particular model. The availability of large computing resources have thereby motivated numerous statistical analyses to infer the information employing different techniques, such as Bayesian analyses or machine learning, focusing on a particular measurement or combining different ones, see e.g.~\cite{Annala:2017llu,Miller:2021qha,Riley:2021pdl,Zhang:2019fog,Lim:2019som,Ferreira:2019bgy,Ferreira:2021pni,Patra:2022yqc,Malik:2022zol,Malik:2023mnx,Beznogov:2022rri,Beznogov:2023jqp, Traversi:2020aaa}.  

The main results can be summarized as follows:

(i) The progressive incorporation of different observations narrows down the cold EOS, but current constraints remain loose and are insufficient to discriminate between models or to determine the composition of high density (and hot) matter. Indeed, the relation between EOS and composition is not free from ambiguity, and very similar EOSs (thus almost indistinguishable mass-radius or mass-$\Lambda$ relations) can be obtained under different hypotheses on the underlying microphysics, giving rise to the so-called ``masquerade'' effect \cite{Blaschke:2018mqw}. We also note that even a very precise knowledge of the NS EOS does not pin down the matter composition without any additional information, such as accurate
measurements on the symmetry energy at high density or possibly complementary information from dynamical NS
observables~\cite{Mondal:2021vzt,Iacovelli:2023nbv}. The composition and in particular a change with respect to a purely nucleonic one considerably affects transport properties and thus NS cooling~\cite{Yakovlev:2004iq} as well as the evolution of oscillation frequencies~\cite{Barta:2019xhg}, but no conclusive measurements exist today, although the recent discovery of three young and cold objects might be interesting in this respect~\cite{Marino:2024gpm};

(ii) In addition to the reduced maximum mass, the onset of new d.o.f.\ in general leads to a sizable reduction of radii and tidal deformabilities for intermediate mass NSs, too. Current constraints can accommodate the presence of additional particles, such as $\Delta$ isobars (e.g. \cite{Ribes:2019kno,Li:2019tjx}) or hyperons \cite{Sedrakian:2022ata,Nandi:2018ami,Biswas:2020puz,Malik:2022jqc,Tolos:2017lgv}. The semi-agnostic modeling of a nucleonic EOS has also been used as a null hypothesis to search for exotic d.o.f., indicating a mild tension between the GW data and ab-initio nuclear physics calculations~\cite{Guven:2020dok}, however, the statistical significance of the discrepancies is currently insufficient to be conclusive.
Overall, current astrophysical data are not constraining enough and remain compatible as well with non-nucleonic as with  purely nucleonic EOSs \cite{Gulminelli:2021pth, Thi:2021jhz, Lim:2019som,Malik:2022zol};

(iii) If the non-nucleonic particles appear via a first-order phase transition, with a density jump at constant pressure, the global properties of compact stars may differ significantly from the nucleonic counterparts. In particular, NSs with the same mass but different compactness dubbed ``twin stars'' can appear in the mass-radius diagram~\cite{Alford:2017qgh,Blaschke:2019tbh} and similarly for tidal deformability~\cite{Paschalidis:2017qmb}. Also in the case of the two-families scenario, in which strange quark stars co-exist with neutron stars \cite{Drago:2015cea}, the existence of stellar objects having the same mass but different radii and tidal deformabilities is predicted \cite{Burgio:2018yix}. Sharp phase transitions thereby lead to very small values of tidal deformabilities and induce discontinuities in the tidal deformability--mass relation~\cite{Damour:2009vw,Postnikov:2010yn,Han:2018mtj,Sieniawska:2018zzj}. Current data are compatible with a phase transition in compact stars~\cite{Christian:2018jyd,Montana:2018bkb,Annala:2019puf,Pang:2023dqj,Takatsy:2023xzf} and several indications could be found on its properties, e.g. it was concluded that a sizable quark core can exist inside a heavy NS ($\sim 2M_{\odot}$) \cite{Annala:2019puf} or that strong phase transitions with a visible jump in the mass-radius relation are ruled out at $1 \sigma$ at densities below 1.7 times saturation density~\cite{Christian:2019qer}. Agnostic studies also found a weak statistical evidence in favor of two stable branches with current data, that is, the existence of hybrid stars with a strong phase transition~\cite{Landry:2020vaw,Essick:2020flb}; 

(iv) Concerning the constraints on empirical nuclear matter parameters, the results from NS observables have almost no effect on low-order parameters, the latter being mainly constrained (in the isoscalar sector) by nuclear physics information. A slight trend to favor ``soft'' EOS with relatively low values for $L_{\rm sym}$, $Q_{\rm sat}$ and $K_{\rm sym}$ has been observed without being conclusive~\cite{Sedrakian:2022ata,Tsang:2019vxn,Thi:2021jhz,Zhu:2022ibs, Huang:2023grj}.

In brief, current constraints remain inconclusive and a larger sample and higher precision measurements of NS properties together with inputs from subatomic physics are needed to advance our understanding of dense matter.



\subsubsection{Reaction rates}\label{subsec:reactionsandtransport}
\label{sec:reactions}
Isolated mature NSs and NSs in the inspiral phase of BNS mergers are typically in near thermal and chemical equilibrium. As a result, knowing the EOS of dense matter is sufficient to relate its microphysical properties to the macroscopic parameters of NSs, such as mass, radius, tidal deformability. However, during the postmerger phase of BNS mergers and in core-collapse supernovae, including also the physics of ejecta and nucleosynthesis, matter is out of chemical (and thermal) equilibrium. In these scenarios, additionally reaction rates governing the various out-of-equilibrium processes are required to model the dynamics of these events and, ultimately, to extract information about subatomic physics from observations. This subsection provides a brief overview of the current state of the key ingredients needed to describe out-of-equilibrium processes. We will omit details about the precise implementation of these effects since there are many different ways to treat them, ranging from a simple source of dissipation within a hydrodynamics description to a full transport coupled to hydrodynamics via conservation equations (particle numbers, energy and momentum). The current incomplete understanding of these processes within dense matter and degeneracies with the EOS and other microphysical parameters limits our ability to draw definitive conclusions from future observations and will need to be improved to optimize the science returns with ET and multimessenger observations. 

\paragraph{Neutrino rates.}
\label{sect_nu_viscosities}

Neutrinos interact with matter through weak interactions. They reduce the lepton number of the system and affect the local cooling and heating via energy- and momentum transfer. Neutrinos play a central role in the dynamics of CCSN and largely determine the neutron-to-proton ratio in the ejecta of CCSN and BNS mergers, and thus the conditions for (heavy element) nucleosynthesis. Current and future neutrino observatories such as Super/Hyper-Kamiokande are capable of detecting neutrinos from an event in the Galaxy and its close environment. Understanding neutrinos in CCSN and BNS mergers is thus important for a better understanding of the multi-messenger emission by these transient events. 
Neutrino interactions are most important in the vicinity of the neutrinosphere, i.e., in the transition region from trapped neutrinos close to equilibrium to free streaming neutrinos which only feebly interact with matter. Although the exact location of the neutrinosphere depends on (anti-)neutrino energies and precise thermodynamic conditions, due to the strong variation of the reaction rates with temperature, the latter is often taken as  indicator for the neutrinosphere with values in the range of a few MeV relevant for CCSN and BNS mergers.

As a first approximation, the effect of neutrino absorption and creation processes can be included via its contribution to bulk viscous dissipation. Bulk viscosity from weak interactions is expected to be effective in damping post-merger density oscillations. The treatment of neutrino dynamics via a kinetic equation allows to include their reactions with the environment via opacities~\cite{Sawyer:1978qe}. 
These opacities can be utilized in various numerical schemes without assuming specific properties of the (anti-)neutrino distribution function, as is self-consistently determined from the transport equation.
A full transport approach also enables the inclusion of other neutrino reactions such as scattering and pair creation via neutral current interactions. In addition, neutrino scattering processes could in principle provide viscosity to influence turbulent flow. Due to the large neutrino mean free paths, however, it is generally assumed that the magnetic field is much more efficient in this respect, see e.g.~the discussion in ~\cite{Radice:2024gic}. It should be noted that neutrinos influence matter properties locally, e.g. the temperature and composition, and its detailed impact on turbulence is not yet fully understood.
 
\subparagraph{Bulk viscosities.}
The main weak reaction for bulk viscosity, first suggested in the study of cold matter~\cite{Sawyer:1989dp}, arises from direct Urca processes describing neutron decay $n \to p + e + \bar{\nu}_{e}$, where $\bar{\nu}_{e}$ is the electron antineutrino, and the corresponding electron/positron capture or proton decay reactions. Recent research has extensively explored its contribution to bulk viscosity in both neutrino-trapped and transparent regimes in a hot and dense environment~\cite{Alford:2019qtm,Alford:2019qtm,Alford:2020pld,Alford:2021lpp,Alford:2020lla,Alford:2023uih}. In addition, muonic Urca~\cite{Alford:2023uih} and modified Urca~\cite{Suleiman:2023bdf}, which involves bystander particles that contribute to energy-momentum balance, may contribute. 
Results for the
bulk viscosity and damping timescales that cover the entire range of temperatures and densities are available 
in~\cite{Alford:2020lla,Alford:2022ufz} showing that bulk viscosity is most relevant at $T\sim 5$~MeV where the transition between the trapped to untrapped neutrino regimes occurs. 
Bulk viscosities can be significantly enhanced due to resonant behavior \cite{Alford:2010gw} and can arise due to phase conversion dissipation~\cite{Alford:2014jha} in NSs that also contain non-nucleonic matter. 

\subparagraph{Neutrino opacities within a transport approach.}

Approximating the effect of neutrino interactions via their contribution to (bulk) viscous dissipation implies averaging over neutrino phase space assuming free streaming. By contrast, neutrino transport allows to follow the dynamics throughout the system and thus requires information on the energy and angular dependence of the reaction rates. The importance of accurate neutrino transport in supernova simulations has spurred numerous studies of the relevant reactions. Nevertheless, it remains challenging to obtain reliable predictions. On the one hand this is due to the difficulties in determining the nuclear (strongly interacting) matrix element and on the other hand due to the computational cost of the high-dimensional phase space integration. The former clearly needs information beyond the EoS, starting with detailed information about the composition.

Concerning the different reactions, the direct Urca type reactions on nucleons contributing to bulk viscosity are in most cases the dominant ones for (anti-)neutrino absorption and creation and thus to determine the neutron to proton ratio, information lost within a viscous description. Opacities for these processes have been extensively studied in the literature, applying different levels of approximation for both matrix element and phase space integration, see e.g.~\cite{Bruenn:1985en,Reddy:1997yr,MartinezPinedo:2012rb,Roberts:2016mwj,Oertel:2020pcg}. At high densities, the predictions can vary by more than one order of magnitude, and depend strongly on the treatment of nuclear excitations. Under certain conditions, reactions on other particles such as hyperons or pions~\cite{Fore:2019wib} or in quark matter~\cite{Iwamoto:1982zz,Jarvinen:2023xrx}, or involving muons as charged leptons~\cite{Fischer:2018kdt} can contribute as well.
In the low-temperature limit, the direct Urca reactions are kinematically suppressed and the modified Urca processes which involve an additional particle for energy-momentum conservation must be considered, see e.g.~\cite{Roberts:2012um,Alford:2018lhf,Suleiman:2023bdf}. 

For neutrino-nucleon scattering kernels, different approximations for treating the nuclear interactions in dense matter have been considered, see e.g.~\cite{Bruenn:1985en,Reddy:1997yr,Thompson:2002mw,Horowitz:2006pj,Vidana:2022ket}. In this case the level of approximation not only affects the opacity, but also influences the scattering angle distribution~\cite{Duan:2023amg}. 
In contrast to the modified Urca processes, the corresponding neutral current process, the Bremsstrahlung reaction, has received some attention in dense and hot matter~\cite{Sigl:1997ga,Hannestad:1997gc,Sedrakian:2000kc,Hanhart:2000ae,Bartl:2014hoa,Guo:2019cvs} since it is known to influence neutrino spectra in CCSN.  Electron capture (EC) on nuclei plays an important role during the late stages of stellar evolution and the pre-bounce phase of CCSN, see e.g.~\cite{Janka:2006fh}. Total EC rates are influenced both by the nuclear distribution and by the rates on individual nuclei, both of which suffer from uncertainties. The former is very sensitive to nuclear masses and for the latter, although microscopic
calculations are available for a large number of different nuclei, see~\cite{Langanke:2020gbk} and references therein, the nuclei identified in several studies~\cite{Sullivan:2015kva,Titus:2017fln,Pascal:2019awl} as having the highest impact lie
outside the region where microscopic calculations exist and parameterizations~\cite{Langanke:2003ii,Raduta:2016hca} are often used. Theoretical and experimental efforts are underway to improve the situation and obtain reliable rates for all relevant nuclei~\cite{Titus:2019pcw,Dzhioev:2019qhq,Giraud:2021fvz,Litvinova:2020ebb}.
Nuclear abundances also influence the neutrino opacity via
neutrino-nucleus scattering~\cite{Furusawa:2017esp}.
Standard rates for these reactions including in addition purely leptonic processes and plasmon decay are available in publicly accessible libraries like \textsc{NuLib}~\cite{OConnor:2014sgn} and \textsc{Weakhub}~\cite{Ng:2023syk}. 

\paragraph{Nucleosynthesis and reaction rates.}
\label{sect_nuclEOSynth}

As detailed in section~\ref{ssec:multimessenger}, compact binary mergers involving at least one NS have been shown, both observationally and theoretically, to be a key astrophysical site for the production of neutron-rich nuclei by the rapid neutron-capture process (or r-process). 
Despite the recent success of nucleosynthesis studies for NS mergers, the estimated abundances and decay heat in the ejecta are still affected by a variety of uncertainties, including those associated with the complex nuclear physics description of exotic neutron-rich nuclei. R-process nucleosynthesis calculations require  a reaction network consisting of about 5000 species from protons up to $Z \simeq 110$ lying 
between the valley of $\beta$-stability and the neutron drip line \cite{Arnould07,Cowan21}. All charged-particle fusion 
reactions on light and medium-mass elements that play a role when the nuclear statistical equilibrium freezes out need to be 
considered, in addition to radiative neutron captures and photodisintegrations. On top of these reactions, $\beta$-decays as well as $\beta$-delayed neutron emission probabilities and $\alpha$-decay rates need to be taken into account, but also fission processes, such as neutron-induced, spontaneous, $\beta$-delayed and photofission, together with the corresponding fission fragment distribution for all fissioning nuclei.  All rates should ideally be based on experimental information whenever available, but since only an extremely tiny amount of data concerning the neutron-rich nuclei produced during the r-process are known experimentally, theoretical models are fundamental. 

For the estimate of reaction and decay rates of relevance in r-process simulations, the necessary nuclear ingredients (properties of cold and hot nuclei, nuclear level densities, optical potentials, $\gamma$-ray strength functions, fission properties, $\beta$-strength functions) should ideally be derived from {\it global}, {\it universal} and {\it microscopic} models \cite{Goriely15b}. 
Nowadays, microscopic or semi-microscopic mean-field-type models can be tuned to the same level of accuracy as the phenomenological models, and therefore could replace the phenomenological inputs in astrophysical simulations \cite{Hilaire16}. 
Some details about existing models to estimate the key ingredients to nucleosynthesis reaction network calculations are given below. They concern nuclear masses, $\beta$-decay and reaction rates.

\subparagraph{Nuclear masses.}

Among the properties of ground state matter, the atomic mass is the most
fundamental quantity (for a review, see e.g. \cite{Lunney03}) and governs the r-process path through the competition between inverse channels like the dominant (n,$\gamma$) and ($\gamma$,n) reactions. Uncertainties in the mass predictions may significantly affect the r-process predictions and consequently the ejecta composition and decay heat resulting from NS merger  \cite{Kullmann23}.
When not available experimentally \cite{Wang21}, these quantities need to be extracted from a mass model which
aims at reproducing measured masses as accurately as possible, i.e typically with a root-mean-square (rms) 
deviation smaller than 800~keV. Modern mass models not only try to reproduce experimental masses and mass differences, but also charge radii, quadrupole moments, giant resonances, fission barriers, shape isomers, infinite nuclear matter properties, and more~\cite{Lunney03,Pearson00}.

Nowadays, nuclear structure properties can be accurately determined within the (relativistic or non-relativistic) mean-field model \cite{Bender03}. This is a variational approach in which the trial wavefunction of the nucleus has the form of a Slater determinant of single-particle wavefunctions. The energy is then calculated by minimizing the expected value of the Hamiltonian with respect to arbitrary variations in the single-particle wavefunctions. Since the mean field itself depends on these wavefunctions, the process has to be reiterated until an acceptable level of self-consistency
is reached. Such non-relativistic Hartree-Fock-Bogolyubov  calculations based on the Skyrme zero-range \cite{Goriely16a,Ryssens22} or Gogny finite-range \cite{Goriely09a} effective interactions have proven their capacity to describe experimental data, provided the interaction parameters are directly fitted to the complete set of known masses. 
The accuracy achieved by relativistic mean field  models also still needs to be improved, as the most successful interaction still leads to rms deviations larger than about 1.2~MeV. Other global mass models have been developed, essentially within the so-called liquid drop or droplet approach, where the binding energy is macroscopically described by the leading volume and surface terms as well as a symmetry energy (also with a volume and surface contribution) and a Coulomb energy. Microscopic corrections to account for quantum shell and pairing correlation effects need to be added to the liquid drop part \cite{Lunney03}. Despite its success, this macroscopic-microscopic approach remains unstable with respect to parameter variations, especially when approaching the neutron drip line, and suffers from major shortcomings, such as the incoherent link between the macroscopic part and the microscopic correction or the instability of the shell correction \cite{Lunney03,Pearson00}. 

\subparagraph{$\beta$-decay rates.}

$\beta$-decay rates are fundamental in much of nuclear astrophysics  \cite{Arnould20,Suzuki22}, particularly so in r-process nucleosynthesis \cite{Arnould07,Cowan21} for which they set the timescale of the nuclear flow and consequently of the production of the heavy elements \cite{Kullmann23}. Most of the nuclei involved during the r-process neutron irradiation have yet to be discovered, with $\beta^-$-decay half-lives known only for about 1200 nuclei \cite{Kondev21}.
A proper prediction of r-abundances in any r-process site calls for $\beta$-decay rates to be estimated within a factor smaller than typically 1.5. However, only a restricted number of global models of $\beta$-decay rates are available for nucleosynthesis applications \cite{Moller03,Klapdor84,Marketin16,Ney20,Minato22} and 
 deviations between their predictions can reach factors of 100 for many neutron-rich regions. In particular, for very heavy $Z\ga 82$ nuclei, as well as along the isotonic chains corresponding to closed neutron shells ($N=50, 82, 126, 184$), responsible for the formation of the r-process peaks observed in the solar system, non-negligible differences can be observed, leading to different estimated r-process peak structures \cite{Goriely15c}.

More effort is needed in the future to include not only the contribution of the forbidden transitions \cite{Marketin16,Ney20} but also deformation effects, as the majority of nuclei is expected to be deformed \cite{Martini14}. Recent studies within the fully self-consistent proton-neutron Quasi-particle random phase approximation (QRPA) model using the finite-range Gogny interaction have now also consistently taken axially symmetric deformations into account \cite{Martini14}, but forbidden transitions remain to be included. The inclusion of finite-temperature effects as well as the phonon coupling has also been shown to give rise to a redistribution of the QRPA strength and significantly impact the $\beta$-decay half-lives of neutron-rich nuclei \cite{Litvinova20}. Inclusion of particle-vibration coupling effects has also been shown to redistribute the Gamow-Teller strength and impact the $\beta$-decay half-lives of neutron-rich nuclei significantly \cite{Robin19}. 
Finally, note that on the basis of the $\beta$-decay strength, the $\beta$-delayed processes, including neutron emission and fission for the heaviest species, need to be derived. Detailed calculations on the basis of statistical reaction codes, like TALYS \cite{Koning:2023ixl}, can account for the competition of the various open channels (neutron, photon, fission) in the daughter nucleus. Reaction models still need to be better exploited to estimate the probability for such $\beta$-delayed processes.

\subparagraph{Neutron capture rates.}

If an $(n,\gamma)\rightleftarrows (\gamma,n)$ equilibrium is established during the r-process, the abundances and decay heat in the ejecta of NS mergers remain rather insensitive to the adopted reaction rates. However, even in this case, i.e. to prove that an $(n,\gamma)\rightleftarrows (\gamma,n)$ equilibrium is achieved, it is necessary to estimate the rates and perform network simulations to show that neutron captures and photoneutron emissions are faster than $\beta$-decays in the conditions prevailing during NS mergers. If an $(n,\gamma)\rightleftarrows (\gamma,n)$ equilibrium is not established, as expected in particular at late times when reactions freeze out, the abundances become directly sensitive to the estimated rates.

Most of the low-energy cross section calculations for nucleosynthesis applications are based on the statistical model of Hauser-Feshbach.  The statistical model has proven its ability to predict cross sections accurately for medium- and heavy-mass nuclei. However, this model suffers from uncertainties that essentially stem from the predicted nuclear ingredients describing the properties of the nuclear ground and excited states, the $\gamma$-ray strength function and the optical potential. Nuclear level densities typically affect rates by a factor of 10 with a strong odd-even effect (i.e. between systems with an even or odd number of nucleons) according to the way the pairing interaction is treated. The $\gamma$-ray strength function may impact the prediction of the rate up to a factor of 100, in particular depending on the description of the low-energy tail of the giant $E1$ resonance, or if the low-energy $M1$ component is included, both for the scissors mode and its low-energy enhancement related to the de-excitation mode (also referred to as the $M1$ upbend) \cite{Goriely18a,Goriely19}. Finally, the optical potential is known to have a negligible impact in the standard case of radiative neutron captures. However, a reduction of the imaginary component of the neutron-nucleus potential (that takes into account the absorption of the reaction flux from the elastic channel) may have a drastic impact in reducing the capture cross section by neutron-rich nuclei.
More details on our capacity to reliably predict all these ingredients can be found in \cite{Goriely15b,Arnould20,Goriely18a,Goriely07b,Goriely08b,Koning08,Capote09}. 

In addition, at the very low energies of astrophysical interest for light or exotic neutron-rich nuclei for which few, or even no resonant states are available, the capture reaction is known to be possibly dominated by direct electromagnetic transitions to a bound final state rather than through a compound nucleus intermediary. This direct contribution to the neutron capture rate can be 2--3 orders of magnitude larger than the one obtained within the statistical approach traditionally used in nucleosynthesis applications \cite{Goriely15b,Xu12,Xu14,Sieja21}. Significant uncertainties still affect the predictions of the direct contribution. These are related to the determination of the nuclear structure ingredients of relevance, i.e.,  the nuclear mass, spectroscopic factor, neutron-nucleus interaction potential and excited level scheme. A special emphasis needs to be put on determining the low-energy excitation spectrum with all details of the spin and parity characteristics. This can be deduced from a nuclear level density model, but not a statistical approach. Further efforts are needed to improve the prediction of such nuclear inputs within reliable microscopic models. The transition from the compound to the direct nucleus mechanism  when only a few resonant states are available also needs to be tackled in a more detailed way, for example within the Breit-Wigner approach \cite{Rochman17}.

In view of all the above-mentioned difficulties and complexities of the relevant nuclear physics, the estimate of nuclear masses, $\beta$-decay and neutron capture rates of exotic neutron-rich nuclei need to be further improved, both experimentally and theoretically, in order to be able to accurately estimate the ejecta composition and decay heat released by compact binary objects after merger. Future kilonova light curve and spectra observations, triggered by ET detection of NS-NS and NS-BH systems, will provide constraints on the r-process nucleosynthesis provided nuclear physics uncertainties are significantly reduced.

\subsection{Prospects for constraints on microphysics with ET data}
\label{sec:constraints}

We expect that in the near future our current understanding of the various fundamental phenomena related to the astrophysics of NSs will be improved thanks to a new generation of telescopes and advanced GW detectors. 
New observational data both in pre- and post-merger phases will be of great help for understanding their internal composition, thermal effects, and  nucleosynthesis. This will significantly contribute to insights in the internal structure of compact objects, dynamics of their mergers, and explosion mechanisms of CCSNe. In this context, an important line of work for ET will be to assess the possibility of finding a phase transition to ``exotic'', i.e., non-nucleonic, matter inside NSs. In this section wediscuss the capabilities of ET. In section~\ref{ssec:ConstraintsonlowT} we review the possibilities to obtain constraints on low temperature matter, in section~\ref{ssect:constraintsmicro} on the postmerger phase of a BNS or on hot matter from CCSN, respectively, and in section~\ref{ssec:multimessenger} on multi-messenger aspects. 
For the following discussion we assume general relativity as theory of gravity; see section~\ref{sec:degeneracies} for a discussion of the impact of alternative theories on the subatomic physics outcome.

\subsubsection{Constraints on low-temperature microphysics}
\label{ssec:ConstraintsonlowT}

\paragraph{Compact Binary Mergers (NS-NS and NS-BH).}\label{subsec:NSmergers}



As discussed in section~\ref{ssec:current-constraints}, current measurements are
insufficient to answer fundamental open questions on the properties of dense baryonic matter and the occurrence of phase transitions, possibly to deconfined quark matter in the core of NSs. With current detectors, we expect at most ten to hundred NS-NS or NS-BH mergers per year~\cite{Colombo:2023une,KAGRA:2013rdx}, whereas ET will enable the detection of up to $10^5$ BNS mergers per year \cite{Maggiore:2019uih}. The enhanced sensitivity will also lead to higher precision measurements of tidal deformability and sub-dominant matter effects during inspiral. In addition, in mixed (NS-BH) binaries, tidal forces on the NS can become so strong that it gets tidally disrupted, which leads to a distinctive sudden shutoff of the GW signal. The disruption occurs at a characteristic frequency in the $O({\rm kHz})$ range that depends primarily on the NS compactness, BH mass and spin~\cite{Shibata:2007zm,Foucart:2012nc,Pannarale:2015jia} and thus yields additional, clean information on the cold EOS, though spin precession may complicate the identification of the shut-off~\cite{Kawaguchi:2017twr}. ET will be able to detect NS-BH binaries out to redshifts of $\sim 20$~\cite{Maggiore:2019uih,Iacovelli:2022bbs}, which leads to a noticeable shift in the detected GW frequencies. For a fraction of these observable events, the NS will be tidally disrupted, which would only be detectable for events with a sufficiently high signal-to-noise ratio (SNR). 

In addition, information on microphysics in and beyond the EOS could come from the characteristic oscillation modes excited by the tidal field of the companion in a binary system, or by other mechanisms in isolated NSs (see e.g. ref.~\cite{Barta:2019xhg}). The most relevant modes from the observational point of view and to obtain spectroscopic information about the NS interior are the fundamental ($f$), 
the gravity ($g$), and $r$ modes that lead to imprints in the GW signal. 
The $f$-modes have the strongest tidal couplings and their behavior below resonance is the primary contributor to the tidal deformability discussed above. For slowly rotating NSs, these modes have frequencies in the kilo-Hertz range, making their effects most significant near the end of the inspiral in circular orbits. However, high spin can lower the mode frequency, allowing these excitations to occur earlier in the inspiral, potentially even reaching resonance before the plunge or merger phase (see, e.g., \cite{Ma:2020rak, Kuan:2023qxo}). Gravity $g-$modes and related classes of interfacial modes are directly associated with steep composition gradients in the NS, and arise, for instance, from a strong phase transition, with their frequencies and amplitudes containing direct information about the properties of the transition. The $r$-modes have frequencies proportional to the spin of the NS and may thus become excited during the early inspiral. Although their effects could be relevant for ET, their influence will likely be subdominant to $f$-mode effects due to the significantly weaker tidal coupling~\cite{Flanagan:2006sb,Landry:2015cva}. The low-frequency sensitivity of next-generation detectors will be key to capture this information in loud events. It is also important to note that $f$- and $r$-mode excitations have opposite effects on the merger timing, with $f$-modes advancing and $r$-modes delaying the merger \cite{Flanagan:2006sb,Gupta:2020lnv}.
In addition, various kinds of spin-tidal couplings~\cite{Pani:2015nua}  will enrich the subdominant GW signatures during binary inspirals. Moreover, effects of viscosity may also play a role, in particular in conjunction with tidal heating due to mode excitation~\cite{Arras:2018fxj,Celora:2022nbp,Kantor:2024wmd,Saketh:2024juq}. Together, these effects can influence the oscillation spectra~\cite{Andersson:2019mxp,Kuan:2022bhu,Kuan:2023kif,Counsell:2023pqp} and the tidal response function~\cite{Ripley:2023qxo,Ripley:2023lsq,HegadeKR:2024slr}. 
Modes can also play an important role in single NSs for continuous gravitational wave (CGW) signals and in the proto-NSs formed in SNe, as will be discussed in sections~\ref{sssec:continousGW} and \ref{subsec:GWCCSN}, respectively. 

As discussed above, the connection between astrophysical observations and microphysical properties of NSs requires both microscopic and macroscopic (global simulation) modeling, where the specific microphysics inputs depend on the astrophysical scenario explored. 
If one restricts the study to cold, non-rotating, and unmagnetized NSs, which are the subject of this section, this connection mainly relies on the knowledge of the EOS [pressure versus energy density relation $P(\varepsilon)$] of cold (baryonic) matter in $\beta$-equilibrium (see section~\ref{sec:microphysics}). 
For static properties of NSs, the hypothesis of validity of general relativity (see section~\ref{subsec:newphysics} for a discussion about possible consequences when relaxing this assumption) suffices to guarantee a direct correspondence between global properties and the corresponding EOS. 
Therefore, it is, in principle, possible to reverse the problem and associate to simultaneous observations of the $M$-$\Lambda$ relation of several NSs the underlying EOS. The measurement of mass and radius of several NSs with high precision as projected by X-ray telescopes, successors of NICER, are very promising in this respect, too (see e.g.~\cite{Watts:2016uzu} for a review), but suffer from larger modeling uncertainties. 
However, intriguing degeneracies should be noted in this context. Very similar $M$-$R$ relations can be obtained under different hypotheses on the underlying microphysics, giving rise to the so-called “masquerade” effect~\cite{Alford:2004pf}, where hybrid stars involving a quark matter core can be difficult to distinguish from those comprising only hadronic matter.  Another potential degeneracy are so-called tidal deformability doppelgängers~\cite{Raithel:2022aee}, which have been found in sets of parameterized EOS models when allowing for a strong first-order phase transition at low density of $1-2 n_{\rm sat}$. Such doppelgängers can have similar tidal deformabilities (as well as main postmerger frequencies) that are indistinguishable within the statistical uncertainties achievable by current GW detectors but different radii. The use of more restrictive prior information from nuclear physics and joint EOS interpretations with astrophysical measurements of NS radii could reduce these degeneracies.  

More and higher-precision measurements such as those expected from ET will contribute to reduce the error bars, thus tightening the constraints on the EOS~\cite{Branchesi:2023mws}. In this section, we discuss constraints on the low-temperature microphysics that can be obtained from (future) binary-merger GW data. They concern information inferred from the pre-merger phase, when matter can still be assumed as `cold', while constraints on the finite-temperature EOS, which is relevant for the post-merger phase or CCSN, will be discussed in sections~\ref{ssec:postmerger} and \ref{subsec:GWCCSN}, respectively. 

\subparagraph{Constraints on hadronic EOS parameters and composition.}

Several studies have been performed highlighting the potential of next-generation GW observatories, such as ET, for constraining the NS EOS. First of all, those observatories will not only detect many events but, due to higher sensitivity, also more precisely determine the component masses and the mass-weighted tidal deformability $\tilde{\Lambda}$; see, e.g., 
 for recent works refs.~\cite{Gamba:2020wgg,Puecher:2022oiz,Branchesi:2023mws,Iacovelli:2023nbv,Puecher:2023twf}.\footnote{The mass-weighted tidal deformability is defined as~\cite{Wade:2014vqa} 
\begin{equation}\label{deftildeLambdadiv6}
\tilde{\Lambda} = \frac{8}{13} \left[(1+7\eta-31\eta^2)(\Lambda_1 + \Lambda_2) + \sqrt{1-4\eta}(1+9\eta-11\eta^2)(\Lambda_1 - \Lambda_2)\right]\, ,
\end{equation}
where $\Lambda_1$ and $\Lambda_2$ are the tidal deformabilities of the two component stars, and $\eta=m_1m_2/(m_1+m_2)^2$ is the symmetric mass ratio. The usefulness of $\tilde{\Lambda} $ is that it
is the combination that enters the inspiral waveform at 5\,PN order, where tidal effects first appear.
 }
Assuming a relatively broad mass distribution in NS binaries, this leads to a determination of $\tilde\Lambda$ as function of the two masses, which in turn allows obtaining the mass--tidal deformability relation. This can be translated into constraints on the NS EOS and hence 
NS radii. The capability of ET to access lower frequencies can also contribute to increasing the precision of the parameter estimation, although this effect is more important for determining the chirp mass and the mass ratio than on the measurement of the tidal deformability \cite{Branchesi:2023mws}.

\begin{table}[t]
    \centering
    \begin{tabular}{|c|c|c|c|c|}
    \hline
         $N_{ET}$ & \multicolumn{2}{c}{$\Lambda_{1.4M_{\odot}}$}  &  \multicolumn{2}{c}{$\Lambda_{2.0M_{\odot}}$} \\
         \hline
          & BSk24 & Nl3$\omega\rho$ & BSk24 & Nl3$\omega\rho$\\
          \hline
          10 & $550.590^{+39.262}_{-36.078}$ & $914.771^{+26.206}_{-33.372}$ & $61.438^{+11.460}_{-11.137}$ & $108.683^{+12.474}_{-10.709}$ \\
          \hline
          50 & $526.677^{+19.882}_{-19.022}$ & $935.696^{+1.357}_{-4.151}$ & $51.473^{+7.854}_{-6.402}$ & $114.872^{+0.084}_{-0.084}$ \\
          \hline
          100 & $518.230^{+22.086}_{-23.094}$ & $934.407^{+0.069}_{-2.862}$ & $45.306^{+6.521}_{-8.030}$ & $114.287^{+0.670}_{-1.116}$ \\
          \hline
          500 & $513.879^{+1.405}_{-2.147}$ & $931.547^{+0.002}_{-0.002}$ & $37.722^{+0.445}_{-0.780}$ & $114.957^{+0.0003}_{-0.0003}$ \\
          \hline
          1000 & $505.346^{+9.939}_{-4.168}$ & $931.545^{+5e-6}_{-5e-6}$ & $38.893^{+0.733}_{-1.616}$ & $114.957^{+5e-7}_{-5e-7}$ \\
          \hline
          2000 & $511.192^{+0.540}_{-0.540}$ & $931.545^{+4e-8}_{-4e-8}$ & $37.170^{+0.228}_{-0.228}$ & $114.957^{+3e-9}_{-3e-9}$ \\
          \hline
          5000 & $512.294^{+0.528}_{-0.528}$ & $931.545^{+1e-13}_{-1e-13}$ & $37.293^{+0.349}_{-0.349}$ & $114.957^{+0}_{-0}$ \\
          \hline
    \end{tabular}
    \caption{Mean and $1\sigma$ statistical uncertainties on tidal deformability for $1.4M_{\odot}$ and $2.0M_{\odot}$ NSs with varying number of simulated detections using BSk24 and Nl3$\omega\rho$ EOS models for injection, as depicted in left panel of figure~\ref{fig:lambda_detec}.}
    \label{tab:uncertainties_tidal}
\end{table}

The added value of multiple detections with ET is shown in the left panel of figure~\ref{fig:lambda_detec} (adapted from ref.~\cite{Iacovelli:2023nbv}), where the extracted value of the tidal deformability for a $1.4 M_{\odot}$ and a $2 M_{\odot}$ NS is shown for an increasing number $N_{\rm ET}$ of detections with ET, see table~\ref{tab:uncertainties_tidal} for the corresponding mean values and $1\sigma$ uncertainties. A uniform distribution for NS masses between $1.1M_\odot$ and the maximum mass for each EOS has been assumed and the events are chosen randomly among the events leading to a $\tilde\Lambda$ measurement from the simulation of detected BNS merger events with ET over one year from ref.~\cite{Iacovelli:2023nbv}. Compared to the total number of $\sim 10^5$ BNS merger detections expected par year with ET, those with $\tilde\Lambda$ measurement correspond --depending on the injected EOS and the waveform model used-- to roughly 60-70\% of the total number. For a discussion of the influence of different detector configurations on these expected detections, see section~\ref{sec:div9_infer_intrinsic_par_golden_binaries}.

Note that the capability to determine the tidal deformability of a star with a given mass, and thus to infer the NS EOS and radius, will depend on the mass distribution of the measured events --detections of events across a wide range of masses will better constrain the EOS for a large range of densities. 
For EOS inference, a nucleonic meta-modeling approach has been employed within a Bayesian framework \cite{Carreau:2019zdy,Thi:2021jhz}. 
With $\gtrsim 500$ detections, the underlying NS EOS can be narrowed down quite accurately irrespective of the different proposed designs (2L shape or triangle) of ET. Without doppelgängers, NS radii for the fiducial masses can also be determined from this analysis~\cite{Iacovelli:2023nbv}. To be precise, in this case the uncertainty on radius and tidal deformability is reduced from (1-2)\% for $N_{ET} = 10$ to less than $ 10^{-3}$\% for $N_{ET} = 5000$. This is in agreement with other studies, e.g., that in ref.~\cite{Warren:2019lgb}, where it was shown that a few loud events within a network of next-generation observatories are sufficient to determine radii with a precision of $< 100$~m.  It should be noted that the difference of applying a unified crust model or a fixed reasonable unique crust does not induce any noticeable difference in the above analysis if the latter is matched consistently at the corresponding crust-core transition and reasonably well describes the crust~\cite{Davis:2024nda} even for ET. This result might seem astonishing on a first sight in view of the large differences obtained for the predicted radii for some EOS models between unified and unique crust~\cite{Fortin:2016hny,Suleiman:2023bdf}.  The reason is probably that recent constraints from nuclear theory, in particular from $\chi$EFT, very strongly constrain the EOS in the crust region~\cite{Davis:2024nda}. 

Most of these constraints can be derived from the measurement of BNS systems; however, NS-BH systems can also provide valuable information about the EOS. Those constraints would mainly arise from observations of NS-BH systems in which the objects have comparable masses and/or the BH is highly spinning, see e.g.~\cite{Lackey:2013axa,Sarin:2023tgv}.

Concerning the underlying nuclear matter parameters, it should be noted that there is some degeneracy and that different combinations of parameters lead to the same EOS for $\beta$-equilibrated NS matter, and thus, measuring the tidal deformability alone does not determine all nuclear matter properties~\cite{Mondal:2021vzt,Iacovelli:2023nbv,Imam:2023ngm}. Similar conclusions have been drawn for precise NS radius measurements~\cite{Xie:2020tdo}. However, with some complementary information either from independent observations or from nuclear physics experiments or theory, ideally on symmetric (i.e., same number of neutrons and protons) matter, nuclear matter properties can be determined very precisely. To cite some examples, within a given framework for the nuclear Hamiltonian, refs.~\cite{Sabatucci:2022qyi,Rose:2023uui} found that next-generation observatories have the potential to constrain the strength of the three-nucleon interaction. ‌‌Complementary information on the low-temperature EOS from the inspiral phase could also come from combined measurements of the tidal deformability and quadrupolar crust-core interface $i$-modes through multi-messenger astronomy (specifically, through resonant shattering flares).
Indeed, even if $i$ modes are expected to be weak, their detection might be within reach for ET, and ref.~\cite{Neill:2022psd} suggest them as a possible way to constrain the empirical nuclear matter parameters with a single loud event being sufficient. 
As discussed above, exotic hadrons may also be present. The GW signatures of a strong phase transition that could reveal their presence is discussed below; in addition they would likely also impact viscosities, which may be constrained from signatures in GWs from binary inspirals, e.g.~\cite{HegadeKR:2024slr, Ripley:2023lsq}. 

Altogether, these studies demonstrate the unprecedented potential of detectors such as ET to determine from the inspiral of BNS mergers the underlying NS EOS, and help to constrain NS structure and properties of dense matter.

\subparagraph{Comparison  between triangular and 2L-configurations.}

Given the current situation in which the final  choice for ET geometry has not yet been made, it is important to also consider the impact of different designs and configurations on the science return regarding the EOS inference; since the parameters encoding most information about the EOS are the masses and tidal deformabilities, in the following we will focus on these when discussing  the difference between a triangular ET with 10km arms vs.\ 2 L-shaped detectors with 15km arms; a detailed comparison can be found in~\cite{Branchesi:2023mws}. We also refer to chapter~\ref{section:div9} for similar studies employing a Fisher matrix approach. 

\begin{figure*}[t]
    \centering
    \includegraphics[width=\textwidth]{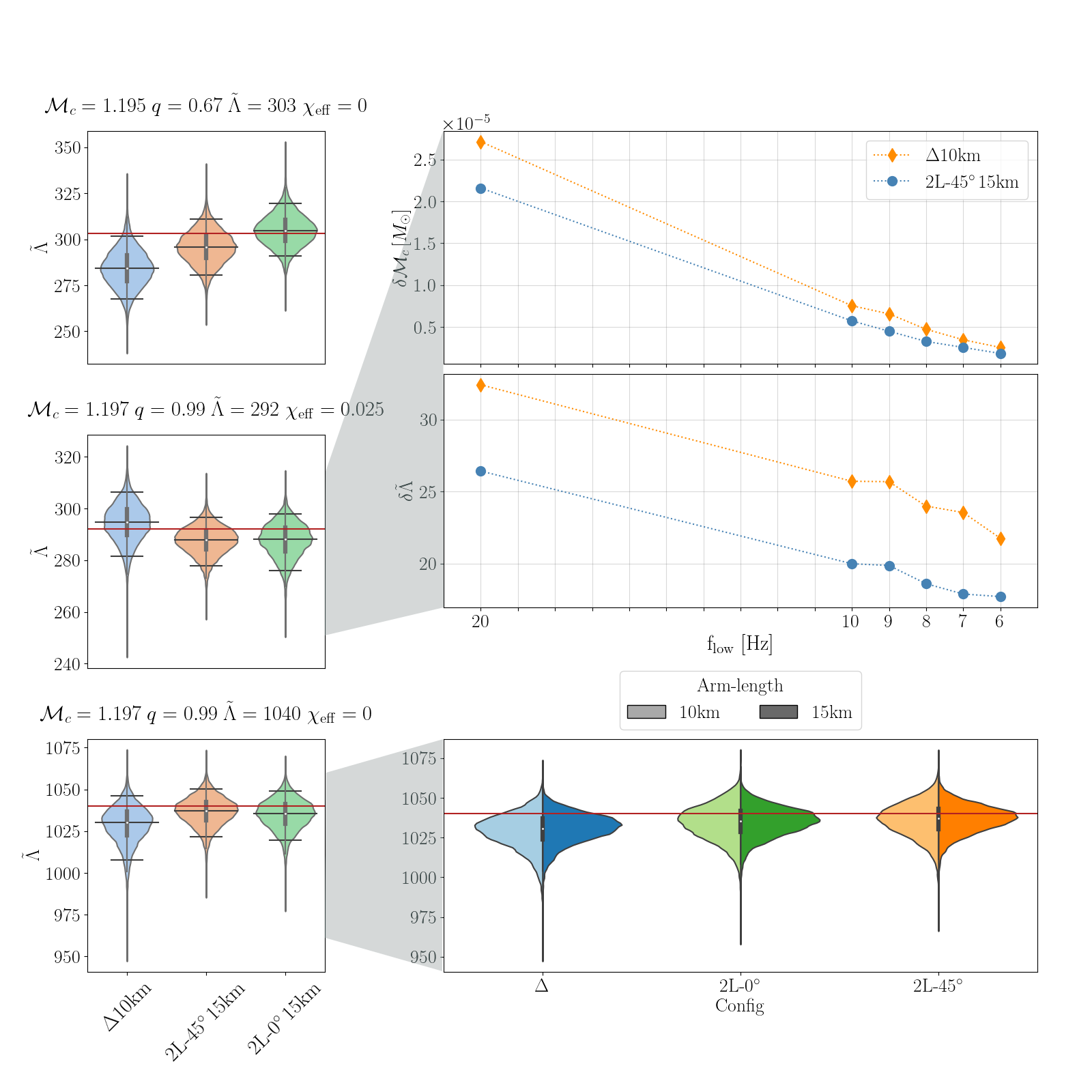}
    \caption{Left: $\tilde{\Lambda}$ posteriors recovered from the simulated signals for three different sources, whose parameters are reported on top of each panel, for three ET configurations; the horizontal red line indicates the injected value. Right - top panel: width of the $90\%$ confidence interval of the $\mathcal{M}_c$ and $\tilde{\Lambda}$ posteriors obtained for the source in the middle on the left-hand side panels, when performing the analysis with different starting frequencies $f_{\rm low}$; the different markers and colors correspond to two different detector configurations. Right - bottom panel: $\tilde{\Lambda}$ posteriors recovered for the source on the bottom panel on the left-hand side, with the red line showing the injected value; th different colors correspond to the different ET geometries, with the lighter shade representing a 10~km arm-length configuration, and the darker ones a 15~km arm-length one.}
    \label{fig:summary_pe}
\end{figure*} 

\begin{figure*}[htb]
    \centering
    \includegraphics[width=\textwidth]{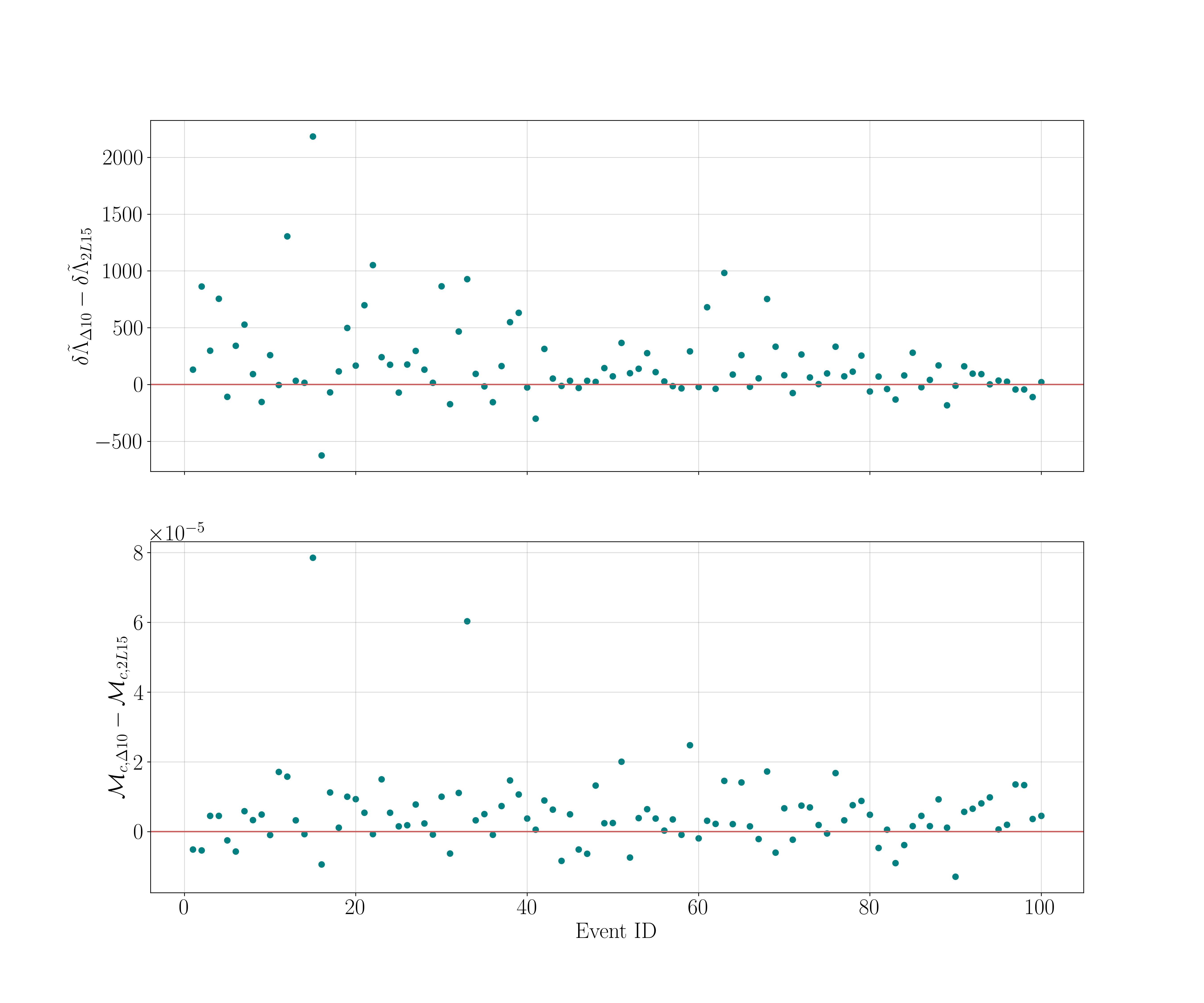}
    \caption{Difference between the width of the $90\%$ confidence interval obtained with a triangular, 10~km arm-length detector and a configuration including two L-shaped interferometers with 15~km arm-length, for the posterior recovered for $\tilde{\Lambda}$ (top panel) and $\mathcal{M}_c$ (bottom panel), over a catalog of simulated signals for 100 different sources. The red line marks zero, meaning that points above (below) the line show a wider posterior for the $\Delta$ (2L) configuration.}       
    \label{fig:more_pe_comparison}
\end{figure*}

\begin{figure}[t]
	\centering
		\includegraphics[width=\textwidth]{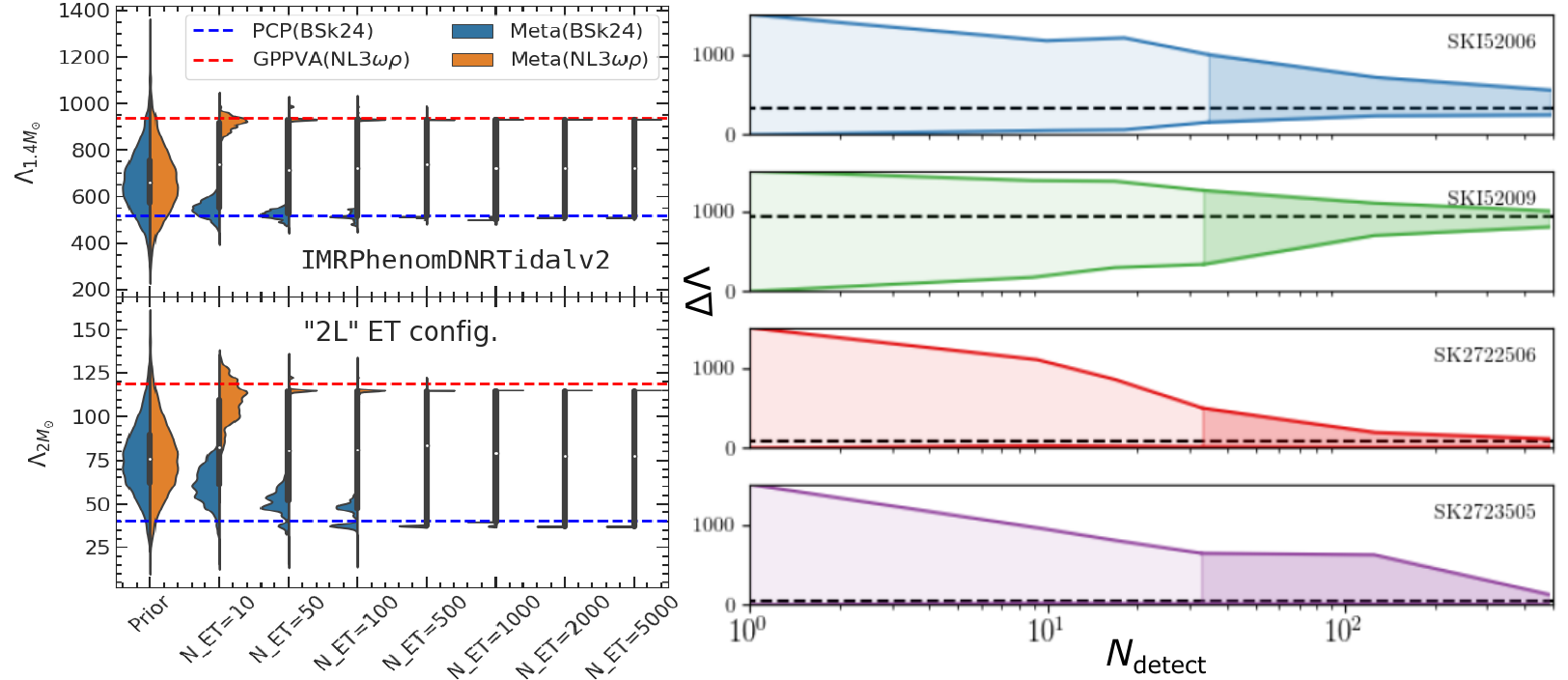}
		\caption{Left panel: Increased precision obtained from multiple ET detections of the tidal deformability for a 1.4$M_{\odot}$ and 2.0$M_{\odot}$ NS using a nucleonic meta-modeling technique (violin shapes); for comparison the prediction of two selected EoSs is shown with horizontal lines. Right panel: Reduction in the uncertainty on the tidal deformability with increasing number of detection. Figures adapted from ref.~\cite{Iacovelli:2023nbv} (left) and ref.~\cite{Landry:2022rxu} (right panel).}
		\label{fig:lambda_detec}
\end{figure}

Figures~\ref{fig:summary_pe} and~\ref{fig:more_pe_comparison} show an example of how accurately we expect to infer these quantities from complete PE runs with ET. Figure~\ref{fig:summary_pe} shows a summary of the results presented in Refs.~\cite{Puecher:2023twf,Branchesi:2023mws}, to which we refer for more details. The panels on the left-hand side show the posteriors obtained for $\tilde{\Lambda}$ from mock signals created for three different sources, all placed at a distance of 100~Mpc, with three possible ET configurations: a triangle with 10~km arm-length ($\Delta$), placed in Sardinia, or two L-shaped interferometers with 15~km arm-length, one in Sardinia and one in Limburg, with either aligned ($2L - 0^\circ$) or misaligned ($2L - 45^\circ$) arms; in all cases the sensitivity for the xylophone configuration in cryogenic mode was considered. For all sources, all configurations recover $\tilde{\Lambda}$ very well. The posteriors for $\Delta$ seem slightly wider than the 2L ones. In order to check whether this is due to the geometry or the arm-length, for one of the sources we repeated the analysis for all the configurations with 10~km and 15~km arm-length. As shown in the bottom panel on the right-hand side of figure~\ref{fig:summary_pe}, if we compare detectors with the same arm-length, the obtained precision on $\tilde{\Lambda}$ is the same. Finally, the top panels on the right-hand side show the width of the $90\%$ confidence interval of the chirp mass $\mathcal{M}_c$ and mass-weighted tidal deformability $\tilde{\Lambda}$ posteriors for the analysis performed at different starting frequencies $f_{\rm low}$, highlighting how, irrespective of the configuration, the increased sensitivity of ET at lower frequencies will help in obtaining more and more accurate measurements of these parameters. 
Figure~\ref{fig:more_pe_comparison}, instead, compares how accurately we measure $\tilde{\Lambda}$ for high-SNR simulated signals from a catalog of 100 sources with a triangular or two L-shaped detectors, showing the difference between the $90 \%$ confidence interval obtained from the analysis considering one triangular detector with 10~km arm-length (placed in Limburg) or two L-shaped, misaligned detectors, with 15~km arm-length (one placed in Sardinia and one in Saxony). As a caveat, these inference runs were not performed as a ``standard" PE run, since the work for which they were produced included six additional free parameters to model the tidal phase in a model agnostic way in the Bayesian inference, which causes the posteriors to be wider than we would have with regular PE. However, figure~\ref{fig:more_pe_comparison} shows clearly how in most cases the 2L configuration recovers more accurate posteriors for both $\tilde{\Lambda}$ and $\mathcal{M}_c$.

\subparagraph{Constraints on a potential phase transition.}
\phantomsection \label{refhereforphasetransitions}



As discussed in section~\ref{ssec:current-constraints}, there are clear features in the mass--tidal deformability relation which are a signature of a phase transition. To identify them, or to break possible degeneracies, the resolution and statistics achievable with next-generation observatories is required. We note that they only depend on the thermodynamic characteristics of the phase transition and not on the details of the two adjacent phases. Thus, although the most studied case is a transition from hadronic to deconfined quark matter within NS, a detection of a phase transition would not be a proof of a quark matter core. As discussed below, detecting signatures from $g$-modes directly associated to the transition could supplement additional information, however, more work is needed to fully explore the information contained in their GW signatures. 

The detectability of a phase transition from the inspiral of a BNS merger with next-generation observatories has been the subject of numerous works recently. For instance, based on the breakdown of quasi-universal relations fitted to purely hadronic EOS in \cite{Chatziioannou:2019yko} and a comparison of the inferred radius in \cite{Chen:2019rja} together with a collection of simulated events for a few purely hadronic and hybrid EoS, the authors suggest that $\sim 50-100$ detections can be sufficient to distinguish different EoSs. Similar findings, showing that, under some conditions, a relatively small number of events -- reachable with ET with a single week of operations -- could be sufficient to infer on a first-order phase transition,  have been obtained in others studies using Bayesian non-parametric EOS inference~\cite{Pang:2020ilf,Landry:2022rxu}. This conclusion relies on the fact that the uncertainty on the tidal deformability is significantly reduced if multiple BNS mergers are detected, as shown in the right panel of Fig. \ref{fig:lambda_detec} (adapted from ref.~\cite{Landry:2022rxu}).

The most complete investigation to date, which was based on representing the full variation of the nucleonic EOSs and phase transitions within a meta-modeling approach (see section~\ref{sec:eos}), concluded that from a single loud event of BNS merger up to a luminosity distance $\lesssim 300$~Mpc, a phase transition at relatively low density (around twice saturation density), can be successfully identified \cite{Mondal:2023gbf}. 
 In this study, the phase-transition signal was quantitatively extracted through Bayes factors, assuming the simulated precision in the tidal deformability that would be achieved by ET.  


Moreover, using a large sample of hybrid EOSs with a first-order phase transition to deconfined quark matter, ref.~\cite{Bauswein:2020xlt} showed that the detection of a system with a total binary mass equal to the threshold mass for prompt black hole formation, together with the (combined) measurement of the dimensionless mass-weighted tidal deformability $\tilde{\Lambda}$ and the threshold mass, could reveal the possible appearance of a phase of deconfined quark matter in NSs. With the large number of events expected with ET, such a detection seems possible.  

In addition, phase transitions modify the spectrum of oscillation modes of the NS. In particular, the principal core $g$-mode in hybrid stars containing quark matter has an unusually large frequency range compared to ordinary NSs of the same mass~\cite{Jaikumar:2021jbw}; as such, for binary mergers involving at least one hybrid star, the fraction of tidal energy pumped into resonant $g$-modes can exceed the corresponding energy of a normal NS by a factor of 2 to 3~\cite{Jaikumar:2021jbw}. Likewise, the resonant tidal excitation of $g$-modes due to a transition from nucleonic to hyperonic matter can potentially be measurable with detectors such as ET~\cite{Yu:2017cxe}.

In conclusion, these studies highlight the enormous potential of ET to answer open questions such as the occurrence of phase transitions in NSs and the need to complement this information by other observables to fully understand not only the dense matter EOS in $\beta$-equilibrium, but its inner structure and behavior for other conditions. 

\paragraph{Continuous GWs.} 
\label{sssec:continousGW}

\subparagraph{Neutron star mountains.}

Rotating NSs that support long-lived, non-axisymmetric deformations, colloquially known as \emph{mountains}, have long been considered interesting sources of continuous GWs. 
Even though the amplitude from such a source is likely to be weak, one may hope to integrate over long-lasting continuous waves to gain in signal-to-noise ratio. Considerable effort has gone into modeling and searching for GWs from deformed rotating NSs. This effort has not yet been rewarded with success. Searches have only provided upper limits on the size of the involved deformations. However, we are beginning to probe the interesting parameter regime and with the improvements in detector sensitivity and analysis techniques anticipated in the era of next-generation instruments, there are good reasons to expect observations, see e.g. \cite{Branchesi:2023mws}. This 
motivates a closer discussion of the physics involved in the problem and the observational effort associated with it.

Why should we expect NSs to support mountains in the first place? The answer is quite simple, even though the precise details are not. 
The formation history of a NS 
starts from the remnant of a supernova. 
A newly born NS will be hot and rapidly rotating, as it retains some of the angular momentum of the progenitor star (or is given a ``kick'' during the collapse process \cite{Spruit:1998sg}). As the star cools, it becomes energetically favourable to form a crystal lattice at low densities and the crust forms. The NS may then, at some point in its lifetime, exhibit starquakes/glitches or accrete from a companion star. These processes are expected to build up (or release) strain in the crust and cause the star to change shape. Intuitively, it seems inevitable that a typical NS will be deformed in a non-axisymmetric fashion and therefore \emph{should} emit GWs. 
 

The general problem of deformed NSs has been reviewed in \cite{Jones:2001ui, Prix:2009oha,Lasky:2015uia,Glampedakis:2017nqy}.
These discussions include mountains, but also long-lived oscillation modes, like the $r$-modes (discussed later)  which may be excited and radiate GWs. The problem of detecting the continuous GW signals from NSs is reviewed in \cite{Palomba:2011icw, Sieniawska:2019hmd, Tenorio:2021wmz, Piccinni:2022vsd, Wette:2023dom,Riles:2022wwz, Haskell:2023yrv, Jones:2024npg}.

In general relativity (and if we ignore internal dynamics) an isolated body must spin and be deformed away from axisymmetry in order to radiate GWs---there must be a time-varying quadrupole moment \cite{Maggiore:2007ulw}. The fact that 
NSs have solid crusts close to the surface enables them to support long-lived, non-axisymmetric deformations. These deformations can be quantified in terms of the principal moments of inertia, $I_j$.
For a rigid uniformly rotating star with angular velocity $\Omega$ about the $z=x^3$-axis, an estimate of the strain amplitude is: 
\begin{equation}
    h_0 = \frac{4 G}{c^4} \frac{\epsilon I_3 \Omega^2}{d} \approx 10^{-25} \left( \frac{{10}{\text{kpc}}}{d} \right) \left( \frac{\epsilon}{10^{-6}} \right) \left( \frac{I_3}{10^{45}\,\text{g\,cm}^2} \right) \left( \frac{\nu}{{500}\,{\text{Hz}}} \right)^2,
    \label{eq:StrainAmplitude}
\end{equation}
where $\nu = \Omega / 2 \pi$ is the star's spin frequency and we have defined the \textit{ellipticity} of the star
as
\begin{equation}
    \epsilon \equiv \frac{I_2 - I_1}{I_3}.
    \label{eq:Ellipticity}
\end{equation}
The ellipticity is a dimensionless measure of the star's quadrupolar deviation from symmetry about the ${z}$-axis. 
From \eq{eq:StrainAmplitude} we see that, even for a reasonably nearby and rapidly rotating source, the GW strain is extremely weak. 
However, if the source can be observed for a sufficiently long period of time, the strain data measured at the detector can be folded over itself to improve the signal-to-noise ratio (roughly as the square root of the observing time for fully coherent searches or as the fourth root of the observing time for semi-coherent searches \cite{Riles:2022wwz}). We also see that the detection problem is degenerate. We need, for example, some insight into the moment of inertia, whose value is uncertain up to a factor of 3-5, depending on the NS mass and EOS \cite{Johnson-McDaniel:2012wbj}. 

\begin{figure}[t]
	\centering
		\includegraphics[width=0.6\textwidth]{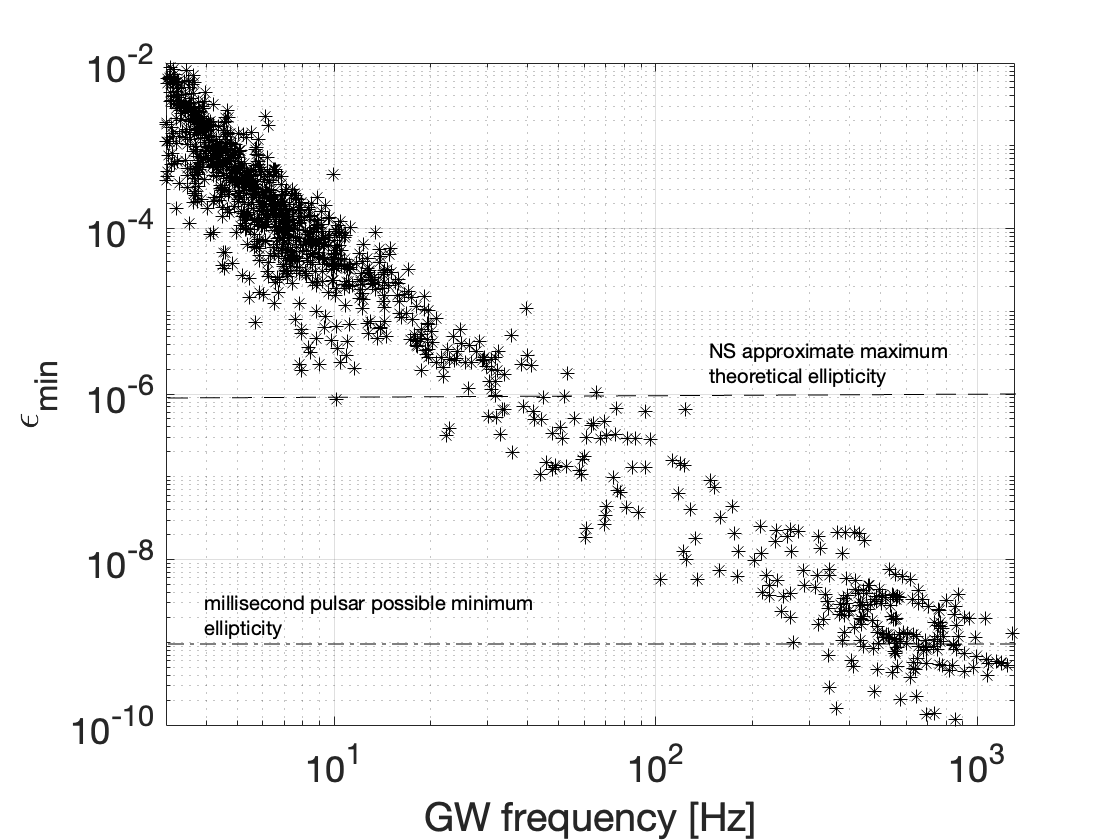}
        \includegraphics[width=0.6\textwidth]{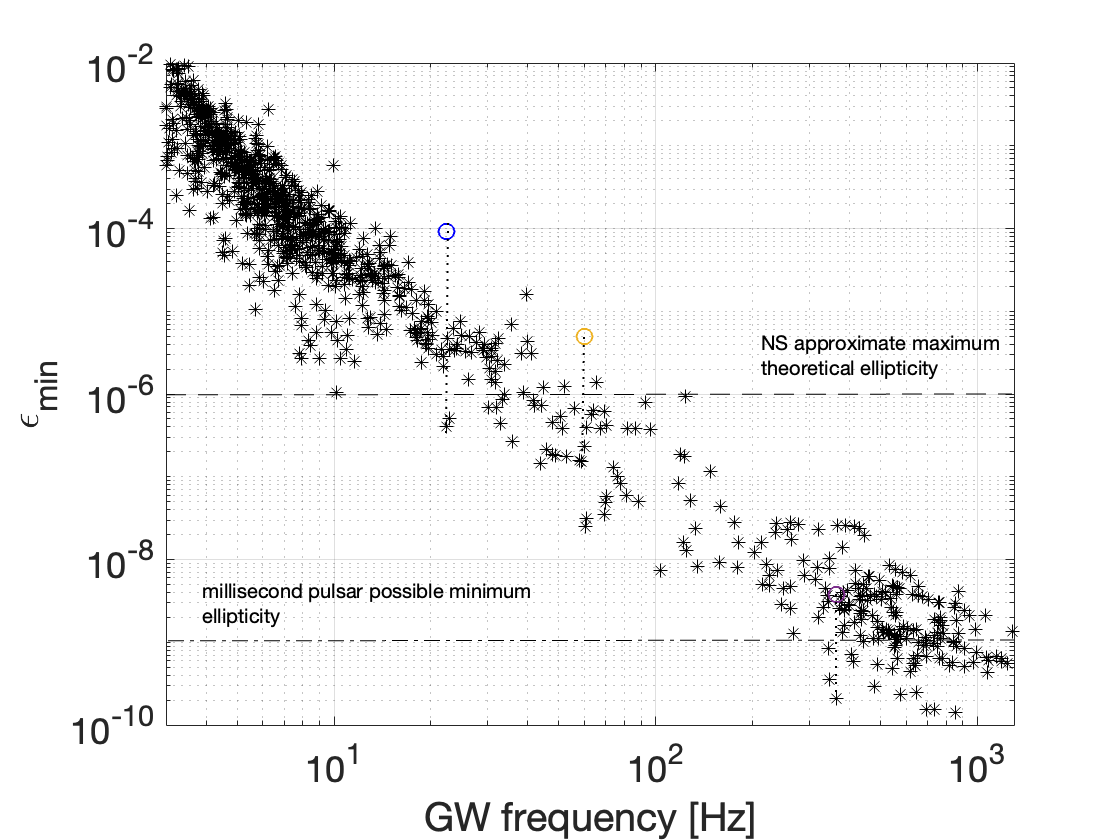}
		\caption{Smallest detectable ellipticity for a search for CWs from known pulsars using a network of two L-shaped ET detectors, with 15 km arms (top plot) or a single triangular detector (bottom plot), with 10 km arms. In both cases, an observation time of three years, with duty cycle 85$\%$, has been considered. The horizontal dashed line roughly indicates NS theoretically predicted maximum ellipticity, see discussion in the main text. The horizontal dashed-dot line indicates a suggested possible minimum ellipticity of observed millisecond pulsars \cite{Woan:2018tey}. For comparison, the three circles indicates upper limits obtained in O3 LIGO-Virgo run for pulsars Vela ($f\simeq 22.38$ Hz), Crab ($f\simeq 59.89$ Hz) and J0711-6830 ($f\simeq 364.23$ Hz).}
		\label{fig:cw_pulsars}
\end{figure}

In an all-sky search for unknown GW sources, i.e., those sources for which no corresponding EM signal is observed, a signal with some given amplitude will not allow us to infer $\epsilon$ and  constrain the relevant nuclear physics, e.g. the crustal strength, unless additional information, e.g., the distance, is available. 
Instead, focusing on known NSs for which an estimated distance and an observed spin-down rate is available, 
we can work out the required energy loss. Assuming that the energy is entirely radiated through GWs we have an upper limit on the NS mountain. This estimate---the so-called spin-down limit---allows us to quantify to what extent known pulsars could in principle be detectable GW sources. figure~\ref{fig:cw_pulsars} provides an illustrative example, showing the minimum required ellipticity for known pulsars and two specific ET configurations. The estimates do not change much among the two possible instrument designs, see section~\ref{section:div7} for a more detailed discussion.
However, we know that some energy is lost as electromagnetic radiation because we observed these NSs in the first place. Moreover, in the relatively rare cases where the braking index\footnote{The braking index is the exponent of the power law describing the time evolution of the NS spin frequency, $\dot{\nu}\propto -\nu^n$, see section~\ref{section:div7} for more details.} $n$ can be inferred from the spin evolution, we know that it is not close to the $n=5$ expected for a NS mountain. This indicates that, while GW losses may contribute, they do not dominate in these systems.

How large  should we expect the mountains to be? 
The size of a NS mountain depends on unknown nuclear physics and the star's evolutionary history. To get an estimate, we 
focus on the physics of a deformed crust: its shear modulus and breaking strain. 
An energetics estimate (see e.g.~\cite{Gittins:2024zbg} for details) leads to $
    \epsilon_\text{max} \approx 2\times 10^{-7} \left( \bar{\sigma}_\text{max}/0.1 \right) $,
where there is some degree of uncertainty around how large the crustal breaking strain $\bar{\sigma}_\text{max}$ may be. Molecular-dynamics simulations for high-pressure Coulomb crystals suggest that the lattice is remarkably strong; $\bar{\sigma}_\text{max} \approx 0.1$ \cite{Horowitz:2009ya}. For comparison, terrestrial solids lie in the range $10^{-4} \leq \bar{\sigma}_\text{max} \leq 10^{-2}$. For a NS with radius $R \sim {10}$km, this would lead to deformations of the size $ \epsilon_\text{max} R \sim {0.1}$~cm. 
Going beyond this simple estimate by also considering the stellar interior, the star's relaxed shape, 
and assuming that the crust is maximally strained at every point \cite{Ushomirsky:2000ax} yields an estimate for the maximum allowed ellipticity
%
for a $M = 1.4 M_\odot$ star (with the DH(SLy4) equation of state \cite{Douchin:2001sv}) of~\cite{Johnson-McDaniel:2012wbj}
\begin{equation}
 \epsilon_\text{max} \approx 3\times 10^{-6} \left( \frac{\bar{\sigma}_\text{max}}{0.1} \right).
\end{equation}
The result differs by a factor of a few from the simple energetics argument.
However, it was noted in \cite{Haskell:2006sv}, and later discussed in more detail in \cite{Gittins:2020cvx}, that demanding the crust to be at breaking strain throughout forces the NS into a shape that violates physical boundary conditions. 
The question then becomes how close a real system can get to this idealized, unphysical state. A useful step towards the answer was taken in \cite{Gittins:2020cvx}, where a fiducial deforming ``force''  was introduced in the problem (see also \cite{Morales:2022wxs}) that contains information about the non-spherical shape of the crust if it were relaxed.
 This force plays no role in supporting the mountain; the deformed shape of the star is self-consistently supported by crustal strains and represents a solution to the elastic perturbation equations. This approach results in smaller predicted mountains but is closer to realistic evolutionary scenarios and an understanding of how deformations actually arise. 

Further work is needed on several fronts in order to make reliable computations of how  NS mountains form and which is their typical size. Firstly, going beyond the assumption of a purely elastic crust we probably need to take into account some form of plastic behavior 
\cite{Smoluchowski:1970zza,Jones:2002bu,Chugunov:2010ac}. In a  plastic flow the crust may retain some of the strain beyond the yield point and as a result be able to develop larger quadrupole moments. 
Secondly, realistic mountain models need to involve evolutionary aspects (some of which will be discussed in section~\ref{section:div7}) and we may seek guidance from electromagnetic observations, as in the case of the spindown upper limit argument. For example,  
 it has been suggested that the observed population of millisecond pulsars is consistent with a \emph{minimum ellipticity} of $\epsilon \approx 10^{-9}$ \cite{Woan:2018tey}.

Accreting NSs have long been considered promising GW emitters \cite{Papaloizou:1978zz,Wagoner:1984pv,Bildsten:1998ey,Andersson:1998qs}. 
Accreting gas is expected to 
transfer angular momentum to the NS \cite{1982Natur.300..728A,1982CSci...51.1096R} and 
the magnetic field will direct the matter flow towards the star's poles. Assuming the magnetic poles are misaligned with the rotation axis, 
the star will naturally develop a mass asymmetry. This argument may be somewhat simplistic, but it provides key motivation for detailed studies discussed in section~\ref{section:div7}.  Specifically, 
ref.~\cite{Bildsten:1998ey} proposed that the observed spin frequencies in low-mass X-ray binaries could be explained by the accreted matter heating the crust giving rise to temperature-sensitive nuclear reactions involving electron captures. Hotter regions of the crust would have these reactions at lower pressures, so the density variations would occur at higher altitudes in these regions. This naturally generates a quadrupole moment,  commonly referred to as a \textit{thermal mountain}. There have  been further developments of this work, mostly in an effort to provide more detail to the modeling \cite{Ushomirsky:2000ax,Singh:2019dgy,Osborne:2019iph,Hutchins:2022chj}. These models require
a precise understanding of the composition of the NS crust (see for example \cite{Fantina:2019lbd}).

The most recent search for continuous GWs from known pulsars using LIGO-Virgo O3 data ~\cite{LIGOScientific:2021hvc} pushed the upper limit on the signal strain amplitude below the  \textit{spin-down limit} for 23 of the target systems. Among these, the inferred upper limits on the ellipticity for J0437-4715 and J0711-6830 are $8.5\times 10^{-9}$ and $5.3\times 10^{-9}$, respectively; only a factor of few larger than estimate from \cite{Woan:2018tey}. Future prospects with  ET are summarized in figure~\ref{fig:cw_pulsars}, which shows the smallest detectable ellipticity for a search over three years (85$\%$ duty cycle) with a network of two L-shaped detectors, each with 15 km arms (top plot) or a single triangular detectors, with 10 km arms (bottom plot).  

These estimates show that there are hundreds of pulsars for which ET will be able to, if not detect a signal, set constraints below the maximum ellipticity predicted by theory. From the figure, we also see that for tens of high frequency millisecond pulsars, ellipticities below $10^{-9}$ will be probed. Based on these predictions and  the available theoretical results, the detection of at least one continuous GW signal by ET seems likely. What information can be drawn from a detection is not a completely solved problem and is a hot topic in the field \cite{Haskell:2023yrv}.
In particular, we need to better understand systematics and degeneracies, see for example \cite{Sieniawska:2021adr} for a relevant discussion.

For a source with an estimated distance, the measure of the signal amplitude and frequency would provide a measure of the $l=m=2$ quadrupole moment \cite{Owen:2005fn}
\begin{equation}
    Q_\mathrm{22}=\sqrt{\frac{15}{8\pi}}\, I_\mathrm{zz}\epsilon.
\end{equation}
This may permit us to exclude EOS models that do not allow quadrupole moments as large as the measured one. Neutron star distance measures typically come from EM observations, like for many known pulsars, but purely GW methods based on the parallax effect have been proposed in \cite{Seto:2005gd,Sieniawska:2022bcn} for EM-silent sources. The results suggest that relative distance errors $\Delta d/d\sim (1-10)\%$ can be obtained with ET over an observation time of 1-3 years, for source distance in the range 0.1-1 kpc and ellipticities in the range $10^{-9}-10^{-7}$. In general, we expect the distance error to dominate the errors in other GW-derived parameters, like the signal strain amplitude, the frequency, and its derivatives. If the GW contribution to the spin-down is dominant or can be estimated, a measure of the star's moment of inertia, with relative error $\Delta I_\mathrm{zz}/I_\mathrm{zz}\approx 10\%$ may be possible \cite{Sieniawska:2022bcn}. An independent measure of the star's mass or radius would in any case be necessary to constrain the EOS, see, e.g., \cite{Bejger:2005jy,Lattimer:2004nj,Raithel:2016vtt,Greif:2020pju}. A  quantitative model of how  accurately the EOS could be determined assuming the observation of one or more continuous GW signals, possibly in combination with results from EM facilities, remains to be worked out. It is clearly a very important step to be taken.   

\subparagraph{The magnetic field.}
\label{sec:Magnetic}

If we focus on the issue of why mountains form in the first place, then it is natural to consider the star's magnetic field. A dipole magnetic field breaks the (essentially spherical) symmetry of the star, deforms the star \cite{Chandrasekhar:1953zz}, and is expected to lead to some level of GW emission. In fact, the internal magnetic field sets a natural lower limit on NS deformations. However,  
magnetic fields in NS interiors remain poorly understood. Even for the external field some studies suggest the presence of complex magnetic-field structures beyond the standard dipole assumption; see~\cite{Riley:2019yda,Bilous:2019knh,Miller:2019cac,Riley:2021pdl,Miller:2021qha,Vinciguerra:2023qxq}.

For the magnetic field, an argument  comparing the energy stored in the magnetic field to the gravitational potential energy leads to \cite{Haskell:2007bh}
\begin{equation}
    \epsilon \sim \frac{B^2 R^3}{G M^2 / R} \approx 10^{-12} \left( \frac{R}{10\,\text{km}} \right)^4 \left( \frac{1.4 M_\odot}{M} \right)^2 \left( \frac{B}{10^{12}\,\text{G}} \right)^2,
    \label{eq:MagneticEllipticity}
\end{equation}
where $B$ represents the magnetic-field strength. 
As in the case of the crust, the NS's self-gravity is very strong compared to the magnetic field so the expected deformations are (very) small. However, this estimate of the field strength is based on the \textit{external} magnetic field inferred from pulsar timing. The interior field, which will be more relevant for the shape of the star, is highly uncertain and presents (quite predictably) a complicated problem. The main theoretical challenge is our present inability to construct stable magnetic NS models. Both analytical \cite{1956ApJ...123..498P,1973MNRAS.161..365T} and numerical studies \cite{Braithwaite:2005su,Braithwaite:2007fy} indicate that purely poloidal and purely toroidal magnetic fields are unstable on dynamical timescales. Mixed fields, where the toroidal field threads the closed-field-line region of the poloidal component, the so-called twisted-torus configuration, may be stable \cite{Braithwaite:2005xi,Akgun:2007ph,Ciolfi:2009bv}, but this result appears to be EOS dependent \cite{Lander:2012iz,2015MNRAS.447.1213M}. Much of our present understanding involves idealized assumptions such as ideal magnetohydrodynamics and barotropic equations of state. It could also be that a real NS is not always in equilibrium---the magnetic field may be gradually evolving, and this could be important \cite{Moraga:2023xsl}. 
Moreover, the high-density NS interior is expected to form a large-scale superconductor \cite{1969Natur.224..673B}. In that case, estimates suggest that a purely toroidal field leads to the ellipticity scaling as \cite{Cutler:2002nw,Glampedakis:2012qp}
$
    \epsilon \sim 10^{-9} \left( B/(10^{12}\,\text{G}) \right) \left( H_\text{c}/(10^{15}\,\text{G}) \right)$,
where $H_\text{c}$ is the critical field strength (above which the superconductivity is broken). Again, this estimate is highly uncertain.
In principle, a strong magnetic field, like that of a magnetar (of order $10^{15}$~G), may enable the star to develop a large ellipticity, of order that expected from crust deformations \cite{Bocquet:1995je}, see section~\ref{section:div7} for discussion. This is yet another degeneracy of the continuous GW problem. 

The magnetic field also comes into play in accreting systems.
In fact, an accreting NS is expected to develop an asymmetry through the magnetic field confining the accreted matter \cite{Brown:1997ji,Payne:2003vy,Melatos:2005ez,Payne:2007kn,Vigelius:2008fv,Vigelius:2009eg,Wette:2009gj,Priymak:2011zv,Fujisawa:2022dzp}. Accretion will interact with the magnetic field and compress it, such that there will be locally strong fields on the star's surface. Estimates suggest that this can lead to slightly larger deformations than those sourced by the background magnetic-field configuration, but (again) precise modeling is difficult. The accretion problem will be discussed in more detail in section~\ref{section:div7}.


\subparagraph{Unstable \textit{r}-modes.}

Neutron stars host a  rich spectrum of oscillation modes \cite{1988ApJ...325..725M}. These oscillations are closely connected to the physics of the stars. Essentially, each characteristic of the star, e.g., the density, composition gradients, rotation has a corresponding mode family, e.g., the \textit{p}-modes, \textit{g}-modes, inertial modes. The mode properties are intrinsically linked to the properties of the nuclear matter.
An excited non-axisymmetric oscillation mode will emit GWs, presenting an opportunity to probe dense nuclear matter through observations. However, the oscillatory motion will be damped by the radiation and viscosity in the stellar fluid (which is sensitive to nuclear transport coefficients) and this damping may be quite rapid. For this reason, it is natural to consider instabilities that  cause the oscillations to grow, such as the \textit{Chandrasekhar-Friedman-Schutz (CFS) instability} \cite{Chandrasekhar:1970pjp,Friedman:1978hf}.

The CFS instability arises in rotating stars. The idea is as follows: There are two frequencies associated with an oscillation of a spinning star---the frequency as measured by an observer co-rotating with the star and the frequency according to an inertial observer. Suppose the mode is travelling in retrograde to the star's rotation such that the inertial frequency is smaller than the co-rotating frequency. If the star spins fast enough, it will drag the mode along with its rotation such that the inertial observer will see the mode travelling in the prograde direction. The emitted gravitational radiation comes from the non-axisymmetric motion of the star in the inertial frame. The GWs will carry positive angular momentum away from the star, which will be subtracted from the mode. Since this mode is travelling in retrograde in the rotating frame, it carries negative angular momentum and the radiation will cause its amplitude to grow. This, in turn, will increase the amplitude of the radiation, leading to an instability.

Many families of NS oscillation modes contain members that, for sufficiently fast rotation rates, satisfy the CFS instability criterion (for reviews, see \cite{Andersson:2000mf,Andersson:2002ch}). The (currently) most promising candidates for the instability are the inertial \textit{r}-modes \cite{Papaloizou:1978zz,1981A&A....94..126P,1982ApJ...256..717S}. The \textit{r}-modes are generically unstable in perfect fluids \cite{Andersson:1997xt,Friedman:1997uh} because they are retrograde in the rotating frame but prograde to an inertial observer at any rate of rotation.

In reality, NSs are more complex than perfect fluids. Notably, they exhibit viscosity, which damps the fluid motion. To assess the actual relevance of the \textit{r}-mode instability, one must compare the growth timescale associated with the GW emission with the damping timescales that arise due to, in the first instance, shear and bulk viscosity  \cite{1989nos..book.....U,1991ApJ...373..213I}. However, most current estimates of these timescales  rely on Newtonian gravity. The relativistic \textit{r}-mode problem is complicated (and may involve a continuous spectrum \cite{Kojima:1997vv,Beyer:1999te}). 
Nevertheless, the Newtonian estimates provide important qualitative insights, while we await the  developments required to understand the relativistic problem better.

Calculations of the competing timescales demonstrate the existence of an \textit{instability window}. The details depend on the state and composition of the nuclear matter. Broadly, dissipation due to shear viscosity suppresses the \textit{r}-mode at low temperatures, while the bulk viscosity dominates when the star is hot. There exists critical rotation rates at which each viscosity balances the radiation-reaction growth. The instability window maps out a region of parameter space in temperature and stellar spin, where the \textit{r}-mode is unstable and will grow. In order to model the associated GW signal, we also need to understand the amplitudes the unstable modes may reach. This is a difficult problem as it requires nonlinear hydrodynamics (the coupling of modes, touching upon turbulence) \cite{Arras:2002dw}. Given our relative ignorance of the nuclear physics and our inability to calculate the relativistic \textit{r}-modes \cite{Andersson:2022vta}, we do not have a precise description of the instability window. Neither do we complete understand the nonlinear fluid aspects. Nonetheless, it is possible that \textit{r}-modes limit the spin frequencies of hot, newly born pulsars \cite{Lindblom:1998wf,Andersson:1998ze} and more mature accreting systems \cite{Bildsten:1998ey,Andersson:1998qs}. 

\begin{figure}[t]
	\centering
		\includegraphics[width=0.49\textwidth]{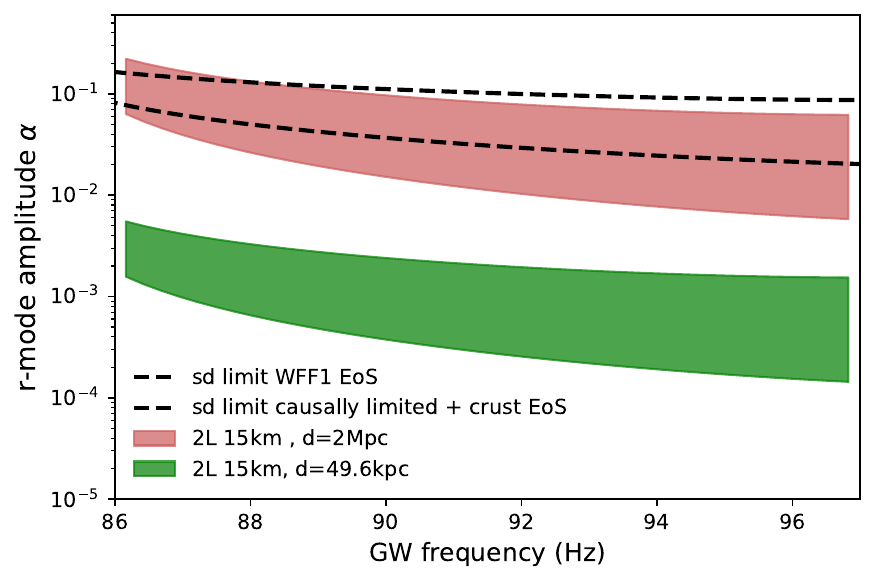}
        \includegraphics[width=0.49\textwidth]{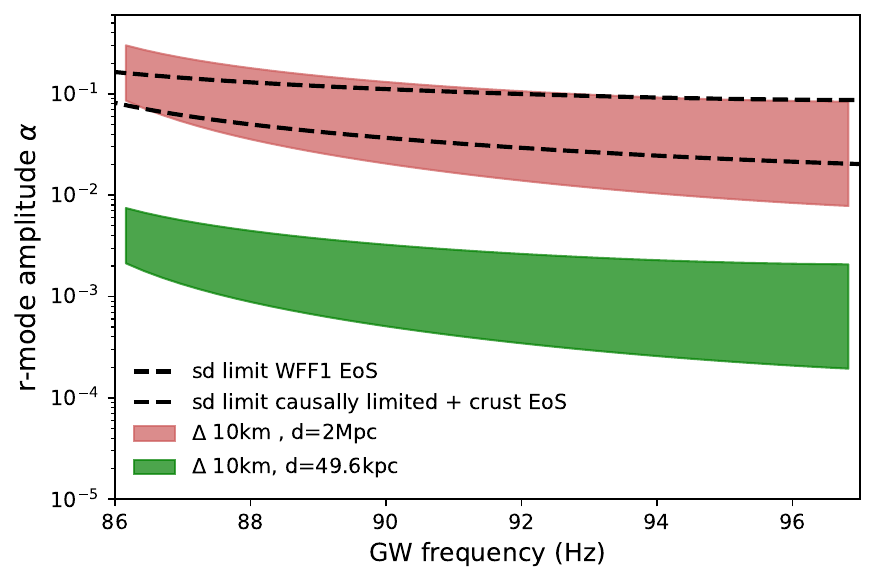}
		\caption{Constraints on the \textit{r}-mode amplitude $\alpha$ using a network of two L-shaped ET detectors, with 15 km arms (left plot) and the triangle configuration, with 10 km arms (right plot), over three years, assuming in both cases a detector duty cycle of $85\%$. The considered parameter space corresponds to that of pulsar J0537-6910. The two colored bands corresponds to a range of EOS, from a stiff causally limited EOS with crust \cite{Haskell:2018nlh} to a soft WFF1 EOS \cite{Idrisy:2014qca}, assuming a distance $d=49.6$ kpc (green band), which is the distance of pulsar J0537-6910, and a hypothetical source emitting at the same frequency and with distance $d=2$ Mpc (red band). The two dashed lines define the spin-down limit for the same range of EOS.}
		\label{fig:alpha_limit}
\end{figure}

From an observational point of view, the GW signal emitted by excited \textit{r}-modes is continuous, with a frequency $\approx (4/3)\nu$, where $\nu$ is the star rotation frequency. The precise value of the GW frequency $f$ actually depends on the EOS, and has been parametrized in \cite{Caride:2019hcv} (making  reasonable assumptions) as
\begin{equation}
    f = A\nu -B\left(\frac{\nu}{\nu_K}\right)^2\nu,
    \label{eq:frmode}
\end{equation}
where $\nu_K$ is the Keplerian frequency and the two parameters $A,~B$ depend on the EOS. For an astrophysically motivated set of EOS considered in \cite{Caride:2019hcv} we have
$1.39 < A < 1.57$ and 
$ 0 < B < 0.195.$
If the GW emission is completely due to \textit{r}-modes, the corresponding ``spin-down limit'' strain amplitude is given by
\begin{equation}
    h_\mathrm{sd}=\sqrt{\frac{10GI_\mathrm{zz}}{c^3}\frac{\nu|\dot\nu|}{f^2}}\,\frac{1}{d}\, ,
\end{equation}
where $d$ is the source distance. The corresponding \textit{r}-mode amplitude is
\begin{equation}
\alpha_\textit{sd}=\sqrt{\frac{5}{8\pi}}\frac{c^5}{G}\frac{h_\mathrm{sd}}{(2\pi f)^3}\frac{d}{MR^3\tilde{J}}\, .
\label{eq:alphasd}
\end{equation}
This quantity, which depends on the  EOS through the relation $f(\nu)$, the star's mass and radius values, and the dimensionless canonical angular momentum of the mode, $\tilde{J}$ (which is weakly dependent on the EOS) can be compared to a given search sensitivity, that is to the minimum detectable \textit{r}-mode amplitude. In case of a detection, parameters $A$ and $B$ of \eq{eq:frmode} could be estimated, constraining the range of possible NS EOS. Even in case of non-detection, setting an upper limit below the spin-down limit allows to constrain the mode amplitude, giving some insight on the excitation mechanism. 

A recent search for \textit{r}-mode emission from pulsar J0537-6910 has been conducted using LVK O3 data \cite{LIGOScientific:2021yby}. This is a frequently glitching X-ray pulsar, and the search was motivated by the observation that spin evolution in the inter-glitch period could be driven by \textit{r}-mode oscillations \cite{Andersson:2017fow}. The search set upper limits around or below (depending on the assumed EOS) the spin-down limit over the narrow frequency band 86-97 Hz, beginning to constrain the range of mode amplitudes predicted by theoretical models.  

Figure~\ref{fig:alpha_limit} shows 
the predicted constraints on the \textit{r}-mode amplitude, $\alpha$, considering two possible ET configurations, one consisting of two L-shaped ET detectors, each with 15 km arms, and one consisting of a single triangular observatory, with 10 km arms. In both cases, a semi-coherent search over three years (with a detector duty cycle of $85\%$), with each data segment of duration three months, is considered. Moreover, we assume a search aimed at a source with the same parameters (spin frequency and spin-down) as J0537-6910 at two possible distances, $d=49.6$ kpc (the distance of J0537-6910) and a hypothetical distance of $d=2$ Mpc. For each assumed distance we plot constraints corresponding to a range of EOS, going from a causally limited stiff EOS in the core, connected to a crustal EOS \cite{Haskell:2018nlh}, to the observationally motivated soft WFF1 EOS \cite{Idrisy:2014qca}. These estimates are compared to the range of spin-down values, computed from \eq{eq:alphasd} taking $M=1.4M_\odot,~R=12$km and $\tilde{J}=0.0164$, the values for a $n=1$ polytrope \cite{Owen:1998xg}.  For a galactic source like J0537-6910, ET would be able to probe $\alpha \in[10^{-4},~10^{-2}]$, possibly allowing us to test various proposed mechanisms for saturating the \textit{r}-mode instability. A hypothetical source emitting gravitational waves at the spin-down limit, i.e., with $\alpha\approx 10^{-2}-10^{-1}$, would be detectable at a distance of about 2 Mpc.




\subsubsection{Constraints on microphysics at finite temperature}
\label{ssect:constraintsmicro}

\paragraph{Postmerger GWs.}
\label{ssec:postmerger}

In general, for numerous parameter combinations of the individual masses of the stars, BNS mergers will give rise to short-lived or long-lived remnants that could emit a GW signal also after the merger before the collapse to a black hole~\cite{Piro:2017zec}. In fact, the post-merger phase is strongly affected by the evolution of nuclear matter at high densities with temperatures reaching values of up to $\approx 100$ MeV; therefore, higher densities than during the inspiral and thermal effects in nuclear matter can be tested, see e.g.~\cite{Dietrich:2020eud, Bernuzzi:2020tgt} for recent reviews. 

\begin{figure}[t]
	\begin{center}
		\includegraphics[width=0.53\textwidth]{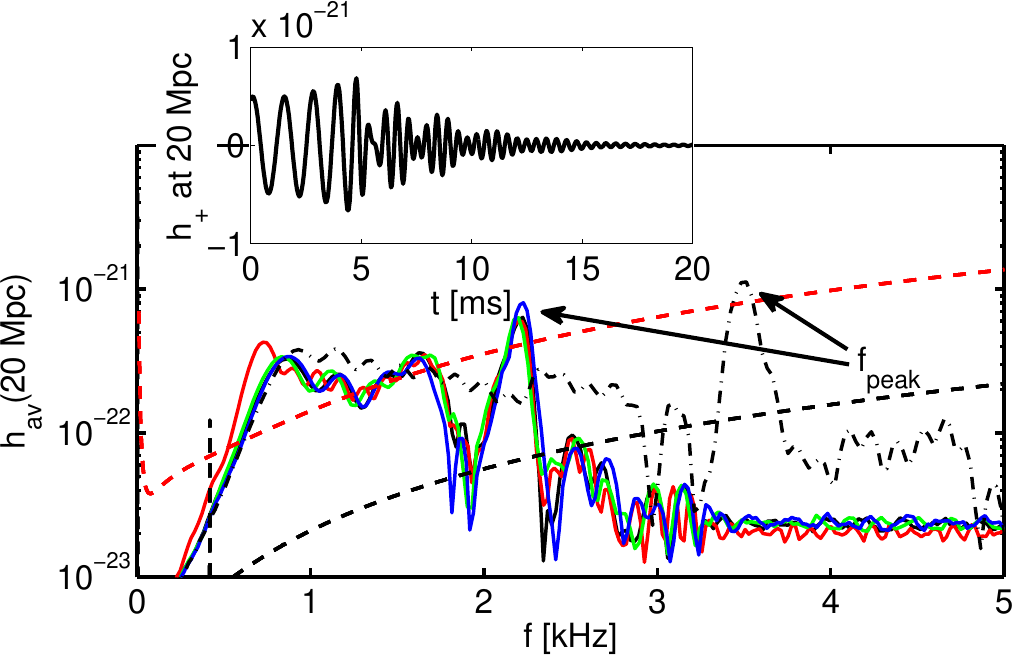}
        \includegraphics[width=0.4\textwidth]{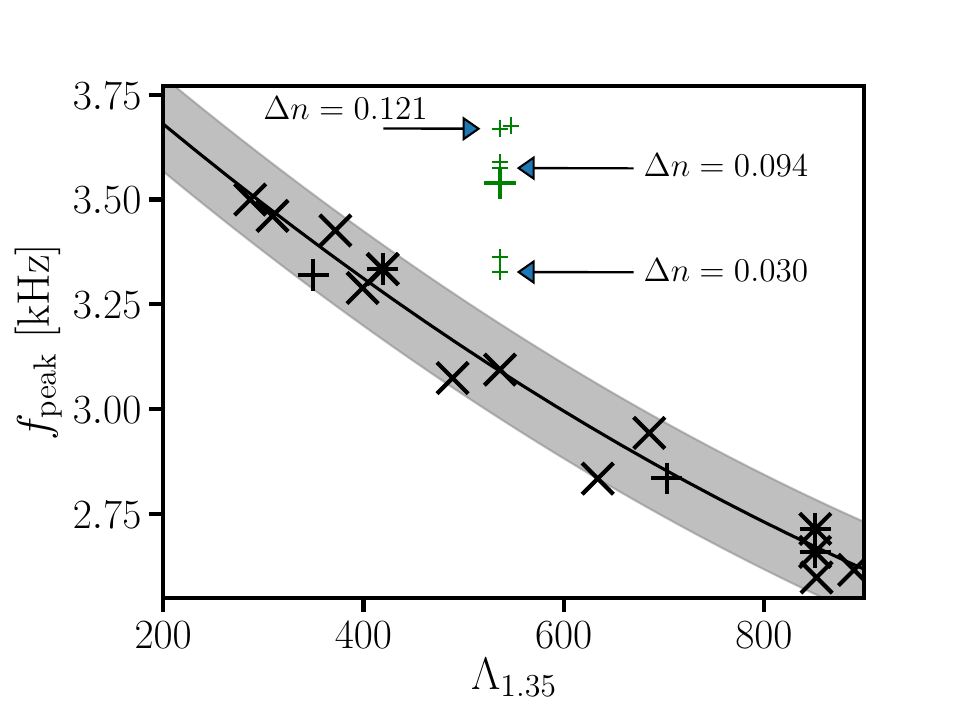}
		\caption{Left panel: Orientation-averaged spectra of the GW signal
  for different EOSs and the Adv LIGO (red dashed) and ET (black dashed) sensitivity curves. The inset shows the GW amplitude of the $+$ polarization at a distance of 20~Mpc for one of the EOSs. Figure from ref.~\cite{Bauswein:2011tp}. Right panel: Peak frequency of the postmerger GW emission as a function of tidal deformability $\Lambda$ for a $1.35 M_\odot - 1.35 M_\odot$ NS-NS merger. Black symbols are for purely hadronic EOSs, while green symbols are for EOSs which include a phase transition to quark matter; characterized by a density jump $\Delta n$ given $\rm fm^{-3}$. The solid curve shows a fit for the purely hadronic EOSs. Figure taken from ref.~\cite{Bauswein:2018bma}.}
		\label{figure_Bauswein:2011tp}
	\end{center}
\end{figure}

In~\cite{Bauswein:2011tp}, the authors used numerical-simulations to derive a correlation between the GW frequency in the post-merger phase and the EOS of NS matter. Since then, numerous studies have followed and revealed several possible relations between inspiral and postmerger properties. Specifically, the main characteristic peak in the GW amplitude spectral density was found to be correlated with the radius of the maximum-mass TOV solution or other quantities, such as the tidal deformability of the stars, which are directly accessible through GW observations of the inspiral. 
The left panel of figure~\ref{figure_Bauswein:2011tp} (taken from~\cite{Bauswein:2011tp}) shows the typical GW spectrum one can expect from the post-merger phase of a BNS merger leading to a hypermassive or supramassive NS remnant. 

As seen in figure~\ref{figure_Bauswein:2011tp}, the post-merger signals are emitted at high frequencies ($\approx 2-4$ kHz), and therefore, their detection by the current LIGO-Virgo-KAGRA interferometers is limited by their sensitivity, and no detection has been reported so far (see~\cite{LIGOScientific:2017fdd,LIGOScientific:2018urg} for a search in the case of GW170817). The higher sensitivity of the ET detector at such high frequency could instead allow for a detection of such signals~\cite{Breschi:2022ens}. Unfortunately, at the moment, we are still lacking accurate enough approximants that could be used in parameter estimation, but some models have been suggested, e.g.,~\cite{Puecher:2022oiz,Easter:2018pqy, Tsang:2019esi, Breschi:2019srl, Soultanis:2021oia, Breschi:2022xnc}, as well as agnostic data analysis techniques that could be used without templates, e.g., \cite{Chatziioannou:2017ixj, Tringali:2023ray}. However, a current issue is that different numerical-relativity codes, even if producing similar results from a qualitative point of view, still produce quantitative differences in the post-merger phase for the same BNS systems~\cite{Espino:2022mtb}; see~section~\ref{Sec:Simulation_Uncertainty}. In the following, we will outline some of the science targets or studies doable with future postmerger detections. 

In general, it is also worth highlighting that the use of EOS-insensitive relations (e.g., refs.~\cite{Takami:2014tva,Bernuzzi:2014kca,Yagi:2016bkt,Bauswein:2017aur}) allows not only for the reduction of the d.o.f.\ in waveform modeling and enables Bayesian studies, e.g., \cite{Breschi:2021xrx,Breschi:2019srl,Wijngaarden:2022sah,Puecher:2022oiz}, or to search for possible phase transitions, it also helps to perform pre/post-merger consistency tests~\cite{Tsang:2019esi,Breschi:2023mdj}; as described below. 
However, these relations are always derived based on a selected set of EOSs, which do not cover all possible scenarios, e.g., the presence of phase transition could alter derived relations and leads to systematic biases. This emphasizes the need for further studies in the coming years to ensure that the full potential of postmerger detections is revealed. 

\begin{figure}[t]
	\begin{center}
		\includegraphics[width=0.6\textwidth]{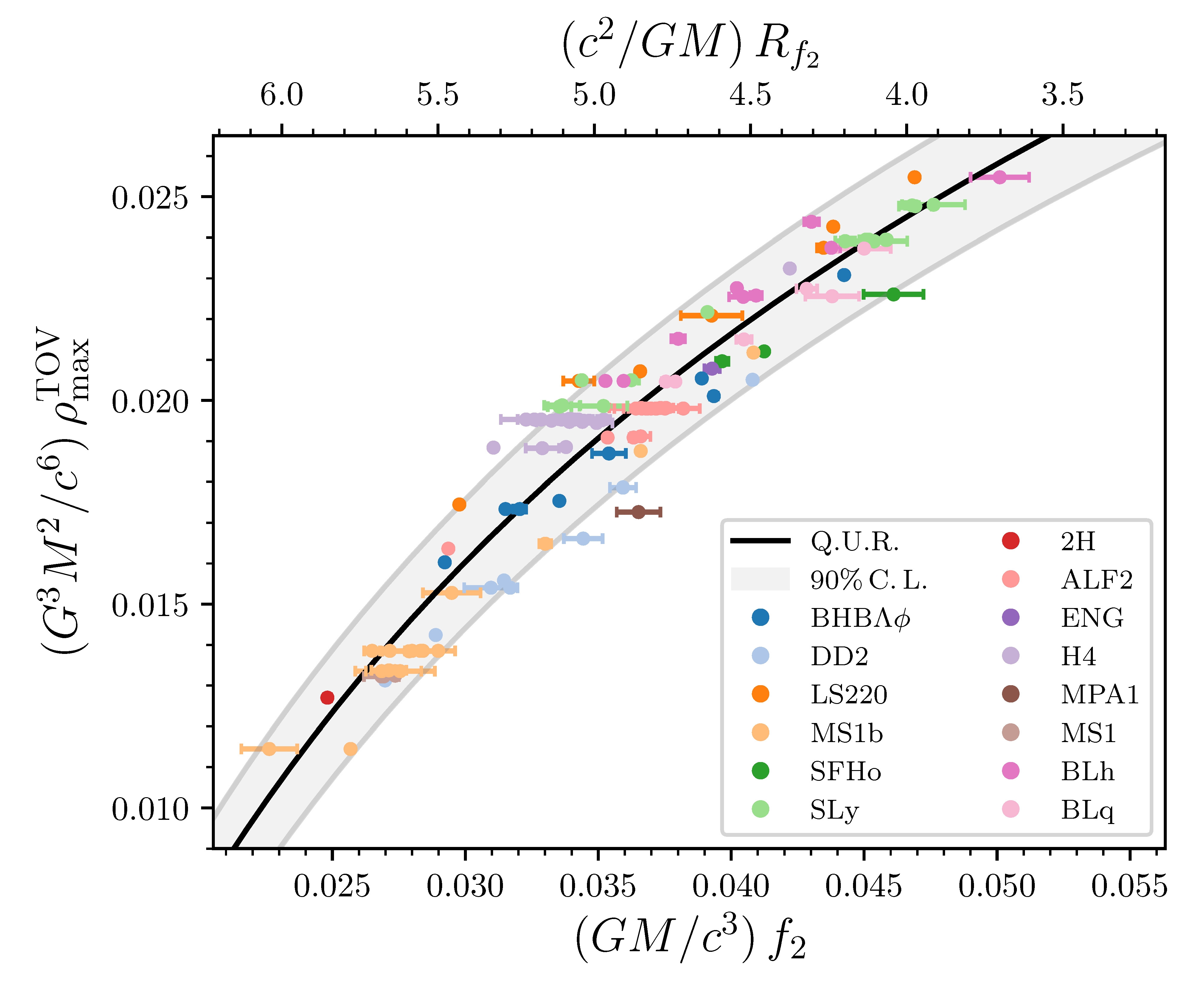}
		\caption{Empirical relation (black line) for the maximum central density $\rho^{\text{TOV}}_{\text{max}}$ of a non-rotating NS as function of the postmerger peak frequency $f_{2}$ and the Keplerian radius $R_{f_{2}}$. The colored markers show the data extracted from 289 numerical-relativity simulations with 14 EOSs, whereas the shadowed area indicates the $90\%$ credibility region of the fit. Figure is adapted from ref. \cite{Breschi:2021xrx}.}
		\label{fig:relation.rhomax}
	\end{center}
\end{figure}

\subparagraph{Determining NS properties for maximum-mass TOV stars.} 
Post-merger frequencies are expected to reveal information on remnants with up to two times the original component's mass reaching $\sim (3-6)\rho_{0}$, with $\rho_{0}$ denoting as before the nuclear saturation density. For those post-merger remnants that do not promptly collapse into black holes, NR simulations predict loud GW transients with a characteristic main peak frequency. The postmerger main peak frequency $f_{2}$ can be phenomenologically associated with an effective remnant radius using Kepler's law \cite{Breschi:2021xrx}.
Even though this quantity has no direct physical interpretation in terms of the remnant properties, it correlates with the maximum central density $\rho^{\text{TOV}}_{\text{max}}$ of a non-rotating equilibrium NS with a weak dependence on the EOS.
This relation $\rho^{\text{TOV}}_{\text{max}}(R_{f_{2}})$ is illustrated in figure~\ref{fig:relation.rhomax}. Hence, to a good approximation, the maximum density $\rho^{\text{TOV}}_{\text{max}}$ can be inferred from a full-spectrum BNS GW observation, including the postmerger; where some further work is required to robustly check these relation shown in figure~\ref{fig:relation.rhomax} in the presence of a strong phase transition and more realistic microphysics. Hence, a postmerger detection can generally impose bounds on the maximum NS mass $M^{\text{TOV}}_{\text{max}}$ and several postmerger detections could enable a slightly more accurate measurement \cite{Bauswein:2014qla} complementary to that from an inspiral EOS inference \cite{LIGOScientific:2018cki,Landry:2018prl,Essick:2019ldf,Breschi:2019srl}. The minimum NS radius $R^{\text{TOV}}_{\text{max}} = R(\rho^{\text{TOV}}_{\text{max}})$, that is the radius of a NS at maximum density $\rho^{\text{TOV}}_{\text{max}}$, is crucial information to constrain the mass-radius diagram for NSs. It can be extracted from quasiuniversal relations between the postmerger peak frequency \cite{Bauswein:2014qla} and measured with an uncertainty of $\sim 2$ km at postmerger signal-to-noise ratio of 10 \cite{Breschi:2019srl,Breschi:2021wzr}; see~\cite{Bauswein:2015yca,Takami:2015gxa,Tsang:2019esi,Puecher:2022oiz}. 
It is worth noting that, in the presence of strong phase transitions, and additional physics such as viscosities, the accuracy of the employed quasi-universal relations might decrease, and constraints might be weaker. 
Imposing stronger constraints on the EOS from BNS mergers will be feasible by combining GW observations with other messengers, e.g., \cite{Bauswein:2017vtn,Radice:2017zta,Breschi:2021tbm}. 

\subparagraph{Search for Phase Transitions and pre/post-merger consistency tests.} The higher densities and temperatures produced after merger may also trigger phase transitions. Such phase-transitions can significantly modify the dynamics and hence the GW emission, so that in some cases, the post-merger GW frequency would not match anymore with previously derived predictions~\cite{Bauswein:2018bma,Most:2018eaw} (e.g., see right panel of figure~\ref{figure_Bauswein:2011tp}). As mentioned, in the absence of phase transitions, the characteristic post-merger GW frequency can be correlated with the tidal deformability $\Lambda$, inferred during the inspiral, e.g.,~\cite{Takami:2014tva,Bernuzzi:2015rla}. Combining inspiral and post-merger GW detections with ET could therefore allow for the detection of phase transitions (e.g., see~\cite{Wijngaarden:2022sah}). At the moment, simulations accounting for phase transitions have focused mainly on two possible scenarios: the softening of the EOS due to the appearance of hyperons (e.g., \cite{Radice:2016rys}) and the transition to deconfined quark matter (e.g., \cite{Bauswein:2018bma,Most:2018eaw,Most:2019onn,Blacker:2020nlq,Prakash:2021wpz}). It has also been pointed out that unless the phase transition is `strong', it could be difficult to disentangle the deviations from other uncertainties, e.g.~\cite{Prakash:2023afe}, and the shift in the peak could also be due to non-convex dynamics \cite{Rivieccio:2024sfm}, both of which are areas for further work. The onset of new d.o.f.\, such as hyperons and quarks, may also impact viscosities, potentially leading to further shifts of the peak frequency, and manifest in sub-dominant oscillations of post-merger objects, e.g.,\cite{Blacker:2020nlq,Counsell:2023pqp,Chabanov:2023abq}.




\subparagraph{Thermal Effects.} Due to the high temperatures reached after the merger, thermal effects in the EOS impact the GW emission in the post-merger and hence become detectable by ET~\cite{Bauswein:2010dn, Fields:2023bhs, Raithel:2023gct, Villa-Ortega:2023cps}. Different studies have demonstrated that they alter the GW spectrum and shift the dominant post-merger frequency by amounts ranging from 60 to 200 Hz, depending on the underlying EOS~\cite{Raithel:2023zml}. 
Hence, future detections with ET allowing precise measurements of the cold part of the EOS during the inspiral could be complemented by post-merger observations to infer the thermal properties of the EOS~\cite{Fields:2023bhs,Pesios:2024bve}.
However, to enable extracting accurate information about the finite temperature EOS requires further work on analyzing quasi-universal relations and other observables in settings that go beyond the approximate treatment of thermal effects within the EOS often used in current studies, which requires a large community effort due to the complexity and computational costs of numerical-relativity simulations. 

\subparagraph{Influence of Magnetic Fields.} 

Another possibly important effect is due to the very large magnetic fields that may be formed after merger~\cite{Kiuchi:2014hja,Giacomazzo:2014qba,Kiuchi:2015sga,Kiuchi:2017zzg}. In this case, while many simulations suggest that the post-merger GW frequency should not be significantly affected~\cite{Radice:2017zta,Ciolfi:2017uak,Ciolfi:2019fie}, unless large magnetic field strengths are present before the merger~\cite{Bamber:2024wqr,Tsokaros:2024wgb}, others claim that the GW amplitude and luminosity may be suppressed~\cite{Shibata:2017xht}, making a detection more difficult, or that the frequency shift of the post-merger signal could be high enough to mask effects due to phase transitions in the EOS~\cite{Tsokaros:2024wgb}. This is because magnetic turbulence acts as an effective viscosity on the fluid, redistributes angular momentum, and may accelerates collapse to BH.

\begin{figure}[t]
	\begin{center}
		\includegraphics[width=0.5\textwidth]{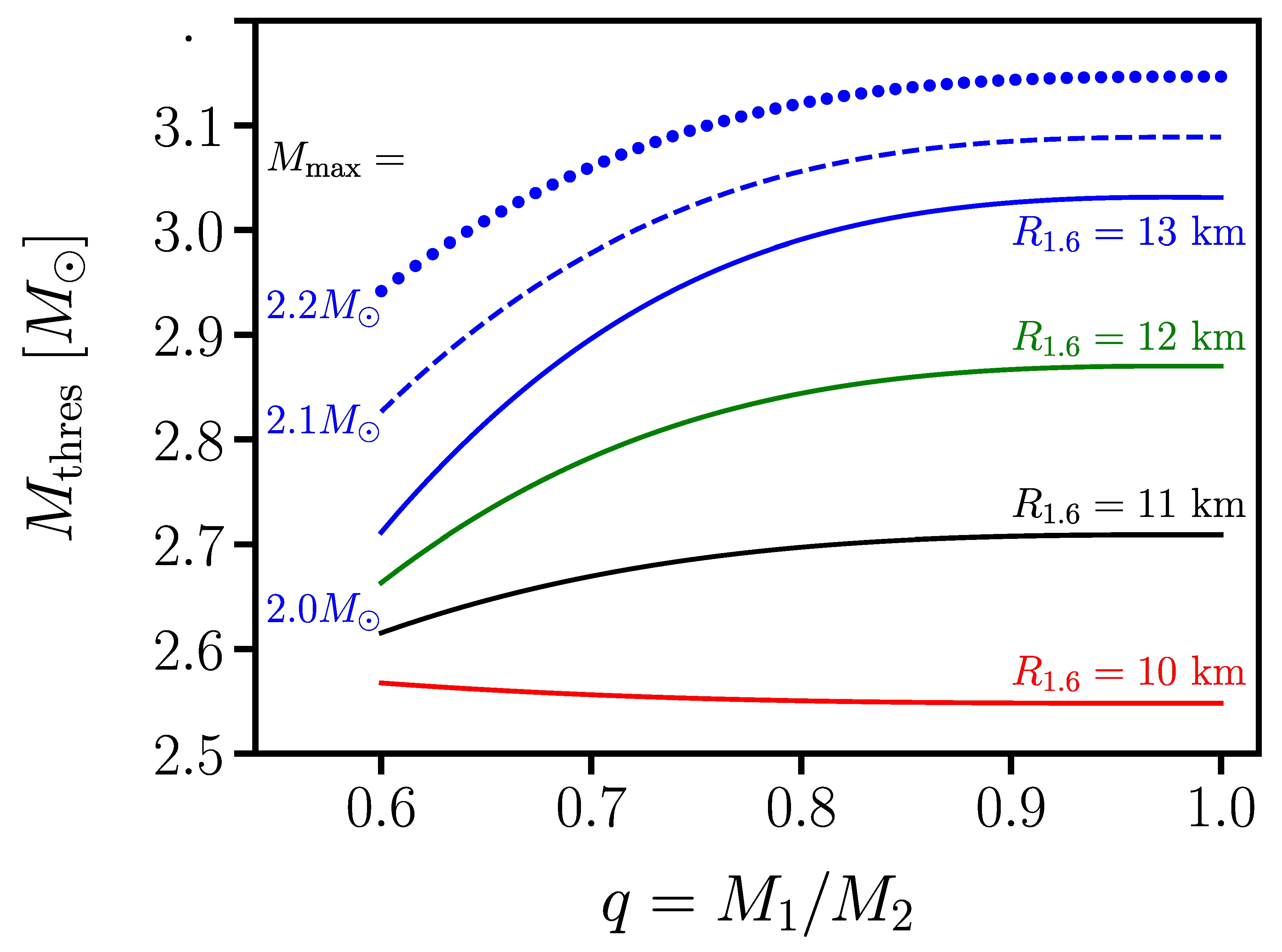}
		\caption{Threshold binary mass for prompt collapse as a function of mass ratio for different neutron star radii. Solid curves assume a fixed maximum mass of $M_{\text{max}} = 2 M_{\odot}$ but different NS radii. Blue curves show $M_{\text{thres}}(q)$ for a	fixed radius $R_{1.6} = 13$ km but with $M_{\text{max}}$ being $2.0 M_{\odot}$ (solid), $2.1 M_{\odot}$ (dashed) and $2.2 M_{\odot}$ (dotted). Figure adapted from ref. \cite{Bauswein:2020xlt}.}
		\label{fig:mtres.for.q}
	\end{center}
\end{figure}

\subparagraph{Inference of prompt black-hole formation.} Constraints on the EOS may also be provided by inferring the threshold binary mass for prompt black-hole formation in NS mergers from analyzing the GW signal radiated in the inspiral phase. The merging of two NSs may lead to either the formation of a BH or of a NS remnant \cite{Shibata:2005xz,Shibata:2006nm}. A direct gravitational collapse occurs if the binary's total mass $M_{\text{tot}}$ is higher than some threshold mass $M_{\text{thres}}$ beyond which the forming remnant cannot be stabilized. For lower $M_{\text{tot}}$, rapid differential rotation and thermal pressure support the central object against prompt collapse even if $M_{\text{tot}}$ is larger than the maximum mass of non-rotating NSs. Such systems may undergo a delayed collapse to a BH due to the reduction of angular momentum and cooling \cite{Bauswein:2020xlt}. The merger outcome is predominantly determined by the ratio of the total mass $M_{\text{tot}}$ and the threshold mass $M_{\text{thres}}$~\cite{Hotokezaka:2011dh,Bauswein:2013jpa}. The system with the lowest $M_{\text{tot}}$ and observational indications for a prompt collapse gives an upper limit on $M_{\text{thres}}$, whereas a detection with characteristics excluding a prompt collapse limits $M_{\text{thres}}$ from below. In general, the dynamics of NS mergers depend on the incompletely known EOS of high-density matter \cite{Ozel:2016oaf,Oertel:2016bki} and other parameters such as the mass ratio or the spin of the merging stars~\cite{Tootle:2021umi,Kolsch:2021lub}. Over the last years, numerous phenomenological relations for $M_{\text{thres}}$ have been derived, e.g.,~\cite{Bauswein:2013jpa,Agathos:2019sah,Koppel:2019pys,Bauswein:2020xlt,Kolsch:2021lub}. 
figure~\ref{fig:mtres.for.q} shows one such relation, where the threshold mass is lined to the mass ratio and the typical radius of a NS with a mass of $1.6M_\odot$, $R_{1.6}$. 
However, further work is needed to verify the robustness of the relations, for example, ref.~\cite{Perego:2021mkd} finds that the threshold mass does not have a universal behavior under EOS variations for $q\neq1$ as the collapse is induced by the tidal disruption and accretion of the secondary star \cite{Sekiguchi:2015dma,Lehner:2016lxy,Bernuzzi:2020tgt}. However, \cite{Perego:2021mkd} demonstrate that observations of prompt collapse thresholds, either from binaries with two different mass ratios or with one mass ratio but combined with the knowledge of the maximum NS mass or compactness, will constrain the incompressibility at the maximum NS density $K_{\rm max}$ to within tens of percent. Hence, measuring the maximum mass in combination with the other systems' parameters directly helps us to constrain important information about the EOS.

\subparagraph{Accuracy in the measurement of postmerger properties.} Recent investigations have attempted to estimate the accuracy with which the post-merger signal could be measured by next-generation GW observatories. For example, in~\cite{Clark:2015zxa} the authors estimated an error in the measure of $f_{\rm peak}$ (sometimes also referred to as $f_2$) of $\delta f_{\rm peak}\approx 139$~Hz at a horizon distance $d_{\rm hor}\approx 267$~Mpc for ET (corresponding to a source with a signal-to-noise ratio of 5 in the post-merger). The same error could be achieved by Advanced LIGO at $d_{\rm hor}\approx 30$~Mpc. The detection rate for the post-merger signal was estimated to be $\approx 1$ every $100$ years for Advanced LIGO and $\approx 3$ per year for ET. 
Combining the merger and post-merger signal, ET could also detect phase transitions in the post-merger, assuming the phase transition causes a shift in the $f_{\rm peak}$ predicted from the inspiral measured $\Lambda$ (solid line in right panel of figure~\ref{figure_Bauswein:2011tp}) of at least $455$ Hz (assuming a signal-to-noise ratio of at least 10 in the post-merger~\cite{Prakash:2023afe}).
Due to the limited number of post-merger detections per year, it will probably be necessary to stack multiple detections together in order to increase the accuracy in the determination of the post-merger EOS. One year of observations may indeed not be enough to distinguish between different EOSs in the post-merger phase~\cite{Martynov:2019gvu}. For example, in~\cite{Yang:2017xlf} it was investigated the possibility of using both power and coherent mode stacking methods and it showed that it could be possible to measure the post-merger frequency and the NS radius with an accuracy of $\approx 4 \%$ or better (depending on the EOS) when combing together $30$ post-merger detections (i.e., with an error of $\approx 20 $ Hz on the post-merger peak frequency and less than $\approx 100$ m in the NS radius). Such an accuracy should be sufficient to detect phase transition in the post-merger phase, since it may be able to show a difference between the measured post-merger frequency and the one expected from the EOS measured during the inspiral. We need to make clear that studies are still very limited at the moment, since the larger sensitivity of ET in the post-merger phase will require more accurate theoretical predictions. NR simulations, for example, sometimes employ approximate modeling of finite temperature effects instead of a fully tabulated finite temperature EOS (which is more computationally expensive). While current detectors are not sensitive enough to be impacted by such approximations, ET could be affected for sources up to $\approx 44$~Mpc in case of softer EOSs, and up to $\approx 113$~Mpc for stiffer EOSs~\cite{Raithel:2023gct} (see also~\cite{Villa-Ortega:2023cps, Miravet-Tenes:2024vba}).\\

\paragraph{GWs from core-collapse supernovae and proto-neutron stars.}
\label{subsec:GWCCSN}


CCSN probe the EOS of supranuclear-dense matter in complementary regimes of temperature and density to BNS mergers. CCSNs are triggered when exothermic nuclear burning ceases at the center of a 
star with a mass between $8 \Msun$  and $100 \Msun$. At this time, the hydrostatic equilibrium is lost and the core (iron) begins to collapse from its original radius of a few thousands of km. The collapse of the core continues until it reaches nuclear densities, at which point the repulsive forces between nucleons cause the collapse to decelerate rapidly. This results in a partial rebound of the inner core, known as the core bounce, which generates a shock wave that propagates outward. 
The collapse is then halted at the formation of a PNS. At an initial mass and radius of $M_{\mathrm{PNS}} \sim 0.6 \Msun$ and $R_{\mathrm{PNS}} \sim 50 \,
\mathrm{km}$, this object has a moderately relativistic
gravitational field, which gets stronger when the newly formed PNS accretes additional matter and contracts.  
  
During this phase, processes such as convection, rotation, and magnetic field interactions, generate non-spherical flows in the core that could produce gravitational radiation. 
These flows also contribute to all of the mechanisms that have been proposed to initiate a CCSN explosion after the prompt shock has failed to break out of the core. In most stars, they support neutrino-driven explosions by enhancing the efficiency with which neutrinos transport energy from the PNS to its surrounding layers~\cite{Janka:1996tu,Janka:2012wk,Vartanyan:2023sxm}.  In a subset of cores, high rotational and magnetic energies drive the explosion in a predominantly anisotropic geometry \cite{Wheeler:2001me,Akiyama:2002xn}.  Even if none of these mechanisms succeeds at launching a CCSN and instead accretion causes the PNS to collapse to a black hole (BH), spherical symmetry will be broken to a considerable degree~\cite{Cerda-Duran:2013swa,Burrows:2023nlq}. The GW signal is determined by the dynamic processes in the collapsing core and PNS, exhibiting a range of features with varying amplitudes and frequencies that are sensitive to 
the stochastic 
source dynamics
and the EOS of nuclear matter.

\begin{figure}[t]
    \centering
    \includegraphics[width=0.6\linewidth]{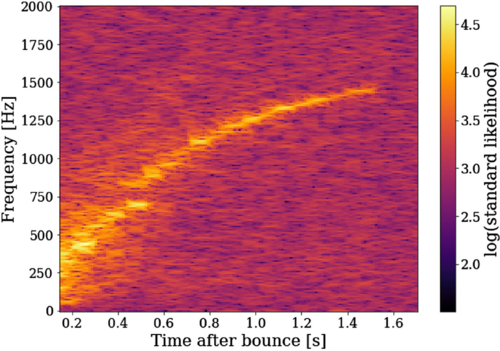}
    \caption{{Time-frequency map of the GW signal of a three-dimensional model of the collapse of a star with $20 \, M_\odot$ showing a PNS oscillation mode whose frequency rises continuously from the time of core bounce ($t = 0$). Figure taken from \cite{Bruel:2023iye}.}}
    \label{fig:CCSN_GW_emission}
\end{figure}

Information on the EOS in the signal comes from hydrodynamic oscillation modes of the PNS whose frequencies have a direct relation with the structure of the PNS, in particular its mass and radius.
The time evolution of the mode spectrum would yield valuable insights into the EOS 
through GW asteroseismology of PNSs.
No such detection has been achieved so far, but progress in theoretical modeling along two lines of work has laid the ground for future observations. As the relevant PNS oscillations are relatively weak perturbations of a slowly varying object close to equilibrium, they can be described as solutions of the linearized hydrodynamic equations \cite{Torres-Forne:2017xhv,Torres-Forne:2018nzj,Sotani:2016uwn,Sotani:2021ygu}. This approach is computationally inexpensive and thus well suited to scanning the large parameter space of nuclear and stellar physics. It cannot, however, account for the large-scale dynamics of the core, including, e.g., the development of a CCSN explosion, the rate at which mass is accreted onto the PNS, its internal dynamics such as convection, and it cannot predict how and to which amplitude the possible modes are excited. These questions can be addressed by large-scale, multi-dimensional simulations of the entire core that capture \mbox{(magneto-)}hydrodynamics, relativistic gravity, neutrino transport, and nuclear reactions. Such models produce detailed synthetic GW signals but can be very expensive. Thus, only a limited number of them are available \cite{Mosta:2014jaa,Lentz:2015nxa, Andresen:2016pdt,Burrows:2019rtd,Burrows:2019zce,Kuroda:2020bdq,Powell:2020cpg,Obergaulinger:2020cqq,Vartanyan:2021dmy}. Figure \ref{fig:CCSN_GW_emission} shows an example result of three-dimensional simulation as a time-frequency map of the GW signal from a $20 \, M_\odot$ star. The figure depicts the continuous rise of the PNS oscillation mode frequency starting from core bounce ($t = 0$). 

Perturbation analyses aimed at uncovering the relationship between GW emission and the physical parameters of progenitors during PNS formation and evolution have shown that the time-frequency evolution of the primary emission mode corresponding to the $g$-mode or $f$-mode, closely relates PNS properties such as surface gravity and average density~\cite{Torres-Forne:2019zwz}. Additionally, advanced numerical simulations of CCSN have established empirical relationships that link oscillation modes to intrinsic system attributes and are largely independent of the EOS or progenitor mass~\cite{Torres-Forne:2019zwz}. However, factors like progenitor mass and the EOS for dense matter can influence GW signals, leading to further empirical relationships between the supernova GW frequency and the average density of the PNS \cite{Sotani:2016uwn}. While most of these relationships are largely EOS-independent, certain PNS $g$-modes, especially those with decreasing frequencies over time, exhibit a clear EOS dependence \cite{Jakobus:2023fru}. 
Building on these insights, a recent investigation has compiled a comprehensive dataset of over 1000 exploding CCSNe models spanning various progenitor masses, metallicities, and nuclear EOS~\cite{Wolfe:2023rfx}. This work identified two main frequencies in a range of order 1 kHz, with the lower/higher one associated to the early/late stage of the signal that are sensitive to the EOS, and with the latter being correlated with the surface gravity of the remnant.  

Although the frequency range of GW emissions from CCSNe is within the reach of current detectors, the detection range for slowly or non-rotating CCSNe is limited to approximately 5 kpc~\cite{Szczepanczyk:2021bka,Abdikamalov:2020jzn}. 
If the progenitor core exhibits very rapid rotation, as seen in only 1\% of electromagnetically observed events~\cite{Li:2010kc,Chapman:2007vp}, the distance increases up to 50 kpc. The estimated occurrence rate for CCSNe within the Milky Way is about 1 to 2 per century~\cite{Rozwadowska:2020nab}, making detection challenging for current detectors.
For ET, the anticipated event rate is about 0.5 events per year~\cite{Ando:2005ka}, with a detection range extending up to (2–4)~Mpc for neutrino driven explosion and up to (5–30)~Mpc for rapidly rotating progenitors~\cite{Gossan:2015xda}. For a detailed assessment of the observed and expected rates of CCSNe in the local Universe and the Milky Way, see section~ \ref{ccsnrates}.

\begin{figure}[t]
\centering
\includegraphics[width=0.36\columnwidth,angle=-90]{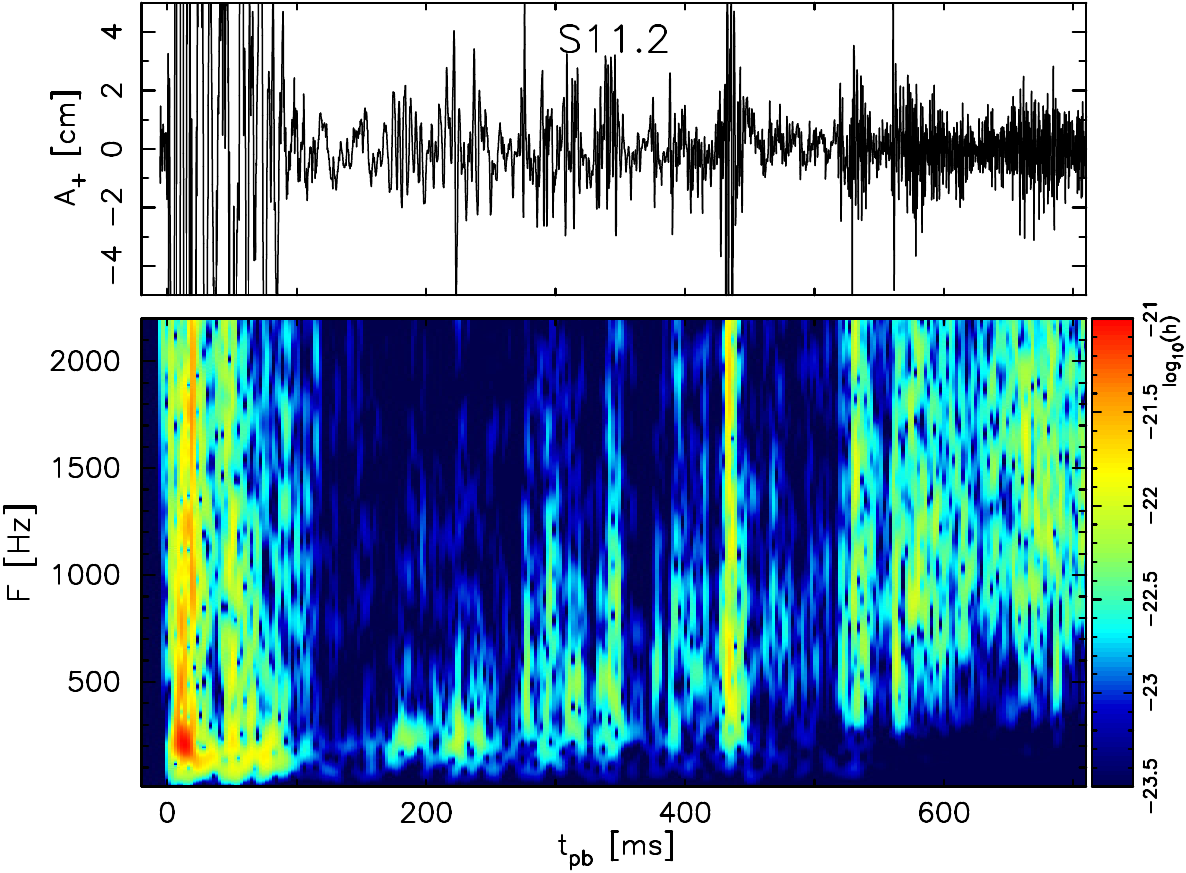}\label{fig:GWa}
\includegraphics[width=0.36\columnwidth,angle=-90]{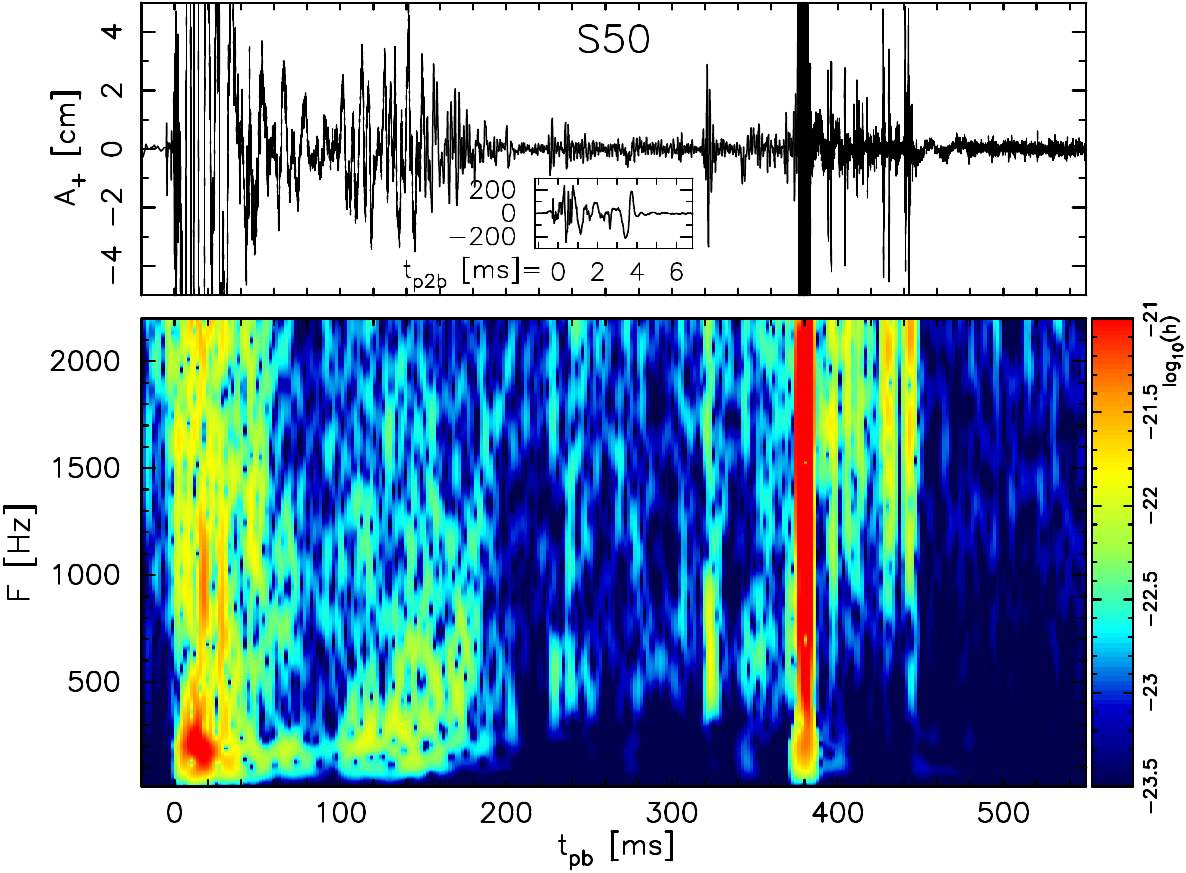}\label{fig:GWb}
\caption{Post-bounce evolution of the gravitational waveform $A_+$ (top panels) and the corresponding spectrogram (bottom panels) for two representative models, neutrino-driven supernova explosion of a 11.2~M$_\odot$ progenitor model (S11.2) and 50~M$_\odot$ model (S50), the latter featuring a QCD phase transition and associated supernova explosion onset at around 376~ms post bounce, which is accompanied by a sudden rise of the GW amplitude. For S50, a magnified view of the gravitational waveform is shown in the inlay of the top panel with respect to the second bounce time, $t_{\rm p2b}$. Figure taken from ref. \cite{Kuroda22}.
\label{fig:GW}}
\end{figure}

\begin{figure}[t]
\centering
\includegraphics[width=0.75\columnwidth,angle=0]{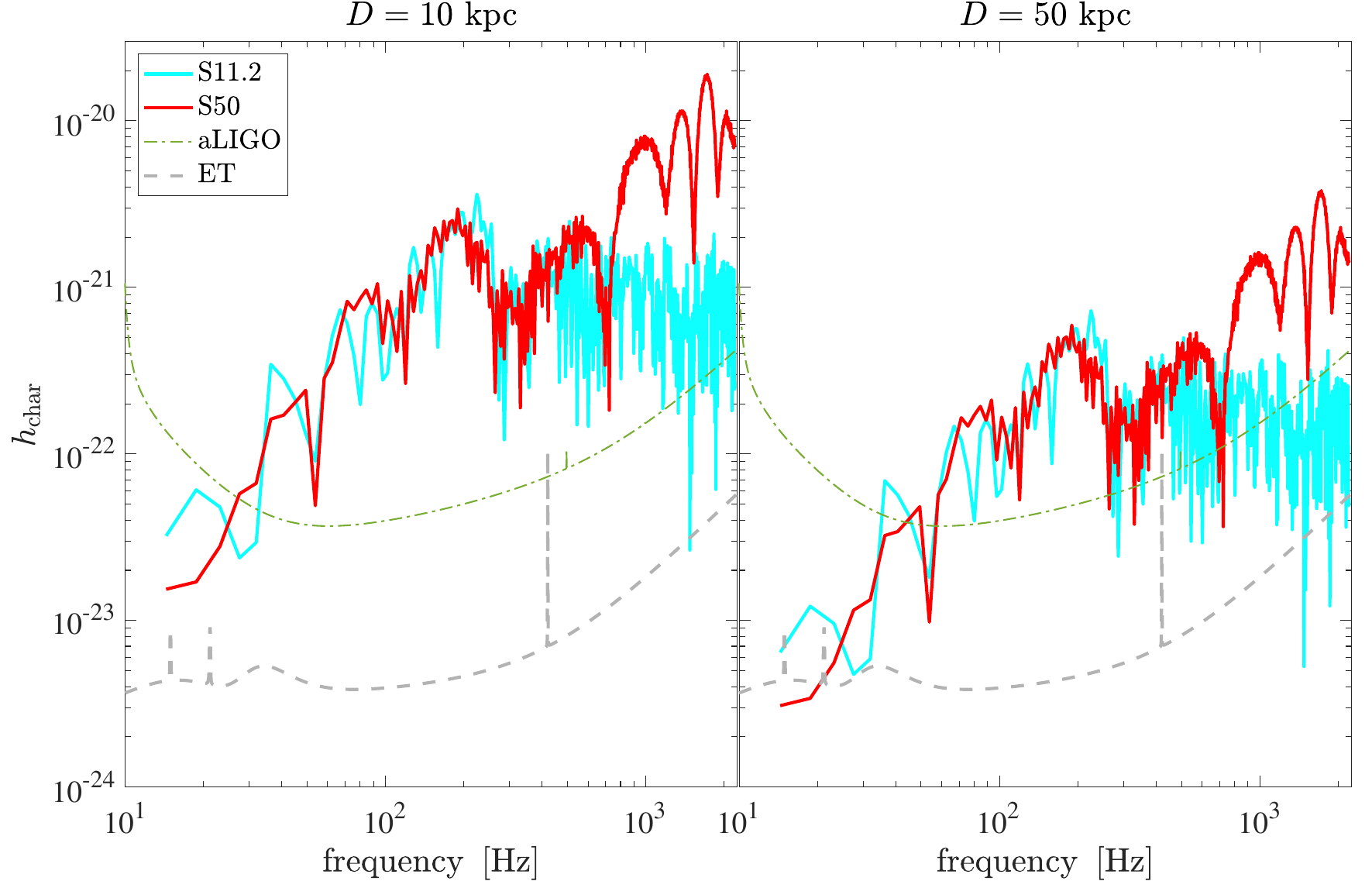}
\caption{Characteristic GW spectral amplitudes, $h_{\rm char}$, for the two models S11.2 (light blue solid lines) and S50 (red solid lines) shown in figure~\ref{fig:GW}, assuming a source distance of 10~kpc (left panel) and 50~kpc (right panel). The noise amplitudes of aLIGO (green dashed lines) and  ET (grey dashed lines)  are plotted as references. Figure reproduced based on data from ref.~\cite{Kuroda22}.
\label{fig:GWstrain}}
\end{figure}

In the case of QCD phase transitions, the GW emission due to the sudden contraction of the PNS results in an abrupt rise in the amplitude by more than a factor of 10, followed by a model-dependent ring-down phase that can last for several hundreds of milliseconds. This unique GW emission, as illustrated in figure~\ref{fig:GW} (right panel), makes the QCD-driven supernova scenario well distinguishable from a canonical neutrino-driven supernova explosion (see the characteristic GW strain $h_{\rm char}$ in figure~\ref{fig:GWstrain}, comparing S11.2 and S50). The detailed GW analysis of these models demonstrates great potential for their detection in galactic events, with characteristic GW spectral amplitudes more than an order of magnitude above those of normal supernovae in the frequency range above about 1 kHz.

Associated with the GW signal from the QCD phase transition is the emission of a non-standard neutrino burst, dominated by $\bar\nu_e$ and contains information about bulk properties of the EOS \cite{Largani:2023oyk}. 
Despite the present uncertainties due to yet-incompletely understood neutrino flavour evolution, this non-standard $\bar\nu_e$ burst is detectable at the currently operating generation of water-Cherenkov detectors \cite{Dasgupta10, Fischer20}. Such QCD-driven massive star explosions also feature $r$-process nucleosynthesis \cite{Fischer20}.

\subsubsection{Nucleosynthesis and multi-messenger signals}
\label{ssec:multimessenger}

In addition to being primary targets for ET,
compact binary mergers and CCSNe are key players in galactic chemical evolution, see, e.g., \cite{Thielemann:2018bwk,Cowan21,Matteucci:2021hhd,Arcones:2022jer}.
Moreover, the variety of scales and processes involved in these events produce radiation of different kinds, 
qualifying them
as genuinely multimessenger events.
In the strong field regime, the heating of dense matter 
boosts weak reactions and produces intense neutrino radiation (with peak luminosities $\sim 10^{53}{\rm erg~s^{-1}}$) in all neutrino flavors, lasting for seconds and influencing the ejecta composition, see, e.g., \cite{Eichler:1989ve,Rosswog:2003rv,Janka:2006fh,Burrows:2012ew}.
At the same time, several mechanisms involving strong magnetic fields, neutrinos, and dynamical spacetime can produce a relativistic jet that drills inside the ejecta or through stellar layers, and possibly breaks out.
Once ejected matter has expanded, it becomes transparent enough to emit electromagnetic (EM) radiation over the entire  spectrum. In particular, the jet is expected to produce the non-thermal emissions powering the prompt and afterglow emission of gamma-ray bursts (GRBs). According to the present paradigm, compact binary mergers produce short GRBs, while the long ones are produced by the core-collapse of fast rotating, highly magnetized stellar cores (collapsars).
Within the same scenarios, the decay of freshly synthesised radioactive nuclei powers the light curves and spectra that characterizes CCSNe and kilonovae.

Nuclear physics represents a key ingredient for the prediction and interpretation of the multimessenger signals from CCSNe and compact binary mergers, as well as of their nucleosynthesis signature. With its unprecedented sensitivity and sky localization capability (especially, if part of a larger network of detectors, not only for GWs, but also for EM radiation and neutrinos) ET will enable us to maximize the outcome of joint multimessenger analysis, see also section~\ref{section:div5} and  section~\ref{section:div7}. 
In particular, it will directly connect the strong field dynamics, where the properties of dense nuclear matter are at play and shape the GW and neutrino signals, with its imprint in post-explosion/post-merger observables, including the EM emissions, as well as the nucleosynthesis yields.

\paragraph{Nucleosynthesis in Core-Collapse Supernovae and the r-process.}
\label{subsubsec:CCSN nucleosynthesis}

\begin{figure}[t]
\centering
\includegraphics[width=0.49\columnwidth,angle=0]{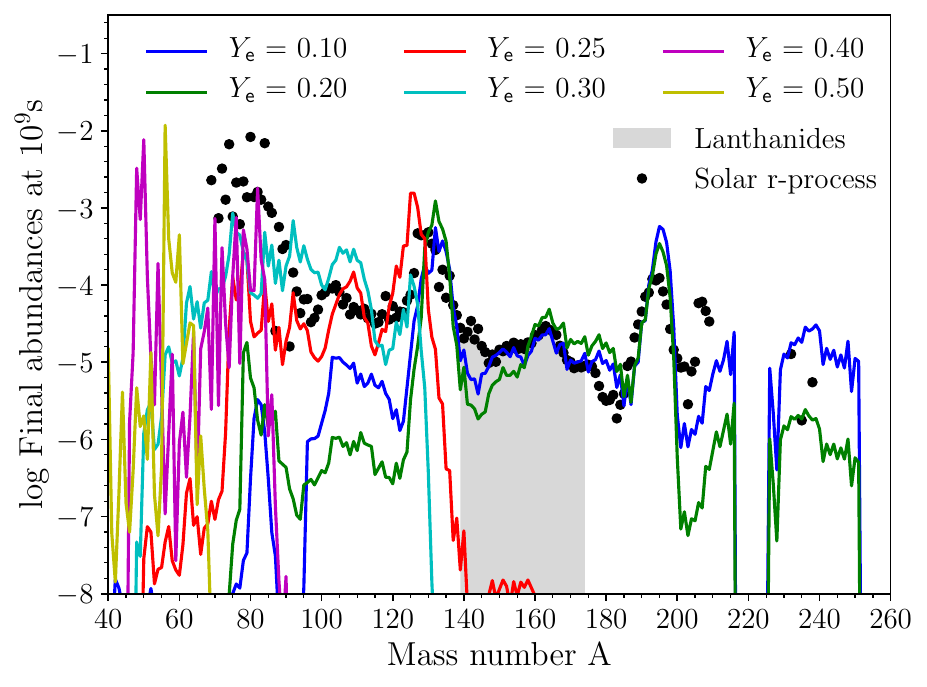}
\includegraphics[width=0.49\columnwidth,angle=0]{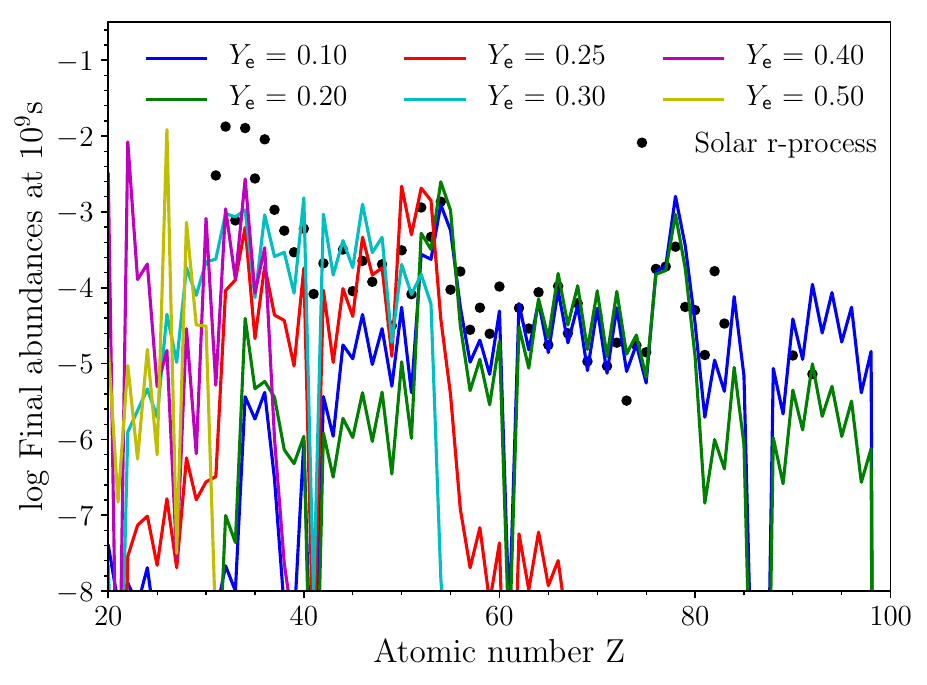}
\caption{Abundances obtained for a fluid element of entropy $s \approx 10 k_{\rm B}{\rm~baryon^{-1}}$, expansion timescale $\tau \approx 10~{\rm ms}$ and for different initial $Y_e$ are presented as a function of the mass number $A$ (left) and of the atomic number $Z$ (right).
For $Y_e \lesssim 0.25$, 
r-process nucleosynthesis produces all heavy elements between the second (Te-I-Xe region) and third (Ir-Pt-Au region) r-process peaks, including lanthanides \cite{Lippuner:2015gwa}. If $Y_e \lesssim 0.15$, also actinides are produced \cite{Lippuner:2015gwa,Giuliani:2019oot}.
The production of elements between the first (Se-Br-Kr region) and the second r-process peaks requires $0.25 \lesssim Y_e \lesssim 0.4 $. Figure from \cite{Perego:2021dpw}.
\label{fig:r_process_abundances}}
\end{figure}

The general picture of nucleosynthesis in standard CCSNe is nowadays well-established \cite{Woosley:2002zz}. 
Extensive calculations were usually performed by means of  parameterized spherically symmetric models, e.g., \cite{Sukhbold:2015wba,Curtis:2018vkh}, including also the study of
correlations with the EOS of dense matter \cite{Ghosh:2021bjl}.
Three-dimensional models featuring nucleosynthesis calculations
have also become available \cite{Hammer:2009cn,Wongwathanarat:2014yda,Vartanyan:2018iah,Sandoval:2021hnk}. 
Their outcome was compared with observations from CCSN remnants and elemental abundances in the Solar System, in the interstellar medium and in stars of our Galaxy, as well as of nearby galaxies.
The largest uncertainties 
still concern iron-group and heavy element formation in the innermost ejecta, mostly powering the EM luminosity ($\sim 10^{41-42}\,{\rm erg~s^{-1}}$).
The silicon shell of the progenitor star is expected to experience explosive Si-burning that has $^{56}$Ni as the major product. Ejecta coming from deeper layers 
first photodissociate into neutrons, protons, and $\alpha$-particles, while the build-up of heavier nuclei occurs during the subsequent expansion.
The nucleosynthesis outcome depends mostly on the entropy and on the initial 
relative abundance of free neutrons and protons (quantified by the electron fraction, $Y_e$) \cite{Hoffman:1996aj}.
For $Y_e \sim 0.5$, the production of iron-group nuclei can occur.
The production of heavy elements beyond the iron-group follows a different path, 
called rapid neutron capture process (r-process) nucleosynthesis, which requires the presence of neutron rich conditions ($Y_e < 0.5$), see e.g. \cite{Cowan21,Perego:2021dpw}. 
Figure~\ref{fig:r_process_abundances} shows the abundances obtained for a representative low entropy fluid element ($s \sim 10\, k_{\rm B}\, {\rm baryon}^{-1}$) and for different initial $Y_e$.
For low entropy material, matter with $Y_e \gtrsim 0.4$
produces iron-group nuclei and only weak r-process elements.
However, in the case of high entropy material ($s \gtrsim 100 \, k_{\rm B}\; {\rm baryon^{-1}}$), an $\alpha$-rich freeze-out occurs and,
even for marginally neutron-rich conditions ($Y_e \lesssim 0.5$),
the synthesis of heavy elements proceeds due to the capture of free neutrons on the few available seed nuclei.
 The electron fraction is set by the interaction between neutrinos and matter. 
The initial neutrino burst is emitted a few milliseconds after core bounce and followed by accretion-powered thermal neutrino emission 
characterized by a luminosity $L_{\nu} \sim 10^{52-53} \,{\rm erg~s^{-1}}$ and mean energies $\langle E_{\nu} \rangle \sim $10-20 MeV, with heavy flavor neutrinos hotter than electron ones, due to their deeper decoupling from matter, see e.g. \cite{Janka:2017vlw}.
In the case of a successful explosion,
the neutrino luminosities decrease exponentially over the neutrino diffusion timescale in the subsequent PNS cooling phase. At the same time, the mean energies decrease down to a few MeV and equalize due to the different neutrinospheres approaching each other as the PNS contracts \cite{Fischer:2009af,Hudepohl:2009tyy}.
In the context of neutrino-delayed explosion mechanism,
neutrinos emitted by the new born PNS interact with matter just behind the shock through 
reactions like
$p + \bar{\nu}_e \rightarrow n + e^+$
and $n + \nu_e \rightarrow p + e^-$.
The precise balance between these reactions determines the equilibrium $Y_e$ in the innermost ejecta, 
e.g., \cite{Qian:1996xt,Martinez-Pinedo:2013jna,Perego:2015zca,Martinez-Pinedo:2017ksl}.

The initial entropy, expansion timescale and $Y_e$ 
are ultimately related to
the explosion dynamics (i.e., the explosion mechanism and the shock strength) and to the neutrino luminosities.
High-entropy neutrino-driven winds from the nascent PNS were long considered promising r-process nucleosynthesis sites \cite{Hoffman:1996aj}. However, recent simulations 
showed that the ejecta of standard CCSNe are not neutron-rich enough to produce robust and strong r-process nucleosynthesis \cite{Martinez-Pinedo:2012eaj,Roberts:2012um,Arcones:2012wj}. Only a partial and weak r-process can be realized in some classes of CCSNe, e.g. electron-capture SNe and CCSNe near the low-mass end \cite{Wanajo:2017cyq}.
At the same time, the lightest heavy elements, including Sr, Y, Zr, could be produced in neutrino-driven winds \cite{Arcones:2010dz}, depending
on the nuclear physics input, see also section~\ref{sect_nuclEOSynth}.
Another argument against the occurrence of full r-process nucleosynthesis in standard CCSNe is the fact that at low-metallicity the abundance of r-process elements is characterized by a large scatter. The latter points to their production in relatively large amounts, but in rare events \cite{Hotokezaka:2015zea}. 
On the other hand, special classes of rare SNe, such as magneto-rotationally driven supernovae \cite{Winteler:2012hu,Mosta:2017geb,Reichert:2024vyd} or collapsar \cite{Fujimoto:2006ek,Siegel:2018zxq},
may harbor suitable conditions in their innermost ejecta to provide strong r-process nucleosynthesis yields. Their existence is, however, still debated, as well as their viability as r-process nucleosynthesis sites, together with their dependence on relevant stellar parameters, such as metallicity, rotational velocity and core inhomogeneities.

The detection of GWs from nearby CCSNe and from long GRBs by ET, especially if combined with the multi-messenger detection of line elements in EM spectra or of thermal neutrinos from the exploding core (see below), could improve our understanding of the nucleosynthesis in the innermost ejecta, mostly by shedding light on the explosion mechanism.
In the  case of a nearby long GRB, the detection of EM signatures from r-process elements in the associated SN could provide a smoking gun evidence for the existence of rare classes of SNe contributing to r-process nucleosynthesis. In the case of an extremely close (yet unlikely) event ($\lesssim $ a few Mpc), the coincident detection of GWs, more plausible in the case of strong asymmetries in the central engine, could better clarify the nature of the progenitor.
Such a discovery would release some of the present tensions in galactic chemical evolution, e.g. \cite{Cote:2018qku}.

\paragraph{Multimessenger detections of core-collapse Supernovae.}

\begin{figure}[t]
\centering
\includegraphics[width=0.45\columnwidth,angle=0]{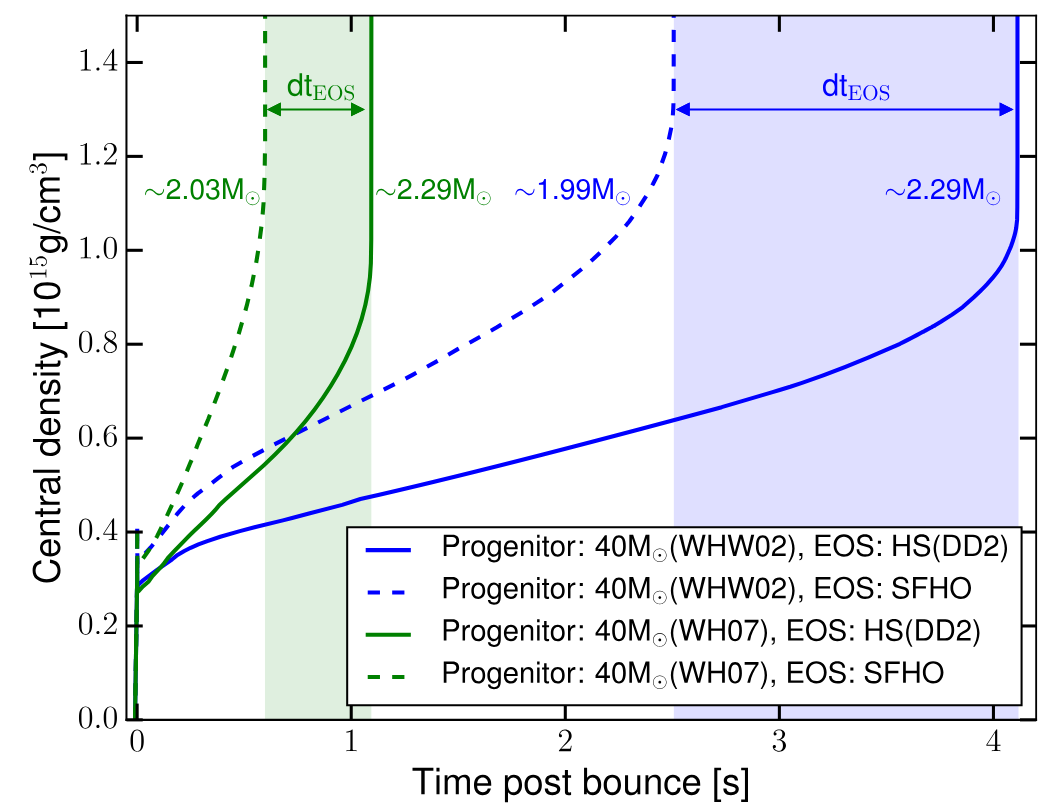}
\includegraphics[width=0.54\columnwidth,angle=0]{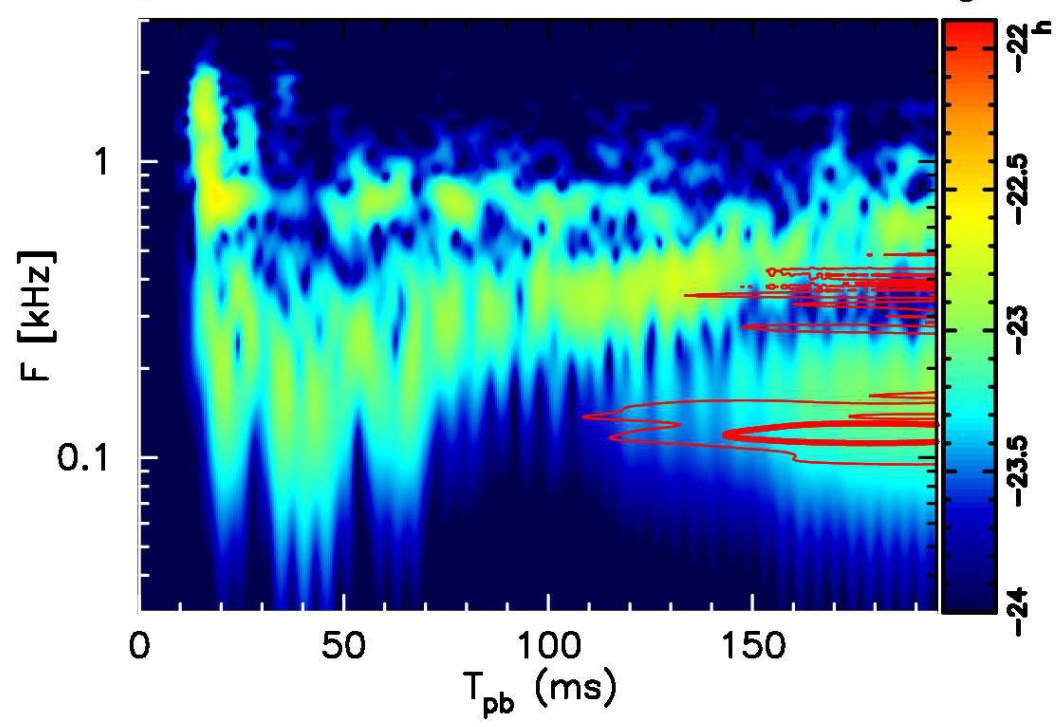}
\caption{Left: Evolution of the maximum density in CCSNe producing a BH. For a given progenitor, the nuclear EOS affects significantly the collapse timescale. Right: 
Color-coded GW spectrum from a 15$M_{\odot}$ CCSN simulation using the SFHx EOS.
The red curves are contours (only for post-bounce times larger than 100~ms) of the $\bar{\nu}_e$'s spectra.
The observer’s direction
is fixed along the $z$-axis for a source at a distance of $d = 10$ kpc. The overlap observed for this model between the two spectra is evident. 
Figures adapted from \cite{Ebinger:2018fkw} (left) and \cite{Kuroda:2017trn} (right).
\label{fig:CCSN_correlations}}
\end{figure}

In the case of a Galactic CCSN, whose rate is expected to be between 1-2 per century, a multimessenger detection could involve GWs, neutrinos and EM radiation.
With optimal GW filters and for cases involving strong asymmetries, the ET detection horizon could reach up 30-500 kpc, i.e., comparable or even larger than the distance of the Large Magellanic Cloud (50kpc) \cite{Halim:2021yqa}. 
The sensitivity of next-generation GW observatories such as ET will be necessary to distinguish between standing accretion shock instability (SASI) and convection-induced modulations in detected GW signals \cite{Abdikamalov:2020jzn}, see also section~\ref{subsec:GWCCSN}.
For a galactic event, future neutrino detectors such as Hyper-Kamiokande, DUNE or JUNO, will detect almost one hundred thousand neutrinos within a few seconds, allowing to investigate in great detail the thermal neutrino emission. Even pre-collapse neutrinos could be observed for a few seconds, probing the latest stellar evolution phases.
However, even for a Galactic CCSN, it is not obvious that all three radiation types could be easily detected: while the EM emission could be observable up to redshift $z\sim 1$, if the explosion happens in the Galactic plane and on the other side of the Galaxy, its EM emission will be likely obscured by dust.
On the other hand, GWs and neutrinos are largely unaffected by absorption processes within the Galaxy, and their combination can very precisely indicate the time of the explosion and its sky localization \cite{Nakamura:2016kkl,SNEWS:2020tbu,Halim:2021yqa}. 
Moreover, if a massive progenitor star directly collapses into black holes without bright explosion, a failed SN occurs. Such a SN will have a very faint EM signature \cite{Lovegrove:2013ssa}, while it could have significant GW and neutrino emissions.
By measuring the neutrino and GW emission during the first few seconds, a sharp end in both emissions should indicate the time of BH collapse, a feature which is possibly distinguishable from the drop in the neutrino emission due to the explosion.
The collapse timescale, in combination with information on the progenitor structure, can set constraints on the EOS of nuclear matter, since the latter sets the density at which the collapse occurs \cite{Ebinger:2018fkw}, as visible in the left panel of figure~\ref{fig:CCSN_correlations}.
Additionally, the detection and very precise localization of CCSNe in GWs and neutrinos, but very faint in EM radiation, can help mitigate the apparent lack of SNe from stars with initial masses  $\gtrsim 17 M_{\odot}$, known as `red supergiant problem' \cite{Smartt:2008zd}.


The nuclear EOS and several related quantities, including the symmetry energy, its slope and the nuclear incompressibility at saturation density, can significantly affect the GW \cite{Andersen:2021vzo,Morozova:2018glm,Murphy:2024uar}, neutrino \cite{Roberts:2011yw,Fischer:2013eka,Couch:2012gh,Schneider:2019shi,Nakazato:2021yjv} and EM emissions from CCSNe through the explosion dynamics and the PNS evolution.
For example, a softer EOS is expected to produce a more compact and faster contracting PNS, increasing both the neutrino luminosities and mean energies, as well as the GW frequency and its evolution rate.
Additionally, the progenitor variety (e.g., mass, rotation, magnetic field) is expected to produce an intrinsic variability in all emissions, see e.g. \cite{OConnor:2010moj,Mezzacappa:2022xmf,Vartanyan:2023sxm}.
Thus, relevant constraints on nuclear physics require a robust way to disentangle the dependence on the progenitor from the one on the nuclear physics, also breaking possible degeneracies.
A multimessenger detection of GWs and neutrinos 
could help to achieve this, for instance, through correlations between the two types of emission, 
which are also related to the progenitor mass and compactness, remnant mass and explosion energy \cite{Warren:2019lgb}.
Moreover, the presence of SASI is expected to induce modulations both in the neutrino and GW signals,
as visible the right panel of figure~\ref{fig:CCSN_correlations},
while in the case of explosions dominated by neutrino-driven convection, such a correlation could be barely seen~\cite{Kuroda:2017trn}.
Since strong SASI activity is observed in the case of more compact stellar cores, such a detection could favor softer nuclear EOSs. In the case of rapidly collapsing cores,
the gravitational collapse can lead to the onset of the low $T/W$ instability within the PNS \cite{Ott:2005gj}, also producing correlated signatures in both the GW and neutrino emissions~\cite{Shibagaki:2020ksk}.
The strength and configuration of the magnetic field could also influence the $T/W$-related emissions as well as
their multimessenger detection prospects \cite{Bugli:2022mlq}.

In the case of extra-galactic CCSNe, 
GW detections by ET could reach such horizons only if strong axisymmetric instabilities are present, for example when the remnant is a BH surrounded by a very massive accretion disk.
The joint detection of high energy photons, neutrinos and GWs from a nearby long GRB could help discriminate between different central engines, e.g. an accreting BH  from a magnetar, and could set constraints on the nuclear EOS through inference of the maximum NS mass.
While ET is expected to boost horizon and sensitivity with respect to current detectors, the observational prospects will however be small, since even for an optimistic GW horizon at the edge of the Local Group ($\lesssim 3$Mpc), the expected detection rate is of the order of a few every $10^5$ years due to the low long GRB explosion rate in the local universe \cite{Ghirlanda:2022edk}.
Only in very optimistic scenarios could  ET be able to provide a multimessenger detection within the Virgo Cluster (i.e., $\sim$16 Mpc) in the time-frame of several years \cite{VanPutten:2017wxo}.
Recently, other possible multimessenger signatures related to collapsar and long GRBs were explored. 
These include the GW and EM signature produced by the collapse of a very massive He core (above the pair-instability mass gap), possibly also ejecting r-process elements. 
In this scenario, ET could be able to detect multiband GWs of 0.1--50 Hz from non-axisymmetric instabilities in self-gravitating massive collapsar disks out to hundreds of Mpc~\cite{Siegel:2021ptt}.
The resulting EM transient would include a long GRB and a quasi-thermal emission, with intermediate properties between a kilonova and a supernova, and r-process signatures.
Another scenario related to CCSNe and long GRBs involves a choked jet, where the relativistic jet fails to break through the stellar envelope, driving a quasi-spherical shock into the surrounding circumstellar material. In this case, high-energy neutrinos are expected to be produced over broader angles compared to a successful jet enhancing the chances of observing both GWs and high-energy neutrinos from nearby Type II CCSNe, which could also be associated with low-luminosity GRBs \cite{Meszaros:2001ms,Ando:2005xi,Senno:2015tsn,Denton:2017jwk,Fasano:2021bwq}.

\paragraph{Ejecta, nucleosynthesis and kilonova emission from compact binary mergers.} 

\begin{figure}[t]
\centering
\includegraphics[width=0.49\columnwidth,angle=0]{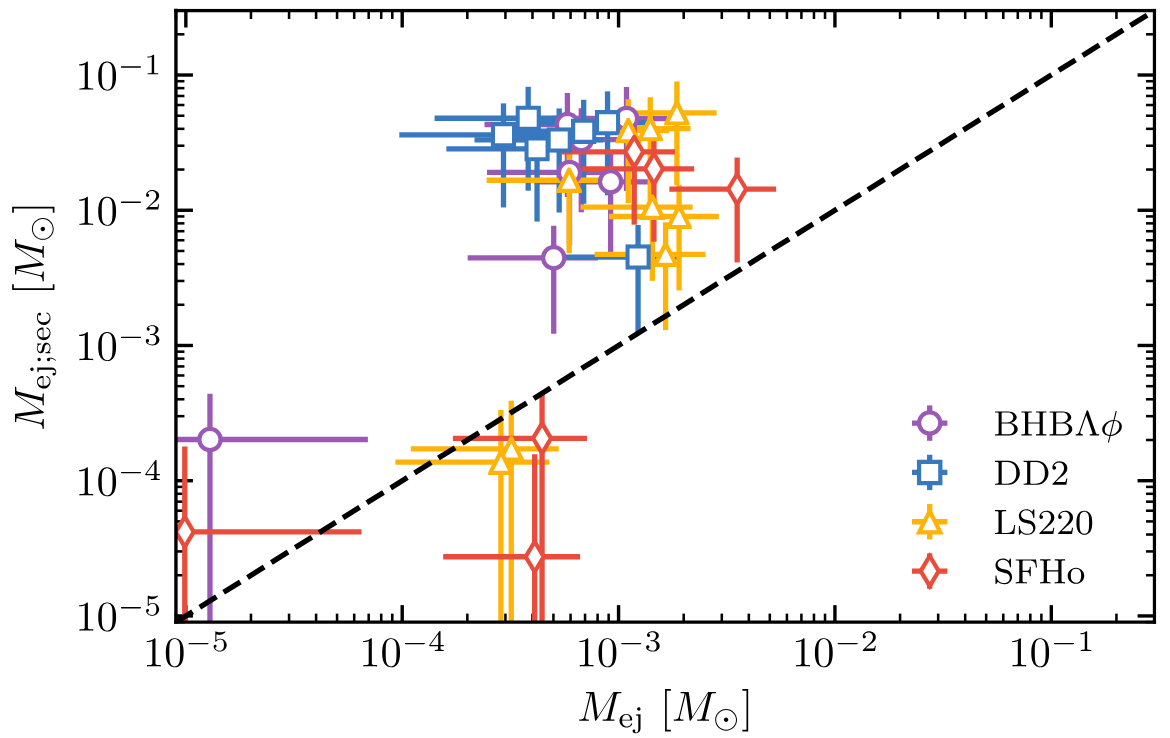}
\includegraphics[width=0.49\columnwidth,angle=0]{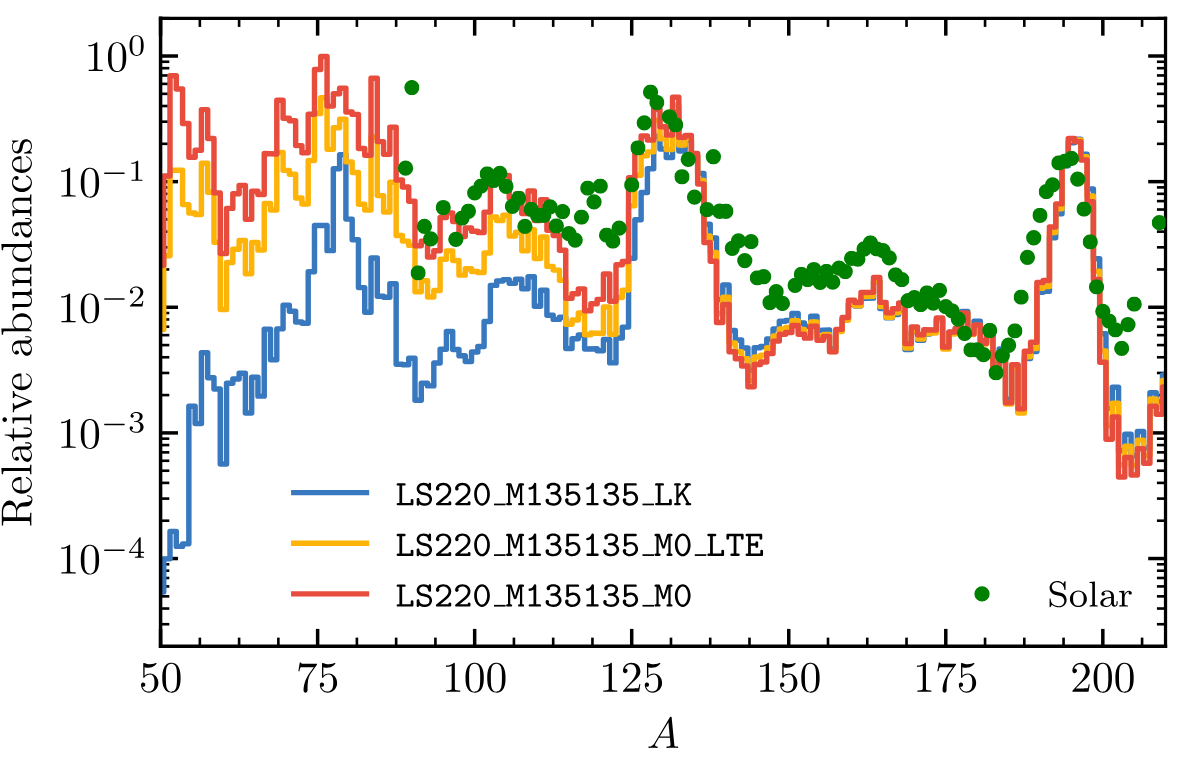}
\caption{Left: Dynamical ejecta mass ($x$ axis) VS secular ejecta mass ($y$-axis, estimated as 20\% of the disk mass) extracted from a large set of BNS merger simulations. Secular ejecta are dominant over the dynamical ones. Different colors show the impact of the nuclear EOS. Right: r-process nucleosynthesis from different simulations of the same BNS merger. Simulations featuring neutrino absorption in optically thin conditions (red and orange lines) produce all r-process nuclei, while the simulation not including it (blue line) only strong r-process nucleosynthesis. Figures adapted from \cite{Radice:2018pdn}.
\label{fig:r_process_BNS}}
\end{figure}

Compact binary mergers involving at least one NS have been long considered promising r-process nucleosynthesis sites \cite{Lattimer:1974slx,Eichler:1989ve,Rosswog:2012wb}. 
Thermal neutrino emission from them shares many similarities with the one of CCSNe. This is due to the similar conditions of matter inside a PNS and in a merger remnant, with possibly larger temperatures and higher densities in the case of BNS mergers \cite{Perego:2019adq}.
Additionally, accretion of matter from the disk or torus around the central object provides accretion-powered neutrino luminosity. This becomes the main source of neutrinos in the case of NS-BH mergers or for BNS merger whose remnant has collapsed to a BH.
While the mean neutrino energies are comparable to the ones observed in CCSNe, the neutrino luminosities are possibly larger at peak. A valuable difference is represented by the initial dominance of $\bar{\nu}_e$ over $\nu_e$, due to the tendency of cold $\beta$-equilibrated matter to leptonize, once decompressed and heated, see e.g., \cite{Foucart:2015gaa}.

Extensive simulation campaigns showed that different ejection mechanisms and neutrino irradiation operate on different timescales, contributing with different components to the total ejecta \cite{Radice:2020ddv,Bernuzzi:2020tgt}, as visible in the left panel of figure~\ref{fig:r_process_BNS}.
During merger, a mass ${\sim}{10^{-4}-{10^{-2}}}\,\Msun$ of neutron rich material is expelled
on dynamical (GW) timescales, i.e., within 5--10 ms after merger, and with characteristic velocities of $\sim (0.1-0.3) c$, see  e.g., \cite{Rosswog:1998hy,Hotokezaka:2012ze,Bauswein:2013yna,Wanajo:2014wha,Radice:2018pdn,Nedora:2020hxc}. Two main mechanisms are responsible for these dynamical ejecta: the tidal interaction between the two merging objects \cite{Rosswog:1998hy,Radice:2016dwd} and the shock developing at the collisional interface \cite{Hotokezaka:2012ze,Bauswein:2013yna,Sekiguchi:2016bjd,Radice:2018pdn}. The former is more relevant in the case of NS-BH mergers, asymmetric BNS mergers and stiff nuclear EOSs. The latter is the dominant mechanism in the case of equal mass BNSs, and leads to mass mainly being expelled at the first bounce of the two NS cores \cite{Radice:2018pdn}. In the case of soft nuclear EOSs, the merging NSs are more compact, the merger is more violent, while the remnant reaches higher temperatures. A small fraction of the shocked component can reach high-speeds, up to $\sim0.8c$ \cite{Bauswein:2013yna,Radice:2018pdn}; this fast-tail has a non-trivial dependence on the EOS \cite{Nedora:2021eoj} and is relevant for the kilonova afterglow \cite{Nakar:2011cw}.
The merger remnant can unbind a total mass of 
$\sim 10^{-2} \Msun$ during its evolution on viscous timescales of $(0.1-1)$~s throughout various wind mechanisms (spiral-wave and disk winds). Binaries with short-lived NS remnants and/or with mass ratio $\gtrsim1.4$ (including NS-BH) produce remnant BH with massive hyperaccreting disks whose evolution is dominated by viscous and thermal effects. Turbulent viscosity of magnetic origin, neutrino absorption or EM interaction with a strong, large-scale magnetic field have been proven to be very efficient in unbinding a significant fraction (between 10 and 40\%) of the accretion disk resulting from the merger dynamics, see e.g. 
\cite{Fernandez:2013tya,Siegel:2014ita,Ciolfi:2020wfx,Fahlman:2022jkh,Kiuchi:2023obe}.
At a quantitative level, it is important to notice that the amount of ejected matter, its composition and, in the case of BNS mergers, the lifetime of the remnant are influenced by both intrinsic binary parameters, as the total mass and the mass ratio, see, e.g., \cite{Bernuzzi:2020txg,Kruger:2020gig}, and nuclear EOS-related properties, such as the reduced tidal deformability, the stellar compactness or the maximum mass for a non-rotating NS, 
see e.g. \cite{Dietrich:2016fpt,Radice:2018pdn,Barbieri:2019kli,Nedora:2020qtd,Kruger:2020gig}.

Due to the variety of ejection processes and timescales, as well as to the anisotropic neutrino emission that characterizes merger remnants, the ejecta feature a broad range of electron fractions \cite{Just:2014fka,Wu:2016pnw}.
The tidal ejecta are very neutron rich ($Y_e \sim 0.1$) and cold, while the shocked ejecta are reprocessed to higher $Y_e$ by capture processes and neutrino irradiation from the NS remnant \cite{Wanajo:2014wha,Perego:2017wtu}. The electron fraction in shocked ejecta can span a wide range of values, $Y_e\sim0.1-0.4$, with the largest $Y_e$ obtained at high latitudes where the neutrino fluxes are more intense. The inclusion of neutrino absorption has a direct impact on the r-process nucleosynthesis, as visible in the right panel of figure~\ref{fig:r_process_BNS}.
The lifetime of a BNS merger remnant and, more in general, the presence of a massive NS or a BH in the center possibly affect the disk wind ejecta composition due to stronger or weaker neutrino irradiation \cite{Perego:2014fma,Lippuner:2017bfm,Nedora:2019jhl}. 

The observational imprint of r-process elements in compact binary mergers is the kilonova EM transient \cite{Li:1998bw,Metzger:2010sy,Metzger:2019zeh}, see also section~\ref{section:div4}.
The peak time and the spectral evolution of the light curves are influenced by the amount of the ejecta, as well as by its composition. In particular, if the r-process nucleosynthesis results in the synthesis of lanthanides or actinides, the resulting opacity is expected to increase by one or two orders of magnitudes, due to the opening of the electron $f$-shell \cite{Roberts:2011xz,Kasen:2013xka,Tanaka:2013ana}. The peak in the emission is then shifted toward later times and the spectrum becomes redder following the adiabatic cooling due to matter expansion.
Kilonova modeling is extremely challenging, since it requires the solution of the photon radiative transfer problem in a fast expanding medium of ionized matter. The latter is composed by radioactive elements characterized by millions of atomic transitions whose strengths are poorly known. Moreover, the ejecta emerging from the collisions are characterized by a non-trivial geometry, due to the presence of different ejecta components, possibly characterized by a radial and angular dependence in the most relevant properties (mass distribution, expansion velocity, nucleosynthesis yields, etc), which are far from being isotropic or radially homogeneous, see e.g. \cite{Perego:2017wtu}. Sophisticated radiative transfer models have shown their ability in reproducing, at least at a qualitative level, the spectral features observed in AT2017gfo, e.g. \cite{Kasen:2017sxr,Wollaeger:2017ahm,Tanaka:2017qxj,Bulla:2019muo,Shingles:2023kua}. 
They are mandatory to identify the presence of specific elements through distinct spectral features. A possible compromise for efficient line identification is represented by radiative transfer codes that model only the effects of the atmosphere on the photospheric emission, see e.g. \cite{Watson:2019xjv}, while efficient kilonova models necessary for Bayesian analysis can rely on kilonova surrogate models of radiative transfert simulations \cite{Coughlin:2018miv} or simplified semi-analytical solution of the radiative transfer equations \cite{Ricigliano:2023svx}.

\paragraph{Multimessenger detections in compact binary mergers.}

\subparagraph{Gravitational waves and neutrinos.} In the case of NS-BH or BNS mergers, a multimessenger detection possibly includes neutrinos and EM radiation, in addition to GWs.
Thermal neutrino luminosities from BNS mergers were analyzed and compared with the corresponding GW luminosities \cite{Cusinato:2021zin}.
In the case of non-promptly collapsing remnants, both emissions seem to correlate with the mass-weighed tidal deformability $\tilde{\Lambda}$,  \eq{deftildeLambdadiv6}, with larger $\tilde{\Lambda}$ providing weaker luminosities.
Both  emissions show $\sim $1 kHz modulations associated to the violent radial bounces of the remnant’s core, ultimately depending on the nuclear EOS properties.
This kind of analysis would require a Galactic merger, which is extremely unlikely ($\sim 10^{-5} {\rm yr}^{-1}$). The detection of one event within the Local group ($\lesssim 3$Mpc) could still results in $\mathcal{O}(10)$ neutrinos in future neutrino detectors.
A more significant probability to observe thermal neutrinos in combination with GW signals from next-generation GW observatories will be realized only when megaton-scale $\nu$ detectors will be available. The impact of ET on these searches will be mostly limited to set triggers and alerts for well localized neutrino searches. However, the access to the post-merger signal, which is very likely in the case of nearby events, could better constrain the neutrino emission and the related nuclear physics, even by setting limits in the case of a lack of detections.

A more promising perspective is the multimessenger detection of GWs and non-thermal neutrinos. High-energy ($>$ TeV) non-thermal neutrinos are expected to be produced in gamma-ray bursts (GRBs) via photo-hadronic interactions between GRB photons and relativistic protons. During the prompt emission, these interactions lead to the production of pions, which decay into neutrinos. Additionally, during the extended emission phase, where the Lorentz factor of the jet becomes lower, the efficiency of meson production is enhanced, increasing the expected neutrino flux. For relatively nearby events ($<$ 100 Mpc), the next-generation neutrino observatory, such as IceCube-gen2, will have the sensitivity required to detect neutrinos associated with ET detections of BNS or NSBH mergers \cite{Kimura2017ApJ}. Another particularly interesting scenario occurs when
a BNS merger results in the formation of a millisecond magnetar remnant surrounded by a low-mass ejecta shell. In this case, a fraction of the magnetar’s rotational energy could be deposited in a pulsar wind nebula, accelerating ions to ultra-high energies. After $\sim \mathcal{O}(1)$ day, pion production in the resulting ion collisions could produce high-energy neutrinos.
The resulting signal may be detectable in conjunction with GWs for individual mergers out to $\sim$100 Mpc by next-generation neutrino telescopes such as IceCube-Gen2 \cite{Fang:2017tla}. 
Within such distances, ET could be able to detect the post-merger signal with a high enough signal-to-noise ratio to confirm
the presence of a long-lived NS remnant and to set constraints on the nuclear EOS from the joint detection.


\subparagraph{Gravitational waves and EM emission.} 
The coincident detection of GWs and EM radiation has already disclosed its potential in constraining the nuclear EOS in the case of GW170817 and of its EM counterparts (GRB170817A and AT2017gfo). The reason is that the properties of the GW and EM emissions are influenced by nuclear physics in many different and often correlated ways, see, e.g., section~\ref{sec:constraints} for the impact on the GW signal.

The contribution of ET to these multimessenger detections will be manifold, see section~\ref{section:div4}. First of all, ET will enlarge the detection horizon, and for longer signals, such as those from BNS mergers, ET will provide accurate sky localization even when operating as a single observatory.
Section~\ref{section:div4} demonstrates that the GW+kilonova detectable binaries are tens to a few hundreds per year by a telescope reaching an AB magnitude of around 26 in the g band and following up GW signals with a sky-localization smaller than $100\,\mathrm{deg}^2$. By exploring realistic observational strategies of ET operating with the Vera Rubin Observatory, ref.~\cite{Loffredo:2024gmx} evaluates 10 to 100 detections per year and a detection increase of an order of magnitude when ET operates with Cosmic Explorer. The impact of the EOS of nuclear matter on the GW+KN detection rate is found to be $\lesssim 20\%$ level for ET observing as a single detector and $\lesssim 40\%$ for the ET+CE network. While for kilonova emission, the horizon of joint detections is limited to $z \sim 0.5$, the larger horizon for GW+GRB detections will add tens to a few hundred joint GW+GRB detections per year (see section~\ref{section:div4} and ref.~\cite{Ronchini:2022gwk}).   
The resulting number of detections will be large enough to provide a significant exploration of the binary parameter space, probing different post-merger scenarios and the involved nuclear physics. The capability to localize compact binary merger events in the sky will allow deep and temporally extended follow-up campaigns, whose results will be detailed light curves and spectra from which the ejecta properties and composition could be accurately inferred (at least for the closer ones).

Second, ET will help break the degeneracies between the nuclear physics effects and the influence of the binary parameters on the post-merger dynamics and observables.
In particular, for sufficiently nearby events, ET will allow a detailed GW parameter estimation to determine with high accuracy the properties of the merging objects and their tidal deformability, see section~\ref{ssec:ConstraintsonlowT}. In this way, the kilonova features and, if available, the post-merger GW signal can be more directly and firmly related to the post-merger dynamics and to the ejecta composition, and from that to the underlying nuclear physics.
For example, post-merger models differing only in the high density and high temperature behavior of the EOS or featuring the appearance of hyperons or quarks (usually producing an EOS softening) could indeed be tested by their effects on the post-merger observables (for example, through their impact on the remnant lifetime or on the amount of ejected matter).
In this respect, it is important to remark that the coincident detection of kilonova provides complementary information with respect to the GW signal alone since, even in the case of a long-lived remnant, the GW emission is expected to subside after a few tens of ms, while matter ejection and neutrino irradiation can last seconds and leave their fingerprints in the kilonova emission.

Third, in the case of a joint GW and kilonova detection, detailed modeling of the matter ejection and kilonova emission based on robust binary parameters extracted from the GW inspiral signal can provide constraints on the ejecta mass, on the nuclear heating rate and on the nucleosynthesis yields required to explain the observed light curves and spectra.
However, starting from the same ejecta properties, different mass models can introduce up to a factor 10 of uncertainty in the kilonova brightness \cite{deJesusMendoza-Temis:2014owk,Rosswog:2016dhy}. It is thus clear that such a kind of analysis has the potential to determine the properties of the exotic nuclei involved in the r-process, see section~\ref{sect_nuclEOSynth}.


Another relevant scenario for which next-generation GW observatories could be key is the one involving promptly collapsing BNS mergers. The latter have the potential of probing nuclear properties at the highest densities, especially if the threshold for prompt collapse is determined for different mass ratios, see, e.g., \cite{Perego:2021mkd} and \ref{ssec:postmerger}.
GW telescopes such as ET have the potential of directly probing the process of BH formation in promptly collapsing or short lived events \cite{Dhani:2023ijt}. For the same events, the detection of the corresponding kilonova could discriminate between different mass ratios or different remnant lifetime: indeed, unequal mass mergers and short-lived remnant are expected to produce more ejecta and, thus, brighter kilonovae, while equal mass mergers tend to eject very small amount of ejecta \cite{Bauswein:2013yna,Bernuzzi:2020txg,Kolsch:2021lub}.



The joint detection of GWs and of a short GRB signal (prompt or afterglow emission) has the potential of shading light on the nature of the central GRB engine. 
In particular, assuming that the post-merger signal could be detected up to $\sim$100~Mpc by ET, the expected joint detection rate involving a post-merger signal can be estimated to be $\sim 0.03-0.6\, {\rm yr}^{-1}$~\cite{Salafia:2023sjx,Colombo:2022zzp}. Such a detection could directly discriminate between the magnetar and the BH central engine scenarios for the production of a relativistic jet.
Additionally, the precise localization of compact binary mergers allowed by ET can also favor the discovery of faint emissions from chocked jets. 
The precise knowledge of the central engine, combined with the measured rates of successful short GRBs and of merging binaries by GW observations, can provide
independent constraints on the maximum NS mass and, from that, on the EOS of nuclear matter, see, e.g.,~\cite{Piro:2017zec,Margalit:2017dij}.
Recent detection of nearby ($z \lesssim 0.1$) long GRBs with kilonova-like signatures have questioned the long-standing paradigm of long-GRBs always originating from core collapses, see, e.g., \cite{Mei:2022ncd,JWST:2023jqa,Rastinejad:2022zbg}.
These observations points on the possibility for compact mergers to produce long GRBs. Since ET will be able to detect compact mergers up to an horizon of $z=3-4$~\cite{Branchesi:2023mws}, in the case of long GRB with kilonova-like signatures, it has the capability to firmly confirm this scenario.

\paragraph{Data analysis and modeling efforts.}

\begin{figure}[t]
\centering
\includegraphics[width=0.54\columnwidth,angle=0]{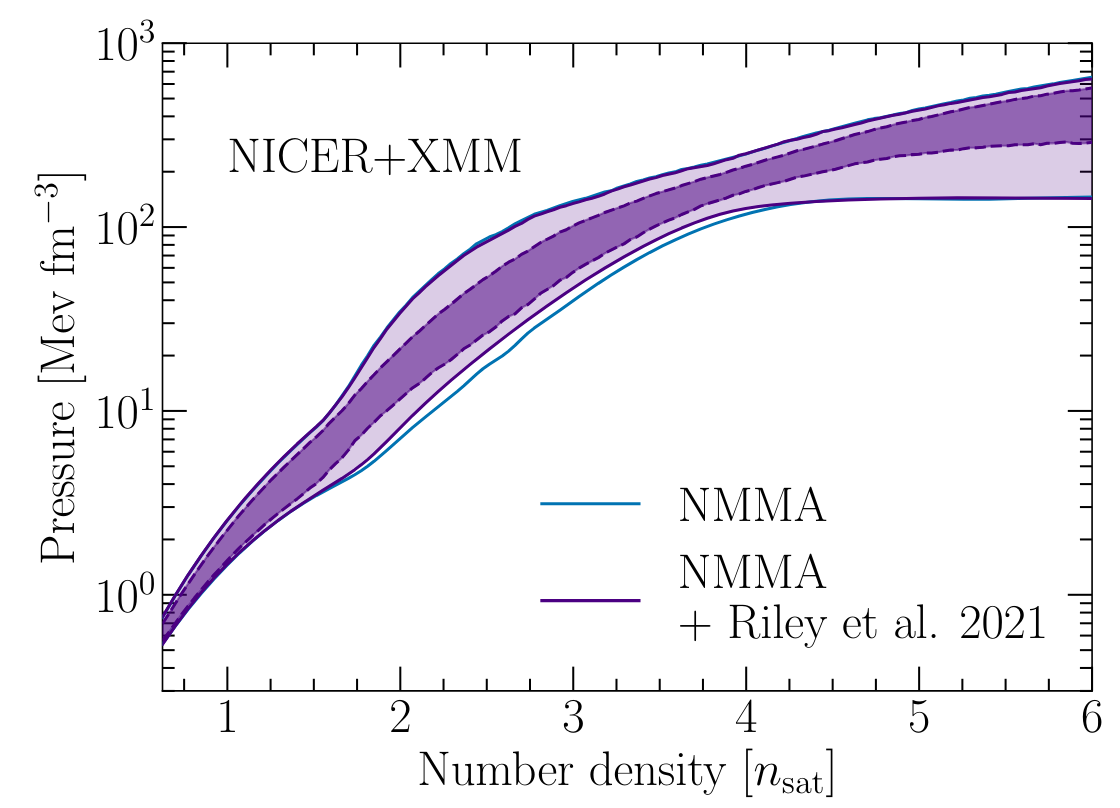}
\includegraphics[width=0.45\columnwidth,angle=0]{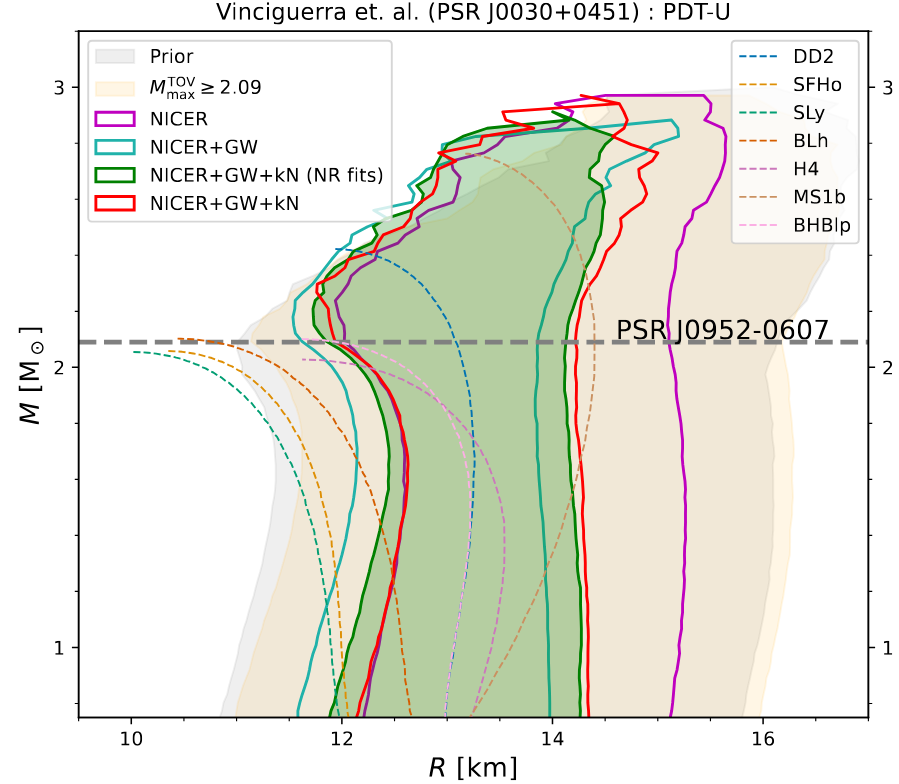}
\caption{Examples of constraints on the nuclear EOS obtained by joint multimessenger analysis combining astrophysical
observations of pulsars, NICER measurements, 
GW170817 and GW190425, kilonova AT2017gfo and GRB170817A modeling.
Left: Posteriors for the pressure as a function of number density with (purple) and without (blue) NICER and XMM observations of PSR J0740+6620.
Right: Posteriors in the $M$-$R$ diagram of the GW-only (blue), joint (red), and NR-informed joint analysis (green).
Figures adapted from \cite{Pang:2021jta} (left) and \cite{Breschi:2024qlc} (right).
\label{fig:MM_bayesian_approaches}}
\end{figure}

The rigorous analysis of multi-messenger signals 
requires a Bayesian approach.
Constraints on the EOS of nuclear matter and on specific NS properties (for example, the maximum mass of a non-rotating NS or the radius of a 1.4 or 1.6 $\Msun$ NS) were obtained by combining parameters extracted from the inspiral signal of GW170817 and GW190425, detailed models of the EM counterparts, additional astrophysical and nuclear physics constraints (e.g., maximum observed NS masses, NS radii as obtained by the NICER mission or data extracted from heavy-ion collisions) and non-trivial relations connecting the remnant and ejecta properties to the nuclear EOS (as obtained by detailed compact binary merger simulations), see, e.g., \cite{Radice:2017lry,Annala:2017llu,De:2018uhw,Radice:2018ozg,Coughlin:2018fis,Raaijmakers:2019dks,Capano:2019eae,Essick:2020flb,Dietrich:2020efo,Breschi:2021tbm,Pang:2021jta,Huth:2021bsp}.
Even the non-detection of EM counterparts, if not due to insufficient accuracy in the sky localization or EM follow-up coverage, can provide constraints on the nuclear EOS, as in the case of GW190425, see, e.g., \cite{Dudi:2021abi,Camilletti:2022jms,Radice:2023zfv}.
In the case of (i) large correlations between the parameters describing the different observations and (ii) systematics in the modeling, the analysis of different data coming from a single source benefits from a joint and coherent analyses. 
In this kind of analysis, single messenger likelihoods are joined and a combined sampling of the full posterior probability distribution is performed.
This approach was recently implemented and presented in refs.~\cite{Biscoveanu:2019bpy,Pang:2022rzc,Breschi:2024qlc}.
It allows joint and coherent analyses of multimessenger signals, including GWs, kilonovae/supernovae and GRBs, possibly taking into account quantitative relations inferred by large sets of sophisticated simulations connecting the source intrinsic properties (for example, the masses of two merging neutron stars) to the ejecta properties, 
nuclear-physics constraints at low densities and multiband observations of isolated NSs, as visible in figure~\ref{fig:MM_bayesian_approaches}. 

It is thus clear that 
the full exploitation of future multimessenger observations involving GW detections from ET will rely on detailed and reliable models of the sources that contains detailed nuclear physics. This will require a great community effort within the next ten years. The ambitious goal will be to connect the strong field dynamics that characterizes the emission of GWs and neutrinos, as well as the expulsion of matter, to the resulting light curves and spectral features that allow to quantify the amount of ejecta and its properties through the identification of elements.
For standard CCSNe, three-dimensional, self-consistent exploding simulations, characterized by a high level of fidelity in the neutrino transport, are now available. Future models should also include consistent nucleosynthesis yields predictions and extend up to shock breakout.
The inclusion of magneto-hydrodynamics effects in CCSN models to predict their imprint will also be key. In case of a sufficiently close-by CCSN, jet-like features in the GW, neutrino and electromagnetic signals can be a smoking gun for the viability of magneto-rotationally driven SNe.
Reliable collapsar simulations, addressing both the formation of the relativistic jet and the nucleosynthesis of the innermost ejecta, will be also necessary to probe the central engine of gamma-ray burst related to SNe.
In the case of compact binary mergers, the ambitious goal will be the production of reliable end-to-end models, i.e., models that seamlessly connect the GW inspiral signal to the emission of EM radiation that characterizes the gamma-ray burst and the kilonova, as well as their afterglows \cite{Fernandez:2016sbf,Perego:2020evn,Just:2023wtj}.
Open challenges here are represented by the inclusion of detailed neutrino transport and microphysics in merger models (see also section~\ref{Sec:Simulation_Uncertainty}), the modeling of the central engine producing the relativistic jet, as well as its interaction with the surrounding medium. In terms of kilonova emission, large effort will be devoted to the calculation of opacities and spectral features that can unambiguously identify the presence and the amount of specific elements in the ejecta.

A great challenge ahead is also represented by the inclusion of neutrino oscillations in models of CCSNe and compact binary mergers. While the effect of vacuum and matter oscillations is presently accounted for the determination of the expected neutrino signal, the role of neutrino-induced oscillations, their impact on the remnant dynamics and on the nucleosynthesis are possibly significant, but still under investigation, see, e.g., \cite{Wu:2014kaa,Mirizzi:2015eza,Wu:2017drk,Zhu:2016mwa,Frensel:2016fge}.

\subsection{Uncertainties and degeneracies in our measurements} 
\label{sec:degeneracies}
To properly extract the information from GW observations, one has to cross-correlate the observational data to theoretical models. In this section, we discuss the impact that various uncertainties in GW modeling and theoretical assumptions may have on the interpretation of ET data for subatomic physics. For a detailed discussion about waveform modeling, we refer to~section~\ref{section:div8}. 

\subsubsection{Impact of waveform-model uncertainties}
\label{ssec:waveforms}

\paragraph{Inspiral -- Merger.}

During the inspiral, characteristic matter signatures in the GW signal arise primarily due to spin and tidal effects, including dynamical tides associated with the excitation of the star's internal oscillation modes, and, in NS-BH systems, also the possible tidal disruption of the NS. 
Waveforms describing the inspiral of compact binaries with at least one NS are commonly built by augmenting a high-fidelity point-mass, i.e., binary BH (BBH), baseline with matter-induced corrections. A more detailed overview of the current state-of-the art waveform models for BNS and BHNS can be found in section~\ref{section:div8}. 

Spin effects that depend on the EOS first enter the GWs through the spin-induced quadrupole moment with the same scaling in frequency as post-2-Newtonian (2PN) point-mass effects. 
For tidal effects, the dominant adiabatic (static) contribution first enters the GW phase with a scaling in frequency similar to a 5PN term~\cite{Flanagan:2007ix} and higher order corrections to these effects are currently known to relative 2.5PN order~\cite{Vines:2011ud,Henry:2020ski}, while non-spinning point-mass effects are currently known completely up to 4PN order. 
More recently, calculations within the effective-one-body framework as well as PN theory have been carried out to start incorporating dynamical tidal effects from the $f-$modes into BNS waveform models~\cite{Hinderer:2016eia, Steinhoff:2016rfi, Pitre:2023xsr,Kuan:2023qxo, Schmidt:2019wrl,Abac:2023ujg}. Even though these effects are most prominent at a GW frequency of around $800$~Hz, they accumulate over a wide range of frequencies and, for a detector like ET, neglecting these effects can lead to systematic biases in the inferred NS radius by more than $1$~km \cite{Pratten:2021pro,Pratten:2019sed,Williams:2022vct}, i.e., they have to be incorporated to achieve the estimated accuracies discussed in section~\ref{subsec:NSmergers}, which only account for statistical uncertainties. In addition, other NS modes such as $r$-modes and $g$-modes can become resonantly excited much earlier in the inspiral and accumulate significant information, see e.g.~\cite{Flanagan:2006sb,Lai:2006pr,Poisson:2020eki,Gupta:2023oyy,Yu:2017cxe,Passamonti:2020fur,Kuan:2021jmk,Xu:2017hqo} for recent work and further references. Moreover, nonlinear tides may also play a non-negligible role~\cite{Yu:2022fzw}. Spins further affect tidal phenomena, e.g., they lead to effective shifts in the resonance frequencies~\cite{Ma:2020rak,Steinhoff:2021dsn} and new interactions characterized by spin-tidal Love numbers~\cite{Castro:2022mpw,Abdelsalhin:2018reg}. Other physics may also play a role, e.g., the NS viscosity~\cite{Ripley:2023qxo,HegadeKR:2024agt}; see section~\ref{sect_nu_viscosities}, and a nonnegligible temperature induced by mode excitations as estimated in~\cite{Arras:2018fxj}; see~section~\ref{ssec:postmerger}. However, such consequences remain to be explored in more realistic settings. Additionally, the effects of eccentricity could alter the dynamics, in turn impacting tidal effects~\cite{Wang:2020iqj,Yang:2019kmf} and thus influencing the inference of EOS physics, as remains to be quantified.  
In summary, incomplete knowledge of the point-mass baseline model and more realistic physical effects can have a significant impact on the measurement of tidal parameters. The implications for ET remain to be comprehensively analyzed once theoretical models are further developed. The main theoretical tools to understand and model the currently missing effects are in hand, but further work is needed to determine which phenomena are essential for the majority of NS binary signals when incorporating more realistic physics. 

Currently, shortcomings in the analytical knowledge of the point-mass baseline are reduced by including phenomenological corrections to incorporate non-perturbative information from NR simulations and in some cases also more theoretical knowledge from other channels such as gravitational self-force. 
Importantly, while the current BBH baseline models such as \texttt{IMRPhenomXPHM} \cite{Pratten:2020ceb,Garcia-Quiros:2020qpx} and \texttt{SEOBNRv5PHM} \cite{Pompili:2023tna,Ramos-Buades:2023ehm} are found to be sufficiently accurate across a large part of the parameter space for current-generation GW detectors, systematic differences due to modeling choices are already becoming apparent in more challenging parts of the parameter space, e.g.~\cite{Dhani:2024jja}. Shortcomings in these baselines impact the inference of all source parameters and hence the EOS reconstruction from global NS properties, including the masses. 
Moreover, current models that include matter effects (e.g. \texttt{NRTidalv3}~\cite{Abac:2023ujg}, \texttt{SEOBNRv4T}~\cite{Hinderer:2016eia, Steinhoff:2016rfi}, \texttt{TEOBResumS}~\cite{Akcay:2018yyh}; see~section~\ref{section:div8} for details on these models) remain even more limited than those for BBHs in both the parameter space covered and the additional physics included, such as the magnitude of spins, spin precession effects due to the misalignment between the component spins and the orbital angular momentum, orbital eccentricity, and higher modes. Spin-precession effects have been shown to be essential to avoid biases in the measurement of the source masses~\cite{Pratten:2020igi,Chatziioannou:2014coa}, which in turn impacts the inference of the mass-radius relationship, and higher-order modes will be especially important for NS-BH systems due to the larger mass ratio of these class of objects.
Any such shortcoming further impacts the measurements of tidal parameters. The extent of these impacts remains to be quantified, however, existing estimates already indicate the need for advances in modeling to include such effects. 


To capture the tidal imprints in the waveform phasing during the late stages of the inspiral, phenomenological expressions informed by NR simulations have been developed. However, their tuning is limited in mass ratio, spin magnitude, and orientation, as well as the dominant quadrupolar mode and BNS systems. Recent progress has been made by extending the calibration range to more asymmetric binaries and incorporating more analytical information into the structure of the model, e.g., \cite{Abac:2023ujg}. However, ET will be sensitive to even more subtle matter effects, as discussed above, that could alter the morphology of the signal and have not yet been included in such state-of-the-art models.


An additional tidal signature that occurs in NS-BH binaries is the complete or partial disruption of the NS, which also gives rise to an electromagnetic counterpart. While disruption preferentially occurs for binaries with similar masses, large BH spins and stiffer EOS, the recent observation of GW230529~\cite{LIGOScientific:2024elc} suggests that such events may be more common than previously expected. Currently, tidal disruption is only treated phenomenologically and modelled as a sudden dropoff in the waveform amplitude \cite{Thompson:2020nei,Matas:2020wab,Gonzalez:2022prs}. The disruption frequency and its main parameter dependencies have been explored in full NR simulations, which are challenging to carry out, as explained in detail in section~\ref{section:div8}, and have been limited in their parameter space coverage. Under the assumption of a perfect model, the improved detector sensitivity alone will allow us to determine the disruption frequency within $\sim 100$~Hz, which translates into an NS radius measurement accuracy of $\sim 1$~km \cite{Clarke:2023rrm}, with much tighter constraints possible when combining this with the information on tidal deformability from the inspiral phasing.

\paragraph {Post-Merger}

Due to the complex multi-scale physics involved, the modeling of the post-merger GW emission is very challenging and involves extreme densities, temperatures, and magnetic field strengths. We also note that, as discussed in section~\ref{ssec:postmerger}, only a fraction of BNS mergers are expected to exhibit such rich post-merger behavior, while others will promptly collapse to a BH.
Understanding and modeling the post-merger phase of the binary evolution requires NR simulations. However, NR simulations are limited by uncertainties from numerical inaccuracies (which increase in the presence of shocks) and incomplete descriptions of the microscopic physics. 
Given the high temperatures and densities after the merger and the possibility of increased magnetic field strength by orders of magnitude during the merger, the physical inputs, which are of interest when constraining the EOS, alter the postmerger GW signal noticeably more than during the inspiral. For example, the emission of neutrinos reduces the remnant's temperature and leads to an earlier BH formation. Similarly, a possible quark-hadron phase transition (that might appear in the post-merger phase) might noticeably change the compactness of the remnant and, therefore, shift the emission frequency of the post-merger GW signal, e.g.,~\cite{Bauswein:2018bma, Most:2018eaw, Weih:2019xvw}. All of these influences must be quantified to extract information from the postmerger GW signal reliably; 
see section~\ref{Sec:Simulation_Uncertainty}.

Up to now, there have been several attempts to construct postmerger models, e.g.,~\cite{Takami:2014tva,Clark:2015zxa,Bose:2017jvk,Easter:2018pqy,Breschi:2019srl,Tsang:2019esi,Soultanis:2021oia,Breschi:2022ens,Breschi:2022xnc,Puecher:2022oiz}, by capturing the most important characteristic frequencies from the postmerger remnant. While our understanding of the main emission frequencies continuously increases, existing estimates still have uncertainties of about 100~Hz. These differences are of the same order as differences in the postmerger spectra due to interesting physical processes; for example, it has been outlined~\cite{Vijayan:2023qrt} that pion production could lead to a frequency shift of about 150~Hz. 
To further improve our understanding of the postmerger spectrum, one would need a larger set of NR simulations with well-quantified error bars across a large range of the BNS parameter space and with accurate modeling of microphysical processes. Given the enormous complexity of physics and scales involved, and despite recent progress on novel codes (e.g.~\cite{Legred:2023zet}), it is likely that computing thousands of NR waveforms with sufficient resolution, physical realism, and parameter space coverage to build accurate waveforms (e.g. via surrogates) for parameter inference of BNS mergers will remain out of reach for the next decade.  

However, it is possible to instead adopt a data-driven approach, for instance, by using a morphology-independent method to reconstruct the postmerger GW signal~\cite{Chatziioannou:2017ixj, Wijngaarden:2022sah}. Such analysis strategies have the advantage that information of the GW spectrum can be directly extracted without reliance on models. However, the interpretation of these results for the underlying source physics still requires theoretical models.

Irrespective of the data analysis approach, extracting the main emission frequencies from the postmerger remnant will remain challenging even with ET and will only be possible for BNS sources with a minimum postmerger signal-to-noise ratio of about $\sim 5$~\cite{Tsang:2019esi}. Depending on the properties of the source, this corresponds to distances of less than 100 Mpc. Hence, only a few of the closest BNS systems will provide reliable information on postmerger physics. For these sources, the properties of the progenitors would be measured by ET with high accuracy from the inspiral physics, which will aid in the analysis of the postmerger signals.

\subsubsection{Uncertainties in simulations and microphysics input}
\label{Sec:Simulation_Uncertainty}


Due to the complexity of Einstein's field equations, NR simulations are an essential tool to improve our understanding of the merging of BNS and NS-BH systems~\cite{Rezzolla:2013,Shibata:2016,Dietrich:2020eud}. However, despite continuous progress and simulations with increasing accuracy, e.g., \cite{Radice:2013hxh,Hotokezaka:2015xka,Kiuchi:2017pte,Bernuzzi:2016pie,Dietrich:2018phi,Kiuchi:2019kzt,Doulis:2022vkx}, these simulations have different sources of errors and uncertainties which have to be taken into account when using numerical-relativity simulations to extract information from future observations of NS mergers. Uncertainties in such simulations arise, on the one hand, due to numerical approximation when solving the GR field equations together with those of general relativistic hydrodynamics and on the other, due to the incorporation of microphysical processes, such as neutrino and photon radiation, magnetic-field amplification, or viscous effects.
 In the following, we briefly discuss the dominant sources of uncertainty. 

\paragraph{Numerical Discretization.}

In contrast to the simulation of black-hole spacetimes, simulations of supernova explosions or compact binaries including at least one NS, not only need to find an accurate description for the simulation of the spacetime but also have to deal with general-relativistic hydrodynamics. 
This adds to the complexity that at the surface of the NS, the shocks and discontinuities have to be described properly.
Until now, most numerical codes used for such simulations use either finite differencing or spectral methods for the spacetime, and high-resolution shock-capturing schemes based on finite differencing or finite-volume methods for the hydrodynamical equations. 
Irrespective of the numerical methods employed, the discretization typically introduces an error that has to be carefully checked by computing numerous resolutions. In the past, several groups have assessed the uncertainty of their simulations and found errors in the phase of the extracted gravitational-wave signal of about $\mathcal{O}(0.1) \, \rm rad$ up to $\mathcal{O}(1)\, \rm rad$  accumulated over the last $10$ to $20$ orbits before the merger~\cite{Radice:2013hxh,Bernuzzi:2016pie,Dietrich:2018upm,Foucart:2018lhe,Kiuchi:2019kzt,Doulis:2022vkx}. This typically leads to mismatches that are larger than the target accuracy required for ET. 
However, recent progress has also been made on using discontinous-Galerkin methods~\cite{Bugner:2015gqa,Deppe:2021bhi} and developing exascale computing codes that will be a promising tool in the upcoming years, e.g., \cite{DUMBSER2020109088}. In addition, alternatives to the ordinary high-resolution shock-capturing schemes and Riemann solvers have been proposed to reduce the uncertainty without increasing the computational footprint of the simulations~\cite{Guercilena:2016fdl,Doulis:2022vkx}.
Overall, it will be crucial to explore alternative techniques to ensure efficient use of the computational hardware. Without such significant progress, based on the accuracy improvements obtained over the last decade, e.g.,~\cite{Baiotti:2011am,Haas:2016cop,Kiuchi:2019kzt}, the accuracy targets will not be reached, which -- as discussed in section~\ref{section:div8} -- will limit the development of improved GW models. 

Moreover, uncertainties related to the finite volume of the simulation domain, though typically smaller than discretization errors, might also impact ET science. Most notably, the finite domain size leads to an extraction of the GW at a finite distance from the emitting system. At these distances, the linear approximation typically employed to describe GWs might not fully apply, and additional errors are introduced. Polynomial extrapolation employing different extraction radii or next-to-leading order information about the GW falloff behavior can improve on this, e.g.,~\cite{Bernuzzi:2016pie}. While increasing the accuracy of the simulation and the extracted GW, this increased accuracy will likely still be too low. Hence, one has to expect that all future simulations aiming to produce accurate simulations for developing new waveform models employed in the ET era will have to use Cauchy Characteristic Extraction~\cite{Bishop:1996gt}, a method that ensures that GWs propagate to future null infinity where they can be properly and accurately extracted. 

\paragraph{Matter-composition and simulating neutrino radiation.}

In addition to the uncertainty introduced by numerical discretization, our limited knowledge about the input physics is causing uncertainties in NR simulations. Among the most important unknown effects are composition effects of the EOS describing the material inside the NSs. 
Certainly, an understanding of the EOS and composition is among the science drivers for ET, and hence, one can not assume to have complete knowledge about these aspects. However, this might cause substantial problems, e.g., there have been numerous studies in the field outlining the existence of quasi-universal relations connecting postmerger properties to the binary properties, see~section~\ref{ssec:postmerger}, but these relations might break in the presence of a strong phase transition, e.g.,~\cite{Most:2018eaw,Bauswein:2018bma}. Furthermore, the influence of pions or muons on the merger dynamics has also not been understood in detail~\cite{Gieg:2024jxs,Kim:2024zjb}, and the role of viscous or out-of-equilibrium effects have to be quantified in greater depth, e.g.,~\cite{Most:2022yhe,Chabanov:2023abq}. Clearly, such studies are of interest to be prepared for upcoming ET discoveries so that observational data can be correctly interpreted, and the community has recently started to make increasing progress in this direction - see section~\ref{sec:microphysics} and section~\ref{subsec:NSmergers} for more detailed discussions. 

Another issue regarding the correct simulation of the composition is the inclusion of neutrino interactions. The correct approach for including neutrinos in NR simulations would require solving equations for relativistic neutrino transport. However, these equations, coupled with the equations of general relativity and general relativistic hydrodynamics, are computationally too expensive to be solved directly. The reason is that, in addition to three spatial and one-time dimensions, three dimensions in the neutrino momentum space would need to be resolved. This 3+3+1-dimensional parameter space is too large considering the existing (and likely also upcoming) high-performance computing facilities. Hence, the community is still focusing on finding good approximations of the relativistic Boltzmann Equations; see~\cite{Foucart:2022bth} for a recent review.

\begin{figure}[t]
\centering
\includegraphics[width=0.5\columnwidth]{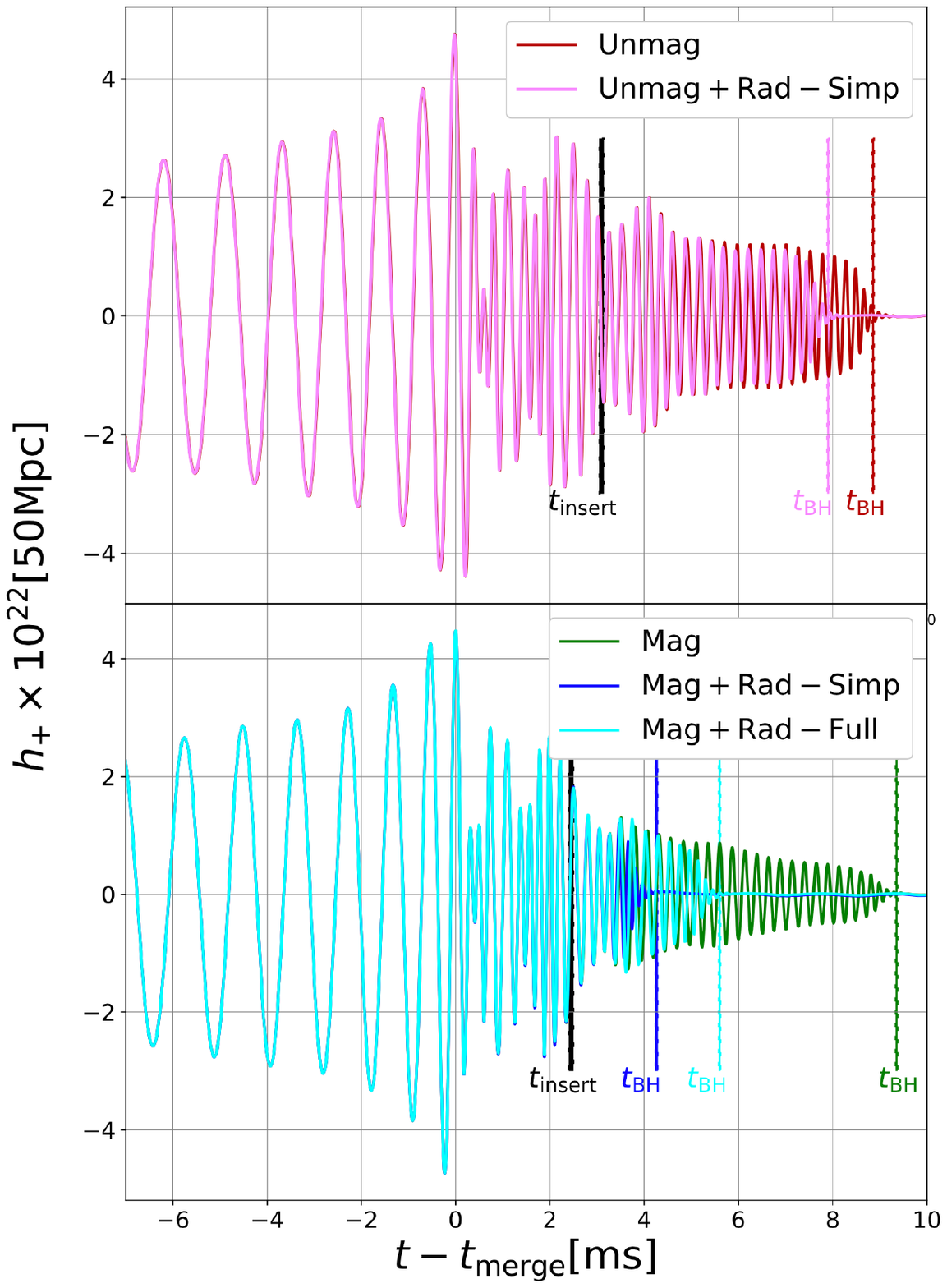}
\caption{\label{Sun_etal_fig8} Influence of magnetic field and neutrino emission in the post-merger phase of a specific BNS simulation. Vertical lines mark the neutrino insertion (Mag + Rad cases), and BH formation. 
Figure from \cite{2022PhRvD.105j4028S}. }.
\end{figure}

\paragraph{Simulating Magnetic fields.}

Another possible accuracy limitation of current simulations is the treatment of magnetic fields. While studies such as the results shown in figure~\ref{Sun_etal_fig8}, where the upper (lower) panel corresponds to a system without (with) magnetic fields, have shown that, for most scenarios, the magnetic field strength is insufficient to cause a large difference during the inspiral dynamics, the postmerger evolution will be affected by the magnetic fields. This is due to magnetic field amplification during the merger through the Kelvin-Helmholtz or the magnetorotational instability~\cite{Rosswog:2015nja,Balbus:1991ay,Siegel:2013nrw,Giacomazzo:2010bx,Rezzolla:2011da,Price:2006fi}. After merger, magnetized winds and jetted outflows can be launched by the now dynamically relevant magnetic field for both black hole and NS remnants~\cite{Blandford:1977ds,Siegel:2017,Moesta:2020}. This has a direct impact on the electromagnetic counterparts signals expected from these events, see section~\ref{ssec:multimessenger}. 
A challenge in current state-of-the-art general-relativistic magnetohydrodynamics simulations of NS mergers is to resolve the turbulent magnetic field amplification and the resulting nonlinear small- and large-scale dynamo processes. High-resolution simulations showed that the length scales driving small-scale dynamos in the early merger are clearly unresolved ~\cite{Kiuchi:2014hja,Kiuchi:2017zzg} and at least two orders of magnitude below the currently highest resolution simulations available but that resolving the magnetorotational instability in the later postmerger evolution can lead to large-scale magnetic field generation~\cite{Kiuchi:2023}. Alternative approaches to incorporate the unresolved merger and early post-merger sub-grid scale effects are underway, e.g., general-relativistic large-eddy simulation schemes, e.g.,~\cite{Radice:2017zta,Radice:2020ids,Vigano:2020ouc,Aguilera-Miret:2020dhz,Celora:2021fry}. However, such approaches are challenging, and most of the proposed methods are coordinate-dependent and do not provide a covariant formulation; see~\cite{Celora:2024rif} for recent progress. Hence, further developmental work on both the theoretical and computational side is needed to ensure that magnetic fields are properly modelled, which requires a better description of turbulent flow and small-scale effects. 
Additionally, beyond-ideal general relativistic magnetohydrodynamics (GRMHD) effects play a key role in how particles are accelerated in jetted outflows from BNS remnants. Hence, such effects have to be modeled to systematically connect GRMHD BNS simulations to their potential short-GRB signatures. While there are a small number of resistive GRMHD simulations~\cite{Palenzuela:2013,Dionysopoulou:2015tda}, more work is needed to develop resistive GRMHD simulation capabilities with all the necessary nuclear and neutrino physics included. Effects from the nuclear stress-energy tensor, such as bulk viscosity, may also play an important role and have only just begun to be included. This is partly due to the complexity of finding suitable and stable formulations of the viscous GRMHD equations that are numerically feasible, e.g.,~\cite{Most:2021zvc}. Including all of these beyond-ideal effects will require significant development efforts from the community.

\subsubsection{Degeneracies with modified gravity and Beyond-Standard-Model physics}\label{subsec:newphysics}

Determining the EOS from GW data requires assuming an underlying theory of gravity. Even though most of the existent analyses assume General Relativity (GR), one can plausibly argue that its modification might be necessary and convenient in cosmology or help in building a consistent quantum--gravity theory (see section~\ref{section:div1} for further detail). Section~\ref{par:DegenEoSGravity}  addresses the influence of modified gravity on the structure of NSs and possible degeneracies with effects coming from the EOS uncertainties \cite{Shao:2019gjj}. Even if GR is adopted, other new physical effects, e.g., from the presence of dark matter, could impact the inference of the subatomic EOS from measurements. Such effects of Beyond the Standard Model (BSM) non-QCD matter are discussed in Secs.~\ref{subsec:BSMandEOS} and \ref{subsec:BSMandEOS_merger}. We note that much of the material discussed in this section is an active and relatively new area of research. Consequently, the examples discussed below mainly serve to illustrate possible phenomena that have been uncovered and point out potential degeneracies with changes in the EOS in qualitative terms, while detailed studies of such degeneracies and implications for data analysis strategies remain an open problem.

\paragraph{Degeneracy of EOS uncertainties with gravity beyond GR.}\label{par:DegenEoSGravity}

\begin{figure}[t]
\includegraphics[width=0.46\textwidth]{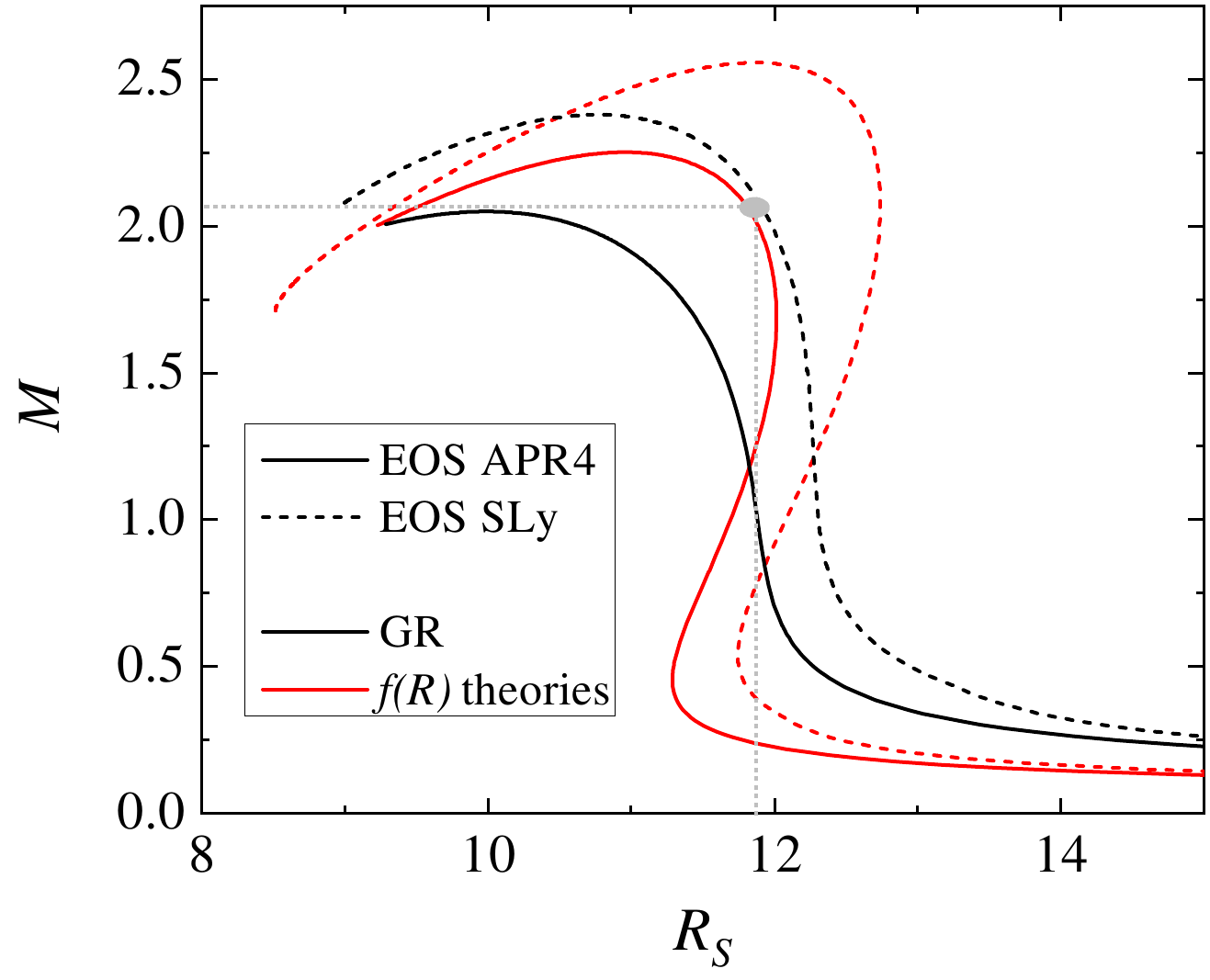}
\includegraphics[width=0.55\textwidth]{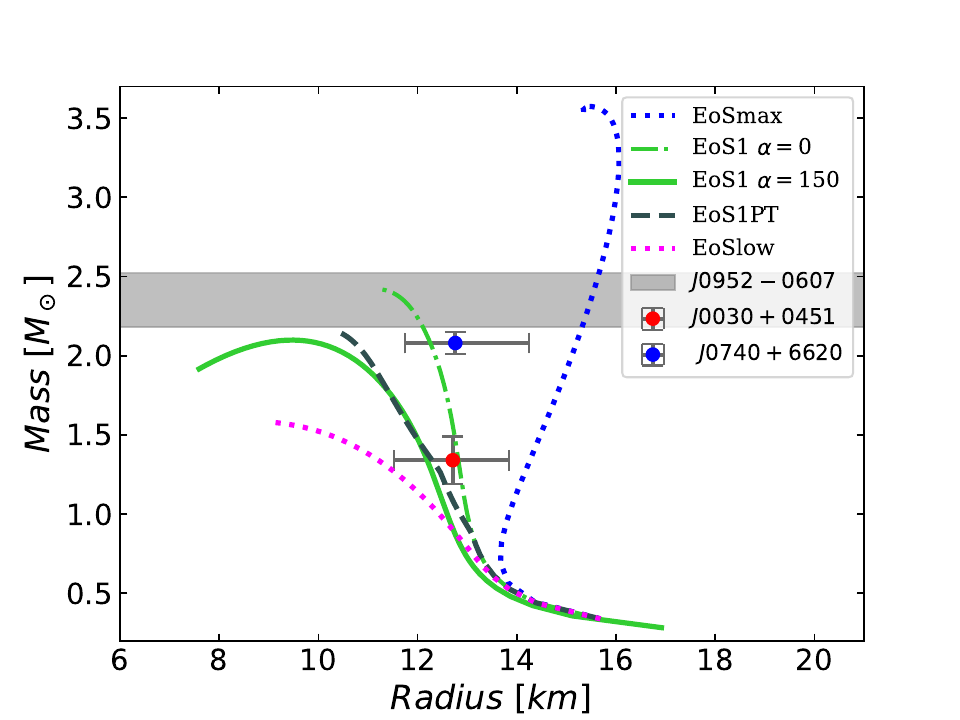}
\caption{\label{fig:degeneracy_FR}
Left: The $M(R)$ curve for two EOS and two theories of gravity -- GR and metric $f(R)=R+\alpha R^2$ gravity. Figure adapted from \cite{Yazadjiev:2014cza}.  
Right: $M(R)$ curves from a typical EOS (green dashed line) modified by either a phase transition (black dashed line) or Palatini $f(R)$ gravity (solid green line). Superposed are example current and near-term pulsar measurement uncertainties~\cite{Romani:2022jhd,Raaijmakers:2019qny,Salmi:2024aum}. Figure adapted from \cite{Lope-Oter:2023urz}.
}
\end{figure}

The degeneracies between EOS uncertainties and gravity beyond GR are far-reaching, covering both the equilibrium properties of NSs as well as their dynamics. Perhaps the best-known example is the degeneracy in the mass-radius diagram. This is demonstrated in figure~\ref{fig:degeneracy_FR} (left panel), which represents $M(R)$ dependencies for two EOSs within GR and for higher-curvature theories of gravity where one substitutes $R\to f(R) $ in the gravitational action, with the choice $f(R)=R+\alpha  R^2$~\cite{Starobinsky:2007hu} and $\alpha$ a parameter.\footnote{We stress that the choice of a particular theory beyond GR, and the values of its parameters, are chosen mainly for illustrative purposes and better visualization.} For example, a hypothetical simultaneous measurement of the mass and radius of a NS as indicated by a grey dot might signal either that the true EOS is SLy or that we actually have a GR modification and the correct EOS is APR4. Such degeneracies are very common, and they reduce our ability to disentangle the two effects. Hence, extracting an EOS from the $M(R)$ diagram alone, which is in one-to-one correspondence in GR, becomes extremely difficult if alternative theories of gravity are considered. 

\begin{figure}[t]
\centering
\includegraphics[width=0.7\columnwidth]{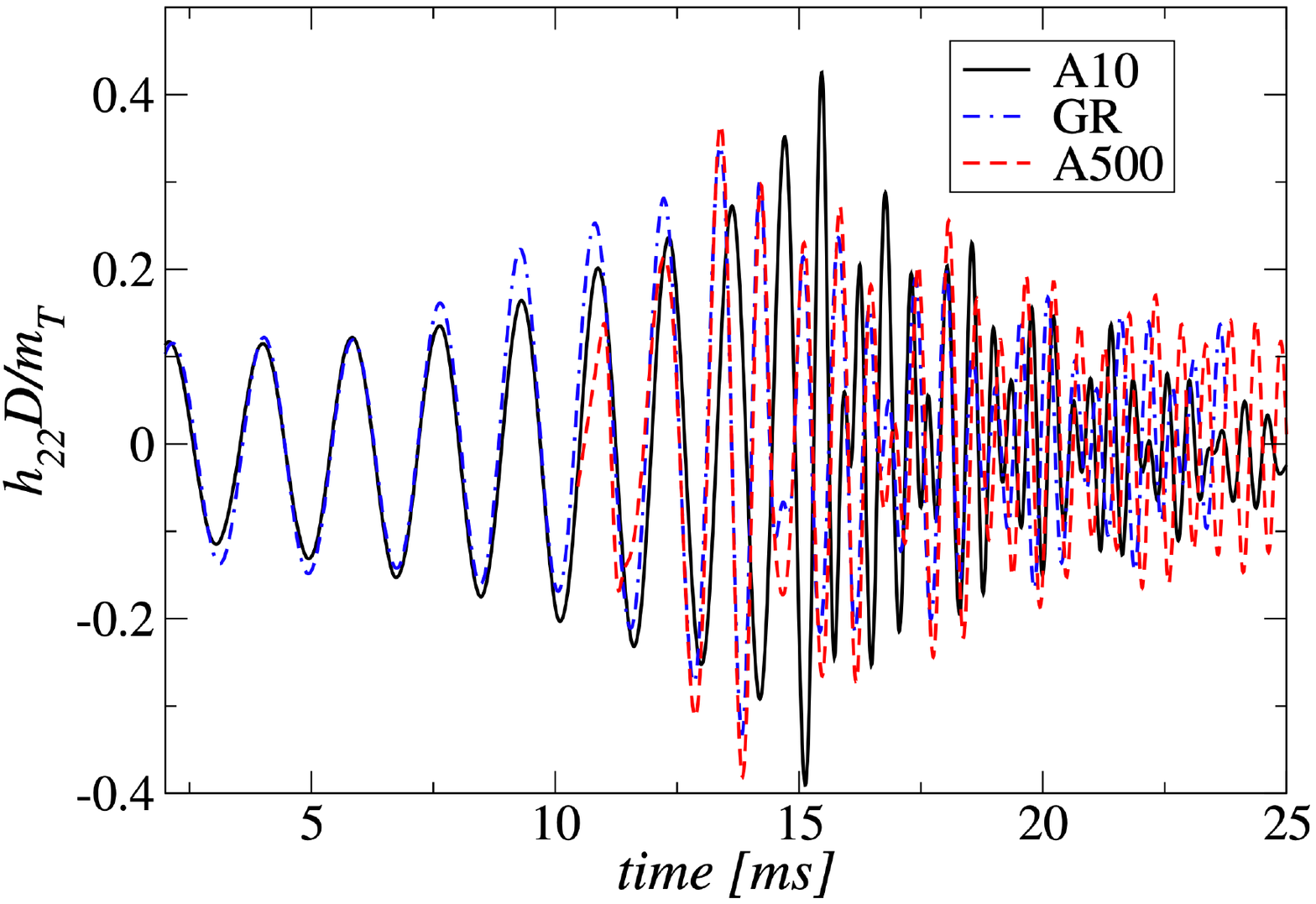}
\caption{\label{Sagunski_etal_fig7} Influence of different values of the coupling constant $\alpha$ of $f(R)=R+\alpha R^2$ gravity on the post-merger GW signal (A10 computed with $\alpha = 21.8\, {\rm km}^2$, A500 with $\alpha = 1090.3\,{\rm km}^2$). (Time shifted so that the peak strain is simultaneous to that in GR). 
Figure from \cite{2018PhRvD..97f4016S} with APS permission. 
\label{fig:modGRstrain}
}
\end{figure}
Another example is represented in the right panel of figure~\ref{fig:degeneracy_FR}. The mass-radius diagram shows how a typical EOS (dashed green line~\cite{Oter:2019rqp}) is modified by (a) adding a first-order phase transition, visible as a kink,  at around $R=13$ km in the black-dashed line, or (b) using the modified-gravity Palatini $R+\alpha R^2$~\cite{Lope-Oter:2023urz} TOV equations with a parameter $\alpha=150$ (solid green line). 
The two modifications would be practically indistinguishable over most of the range without extremely precise data (two data bars are shown to understand current precision in this diagram). It is interesting that such complicated phenomena like kinks, i.e. nonanalyticities, points of discontinuous derivatives in the EOS due to the latent heat of the phase transition~\cite{Lope-Oter:2021mjp} in subatomic matter, can be mimicked by modified gravity. Another interesting example of a non-trivial degeneracy is the phase transition between two equilibrium states of NSs caused by the formation or the complete dissolution of a scalar condensate above a certain threshold condition~\cite{Kuan:2022oxs,Doneva:2023kkz}. This process shares many similarities with a potential first-order phase transition between normal nuclear matter and deconfined quark matter in the stellar cores. 
As discussed in subsection~\ref{subsec:NSmergers}, ET has the potential to determine NS radii down to 0.1 km. Even falling short of that goal would be extremely useful in this diagram, given that current uncertainties are over $\times 10$ times larger~\cite{LIGOScientific:2018cki}. We refer to~\cite{Yazadjiev:2018xxk} for the equivalent tidal deformability vs. mass diagram, which is the quantity directly relevant for GW measurements, while the radius is only inferred from the resulting preferred EOSs.

In addition to modifying properties of individual NSs and their binary inspirals, the theory of gravity also impacts the postmerger phase, as shown for the strain in figure~\ref{fig:modGRstrain}. 
\begin{figure}[t]
\centering
\includegraphics[width=0.9\columnwidth]{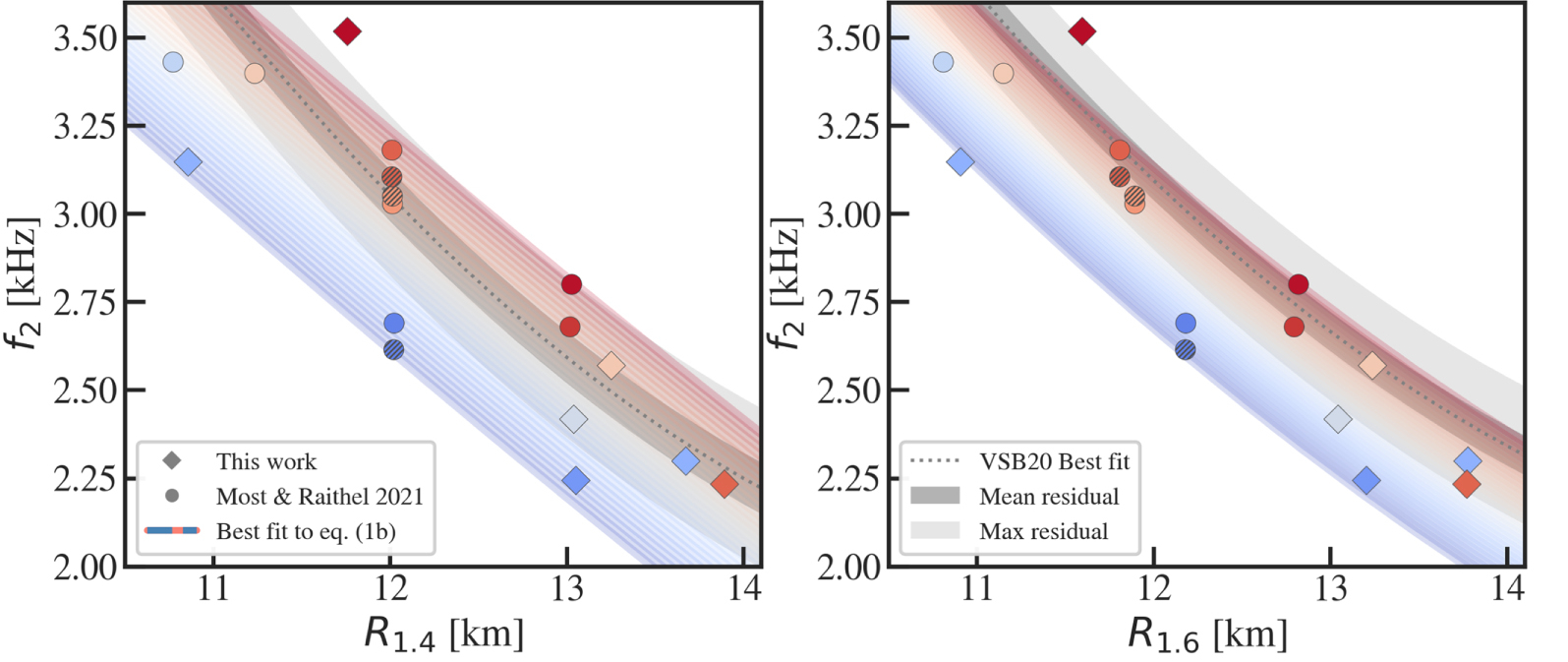}
\caption{\label{fig:Universalbreakdown} The universal relation for the peak frequency of the remainder following a neutron-star binary merger breaks down even in GR if the band of EOS used is broad enough: full tests of the gravity theory will need good control of the EOS from other sources. 
Reproduced from \cite{Raithel:2022orm} under Creative Commons 4.0 Attribution license. 
}
\end{figure}
A popular method to test the theory of gravity without having to worry about the underlying EOS is the use of universal relations \cite{Silva:2020acr,Yagi:2016bkt,Maselli:2013mva} that correlate quantities, like the dominant frequency following a binary merger and the radius of a cold spherically symmetric NS in $\beta$-equilibrium. These relations generally no longer hold in theories beyond GR. However, even within GR, such relations tend to be empirical and based on computing with a swath of EOS, whereas they sometimes fail when checked with a broader choice of EOS (colored symbols in the example of figure~\ref{fig:Universalbreakdown}, that are out of the initially
predicted grey band). Hence, a violation of quasi-universal relations could hint at multiple
things: (i) interesting nuclear physics such as phase transitions, (ii) modifications of GR, (iii) too uncertain NR simulations employed to derive the quasi-universal relations, or (iv) a non ideal choice of variables for the quasi-universal relation. The latter is actually the case for the example shown in figure~\ref{fig:Universalbreakdown}, where a rescaling of the frequency and the radius with the NS mass leads to much tighter and better fulfilled relations for a large set of NR data~\cite{Breschi:2022xnc}.

Moreover, similarly to the case of a BNS merger, a close  (within 10 kiloparsecs) supernova event's $f$-mode is expected to be reconstructible by ET with 90\% accuracy~\cite{Afle:2023mab} in the frequency measurement. While we should be able to extract information about the EOS from it, once again, modifications to GR can produce degeneracies as outlined in ref.~\cite{Blazquez-Salcedo:2022pwc}.

Despite the mentioned matter-gravity degeneracies, interesting features in modified gravity can sometimes be solely attributed to the generalization of Einstein's equations. One such example is the existence of additional polarizations of the gravitational waves not present in GR, such as a scalar  ``breathing mode'', a scalar longitudinal mode, and any vector modes \cite{Will_2018}. Special attention has been paid to the ``breathing mode'', present in certain classes of scalar-tensor models (such as Brans-Dicke theory). It can lead to interesting scalar-field induced GW signals,  potentially detectable with ET~\cite{Zhang:2017sym,Sperhake:2017itk,Kuan:2022oxs}.

\paragraph{Dark matter and other BSM physics.} \label{subsec:BSMandEOS}

Models of the galactic dark matter (DM) halo (e.g.~\cite{Quintana:2022yky}) suggest that many NSs may not be in vacuum but rather in a DM environment. As a consequence, NSs could accumulate a nonnegligible amount of DM within their interiors and vicinity, depending on the nature of the DM. Studies of the capture cross-section for DM particles such as~\cite{Anzuini:2021lnv} imply that it is possible that NSs could efficiently thermalize and accumulate enough DM to affect their macroscopic astrophysical properties.

Depending on its nature, quantum numbers, and particles' mass, the effects of DM on the NS include modification of mass, radius, tidal deformability, kinematic properties, thermal evolution, x-ray pulse profile, etc~\cite{Karkevandi:2021ygv, Rutherford:2022xeb,Tolos:2015qra,Dengler:2021qcq}.
As the nature of DM is unknown, proposed model DM particles span a broad mass
range, from ultralight ($<$ 1 keV) to heavy ($\gtrsim$ GeV)~\cite{Baryakhtar:2022hbu}. Here, selected DM candidates and their impact on compact stars are briefly discussed, and we refer to~section~\ref{section:div1} for more details.

An observable of special note is the NS cooling behavior \cite{_ngeles_P_rez_Garc_a_2022}, where DM can lead to a higher temperature plateau for old NSs. Such indications of a nonvanishing DM fraction in some NSs is complementary to signatures of matter properties in the GW signal \cite{Duvieretal}. Recent work~\cite{Mariani_2023} has also considered the efficiency of scalar-vector field theories describing the DM sector to reproduce current astrophysical measurements. In this way it is possible to further constrain the available DM parameter space complementarily to other existing bounds.

The impacts of DM on NS structure depend on the DM properties. Very weakly interacting dust-like DM acts by increasing the net energy density of the star without increasing the pressure, and thus at a given mass, the star contracts and the tidal deformability drops as $R^5$. The effect of such DM is akin to a softening of the EOS. Moreover, the presence of DM has been found to be potentially capable of triggering a phase transition to a quark star~\cite{Herrero:2019esf}. This arises because the added DM increases the compactness of the NS at a given mass, which could make it energetically favorable to change at least part of the subatomic matter to a non-nucleonic phase.

\begin{figure}[t]
\centering
\includegraphics[width=0.5\columnwidth]{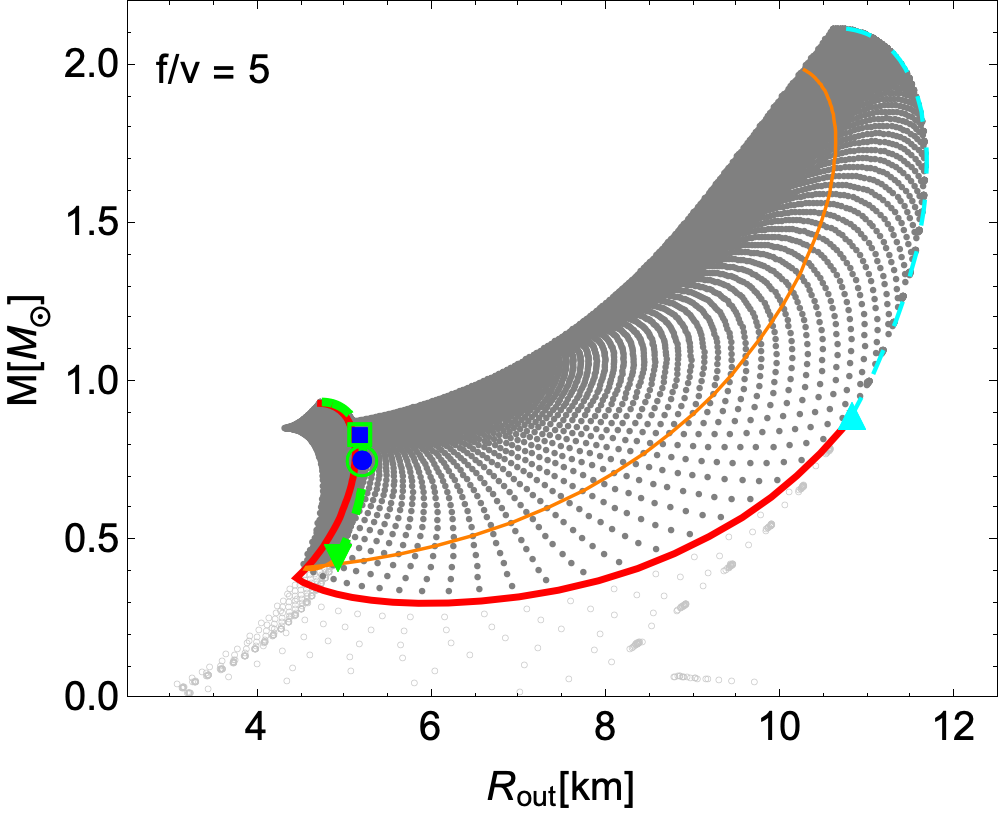}
\caption{If mirror DM is added in varying quantities to ordinary NS matter, the admixture is equivalent to having a one-parameter family of EOSs, and observable to observable diagrams such as $M(R)$ or $\Lambda(M)$ are now populated in two dimensions, with pure NS matter providing the (cyan online) upper right edge of the figure.\label{fig:MirrorDM}
Reproduced from~\cite{Hippert:2022snq} with permission from APS.}
\end{figure}

If DM belongs to a mirror sector, where each ordinary particle is assumed to have a corresponding dark partner, it generally does not equilibrate with ordinary matter via the strong force, but nevertheless accrues in NSs due to gravitational interactions. This leads to matter in NSs being characterized by two pressures and two energy densities for each of the components (normal and mirror). Consequently, observable properties such as the $M$-$R$ curves shown in figure~\ref{fig:MirrorDM} depend not only on the nuclear EOS but additionally on the dark matter parameters, i.e. $M(R)$ becomes an envelope of 
a neutron+DM $M(R)$ region in a two-dimensional plane~\cite{Hippert:2022snq}. Similar results apply to the tidal deformability vs. mass curves. The width of the ``line'' in tidal vs. mass or radius vs. mass diagrams can be used to constrain DM properties in this scenario. 

\begin{figure}[t]
\centering
\includegraphics[width=0.49\columnwidth]{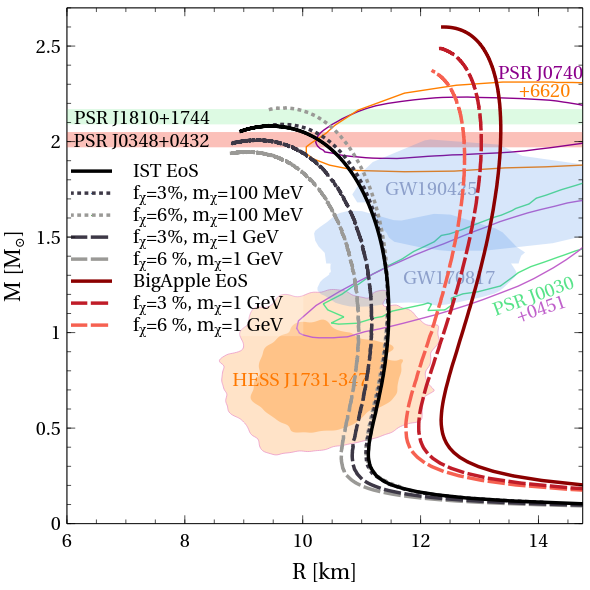}
\includegraphics[width=0.49\columnwidth]{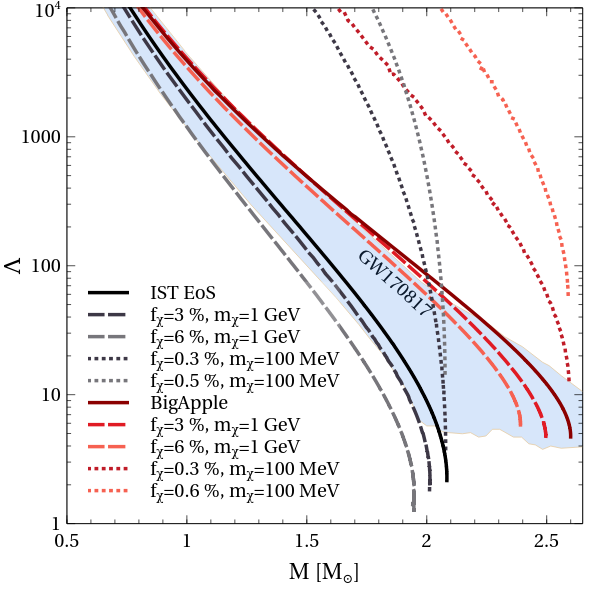}
\caption{Modifications of the mass-radius relation (left panel) and tidal deformability of NSs with a fraction of DM, $f_{\chi}$, and DM particle's mass, $m_{\chi}$. To address the uncertainties of the baryonic matter EOS, the soft IST EOS~\cite{Sagun:2020qvc} (black solid curve) and stiff BigApple EOS~\cite{Fattoyev:2020cws} (dark red solid curve) are considered. The 1$\sigma$ constraints from GW170817, GW190425, the NICER measurements of PSR J00030+0451, and PSRJ0740+6620, as well as mass measurements of heavy radio pulsars (PSR J1810+1744, PSR J0348+0432) are plotted. The 1$\sigma$ and 2$\sigma$ contours of HESS J1731-347~\cite{Doroshenko:2022} are shown in dark and light orange. Figure adapted from~\cite{Giangrandi:2022wht, Sagun:2023rzp}}.
\label{fig:DMeffect}
\end{figure}

Another proposed BSM phenomenon is an axion cloud surrounding NSs~\cite{Noordhuis:2023wid}. While the motivation for axions in the context of the Strong-CP Problem is currently being debated~\cite{Ai:2020ptm,Schierholz:2024var,Nakamura:2021meh},
many kinds of light DM particles can create an extended halo around the star that increases the total gravitational mass. By contrast, heavy DM tends to condense in a core, leading to a smaller radius and tidal deformability compared to a pure baryonic star due to the additional gravitational pull. Figure~\ref{fig:DMeffect} illustrates these two scenarios considering the asymmetric fermionic DM that interacts with the visible sector only through gravity~\cite{Ivanytskyi:2019wxd}. The outermost radius of the two configurations are different, as is the tidal deformability. Consequently, DM distributed in a halo/core could mimic an apparent stiffening/softening of strongly interacting matter EOS and spoil astrophysical constraints at high densities~\cite{Giangrandi:2022wht,Barbat:2024yvi}. This degeneracy between the effect of DM and strongly interacting matter properties at high densities could potentially lead to misleading conclusions and remains to be studied in detail. 

The contribution of the next-generation GW telescopes is important in breaking a possible degeneracy between the effects of DM and strongly interacting matter at high densities. Along with advances in subatomic theory and experiment, astrophysical observations of NSs, and their thermal evolution require simultaneous efforts in setting bounds on the microscopic DM parameters~\cite{Avila:2023rzj}. Thus, for popular particle-like DM models with mass $m_{\rm DM}\in(1,10^4)$ GeV, accounting for the impact of DM on the star sequence leading to $\rm NS\to BH $ collapse should allow constraining the nucleon-DM cross-section $\sigma_{N-\rm DM}$ to the $10^{-48}$ to $10^{-50}$ cm$^2$ range~\cite{Singh:2022wvw}. This approach is competitive with other experiments, even below the solar system neutrino floor. Therefore, we emphasize the importance of combining multi-messenger data on NS properties with microscopic nuclear input and DM distribution in the target galaxy to address this question.

\paragraph{The effect of dark matter on the merger dynamics.} \label{subsec:BSMandEOS_merger}

Binary NS coalescences provide a new and sensitive indirect way of constraining the properties of DM, as the GW signals
contains information on their environment, where effects such as dynamical friction or accretion could change the signals away from those in vacuum. The environments of merging binaries could be different depending on the particular galaxy, the distance to the centre, and the shape~\cite{Quintana:2022yky} of the galactic DM halo. Studies of the impact of DM on the post-merger phases of the binary NS system showed that DM cores can lead to additional peaks in the post-merger GW spectrum~\cite{Ellis:2017jgp} (and formation of a one-arm spiral instability~\cite{Bezares:2019jcb, Bauswein:2020kor}). 
The extra spectral peak (or two for unequal masses), which occurs at higher frequency than the main post-merger peak associated to the rotation of the hypermassive remnant,  provides a possible way to disentangle DM from nuclear EOS effects. It is found to arise from the fact that, while the normal matter components of two NSs merge, the compact DM cores are still orbiting within the remnant and take more time to coalesce~\cite{Ellis:2017jgp}. The location and features of the spectral peaks are expected to correlate with properties of the merging DM cores~\cite{Ellis:2017jgp}, and may enable inferring their size. Constraining the DM core size then allows to more confidently extract the information about the nuclear matter. 
Moreover, simulations have indicated that 
due to the absence of non-gravitational interactions, the DM cores and the surrounding post-merger matter likely exhibit distinct rotational frequencies immediately after the merger. Over time, these frequencies might synchronize through gravitational angular momentum transfer, including tidal effects~\cite{Hippert:2022snq}. The first two-fluid NR simulations of NSs mixed with mirror DM, concluded that DM might prolong the inspiral phase as the amount of DM increases as well as decrease the kilonova ejecta~\cite{Emma:2022xjs}.

\begin{figure}[t]
    \centering
    \includegraphics[width=\linewidth]{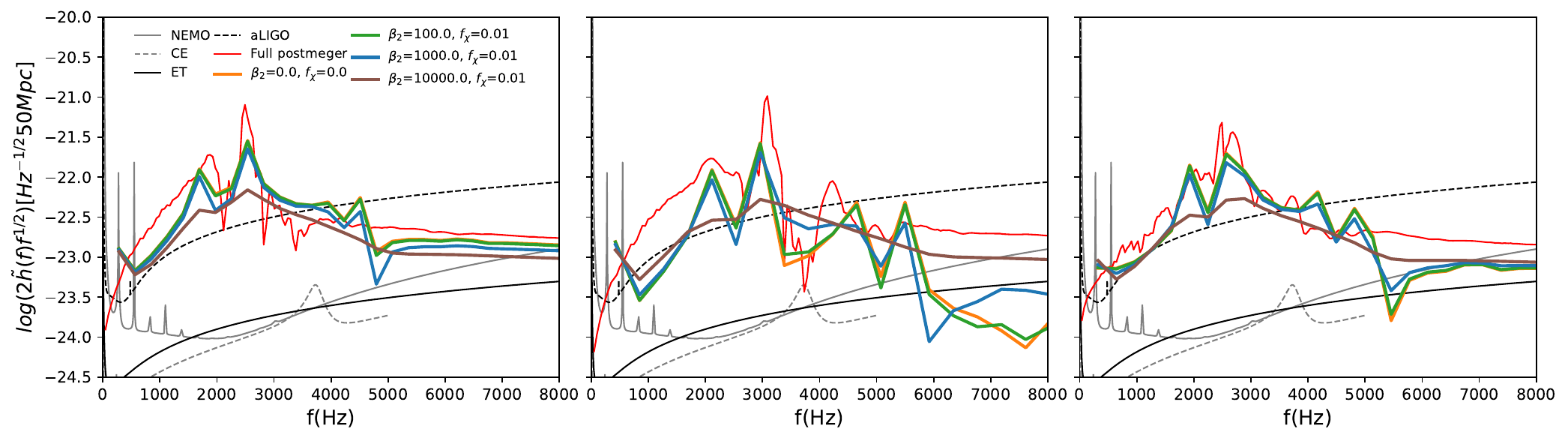}
    \caption{
Spectral density computed for the full post-merger data (thin red line) and for the early post-merger interval (thick colored lines) obtained from variation of DM mass fraction $f_\chi$ and viscosity parameter $\beta_2$ defined as the coefficient of an effective velocity-dependent dissipative force $\sim \beta_2 v^2$.
We also plot sensitivity curves of the NEMO, CE, ET, and Advanced LIGO detectors (black and grey curves).  CoRe database \cite{dietrich2018coredatabasebinaryneutron} sets THC005, THC0032, and THC0040 appear in the left, center, and right plots, respectively.  The damping of the peak amplitude due to DM can be seen. Distance fixed to $d = 50$~Mpc.
}
    \label{FIGdampX}
\end{figure}

Recently, DM models deployed in NS physics also have incorporated possible DM self-interactions, as suggested by galactic dynamics, 
as DM most likely flows with an effective viscosity~\cite{Duvieretal}. 
In the power spectral distribution,
this damps the peak amplitudes at the frequencies of the (expected) main modes in the power-spectrum distribution,
see figure~\ref{FIGdampX} for selected CoRe database BNS merger simulations. ET, among other detectors, will better constrain whether such damping is at work.

To have a comprehensive understanding of the effect of DM, it is necessary to perform binary NR simulations and analyze kilonova ejecta~\cite{Ruter:2023uzc}. These simulations should encompass various DM candidates, particle masses, interaction strengths, and fractions. Furthermore, a thorough investigation should include a comparison with GW and electromagnetic signals.\\

To conclude this subsection, we stress that extracting an EOS or other properties of nuclear matter from ET data is always subject to a hypothesis about the gravity theory. To prevent misleading conclusions, the possible degeneracies need to be studied in detail, in particular: (a) impact of exotic matter (whose effect is to soften/stiffen the EOS, modify the merger outflow, kilonova ejecta, and electromagnetic counterpart) and (b) degeneracies with modifications of General Relativity (which needs to be simultaneously tested with NS data).

\subsection{Executive summary}
\label{sec:conclusion}

As exposed throughout this section, the science potential for subatomic physics with GWs from next-generation observatories such as ET will be immense and is poised to enable groundbreaking advances in the fundamental understanding of the structure of NSs, dense hadronic matter, and the QCD phase diagram. Among others, copious detections of GWs from BNS and NSBH systems are guaranteed. Rapid progress is underway to face computational and modeling challenges and those for interdisciplinary interactions, such as optimally benefiting from the opportunities offered by GW astronomy. Here we will summarize the main points discussed in this section, on the areas where ET's capabilities will lead to advances in our understanding, the main challenges to be overcome and potential synergies with other facilities. 
\begin{highlightbox} {ET observables for probing subatomic physics}
Different observable GW sources will probe different aspects of subatomic matter: 
\begin{itemize}
\item Isolated mature NSs, and NSs in the inspiral phase of BNS mergers, can be considered as cold and in chemical ($\beta$-)~equilibrium. From the inspiral phase of such mergers and the tidal disruption in NS-BH systems, NS properties such as radius and tidal deformability  along with the underlying EOS for cold $\beta$-equilibrated matter can be recovered with great accuracy with a limited number of detections improving considerably on current knowledge, see section~\ref{ssec:ConstraintsonlowT}. A potential phase transition in dense NS matter is detectable for an onset at low densities from a few loud events, which will  potentially also enable measuring subdominant matter effects such as those from characteristic oscillation modes.
\item In a post-merger remnant, matter is heated up and higher densities are reached such that signals from the post-merger phase probe dense and hot matter. The complicated dynamics and the fact that matter is out of chemical equilibrium renders the interpretation of GW signals from this phase more difficult than during inspiral. Nevertheless, as discussed in section~\ref{ssec:postmerger}, the detection of such a signal in combination with the associated inspiral allows to obtain additional information on the EOS, for example on a potential phase transition at high densities. \item The GW signal from a CCSN is highly sensitive to the progenitor properties and the supernova dynamics. Nevertheless, GW asteroseismology of the newly born PNS can reveal  properties such as its surface gravity and some features of hot and dense matter, see section~\ref{subsec:GWCCSN} for more details. 
\item   ET will be able to  probe the deformations of hundreds of pulsars  either detecting a signal, or setting constraints below the maximum ellipticity predicted by theory, and for tens of high frequency millisecond pulsars ET will probe
ellipticities below $10^{-9}$;  see section~\ref{sssec:continousGW} and section~\ref{section:div7}. 

 \item In combination with other observables, e.g., electromagnetic counterparts of BNS or NSBH mergers, there will be additional information to place constraints on the EOS as well as on the formation of heavy elements. In particular, the improved sky-localization of future GW signals observed by ET will provide important information for follow-up searches of electromagnetic counterparts; see~section~\ref{ssec:multimessenger} and section~\ref{section:div4}.
\end{itemize}
\end{highlightbox}

\begin{highlightbox}{Challenges}
In order to take full advantage of this scientific potential for nuclear physics with ET, important theoretical, numerical, and computational challenges remain to be resolved. 
\begin{itemize}
\item 
A key feature of GW measurements is that they are based on cross-correlating theoretical models with the data to  make the link between the signals and the fundamental source physics. Models for NS binary systems that include more realistic physics are urgently needed for data analysis; the specific aspects requiring prioritized efforts for subatomic physics  were highlighted throughout this section. This also requires key inputs from nuclear physics on the possible matter properties beyond the pressure-density relationship, such as finite temperature effects and transport coefficients (see section~\ref{sec:reactions}), from studies that follow a first-principle methodology as far as possible. Interactions between the GW and nuclear-physics communities are essential to communicate what information is required and what formats of the results are most useful as inputs for numerical simulations and analytical studies. Efforts are also necessary to develop suitable waveforms for data analysis that include NS matter effects (see section~\ref{ssec:waveforms}). 
\item On the computational side, the complex merger process in BNS events requires NR simulations with high computational cost on supercomputers. At present, they remain limited in length, resolution, parameter space coverage, and microphysical realism, see the discussion in section~\ref{Sec:Simulation_Uncertainty}. NS binary signals for ET will linger for several hours in the sensitive band, which also presents a computational challenge for data analysis. With current methods and infrastructures, it will become prohibitively expensive. New tools such as deep learning could help address this critical issue; see~section~\ref{section:div10}.

\item In this context, it should also be noted that the extraction of nuclear matter properties is simultaneous with constraining modifications of GR and exotic (perhaps dark matter) components of NSs, which can mask or distort the subatomic effects and that further work is required to understand resulting degeneracies and develop strategies to mitigate them, as discussed in section~\ref{subsec:newphysics}. A better and independent understanding of the underlying nuclear physics from theory and experiment will  open new avenues for tests of gravity, cosmology, and explorations of dark matter with GWs from NS events. 
\end{itemize}
\end{highlightbox}

\begin{highlightbox}{Synergies with other facilities}
In addition, to optimize the science payoffs with GWs for nuclear
physics requires strengthening the efforts on interdisciplinary
connections with various other probes from astrophysical observations, 
and nuclear physics theory and experiment, see section \ref{section:div5} for more details.  As an example, as discussed in section~\ref{ssec:ConstraintsonlowT}, from the inspiral of BNS mergers, the cold $\beta$-equilibrated EOS can be determined accurately, but without additional information it is difficult to determine matter composition or properties of symmetric nuclear matter. Only a combination of data from different probes will enable pinning
down the forces acting in dense matter. In particular:
\begin{itemize}
\item 
Heavy-ion
collisions at lower energies probe the EOS of warm dense
matter and yield valuable information in a regime of density and
isospin asymmetry different from NSs; data on nuclear masses,
excitations, resonances, and the determination of the neutron skin
thickness probe properties of nuclear interactions in yet other complementary
realms than NSs. Likewise, it will be important to exploit
experiments to gain insights into the complex physics of kilonovae, for instance, by measuring
masses of exotic nuclides and synthesizing neutron-rich nuclei with
rare isotope beam facilities. 

\item New electromagnetic telescopes will also come online in the next
years, for some of which an important objective is to observe
counterparts to GW events in different wavelengths. In addition, X-ray
telescopes, such as e.g., NICER or the planned Strobe-X, and
radio-timing, e.g., with SKA and its precursors, will provide
information on NS masses and radii complementary to BNS and NSBH
inspirals. Multi-messenger observations of (P)NS cooling will allow us to gain valuable insight into the composition of the dense (and hot) matter in these compact stars.
\end{itemize}
\end{highlightbox}


\section{Stellar collapse and rotating neutron stars}\label{section:div7}

Stellar collapses are among the most powerful astrophysical phenomena, in which extreme temperatures and densities are reached. They are characterized by complex physical processes, involving hydrodynamics, magnetic fields, neutrinos, nuclear physics, electromagnetic and gravitational radiation.
The final remnant of a stellar collapse is a compact object, namely a neutron star or a black hole. 

Although some aspects are still not totally clear, stellar collapses are unanimously considered as potential sources of gravitational-waves (\acrshort{gw}s). GWs are associated both to the collapse itself, in the form of transient emission, and to the collapse aftermath in the form of persistent emission from an asymmetric neutron star, isolated or part of a binary system. Further possible channels of GWs include various kinds of dynamical instabilities, which can set in during the early stages of the remnant life, and glitches, likely associated to cracks in the neutron star crust or to the interplay among the crust and the inner superfluid. 

In this section, we describe the main aspects of the collapse process and of the resulting neutron star, focusing attention on the associated GW emission in view of the unprecedented sensitivity promised by Einstein Telescope. Benefits of the multi-messenger approach, in which information from electromagnetic, neutrino and GW observatories are combined, are also stressed both in terms of improved detection capabilities and in terms of science return.

\subsection{Introduction}
In this era of \acrshort{gw}s as new and exciting astrophysical messengers, stellar collapses and the resulting neutron star (\acrshort{ns}) will play a crucial role. Their emission encompasses all the available multi-messenger tracers: electromagnetic (\acrshort{em}) waves, cosmic rays, neutrinos, and \acrshort{gw}s. They are unique laboratories where the most extreme phases of matter can be studied: not only probing extreme regimes of gravity and electromagnetism but also the strong and weak interactions can be studied in regimes that there is no hope to explore on Earth. To interpret at the best future \acrshort{gw} detections, it is necessary to consolidate our understanding of the stellar collapse mechanisms and of the different \acrshort{ns} classes, their \acrshort{em} and neutrino signatures, rates and the expected \acrshort{gw} characteristics. 

Concerning \acrshort{gw} emission, stellar collapses are predicted to emit short bursts of radiation, with duration from a few milliseconds to a few seconds, associated to the core bounce and following explosion, with a more structured emission due to the excitation of oscillation modes of the newly formed proto-neutron star (\acrshort{pns}). 
\acrshort{ns}s have been predicted to emit short \acrshort{gw} transients in response to sudden changes in the star's mass quadrupole, like glitches, and persistent signals due to long-lived deviations from an axi-symmetric shape. Young \acrshort{ns}s could also be subject to the excitation of various kinds of oscillation modes triggered by dynamical instabilities or following fast processes, like crust cracks or rearrangements of the star's magnetic field.  

Overall, stellar collapse events detectable by \acrshort{et} will belong to the Milky Way or the local group of galaxies (i.e. within a few Mpc), with the possible exception represented by long-transient emission from fast spinning newborn magnetars endowed with a super-strong inner magnetic field, which could be detectable up to a distance of a few tens of Mpc. 

In this section, we will briefly discuss the different data analysis techniques required to detect the \acrshort{gw}s emitted by such sources.Depending on their characteristics, different data analysis algorithms are needed to extract the signals from detector data. In general, the development of suitable data analysis methods must conform, as much as possible, to three keystones: \textit{sensitivity}, \textit{robustness} and \textit{computational efficiency}. On one hand, search algorithms must be as sensitive as possible to the specific kind of \acrshort{gw} signal being searched. On the other hand, these algorithms must often balance search sensitivity to generic, unmodeled signal features with robustness against detector noise artefacts.In addition, analyses computational cost also play a relevant role and must be taken into account during method development.

Since in this section we deal with the outcomes of the stellar collapse, we begin in subsection~\ref{sec:collapse} to recall the main evolutionary path of stars toward the collapse and discuss the limiting masses of stars that explode and those that collapse directly to a black hole (BH). We also discuss the minimum mass of stars that enter the pulsation pair instability exploding, therefore, as Pair Instability Supernovae leaving no remnant. We also address the dynamics of the explosion of core collapse supernovae, the current status of their observations and the expected \acrshort{gw} and neutrino emission from these objects. We provide the observed and expected core collapse supernova rates (CCSNR) and in particular the expected rate in the Milky Way, which is of key importance to plan the observational strategy for the current and next generation \acrshort{gw} and neutrino detectors.

In subsection~\ref{sec:neutron}, we outline \acrshort{ns}s' main features, in terms of structure, and emission of \acrshort{em} radiation, going from steady beamed radio or gamma emission of pulsars to burst-like highly energetic X- or gamma-ray flares in young magnetars. Moreover, we describe the main mechanisms through which \acrshort{ns}s are expected to emit \acrshort{gw}s. We are here mainly interested in isolated \acrshort{ns}s, or \acrshort{ns}s in stellar binary systems like Low Mass X-ray Binaries (LMXBs) and pulsars in binary systems, in which the \acrshort{gw} emission is powered by the \acrshort{ns} rotational energy rather than the orbital (as  for coalescing binaries, discussed in section 6). 

Subsection~\ref{sec:observation} is devoted to a discussion of \acrshort{gw} detection prospects of the aforementioned sources and to outline the role of \acrshort{em} and neutrino future facilities in the context of multi-messenger astronomy. A brief classification of the main search categories is also presented for both stellar collapses and rotating \acrshort{ns}s \acrshort{gw} signals. Moreover, some basic information on the main analysis methods is given, stressing possible open points to be tackled in view of \acrshort{et} and promising routes for improvement.       

\subsection{Stellar collapse}\label{sec:collapse}

\subsubsection{Stellar evolution toward stellar collapse}\label{stelev}

In the general picture of stellar evolution stars with initial mass $\rm M \gtrsim 10~M_\odot$ never experience a significant electron degeneracy in the core during all their nuclear burning stages; therefore, they evolve to higher and higher temperatures, fusing heavier and heavier nuclei until an iron core is formed. The iron core, then, becomes unstable, collapses and through a sequence of events drives the explosion of the star as a core collapse supernova (\acrshort{ccsn}) (see section \ref{corecollapse}). However, as the initial mass increases the explosion becomes harder and it is expected that above a critical initial mass the stars, at the end of their evolution, will collapse directly to a \acrshort{bh}, producing a so-called failed explosion. As the progenitor mass increases even more, the core of these stars becomes unstable for pair creation, it collapses and eventually drives a pair instability supernova (\acrshort{pisn}) that leaves no remnant. The minimum masses of the stars that produce a successful explosion, $M_{\rm CCSN}$, a direct collapse (\acrshort{dc}) to a \acrshort{bh}, $M_{\rm DC}$, and a \acrshort{pisn}, $M_{\rm PISN}$, depend in general on the initial metallicity and rotation velocity. In general $M_{\rm CCSN}<M_{\rm DC}<M_{\rm PISN}$.

\begin{figure}[t]
\centering
\includegraphics[scale=0.3]{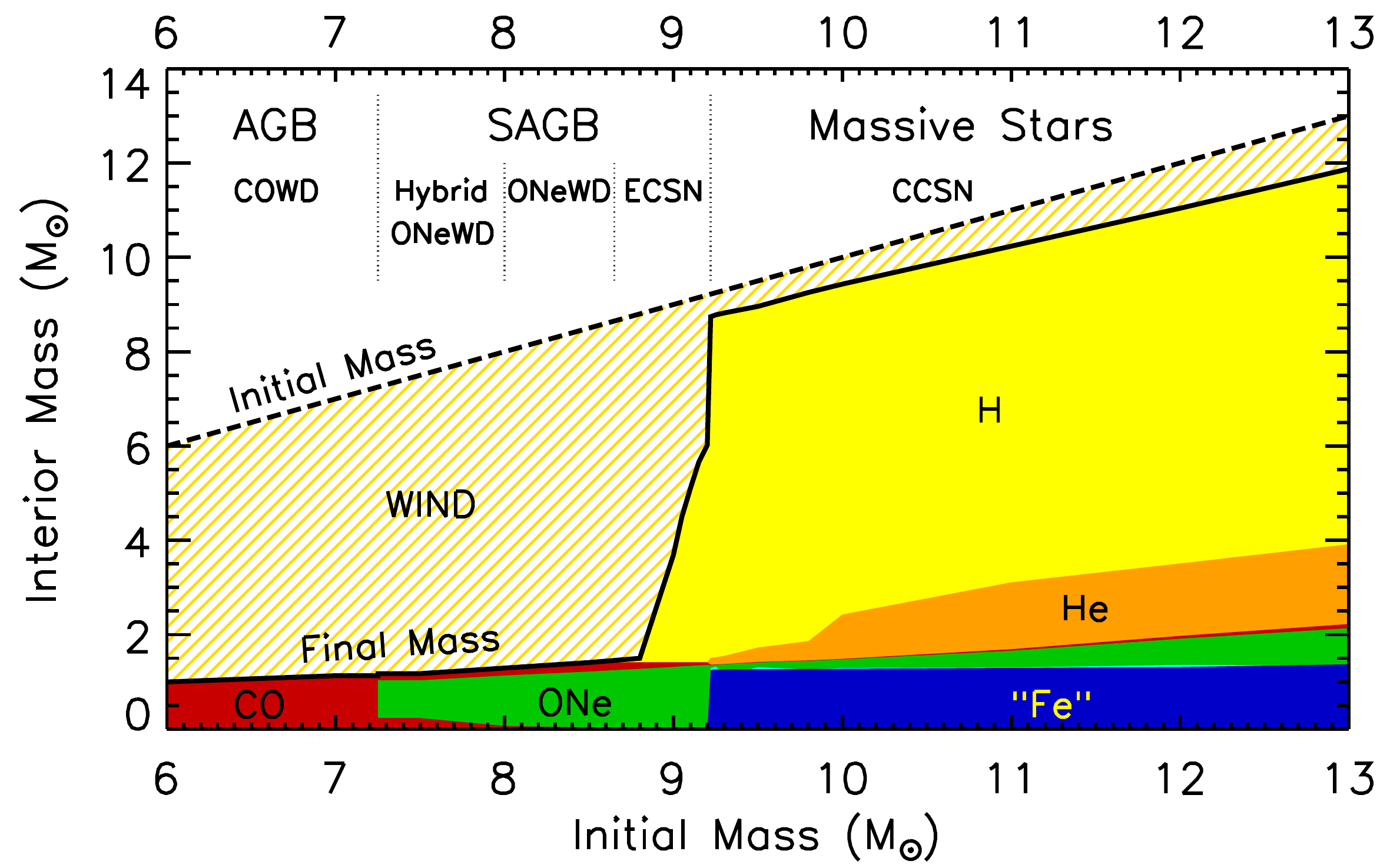}
\caption{Final fate of stars in the mass range $(7-13)\, M_\odot$ according to \cite{Limongi:2023bcg}.}
\label{fig:last_sagb_massive}
\end{figure}

According to the most recent studies \cite{Limongi:2023bcg}, stars with mass $M\lesssim 7.5~M_\odot$ evolve along the Thermally Pulsing Asymptotic Giant Branch (\acrshort{tpagb}) phase and end their life as CO White Dwarfs (\acrshort{cowd}s). Stars with $M \gtrsim 9.2~M_\odot$, on the contrary, evolve through all the major stable nuclear burning stages and eventually explode as \acrshort{ccsn} leaving a compact remnant, i.e. either a neutron
star or a \acrshort{bh}. Therefore, according to this study $M_{\rm CCSN}\sim 9.2\, M_\odot$. However, the minimum mass that produces an explosion cannot be set to such a value because stars in the range $\sim (7.5-9.2)~M_\odot$ have a much more complex evolution and a fraction on them can eventually explode as electron capture supernovae (\acrshort{ecsn}e). In particular, as a result of the carbon burning phase, stars in this mass range develop a degenerate ONeMg core and enter the thermally pulsing phase. These stars are usually called Super Asymptotic Giant Branch (\acrshort{sagb}) stars. Among them, the most massive ones, $\sim (8.5-9.2)\, M_\odot$, achieve central densities close to the threshold density for the activation of the electron capture (\acrshort{ec}) $\rm ^{20}Ne(e^{-},\nu)^{20}Fe$. Once such an electron capture becomes efficient, deleptonization and thermonuclear instability develop and may eventually lead to the \acrshort{ecsn}.  
The final fate of such stars however is not yet completely well established since it depends on the competition between the
energy released by the nuclear burning front and the loss of pressure
due to the deleptonization occurring in the central zones. If the
energy released by nuclear burning prevails, degeneracy is removed and
the star explodes as a thermonuclear \acrshort{ecsn} \cite{Miyaji80,Isern91,Jones:2016asr,NomotoLeung2017}.  If, on the contrary, the deleptonization dominates, the collapse cannot be
halted and the star collapses into a \acrshort{ns} (core collapse \acrshort{ecsn}) \cite{Miyaji80,Nomoto87,Kitaura:2005bt,Fischer:2009af,Jones:2016asr,Zha:2019rpq,Zha:2021bev}. Which outcome is realized from \acrshort{ecsn}e depends on both the details of the modeling of the presupernova evolution and explosion. For all these reasons the minimum mass that can produce an explosion as an \acrshort{ecsn} cannot be identified with certainty of precision and we can only say that it is in the range $M_{\rm ECSN}\sim (8.5-9.2)\, M_\odot$. A summary of the evolution and final fate of stars in the mass interval between 7 and 13 $M_\odot$ is shown in figure~\ref{fig:last_sagb_massive}.

For sufficiently high masses (with $M>M_{\rm CCSN}$), the binding energy of the stars are so large that they do not explode but, rather, collapse directly to a \acrshort{bh}.
The determination of the minimum mass that produces a direct collapse to a \acrshort{bh} ($M_{\rm DC}$) is quite difficult because it depends on the details of the explosion and in particular on the amount of fallback of material and mixing occurring at late times. After about three decades of research, the current reference model for \acrshort{ccsn}e is based on the ``delayed shock mechanism'' or ``neutrino-heating mechanism'', which was originally proposed by Wilson and Bethe \cite{Bethe:1985sox,Herant:1994dd,Burrows:1995ww,Janka:1996tu,Kotake:2005zn,Janka:2016fox,Janka:2017vcp,Burrows:2019rtd}. However, state-of-the-art 3D simulations of the explosion, which include the most sophisticated treatment of neutrino transport, cannot follow the explosion at late times ($t\gtrsim 10$~s), when most of the fallback should occur, because of the tremendous cost of computer time. As a consequence these simulations cannot accurately predict the
the final mass of the compact remnant and whether the supernova (\acrshort{sn}) will be successful or failed, in the entire \acrshort{sn} progenitor mass range \cite{Blondin:2002sm,Suwa:2009py,Muller:2011yi,Suwa:2012xd,Lentz:2012xc,Summa:2015nyk,Lentz:2015nxa,Bruenn:2014qea,Burrows:2019zce,Mezzacappa:2020oyq,Bollig:2020phc}.

\begin{figure}[t]
\centering
\includegraphics[scale=0.33]{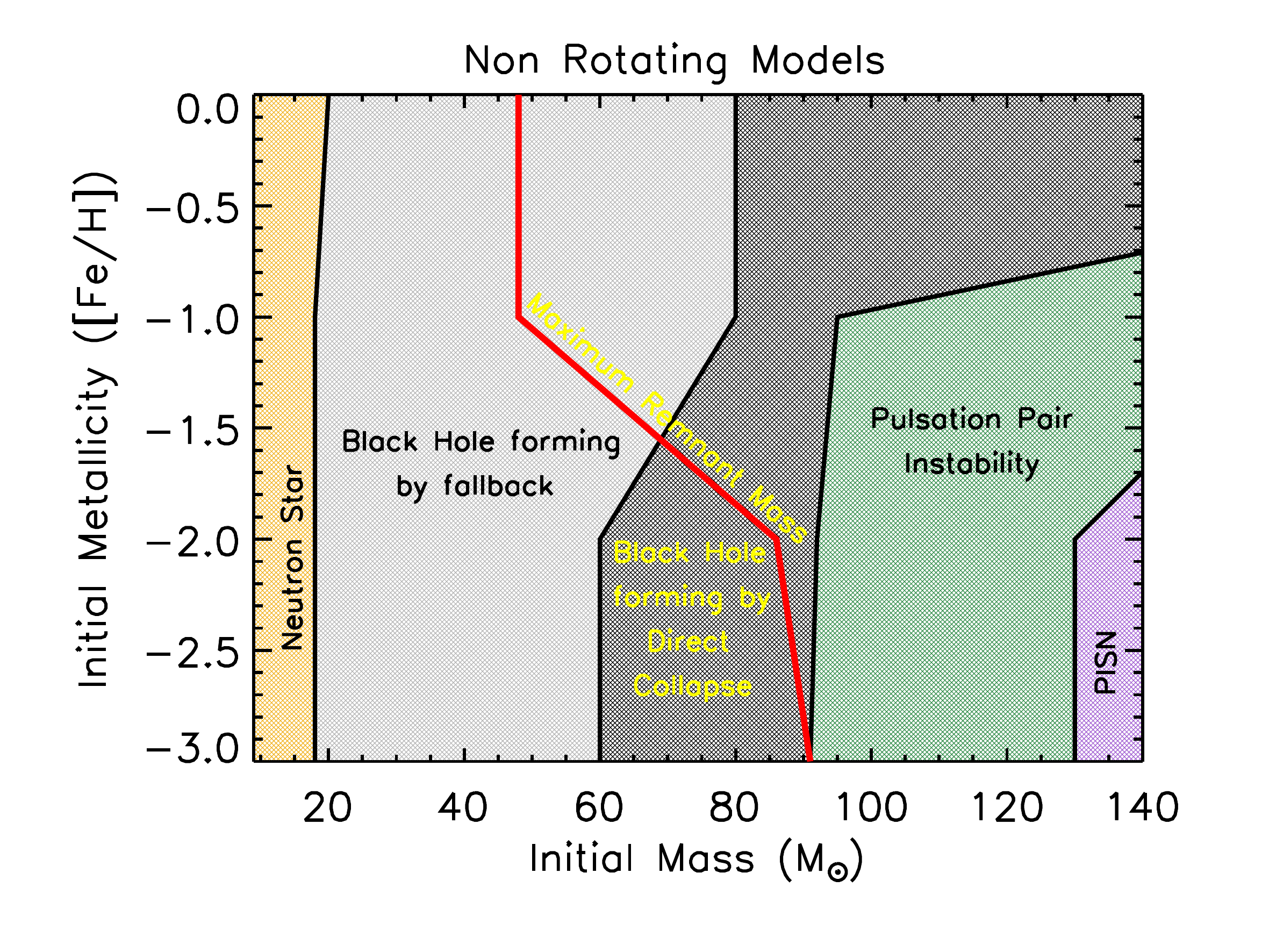}
\caption{Final fate  and remnant masses predicted for non rotating stars as a function of the initial mass and initial metallicity. The red line marks the predicted maximum remnant mass.}
\label{fig:summanorot}
\end{figure}

\begin{figure}[t]
\centering
\includegraphics[scale=0.33]{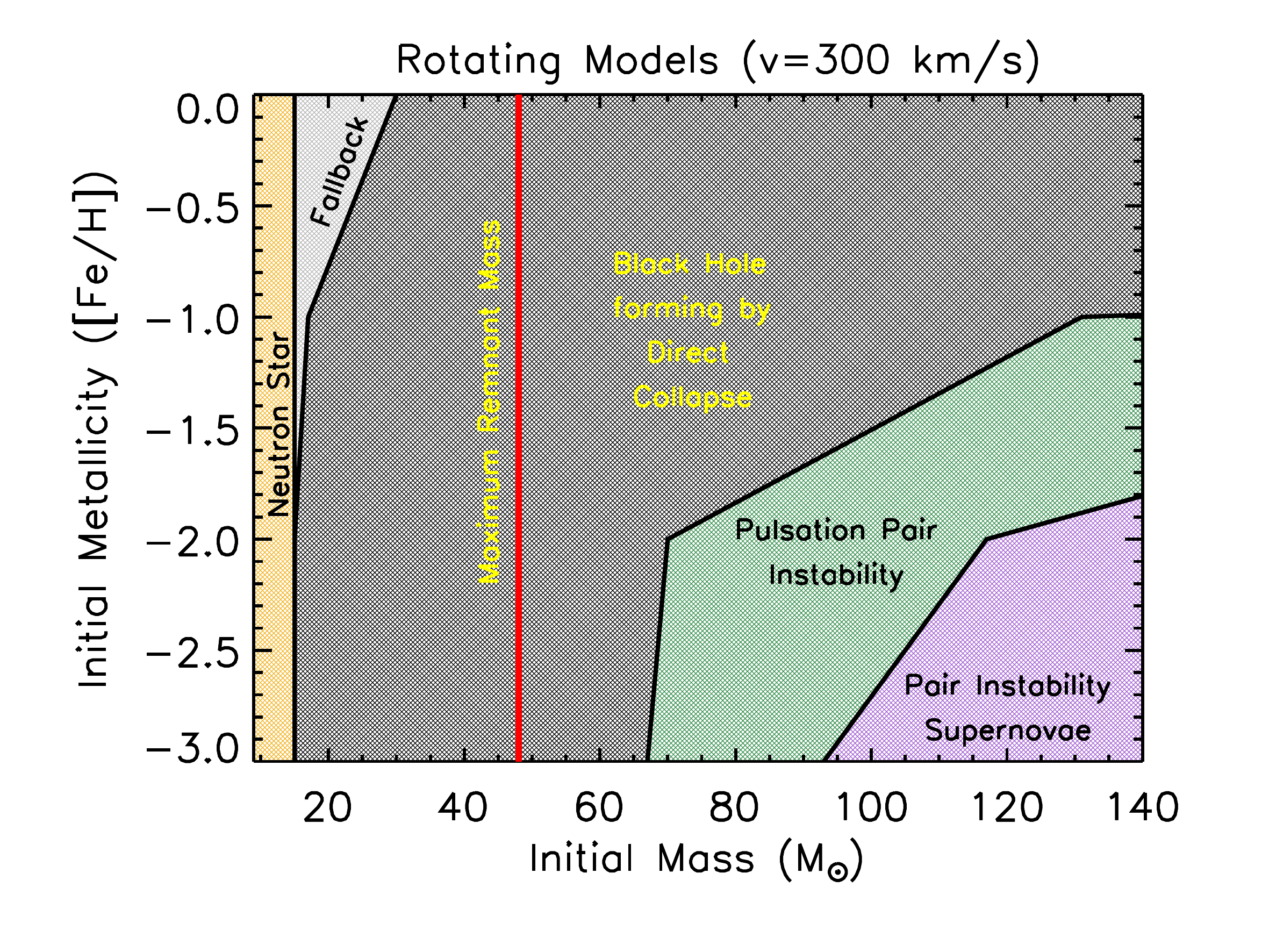}
\caption{Final fate  and remnant masses predicted for stars with initial rotation velocity of 300 km/s as a function of the initial mass and initial metallicity. The red line marks the predicted maximum remnant mass.}
\label{fig:summarot}
\end{figure}

A simplified approach for studying the explosion of massive stars is to artificially stimulate the \acrshort{sn} explosion by injecting some amount of energy at an arbitrary mass coordinate in the progenitor model. This extra energy induces the formation of a shock wave whose propagation through the mantle is followed by means of a hydro code. Different groups adopted different schemes to inject the extra energy: a thermal bomb \cite{Thielemann+90}: a piston \cite{Woosley:1995ip}: a kinetic bomb \cite{Limongi:2003ui}: thermal energy due to a calibrated neutrino luminosity \cite{Ugliano:2012fvp}. Whichever is the way in which the energy is deposited into the presupernova model, the typical evolution of the explosion involves (1) an initial remnant; (2) a shock wave propagating through the mantle and inducing compression, heating, explosive nucleosynthesis and expansion of the shocked material; (3) a fallback of the deepest layers of the expanding mantle because of the gravitational field of the compact initial remnant; (4) a mass cut which is the mass separation between the ejecta, which will have a given final kinetic energy, and the final remnant, which in general will be larger than the initial one. The main advantage of this approach is that the computing cost is relatively small. Thus, it is possible to follow the evolution of the shock over longer timescales and to estimate the amount of fallback and the final remnant mass. The 1D approach has been extensively used to study the chemical composition of the ejecta produced by \acrshort{sn} explosion, especially in the context of explosive nucleosynthesis. A recent calculation of the explosion of an extended set of presupernova models in the range $ (13 -120)~M_\odot$, with initial metallicities corresponding to ${\rm [Fe/H]}=0,\,-1,\,-2,\,-3 $ and initial rotation velocities $v=0,~300~{\rm km/s}$, in the framework of the thermal bomb, has shown that: (1) for non rotating models the threshold main sequence masses above which the stars undergo a direct collapse to \acrshort{bh} are  $M \simeq 60~M_\odot$ for ${\rm [Fe/H]}=-2,~-3$ and $M \simeq 80~M_\odot$ for ${\rm [Fe/H]}=0,\,-1$; 
(2) for stars with initial rotation velocity $v=300~{\rm km/s}$ the transition from a successful \acrshort{sn} explosion to a direct collapse to \acrshort{bh}s occurs for initial main sequence masses of $\simeq 30~M_\odot$, $\simeq 17~M_\odot$, $\simeq 15~M_\odot$, and $\simeq 15~M_\odot$, for ${\rm [Fe/H]}=0, ~1, ~2, ~3$, respectively. 
A summary of these results are shown in figures \ref{fig:summanorot} and \ref{fig:summarot}.

Above a given threshold value of the initial mass, or more correctly of the CO core mass, after core He burning the stars evolve at sufficiently high temperature and low density (i.e. high central entropy) that electron-positron pairs are created copiously, converting internal energy into rest mass of the pairs without contributing much to the pressure. When the star enters this stage the equation of state softens, the structural adiabatic index decreases below 4/3 and the core becomes unstable. This is the ``pair-instability'' \cite{Fowler:1964zz,Barkat:1967zz,RakavyShaviv67}. According to \cite{Woosley:2016hmi}, if the mass of the CO core is larger than $\sim 53~M_\odot$, this instability results in a dynamical collapse of the core that triggers explosive oxygen and silicon burning.  These nuclear burning produce enough energy to revert the collapse leading to the explosion of the star as a \acrshort{pisn}. Unlike other types of supernovae, the explosion mechanism of a \acrshort{pisn} is well understood, without complications like mass cut or fallback. If, on the contrary, the mass of the CO core is in the range $\sim (35-53)~M_\odot$, the nuclear burning are not sufficiently energetic to disrupt the star but, rather, they induce a series of pulsations. This is the ``pulsation pair-instability''. After a number of pulsations the H rich envelope is almost completely lost and the core of the star resumes the classical evolution of a typical massive star eventually collapsing to a \acrshort{bh} (\acrshort{ppisn}). In general the limiting masses that eventually become \acrshort{ppisn} or \acrshort{pisn}, depend on the initial metallicity and initial rotation velocity. According to recent studies \cite{Limongi:2018qgr}, for solar metallicity models, regardless of the initial velocity, the mass loss is efficient enough that the CO core never becomes massive enough to enter the pair instability regime. For non-rotating models of lower metallicities, on the contrary, the mass loss is almost negligible, in this way $M_{\rm PPISN}=95,~92,~91~M_\odot$ and $M_{\rm PISN}=163,~130,~130~M_\odot$, for metallicities ${\rm [Fe/H]}=-1,\,-2,\,-3$, respectively. The effect of rotation is that of increasing the CO core mass due to the rotation driven mixing, as a consequence for rotating models with initial velocity of $\rm ~300~km/s$ the limiting masses are $M_{\rm PPISN}=131,~70,~67~M_\odot$ and $M_{\rm PISN}=235,~117,~93~M_\odot$, for metallicities ${\rm [Fe/H]}=-1,\,-2,\,-3$ (see figures 
\ref{fig:summanorot} and \ref{fig:summarot}). 

The most massive \acrshort{bh} ($M_{\rm BH,max}$) that can be formed corresponds to the mass of the most massive stars that collapses directly to a \acrshort{bh}. For non-rotating stars $M_{\rm BH,max}\simeq 91~M_\odot$ and is the result of the direct collapse of the most massive star at ${\rm [Fe/H]}=-3$ not undergoing \acrshort{ppisn}e (red line in figure~\ref{fig:summanorot}). In the case of rotating stars (with initial rotation velocity of 300 km/s), because of the enhanced mass loss driven by rotation, even if more stars collapse directly to a \acrshort{bh}, $M_{\rm BH,max}$ is lower than for non-rotating stars. In particular, in the rotating case $M_{\rm BH,max}\simeq 47~M_\odot$ and is the result of the direct collapse of a star undergoing \acrshort{ppisn}e, according to the \acrshort{ppisn}e models of \cite{Woosley:2016hmi}.


An important aspect of the progenitors of a \acrshort{ccsn}e is the presence of an extended H-rich envelope. The basic classification of \acrshort{ccsn}e  in Type IIP, Ib and Ic, in fact, is based on the presence of hydrogen in their ejecta (see section \ref{SNObs}). During the presupernova evolution, stars with initial mass larger than $M\sim 20~ M_\odot$ lose a substantial amount of mass. Stars above a critical mass, then, lose their entire H-rich envelope and become Wolf-Rayet (\acrshort{wr}) stars. The limiting masses that form the various kinds of \acrshort{wr} stars (the classification of a \acrshort{wr} star is based on the chemical composition of their atmospheres) depend on the interplay between the evolutionary times and the efficiency of mass loss. A detailed survey of massive stars in a wide range of initial masses, metallicities and rotation velocities \cite{Limongi:2018qgr} has shown the following results. At solar metallicity, non-rotating stars populate the Red Super Giant (\acrshort{rsg}) branch up to a luminosity 
${\rm Log}(L/L_\odot) \sim 5.7$, which corresponds to a mass of $\rm \sim 40~M_\odot$. More massive stars lose a major fraction of their H-rich envelope because they overcome the Eddington luminosity before reaching the \acrshort{rsg} branch and hence turn back blueward where they spend all their core He-burning lifetime. Stars in the mass range $\rm \sim 20-40~M_\odot$ leave the \acrshort{rsg} branch during the core He burning and evolve towards a Blue Super Giant (\acrshort{bsg}) configuration, eventually becoming \acrshort{wr} stars. Stars with initial mass  $M\lesssim 15~M_\odot$, on the contrary, remain \acrshort{rsg} until their final explosion. Accordingly, the expected maximum mass for the progenitors of SNIIP is $M_{\rm IIP} \sim 17~M_\odot$ (which corresponds to a maximum luminosity of ${\rm Log}(L/L_\odot) \sim 5.1$). Moreover, stars with initial mass $M\gtrsim 25~M_\odot$ loose their entire H-rich during the \acrshort{rsg} phase and therefore are expected to explode as SNIb. Stars in the range $\sim (20-25)~M_\odot$ retain a modest amount of H in the envelope at the presupernova stage and therefore should explode as SNIIb. Since the transition SNII/SNIIb/SNIb is not yet clear either from the theoretical and observational side, we can consider only one limiting mass $M_{\rm IIP} \sim 17~M_\odot$ such that stars with $M<M_{\rm IIP}$ explode as SNII while stars with $M>M_{\rm IIP}$ explode as SNIb. These limiting masses are in fair agreement with some observed properties of massive stars: (a) $M_{\rm IIP} \sim 17~M_\odot$ is compatible with the lack of detected progenitors (of SNIIP) with luminosities higher than ${\rm Log}(L/L_\odot) \sim 5.1$, as reported in  \cite{Smartt:2015sfa} - therefore this result constitutes a natural possible explanation of the so called ``\acrshort{rsg} problem''
\cite{Smartt:2009zr}; (b) in a scenario in which all the stars with $M>25~M_\odot$ fail to explode and collapse directly to a \acrshort{bh}, no SNIb is expected at luminosities larger than ${\rm Log}(L/L_\odot) \sim 5.6$ - this result is at least not in contradiction with the 14 SNIb progenitors with no detection reported in \cite{Smartt:2015sfa}, moreover, the expected amount of mass ejected by stars exploding as SNIb is of the order of $M_{\rm ejecta}\sim 6~M_\odot$, i.e., in rather good agreement with the ejected masses of $ (3-4)~M_\odot$ estimated from the SNIb light curve fitting (see, e.g., \cite{Dessart:2016fun}).
The effect of rotation is in general that of reducing both the maximum mass of stars that evolve, with a significant fraction of their lifetime, as \acrshort{rsg} and the minimum mass that becomes \acrshort{wr}. In particular, solar metallicity models, with initial rotation velocities $v\geq 150~\rm km/s$, are expected to populate the \acrshort{rsg} up to a luminosity of the order of ${\rm Log}(L/L_\odot) \sim 5.5$ and then to explode as SNIb ($M_{\rm IIP}\sim 13~M_\odot$). 

At metallicity corresponding to ${\rm [Fe/H]}=-1$, in the absence of rotation, the strong decrease of the mass loss raises the maximum mass that settles on the \acrshort{rsg} branch up to $\sim 25-30~M_\odot$. The more massive ones, on the contrary, do not reach the \acrshort{rsg} branch before the explosion because these stars turn towards their Hayashi track on a nuclear timescale. Therefore, at this metallicity, without rotation, \acrshort{rsg} SNIIP progenitors are predicted with masses as high as $\sim (25-30)~M_\odot$;  above this value stars are expected to explode as \acrshort{bsg}s SNIIP.
Given the very low amount of mass lost at this metallicity, only stars with initial mass $M \geq 80~M_\odot$ may become \acrshort{wr}, therefore in this case $M_{\rm IIP} \sim 80~M_\odot$. At lower metallicities $\rm ([Fe/H] \leq -2)$ also the less massive stars turn to the red on a nuclear timescale so that basically no stars are expected to populate significantly the \acrshort{rsg} branch at these metallicities. This does not imply that no \acrshort{rsg} SNIIP are expected, because stars in the range $\sim (13-25)~M_\odot$ reach the \acrshort{rsg} branch towards the end of the core He burning. Given the extremely low amount of mass lost by these stars, all of them explode as SNIIP (blue or red supergiants). The main effect of rotation at these (low) metallicities is that of pushing the stars towards the \acrshort{rsg} branch. As a consequence, low metallicity rotating stars with $M\lesssim (25-30)~M_\odot$ spend a considerable amount of time on the \acrshort{rsg} branch where they eventually explode as \acrshort{rsg} SNIIP.  The more massive ones, on the contrary, approach their Eddington luminosity when their surface temperature drops to ${\rm Log}\, \[T_{\rm eff}~({\rm K})\]\sim 3.8-4.0$, lose a huge amount of mass and turn to the blue becoming \acrshort{wr} stars. As a consequence, at subsolar metallicities the limiting mass between SNIIP and SNIb for rotating models is in the range $M_{\rm IIP} \sim (25-30)~M_\odot$. Though no progenitor for SNIIP as massive as $(25-30)~M_\odot$ has been detected yet, the present results imply the existence of these massive \acrshort{rsg} progenitors (and SNIIP) at low metallicities and we expect their detection in the next future.

A summary of the limiting masses discussed in the previous paragraph is reported in table~\ref{tab:my_label}.

\begin{table}[ht]
    \centering
    \begin{tabular}{c|c|c|c|c|c|c|c}
    Metallicity  &  Velocity   & $M_{\rm CCSN}$ & $M_{\rm IIP}^{**}$ & $M_{\rm DC}$ & $M_{\rm BH,max}$ & $M_{\rm PPISN}$ & $M_{\rm PISN}$ \\
    ${\rm [Fe/H]}$ &  (km/s)     & $(\rm M_\odot$)    & $\rm (M_\odot$)        & $\rm (M_\odot)$  & $\rm (M_\odot)$      & $\rm (M_\odot)$     & $\rm (M_\odot)$    \\
    $~0$         &      0      &     $9.2^{~}$      &        17              &      80          &          47          &       -             &         -          \\
    $~0$         &  $300$      &     $9.2^*$        &        13              &      30          &          47          &       -             &         -          \\
    -1           &      0      &     $9.2^*$        &        80              &      80          &          47          &       95            &        163         \\
    -1           &  $300$      &     $9.2^*$        &        25              &      17          &          47          &      131            &        235         \\
    -2           &      0      &     $9.2^*$        &    $>120$              &      60          &          89          &       92            &        130         \\
    -2           &  $300$      &     $9.2^*$        &        25              &      15          &          47          &       70            &        117         \\
    -3           &      0      &     $9.2^*$        &    $>120$              &      60          &          91          &       91            &        130         \\
    -3           &  $300$      &     $9.2^*$        &        25              &      15          &          47          &       67            &         93         \\
    \end{tabular}
    \caption{Summary of the limiting masses. Values marked with   $^*$ are obtained for the non rotating solar metallicity case; values marked with $^{**}$ are obtained during the presupernova evolution and hence do not take into account the stars that directly collapse to a \acrshort{bh}. }
    \label{tab:my_label}
\end{table}

\subsubsection{Modelling, explosion mechanisms, and dynamics}\label{corecollapse} 

Besides the aforementioned stellar models our understanding of the
dynamics of stellar core collapse owes a great deal to a growing set
of ever more sophisticated numerical models (for recent reviews and works,
see, e.g., \cite{Burrows:2012ew,2017hsn..book.1095J,2017hsn..book.1671K,OConnor:2018tuw,Muller:2020ard,Mezzacappa:2020oyq,Burrows:2020qrp}).  The phase during which
the \acrshort{gw}s detectable by ground-based interferometers such as \acrshort{et} are
emitted is only a short window of a few seconds in the wider sequence
of events that transforms a star in hydrostatic equilibrium into a
compact object.  It is dominated by a very specific combination of
input physics that differs from the stages before and after and thus
calls for highly specialized methods for simulation and analysis.
Primary ingredients for modelling are multi-dimensional and
multi-scale (magneto-)hydrodynamics, relativistic gravity, the
equation of state (EOS) describing matter at densities up to the
supernuclear regime, reactions between nuclei and between matter and
neutrinos, and the transport of neutrinos from an opaque core to the
transparent stellar envelope.

The main drivers of computational costs come from the neutrino sector  \cite{Mezzacappa:2020oyq}
where the direction dependence of the neutrino distribution and the
energy dependence of the interactions with matter have to be taken
into account, thus increasing the dimensionality of the problem beyond
the usual three spatial and one temporal dimensions.  Ideally, the
Boltzmann transport equation for the neutrino distribution function in
seven-dimensional phase space coupled to the
equations of fluid dynamics would have to be solved.  Additionally, the development of
small-scale turbulence and the many dynamical time scales over which
the cores have to be simulated increase the effort.  Therefore, groups
working on the field can perform state-of-the-art simulations at moderate
rates of typically a few models per year  (see, e.g., \cite{Burrows:2020qrp} for a
particularly extensive set of 3d models).  Thus, many studies addressing
specific questions on, e.g., the nature of the \acrshort{gw} emission, employ
approximations in one or several of these fields such as simplified
neutrino schemes, e.g., expansions of the Boltzmann equations in
angular moments, or a reduced dimensionality.  Of particular interest
for studying the \acrshort{gw} signal might be that the moderate compactness of
the cores and the lack of backreaction of the \acrshort{gw} emission on the
dynamics make meaningful results possible with post-Newtonian
approximations, including calculating waveforms with the quadrupole
formula, instead of a fully general relativistic treatment  \cite{Blanchet:1989fg,Mueller_Janka__1997__aap__Gravitational_radiation_from_SN_convection,Scheidegger:2007nk}.

Beyond the technical difficulties related to computational costs,
modellers have to contend with several physical uncertainties.  A
principal concern is the possible impact of neutrino flavor
oscillations for which the lack of tight experimental and theoretical
constraints still leaves a large parameter space to be explored   \cite{Mirizzi:2015eza,Richers:2022zug,Tamborra:2020cul,Capozzi:2022slf,Fischer:2023ebq}.
Further microphysical uncertainties come from the nuclear EOS, in
particular the potential transition to hyperons or quark matter at very
high densities \cite{Janka:2022krt}.

The progenitor star arrives to the point of core collapse after having
evolved through subsequent burning stages starting on the main
sequence with the fusion of hydrogen to helium \cite{Woosley:2002zz,Smartt:2009zr,Langer__2012__ARAA__Presupernova_Evolution_of_Massive_Single_and_Binary_Stars,Limongi:2017ggo,Alsabti:2017ahu,Limongi:2023bcg} (see also section \ref{stelev}).  By then it has formed
an onion-shell structure with an inert iron core of about a mass of
$1.5 \, \msol$ and a few 1000 km radius surrounded by layers composed
of lower-mass elements.  Once the core loses stability, it collapses
within a few 100 ms to a \acrshort{pns} with a radius of
initially around 50 km.  While this qualitative picture is the
same for all progenitors, with the possible exception of the
supermassive stars formed at Cosmic dawn \cite{Klessen:2023qmc}, the precise structure of the
core is still not fully understood.  Important uncertainties arise
from the necessary approximations made to model the evolution up to
collapse.  Spherically symmetric hydrostatic modelling is still and
will for the foreseeable future remain the workhorse of stellar
evolution theory.  It can by design incorporate effects such as
convection and other mixing processes, rotation, magnetic fields, or
binary interaction only approximately.  Modifications of the stellar
profiles caused by these processes, however, can have crucial impact
on the post-collapse evolution.  Due to progress in numerical methods and computing power, traditional
stellar evolution models are now supported by multi-dimensional MHD simulations
of short episodes of, e.g., convective phases immediately prior to
collapse, which makes it possible to test and calibrate approximate descriptions used
to describe these effects \cite{Muller:2020ard}.

The \acrshort{pns}, composed of highly incompressible matter at supernuclear
density, is close to hydrostatic equilibrium.  Hence, the infall of
surrounding shells of matter is abruptly stopped at its surface.
These layers bounce back in the form of an outward moving shock wave.
An intense burst of neutrinos streams out of the cores during the few
ms that \acrshort{pns} formation takes and carry away the largest fraction of the
energy released in the gravitational collapse.  Thus weakened, the
shock wave fails to leave the core and turns into an accretion shock
located at a radius of between 100 and 200 km, having traversed less
than $\approx 1\, \msol$ of core matter.  The exterior layers, several $\msol$
of material, keep falling supersonically toward the center.  If no
additional processes entered the picture, the accretion would cause
the \acrshort{pns} to grow in mass until it collapses to a \acrshort{bh} in what amounts to
a failure of the prompt explosion mechanism \cite{Janka:2006fh}.

Neutrinos are produced thermally in the hot and dense \acrshort{pns} and by the
gas accreted onto it.  They are in local thermodynamic equilibrium with
matter inside the bulk of the \acrshort{pns} and start to decouple at the
neutrinosphere marking its surface.  They traverse a semi-transparent
region extending up to the shock wave before freely streaming to
infinity.  Their mean energy, set by the temperatures at the
neutrinosphere, exceeds that of the post-shock matter.  Thus,
reactions with matter will, on average, heat the gas at the expense of
the neutrinos.  If the energy transfer in this gain layer is
sufficiently rapid compared to the velocity at which matter falls onto
the \acrshort{pns}, it can unbind the gas from the gravitational pull of the \acrshort{pns}
and thus cause an explosion marked by the shock revival and runaway to
the stellar surface \cite{Burrows:1993pi}.  The balance between the infall and the
infrequent neutrino-matter reactions is too close to determine the
outcome analytically.  Detailed numerical models with accurate
neutrino transport, however, show that this delayed neutrino-driven
mechanism usually does not succeed in reviving the shock.

\begin{figure}[t]
    \centering
    \includegraphics{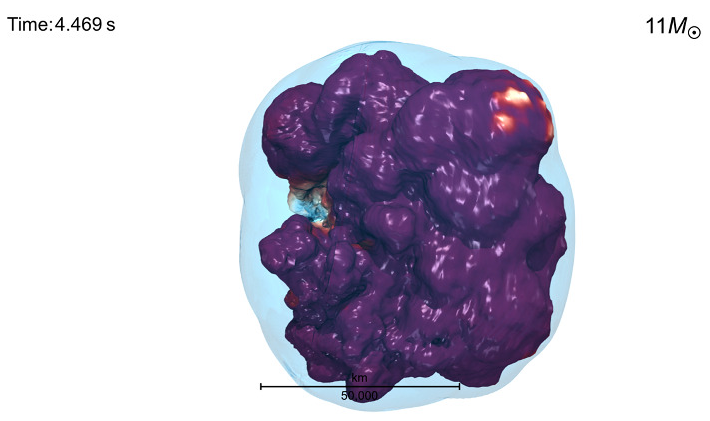}
    \includegraphics{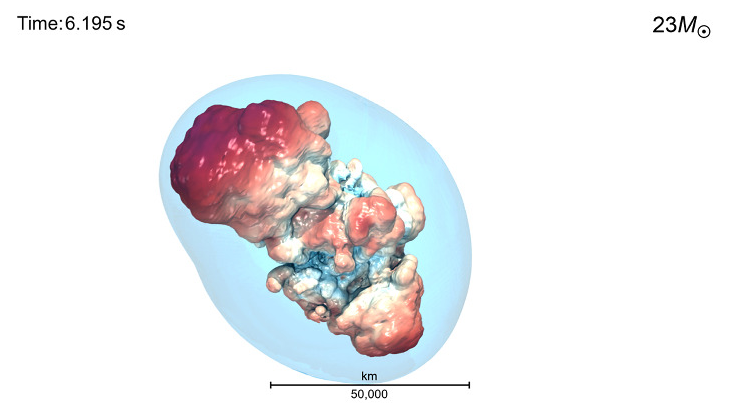}
    \caption{Morphology of the ejecta of two three-dimensional models of neutrino-driven SNe for progenitors of initial masses of $11$ and $23 \, \msol$ \cite{Burrows:2024pur}: isosurfaces of $10\%$ abundance of ${}^{56}\mathrm{Ni}$ colored by the Mach number of the radial velocity and, as blue veil, the shock surface.}
    \label{fig:3dSNe-Burrows2024}
\end{figure}

Multi-dimensional models show an increased heating efficiency because
the fluid, no longer bound to travel on purely radial trajectories,
can dwell for a longer time in a place where it is exposed to the
neutrino flux, e.g., because it is caught up in a vortical flow \cite{Radice:2017kmj}.  In
the gain layer, such eddies are a natural consequence of the
convective instability induced by neutrino heating \cite{Foglizzo:2006fu}.  Further
non-radial motions can be caused by the standing accretion shock
instability (SASI) that leads to large-scale deformations of the shock
and the enclosed matter \cite{Blondin:2002sm,Foglizzo:2005xr}.  If these effects manage to revive the shock
wave, it will start to propagate outward and eject in a very aspherical manner parts of the
stellar material enriched with heavy elements produced before or
during the SN (see figure~\ref{fig:3dSNe-Burrows2024}).  After breaking out of the stellar surface, a
bright SN starts to shine electromagnetically, powered by the
thermal energy of the gas and by the decay of newly produced
radioactive isotopes.

The consequences of the multi-dimensional nature of the event go
beyond the mechanism of shock revival.  The hydrodynamic instabilities
can transfer linear and angular momentum from the envelope to the \acrshort{pns},
thus giving it a non-zero kick velocity and spin \cite{Blondin:2006yw,Nordhaus:2011aa,Wongwathanarat:2012zp,Nakamura:2019snn,Coleman:2022lwr,Janka:2021deg}.  Additionally,
convection is also excited in the \acrshort{pns} where it enhances the neutrino
emission (e.g., \cite{Janka:2017vlw,Tamborra:2018upn,Nagakura:2019tmy}).  The simultaneous presence of downflows and outflows behind
the shock means that the \acrshort{pns} can continue to accrete even while the
explosion is already proceeding.

That this hydrodynamically aided neutrino-driven mechanism can work in
principle has been shown in two-dimensional axisymmetric and in fully
three-dimensional simulations independently with many different
numerical approaches (e.g., \cite{Cabezon:2018lpr}).  Irrespective of currently open or discussed
issues such as the explosions conditions depending on the progenitor
properties, the role of different instabilities and of turbulence, the
origin of the spin and kick velocity of the \acrshort{ns} formed from the \acrshort{pns},
and of the difficulties to match simulations to an observed event, the
current generation of models seem to offer a viable explanation for a
large fraction of observed SNe (e.g., \cite{Bollig:2020phc,Burrows:2020qrp}).  They also indicate that this
mechanism might have difficulties to produce more extreme explosions,
in particular ones with above average kinetic (i.e.,
$E_{\mathrm{kin}} \gg 10^{51} \, \mathrm{erg}$) or \acrshort{em}
energies, relativistic outflow velocities, or particular
nucleosynthetic yields.

The main limitation of the neutrino-driven mechanism is the very loose
coupling of the energy carried by neutrinos to the post-shock gas.
The magneto-rotational explosion mechanism overcomes this restriction
by invoking an alternative energy source for the powering outflows in
the form of the rotational energy of a \acrshort{pns} spinning with periods in
the ms range.  The energy thus available is at
$E_{\mathrm{rot}} \sim \mathcal{O}(10^{52}) \, \mathrm{erg}$
sufficient for events on the hypernova branch.  Sufficiently strong
magnetic fields then act as very efficient conduits for transporting
angular momentum outward and hence launching outflows at the cost of
extracting rotational energy and spinning down the \acrshort{pns} \cite{Thompson:2004wi,Thompson:2004if}.
Multi-dimensional simulations have demonstrated the viability of this
mechanism (recent work, e.g., \cite{Mosta:2014jaa,Kuroda:2020bdq,Obergaulinger:2020cqq,Powell:2022ljv}).  Apart from the high energies that can be reached, it
differs from the standard mechanism by the potential to accelerate
mildly relativistic jet-like outflows preferentially along the
rotational axis.  Apart from observed explosions with such
characteristics, the existence of magnetars, i.e., a population of
young \acrshort{ns}s with extremely strong magnetic fields, suggests
that the magneto-rotational mechanism works for a non-zero, yet so far
poorly determined, fraction of stars.  The uncertainty on the rates
comes from our lack of knowledge about the internal rotation rates of
potential progenitors and from the unsettled question as to how the
cores reach the required strong magnetic fields.  Compression during
the infall and differential rotation will certainly increase the
pre-collapse field strength, and additional amplification mechanisms
such as convection and the magnetorotational instability can boost the
field further \cite{Akiyama:2002xn,Mosta:2015ucs}.  The latter, however, are connected to small-scale
turbulence and, thus, not easy to follow in numerical simulations with
limited resolution.

Other proposed explosion mechanisms rely on tentative modifications of
nuclear physics at a few times the saturation density \cite{Janka:2022krt,Fischer:2020xjl,Jakobus:2022ucs}.  A phase
transition from ordinary matter to a state composed of $u,d,s$ quarks
can liberate enough energy to launch a successful explosion
accompanied by a sudden increase of the \acrshort{pns} density and a burst of
neutrino emission.  Such a phase transition could power an explosion
even if none of the other mechanisms worked, as in that case the \acrshort{pns}
is bound to accrete large amounts of matter and contract to reach the
threshold density.  While the nuclear physics input is uncertain and
somewhat speculative, simulations, so far mostly in spherical
symmetry, have shown the principle viability.  This result can be
turned around to use searches for observations compatible with such an
explosion type to put constraints on the properties of nuclear matter \cite{Fischer:2021jfm}.

In the absence of shock revival and jet formation, and sometimes even
going together with them, the \acrshort{pns} can, by accretion of enough matter,
transform into a \acrshort{bh}.  Even then, explosions can occur in the presence
of rapid rotation.  In the collapsar scenario \cite{MacFadyen:1998vz,MacFadyen:1999mk}, an accretion disk would
form orbiting the \acrshort{bh} inside the core, i.e., composed of matter of
subnuclear, but nonetheless very high density.  Magnetic fields
generated in the disk and/or neutrinos emitted by it can extract a
large fraction of the gravitational energy released in the accretion
process, and power collimated jets which, for sufficiently low ambient
densities, can have highly relativistic velocities \cite{Metzger:2010pp}.  As shown in multi-dimensional simulations,  the outcome of such an
evolution can be  a long
gamma-ray burst.

In most of the successful explosions, by far most of the released energy is
carried away by neutrinos ($E_{\nu} \sim 10^{53} \, \mathrm{erg}$)
over a period of several seconds.  The kinetic energy of the ejecta
is usually quoted with a canonical value of $E_{\mathrm{kin}} \sim
10^{51} \, \mathrm{erg}$, but has a large scatter.  The
\acrshort{em} emission tends to be lower by  roughly two more orders of
magnitude.  There is a basic consistency between observations and
theory/simulations on these figures.  For the energy emitted in \acrshort{gw}s,
the upper limits from the non-detections so far are not very tight.
Theoretical results indicate that the \acrshort{gw} emission is again several
orders of magnitude less and thus dynamically insignificant.  By
probing the wide range of non-spherical motions described above, in
particular \acrshort{pns} oscillations, rotation, and aspherical ejecta
expansion, \acrshort{gw}s nonetheless offer an intriguing possibility to directly
learn about the dynamics in the core.




\subsubsection{GW and neutrino emission}\label{gwnuemission} 
\label{sec:CCSNeMM}

Neutrinos (for recent reviews, see, e.g., \cite{Janka:2017vlw,Mirizzi:2015eza,Muller:2019upo}) and \acrshort{gw}s share many characteristics as the only direct
observables of the dynamics in the stellar core.  However, they probe
different and complementary aspects of the physics.  Neutrinos are
produced as a consequence of the release of gravitational and thermal
energy and carry most of it from the core (see section~\ref{corecollapse}).  
This property makes them
primary probes of the energetics of the collapse and explosion.  \acrshort{gw}s,
on the other hand, are only emitted in proportion to the breaking of
spherical symmetry.  
Hence, not all processes necessarily leave an
observable imprint in the \acrshort{gw} signal.  However, as outlined above, all
cores will eventually develop pronounced anisotropies and thus produce
\acrshort{gw}s.  In line with these different emission properties, the methods
used for detecting and analysing neutrinos and \acrshort{gw}s each have their own
characteristics.  For a sufficiently nearby explosion, neutrino
detectors would collect large numbers of neutrinos, enabling us to
reconstruct the neutrino light curve and the energy spectrum of,
ideally, more than one flavour.  For the \acrshort{gw} signal, a time-frequency
analysis via spectrograms turns out to be a powerful tool which might
reveal details about, e.g., oscillations in the \acrshort{pns}.  Reconstructing
the dynamics from future detections of this kind, however, is an
intricate procedure and requires a detailed understanding of the physics.

The thermal production of large neutrino fluxes does not start with
core collapse.  During the last phases of nuclear burning, massive
stars cool predominantly via neutrinos rather than photons \cite{Limongi:2000km}.  Hence,
detecting the characteristic increase in neutrino emission as the star
evolves though its last burning stages would provide a useful early
warning months before collapse ensues \cite{Super-Kamiokande:2022bwp}.  However, given the
luminosities and the relatively low (sub-$\mathrm{MeV}$) neutrino energies, it
would only be feasible for nearby stars.  At this stage, the
compactness of the star and the, e.g., convective, velocities are too low for any
detectable \acrshort{gw}s to be produced.

\begin{figure}[t]
    \centering
    \includegraphics[width=0.8\linewidth]{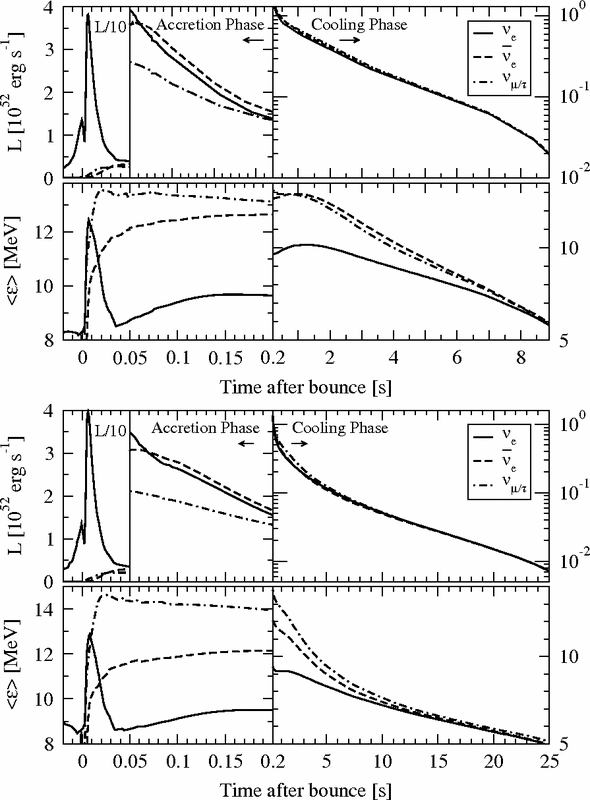}
    \caption{Time evolution of neutrino luminosities $L$ and mean energies $\langle \varepsilon \rangle$ for a typical long-time simulation \cite{Hudepohl:2009tyy}.}
    \label{fig:SNnuLC}
\end{figure}

For the subsequent core collapse and explosion itself (addressed in general in section~\ref{corecollapse}), figure~\ref{fig:SNnuLC} shows the evolution of the luminosities and the mean energies of neutrinos observed at infinity obtained using a spherically symmetric simulation \cite{Hudepohl:2009tyy}. The different phases indicated in the panels will be outlined in the following. 

As the density and temperature of the core increase during collapse,
the neutrino emission rises, in particular via the electron-capture
process $e^- + p^+ \to n + \nu_e$.  The rise is briefly interrupted
when the core becomes optically thick at a central density
$\rho_{\mathrm{c}} \sim 10^{12} \, \mathrm{g} \mathrm{cm}^{-3}$.  From
this moment on, the properties of the emitted neutrinos are set to
large degree by the conditions at the newly developed neutrinospheres \cite{Raffelt:2001kv}.
The collapse proceeds further until the \acrshort{pns} is formed.  The outward
moving shock wave releases a large fraction of the trapped neutrinos
in the intense deleptonisation burst in which the luminosity of
$\nu_e$ exceeds for a few ms values
$L_{\nu_e} \gtrsim 3 \times 10^{53} \, \mathrm{erg}/\mathrm{s}$.  This 
very distinctive electron neutrino burst occurs in all stellar core
collapse events.  Its detection would serve as a very clear indication
of the moment of bounce \cite{Yokozawa:2014tca}.  As the height of the burst varies only slightly
between stars, the count rates in a detector could furthermore be used
to estimate the distance to the event.

The collapse is mostly spherical and by itself does  not generate any
significant \acrshort{gw}s.  It can, however, amplify pre-existing asphericities
caused, e.g., by rapid rotation of the core.  In that case, the
ensuing rapid anisotropic acceleration of the gas at bounce, followed
by the decaying oscillations as the \acrshort{pns} settles into its new
quasi-equilibrium, produces a signal composed of a short (few ms), large
spike, and a series of lower-amplitude oscillations thereafter.  The
amplitude of the spike and the frequencies at which it is emitted can
be used as probes for the rotational energy of the core as shown,
e.g., in ref.~\cite{Richers:2017joj}.

While $L_{\nu_e}$ declines after the burst, the production of the
other flavours increases such that, after a few tens of ms, all
flavours are emitted at roughly similar rates.  Their spectra are
approximately, though not exactly, thermal \cite{Janka1989,Keil:2002in,Tamborra:2012ac}, with mean energies close
to the gas temperature at the neutrinospheres.  The ordering of the
neutrinosphere radii, $R_{\mathrm{ns}, \nu_e} > R_{\mathrm{ns},
  \bar{\nu}_e} > R_{\mathrm{ns}, \nu_{\mu,\tau}}$, reflects the
hierarchy of neutrino-matter interaction rates from strongest to weakest
coupling and translates into an inverse order of the mean energies
$\bar{e}_{\nu_e} < \bar{e}_{\bar{\nu}_e} < \bar{e}_{\nu_{\mu,\tau}}$.

Two main sources feed the neutrino emission: the thermal energy of
the hot \acrshort{pns} and the accretion of matter falling toward it \cite{Mirizzi:2015eza,Muller:2014rpb}.  The former
leads to a luminosity that is similar for all flavours and decays on the Kelvin-Helmholtz timescale of
the \acrshort{pns}, typically several seconds  \cite{Burrows:1984zz}.  The total energy emitted until
the transformation of the \acrshort{pns} into a cold \acrshort{ns} should approximately
correspond to the total gravitational binding energy.  That, in turn,
is, for a given nuclear EOS, a function of the \acrshort{ns} mass.  Hence,
observing the neutrino emission for a sufficiently long time could
provide an estimate of the mass of the newly born \acrshort{ns}.  The cooling \acrshort{pns}
emits all flavours at similar rates.  

Matter keeps falling onto the \acrshort{pns} at least until an explosion sets in,
and, in the case of very asymmetric explosions, potentially for much
longer.  The gravitational binding energy released in the process
powers the second main component of the neutrino luminosity, which
thus is a possible probe for the instantaneous accretion rate.  It
occurs mostly near the \acrshort{pns} surface where the matter settles into
equilibrium.  In the corresponding lower density range,
$\nu_{\mu, tau}$ couple too weakly to matter to be emitted at
significant rates.  Consequently, accretion contributes mainly to the
emission of $\nu_{e}$ and $\bar{\nu}_{e}$ \cite{Mirizzi:2015eza}.  A pronounced drop in the
luminosities of these flavours would thus be signature of the end of
mass accretion, signalling in many cases a successful explosion.

Besides these long-term trends, the neutrino luminosities can show
fluctuations on a range of time scales down to a few milliseconds.
Additionally, the nonspherical geometry of the core leads to
significant variations with emission directions.  Given sufficiently
high count rates, the former can be discovered in a detector equipped
with a sufficiently fast readout system and with the help of
time-frequency analysis techniques.  The latter is harder to observe
as any detector is limited to a single line of sight.

The \acrshort{sasi}, which is only excited before the explosion sets in while the
shock stalls at a low radius, leaves an imprint in the form of a
modulation of the emission at frequencies around
$\mathcal{O}(100) \, \mathrm{Hz}$ \cite{Lund:2012vm,Tamborra:2013laa,Tamborra:2014hga,Walk:2018gaw,Kuroda:2017trn}.  An eventual detection could offer
a way to infer the radius of the shock wave via its influence on the
period of \acrshort{sasi} spiral waves or sloshing motions.  Additional
variations at lower frequencies can come from the dynamics of the
convection in the gain layer.

The neutrino emission is never perfectly spherical, but usually
develops a large-scale anisotropy, in particular after the explosion
is launched.  Aside from this effect, 
three-dimensional simulations have revealed a large-scale asymmetry in
the emission of $\nu_e$ and $\bar{\nu}_e$, the lepton
number emission self-sustained asymmetry (LESA, \cite{Tamborra:2014aua,Janka:2016fox,Walk:2019ier,Lin:2019wwm}).  It consists of a
dipolar pattern in which the emission in one direction is dominated by
$\nu_e$ and in the opposite direction, by $\bar{\nu}_e$.  This
geometry corresponds to a similar dipolar anisotropy of the electron
fraction in the \acrshort{pns}.  The dipole direction is random, but changes only
slowly in time.  

Further variations of the neutrino luminosities and mean energies can
be observed if the rotation of the core periodically moves hotter and
cold spots on the neutrinospheres across the line of sight.
Furthermore, rapid rotation can cause a flattening of the core.  The
\acrshort{pns} and with it the neutrinospheres are situated at lower radii along
the rotational axis than in the equatorial plane.  As a consequence,
the polar neutrino fluxes can be enhanced with respect to those at equatorial
directions by several tens of percent.

Potential post-bounce episodes in which gravitational binding energy
is released on short times are reflected in bursts of the neutrino
emission \cite{Dasgupta:2009yj,Fischer:2010wp}.  They can occur if the contraction of the core leads to a
regime in which phase transitions of the nuclear matter change the
adiabatic index of the EOS and, thus, cause the \acrshort{pns} to rapidly settle
into a new quasi-equilibrium.  These bursts tend to be lower and
slower than the deleptonisation burst.  Futhermore, they would not be
visible in all flavours, not only the $\nu_e$.  If the \acrshort{pns} accretes
enough mass for collapse to a \acrshort{bh} to be inevitable, a similar rise of
the neutrino emission occurs \cite{Mirizzi:2015eza}.  In this case, however, it is followed
by a very rapid dimming marking the moment of \acrshort{bh} formation.

We note that most results on the neutrino emission are based on models
that take into account neutrino emissions in a simplified manner at
best.  Thus, they may miss an important ingredient in the dynamics and
the emission of the core itself as well as the propagation of the
neutrinos to the detector.  Collective oscillations could, e.g., swap
neutrino flavors among each other, thus complicating the
interpretation of a detection.  The physical input such as neutrino
masses and mixing angles required for describing the oscillations are
still uncertain, and numerical methods adequate for long-term,
large-scale simulations are under development \cite{Johns:2019izj,Grohs:2023pgq,Nagakura:2021hyb,Richers:2022bkd,Ehring:2023lcd,Ehring:2023abs}.




Although at first sight the \acrshort{gw} signal from CCSN may seem completely stochastic, in reality is rich in different features that carry information about the emitting object, mainly the nascent \acrshort{pns}. The main burst has a typical duration of $\sim 1$~s, starting at the core bounce.  A number of numerical simulations have been performed focused in the analysis of the \acrshort{gw} signal for the case of non-rotating progenitors both in 2D \cite{Murphy:2009dx,Mueller:2012sv,Yakunin:2010fn,Kuroda:2015bta,Kuroda:2017trn,Pan:2017tpk,Morozova:2018glm,Andersen:2021vzo} and 3D \cite{Andresen:2016pdt,OConnor:2018tuw,Powell:2018isq,Radice:2018usf,Mezzacappa:2020lsn,Mezzacappa:2022xmf,Vartanyan:2023sxm}. For the case of rotating progenitors there is also extensive work by means of 2D  \cite{Cerda-Duran:2013swa,Andersen:2021vzo,Jardine:2021fsf} and 3D simulations  \cite{Scheidegger:2010en,Andresen:2018aom,Powell:2020cpg,Pan:2020idl,Bugli:2022mlq}.
The different features present in the signal are described next:

\begin{figure}[t]
    \centering
    \includegraphics[width=0.8\linewidth]{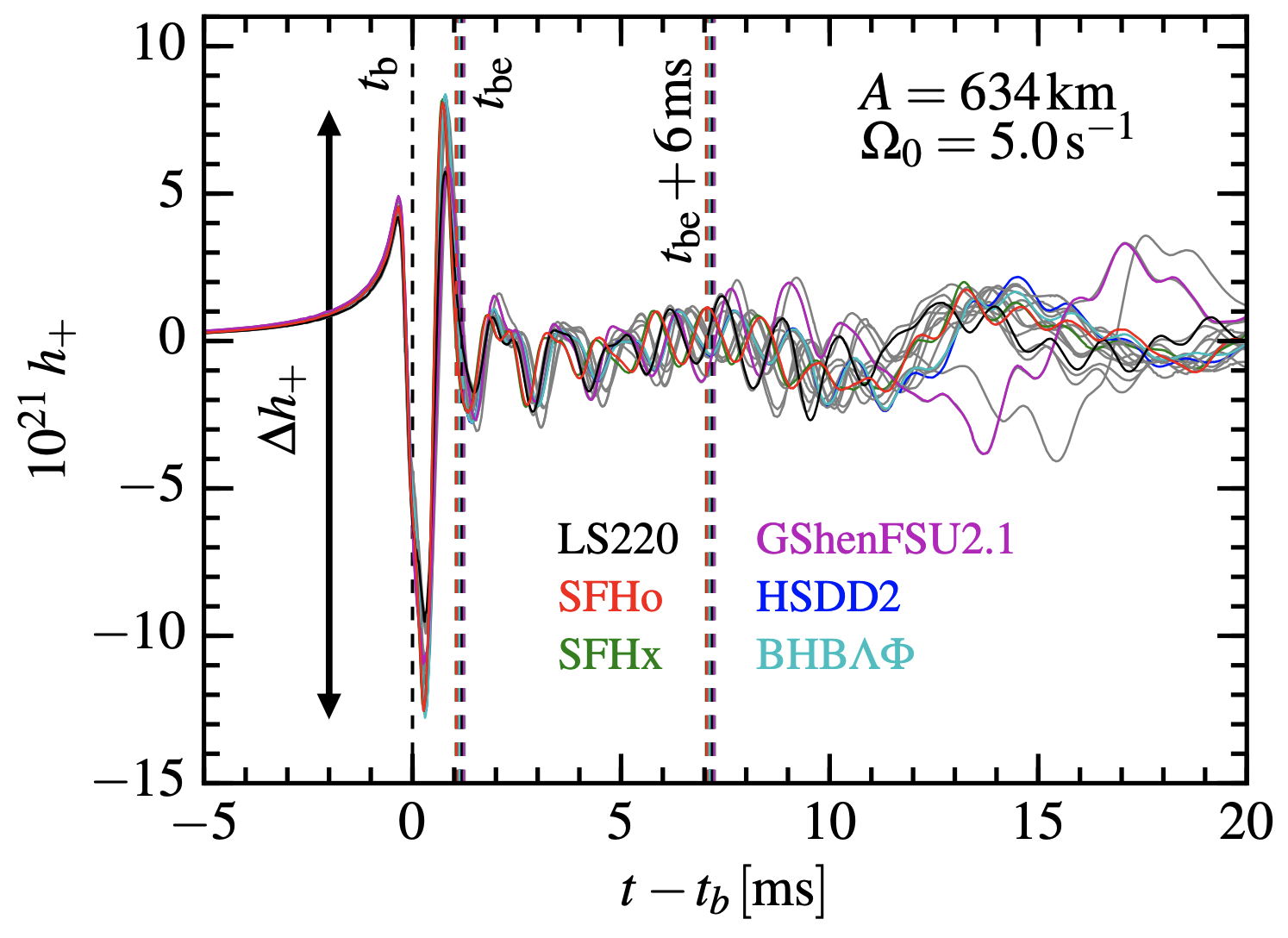}
    \caption{ {\it Bounce \acrshort{gw} signal}: strain during the first few ms of the collapse of a fast rotating progenitor computed for different EOS. The signal is cheracterized by the peak to peak amplitude, $\Delta h_+$, and the frequency $f_{\rm peak}$ of the post bounce oscillations up to $\sim 6$~ms \cite{Richers:2017joj}.}
    \label{fig:BounceGW}
\end{figure}

\begin{figure}[t]
    \centering
    \includegraphics[width=0.8\linewidth]{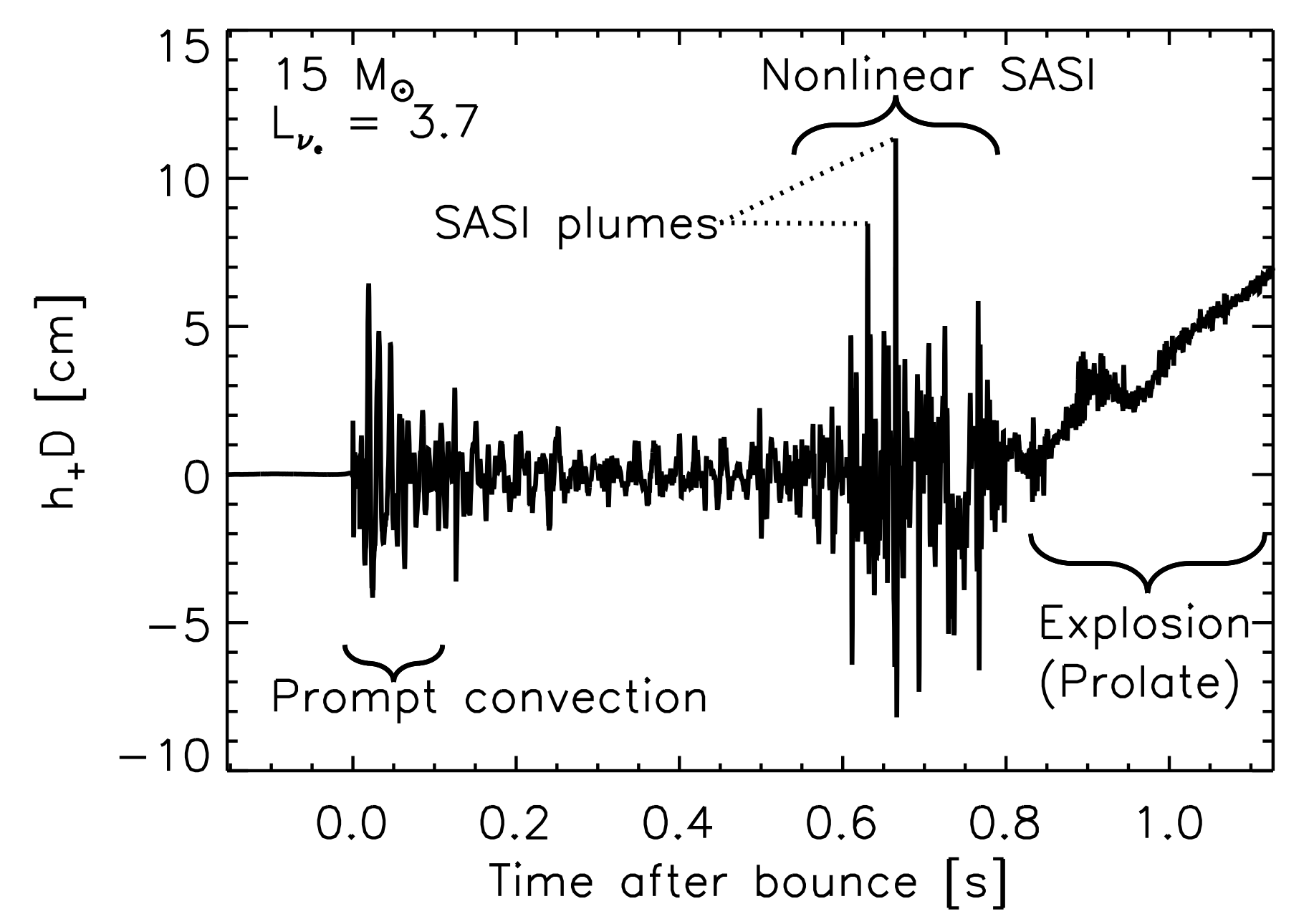}
    \caption{Typical strain during neutrino-driven SNe showing the different features present~\cite{Murphy:2009dx}.}
    \label{fig:NeutrinoDrivenSignalMurphy}
\end{figure}

{\it Bounce signal}: If rotation is present, the collapsing core is not perfectly spherically symmetric but deformed by rotation. The rotationally-induced quadrupolar deformation has its maximum time variation at the time of bounce, which produces a characteristic signal in coincidence with the bounce followed by a few oscillations in a timescale of a few ms (see figure~\ref{fig:BounceGW}). The dependence of the waveform on the rotation rate, the rotational profile,  the equation of state and the  progenitor mass has been extensively studied by \cite{Dimmelmeier:2008iq,Abdikamalov:2013sta,Richers:2017joj}. Numerical simulations show that the signal only has plus polarization and is characterized by its amplitude $\Delta h$ (at a given source distance) and its peak frequency, related to the post-bounce oscillations.  \cite{Richers:2017joj} found that $\Delta h$ is proportional to the ratio of rotational-kinetic energy to potential energy $T/|W|$. This relation breaks for sufficiently large rotation rates, $T/|W|>0.06$, due to the effect of centrifugal forces in the dynamics, reaching a maximum value of about $3\times10^{-20}$ at $10$~kpc. In the same regimen, they found that the peak frequency, which is in the range $100-1000$~Hz, scales with the square root of the mean density in the \acrshort{pns}. This dependencies open the possibility of measuring the rotation rate and putting constrains on the EOS. Unfortunately, this feature is only present for fast rotating progenitors that are expected to be only a small fraction of all collapse events. 

\begin{figure}[t]
    \centering
    \includegraphics[width=0.8\linewidth]{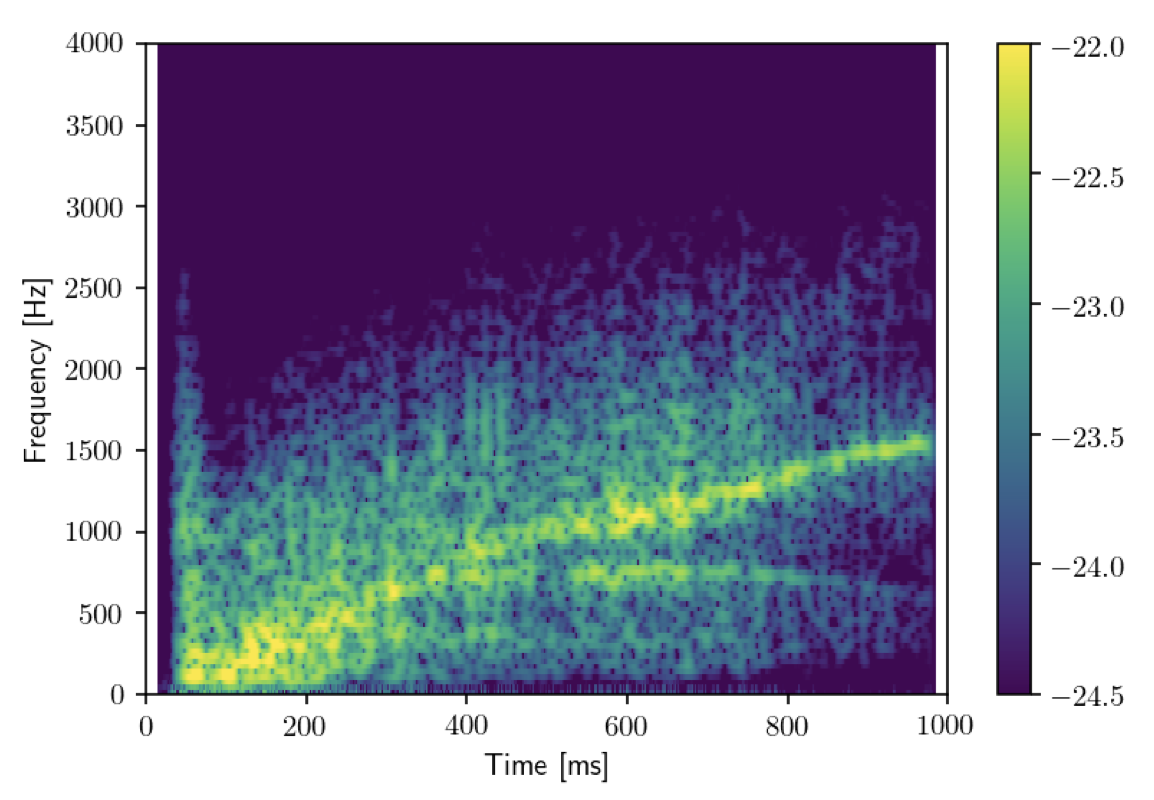}
    \caption{Spectrogram of a typical core-collapse simulation showing the tracks of different oscillation modes of the \acrshort{pns} \cite{Torres-Forne:2018nzj}.}
    \label{fig:NeutrinoDrivenSignal}
\end{figure}

{\it Prompt convection}: About $10$~ms after bounce symmetry of the \acrshort{pns} breaks due to the development of convection. In some cases the conditions in the \acrshort{pns} after bounce can trigger a strong episode of convection that produces a \acrshort{gw} signature lasting for at most $\sim 50$~ms \cite{Murphy:2009dx,Marek:2008qi,Yakunin:2010fn} (see figure~\ref{fig:NeutrinoDrivenSignalMurphy}). Typical frequencies are $100-200$~Hz and the strain amplitude is about $10^{-22}$ at $10$~kpc. In case it is present, this phase can be followed by a quiescent phase, with lower \acrshort{gw} amplitude, lasting for, at most, $200$~ms.

{\it \acrshort{pns} oscillations}: The dominant \acrshort{gw} emission mechanism in neutrino-driven supernovae are \acrshort{pns} oscillations sustained by the instabilities developing at the \acrshort{pns} region \cite{Murphy:2009dx,Mueller:2012sv,Cerda-Duran:2013swa}, which are observed generically in all numerical simulations. The dominant component of the signal (the so-called ramp-up mode) is particularly easy to identify in the spectrogram of the \acrshort{gw} signal 
where it appears as an arch with increasing frequency starting at about $100-500$~Hz after bounce and raising to $500-2500$~Hz after $0.6$~s (see figure~\ref{fig:NeutrinoDrivenSignal}). Most of the energy of the signal is emitted in the first $0.2-1$~s after bounce, with a strain amplitude in the range $\sim10^{-23}-10^{-22}$ for non rotating progenitors. For rotating progenitors the strain increases with rotation up to values as high as $\sim10^{-21}$ for a source located at $10$~kpc. By comparing to the oscillation modes of the \acrshort{pns}, computed by solving the eigenvalue problem associated to a spherical background, it has been shown that this dominant component of the signal is the result of the excitation of a particular oscillation mode of the \acrshort{pns} \cite{Torres-Forne:2017xhv,Morozova:2018glm,Torres-Forne:2018nzj,Sotani:2019ppr,Sotani:2020eva,Westernacher-Schneider:2019utn}. The nature of this mode is unclear and has been proposed to be a g-mode \cite{Mueller:2012sv,Cerda-Duran:2013swa,Torres-Forne:2017xhv,Torres-Forne:2018nzj,Torres-Forne:2019zwz}, a f-mode \cite{Morozova:2018glm,Sotani:2019ppr,Sotani:2020eva,Rodriguez:2023nay} or some combination of both \cite{Vartanyan:2023sxm}. The frequency of the dominant mode depends on the properties of the \acrshort{pns} at each time. Have been proposed universal relations, independent of the EOS or the progenitor model used,  relating this frequency with the surface gravity ($M/R^2$), the mean density ($\sqrt{M/R^3}$) and the compactness ($M/R$) of the \acrshort{pns} \cite{Torres-Forne:2019zwz, Sotani:2020dnh,Sotani:2021ygu, Mori:2023vhr}. These relations allow  developing \acrshort{gw} asteroseimology by measuring the \acrshort{pns} properties from the frequency evolution of the dominant mode in the \acrshort{gw} spectrogram. The development of this kind of relations for the case with rotation is something important to consider for the future.

Additionally, some simulations show a power gap in the spectrograms at about $1000-1300$~Hz \cite{Morozova:2018glm,OConnor:2018tuw,Andersen:2021vzo,Bruel:2023iye,Vartanyan:2023sxm,Zha:2024muc}, breaking the dominant component in two. This gap has been attributed either to an avoided crossing between g and p-modes \cite{Morozova:2018glm,Vartanyan:2023sxm} or to to a minimum of the eigenfunctions of the \acrshort{pns} oscillation modes \cite{Zha:2024muc}. \cite{Wolfe:2023rfx} have proposed the possibility of using the exact location of this gap to constraint the EOS of dense matter. 

Apart from the dominant component there are some weaker modes excited that could observable and linked to the properties of the \acrshort{pns}. 

\begin{itemize}

\item {\it \acrshort{sasi}} induces a periodic signal (see figure~\ref{fig:NeutrinoDrivenSignal}) that can be also seen inf \acrshort{gw} spectrograms
\cite{Murphy:2009dx,Marek:2008qi,Mueller:2012sv,Cerda-Duran:2013swa}. Although there is some basic understanding of the \acrshort{sasi} through mode analysis (e.g. \cite{Foglizzo:2006fu}), detailed calculations based of simulations are still missing. If we use p-modes as a proxy for \acrshort{sasi} modes (they behave similarly and share properties) one can conclude that the frequency of these modes could be used to determine the shock radius \cite{Torres-Forne:2019zwz}.

\item A {\it g-mode with decreasing frequency} 
\cite{Morozova:2018glm,Torres-Forne:2019zwz,Andersen:2021vzo, Jakobus:2023fru,Zha:2024muc,Jakobus:2024cwx}
has been identified starting at $\sim 0.3$~s post-bounce at a frequency of $\sim 600-800$~Hz that decreases with time to values as low as $200$~Hz. This mode has been proposed to be a g-mode of the \acrshort{pns} core and its frequency seems to be highly sensitive to the EOS \cite{Jakobus:2023fru,Jakobus:2024cwx}.

\item For fast rotating models, the {\it quasi-radial mode} contribute to the \acrshort{gw} emission \cite{Cerda-Duran:2013swa}. The frequency of this mode goes to zero if the \acrshort{ns} becomes unstable to collapse, therefore it can be used to infer the formation of a \acrshort{bh} \cite{Cerda-Duran:2013swa,Torres-Forne:2018nzj}.

\end{itemize}

{\it Long-term \acrshort{pns} convection}: Even in the case of a successful explosion, the cooling \acrshort{pns} is still convectively unstable for timescales of $10-50$~s. This convection can produce a weak ($<10^{-24}$) broad-band signal in the range $100-1000$~Hz \cite{Raynaud:2021cgu}. If fast rotation is present inertial modes can develop, with frequencies scaling with the rotational frequency of the \acrshort{pns}. The detection of these inertial modes is an interesting oportunity for asteroseismology. However, this signal can be suppressed if a strong dynamo is developed, case in which a low frequency signal ($<100$~Hz) may apear instead.

{\it Triaxial instabilities}: A sufficiently fast rotating \acrshort{pns} can develop triaxial instabilities, typically a bar, that breaks axisymmetry and induces a strong \acrshort{gw} emission with a frequency related to the rotation of the bar. The presence of differential rotation lowers the threshold of the instability allowing for the formation of bars a moderate rotation ($T/|W|>0.01$). This is the so called Low-$T/|W|$ instability and has been observed in a number of simulations \cite{Ott:2005gj,Scheidegger:2010en,Kuroda:2013rga,Shibagaki:2019mlq,Pan:2020idl,Takiwaki:2021dve,Bugli:2022mlq}. It leaves a very characteristic periodic oscillation with increasing frequency in the range $400-1000$~Hz and a strain amplitude up to about $10^{-21}$. Once formed it is unclear how long this bars can survive since it could be destroyed by the presence of turbulent viscosity, driven by the presence of magnetorotational instabilities. In any case, this feature is only present for fast-rotating progenitors which are likely a small subset of all collapsing cores.

{\it Memory}: The asymmetric expansion of the exploding shock and anisotropic neutrino emission produces a low-frequency ($1-10$~Hz) component of the \acrshort{gw} signal \cite{Murphy:2009dx,Marek:2008qi,Yakunin:2010fn,Richardson:2021lib} that carries information about global asymmetries of the system (see figure~\ref{fig:NeutrinoDrivenSignal}).

\vspace{2mm}\noindent
Finally, we discuss other parameters that have an impact on the \acrshort{gw} signal:

\begin{itemize}

\item The main impact of the {\it progenitor star} is through the structure of the iron core at the time of collapse, although for most stars the \acrshort{gw} signal is qualitatively similar. Even for some extreme progenitors such that ultra-stripped helium stars, with masses as low as $3.5 M_\odot$, produce \acrshort{gw} signals comparable to other SN cases \cite{Powell:2018isq}. Special mention require the case of low mass progenitors ($9-10 M_\odot$) which explode easily and produce very short \acrshort{gw} bursts of $0.1-0.35$~s (e.g. \cite{Radice:2018usf,Mezzacappa:2022xmf}). 

\item {\it Black hole formation:} The signal is qualitatively similar for the case of successfully and  failed explosions because the typical timescale in which the explosion develops is comparable or longer than the \acrshort{gw} emission (see e.g. \cite{Cerda-Duran:2013swa}). The ring-down of the forming \acrshort{bh} is expected to leave a very short ($\sim0.1$ ms) \acrshort{gw} burst  with a frequency of $\sim 5$~kHz \cite{Kokkotas:1999bd}.

\item The consideration of EOS with {\it QCD phase transitions} \cite{Zha:2020gjw,Kuroda:2021eiv,Zha:2022pnc} may leave an imprint in the \acrshort{gw} signal, namely a secondary \acrshort{gw} short burst (a few ms) at late times with strain amplitude $\sim 10$ times the main burst and frequencies in excess of $1$~kHz

\end{itemize}

\subsubsection{Observations: state of the art}\label{SNObs}
Although most \acrshort{sn}e probably are not strong sources of \acrshort{gw}s, some cases, such as rapidly rotating hypernovae associated with Long Gamma Ray Bursts (\acrshort{lgrb}s), magnetar progenitors or asymmetric \acrshort{ccsn}e, are. Furthermore, the existence in binary systems of compact objects such as  \acrshort{ns} and \acrshort{bh}, whose merger has been detected by \acrshort{gw}s, implies that core-collapse supernovae have occurred before.

The characteristics of a \acrshort{sn}, including its outcome, rest upon several factors, notably the evolutionary stage of the star that acts as the \acrshort{sn} progenitor at the moment of its explosion, as well as its chemical composition and, possibly, its angular momentum \cite{Arnett:1982ioj} (see also section \ref{stelev}). The observational heterogeneity observed in supernovae introduces a significant challenge in making precise theoretical predictions regarding the expected outcomes.

In the initial fundamental classification of \acrshort{ccsn}e, we already distinguish between Type II \acrshort{sn}e (\acrshort{snii}), which exhibit the residual hydrogen from the stellar envelope, and Type Ib \acrshort{sn}e (\acrshort{snib}, He-rich) or Type Ic \acrshort{sn}e (\acrshort{snic}, He-poor), characterized by the absence of hydrogen (e.g. \cite{Gal-Yam:2016yms}). 
\acrshort{snii}e are classified into two photometric subclasses based on their light curves: those displaying prominent plateaus (\acrshort{sniip}) and those with linearly declining curves (\acrshort{sniil}). Nowadays, the gap between the two classes has been filled with a continuous distribution of \acrshort{sn}e with different light-curve slopes (e.g. \cite{Anderson:2014hta}).
Additionally, two other main \acrshort{ccsn} subclasses are recognised according to their spectra. 
The interacting \acrshort{sn}e which are any \acrshort{sn} type (it includes the core-collapse subfamilies IIn, Ibn, Icn) with fast ejecta and sufficient energy to interact with a slower and denser circumstellar material (\acrshort{csm}; \cite{Smith:2016dnb}), and the Type IIb SNe (\acrshort{sniib}), whose spectra metamorphose in the following week from hydrogen-rich to helium-dominated, similar to those of the \acrshort{snib} (\cite{1988AJ.....96.1941F}). Since then, numerous additional SN groups have been identified, some of which we will discuss below. 

Together the \acrshort{sniip} and \acrshort{sniil} represent the majority of \acrshort{ccsn}e (considering a volume-limited rate in the local Universe; \cite{Li:2010kc}). Almost 9\% is formed by interacting \acrshort{sn}e. Instead, the intermediate grouping of \acrshort{sniib} and the H-poor \acrshort{sn}e (\acrshort{snib}, \acrshort{snic}) constitute the remaining $\sim$ 37\%. \cite{Cappellaro:2015qia} find similar rates for \acrshort{ccsn} groups considering a redshift range $0.15 < z < 0.35$ (see also section \ref{ccsnrates}).

The large observational heterogeneity present in the \acrshort{sn}e is closely related to the progenitor star's evolutionary state at the moment of the explosion. Here, a key parameter to predict the final fate of star evolution is the initial mass. Several indirect techniques to probe the progenitor have been utilized over the years, but the most straightforward method involves pinpointing the star that underwent an explosion. The search for such progenitors demands precise spatial resolution and accurate astrometry in both pre- and post-\acrshort{sn} explosion images. Source confusion in densely populated environments poses a significant challenge, especially for more distant objects. Therefore, the majority of these identifications have been achieved by examining archival images captured by the Hubble Space Telescope (\acrshort{hst}; e.g. \cite{Elias-Rosa:2016etj}) and recently by the James Webb Space Telescope (\acrshort{jwst}; e.g. \cite{VanDyk:2023sgs}) before the explosion occurred.

In recent years, it has derived a limited range from 8 to 17 $M_{\odot}$ for the progenitor mass of \acrshort{sn}e H-rich. As theoretically expected (see section~\ref{stelev}) these seem to have exploded in the red supergiant phase. However, other surprising cases of massive stars exploding unexpectedly as yellow hypergiants (e.g. \cite{Maund:2011mu,VanDyk:2011tj}) or luminous blue variables \cite{Gal-Yam:2006kfs,2018ApJ...860...68E} have been observed. 
Furthermore, evidence for enhanced mass loss on timescales of years to days before a star undergoes core collapse \cite{Yaron:2017umb}, as well as stars exploding within shells of circumstellar dust \cite{Kilpatrick:2023pse} has been accumalated. 

A well-known case of H-rich \acrshort{sn} whose progenitor was not found to be a \acrshort{rsg} is the peculiar \acrshort{sniip} SN~1987A in the Large Magellanic Cloud. Sk -69$^{\circ}$ 202, a blue B3I supergiant star, was identified at the \acrshort{sn} position on pre-\acrshort{sn} ground-based images (\cite{1987Natur.328..318G,1987ApJ...323L..35S,1987Natur.327...36W}), with a mass of $\sim$ 20 $M_{\odot}$ (\cite{1991ApJ...383..295A,1987ApJ...318..664W}). SN~1987A is the \acrshort{sn} know well as the first example of a multi-messenger event observed through neutrinos and photons. This \acrshort{sn} showed a light curve that did not resemble any of those observed up to that moment, challenging many of the scientific models of the time. It showed that its explosion was not as symmetric as previously thought (\cite{Arnett:1989tnf,McCray:1993ga}). It was also possible to analyse the dust released by a \acrshort{sn} (e.g., \cite{1989LNP...350..164L,1989Natur.337..533R,1989Natur.339..123F,1991supe.conf...82L,Indebetouw:2013sva} and detect for the first time neutrinos created just when the core of the massive star exploded as a \acrshort{sn} (see, e.g., \cite{Scholberg:2012id}). The latest discovery provided by SN~1987A came recently with observations from the ALMA radio interferometer and \acrshort{jwst}, confirming that in the central parts of the \acrshort{sn} remnant, there is a cloud of dust that is very hot and brighter than its surroundings. This discovery suggests that the explosion remnant was a \acrshort{ns} currently hidden within the dust cloud (\cite{Zanardo:2014uoa,Matsuura:2017dzh,Fransson:2024csf}).

H-poor \acrshort{ccsn}e could instead arise either from more massive single-progenitor stars than hydrogen-rich \acrshort{sn} one ($\geq$ 20 $M_{\odot}$) or from lower mass stars $8M_{\odot} < M < 30 M_{\odot}$ in a binary system \cite{2017hsn..book..693V}. In binary systems, they are expected to be stripped of mass by their companion stars, however, this process alone is insufficient to completely remove the hydrogen envelope of a star and they must evolve into a helium giant phase to shed the remaining mass \cite{Eldridge:2013tn}. 
The difficulty in detecting them is because these \acrshort{sn}e tend to occur near complex, crowded environments (see, e.g. \cite{Anderson:2012px}). Despite our best efforts and a bunch of upper limits \cite{Eldridge:2013tn}, there are only two tentative identifications of progenitor stars for \acrshort{snib} (iPTF13bvn; e.g. \cite{Eldridge:2016wlv} and SN~2019yvr; \cite{Kilpatrick:2021hup}), and a potentially hot and luminous candidate progenitor for a \acrshort{snic} (SN 2017ein; \cite{Kilpatrick:2018lil,VanDyk:2018dhz,Xiang:2018xii}).

It is worth noting that both single-star and binary systems scenarios are used in the current search for \acrshort{ccsn} progenitors since most massive stars ($\sim$ 70$\%$) live in interacting multiple systems \cite{Sana:2012px}. From a different perspective, the abundance of binary systems implies that binary interaction plays a predominant role in the evolution of massive stars, giving rise to phenomena such as stellar mergers and gamma-ray bursts.

\begin{figure}[t]
\centering
\includegraphics[width=0.9\textwidth]{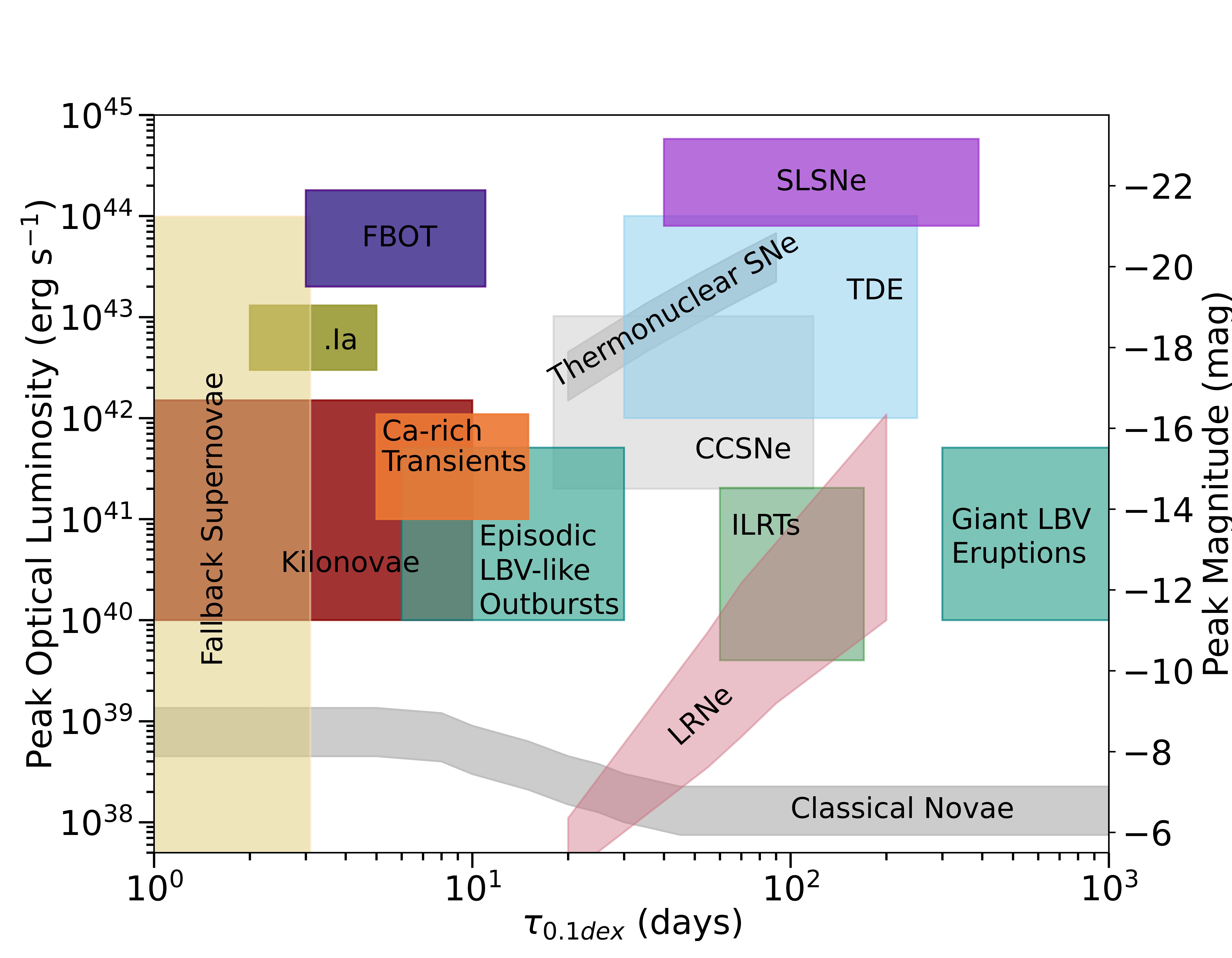}
\smallskip
\caption{Relationship between peak optical luminosity and characteristic time (defined as the duration for a luminosity decline of 0.1 dex) for optical transients within the local Universe. Different regions, denoted by various colours and shapes, indicate the typical location of some representative classes of transients such as SuperLuminous Supernovae (\acrshort{slsn}e), Tidal Disruption Events (\acrshort{tde}s), \acrshort{ccsn}e, Luminous Blue Variables (\acrshort{lbv}s), Intermediate Luminosity Red Transients (\acrshort{ilrt}s), Luminous Red Novae (\acrshort{lrn}e), and others. Figure adapted from \cite{Cai:2022uqa}.}
\label{fig_secCCSNeobserv}
\end{figure}

Since the initial \acrshort{sn} classification, further \acrshort{sn} types have been recognized (see figure~\ref{fig_secCCSNeobserv}), such as the Broad-Line Type Ic supernovae (\acrshort{snicbl}) linked to high-energy gamma-ray bursts (\acrshort{grb}s; \cite{Woosley:2006fn}) and the Super Luminous Supernovae (\acrshort{slsn}; \cite{Gal-Yam:2012ukv}). 

\acrshort{snicbl} are a subclass of \acrshort{snic} characterised by extreme velocity dispersion in their spectra at and before the peak ($\sim$ 19000 km s$^{-1}$; \cite{Modjaz:2015cca}), causing the spectral lines to appear broader than in other types of supernovae. They are often found in association with \acrshort{lgrb}s and are then believed to be linked to the deaths of rapidly rotating massive stars, which can produce powerful jets of material and radiation (e.g. SN~1998bw, located in the error region of the GRB980425; \cite{Galama:1998ea}). However, there are relativistic \acrshort{snicbl} without a detected \acrshort{grb} (e.g. SN~2009bb; \cite{Pignata:2010ap}). 
\acrshort{snicbl} similar to those associated with \acrshort{grb}s are intrinsically rare, being constrained  its rate to $\leq$ 950 Gpc$^{-3}$ yr$^{-1}$ (\cite{Corsi:2022tyl}). 

The \acrshort{slsn}e can be hundreds of times brighter than SNe found over the last decades (M$_V > -21$ mag), having an energetic not explained by the standard core-collapse and neutrino-driven mechanisms (see, e.g., \cite{Nicholl:2021pou}). 
Since the first reported \acrshort{slsn}, SN~2005ap (\cite{Quimby:2007tb}), we have built a catalogue of approximately 100 such events, a number steadily growing thanks to ongoing sky surveys. This sample of events has allowed us to classify \acrshort{slsn}e into two main types based on the presence or absence of hydrogen in their spectra: \acrshort{slsn}-I and \acrshort{slsn}-II. While \acrshort{slsn}e are rare compared to more common \acrshort{sn} types, with \acrshort{slsn}-II being less frequent than hydrogen-poor \acrshort{slsn}e, they are nonetheless significant cosmic phenomena. The volumetric rate of \acrshort{slsn}e, including both hydrogen-rich and hydrogen-poor events, is estimated to be between 0.4 and 1 event per several thousand supernovae (e.g., \cite{Frohmaier:2020pec}).
Standard \acrshort{sn} progenitors typically synthesize about  $0.1\, M_{\odot}$ of $^{56}$Ni, while \acrshort{slsn}e demand 
$\geq (1 - 10) M_{\odot}$ of this isotope (see, e.g. \cite{Umeda:2007wk,Dessart:2012vc}). Therefore, \acrshort{slsn}e seem to favor alternative explosion scenarios to the traditional SN energy sources of radioactive decays of $^{56}$Ni and $^{56}$Co. It is proposed that the additional energy is provided, for example, by a shock between the \acrshort{sn} ejecta and \acrshort{csm} shed by the progenitor star before its explosion (see, e.g., \cite{Woosley:2007qp}) or by the radiation from a newly-formed, highly magnetized \acrshort{ns} (a magnetar) deposited into the ejecta (see, e.g., \cite{Woosley:2007qp,Kasen:2009tg}). However, no models yet provide an excellent and unique interpretation of the observations.

The peak luminosities and radiant energies of \acrshort{snicbl} and \acrshort{slsn}e, which exceed those of typical supernovae by several orders of magnitude, lead to the hypothesis that the progenitor stars are more massive than typical \acrshort{ccsn} progenitors (e.g., \cite{Taddia:2018myd,Blanchard:2020wpn}). The spectroscopic analysis, particularly during the nebular phase, and the study of their host galaxies suggest that \acrshort{slsn}e and \acrshort{snicbl} associated with long GRB  might share similar progenitors and/or explosion mechanisms (\cite{Nicholl:2021pou}). Although there is no direct observational evidence, magnetar-driven models which involve the energy injection from a rapidly spinning magnetar (see below and \cite{Woosley:2009tu,Kasen:2009tg}) successfully explain the observed light curves of both \acrshort{sn} families (e.g., \cite{Wang:2016yel,Dessart:2012vc,Inserra:2013ila,Wang:2019qob,Jerkstrand:2016uyh}).
In addition, jets launched by the central engine in these events are widespread (e.g., \cite{Lazzati:2011ay,Gilkis:2015adr,Barnes:2017hrw,Soker:2017dec,Eisenberg:2022law,Soker:2022vdg}). The possible magnetar formed at the collapse of a massive star slows down quickly due to its magnetic field, sometimes causing it to emit focused jets that can punch through the star's outer layers. 
These jets can also provide enough energy to cause the star to explode as a \acrshort{sn}.
Furthermore, a jet will introduce asymmetries in the ejecta that may explain peculiar features also depending on the viewing angle. Indeed, \acrshort{sn} observations suggest that many, if not most, \acrshort{ccsn} explosions exhibit asymmetric features (e.g., \cite{Wang:2008bd,Taubenberger:2009kx,Chornock:2010ay,Inserra:2016vbq,Saito:2020lfu}). 

As previously described the more massive stars should end their lives as failed, or dark, \acrshort{sn}e (see section \ref{stelev}). 
These events occur when, after the initial phase of the explosion, 
the mantle of the star collapses back onto the core and the consequence is the formation of a \acrshort{bh} without the occurrence of a visible \acrshort{sn} (e.g., \cite{Heger:2002by,OConnor:2010moj,Ertl:2015rga,Sukhbold:2015wba}). Models suggest that about $(10-30)\%$ of core collapses lead to failed \acrshort{sn}e, supported by the lack of higher mass \acrshort{sn} progenitors (\cite{Kochanek:2008mp,Smartt:2008zd}, but see \cite{Davies:2020iga,2020MNRAS.493.4945K}).  Some efforts have been made to identify failed \acrshort{sn}e using telescopes to search for disappearing \acrshort{rsg}s without an associated \acrshort{sn} (e.g., \cite{Kochanek:2008mp,Gerke:2014ooa,Adams:2016hit,Neustadt:2021jjt}). The first candidate discovered was N6946-BH1 \cite{Gerke:2014ooa,Adams:2016ffj}. It was a red source consistent with a mass  $(18-25)\,M_{\odot}$ \acrshort{rsg} star, which disappeared after a long-duration weak transient. Later, ref.~\cite{Basinger:2020iir} did not find any optical counterpart and only low-luminosity near-IR and no mid-IR counterparts compared to the estimated progenitor luminosity. Recently, \cite{Kochanek:2023aob} have found a very red, dusty source at the candidate location in the mid-IR with data from the \acrshort{jwst}. Overall, it has been estimated that failed \acrshort{sn}{sn}e occur at a rate of one every 300--1000 years (see, e.g., \cite{Adams:2013ana}) in the Milky Way galaxy and its satellites. More recently, \cite{{Neustadt:2021jjt}} quantified the fraction of core collapses that fail to produce \acrshort{sn}e at $0.16^{+0.23}_{-0.12}$ (at 90\% confidence level).  

Failed supernovae could theoretically produce \acrshort{gw}s; however, these waves may be weak due to a lower level of asymmetry compared to more energetic supernovae and could have lower frequency emissions since the core collapse often leads directly to the formation of a black hole without significant shock waves or turbulence (e.g., see \cite{Ott:2010gv} and \cite{Cerda-Duran:2013swa}). As a result, detecting these \acrshort{gw}s is challenging for current observatories like LIGO and Virgo. However, third-generation detectors, such as the Einstein Telescope, may detect them.


\subsubsection{Observed rates and expected rates in the local Universe}\label{ccsnrates} 

\paragraph{The observed \acrshort{ccsn} rate in the local Universe.}
The history of \acrshort{sn} discovery and the measurement of the \acrshort{sn} rate is a fascinating evolution from serendipitous observations to systematic surveys.
In the past, nearby \acrshort{sn}e  were all discovered serendipitously by amateurs and astronomers. 
The first significant organized effort to search for \acrshort{sn} events was initiated by F. Zwicky in 1942 with the Schmidt telescope on Palomar.
Zwicky  was able to obtain the first estimate of the \acrshort{sn} rate based on a small sample of 19 \acrshort{sn} events. 
In 1964 Rosino began a systematic search for \acrshort{sn}e  at Asiago Observatory discovering 30 \acrshort{sn}e.
Since Zwicky and Rosino pioneering efforts there have been numerous attempts to estimate the local \acrshort{sn} rate. However, it was not until the 1980s and early 1990s that the measurement of the local SN rate became more accurate. During this period, the sample size of observed \acrshort{sn}e grew substantially, and improvements in observational techniques allowed astronomers to make more precise estimates ~\cite{Cappellaro:1999qy,Leaman:2010kb}.
ref.~\cite{Cappellaro:1999qy} combined data of five optical surveys in the nearby Universe using wide-field telescopes and photographic plates, collecting 67 \acrshort{ccsn}e in a reference sample of about 10000 galaxies. More recently, the Lick Observatory Supernova Search (\acrshort{loss}) at the Katzman Automatic Imaging Telescope equipped with a digital camera measured the rate for each \acrshort{sn} type monitoring a sample of 14882 galaxies up to $z\leq 0.035$  and discovering 106 \acrshort{ccsn}e within 60 Mpc and 440 \acrshort{ccsn}e in the overall galaxy sample \cite{graur2017loss}.

\begin{figure}[t]
    \centering
    \includegraphics[width=0.9\textwidth]{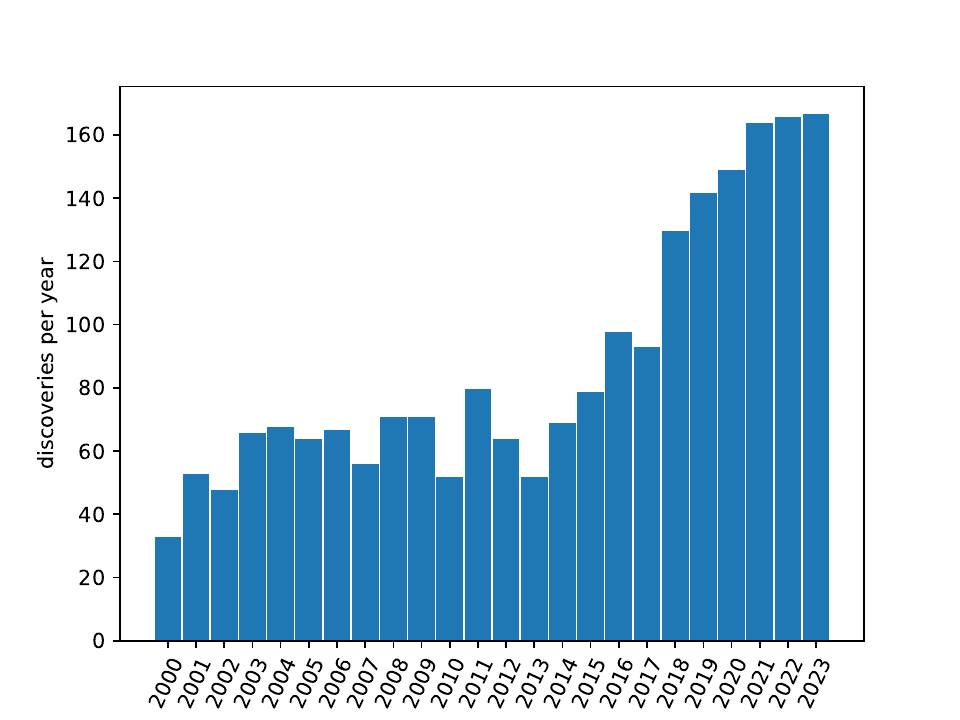}
    \caption{The yearly distribution of \acrshort{ccsn}e with spectroscopic classification from 2000 to 2023 within 100 Mpc.}
    \label{discovery}
\end{figure}

The landscape of optical transient searches is changed considerably in the last decade with the advent of  wide-field  all-sky surveys (e.g. Catalina Real-Time Survey, Palomar Transient Factory, Rapid Response System-Pan-STARRS, All Sky Automated Survey for SuperNovae) monitoring the local Universe with a regular cadence and a deeper limiting magnitude compared to past searches.
The significant increase in the number of \acrshort{sn} discoveries over
the past decade, driven by new \acrshort{sn} surveys (as shown in figure~\ref{discovery}), is revealing a remarkable diversity in \acrshort{sn} types. This progress has enabled the discovery of new classes of stellar explosions as such as \acrshort{slsn}e and Intermediate Luminosity Optical Transients (\acrshort{ilot}s) (see section~\ref{SNObs}).

In particular, from 2017 the  Zwicky Transient Facility (\acrshort{ztf}) \cite{bellm2018zwicky} at  48 inch Samuel Oschin Telescope (Schmidt-type) at the Palomar Observatory scans the entire Northern sky every two days with a median single visit depth (5 sigma) of 20.4 mag in R band and the Asteroid Terrestrial impact Last Alert System (\acrshort{atlas}) \cite{tonry2018atlas} which consists of four telescopes (two in Hawaii, one in Chile, one in South Africa) automatically scans the whole sky several times every night with a median depth of 19.5 mag.               

Assuming that modern \acrshort{sn} searches are very close to guarantee a continuous full-sky  monitoring and an almost complete census of \acrshort{sn} explosions in the local Universe we can easier estimate the \acrshort{ccsn} 
rate (\acrshort{ccsnr})  counting the \acrshort{ccsn}e reported in transient discovery repositories and  archives (the Asiago \acrshort{sn} catalogue\footnote{http://sngroup.oapd.inaf.it/\cite{Barbon2008}}, Transient Name Server\footnote{https://www.wis-tns.org} and Bright \acrshort{sn}e catalogue\footnote{https://www.rochesterastronomy.org/snimages/}). 

We analysed the sample of transients discovered in different volumes of the Local Universe (within a radius of 10, 50 and 100 Mpc) in two time intervals, from 2000 to 2023 and from 2018 to 2023, the beginning of \acrshort{ztf} and \acrshort{atlas} surveys.
The number of \acrshort{ccsn}e discovered from 2000 to 2023 is 27, 680  and  2030  within 10, 50 and 100 Mpc, respectively, yielding rates of  1$^{+1.7}_{-0.6}$, 28$\pm 1$, and 85$\pm 2$ events per year. 
If we consider the temporal range from 2018 to 2023, we found 229, and  916 \acrshort{ccsn}e detected within 50, and 100 Mpc, respectively, corresponding to rates of 38$\pm 3$, and $153\pm 5$ events per year.
The results of our analysis are presented in 
table~\ref{tab:Tab1Div7}. We emphasize that the observed number of \acrshort{ccsn}e provides a stringent lower limit on the intrinsic number of \acrshort{ccsn}e exploded in each volume. 

\begingroup
\renewcommand{\arraystretch}{1.5} 
\begin{table*}[t]
\centering
\resizebox{\textwidth}{!}{
\begin{tabular}{{|l|cc|cc|cc|}}
   \hline
     \multirow{2}{*}{} &
      \multicolumn{2}{c|}{\textbf{10 Mpc}} &
      \multicolumn{2}{c|}{\textbf{50 Mpc}} &
      \multicolumn{2}{c|}{\textbf{100 Mpc}} \\
     & \acrshort{sfr}     &      \acrshort{ccsnr}& \acrshort{sfr}&      \acrshort{ccsnr}& \acrshort{sfr}     &      \acrshort{ccsnr}  \\
       & ($M_{\odot}$ yr$^{-1}$)    &   (yr$^{-1}$) &   ($M_{\odot}$ yr$^{-1}$) &  (yr$^{-1}$) &  ($M_{\odot}$ yr$^{-1}$ )    &  (yr$^{-1}$) \\
      \hline
      \hline
\acrshort{loss} &  & 0.3$\pm{0.04}$   &  & 37.0$\pm 6$ &  & 295$\pm46$ \\
  \hline
Kennicutt et al. & 87$\pm{4}$ & 0.40$\pm{0.02}$   & &  &  & \\ 
Lee et al. &  123$\pm{8}$  & 0.6$\pm{0.04}$    & &  &  & \\
Bothwell et al. & 75$\pm{5}$ & 0.4$\pm{0.02}$  & 9420$\pm{602}$ & 45$\pm{3}$  & 75360$\pm{4814}$ & 362$\pm{23}$ \\
 Hopkins $\&$ Beacom & 65 & 0.3  & 8836 & 42  & 76121 & 365  \\
 Madau $\&$ Dickinson & 63& 0.3  & 8059 & 39  & 66568 & 319   \\
  \hline

 Observations &  & 1.1$^{+1.7}_{-0.6}$ &   & 38.0$\pm{3}$  &   & 153$\pm{5}$ \\
\hline
\end{tabular}
}
\caption{Our estimates of the expected \acrshort{ccsn}e per year within 10 Mpc, 50 Mpc and 100 Mpc, respectively. These estimates are derived from the volumetric rate measured by \acrshort{loss}, several measurements of the total \acrshort{sfr} in the same volumes, and our estimates of the observed number of \acrshort{ccsn}e per year based on discovery reports and archives.}\label{tab:Tab1Div7}
\end{table*}
\endgroup

The incompleteness of \acrshort{ccsn} sample collected from archives arises from several factors: the effective search area in the sky, the
fraction of intrinsically faint \acrshort{ccsn}e that are  missed by local \acrshort{sn}e surveys due to their limiting magnitude, the fraction of \acrshort{ccsn}e obscured by dust that remain undetected by optical searches, and  the fraction of transients lacking classification.  


To estimate the fraction of faint \acrshort{ccsn}e missed by modern surveys, we assumed an average limiting magnitude of 20 mag for the \acrshort{sn} searches which corresponds to absolute magnitudes of $-10$~mag, $-13$~mag, and $-15$~mag, at distances of 10, 50 and  100 Mpc, respectively.
The cumulative luminosity function of \acrshort{ccsn}e measured by \acrshort{loss} survey \cite{Li:2010kc} indicates that approximately 20$\%$ of \acrshort{ccsn}e are fainter than $-15$~mag and about $5\%$ of
\acrshort{ccsn}e are fainter than $-13$~mag. We expect that local \acrshort{sn} searches are nearly complete within 50 Mpc, but miss about 20$\%$ of \acrshort{ccsn}e that explode within 100 Mpc. 

The number of nearby \acrshort{ccsn}e missed due to dust extinction remains largely uncertain. Dust extinction can significantly reduce the number of detectable \acrshort{sn}e, especially in starburst galaxies, where the dust content is higher than in normal galaxies.

\acrshort{sn} searches in near-IR wavelength range 
using adaptive optics have yielded a number of \acrshort{sn}e in nuclear region of the  Ultra
Luminous InfraRed Galaxies (\acrshort{ulirg}s)  with a broad range of
host galaxy extinctions (for more details, see
\cite{Kankare:2021fzf}).  
ref.~\cite{Mannucci:2007ex} monitored 46 \acrshort{lirg} at near infrared wavelengths to search for obscured \acrshort{sn}e, and detected
only about $(10-30)\%$ of the expected \acrshort{sn}e for far infrared luminosity of the galaxies. Ref.~\cite{mattila2012core}, following the approach of \cite{Mannucci:2007ex} and exploiting high resolution adaptive optics, found that the fraction of \acrshort{sn}e missed due to high dust extinction has an  average local value of approximately $20\%$. Results from the Spitzer InfraRed
Intensive Transients Survey (\acrshort{spirits}) which monitored a large number of very nearby ($< 35$ Mpc) galaxies between 2014 and 2018 suggest that the fraction of \acrshort{ccsn}e in nearby galaxies missed
by optical surveys could be as high as $38.5\%$ \cite{Jencson:2019jrg}. More recently, ref.~\cite{Fox:2021zsm} performed a
systematic search at $3.6\, {\rm \mu m}$ for dust-extinguished \acrshort{sn}e in the nuclear regions of forty \acrshort{lirg}s
within 200 Mpc, resulting in the detection of 5 \acrshort{sn}e that had
not been discovered by optical searches.
However, the spatial resolution of Spitzer was not optimal for searching for obscured \acrshort{sn}e
within the nuclear regions of \acrshort{ulirg}s. 
In contrast, high-resolution ground-based radio surveys have successfully discovered nuclear \acrshort{sn}e with rates consistent with
the galaxy IR luminosities. Nevertheless, not all \acrshort{sn}e are are sufficiently bright at
radio wavelengths to allow a complete survey with the radio telescopes currently available to us \cite{pereztorres}.

We also estimated the fraction of transients without spectroscopic classifications and found an average value of  $35\%$ within both 50 and 100 Mpc.
A portion of these transients are likely \acrshort{ccsn}e but accurately 
determining their number remains challenging.
 
 By combining  the fraction of faint \acrshort{ccsn}e,  dust obscured \acrshort{ccsn}e and  \acrshort{ccsn}e without a spectroscopic classification, we infer that about $40\%$ of the \acrshort{ccsn}e  exploded within 100 Mpc remain undetected.

Our estimate of the observed number of \acrshort{ccsn}e in each  volume can be compared with the expected number  based on the \acrshort{ccsnr} measured by the systematic \acrshort{sn} search \acrshort{loss}.
The \acrshort{ccsnr} measurements from \acrshort{loss} have been normalised per unit $B$ and $K$ band luminosity, and their conversion  into a volumetric rate requires knowledge of the local luminosity density for galaxies of different Hubble types. 
ref.~\cite{Li:2010kc}, adopting the following luminosity densities for early type and late type galaxies from \cite{kochanek2001k},
    \begin{equation}
        j_{\rm early\, type}=(2.25\pm0.36) \times 10^{8}\, L_{\odot} {\rm Mpc}^{-3} \, ,
        \label{eq:jearly}
    \end{equation}
    \begin{equation}
        j_{\rm late\, type}=(2.96\pm0.42) \times 10^{8}\, L_{\odot} {\rm Mpc}^{-3} \, ,
        \label{eq:jlate}
    \end{equation} 
obtained a volumetric rate of $0.258 \pm 0.044 \times 10^{-4}$  Mpc$^{-3}$yr$^{-1}$ for \acrshort{snibc}, and of $0.447 \pm 0.068 \times 10^{-4}$ Mpc$^{-3}$yr$^{-1}$ for \acrshort{snii}. 
The total \acrshort{ccsnr}  ($0.7 \pm 0.11 \times 10^{-4}$  Mpc$^{-3}$yr$^{-1}$), when multiplied by the volumes within 10 Mpc, 50 Mpc and 100 Mpc, gives  $0.3\pm 0.04$ , $37 \pm 6$  and $295 \pm 46$ \acrshort{ccsn}e  per year, respectively.  

Our estimate of the observed number of \acrshort{ccsn}e per year  within 10 Mpc is about a factor four higher than the expected value from the \acrshort{loss} volumetric rate.
The discrepancy can be explained by the observed galaxy overdensity within 10 Mpc \cite{karachentsev2004catalog} which results in a higher luminosity density compared to that assumed in \cite{Li:2010kc} for estimating the volumetric \acrshort{ccsnr}, as described by \eqs{eqCCSNRz}{eq:KCCdiv7} below.
The expected \acrshort{ccsn} number from \acrshort{loss} volumetric rate matches the observed \acrshort{ccsn}e within 50 Mpc  while
is about a factor two higher within 100 Mpc.

\paragraph{The expected \acrshort{ccsn} rate from \acrshort{sfr} measurements.}

Due to the short lifetime of their progenitor stars,  the \acrshort{ccsnr} closely follows the current \acrshort{sfr} in a stellar system, making it possible to infer the \acrshort{ccsnr} in a galaxy sample by adopting a progenitor scenario and measuring the total \acrshort{sfr} in the sample.
If we  assume that  the time interval between the formation of the \acrshort{ccsn} progenitors and their explosion (delay time) is negligible, and that the \acrshort{sfr} remains constant during this period, we obtain a linear relation between \acrshort{ccsnr} and the \acrshort{sfr} $\psi(z)$,
 \begin{equation}\label{eqCCSNRz}
{\rm CCSNR}(z) = K_{CC}\psi(z)\, .   
 \end{equation}
Assuming that all stars with suitable masses  ($m_{l}^{CC}-m_{u}^{CC} $) become \acrshort{ccsn}e,  the scale factor  is given by:
\begin{equation}\label{eq:KCCdiv7}
    K_{CC}=\frac{\int_{m_{l}^{CC}}^{m_{u}^{CC}}\phi(m)dm}{\int_{m_{l}}^{m_{u}}m\phi(m)dm}
\end{equation}
where $\phi(m)$ is the initial mass function (\acrshort{imf}), and $m_{l}-m_{u}$ is the mass range for the \acrshort{imf}.

The estimate of $K_{CC}$ requires knowledge of the presupernova evolution of massive stars and of their final fate and depends on: (1) the value of the minimum mass that explodes as a \acrshort{ccsn} ($M_{\rm CCSN}$); (2) the minimum mass that produces a failed (and therefore undetectable) SN $M_{\rm DC}$; (3) the limiting masses that form the various types of Wolf-Rayet stars ($M_{\rm IIP}$) (see section~\ref{stelev}). 
Since we are focusing on the Local Universe we can, as a first approximation, assume the limiting masses for the solar metallicity reported in table~\ref{tab:my_label} at the beginning of this section.
Adopting a Salpeter \acrshort{imf} in the range $(0.1-100)\, M_{\odot}$  \cite{Salpeter:1955it} and a progenitor mass range of  $(9 - 25)\, M_{\odot}$ for \acrshort{ccsn}e, we obtained:
\begin{equation}
K_{CC}= 4.8\times 10^{-3} M_{\odot}^{-1}\, .
\end{equation}
The luminosity of a galaxy is a sensitive tracer of its stellar population, allowing a direct connection between luminosity and instantaneous \acrshort{sfr} when the observed emission originates from short-lived stars or brief phases of stellar evolution.
Measurements of the \acrshort{sfr} in the local Universe are based on H$\alpha$, Far Ultraviolet (\acrshort{fuv}) and Total Infrared (\acrshort{tir}) galaxy luminosities. The total \acrshort{sfr} in a given volume can be estimated using two different approaches: summing the \acrshort{sfr} of each galaxy within this volume or extrapolating  at $z\sim0$ a model of the cosmic Star Formation Density (\acrshort{sfd}) based on measurements in a larger range of redshift.

\vspace{2mm}\noindent
\textit{The total \acrshort{sfr} in a galaxy sample}. 
We estimated the total \acrshort{sfr} within a volume of radius 10 Mpc by summing the measurements based on  H$\alpha$ \cite{Kennicutt:2008ce}, Ultraviolet (\acrshort{uv}) and \acrshort{tir} luminosity \cite{Lee:2010gj} for a sample of 398 nearby galaxies from the 11 Mpc H$\alpha$ UV Galaxy Survey (\acrshort{11hugs}) and Spitzer Local Volume Legacy programs \cite{Kennicutt:2008ce} as in \cite{Botticella:2011nd}.
The total \acrshort{sfr} based on L$_{H\alpha}$ is lower than that obtained from L$_{FUV}$ and \acrshort{tir} luminosities, likely due to the attenuation correction. A detailed analysis of these differences can be found in \cite{Botticella:2011nd} and in \cite{Relano:2012ge}. 

By adopting the \acrshort{sfr} based on L$_{H\alpha}$, we found that the expected \acrshort{ccsn} number within 10 Mpc is 0.4 events per year, whereas when adopting the \acrshort{sfr} based on L$_{FUV}$ and total IR luminosity, it is 0.6 events per year.
For volumes with radii of 50 Mpc and 100 Mpc, we used the \acrshort{sfr} estimate in \cite{Bothwell:2011rk}, based on measurements from a large ($>10 000$) sample of galaxies selected from the Imperial IRAS Faint Source Catalogue Redshift Database (\acrshort{iifscz}) catalogue, as well as the GALEX All-Sky Imaging Survey, augmented by the \acrshort{11hugs} dataset to include systems with low \acrshort{sfr}, low metallicity and low dust content.
This \acrshort{uv} and \acrshort{ir} dataset covers the full range of \acrshort{sfr} behaviour, ensuring that both obscured and unobscured star formation are represented. It provides a \acrshort{sfd} of $0.025 \pm 0.0016\, M_{\odot} {\rm yr}^{-1} {\rm Mpc}^{-3}$ at $z\sim0$.
We used a scaling factor of 0.72 to adjust the \acrshort{sfd} to our cosmological model yielding $(0.018 \pm 0.015)\, M_{\odot}{\rm yr}^{-1} {\rm Mpc}^{-3}$. 

By adopting the \acrshort{sfr} estimate from \cite{Bothwell:2011rk}, we derived the expected number of \acrshort{ccsn}e as $0.4\pm0.02$ events per year within 10 Mpc, $45\pm3$ events within 50 Mpc, and $362\pm23$ events per year within 100 Mpc, as shown in table~\ref{tab:Tab1Div7}. The expected number of events within 10 Mpc is consistent with the value obtained by adopting the \acrshort{sfr} measurements from \cite{Kennicutt:2008ce,Lee:2010gj}. 

\vspace{2mm}\noindent
\textit{The cosmic \acrshort{sfd} models}.
ref.~\cite{Hopkins:2006bw} mapped the evolution of the cosmic \acrshort{sfd} by collecting \acrshort{fir} measurements from the Spitzer and \acrshort{uv} measurements from the SDSS GALEX  and the COMBO17 and Hubble Ultra Deep Field project and H$\alpha$ derived measurement at low redshift. They performed parametric fits to the \acrshort{sfd} using the form of ref.~\cite{cole20012df},  
\begin{equation}
\psi(z) =  \frac{(a+bz)h}{1+(z/c)^{d}}\,\, M_{\odot} {\rm yr}^{-1} {\rm Mpc}^{-3}\, .
\end{equation}
The coefficients of the  parametric form used in \cite{Hopkins:2006bw} depend on the choice of the \acrshort{imf} and are $a=0.017$, $b=0.13$, $c=3.3$ and $d=5.3$ for a Salpeter \acrshort{imf} A with a high-mass power-law slope of 1.35 \cite{baldry2003constraints}. We scaled these parameters to a Salpeter \acrshort{imf} using a factor of 0.77.

Madau $\&$ Dickinson (2014)\cite{Madau:2014bja}  based their modeling of the cosmic \acrshort{sfd} evolution on a limited set of post-2006 galaxy surveys. This set includes only surveys that measured \acrshort{sfr}s from rest-frame \acrshort{fuv} or \acrshort{mir} and \acrshort{fir} data from Spitzer or Herschel. In some cases, this set also incorporates older measurements for local luminosity densities from IRAS or GALEX, as well as GALEX-based measurements at higher redshifts. They adopted a fitting function that is a double power law in $(1 + z)$ defined as:
\begin{equation}
\psi(z)= 0.015 \frac{(1+z)^{2.7}}{1+[(1+z)/2.9]^{5.6}}\,
M_{\odot} {\rm yr}^{-1} {\rm Mpc}^{-3}\, .
\end{equation}
We extrapolated both the parametric form from \cite{Hopkins:2006bw} and from \cite{Madau:2014bja} to $z\sim0$ in order to estimate the expected \acrshort{ccsnr} in the Local Universe.

We found that the expected number of \acrshort{ccsn}e from \acrshort{sfd} measurements is 0.3 events per year within 10 Mpc when adopting  \acrshort{sfd} values from both  \cite{Hopkins:2006bw} and \cite{Madau:2014bja}.  Within of 50 Mpc, we expect 42 and 39 \acrshort{ccsn}e  per year,  corresponding to the \acrshort{sfd} values from  \cite{Hopkins:2006bw} and \cite{Madau:2014bja}, respectively. And within 100 Mpc, the expected number of \acrshort{ccsn}e is 365 and 319 events per year, respectively. These results are in good agreement with those obtained from \acrshort{sfd} measurements in \cite{Bothwell:2011rk}. See table~\ref{tab:Tab1Div7} for a detailed summary.

\paragraph{Comparison between observed and expected \acrshort{ccsn}e.}
The results of the comparison between the expected \acrshort{ccsn} numbers  from \acrshort{sfr} measurements assuming a mass range of  $(9 - 25)\,   M_{\odot}$ for \acrshort{ccsn} progenitors and the observed \acrshort{ccsn} numbers for each volume we analysed are presented in table \ref{tab:Tab1Div7} and  shown in figure~\ref{fig:100}.

\begin{figure}[t]
\minipage{0.5\textwidth}
  \includegraphics[width=\linewidth]{figures/figures_div7/10Mpc_v2.pdf}

\endminipage\hfill
\minipage{0.5\textwidth}
  \includegraphics[width=\linewidth]{figures/figures_div7/50Mpc_v2.pdf}

\endminipage\hfill
\minipage{0.5\textwidth}%
  \includegraphics[width=\linewidth]{figures/figures_div7/100Mpc_v2.pdf}

\endminipage
\caption{Comparison between the observed number of \acrshort{ccsn}e and expected number obtained using different approaches within 10, 50 and 100 Mpc. For the expected \acrshort{ccsnr} from the cosmic \acrshort{sfd}, we adopted the parametric form from ~\cite{Madau:2014bja} as a reference.}\label{fig:100}

\end{figure}

Our estimate of the observed number of \acrshort{ccsn}e within 10 Mpc is 2 to 3 times higher than our estimate of the expected number from the  total \acrshort{sfr} values by  \cite{Kennicutt:2008ce} and \cite{Lee:2010gj}. It is important to note that all the host galaxies of \acrshort{ccsn}e discovered within the last 24 years in this volume are included in the galaxy catalogues selected by \cite{Kennicutt:2008ce} and  \cite{Lee:2010gj}.
ref.~\cite{Botticella:2011nd}, analysing the \acrshort{ccsn}e discovered within 11 Mpc from 1998 to 2010, measured a \acrshort{ccsnr} of $1.1^{+0.4}_{-0.3}$ events per year. By comparing this result with the expected \acrshort{ccsnr} from \acrshort{11hugs} \acrshort{sfr} measurements, they  concluded that the measurements from L$_{\alpha}$ may under-estimate the total \acrshort{sfr}  by nearly a factor of two. Furthermore, our estimate of the observed number of \acrshort{ccsn}e per year  within 10 Mpc is about a factor four higher than  our estimate of the expected number based on the extrapolation $z\sim0$ \acrshort{sfd} models, a discrepancy attributed to the observed galaxy over-density within 10~Mpc \cite{karachentsev2004catalog}.

The expected numbers of \acrshort{ccsn}e from different \acrshort{sfd} measurements \cite{Bothwell:2011rk,Hopkins:2006bw,Madau:2014bja} are in good agreement  with the observed number within 50 Mpc. However, they all exceed the observed rates by a factor of two within 100 Mpc.

These results are consistent with those obtained from the comparison with the expected values derived from \acrshort{ccsnr} by \acrshort{loss} survey.



\paragraph{Observed and expected \acrshort{ccsnr} in the Milky Way.}
The estimate of \acrshort{ccsnr} in the Milky Way (\acrshort{mw}) is crucial for planning the observational strategy of current  and next generation \acrshort{gw} and neutrino detectors.
Detecting a neutrino burst, a \acrshort{gw} signal, shock breakout emission and gamma rays from a Galactic \acrshort{ccsn} will provide direct and significant constraints on \acrshort{ccsn} explosion mechanism.
The most recent recorded \acrshort{sn} explosion in our Galaxy occurred in 1604. However, the study of young \acrshort{sn} remnants suggests that at least two \acrshort{sn} events have occurred in the past 400 years in our Galaxy. The expansion age of the Cassiopeia A remnant (G111.7-2.1), approximately 325 yr, indicates that the \acrshort{sn} explosion took place around AD 1681 $\pm$ 19 \cite{Fesen:2006zma}. Additionally, the most recently discovered \acrshort{sn} remnant G1.9+0.3 has an age of $\sim 110-180$ years, making it the youngest-known remnant in our Galaxy with an explosion date placed in the second half of the 19$^{\rm th}$ century \cite{Luken:2019zbs,green1984two,reynolds2018evolution}.
A significant fraction of \acrshort{sn} events in the \acrshort{mw} remains undetected in the optical range due to heavy dust absorption in
the Galactic disk. Despite advancements in observational capabilities across other wavelengths (e.g., radio telescopes), and the advent of neutrino and \acrshort{gw} detectors, no other \acrshort{sn}e have been discovered since SN1604.
 \begin{figure}[t!]
     \centering
     \includegraphics[width=13 cm]{figures/figures_div7/historicalsn.pdf}
     \caption{The distribution of the historical \acrshort{sn}e and \acrshort{sn} remnants in the galactic coordinate system is shown. \acrshort{ccsn}e  (SN1054 and SN1181) are showed as red stars, \acrshort{snia} (SN1006,SN1572,SN1604) as green stars, and the \acrshort{sn} remnants from the Green catalogue \cite{Green:2019mta} as blue dots.}
     \label{fig:enter-label}
 \end{figure}
Different approaches can be adopted to infer the Galactic \acrshort{ccsnr} based on both direct and indirect evidences:
\begin{itemize}
     \item the number of historical records of  \acrshort{sn}e combined with spatial distribution models for Galactic  stellar populations and dust, 
     \item the \acrshort{ccsnr} measurements for galaxies  with morphological type and luminosity similar to these of the \acrshort{mw}, 
 \item  probes of \acrshort{ccsn} explosions, such as the census of massive stars, the birth rate of \acrshort{ns}s, the age distribution of \acrshort{sn} remnants, the $^{26}$Al distribution in the Galaxy. 
 \end{itemize}
These estimates are summarized in table~\ref{Tab3}. 

\vspace{2mm}\noindent
\textbf{\acrshort{ccsnr} from historical records and Galaxy models}.
Historical records indicate that five \acrshort{sn} events have occurred in the Milky Way over the past millennium: SN1006 (\acrshort{snia}), SN1054 (\acrshort{ccsn}), SN1181 (\acrshort{ccsn}), SN1572 (\acrshort{snia}) and SN1604 (\acrshort{snia}).  Their apparent magnitudes range from  $-4$  to 0 mag (SN1181) \cite{1977hisu.book.....C} and their observability timescales span from 6 months (SN1181) to 3 years (SN1006) \cite{green2017historical}. \acrshort{sn} 1054 and 1181 have been classified as \acrshort{ccsn}e due to the presence of a pulsar in their remnants: SN1054 is thought to be a IIn-P \cite{Smith:2013gya}, while SN1181 classification 
remains uncertain, with several subtypes proposed, including (\acrshort{snib}, \acrshort{snic}, \acrshort{sniil}, and possibly an underluminous \acrshort{sniip}) \cite{Kothes:2013hka}.

\begingroup
\renewcommand{\arraystretch}{1.5} 
\begin{table}[t]
    \centering
    \begin{tabular}{|l|l|l|}
    \hline
    Method & CCSNR (100 yr)$^{-1}$) & Reference\\
    \hline
    \hline 
        
        Galaxy models &2.1 & \cite{Tammann:1994ev} \\ &$3.2^{+7.3}_{-2.6}$ &\cite{Adams:2013ana} \\
        
     &$1.4^{+1.6}_{-0.9}$  &  \cite{Murphey:2020unk}   \\
        \hline
         
         CCSNR measurements &1.7 $\pm$1.1   & \cite{Cappellaro:1993ns} \\
        
         &2.4-2.7$\pm$ 0.9 & \cite{van1994rediscussion} \\
        
        &2.30 $\pm 0.48$ &  \cite{Li:2010kc} \\
        \hline
        Counts of massive stars 
        &1 - 2 & \cite{Reed:2005en}\\
        
        \hline
        NSs & 7.2 &   \cite{Rozwadowska:2020nab}\\
        \hline
        SN remnants & 0.43 &\cite{Leahy:2018pgb}\\
        \hline
        $^{26}$Al distribution & 1.9 $\pm$ 1.1 & \cite{Diehl:2006cf} \\
        \hline
        Combination of different methods &$1.63 \pm 0.46$ & \cite{Rozwadowska:2020nab}   \\
        \hline
        
    \end{tabular}
    \caption{The expected  \acrshort{ccsnr} per century in the Milky Way estimated with different proxies.}
    \label{Tab3}
\end{table}
\endgroup

The first attempt to estimate the Galactic \acrshort{ccsnr}  \cite{Tammann:1994ev} was made by combining rate measurements from other galaxies and from the historical \acrshort{sn}e in the Galaxy. This approach yielded a Galactic \acrshort{sn} rate  of $2.5 (100\, \rm yr)^{- 1}$  and \acrshort{ccsnr} of $2.1 (100\,\rm yr)^{-1}$, assuming that about $85\%$ of all \acrshort{sn}e are core-collapse events. 

Ref.~\cite{Adams:2013ana} presented a new measurement of the Galactic \acrshort{ccsnr} by exploiting advancements in models of Galactic stellar populations  and modern dust maps.  They performed a Monte Carlo simulation to determine the position and dust extinction of Galactic
\acrshort{sn}e modeling both progenitor and dust distributions with a double-exponential function and adopting new estimates of the  Galactic scale lengths, heights and total extinction along any line of sight.
The resulting cumulative apparent magnitude distribution for Galactic \acrshort{ccsn}e, derived  under  the assumption of  a Salpeter \acrshort{imf} and a range mass for \acrshort{ccsn} progenitors of $8-100 M_\odot$, when combined  with the number of historical \acrshort{sn}e suggests a value of  $3.2^{+7.3}_{-2.6} $ (100yr)$^{-1}$ for Galactic \acrshort{ccsnr}.  

More recently, ref.~\cite{Murphey:2020unk} conducted a similar study analysing the \acrshort{sn} visibility by the naked eye in our Galaxy. They adopted the spatial distributions of \acrshort{sn} and dust by \cite{Adams:2013ana} and modeled the sky distributions for naked eye \acrshort{sn}e using axisymmetric models of \acrshort{mw}. Their fiducial model was compared to the location of historically \acrshort{sn}e, which all avoid the the high-probability region (95$\%$), likely due to spiral arm effects. The authors calculated  the fraction of \acrshort{sn}e visible to the naked eye and found that about $33 \%$ of \acrshort{ccsn}e are visible at a magnitude limit of m$_{V,lim}=-2$ mag.
The number of historical \acrshort{sn}e within the observable fraction suggests an underlying  \acrshort{ccsnr} of $1.4^{+1.6}_{-0.9}\,  (100{\rm yr})^{-1}$. \\

\vspace{2mm}\noindent
\textbf{\acrshort{ccsnr} from measurements in the local Universe}.
Historically, the \acrshort{ccsnr} in the Local Universe has been estimated for samples of galaxies with various Hubble types and sizes and has been normalized to B-band or K-band  galaxy luminosities. The Galactic \acrshort{ccsnr} can be inferred by adopting a morphological classification and luminosity for \acrshort{mw}, along with the corresponding value of \acrshort{ccsnr} measured in the local Universe. 

Ref.~\cite{Cappellaro:1993ns}, exploiting  their \acrshort{ccsnr} measurements for a sample of more 10000 galaxies and assuming that the \acrshort{mw} is a type Sb$\pm$ 0.5 galaxy with a luminosity of $(2.0 \pm 0.6) \times 10^{10} L_{B,\odot}$ \cite{vanderkruit1987}, obtained a value of $(1.7 \pm 1)\, (100\, {\rm yr})^{-1}$ for the Galactic \acrshort{ccsnr}.

Ref.~\cite{van1994rediscussion} estimated the \acrshort{ccsnr} for a sample of 800 galaxies from the Shapley-Ames catalogue and \acrshort{sn} discoveries by \cite{Evans:1989}. Assuming that the \acrshort{mw} is a Sbc galaxy, with properties similar to the Sab-Sd galaxies for which they estimated the rates, and a B-band luminosity of $(2.6 \pm 0.6)\, \times 10^{10} L_{B,\odot}$ \cite{van1994rediscussion}, they obtained a Galactic \acrshort{sn} rate of $(3.0\pm 1.0) (100 {\rm yr})^{-1}$ and, by assuming that \acrshort{ccsn}e account for $80\% - 90\%$ of all \acrshort{sn}e,
they derived  a \acrshort{ccsnr}  in the range $[2.4-2.7]\pm0.9$ per $100 {\rm yr}$.

Ref.~\cite{Li:2010kc} estimated the \acrshort{ccsnr} per unit luminosity for a galaxy sample monitored by \acrshort{loss}. They assumed that the \acrshort{mw} has the average size of a Sbc and a B-band luminosity by \cite{1987ASIC..207...27V} and derived a Galactic \acrshort{ccsnr} of $(2.30\pm 0.48) \,(100 {\rm yr})^{-1}$. 

These results, based on similar assumptions about the morphological classification and luminosity of the \acrshort{mw},  are in good agreement within the reported uncertainties. Considering the uncertainties in the Hubble type and luminosity of the \acrshort{mw} we can conclude that these measurements have an uncertainty approximately a factor of two.\\

\vspace{2mm}\noindent
\textbf{ \acrshort{ccsnr} from the local massive star birthrate}.
An estimate of \acrshort{ccsnr} can be inferred by extrapolating a direct census of  massive stars within solar neighborhood. 

Ref.~\cite{Reed:2005en} estimated the local massive-star birthrate from a sample of  420 O3-B2 dwarfs within 1.5 kpc of the Sun to a limiting magnitude of 8 mag in V band. 
To extrapolate the local birthrate to the entire Galaxy, they adopted a two-component “disk + hole” model  for the 
stellar density distribution, along with models for extinction and main-sequence star lifetime. 
The authors  found that the massive-star birthrate is on average 7.86 stars kpc$^{-3}$ 10$^{-6}$ yr$^{-1}$ and $176$ stars kpc$^{-3}$ 10$^{-6}$ yr$^{-1}$ on the Galactic-plane and  concluded that many more galactic OB stars likely remain to be discovered. 

Assuming a mass cutoff  of 10 $M_{\odot}$ for \acrshort{ccsn} progenitors, the  resulting \acrshort{ccsnr} ranges from  1 to 2 (100yr)$^{-1}$ depending on the values of the model parameters. 

Ref.~\cite{Hohle:2010yp} compiled a sample of massive stars (O and early B type) within
3 kpc detected in 2MASS \cite{2003tmc..book.....C}  and observed by Hipparcos \cite{Perryman:1997sa} with parallax and precise photometry.
Luminosities and temperatures of these stars were used to estimate the masses and ages adopting mass tracks and isochrones from different sources, such as 
\cite{Bertelli:1994zz,claret2007,Schaller:1992nr}. The authors identified 759 stars with masses greater than $8 M_{\odot}$ and inferred a \acrshort{ccsnr} of 21.3$\pm$4.7 per Myr within 600 pc from the Sun. 

\vspace{2mm}\noindent
\textbf{\acrshort{ccsnr} from \acrshort{ns}s birthrate}.
The census of \acrshort{ns}s, considering their various manifestations, can also provide an estimate of the \acrshort{ccsnr}, as they are the outcomes of core collapse events for the progenitor mass range assumed in our analysis.

Ref.~\cite{Keane:2008jj} reviewed the \acrshort{ns} birthrate estimates in the literature by analysing different classes of \acrshort{ns}s: magnetars, radio pulsars, Rotating Radio Transients (\acrshort{rrat}s), X-Ray Dim Isolated Neutron Stars (\acrshort{xdins}s) and X-ray binaries. 
For the magnetars, the authors used an overall estimate derived from two different age estimation methods: i) spin-down age estimates \cite{Kouveliotou:1998ze}, ii) \acrshort{sn} remnant age estimates \cite{1995A&A...299L..41V}. 

For radio pulsars, they used estimates from three different studies  \cite{Faucher-Giguere:2005dxp,Lorimer:2006qs,Vranesevic:2003tp}. 
The  birthrate of \acrshort{xdins}s was estimated by \cite{Gill:2007ra} from a sample of seven \acrshort{xdins}s detected by the ROSAT All-Survey. Finally, the authors assumed a direct proportionality between the pulsars and the \acrshort{rrat}s birthrate, as in \cite{Popov:2006ma}. 

If we consider the average of the birthrate values presented by \cite{Keane:2008jj}, based on 
the Galactic electron density model NE2001\cite{Rozwadowska:2020nab}, we obtain a \acrshort{ccsnr} of $7.2$ (100yr)$^{-1}$.  This value is higher than the other estimates and suggests that there may be an excess of \acrshort{ns}s in the
Galaxy, known as the \acrshort{ns} ‘birthrate problem’.

\vspace{2mm}\noindent
\textbf{\acrshort{ccsnr} from the \acrshort{sn} remnant birthrate}.
Approximately 300 \acrshort{sn} remnants have been observed in our Galaxy through their radio emission \cite{Green:2019mta}. However, only a small number have been physically characterized, including
determination of evolutionary state, explosion energy and age as discussed in 
\cite{Leahy:2020jpk}.
The authors developed a set of spherically symmetric \acrshort{sn} remnant models based on hydrodynamic calculations and used them to estimate age, explosion energy, and circumstellar medium density for a sample of 43 Galactic \acrshort{sn} remnants whit known distances and temperatures, assuming spherically symmetric evolution of the \acrshort{sn} remnant. 

The distribution of ages yields a \acrshort{sn} birthrate of 1/260 yr for the sample of 43 \acrshort{sn} remnants and of 1/160 yr including 15 Galactic \acrshort{sn} remnants from \cite{Leahy:2018pgb}. A final correction for the incompleteness of
X-ray observations of \acrshort{sn} remnants increases the birth rate to $\sim$1/35 yr,which corresponds to a  \acrshort{ccsnr} of $\sim 0.43$ (100yr)$^{-1}$. 
This result is significantly lower than other \acrshort{ccsnr} estimates, likely due to the incompleteness of the \acrshort{sn} remnant sample.

\vspace{2mm}\noindent
\textbf{ \acrshort{ccsnr} from $^{26}$Al distribution in the \acrshort{mw}}.
Ref.~ \cite{1977Icar...30..447C} suggested that the injection of elements into the interstellar medium, such as $^{26}$Al, may originate from \acrshort{ccsn}e. 
Ref.~\cite{Diehl:2006cf} analysed the $\gamma$-ray emission from $^{26}$Al decays to derive its distribution using INTEGRAL data. By mapping the isotope, they found an irregular distribution, primarily concentrated along the plane of our Galaxy. This result strongly supports the hypothesis that \acrshort{ccsn}e are $^{26}$Al emitters, as \acrshort{ccsn}e are mainly located in this region. 
The  mean mass of $^{26}$Al ejected in the Galaxy  has been estimated by combining observations with the theoretical models. The $^{26}$Al yield was inferred from the total mass, assuming a Scalo \acrshort{imf} (i.e. $\phi(m) \sim m^{-2.7}$) \cite{scalo1986stellar}, the isotope lifetime (1.04 Myr) and a progenitor mass range $(10-120) \, M_{\odot}$. 
By correlating the observed $^{26}$Al $\gamma$-ray emission with the expected $^{26}$Al yield \cite{Limongi:2006xu} they estimated the \acrshort{ccsnr} to be $(1.9\pm1.1)\, (100 {\rm yr})^{-1}$.

\vspace{1mm}\noindent
\textbf{  \acrshort{ccsnr}  combining different proxies}.
Ref.~\cite{Rozwadowska:2020nab} obtained a more precise estimate of the Galactic \acrshort{ccsnr} by combining various measurements from the literature  over the past twenty years with direct information on the \acrshort{ccsnr} from historical \acrshort{sn}e  in the Local Group. Their first estimate is  based on the following measurements:
 i) counts of massive stars (\acrshort{ccsnr}=$(1.5\pm 0.5)$(100yr)$^{-1}$ \cite{Reed:2005en}), ii) the \acrshort{ns} birthrate (\acrshort{ccsnr}=$7.2$ (100yr)$^{-1}$) \cite{Keane:2008jj}, iii) \acrshort{ccsnr} measurement from a large galaxy sample by \cite{Li:2010kc} (\acrshort{ccsnr}= $(1.95\pm 0.41)$(100yr)$^{-1}$), iv) Galactic enrichment of $^{26}$Al (\acrshort{ccsnr}=$(1.9\pm1.1)$ (100yr)$^{-1}$ \cite{Diehl:2006cf}). By multiplying the likelihood functions of these different estimates,  the authors obtained a \acrshort{ccsnr} of 
$(1.79\pm 0.55)\, (100 {\rm yr})^{-1}$.

The second estimate of the galactic \acrshort{ccsnr} is based on: i) the Galactic \acrshort{ccsnr} from historical records adopting the fraction of visible \acrshort{ccsn}e by \cite{Adams:2013ana}, ii) the \acrshort{ccsnr} from neutrino observations of the \acrshort{mw}, iii) \acrshort{ccsnr} in M31 galaxy, iv) \acrshort{ccsnr} in other galaxies within the Local Group. 

By combining these four likelihood functions, a range of $[0.66-2.04]\, (100 {\rm yr})^{-1}$ was obtained.
Finally, by incorporating all available information, the first combination of measurements and direct data on the \acrshort{ccsnr} in the local Group, the authors performed a global fit, resulting in a \acrshort{ccsnr} of $(1.63 \pm 0.46)\,  (100 {\rm yr})^{-1}$. 



\subsection{Neutron stars}
\label{sec:neutron}
Neutron stars are the end product of the collapse of progenitor stars with mass in the range $\approx [9,~20]M_\odot$. See section~\ref{sec:collapse} for a detailed discussion regarding theory and observations related to star collapses.  


\subsubsection{Observed NS population}

Neutron stars manifest themselves in a bewildering diversity, and over the whole \acrshort{em} spectrum, from radio to TeV energies. The observed \acrshort{ns}s population is dominated by isolated radio or gamma pulsars, which beamed emission is powered by their large rotational energy, with about 2500 objects observed\footnote{v. 2.4.0, \url{https://www.atnf.csiro.au/research/pulsar/psrcat/}}. Then, there are about 1500 systems consisting of \acrshort{ns}s in binary systems, showing different emission phases, from accreting X-ray binary systems to rotational powered radio pulsars orbiting low and high mass stars, or transitional pulsars that oscillate between direct matter accretion and quiescent phases where the radio pulsar switches on. 
However, in the last decades, several extreme and puzzling sub-classes of \acrshort{ns}s were discovered: Magnetars, X-ray Dim Isolated Neutron stars (\acrshort{xdins}s), or Central Compact Objects (CCOs), all somewhat hosting extreme magnetic fields (in the case of CCOs buried by fall-back accretion after their SN explosion). A recent brief review can be found in \cite{Borghese:2020emz}.

The majority of \acrshort{ns}s show a measurable slow decrease of their rotational frequency $\nu_s$, the \textit{spin-down} $\dot{\nu}_s$, consequence of rotational energy losses due to several possible mechanisms, including emission of \acrshort{em} radiation, particle wind and, possibly, \acrshort{gw}s. The star's rotational energy loss rate, called \textit{spin-down luminosity} $L_\mathrm{sd}$ , can be related to the spin-down by \cite{2004hpa..book.....L}
\begin{equation}
    -\dot{E} \equiv L_\mathrm{sd} = -4\pi^2I_3\dot{\nu}_s\nu_s
    \label{eq:emlum}
\end{equation}
where $I_3$ is the star moment of inertia with respect to the star's rotation axis. Without loss of generality, we can express the spin-down rate as a power law in terms of rotational frequency, 
\begin{equation}
\dot{\nu}_s = -K\nu_s^{n}\, ,
\label{eq:pdot}
\end{equation}
where the coefficient $K$ and the \textit{braking index} $n$ depend on the energy loss mechanism. For instance, $n=1$ describes pulsar wind, $n=3$ corresponds to magnetic dipole radiation, $n=5$ to quadrupole gravitational radiation and $n=7$ to \textit{r-mode} \acrshort{gw} 
 emission. 
For the typical observed spin-down rates, the star's time-dependent rotation period, $P=1/\nu$, deriving from integration of \eq{eq:pdot}, can be described by a Taylor expansion around the value $P_0$ at a reference time $t_0$,
\begin{equation}
    P(t) = P_0 + \dot{P}(t-t_0) + \ddot{P}\frac{(t-t_0)^2}{2} + \ldots \, .
\end{equation}
The set of values $(\dot{P},~\ddot{P},~\ldots )$ are called spin-down parameters and can be in principle measured from the pulsar timing. From a measure of the first two parameters, an estimation of the \textit{braking index} can be obtained from the time derivative of \eq{eq:pdot},
\begin{equation}
    n = 2-\frac{P\ddot{P}}{\dot{P}^2}\, .
\end{equation}
In practice, only for a very small number of pulsars the measure of $\ddot{P}$ (which is typically affected by timing noise) is sufficiently accurate to provide a reliable measure of the braking index, see e.g. \cite{Archibald:2016hxz}.
\begin{figure}[t]
\centering
\includegraphics[scale=0.35]{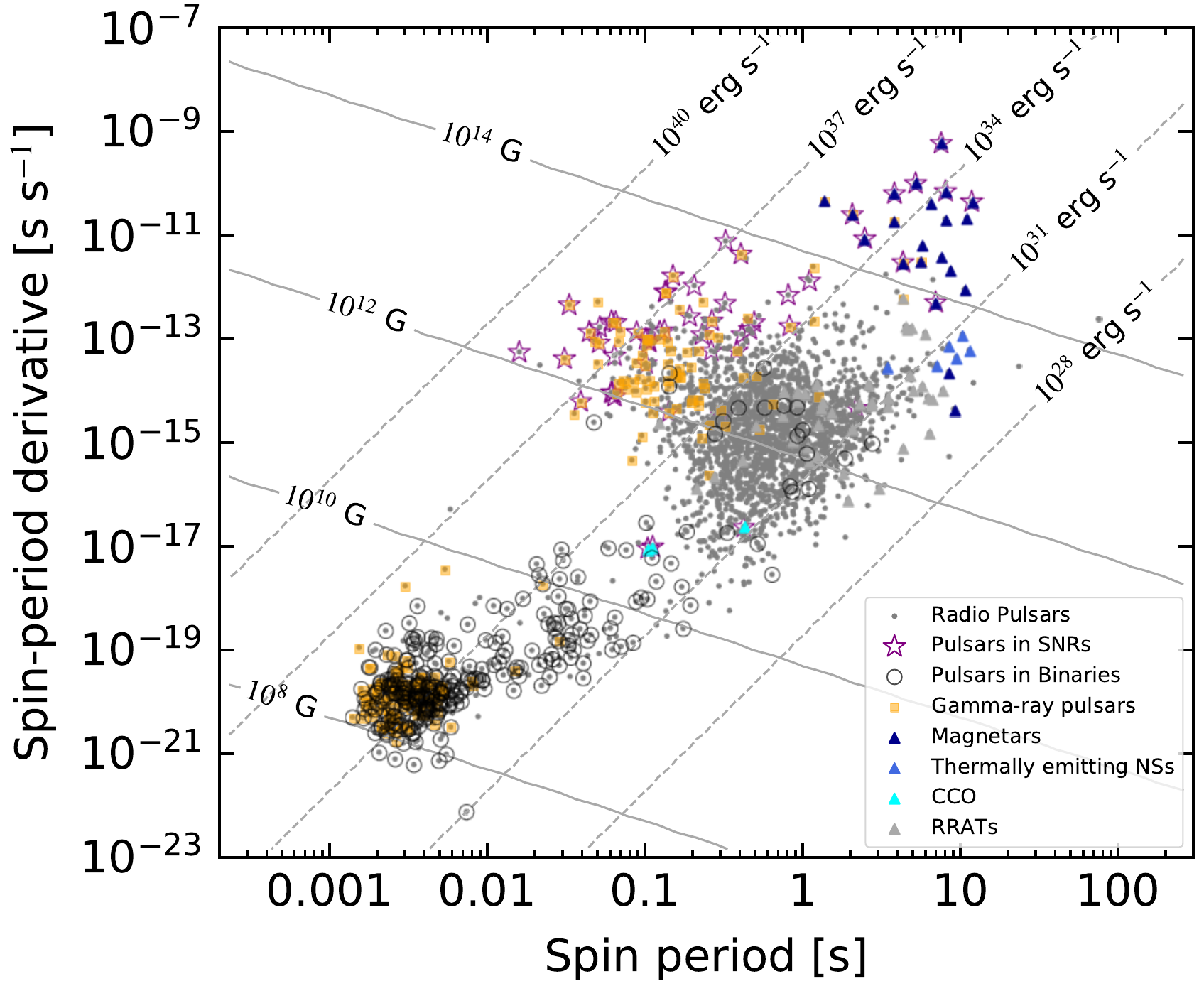}
\caption{Period versus period derivative of different classes of pulsars. Lines of constant estimated B-field are also shown, see\eq{eq:Bppdot}, as well as lines of constant luminosity, see \eq{eq:emlum}. Data taken from \cite{Manchester:2004bp,CotiZelati:2017rgc}.}
\label{fig:p_pdot}
\end{figure}

\begin{figure}[t]
\centering
\includegraphics[scale=0.35]{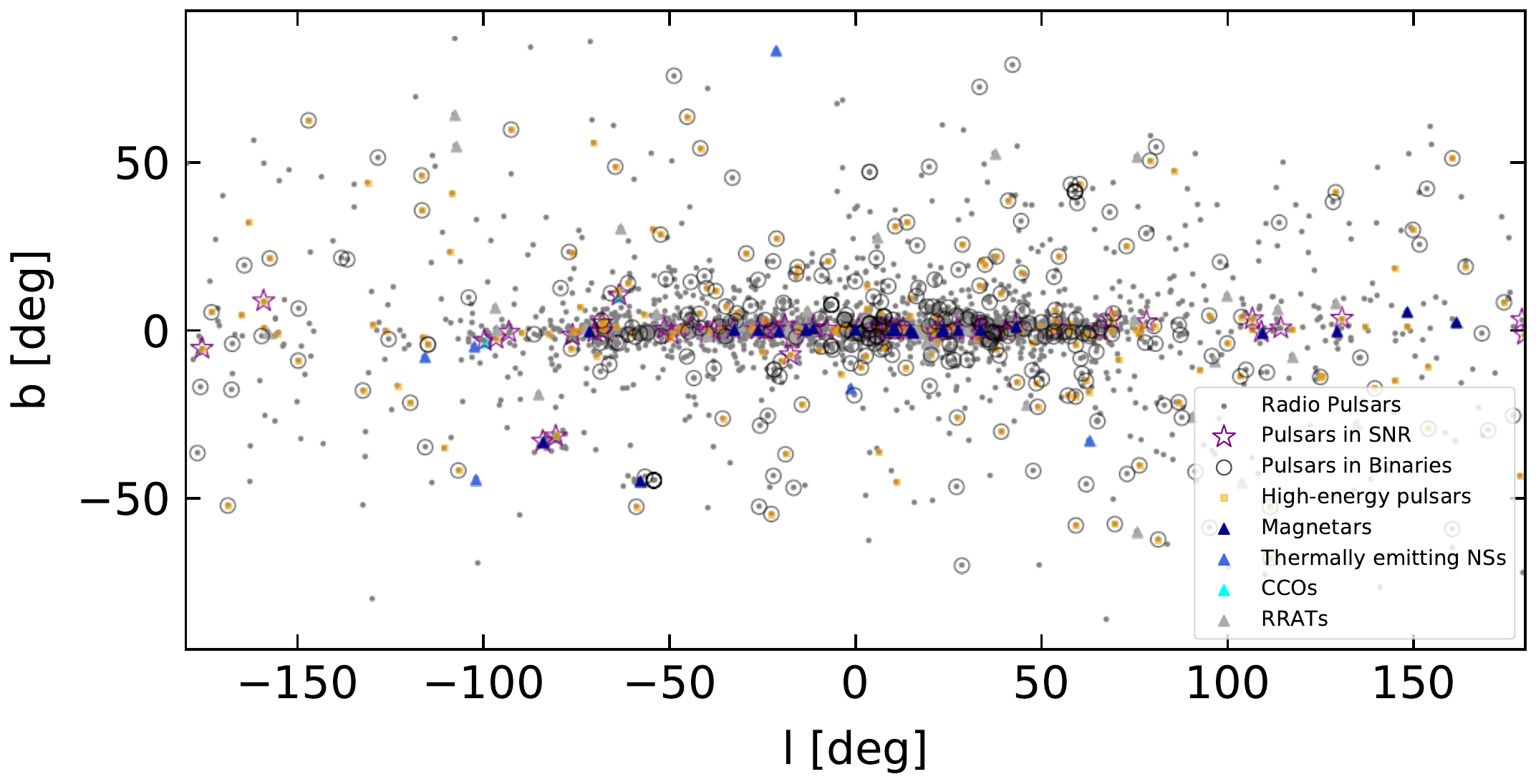}
\caption{Position in the sky (galactic coordinates) of the various classes of pulsars shown in figure~\ref{fig:p_pdot}. Data taken from \cite{Manchester:2004bp,CotiZelati:2017rgc}.}
\label{fig:pulsar_position}
\end{figure}

From \eq{eq:pdot}, and assuming spin-down is dominated by the emission of magnetic dipole radiation ($n=3$), an expression of the magnetic field at the surface can be obtained:
\begin{equation}
    B_s = 3.2\times 10^{19}\sqrt{P\dot{P}}~G\, ,
    \label{eq:Bppdot}
\end{equation}
where the period $P$ is expressed in seconds.
The various \acrshort{ns}s sub-populations can be effectively represented in the so-called $P-\dot{P}$ diagram, see figure~\ref{fig:p_pdot}, where each known \acrshort{ns} for which rotation period and the first spin-down parameter have been measured, corresponds to a point. In general, different sub-populations mostly occupy a different region in the plane. In the plot, lines of constant spin-down luminosity (\eq{eq:emlum}) and B-field (\eq{eq:Bppdot}) are also indicated. The sky distribution (in galactic coordinates) of these objects is shown in figure~\ref{fig:pulsar_position}. 

Rotating \acrshort{ns}s can be sources of long duration \acrshort{gw}s, if their shape deviates from axi-symmetry with respect to the rotation axis. The two basic emission mechanisms (the so-called `mountains' and `r-modes') have been introduced in section~\ref{section:div6}. Moreover, they can also be source of burst-like \acrshort{gw}s, following e.g. pulsar glitches. A few specific emission mechanisms, not covered in section~\ref{section:div6}, will be discussed in section~\ref{sec:nscw}, while a more general discussion about ET detection prospects will be done in section~\ref{sec:Prospects:CW}. 

\paragraph{Galactic radio pulsars.} \label{par:Galactic radio pulsars}


Pulsars -- a contraction of `pulsating radio source', as they were initially observed at radio wavelengths -- are rapidly rotating \acrshort{ns}s with a plasma-filled magnetosphere that emits beamed radiation at many wavelengths. Due to a conspicuous particle acceleration in their magnetosphere undergoing a series of pair production cascades, ultimately a beamed synchrotron emission is observed, see e.g. \cite{Philippov:2020jxu}. Pulsars also emit their \acrshort{em} signal in the optical, X-ray and gamma-ray bands, often also due to curvature emission and/or surface thermal emission \cite{Harding:2021wum}.
    
Pulsars are extremely interesting objects due to their intrinsic regular \acrshort{em} emission, as some of the known millisecond pulsars rival the stability of atomic clocks. The possibility of tracking their rotation rate via pulsar timing techniques to a high level of precision opens a wealth of options, from precision tests of general relativity in the strong field regime \cite{Ding:2021bhk} to inferring dense matter properties by studying their timing instabilities \cite{amp:2023vpd} or estimating pulsar masses via Shapiro delay \cite{Demorest:2010bx} and using millisecond pulsars as a pulsar timing array, i.e. a galactic-scale interferometer for \acrshort{gw}s~\cite{EPTA:2023fyk,NANOGrav:2023gor}. 

\paragraph{Pulsar glitches}\label{sec:glitches}

The slow and predictable spin-down of most young pulsars suffers from sporadic timing irregularities that spoil their perfect timing. 
Abrupt increases in a spinning-down \acrshort{ns}'s observed rotation frequency -- a phenomenon known as a `glitch' -- are commonly observed in most young canonical pulsars (notably, the Vela and Crab pulsars among others) and, more exceptionally, in some old millisecond pulsars, see e.g. \cite{Manchester:2017hyr,Antonopoulou:2022rpq}. 
Since glitching rotation-powered pulsars are isolated, the glitch phenomenon must originate from their interiors. Accordingly, the question of where the apparent increase in angular momentum comes from is directly linked to the understanding of the inner structure and transport properties of \acrshort{ns} interiors \cite{amp:2023vpd}. 
The exact mechanisms responsible for glitches is still not exactly understood, but it is likely to involve an interplay between the \acrshort{ns}'s elastic crust and the superfluid that lies within the inner crust and outer core. 
Shortly after the first glitches were detected in the Vela and Crab pulsar, it was suggested that the observed spin-up is caused by a decrease in the \acrshort{ns}'s moment of inertia -- a consequence of the conservation of angular momentum -- as the solid crust suddenly reduces to a less oblate shape. This hypothesis is known as the `starquake model'.
However, the observed quasi-exponential post-glitch relaxation of the rotation frequency to the pre-glitch state cannot be easily explained only in terms of starquakes. Therefore, the long post-glitch relaxation timescale (of the order of days to months) was linked to the presence of superfluid neutrons in the interior of glitching \acrshort{ns}s \cite{Haskell:2015jra,amp:2023vpd}. 
Superfluidity is also likely to be involved in timing noise \cite{Antonelli:2022gqw}, a random and long-timescale wandering of the rotation frequency around its secular value, that is frequently seen in canonical and millisecond pulsars. Besides giving us a glimpse of the interior of \acrshort{ns}s, glitches and timing noise can also impact the search for \acrshort{cw} from \acrshort{ns} mountains, as they spoil the perfect coherence of the periodic \acrshort{gw} signal. Finally, pulsar glitches are expected to be accompanied by some degree of \acrshort{gw} emission, both burst-like emissions and longer-lived signals \cite{Yim:2020trr,Yim:2022qcn}, as recently reviewed in \cite{Haskell:2023exo}. See \ref{sec:cwglitch} for a brief summary on \acrshort{gw} emission from pulsar glitches.

\paragraph{Magnetars, XDINs and CCOs.}
\label{sec:magnetars}
    
Magnetars are isolated X-ray pulsars whose emission is thought to be powered by the decay and instabilities of their extreme magnetic fields, typically $B$ $\sim$ 10$^{14}$--10$^{15}$~G at the surface (see \cite{Esposito2021} for a review; stars in figure~\ref{fig:p_pdot}). At the time of writing, about 30 sources are listed as magnetars. 
With spin periods in the 1--12\,s range and relatively large spin-down rates, these objects have a persistent X-ray luminosity of $L_{\rm X}$ $\sim$ 10$^{31}$--10$^{36}$\,erg\,s$^{-1}$ generally larger than their rotational energy loss rate. 
The birthmark of these isolated \acrshort{ns}s is the unpredictable and variable bursting activity in the X-/gamma-ray bands on different time scales, from a few milliseconds to hundreds of seconds. In particular, short bursts are the most common, with peak luminosities in the range $\sim$ 10$^{36}$--10$^{43}$\,erg\,s$^{-1}$. 
Burst durations span over two orders of magnitude, ranging from a few milliseconds to a few seconds. They can occur sporadically or cluster in time, and it is not possible to predict their appearance and which source is about to burst. 
The brightest bursts, also called intermediate bursts, sometimes show a tail lasting up to several thousand seconds that might show flux modulation at the \acrshort{ns} spin period \cite{Kaspi:2017fwg}. The most energetic events associated with magnetars are the giant flares. So far, only three have been observed, each from a different source, implying that giant flares are rare. 
Their properties are very similar: all of them started with a short ($\sim 0.1-0.2$~s) spike of $\gamma$-rays, with emission detected up to a few MeVs that reached a peak luminosity $\geq$10$^{44}$--10$^{45}$\,erg\,s$^{-1}$ (and $\geq$10$^{47}$\,erg\,s$^{-1}$ in one case \cite{Mereghetti:2023iwx}). The initial flashes were followed by hard X-ray tails strongly modulated at the spin period and observed to decay in a few minutes.
    
These flaring events often indicate that the magnetar has entered an active phase, commonly referred to as an outburst. During an outburst, the persistent X-ray flux increases by up to three orders of magnitude above the quiescent level. Then, it usually relaxes back to the pre-outburst level on time scales spanning from weeks to months/years. The outbursts are most likely driven by magnetic stresses, which result in elastic movements of the \acrshort{ns} crust and/or rearrangements/twistings of the external magnetic field, with the formation of current-carrying localized bundles. 

\acrshort{xdins}s are a small group of seven nearby (hundreds of parsecs), thermally emitting, isolated \acrshort{ns}s (squares in figure~\ref{fig:p_pdot}). They are radio quiet while relatively bright in the X-ray band, and five out of the seven have spin periods similar to the magnetars. Moreover, their magnetic fields seem to be comparable to the electron critical field. At variance with the majority of X-ray emitting pulsars, \acrshort{xdins}s X-ray spectra are rather perfect black bodies ($kT\sim0.1$ keV, eventually with some broad lines), which make them the key objects for a large branch of research on the \acrshort{ns} equation of state, and the cooling properties of the \acrshort{ns} crust and core.

Central compact objects (CCO) consists of radio quiet thermally emitting X-ray sources ($kT\sim 0.2 - 0.5$~keV), located at the center of luminous shell-like SN remnants (triangles in figure~\ref{fig:p_pdot}). The association with SN remnants implies ages at most of a few tens of kyrs. Three CCOs show X-ray pulsations with periods in the $0.1-0.4$\,s range, but with small spin-down rates. The rotational energy loss of the three pulsed CCOs is certainly too small to give a detectable contribution to their observed X-rays. Furthermore, they are often called “anti-magnetars”, because of their extremely low inferred dipolar magnetic fields ($B_\mathrm{dip} \sim10^{10-11}$\,G). The spin-down characteristic ages of these three CCOs exceed by orders of magnitude the ages of their associated SN remnants, indicating that they were either born with spin periods very close to the current values, or underwent strong accretion during the fallback phase soon after the SN explosion, which might have buried a strong magnetic field inside. Furthermore, the large spin modulated X-ray emission (i.e. $\sim 60\%$ in Kes 79 SN remnant) and the evidence for bright and tiny (a few km radii) thermally emitting hot spots in most of them, are difficult to explain in a weakly magnetized \acrshort{ns} scenario.
Magnetars, \acrshort{xdins}s and CCOs are examples of different stages of the evolution of strongly magnetic \acrshort{ns}s: the first ones being young, the second their older peers and CCOs young systems but having witnessed strong accretion at birth. 

Strongly magnetized \acrshort{ns}s are potentially relevant sources of \acrshort{gw}s. As will be discussed in more detail later, burst-like emission of \acrshort{gw}s can happen in association with outburst (section~\ref{sec:burst_prospects}), while magnetars can emit ``long-transient'' signals during the early phases of their life due to the distortion of the star's shape induced by the strong inner magnetic field (if not aligned with the rotation axis), as discussed in section~\ref{sec:cwmagn} and \ref{sec:Prospects:CW}.

\paragraph{Low Mass X-ray Binaries.} 
\label{sec:lmxb}
Low Mass X-ray Binaries (LMXBs) with \acrshort{ns} accretors are crucial systems in astrophysics, where a \acrshort{ns} accretes material from a companion star that is less massive than the Sun. These systems are significant sources of X-rays, as the material from the companion star spirals in towards the \acrshort{ns}, heating up and emitting high-energy radiation in the process. In the majority of Low Mass X-ray Binaries (LMXBs), the \acrshort{ns} undergoing accretion experiences a rapid increase in its spin rate to millisecond durations, a process often referred to as \textit{pulsar recycling}. This acceleration is attributed to the transfer of mass from the companion star, which carries substantial specific angular momentum, thus enhancing the spin of the \acrshort{ns} through accretion torques~\cite{Patruno:2012ab}. When the period of mass transfer concludes, the \acrshort{ns}, now spinning rapidly, has the potential to reactivate as a ``recycled'' millisecond radio pulsar. This reactivation typically occurs once the accretion ceases, allowing the swiftly spinning \acrshort{ns} to emerge as a radio millisecond pulsar.
This scenario has somehow blurred in recent years with the discovery of a class of LMXBs known as \textit{transitional pulsars} which have shown the characteristics of both accreting millisecond pulsars and radio millisecond pulsars \cite{Papitto:2013hza,Ambrosino:2017trt}. 
    
Neutron stars in LMXBs with spin periods in the millisecond range are of particular interest for the study of continuous \acrshort{gw}s. As these dense objects accrete mass, they can develop asymmetries in their mass distribution, leading to the emission of continuous \acrshort{gw}s as they spin \cite{Bildsten:1998ey}, see section~\ref{sec:cwlmxb} for more details. 
    


\subsubsection{GWs from rotating NSs} 
\label{sec:nscw}
We start this section with a brief summary of the main mechanisms and features of long-lived \acrshort{gw} emission (\acrshort{cw}, Continuous Waves) from rotating \acrshort{ns}s, deferring to section~\ref{section:div6} for more details. Later we provide a more in-depth discussion for a few specific sources not  covered in section~\ref{section:div6}.  An excellent review on \acrshort{gw} emission from rotating \acrshort{ns}s is given in \cite{Lasky:2015uia}. Detection prospects with ET will be presented in section~\ref{sec:Prospects:CW}.

If a rotating \acrshort{ns} is deformed in a non-axisymmetric way, it produces \acrshort{cw}s. 
There are essentially two main mechanisms that can be at play: a static deformation of the star, a so-called `mountain', which creates (to leading order) a mass quadrupole, that is then swept around by rotation, or modes of oscillation of the \acrshort{ns}, that can create time-varying mass or current quadrupoles. The mountain is supported by either crustal rigidity or the star's magnetic field \cite{Sieniawska:2019hmd}, as the field evolves due to the Hall effect \cite{Suvorov:2016hgr}.
Theoretical estimates suggest the crust could sustain deformations, in terms of the star's ellipticity $\epsilon$, which measures the degree of asymmetry with respect to the rotation axis, of up to $\epsilon\approx 10^{-5}$ \cite{Horowitz:2009ya} (although relativistic effects may reduce the ellipticity by more than an order of magnitude \cite{Gittins:2021zpv}). In the case of compact stars with exotic EOS, due for instance to the presence of hybrid quark-baryon mixtures or meson-condensates, significantly higher ellipticities, up to $10^{-4}-10^{-3}$, could be reached \cite{Owen:2005fn}. See section~\ref{section:div6} for a detailed discussion on possible mechanisms for mountains formation and the role of the star's EOS. The strain amplitude of the \acrshort{cw} emitted by an asymmetric \acrshort{ns} rotating  around one of its principal axes of inertia is given by \eq{eq:StrainAmplitude} which, for convenience, we re-write here:
\begin{equation}
    h_0 \simeq 10^{-25} \left( \frac{{10}{\text{kpc}}}{d} \right) \left( \frac{\epsilon}{10^{-6}} \right) \left( \frac{I_3}{10^{45}\,\text{g\,cm}^2} \right) \left( \frac{\nu}{{500}\,{\text{Hz}}} \right)^2,
    \label{eq:StrainAmplitude_div7}
\end{equation}
where $\nu=2\nu_s$ is the \acrshort{gw} signal frequency, $I_3$ is the star's momentum of inertia with respect to the rotation axis and $d$ is the source distance from the detector.
For the standard case of a star emitting at a single harmonic emission from the $l = m = 2$ mass quadrupole mode i.e. with a \acrshort{gw}  frequency at twice the pulsar rotation frequency, assuming all the observed source spin-down is due to the emission of \acrshort{cw}s brings to the so-called \textit{spin-down limit} on the star ellipticity, given by 
\begin{equation}
\label{eq:eps_sd}
\epsilon_{\rm sd} = \left(\frac{5c^5}{32 \pi^4 GI_{3} }\frac{|\dot{\nu}|}{\nu^5}\right)^{1/2},
\end{equation} 
where $\dot{\nu}$ is the signal frequency first-time derivative.

\Eq{eq:StrainAmplitude_div7} shows that \acrshort{cw} from ``standard'' spinning \acrshort{ns}s are very faint even for galactic sources. On the other hand, signal persistency can be exploited to build signal-to-noise ratio (\acrshort{snr}) by analyzing long chunks of data. This has significant implications in the way in which the analyses must be carried on, because some secular effects, like the Doppler effect induced by the detector motion, the source intrinsic spin-down and the sidereal modulation due to the detector time-dependent response, affect the signal at the detector and must be properly taken into account, see section~\ref{sec:Prospects:CW} for more details.

In addition, as the magnetic and rotational axis are not, in general, aligned, a magnetic mountain may emit not only via the $Q_{22}$, but also the $Q_{21}$ (i.e. $l=2, m=1$) multipole moment, thus leading to emission not only at $\nu=2\nu_s$, but also at $\nu=\nu_s$ (see \cite{Riles:2022wwz} for a review).

Newborn millisecond magnetars (section~\ref{sec:magnetars}) may sustain large ellipticities, due to the strong inner magnetic field, and this can lead to a significant \acrshort{gw} emission, in principle detectable up to several Mpc with ET.  

Accreting \acrshort{ns}s (section~\ref{sec:lmxb}), like LMXBs (e.g. Sco-X1), may develop crustal mountains as a consequence of the matter accretion process and reach spin equilibrium making them potential interesting emitters of \acrshort{cw}.

When it comes to modes of oscillation, 
the most relevant for our discussion are the $f$-mode and the $r$-mode. 
The $f$-mode frequency $\nu$ is in the kHz region, and proportional to the average density of the star, $\nu\propto\sqrt{M/R^3}$. The mode can be excited efficiently by transient events, for instance pulsar glitches (section~\ref{sec:glitches}), leading to copious \acrshort{gw} emission, which also leads to rapid damping, on a timescale $\tau\propto 
 \nu^{-6} R^{-5}$, which for typical \acrshort{ns} parameters is less than a second. However, if both frequency and damping timescale are measured, this leads to a constraint on the mass and radius of the star, and therefore on the EOS of dense matter \cite{Andersson:1997rn}. 

 The $r$-mode, as discussed in section~\ref{section:div6}, is a toroidal mode of oscillation for which the restoring force is the Coriolis force, and can be driven unstable by \acrshort{gw} emission, growing to large amplitudes that make it an interesting mechanisms for current and future searches (see \cite{Haskell:2015iia} for a review).
 
 In the following subsections we focus 
 on three scenarios potentially relevant for the emission of \acrshort{gw} from spinning \acrshort{ns}s, which have not been discussed in detail in section~\ref{section:div6}: accreting \acrshort{ns}, newborn millisecond magnetars and pulsar glitches.

\paragraph{Spin equilibrium in LMXBs.}
\label{sec:cwlmxb}
Accreting \acrshort{ns}s in LMXBs are a promising source of \acrshort{cw}s. Observations
suggest that these systems rotate well below the break-up limit,\footnote{The break-up limit is the star's maximum rotation frequency above which the centrifugal force overcomes the gravitational force causing the star's break-up \cite{Fattoyev:2018vbj}.} implying that some kind of speed limit may be enforced, and the already discussed r-modes could provide an explanation.
Alternatively, the observed spin frequencies of accreting \acrshort{ns}s in LMXBs could be determined by the balance between spin-up torques due to accretion and spin-down torques due to \acrshort{gw} emission from mountains created by the accreted material in the crust (crustal mountains) \cite{Bildsten:1998ey}.

This possibility is supported by X-ray measurements of the spins of \acrshort{ns}s in LMXBs. The spin-distribution is, in fact, cut-off well before any theoretically conceivable break up frequency for the star \cite{Haskell:2018nlh}, and the cutoff appears to be sharp, with evidence for two populations: a slower, more widely distributed population, and a faster, narrower population with a cutoff around 700 Hz \cite{Patruno:2017oum}. Theoretical population synthesis models show that such a feature is better modeled by the presence of \acrshort{gw} torques which stall the spin-up above approximately 700 Hz \cite{Gittins:2018cdw}. 

The \acrshort{gw} amplitude that would give spin-equilibrium in observed LMXB systems is thus a useful benchmark for \acrshort{gw} searches, and future observations with ET will allow to thoroughly investigate this scenario. For the  simplest accretion torque model, that assumes accretion at the surface of a $M=1.4 M_\odot$ and $R=10$ km \acrshort{ns}, this amplitude, as a function of the observed X-ray flux $F_x$ reads
\begin{equation}
    h_\text{eq}\approx 5.48\times 10^{-27} \left(\frac{F_\mathrm{x}}{10^{-8}\;\mbox{erg cm$^{-2}$ s$^{-1}$}}\right)^{1/2}\left(\frac{\nu_s}{300\;\mbox{Hz}}\right)^{-1/2}\, ,
    \label{eq:accreting}
\end{equation}
where $\nu_s$ is the spin frequency of the star at equilibrium. Recent searches for \acrshort{gw} signals in O2 and O3 data \cite{Zhang:2020rph, LIGOScientific:2022enz} from the brightest (in X-ray) LMXB, Sco X-1, have in fact surpassed this limit, and thus put constraints on the magnetic field of the star (as we discuss in the following paragraph) and the EOS. The current sensitivity of \acrshort{cw} searches has thus started to put astrophysical constraints on these sources, and the increased sensitivity of next generation detectors, such as ET, will allow to fully probe quantitative details of the physics of these systems.

In accreting \acrshort{ns}s, the field structure evolves due to infalling matter \cite{Melatos:2005ez}. This is particularly interesting for multimessenger observations, as the magnetic mountain may lead to cyclotron resonance scattering features (CRSFs) in the X-ray spectrum of an accreting \acrshort{ns} \cite{Priymak:2014dqa}. Given a \acrshort{gw} detection, X-ray observations may thus offer the means to understand which kind of mechanisms is producing the mountain on a \acrshort{ns} \cite{Haskell:2015psa}.
Note that in the presence of a magnetic field, the disc will be magnetically truncated, and the spin-equilibrium formula above, \eq{eq:accreting}, no longer applies, although it is still a useful order of magnitude estimate (see \cite{LIGOScientific:2022enz} for a more detailed discussion of the topic).

\paragraph{Newborn millisecond magnetars.}
\label{sec:cwmagn}
As previously discussed, magnetars are believed to be born rapidly rotating (with millisecond periods), from the collapse of massive stars, accretion induced collapse of white dwarfs, or from a \acrshort{ns}-\acrshort{ns} merger. 
Two main fundamental mechanisms have been proposed to determine long-duration \acrshort{gw}s from newborn millisecond magnetars.  
The large magnetic field and likely presence of fallback accretion, can lead to the creation of a mountain, which couple with the rapid rotation rate, can lead to strong \acrshort{gw} emission from these systems. The braking index of a number of systems is, in fact, consistent with $n=5$ \cite{Sarin:2020mvf}, corresponding to spin-down due to a \acrshort{gw} mountain, which for the estimated durations of up to $50$ s, would require an ellipticity of the order of $\epsilon\approx 10^{-3}$ \cite{Gao:2015xle}, consistent with theoretical models for fallback accretion, and which could be detected by next generation detectors at distances of 1 Mpc (see e.g. \cite{DallOsso:2021xbv, Sur:2020imd}).
Alternatively, the strong inner magnetic field can significantly distort the star, up to $\epsilon\approx 10^{-4}-10^{-3}$, and trigger long-lasting \acrshort{gw} emission \cite{2001A&A...367..525P}, in particular in presence of a toroidal field which induces a prolate deformation that maximizes \acrshort{gw} emission (through the so called \textit{spin-flip} mechanism) \cite{Cutler:2002nw,Stella:2005yz,DallOsso:2018dos}. Such \acrshort{gw} signal, with duration from several hours to a few days, could be detected by ET up to 5-10 Mpc with reasonable data analysis approaches, see section~\ref{sec:Prospects:CW} for more details. 

In the case of millisecond magnetars produced by \acrshort{ns}-\acrshort{ns} mergers, there is evidence that a meta-stable \acrshort{ns} may be produced in some cases, and power the X-ray plateau observed in some short gamma-ray bursts \cite{Rowlinson:2013ue}. A magnetar central engine has been also proposed for short gamma-ray burst with extended emission, see e.g. \cite{Bucciantini:2011kx,Lu:2020gse}. See \cite{Bernardini:2015coa} for a review on the connection among gamma-ray bursts and magnetars. If the star is supra-massive, i.e. has a maximum mass that exceeds that of a non-rotating, stable, \acrshort{ns}, it will eventually collapse to a \acrshort{bh} as it slows-down, and a measurement of the collapse time, from combined \acrshort{em} and \acrshort{gw} observations, could thus constrain the equation of state of dense matter \cite{Lasky:2013yaa}. 

\paragraph{Oscillations and mountains after pulsar glitches.}
\label{sec:cwglitch}
Pulsar glitches, the sudden spin-up events discussed in section~\ref{sec:glitches},
are also likely to excite \acrshort{gw} emission. Key to this discussion is an understanding of the superfluid interior of \acrshort{ns}s which, when rotating, form an array of quantum vortices that carry the angular momentum of the fluid. Each vortex essentially acts as a quantum of angular momentum. The observed rise in the pulsar's spin from a glitch occurs rapidly (the best current upper limits indicating it is over in less than a minute), and is thought to be associated with the sudden outward motion of quantum vortices in the superfluid interior of the star \cite{Antonopoulou:2022rpq, amp:2023vpd}. The rearrangement of the vortex configuration is likely to be non-axisymmetric and can lead to the excitation of oscillation modes during and after a glitch. 
In particular, the f-mode is likely to be excited efficiently and therefore lead to a short-duration burst-like signal, which is damped on short timescales comparable to that of the rise \cite{Keer:2014uva, Ho:2020nhi, Yim:2022qcn}.

On longer timescales, associated with those of the post-glitch relaxation (hours to months), transient mountains may be created in the crust of the \acrshort{ns}, leading to continuous signals, and leaving an imprint also on the evolution of the spin-down rate of the pulsar, as measured from \acrshort{em} timing \cite{Yim:2020trr}. See \cite{Haskell:2023exo} for a review of \acrshort{gw} emission mechanisms associated with glitches.
Finally, on even longer timescales, glitch-induced intermittent emission of \acrshort{gw}s leads to a fluctuating braking torque that can drive timing noise, a slow modulation of a pulsar's period that is generally consistent with a red noise process \cite{Antonelli:2022gqw,Meyers:2021myb}.

A benchmark for searches for signals following glitches can be obtained by 
expressing the energy associated with the event in terms of the glitch parameters \cite{Yim:2020trr},
\begin{equation}
    E_{sd} = 4 \pi^2 Q I_3 \nu_s \Delta \nu_s\, ,
\end{equation}
where $Q$ is the healing parameter which quantifies the fraction of the total spin-up that recovers. The case $Q=1$ corresponds to the situation in which all the energy associated with the event is radiated away as \acrshort{gw}s \cite{Yim:2024eaj}, both for burst-like and continuous emission.



Furthermore, one can also predict the maximum \acrshort{gw} amplitude $h_0$ under certain models. If non-axisymmetric $f$-modes are excited, which carry angular momentum and therefore affect the \acrshort{ns}'s spin, one expects \cite{Yim:2022qcn}
\begin{equation}
h_0 \approx 6.8 \times 10^{-24}  \left(\frac{\alpha}{1\times10^{-6}}\right) \left(\frac{M}{1.4~\text{M}_\odot}\right)^{2} \left(\frac{R}{10~\text{km}}\right)^{-1} \left(\frac{d}{1~\text{kpc}}\right)^{-1}~,
\end{equation}
which assumes an $l = 2$, $m=2$ (retrograde) $f$-mode has been excited with an amplitude $\alpha \sim \delta r / R$, where $\delta r$ is the radial displacement caused by the oscillation. A different parametrization is found in \cite{Ho:2020nhi} for $f$-modes that are excited by the glitch but are not coupled to rotation (axisymmetric oscillations).

Transient mountains have an evolution that follows the post-glitch exponential recovery of the spin-down rate. As expected, it leads to an exponentially decaying amplitude \cite{Yim:2020trr}
\begin{equation}
h_0(t) = \sqrt{-\frac{5}{2} \frac{G}{c^3} \frac{I_3}{d^2} \frac{\Delta \dot\nu_\text{s,t}}{\nu_s}} \, e^{-t/(2\tau_\text{EM})}\, ,
\end{equation}
where $t$ is the time after the glitch, and $\Delta \dot\nu_\text{s,t}$ is the transient change in the spin-down rate.
It is notable that the recovery timescale of the \acrshort{gw} strain is twice that of the glitch recovery timescale observed in the \acrshort{em} band, i.e.~$\tau_\text{GW} = 2 \tau_\text{EM}$, which is on the order of weeks to months. Putting in typical values, we find an initial \acrshort{gw} amplitude of
\begin{equation}
h_0(0) \approx 2.5 \times 10^{-26} \left(\frac{d}{1~\text{kpc}}\right)^{-1} \left(\frac{\nu_s}{10~\text{Hz}}\right)^{-\frac{1}{2}} \left(\frac{-\Delta \dot\nu_\text{s,t}}{10^{-14}~\text{Hz~s}^{-1}}\right)^{\frac{1}{2}}~.
\end{equation}
Although the transient mountain amplitude is smaller than the $f$-mode case by a few orders of magnitude, the fact that they last much longer results in more signal accumulating which improves detectability prospects. 

Searches in O3 data did not detect a signal, and all observational upper limits were all above these theoretical upper energy and spin-down limits \cite{LIGOScientific:2021quq}. Nevertheless, for several pulsars they were close to the theoretical expectation: several glitching pulsars would be targets for next-generation detectors, which could also help us detect glitches that are not observed electromagnetically, and allow for \acrshort{em} follow-up of such events \cite{Haskell:2023exo, Yim:2024eaj}.

\subsection{ET observational prospects}
\label{sec:observation}

\subsubsection{Bursts}

\label{sec:burst_prospects}

\paragraph{CCSN.}\label{sect:CCSNdiv7}
The case of \acrshort{ccsn} \acrshort{gw} detection deserves special attention. The stochastic nature of its \acrshort{gw} signal makes it impossible the use of matched-filtering techniques. The main reason behind is that, unless the case of binary mergers, the phase is stochastic and determined by the instabilities developing in the \acrshort{pns}, such as convection, \acrshort{sasi} and turbulence (see section~\ref{gwnuemission}). Therefore, excess-power search algorithms are the basic tool that are used in \acrshort{ccsn} signal searches with current detectors. This will remain the case for third generation detectors, as well as the fact that searches are carried out differently depending on the observation or not of another messenger: neutrinos and/or the light of the SN. As stated in section~\ref{SNObs} a certain fraction of \acrshort{ccsn}e are observed by telescopes. This has the advantage of a precise sky location, less than few arcminutes, an accuracy smaller than what neutrinos detectors can achieved \cite{Scholberg:2012id}. When neutrinos are detected, the time of the collapse is known within ${\cal O}(10)$~ms \cite{Hansen:2019giq, Mirizzi:2015eza, Scholberg:2012id}. In both cases, this allows to conduct a targeted search that enhances the search performance. 

In {\it targeted searches}, when only the \acrshort{em} counterpart is detected (this corresponds to targeted searches conducted by the \acrshort{lvk} collaboration \cite{LIGOScientific:2016jvu,LIGOScientific:2019ryq,Szczepanczyk:2023ihe}) the source location is known to good accuracy (less than few arcminutes depending on the telescope angular resolution). The source sky location is thus fixed in the search. The time of the event can be bounded within $\sim 2-10$\, days. The remaining sources of uncertainty of this time bound, the so called ``on source window'' (OSW), are the determination of the time at which the shock breaks out of the star (in most of the cases it is not observed directly and the light-curve has to be modeled backwards in time to estimate it) and the time it takes for the shock to cross the star. It depends mostly on the size of the star and can be modeled if there are observations of the progenitor star prior to the explosion \cite{Szczepanczyk:2023ihe}.

The detection of the neutrino counterpart is less likely than for the \acrshort{em} emission because of the horizon distance which is limited to $100$~kpc with current neutrino detectors (e.g. super-Kamiokande, \cite{Super-Kamiokande:2007zsl}) and to $\sim 1$~Mpc for next generation detectors (e.g. Hyper-Kamiokande, \cite{Hyper-Kamiokande:2021frf}) that will be available at the time of \acrshort{et}, see section~\ref{sec:mm}. The main advantage of the neutrino is that it reduces the OSW to a few ms. This bound takes into account the delay between the collapse and the neutrino burst emission (see 
section~\ref{sec:CCSNeMM} for more details). Besides, neutrino detectors can also determine the host galaxy, but it is also very likely that the \acrshort{ccsn} sky location be precisely identified thanks to \acrshort{em} observations. 
An OSW of a few seconds is extremely powerful as it allows to reduce the probability of a false alarm. Because the parameter space is also reduced (known sky position and reduced OSW) the false alarm rate is reduced. This permits to set a detection threshold lower than for the case one does not know when and where the collapse happened.
It also allows  carrying out Bayesian inference for unmodeled \acrshort{gw} burst signal. 

When no \acrshort{em} nor neutrino counterpart is found, {\it all-sky/all-time} generic burst searches are considered to detect \acrshort{ccsn}e. \acrshort{em} counterparts could be absent or be very dim for the case of \acrshort{bh} forming core-collapse events, which nevertheless have a \acrshort{gw} signature very similar to regular \acrshort{ccsn}e. Additionally, galactic disk dust obscuration may prevent the \acrshort{em} observation of some \acrshort{ccsn}e in both our galaxy and distant galaxies. Regarding a neutrino counterpart, the horizon distance will remain limited to $\sim 1$~Mpc for next generation detectors. This means that, for sufficiently distant supernovae, there may not be any detectable neutrino counterpart (see section~\ref{sec:CCSNeMM} for  more detailed discussion).

In the \acrshort{et} era, we expect that different families of search pipelines will be available. Among them, excess power techniques search pipelines which implement time-frequency techniques such as wavelet transform (cWB, \cite{Klimenko:2015ypf}) or spectrogram (X-pipeline, \cite{Sutton:2009gi}) are targetting \acrshort{ccsn} signals. Cross-correlation techniques, implemented in cWB and PySTAMPAS \cite{Macquet:2021ttq} are also promising to better catch up the complex \acrshort{gw} waveform which is expected to be very widely spread in frequency. Finally, Bayesian algorithm using a superposition of wavelets as implemented in Bayeswaves \cite{Cornish:2020dwh} is also adapted to detect \acrshort{ccsn} \acrshort{gw} emission.

\begin{figure}[t]
\centering
\includegraphics[scale=0.5]{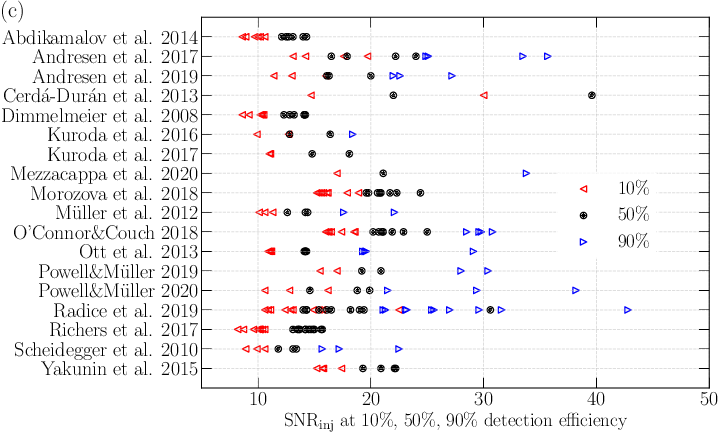}
\caption{Sensitivity of the cWB search pipeline to different \acrshort{ccsn} waveform models as function of the signal \acrshort{snr} using simulated O5 LIGO Hanford and Livingston data (from \cite{Szczepanczyk:2021bka}). The search sensitivity is defined as the average detection efficiency percentage.}
\label{fig:cWB}
\end{figure}

Carrying out a {\it targeted} or {\it all-sky/all-time} search implies different pipeline configuration and tuning but the intrinsic detection performance depends mainly on the way data from multi-detectors are processed to disentangle signal from noise.
Current pipelines sensitivity to \acrshort{gw} emission strongly depends on the total amount of \acrshort{gw} energy and the complexity of the signal. \acrshort{gw} burst at the bounce due to fast rotating progenitor, convection/SASI and \acrshort{pns} fundamental mode excitation correspond to very different frequency bands. This results into a still rather large loss of the \acrshort{gw} signal. For instance, for the current network of \acrshort{gw} detectors, cWB sensitivity requires signals with a minimal \acrshort{snr} amplitude of $\sim$ 20 as shown in figure~\ref{fig:cWB} (from ref.~\cite{Szczepanczyk:2021bka}). 
Assuming optimal glitch characterization and data quality, there is potential to improve algorithms, enabling more efficient searches with a lower detection SNR threshold compared to that used for data taken by current detectors. 

\begin{figure}[t]
\minipage{0.5\textwidth}
  \includegraphics[width=\linewidth]{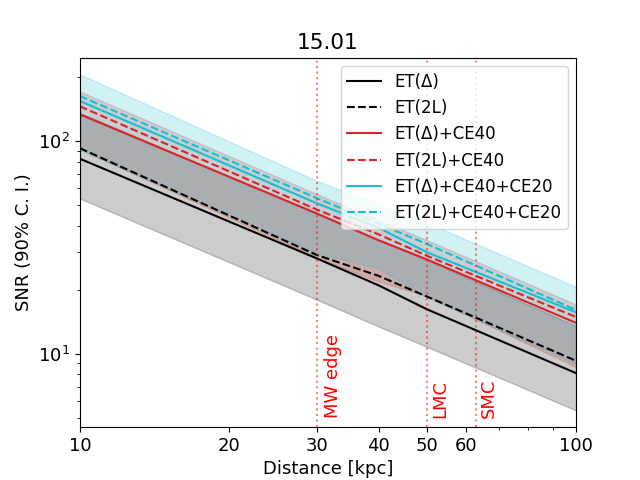}
  
\endminipage\hfill
\minipage{0.5\textwidth}
  \includegraphics[width=\linewidth]{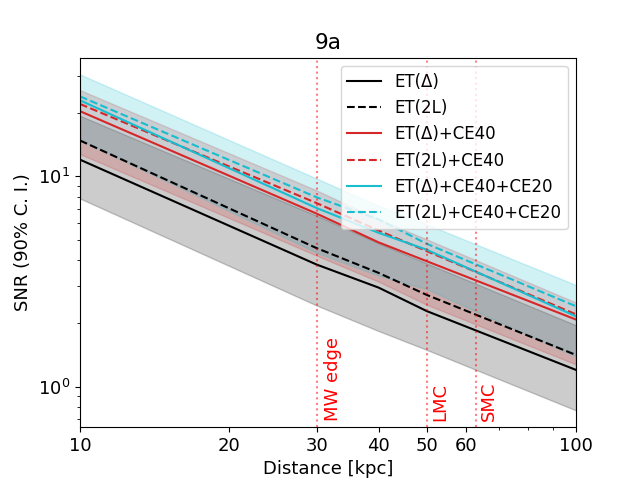}

\endminipage\hfill
\caption{Expected SNR as a function of distance for the 15.01 (left panel) and 9a (right panel) models corresponding to a progenitor mass of 15.01~M$_\odot$ and 9~M$_\odot$ \cite{Vartanyan:2023sxm}, respectively. Solid lines represent the median SNR for \acrshort{et} 10 km triangle, while dashed lines indicate the median SNR for \acrshort{et} 15 km 2L considering \acrshort{et} operating as a single observatory or in a network with CE(40 km), and CE(40 km)+CE(20km).  Colored bands indicate the 90\% confidence interval. The vertical lines indicate reference distances; the Milky-Way edge, the large and small Magellanic clouds.}
\label{fig:CCSN_horizon}
\end{figure}

The estimated detection horizon for neutrino-driven \acrshort{ccsn} in 2nd generation detectors is about $10$~kpc, while for fast rotating progenitors it can reach $100$~kpc \cite{Szczepanczyk:2023ihe}. Yet, these numbers hide lots of variation depending on the progenitor properties and which simulation code is used (see section \ref{sec:collapse}). Moreover, as already mentioned, the current generation of search pipelines is still sub-optimal for complex \acrshort{ccsn} waveforms.  
For third generation detectors the performance of current search pipelines (cWB, X-pipeline, Bayeswave, PySTAMPAS) has not been tested in detail. However, given the increase of about a factor $10$ in sensitivity at all frequencies in \acrshort{et} with respect to 2nd generation detectors, the expectation is an increase of a factor $10$ in the detection range. This is consistent e.g. with the estimates of \cite{Powell:2020cpg} that show that, for a fixed \acrshort{snr}, a detectable source would be located typically a factor $10$ further away for \acrshort{et} than for LIGO/Virgo/KAGRA detectors. Moreover, there is a possibility to gain a factor $\sim$ 2 by developing improved search pipeline fully targeted on electromagnetic \acrshort{ccsn} signals\footnote{E.g., the sensitivity of optically targeted search for gravitational waves emitted by \acrshort{ccsn} is approximately a factor of 2 better than the all-sky, all-time burst search in O2~\cite{LIGOScientific:2019ryq,LIGOScientific:2019ppi}}. 
Another important aspect of the \acrshort{ccsn}e search with \acrshort{et} concerns the exact configuration of the \acrshort{et} detector. All cited sensitivity performances assume that the false alarm rate is reduced by the coincidence/coherence of a non-colocated multi-detectors search. A detailed study, assuming data from three colocated detectors, remains to be done to assess the \acrshort{et} performance. In general, however, we expect such configuration should reduce the search distance reach to \acrshort{ccsn} signals, due to the  randomness of the polarization.  
These semi-quantitative considerations are confirmed by the following results, related to a simplified simulation to compute the horizon of supernova detection using GWFish \cite{Dupletsa:2022scg}. Since \acrshort{ccsn} signals are stochastic, a Fisher matrix approach is not fully usable. Despite losing out on parameter estimation, all the functionalities of the code can be exploited to compute the \acrshort{snr} for any observatory network, source distance and position in the sky. We use waveforms based on the models developed by \cite{Vartanyan:2023sxm}, which provides 11 3D supernova GW emission simulations from 9 to 23~M$_\odot$ progenitors. For distances covering the Milky-way and up to 100 kpc, \acrshort{snr} distributions are obtained by injecting 400 source location (RA and Dec) randomly selected for each combination of distance, waveforms and observatory network. Figure \ref{fig:CCSN_horizon} shows the \acrshort{snr} distributions as a function of distance for the waveforms corresponding to a progenitor mass of 15.01~M$_\odot$ and 9~M$_\odot$, which represent the best and worst \acrshort{gw} emission scenario in \cite{Vartanyan:2023sxm}.
In the best scenario \acrshort{et} will be able to detect a \acrshort{ccsn} up to the Small Magellanic Cloud with \acrshort{snr} $> 10$, 
while in the worst scenario the \acrshort{ccsn} detections will be limited to the Milky-Way. \acrshort{et}  observing in network  with CE(40 km) would increase the distance reach of about 70\%,  while a network including \acrshort{et}, and CE(40 km)  and CE(20 km)  would double it. \acrshort{et} 15 km 2L performs better than \acrshort{et} 10 km triangle with a horizon increase of around 20\%, while the increase is limited to within 10\% when \acrshort{et}  operates in a global network.

When considering a galactic SN, the improved sensitivity of \acrshort{et} could open the possibility of detecting components of the signal that are more challenging to detect. For example contributions that are weaker than the main emission mechanism, such as \acrshort{sasi} \cite{Lin:2022jea}, long term \acrshort{pns} convection \cite{Raynaud:2021cgu} or quasi-radial oscillations \cite{Cerda-Duran:2013swa}. Also components with low frequencies, such as \acrshort{gw} memory effects \cite{Richardson:2021lib}, present a detection challenge, although there are chances of being detectable with current detectors \cite{Richardson:2024wpp}. This is in particular the case of the \acrshort{gw}s from anisotropic emission of neutrinos which is ignored in many \acrshort{ccsn} simulations. This low-frequency signal is rather high amplitude and detection prospects for \acrshort{et} should be enhanced \cite{Powell:2024nvv, Mueller_Janka__1997__aap__Gravitational_radiation_from_SN_convection}. This also means that the \acrshort{et} sensitivity improvement must be accompanied with new methodological developments such that a fully \acrshort{ccsn} dedicated search pipeline is ready for the time of \acrshort{et}. Below we list some of these possible methodological developments. 

{\it Machine learning methods:} Searches based on excess power detection have the advantage that they are model independent. However, this may also limit the detection capability when searching for specific waveforms, such as those from supernovae. For example, cWB is able to detect SN signals with a typical \acrshort{snr} in the range 15-25 \cite{Szczepanczyk:2021bka}. This is a relatively high \acrshort{snr} when compared to binary mergers, for which matched filtering techniques can easily detect sources with a \acrshort{snr} $\sim 8$~\cite{KAGRA:2021vkt}. 
Given that template matching is not possible in this case, machine learning techniques have been proposed by a number of authors \cite{Astone:2018uge,Chan:2019fuz,Cavaglia:2020qzp,LopezPortilla:2020odz,Mukherjee:2021xdt,Lopez:2021ikt,Antelis:2021qak,Iess:2023quq,Lagos:2023qli,Powell:2023bex,Perez2023,Nunes:2024fig} as an alternative for \acrshort{ccsn}e. These works are part of a more general effort of introducing ML techniques in \acrshort{gw} searches \cite{Cuoco:2020ogp}. In many ML detection pipelines there is a first step in which a neural network is trained, either using time series or time-frequency data, to recognize data in which a signal is present from data in which only noise is present. Once the neural network has been trained, it can be used as a detection pipeline. The properties (efficiency, false alarm rate) can be tested by evaluating its capabilities on test data. In general, ML algorithms seem to perform similar to other excess-power techniques; in many cases they are capable of detecting signals with \acrshort{snr}$\sim 15$. However, a direct comparison with current excess-power techniques (e.g. cWB) is currently not available, so it is still difficult to evaluate if they provide any advantage. Nevertheless, the ever increasing sophistication of ML techniques in a field that is fast moving makes them very promising for the future and should be explored further for \acrshort{et}. Additionally, ML techniques are typically faster than traditional techniques and this could produce a critical advantage over traditional searches for the \acrshort{et} era, where the number of events is expected to increase significantly (although not necessarily in the \acrshort{ccsn} case). 

The main limitation of ML techniques is the necessity to train neural networks with known examples of signals. In the case of \acrshort{ccsn}e, the total number of \acrshort{gw} signals available in the literature computed from numerical simulations is of the order of $\sim 100$, when considering only models with sufficient degree of realism (three-dimensional simulations with accurate neutrino transport). This amount is clearly insufficient to train any neural network, for which one typically needs 10-100 thousand examples. This number is not expected to increase significantly in the future given that is limited by the computational cost of the numerical simulations (millions of CPU hours). To address this problem different strategies have been followed: i) \cite{Astone:2018uge,LopezPortilla:2020odz,Powell:2023bex} used phenomenological templates to train neural networks, that mimic the temporal and frequency evolution of \acrshort{ccsn} waveforms but are inexpensive to produce ($\sim$ms). Those can be generated in any amount necessary to train the network and the trained networks are still capable of detecting signals from numerical simulations. ii) Dimensionality reduction allows to reduce the information contained in waveforms to a minimum set of parameters that can be used to train a simple neural network. One possible approach is the use of principal component analysis (see e.g. \cite{Heng:2009zz,Rover:2009ia,Powell:2016wke,Suvorova:2019ebd} for other uses in the context). In this case the number of waveforms necessary to train the network could be reduced significantly. iii) In combination with the strategies above (and in many cases as the only strategy) one may use only a few numerical simulations and perform tens of thousands of injections at different sky locations and source orientations. Given that the strain at the detector changes depending on these parameters, the variability introduced could provide sufficient variety of examples to the network to have a proper training. However, the use of this strategy alone may lead to overfitting, i.e. networks that work very well for a small set of known waveforms but do not generalize properly when used on new waveforms. An alternative approach is the use of ML techniques in combination with current excess-power techniques as a post-processing step to improve the detection statistics \cite{Cavaglia:2020qzp,Mukherjee:2021xdt,Antelis:2021qak,Perez2023}. This approach is currently being used for the analysis of the O4 \acrshort{lvk} data as a modification of the cWB and PySTAMPAS pipelines that allows to improve the false alarm rate of the detected events. 

{\it Waveform reconstruction}: Once the signal has been detected one may attempt to reconstruct the signal based on a model independent algorithm. For the case of \acrshort{ccsn}e this has been developed with cWB \cite{Szczepanczyk:2021bka}, MuLaSECC \cite{Mukherjee:2021xdt}, Bayeswave \cite{Raza:2022kcs} and an ensemble of empirical mode decomposition \cite{Yuan:2023umh}. To get a waveform overlap better than $90\%$ in the reconstructed signal one needs typically \acrshort{snr} values of $\sim 40-100$.

{\it Model classification}: The detection of a \acrshort{gw} signal from a \acrshort{ccsn} might provide an insight on the mechanism driving the explosion. To this purpose a number of data analysis methods have been developed with the aim at classifying \acrshort{ccsn}e signals in categories, mostly between neutrino driven SN and magneto-rotational SN. Refs.~\cite{Logue:2012zw,Powell:2016wke,Powell:2017gbj} have developed a classification pipeline (Supernova Model Evidence Extractor, SMEE) based on Principal Component Analysis of the signal, capable of performing Bayesian model selection. They showed that this classification is possible for sources up to the LMC with second generation detectors. Ref.~\cite{Saiz-Perez:2021bce} used supervised dictionary-learning for the task obtaining a performance comparable to SMEE. Machine learning has also been used in this context \cite{Chan:2019fuz,Iess:2023quq}. Ref.~\cite{Powell:2023bex} performed a comparison of different classification methods including all the mentioned above. For \acrshort{et},  they conclude that for signals with \acrshort{snr} in the range $25-40$ is possible to classify between neutrino-driven and magneto-rotational. In their work they also include a class for non-exploding models. They show that it is generally difficult to distinguish between neutrino-driven and non-exploding models, unless the \acrshort{snr} is high and \acrshort{sasi} is strong. A different approach to distinguish the presence of strong rotation is to check the presence of circular polarization in the Stokes parameters of the signal \cite{Hayama:2016kmv,Chan:2020gnx}. They have shown that circular polarization should be detectable up to $5$~kpc with current detectors. Another feature that can benefit of classification algorithms is the presence of \acrshort{sasi}. Ref.~\cite{Lin:2022jea}
used the modified constrained likelihood method, in combination with neutrino detector data, to detect the presence of \acrshort{sasi}, reaching a detection distance of about $5$~kpc with current \acrshort{gw} and neutrino detectors.

{\it \acrshort{pns} asteroseismology:} Among the most promising perspectives in \acrshort{ccsn}e observations by \acrshort{gw}s is the detection of oscillation modes of the \acrshort{pns}. \acrshort{pns} asteroseismology can allow measuring the properties of the nascent \acrshort{ns} through the universal relations existing between the frequency of the modes and combinations of mass and radius of the \acrshort{pns} (see discussion in section~\ref{sec:CCSNeMM}). Refs.~\cite{Bizouard:2020sws,Powell:2022nrs,Bruel:2023iye} have shown that is possible to reconstruct the time evolution of the \acrshort{pns} surface gravity ($M_{\rm PNS}/R_{\rm PNS}^2$) from \acrshort{gw} data alone. For \acrshort{et}, inference is possible up to hundreds of kpcs \cite{Bruel:2023iye}. 
Additionally, ref.~\cite{Afle:2023mab} has shown that, with third generation detectors, it will be possible to reconstruct the frequency track of the dominant oscillation mode with a $10\%$ accuracy for a galactic SN. Instead of trying to reconstruct the full track of the oscillation mode, some work has also be done trying to infer the frequency and slope of the dominant mode after bounce  using machine learning techniques~\cite{Lagos:2023qli}. The use of other oscillation modes, beyond the dominant emission, is still rather unexplored, mainly because they are weaker. Those modes could contain complementary information to the dominant mode and are certainly a target for \acrshort{et} in case of a galactic SN.  In this context, ref.~\cite{Lin:2022jea} explored the possibility of measuring the frequency and duration of the \acrshort{sasi} component. Many of the algorithms mentioned require the reconstruction of the frequency track of the modes in spectrograms. Two aspects of these analysis could be be improved in the future: i) the error associated to the tracking algorithm itself is poorly understood and could be the dominant source of error in the inferred parameters, in particular for weaker features such as \acrshort{sasi}. A possible solution could be the use of Bayesian inference methods, but also the use of error propagation techniques such as Gaussian processes. ii) Tracking algorithms might depend on the way the spectrogram is constructed. Alternative methods such as the Hilbert-Huang transform \cite{Takeda:2021hmf}, the S-method \cite{Kawahara:2018xjl} or the Prony method (\cite{Maione:2017aux} applied  to binary \acrshort{ns} mergers (\acrshort{bns}) extract directly the frequency evolution of different components in the signal and could be of use here.

{\it Bounce signal and rotation:} In the case of fast rotating progenitors the signal at bounce is very characteristic, linearly polarized and its amplitude is related to the rotation rate of the \acrshort{pns} (see section~\ref{sec:CCSNeMM} for a longer discussion). A variety of methods have been explored to try to determine the rotation rate of the \acrshort{pns} using the bounce signal, including Principal Component Analysis \cite{Rover:2009ia,Abdikamalov:2013sta,Engels:2014nua}, coherent network analysis \cite{Hayama:2008ws}, Bayesian parameter inference \cite{Afle:2020zgd,Pastor-Marcos:2023tcc}, maximum likelihood estimator \cite{Villegas:2023wsu} and neural networks \cite{Nunes:2024fig}. One of the main drawbacks of this approach is that the rotation rate is degenerated with the distance and source inclination. Although the distance can be easily estimated if an \acrshort{em} counterpart is present, the inclination angle is more challenging and specific methods should be developed (see discussion in \cite{Pastor-Marcos:2023tcc}).
Additionally, the frequency of the bounce signal is related to the density in the \acrshort{pns} and might be used to constraint the EOS of dense matter \cite{Rover:2009ia,Edwards:2020hmd,Afle:2020zgd,Mitra:2022nyc}. 

{\it Collapsars:}  Recent numerical simulations reported in  \cite{Gottlieb:2024dfw} suggest that cooled accretion disks formed during the collapse of a star with high mass and with sufficient angular momentum in the core, the so-called collapsars, can produce strong, coherent gravitational wave (GW) signals via Rossby wave instabilities. These GW emissions, peaking around 100 Hz, fall within the most sensitive frequency band of Einstein Telescope. With predicted signal amplitudes and event rates much higher than those of core-collapse supernovae, collapsar disks represent one of the most promising new classes of burst-type GW sources for third-generation detectors such as the Einstein Telescope.

\begin{figure}[t]
	\centering
	\includegraphics[width=0.9\textwidth]{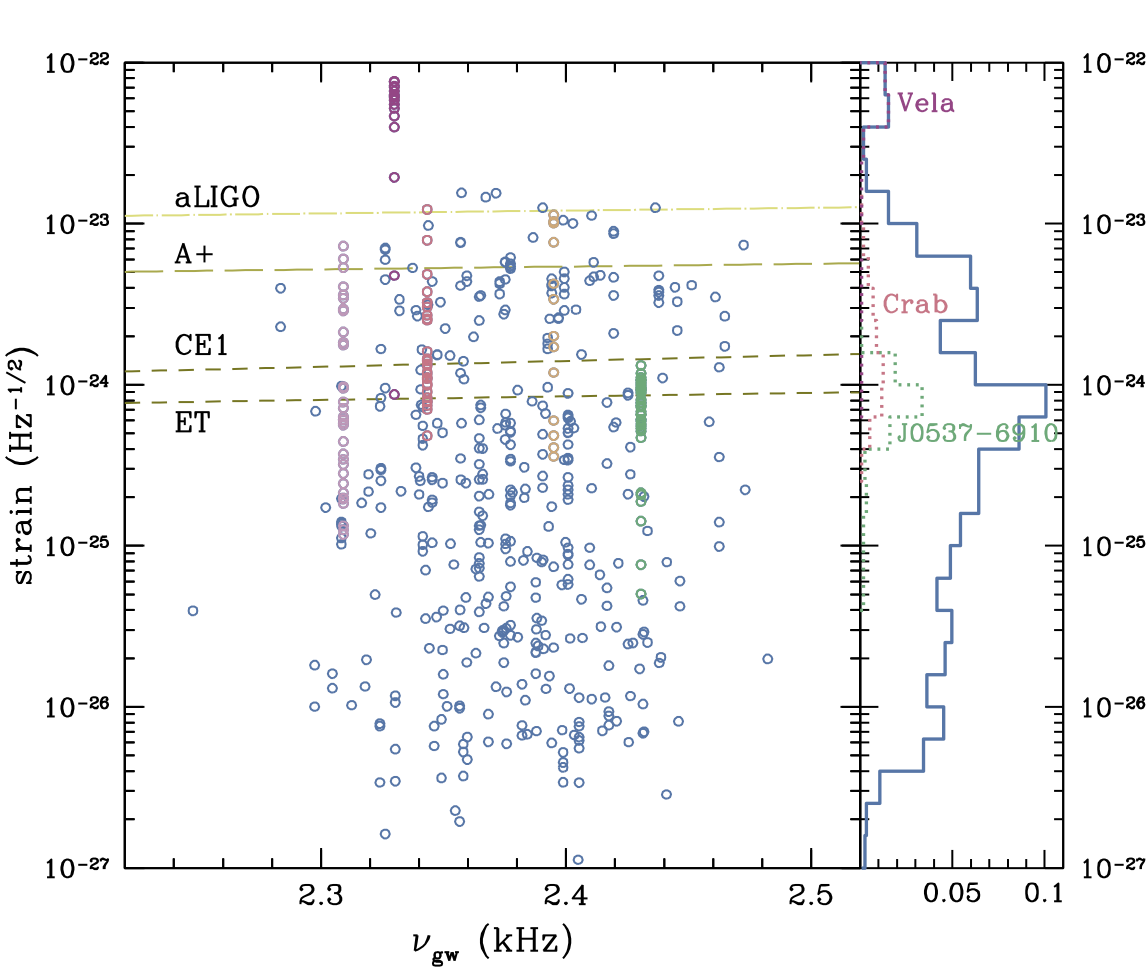}
		\caption{Each circle represents a value of $h_0\sqrt{\tau_\text{GW}}$ for $f$-modes excited by glitches of different pulsars and the sensitivity of several \acrshort{gw} detectors are plotted as horizontal dashed lines. Points that lie above the sensitivity curve have signal-to-noise ratio estimates greater than 1. Typically, one would require a signal-to-noise greater than $\sim$10 before a detection is classified as astrophysical. The ET sensitivity curve has been computed assuming the 2L 15 km configuration. The triangular 10 km configuration curve is not be distinguishable on the plot scale. Adapted from \cite{Ho:2020nhi}.}
		\label{fig:f_mode_prospects}
\end{figure}

\paragraph{Pulsar glitches.} 
As mentioned in section~\ref{sec:cwglitch}, we also expect burst signals from pulsar glitches. It is believed that pulsar glitches can excite the $f$-mode which in turn can radiate away \acrshort{gw}s in a burst-like way \cite{Keer:2014uva, Ho:2020nhi, Yim:2022qcn}. One can calculate the maximal energy available for \acrshort{gw} radiation and relate it to the signal-to-noise ratio that a \acrshort{gw} detector would detect. From this energy-based argument, one finds that next-generation detectors should be able to detect signals from $f$-modes excited by pulsar glitches, as seen in figure~\ref{fig:f_mode_prospects} \cite{Ho:2020nhi, Yim:2022qcn, Wilson:2024tcc}. Since the exact $f$-mode frequency depends on mass and radius, a successful detection offers the exciting prospect of probing the elusive EOS \cite{Andersson:1997rn, Wilson:2024tcc}. There is also the possibility to detect burst signals from small spin-up and spin-down events, which are similar to glitches (and antiglitches) but are much smaller in magnitude, and it is thought that the succession of these events could potentially contribute towards timing noise \cite{Yim:2022qcn}. Although it may be difficult to detect \acrshort{gw}s from individual spin-up or spin-down events, the accumulation of signal power from several events could yield a detection with next generation detectors. However, before this becomes a reality, more work needs to be done on stacking signals together so as to ensure maximal extraction of signal from the data.

\paragraph{Magnetar bursts.}
As discussed in section~\ref{sec:magnetars}, some of the magnetars have a flaring activity. \acrshort{gw} emission associated with these bursts is the subject of a targeted search, since we know both the source sky location and the time of the burst with a very good accuracy. Few known magnetars have a regular flaring activities and \acrshort{lvk} data are regularly searched with standard unmodeled search pipelines. The search sensitivity is rather limited. For instance in \cite{LIGOScientific:2022sts} the search was sensitive to the ratio of the \acrshort{gw} energy to \acrshort{em} energy of $3\times 10^3$\, for SGR 1935+2154. The main difficult of this search is the weakness of the \acrshort{gw} bursts, whose energy is expected to be a small fraction of the \acrshort{em} burst energy. To enhance the \acrshort{snr}, stacking methods have been proposed such as \cite{Kalmus:2009uk}. The \acrshort{snr} enhancement scales with the number of flares. With \acrshort{et}, the search sensitivity for a single burst search will be 10 times higher. It is expected that with stacking methods we could be sensitive to \acrshort{gw} energy weaker than the \acrshort{em} energy. The repeated flaring activity from individual  magnetars would also have characteristic signatures in terms of the energy distribution of the flares, and the waiting times between  individual events~\cite{Dubath:2004vv}.

Yet, magnetar giant flares are maybe the most promising source of \acrshort{gw}s from magnetar flaring activities. Those events are rarer but more energetic. With current generation detectors, searches are sensitive to the ratio of the \acrshort{gw} energy to \acrshort{em} energy of the order of $0.1$ for magnetar giant flare at 10 kpc. With \acrshort{et} we could probe this ratio in the regime of $10^{-5}- 10^{-4}$ for a galactic magnetar giant flare \cite{Macquet:2021eyn}.


\subsubsection{Continuous waves and long transients}
\label{sec:Prospects:CW}

In general, the class of \acrshort{cw} comprises all deterministic \acrshort{gw} signals that last much longer than typical transient signals, with duration from thousands of seconds to years. No \acrshort{cw} has been detected so far, although notable 
 upper limits on the star's ellipticity  have been established, see e.g. \cite{Riles:2022wwz,Piccinni:2022vsd,Tenorio:2021wmz} for recent reviews. The prototypical source of \acrshort{cw}s is a spinning \acrshort{ns}, asymmetric with respect to the rotation axis, that emits a quasi-monochromatic \acrshort{gw}s whose frequency slowly changes over time. For such sources, the long signal duration implies that the frequency modulation due to the Doppler effect, caused by the motion of the Earth, and the sidereal modulation due to the detector directional response, play a relevant role. 
 Doppler modulation is given by the following equation, which relates the emitted signal frequency $\nu_{0}$ to the frequency at the detector $\nu$:
 \begin{equation}
     \nu(t) = \nu_{0}\left(1+\frac{\vec{v}(t)\cdot \hat{n}}{c}\right),
     \label{eq:dopp}
 \end{equation}
 where $\vec{v}$ is the detector time-dependent velocity and $\hat{n}$ is the versor toward the source position in the solar system barycenter reference frame. The sidereal modulation, on the other hand, implies that the received signal is a linear combination of the wave plus and cross components,
 \begin{equation}
h(t)=F_+(t) h_+(t)+F_{\times}(t) h_\times(t),
\label{eq:sider}
 \end{equation}
 where the detector beam-pattern functions, $F_+,~F_\times$, are a combination of periodic functions with period of one and half sidereal day, with coefficients depending on the detector location and orientation and the position and polarization angle of the source. Such amplitude modulation, for signal lasting more than about one sidereal day, produces a splitting of the signal power at the five frequencies $\nu\pm \vec{k}\nu_\mathrm{sid}$, with $\vec{k}=-2,...+2$, being $\nu_\mathrm{sid}\simeq 1.1606\times 10^{-5}$ Hz the Earth sidereal frequency. 
 The overall observed phase evolution then depends on the intrinsic signal frequency, frequency time derivatives, and source sky position and polarization. If the source is located in a binary system, there is a further frequency modulation caused by the source orbital motion, usually described by five unknown Keplerian parameters. The frequency shift resulting from the source binary orbit is typically more significant, and the computational challenge of searching for the additional five parameters describing the binary orbit is substantial~\cite{Leaci:2015bka,Leaci:2016oja}. Hence, these Doppler effects must be accurately taken into account at the analysis level, to prevent the signal power from being spread across many frequency bins, resulting in a significant \acrshort{snr} loss.


Various techniques have been devised to search for \acrshort{cw} signals, each reflecting diverse approaches to detect these persistent signals. These methods prioritize either the pipeline sensitivity or robustness against signal deviations from the assumed model while keeping the computing cost still accessible, see recent reviews \cite{Tenorio:2021wmz,Riles:2022wwz}.

Matched filtering stands out as the optimal search method for these signals when the waveform is completely understood and under the assumption of Gaussian noise. Typically matched filter is applied in known pulsar searches, where all source parameters are identified. These parameters may be obtained from \acrshort{em} observations (typically radio, gamma, or X-rays), providing information such as the source's sky position, its rotational parameters and Keplerian parameters for \acrshort{ns}s in binary systems. In such case, fully coherent targeted searches are conducted. Various implementations of matched filtering techniques are available, utilizing different detection statistics or algorithms. See \cite{LIGOScientific:2021hvc} for the latest O3 results using LIGO and Virgo data.

Also searches covering a larger parameter space, with a corresponding increase in the computational cost,
are regularly conducted. For instance, narrow-band searches expand a bit the parameter space around the \acrshort{em}-inferred parameters of known pulsars to accommodate potential small deviations, such as fractions of a Hz, between the \acrshort{em}-inferred rotational and \acrshort{gw} frequencies \cite{LIGOScientific:2021quq}.
Cases of interesting targets or specific sky regions, where the sky position is assumed as known, are referred to as directed searches. In this case,  the star's rotational parameters are often unknown, as in the case of SN remnants or the galactic centre \cite{LIGOScientific:2021mwx, KAGRA:2022osp}. This class also includes accreting \acrshort{ns}s, like LMXBs, among which Sco-X1 has a prominent role. It is the most luminous system and potentially, if the torque balance condition is met, it could generate a large \acrshort{gw} strain, see \eq{eq:accreting}. Even though the sky position is known, the search for \acrshort{cw}s from LMXBs is complicated by the need to cover a large frequency range, and by the possibility that matter accretion from the companion star induces fluctuations in the spin period (\textit{spin wandering}) which may significantly affect the search sensitivity. Recent searches of LIGO-Virgo-KAGRA data started to constrain the torque balance limit over some frequency range \cite{LIGOScientific:2022enz,Whelan:2023bha}. 

All-sky searches for \acrshort{ns}s lacking \acrshort{em} counterparts have been developed to explore as much parameter space as possible. Given their computational limitations, these searches employ hierarchical semi-coherent methods, trading off some sensitivity for computational feasibility. Compared with known pulsar searches, most searches are robust against possible deviations from the assumed emission models despite the lowest sensitivity achieved. The latest LIGO-Virgo-KAGRA results of such all-sky searches can be found in \cite{KAGRA:2022dwb}. 


Besides conventional continuous waves, efforts have also been made to detect long transient signals (often called ``transient continuous waves'' - \acrshort{tcw}), which are shorter in duration  - typically O(minutes-days) - but not as short as burst signals. The signals can be modeled or unmodeled and that choice determines whether a semi-coherent or a cross-power accumulation (incoherent) search is used. Triggers of long transients include pulsar glitches \cite{Haskell:2023exo} 
and magnetar flares \cite{Kashiyama:2011fs} and, so far, searches for coincident signals have not resulted in detection, leading to upper limits on the \acrshort{gw} strain as a function of signal duration or upper limits in the root-sum-squared strain \cite{LIGOScientific:2021quq, LIGOScientific:2022sts}. 
Newborn fast-spinning magnetars, produced in core collapses or merger of \acrshort{ns} binary systems, see section~\ref{sec:cwmagn}, represent one potentially very interesting source of \acrshort{tcw}s. The strong inner magnetic field of magnetars can induce a distortion of the star's shape which, if not aligned with the rotation axis, causes a powerful emission of \acrshort{tcw}s \cite{2001A&A...367..525P,DallOsso:2008kll,DallOsso:2018dos}.
The downside is that a strong magnetic field would contribute to a rapid spin-down of the star due to \acrshort{em} radiation, therefore such \acrshort{gw} emission will only take place over a timescale of hours or days. Some fraction of \acrshort{ns}s, probably higher than 10$\%$, are likely to be born as magnetars following the collapse of a massive progenitor star \cite{Kouveliotou:1998ze,Woods:2004kb,Beniamini:2019bga}. A search for \acrshort{gw}s from a possible long-lived remnant of GW170817 has been conducted in O2 data, using four different semi-coherent methods \cite{LIGOScientific:2018urg}. A horizon distance of less than $\sim$0.5 Mpc has been reached, about two orders of magnitude smaller than the estimated remnant distance ($\sim 40$Mpc). Even considering magnetars formed in stellar collapses, which may happen once every few years within 5-10 Mpc \cite{Beniamini:2019bga,DallOsso:2021xbv}, it is clear that the sensitivity improvement provided by \acrshort{et} will need to be accompanied by a significant increase in the analysis pipeline sensitivity. 

All these sources share the characteristic that, once detected, they can be monitored over extended periods, transforming them into a valuable laboratory for fundamental physics and astrophysics. Additionally, the persistence of the signal helps to reduce virtually to zero the false alarm probability for any candidate. If a signal is present in a dataset, even with low significance, it will reappear in new datasets, increasing in significance until a detection can be confidently claimed (though this is not always true for \acrshort{tcw}s). Moreover, \acrshort{cw} source parameters can be measured with great precision, as uncertainties decrease with longer observation times. Finally, this long-term observation can potentially reveal minute deviations from assumed models, such as non-standard polarizations indicating deviations from General Relativity (see \cite{Will:2014kxa}).

However, \acrshort{cw} signals are predicted to have small amplitudes compared to typical CBC signals and searching for them can be computationally intensive. For this reason, developing more sensitive, robust, and computationally efficient algorithms is a highly active area of research, progressing alongside upgrades to current detectors and R\&D for future detectors like \acrshort{et}.



In the following, we discuss \acrshort{et} detection prospects for different classes of \acrshort{cw}/\acrshort{tcw} sources, in some cases relying on \cite{Branchesi:2023mws} and in other cases presenting new results.

Concerning searches for known pulsars, we report in  table~\ref{tab:pulsar_eps} the number of potentially detectable sources (among those contained in the ATNF catalogue\footnote{v. 2.4.0, \url{https://www.atnf.csiro.au/research/pulsar/psrcat/}}), for both the 10 km triangle and the 15 km 2L configurations. As mentioned above, known pulsars are searched for using a fully coherent method based on matched filtering. Here we assume taking one year of data at a duty cycle of 85\%.  We consider three different scenarios for the ellipticity, $\epsilon_1=\epsilon_{sd}$ (second column), meaning the sources emit at their spin-down limit, i.e. all their spin-down is due to the emission of \acrshort{gw}s, see \eq{eq:eps_sd}. 
The second scenario (column three) assumes an ellipticity $\epsilon_2$ equal to the minimum between the spin-down limit value $\epsilon_{sd}$ and $10^{-6}$, which is of the order of the maximum theoretical ellipticity a \acrshort{ns} with a standard EOS could sustain before breaking off \cite{Morales:2022wxs}. Finally, we also consider the case $\epsilon_3$ of minimum ellipticities between $\epsilon_{sd}$ and $10^{-9}$ (fourth column), which some works suggest as a possible ellipticity value for millisecond pulsars \cite{Woan:2018tey,Soldateschi:2021hfk}.
For each case, we also report the median value of the ellipticity for the potentially detectable sources. Due to the implicit averaging over time of \acrshort{cw} searches, the results depend very weakly (within less than 1$\%$) on the relative orientation of the two detectors.

It is important to note that $\epsilon_2$ and $\epsilon_3$ are more realistic estimates for the ellipticities of pulsars. Most potentially detectable sources are millisecond pulsars or young, fast-spinning pulsars with frequencies exceeding a few tens of Hertz. For these frequencies, the detectors' low-frequency sensitivity is not particularly relevant. 

\begin{table}[t!]
   \centering
    \begin{tabular}{|c|c|c|c|}
    
\hline
    Configuration  &   $n_1~(\epsilon_{\rm med})$ & $n_2~(\epsilon_{\rm med})$ & $n_3~(\epsilon_{\rm med})$ \\ \hline
    \hline
   $\Delta$ 10km & 866 ($1.3\times 10^{-4}$) & 180 ($4.4\times 10^{-9}$) &  19  ($7.5\times 10^{-10}$)\\
   2L 15km & 959  ($1.2\times 10^{-4}$) & 206   ($4.2\times 10^{-9}$) &  29 ($8.1\times 10^{-10}$) \\
   \hline
    \end{tabular}
    \caption{\small Expected number of detectable sources, including binaries, by two different detector configurations (10 km triangle and 15 km 2L), assuming three different conditions for the ellipticity:
    $\epsilon_1=\epsilon_{sd}$, $\epsilon_2=min(\epsilon_{sd},~10^{-6})$, $\epsilon_3=min(\epsilon_{sd},~10^{-9})$, assuming a total observation time $T_\mathrm{obs}=1$ year and a duty cycle of 85$\%$. The spin-down limit on ellipticity, $\epsilon_{sd}$, is given by \eq{eq:eps_sd}. For each case, we give in parentheses the median $\epsilon_{\rm med}$ values of ellipticity for detectable signals.\\} 
    \label{tab:pulsar_eps}
\end{table} 
For the most conservative scenario, the typical ellipticity values for detectable sources are of the order of few times $10^{-10}$, corresponding to ``mountains'' on the star's surface that are not taller than a few microns.

\begin{figure}[t]
	\centering
		\includegraphics[width=0.9\textwidth]{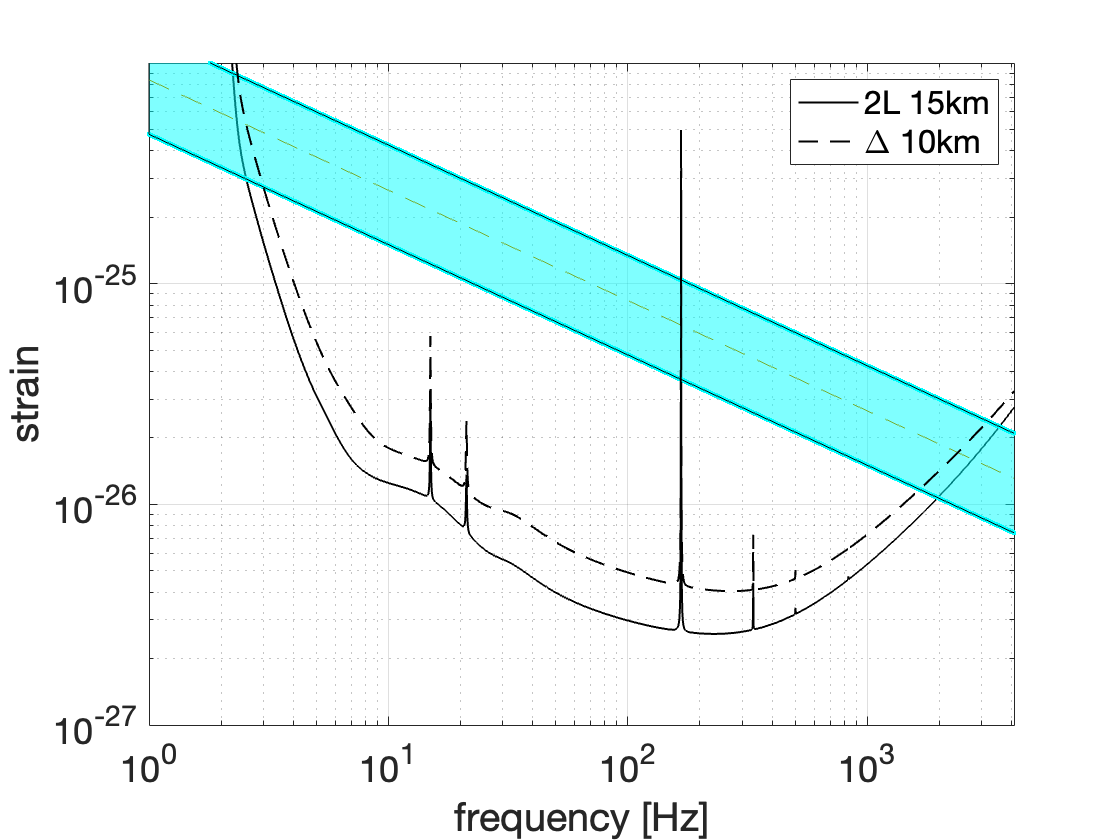}
		\caption{Estimated \acrshort{et} sensitivity (solid line: 2L 15 km configuration, dashed line: triangular 10 km configuration - 1 year of data, duty cycle 85$\%$, drift time of 10 days) to \acrshort{cw} emission from Sco-X1 using the semi-coherent Viterbi algorithm \cite{Sun:2017zge}. The transversal orange dashed line represents the standard torque balance limit (with an average over the star rotation axis inclination), assuming surface accretion. The cyan band indicates torque balance strains from circular polarization (lower) to linear polarization (upper).}
		\label{fig:scox1}
\end{figure}

Moving to accreting \acrshort{ns}s, \acrshort{et} will significantly improve the perspectives for  detection of \acrshort{cw}s from LMXBs (such as Sco-X1) over a broad frequency range, even if the emission occurs at an amplitude below the torque balance limit. Figure~\ref{fig:scox1} shows the expected sensitivity of a semi-coherent search for Sco-X1 using the Viterbi algorithm, able to take into account random frequency fluctuations \cite{Sun:2017zge} (both with the 2L 15km detector configuration and the 10km triangle configuration, 1 year of observation time, duty cycle of 85$\%$ and a drift time of 10 days, compatible with the characteristic time-scale of the random walk inferred from accretion-driven fluctuations in the observed x-ray flux of Sco X-1 \cite{Mukherjee:2017qme}), compared to an estimate of the torque balance limit (assuming the accretion occurs at the star's surface), given by \eq{eq:accreting}.
However, as already noted in section~\ref{sec:cwlmxb}, the torque balance limit is rather uncertain, potentially varying by up to an order of magnitude due to factors such as the star magnetic field topology, viscous heating, X-ray emission efficiency, and the effects of radiation pressure \cite{LIGOScientific:2022enz}. Searches for Sco-X1, and other LMXBs, would receive a significant sensitivity boost if more accurate measures of the main parameters were available from \acrshort{em} observations \cite{Riles:2022wwz}.    

\begin{figure}[t]
	\centering
		\includegraphics[width=0.9\textwidth]{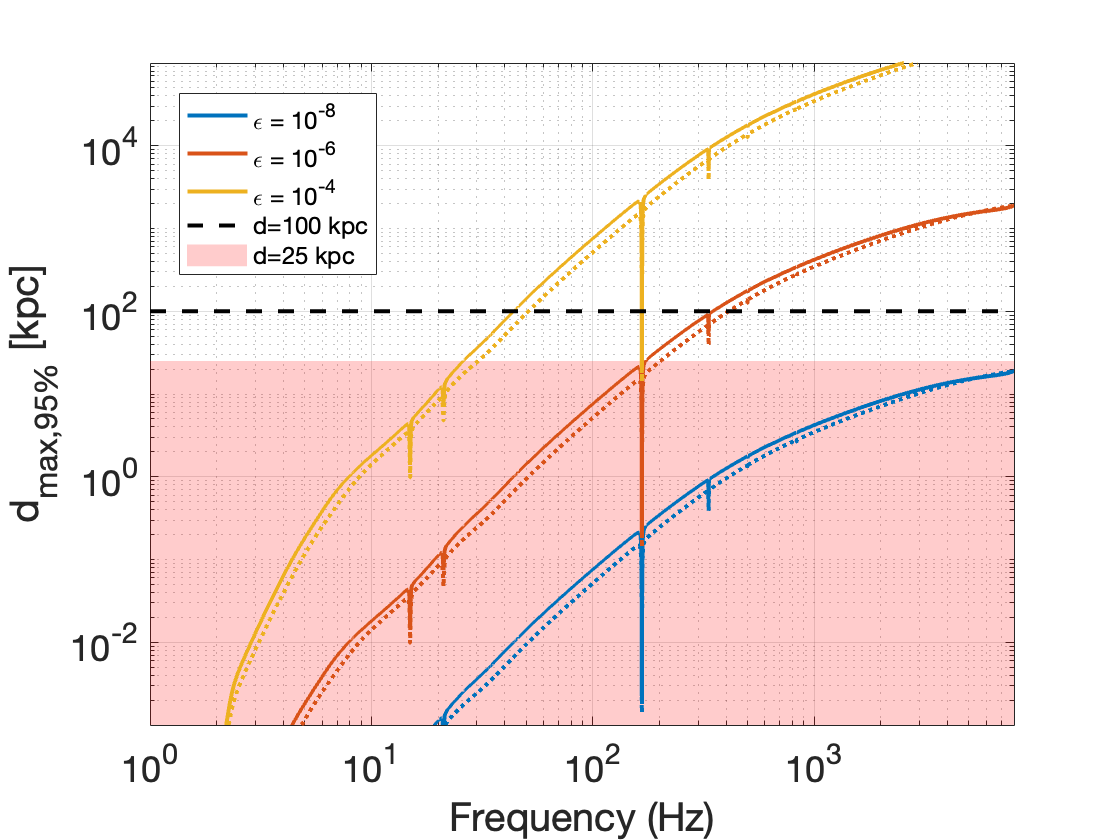}
		\caption{Maximum distance reached for an all-sky search of rotating \acrshort{ns}s, based on the FrequencyHough transform. The shaded area roughly covers the Milky Way extension (set at 25 kpc). The black dashed line represents a distance of 100 kpc. We assumed three plausible ellipticity values (and a realistic analysis setup, see main text). Two configurations, one consisting of two L-shape 15km arms detectors (continuous lines) and one with a triangular 10 km arms detector (dashed lines), are considered. In both cases, one year of observation time and a duty cycle of 85$\%$ is assumed.}
		\label{fig:d_allsky}
\end{figure}

Concerning all-sky searches for \acrshort{ns}s without an \acrshort{em} counterpart, figure~\ref{fig:d_allsky} shows the astrophysical reach of the semi-coherent FrequencyHough pipeline \cite{Astone:2014esa}, routinely used in current \acrshort{cw} searches, assuming the \acrshort{et} design noise curve $S_n(f)$. 
The astrophysical reach can be obtained by recasting the 95\% confidence level pipeline's sensitivity $h_{\rm min, 95\%}$ in terms of the distance $d_{\rm max, 95\%}$ as a function of the ellipticity by inverting \eq{eq:StrainAmplitude_div7}. Such distance is given by: 
\begin{equation}
d_{\rm max, 95\%}= \frac{4\pi^2G\epsilon I_{3}\nu^2}{h_{\rm min, 95\%} c^4} \approx \frac{4\pi^2G\epsilon I_{3}\nu^2}{c^4}\frac{N^{1/4}}{4.97}\sqrt{\frac{T_{\rm FFT}}{S_n(f)\left(CR_{\rm thr}+1.6449\right)}}\, .
\label{eq:distance_allsky}
\end{equation}
This expression assumes that the detector noise is Gaussian; it depends on $N$, the number of data segments of duration $T_{\rm FFT}$ (coherence time) to be incoherently combined, and on the Critical Ratio threshold used to select outliers, ${\rm CR}_{\rm thr}$. For figure~\ref{fig:d_allsky} we assume a data-taking period of $T_{\rm obs}$=1 year, $T_{\rm FFT}$=10 days, hence $N=T_{\rm obs}/T_{\rm FFT}$, and a duty cycle of 85\%.  The Critical Ratio threshold is set equal to 4. A standard value $I_3=10^{38}~\rm kg~m^2$ has been taken for the star's moment of inertia. Two configurations, one consisting of two L-shape 15km arms detectors and one with a triangular 10 km arms detector, are considered. With this search setup we can reach the center of the galaxy (at about 8 kpc) for sources emitting at frequencies above 100 Hz and ellipticities larger than $10^{-6}$. 
For ellipticities of the order of $10^{-8}$, \acrshort{et} would be able to detect galactic sources at a distance larger than 100 pc, provided they emit signals at frequencies higher than 100 Hz (and up to about 20 kpc for neutron stars spinning near the break-up limit). 

Prospects of detecting \acrshort{tcw}s from pulsar glitches with \acrshort{et} are promising, with up to 40 percent of all glitches from pulsars in our Galaxy being detectable by ET (with \acrshort{snr} $> 10$) if one assumes all the glitch energy goes into generating \acrshort{gw}s \cite{Yim:2020trr, Moragues:2022aaf, Yim:2024eaj}. This result comes from the analysis of historic glitch data. Looking specifically at the glitches of the Crab and Vela pulsars, one expects signal-to-noise ratios of $\sim50$ and $\sim1000$, respectively, for \acrshort{et} \cite{Yim:2020trr}. This raises the question of whether these types of signals could be detectable by current detectors. Calculations show this is indeed possible, if a glitch were coincident with one of the observing runs, like the most recent April 2024 Vela glitch \cite{Yim:2024eaj}. The assumption that all the glitch energy goes into the production of \acrshort{gw}s is an optimistic assumption. However, the non-detection of a supposedly detectable signal would give an upper limit on how much glitch energy can be radiated away as GWs, much like the upper spin-down limit for quasi-infinite \acrshort{cw}s.

\begin{figure}[t]
	\centering
		\includegraphics[width=0.9\textwidth]{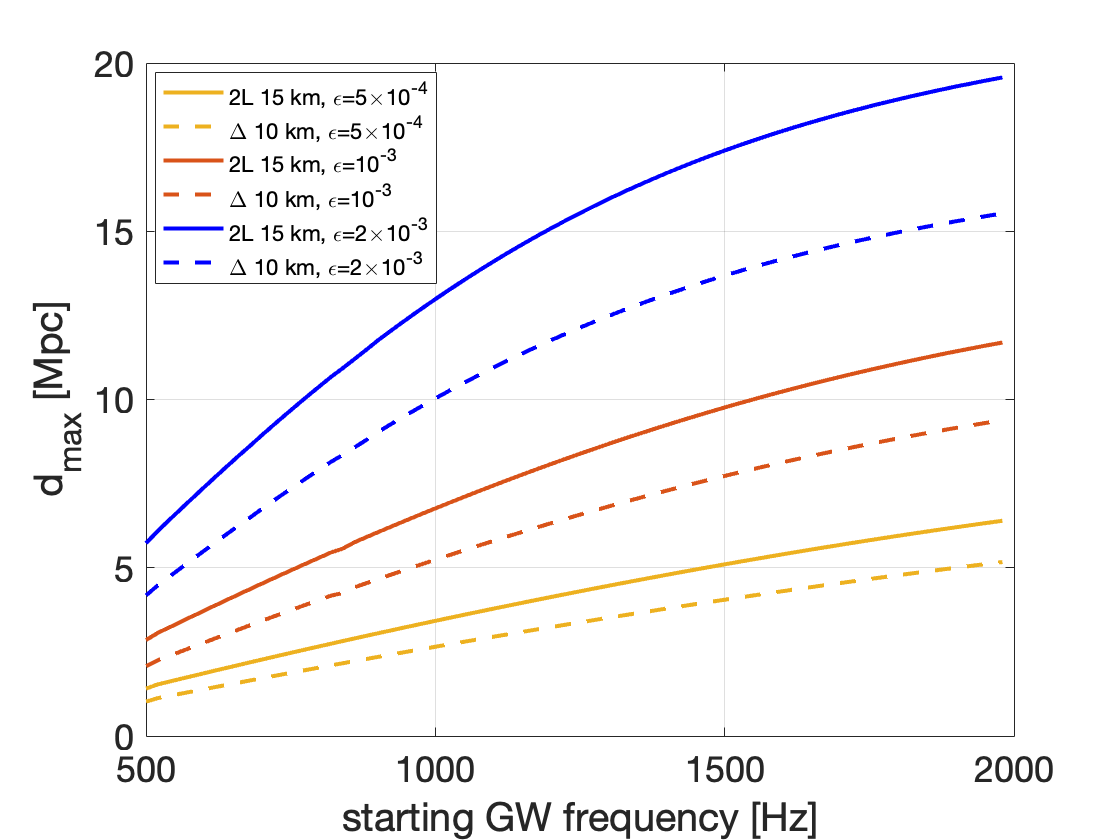}
		\caption{Maximum distance at which \acrshort{tcw} emission from a newborn magnetar could be detected by \acrshort{et} for three different plausible values of the ellipticity (and realistic analysis setup, see text for more details). Two configurations, consisting of two L-sphape 15km arms or one triangular 10 km arms detectors, are considered.}
		\label{fig:dmax_magnetar}
\end{figure}

In relation to the search for \acrshort{tcw} signals emitted by newborn magnetars, figure~\ref{fig:dmax_magnetar} shows the distance reach with \acrshort{et} for a method based on the Generalized FrequencyHough (GFH) \cite{Miller:2018rbg}, assuming data segments of duration 100 seconds are combined over an observational window of 5 hours, as a function of the initial signal frequency and the ellipticity induced by the inner magnetic field. Both the 10 km triangle and the 15 km 2L detector configurations are considered.
As already said, reaching distances of $5-10$~Mpc, for moderately large ellipticities, would correspond to an expected rate of one newborn magnetar every few years.    
While current methods are limited to segment duration of few seconds, an intense effort toward more sensitive approaches is being carried on, so reaching durations of hundreds of seconds is considered very likely by the time \acrshort{et} will start taking data. 
In parallel to the improvement of classical semi-coherent methods, new methodologies based on ML techniques are also being explored, and have been shown to have a comparable sensitivity at a lower computing cost \cite{Miller:2019jtp,Modafferi:2023nzt,attadio2024neuralnetworksmethodsearch}. The use of cutting-edge image processing techniques looks very promising too, see e.g. \cite{Pierini:2020ulw}, as a way to pre-process data in order to increase the initial \acrshort{snr}. Overall, this is a field in which significant progresses are expected in coming years. 
In parallel, adopting a multi-messenger approach could significantly boost chances of detection. Indeed, as shown e.g. in \cite{Menon:2023sfa}, \acrshort{em} observations of the shock breakout emitted during the collapse of massive progenitor stars can provide crucial information about the newly born magnetars, thus allowing to make deeper searches over a restricted parameter space. 

\subsubsection{Synergies with neutrino and electromagnetic observatories}
\label{sec:mm}
Concurrent multi-messenger observations of a core-collapse supernova would address many long-standing issues regarding the physical processes during stellar explosions, see e.g. \cite{Fryer:2023ehc}. Historically, the first multi-messenger observations of a \acrshort{sn} were obtained for SN 1987A in the \acrshort{lmc}. 
The detection of 25 MeV neutrinos shortly before the optical breakout definitively proved that the supernova began with the collapse of a stellar core into a compact remnant \cite{Hirata1987}. 
This nearby event allowed for unique multi-wavelength observations, including the identification of the progenitor as a massive star \cite{1987Natur.328..318G} and the detection of gamma-rays from the radioactive decay of $^{56}$Ni \cite{1996ssr..conf..201B} much earlier than expected. This finding is inconsistent with a model of spherically symmetric ejecta expansion.
SN1987A was both radio-loud and aspherical, featuring relativistic radio jets and possible \acrshort{ns} or \acrshort{bh} remnant. Initially, the black hole scenario was supported by the absence of a neutron star detection, however, recent \acrshort{jwst} \cite{Gardner:2006ky} observations of high-ionization emission lines in the center of the SN 1987A remnant appear to have resolved this ambiguity indicating the presence of a source of ionizing photons, probably a \acrshort{ns} \cite{Fransson:2024csf}.

The detection of \acrshort{gw}s  from a nearby core-collapse would be immensely valuable. \acrshort{gw}s  are direct probes of the angular momentum in the collapsing stellar core and of convection processes in both the proto-neutron star and the convective engine region. They can also originate from highly asymmetric neutrino emission, and thus, even upper limits on \acrshort{gw} emission would add significant value to concurrent neutrino detections. See discussion in section~\ref{sec:CCSNeMM}.


There are two observing approaches that need to be implemented to ensure the synergy of multi-messenger probes.

The first one is to secure detailed multi-wavelength \acrshort{em} follow-up immediately after the detection of a burst of \acrshort{gw}s  or neutrinos. The infrastructure to guarantee this kind of observation strategy is already in place, including the rapid alert system of the current generation of \acrshort{gw} interferometers and neutrino detectors, as well as a wide network of \acrshort{em} facilities dedicated to target-of-opportunity observations \cite{LIGOScientific:2017ync} for these kind of astrophysical objects.

The second approach is instead driven by the discovery of a nearby core-collapse supernova observed as a transient \acrshort{em}. The search for gravitational waves and neutrinos can be triggered in the time window and sky position of the EM transient,  allowing for sub-threshold detections or, alternatively, for more stringent upper limits (e.g. \cite{Szczepanczyk:2023ihe}). In this scenario, it is crucial to assess the ability of \acrshort{em} observations to constrain the time of explosion.



\subsubsection*{Neutrinos Detectors}
\label{sec:neutri}

As discussed in sections \ref{corecollapse} and \ref{gwnuemission}, neutrinos play a fundamental role in the physics of a \acrshort{ccsn}. 
Thanks to their weak interaction cross section, they immediately escape from the innermost regions of the collapsing core just after the core bounce. 
Therefore, their detection provides the time of the bounce within few milliseconds, allowing to trigger the search for \acrshort{gw}s within an extremely precise time window \cite{Pagliaroli:2009qy}. 
The detection of neutrino emission is guaranteed for a Galactic \acrshort{sn} by a worldwide network of sensitive detectors able to provide fast alerts with a low false alarm rate. A cooperative effort between the neutrino detection experiments, known as the Supernova Neutrino Early Warning System (SNEWS) \cite{SNEWS:2020tbu}, has been established to exploit the combined potential of the neutrino detectors in operation around the globe. Through a coincidence-based approach,  the detection significance is increase by combining the observations of several neutrino detectors. The same approach can work for a combined network of \acrshort{gw}s and neutrino detectors, with an overall increase of the \acrshort{sn} detection capability\cite{Halim:2021yqa}.
 
The neutrino signal can be used to localize 
the \acrshort{sn} 
in the sky thanks to specific detection channels due to anisotropic neutrino interactions \cite{Tomas:2003xn} or triangulation techniques in a network \cite{Beacom:1998fj}. In particular, the neutrino elastic scattering (ES) on electrons shows a strong correlation with the direction of the neutrino flux. For a \acrshort{sn} distance of 10~kpc, the current Super-K, doped with Gadolinium, is expected to have a pointing accuracy of $3.3^{\circ} - 4.1^{\circ}$.    

Current and next-generation neutrino detectors have the capabilties to make high-statistics observations of the neutrino burst (see e. g. \cite{Scholberg:2012id} for a review), 
enabling the characterization of the neutrino light curves to study the effects of 
the standing accretion shock instability \cite{Lund:2010kh}, the lepton number emission self-sustained asymmetry \cite{Tamborra:2014aua}, or to determine the stellar core compactness \cite{Horiuchi:2017qlw} or the mass and radius of the resulting neutron star \cite{Nakazato:2020ogl}.
The detection horizon of the current generation of neutrino detectors includes the Milky Way and the Magellanic Clouds, while the next generation of neutrino detectors can reach Andromeda.

As a leading example, Hyper-Kamiokande (HK) \cite{Hyper-Kamiokande:2018ofw} with a fiducial mass of 187 kton of pure water is expected to observe more than 50 000 events in a temporal window of a few tens of seconds for a supernova distance of 10 kpc. 
With this statistic, the pointing capability of HK trough ES can reach 
$1^{\circ}-1.3^{\circ}$. 
For a \acrshort{ccsn} at the Andromeda distance (700 kpc) the number of events is still enough (tens) to identify the neutrino burst with high significance. HK is under construction near the town of Kamioka in Japan's Gifu Prefecture, and data-taking is scheduled to start in 2027.
Shortly, also the Jiangmen Underground Neutrino Observatory (JUNO), a 20 kton liquid scintillator detector, will start the data-taking. JUNO is expected be sensitive to  \acrshort{ccsn}e within $\sim300$ kpc.

While HK and JUNO are sensitive primarily to the scattering of $\bar{\nu}_e$ on protons, by the inverse beta decay (IBD) channel, the Deep Underground Neutrino Experiment (DUNE) \cite{DUNE:2020zfm}, made up of four 10 kton liquid argon time-projection chambers, will be sensitive to $\nu_e$ via $\nu_e$ charged-current interactions on argon nuclei. DUNE will provide high statistic and good reconstruction within a horizon of $\sim 100$ kpc.


In addition to the \acrshort{sn} neutrinos accompanying the star collapse, during the last years, the possibility of detecting pre-supernova neutrinos has also been investigated. 
Pre-supernova neutrinos are neutrinos of $\sim 0.1 - 5$ MeV emitted during the advanced nuclear burning stages of a massive star years/days before the onset of the core collapse \cite{Limongi:2000km,Mukhopadhyay:2020ubs}. As the central temperature and density increase during the last evolution stages, the $\nu_\alpha\bar{\nu_\alpha}$ pair production by 
electron-positron pair annihilation becomes the dominant mechanisms of energy loss. When the core silicon burning begins, the neutrino luminosity and average energy increase and the neutrino signal becomes potentially detectable in the hours prior to the collapse. In particular, detection of $\bar{\nu_e}$ through inverse beta decay allows the clear identification of such signal from nearby \acrshort{sn}e. 
The current generation of neutrinos detectors, such as KamLAND, Borexino, SNO+, Daya Bay and SuperKamiokande are already able to detect pre-supernova neutrinos in case of a \acrshort{sn} at a distance similar to that of Betelgeuse (see \cite{Kato:2020hlc} for a review).
The upcoming HK~\cite{Hyper-Kamiokande:2018ofw}, DUNE~\cite{DUNE:2020zfm} and JUNO~\cite{JUNO:2023dnp} will be able to reach $\sim1$ kpc away from Earth. 

Finally, the next-generation long-string water detectors, IceCube Gen-2 \cite{IceCube-Gen2:2020qha} and KM3NeT~\cite{KM3NeT:2021oaa}, will bring improved burst timing for a Galactic  \acrshort{ccsn}. These detectors primarily aim to detect GeV-PeV neutrinos, but are also sensitive to SN MeV neutrinos, searching for excess over the background expectation. Combining signals form different detectors and using only the IBD channel (IceCube, KM3NeT, HK and JUNO), triangulation enables to provide a sky-localization uncertainty of $\sim 70 \text{ deg}^2$  (at $1\sigma$ level) for a \acrshort{sn} distance of 10 kpc \cite{Coleiro:2020vyj}. The main drawback of detecting MeV neutrinos is the low rate of SN events within the Local Group of galaxies, approximately one event every 30 years. However, recently, 
new scenarios have expanded the detection horizon for high-energy neutrinos from kpc to 10-15 Mpc, making their detection a promising scientific case.  In recent years, an older hypothesis regarding the jet productionin CC-SN events has regained attention, particularly due to studies of long-duration \acrshort{grb}s. This hypothesis \cite{2019ApJ...871L..25P,Soker:2022vdg} suggests that jets may play a crucial role not only in the explosion of massive \acrshort{grb} progenitors but also in other \acrshort{sn}e. The idea is that jets might be produced and launched immediately after the core collapse. If the progenitor has lost its hydrogen and helium envelopes, the relativistic jet emerge from the stellar envelope resulting in a type Ic supernova associated with a long-duration \acrshort{grb}. Conversely, if the progenitor retains the hydrogen and helium envelopes, the jet is ``choked". It remains trapped within the huge envelopes, forming a ``cocoon" where significant energy is deposited. This energy is comparable to that of standard \acrshort{ccsn} and \acrshort{grb}-SNe. Supporting evidence for this scenario includes the 2019 first detection of a jet cocoon in \acrshort{sn} 2017iuk \cite{Izzo:2019akc}. Furthermore, internal shocks within the jet can accelerate protons, leading to TeV neutrino production, detectable with IceCube and Km3Net experiments~\cite{Guetta:2023mls}. A similar scenario for high-energy neutrino production, characterized by similar horizon distance, has been elaborated by \cite{Murase:2010cu} (see also \cite{Kheirandish:2022eox,Tsuna:2022llm}). These authors consider the collisions of the \acrshort{sn} ejecta with massive CSM shells as potential sources of accelerated protons that may interact with the protons present in the dense CSM producing high-energy neutrinos. 

\subsubsection*{Electromagnetic Facilities}

As extensively discussed in section~\ref{section:div4}, the observation of the first GW signal from a BNS merger
GW170817, together with the detection of EM signals across the entire electromagnetic spectrum, from gamma rays to the radio, marked the birth of multi-messenger astronomy including GWs, and highlighted the important scientific breakthrough achieved thanks to the synergy between GW detectors and EM facilities.

The enhanced sensitivity and broader frequency range of the new generation of \acrshort{gw} detectors, such as \acrshort{et}, will significantly extend the range of detectable distances and source masses, while finally enabling the detection of \acrshort{gw} signals from a wider variety of astrophysical sources, including nearby \acrshort{ccsn}e~\cite{Gossan:2022}. 
Even in the context of these new accessible sources, multi-messenger astronomy combining \acrshort{gw} and electromagnetic (EM) signals is imperative to fully understand the underlying physics of these phenomena.
Thus, the operation of efficient, deep, and wide-field transient dedicated  \acrshort{em} surveys — such as those currently running and the more sensitive ones planned for the near future —are of great importance. 
These wide-field surveys will continuously scan the sky across multiple EM bands. By employing specific observing strategies with the appropriate sensitivity and cadence, optimized for the expected properties of EM counterparts, these survey will be crucial for both detecting \acrshort{em} counterparts to \acrshort{gw} events and identifying the best candidates for subsequent characterization by dedicated \acrshort{em} resources.

\vspace{1mm}\noindent
{\bf EM prompt follow-up}.
The optimal strategy for prompt \acrshort{em} follow-up of \acrshort{gw} events relies on several steps. The initial search for the \acrshort{em} counterpart using wide-field facilities enables to cover the sky localization provided by the \acrshort{gw} detectors and to identify EM counterpart candidates. This step needs then to be followed by the selection and characterization of the most promising candidates among the large number of contaminant detected transients. While the wide-field facilities can search for the EM counterparts of GW events both in survey mode or pointing the target through Target of Opportunity (ToO) observations, the characterization of the most promising candidates requires the use of dedicated ToO.
As the wide-field survey will be more and more sensitive, the expected rate of detected transients 
will significantly increase making the selection of the candidate counterparts of GW events more difficult and requiring a larger amount of characterization resources. 


In the case of \acrshort{ccsn}e, as discussed in the previous sections, the detection of GW signal from the core-collapse by ET will be limited to the local Universe with a relatively low expected rate of detections.
Thus, a reverse observing approach could become  more effective; the \acrshort{em} detection of a nearby  \acrshort{ccsn} can be used as a trigger to search for an\acrshort{gw} signal in the same sky position and in temporal coincidence with the core collapse time extrapolated from the EM signal.  In this scenario as well, efficient  \acrshort{em} facilities that can promptly detect transients, issue alerts, and conduct characterization follow-ups are required.


\vspace{1mm}\noindent
{\bf Wide field transient surveys}.
The availability of increasingly efficient ground-based facilities specifically dedicated to transient phenomena has greatly driven the evolution of time-domain astronomy over the past decade. In particular, we have witnessed the development of several wide-field optical transient surveys, such as the \acrshort{asassn}  \cite{Shappee:2013mna}, \acrshort{atlas} \cite{Rest:2018amw}, the Panoramic Survey Telescope and Rapid Response System (\acrshort{panstarrs} or PS) \cite{Kaiser:2002zz}, the Palomar Transient Factory (PTF/iPTF) \cite{Law:2009ys}, and, more recently, \acrshort{ztf} \cite{Masci:2018neq}. These surveys have played a crucial role in making the optical and near-infrared (NIR) bands the primary channels for the discovery of transients.

In the past, the discovery of such phenomena was mainly a serendipitous process. However, with the advent of these synoptic all-sky surveys, it has become the result of systematic, repeated scanning of the sky over days to weeks. Thanks to dedicated and efficient observing strategies, these once rare events have become more common, and a number of new, exotic, and exciting phenomena have started to unveil their existence.

Each survey is characterized by different scientific goals and associated modes of operation and properties, such as sensitivity, cadence of repeating observations in the same area of the sky, slewing time, 
filters provided by the different cameras and telescopes employed. For instance, \acrshort{asassn} is a long-term project specifically dedicated to the study of supernovae (SNe). It consists of a network of small, wide-field telescopes distributed across both hemispheres, allowing it to survey the entire night sky with a cadence of approximately 2-3 days. It reaches a magnitude limit of about 18.5 each night in the 470-700 nm wavelength range~\cite{Shappee:2013mna}.

\acrshort{panstarrs} and \acrshort{atlas} are mainly dedicated to the study of Near-Earth Asteroids (NEOs). Both consist of a system of two telescopes located in the Hawaiian Islands, with diameters of 1.8 meters and 0.5 meters, respectively. Their observing strategies are slightly different: \acrshort{panstarrs} provides observations with a cadence of about 7 days in the 400-1000 nm band, reaching a maximum depth of approximately 24 magnitude, while \acrshort{atlas} provides a 2-day cadence in the 420-975 nm band, reaching around 19 magnitude \cite{Rest:2018amw,Kaiser:2002zz}. Although designed for NEO monitoring, both surveys have proven to be extremely useful in the discovery and follow-up of transient sources of various natures. They systematically report new transient discoveries and publish updated light curves. In particular, one of the two \acrshort{panstarrs} telescope, Pan-STARRS1 (PS1), provides deep reference imaging for the northern sky in Sloan filters. Reference images are essential for the image subtraction technique. This image processing method is widely used in transient searches and consists on the subtracting images of the same region of sky taken at different times to detect new transients.

The Zwicky Transient Facility (\acrshort{ztf}) has increased the transient discovery rate by an order of magnitude.  \acrshort{ztf} represents the forefront  of what can be achieved with a single 1-meter class survey telescope. It consists of a wide-field imager (with a 47 square degrees field of view) mounted on the 1.22-meter Palomar Oschin Schmidt telescope, currently scanning the entire Northern sky with a 3-day cadence in the 400-900 nm band, and reaching a depth of approximately 21 magnitude. Additionally, an integral field unit spectrograph on the 1.52-meter Palomar telescope is optimized for the spectral classification of transients brighter than 19 magnitude. The significant increase in detected transients has highlighted the urgent need for efficient automated tools to analyse the data and provide the community with alerts.

Another project of great interest is the Gravitational-wave Optical Transient Observer (GOTO) \cite{Steeghs:2021wcr}, which is specifically dedicated to the search for optical transients associated with \acrshort{gw} detections, maximising the coverage of wide sky areas. GOTO consists of an array of wide-field optical telescopes with multiple 40-cm reflectors on a single robotic mount. The project includes a node located on La Palma island (GOTO-N) and one located in Australia (GOTO-S), ensuring northern and southern hemisphere coverage. 
This configuration results in an instantaneous field of view of up to 80 square degrees at each site. Two observing strategies are available: sky-survey with a cadency of 2-3 days and  target trigger mode. Each unit is equipped with Baader filters, covering the 380-700 nm range down to a magnitude of approximately 20.5. 
In line with the goal of monitoring large regions of the sky, new optical system concepts, such as Mezzocielo \cite{2022SPIE12182E..0HR}, are being developed to enable continuous monitoring of the entire sky, aimed at detecting the prompt and early optical emission of transients such as GRBs and SNe.

Beyond the substantial advancements expected from the next generation of \acrshort{gw} detectors such as ET, we are also entering a new era in time-domain astronomy. A revolutionary optical/NIR facility, the Vera Rubin Observatory is in the process of starting the Large Synoptic Survey Telescope (LSST) \cite{LSSTScience:2009jmu}. 
The VRO-LSST will offer a unique combination of a large effective aperture (6.7 meters) and a wide field-of-view (9.6 square degrees) imaging camera equipped with six broad-band filters (covering 0.35 to 1.1 microns), reaching a total point-source depth in r-band of 27.5.The VRO-LSST is expected to detect a few million transients per night, among them core-collapse SNe, kilonovae and GRB oprphan afterglow which can trigger the GW search. 

The combination of deep sensitivity, wide field-of-view and rapid slewing time also makes VRO ideally suited for the \acrshort{gw} follow-up. In particular, VRO will be able to cover well-localized \acrshort{gw} regions with just a few pointings and achieve deep observations with relatively short integration times. This means VRO has the potential to detect and identify \acrshort{em} counterparts to \acrshort{gw} sources rapidly and effectively, especially at the large distances where such counterparts (with an absolute magnitude in the optical of M$\sim$-16 mag) are expected to be too faint for all the other wide field-of-view telescopes. Ref.~\cite{Loffredo:2024gmx} discusses in detail the perspectives of the VRO observing in synergy with ET to detect the optical counterparts associated with \acrshort{bns}. When operating in synergy with   \acrshort{et} and other available multi-wavelength facilities, VRO will open enormous opportunities for new discoveries, significantly advancing our understanding of the transient universe.

In other EM bands, particularly the ultraviolet and soft X-ray bands, wide-field missions are expected to detect the first emission from CCSNe caused by shock breakout. For example, the upcoming Ultraviolet Transient Astronomy Satellite (ULTRASAT) \cite{Ben-Ami2022}, with an unprecedentedly large field of view of $204 , \text{deg}^2$, is expected to collect the early UV light curves for hundreds of CCSNe. Nearby, low-luminosity long GRBs, CCSN shock breakout signals, and soft gamma repeaters are promising target signals for detection by the Transient High Energy Sky and Early Universe Surveyor (THESEUS) \cite{Amati:2017npy}.





\noindent

\vspace{1mm}\noindent
{\bf Candidate selection/characterization}.
The wide-field optical surveys previously described are fundamental resources for the prompt discovery and for alerting the community to their detection. Many of these surveys also provide updated multi-band optical light-curves, which are invaluable for investigating the properties of transients, facilitating initial classification, and identifying promising candidates. Then, prompt spectroscopic observations of the detected transients enable a more precise classification necessary to select interesting candidates among contaminants. This step is essential for the subsequent characterization through multi-wavelength follow-up campaigns dedicated to selected events. 
For nearby  \acrshort{sn}e, the combination of prompt identification and early photometry and spectroscopic follow-up allows the transient to be correctly classified as an \acrshort{ccsn}e and the initial light curve to be accurately sampled, which is particularly important for extrapolating the core collapse time and precisely defining the time window for searching for \acrshort{gw} and neutrino signals.

Among the available ground-based facilities and projects currently dedicated to transient spectroscopic observations, the Public ESO Spectroscopic Survey of Transient Objects (PESSTO) \cite{Smartt:2014rpa}  and its extension ePESSTO+ have proven to be extremely useful and productive. ePESSTO+ targets supernovae and optical transients brighter than 20 mag. 
Since its starting operations in 2012, the ESO New Technology Telescope (NTT) EFOSC2 (optical) and SOFI (NIR) spectrographs have been systematically used to classify transients selected from publicly available surveys and to select interesting science targets for detailed follow-up campaigns. To give some numbers, spectra for 2138 objects were provided during the first 2.5 years of ePESSTO+ operations (from April 2019 to October 2021) . Spectroscopic classification reports along with the corresponding optical spectra are publicly available within 24 hours,  while follow-up spectra are delivered to the community after one year through dedicated public data release events. While this project is approaching its end, a more efficient instrument is being installed at the NTT to replace SOFI: the Sons of X-shooter (SoXS) \cite{schipani22}.

SoXS is a broad-band, high-efficiency, medium-resolution spectrograph expected to be operational in 2025. It is capable of simultaneously observing a spectral range of 350-2000 nm with a resolution of about 4500, thanks to its two-arm configuration (UV-VIS+NIR). Its acquisition camera is equipped with the u, g, r, i, z, and Y filters and features a 3.5'×3.5' field of view. This facility is fully 
dedicated to the classification and follow-up of transient sources, as well as to supporting wide-field transient surveys.

The next-generation of 30-40 m class telescopes comprises facilities such as the Extremely Large Telescope (ESO-ELT) \cite{2007Msngr.127...11G}, the Giant Magellan Telescope (GMT) \cite{2012SPIE.8444E..1HJ} and the Thirty Meter Telescope (TMT) \cite{2013JApA...34...81S}. Their instruments will revolutionize the study of gravitational wave sources by providing unprecedented capabilities for spectroscopic and imaging observations. Their advanced instruments will enable for example the direct detection of spectral signatures from the heaviest elements at their production sites, track the temporal evolution and spectral features of faint emissions, and constrain the physics of matter ejection in compact object mergers. Because of their narrow fields of view, they will only come into operation when a well-localized counterpart is found.

Another proposed next-generation project is the Wide-field Spectroscopic Telescope (WST) \cite{WST:2024rai}, a 10-meter class telescope dedicated to wide-field spectroscopic surveys. It aims to combine a large field of view (approximately 3 square degrees) with a high-multiplex multi-object spectrograph (MOS) operating in both low and high-resolution modes (R4000 and R40000, respectively) across a wavelength range of 370-970 nm, , as well as a giant panoramic central integral field spectroscopy (IFS) unit. This combination will greatly enhance our ability to conduct comprehensive transient and multi-messenger observations.




\vspace{1mm}\noindent
{\bf \acrshort{em} facilities to observe isolated neutron stars}.
EM observations of isolated neutron stars has uncovered a wide variety of these fascinating objects, each distinguished by unique observational properties and emission mechanisms. Extensive observations and theoretical modeling have significantly advanced our understanding of rotation-powered pulsars, magnetars, XDINSs and CCOs. Each class offers valuable insights into the underlying physics of neutron stars, including their magnetic fields, rotational dynamics, emission processes, and evolutionary paths.

Looking ahead, the Square Kilometre Array (SKA) with its unprecedented sensitivity will make it possible the discovery of many more pulsars and detailed studies of their properties. Large-scale radio surveys with the SKA and other telescopes will be critical in uncover new, peculiar neutron stars, such as long-period radio sources, and understanding their full population diversity. 

As powerful gamma-ray emitters, isolated neutron stars are one of the primary targets of the Cherenkov Telescope Array Observatory which aim to probe the (very) high-energy emission mechanisms at work in these sources. For neutron star in X-ray binaries, satellites such as NewAthena with the X-ray Integral Field Unit (X-IFU), a cryogenic imaging spectrometer covering the 0.2 to 12 keV energy range,  will observe absorption and emission lines from the outflow to probe the physics of accretion.
Overall, the study of isolated neutron stars is poised for significant advancements, with each discovery offering new opportunities for a deeper understanding of these enigmatic objects.



\noindent








\vspace{1mm}\noindent
{\bf \acrshort{em} characterisation of progenitor population and nearby SNe}.
While waiting for the fortunate occurrence of a nearby core collapse SN and the possible detection of a related \acrshort{gw} signal there are two areas where \acrshort{em} surveys can contribute to better define the \acrshort{gw} search context, namely:
i) the characterisation of massive star population in the Milky Way and nearby galaxies and ii) the detailed studies of bright, nearby core-collapse events.

\subparagraph{Massive stars population}

Surveys of massive stars in our Galaxy are providing a view on the properties of the progenitors of core-collapse SNe (e.g. \cite{Apellaniz:2010vn}).  
One of their most important finding is that most massive stars are born in binary or, in general, multiple systems, and about three-quarters of them are expected to undergo interactions during their evolution, either by exchanging mass or merging  \cite{Sana:2012px}. This is expected to perturb their evolution, explosion mechanism and remnant properties with respect to the outcomes of an isolated star. 
Indeed, as emphasised in a recent review \cite{Marchant:2023wno}, binary evolution provides a rich set of post-interaction products such as stripped stars, rapidly rotating accretors, mergers and X-ray binaries. It is claimed that binary systems provide the favored formation channel of stripped envelopes, exotic supernovae, and double-degenerate binaries, which are best-known sources of \acrshort{gw} emission. In this case, it is important to note that such systems, particularly double-degenerate binaries, will be detectable by the Laser Interferometer Space Antenna (LISA). 
A major contribution to the characterisation of galactic massive binaries is expected from GAIA starting from the fourth data release (DR4), that is due by the end of 2025.

Then, as discussed in section~\ref{stelev} wind and rotation are the other key ingredients in determining the evolution of massive stars and their final remnant mass. Theoretical models are able to describe the effects of these parameters but there are large uncertainties on the initial conditions and therefore observational constraints are needed.

The measurements of rotational velocities in massive stars of the Milky Way and nearby galaxies indicate that in most cases they are relatively slow rotators, with a rate less than 20\% of their break up velocity. There is however a sizeable population of fast rotators that it is attributed to the outcome of close binary interaction  \cite{Ramirez-Agudelo:2013mpa}. Interestingly, there is the tendency of fast rotators to occur at low metallicity \cite{Schootemeijer:2022} that intriguingly matches the fact that long GRBs also occurs in metal-poor galaxies.

For what concern mass loss, there is a well know problem that empirical estimates of rates for OB stars gives much lower values than predicted by theory (the weak wind problem) \cite{Marcolino:2009mb}. This weak winds is not sufficient to remove the hydrogen envelope, 
that could allow for the formation of massive \acrshort{bh} even for moderately sub-solar metallicity  \cite{Ramachandran:2019xyz}. On the other hand, this scenario would not lead to a Wolf-Rayet phase that therefore requires some other formation mechanism. There are evidences that for very massive stars mass (above $(40-50) M_\odot$) loss may be dominated by ejection in brief, powerful eruption in the so-called luminous blue variable (LBV) phase~\cite{Smith:2006kx}.

A long debated issue is the shape and the universality of the initial mass function (IMF). This is important because, as reminded in a previous section, IMF is the link between the  star formation rate and the number of expected CCSNe. The standard assumption  is that of a universal IMF, typically with a power law  shape, although the exact parameters are debated (see section~\ref{ccsnrates}).

Yet, there is evidence that in starburst region the IMF can be “top-heavy”, implying a higher fraction of massive star for a given SF \cite{Weidner:2010md}.  The study of the 30 Doradus star forming region in the Large Magellanic cloud shows indeed that the IMF  contains 1/3 more stars above $30\,M_\odot$ than predicted by a standard Salpeter IMF \cite{Schneider:2018}. If this is confirmed, it would imply a rate of SNe from very massive stars a factor two higher than for a standard IMF, for a given SFR.


\subparagraph{Nearby core collapse}

It is important to continue to secure detailed \acrshort{em} observations of nearby core collapse to be used  to constrain the explosion properties, mass loss history, amount of asymmetry, binary companion and compact remnant properties. 

We notice that the peculiar properties of the two closest SNe discovered in modern times, SN~1987A  \cite{Podsiadlowski:1989aaa,Menon:2017hva} and SN~1993J \cite{Maund:2004cp}
were attribute to close binary evolution that is consistent with the global properties of massive star population mentioned above. Also, recent spectroscopic observations of the remnant of SN~1987A with \acrshort{jwst} were explained with the contribution of energy injection by a cooling \acrshort{ns} or a pulsar wind nebula and the observed line velocity shift evidence for a \acrshort{ns} natal kick \cite{Fransson:2024csf}.

For nearby SNe it is sometimes possible to identify the progenitor in pre-explosion images. Based on these studies, it appears that there is an upper limit for the maximum mass for SN II-P of $16.5 \pm 1.5\,{\rm M}_\odot$, that is in fair agreement with the theoretical predictions, while the majority of stars with mass above $\sim 20 {\rm}_\odot$ would explode as SNIb/Ic or would collapse directly to a black-hole  (see section~\ref{stelev}). In the first case the progenitor of the SN is difficult to detect in the pre-explosion images while in the second case the explosion remains undetected \cite{Smartt:2009zr}.
Therefore, more data are needed. 

A significant contribution in this field is expected by VRO-LSST that, with its deep, multicolor  monitoring of the southern sky would allow a better constraints of the progenitor on nearby SNe. At the same time, the survey will allow for a systematic search of failed SNe, monitor stars that disappear rather than for bright SNe. There have been already some attempts of this kind of search that, however, were not yet able to provide definite evidences \cite{Adams:2016hit}.

\subsection{Executive Summary}
\acrshort{gw} radiation from \acrshort{ccsn} and from rotating \acrshort{ns} has not been detected so far by current detectors, due to the predicted signal weakness.
In this chapter it has been shown that ET, considering foreseeable advances in data analysis and computing techniques, has significant chances to detect such signals. Relevant sources are in the Milky Way or in the local group of Galaxies, with a few notable exceptions, like newborn magnetars, for which \acrshort{gw} radiation could be detected up to tens of Mpc. 

\begin{highlightbox}{GWs from CCSNe}
\begin{itemize}
\item
The rate of \acrshort{ccsn}e within 10 Mpc is about 1 per year and the predicted \acrshort{ccsn} rate within the Milky way is $\sim 2$ to 3.2 per 100 years.
\item
ET is expected to have a \acrshort{ccsn} detection horizon 10 times further than the 2nd generation detectors, which are expected to detect neutrino-driven CCSN up to about 
10~kpc and fast rotating progenitors up to about 100~kpc. There is a potential factor of $\sim 2$ gain in horizon distances from future improvement in search techniques.
\item
Observed \acrshort{ccsn} signals will provide insight into the explosion mechanism and, for galactic \acrshort{ccsn}e, the frequency of the dominant oscillation gravitational wave mode, for example from g-mode excitations of newly formed proto-neutron star, can be tracked with an uncertainty of 10$\%$.
\item Neutrino and/or electromagnetic counterparts will provide excellent, arcsecond sky localisation information for \acrshort{ccsn}e. Neutrino counterparts will potentially constrain the on-source window for gravitational wave signals to a few milliseconds while electromagnetic observations that estimate the shock breakout time will lead to on-source windows ranging from seconds to hours, depending on the progenitor mass. 
\item 
ET can also detect transient gravitational wave signals from isolated \acrshort{ns}s, for example associated with pulsar glitches and magnetar bursts. It is likely that ET will observe f-mode excitations resulting from pulsar glitches, for pulsars such as Vela. For magnetar bursts, ET will probe the regime where the ratio of gravitational wave to electromagnetic energy is $10^{-5}$ to $10^{-4}$ for a galactic giant flare. The observation of gravitational waves from less energetic flares and pulsar glitches relies on the successful implementation of stacking methods to accumulate signal-to-noise ratio from a population of bursts/glitches.
\item The role of  \acrshort{em} surveys, which involve efficient, deep, wide-field instruments with planned observing strategies based on \acrshort{gw} counterpart properties, along with continuous sky scanning across different bands, will be crucial for detecting the shock breakout and early emission from nearby CCSNe, magnetar flares,  nearby low-luminosity long GRBs. These EM-detected transients will enable to trigger a more efficient GW search.
\item \acrshort{em} follow-up of GW burst triggers will require initial wide-field searches with the appropriate cadence, followed by the selection and characterization of promising candidates through dedicated spectroscopic follow-up. The high rate of unrelated transient discoveries will require careful optimization of data analysis and coordinated planning for the efficient use of observational resources.
\item Wide field of view, sensitive and fast optical telescopes such as the VRO dedicated to GW-related transients providing global coverage are critical in the ET era.  Spectroscopic facilities like SoXS (an upcoming broad-band spectrograph at NTT), ELT, and WST will play key roles in transient classification.

\end{itemize}
\end{highlightbox}

\begin{highlightbox}{\acrshort{gw}s from rotating \acrshort{ns}s}
\begin{itemize}
\item ET will detect persistent emission from several known \acrshort{ns}s rotating at high frequency (namely, millisecond pulsars) if their ellipticity is larger than a few times $10^{-10}$, corresponding to ``mountains’’ a few micron high. Such values are orders of magnitude below the maximum theoretical ellipticity that neutron stars can sustain and is comparable to the minimum ellipticity millisecond pulsars could have, as suggested on the basis of their observed properties.
\item Accreting \acrshort{ns}s, like Sco-X1, are promising sources of persistent \acrshort{gw}s. ET will be able to probe the standard torque balance limit scenario. Actually, ET would be sensitive to the \acrshort{gw} emission even considering more conservative signal amplitude estimations, which take into account uncertainties in the star magnetic field topology, viscous heating, X-ray emission efficiency, and the effects of radiation pressure.
\item In all-sky surveys, ET will be able to detect signals emitted by \acrshort{ns}s rotating at frequency larger than about 200 Hz, up to the distance of the galactic center, provided their ellipticity is larger than about $10^{-7}$. Slower \acrshort{ns}s, with spin frequency of 20-30 Hz, would be detectable in the solar neighborhoods (with distance up to about 500 pc).
\item Newborn millisecond magnetars, endowed with a super-strong internal magnetic field, could have significantly larger deformation. Therefore, they are expected to emit \acrshort{gw} signals with durations of hours-days detectable by ET up to 5--10 Mpc, for moderately large ellipticities up to $10^{-4}-10^{-3}$ (and up to about 15--20 Mpc for more extreme ellipticities).
\item A multi-messenger approach, exploiting observations in the electromagnetic band by future facilities, will significantly help in the endeavor of detecting \acrshort{gw} emission from rotating \acrshort{ns}s, as it would allow to restrict the parameter space (e.g., providing accurate sky localization and decent time of formation for magnetars, and accurate measure of the binary parameters for accreting systems). Therefore, a deeper (i.e., more sensitive) search could be performed at fixed computing cost. The search for a possible \acrshort{em} counterpart would be also a natural step to confirm the detection of a \acrshort{gw} source in all-sky searches.
\item 
The detection of a persistent signal from a rotating \acrshort{ns} would open a new window on the observation of our Universe, as the long signal duration would allow to measure the star’s parameters with extremely high accuracy, and to spot possible tiny deviations from the models (for instance, violation of General Relativity which would leave an imprint on the signal polarization). Constraining the \acrshort{ns} internal structure is another important possibility that can be achieved, although accurately determining the equation of state is not that simple due to the presence of degeneracies, and careful studies are needed on this topic.   

\end{itemize}
\end{highlightbox}

\section{Waveforms}\label{section:div8}

\subsection{Introduction}
\label{sec:introdiv8}

The upcoming era of precision gravitational-wave astronomy enabled by ET requires unprecedented knowledge of gravitational waveforms of modelled sources to achieve many of its science goals and to maximize its science output. Waveforms that are physically detailed, numerically accurate and computationally fast to generate, underpin much of ET's data analysis challenges: detection by matched filtering, parameter estimation via comparison of measured data with theoretical expectations of General Relativity, and tests of deviations of the data from the predictions of General Relativity.

The Waveforms Division of the ET Observational Science Board aims to provide the waveform models needed for ET to reach its science goals. Despite more than a hundred years in semi-analytic and numerical developments since the birth of General Relativity, major advances will be required in all areas of waveform modeling to achieve the ultimate scientific goals of ET, as will be detailed below.
In this section we present the current state-of-the-art, the main open challenges and the important next steps in all domains of waveform modeling for ET: waveform systematics and accuracy requirements for third generation detectors, established and novel techniques for waveform modeling, waveform models for specific sources and waveform acceleration techniques. 

ET has the potential to make extremely precise measurements. To enable this potential, however, requires sufficiently accurate waveform models, which encompass all relevant physical effects. Waveform systematics and accuracy requirements are discussed in
section~\ref{sec:accuracy}, with the overall conclusion 
  that waveform
  models need further improvements to avoid systematic biases due to
  modeling errors, with the precise accuracy needs being a topic of ongoing
  scientific studies.  Moreover, the large event rate and the larger
  frequency band of ET may uncover GW source populations in other
  parts of parameter space, so that enlarged parameter space coverage is important (e.g.\ mass-ratio, spins, eccentricity).

The state-of-the-art and latest advances in techniques of waveform modeling are outlined in section~\ref{sec:tools}. There are currently four techniques being used to formulate waveforms within the framework of Einstein gravity: numerical relativity, weak-field expansions, gravitational self-force and inspiral-merger-ringdown models. In addition, some techniques have been extended beyond General Relativity, although they remain largely underdeveloped. Each of these techniques has its own current limitations, and all need significant improvements to be able to deliver the scientific promise of ET.

Section~\ref{Sec:models} discusses waveform models for the expected sources of ET,  which are built utilizing results of the modeling techniques discussed in section~\ref{sec:tools}. Models need to be formulated for compact binary mergers involving black holes and neutron stars or yet undetected but theoretically studied exotic compact objects, for core-collapse supernovae, for galactic binaries such as double white dwarf systems, but also for potentially detectable systems such as hyperbolic encounters and triple systems. Waveform modeling beyond vacuum General Relativity is necessary to test GR and infer the environment of targeted sources. Waveform models of dark matter environments and alternative theories of gravity have been formulated both as theory-specific models, or as parametric deviations of vacuum General Relativity in an effective theory framework.

ET data analysis will not only require faithful and accurate waveforms but also fast generation of waveforms, which allows for efficient determination of the source's properties. The waveform acceleration techniques discussed in section~\ref{sec:acc} focus on methods used to improve the efficiency of evaluation of individual waveforms. Current techniques, such as reduced-order-models or surrogates, which might include an architecture built using machine learning, are trained to be accurate for 2G detectors, but further refinements or even new tools will be needed for ET.

\subsection{Waveform systematics and accuracy requirements for 3G}
\label{sec:accuracy}

With increasing signal-to-noise ratio of a GW signal, the statistical
uncertainty in the posteriors of parameter estimation becomes smaller,
whereas the impact of systematic errors is independent of SNR.
Therefore, control of systematic errors ---like inaccuracies in
waveform models or decisions to simplify the treatment of certain
physical effects--- becomes ever more important as the sensitivity of
detectors increases.  The high sensitivity of ET, where the loudest
events detected will likely have SNRs~$\sim
10^3$~\cite{Borhanian:2022czq, Dupletsa:2022wke,Iacovelli:2022bbs,Branchesi:2023mws}, allows astonishingly
precise measurements, but in turn places high demands on the waveform
models.  The high sensitivity, broader frequency bands and the higher
event rates in future detectors also imply that the detectors will be
sensitive to signals from a larger part of parameter space
(e.g. binaries with more unequal masses).  Therefore, not only do
waveform models need to be accurate enough in those parts of parameter
space where particularly high SNRs are expected, they also need to
maintain sufficient accuracy across the entire parameter space of
conceivable sources and all physical effects potentially measurable.

\subsubsection*{State-of-the-Art}

Studies of waveform accuracy fall into several categories,
  which follow different trade-offs between the ease of the study and
  reliability of results. We will first discuss easy, yet less reliable techniques, and subsequently more reliable, but computationally costly approaches.

A simple way to approach the problem is through \textit{indistinguishability} criteria between waveforms: inequalities which -- when statisfied -- ensure that a model waveform is indistinguishable from a target (exact) waveform when performing measurements with a chosen detector.
    These criteria \cite{Flanagan:1997kp,Cutler:2007mi,Vallisneri:2007ev,Lindblom:2008cm, Lindblom:2009ux,Damour:2010zb, Baird:2012cu, Chatziioannou:2017tdw} are phrased in terms of the noise-weighted inner product between two waveforms $h_1$ and $h_2$,
    \begin{equation}
        (h_1|h_2) \equiv 4 \operatorname{Re} \int_0^\infty \frac{h_1(f)\bar{h}_2(f)}{S_n(f)} df,
    \end{equation}
    where $S_n(f)$ is the power spectral density, the overlap $\mathcal{O} \equiv (s|h)/\sqrt{(s|s)(h|h)}$ and the mismatch $\mathcal{M}\equiv 1-\mathcal{O}$.
A typical version of the criteria reads that, given a signal $s$ and a
 model waveform $h$, if
\begin{equation}
    \label{eq:crit}
    \mathcal{M}
    < \frac{D}{2 \rho^2} \, ,
\end{equation}
then $h$ cannot be distinguished  from $s$ beyond statistical uncertainty.
Here $\rho=\sqrt{(s|s)}$ is the signal to noise ratio (SNR) and when calculating $\mathcal{M}$, the waveform model $h(\vec\theta)$ is to be evaluated at the same parameters $\vec\theta$ as the signal.  As formulated in~\cite{Baird:2012cu}, $D$ is the chi-square value, for a given number of degrees of freedom, and a chosen confidence interval in the measurement.  Some authors replace the chi-square by the number of intrinsic parameters in the model~\cite{Chatziioannou:2017tdw}, which is the chi-square value for 1-sigma confidence.

If the waveform model is approximate, which is typically the
  case, the mismatch in eq.~(\ref{eq:crit}) should be replaced by
  $\mathcal{FF}-\mathcal{O}$~\cite{Vallisneri:2007ev,Toubiana:2024car}, where the fitting factor $\mathcal{FF}$ between the
  signal $s$ and the waveform model $h(\theta)$ is given by
  $\mathcal{FF}\equiv \max_{\vec{\theta}} \mathcal{O}(s,
  h(\vec{\theta})$.  This replacement weakens the accuracy requirements by allowing waveform errors orthogonal to the waveform manifold, which do not cause parameter estimation biases.
Indistinguishability criteria can also be generalized to multiple detectors, where  $\mathcal{M}$ and $\rho$ are replaced by suitable SNR-weighted averages across detectors~\cite{Toubiana:2024car}.  Finally, the true signal $s$ is generally not known; in that case one often substitutes a second waveform model for the signal $s$, and thus takes the difference between two waveform models for the same class of signal as indicative of waveform error (see, e.g., ~\cite{Hu:2022rjq}).

\begin{figure}[t!]
	\centering
    \includegraphics[width=0.8\textwidth]{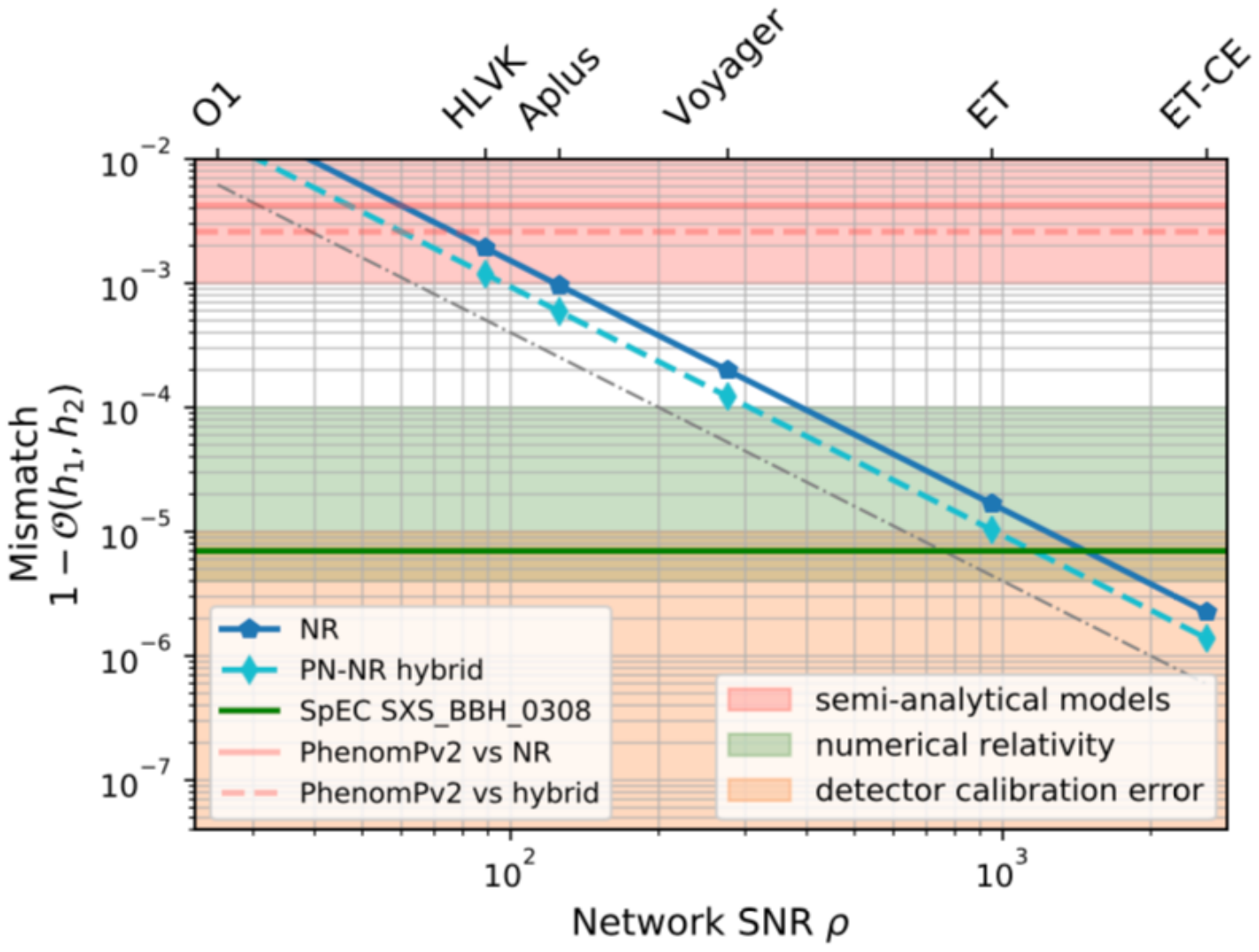}
\caption{Predicted waveform accuracy requirements for second and third generation ground based detector networks (from \cite{Purrer:2019jcp}). 
}
\label{figurePH}
\end{figure}

While easy to evaluate, indistinguishablity criteria impose a (too) strong requirement on waveform models. They provide a sufficient criterion, essentially requiring waveforms to be accurate enough that no possible measurement in the infinite dimensional space of potential waveforms would ever show a systematic bias. For instance, at  SNRs~$\sim 10^3$, they imply waveform models would need a mismatch~$\mathcal{M}\sim 10^{-5}-10^{-6}$~\cite{Purrer:2019jcp} (see figure~\ref{figurePH}).
In practice, one is often only interest in measuring a subset of parameters, and only desires those to be unbiased. \textit{Injection studies} can be used to gauge the actual biases in such measurements. In an injection study a waveform from one model or from numerical relativity is injected in simulated detector noise, and its parameters are estimated with a different waveform model through Bayesian inference. Such studies find~\cite{Purrer:2019jcp} that in some cases where the indistinguishablity criteria suggest biases may occur, none of the typically measured parameters show any bias.

Injection studies can be computationally intensive, especially at high
SNR values.  The linear signal approximation provides a compromise
between rigor and computational cost.  One assumes that in the
vicinity of a signal to be studied, the waveforms change only linearly
with model parameters; this leads to the Fisher information matrix
analysis for statistical uncertainity, which is combined with an
estimate of the systematic bias from the maximum likelihood
point~\cite{Flanagan:1997kp,Cutler:2007mi}.  With careful alignment
between distinct waveform models~\cite{Dhani:2024jja} and
modifications to avoid posteriors to exceed physical
bounds~\cite{Cutler:1994ys,Poisson:1995ef,Berti:2004bd}, the linear
signal approximation can yield estimates of systematic parameter estimation (PE) biases
across parameter space.
One such analysis~\cite{Kapil:2024zdn} (restricted to aligned spin
  systems using only the (2,2) waveform mode) reports that current
  waveform models would result in significant PE bias in several hundreds events from
  a population of 3000 events with SNR $>100$.  Another analysis~\cite{Dhani:2024jja} (for precessing binaries using waveform models including higher modes) concludes that as much as 40\% of binaries with high spins would lead to biased PE results, and further illustrates the severity of potential biases through three case studies performed with full Bayesian PE.

Waveform modeling errors increase with magnitude of spin and
  with mass ratio, so that PE biases will be more common at large
  spins and unequal masses, i.e. precisely for less common, but perhaps particularly interesting binary parameters at the edge of the astrophysical distributions.
      


\vspace{2mm}\noindent    
\subsubsection*{\bf Main Open Challenges}
Injection studies are mainly a test of whether a particular waveform model is ``good enough''.
Such studies~\cite{Owen:2023mid, Ramos-Buades:2023ehm} tell us that present day state-of-the-art models may produce biases even for certain possible events in ongoing observation campaigns, and publications of the LVK collaboration regularly and carefully discuss waveform systematics, both for GW catalogs~\cite{LIGOScientific:2018mvr,LIGOScientific:2020ibl,KAGRA:2021vkt} and for individual events~\cite{LIGOScientific:2016ebw,LIGOScientific:2017bnn,LIGOScientific:2018hze,LIGOScientific:2020aai,LIGOScientific:2020stg,LIGOScientific:2020ufj,LIGOScientific:2020zkf,LIGOScientific:2021qlt,LIGOScientific:2024elc}.
  
What they typically do not tell us is how much better a model needs to be if it is found lacking. Properly quantifying the needed improvement in modeling remains an open challenge.
Ideally, such quantification should come in terms that are meaningful to first principles calculations of the waveforms: How many orders are needed in post-Newtonian  expansions? What resolution is needed in numerical relativity  simulations? How many post-adiabatic orders need to be included in the self-force formalism? Some efforts have been made to start answering these questions in the context of post-Newtonian~\cite{Owen:2023mid}, numerical relativity~\cite{Ferguson:2020xnm}, and self-force~\cite{vandeMeent:2020xgc,Wardell:2021fyy, Burke:2023lno} calculations. In most case these still come with limitations such as restricting to quasicircular (possibly non-spinning) binaries.

One of recurring challenges is that accuracy standards typically require a comparison to a ``true'' signal. The typical solution adopted is to use a waveform from a similarly (or more) accurate, but different, model as a stand-in for the ``true'' model. This is in many cases a reasonable solution, but does come with the risk of missing some systematics.
A relative new challenge for future detectors comes from the increased chance of observing overlapping signals.  Ref.~\cite{Antonelli:2021vwg} shows that the incorrect subtraction of a loud signal due to modeling error can lead to significant bias in the parameter estimation of a weaker overlapping signal in 3G detectors.
A final challenge is the step from quantifying biases of individual PE
results (as discussed above) to biases in studies that combine many GW
signals, like population inference, measuring cosmological parameters,
tests of General Relativity or measurements of the equation of state,
continuing and refining the studies presented in \cite{Dhani:2024jja,Kunert:2021hgm,Kunert:2024xqb,Moore:2021eok}.
It is also important to quantify the impact on each of ET's science objectives for the situations in which waveform models do not fully satisfy the  accuracy requirements.

Integrating waveform modeling error estimates directly into waveform models may also provide a method to mitigate biases due to waveform systematics. Proposed methods include using Gaussian Process Regression to interpolate either waveform residuals or directly NR waveforms~\cite{Moore:2014pda, Gair:2015nga, Moore:2015sza, Doctor:2017csx, Williams:2019vub, Andrade:2023sal, Khan:2024whs}, applying frequency-dependent amplitude and phase corrections similar to those used for detector calibration uncertainty~\cite{Read:2023hkv}, or marginalizing over the uncertainty of calibration parameters in the models~\cite{Pompili:2024yec} (see also~\cite{Owen:2023mid}). In these approaches, one can account for waveform modeling uncertainties by marginalizing over these additional degrees of freedom during the inference process. While this may lead to parameter estimates with reduced precision (e.g. wider posterior distributions), it should improve their robustness, ensuring they remain reliable even in the presence of systematic modeling errors. However, most studies incorporating error estimates into waveform models have been proof-of-principle, and assessing the impact of these approaches on the analysis of realistic signals observable with ET remains to be done.

\subsubsection*{Important Next Steps}

In the absence of a clear goal post of how much waveform models need
to improve to meet ET's science goals, it is important to continue
monitoring potential systematic biases due to modeling error through
the use of injection studies, as improvements to waveform models are
made.

Of high importance is an improved understanding of
  systematics, e.g., with improved distinguishability estimates that
  are not conservative, and with specific studies on the accuracy
  requirements of the separate ingredients of models (PN, NR, self
  force), and how they relate to the accuracy of final models.

Of course, this also raises the question, how much systematic
uncertainty can be tolerated? The traditional answer to this, is that
in order to maximize the scientific pay-off from investments in better
detectors, we should strive to always have the systematic uncertainty
due to modeling error be smaller than the statistical uncertainty in
the observations. This may be a too conservative requirement, and it
may be possible to reach some science goals with suboptimal waveforms.
This should be considered for each science goal.

\subsection{Techniques for waveform modeling: Current state and advances}
\label{sec:tools}

Several individual calculation techniques provide important input into
waveform models.  The subsections here keep track of developments in
these areas.

\subsubsection{Numerical Relativity}
\label{sec:NR}


General relativistic  simulations of binary black hole (BBH), binary neutron star (BNS), and neutron star-black hole (NSBH) systems have  rapidly evolved over the last decades.
Simulations of BNS~\cite{Shibata:1999wm} and BBH~\cite{Pretorius:2004jg,Campanelli:2005dd,Baker:2005vv} mergers have been possible for $\sim 15-20$ years. Today, thousands of high-accuracy BBH GW signals have been produced. Simulations have also been performed using alternative theories of gravity, while more advanced models for the dense matter at the core of neutron stars and the inclusion of magnetic fields, neutrino radiation transport, or non-ideal fluid dynamics in simulations allow us to study the merger and post-merger evolution of BNS and NSBH mergers with increasing realism. These simulations play a crucial role in testing, calibrating, and developing the models used by GW observatories, as well as in the analysis of electromagnetic (EM) signals following these mergers. Nevertheless, the accuracy of the simulations, their coverage of the available parameter space, and their approximate treatment of important physical process are insufficient for the expected needs of ET.

\vspace{2mm}\noindent
{\bf State-of-the-Art}

\subparagraph{Formalism.} Most GR simulations use one of two formulations of Einstein's equations known to be amenable to numerical evolution: the Generalized Harmonics (GH) formalism \cite{Friedrich1985,
  Garfinkle:2001ni, Pretorius:2004jg,Lindblom:2005qh} and the Baumgarte-Shapiro-Shibata-Nakamura (BSSN) formalism \cite{Baumgarte:1998te,Shibata:1995we}
  (and its Z4c~\cite{Bernuzzi:2009ex, Hilditch:2012fp} and ccZ4~\cite{Alic:2011gg,Alic:2013xsa} extensions). These methods have been improved over the years through appropriate gauge choices \cite{Bona:1994dr,Alcubierre:2002iq,vanMeter:2006vi,Gundlach:2006tw,Szilagyi:2009qz,Chen:2021rtb,Rosato:2021jsq}. Black holes are evolved either by excising a region inside their horizon~\cite{Pretorius:2004jg,Scheel:2006gg}, or using the {\it puncture} method~\cite{Campanelli:2005dd,Baker:2005vv,Hannam:2006vv,Bruegmann:2006ulg,Etienne:2024ncu}. Most simulations are performed with high-order finite difference methods~\cite{Campanelli:2005dd,Baker:2005vv,Husa:2007hp,EinsteinToolkit:2022_11} or spectral methods~\cite{Boyle:2019kee}. For neutron stars, the GR equations of fluid dynamics are usually evolved in the Valencia formalism~\cite{1997ApJ...476..221B} using high-order shock capturing finite difference~\cite{Radice:2013xpa,Bernuzzi:2016pie} or finite volume~\cite{Most:2019kfe} methods. A Lagrangian formulation of the fluid equations has also been developed \cite{Rosswog:2020kwm}. An important aspect of these simulations, particularly for the analysis of high-frequency GWs, is the assumed equation of state (EoS) for dense neutron rich matter. Simulations now routinely use tabulated EOS derived from nuclear physics models (e.g.~\cite{Sekiguchi:2011mc,Deaton:2013sla,Palenzuela:2015dqa,Gonzalez:2022mgo}), as well as simpler analytical models~\cite{Read:2008iy,Lindblom:2010bb,Foucart:2019yzo} that can however be extended to include temperature and composition effects~\cite{Raithel:2019gws}. Magnetic fields have been evolved using ideal magnetohydrodynamics (MHD) \cite{Anderson:2008zp,Chawla:2010sw,Giacomazzo:2010bx,Etienne:2010ui,Kiuchi:2014hja}, resistive MHD \cite{Dionysopoulou:2015tda,Shibata:2021bbj}, and the force-free formalism \cite{Ponce:2014sza}. 
Neutrinos have been taken into account using order-of-magnitude leakage schemes~\cite{Ruffert1996,Rosswog:2003rv},  approximate transport (moment formalism)~\cite{1981MNRAS.194..439T,shibata:11,Foucart:2016rxm,Radice:2021jtw}, and Monte-Carlo methods~\cite{Foucart:2020qjb,Miller:2019dpt}. Formalism for the evolution of theories of gravity beyond GR have recently been developed for use in numerical simulations with \cite{Barausse:2012da,Shibata:2013pra} and without matter \cite{Witek:2020uzz,Okounkova:2022grv}.
Combining all of these important ingredients into sufficiently accurate simulations across the entire available parameter space remains however beyond our current abilities.

\subparagraph{Initial data.} GR simulations require solving the Einstein constraint equations on an initial spatial slice. In the case of compact binaries involving black holes and neutron stars, this initial data problem admits its formulation in terms of elliptic equations \cite{Pfeiffer:2002iy}, which in the case of neutron star mergers are solved iteratively to obtain initial conditions in hydrostatic equilibrium \cite{Taniguchi:2006yt}. The community has put significant effort in developing numerical codes to solve elliptic problems. 
Most of the state-of-the-art codes employ spectral methods to discretize the elliptic equations~\cite{Pfeiffer:2002wt,Ansorg:2004ds,Ansorg:2005bp,lorene,Grandclement:2006ht,Taniguchi:2006yt,Foucart:2008qt,Grandclement:2009ju,Ossokine:2015yla,Tacik:2015tja,Dietrich:2015pxa,Tichy:2019ouu,Papenfort:2021hod,Rashti:2021ihv,Vu:2021coj}, with the exception of \texttt{COCAL}~\cite{Uryu:2011ky,Tsokaros:2015fea,Uryu:2019ckz} and \texttt{NRPyElliptic}~\cite{Assumpcao:2021fhq} which employ finite-difference methods.  
The cost of solving the initial data problem is typically negligible compared to the cost of evolution. 
In addition, there are issues relating to transients arising from the relaxation of the initial data to a quasi-stationary geometry \cite{Lovelace:2008hd,Varma:2018sqd} (often referred to as ``junk radiation'') and gauge relaxation~\cite{Hannam:2006vv,Etienne:2024ncu}.

\begin{figure}
	\centering
    \includegraphics[width=0.98\textwidth]{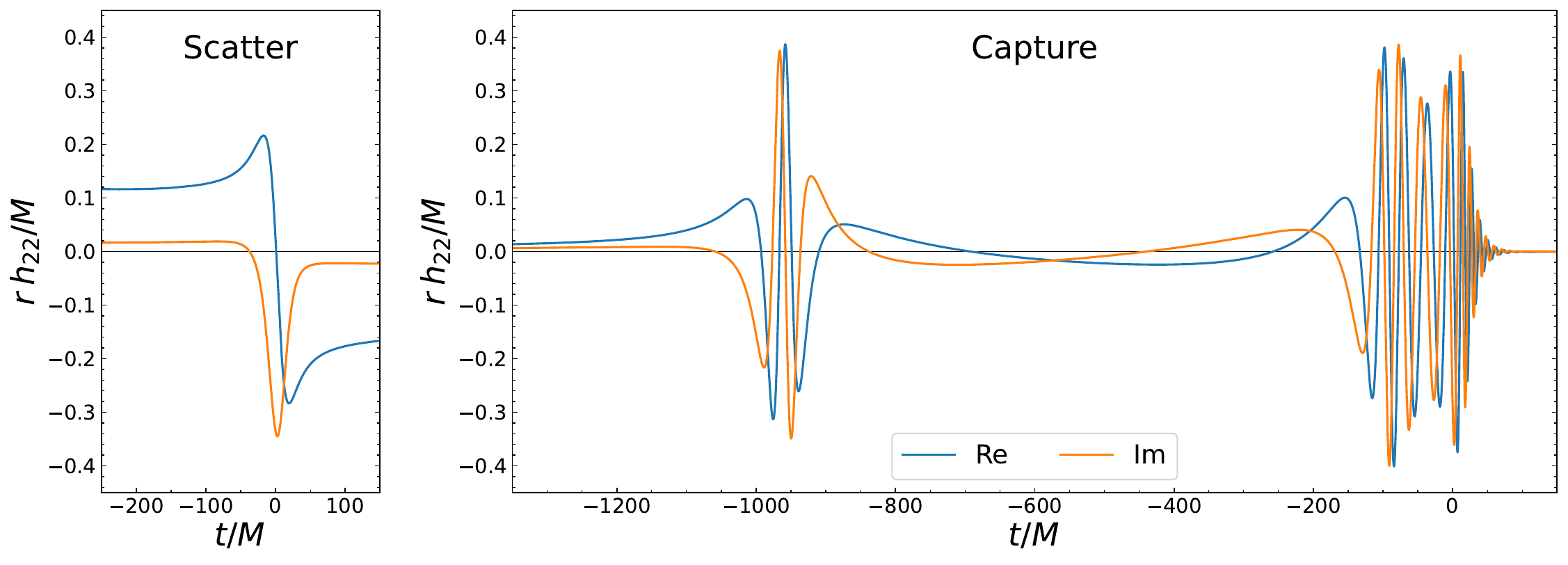}
    \caption{
            Exemplary waveforms of hyperbolic BBH encounters where the two BHs 
      are initially unbound. The left panel shows a scattering event
      where the BHs remain unbound, changing their direction of motion by $\sim 90^\circ$. The configuration shown at the right loses enough energy at the first interaction
      to be captured into a bound, eccentric binary that subsequently
      merges.  Both systems start with equal-mass, non-spinning BH,
      with data from the SXS waveform catalog~\cite{sxscatalog}.  Left
      SXS:BBH:3999 (initial energy $E_i=1.023$ and angular
      momentum $J_i=1.600 M$).  Right SXS:BBH:4000 ($E_i = 1.001$
      and $J_i=1.005 M$).
    }
    \label{fig:HyperbolicBBH}
\end{figure}

\subparagraph{BBH simulations.} 
One of the key motivations driving BBH simulations is the accurate modeling of the GW signal, especially through the merger and ringdown where NR information is essential. The semi-analytical waveform models used
to analyze GW data
(see section~\ref{sec:imr} for details) are calibrated and validated against entire catalogs of NR simulations~\cite{Mroue:2013xna,Husa:2015iqa,Jani:2016wkt,Boyle:2019kee,Ramos-Buades:2019uvh,Pratten:2020fqn,Healy:2022wdn,Joshi:2022ocr,Ramos-Buades:2022lgf}, emphasising the central importance of NR in modern GW astronomy. Models have also been developed based solely on simulations~\cite{Varma:2018mmi, Varma:2019csw, Yoo:2022erv}. The parameter space coverage of BBH simulations can be broadly split between quasi-circular binaries and eccentric binaries, including hyperbolic encounters. Significant attention has been given to modeling the 7-dimensional non-eccentric parameter space (mass ratio $q=m_1/m_2$ and black hole spins).
For aligned-spin binaries, current catalogs extend up to mass ratios  $q = 18$ and spins $|{\boldsymbol{\chi}}| \lesssim 0.8$ \cite{Husa:2015iqa,Jani:2016wkt,Boyle:2019kee,Pratten:2020ceb,Healy:2022wdn},
whereas for precessing binaries, the vast majority of simulations cover a restricted parameter space up to $q \sim 4$ and $| \boldsymbol{\chi}_i | \lesssim 0.8$ \cite{Jani:2016wkt,Boyle:2019kee,Healy:2022wdn}, though see \cite{Hamilton:2021pkf} which presented simulations sufficient to model single-spin binaries up to $q=8$. 
For spins $\gtrsim 0.8$, it becomes increasingly difficult to resolve the apparent horizons, which become smaller and increasingly deformed in the coordinate systems used in NR.  Nevertheless, individual simulations have reached spins as high as 0.998~\cite{Scheel:2014ina,Zlochower:2017bbg,Boyle:2019kee}.
For large mass ratio, computational cost increases because the decreasing size of the smaller BH forces the time-step to decrease as $\sim 1/q$, whereas the physical evolution time increases as $\sim q$.  Combined, these two effects increase computational cost at least as fast as $\sim q^{2}$ \cite{Dhesi:2021yje,LISAConsortiumWaveformWorkingGroup:2023arg} for high mass ratio binaries covering a certain frequency band.
Individual simulations have reached mass ratio up to $q = 1000$ \cite{Lousto:2020tnb,Rosato:2021jsq,Lousto:2022hoq}, and an alternative framework to tackle intermediate mass ratio binaries based on a worldtube excision method was presented in \cite{Dhesi:2021yje,Wittek:2023nyi,Wittek:2024gxn}, but long and accurate numerical waveforms at such high mass ratios remain beyond our current reach. 
For eccentric and hyperbolic binaries, preliminary catalogues are typically limited to $q \leq 6$, spins $|{\boldsymbol{\chi}}| \lesssim 0.8$, and $e \lesssim 0.8$ \cite{Ramos-Buades:2019uvh,Gayathri:2020coq, Gamba:2021gap, Joshi:2022ocr, Bonino:2024xrv}; non-spinning eccentric simulations up to $q = 10$ were presented in \cite{Ramos-Buades:2022lgf}.  Exemplary waveforms for hyperbolic encounters are shown in figure~\ref{fig:HyperbolicBBH}.

\subparagraph{NSBH/BNS simulations.} 
Simulations of  NSBH binary and of BNS are not as accurate as BBH simulations. Waveforms in GW databases~\cite{Dietrich:2018phi,Foucart:2018lhe,Kiuchi:2019kzt,Gonzalez:2022mgo} 
typically capture the impact of the finite size of the neutron star with $
 \gtrsim$ a few parts of a radian error at merger, accumulated over $\sim 10$ orbits.
 For BNS systems, non-spinning low-mass binaries ($M_{\rm binary}\lesssim 3M_\odot$) in quasi-circular orbits have been studied most extensively, though more asymmetric, eccentric, higher mass systems and/or spinning neutron stars have also been simulated (see e.g.~\cite{Baiotti:2016qnr,Paschalidis2017,Shibata:2019wef} for reviews). For NSBH systems, most simulations consider low mass or high spin
black holes in quasi-circular orbit with a low-mass neutron star (as
this configuration leads to EM signals), with the rest of the
parameter space more sparsely covered (see
e.g. \cite{Foucart:2020ats,Kyutoku:2021icp} for reviews).
  Specifically, 
  simulations have focused on the features of tidal disruption in the GW
  amplitude~\cite{Kyutoku:2010zd,Kyutoku:2011vz,Kawaguchi:2015bwa,Duez:2009yy,Foucart:2018lhe,Ruiz:2021gsv,Khamesra:2021duu}.  This has enabled predictions of the outcome of a NSBH merger based on the
  mass ratio, $|{\boldsymbol{ \chi}}_{\rm BH}|$, and the NS
  compactness for circular inspirals and negligible NS
  spin~\cite{Foucart:2012nc,Kruger:2020gig}, the shutoff
  frequency~\cite{Pannarale:2015jia}, and the properties of
  remnants~\cite{Zappa:2019ntl,Hayashi:2020zmn,Hayashi:2022cdq,Hayashi:2021oxy,Most:2021ytn}. Simulations
  have also explored effects of spin
  precession~\cite{Kawaguchi:2015bwa,Foucart:2020xkt}, and
  eccentricity~\cite{Stephens:2011as, Gold:2011df}, which complicate
  the identification of the tidal disruption shutoff.

Simulations provide us with a decent qualitative understanding of BNS/NSBH mergers, as well as quantitative predictions for the remnants/outflows of NSBH mergers~\cite{Pannarale:2013jua,Kawaguchi:2016ana,Foucart:2018rjc, Zappa:2019ntl, Gonzalez:2022prs}, the threshold for collapse of BNS merger remnants~\cite{Bauswein:2013jpa, Agathos:2019sah}, and its post-merger GW frequency~\cite{2015PhRvD..91f4001T, Breschi:2022xnc} (though magnetic fields and/or exotic matter may modify the latter~\cite{Bauswein:2018bma,Raithel:2022orm}). Quantitative predictions for many other properties of the post-merger remnant remain unreliable (see e.g. \cite{Nedora:2020qtd,Hayashi:2020zmn,Henkel:2022naw}). Simulations of BNS~\cite{Kiuchi:2015sga} and NSBH~\cite{Kiuchi:2015qua} have demonstrated the rapid growth of magnetic fields during merger, but demonstrating convergence of this effect remains difficult even the highest resolution simulations. This is problematic as magnetic fields can drive collimated outflows~\cite{Rezzolla:2011da,Paschalidis:2014qra,Ruiz:2016rai} and disk winds~\cite{2018ApJ...858...52S,Christie:2019lim}. Neutrino absorption, for its part, can greatly impact nucleosynthesis and the properties of post-merger EM signals~\cite{2013ApJ...775...18B,Wanajo:2014}. Self-consistent simulations of the merger and post-merger evolution are only becoming available in 2D~\cite{Fujibayashi:2022ftg} and 3D \cite{Hayashi:2022cdq}, and will certainly play an important role in our understanding of EM signals.

\subparagraph{Core-collapse.} Historically, core-collapse supernova (CCSN) simulations have included significantly more advanced microphysics (neutrino physics, dense matter EoS, nuclear reactions) than BNS simulations (see the discussion in
section~\ref{sec:collapse} and, e.g., refs.~\cite{Janka:2016fox,Fischer:2017zcr,Muller:2020ard} for reviews of CCSN simulations and microphysics), while making stronger approximations when modeling gravity and/or imposing symmetries on the system -- though as the importance of GR and 3D effects (e.g. turbulence) in CCSNe and the need for improved microphysics in BNS mergers became more obvious, technology transfers between these two fields has become more important. In CCSNe, GW emission is due to the oscillation of the post-collapse neutron star excited by 3D hydrodynamics instability, the asymmetric emission of neutrinos, and/or larger scale deformation of the forming neutron star in rapidly rotating systems~\cite{Abdikamalov:2020jzn,Vartanyan_2020}, thus potentially providing us information about the properties of dense matter and the rotation rate of the system~\cite{Afle:2020zgd}. GWs from CCSNe can only be seen for close-by objects~\cite{Gossan:2015xda}, yet provide us with a unique opportunity to combine neutrino and GW observations to better understand nuclear physics at high density. See  section~\ref{sect:CCSNdiv7} for a discussion of the corresponding observational prospects at ET.

\vspace{2mm}\noindent 
{\bf Main Open Challenges}

\subparagraph{BBH simulations.}
The main open challenges in extending the parameter space coverage of BBH simulations are producing accurate, long-duration simulations at high mass ratios and high spins where one has to resolve multiple length and time scales associated to both the large and small BHs. Implicit time-stepping may help with this issue~\cite{Lau:2011we}, but is costly and not used in current state-of-the art simulations. The BH excision used in multi-domain pseudospectral codes such as SpEC \cite{Scheel:2006gg,Szilagyi:2009qz,Buchman:2012dw} becomes increasingly complicated for highly distorted BH horizons and can require fine-tuning of numerical algorithms \cite{Hemberger:2012jz,Scheel:2014ina}.
Finite-difference codes based on the moving puncture framework are numerically more robust, but less computationally efficient at comparable mass ratios. High mass ratios ($q > 10$) are computationally challenging for all current codes, and it is currently not possible to study the high mass ratio regime with anywhere near the detail that has been achieved for comparable masses. Developing methods to do this is one of the major unsolved problems for BBH simulations.
In addition, high-spin simulations with BH spins $\chi\gtrsim 0.85$ constitute a very small portion of current catalog of NR simulations \cite{Jani:2016wkt,Boyle:2019kee,Lousto:2022hoq}. This is because standard initial data impose conformal flatness \cite{Brandt:1997tf,Caudill:2006hw}, which limits spins to $\chi < 0.93$ \cite{Gleiser:1997ng,Dain:2002ee,Dain:2008ck,Lovelace:2008tw}, and high-spin simulations~\cite{Lovelace:2008tw, Lovelace:2010ne, Lovelace:2014twa, Zlochower:2017bbg, Healy:2017vuz} are more challenging.

Besides difficulties with the numerical methods for solving initial data, the choice of adequate initial parameters 
is also an open problem. While there has been a lot of work to develop methods to estimate initial parameters for quasi-circular evolutions \cite{Pfeiffer:2007yz,Husa:2007rh,Buonanno:2010yk,Tichy:2010qa,Purrer:2012wy,Ramos-Buades:2018azo,Ramos-Buades:2019uvh,Habib:2020dba}, the choice of parameters for the most generic binaries such that the simulations have the desired values of eccentricity, radial phase as well as time to merger is still fairly unexplored \cite{Hinder:2008kv,Ramos-Buades:2019uvh,Islam:2021mha,Ramos-Buades:2022lgf,Ciarfella:2022hfy}. Covering the parameter space of eccentric binaries will be even more challenging, particularly for nearly hyperbolic encounters with long time intervals between periastron passages.

\subparagraph{NSBH/BNS Mergers.} Simulations involving matter face important challenges to reach the level of accuracy and physical realism required to serve ET science. In order to make sure that GW models do not bias our measurements of the neutron star EoS, pre-merger GW models with significantly improved phase accuracy are needed (see e.g.~\cite{Purrer:2019jcp,Read:2023hkv}). 
This is a steep challenge for current numerical codes which have shown 3rd \cite{Radice:2013xpa}
or 4th order convergence~\cite{Doulis:2022vkx} so far for simple EoS, and have reduced accuracy for more complex EoS~\cite{Foucart:2019yzo,Raithel:2022san}. 
Additonally, while schemes enabling 4th order convegence, such as the entropy-based flux-limiting scheme of ref.~\cite{Doulis:2022vkx}, mark a significant advancement towards ET accuracy requirements, such simulations remain limited in number.
Simulations have begun exploring phase transitions~\cite{Prakash:2021wpz,Espino:2023llj,2023PhRvD.107b4025U} and strange matter~\cite{2010PhRvD..81b4012B,Sekiguchi:2011mc} in neutron stars, as well as boson stars~\cite{Palenzuela:2007dm,Bezares:2022obu} but a more systematic understanding of the impact of more advanced nuclear physics, MHD, and neutrinos on post-merger GW signals will be needed in preparation for ET.
Recent simulations relying on subgrid models~\cite{Vigano:2020ouc} as well as extremely high-resolution simulations of merger remnants~\cite{Kiuchi:2023obe} and post-merger accretion disks~\cite{Gottlieb:2023est} have been able to capture the growth of magnetic fields from relatively low initial magnitude, yet such simulations remain rare and expensive, and even now the process leading to the production of collimated jets and gamma-ray bursts remains an open question. The cooling of the remnant and heating of the outflows by neutrinos, as well as composition changes in the outflows, is captured at the $\sim 10\%$ level by the best existing transport schemes~\cite{Foucart:2020qjb,Radice:2021jtw}, but with the exception of the post-merger evolutions in~\cite{Miller:2019dpt,Hayashi:2021oxy,Curtis:2022hjy}, only simpler treatment of neutrinos have been used in conjunction with high-resolution MHD simulations~\cite{Palenzuela:2015dqa,Hayashi:2022cdq}, limiting our ability to study the impact of neutrinos on outflow composition, jet formation, and bulk viscosity. Performing sufficiently long simulations of the post-merger remnant with adequate microphysics across the entirety of the relevant parameter space thus remains an important issue for ET. More generally, it is worth noting that the exact impact of currently missing physics (nuclear physics, neutrino physics, non-ideal MHD,...) on the observable properties of neutron star mergers remains unknown, greatly complicating the task of determining what is actually measurable in these systems.

\subparagraph{CCSNe simulations.} The simulations of CCSNe suffer from some of the same issues as NSBH/BNS mergers: the need to perform expensive 3D simulations with advanced microphysics across a high-dimensional parameter space (of both initial conditions and input nuclear physics). In addition, the outcome of CCSNe is impacted by the late-stage 3D evolution of massive stars, and the presence of inhomogeneities in the pre-collapse remnant. Generating realistic initial conditions for these simulations is thus a complex and important question, that itself can only be solved through 3D hydrodynamics simulations of massive stars including a detailed modeling of nuclear burning.

\paragraph{Important Next Steps}

\subparagraph{Code development.} The previous sections should make it clear that there are two important aspects of NR simulations that need to greatly improve for ET: accuracy (specifically for inspirals) and physical realism (nuclear physics, MHD, neutrinos, for simulations with matter; gravity beyond General Relativity). This will require NR to leverage current and upcoming exascale computing facilities, and has driven the development of next-generation codes, including the Disontinuous Galerkin code SpECTRE~\cite{Kidder:2016hev,Moxon:2021gbv,Vu:2021coj,deppe_nils_2021_5083825}, the octree-refined wavelet code DendroGR \cite{Fernando:2022php} and finite difference codes GRAthena++\cite{Daszuta:2021ecf}, CarpetX (part of the Einstein Toolkit~\cite{EinsteinToolkit:2022_11}, the most used open-source NR code), and AMReX \cite{Peterson:2023azu}; as well as continued development of existing codes already used for near exascale simulations (e.g. SACRA~\cite{Hayashi:2022cdq}). Testing and development of these (and other future) codes is an important step to allow the NR community to provide ET with simulations of the required quality.

\subparagraph{Understanding current accuracy and future requirements.}
Recent studies investigating the accuracy necessary to perform unbiased parameter estimation with next generation detectors have highlighted the need of orders of magnitude improvements for both waveform models and NR simulations~\cite{Purrer:2019jcp, Gamba:2020wgg}.
These estimates, however, are largely based on simple comparisons between simulations performed at different resolutions, and often do not account for other possible sources of error, such as systematic differences between NR codes, the integration from Newman-Penrose scalar $\Psi_4$ to GW strain~\cite{Reisswig:2010di}  (challenging for highly eccentric or hyperbolic systems), extrapolation of the GWs to infinity and -- when considering long inspirals -- EOB-NR or PN-NR hybridization.
The quoted requirements further depend on the parameters of the binary, and may become more stringent in the more extreme regions of the parameter space.
In order to drive waveform development it is then fundamental to develop a framework to fully characterize the sources of error that currently affect our simulations, allowing important insights on the determination of accuracy requirements for ET.

\subparagraph{BBH binaries.} For BBH simulations, a dual push towards simulations that can match the accuracy requirements discussed in the previous section and an expansion of the covered parameter space to high-spin, high mass ratios, and eccentric systems are likely to be needed to satisfy ET requirements. 

\subparagraph{NSBH/BNS binaries.} Matter simulations should greatly benefit from next-generation codes aimed to leverage exascale computational resources. The development of those codes may however not be sufficient to allow simulations to reach the level of accuracy desirable for ET science. Subgrid models capable of capturing the growth of magnetic fields in mergers may remain necessary even in next-generation codes. While progress has been made in that direction~\cite{Radice:2017zta,Shibata:2017jyf,Palenzuela:2022kqk}, whether these models can reasonably match the results of a MHD simulation remains an open question. Larger parameter studies of merger and post-merger evolution including magnetic field effects, neutrinos, and a range of potential equations of state (including exotic physics that can be probed with ET) are also necessary for post-merger GW and EM modeling, but computationally expensive. On shorter time scales, continued development of existing post-merger GW models for BNS systems, especially with improved/more exotic nuclear physics, and the improvement of the qualitative disruption models currently used by NSBH models, could benefit from large parameter space studies with existing codes including the range of equations of state currently considered in nuclear physics.
Finally, it will be crucial to better understand the impact of currently missing physics in simulations, and determine more accurately the level of realism that is desired in order to reach a desired accuracy level in predictions for the GW and EM signals powered by these mergers.

\subsubsection{Weak-field Expansions}
\label{sec:weak}


{\bf State-of-the-Art}
\vspace{2mm}

\noindent
Parallel to the numerical methods presented in the previous section, \emph{analytical} frameworks are also used to derive gravitational waveforms.
In this section, we discuss the class of \emph{weak-field} methods, that are valid when the separation of the two object is large (typically during the inspiral phase).
Those methods rely on post-Newtonian (PN, low velocity) and post-Minkowskian (PM, large separation) expansions: the first one is an expansion in $(v/c)^2$ where $v$ is the relative velocity of the two bodies and $c$ the light speed; the second one, an expansion in $Gm/rc^2$ where $G$ is Newton's constant, $m$ is the total mass and $r$ the separation. 
In a bound and virialized state, the two expansions are correlated, but they stay independent in the scattering problem, which has lately attracted more and more attention as an useful tool to study the two-body gravitational interaction. For recent reviews, see notably~\cite{Buonanno:2014aza, Blanchet:2013haa, Goldberger:2007hy, Foffa:2013qca,Porto:2016pyg,Levi:2018nxp}.
Note that another analytical framework is discussed in the next section: the Gravitational Self-Force, that is able to span the whole history of the binary, but valid only for a large mass ratio.

The key observables delivered by weak-field methods are the GW phase and amplitude of the waveform. For bound systems, the GW phase is computed \emph{via} the energy flux-balance equation, expressing the loss of energy by GW emission ${\frac{\mathrm{d}E}{\mathrm{d}t} = - \mathcal{F}_\text{GW}}$
where $E$ is the (conservative) binding energy and $\mathcal{F}_\text{GW}$, the flux of radiation emitted.
Once both quantities are derived, the GW phase is obtained by a simple integration, and the amplitude is extracted from the moments that enter the flux.
For non-spinning compact binary systems in quasi-circular orbits, complete results for the phase and amplitude are respectively known at 4.5PN~\cite{Blanchet:2019zlt,Blanchet:2023bwj} and 3.5PN~\cite{Henry:2022ccf}. For binaries with spinning but non-precessing objects, the phase has been derived at 4PN~\cite{Cho:2022syn}, and the amplitude, up to 3.5PN in the case of aligned spins~\cite{Henry:2022dzx}.
Once precession is added, and while higher-PN order information exists for generic orbits~\cite{Chatziioannou:2012gq}, the accuracy is degraded to 2PN for the phase, and 1.5PN for the amplitude to get ready-to-use waveforms on precessing quasi-circular orbits~\cite{Arun:2008kb}.
The contribution of the quadrupole-monopole interaction due to a rotating star’s oblateness which enters at 2PN order has been computed at leading-order in the radiation losses for generic orbits~\cite{Poisson:1997ha,Gergely:2002fd}.
As for the tidal effects, that enter at 5PN, their contribution in the phase is known at 2.5PN relative order~\cite{Henry:2020ski}, and their coupling with spins, at the leading 6.5PN order~\cite{Abdelsalhin:2018reg}. 

For the conservative sector of non-spinning objects, three different techniques have been used to compute the binding energy at the 4PN level of precision: the canonical Hamiltonian formulation of GR~\cite{Jaranowski:2013lca, Jaranowski:2015lha, Damour:2014jta, Damour:2015isa, Damour:2016abl}, the Fokker Lagrangian~\cite{Bernard:2015njp, Bernard:2016wrg, Bernard:2017bvn, Marchand:2017pir, Bernard:2017ktp} and the EFT~\cite{Goldberger:2004jt,Foffa:2012rn, Galley:2015kus, Porto:2017dgs, Foffa:2019rdf, Foffa:2019yfl, Blumlein:2020pog} approaches, yielding physically equivalent results. The equations of motion at 4.5PN order, resulting from the 2PN corrections of the radiation-reaction effect, are also known~\cite{Leibovich:2023xpg}. The next level of precision, at 5PN order, is in a mature stage \cite{Blumlein:2020pyo,Porto:2024cwd}, with partial results also at 6PN~\cite{Bini:2020nsb} and 7PN~\cite{Blanchet:2019rjs} also known. Other methods have also been traditionally applied in the literature to derive the conservative dynamics, but have not (yet) attained the 4PN order, such as the surface-integral approach at 3PN~\cite{Itoh:2003fy}, or the extended-body framework at 2.5PN order~\cite{Kopeikin:1985}.

For radiative effects, these are expressed in terms of multipole moments~\cite{Thorne:1980ru} (which are also used to compute the amplitude), that generalize the Einstein's quadrupole formula~\cite{Einstein:1918btx}.
Using the post-Newtonian-Multipolar-post-Minkowskian algorithm~\cite{Blanchet:2013haa}, the relevant (spin-independent) moments have been derived to 3.5PN~\cite{Blanchet:2001ax,Faye:2012we,Faye:2014fra}. The 4PN moments, leading to the phase and amplitude at the same order,  are also known~\cite{Faye:2014fra,Blanchet:2008je,Henry:2021cek}, notably the quadrupole moment which requires a proper regularization scheme~\cite{Marchand:2020fpt,Larrouturou:2021dma,Larrouturou:2021gqo}, and including non-linear interactions~\cite{Blanchet:2022vsm,Trestini:2022tot,Trestini:2023wwg}. For quasi-circular orbits, the full 4.5PN flux has also been computed~\cite{Marchand:2016vox,Blanchet:2023sbv}, and the 4.5PN sector has been confirmed by an independent PN reexpansion of resummed waveforms~\cite{Messina:2017yjg}.
The 4.5PN phase for quasi-circular orbits have also been derived~\cite{Blanchet:2023bwj}.
Two other methods are also in use, the EFT approach developed to 3PN order in the radiative sector~\cite{Leibovich:2019cxo,Amalberti:2024jaa}, and the direct integration of the relaxed equations, at 2PN~\cite{Will:1996zj}.

Note that in order to derive the observable phase, one has also to take into account the absorption of the wave by the black holes horizons, see~\cite{Poisson:1994yf,Tagoshi:1997jy,Alvi:2001mx,Porto:2007qi,Chatziioannou:2012gq,Saketh:2022xjb}.

For the case of spinning bodies, extensive dynamical studies have been initially developed in the framework of the PN approximation~\cite{Kidder:1995zr,Gergely:2014vea,Gergely:1998sr,Gergely:1999pd,Faye:2006gx,Blanchet:2006gy,Bohe:2013cla,Bohe:2012mr}, while significant present knowledge of the dynamics has been achieved through the EFT approach~\cite{Goldberger:2004jt,Porto:2005ac,Porto:2006bt,Porto:2008tb,Porto:2008jj,Porto:2016pyg,Levi:2018nxp}. In the conservative sector, the binding energy is known up to 5PN order~\cite{Bohe:2015ana,Levi:2022rrq}, whereas the radiative multipoles and GW phase at the 4PN level of accuracy for aligned spins were derived in~\cite{Cho:2021arx,Cho:2022syn,Henry:2022dzx} (see also \cite{Bohe:2012mr,Bohe:2013cla,Marsat:2013caa} for alternative derivations up to 3.5PN order). The radiation-reaction force has also been computed up to 4.5PN order in \cite{Maia:2017gxn,Maia:2017yok}. Spin-dependent effects in the amplitude have also been derived to 3.5PN order~\cite{Porto:2012as,Mishra:2016whh,Henry:2022dzx}. Note also that leading PN order results for the binding energy, the GW flux and GW modes were obtained to all orders in spins~\cite{Siemonsen:2017yux}.

To get inspiral waveforms for compact binaries in eccentric orbits, one commonly relies on the quasi-Keplerian parametrization (QKP), a semi-analytic representation of the perturbative, slowly precessing post-Newtonian motion. The conservative motion as well as instantaneous and hereditary contributions to the secular evolution of the orbital elements are known to 3PN order, in both modified harmonic and ADM-type coordinates \cite{Damour:2004bz,Memmesheimer:2004cv,Arun:2007rg,Arun:2007sg,Arun:2009mc} and to 4PN order in the ADM-type coordinates only~\cite{Cho:2021oai}.
The complete 3PN-accurate GW amplitudes from non-spinning eccentric binaries have been derived using the multipolar post-Minkowskian formalism, including all instantaneous, tail and non-linear contributions to the spherical harmonic modes \cite{Mishra:2015bqa, Boetzel:2019nfw, Ebersold:2019kdc}.

For non-precessing spinning eccentric binaries, the effects of spin-orbit and spin-spin couplings on the binary evolution and gravitational radiation have been worked out to leading order in QKP~\cite{Klein:2010ti}. Subsequent work has aimed to extend the treatment of spins in QKP to higher PN orders~\cite{Tessmer:2010hp, Tessmer:2012xr}. The waveforms modes to next-to-leading (3PN) order for aligned spins and eccentric orbits have been derived in~\cite{Khalil:2021txt,Henry:2023tka}, including tail and memory contributions.
Regarding precessing eccentric systems, the inspiral dynamics of spinning and precessing compact binaries with arbitrary eccentricities, including the mass quadrupole-monopole interaction, has been obtained at leading order~\cite{Gergely:2009zz,Gergely:2010mu}.
Gravitational waveforms have been obtained to leading order in the precessing equation~\cite{Klein:2021jtd}. A fully analytical treatment has also been proposed to leading order order in the spin-obit and spin-spin couplings, including higher harmonics~\cite{Paul:2022xfy}.

For compact binaries evolving on hyperbolic trajectories (or scattering case),\footnote{Note that orbits dubbed chameleon which alternate between hyperbolic and highly eccentric due to spin effects were also found~\cite{Gergely:2014vea}.} the knowledge of the scattering angle or gravitational waveform would, in principle, not be relevant for the study of bound orbits. Nevertheless, such configurations have recently gained much interest in the context of the Post-Minkowskian expansion, notably after the connections to the EOB potentials discussed in \cite{Damour:2016gwp}, as well as the (boundary-to-bound) map that links scattering data to observables for elliptic orbits~\cite{Kalin:2019rwq,Kalin:2019inp,Cho:2021arx,Dlapa:2023hsl,Dlapa:2024cje}. As, for instance, the scattering angle contains terms computed to all orders in the relative velocity, one can extract valuable high-precision information on the bound dynamics from the knowledge of the hyperbolic case. The current state-of-the-art for the total relativistic impulse is at 4PM for both conservative and radiative sectors, for non-spinning ~\cite{Dlapa:2021npj,Dlapa:2021vgp,Dlapa:2022lmu}, and spinning \cite{Jakobsen:2023hig,Jakobsen:2023ndj} objects, and also partial results in the sector at 5PM (and first order in the SF expansion) \cite{Driesse:2024xad,Driesse:2024feo}, 
obtained by means of a Post-Minkowskian version \cite{Kalin:2020mvi,Liu:2021zxr,Kalin:2022hph,Dlapa:2023hsl,Mogull:2020sak,Jakobsen:2021zvh} of the worldline EFT approach \cite{Goldberger:2004jt,Porto:2005ac}. Other methods have  been developed, such as the \textit{scattering amplitudes} approach, that have also completed the conservative \cite{Bern:2021dqo,Bern:2021yeh} and dissipative \cite{Damgaard:2023ttc} (non-spinning) impulse at 4PM order, while yielding several results at low PM orders but high orders in the spins \cite{Aoude:2022thd}. The derivation of the waveform has also received attention, and was computed by several groups using scattering amplitudes methods to 3PM order \cite{Bini:2024rsy,Georgoudis:2024pdz,Brandhuber:2023hhy}. Other relevant results include the inclusion of various types of tidal effects \cite{Kalin:2020lmz,Saketh:2023bul,Jakobsen:2023pvx}.

\vspace{2mm}\noindent
{\bf Main Open Challenges}

\vspace{2mm}

\noindent
For non-spinning systems on bound orbits, the current frontier is the expression of both the energy and the flux at 5PN order. The derivation of the binding energy and equations of motion at 5PN has been tackled in two different frameworks: the so-called ``Tutti-Frutti'' (TF) Hamiltonian approach~\cite{Bini:2020nsb,Bini:2020hmy} and EFT methodologies \cite{Blumlein:2020pyo,Blumlein:2021txe,Almeida:2022jrv,Almeida:2023yia,Porto:2024cwd} (see also \cite{Henry:2023sdy}). While the TF Hamiltonian still contains two unknown parameters at 5PN, the EFT derivation is in principle complete and includes the subtle tail and memory effects (where the GWs emitted are back-scattered on the geometry and earlier radiation, respectively, and reabsorbed by the system), as well as contributions quadratic in the (linear) radiation-reaction forces. The subtleties involved in the computation led to an initial disagreement between the results at 5PN in \cite{Blumlein:2021txe,Almeida:2022jrv,Almeida:2023yia} and the overlap with the 4PM scattering dynamics in \cite{Dlapa:2021vgp,Dlapa:2022lmu}. This has been recently resolved in \cite{Porto:2024cwd}. Yet, the identification of the total conservative part of the full dynamics, in particular the determination of the two unknowns in the TF formalism, remains an open quest. We should also emphasize that the inclusion of spin effects is relatively straightforward within the EFT approach \cite{Porto:2016pyg,Levi:2018nxp}, and therefore the resolution of these discrepancies can also ultimately lead to higher order waveform models including spin contributions.

Other progresses in the case of bound orbits include the determination of the flux and waveform for generic precessing spins on eccentric orbits, beyond the slowly-precessing and/or small eccentricity assumptions.
The post-circular (PC) formalism \cite{Yunes:2009yz} provides a method to recast the time-domain response function $h(t)$ into a form that permits an approximate fully analytic Fourier transform in the stationary phase approximation (SPA), under the assumption that the eccentricity is small, leading to non-spinning, eccentric Fourier-domain inspiral waveforms. It has been extended to 3PN order~\cite{Tanay:2016zog, Moore:2016qxz}, and has included previously unmodeled effects of periastron advance~\cite{Tiwari:2019jtz}.
However such an approach is limited by the necessary expansion in small eccentricity. Newer models aim for validity in the range of moderate to high eccentricities, by utilizing numerical inversions in the SPA and resummations of hypergeometric functions to solve orbital dynamics~\cite{Moore:2018kvz, Moore:2019xkm}.
A semi-analytic frequency-domain model for eccentric inspiral waveforms in the presence of spin-induced precession has been developed with the help of a shifted uniform asymptotics technique to approximate the Fourier transform, a numerical treatment of the secular evolution coupled to the orbital-averaged spin-precession, and relying on a small eccentricity expansion~\cite{Klein:2018ybm}. However, the full treatment of precession at arbitrary eccentricities still remains to be done beyond leading orders.

Finally, the proper PN treatment of the dynamics of magnetized binaries (e.g. composed of neutron stars) has also been studied~\cite{Vasuth:2003yr,Henry:2023guc}.

When it comes to scattering results, and their relevance to the study of bound orbits, there are two major obstacles. Firstly, a computational hurdle, in the form of a significant increase in complexity in the Feynman integration problem, see the discussions in e.g. \cite{Dlapa:2023hsl,Driesse:2024xad}; and secondly, in circumventing the presence of nonlocal-in-time effects \cite{Damour:2014jta,Galley:2015kus,Bini:2020nsb,Bini:2020hmy} that are not directly amenable to a boundary-to-bound continuation to elliptic orbits \cite{Cho:2021arx}. 

\vspace{2mm}\noindent
{\bf Important Next Steps}

\vspace{2mm}
\noindent
Pushing the weak-field results to high accuracy seems an important goal for at least three reasons.
First, it may be necessary for the precise parameter estimation of low-mass binaries (that will be observed in their inspiral phase during a long time).
Second, high precision waveforms are very beneficial for the calibration of numerical waveforms, and are crucial for testing the second-order self-force computations, as done e.g.  in~\cite{Warburton:2021kwk}.
Last, as the difference between canonical and exotic compact object enter at high orders (at least 5PN, due to the so-called ``effacement principle"~\cite{Damour:1982wm}), high precision can be crucial to clearly discriminate between those.

Another important frontier for the ET science is not only high accuracy, but also the inclusion of physical effects, such as eccentricity, that are poorly understood, even at low precision.
For instance, most of (if not all) the compact objects that ET will observe will have nonzero spins, yet our current waveforms are usually derived under the much simplified assumption of aligned-spin and non-precessing, quasi-circular orbits, which the compact binaries are definitively not expected to necessarily follow. Moreover, neutron stars bear strong magnetic fields, that will act on the trajectories, and thus on the GW emission~\cite{Bourgoin:2021yzf}.
Furthermore, all these effects couple to each other, and to the eccentricity and precession of the orbit. Yet, even though they may play a crucial role in the data analysis of ET, only a handful of such couplings have been studied and at very low levels of accuracy. A greater effort is needed to  include all relevant physical effects and respective couplings in the computation of analytical waveforms.

\vspace{2mm}\noindent
{\bf Recent Progress}

\vspace{2mm}
\noindent
Recently, the comparison between the 4.5PN and 2SF (see section~\ref{sec:GSF}) results has been performed and it proved to be in very good agreement, even in the late-inspiral regime of PN when non-linearities become important~\cite{Warburton:2024xnr}. Notable progress has also been achieved in the realm of both the PN and PM expansions. In particular, the discrepancy between the previous 5PN results in \cite{Blumlein:2021txe,Almeida:2022jrv,Almeida:2023yia} and the complete 4PM knowledge of the impulse derived in \cite{Dlapa:2022lmu,Dlapa:2023hsl,Jakobsen:2023hig}, has been resolved \cite{Porto:2024cwd}. Furthermore, a first attempt at the complete fourth-loop level of precision in our understanding of scattering events was recently done in \cite{Driesse:2024xad}, at first order in the SF expansion. Novel developments, in particular in the Feynman-integration frontier, will be sorely needed to continue forward in this realm. At the same time, the separation of local- and nonlocal-in-time effects was recently performed in \cite{Dlapa:2024cje} in the context of unbound/bound dynamics at 4PM order. The results reported in \cite{Dlapa:2024cje} are readily applicable to waveform models for elliptic-like orbits, incorporating an infinite tower of velocity corrections. New ideas are also needed here to enable the direct application of scattering result to the dynamics of bound systems at higher orders.

\subsubsection{Gravitational Self-Force}
\label{sec:GSF}


\vspace{2mm}\noindent
{\bf State of the art}

\subparagraph{Formalism.} GSF theory is applicable to binaries with small mass ratios, for which $\eps := m_2/m_1\ll 1$. The binary's spacetime metric is expanded in powers of the mass ratio, 
\begin{equation}
g_{\alpha\beta}=g^{(0)}_{\alpha\beta}+\eps h^{(1)}_{\alpha\beta}+\eps^2 h^{(2)}_{\alpha\beta}+\ldots,
\end{equation}
where the background metric $g^{(0)}_{\alpha\beta}$ describes the spacetime of the primary object, typically taken to be a Kerr BH, and the corrections $h^{(n)}_{\alpha\beta}$ are perturbations created by the presence of the secondary object, which may be a BH or a material body. At zeroth order, the secondary moves on a geodesic of $g^{(0)}_{\alpha\beta}$. At subleading orders, the perturbations $h^{(n)}_{\alpha\beta}$ exert a self-force on the secondary, accelerating it away from geodesic motion~\cite{Poisson:2011nh,Harte:2014wya,Pound:2015tma,Barack:2018yvs,Pound:2021qin}. The emitted waveform is then contained in the perturbations $h^{(n)}_{\alpha\beta}$. Calculations involving the $n$th-order perturbation $h^{(n)}_{\alpha\beta}$ are referred to as $n$SF. 

Equations of motion for the secondary and field equations for the perturbations $h^{(n)}_{\alpha\beta}$ have been derived through second order in $\eps$ in a generic background spacetime~\cite{Mino:1996nk,Quinn:1996am,Detweiler:2000gt,Detweiler:2002mi,Gralla:2008fg,Pound:2009sm, Gralla:2011zr,Harte:2011ku,Detweiler:2011tt,Pound:2012nt,Gralla:2012db,Harte:2014wya,Pound:2017psq,Upton:2021oxf}. When specialized to a BH background, the field equations can be solved using methods of BH perturbation theory~ \cite{Regge:1957td,Zerilli:1970wzz,Teukolsky:1973ha,Wald:1973wwa,Chrzanowski:1975wv,Wald:1978vm,Kegeles:1979an,Campanelli:1998jv,Sasaki:2003xr,Martel:2005ir,Barack:2005nr,Brizuela:2009qd,Green:2019nam,Pound:2021qin,Dolan:2021ijg,Spiers:2023cip,Spiers:2023mor}.

\subparagraph{Waveforms.} Waveform-generation schemes in the GSF approach have traditionally focused on the inspiral phase, and waveforms are now typically generated within a multiscale framework~\cite{Chua:2020stf,Miller:2020bft, Pound:2021qin,Hughes:2021exa,Katz:2021yft, Isoyama:2021jjd}, which exploits the fact that a small-mass-ratio inspiral occurs slowly, over $\sim 1/\eps$ orbits. The evolution is written in terms of a set of ``slow'' parameters $J_A$ (the secondary's orbital energy, angular momentum, and Carter constant, the primary's mass and spin), which evolve on the slow time scale $\sim m_1/\eps$, and a set of ``fast'' phase variables $\psi^A$, which evolve on the orbital time scale $\sim m_1$. In terms of these variables, the system's evolution is governed by equations of the form~\cite{Hinderer:2008dm,VanDeMeent:2018cgn,Miller:2020bft,Pound:2021qin}
\begin{align}
\frac{d\psi^A}{dt} &= \Omega_{(0)}^A(J_B)+\eps\Omega_{(1)}^A(J_B)+O(\eps^2),\label{GSFeq:psidot}\\
\frac{dJ_A}{dt} &= \eps \left[G^{(0)}_A(J_B) + \eps G_A^{(1)}(J_B) + O(\eps^2)\right],
\label{GSFeq:Jdot}
\end{align}
where $\Omega^A_{(0)}=(\Omega^r_{(0)},\Omega^\theta_{(0)},\Omega^\phi_{(0)})$ are the three independent frequencies of geodesic orbits in Kerr spacetime~\cite{Schmidt:2002qk,Mino:2003yg,Fujita:2009bp,Grossman:2011ps,Pound:2021qin}.
Each $\ell m$ mode of the perturbations $h^{(n)}_{\alpha\beta}$ (and therefore of the emitted waveform) is then a sum of slowly varying amplitudes $h^{(n)}_{\ell m \omega_{k}}(J_A)$ multiplied by rapidly evolving phase factors:
\begin{equation}\label{GSFeq:waveform}
h^{(n)}_{\ell m} = \sum_{k_A}h^{(n)}_{\ell m \omega_{k}}(J_B)e^{ik_A \psi^A},
\end{equation}
where $k_A=(k_r,k_\theta,k_\phi)$ are integer constants and $\omega_k:=k_A\Omega^A_{(0)}$. Concretely, $\psi^A=(\psi^r,\psi^\theta, \psi^\phi)$ are the phases of the secondary's radial, polar, and azimuthal motion around the primary, such that the ratios $\dot\psi^\phi/\dot\psi^r$ and $\dot\psi^\phi/\dot\psi^\theta$ represent precession of periapsis and precession of the orbital angular momentum around the central BH's spin axis. If the secondary is spinning, then eqs.~\eqref{GSFeq:psidot}--\eqref{GSFeq:waveform} are straightforwardly extensible to include its spin precession phase and slowly evolving parameters associated with its spin components~\cite{Drummond:2022xej,Drummond:2022efc,Skoupy:2023lih,Witzany:2023bmq}. These effects modify the phasing at 1PA (see below).

The multiscale structure of eqs.~\eqref{GSFeq:psidot}--\eqref{GSFeq:waveform} conveniently splits waveform generation into offline calculations and online ones. In the offline stage, $h^{(n)}_{\ell m\omega_k}$, $\Omega^A_{(n)}$, and $G^{(n)}_A$ are  pre-computed by solving the field equations (formulated in multiscale form) on a grid of parameter values $J_A$. Using this stored offline data, the online stage solves eqs.~\eqref{GSFeq:psidot} and~\eqref{GSFeq:Jdot} and evaluates the sum~\eqref{GSFeq:waveform} to obtain a waveform. The Fast EMRI Waveforms (FEW) package has made this online stage sufficiently rapid for gravitational-wave data analysis~\cite{Chua:2020stf,Katz:2021yft}.

A leading-order waveform in this construction uses the slowly evolving geodesic frequencies $\Omega^A_{(0)}(J_B)$ together with the dissipative driving forces $G^{(0)}_A(J_B)$; this is referred to as adiabatic order (0PA). Subsequent orders in eqs.~\eqref{GSFeq:psidot}--\eqref{GSFeq:Jdot} are referred to as $n$th post-adiabatic order ($n$PA). In the remainder of this section, we summarise the state of the art at each order as well as progress to cover regions of phase space where the multiscale expansion breaks down.

Before proceeding to that summary, we note that alternatives to the multiscale expansion have also been formulated, though never implemented beyond 0PA order. Most prominently, waveforms have been generated using an iterative scheme in which the linearized field equations are solved in the time domain with an adiabatically inspiraling secondary~\cite{Sundararajan:2008zm,Barausse:2011kb,Taracchini:2014zpa,Harms:2014dqa,Rifat:2019ltp}. This method does not immediately provide rapid waveform generation, due to the cost of solving the field equations in the online step, but the resulting waveforms can be used to build a surrogate model~\cite{Rifat:2019ltp,Islam:2022laz}.

\subparagraph{0PA waveforms.} Leading-order (0PA) waveforms require the driving forces $G^{(0)}_A(J_B)$, which are calculated from the radiative pieces of the modes $h^{(1)}_{\ell m\omega_k}$ (1SF). Decades of work have enabled efficient calculations of these inputs~\cite{Press:1973zz,Teukolsky:1974yv,Galtsov:1982hwm,Nakamura:1987zz,Cutler:1994pb,Tagoshi:1995sh,Mano:1996vt,Mino:1997bx,Hughes:1999bq,Finn:2000sy,Hughes:2001jr,Glampedakis:2002ya,Mino:2003yg,Martel:2003jj,Sasaki:2003xr,Poisson:2004cw,Drasco:2005is,Drasco:2005kz,Sago:2005gd,Sago:2005fn,Mino:2005an,Hughes:2005qb,Sundararajan:2007jg,Ganz:2007rf,Sundararajan:2008zm,Hinderer:2008dm,Fujita:2009us,Fujita:2010xj,Pan:2010hz,Zenginoglu:2011zz,Fujita:2012cm, Isoyama:2013yor,Fujita:2014eta, Shah:2014tka,Harms:2014dqa,Sago:2015rpa,Gralla:2016qfw,Isoyama:2018sib,Burke:2019yek, Chua:2020stf,Fujita:2020zxe,Munna:2020juq,Munna:2020som}, and 0PA waveforms can now be computed for generic orbits around a Kerr BH, both numerically~\cite{Hughes:2021exa} and analytically (in a weak-field expansion)~\cite{Isoyama:2021jjd}. However, substantial effort is still required to populate the parameter space with sufficient offline data.

\subparagraph{1PA waveforms.} First-post-adiabatic waveforms (1PA) require the corrections $\Omega_{(1)}^A(J_B)$, which are calculated from the time-symmetric pieces of $h^{(1)}_{\ell m\omega_k}$ (1SF), together with the driving forces $G^{(1)}_A(J_B)$, which are calculated from the radiative pieces of $h^{(2)}_{\ell m\omega_k}$ (2SF). 

Practical methods of solving the 1SF field equations and calculating the necessary inputs for $\Omega_{(1)}^A$ have been developed over the past two decades~\cite{Barack:1999wf,Barack:2001gx,Barack:2002bt,Barack:2002mh,Barack:2005nr,Barack:2007tm,Barack:2007we,Vega:2007mc,Casals:2009zh,Barack:2010tm,Keidl:2010pm,Shah:2010bi,Dolan:2012jg,Akcay:2013wfa,Pound:2013faa,Merlin:2014qda,Osburn:2014hoa,Isoyama:2014mja,Wardell:2014kea,Wardell:2015ada,vandeMeent:2015lxa,vandeMeent:2016pee,Merlin:2016boc,vanDeMeent:2017oet}, leading to 1SF calculations for generic bound orbits in Kerr spacetime~\cite{vandeMeent:2017bcc}. These results have been incorporated into inspiral and waveform models covering much of the parameter space~\cite{Warburton:2011fk,Lackeos:2012de,Osburn:2015duj,VanDeMeent:2018cgn,Lynch:2021ogr,Lynch:2023gpu}. 

Further development of 1SF methods is ongoing~\cite{Toomani:2021jlo,Dolan:2021ijg,PanossoMacedo:2022fdi,Barack:2017oir, Hughes:2005qb, Hughes:2006pm,Long:2021ufh,Canizares:2008dp,Canizares:2010hu,Canizares:2011fd,Field:2009kk,Diener:2011cc,Markakis:2014nja,DaSilva:2023xif,OBoyle:2022yhp,Markakis:2023pfh,OBoyle:2023jqo,Casals:2013mpa, Wardell:2014kea, Casals:2019heg,Yang:2013shb,OToole:2020ejc}, but the frontier in GSF theory is now the calculation of $G^{(1)}_A(J_B)$, which involves solving the radiative part of the 2SF field equations. Progress in 2SF formulations~\cite{Rosenthal:2006iy,Detweiler:2011tt,Pound:2012nt,Gralla:2012db,Pound:2012dk,Pound:2015fma,Pound:2017psq,Upton:2021oxf} and in practical implementation techniques~\cite{Pound:2014xva,Warburton:2013lea,Wardell:2015ada,Pound:2015wva,Miller:2016hjv,Miller:2017coe,Pound:2021qin,Durkan:2022fvm,Spiers:2023mor} has recently led to the first 2SF calculations and complete 1PA waveforms~\cite{Pound:2019lzj,Warburton:2021kwk,Wardell:2021fyy} in the case of quasicircular orbits around a Schwarzschild BH.

\subparagraph{Secondary spin and finite-size effects.} The secondary's spin enters the multiscale expansion at 1PA order~\cite{Mathews:2021rod}. There has been a wealth of work on this topic in recent years~\cite{Mino:1995fm,Tanaka:1996ht,Han:2010tp,Harms:2015ixa,Harms:2016ctx,Warburton:2017sxk,Lukes-Gerakopoulos:2017vkj,Nagar:2019wrt,Akcay:2019bvk,Zelenka:2019nyp,Witzany:2019nml,Piovano:2020zin,Piovano:2021iwv,Skoupy:2021asz,Skoupy:2022adh}, and the community is now able to calculate these effects for generic binary configurations~\cite{Drummond:2022xej,Drummond:2022efc,Skoupy:2023lih}. 

Little work~\cite{Steinhoff:2012rw,Rahman:2021eay} has been done to incorporate the secondary's higher moments into evolution models because their effects are suppressed by $\eps^\ell$~\cite{Pound:2015tma} and by even higher powers of $\eps$ in the case of tidally induced moments~\cite{Binnington:2009bb}. However, these effects might be important for IMRIs and highly deformable secondaries~\cite{Rahman:2021eay} including certain exotic compact objects and white dwarfs~\cite{Yang:2022tma}.

\subparagraph{Transient resonances.} The multiscale expansion~\eqref{GSFeq:psidot}--\eqref{GSFeq:waveform} breaks down during transient resonances, when two or more of the orbital frequencies have an approximately rational ratio~\cite{Tanaka:2005ue,Levin:2008mq,
Apostolatos:2009vu,Flanagan:2010cd,Brink:2013nna,Ruangsri:2013hra,Brink:2015roa,Berry:2016bit,Mihaylov:2017qwn,Lukes-Gerakopoulos:2021ybx}. Passage through such a resonance can lead to a large, 0.5PA contribution to the waveform phase, i.e.\ only $\mathcal{O}(\epsilon^{1/2})$ below the leading order in eqs.~\eqref{GSFeq:psidot}--\eqref{GSFeq:waveform}. There is ongoing effort to model these effects and incorporate them into evolution schemes~\cite{Gair:2011mr,Flanagan:2012kg,vandeMeent:2013sza,Isoyama:2013yor,Lewis:2016lgx,Isoyama:2018sib,Isoyama:2021jjd,Lukes-Gerakopoulos:2021ybx,Nasipak:2021qfu,Nasipak:2022xjh,Gupta:2022fbe}.

\subparagraph{Merger and ringdown.} The multiscale expansion also breaks down as the secondary approaches the last stable orbit, where it gradually transitions into a rapid plunge into the primary~\cite{Buonanno:2000ef,Ori:2000zn,Sundararajan:2008bw,Kesden:2011ma,Taracchini:2014zpa,Apte:2019txp,Burke:2019yek,Compere:2019cqe,Compere:2021zfj}. Using alternative expansions across the transition and through the plunge, a number of groups have generated leading-order inspiral-merger-ringdown waveforms in some scenarios~\cite{Barausse:2011kb,Taracchini:2014zpa,Harms:2014dqa,Folacci:2018cic,Rifat:2019ltp,Rom:2022uvv,Islam:2022laz}. Refs.~\cite{Compere:2021iwh,Compere:2021zfj} have begun to extend this progress toward 1PA/2SF order.

\vspace{2mm}\noindent
{\bf Main Open Challenges}

\vspace{2mm}

\noindent
Currently, there are two main challenges in GSF calculations. The first is expanding their coverage of parameter space: (i) completing the coverage of 0PA waveform models and 1PA spin contributions for generic orbital and spin configurations around a Kerr BH, (ii) extending the 1PA waveform model to include generic spin on the primary, orbital inclination, and eccentricity. The second major challenge is to extend waveform models to include the merger and ringdown; at present, complete models that extend beyond the inspiral phase are only available at leading order in the mass ratio and only for quasicircular, nonspinning binaries.

\subparagraph{0PA waveforms for generic orbits in Kerr spacetime.} A fast 0PA waveform model is available for generic orbits in Kerr spacetime, but only in the large-separation regime~\cite{Isoyama:2021jjd}. At small separations, below $\sim 10$ Schwarzschild radii, the requisite tools are available, but only a small number of waveforms have been produced~\cite{Hughes:2021exa}. Populating the parameter space with the necessary offline data will require highly accurate and efficient interpolation of the forcing terms $G^{(0)}_A$.

\subparagraph{1PA waveforms beyond the quasicircular, nonspinning case.} At present, the only 1PA waveform model is restricted to quasicircular inspirals into nonspinning primaries~\cite{Wardell:2021fyy}. Numerous challenges arise in going beyond this case.

Many of the challenges in going beyond this case stem from singularities that appear in 2SF calculations. Because the secondary is modelled as a point particle (or a puncture in the spacetime~\cite{Pound:2015tma}), strong singularities appear at its location. Correctly capturing these singularities is difficult using standard methods of BH perturbation theory: common gauge choices lead to pathological singularities extending away from the particle's position~\cite{Toomani:2021jlo}; and decompositions into $\ell m\omega$ modes suffer from intransigently slow convergence in the particle's neighbourhood~\cite{Miller:2016hjv}. All of these problems have been surmounted or bypassed in 1SF calculations, but existing 1SF methods fail in 2SF calculations, where gauge singularities and slow convergence become more problematic. 2SF calculations additionally suffer from an infrared breakdown of the multiscale expansion at large distances and near the primary's event horizon~\cite{Pound:2015wva}. Methods currently used to overcome these problems are cumbersome, expensive, and tailored to the spherical symmetry of a Schwarzschild background spacetime.

Extending 1PA waveforms to generic orbits around generic Kerr BHs will require several advances: improved computational efficiency of current methods; development of new approaches in BH perturbation theory, particularly in practical methods of working with better-behaved gauges; and development of new tools to improve or avoid slow convergence.


\subparagraph{Secondary spin and finite-size effects.} After rapid recent progress, the next obstacle in calculating 1PA secondary-spin effects is to find efficiently evaluated flux-like expressions for the evolution of the spin-corrected Carter constant and the spin component parallel to the total orbital angular momentum~\cite{Skoupy:2023lih}. The remaining challenge will then be the same as in calculating 0PA effects: covering the full parameter space of generic, inclined orbits with generic, precessing spins. Significantly more work is required to assess the impact of the secondary's quadrupole and higher moments in the range of mass ratios where GSF models are relevant for ET and (if needed) to incorporate these effects into GSF waveform models.

\subparagraph{Transient resonances.} Several strategies have now been developed to evolve through orbital resonances, although more work is likely required to achieve the same waveform-generation speed as in the non-resonant case. Currently, the largest potential concern related to resonances is the cumulative loss of phase accuracy in the case of multiple resonances: each successive resonance causes an $O(\epsilon^{1/2})$ growth in phase error, meaning a 1PA model will have an $O(\epsilon^{0})$ error if an inspiral passes through two significant resonances. Further study is required to determine the absolute magnitude of this error and its relevance for ET sources, as well as to develop mitigation strategies. 

\subparagraph{Merger and ringdown.} GSF calculations were historically motivated by EMRIs, for which the merger and ringdown contain negligible SNR. However, for small-$\epsilon$ signals detectable by ET, including the merger will likely be crucial. At present, the merger has only been modelled at leading order in the mass ratio. These calculations must be extended beyond leading order in such a way that they remain compatible with fast waveform generation.

\vspace{2mm}\noindent
{\bf Important next steps}

\vspace{2mm}

\noindent
There are several obvious next steps in the GSF modeling program. 

The first, simplest step is to build a complete 0PA waveform model covering the space of generic inspirals. This means computing (and interpolating) data for $G^A_{(0)}$ and $h^{(1)}_{\ell m\omega_k}$ across the space of inclined, eccentric orbits in Kerr spacetime. Similarly, another next step will be to cover the parameter space with 1PA data proportional to the secondary spin. At 1PA order these spin contributions add linearly onto a nonspinning-secondary waveform model, making this step relatively straightforward (though more demanding than the calculation of 0PA data).

Extending the 1PA waveform model toward generic inspiral configurations will be more challenging. A major step will be to include the primary spin, which means performing 2SF calculations in Kerr spacetime. There are several possible routes toward this goal~\cite{Toomani:2021jlo,Osburn:2022bby,Spiers:2023cip,Dolan:2023enf}. Such calculations can then be extended to spherical orbits with spin-orbit precession and finally to fully generic orbits with eccentricity.

Another important, largely independent step will be to extend inspiral waveform models to include the final merger and ringdown. This can begin with the simplest case of quasicircular, nonspinning binaries by further developing the expansions across the transition to plunge and plunge, extending the existing leading-order treatments.

Depending on the difficulty of completing these steps, high-risk, high-gain approaches could also be pursued. In particular, work on scattering orbits in GSF theory~\cite{Hopper:2017iyq,Hopper:2017qus,Long:2021ufh,Barack:2022pde,Long:2022sdq,Barack:2023oqp,Whittall:2023xjp} might provide an independent avenue toward obtaining GSF data for bound orbits~\cite{Gonzo:2023goe}. 

Synergies with other solution techniques like weak-field expansions (section~\ref{sec:weak}) or numerical relativity (section~\ref{sec:NR}) will provide very useful external checks on all calculations.
In tandem, a critical task will be to assess the required accuracy and range of parameters that are relevant for ET sources, which will influence the prioritization of other steps. For example, for ET sources it might be more beneficial to target specific GSF calculations that can be used to improve EOB models, rather than focus on complete inspiral-merger-ringdown 1PA waveform models across the full parameter space. 

\subsubsection{Inspiral-Merger-Ringdown Models}
\label{sec:imr}

\vspace{2mm}\noindent
{\bf State-of-the-Art}
\vspace{2mm}

\noindent
The analytical and numerical methods presented in the previous sections can be combined to create reliable inspiral-merger-ringdown (IMR) models.
In particular, we will focus on the three most fruitful approaches: NR surrogates, effective-one-body (EOB) models, and phenomenological (Phenom) approximants.

As previously discussed, NR simulations represent the most accurate GR solution of binary systems but are still too slow to be directly used in parameter estimation.
Thus, available NR waveforms have been interpolated to build surrogate 
models \cite{Field:2013cfa, Blackman:2017dfb, Varma:2019csw, Tiglio:2021ysj}, 
which are very accurate but are currently limited in waveform length and parameter space coverage.
State-of-the-art NR surrogates include
\texttt{NRSur7dq4}~\cite{Blackman:2015pia,Blackman:2017pcm,Varma:2019csw}, which
can handle generically spinning
BBHs but is restricted to mass ratios $q\leq 4$, and $\sim 20$ orbits before
merger, and \texttt{NRHybSur3dq8}~\cite{Varma:2018mmi}, which can handle mass
ratios $q\leq 8$ and includes the early inspiral through PN/EOB hybridization,
but is restricted to nonprecessing BBHs. Recent efforts have been made to
include extreme spins~\cite{Walker:2022zob}, high mass ratios~\cite{Yoo:2022erv,
Islam:2022laz,Islam:2023mob}, GW memory~\cite{Yoo:2023spi} and even eccentricity
effects~\cite{Islam:2021mha}.

NR simulations are also a key ingredient in the construction of semi-analytical waveform models.
They are necessary to correctly model the plunge, merger and post-merger waveforms and provide a much needed validation of analytical approximations.
These latter, on the other hand, constitute a very useful complement to NR simulations, being able to deal with long inspirals and a wide range of different physical effects. 

An important step in the improvement of analytical approximations was the introduction of the EOB approach~\cite{Buonanno:1998gg,Buonanno:2000ef,Damour:2000we,Damour:2001tu}, which provides a way to map the general relativistic two-body problem onto an effective one-body system where a test particle moves in an external effective metric. 
This metric is a deformation of the Schwarzschild (or Kerr) one, with the symmetric mass ratio $\nu = (m_1 m_2)/M^2$ as a deformation parameter.
The continuous link between the test-mass case, where GR is exactly solvable, and the equal-mass one, allows to meaningfully resum perturbative information, improving its accuracy during the late stages of the inspiral and plunge.
Currently, these semi-analytical models are built using (mostly) PN series 
resummed through the EOB prescription and calibrated to NR information \cite{Nagar:2018zoe,Cotesta:2018fcv,Pompili:2023tna}.
Importantly, this framework does not allow to model the postmerger and ringdown
waveform per se.
Hence, this part is generally modeled using NR fits and a smooth transition between the inspiral 
and postmerger waveforms is required~\cite{Pan:2011gk,Taracchini:2012ig,Taracchini:2013rva,Damour:2014yha,DelPozzo:2016kmd}.

Current state-of-the-art EOB models such as the \TEOBResumS{}~\cite{Damour:2014sva,Nagar:2018zoe,Gamba:2021ydi} and \texttt{SEOBNR} families~\cite{Barausse:2009xi,Cotesta:2018fcv,Ossokine:2020kjp,Pompili:2023tna,Ramos-Buades:2023ehm} take into account tidal effects, higher harmonics and spin-induced precession.
Significant progress has been made in the modeling of non-circular binaries~\cite{Chiaramello:2020ehz,Nagar:2021gss,Nagar:2021xnh,Placidi:2021rkh,Khalil:2021txt,Ramos-Buades:2021adz,Ramos-Buades:2022lgf,Gamboa:2024imd,Gamboa:2024hli}.
Additionally, the framework's flexibility allows to straightforwardly include analytical contributions computed in different approximations such as the GSF~\cite{Antonelli:2019fmq,Albertini:2022rfe,Albertini:2022dmc,vandeMeent:2023ols} and PM~\cite{Antonelli:2019ytb,Khalil:2022ylj,Damour:2022ybd,Buonanno:2024vkx, Buonanno:2024byg} ones. 
EOB models allow complete control over the underlying physics, making it easy to modify them, 
for example to take into account beyond-GR effects~\cite{Julie:2017pkb,Khalil:2018aaj,Julie:2018lfp,Bernard:2018hta,Julie:2019sab,Jain:2022nxs,Julie:2022qux,Jain:2023fvt}.
The models generally rely on solving a system of ordinary differential equations, which provides a highly accurate description of the binary dynamics, but can also make waveform generation computationally expensive.
This drawback can be mitigated, 
at the cost of a slight reduction in accuracy,
through the building of surrogate models~\cite{Lackey:2018zvw,Cotesta:2020qhw,Schmidt:2020yuu,Khan:2020fso,Riemenschneider:2021ppj,Thomas:2022rmc}, by using a post-adiabatic approximation during the early inspiral \cite{Nagar:2018gnk,Riemenschneider:2021ppj,Mihaylov:2021bpf}, or through a numerical implementation of the stationary phase approximation \cite{Gamba:2020ljo}. 
As EOB models are based on physical equations of motion, we can use a broader range of observables to test the consistency and accuracy of the resulting waveform model, e.g. periastron advance~\cite{LeTiec:2011bk,Hinderer:2013uwa}, binding energy~\cite{Damour:2011fu,Nagar:2015xqa,Nagar:2019wds,Ossokine:2017dge,Khalil:2020mmr,Khalil:2022ylj}, or their scattering angles~\cite{Damour:2016gwp,Damour:2017zjx,Khalil:2022ylj,Nagar:2020xsk,Damour:2022ybd,Rettegno:2023ghr,Albanesi:2024xus,Swain:2024ngs}.

Another successful approach to waveform generation consists in 
building phenomenological approximants~\cite{Ajith:2007qp, Ajith:2009bn,Santamaria:2010yb, Hannam:2013oca, Khan:2015jqa, London:2017bcn,  Khan:2018fmp, Khan:2019kot, Pratten:2020ceb, Pratten:2020fqn, Garcia-Quiros:2020qpx, Hamilton:2021pkf, Thompson:2023ase}.
These focus predominantly on directly modeling the GW signal 
without solving the equations of motions for the binary system, thereby significantly reducing the computational cost and hence enabling fast yet accurate analyses. 
In order to do so, they separately model the amplitude and phase
of the waveforms and split them into three distinct regions:
(i) the inspiral region is modeled using a high-order pseudo-PN series typically calibrated 
to EOB models; 
(ii) the intermediate region and (iii) the merger-ringdown part. The latter two are modeled using different functional forms and are both calibrated to NR simulations.
Spin-induced orbital precession is included via an SO(3) frame transformation~\cite{Apostolatos:1994mx,Buonanno:2002fy, Schmidt:2010it, Boyle:2011gg} by separately modeling the multipolar modes in a co-precessing frame and a time-dependent rotation that encapsulates the orbital precession~\cite{Schmidt:2012rh}. 
Test-mass waveforms are also sometimes used to extend the models' validity 
beyond the available comparable-mass NR simulations. 

State-of-the-art phenomenological approximants, such as the \texttt{PhenomX} family~\cite{Pratten:2020fqn,Pratten:2020ceb, Garcia-Quiros:2020qpx, Thompson:2023ase}, are able to generate 
multipolar waveforms for spin-precessing systems and are reliable 
also for intermediate mass-ratios in the aligned-spin case.
Phenomenological models can also be directly generated in the time-domain, see e.g. the \texttt{PhenomT} family~\cite{Estelles:2020osj,Estelles:2020twz,Estelles:2021gvs}, which can be helpful to include precession and eccentricity effects. 
The phenomenological approximants are known to be less faithful than some of the EOB models in regions of the parameter space~\cite{Pratten:2020fqn,Garcia-Quiros:2020qpx,Albertini:2021tbt, Pompili:2023tna}, but can be several times faster than surrogate models for EOB~\cite{Gamba:2021ydi,Garcia-Quiros:2020qpx}.

EOB, phenomenological and NR surrogate waveforms can be extrapolated beyond their calibration range, but their accuracy and validation strongly depend on the availability of NR simulations across the parameter space.

For sufficiently small mass ratios, the situation is somewhat different: GSF theory (see section~\ref{sec:GSF}) by itself can provide sufficiently accurate IMR waveforms, and it may be possible to make these sufficiently rapid to meet data analysis requirements by extending~\cite{Compere:2021zfj} the acceleration methods used for inspiral waveforms~\cite{Katz:2021yft}. Alternatively, existing, slow GSF IMR simulations can also be used to build surrogates~\cite{Rifat:2019ltp,Islam:2022laz} analogously to NR surrogates. For nonspinning binaries, surrogates such as the ones developed in refs.~\cite{Rifat:2019ltp, Islam:2022laz}, have also been made highly accurate even for comparable masses through simple, NR-calibrated rescalings of time and amplitude, although it remains to be seen whether such a simple method can be successfully adapted for non-circular and/or precessing systems.

\vspace{2mm}\noindent
{\bf Main Open Challenges}
\vspace{2mm}

\noindent
The sensitivity of future detectors such as ET imposes a strong requirement on the accuracy of IMR models for compact binary coalescences (see section~\ref{sec:accuracy}).
Much work is still needed to meet such stringent accuracy requirements for higher harmonics, large mass ratios and extremal black hole spins.
Spin-induced precession adds another layer of complications, which concerns the modeling of the multipolar co-precessing waveforms~\cite{Ramos-Buades:2020noq}, the precession angles~\cite{Hannam:2013oca, Chatziioannou:2017tdw, Ramos-Buades:2023ehm}, a dedicated ringdown model~\cite{Hamilton:2021pkf} and modeling the multipole asymmetry in precessing systems that leads to large out-of-plane recoils~\cite{ Varma:2019csw, Thompson:2023ase, Ghosh:2023mhc}.
Another open but crucial issue is the inclusion of non-circular effects in semi-analytical models.
The inclusion of effects due to orbital eccentricity into waveform models have recently started to gain traction~\cite{Moore:2019xkm, Nagar:2021gss, Ramos-Buades:2021adz,Gamboa:2024imd,Gamboa:2024hli}.
In the future, complete models that accurately model the GWs from fully generic systems that are both eccentric and spin-precessing will be required~\cite{Klein:2021jtd, Liu:2023ldr, Gamba:2024cvy}. 
This will demand improved analytical information but also coincides with the lack of NR simulations covering such a high-dimensional parameter space. Work will be required in producing accurate and long NR simulations that include both effects.

At the same time, the increased number of detections and observation of signals for a longer duration, will be even more demanding on the waveform generation time of any model.
This introduces several challenges for EOB models, which require solving a coupled system of ODEs. 
However, there has been significant work on techniques to amortize the cost of EOB models.
For example, the post-adiabatic approximation~\cite{Nagar:2018gnk}, used to quickly generate waveforms in the co-precessing frame, has not yet been extended to take into account eccentricity effects. The building of very fast surrogates seems to be necessary, possibly making use of machine learning techniques (see section~\ref{sec:acc}).
Phenomenological models on the other hand are already several orders of magnitude less computationally expensive than other models: Their PN-based inspiral prescription allows them to be evaluated to arbitrarily low frequencies (unlike surrogates) and the closed-form analytical expressions make them very fast to evaluate. This makes them particularly suitable for efficient parallelization with GPUs, as well as the use of non-uniform time or frequency grids for a further reduction in evaluation cost.
However, the accuracy of phenomenological models must be improved, raising a number of important challenges in the construction of generic waveform models and overcoming known limitations, such as those imposed by the stationary phase approximation and its extensions.

\vspace{2mm}\noindent
{\bf Important Next Steps}
\vspace{2mm}

\noindent
The first step to improve current models would be a thorough exploration of the parameter space with NR simulations.
This implies, e.g., performing long high-resolution simulations for high-mass ratio systems.
Numerical simulations are also necessary to build post-merger templates for BNS and BHNS systems~\cite{Breschi:2019srl,Thompson:2020nei,Matas:2020wab,Steinhoff:2021dsn,Breschi:2022xnc,Puecher:2022oiz}.
A larger number of eccentric and/or precessing binaries not only would help in building dedicated NR surrogates, but would constitute a needed benchmark to inform and validate semi-analytical models.

At the same time, high-order analytical contributions, will be useful in building fast (inspiral) waveforms for the most challenging systems.
GSF and PM contributions can complement the more common PN terms in different corners of the parameter space.

\subsubsection{Alternative Theories of Gravity}
\label{sec:altth}

\vspace{2mm}\noindent
{\bf State-of-the-Art}
\vspace{2mm}

\noindent
The modeling of gravitational waves (GWs) in alternative theories of gravity faces similar challenges to those in General Relativity (GR), since in the end the same level of accuracy is required.
This section focuses on numerical and analytical techniques for making predictions in these theories of gravity, with particular emphasis on the challenges that arise when generalizing the methods presented in sections~\ref{sec:NR}--\ref{sec:imr}  to alternative theories.
We focus on GWs from compact binary systems, the most promising probes of strong-field gravity, with the notable exception of GWs that might be emitted by dark matter clouds
(see section~\ref{section:div1} for a more detailed discussion) or supernovae.

The modeling of GWs from compact binaries starts at the smallest length scale in the EFT treatment, i.e. for isolated compact objects.
Single-object solutions are needed to provide initial data for numerical simulations, but also as input to analytical methods that match single objects to effective point particles. For example, to obtain e.g.~their spin-induced multipoles~\cite{Laarakkers:1997hb, Pappas:2012ns} or tidal deformabilities~\cite{Hinderer:2007mb, Damour:2009vw, Binnington:2009bb}.
In the presence of additional fields, such as in scalar-tensor theories, significant complications can arise in calculating such quantities~\cite{Pappas:2014gca, Pappas:2018csu, Pani:2014jra, Yazadjiev:2018xxk, Brown:2022kbw, Creci:2023cfx}.
For example, new families of tidal deformations emerge, such as the scalar tidal ones, and their calculation is nontrivial both theoretically and numerically~\cite{Creci:2023cfx}.
However, self-consistent calculations of multipole moments and tidal deformabilities are still largely lacking in many modified theories of gravity, such as the effective field theories, so further development is needed.

The post-merger object, either a black hole or a neutron star, can be also well modeled as a perturbed single object, simplifying the calculations of the emitted gravitational wave signal.
The study of the compact object quasinormal modes (QNMs) in modified theories of gravity is conceptually very similar to the GR case.
However, there can be serious technical challenges due to the fact that the field equations of the modified gravity theories are as a rule more complicated and contain more degrees of freedom.
Systematic studies of the QNM beyond GR  were mainly performed for static black holes and static neutron stars within the (extended) scalar-tensor theories~\cite{Blazquez-Salcedo:2018pxo, Blazquez-Salcedo:2016enn, Blazquez-Salcedo:2017txk, AltahaMotahar:2019ekm, Staykov:2021dcj, Blazquez-Salcedo:2020caw, Blazquez-Salcedo:2020ibb, Kruger:2021yay}.
Those studies show that, in the general case, the QNM in modified gravity are different from those in GR, the well-known GR isospectrality is violated and there can exist new type of (scalar) QNMs absent in GR.
Studies of QNM without approximations for rotating beyond-GR black holes and neutron stars are practically lacking.
Available studies are based on approximate techniques using perturbations in the spin and/or in the coupling constants~\cite{Wagle:2021tam, Pierini:2022eim}.
Ideas for different approaches to the QNM calculation, like spectral decompositions without the need to separate the radial and angular dependence~\cite{Chung:2023zdq}, are also being developed, but much further work is needed in this direction in order to calculate the QNM frequencies of beyond-Kerr black holes close to the extremal limit.
Also, analytic solutions for stationary BHs in alternative theories can be developed as a perturbation series in beyond-GR coupling constants, which then in principle allows to express their linear perturbations as an extension of the (relatively simple) GR master equations, see also refs.~\cite{Hui:2022vov, Li:2022pcy}.

The late inspiral and merger of a compact binary involves strong and highly dynamical fields that are generally accessible only to numerical-relativity simulations.
It has taken decades to get the GR field equations into a form that is numerically well-posed, and unfortunately there is still no approach that works for generic beyond-GR theories without compromises; usually the GR modifications cannot become too large.
One possibility is to formulate the theory as GR with a modified matter source, as in scalar-tensor theories by going to the Einstein frame, or for studies of higher-dimensional gravity~\cite{Zilhao:2010sr}.
Another possibility is a reduction of the derivative order in the field equations within an effective-field-theory framework~\cite{Okounkova:2017yby, Okounkova:2019dfo, Witek:2020uzz, Silva:2020omi}.
The generalized harmonic gauge approach to numerical simulations can also be modified or adapted~\cite{East:2020hgw, Corman:2022xqg, AresteSalo:2022hua}.
Finally, one can try to find modifications with the sole purpose of ``fixing" the field equations (making them well-posed), while making sure that these modifications are of higher order within an effective-field-theory power counting~\cite{Cayuso:2020lca, Bezares:2020wkn, Franchini:2022ukz}.

The only alternative to these numerical-relativity methods with access to the strong-field regime is an expansion in the mass ratio: the gravitational self-force program. Self-force models benefit from several simplifications in the most commonly considered alternatives to~GR. Corrections to GR and coupling of gravity to new fields typically manifest themselves at high curvature, meaning (i) their impact on the primary black hole is typically suppressed by a power of the mass ratio relative to their impact on the secondary object, and (ii) coupling of new fields to gravity is also suppressed by a power of the mass ratio. Detailed scaling arguments show that in a broad class of theories, corrections to the primary can then be treated perturbatively; new fields entirely decouple from gravity at leading, 0PA order; and standard GSF modeling methods can be applied~\cite{Spiers:2023cva}. These simplifications have enabled a recent program to develop complete GSF waveform models at 0PA order in large classes of theories beyond GR~\cite{Maselli:2020zgv,Maselli:2021men,Barsanti:2022ana,Barsanti:2022vvl}. 
Note, however, that such a simplification only applies when the coupling to the new physics is dimensionfull, which is not the case in certain theories, such as Einstein-Aether gravity.
Numerous earlier studies also carried out partial or semi-relativistic calculations beyond GR at 0PA order (see, e.g.,~\cite{Sopuerta:2009iy,Pani:2011xj,Cardoso:2011xi,Yunes:2011aa,Canizares:2012is,Blazquez-Salcedo:2016enn}), and the first-order GSF equations of motion have been derived in a class of scalar-tensor theories~\cite{Zimmerman:2015hua}. All concrete calculations to date have been restricted to the inspiral phase prior to merger, but there does not appear to be an obstacle (in principle) to handling the merger phase with GSF methods as in GR.

As discussed in section~\ref{sec:weak}, the early inspiral can be modeled in a weak-field and slow-motion (or post-Newtonian) approximation, and a scattering for large impact parameter in a weak-field (or post-Minkowskian) approximation.
For simple scalar-tensor theories, there are essentially no new technical obstacles in performing this calculation, compared to the GR computations discussed in 
section~\ref{sec:weak}, as demonstrated by high-order calculations of the binary dynamics in these theories~\cite{Bernard:2018ivi, Bernard:2018hta, Julie:2019sab}; for the waveform see refs.~\cite{Bernard:2022noq, Sennett:2016klh}, and for scattering ref.~\cite{Barack:2023oqp}. In Einstein-Aether theories, preliminary results were derived in order to obtain the equations of motion up to 2.5PN order~\cite{Taherasghari:2023rwn}.
However, in most alternative theories that introduce higher derivative terms, there are complications and one should consider the weak-field expansion only in a certain EFT regime of validity~\cite{Endlich:2017tqa}.
An essential input is an effective point-particle model representing the compact object, which is matched to solutions of the individual objects described above.
While this presents some challenges, it is conceptually not that different from GR, with notable examples being the description of spontaneous scalarization in a wide range of theories~\cite{Julie:2022huo} and also dynamical scalarization~\cite{Khalil:2022sii, Khalil:2019wyy, Sennett:2017lcx, Sennett:2016rwa, Sampson:2014qqa, Palenzuela:2013hsa, Sampson:2013jpa}.

Finally, predictions from the above techniques must be combined into complete inspiral-merger-ringdown models, as explained in section~\ref{sec:NR}.
For effective-one-body waveform models, no immediate technical problems should arise in this process, and so far most efforts focused on theories with an additional scalar field~\cite{Jain:2022nxs, Julie:2022qux,Julie:2024fwy}. 
Aside from predictions in specific alternative theories, a complementary agnostic approach is to introduce phenomenologically parametrized deviations from GR directly in the waveform, as we will discuss in section~\ref{sec:waveforms_alttheories}.

\vspace{2mm}\noindent
{\bf Main Open Challenges}
\vspace{2mm}

\noindent
As explained above, numerical-relativity approaches to beyond-GR theories are often problematic when the modifications become large in the strong-field regime.
One reason may be that numerical stability breaks down, or that one has to modify the equations without a clear physical interpretation.
Similarly, the gravitational self-force program makes crucial use of the separation of the BH perturbation equations into master equations, but the separability cannot be achieved generically when the beyond-GR modifications are large, i.e.~no expansion in the beyond-GR coupling constants is possible (and BH solutions may even only exist numerically).
That is, while the techniques for dealing with the strong-field regime are somewhat limited to small modifications relative to GR, the modifications to GR are in fact not necessarily small, especially in the strong-field regime.
Resolving this conundrum is the major open challenge for waveform modeling techniques beyond GR.

The matching of single-object solutions to the effective point-particle models used in the weak-field approximations (and for the secondary object in GSF models) also remains a challenge.
Even in GR, obtaining the relativistic (tidal) response to external fields is conceptually controversial (see e.g.~refs.~\cite{Gralla:2017djj, Creci:2021rkz}), especially when the external fields are dynamical.
Fortunately, this is mostly relevant for higher orders in the perturbative weak-field expansion, with the notable exception of dynamical scalarization effects~\cite{Khalil:2022sii} and screening mechanisms~\cite{Babichev:2013usa}.

\vspace{2mm}\noindent
{\bf Important Next Steps}
\vspace{2mm}

\noindent
An important next step is to advance the development of numerical relativity in beyond-GR theories of gravity.
This may involve improvements in numerical algorithms, but it may also involve rethinking the whole approach to strong-field modifications, e.g.~from an effective-field-theory perspective~\cite{Coates:2023swo}.
These efforts on the numerical-relativity side should be coordinated with the efforts on weak-field approximations, solutions for single compact objects, and the gravitational self-force, also considering which modifications of GR are best motivated and compatible with contemporary experimental constraints.

\subsection{Waveform Models for Specific Sources}
\label{Sec:models}

Results of the individual techniques for waveform modeling are combined into complete waveform models for concrete GW sources.  This section covers the various models that exist for each of the specific sources and stress the directions where major progresses are required. 

\subsubsection{Binary black holes}
\label{sec:bbh}


\vspace{2mm}\noindent
{\bf State-of-the-Art}
\vspace{2mm}

\noindent
The main families of inspiral-merger-ringdown (IMR) waveform models commonly used in gravitational-wave data analysis are phenomenological models, the effective-one-body (EOB) framework, and numerical relativity (NR) surrogate models. The first two families are semi-analytical models predicated on a range of analytical information including the post-Newtonian (PN) framework  and gravitational self-force. Information on the strong-field regime is incorporated by calibrating the semi-analytical expressions against NR simulations. The NR surrogates, however, effectively interpolate NR simulations across the parameter space to directly reproduce the input NR information. 

The PhenomX family is the latest generation of phenomenological waveform models which were first proposed in \cite{Ajith:2007kx,Ajith:2007qp,Ajith:2007xh}. 
PhenomXAS models the dominant $\ell = m = 2$ mode for aligned-spin binaries \cite{Pratten:2020fqn}, offering nearly two orders of magnitude improvement in the mismatch of the model against EOB-NR hybrids relative to earlier generations. 
The number of NR simulations used to calibrate the model has increased dramatically to $\sim 400$ simulations. 
In addition, waveforms in the test-particle limit provided through numerical solutions to the $2+1$ Teukolsky equation \cite{Harms:2013ib,Harms:2014dqa,Harms:2016ctx}, and are used to extend the calibration of the model to extreme mass ratios. 
The aligned-spin NR simulations cover mass ratios up to $q = 18$ and spins up to $| \boldsymbol{\chi} | \sim 0.8$. 
In the equal-mass limit, high-spin simulations produced by the SXS collaboration~\cite{Boyle:2019kee} allow a calibration up to $| \boldsymbol{\chi} | \sim 0.99$. 
An extension to incorporate subdominant multipoles calibrated against NR was provided by PhenomXHM, using the same catalog of NR simulations and test-particle waveforms. The PhenomX family also benefits from including recently derived higher order PN information. 
The phasing model incorporates non-spinning corrections to 3.5PN order \cite{Buonanno:2009zt,Blanchet:2013haa}, spin-orbit corrections to 3.5PN \cite{Bohe:2013cla,Bohe:2015ana}, spin-orbit tail corrections at 4PN \cite{Marsat:2013caa}, quadratic-in-spin-spin corrections to 3PN \cite{Poisson:1997ha,Mikoczi:2005dn,Arun:2007rg,Bohe:2015ana}, and cubic-in-spin corrections at 3.5PN \cite{Marsat:2014xea}. 
PhenomX also includes the recently derived non-spinning corrections at 4PN to the equations of motion for compact binaries \cite{Damour:2014jta,Bernard:2015njp,Bernard:2016wrg,Damour:2017ced,Marchand:2017pir,Jaranowski:2015lha} and a higher non-linear tail effects associated to quartic non-linear interactions that were derived from first principles in the multipolar post-Minkowski (MPM) framework \cite{Marchand:2016vox}, see also \cite{Messina:2017yjg} for an independent derivation from the PN re-expansion of the factorized and resummed EOB fluxes.

The aligned-spin models incorporate precession following the paradigm introduced in \cite{Schmidt:2010it,Boyle:2011gg,OShaughnessy:2011pmr}, in which the co-precessing frame modes can be well approximated by aligned-spin modes with an augmented merger-ringdown \cite{Schmidt:2012rh}. 
The subsequent model, PhenomXPHM \cite{Pratten:2020ceb}, incorporates multiple prescriptions for precession, including a single-spin next-to-next-to-leading-order  post-Newtonian description \cite{Hannam:2013oca,Bohe:2016gbl}, a prescription based on a multiple scale analysis \cite{Kesden:2014sla,Gerosa:2015tea,Chatziioannou:2016kem,Chatziioannou:2016ezg} that captures double-precessing spin effects, and the option to numerically integrate the SpinTaylor equations for the precession dynamics~\cite{Colleoni:2023czp}.
Most recently, PhenomPNR introduced a model for the precession dynamics and co-precessing frame modes that are calibrated against a suite of 40 single-spin precessing NR simulations up to $q = 8$ and spins $|\boldsymbol{\chi} \lesssim 0.8|$. This includes the modeling of mode asymmetries through the antisymmetric contribution to the dominant coprecessing-frame signal multipole \cite{Ghosh:2023mhc}.

A family of time domain phenomenological models, PhenomT \cite{Estelles:2020osj,Estelles:2020twz,Estelles:2021gvs}, offers a complementary development to PhenomX. 
A key motivation for this is to simplify insight into modeling precession and eccentricity, which can be non-trivial in the frequency-domain and often invokes approximations such as the stationary phase approximation to analytically approximate the Fourier transformation.
The domain of validity of PhenomT is comparable to the PhenomX family of models, being calibrated against a similar suite of NR simulations. 

The EOB framework, first introduced in \cite{Buonanno:1998gg,Buonanno:2000ef}, provides a means to map the general relativistic two-body problem to that of a test particle moving in some effective spacetime metric, typically a deformed Kerr or Schwarzschild spacetime. 
The two main state-of-the-art EOB models are SEOBNRv5 \cite{Pompili:2023tna,Ramos-Buades:2023ehm,Khalil:2023kep,vandeMeent:2023ols} and TEOBResumS \cite{Nagar:2018zoe}. 
The models differ in a number of ways, including the choice of the underlying Hamiltonian, re-summation of PN information, how spin-effects enter the model, and incorporation of information from numerical relativity. The analytical structure of the different EOB models, and their differences, was for example explored in \cite{Rettegno:2019tzh, Khalil:2020mmr}. 

SEOBNRv5, the latest generation of SEOBNR models, introduces notable changes in the underlying structure of the Hamiltonian, which  is based on that of a \textit{test-mass} in a deformed Kerr background \cite{Damour:2001tu,Damour:2008qf,Nagar:2011fx,Damour:2014sva,Balmelli:2015zsa} whereas SEOBNRv4 was based on the Hamiltonian for a \textit{test-spin} in a deformed Kerr background \cite{Barausse:2009aa,Barausse:2011ys}. 
Further improvements include higher-order PN information, the modeling of mode-mixing \cite{Berti:2014fga}, calibration against a larger suite of 442 NR simulations and 13 waveforms from black-hole perturbation theory, and information from second-order GSF in the non-spinning waveform modes and radiation reaction. 
The extension to spin-precession~\cite{Ramos-Buades:2023ehm} builds on previous studies that employed aligned-spin orbital dynamics in the co-precessing frame coupled to PN-expanded precessing-spin equations~\cite{Estelles:2020twz,Akcay:2020qrj,Gamba:2021ydi}, to mitigate the computational expense of solving the equations of motion using the full precessing-spin EOB Hamiltonian~\cite{Pan:2013rra,Ossokine:2020kjp, Babak:2016tgq}.
  SEOBNRv5PHM extends beyond this approach by using, in the co-precessing frame, an EOB Hamiltonian that includes partial precessional effects, in the form of orbit-averaged in-plane spin contributions for circular orbits~\cite{Khalil:2023kep}. Additionally, the model employs PN-expanded evolution equations for the spins and angular momentum that include higher-order PN information and uses a spin-supplementary condition consistent with such a Hamiltonian.  SEOBNRv5PHM achieves more than an order of magnitude improvement in speed with respect to the previous generation SEOBNRv4PHM.
  Building on early attempts to incorporate eccentricity into the EOB framework \cite{Hinderer:2017jcs,Cao:2017ndf,Liu:2019jpg}, there have been a number of parallel developments within the SEOBNR framework \cite{Cao:2017ndf,Ramos-Buades:2021adz,Ramos-Buades:2022lgf,Gamboa:2024hli}. 
SEOBNRv5EHM~\cite{Gamboa:2024hli} is the state-of-the-art model for eccentric, aligned-spin BBHs within this framework.  This time-domain, inspiral-merger-ringdown, multipolar model includes the dominant (2,2) mode and the modes $(2,1),(3,3),(3,2),(4,4), (4,3)$, and it employs two eccentric parameters, namely an eccentricity and a radial anomaly. It is inspired by the previous-generation SEOBNRv4EHM eccentric model \cite{Ramos-Buades:2021adz}, and it is constructed upon the SEOBNRv5HM model for quasi-circular, aligned-spin BBHs \cite{Pompili:2023tna}. SEOBNRv5EHM benefits from analytical results at 3PN order for the eccentricity contributions to the inspiral EOB waveform modes and radiation-reaction force \cite{Khalil:2021txt,Gamboa:2024imd}, employs a quasi-circular description for the merger-ringdown, and targets binaries with eccentricities up to $\sim 0.45$ at 20~Hz.

\begin{figure}
    \centering
    \includegraphics[width=0.98\textwidth]{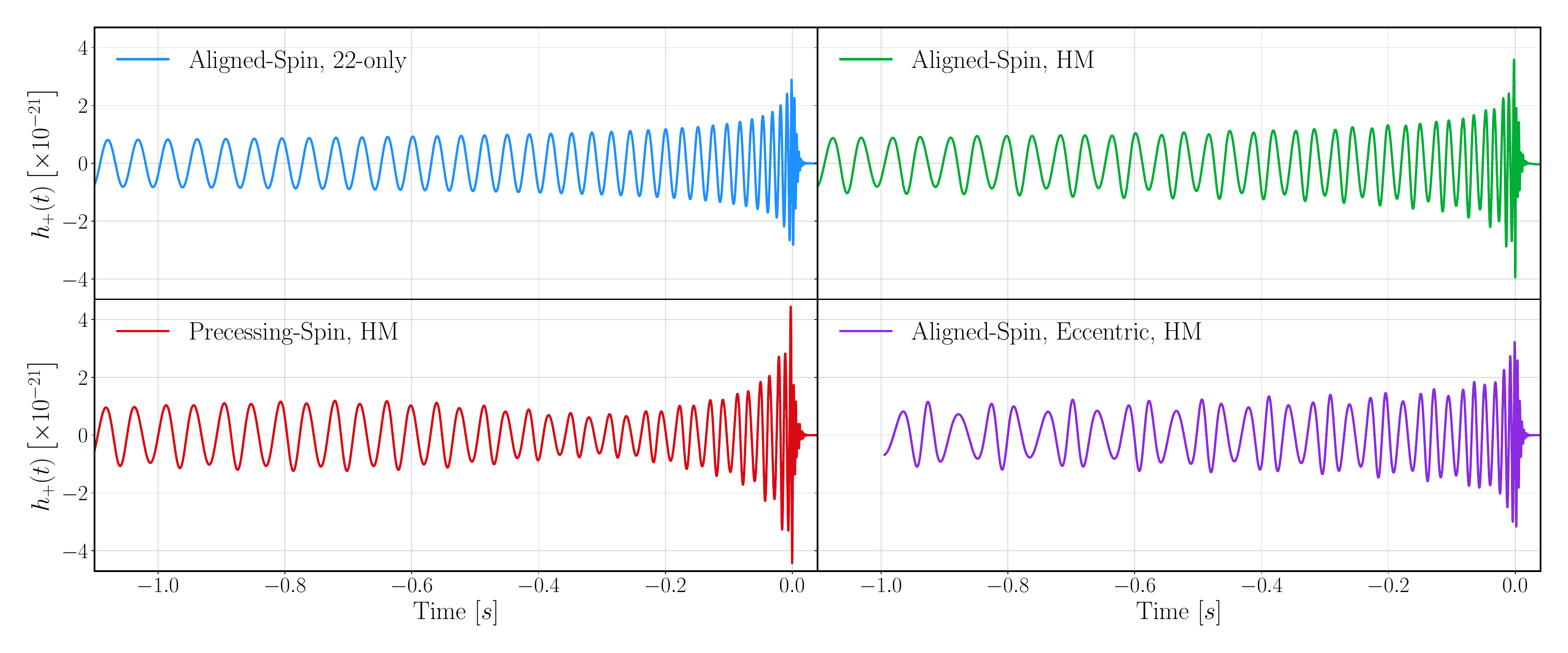}
    \caption{
      Exemplary IMR waveforms of coalescing binary black holes with mass ratio $1:3$ and a total source-frame mass of $60M_\odot$ located at a distance of $100$ Mpc and viewed under an inclination angle of $\pi/3$ relative to the line-of-sight. The initial GW frequency is $20$ Hz. The four panels show waveforms from four different state-of-the-art approximants discussed in the text highlighting the impact of different physical effects on the waveform. Top left: Example of an aligned-spin binary with only the dominant harmonic using the IMRPhenomTPHM approximant. Top right: Example of an aligned-spin binary including higher-order modes using the IMRPhenomXPHM approximant. Bottom left: Example of a spin-precessing binary using the SEOBNRv5PHM approximant. Bottom right: Example of an aligned-spin binary with non-zero orbital eccentricity using the TEOBResumS-Dal\'{i} approximant. We note that the latter waveform is shorter due to the enhanced radiation reaction in eccentric binaries relative to quasi-circular ones.
    }
    \label{fig:BBHComparison}
\end{figure}

The TEOBResumS family currently provides two state-of-the-art models.
The first is the quasi-circular model TEOBResumS-Giotto, which includes detailed modeling of the higher-multipoles and precession. The second is TEOBResumS-Dal\'{i}, which is the extension to generic orbits \cite{Chiaramello:2020ehz,Nagar:2021gss,Nagar:2021xnh,Bonino:2022hkj}. 
The two variants feature different choices in numerous aspects, including analytical information and calibration to NR. 
TEOBResumS-Giotto is informed by a suite of NR simulations in the aligned-spin sector \cite{Nagar:2017jdw,Nagar:2018zoe,Nagar:2019wds,Nagar:2020pcj} and incorporates precession following the paradigm of \cite{Schmidt:2010it,Boyle:2011gg,OShaughnessy:2011pmr,Schmidt:2012rh} by solving a post-Newtonian system of equations \cite{Akcay:2020qrj,Gamba:2021ydi}. 
Radiation reaction and higher multipoles follow an improved factorization and resummation procedure introduced in \cite{Messina:2018ghh,Nagar:2020pcj} with an NR informed merger-ringdown based on \cite{Nagar:2020pcj,Riemenschneider:2021ppj}. 
TEOBResumS-Dal\'{i} includes recent 5PN information 
\cite{Nagar:2023zxh} with radiation reaction on generic orbits being incorporated by dressing the factorized quasicircular modes \cite{Chiaramello:2020ehz,Nagar:2021gss} with the eccentric corrections up to 2PN \cite{Placidi:2021rkh,Albanesi:2022ywx}.  
Exemplary waveforms of coalescing binary black holes from the Phenom and EOB waveform families are shown in figure~\ref{fig:BBHComparison}.

For a detailed overview of current progress within the PN framework, see section~\ref{sec:weak}. Here we highlight recent progress in modeling eccentric binaries. 
Key steps towards building non-spinning eccentric waveform models in the PN framework included deriving the quasi-Keplerian equations for the orbits \cite{Damour:1985de,Schafer1993:wex,Memmesheimer:2004cv,Cho:2021oai}, the evolution equations for the orbital terms \cite{Damour:2004bz,Konigsdorffer:2006zt,Arun:2007rg,Arun:2007sg,Arun:2009mc}, and the derivation of the waveform amplitudes \cite{Mishra:2015bqa,Boetzel:2019nfw,Ebersold:2019kdc}. 
Spin effects were subsequently incorporated in \cite{Wex:1995sp,Konigsdorffer:2005sc,Keresztes:2005tp,Majar:2008zz,Cornish:2010cd,Klein:2010ti,Csizmadia:2012wy,Ireland:2019tao,Paul:2022xfy}. 
Post-circular or small-eccentricity expansions have proven to be particularly successful, and have been used to derive analytical expressions for eccentric binaries in the Fourier-domain \cite{Yunes:2009yz,Tanay:2016zog,Loutrel:2016cdw,Loutrel:2018ydu,Klein:2018ybm,Moore:2019xkm,Klein:2021jtd}. 
Early efforts at modeling both precession and eccentricity include \cite{Klein:2018ybm}, which used a small-eccentricity expansion of the waveform amplitudes to separate the periastron precession timescale from the orbital timescale. 
This was extended in \cite{Klein:2021jtd}, building on the multiple scale analysis (MSA) framework, to introduce a system of equations for precessing binaries on elliptic orbits that can be integrated on the radiation-reaction timescale. 
The validity of the model was extended to moderate eccentricities $(e \lesssim 0.8)$ \cite{Arredondo:2024nsl}.
In \cite{Morras:2025nlp}, a number of significant improvements were introduced including closed-form analytical expressions for the eccentric amplitudes, higher-order PN information, and a reformulation of the MSA equations to improve numerical stability and robustness. 

The existence of the BH event horizons implies that a fraction of the emitted GWs can be absorbed by the binary objects and lead to a change in the BH masses and spins.
These fluxes are called horizon fluxes or BH absorption to distinguish them from the ones carried out to infinity.
The increased sensitivity of future detectors may be sufficient to  discern such dissipative effects and help to probe the nature of BHs and the existence of event horizons~\cite{Maselli:2017cmm,Datta:2019epe,Datta:2019euh, Mukherjee:2022wws}.
In the case of nonspinning BHs, the horizon flux starts at 4PN~\cite{Tagoshi:1997jy, Poisson:2005pi} relative to the leading quadrupolar flux to infinity, while the same effect appears earlier, at 2.5PN~\cite{Tagoshi:1997jy, Chatziioannou:2016kem, Goldberger:2020fot,Poisson:2005pi, Comeau:2009bz, Poisson:2004cw, Yunes:2005ve,Brown:2007jx,Isoyama:2017tbp}, when spins are considered.
These effects, for quasi-circular spin-aligned binaries, are known up to 11PN in the test-mass limit~\cite{Tagoshi:1997jy,Shah:2014tka,Fujita:2014eta,Kavanagh:2015lva}, while the corresponding generic mass ratio contributions have been computed only up to relative 1.5PN (absolute 4PN) order~\cite{Alvi:2001mx,Chatziioannou:2012gq, Chatziioannou:2016kem,Saketh:2022xjb}.
Although many studies have focused on the inclusion of horizon absorprion in EOB models (see, e.g., refs.~\cite{Yunes:2009ef,Yunes:2010zj,Nagar:2011aa,Bernuzzi:2012ku,Taracchini:2013wfa}), complete state-of-the-art models either do not consider it or include it only at leading order~\cite{Damour:2014sva,Nagar:2018zoe}. 
This situation is broadly similar for the phenomenological waveform models, e.g., \cite{Mukherjee:2023pge, Chia:2024bwc}. 

The NR surrogate models are based on reduced order modeling techniques \cite{Field:2013cfa,Tiglio:2021ysj} and are capable of accurately reproducing NR information without being required to use any analytical approximations or phenomenological model. The first such models were restricted to non-spinning NR simulations up to $q = 10$ including all multipoles up to $\ell = 8$ \cite{Blackman:2015pia}. Subsequent work has extended the framework to aligned-spins \cite{Varma:2018mmi}, precession \cite{Blackman:2017pcm,Varma:2019csw}, extremal spins \cite{Walker:2022zob}, extreme mass ratios \cite{Islam:2022laz,Islam:2023mob}, and eccentric non-spinning binaries \cite{Islam:2021mha}. The NR surrogates inherit any limitation in the underlying NR dataset, including finite duration, finite coverage of the parameter space, and data quality issues. One possible avenue for overcoming the finite duration of NR simulations is to construct PN/EOB-NR hybrids that are subsequently used to build the surrogate model \cite{Varma:2018mmi}. Various metrics have been introduced in order to quantify the optimal placement of simulations for constructing NR surrogates, including greedy algorithms or neural networks \cite{Blackman:2014maa,Blackman:2017pcm,Varma:2019csw,Ferguson:2022qkz}. Surrogate models for other NR observables, including the remnant properties, have also been developed \cite{Gerosa:2018qay,Islam:2021mha,Islam:2023mob}. The current state-of-the-art model is NRSur7dq4 for precessing, quasi-circular binaries \cite{Varma:2019csw}. This model was trained against 1528 NR simulations with mass ratios $q \leq 4$ and spins $|\boldsymbol{\chi}_i| \leq 0.8$, however the model supports extrapolation to $q \sim 6$. The model begins about 20 orbits before merger and includes all multipoles up to $\ell \leq 4$. 

Unlike the methods described above, the gravitational self-force program has historically focused on LISA sources and has hence only recently begun to produce full-fledged waveform models, see section~\ref{sec:GSF}. The Fast EMRI Waveforms (FEW) framework~\cite{Katz:2021yft,Speri:2023jte} currently includes (i) a fully relativistic waveform model for eccentric, nonspinning BBHs at leading, 0PA order, with complete harmonic content, as well as (ii) ref.~\cite{Wardell:2021fyy}'s 1PA model for quasicircular, nonspinning BBHs. For generic systems with eccentricity and precession (but nonspinning secondaries), FEW includes the lower-accuracy ``augmented analytical kludge" model~\cite{Chua:2017ujo}. However, all FEW models are truncated at the end of the inspiral phase, before the final merger, because GSF modeling has focused on LISA EMRI sources, which contain negligible SNR in the final merger. Currently, the only IMR model based on the GSF approach is the nonspinning, quasicircular 0PA surrogate model developed in refs.~\cite{Rifat:2019ltp,Islam:2022laz}.

\vspace{2mm}\noindent
{\bf Main Open Challenges}
\vspace{2mm}

\noindent
One of the key open challenges is to overhaul current paradigms for modeling precessing compact binaries to incorporate in-plane spin-effects in the co-precessing frame modes \cite{Ramos-Buades:2020noq}, precession-induced mode asymmetries \cite{Boyle:2014ioa,Ramos-Buades:2020noq,Kalaghatgi:2020gsq}, and merger-ringdown models that are calibrated to precessing NR simulations. A limitation is that the waveform modes for precessing binaries are only known for circular orbits at leading PN order \cite{Arun:2008kb}, so extensions to higher PN orders and to generic orbits will be important. Recent progress has been made to complete the amplitudes and the spin contributions to the waveform modes up to 3.5PN within the PN framework \cite{Henry:2021cek,Henry:2022ccf,Henry:2022dzx}. 

Other key challenges will be to develop complete IMR waveform models that describe precessing binaries on generic orbits. This will require numerous advances on both the analytical and numerical fronts. Numerical explorations of the precessing, eccentric parameter space are still limited but already display a wealth of phenomena, including resonant cycles \cite{Lewis:2016lgx}. The increased dimensionality introduced by eccentricity is likely to be a challenge for a full calibration against NR. Dimensional reduction strategies informed by analytical and numerical relativity have been remarkably successful \cite{Buonanno:2002fy, Schmidt:2010it,Schmidt:2012rh,Blackman:2014maa,Thomas:2020uqj} and will likely play a key role in modeling precessing eccentric binaries. 

A limitation in further improving the calibration of the BBH models against NR is the lack of high-accuracy, long-duration simulations at large mass ratios $q \gtrsim 18$ and spins $|{\boldsymbol{\chi}}| > 0.8$. This situation is exacerbated by the increased dimensionality for precessing binaries on non-circular orbits. Recent efforts have focused on understanding the domain of validity of small mass-ratio perturbation theory \cite{Ramos-Buades:2022lgf,vandeMeent:2020xgc} as well as a extending the coverage of numerical simulations to higher mass ratios, e.g. \cite{Fernando:2018mov,Lousto:2020tnb,Daszuta:2021ecf,Rosato:2021jsq,Fernando:2022php,Lousto:2022hoq}. In particular, it is anticipated that recent second-order (1PA) gravitational self-force developments help to bridge the gap between extreme and comparable mass ratios \cite{Pound:2019lzj,Warburton:2021kwk,Wardell:2021fyy,Albertini:2022rfe,Albertini:2022dmc}. However, significant work will be required to extend these models to cover the merger phase as well as more generic binary configurations.

\vspace{2mm}\noindent
{\bf Important Next Steps}
\vspace{2mm}

\noindent
A key step will be to push the calibration of precessing, quasi-circular waveform models towards the complete 7D parameter space. This will require a community effort to produce long-duration NR simulations that extend the current catalogs to higher mass ratios. Similarly, producing more aligned-spin simulations at high-mass ratios and high-spins is likely to be key in driving down waveform systematic areas in regions of the parameter space with poor NR coverage, e.g. \cite{Bohe:2016gbl,Pratten:2020fqn,Ferguson:2022qkz}.  

\vspace{2mm}\noindent
{\bf Recent Progress}
\vspace{2mm}

\noindent
In this section we highlight a subset of the developments that make progress towards tackling some of the key open challenges identified above. 

Within the post-Newtonian framework, ref.~\cite{Arredondo:2024nsl} extends the precessing eccentric model of \cite{Klein:2021jtd} to moderate eccentricities $e \leq 0.8$ by introducing a framework to derive the mode amplitudes for arbitrary eccentricity. A new analytical model for precessing binaries in the extreme mass-ratio limit was introduced in \cite{Loutrel:2024qxp}. Recent progress within the GSF paradigm includes the modeling of the transition to plunge \cite{Kuchler:2024esj} and progress towards tackling the second-order self-force problem in Kerr \cite{Bourg:2024cgh, Bourg:2024vre}. 

There have been several developments in the EOB framework. In particular, \cite{Gamba:2024cvy} extends TEOB to generic, non-planar orbits including validation against precessing eccentric NR simulations \cite{Healy:2022wdn}. There have been improvements to the treatment of radiation reaction \cite{Nagar:2024dzj}, incorporation of GSF information into the TEOB framework \cite{Albertini:2023aol}, explorations of the structure of the merger-ringdown in eccentric binaries \cite{Carullo:2023kvj}, and incorporation of higher-order PN information for spin-orbit couplings \cite{Placidi:2024yld}. 
Progress on the inclusion of post-Minkowskian information in waveform models can e.g. be found in~\cite{Buonanno:2024vkx}, which presents a new spinning EOB Hamiltonian for unbound orbits, including the use of a physical power counting scheme and incomplete 5PM results~\cite{Driesse:2024xad}. Post-Minkowskian information was also leveraged to construct a Hamiltonian valid for bound orbits, enhanced with 4PN non-spinning corrections, to build a complete inspiral-merger-ringdown waveform model based on SEOBRNv5 for aligned-spin BBHs~\cite{Buonanno:2024byg}.

\subsubsection{Binary Neutron Stars}
\label{sec:matter}

Binary neutron star (BNS) systems are one of the main targets of next-generation GW detectors, with more than $10^4$ expected events per year with SNRs above $10$, and ${\cal O}(200)$ per year  with SNRs larger than $50$~\cite{Iacovelli:2022bbs,Branchesi:2023mws} (with an uncertainty by one order of magnitude in both directions due to the uncertainty in the local merger rate). Due to the presence of matter effects, GWs from NS binaries are quantitatively different from GWs emitted by BBHs in both the pre- and post- merger stages of the coalescence.

\vspace{2mm}\noindent
{\bf State-of-the-Art}

\subparagraph{Inspiral-to-merger.}  
While the dynamics of BNS systems is effectively indistinguishable from that of BBHs when the two stars are far apart, as the orbits shrink due to the emission of GWs finite-size effects become more and more relevant. The quadrupole moment of the system, which impacts the waveforms at leading order, is modified by the presence of tidal effects~\cite{Flanagan:2007ix,Hinderer:2007mb,Binnington:2009bb,Damour:2009vw,Vines:2010ca,Vines:2011ud,Bini:2012gu,Damour:2012yf,Banihashemi:2018xfb,Henry:2019xhg,Henry:2020pzq,Henry:2020ski}, quadrupole–monopole terms~\cite{Poisson:1997ha,Mikoczi:2005dn,Landry:2018bil,Nagar:2018plt} as well as possible resonant excitations of the stars' modes~\cite{Lai:1993di,Reisenegger:1993jh,Kokkotas:1995xe,Ho:1998hq,Hinderer:2016eia,Dietrich:2017feu,Steinhoff:2021dsn} or nonlinear fluid instabilities at low frequencies~\cite{Venumadhav:2013nla,Weinberg:2013pbi,Weinberg:2015pxa}.
All of these effects can be successfully described in post-Newtonian theory, i.e. they can be modeled by resorting to expansions in the small parameter $(v/c)^2$, where $v$ denotes the typical orbital speed of the binary, assuming slow-motion and adiabaticity.
    
From PN theory it is relatively straightforward to construct families of PN 
approximants.
The most successful PN model is the analytical, frequency domain {\tt TaylorF2} approximant~\cite{Sathyaprakash:1991,Arun:2008kb}. Typically, the 3.5PN description of point-mass effects is augmented by 7.5PN contributions due to tidal effects~\cite{Flanagan:2007ix, Vines:2011ud, Damour:2012yf, Henry:2020ski} and EOS-dependent spin terms up to 3.5PN~\cite{Nagar:2018plt, Dietrich:2019kaq}. Recent work has also seen the development of PN models for dynamical tides associated with the fundamental ($f$-) mode of the neutron star~\cite{Schmidt:2019wrl, Ma:2020rak}.
Although very efficient, PN models are known to become less and less accurate the closer one approaches merger -- where the importance of matter effects grows -- and are not valid beyond contact. 

Phenomenological models, such as the \texttt{NRTidalv*} approximants~\cite{Dietrich:2017aum, Dietrich:2019kaq, Colleoni:2023czp, Abac:2023ujg} and the model of refs.~\cite{Kawaguchi:2018gvj,Narikawa:2019xng}, are obtained by fitting the tidal and spin-squared contributions of the waveform phase and amplitude to a suite high resolution simulations of BNS hybridised with an inspiral waveform. The functional forms chosen to represent the NR contributions are such that these models naturally recover the correct PN limit in the low frequency regime.
Differently from pure PN models, the calibration to hybrid waveforms implies that they are also predictive in the strong field regime close to merger, whilst retaining all of the computational convenience of analytical models. 
    
Effective-one-body models are some of the most accurate waveform models for BNS coalescences. The two main flavours currently available within the EOB framework are the {\tt TEOBResumS} and {\tt SEOBNRv4T} approximants:

{\tt TEOBResumS}~\cite{Nagar:2018zoe} is based on a conservative Hamiltonian that includes gravitoelectric and gravitomagnetic tidal terms up to $\ell=8$ and $\ell=2$, respectively. Quadrupolar $(\ell=2)$ and octopolar $(\ell=3)$ contributions of the former kind are included via a ``GSF-inspired'' resummation~\cite{Akcay:2018yyh}, which enhances the strength of tidal effects close to merger. Spin-squared terms are included through a modification of the centrifugal radius~\cite{Damour:2001tu,Damour:2014sva,Nagar:2018plt} to next-to-next-to leading order. Furthermore, tidal terms are also included in the waveform multipoles with $(\ell, m) \in \{(2,2), (2,1), (3,3), (3,2), (3,1)\}$. 

{\tt SEOBNRv4T}~\cite{Hinderer:2016eia, Steinhoff:2016rfi, Steinhoff:2021dsn} is based on a Hamiltonian with modified potentials that includes resummed PN expressions of the tidally-induced quadrupole $(\ell = 2)$ and octopole $(\ell =3)$ gravitoelectric multipole moments, and spin-squared corrections. Dynamical tidal effects due to the excitation of the fundamental modes of the stars close to the mode eigenfrequency are included via an effective Love number that approximates the frequency-dependence of the response. Recently, spin-tidal couplings due to frame dragging and Coriolis interactions, which result in a spin-induced shift of the resonance frequency, have been derived~\cite{Steinhoff:2021dsn} but are not yet incorporated in the model. The incorporation of dynamical tides leads to an enhancement of the tidal response. 
The dynamical tides are incorporated in the waveform multipoles via the effective response. A frequency-domain surrogate model, ${\tt SEOBNRv4Tsurrogate}$~\cite{Lackey:2018zvw}, is also available.

\subparagraph{Post-merger.}
In contrast to the inspiral-to-merger emission, GWs emitted during the post-merger phase of BNS coalescences cannot be modeled accurately via simple expansions in the binary orbital parameters. 
Typically, if the remnant does not collapse into black hole, 
the post-merger spectrum is characterized by a dominant frequency 
peak and multiple subdominant components that reflect the dynamics of the remnant~\cite{Bauswein:2011tp,Kastaun:2021zyo}, see also the discussion in section~\ref{ssec:postmerger}. 
Numerical relativity simulations illustrate how, on top of gravitational interactions, nuclear reactions and hydro- and thermo-dynamical processes drive the evolution of the post-merger remnant object (see e.g.~\cite{Clark:2015zxa, Radice:2016rys}). 
Therefore, due to the non-trivial morphology of these signals and the various physical processes that contribute to the merger outcomes, post-merger templates are currently calibrated to numerical relativity results~\cite{Breschi:2019srl, Breschi:2022xnc, Hotokezaka:2013iia, Takami:2014tva, Clark:2015zxa, Easter:2018pqy}.
These models capture few, robust features of numerical relativity simulations resorting to quasi-universal relations -- which link the characteristic properties of the signal to the binary masses, spins and (adiabatic) tidal parameters -- and are typically built via superpositions of damped sinusoids or wavelets.
More complicated approaches based on principal component analysis~\cite{Clark:2015zxa}, hierarchical models~\cite{Easter:2018pqy} and conditional variational autoencoders~\cite{Whittaker:2022pkd} are also being investigated. 
Alternatively, one can consider a flexible morphology-independent analysis for detecting the post-merger phase, which still allows the estimation of key features such as the peak frequency of the post-merger spectrum~\cite{Wijngaarden:2022sah}.

While the aforementioned methods are typically applied to short-lived post-merger signals, the merger of two neutron stars can also result in a long-lived unstable or stable neutron star, see also the discussion in sections \ref{sssec:continousGW} and \ref{sec:nscw}. The analysis of such intermediate or long-duration GW signals does not commonly use waveform models but instead utilises other techniques such as cross-power spectrograms~\cite{Thrane:2010ri}, Hidden Markov Model Tracking~\cite{Suvorova:2016rdc, Sun:2018owi} or algorithms based on the Hough transform, e.g.~\cite{Oliver:2019ksl, Miller:2018rbg}. Searching for long-duration post-merger signals spanning hundreds of seconds comes with its own host of challenges, including the treatment of data gaps and sharp features in the noise spectrum such as lines~\cite{LSC:2018vzm}.

\begin{figure}[t]
    \centering
    \includegraphics[width=0.98\textwidth,trim=20 10 20 35]{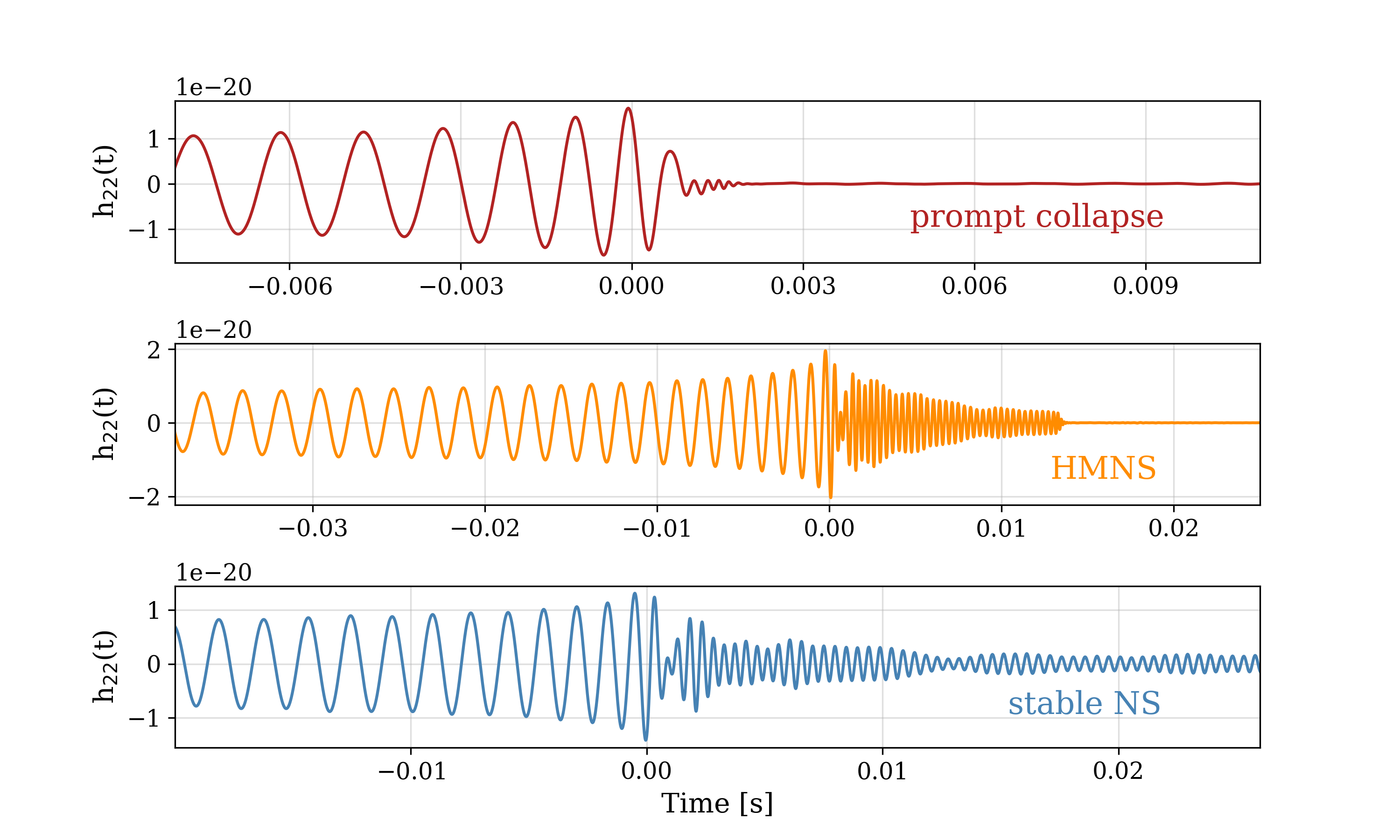}
    \caption{
    Selection of binary neutron star waveforms from NR simulations showcasing three different postmerger phenomenologies: The top panel shows prompt collapse to a black hole; the middle panel shows the formation of a hypermassive neutron star (HMNS), and the bottom panel shows the formation of a stable neutron star. 
    The simulations are BAM:0005~\cite{Dietrich:2016hky}, THC:0084~\cite{Nedora:2020hxc} and BAM:0080~\cite{Bernuzzi:2015rla}, publicly available from the CoRE database~\cite{Gonzalez:2022mgo}.
    }
    \label{fig:BNSComparison}
\end{figure}

\vspace{2mm}\noindent
{\bf Main Open Challenges}
\vspace{2mm}

\noindent
The current selection of BNS NR simulations is sparse and limited in its physics content but post-merger models rely entirely on the availability of such simulations. Moreover, BNS mergers can result in different post-merger scenarios, which are characterised by distinct waveform morphologies. Examples of NR simulations of BNSs with different postmerger outcomes from the CoRE database~\cite{Gonzalez:2022mgo} are shown in figure~\ref{fig:BNSComparison}.
Currently available NR-informed post-merger models are phenomenological in nature~\cite{Breschi:2022xnc} and focus on the dominant harmonic. Going beyond the $(2,2)$-mode will be required in order to identify possible phase transitions~\cite{Espino:2023llj}. Highly accurate NR simulations with smaller errors and improvements in the numerical treatment of effects such as magnetic fields, nuclear reactions and neutrino emission will be crucial.  
NR simulations are not only important to model the post-merger signal but are also needed to inform the description of the late stages of the inspiral and to determine its end. Improving the accuracy and physics content of NR simulations of BNS remain an key open challenge that needs to be tackled urgently.

As BNS signals will be observed for minutes to hours (and up to a day) by ET, any waveform model will need to be computationally highly efficient to provide accurate source parameter estimates and sky localisation information in low latency, which is at odds with the need for improving the model accuracy. Building on existing techniques such as multibanding~\cite{Garcia-Quiros:2020qlt}, reduced-order modeling~\cite{Lackey:2018zvw} or the stationary phase approximation~\cite{Gamba:2020ljo}, and developing novel techniques and tools including the use of AI and machine learning algorithms~\cite{Tissino:2022thn} to reduce the computational cost for generating long-duration signals remains a pertinent challenge.


\vspace{2mm}\noindent
{\bf Important Next Steps}
\vspace{2mm}

\noindent
As discussed in section~\ref{section:div6},  ET promises to deliver some of the most stringent constraints on the properties of neutron star matter, with projected statistical uncertainties in the measurement of the neutron star radius of $50-200$ m~\cite{Huxford:2023qne}. To deliver such a precise measurement, any potential systematic errors must be less than the statistical one. Waveform systematics due to incomplete physics content is one of the dominant sources of error. To achieve the projected measurement accuracy of the neutron star radius, waveform modeling errors must be small enough to induce a dephasing of less than $\lesssim 0.3$ rad~\cite{Chatziioannou:2021tdi}. Currently available waveform models can have typical dephasing errors of up to several radians (depending on the part of the parameter space), and NR simulations typically quote errors of $\sim 1$ rad~\cite{Dietrich:2017feu}.

It has been shown (see e.g.~\cite{Gamba:2020wgg, Pratten:2021pro}) that systematics due to the lack of relevant physical effects pertaining to the presence of matter will need to be minimised in order to make accurate inferences about the nuclear EOS with the next generation of GW observatories.  
There are numerous such effects that are not yet consistently incorporated in current BNS waveform models, including fluid modes other than the $f$-modes, such as Rossby modes or gravity modes, dissipative effects~\cite{Ripley:2023qxo}, and orbital eccentricity, which leads to a distinctly different GW spectrum~\cite{Chirenti:2016xys, Gold:2011df}. 
The availability of analytic tidal-spin information, including self-spin, is currently limited to low PN orders, and an extension of such gravitomagnetic effects to higher order is highly desirable.

As discussed before, while recent progress has been made on modeling the merger and post-merger phases~\cite{Breschi:2022xnc}, no such general model exists across the entirety of the parameter space. As ET will be able to observe the post-merger phase of BNS, which provides a unique opportunity to study the hot nuclear EOS, an improved post-merger model with be required. This necessarily demands further progress in NR simulations. 

Tidal effects and certain beyond-GR effects are degenerate, see section~\ref{subsec:newphysics}. Therefore, missing tidal physics could also lead to a misclassification of a violation of GR. Concrete BNS models in alternative gravity theories such as scalar-tensor or scalar Gauss-Bonnet~\cite{Julie:2022qux} will be advantageous for gauging systematics. 



\subsubsection{Neutron Star -- Black Hole Binaries}
\label{sec:NSBH}


\vspace{2mm}\noindent
{\bf State-of-the-Art}

\subparagraph{Parameter space of NSBH systems.} The parameter space of NSBH systems involves masses in the ranges
$\sim 1 M_\odot<m_{\rm NS} \lesssim 2 M_\odot$,  $m_{\rm BH}\gtrsim$ few $M_\odot$ (though primordial BHs could have lower masses), dimensionless black hole spins $|{\boldsymbol{ \chi}}_{\rm BH}|\lesssim 1 $ and a NS spin frequency up to $f_{\rm NS}\sim 720 $Hz~\cite{Hessels:2006ze}, where the maximum value and the conversion to $|{\boldsymbol{ \chi}}_{\rm NS}|$ depend on the EOS and, with NS spins observed in binary systems, is  generally much lower. 
While most NSBH binaries are expected to have negligible eccentricities in the ET band, some environments may lead to a significant remaining eccentricity~\cite{Fragione:2018yrb,Trani:2021tan}. The physics of NS matter enters mainly through the cold EoS, while effects of temperature, neutrinos, magnetic fields, and viscosity are expected to be less important for the GW signals~\cite{Foucart:2020ats,Kyutoku:2021icp}. Altogether, this makes NSBH systems clean probes of the interplay of dense matter and strong-gravity but also complicates the development of reliable waveform models over this large parameter space. 

\begin{figure}
\centerline{ \includegraphics[width=0.8\textwidth,clip=true,trim=0 3 50 0]{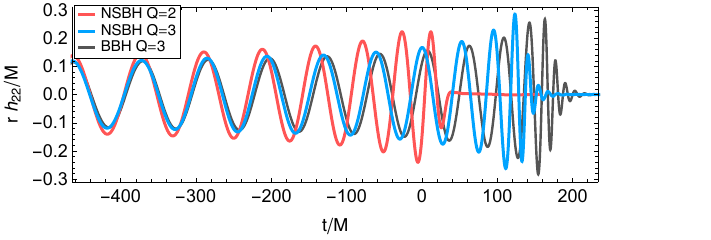}}
 \caption{\textit{Features of NSBH waveforms} from the SXS catalog~\cite{sxscatalog} for nonspinning systems, which were aligned at early times. The case with mass ratio $Q=2$ (red curve) clearly shows the shutoff due to tidal disruption. For $Q=3$, the NSBH waveform (blue curve) is similar to the corresponding BBH case (dark grey curve) up to a small dephasing due to tidal effects and details in the evolution near the plunge-merger (not visible on this scale here). }
 \label{fig:NSBH_waveforms}
 \end{figure}

\subparagraph{Features of the GWs.} NSBH inspiral waveforms have a rich structure as the moderate to large mass ratio increases the power in higher modes of the GWs, and spin precession, eccentricity, and signatures of NS matter (described in section~\ref{sec:matter}) also have an impact. Depending on the mass ratio, the BH spin and the EoS, the inspiral terminates in different ways~\cite{Foucart:2020ats, Kyutoku:2021icp} as illustrated in figure~\ref{fig:NSBH_waveforms}:
\begin{itemize}
\item  {\it Tidal disruption} of the NS leads to a distinct shutoff of the GW signal at $f_{\rm GW}\sim 1-1.5$ kHz, with the exact value depending predominantly on the BH mass, BH spin, mass ratio and NS EOS (see e.g.~\cite{Foucart:2020ats, Kyutoku:2021icp}). During the post-merger phase more matter is accreted onto the remnant, which leads to a dampening of the ringdown signal and is important for multimessenger counterparts. 

\item {\it Plunge} of the NS into the BH, which occurs in a large swath of NSBH parameter space. The GW signals are similar to  BBH and multimessenger counterparts are absent. 

\item {\it Mild disruption}, which represents an intermediate case where part of material is stripped off the NS near merger. The GWs qualitatively resemble those of BBH mergers but with the properties of the final remnant altered by accretion. 
\end{itemize}

\smallskip 

\subparagraph{Numerical relativity simulations.} 
These play a crucial role for accurately modeling the GW signal in the strong-field regime as well as for robustly determining if and at what frequency distruption occurs. A summary of the current state-of-the-art NR simulations of NSBH systems can be found in section~\ref{sec:NR}. 

\subparagraph{Analytical models.} Full waveform models for NSBH coalescences involve a baseline point-mass model for the inspiral with corrections due to matter effects that smoothly connects with a description of the tidal disruption or merger. The purely theoretical models for matter effects during the inspiral apply for any kind of compact object and can be used both for BNS and NSBH systems; differences between these systems only appear in the late stages of the inspiral. The basic features of the GW signals beyond the inspiral due to plunges or tidal disruption have long been qualitatively understood~\cite{Vallisneri:1999nq,Shibata:2009cn,Shibata:2011jka,Foucart:2012nc}, and improved through fits to numerical relativity simulations~\cite{Lackey:2011vz,Lackey:2013axa,Pannarale:2013uoa,Pannarale:2015jka}. Following a first complete model developed in~\cite{Lackey:2013axa}, the current state-of-the-art for NSBH-specific waveform models comprises three models, ordered here in the sequence in which they became available: 

\texttt{IMRPhenomNSBH}~\cite{Thompson:2020nei} is a frequency-domain model for the dominant mode of GWs for aligned BH spins $|\chi_{\rm BH}|\leq 0.5$, circular orbits, and nonspinning NSs with dimensionless tidal deformability up to $\Lambda\leq 5000$. The point-mass phase baseline is~\texttt{IMRPhenomD}~\cite{Husa:2015iqa,Khan:2015jqa} and tidal effects are added through the \texttt{NRTidalv2} model~\cite{Dietrich:2019kaq}, which was calibrated to equal-mass BNS waveforms from the \texttt{BAM} code. The merger model is from~\cite{Pannarale:2015jka} which uses the disk mass fit of~\cite{Foucart:2012nc}, the BBH baseline amplitude of \texttt{IMRPhenomC}~\cite{Santamaria:2010yb}. 
It is calibrated to NR data from the \texttt{SACRA} code for mass ratios $2,3,4,5$, aligned BH spins with $\chi_{\rm BH}=-0.5, 0, 0.25, 0.5, 0.75$, $m_{\rm NS}\sim 1.35M_\odot$ with a few cases of $1.2, 1.45 M_\odot$, and piecewise polytropic EoSs designed to cover the range $\Lambda\sim 100-1000$. The NS compactness appearing in the disk mass predictions is eliminated in favor $\Lambda$ using quasi-universal relations~\cite{Yagi:2016bkt} adapted to ensure a well-behaved BH limit for $\Lambda\to 0$. 

\texttt{SEOBNRv4ROM\_NSBH}~\cite{Matas:2020wab} is a frequency-domain model for the dominant mode of GWs for aligned BH spins of magnitude $|\chi_{\rm BH}|\leq 0.9$, circular orbits, and nonspinning NSs. The point-mass baseline is \texttt{SEOBNRv4\_ROM}~\cite{Bohe:2016gbl}, which was calibrated to NR results 
and to waveforms from a BH perturbation code. Tidal effects are also based on the~\texttt{NRTidalv2} model (see the limitations listed above). The tidal disruption -- plunge model is structurally also based on~\cite{Pannarale:2015jka} with~\cite{Foucart:2012nc} and using quasi-universal relations to eliminate compactness in terms of $\Lambda$. The difference is that the coefficients were re-calibrated for the EOB baseline to the \texttt{SACRA} waveforms (see above) and NSBH waveforms from SpEC for mass ratios $1-8$, $-0.5 \leq \chi_{\rm BH} \leq 0.9$, $m_{\rm NS}\sim 1.2-1.45M_\odot$ and $\Lambda\sim 100 - 1400$.

The NSBH version of \texttt{TEOBResumS-GIOTTO}~\cite{Gonzalez:2022prs} is a time-domain model with accompanying transformation to the frequency-domain using the stationary phase approximation. The model is calibrated to non-precessing \texttt{SACRA} and \texttt{SpEC} waveforms for mass ratios $\lesssim 8$, BH spins of $\chi_{\rm BH}\leq 0.75$, nonspinning NSs and circular orbits. 
Several higher modes and approximate spin precession effects are included. The point-mass baseline is the \texttt{TEOBResumS} model of~\cite{Gamba:2021ydi}, which is based on an EOB Hamiltonian, PN spin precession effects, and tidal effects from~\cite{Bernuzzi:2014owa,Akcay:2018yyh}. The tidal disruption -- plunge model builds on an updated version of predictions for the remnant BH from~\cite{Zappa:2019ntl} and is an NR-informed fit to the differences from a BBH signal. 
We note that the calibration regions of the models quoted above are non-uniformly sampled over the parameter space; see the original references for more details.

\vspace{2mm}\noindent
{\bf Main Open Challenges}
\vspace{2mm}

\noindent
The wide range of parameter space of NSBH systems covers regions that are inherently difficult to model, both analytically and with NR simulations. Waveform models for NSBH remain less developed than for BBH and the following key challenges remain.

\subparagraph{Parameter space coverage and accuracy.} This presents challenges for both numerical and analytical models. NR simulations require capabilities to evolve both an event horizon and the NS over a number of orbits. More unequal mass ratios and higher spins lead to larger differences in the scales involved, making simulations more challenging and computationally expensive. 
Due to the greater complexity of the physics involved in simulations with matter, the numerical errors in the waveforms (e.g. resolution, mass loss) are usually larger than for BBHs and assessing them requires additional efforts. For analytical models, the challenges for the inspiral are similar to those for BNS in section~\ref{sec:matter}, up to a few differences in the very late inspiral: NSBH lack overlapping matter distributions, i.e., while for BNS the outer layers of the stars come in contact before their high-density cores merge, this is not the case for NSBH, which may impact the GW signal during the very last inspiral stages shortly before merger. Detailed studies and quantification of the differences are yet to be carried out, while NR calibration would be one avenue to improve the accuracy of NSBH models during this late stage. 
NSBHs also involve regimes of stronger gravity and higher companion spin. This impacts matter signatures and involves parameters outside the regimes covered by BNS. 
For complete models, a challenge remains to exploit physical insights to reduce the number of calibration parameters and thus improve their robustness. Current NSBH models rely on a large number of fitting parameters in a sparsely sampled parameter space to describe the merger regime and use information from remnant disk masses that is only indirectly related to the GW signals. 
Another challenge is  that existing models are either not entirely independent in their assumptions and inputs, or differ in all aspects at once, thus making it difficult to assess systematic uncertainties. 

\subparagraph{Impact of more realistic NS physics.} While the dominant GW signatures of NS matter in NSBH systems even during tidal disruption depend only on the cold EoS, potential subdominant effects from combined finite temperatures~\cite{Gittins:2024jui}, out-of-equilibrium microphysics, multi-fluids, magnetic fields~\cite{Etienne:2011ea}, or phase transitions in dense matter~\cite{Bauswein:2018bma} cannot be completely excluded. Most estimates and amplification mechanisms have assumed simplified scenarios and individual effects, see section~\ref{sec:matter}, and fully quantifying their effects and potential importance for NSBH waveforms remains an open problem that needs to be addressed.  
 
\subparagraph{Efficiency.} Including additional complexities in the models to improve their accuracy, parameter space coverage, and physical realism generally decreases their efficiency. Recent vigorous progress on deep learning methods and other means to speed up waveform evaluations has made this a separate concern (see section~\ref{sec:acc} for an overview). Still, the high dimensionality and large extent of the NSBH parameter space of future models that will include more realistic physics poses challenges. 

\subparagraph{Flexibility to constrain modified gravity, dark matter and other BSM physics.} While modeling astrophysically expected systems within GR is the priority and the fundamentally required baseline, it is also important to develop models with enough flexibility to test for new physics and quantify potential degeneracies and biases (see also 
section~\ref{section:div1} and section~\ref{subsec:newphysics}). Some studies have begun to explore aspects of this issue but more work is needed, especially NSBH-specific considerations.

\vspace{2mm}\noindent
{\bf Important Next Steps}

\subparagraph{Accuracy requirements.} Dedicated studies to determine the accuracy requirements for NSBH waveform models for ET and other 3G GW detectors are largely missing. The lack of a systematic sampling of the NSBH parameter space with NR limits the accuracy assessment of current models and the determination of systematic errors inherent to them. Limited studies indicate that the calibration of models to NSBH simulations is crucial for SNRs $\gtrsim 70$~\cite{Huang:2020pba} and that higher-order tidal effects such as the $f$-mode excitation can lead to significant biases~\cite{Foucart:2018lhe}. With the recent observation of the NSBH candidate GW230529~\cite{LIGOScientific:2024elc} 
and the attendant increase in event-rate for NSBH systems and associated EM counterparts, understanding the modeling error budget and determining accuracy requirements will be paramount for the next generation. 

\subparagraph{NR simulations.} Further studies of the features of NSBH systems with large BH spin magnitudes, arbitrary spin orientations, appreciable NS spins, and eccentricity are required to improve the modeling. Regarding the coverage of NS masses and EoSs, the range of $\Lambda$ explored to date covers a relatively wide range but also leaves some parts of the parameter space sparsely sampled, for instance small $\Lambda$, as also relevant for higher mass NSs. Studies with varying EoSs have indicated that primarily $\Lambda$ or compactness matters for the waveforms; however, any potential impact of e.g. strong phase transitions in cold NSs on the GWs from the tidal disruption regime remain to be explored. Likewise, further detailed studies of the potential impact of more realistic physics mentioned above will be needed to assess their full impact. In addition, a few case studies for NSBH systems in well-motivated classes of modified gravity theories and in the presence of dark matter are also needed as unexpected nonlinear phenomena could arise, as discovered in BNS~\cite{Barausse:2012da} and BBH~\cite{Elley:2022ept} systems. 
The parameter space of large mass ratios, far from regimes where NS matter effects are discernible, for which the exact boundary still needs to be fully mapped out over the parameter space, does not require NSBH-specific simulations and can be explored with less challenging BBH simulations. The current two-pronged strategy to cover the interesting features might continue to be useful, with shorter simulations exploring the merger and fewer long, high-accuracy waveforms also covering the late inspiral and consistency checks on the mergers.  

\subparagraph{Analytical modeling.} Further work is needed to improve several components of analytical models to cover more of the parameter space with greater accuracy: the BBH baseline, description of matter effects during inspiral, and the modeling of the merger. In addition, models must include or improve the accuracy of higher modes and several independent models are needed to assess the systematic uncertainties.
For the BBH baseline, the important steps are similar as for BBH-specific waveforms discussed in section~\ref{sec:bbh}, and for matter effects in the inspiral similar to BNS covered in section~\ref{sec:matter}. For the latter, a few extra considerations must be taken into account for NSBH, for instance, more realistic descriptions of dynamical tides that include more effects of strong gravity and a potentially high BH spin, NS spins, precession, and higher modes. Although matter effects during the inspiral are expected to be rather small in NSBH systems compared to BNS, their measurement would be a uniquely clean probe of the physics of cold dense matter and its interplay with strong gravity, especially in the case of NS disruption~\cite{Clarke:2023rrm}. Similarly to the inspiral, the analytical structure for modeling tidal disruption modifications away from a BBH waveform requires further refinements, for instance, by making use of physical considerations that are directly related to features in the GWs to develop a more robust underlying structure of the model and its parameter-dependence and thus reduce the number of fitting parameters. Likewise, the description of mild tidal disruptions and mapping from the progenitor parameters to the remnant BH over a wider range of parameter space could be improved with further physical considerations as is done for the BBH case before calibrating to NSBH simulations. 
Moreover, a few analytical models need to also allow for capabilities to discern additional physics, both related to NS matter as well as beyond the standard paradigms of gravity and particle physics.

\subsubsection{Modelled sources beyond binary inspirals}
\label{sec:other}



\paragraph{Early universe sources}

\mbox{}\\

\vspace{0mm}\noindent
{\bf State-of-the-Art}
\vspace{2mm}

\noindent
With the Einstein Telescope, we may discover GW sources that do not originate from standard stellar evolution. As we discussed in detail in section~\ref{sec:early_universe},
other types of structures may have formed in the early universe, such as 
cosmic strings (section~\ref{subsec:cosmic_strings}),
domain walls (section~\ref{sec:domain_walls}) or
primordial BHs (section~\ref{sect:PBHdiv2}). Other exotic compact objects with a possible cosmological origin include quark nuggets \cite{Witten:1984rs,DiClemente:2024lzi},  Q-balls \cite{Coleman:1985ki}, boson stars \cite{Liebling:2012fv}, fermion stars \cite{Lee:1986tr,DelGrosso:2023trq,DelGrosso:2023dmv}, gravastars, and anisotropic stars (for reviews see refs.~\cite{Giudice:2016zpa,Cardoso:2019rvt} and references therein). These
could leave behind a trace of their existence by the production of observable GWs, both in the form of  stochastic GW backgrounds,  and  as individually resolvable events, for which we could reconstruct the waveform.

In this section, we overview the current state-of-the-art and highlight future challenges regarding the GW waveform of sources with early universe origins. More details and signatures of these early universe sources, e.g., in the stochastic GW background and BH mergers, are given in section~\ref{section:div1}, \ref{section:div2} and \ref{section:div3}. We proceed to discuss each source separately.

\subparagraph{Primordial black holes.} The waveform for the coalescence of binaries made of PBHs is the same as for astrophysical binary black hole mergers; however, unlike astrophysical black holes, PBHs could be lighter than a solar mass. Interestingly, though there is no definite detection yet, there are tentative hints of solar and sub-solar mass compact objects in the LVK data \cite{Clesse:2020ghq,Morras:2023jvb,LIGOScientific:2022hai,Prunier:2023cyv}. Finding a single binary of sub-solar mass BHs would be a very strong evidence of new physics in the very early universe and such possibility is within reach of ET \cite{Crescimbeni:2024cwh}. It should be noted that, in the case of masses larger than a solar mass, one needs additional information, e.g. by doing population analysis or looking for the induced GW background counterpart, to distinguish PBH binaries from astrophysical black holes binaries, see the discussion in sections \ref{sect:PBHdiv2} and \ref{sec:PBHsdiv3}.

Current bounds from the LVK collaboration place the abundance of PBHs as dark matter to be less than unity in the subsolar mass range \cite{LIGOScientific:2022hai}, complementing microlensing constraints \cite{EROS-2:2006ryy,Gorton:2022fyb,Petac:2022rio,DeLuca:2022uvz,Mroz:2024mse}.
Remarkably, even though PBHs might comprise less than $1\%$ of the dark matter in the sub-solar mass range, the number of resolvable PBH mergers in $1\,\rm yr$ of observation of ET could still be significant due to the improved strain sensitivity~\cite{Franciolini:2023opt,Crescimbeni:2024cwh}. 
Also, strongly lensed GWs can be used also to constrain the abundance of PBHs in the stellar mass regime $(1-20)\, \Msun$~\cite{Diego:2019rzc}.

Another mechanism that could produce observable GW signals is the close hyperbolic encounter of PBHs, leading to GW bursts. The waveform of hyperbolic encounters has been studied extensively in the literature~\cite{Hansen:1972jt,Turner:1977tm,Turner:1978zz,Capozziello:2008ra,DeVittori:2012da,Bini:2012ji,Damour:2014afa,DeVittori:2014psa,Cho:2018upo,Grobner:2020fnb,Nagar:2020xsk,Hopper:2022rwo,Damour:2022ybd,Rettegno:2023ghr,Caldarola:2023ipo,Roskill:2023bmd}. For graphic examples of the GW waveform see, e.g., figure~2 of ref.~\cite{Garcia-Bellido:2017qal} and figure~2 of ref.~\cite{Teuscher:2024xft}. The peak frequency of the GW burst depends on parameters such as the PBH masses, the speed at which they encounter and the eccentricity of the orbit. Searches for hyperbolic events have been performed using the current LVK data~\cite{Morras:2021atg, Bini:2023gaj}, with no significant candidate so far. Although the standard expectation is that close hyperbolic encounters are rarer than PBH binary mergers, they could be detectable by ET if PBHs constitute a significant fraction of dark matter and are clustered at formation \cite{Garcia-Bellido:2017qal,Morras:2021atg,Dandapat:2023zzn,Caldarola:2023ipo,Bini:2023gaj,Kerachian:2023gsa,Codazzo:2023kcx,Garcia-Bellido:2021jlq}.

\begin{figure}
\centerline{ \includegraphics[width=0.65\textwidth,trim=0 10 0 7,clip=true]{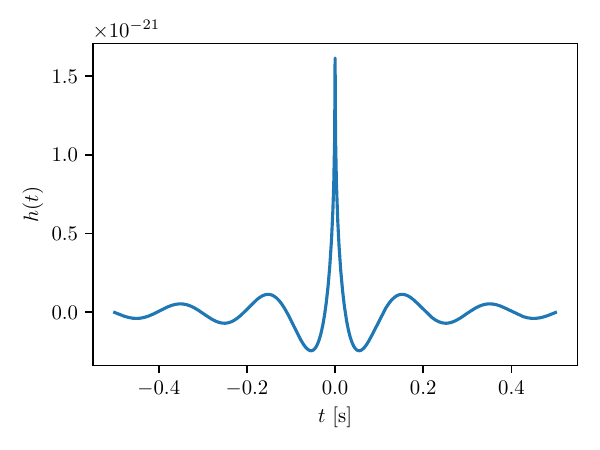}}
 \caption{Cusp strain in time domain computed using eq. (7) of ref.~\cite{Auclair:2023brk}, and fixing $f_{\rm low} = 5\, {\rm Hz}$, $f_{\rm high} = 10^4\, {\rm Hz}$ and for a cusp burst of characteristic amplitude $A = 10^{-21}\,  {\rm s}^{-1/3}$.}
 \label{fig:cosmicstring_waveforms}
 \end{figure}

\subparagraph{Cosmic strings.} As discussed in section~\ref{subsec:cosmic_strings}, the simplest models of strings are parametrized by a single energy scale, $\eta$ that controls their energy density per unit length as well as tension. This scale is usually described by the dimensionless quantity, $G\mu \sim (\eta/M_{\rm pl})^2$. Once formed, cosmic strings start to evolve and oscillate according to their tension. Individual events on a loop, such as cusps and kinks, can generate a strong GW signal. Cusps are points where the string moves momentarily at the speed of light \cite{Turok:1984cn,Blanco-Pillado:1998tyu,Olum:1998ag}, and they typically produce a strong burst of radiation beamed in the direction of motion \cite{Damour:2001bk,Damour:2004kw,Siemens:2006vk}. Kinks, which are discontinuities in the tangent vector for string, and kink-kink collision may also produce GWs, which will be observed as transient bursts \cite{Damour:2001bk,Damour:2004kw}. For an illustration of the typical GW waveform from cosmic strings, see figure~\ref{fig:cosmicstring_waveforms}. Interestingly, bursts may repeat themselves \cite{Auclair:2023mhe} since the dynamics of the Nambu-Goto string predict that loops follow a periodic motion.

A search for these types of bursts has been done using the LVK data in ref.~\cite{LIGOScientific:2017ikf,LIGOScientific:2021nrg}. The lack of any signal compatible with these bursts can then be used to put constraints on the scale of the string.
The capabilities of ET for detecting cosmic string bursts have been discussed in section~\ref{sect:GWburststringsdiv2}, see in particular figure~\ref{fig:cs_burst_rate}, where we see that the absence of detection of GW bursts will  place constraints on the string tension, around $G\mu < 10^{-10} - 10^{-11}$, depending on the string model. For reference, the CMB data imposes a much lower constraint on the string tension, namely, $G\mu \lesssim 10^{-7}$ \cite{Planck:2013mgr}. A similar analysis has also been conducted in the case of the LISA  observatory in \cite{Auclair:2023brk}.

It may also be possible for a cosmic string loop to collapse to a BH and emit GWs.\footnote{This type of process was first described by Hawking as an alternative way to create a primordial black hole \cite{Hawking:1987bn,Hawking:1990tx}.} In this case the  waveform, which must be calculated using full non-linear General Relativity, differs from the  waveform of a burst and of a black hole merger \cite{Helfer:2018qgv,Aurrekoetxea:2020tuw}. Interestingly, there is a candidate of such a new cosmic string waveform in the LVK data \cite{Aurrekoetxea:2023vtp}. It should be noted, though, that the probability of these types of events is likely to be rather small \cite{Polnarev:1988dh,Hansen:1999su}.

\begin{figure}
\centerline{ \includegraphics[width=0.7\textwidth,trim=0 10 0 10]{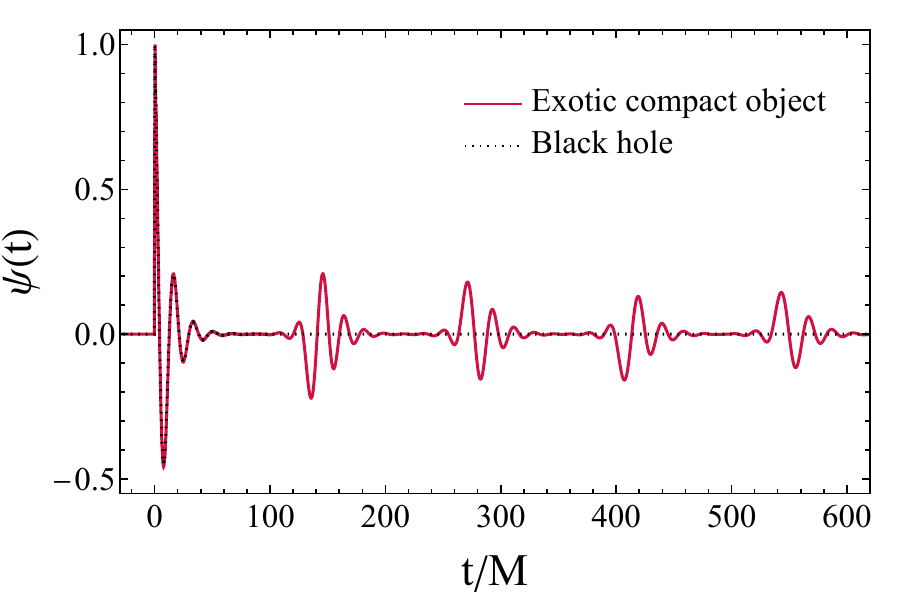}}
 \caption{Ringdown signal featuring echoes, when the remnent of the merger is a horizonless ultracompact object as opposed to a black hole.}
 \label{fig:GW_echoes}
 \end{figure}

\subparagraph{Exotic compact objects.}
Depending on the matter content of the collapsing overdensities in the early Universe, the end state of gravitational collapse might be different from a PBH, and other compact objects can be stable in Beyond Standard Model physics \cite{Giudice:2016zpa,Cardoso:2019rvt}. As in the case of PBHs, sub-solar mass compact objects are possible, depending on the parameters of the theory~\cite{DelGrosso:2024wmy}. 
Such compact objects then emit GWs when in a binary. The resulting GW frequency and waveform is similar to the standard GWs from binaries. However, as discussed in section~\ref{div1_natureofcompact},  any compact object that is not a black hole \cite{Damour:1991yw, Binnington:2009bb, Damour:2009vw,Chia:2020yla} has a non-vanishing tidal deformability \cite{Pani:2015tga,Porto:2016zng, Cardoso:2017cfl}. Thus, the presence of tidal effects in the GW waveform would distinguish exotic compact objects from PBHs as well as stellar BHs~\cite{Crescimbeni:2024cwh,Golomb:2024mmt}. The Love numbers of exotic compact object have been calculated for boson stars \cite{Cardoso:2017cfl,Sennett:2017etc,Mendes:2016vdr}, fermion stars \cite{Diedrichs:2023trk}, 
strange quark stars \cite{Postnikov:2010yn}, gravastars \cite{Pani:2015tga,Cardoso:2017cfl,Uchikata:2016qku}, anisotropic stars \cite{Raposo:2018rjn} and stars with a stiff equation of state at the surface \cite{Cardoso:2017cfl}.
If the remnant of a merger is an ultracompact object without a horizon, the prompt ringdown signal would be nearly the same as the one emitted by a BH. However, the absence of a horizon in the remnant would lead to the emission of a modulated train of GW echoes at later times, as shown in figure~\ref{fig:GW_echoes}.

\vspace{2mm}\noindent
{\bf Main Open Challenges}
\vspace{2mm}

\noindent
GWs from binaries and hyperbolic encounters of sub-solar mass compact objects, such as PBHs and exotic compact objects, are a strong indication of physics beyond the standard model of particles and cosmology. GW bursts, and repeated GW bursts, can be sourced by nearby cosmic string loops, providing hints of grand unified theories. ET may be able to detect such sources with early Universe origin and resolve their GW waveform.

\subparagraph{Challenges on primordial black holes.} To distinguish PBH subsolar binaries from other astrophysical candidates, one should rule out the presence of tidal effects or disruption before ISCO frequency \cite{Crescimbeni:2024cwh,Golomb:2024mmt}. Having reliable waveform models that could capture these effects and are calibrated in the subsolar regime is crucial to rule out the alternative hypothesis of neutron star binaries or more exotic deformable compact objects (e.g., boson star binaries).

Regarding the hyperbolic encounters, the main challenge is to identifying such a GW burst event, specifically distinguishing it from glitch noises, as the waveform is mostly a featureless bump. It is also important to estimate the event rate for a PBH scenario, which strongly depends on clustering. This information is key to distinguishing whether the event is of astrophysical or primordial black hole origin (see section~\ref{sec:PBHsdiv3}). The current estimates are made by introducing a simple clustering parameter~\cite{Garcia-Bellido:2017qal,Garcia-Bellido:2017knh}, but it must connect with factors such as initial non-Gaussianity and the evolution of PBH clusters. Information on the event rate and mass distribution is also crucial for increasing search efficiency.

\subparagraph{Challenges on cosmic strings.} 

The calculation of the number of burst events from a network of strings as well as their waveforms does not take into account the building up of gravitational backreaction on the loop's evolution. This backreaction affects these results since it can alter the average number of cusps per loop as well as their strength. Though information on the shape of the loops may be extracted from numerical simulation, one should also take into account that backreaction may change the shape of loops \cite{Blanco-Pillado:2015ana,Blanco-Pillado:2017oxo}. Furthermore, the number of GW events is directly related to the number density of loops, which at the moment depends on the model \cite{Blanco-Pillado:2013qja,Lorenz:2010sm}.
Regarding the possible misidentification of the GW burst signal with glitches, some authors are considering alternative strategies based on deep learning to identify the cusps signals \cite{Meijer:2023yhn}.

\subparagraph{Challenges on exotic compact objects.} It remains a challenge to build consistent initial conditions for ECOs in order to perform trustable numerical simulations of such binary systems. Without a consistent treatment of initial conditions for ECOs, current numerical simulations are likely to contain systematic issues and biases.


\subparagraph{Core-Collapse Supernovae.}
The detection of GWs from core-collapse supernova explosions (CCSN) is also a primary science goal of the Einstein Telescope.  
The CCSN   GW signal 
\cite{murphy:09,yakunin:10,MuJaMa13,kuroda:14,yakunin:15,kuroda:16,andresen:17,muller2017b,kuroda:17,tk18,2018ApJ...861...10M,radice2019} 
from bounce through supernova explosion and into the 
late-time (many-second) proto-neutron star (PNS) cooling phase (or black hole formation) 
can be decomposed into various stages with characteristic features, frequency spectra, strains 
and polarizations~\cite{2018hayama}. GWs are generated immediately around bounce in core-collapse 
supernovae by time-dependent rotational flattening and prompt post-shock overturning 
convection (lasting tens of milliseconds)~\cite{burrows:95,marek:09}, then during a low phase (lasting perhaps $\sim$100$-$200 milliseconds) 
during which post-shock neutrino-driven turbulence builds, followed by vigorous accretion-plume-energized PNS modal oscillations 
(predominantly f- and g-modes) \cite{murphy:09,2018ApJ...861...10M}. If an explosion ensues, these components are accompanied by 
low-frequency ($\sim$1$-$10 Hz) gravitational-wave ``memory" due to asymmetric emission of neutrinos and 
aspherical explosive mass motions \cite{burrows1996,2020ApJ...901..108V,Mukhopadhyay:2021zbt,2022PhRvD.105j3008R,2022PhRvD.106d3020M}.  
Section~\ref{section:div7} discusses CCSN, and we refer the reader there for more details.

\subsubsection{Waveforms in alternative theories of gravity}
\label{sec:waveforms_alttheories}


In this section we provide a birds-eye-view of the present state of
gravitational waveform modeling of comparable mass binaries in alternative
theories of gravity and in the presence of dark matter clouds.
We will start by reviewing what has been done in the context of post-Newtonian
theory, then numerical relativity, and, at last, black-hole perturbation
theory.
We then survey the progress towards inspiral-merger-ringdown waveform models.
In addition, we discuss theory-agnostic parametrized waveform models (including
their relation to theory-specific predictions).
Before proceeding we must make a disclaimer: the zoo of contenders to general
relativity has a plethora of specimens, and it is impossible to review what we
know about waveform modeling for each theory here.

\vspace{2mm}\noindent
{\bf State-of-the-Art}


\subparagraph{Post-Newtonian theory.}
%
%
Perhaps unsurprisingly, most progress has been made in the context of
scalar-tensor theories of the Bergmann-Wagoner
type~\cite{Bergmann:1968ve,Wagoner:1970vr} (referred to as just ``scalar-tensor theories'' hereafter) which are arguably the
simplest extension of General Relativity.
In virtue of black-hole no-hair theorems (see ref.~\cite{Herdeiro:2015waa} for
a review), the dynamics of black-hole binaries will be identical to what is
predicted by General Relativity (see e.g., refs.~\cite{Berti:2018cxi,Healy:2011ef}),
as long as the background scalar field is spatially homogeneous or time-independent~\cite{Jacobson:1999vr,Berti:2013gfa,Clough:2019jpm}.
However, neutron stars can carry a ``monopole scalar charge'' in these theories
(i.e., the scalar field decays as $1/r$ asymptotically), making neutron-star or
mixed binaries potential test-beds for these theories.
When one or both of the binary components have such a charge, scalar dipolar
radiation can be emitted and thus accelerate the binary inspiral relative to
GR.
Regarding the analytical treatment of the two-body problem in these theories,
the dynamics were studied in
refs.~\cite{Damour:1992we,Mirshekari:2013vb,Lang:2013fna,Lang:2014osa,Bernard:2018hta,Bernard:2018ivi,Bernard:2019yfz,Schon:2021pcv,Bernard:2023eul}
and  waveform generation in refs.~\cite{Sennett:2016klh,Bernard:2022noq}.

The next simplest extension of GR involves the coupling of the scalar field to
quadratic-in-curvature scalars.
Two such theories that have received considerable attention
are Einstein-scalar-Gauss-Bonnet (EsGB) gravity and dynamical Chern-Simons (dCS) gravity.
In the former, a scalar field couples to the Gauss-Bonnet invariant, while in
the latter a pseudoscalar field couples to the Pontryagin density.
No-hair theorems do not apply to these theories, and black holes will in
general carry some ``scalar charge''. In EsGB, this charge is monopolar, while
in dCS the charge is ``dipolar'' (i.e., the pseudoscalar field decays as
$1/r^2$ asymptotically).
The post-Newtonian dynamics in these theories was studied in refs.~\cite{Yagi:2011xp,Loutrel:2018ydv,Julie:2019sab,Shiralilou:2020gah,Shiralilou:2021mfl,Loutrel:2022tbk,Lyu:2022gdr,Julie:2022qux}.

Isolated neutron stars in scalar-tensor theories can undergo a phase transition
known as spontaneous scalarization~\cite{Damour:1993hw}. In a neutron star
binary, one or both binary components can scalarize during the merger.
This effect was first observed in numerical relativity simulations as described
below, and can be formulated in terms of an effective
action~\cite{Sennett:2017lcx,Khalil:2019wyy,Khalil:2022sii}. See also
ref.~\cite{Sampson:2014qqa,Sennett:2016rwa} for an incorporation of these
effects in GW models.

Post-Newtonian calculations were also performed in Einstein-Maxwell-Dilaton theories~\cite{Julie:2017rpw,Julie:2018lfp}, in Lorentz-violating theories~\cite{Yagi:2013ava}, and in effective-field-theory (EFT) of GR~\cite{Sennett:2019bpc,AccettulliHuber:2020dal, Endlich:2017tqa}.

%

\subparagraph{Numerical relativity.}
%
We divide the zoo of alternative theories of gravity into two main categories
that are determined by their type of modification of GR and their challenge for numerical relativity.

The first category represents beyond-GR theories in which scalar or vector fields are minimally coupled to gravity.
It includes scalar-tensor theories or Einstein-Maxwell-dilaton gravity, as well
as dark matter ``clouds'' (represented by massive scalar or vector fields)
around black holes.
In this case, the field equations can be written as a set of wave equations for each field, and standard numerical relativity techniques can be applied. 
%
In scalar-tensor theories, black holes are the same as in GR, but neutron stars
can scalarize. If a coalescing binary that contains at least one neutron star
becomes sufficiently compact, the scalar field can be excited dynamically via
``dynamical scalarization''~\cite{Palenzuela:2013hsa,Shibata:2013pra,Taniguchi:2014fqa,Ponce:2014hha}.
Another example in this group is Einstein-Maxwell-dilaton gravity, for which the
binary coalescences were modeled in ref.~\cite{Hirschmann:2017psw}.
Binary black holes surrounded by a scalar field (or dark matter) condensate
can generate scalar radiation that accelerates the inspiral and causes a gravitational wave phase shift, see, e.g.,~\cite{Bamber:2022pbs,Aurrekoetxea:2023jwk}.

\begin{figure}
\begin{subfigure}{0.5\textwidth}
\centerline{ \includegraphics[width=\textwidth,trim=10 0 80 80]{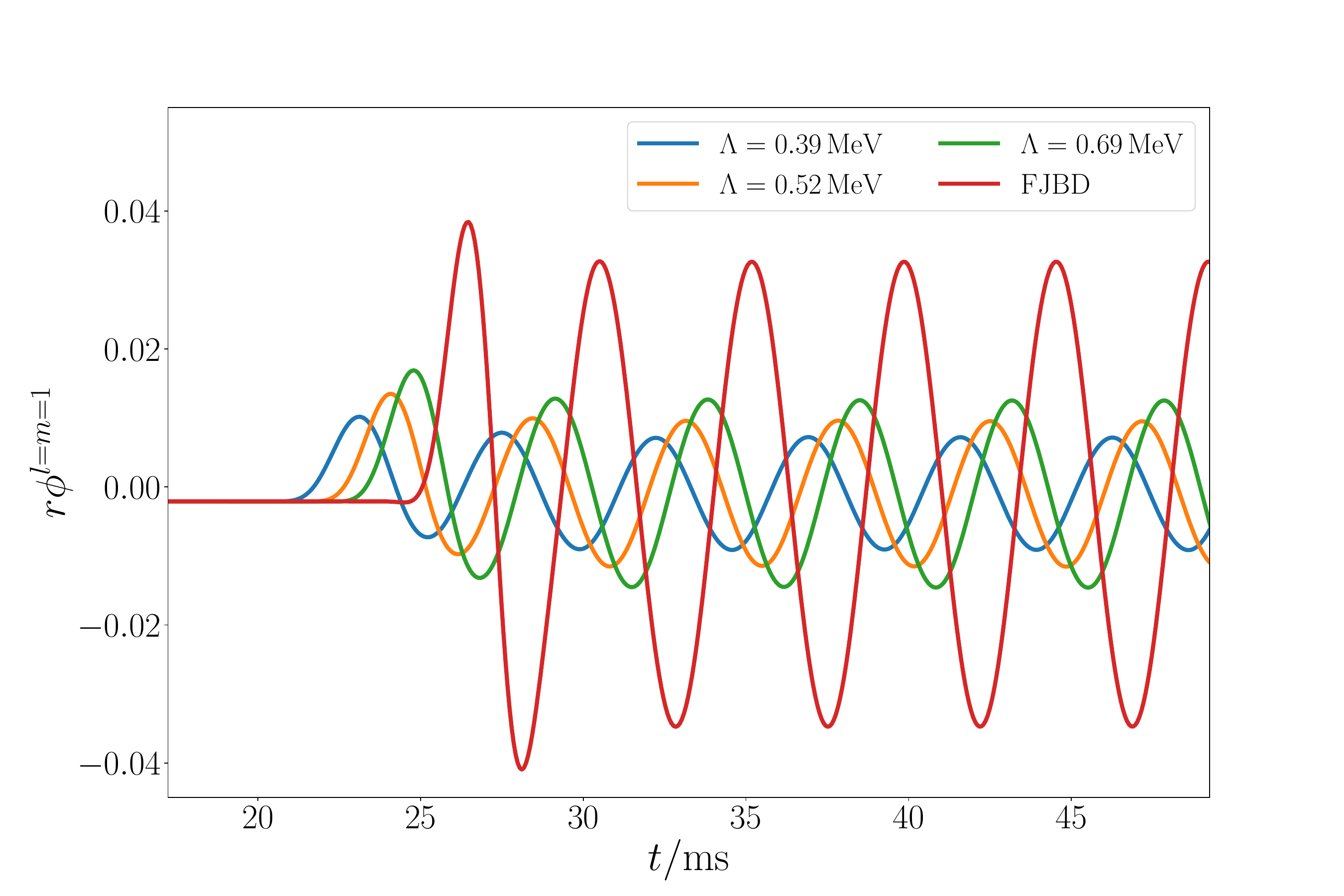}}
 \end{subfigure}
 \begin{subfigure}{0.5\textwidth}
\centerline{ \includegraphics[width=\textwidth,trim=10 0 80 80]{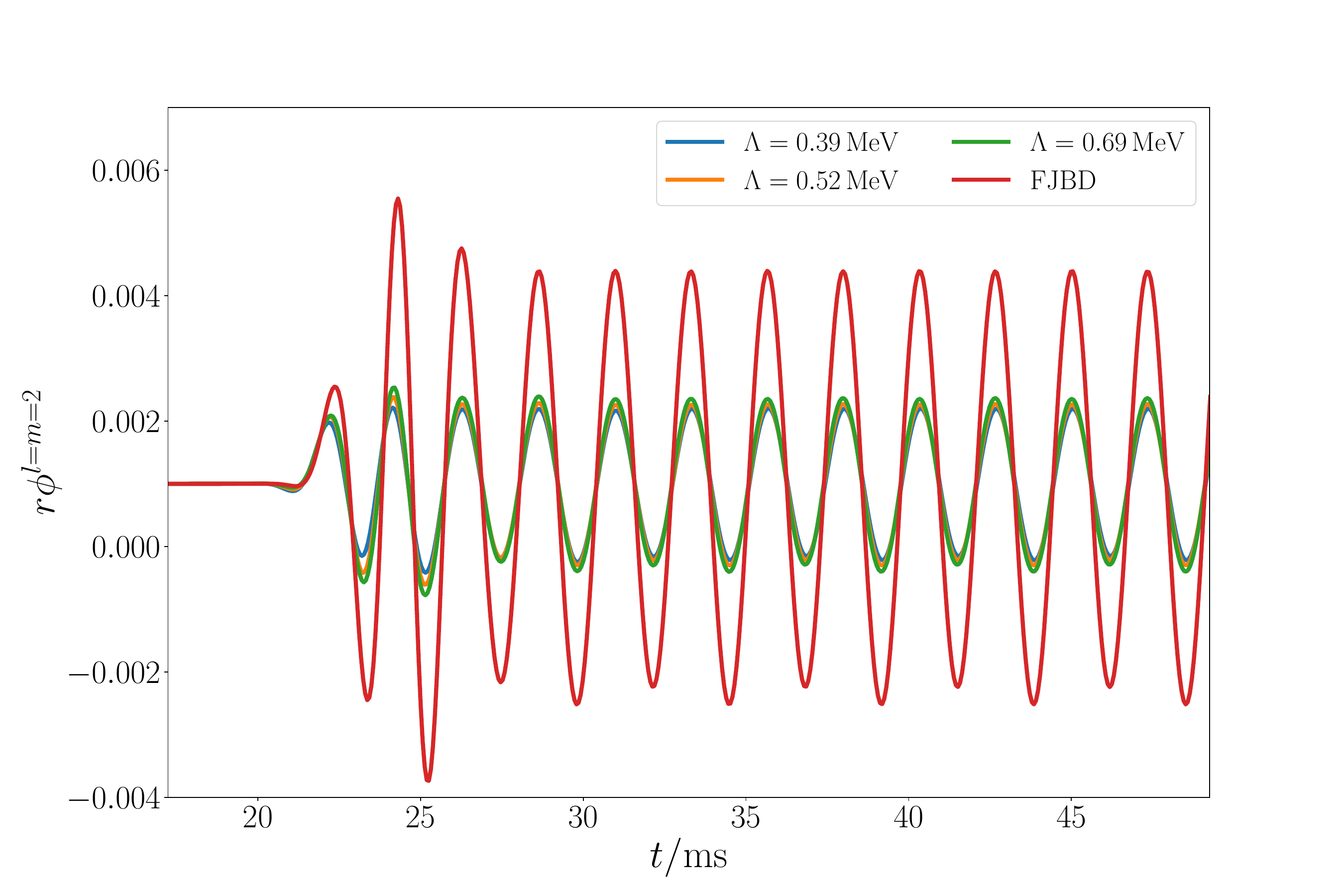}}
 \end{subfigure}
 \caption{Numerical waveforms for the dipole $l = m = 1$ mode (left) and the quadrupole $l = m = 2$ mode (right) of the outgoing scalar radiation emitted by a BH-NS system in $K$-essence theory, for different values of the strong coupling scale $\Lambda$ and extracted at a radius of r = 7383km. The comparison to Fierz-Jordan-Branz-Dicke (FJBD) unscreened scalar-tensor theory have been included for comparison. The first peak (after the transient phase) of the quadrupole waveforms have been aligned for ease of comparison. See ref.~\cite{Cayuso:2024ppe} for details.}
 \label{fig:dipkessence_and_quadkessence}
 \end{figure}

\begin{figure}
\begin{subfigure}{0.5\textwidth}
\centerline{ \includegraphics[width=0.92\textwidth]{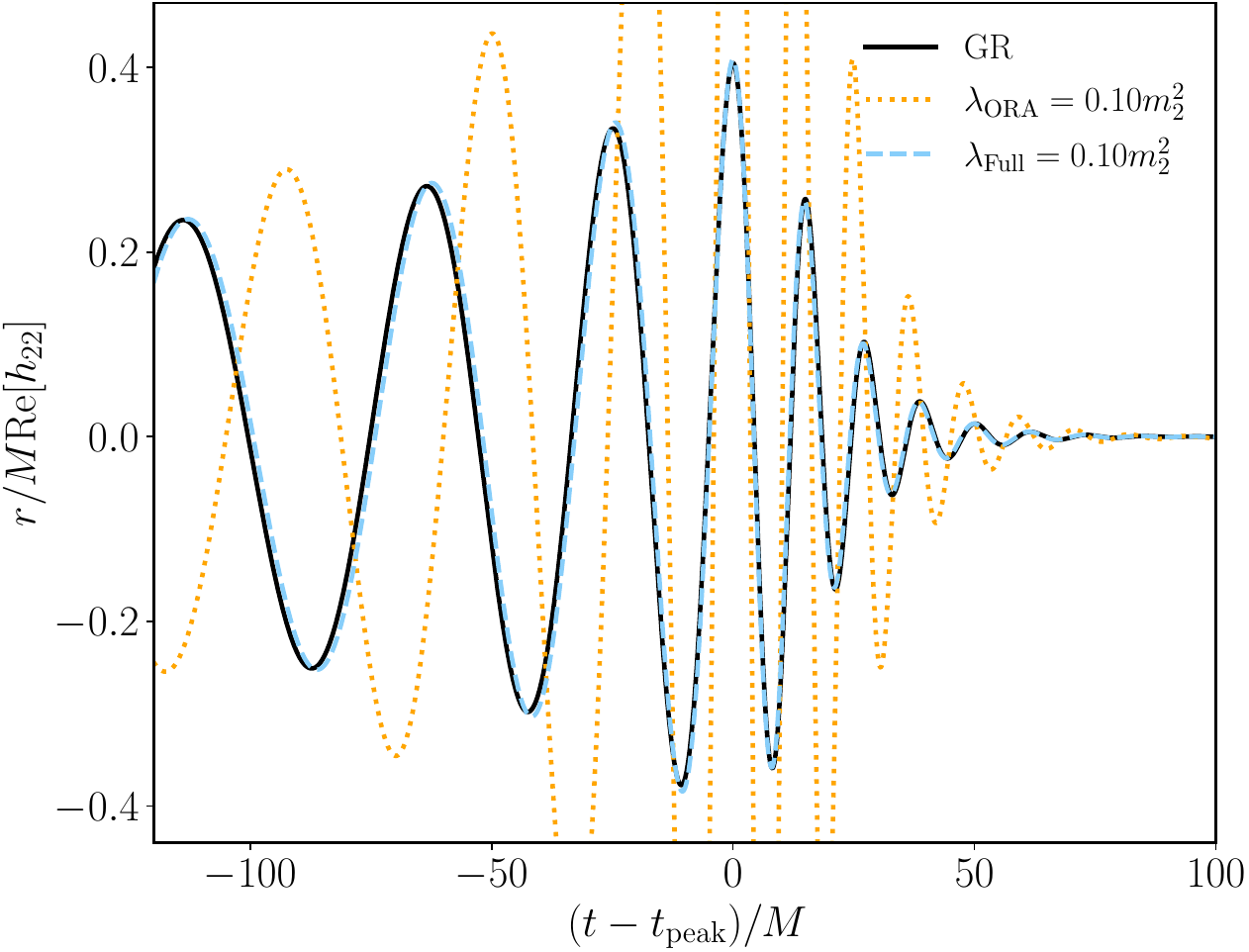}}
 \end{subfigure}
 \begin{subfigure}{0.5\textwidth}
\centerline{ \includegraphics[width=0.92\textwidth]{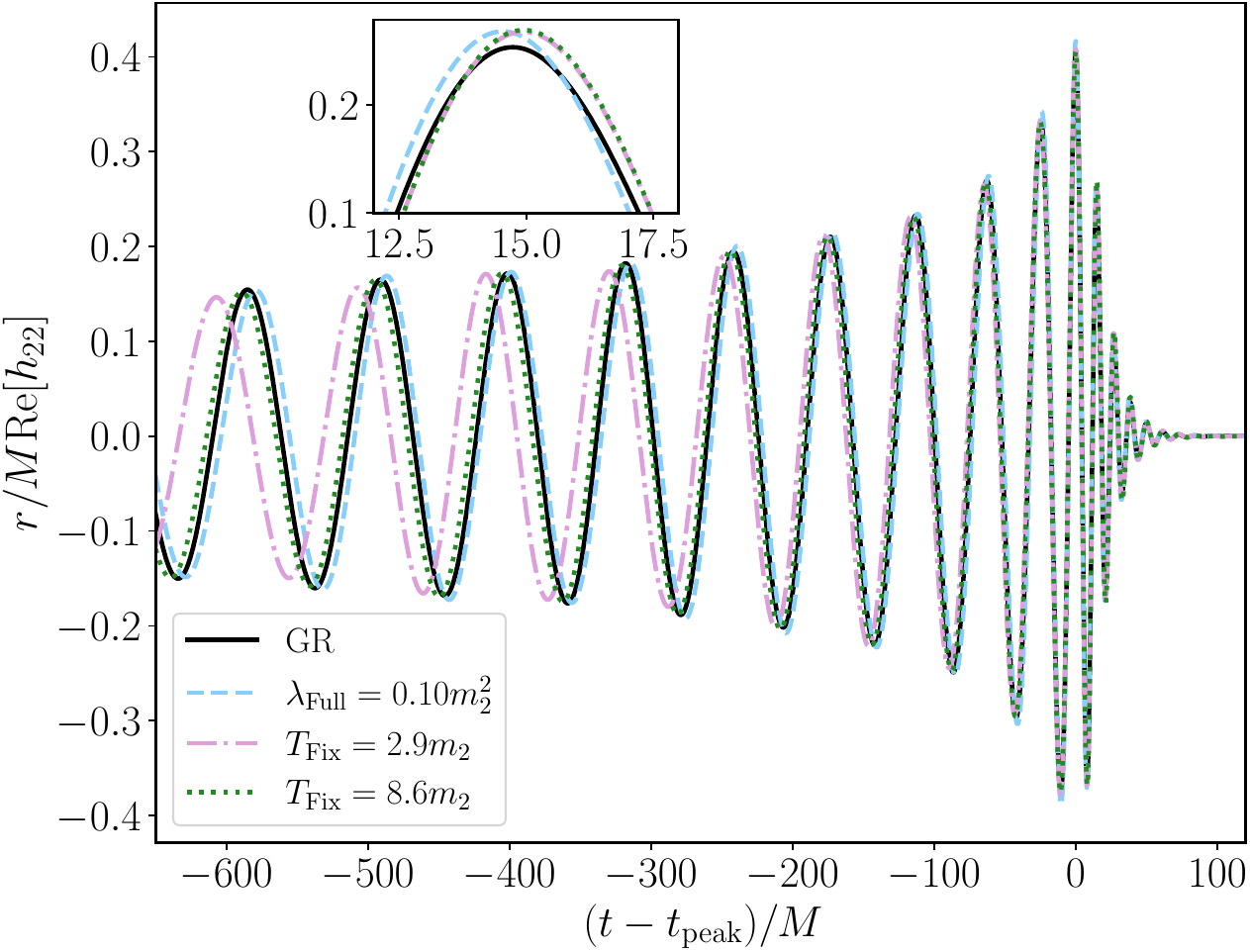}}
 \end{subfigure}
 \caption{Real part of the $l = m = 2$ spherical harmonic of the GW strain obtained in the order-by-order (left) and fixing-the-equations (right) approaches, and compared when solving the full equations for EsGB gravity. The GR waveform is shown for comparison and the waveforms are aligned in phase and time at peak amplitude. See ref.~\cite{Corman:2024cdr} for details.}
 \label{fig:NRORAandfixvsfull}
 \end{figure}

The second group are higher derivative theories of gravity, including the Horndeski class, EsGB and dCS gravity and EFTs of GR.
%
Because of the higher-order derivatives, the bottleneck to performing numerical simulations is the theories' formulation as a well-posed initial value problem.
A straightforward approach is an order-by-order formulation used in the first
simulations of dCS gravity~\cite{Okounkova:2017yby} and EsGB
gravity~\cite{Witek:2018dmd}.
This approach provides insight into the dynamics of the scalar field, and was
used to estimate the GW phase shift or to identify new
nonlinear phenomena like dynamical
(de-)scalarization in black-hole binaries~\cite{Silva:2020omi,Elley:2022ept}.
However, at second order in the expansion, this method suffers from secular
effects~\cite{Okounkova:2019dfo,Okounkova:2020rqw}.
A new formulation using modified general harmonic
coordinates~\cite{Kovacs:2019jqj,Kovacs:2020pns,Kovacs:2020ywu} allowed the
first full simulations of binary black holes, and the computation of the
gravitational radiation, in EsGB
gravity~\cite{East:2020hgw,East:2021bqk,Corman:2022xqg,AresteSalo:2023mmd}. The
third approach is to use a regularization
scheme~\cite{Cayuso:2017iqc,Allwright:2018rut} inspired by the Israel-Stewart
formulation of relativistic viscous hydrodynamics.
This approach has been applied to simulate compact binaries in Horndeski
gravity~\cite{Lara:2021piy,Figueras:2021abd}, in EFTs of
GR~\cite{Cayuso:2023xbc}, as well as in scalar-tensor theories with kinetic screening ($K$-essence)~\cite{Cayuso:2024ppe}, see figure~\ref{fig:dipkessence_and_quadkessence}  for an example of screened scalar field  scalar in a waveform of a BH-NS merger in a quasi-circular inspiral.
Simulations in quadratic gravity were done in ref.~\cite{Held:2023aap}.
A systematic and fully self-consistent comparison of the three different numerical approaches was performed in~\cite{Corman:2024cdr}. Considering quasi-circular binary black hole mergers in shift-symmetric EsGB gravity, they were able to quantify the errors introduced when using the approximate approaches. See figure~\ref{fig:NRORAandfixvsfull} for an illustration of the differences in the GW signal of merging binary black holes obtained with these approaches.

\subparagraph{Post merger.}
%
%
Considerable effort has been made to determine the QNM
spectrum of compact objects in alternative theories of gravity.
This is a technically challenging problem, due to the complexity of the field
equations of alternative gravity theories, the lack of closed form analytical
black-hole spacetimes (and their algebraic symmetries), and the numerical difficulties in solving the
boundary-value problem for the QNM determination, which often involves
coupled systems of equations.
For these reasons, most of the literature has focused on nonrotating (or
slowly-rotating) black holes. Yet, these calculations have taught us a few
lessons.
First, modifications to GR generically cause the QNM spectra of axial and polar
QNMs, which are identical in GR, to become different (``isospectrality
breakdown'').
Second, each extra field introduced by the alternative to GR, will have its own
QNM spectrum.

The most extensive work has been performed in EsGB theories, e.g., in
refs.~\cite{Blazquez-Salcedo:2016enn,Blazquez-Salcedo:2017txk,Blazquez-Salcedo:2020caw,Blazquez-Salcedo:2020rhf,Staykov:2021dcj}.
%
The quasinormal modes of BHs in EsGB theory have also been considered for
rotating solutions at linear~\cite{Pierini:2021jxd} or
quadratic~\cite{Pierini:2022eim} order in the expansion of small spin of the
black hole (see also ref.~\cite{Evstafyeva:2022rve}).
Similar calculations have been made in dCS~\cite{Wagle:2021tam,Srivastava:2021imr},
Horndeski theories~\cite{Tattersall:2019nmh}, and EFTs of GR~\cite{deRham:2020ejn,Cano:2021myl}.


All previous results fall short of being directly applicable to perform
spectroscopy of black-hole remnants of binary coalescences which have
dimensionless spin values  $\simeq 0.7$.
This has motivated the pursuit of a generalization of Teukolsky's formalism
(which in GR is used, e.g., to determine the QNMs of Kerr black holes at arbitrary spin) to beyond-GR
theories. This avenue was explored in refs.~\cite{Hussain:2022ins,Li:2022pcy,Cano:2023tmv}.
See also refs.~\cite{Li:2023ulk,Wagle:2023fwl}.
First calculations of QNMs of rapidly rotating BHs in EFTs of GR were obtained in ref.~\cite{Cano:2023jbk}.
A novel approach based on perturbative spectral expansions \cite{Chung:2023wkd, Chung:2023zdq,Blazquez-Salcedo:2023hwg} has been applied in refs. \cite{Chung:2024ira, Chung:2024vaf} to calculate QNM frequencies of rapidly rotating BHs in EsGB theories.

\subparagraph{Effective-one-body and phenomenological waveform models.}
In order to accurately describe the binary dynamics and gravitational wave
emission up to merger, i.e., in the strong-field regime, the construction of
semi-analytical approximants becomes necessary.
As in GR, it is possible to increase the reliability of post-Newtonian results
by mapping them onto EOB-type of coordinates~\cite{Buonanno:1998gg,Buonanno:2000ef,Damour:2000we}.
The EOB framework imposes a physically motivated resummation of the
perturbative results and grants a better behavior for the inspiral evolution
up to the plunge.
For scalar-tensor and EsGB theories, the explicit form of this mapping was
computed up to 3PN
order~\cite{Julie:2017ucp,Julie:2017pkb,Jain:2022nxs,Julie:2022qux};
see also refs.~\cite{Khalil:2018aaj,Julie:2017rpw} for similar calculations in Einstein-Maxwell-dilaton theories.
The first full IMR gravitational waveform in EsGB theories was also computed within the EOB framework and applied to analyse GW events to place new constraints on the fundamental coupling of the theory~\cite{Julie:2024fwy}. 
The GW phase corrections induced by modification to GR computed
in post-Newtonian theory can be included on a baseline GR frequency-domain
waveform model, as we will see next.

\subparagraph{Parametrized waveform models.}
%
Gravity tests in the weak-field, slow-motion regime can be conveniently performed in the framework of the parametrized post-Newtonian (ppN) formalism~\cite{Will:2014kxa,Shao:2016ezh}.
Likewise, in the strong-field regime for compact binary coalescences, a
parametrized approach was developed in the construction of gravitational
waveforms for beyond-GR effects~\cite{Yunes:2009ke,Cornish:2011ys},
in a theory-agnostic manner.
%
The idea behind parametrized post-Einsteinian (ppE) framework
is to allow for modifications to the frequency-domain gravitational
wave amplitude and phase predicted by GR.
The type and magnitude of these modifications are controlled by ppE parameters
which, when set to zero, return back the GR waveform. One can then let data determine
the most likely values of this parameter, i.e., perform a theory-agnostic inference.
The posteriors on the ppE parameters can then be translated into predictions of specific
theories. The mapping between theory-agnostic and theory-specific modifications to GR
have been worked out mostly for the inspiral part, where post-Newtonian calculation can
be used; see refs.~\cite{Tahura:2018zuq,Berti:2018cxi} for such mappings.
For this reason, although not strictly speaking a limitation of the formalism,
most application of the ppE model have used the \emph{inspiral} part of the
waveform. The ppE has also been extended to include additional gravitational-wave
polarizations~\cite{Chatziioannou:2012rf}, spin-precessing binaries~\cite{Loutrel:2022xok}
and parity-violation effects in the gravitational wave propagation, such as velocity birefringence~\cite{Jenks:2023pmk}.
Similar approaches have been developed in refs.~\cite{Li:2011cg,Mehta:2022pcn} and
are in use by the LVK Collaboration.
Bounds on generic deviations to the GW phase evolution in GR at different PN
orders have been obtained from the GW transient
catalogs~\cite{LIGOScientific:2019fpa,LIGOScientific:2020tif,LIGOScientific:2021sio}.

More recently, a parametrized framework around the \emph{merger and ringdown}
time of black-hole coalescences was developed within the EOB framework in
ref.~\cite{Maggio:2022hre}, building on
refs.~\cite{Brito:2018rfr,Ghosh:2021mrv}; see section~\ref{section:div1} for other ringdown tests
of GR. In parallel, ref.~\cite{Silva:2022srr} developed an EOB waveform model in
which the QNMs are parametrized according to the formalism ParSpec of ref.~\cite{Maselli:2019mjd}. ParSpec is a double expansion in the theory coupling and in the BH spin parameter such that when the coupling goes to zero, one recovers GR, \textit{i.e.} the QNMs for the Kerr solution.
While this approach is built to map theoretical calculations beyond GR, as was done in ref.~\cite{Maselli:2023khq} for EsGB, dCS and some EFT theories of gravity, it can also be used agnostically to find the constraints at each relevant order in the spin~\cite{Maselli:2019mjd,Maselli:2023khq}.
Note that ref.~\cite{Silva:2022srr} also mapped QNM calculations in specific theories (described earlier) to the free parameters of this parametrization.

\vspace{2mm}\noindent 
{\bf Main Open Challenges} 
\vspace{2mm}

\noindent
Below are some important questions to be addressed in the future regarding the construction of beyond GR waveforms in preparation of next-generation gravitational detectors. See also section~\ref{section:div1}, where some of these points are discussed in more details.
\begin{itemize}
\item[--] Will higher-order PN terms affect the constraints derived at leading PN orders on specific gravity theories?
%
%
\item[--] To what extend a more accurate modeling of waveforms beyond GR can help us to constraint the plethora of new physics introduced by modified gravity? Examples of such effects are: extra dimensions, variation of fundamental constants (e.g. Newton's one), modified dispersion relations, violation of Lorentz invariance, etc.
\end{itemize}

\vspace{2mm}\noindent
{\bf Important Next Steps}

\subparagraph{Post-Newtonian theory.}
In the near fututre, it will be crucial to extend current calculations by exploring the effects of spin (including
spin-precession), non-Kerr multipolar structure, orbital eccentricity, and
tidal interactions in the inspiral of beyond-GR theories.

\subparagraph{Numerical relativity.}
So far the first, proof-of-principle calculations of GWs  in
alternative theories of gravity have been attained~\cite{Palenzuela:2013hsa,Shibata:2013pra,Taniguchi:2014fqa,Ponce:2014hha,Hirschmann:2017psw,East:2020hgw,East:2021bqk,Corman:2022xqg,AresteSalo:2023mmd,Lara:2021piy,Figueras:2021abd,Cayuso:2023xbc}. However, more substantial
work is required to chart out the parameter space and to improve the quality of
the waveforms (resolution, eccentricity reduction, etc) that will enable new,
theory-specific tests of gravity.
One can also explore other alternatives to GR, e.g., Lorentz-violating theories,
for which no simulations of compact binaries exist yet. To progress in this direction, several preliminary studies have already been performed.
Thus, a well-posed initial value formulation of Einstein-Aether gravity in 3+1
dimensions was presented in ref.~\cite{Sarbach:2019yso}.
Spherically symmetric scalar field collapse in Einstein-Aether gravity was studied in ref.~\cite{Garfinkle:2007bk}.
A well-posed time evolution formulation of bi-gravity was derived in~\cite{Torsello:2019tgc}.
Simulations of black holes in massive gravity were performed in spherical symmetry~\cite{Kocic:2020pnm,deRham:2023ngf}.

\subparagraph{Post merger.}
It is important to extend the calculation of QNM frequencies, as well as the understanding of their excitation, for rapidly-rotating black holes beyond-GR. Another important point to be studied is to understand how the loss of isospectrality affects the observable ringdown signal.
Also, being able to accurately estimate how much the remnant mass and spin of a BBH coalescence deviate from their GR predictions in beyond-GR theories is crucial.

\subparagraph{Gravitational self-force (GSF).}
In GR, as seen in section~\ref{sec:GSF}, self-force results can be used to improve
our modeling of comparable mass binaries. It is natural to expect that the same
will occur for alternative theories of gravity.
It would also be interesting to develop self-force calculations in
modified gravity theories, see, \textit{e.g.}, refs~\cite{Zimmerman:2015hua,Spiers:2023cva}.

\subparagraph{IMR models.}

Other important questions to be addressed in order to perform meaning-full tests of GR with ET are:
\begin{itemize}
    \item[--] How to ensure that the waveform systematics do not affect the tests of GR?
    \item[--] Could we develop a technique to perform tests of GR with NRsurrogate models?
\end{itemize}

\subsection{Waveform Acceleration Techniques}
\label{sec:acc}


The large expected event rate for 3G detectors  (see section~\ref{section:div9}) drives the need for quicker data analysis techniques including searches and parameter estimation for compact binaries in order to optimise the use of computing resources and to deliver time-critical information such as accurate sky maps or source classification
for follow-up by the scientific community.
The cost of Bayesian inference is typically dominated by the generation of waveforms, and commonly a two-pronged approach is chosen to reduce the computational expense. 

The first class of approaches aims to increase the efficiency of posterior exploration.  This can be achieved through novel techniques to speed up the likelihood evaluation, for example by using heterodyning/relative binning~\cite{Cornish:2010kf, Zackay:2018qdy, Cornish:2021lje, Leslie:2021ssu}, reduced-order quadratures (ROQs)~\cite{Antil:2012wf, Canizares:2013ywa, Smith:2016qas} or multibanding~\cite{Vinciguerra:2017ngf,Morisaki:2021ngj}. One can also opt for novel sampling techniques such as Hamiltonian Monte Carlo that forego the random-walk process of stochastic samplers~\cite{Bouffanais:2018hoz}.  Hardware-based acceleration provides another means to reducing the computational cost.  A final alternative lies in likelihood-free sampling using machine learning (ML) techniques~\cite{Green:2020dnx,Dax:2021tsq,Dax:2022pxd}.


The second class of approaches aims to speed up individual waveform evaluations. This is the approach we will be focusing on here and will be discussed in more detail below. The most commonly used waveform acceleration techniques for IMR models are reduced-order modeling~\cite{Field:2011mf, Field:2013cfa}, multibanding/adaptive grids~\cite{Garcia-Quiros:2020qlt} and waveform decomposition techniques such as singular value decomposition (SVD)~\cite{Cannon:2011rj}. For inspiral-only models, the stationary phase approximation (SPA)~\cite{Droz:1999qx} and Shifted Uniform Asymptotics (SUA)~\cite{Klein:2014bua} are common tools to obtain fast frequency-domain approximants. In recent years, the use of ML alone or in combination with these approaches has gained prominence.

Both types of approaches can be combined to achieve additional savings.

\vspace{2mm}\noindent
{\bf State-of-the-Art}

\subparagraph{Reduced-order models.}

Reduced-order models (ROMs) or surrogates rely on representing waveform models in terms of a small set of basis components. We refer the reader to~\cite{Tiglio:2021ysj} for a review. The earliest work on ROMs~\cite{Field:2011mf} was motivated by the observation that the size of the basis needed to represent a typical waveform model was smaller than the number of time samples in a typical observation and the fact that the representation error of such a basis shows exponential convergence.
ROMs require determination of the basis coefficients as a function of the waveform parameters. This can be done by building an interpolant to obtain the coefficients of that expansion as a function of the waveform parameters~\cite{Cannon:2011rj}, or through empirical interpolation~\cite{Field:2011mf}, or through Gaussian process interpolation~\cite{Doctor:2017csx,Lackey:2018zvw}. For expensive waveform models computational savings can be obtained even when the size of the basis is large, because the typical cost of evaluating the interpolants is much lower than that of calculating the original waveforms. Typical ROMs are between one and three orders of magnitude faster to evaluate and, for waveform models that rely on ODE integration such as EOB models, often provide smoother likelihoods than the original models, helping the sampling process.

The reduced bases on which ROMs are based can be constructed using greedy algorithms~\cite{Field:2011mf}
or by performing SVD decompositions of a set of training waveforms~\cite{Smith:2012du,Cannon:2012gq,Cannon:2011rj}. For simple models, the reduced basis can be built directly with time or frequency domain waveforms. More complex waveforms are decomposed into harmonic modes and a separate reduced basis, with corresponding parameter space interpolant, is constructed for the Fourier domain amplitude and phase of each mode~\cite{Purrer:2014fza,Purrer:2015tud}.

ROMs have been constructed for non-spinning black hole waveform models, e.g., EOBNR \cite{Field:2013cfa}, for SEOBNR aligned-spin black hole models without higher modes~\cite{Purrer:2014fza,Purrer:2015tud,Bohe:2016gbl,Pompili:2023tna}
and for BNS models~\cite{Canizares:2014fya,Lackey:2016krb, Lackey:2018zvw}. 
The first ROMs for higher-mode models without precession were described in~\cite{Marsat:2020rtl,Cotesta:2020qhw} and the first surrogates for a model including both higher modes and precession, SEOBNRv4PHM, were obtained in~\cite{Gadre:2022sed, Thomas:2022rmc}. Recently, a ROM for stellar origin BBH mergers observed with LISA was also proposed that uses a wavelet basis to construct waveforms in the time-frequency domain~\cite{Digman:2022igm}. 

The ROMs described above are based on semi-analytic models and build a reduced basis from a very large set of training waveforms. A second family of ROMs, usually referred to as NR surrogate models, are built in the same way but use NR simulations to construct the reduced basis. Waveform models exclusively based on NR simulations are generally regarded as more accurate than semi-analytic models but they are usually restricted in parameter range due to the limited length of NR simulations. The first such surrogate was built using non-spinning BBH simulations~\cite{Blackman:2015pia} and subsequently extended to aligned spin~\cite{Varma:2018mmi} and then systems with precession~\cite{Blackman:2017dfb, Blackman:2017pcm}. The latter model was the first ROM of any kind for a precessing waveform model. More recent surrogate models have targeted extensions to higher mass ratios~\cite{Varma:2019csw,Rifat:2019ltp} and eccentricity~\cite{Islam:2021mha}. Surrogates have also been built to analyse specific observed events, for example GW190814~\cite{Yoo:2022erv}, by targeting the simulations used to build the surrogate to a narrow region of parameter space.

\textsc{SEOBNRv4PHM\_surr}~\cite{Gadre:2022sed} and the NRSur7dq4 surrogate~\cite{Varma:2019csw} represent the current state of the art in reduced order models. The latter is valid for systems with mass ratio $q \lesssim 4$, arbitrary spin directions and spin magnitudes $|\chi_i| \lesssim 0.8$, and can generate the last $\sim 20$ orbits of the system before final plunge, while the former has a broader parameter space in mass ratio ($q\leq 20$) with the same spin magnitude limits. Work is currently ongoing to extend the NR surrogate model to mass ratios of $q \lesssim 8$.

\subparagraph{Machine Learning.}

Thanks to reduced order modeling, we can represent waveforms in a lower-dimensional space that still contains enough information to perform inference. 
This representation will have some functional dependence on the waveform parameters, which can be reconstructed by a Machine Learning (ML) architecture in a supervised learning setting.
Such framework describes a good fraction of the work on ML applied to waveform modeling \cite{Chua:2018woh,Setyawati:2019xzw,Lee:2021isa,Liao:2021vec,Thomas:2022rmc,Tissino:2022thn,Huerta:2017kez,Lackey:2018zvw,Nousi:2021arn,Fragkouli:2022lpt,Khan:2021czv,Chua:2020stf}.
The distinction between simple fits and ``Machine Learning'' is not well defined, and is typically heuristically made based on the complexity of the architecture that performs the fit. 

Efforts have mostly focused on modeling ET/LVK-band BH binaries in the time domain, without spin \cite{Liao:2021vec,Nousi:2021arn,Barsotti:2021wks}, with only aligned spin \cite{Setyawati:2019xzw,Khan:2020fso,Lee:2021isa,Schmidt:2020yuu,Fragkouli:2022lpt,Khan:2021czv}, with precessing spin \cite{Setyawati:2019xzw,Thomas:2022rmc}, and with eccentricity but no spin \cite{Huerta:2017kez}.
Comparatively little effort has gone into applying a similar procedure to BNS \cite{Lackey:2016krb,Lackey:2018zvw,Tissino:2022thn} and to exotic objects such as Proca stars \cite{Freitas:2022xvg}.

Some effort has also gone into modeling low-frequency sources for LISA \cite{Chua:2018woh,Chua:2020stf}. While these are typically qualitatively different from the ones in the ET band, the techniques employed for LISA may be beneficially applied to ET sources.

A variety of ML models have been applied: Gaussian Process Regression \cite{Huerta:2017kez,Lackey:2016krb,Lackey:2018zvw}, shallow perceptrons \cite{Khan:2020fso,Thomas:2022rmc,Fragkouli:2022lpt,Keith:2021arn,Tissino:2022thn}, deep neural networks \cite{Chua:2018woh,Lee:2021isa,Chua:2020stf}, autoencoders \cite{Nousi:2021arn,Liao:2021vec}, text-processing inspired transformers \cite{Khan:2021czv}, generative adversarial networks \cite{Freitas:2022xvg}, residual networks \cite{Fragkouli:2022lpt}.
Some effort has also gone into comparing different ML architectures and approaches for a specific task \cite{Setyawati:2019xzw,Barsotti:2021wks}.
Most of these architectures, have the advantage of being intrinsically differentiable; this may allow for the direct computation of waveform gradients, which could provide precious information for downstream applications.

It is difficult to compare performances of different models, since the metrics and hardware used may vary among different works.
Acceleration efforts tend to focus on speed-up at moderate fidelity (mismatches around $10^{-2}$ to $10^{-4}$), with typical timing results on the order of single milliseconds per waveform on a CPU \cite{Tissino:2022thn,Thomas:2022rmc,Setyawati:2019xzw}, whereas sub-millisecond performance can be achieved thanks to GPU acceleration \cite{Thomas:2022rmc}.

\vspace{2mm}\noindent
{\bf Main Open Challenges}
\vspace{2mm}

\noindent
Despite the recent burst of interest on waveform acceleration techniques, the current fast waveform surrogates are unlikely to be directly usable in the context of ET and they will need to be updated.
First of all, they will need to meet a strict accuracy requirement, see section~\ref{sec:accuracy}; so far, most of the surrogates available (especially the ML ones) achieve a faithfulness with the training model which is suitable only for current detectors. As we move towards ET, the accuracy of most of the fast surrogates will need to be improved. Since accuracy is often at trade with speed, the conciliation of the two can be challenging.

Moreover, ROMs/surrogate models for ET will need to be able to reach significantly lower waveform starting frequency, which amounts to longer waveforms in the time domain,
{\it and} will need to incorporate realistic physics.
On one hand, pushing the range of validity of a surrogate to lower frequencies may be relatively straightforward, as the binary dynamics asymptotically approach the Newtonian limit. 
On the other hand, incorporating a large number of physical effects into a single surrogate will be a challenge.
So far, surrogates mostly tackled a single physical effect at a time (e.g. precession, eccentricity, tides...). Assembling a more realistic surrogate requires to deal with a larger variance in the training waveforms, meaning that current methods' architectures may prove inadequate.
A ROM will probably employ a large number of basis functions, requiring excessive memory and computational time. A ML model might be able to represent the features of the dataset more efficiently, but very little work has been done to build such model and a change over the state-of-the-art will be probably needed.
It is also worth noting that the increase in the model complexity must not affect the runtime, which is often challenging to achieve.

Studies have shown  that the evaluation of a surrogate on a large time/frequency grid constitutes a large fraction of the overall cost~\cite{Schmidt:2020yuu, Tissino:2022thn}. It is very hard to optimize such cost, unless we depart from the need of a large grid altogether. To this purpose, an effective ET waveform should be coupled with a scheme (such as ROQ or relative binning) to reduce the number of grid points needed for a given data analysis application.

Finally, we note that an effective model should be able to generate a large batch of waveforms in parallel, in order to fully exploit the power of modern computing hardware. While this has been already explored \cite{Khan:2020fso} and might be trivial in modern ML code frameworks, an efficient parallelization scheme remains a challenge that any developer should keep in mind.

\vspace{2mm}\noindent
{\bf Important Next Steps}
\vspace{2mm}

\noindent
While ROMs have proven to be reliable, ML techniques are promising candidates to build the next generation of surrogates, due to their flexibility and ubiquity.
While a rich literature has been developed in the field, to our knowledge no ML surrogate has been developed through production stage (i.e.\ used to make a novel scientific statement) in the current generation of GW observatories. Advancing such models to production quality will be a milestone in their developments, marking a necessary step towards a reliable surrogate for the ET era.

Recent interest in eccentricity has produced a number of models incorporating both the effects of eccentricity and spins \cite{Ramos-Buades:2021adz, Nagar:2021gss,Gamboa:2024hli}.  A spinning-eccentric surrogate has not been developed yet and would mark an important step towards a more physics-complete surrogate model, possibly leading to the development of new ML techniques.

While the use of adaptive rather than uniform grids yields further speed-up for already fast frequency-domain models~\cite{Garcia-Quiros:2020qlt}, this technique is equally applicable to time-domain models such as the \textsc{PhenomT} family and worth exploring for phenomenology-rich models where ROMs/surrogates may incur additional accuracy losses.

Finally, closed-form approximants such as the \textsc{Phenom} family and ML-based surrogates models are particularly amenable to (additional) hardware-based acceleration e.g., through the use of GPUs and parallelisation of the frequency or time array. 

Besides the effort of producing more complete and accurate approximants, it would be beneficial to set a common set of requirements for fast ET waveform approximants, stated in terms of parameter space coverage, generation time and faithfulness. These will help to establish a common target for the developers to reach.

\subsection{Executive summary}
As highlighted throughout this and other sections, gravitational-wave observations with ET will provide unprecedented insights into fundamental physics and astrophysics, including revealing the equation of state of nuclear matter and firmly determining the mass functions and merger rates of black holes and neutron stars throughout the Universe. Delivering many of these key science goals requires highly accurate waveform models of the emitted gravitational-wave signal. Here, we summarize the main points discussed in this section which encompasses the current state-of-the-art of techniques and models and avenues for achieving the accuracy and computational efficiency needed for ET.

\begin{highlightbox}{Waveform accuracy requirements}
\begin{itemize}
    \item Inaccurate waveform models may lead to errors in the
      estimated source parameters with subsequent erroneous
      conclusions about astrophysics, cosmology or fundamental
      physics.
      \item Inaccuracies can arise either through the omission of
        relevant physical effects or through insufficiently precise
        modeling.
      \item The high sensitivity of ET requires that waveform models
        must become significantly more accurate compared to today over broader regions of parameter space of sources.   By how much is under active
        investigation.

\end{itemize}

\end{highlightbox}

\begin{highlightbox}{Waveform models and sources}

\begin{itemize}
    \item Current waveform models for binary systems synthesize different approaches such as numerical relativity (NR), perturbation theory including post-Newtonian and post-Minkowskian expansions and gravitational self-force. Recently, techniques from high-energy physics such as the calculation of scattering amplitudes and effective-field theory have been utilized to derive higher-order perturbative corrections.
    \item The current state-of-the-art waveform models that capture the inspiral, merger and ringdown fall into two categories: NR-calibrated Effective-one-body (EOB) and phenomenological (Phenom) models. These models include higher-order modes and precession, with recent advancements to include orbital eccentricity, and tides.
    \item The main targeted sources for ET are binary systems consisting of black holes and neutron stars, for which models have been developed to high accuracy but further improvements are needed for ET. Possible progresses to be made for BBH require to produce longer NR simulations, that includes higher mass ratios and higher spin systems. For BNS and NSBH, the challenge consists in accounting for more physical effects, especially in the tidal sector and post-merger models. Producing longer duration waveforms will also be key as those signals are expected to be seen for minutes to hours.
    \item Other modeled sources involve hyperbolic encounters, triple systems, core-collapse supernovae and GWs from cosmological origin such as primordial BHs, exotic compact objects or cosmic strings. However, for most of these sources reliable waveform models are not yet readily available.
    \item Waveform models in alternative theories of gravity are also being developed, using both analytical and numerical techniques to describe all three stages of the coalescence: inspiral, merger and post-merger. However, this is still at its infancy and should be further developed to provide reliable waveforms beyond GR that span a large parameter space. Alternative tests of gravity could be performed by the use of parametrized waveform models.
\end{itemize}

\end{highlightbox}

\begin{highlightbox}{Computational efficiency}

\begin{itemize}
    \item ET is expected to observe tens of thousands of gravitational waves from modeled sources per year. Such high event rates require computationally efficient analysis algorithm including searches and parameter estimation. Bayesian inference, in particular, is dominated by the waveform generation cost.
    \item Commonly used techniques to increase the efficiency of waveform models include the development of phenomenological or surrogate models, the use of adaptive grids, singular value decomposition, shifted uniform asymptotics, dimensional reduction and advanced reduced-order modeling. 
    \item Recently, machine learning techniques including Gaussian process regression and neural networks have been applied and are becoming increasingly more prominent to tackle the efficiency problem. 

\end{itemize}

\end{highlightbox}

\section{Tools for assessing the scientific potentials of detector configurations}\label{section:div9}

The goal of this section is to provide an introduction to various software products developed within ET Observational Science Board, with the specific task of releasing tested and publicly available software that allows us to quickly evaluate the effect on the ET science output of possible variations of its design such as sensitivity curve,  geometry, or geographical location.  This allows assessing the scientific potentials of different detector configurations with standardized, official figures of merits (FOMs). 
In this section we will present these tools, 
and illustrate their applications to the production of such FOMs used to investigate the ET science case.
The packages presented here are also meant to provide consistent, tested and rapid tools for the community for first estimations and prompt forecasting.

The content is organized as follows. In section~\ref{sect:Basicdiv9} we introduce basic notions  for detection and parameter estimation: in particular, in section~\ref{sec:div9_basics_detection_pe} we discuss the   detector's response and signal parameters for compact binaries, while in section~\ref{sec:div9_FisherIntro} we discuss the Fisher information matrix formalism. Section~\ref{sec:div9_software:tools} describes software tools for compact binary coalescence (CBC) sources: in particular, the sky location-polarisation-inclination averaged signal-to-noise ratio (SNR) is discussed in section~\ref{sec:div9_skyloc_average_snr}, Fisher matrix pipelines are presented in section~\ref{sec:div9_fm_pipelines},  improvements of Fisher baseline models are discussed in section~\ref{sec:div9_fim_improvements},  and tools for performing inference at the level of astrophysical populations in section~\ref{sect:inferencepopdiv9}. In section~\ref{sec:div9_metrics} we have gathered FOMs related to pattern functions and Earth rotation in section~\ref{sec:div9_pattern_func_earth_rot}, the relation between the distance reach (horizon) and SNR in section~\ref{sec:div9_horizons_snr}, sky localisation and distance performance in section~\ref{sec:div9_skyloc}, and the inference of intrinsic parameters in section~\ref{sec:div9_infer_intrinsic_par_golden_binaries}.
In section~\ref{sec:div9_ringdown} 
we review the conceptual aspects and the implementation  of a Fisher information  code for black hole ringdown spectroscopy.

Section~\ref{sec:div9_stoch_searches} is devoted to GW stochastic background searches. General definitions for the characterization of stochastic backgrounds are collected in section~\ref{sect:CharStocBackdiv9}. In section~\ref{sec:div9_PLSdefinition} we discuss 
the power-law sensitivity curve and we present the tools for computing it for different ET configurations. The methodology and tools for the subtraction of astrophysical backgrounds in the search for cosmological backgroud are discussed in section~\ref{sect:subtractastrobkgdi9}.

Section~\ref{sec:div9_null_stream} describes a package for the use of the null stream in  the triangle configuration.  Section~\ref{sec:div9_conclusions} contains the conclusions, and is followed by the Executive summary.

\subsection{Basic formalism}\label{sect:Basicdiv9}

\subsubsection{Detection and parameter estimation of resolved signals }
\label{sec:div9_basics_detection_pe}

In this subsection we summarize the basic concepts behind 
the detectability and the parameter estimation of 
resolved GW signals,
providing  a minimal and self-consistent discussion of 
the methods used by the software tools discussed in the 
rest of the section.
We refer the reader 
to~\cite{Maggiore:2007ulw,Christensen:2022bxb} for 
exhaustive treatments on the waveform 
models and on the data analysis approaches 
currently used to infer the source properties 
from GW observations.

\paragraph{Matched filtering.}
\label{sec:div9_matched_filtering}

In the presence of a GW signal $h(t)$, the 
output of a detector, $s(t)$, 
can be written as
\begin{equation}\label{eq:s_from_h}
    s(t) =n(t) + h(t) \,,
\end{equation}
where $n(t)$ denotes the detector's noise, which we assume here to 
be Gaussian, stationary, and with zero mean. The statistical 
properties of the noise are then summarized by the one-sided 
power spectral density (PSD), defined as 
\begin{equation}\label{eq:PSD_definition}
    \langle \tilde{n}^*(f)\,\tilde{n}(f^\prime)\rangle = \dfrac{1}{2}\delta(f-f^\prime)\, S_n(f)\,,
\end{equation}
where a tilde denotes Fourier transform from time to 
frequency domain.

Partial knowledge of the form of the signal and of the noise 
properties allows in general to detect $h(t)$ even if its 
amplitude is smaller than $n(t)$.
To assess the signal detectability it is useful to introduce 
a \emph{filter} function $K(t)$, and to define the signal-to-noise 
ratio (${\rm SNR}$) for $s(t)$ as the 
ratio among the expectation value of the filtered detector 
output $\hat{s} = \int dt\,s(t)\,K(t)$ when a signal is 
present,
\begin{equation}
    S = \langle \hat{s}(t)\rangle = \int_{-\infty}^{\infty} dt\, h(t)\, K(t) = \int_{-\infty}^{\infty} df\, \tilde{h}(f)\, \tilde{K}^\ast(f)\,,
\end{equation}
and the root mean square of the filtered output in absence of 
signal:
\begin{equation}
    N^2 = \langle \hat{s}^2(t)\rangle_{h=0} = \dfrac{1}{2}\int_{-\infty}^{\infty} df\, S_n(f) |\tilde{K}(f)|^2\,.
\end{equation}
The optimal filter, or matched filter, is then defined as the one 
which maximizes ${\rm SNR} = S/N$ for a given $h(t)$. This can be 
obtained by defining the scalar product on the waveform 
space,
\begin{equation}\label{eq:innerprod_def}
    (g\,|\,h) = 4\, {\rm Re} \int_0^{\infty}df\,  \dfrac{\tilde{g}^*(f)\,\tilde{h}(f)}{S_n(f)}  \ ,
\end{equation}
such that 
\begin{equation}\label{eq:SN_def_innerprod}
    \dfrac{S}{N} = \dfrac{(u|h)}{(u|u)^{1/2}}\, , 
\end{equation}
where $u(t)$ is a function whose Fourier transform is given by 
\begin{equation}
\tilde{u}(f) = \dfrac{1}{2} S_n(f) \tilde{K}(f)  \, .
\end{equation}
The maximum is then found choosing $\tilde{K}(f) = {\rm const}\,\times \tilde{h}(f) / S_n(f)$, 
which defines the optimal matched filter. By substituting this expression 
in \eq{eq:SN_def_innerprod} we obtain the expression for the 
optimal SNR,
\begin{equation}
    {\rm SNR} = (h\,|\,h)^{1/2} = 4 \int_0^{\infty}df\,  \dfrac{|\tilde{h}(f)|^2}{S_n(f)} \, .
\label{eqn:snr}
\end{equation}
When forecasting the capabilities of a given 
GW detector it is customary to use \eq{eqn:snr} as a 
FOM to assess the detectability of 
source. A source is considered detectable if its
optimal SNR exceeds a certain detection threshold.

\paragraph{Parameter estimation.}
\label{sec:div9_pe}

Once the presence of a signal in a segment of 
the datastream has been assessed, the next step is the reconstruction 
of the source parameters $\vb*{\theta}$. 
%
We adopt a Bayesian perspective, where the information on $\vb*{\theta}$ is fully encoded in the posterior distribution $p(\vb*{\theta}|s)$, which can be expressed as
\begin{equation}\label{eq:posterior}
	p(\vb*{\theta}|s) = \dfrac{{\cal L}(s|\vb*{\theta})\,\pi(\vb*{\theta})}{{\cal Z}(s)}\,,
\end{equation}
where ${\cal L}(s|\vb*{\theta})$ represents the likelihood 
function of $\vb*{\theta}$ given the data, $\pi(\vb*{\theta})$ 
is the prior distribution on the parameters, and 
the normalization ${\cal Z}(s)$ corresponds to the 
evidence
\begin{equation}
	{\cal Z}(s) = \int d\vb*{\theta}\, {\cal L}(s|\vb*{\theta})\,\pi(\vb*{\theta})\,.
\end{equation}
The prior can include physically motivated bounds (e.g. positive masses 
and distances) and all pre-existing information on the 
source parameters. For CBCs, for example, 
a common choice for the luminosity distance is given 
by assuming a uniform prior in the coming volume, to take into account 
the effect of the Universe expansion at larger distances.

The likelihood function ${\cal L}(s|\vb*{\theta})$ depends on the 
statistical properties of the noise and on the specific 
class of sources targeted. 
For resolved sources, and from the assumption of stationarity 
and gaussianity of the noise \eqref{eq:PSD_definition}, 
the probability of having a specific realization of the noise 
$n$ in the frequency domain is given by
\begin{equation}
	p(n) \propto \exp\left\{ -\dfrac{1}{2} \int df \dfrac{|\tilde{n}(f)|^2}{S_n(f)} \right\} = {\rm exp} \left\{-\dfrac{1}{2} (n|n)\right\}\, .
\end{equation}
This can be re-expressed in terms of the data $s$ and of a waveform model for the 
signal $h(\vb*{\theta})$ as 
\begin{equation}\label{eq:gwlik_general}
	{\cal L}(s|\vb*{\theta}) \propto {\rm exp} \Big\{-\dfrac{1}{2} (s - h(\vb*{\theta})|s - h(\vb*{\theta}))\Big\}\ .
\end{equation}

\paragraph{Detector response.}
\label{sec:div9_response}

The function $h(\vb*{\theta})$ in \eq{eq:gwlik_general} describes 
the GW detector response to an incoming 
signal. Such response is not uniform, and 
depends on the geometry of the instrument, 
and on the source position in the sky. 
In order to discuss specifically 
the detector response, in this section we will explicitly 
introduce another set of parameters, denoted by $\vb*{\theta}_p$, characterising the detector shape and position. 
We will also explicitly write the time dependence of the strain.
In summary, we will use the notation $h = h(t,\vb*{\theta},\vb*{\theta}_p)$.
For a GW source with polarization amplitudes 
$h_{+,\times}$, the strain $h$ 
can be decomposed as 
\begin{equation}
    h(t,\vb*{\theta},\vb*{\theta}_p) = h_{+}(t,\vb*{\theta})F_{+}(t, \vb*{\theta}, \vb*{\theta}_p) + h_{\times}(t,\vb*{\theta}) F_{\times}(t, \vb*{\theta}, \vb*{\theta}_p) \, .
\label{eqn:projected_strain_td}
\end{equation}
where $F_{+,\times}(t, \vb*{\theta}, \vb*{\theta}_p)$ are the 
so-called \emph{antenna pattern functions}, that allow 
to project the two polarizations of a GW signal 
$h_{+,\times}(t,\vb*{\theta})$ onto the arms of a detector 
on Earth~\cite{Schutz:2011tw}. 
The parameters that determine the detector's position are   
$\vb*{\theta}_p=\{\lambda,\ \varphi,\ \gamma,\ \zeta\}$, with 
$\lambda$ and $\varphi$ the detector's latitude and longitude,
$\gamma$ the angle formed by the 
arms bisector and East, and $\zeta$ the angle between the detector's arms 
(e.g. $90^{\circ}$ for an L-shaped detector and $60^{\circ}$
for a component of a triangular detector). Intuitively, this projection 
can be seen as the composition of three rotations~\cite{Jaranowski:1998qm}: one from the source's 
frame to a geocentric frame, a second from the geocentric 
frame to a frame centered on the detector on Earth's surface, 
and a final rotation from this frame to the detector's proper coordinates. Denoting by $\phi$ and $\theta$, respectively, the right ascension and  declination of the source sky position,
the antenna pattern functions can be explicitly recast in the form 
\begin{equation}\label{eq:patt_func_expr}
    \begin{aligned}
        F_{+}(t, \theta, \phi, \psi, \vb*{\theta}_p) &= \sin{\zeta}\, \big[a(t, \theta, \phi, \vb*{\theta}_p)\cos{2\psi} + b(t,  \theta, \phi, \vb*{\theta}_p)\sin{2\psi}\big] \, ,\\
        F_{\times}(t, \theta, \phi, \psi, \vb*{\theta}_p) &= \sin{\zeta}\, \big[b(t,  \theta, \phi, \vb*{\theta}_p)\cos{2\psi} - a(t,  \theta, \phi, \vb*{\theta}_p)\sin{2\psi}\big]\, ,        
    \end{aligned}
\end{equation}
with
\begin{equation}
    \label{eq:pattern:functions:2}
    \begin{aligned}
        a(t,  \theta, \phi, \vb*{\theta}_p) &= \dfrac{1}{16}\sin{2\gamma}(3-\cos{2\lambda})(3+\cos{2\theta})\cos{[2(\phi-\varphi-2\pi f_{\oplus}t)]} \\
        &\quad\, -\dfrac{1}{4} \cos{2\gamma}\sin{\lambda}(3+\cos{2\theta})\sin{[2(\phi-\varphi-2\pi f_{\oplus}t)]} \\
        & \quad\, +\dfrac{1}{4}\sin{2\gamma}\sin{2\lambda}\sin{2\theta}\cos{(\phi-\varphi-2\pi f_{\oplus}t)} \\
        &\quad\,-\dfrac{1}{2}\cos{2\gamma}\cos{\lambda}\sin{2\theta}\sin{(\phi-\varphi-2\pi f_{\oplus}t)} \\ & \quad\, + \dfrac{3}{4}\sin{2\gamma}\cos^2{\lambda}\sin^2{\theta} \, ,\\ \\
        b(t,  \theta, \phi, \vb*{\theta}_p) &= \cos{2\gamma}\sin{\lambda}\cos{\theta}\cos{[2(\phi-\varphi-2\pi f_{\oplus}t)]} \\
        &\quad\,+ \dfrac{1}{4}\sin{2\gamma}(3-\cos{2\lambda})\cos{\theta}\sin{[2(\phi-\varphi-2\pi f_{\oplus}t)]} \\
        & \quad\, +\cos{2\gamma}\cos{\lambda}\sin{\theta}\cos{(\phi-\varphi-2\pi f_{\oplus}t)} \\
        &\quad\,+ \dfrac{1}{2} \sin{2\gamma}\sin{2\lambda}\sin{\theta}\sin{(\phi-\varphi-2\pi f_{\oplus}t)}\, ,
    \end{aligned}
\end{equation}
where $f_{\oplus} \simeq 1 \, {\rm day}^{-1}$ is the Earth's rotational frequency. The pattern functions for the three ET configurations considered in this section at a specific time will be shown in section~\ref{sec:div9_pattern_func_earth_rot} below.\footnote{These are obtained by summing in quadrature the contribution for each detector in a given configuration and then extracting the square root, as for the SNR.}

We expect the signal modulation due to the Earth's motion 
to be negligible for the short-lived signals detected by 
2G interferometers. 
This is not the case for 3G detectors, which could observe 
signals from light binaries up to $\sim 1\, {\rm day}$. 

%
%
The explicit dependence on $t$ in 
\eq{eq:patt_func_expr} introduces three 
effects~\cite{Cutler:1997ta, Cornish:2003vj}: (i) an amplitude modulation due to 
the time evolution of the pattern functions, (ii) 
a phase modulation due to the different time dependence 
in $F_{+,\times}$, (iii) a Doppler shift given by the relative motion between the 
source and the detector.\footnote{A similar correction arises for the amplitude, 
although its is expected to be negligible given that 
$f_{\oplus}$ is much smaller than the frequencies relevant 
for ground-based detectors.}

\paragraph{Signals from compact binaries.}
\label{sec:div9_sig_par_cbc}

Until now 
we have made no explicit assumption on the 
specific form of the signal. We now focus
on parameter estimation for CBCs. For binary black holes with masses $m_1$ and $m_2$ (we take $m_1>m_2$) 
on circular orbits, the GW signal depends on 15 parameters~\cite{Maggiore:2007ulw}
\begin{equation}
    \vb*{\theta} = \{{\cal M}_c,\,\eta,\,d_L,\,{\rm RA},\,{\rm dec},\,\iota,\,\psi,\,t_c,\,\Phi_c,\,\vb*{\chi}_1,\,\vb*{\chi}_2\}\,,\label{CBCparams}
\end{equation}
where ${\cal M}_c$ and $\eta$ are the detector-frame 
chirp mass and symmetric mass ratio,\footnote{Alternative parametrizations for 
the mass variables are given in terms of the individual 
component masses $m_{1,2}$ or of ${\cal M}_c$ and  
the mass ratio $q=m_2/m_1$.}, $d_L$ is the luminosity 
distance from the detector, RA and dec are the sky position coordinates, $\iota$ denotes the inclination angle of the binary orbital angular momentum with respect to the line of 
sight,\footnote{Instead of $\iota$ one could use 
the inclination of the total angular momentum with 
respect to the line of sight, $\theta_{JN}$, which coincides 
with $\iota$ in the non-precessing case.} $\psi$ is the 
polarization angle, $(t_c,\Phi_c)$ are the time and phase 
of coalescence, and $\vb*{\chi}_{1,2}$ denote the 
three-dimensional spin vectors of the two black 
holes.\footnote{A commonly used parametrization is 
also given in terms of the individual spin magnitudes 
$\chi_{1,2}$, the tilt angles of the spin vectors with 
respect to the binary orbital angular momentum at a reference frequency $\theta_{1,2}$, the azimuthal angle between the 
total and orbital angular momentum $\phi_{JL}$ and 
the difference in azimuthal angle between spin vectors 
$\phi_{1,2}$, all given at a reference frequency.} 
Note that the spins  contribute, at the leading order, 
with the mass-weighted effective parameter \cite{Damour:2001tu,Racine:2008qv}
\begin{equation}\label{eq:chieff_def}
    \chi_{\rm eff} = \dfrac{m_1\, \chi_{1,z}\, +\, m_2\, \chi_{2,z}}{m_1\,+\,m_2}\,,
\end{equation}
where $z$ conventionally denotes the direction of the 
binary orbital angular momentum.
For coalescing binaries on eccentric orbits, 
the set of parameters \eqref{CBCparams} is complemented 
by two extra quantities, the eccentricity $e$ and the 
mean anomaly $\zeta_M$. 
If $e\ll 1$ the vector $\vb*{\theta}$ includes 
only one new parameter, given by $e_0$ at some 
reference frequency $f_{e_{0}}$.

For binary systems in which at least one of the objects 
is a neutron star, we need to include finite size effects 
due to the star internal composition. The dominant 
contribution\footnote{Finite size effects are 
also encoded by the spin induced quadrupole 
moments \cite{Krishnendu:2017shb}. However 
such contributions depend on the stellar 
spins which are expected to be small for NS 
at the end of the binary evolution.} 
is provided by the tidal parameters which 
encode the stellar deformability properties.
In the case of a double neutron star system, 
the tidal deformability $\Lambda_{1,2}$ enter 
the waveform with the following contribution 
\cite{Wade:2014vqa}:
\begin{equation}\label{eq:tildeLam_def}
    \begin{aligned}
        \tilde{\Lambda} &= \dfrac{8}{13} \Big[(1+7\eta-31\eta^2)(\Lambda_1 + \Lambda_2) + \sqrt{1-4\eta}(1+9\eta-11\eta^2)(\Lambda_1 - \Lambda_2)\Big]\\
        \delta\tilde{\Lambda} &= \dfrac{1}{2} \Bigg[\sqrt{1-4\eta} \left(1-\dfrac{13272}{1319}\eta + \dfrac{8944}{1319}\eta^2\right)(\Lambda_1 + \Lambda_2) \\
        & \qquad+ \left(1 - \dfrac{15910}{1319}\eta + \dfrac{32850}{1319}\eta^2 + \dfrac{3380}{1319}\eta^3\right)(\Lambda_1 - \Lambda_2)\Bigg]\, .\\
    \end{aligned}
\end{equation}
We next turn to the detector response to CBC signals. As discussed in section~\ref{sec:div9_response}, the antenna pattern functions depend on the relative orientation between the instruments and the source, which evolve in time due to the Earth's rotation. This effect is not negligible for long--lasting signals, which are expected at ET.
For those signals, however, the amplitude 
variation during a GW cycle is much slower than the 
corresponding change in the phase, which makes it 
possible to resort to the Stationary Phase Approximation 
(SPA) to compute the waveform in the frequency domain. 
As discussed in \cite{Iacovelli:2022bbs}, the SPA 
can be carried out similarly to the case of 
time-independent pattern functions~\cite{Maggiore:2007ulw}. 
The stationary point $t^\ast (f)$ is
determined by the condition $2\pi f = \dot{\Phi}(t^\ast)$, being $\dot{\Phi}$ the derivative of the time-domain GW phase 
$\Phi$, which in the time-dependent case inherits a dependence on the frequency. 
{In particular, for CBC signals} at lowest Post-Newtonian order we have
%
\begin{equation}\label{tstar}
    t^\ast (f) = t_c - \dfrac{5}{256} \left(\dfrac{G{\cal M}_c}{c^3}\right)^{-5/3} (\pi f)^{-8/3} \left[1 + \order{(\pi f G M_{\rm tot}/c^3)^{2/3}}\right]\,.
\end{equation}
We note that the SPA is applicable to signals from compact binaries and short transients, but not for long-lived signals such as those expected from continuous wave sources.

\subsubsection{Fisher information matrix formalism}
\label{sec:div9_FisherIntro}

With the likelihood function encoding a model 
of the experiment, and a template for 
the expected signal, we can reconstruct the posterior probability in \eq{eq:posterior}. This in general achieved with techniques such as Markov Chain Monte Carlo or Nested Sampling~\cite{Christensen:2022bxb,Ashton:2022grj}. Indeed, an accurate reconstruction of the posterior is crucial for precision science when data are available. 
Such techniques however share an high computational cost.

In the case of characterising a future experiment instead, 
the primary need can shift to having a quick indicator of its parameter estimation capabilities \emph{without having to perform actual injection recovery in simulated noise}. 
The main reason is grounded in the computational cost of such procedure, which nowadays can require $\mathcal{O}(\rm{hours-days})$ for a single event, whose analysis can furthermore require manual tuning, e.g. of the prior range and sampler configuration. 
For instance, in the context of the study of an experiment such as ET, the recent activities of the Observational Science Board~\cite{Branchesi:2023mws} have considered a total of $6\times 2$ configurations for 2 populations of compact objects of order $\mathcal{O}(10^5)$ detected sources each, plus their combinations with Cosmic Explorer. Assuming an optimistic average computation time of $3$ hours per source, this would have required $\sim 150000$ days of computing time for each population. 
In order to reduce the total time to $\mathcal{O}(\rm{days-months})$, one has to be able to forecast the parameter estimation (PE) for a single event in $\mathcal{O}(1-10) \, \rm{s}$. Currently, this is not yet achievable with full Bayesian analysis software on such large catalogs. 
Besides issues in computing time, the modeling of the signal is also in some cases still under active investigation. In particular, including in Bayesian analyses software the effect the rotation of the Earth on long signals while keeping the computational cost under control is not  fully achieved at the present, especially for large catalogs of events (but see~\cite{Nitz:2021pbr,Smith:2021bqc,Wouters:2024oxj} for recent progress).

All the above considerations motivate the adoption of approximations that make the computation for each single event feasible, as explained, in $\mathcal{O}(1-10)$~s. A standard approach is that of the Fisher information matrix (FIM) approximation, that we recall here in the context of GW parameter estimation. We refer the interested reader to  \cite{Cutler:1994ys,Vallisneri:2007ev} for exhaustive treatments.
We assume that both  the true signal $h$ determining the data $s$ as in~\ref{eq:s_from_h}, and the waveform model $h(\vb*{\theta})$ that appears in the likelihood in \eq{eq:gwlik_general}, are obtained from the same waveform approximant, 
so $h = h(\vb*{\theta}_0)$ for some set of true parameters $\vb*{\theta}_0$. 
The FIM aproximation consists in considering an expansion of the template around the true values and retaining only the first derivatives, i.e. $h(\vb*{\theta}) \approx h + \partial_i h (\theta^i - \theta^i_0)$. This is known as \emph{linearized signal approximation} (LSA), and it can be shown to be equivalent to the high-SNR limit~\cite{Vallisneri:2007ev}. This means in particular that the approximation is valid for events at large SNR, which usually drive a significant part of the scientific output of the experiment~\cite{Branchesi:2023mws}. 

Omitting noise-dependent factors, the likelihood in \eq{eq:gwlik_general} can be written in the LSA as 
\begin{equation}\label{eq:lsa_gw_lik}
    -2\log{{\cal L}}(s\,|\,\vb*{\theta}) \appropto (\partial_i h\,|\,\partial_j h)(\theta^i - \theta^i_0) (\theta^j - \theta^j_0)\,,
\end{equation}
i.e. as a multivariate Gaussian distribution with covariance given by the inverse of the FIM
\begin{equation}
\label{eq:fisher_def}
    {\rm Cov}_{ij}^{-1} = \Gamma_{ij} \equiv (\partial_i h\,|\,\partial_j h) = -\eval{\langle\partial_i\partial_j {\log \cal L}(s\,|\,\vb*{\theta})\rangle}_{\vb*{\theta}=\vb*{\theta}_0} \, ,
\end{equation}
where $\langle\,\dots\rangle$ denotes an average over noise realizations with fixed parameters. 
In particular, \eq{eq:fisher_def} shows that one can obtain forecasts for the covariance of the parameters by computing derivatives of the signal around the injected values of the parameters, i.e. without the need of a full injection recovery with simulated noise, which makes the process faster and parallelizable. 

It is useful to recall that the result of \eq{eq:fisher_def} has different complementary interpretations~\cite{Vallisneri:2007ev}. In particular, the inverse of the FIM can be seen as:

\begin{itemize}
    \item the covariance (over noise realizations) of the frequentist maximum likelihood estimator;
    \item the covariance of the Bayesian posterior probability for $\vb*{\theta}_0$ given one observation of $s$ (assuming flat priors on each parameter);
    \item the lower limit for the covariance of any unbiased estimator of $\vb*{\theta}_0$ under repeated data realisations (Cramer-Rao bound). Importantly, this does not trivially bound the variance of Bayesian posterior, so the Fisher forecasts should not be expected to be necessarily more precise than any forecast obtained from a full Bayesian simulation~\cite{Rodriguez:2013mla}.
\end{itemize}

Finally, we comment about the use of priors when adopting the Bayesian perspective. Within the FIM framework it is possible to implement straightforwardly only Gaussian priors on the signal parameters, whose inverse variance can be added to the FIM diagonal elements prior to inversion~\cite{Vallisneri:2007ev}. While practical, this in general does not reflect the actual priors employed in real PE scenarios. In particular, in the context of GW sources hard priors limiting the parameters to the physical region of the domain are used, such as positive masses and distance, and angles typically either in $[0,\pi]$ or in $[0,2\pi]$.
To further improve the reliability of the FIM approach, one possibility is to sample the multivariate Gaussian likelihood in \eq{eq:lsa_gw_lik}, and include such physical priors on the parameters via rejection sampling~\cite{Vallisneri:2007ev}. For details on such procedure, and practical implementations, see appendix~C of \cite{Iacovelli:2022bbs} and \cite{Dupletsa:2024gfl}.

When resorting to approximations, a discussion of their validity is always important.
As already pointed out before, 
many conclusions driven on the performances for ET are based on high-SNR events for which the LSA is valid. 
However, there can be specific cases where the FIM fails to adequately approximate the GW likelihood \emph{independently on the SNR}. 
This is the case, for example, of ill-conditioned Fisher matrices, namely matrices where the condition number, defined as the ratio between the largest and smallest eigenvalues, is larger than machine precision~\cite{Vallisneri:2007ev}. In this case, numerical instabilities in the inversion of the matrix can lead to numerical errors up to $100\%$. More generally, even if the numerics could be kept under control, a large condition number signals a potential breakdown of the validity of the LSA over the whole likelihood surface, i.e. unreliable forecasts for the error~\cite{Vallisneri:2007ev}. In general, there is no clear and numerically quick criterion to assess the validity of the approximation in those cases~\cite{Vallisneri:2007ev,Rodriguez:2013mla}. 
Different software packages adopt different strategies to exclude or regularize Fisher matrices with large condition numbers. We will describe them in the next section and, in particular,
in section~\ref{sec:div9_fim_improvements}, we will introduce recent advancements and software tools conceived to go beyond such limitations.

\subsection{Software tools for CBC sources}
\label{sec:div9_software:tools}

In this section we provide an overview of the main forecast tools for resolved sources. These are publicly available and can be found in particular in the \href{https://gitlab.et-gw.eu/osb/div9/}{OSB division 9 GitLab repository} (see also table~\ref{tab:code_cfr} below for each code's repository).

\subsubsection{Sky location-polarisation-inclination averaged SNR}
\label{sec:div9_skyloc_average_snr} 

As shown in \eq{eqn:projected_strain_td}, 
GW signals can be decomposed as a linear combination 
of the two polarizations $\hplus(t, \vb*{\theta})$ 
and $\hcross(t, \vb*{\theta})$, projected in a detector through 
its unique antenna pattern functions $\Fplus(t, \theta, \phi, \psi, \vb*{\theta}_p)$ and $\Fcross(t, \theta, \phi, \psi, \vb*{\theta}_p)$. 
From \eq{eqn:snr}, or equivalently from 
figure~\ref{fig:pattern_functions_allconf} below, one can 
observe that the set of angles $\{\theta, \phi, \psi\}$ 
affects the loudness of a given GW source, i.e. 
its SNR. Following \cite{Robson:2018ifk}, we can define a 
sky location-polarisation averaged signal in the frequency 
domain as:
\begin{equation}
\begin{split}
    \anglesavg{|\tilde{h}(f)|^2} &= \anglesavg{\,\Fplus^2(t(f))\,}|\tilde{h}_{+}(f)|^2 + \anglesavg{\,\Fcross^2(t(f))\,}|\tilde{h}_{\times}(f)|^2 \\
    & = \mathcal{R}(f) \bigl(|\tilde{h}_{+}(f)|^2 + |\tilde{h}_{\times}(f)|^2 \bigr) \, ,
\end{split}
\end{equation}
where $\mathcal{R}(f) = \anglesavg{\,\Fplus^2(t(f))\,} = \anglesavg{\,\Fcross^2(t(f))\,}$. Angular brackets denote 
the sky location-polarisation average given, 
for a quantity $X$, by
\begin{equation}
    \anglesavg{X} = \frac{1}{4\pi^2} \int_{0}^{\pi} d\psi \int_{0}^{2\pi} d\phi \int_{0}^{\pi} X \sin\theta d\theta \, .
\end{equation}
Assuming that the signal is short-lived in the detector's band, we can drop the frequency dependency 
from $\Fplus$ and $\Fcross$.

As an example, for a detector located at the North Pole (i.e. $\lambda = \pi/2$, $\varphi = 0$ and $\gamma = \pi/4$) the antenna pattern functions in \eq{eq:patt_func_expr} become 
\begin{equation}
\begin{split}
    \Fplus(\theta, \phi, \psi, \zeta) &= \sin\zeta \biggl(\frac{1}{2}(1+\cos^2\theta) \cos2\phi \cos2\psi - \cos\theta \sin2\phi \sin2\psi \biggr) \, , \\
    \Fcross(\theta, \phi, \psi, \zeta) &= \sin\zeta \biggl(\frac{1}{2}(1+\cos^2\theta) \cos2\phi \cos2\psi + \cos\theta \sin2\phi \sin2\psi \biggr) \, ,
\end{split}
\label{eqn:antenna_pattern_specific_case}
\end{equation}
which in turn leads to the average values
\bees
    \mathcal{R} &=& \anglesavg{\Fplus^2} =\anglesavg{\Fcross^2} \nn \\
    & = & \frac{\sin^2\zeta}{4 \pi^2} \biggl[\frac{1}{4} \int_{0}^{\pi} \cos^2 2\psi \, d\psi \int_{0}^{2\pi} \cos^2 2\phi \, d\phi \int_{0}^{\pi} \sin \theta \, d\theta \nn\\
    & &+ \frac{1}{4} \int_{0}^{\pi} \cos^2 2\psi \, d\psi \int_{0}^{2\pi} \cos^2 2\phi \, d\phi \int_{0}^{\pi} \cos^4 \theta \sin \theta \, d\theta \nn\\
    & &+ \frac{1}{2} \int_{0}^{\pi} \cos^2 2\psi \, d\psi \int_{0}^{2\pi} \cos^2 2\phi \, d\phi \int_{0}^{\pi} \cos^2 \theta \sin \theta \, d\theta \nn\\
    & &+ \int_{0}^{\pi} \sin^2 2\psi \, d\psi \int_{0}^{2\pi} \sin^2 2\phi \, d\phi \int_{0}^{\pi} \cos^2 \theta \sin \theta \, d\theta \biggr] \nn \\
    &=& \frac{1}{5}\sin^2\zeta\ .
\ees
Therefore, in particular,
$\mathcal{R}=1/5$  for an L-shaped detector
($\zeta=\pi/2$), and $\mathcal{R}=3/20$  for $\zeta=\pi/3$ as is the case for the detectors nested in a triangle.

\begin{figure}
    \centering
    \includegraphics[width=1.\textwidth]{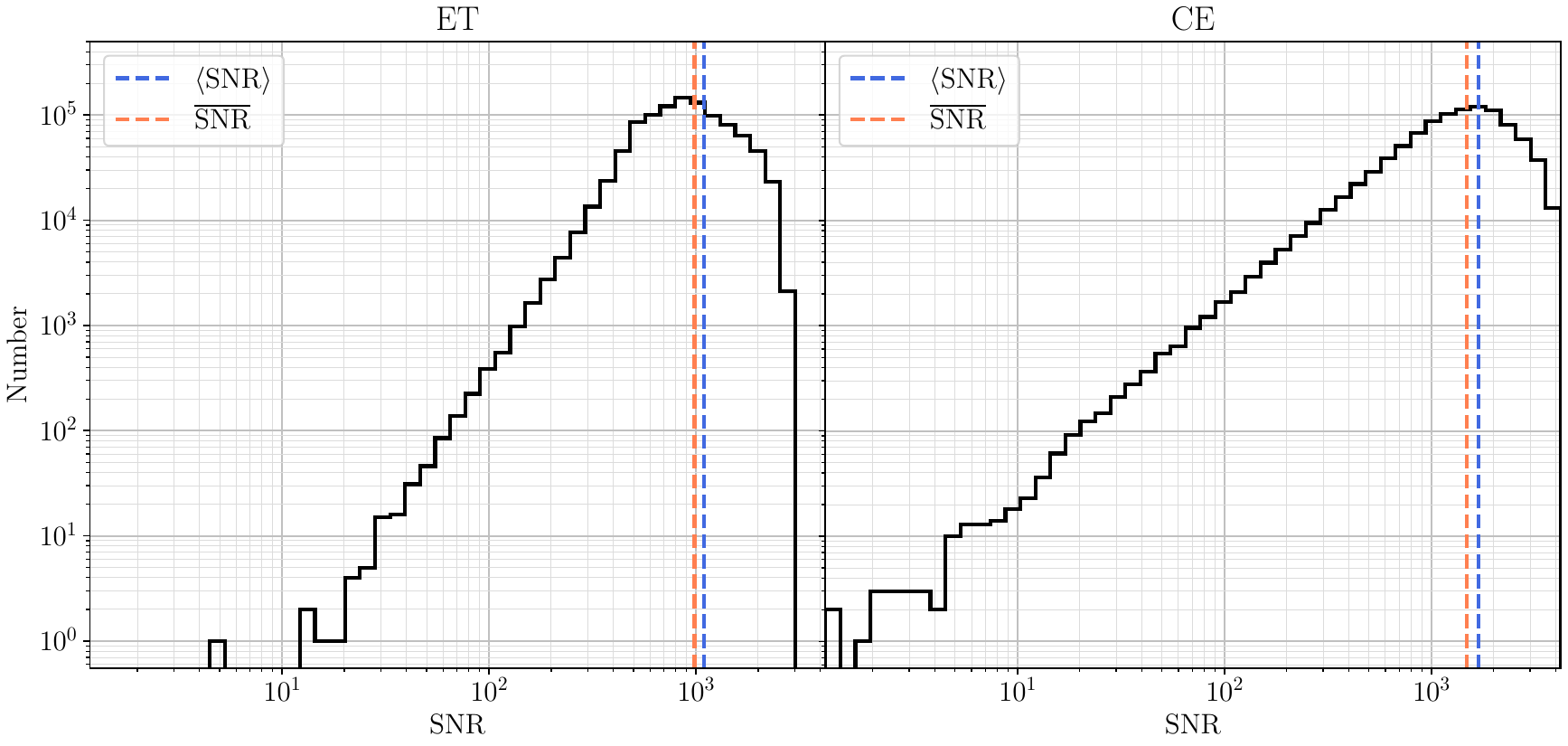}
    \caption{$\snr$ distribution for ET in the triangular configuration (left) and CE (right) considering a GW source with $M_1=36.9\ M_\odot$, $M_2 = 29\ M_\odot$ and located at $z=0.085$ ($d_L=402.4$ Mpc), 
    obtained by selecting a random sets of angles. The mean of the distribution $\overline{\snr}$ approaches the average $\anglesavg{\snr}$ as the size of the samples increases.}
    \label{fig:snr_distribution}
\end{figure}

The angle average can be extended to the inclination 
angle  $\iota$; if higher order modes are neglected, which we assume to be the case here, we have
\begin{equation}
    \anglesavg{|h(f)|^2} = \mathcal{R} A^2(f) \frac{1}{2} \int_{-1}^{1} \biggl(\frac{(1+\cos\iota)^2}{4} + \cos^2\iota \biggr) d\cos\iota = \frac{4}{5} \mathcal{R}A^2(f) \, ,
\end{equation}
where $A(f)$ is the amplitude of the signal. 
Plugging this results into \eq{eqn:snr} 
we obtain the SNR averaged over sky location, polarization and inclination, that we denote by $\anglesavg{\snr}$. The latter corresponds to the 
mean value of a distribution of $\snr$, 
obtained by randomizing over all angles, 
as shown in figure~\ref{fig:snr_distribution}.

\begin{figure}
    \centering
    \includegraphics[trim={0 0 0 25pt},clip,width=1.\textwidth]{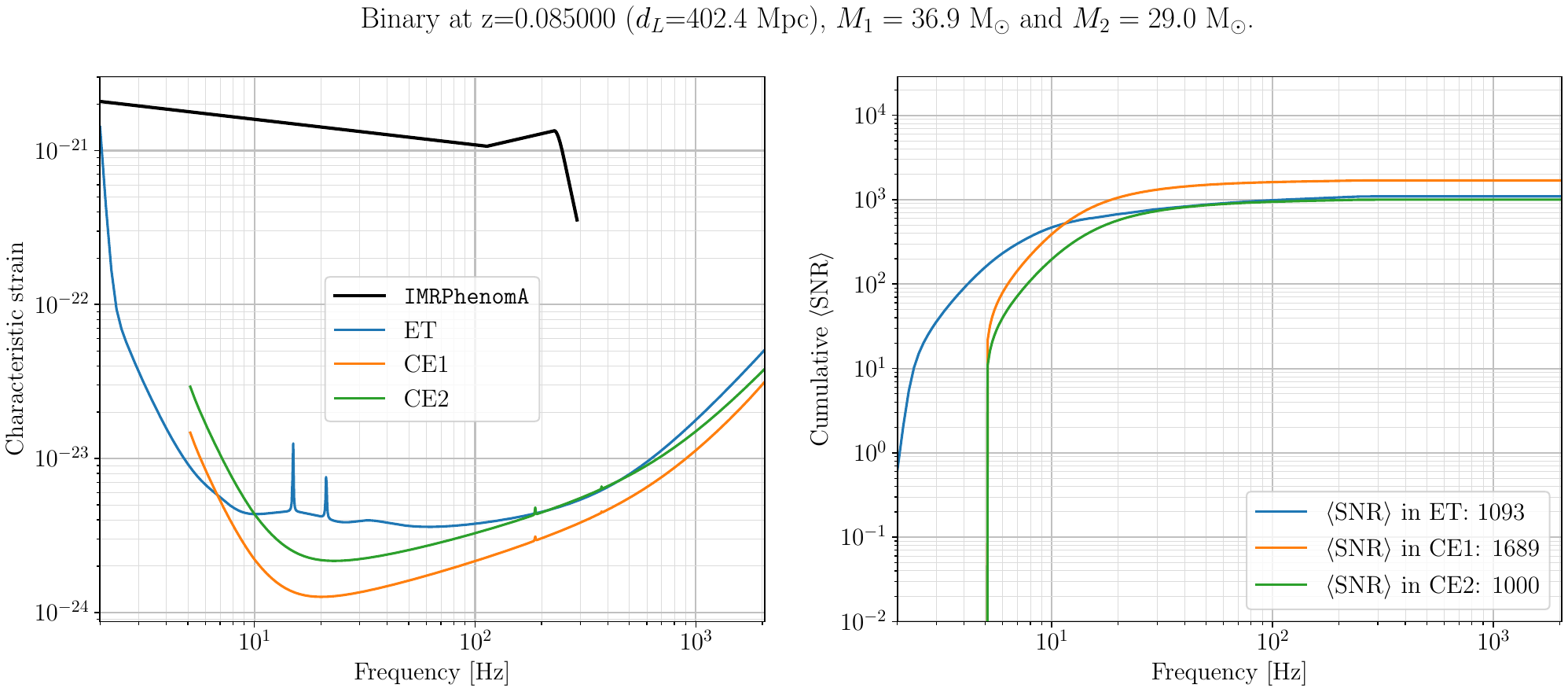}
    \caption{Left panel: Characteristic strain for 
    a sky location-polarisation-inclination averaged GW signal from a binary system composed of $M_1=36.9\ M_\odot$, $M_2 = 29\ M_\odot$ is located at $z=0.085$ ($d_L=402.4$ Mpc), compared against the sensitivity curves of three different 3G detectors (CE sensitivities defined as in footnote~\ref{foot:CEdiv9} on page \pageref{foot:CEdiv9}). Right panel: cumulative $\anglesavg{\snr}$ of the source as a function of frequency for the three detectors. Final values of the SNR are listed in the legend.}
    \label{fig:anglesavg_snr}
\end{figure}

A package to quickly compute the $\anglesavg{\snr}$ of a binary system 
with a given source-frame masses and redshift, for 
a custom number of detectors with 
specific noise curves, is available at \href{https://gitlab.et-gw.eu/osb/div9/tools/-/tree/main/SNR/Sky_loc_pol_inc_avg_SNR?ref_type=heads}{this link}. The code uses as a waveform 
model for the signal obtained with the \texttt{IMRPhenomA} template, 
which includes the inspiral, merger and 
ringdown phases of the coalescence 
\cite{Robson:2018ifk}.
The tool also provides a graphic representation of the characteristic strains, together with the cumulative $\anglesavg{\snr}$ as a function of frequency, as shown 
in figure~\ref{fig:anglesavg_snr}.


\subsubsection{Fisher Matrix pipelines}
\label{sec:div9_fm_pipelines} 

As discussed in section~\ref{sec:div9_FisherIntro}, the Fisher matrix approximation allows us to forecast the parameter estimation capabilities of a given detector network without the computational expenses of a full Bayesian analysis. 

To this scope, five different Fisher matrix pipelines were developed:  GWBench \cite{Borhanian_2021}, GWFast \cite{Iacovelli:2022bbs,Iacovelli:2022mbg} GWFish \cite{Dupletsa:2022scg}, TiDoFM \cite{Chan:2018csa,Li:2021mbo} and GWJulia~\cite{Begnoni:2025oyd}. 
The first four presented above 
are \texttt{Python} based, while the 
latter is developed in \texttt{Julia}. 
All codes but \texttt{TiDoFM} work with 
frequency domain waveforms.
Finally, all codes include the effect of the Earth rotation in the pattern functions. This is relevant for ET given the higher sensitivity and broader frequency band accessible with respect to current--generation detectors, that will allow signals to remain in band up to $\mathcal{O}(1)$ day. Codes in frequency domain use the SPA to account for the time evolution of the antenna
patterns, see \eq{tstar}, while \texttt{TiDoFM} directly encodes the time--dependent response in time domain.

These codes were developed independently and then cross-checked and validated in different ways (see e.g.~\cite{Iacovelli:2022bbs,Branchesi:2023mws}) to ensure the reliability of their results.

In particular, from the computational point of view, the evaluation of a Fisher matrix relies on the following building blocks [see \eq{eq:fisher_def}]:
\begin{itemize}
    \item Waveform approximants
    \item Derivatives
    \item Matrix inversion (to obtain the covariance matrix)
\end{itemize}

In the following, we will give a brief descritpion 
of each code, highlighting the different implementations 
of the core functions. Finally we will discuss 
their comparison.

\paragraph{GWBench.}
\label{sec:div9_gwbench} 

\texttt{GWBench} \cite{Borhanian_2021} uses two packages to compute the partial derivatives of the frequency-domain waveform models necessary for the computation of the FIMs: the package \texttt{numdifftools} enables numerical differentiation of any waveform model written using \texttt{numpy}, such as all the frequency-domain waveform models from LALSimulation \cite{lalsuite} via the \texttt{SWIGLAL} \cite{Wette:2020air} interface, and \texttt{sympy} for analytical differentiation. The latter requires the waveform model to be written with \texttt{sympy} syntax such as the two \texttt{TaylorF2} implementations, with and without tidal corrections, within the \texttt{GWBench} package. To improve the accuracy and speed when using the numerical differentiation option, \texttt{GWBench} further allows the user to use analytical derivative expressions for the  following six waveform parameters: luminosity distance $d_L$, time $t_c$ and phase $\phi_c$ of coalescence, right ascension $\text{RA}$, declination $\text{DEC}$, polarization angle $\psi$. The inversion of the FIMs is handled through the \texttt{mpmath} package at an arbitrary decimal precision which is dynamically adjusted according to the condition number of the FIM.

\paragraph{GWFast.}
\label{sec:div9_gwfast} 

GWFast \cite{Iacovelli:2022bbs,Iacovelli:2022mbg} 
features several state-of-the-art frequency-domain waveform models written in pure \texttt{Python} and cross-checked against their \texttt{C} implementation in the LIGO Algorithm Library \texttt{LAL}~\cite{Iacovelli:2022bbs}, 
namely \texttt{TaylorF2} (both with its tidal and mildly eccentric extensions), \texttt{IMRPhenomD} and \texttt{IMRPhenomHM} (suitable for BBH systems, with the latter including the contribution of higher order harmonics in the signal), \texttt{IMRPhenomD\_NRTidalv2} (suitable for BNS systems), and \texttt{IMRPhenomNSBH} (suitable for NSBH systems)\cite{Iacovelli:2022mbg}.\footnote{See \href{https://github.com/CosmoStatGW/WF4Py}{github.com/CosmoStatGW/WF4Py} for the open-source \texttt{python} implementation of those waveforms.} 
The code also incorporates a wrapper to use the waveform approximants implemented in \texttt{LAL}, as well as the \texttt{TEOBResumSPA} model.
When using pure \texttt{Python} waveforms, the code can handle multiple events at a time even 
on a single CPU (on top of the possibility to 
parallelize the calculations over multiple cores) 
and the derivatives can be computed using the automatic differentiation~\cite{Margossian_2019} routines implemented in the \texttt{JAX} package~\cite{jax2018github}. This allows the code to be particularly efficient when running on large catalogs of events,  
and to get machine-precision estimates of the derivatives, avoiding issues that may arise using finite-differences techniques. On top of this, in order to further gain in speed and accuracy, the derivatives with respect to the luminosity distance $d_L$, sky position angles $\theta$ and $\phi$, polarization angle $\psi$, coalescence time $t_c$, coalescence phase $\Phi_c$ and inclination angle $\iota$ (the latter only in the non-precessing case and when using waveforms containing the fundamental (2,2) mode) are computed analytically. 
If using \texttt{LAL} waveforms, derivatives are computed through the finite-differences-based package \texttt{numdifftools}. The inversion of the FIMs is handled through normalization of the eigenvalues, and the use of the \texttt{mpmath} package, developed for precision arithmetic computation, allowing to partly mitigate issues related to the ill-conditioning of some matrices. It can be performed through Cholesky, Lower-Upper, and Singular Values decompositions, with the Cholesky decomposition being the most stable and default option. 
The recommended criterion for a stable matrix inversion is ensuring that the inversion error should be larger than $0.05$~\cite{Iacovelli:2022bbs}.

\paragraph{GWFish.}
\label{sec:div9_gwfish} 

GWFish is a frequency-domain detector network simulation software \cite{Dupletsa:2022scg}. It accurately models the response of terrestrial, space-based, and lunar GW detectors, which includes the slow variations in the detector response due to the detector motion and working beyond the long-wavelength approximation. Waveform approximants can be taken from LALSimulation \cite{lalsuite} or be based on numerical data.  Networks can consist of mixed detector types, and GWFish puts particular attention to the numerical problems emerging when simulating a network in a common coordinate system with widely separated detectors. This makes GWFish suitable for the simulation of multiband observations. The waveforms can be given in time domain or frequency domain. An important technical choice is that GWFish allows the user to work with ill-conditioned Fisher matrices by projecting on a suitable parameter subspace before inversion of the matrix. This accounts for the fact that ill-conditioned Fisher matrices still contain accurate information about estimation errors for the majority of the parameters, and that ill-conditioning is an artifact of the Fisher-matrix approach irrelevant to the full Bayesian analyses where non-Gaussian likelihoods and prior distributions naturally cure the problem. GWFish implements a module to sample from a Gaussian likelihood with priors, which provides more realistic error estimates whenever priors are important~\cite{Dupletsa:2024gfl}.

\paragraph{TiDoFM.}
\label{sec:div9_tidofm} 

TiDoFM \cite{Chan:2018csa, Li:2021mbo} 
performs all calculations in time domain.
This naturally allows one to track the evolution of the errors on source parameters of a GW event. This feature makes TiDoFM particularly useful for evaluating the prospect for early warning alerts. 
GWs from BNSs may last up to days before the merger. Such lengthy time-domain signals require excessive storage space and operational memory for subsequent computations. To enhance computational efficiency, the time-domain signals are segmented. By default, each segment lasts for 100 seconds (adjustable), wherein we compute the signal-to-noise ratio, antenna pattern, and FM for each segment separately.


\paragraph{GWJulia.}
\label{sec:div9_gwjulia} 

GWJulia is the only code written in  \texttt{Julia}.
It is completely self contained, and does 
not rely on external libraries, increasing its 
computational speed. 
As an example, to calculate the Fisher matrix for an event with the most demanding waveform implemented yet (i.e., \texttt{IMRPhenomXHM}, which includes higher order harmonics) the code takes less than a second for a network of three interferometers. This allows to evaluate the catalog of 1 year of observations in less than two hours (considering 8 CPUs), making feasible to analyze 3G detector catalogs on a laptop. The code employs \textit{automatic differentiation} to perform derivatives in a fast and reliable way, which are based on the \texttt{Julia} package \texttt{ForwardDiff} \cite{Revels:2016for}. 
This is possible thanks to the use of state-of-the-art frequency-domain waveform models written entirely in \texttt{Julia}. 
The code supports \texttt{IMRPhenomXHM}, \texttt{IMRPhenomD} and \texttt{IMRPhenomHM} for BBH (i.e., no tidal deformation allowed), \texttt{IMRPhenomD\_NRTidalv2} for BNS and \texttt{IMRPhenomNSBH} for NSBH. Moreover, the code can perform a full pipeline analysis, including the generation of a catalog of events. 
At any point it is possible to save the results and continue the analysis using other programming languages such as \texttt{Python}. 
To invert the Fisher matrix the Cholesky 
decomposition is used.

\paragraph{Summary and comparison of software tools}
\label{sec:div9_summary_comparison_softtools} 

The key features of the five Fisher matrix tools 
presented above are summarized in table~\ref{tab:code_cfr}. 
We have performed a further comparison among the code performances besides those discussed in~\cite{Iacovelli:2022bbs,Branchesi:2023mws}. 
To this purpose, we have chosen 9 BNS signals, reported in table~\ref{tab:signals}, for which we compare the SNRs and errors obtained with the different codes.
The test is performed using the \texttt{TaylorF2} waveform approximant\footnote{The code \texttt{TiDoFM}, being in time domain, uses instead the \texttt{TaylorT3} waveform model.} with a triangular-shaped ET located in Sardinia, and two Cosmic Explorers located at Hanford and Livingston (see table~1 of \cite{Gossan:2022} for the details of detectors characteristics). The sensitivity curves are \texttt{et\_d.txt} for ET, \texttt{ce1.txt} for CE located at Hanford and \texttt{ce2.txt} for CE located at Livingston.\footnote{The sensitivity curves are public and can be downloaded from \href{https://dcc.ligo.org/LIGO-T1500293/public}{this link}.\label{foot:CEdiv9}} 

Table~\ref{tab:snr_cfr} contains the SNRs.
The result show an excellent agreement between the five pipelines. 
%
Table~\ref{tab:err_cfr} contains the errors obtained from the Fisher matrix. In particular, we choose to display errors on the most relevant parameters for the metrics that will be considered in the rest of the section to compare detector performances, namely detector--frame chirp mass, symmetric mass ratio, luminosity distance, sky localization and spin component of the primary object along the angular momentum.
Despite the difference in implementation, differentiation technique, and programming language, we find excellent agreement for extrinsic parameters, and very good agreement for intrinsic parameters (in which case the different differentiation strategies and waveform implementations play a larger role).

%


\begin{table}
\centering
\scriptsize{
\begin{tabular}{r | l l l l l}
\toprule
\toprule
&\textbf{GWBench} &\textbf{GWFast} &\textbf{GWFish} &\textbf{TiDoFM} & \textbf{GWJulia}\\
\midrule
link & \href{https://gitlab.com/sborhanian/gwbench/-/tree/master}{GWBench} & \href{https://github.com/CosmoStatGW/gwfast}{CosmoStatGW/gwfast} & \href{https://github.com/janosch314/GWFish}{GWFish} & \href{https://github.com/IrisLi135/TiDoFM/tree/main}{TiDoFM} & \href{https://github.com/andrea-begnoni/GW.jl}{GWJulia}\\
domain & Frequency & Frequency & Frequency & Time & Frequency\\
language & \texttt{Python} & \texttt{Python} &\texttt{Python} & \texttt{Python} & \texttt{Julia}\\
waveforms & \texttt{LALSimulation} &self+\texttt{LALSimulation}&\texttt{LALSimulation}+num. & \texttt{Pycbc} & self\\
derivatives & an.+fd.  & an.+AD or +fd. &an.+fd. & num. & AD\\
inversion & \texttt{mpmath} & Cho.,LU,SVD+\texttt{mpmath}& SVD & &Cho.\\
reference & \cite{Borhanian_2021} & \cite{Iacovelli:2022mbg} & \cite{Dupletsa:2022scg} &  \cite{Chan:2018csa, Li:2021mbo} & \cite{Begnoni:2025oyd}\\
\bottomrule
\bottomrule
\end{tabular}
}
\caption{Summary of main characteristics of the five Fisher matrix codes. Abbreviations used are: 
an. for analitical, fd. for finite differences method, num. for numerical, AD for automatic differentiation, Cho. for Cholesky. See text for other pipeline--specific features.
}
\label{tab:code_cfr}
\end{table}
\begin{table}
\centering
\footnotesize
{
\begin{tabular}{r | c c c c }
\toprule
\toprule
\multirow{2}{*}{} &\multirow{2}{*}{$\eta$}  &${\rm dec}$ &$ {\rm RA}$ &$\iota$ \\
\multirow{2}{*}{} &\multirow{2}{*}{}  &[rad] &[rad] &[rad] \\
\midrule
1  &$0.24900$  &$\pi/4$ &$\pi/4$ &$\pi/4$ \\
2  &$0.24900$  &$\pi/4$ &$0$ &$\pi/4$ \\
3 &$0.24900$&$\pi/4$ &$\pi/2$ &$\pi/4$ \\
4&$0.24900$  &$\pi/2$ &$\pi/4$ &$\pi/4$ \\
5 &$0.24900$  &$0$ &$\pi/4$ &$\pi/4$ \\
6  &$0.24900$ &$\pi/4$ &$\pi/4$ &$0$ \\
7 &$0.24900$  &$\pi/4$ &$\pi/4$ &$\pi/2$ \\
8 &$0.24999$  &$\pi/4$ &$\pi/4$ &$\pi/4$ \\
9  &$0.25000$  &$\pi/4$ &$\pi/4$ &$\pi/4$ \\
\bottomrule
\bottomrule
\end{tabular}
}
\caption{Source parameters for the 9 simulated signals 
used to test the agreement between the five 
codes presented in this section. The parameters not reported in the table are kept fixed in all cases to the following values: ${\mathcal{M}}_{c, det}=1.15 M_{\odot}$, $d_L=0.1 \, \rm Gpc$, $t_c = 0.290958$, $\psi=\pi/4 \, \rm rad$, $\Phi_c = \pi/4  \, \rm rad$, $\chi_{1,z} = 0.01$, $\chi_{2,z}=0.005$. }
\label{tab:signals}
\end{table}

\begin{table}
\footnotesize 
\centering
\begin{tabular}{l | c c c c c}
\toprule
\toprule
&\multicolumn{5}{c}{\textbf{SNR} ${\bf \times 10^{-2}}$}\\
\midrule
\midrule
&GWBench &GWFast &GWFish &TiDoFM & GWJulia\\
\midrule
\midrule
%
1 &9.4 &9.4 &9.4  & 9.8 & 9.4 \\
2 &10 &10 &10 & 9.1& 10\\
3 &8.0 &7.9 &7.9& 7.9& 7.9\\
4 &7.4 &7.3 &7.3& 6.7& 7.3\\
5 &7.3 &7.2 &7.2 & 8.7& 7.2\\
6 &13 &13 &13 & 14& 13 \\
7 &2.1 &2.1&2.1 & 3.6 & 2.1\\
8 &9.4 &9.4 &9.4\ & 9.8 & 9.4\\
9 &9.4 &9.4 &9.4 & 9.8& 9.4\\
\bottomrule
\bottomrule
\end{tabular}
\caption{Comparison of SNR calculation by the five 
codes for the signals of table~\ref{tab:signals}. }
\label{tab:snr_cfr}
\end{table}
%
%
\begin{table}
\footnotesize 
\centering
\begin{tabular}{l | c c c c c}
\toprule
\toprule
&\multicolumn{5}{c}{\textbf{Uncertainties}}\\
\midrule
\midrule
&GWBench &GWFast &GWFish &TiDoFM & GWJulia\\
\midrule
\midrule
&\multicolumn{5}{c}{Detector--frame chirp mass (1$\sigma$ relative error $\times 10^7$)}\\
\midrule
%
1 &  5.6 &6.2 &6.9 & 5.9 & 6.2\\
2 & 5.5 &6.0 &6.6 & 5.9 & 5.9\\
3 & 1.4 &6.9 &7.7 & 6.8 & 6.9\\
5 & 7.1 &7.7 &8.5 & 7.2& 7.7\\
7 & 16 &19 &21 & 18& 19\\
\midrule
&\multicolumn{5}{c}{Symmetric mass ratio  (1$\sigma$ error $\times 10^4$)}\\
\midrule
1 & 2.4 &2.7 & 2.6 & 2.6 & 2.3\\
2 & 2.4 &2.6 & 2.5 & 2.5 & 2.6\\
3 & 2.5 &2.9 & 2.8 &  2.8& 2.9\\
5 & 3.1 &3.3 & 3.2 & 3.1& 3.3\\
7 & 7.3 &8.7 & 8.4 &  8.4 & 8.7\\
\midrule
&\multicolumn{5}{c}{Luminosity distance (1$\sigma$ relative error $\times 10^2$)}\\
\midrule
1 & 1.04 &1.1 &1.1 & 1.05 & 1.1\\
2 & 1.6 &1.8 &1.8 & 1.7 & 1.8\\
3 & 0.73 &0.73 &0.73 & 0.67 & 0.73\\
5 & 1.5 &1.5 &1.5 & 1.4 & 1.5\\
7 & 0.53 &0.53 &0.53 & 0.45& 0.53\\
\midrule
&\multicolumn{5}{c}{$90\%$ sky localization $\times 10^3$ $[\rm deg^2]$}\\
\midrule
1 & 4.9 & 4.9 &4.7 & 4.05 & 5.0\\
2 & 6.5 & 6.9  &6.4 & 5.8 & 7.0\\
3 & 6.5 & 6.4  &6.3 & 5.1& 6.4\\
5 & 20 & 26  &18 & 17 & 26\\
7 & 215 & 168  & 161 & 117 & 169\\
\midrule
&\multicolumn{5}{c}{Primary spin, aligned component (1 $\sigma$ error)}\\
\midrule
1 & 0.10 &0.11 &0.11 & - & 0.11\\
2 & 0.10 &0.11 &0.10 & - & 0.11\\
3 & 0.11 &0.12 &0.12 & - & 0.13\\
5 & 0.13 &0.15 &0.14 & - & 0.15\\
7 & 0.34 &0.41 &0.39 & - & 0.42\\
\bottomrule
\bottomrule
\end{tabular}
\caption{Comparison of marginal 1$\sigma$ relative error calculation by the five codes for the signals of table~\ref{tab:signals} for the most representative parameters of the metrics considered in this section. We exclude events 4 and 6 as they have a singular Fisher matrix, and events 8 and 9 as they do not significantly differ from the others. In the last block, forecasts for spin are not shown for \texttt{TiDoFM} as this code does not include their effects. 
}
\label{tab:err_cfr}
\end{table}

\subsubsection{Improvements of Fisher baseline models}
\label{sec:div9_fim_improvements}


As already noticed, in some cases the Fisher Matrix formalism does not provide a good approximation of the likelihood. 

In particular, this happens for ill-conditioned Fisher matrices. While in many science cases these events are only a subset of the total and can be discarded without altering the results, there can be specific situations where this is not the case.

%

A relevant example comes from GW cosmology, for which the distance estimates is essential. For low inclination angles of the binary's angular momentum with respect to the line of sight, distance and inclination become degenerate, which leads to ill-conditioned Fisher matrices (see~\cite{Iacovelli:2022bbs} for a thorough discussion).
On the other hand, events at low inclination are the most relevant for the possible joint detection of a coincident gamma--ray burst which can lead to the determination of the redshift. Hence, dealing accurately with this case is relevant for standard siren cosmology~\cite{Chen:2020dyt,Chen:2023dgw,Mancarella:2024qle}.

In this section, we describe some approaches and software tools aimed at improving the reliability of Fisher--matrix based forecasts and address the issue of ill-conditioned matrices.

 

\subparagraph{GWDALI.}

The code GWDALI provides a Fisher Matrix formalism in which the 
likelihood function is expanded beyond first order 
\cite{deSouza:2023ozp},\footnote{\href{https://github.com/jmsdsouzaPhD/GWDALI/}{github.com/jmsdsouzaPhD/GWDALI/}} and implements 
the \textit{Derivative Approximation for LIkelihoods} (DALI) \cite{Sellentin:2014zta} in the GW context \cite{Wang:2022kia}. The method involves exploring high-order terms in the likelihood expansion, organizing them based on the order of derivatives. This ensures that the likelihood remains positive definite and normalizable. This expansion, up to the third order of derivatives (evaluated at the true parameters $\{\theta^i_0\}$), is given by:
    \begin{equation}\label{eq:DALIlik}
        \begin{split}
		 \log\mathcal{L} \approx &
		 -\left[\frac{1}{2}( \partial_{i}h|\partial_{j}h)_0\Delta\theta^{i}\Delta\theta^{j}\right]
		 \\
		& 
		-\left[\frac{1}{2}(\partial_{i}h|\partial_{j}\partial_{k}h)_0\Delta\theta^{ijk}
		+\frac{1}{8}(\partial_{i}\partial_{j}h|\partial_{k}\partial_{l}h)_0\Delta\theta^{ijkl}\right]  \\
		& 
			-\left[\frac{1}{6}(\partial_{i}h|\partial_{j}\partial_{k}\partial_{l}h)_0 \Delta\theta^{ijkl}
			+\frac{1}{12}(\partial_{i}\partial_{j}h|\partial_{k}\partial_{l}\partial_{m}h)_0 \Delta\theta^{ijklm} \right.  \\
		& 	\left.
			+\frac{1}{72}(\partial_{i}\partial_{j}\partial_{k}h|\partial_{l}\partial_{m}\partial_{n}h)_0 \Delta\theta^{ijklmn}
			\right],
   \end{split}
	\end{equation}
 where $\Delta\theta^i\equiv \theta^i-\theta^i_0$ is the displacement of some parameter $\theta^i$ from the true parameter $\theta^i_0$. The tensors $\Delta\theta^{ij\dots n}$ are defined by displacements products as $\Delta\theta^{ij\dots n}=\Delta\theta^i\Delta\theta^j\dots\Delta\theta^n$. The first term of the expansion is just the Fisher term. The expansion up the second term (second square brackets) is called \textit{Doublet} and the expansion up to the third term (third square brackets) is called \textit{Triplet}. 
 After computing the derivatives entering in \eq{eq:DALIlik} with finite differences techniques, GWDALI extracts samples from the likelihood (\ref{eq:DALIlik}) with standard stochastic samplers, from which the uncertainties can be extracted. 
 
\begin{figure}[t]
    \centering
    \includegraphics[width=.75\linewidth]{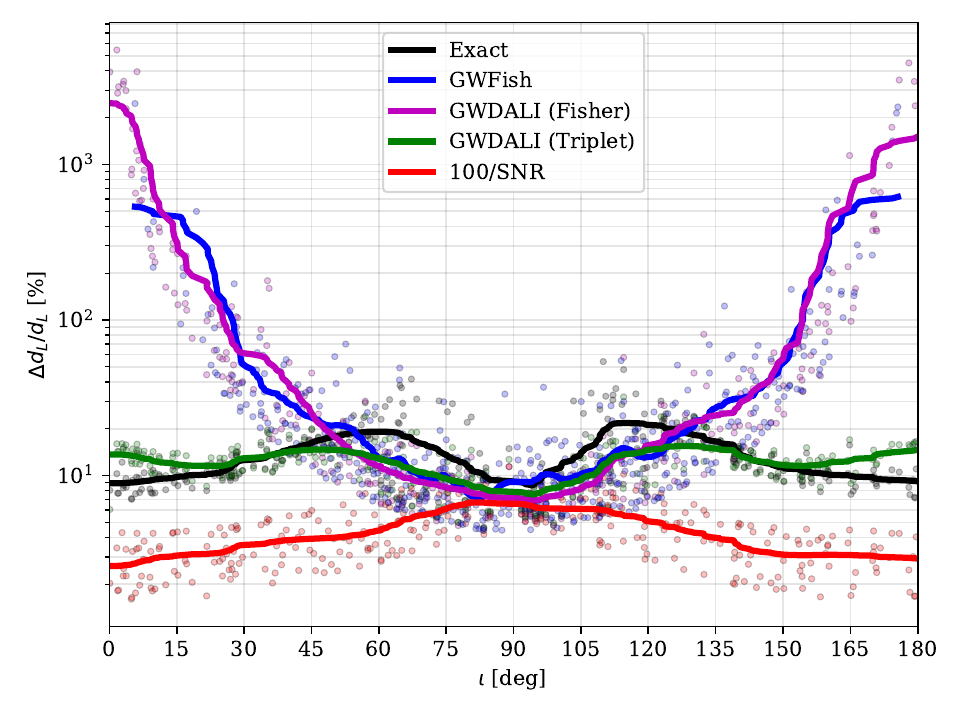}
    \caption{Relative uncertainty of $d_L$ as a function of the inclination $\iota$ for binary neutron stars (BNSs) with total mass $M_{\mathrm{tot}}=3M_{\odot}$. Both $d_L$ and $\iota$ are treated as free parameters. Results are obtained using exact likelihood via nested sampling (black), and GWDALI with Fisher (magenta) and Triplet (green) estimators. These are compared to GWFISH (blue) and to the lower limit predicted by the Fisher matrix when only $d_L$ is a free parameter, leading to $\Delta d_L/d_L \approx 1/\mathrm{SNR}$ (red points). Figure adapted from \cite{deSouza:2023ozp}.}\label{fig:gwfishDLiota}
\end{figure}

 GWDALI allow users to set any detector coordinates, orientations, sensitivities, shapes (opening angle of interferometer arms) as well as waveform approximants (available from \texttt{LALSimulation}), priors, likelihood (exact, Fisher or DALI) and sampler method (available from \textit{Bilby} library).
As an example, figure~\ref{fig:gwfishDLiota} shows the relative uncertainty on the distance $d_L$ versus inclination $\iota$ obtained with GWFish (blue) against the ones obtained with GWDALI Fisher (magenta) and Triplet (green) estimators. For inclinations close to 0 or $180^{\circ}$, the Fisher forecasts become unrealistically large, differently from the GWDALI prediction. For comparison, the red points represent the scaling $\Delta d_L/d_L \approx 1/\mathrm{SNR}$, which is expected from the Fisher matrix when all other waveform parameters are perfectly known. The sampling result obtained using the exact likelihood is shown in black. Note that the implementation of prior distributions greatly improves the parameter-estimation errors even in the Gaussian likelihood approximation as was recently assessed in a comparison of GWFish with GWTC-3 results \cite{Dupletsa:2024gfl}.

\subparagraph{Sampling with prior-informed Fisher approximants.}
Standard Fisher analysis assumes the parameters' priors to be flat and uninformative. This choice is well motivated for high SNR events, for which the Fisher approximation is reliable, since in this case the likelihood function is expected to be sharply peaked so that prior choices become irrelevant.
However, even in this case, correlations among parameters can widen the likelihood beyond the physical range of certain parameters, independently of the SNR value. This is the case for angular parameters, whose range is limited, and for the luminosity distance and masses which are bound to be positive. The inclusion of priors becomes then necessary to restrict the posterior within the physical range of parameters. 
Hard priors can be included in Fisher analyses via rejection sampling from the approximate multivariate likelihood given in \eq{eq:lsa_gw_lik} \cite{Vallisneri:2007ev}. An implementation and a discussion of the role of priors imposed via rejection sampling can be found in~\cite{Iacovelli:2022bbs}, which leverages on \texttt{GWFast} and samples from the LSA likelihood with an algorithm that avoids the inversion of the Fisher matrix, thus also improving the numerical reliability. We refer to table~I of~\cite{Iacovelli:2022bbs} for the default prior choices in \texttt{GWFast}.
The Fisher matrix package \texttt{GWFish} was recently supplemented by a module which returns the prior-informed posteriors~\cite{Dupletsa:2024gfl}. The algorithm - based on importance sampling - is computationally efficient, requiring less than twice the computational time with respect to a Fisher analysis alone. In particular, a truncated version of the likelihood function is sampled with an algorithm developed in ref.~\cite{Botev:2016}, which allows  avoiding regions with vanishing likelihood. The prior-improved Fisher results were compared to the result of the full Bayesian analysis for the GW events observed in the first three observing runs of the LVK collaboration~\cite{LIGOScientific:2018mvr, LIGOScientific:2020ibl, KAGRA:2021kbb}, showing that priors play an important role for improving the reliability of Fisher matrix forecasts for moderate and low--SNR events~\cite{Dupletsa:2024gfl}. We refer to table~II of~\cite{Dupletsa:2024gfl} for the default prior choices in \texttt{GWFish}. 

\subparagraph{Hybrid exact--Fisher likelihood.}
Another possibility to cure singularities in the FIM is to avoid any approximation in some specific directions in the parameter space. This is possible in particular for parameters that determine the amplitude of the GW signal. This is the case for the aforementioned degeneracy between distance and inclination. In this case, a hybrid approach can be used. This was introduced in~\cite{Mancarella:2024qle} which we follow for the rest of the discussion.
%
%
The strategy consists in dividing the parameters $\vb*{\theta}$ into a subset $\bar{\vb*{\theta}}$ for which the LSA is used, and a subset $\vb*{\beta}$ on which the dependence of $\cal L$ is kept exact. In this framework, we have
\begin{equation}
    \begin{aligned}
        & h(\bar{\vb*{\theta}}, \vb*{\beta})\simeq \bar{h}(\vb*{\beta}) +\delta\bar{\theta}^i\ h_i (\vb*{\beta})\,,
        \quad \bar{h}(\vb*{\beta})\equiv h(\vb*{\beta}\,,\bar{\vb*{\theta}}_0)\,,\\
        & h_i(\vb*{\beta})\equiv \left.\partial_{\bar{\theta}_i}h(\vb*{\beta}\,,\bar{\vb*{\theta}})\right|_{\bar{\vb*{\theta}}=\bar{\vb*{\theta}}_0}\,,
    \quad \delta\bar{\theta}^i \equiv \bar{\theta}^i-\bar{\theta}_0^i\,,
    \end{aligned}
\end{equation}
where the subscript $0$ labels the true parameters of the event  and, by introducing the notation $\Delta h(\vb*{\beta})\equiv  \bar{h}(\vb*{\beta})-h$ with $h\equiv h(\vb*{\beta}_0\,,\bar{\vb*{\theta}}_0)$, one obtains the following expression for the likelihood:
\begin{equation}\label{eq:gwlik_hybridfimexact}
-2\log{\cal L}(s|\vb*{\beta\,,\bar{\theta}})\propto
(\Delta h(\vb*{\beta})|\Delta h(\vb*{\beta})) -2\delta\bar{\theta}^i(\Delta h(\vb*{\beta})|h_i(\vb*{\beta}))+\delta\bar{\theta}^i\delta\bar{\theta}^j \Gamma_{ij}(\vb*{\beta})\,.
\end{equation}
In order to compute the likelihood terms that depend on $\vb*{\beta}$, especially $\Gamma_{ij}(\vb*{\beta})$ and $(\Delta h(\vb*{\beta})|h_i(\vb*{\beta}))$, one then needs to assume an explicit form of the signal.
%

\begin{figure}
    \centering
    \includegraphics[width=0.9\textwidth]{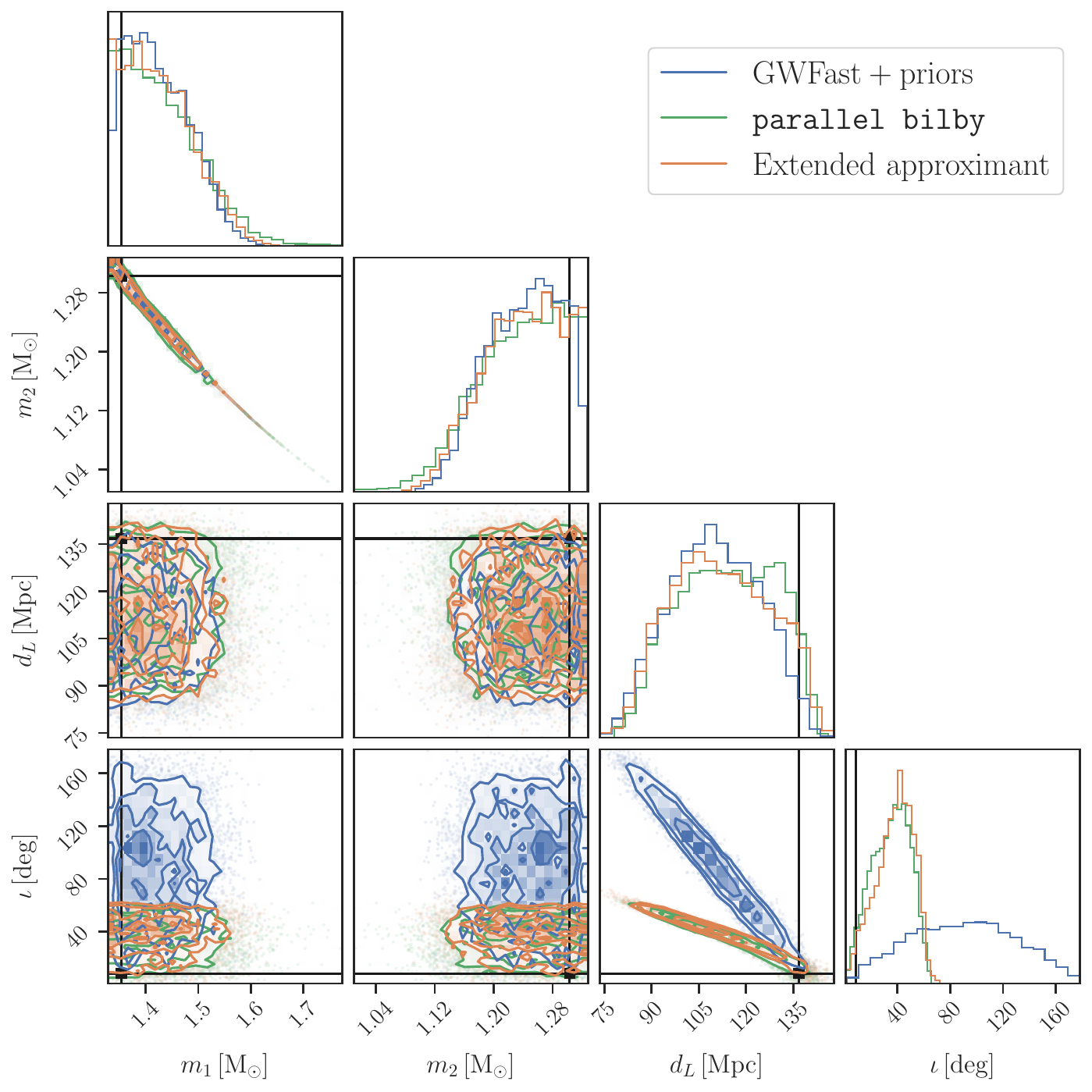}
    \caption{Posterior samples for a face-on BNS systems with ${\rm SNR}\ssim45$ at a LVKI O5 detector network obtained with: the standard FIM computed with \texttt{GWFAST}, with the addition of priors -- in particular, a prior on the inclination angle flat in $\cos \iota$, between -1 and 1 (blue); the extended hybrid likelihood approximant described in the text (orange); and a full Bayesian parameter estimation performed with \texttt{bilby}.}
    \label{fig:corner_hyblik}
\end{figure}

We consider the case $\vb*{\beta} = \{d_L,\,\iota\}$, and we restrict to  the fundamental mode of the GW signal.\footnote{Extending to higher--order harmonics is possible by repeating the procedure for each mode separately.} We define the polarization--decomposed FIM elements as
\begin{equation}
    \Gamma^{ab}_{i j} \equiv \frac{(f^a_{,i}|f^b_{,j})}{d_{L\,0}^2} \, , \quad a,b \in\{+,\,\cross\} \, ,
\end{equation}
with $f^{a,b}$ given by
\begin{equation}
    h(\vb*{\theta})=\frac{c_+(\iota)}{d_L}f^+(\bar{\vb*{\theta}})+i\frac{c_{\cross}(\iota)}{d_L}f^{\cross}(\bar{\vb*{\theta}})\,, \quad c_+(\iota) = \frac{1+\cos^2 \iota}{2}\, , \quad c_{\cross}(\iota) = \cos \iota \, .
\end{equation}
With these definitions, the likelihood can be written as

\begin{equation}
        -2\,\log{\cal L}( {\cal D}_{\rm GW} | d_L \,,\iota\,, \bar{\vb*{\theta}}) \propto \left(\frac{ d_{L\,0}}{d_L}\right)^2\Bigg[\tilde{\delta}d^a_L \Gamma^{ab}_{d_L d_L}\tilde{\delta}d^b_L 
        +(\tilde{\delta}\bar{\theta}^a_i\  \tilde{\delta} d^b_L +\tilde{\delta}\bar{\theta}^b_i\  \tilde{\delta} d^a_L)\Gamma^{ab}_{i d_L}+\tilde{\delta}\bar{\theta}^a_i\Gamma_{ij}^{ab}\tilde{\delta}\bar{\theta}^b_j\Bigg]\,,
\end{equation}
with
\begin{equation}
    \begin{split}
        &\Gamma^{ab}_{i d_L} \equiv \frac{(f^a_{,i}|f^b)}{d_{L\,0}^3}\,,\quad \Gamma^{ab}_{d_L d_L} \equiv \frac{(f^a|f^b)}{d_{L\,0}^4}\,,\\
        &\tilde{\delta}\bar{\theta}_i^{+\,,\cross}\equiv \delta\bar{\theta}_i \cdot c_{+\,,\cross}(\iota)\,, \\
        &\tilde{\delta}d_L^{+\,,\cross}\equiv \delta d_L\cdot c_{+\,,\cross}(\iota_0) -d_{L\,0} [ c_{+\,,\cross}(\iota)-c_{+\,,\cross}(\iota_0)]\,. 
    \end{split}
\end{equation}
An example of the application of this approximant is shown in figure~\ref{fig:corner_hyblik}, where we plot the posterior samples for a face-on BNS system with ${\rm SNR} \simeq 45$ observed by a 2G detector network consisting of LIGO, Virgo, KAGRA and LIGO India at design sensitivity.

The contours obtained by a full Bayesian PE with \texttt{parallel bilby}~\cite{Ashton:2018jfp,Smith:2019ucc} are reported in green.
In blue we show the samples obtained by a simple FIM enhanced by the inclusion of priors, which fails to accurately reproduce the likelihood in the $d_L-\iota$ plane. The extended approximant (orange samples) instead is able to correctly capture the posterior shape. In both cases a good agreement in the mass plane is found.
It should also be noted that the overall computational cost of this procedure is not much higher than the one of two FIMs, plus the time to extract samples from the posterior, that can be made quick enough with a smart initialization of the chains informed by the knowledge of the Fisher matrix. Indeed, the most computationally expensive part is the evaluation of derivatives, which in this case can however be reduced to the computation of those of the two polarizations separately.

The formalism developed here can be further extended to other parameters. The expression of the likelihood for $\vb*{\beta} = \{d_L,\,\iota,\,\psi\}$ is given in appendix~B of \cite{Mancarella:2024qle}, and further work to include the sky--position parameters is ongoing.

\subsubsection{Inference at the population level}\label{sect:inferencepopdiv9}

The Fisher matrix codes described in section~\ref{sec:div9_software:tools} enable the estimation of the uncertainties on the single-event parameters $\vb*{\theta}$ that characterize each detected compact binary system. 
However, given the tremendous detection rate expected for ET, the main science driver for astrophysics and cosmology is expected to be the study of the properties of the underlying population from which individual events are drawn, rather than individual events on a one-by-one basis.
This is the goal of population studies in GW astronomy~\cite{LIGOScientific:2018jsj,LIGOScientific:2020kqk,KAGRA:2021duu,Callister:2024cdx}.

In this section we introduce the formalism and implementation of a forecast tool for population studies, based on the Fisher matrix approximation. 
The underlying mathematical framework has been developed in ref.~\cite{Gair:2022fsj}, where was first derived a Fisher matrix formalism for the population problem. 
An open--source implementation tailored to 3G observatories was then developed in ref.~\cite{DeRenzis:2024dvx} and is available as a module of the package \texttt{GWFast}~\cite{Iacovelli:2022mbg}.\footnote{\url{https://github.com/CosmoStatGW/gwfast/tree/master/gwfast/population}}

\subparagraph{From single-event to population-level inference.}
The properties of a population of sources are described by a function $p_{\rm pop}(\vb*{\theta}|\vb*{\lambda})$ encoding the probability for each source to have parameters $\vb*{\theta}$, given hyperparameters $\vb*{\lambda}$ describing the population. The latter can be, for example, the spectral index of the mass distribution, cutoffs for the minimun/maximum mass allowed for compact objects, or the slope of the redshift distribution. All these are related to the processes governing the formation and evolution of the binaries, therefore their knowledge is of enormous importance for astrophysics.
The goal of population studies is inferring the hyperparameters $\vb*{\lambda}$ by propagating the uncertainty on the single--event parameters and correcting for selection effects, namely, the fact that not all sources are equally likely to be observed (an effect also known as Malmquist bias).

The problem can be modeled as a inhomogeneous Poisson process in presence of selection bias.
The likelihood for observing an ensemble of $N_{\rm det}$ statistically independent, non overlapping events drawn from the population model $p_{\rm pop}(\vb*{\theta}|\vb*{\lambda})$, with data denoted by $\{s\}$, is given by~\cite{Loredo:2004nn,Mandel:2018mve,Vitale:2020aaz}
\begin{align}
\mathcal{L}\left(\{s\}  |   {\vb*{\lambda}}\right) \propto P_{\rm det}({\vb*{\lambda}})^{-N_{\rm det}}\prod_{k=1}^{N_{\rm det}} \int \mathcal{L}(s_k  |   \vb*{\theta}) \, p_{\rm pop}(\vb*{\theta} | \vb*{\lambda}) \, \dd \vb*{\theta}\,,
\label{eq:populationLikelihood}
\end{align}
where 
$\mathcal{L}(s_k  | \vb*{\theta})$ is the single-event likelihood, see \eq{eq:gwlik_general}.
The quantity $P_{\rm det}({\vb*{\lambda}})$, encoding the effect of selection bias, is the fraction of detectable events in the Universe, defined as
\begin{equation}
P_{\rm det}(\vb*{\lambda}) = \int P_{\rm det}(\vb*{\theta}) \, p_{\rm pop}(\vb*{\theta} |   \vb*{\lambda}) \, \dd \vb*{\theta} \leq 1\,,
\label{eq:pdet_lambda}
\end{equation}
where $P_{\rm det}(\vb*{\theta})$ is the probability to observe a compact binary merger with parameters $\vb*{\theta}$.
Note that \eq{eq:populationLikelihood} assumes marginalization over the overall event rate $R$, with a scale-invariant prior $\propto 1/R$.

While codes are available to perform population analyses using hierarchical Bayesian inference for current detectors~\cite{Talbot:2024yqw,Mastrogiovanni:2023zbw}, their scaling to the amount of data expected at 3G observatories remains to date an open problem, both for the computational cost and -- perhaps more worrisome -- for the need of computing the likelihood, and in particular the selection bias, at a sufficient level of accuracy (see e.g.~\cite{Talbot:2023pex}).

\subparagraph{Fisher matrix approximation for population studies.}
Similarly to the case of single events, the above considerations motivate the development of approximation schemes to deliver forecasts for 3G observatories.
In particular, it is possible to write a ``hyper-Fisher matrix" approximating the likelihood~(\ref{eq:populationLikelihood}). This relies on two steps of approximations~\cite{Gair:2022fsj,DeRenzis:2024dvx}: (i) the linear signal approximation for the single-event likelihood $\mathcal{L}(s|\vb*{\theta})$, as in the standard Fisher expansion,  and (ii) the assumption of a large number of events, namely $N_{\rm det} \rightarrow \infty$.
Following these assumptions, the hyper-Fisher matrix, denoted by $\Gamma_\lambda$, can be written as the sum of five contributions, in the form\footnote{Note that in this notation the symbol $\Gamma_\lambda$ indeed indicates the population Fisher matrix and not its single-event contribution as in ref.~\cite{Gair:2022fsj}. This means that the square root of the covariance matrix $(\Gamma^{-1}_{\lambda})_{ii}^{\nicefrac{1}{2}}$ directly provide the uncertainties on $\vb*{\lambda}$ for a catalog composed of $N_{\rm det}$ events. 
See ref.~\cite{DeRenzis:2024dvx} for more details.}

\begin{equation}
\Gamma_{\lambda,ij} =\frac{N_{\det}}{P_{\rm det}({\vb*{\lambda}})} \, \Big( \Gamma_{\mathrm{I}} +\Gamma_{\mathrm{II}} +\Gamma_{\mathrm{III}} +\Gamma_{\mathrm{IV}} +\Gamma_{\mathrm{V}} \Big)_{ij}\,,
\label{eq:5terms}
\end{equation}
where 
\begin{flalign} 
&\Gamma_{\mathrm{I},ij}\!=-\! \int \! \frac{\partial^2 \!\ln [p_{\rm pop}(\vb*{\theta_0}  |  \lambda)p_{\rm det}^{-1}(\lambda)]}{\partial\lambda_i \partial\lambda_j}\Bigg|_{\vb*{\lambda}=\vb*{\lambda_0}}  \!\!\!\!p_{\rm det}(\vb*{\theta_0}) p_{\rm pop}(\vb*{\theta_0}  |  \vb*{\lambda_0}) \dd \vb*{\theta_0} \label{eq:termI}\,, \\
&\Gamma_{\mathrm{II},ij} \!= \frac{1}{2} \! \int \! \frac{\partial^2\! \ln {\rm det}(\Gamma_\theta\!+\!H)}{\partial\lambda_i \partial\lambda_j} \Bigg|_{\vb*{\lambda}=\vb*{\lambda_0}} \!\!\!\! p_{\rm det}(\vb*{\theta_0}) p_{\rm pop}(\vb*{\theta_0}  |  \vb*{\lambda_0}) \dd \vb*{\theta_0} \label{eq:termII} \,,\\
& \Gamma_{\mathrm{III},ij} \!=- \frac{1}{2} \! \int \! \frac{\partial^2\!\left[(\Gamma_\theta\!+\!H)^{-1}_{kl}\right] }{\partial\lambda_i \partial\lambda_j} \Bigg|_{\vb*{\lambda}=\vb*{\lambda_0}}  \!\!\!\!\Gamma_{\theta,kl} \, p_{\rm pop}(\vb*{\theta_0}  |  \vb*{\lambda_0}) \dd \vb*{\theta_0}\label{eq:termIII}  \,,\\
& \Gamma_{\mathrm{IV},ij} \!=- \! \int \! \frac{\partial^2 \!\left[ P_k(\Gamma_\theta\!+\!H)^{-1}_{kl}\right]}{\partial\lambda_i \partial\lambda_j} \Bigg|_{\vb*{\lambda}=\vb*{\lambda_0}} \!\!\!\!D_{l} \, p_{\rm pop}(\vb*{\theta_0}  |  \vb*{\lambda_0}) \dd \vb*{\theta_0} \label{eq:termIV}\,,\\
&\Gamma_{\mathrm{V},ij} \!=- \frac{1}{2}\! \int \! \frac{\partial^2 \!\left[ P_k (\Gamma_\theta\!+\!H)^{-1}_{kl} P_l \right]}{\partial\lambda_i \partial\lambda_j}  \Bigg|_{\vb*{\lambda}=\vb*{\lambda_0}} \!\!\!\! p_{\rm det}(\vb*{\theta_0}) p_{\rm pop}(\vb*{\theta_0}  |  \vb*{\lambda_0}) \dd \vb*{\theta_0}\label{eq:termV} \,.
\end{flalign}
In the above equations, $\vb*{\lambda_0}$ denotes the fiducial population hyperparameters, $\Gamma_\theta$ is the single-event Fisher matrix defined in~\eq{eq:fisher_def}, and we defined
\begin{align}
P_i&=\dfrac{\partial \ln p_{\rm pop}(\theta|\lambda)}{\partial \theta_i}\bigg|_{\vb*{\theta}=\vb*{\theta_0}} \,,
\\
H_{ij}&= -\dfrac{\partial^2 \ln p_{\rm pop}(\theta|\lambda)}{\partial \theta_i \partial \theta_j}\bigg|_{\vb*{\theta}=\vb*{\theta_0}} \,,
\\
D_{i}&= \dfrac{\partial P_{\rm det}(\theta)}{\partial\theta_i}\bigg|_{\vb*{\theta}=\vb*{\theta_0}}= \dfrac{\partial P_{\rm det}}{\partial\rho}\dfrac{\partial\rho}{\partial\theta_i}\bigg|_{\vb*{\theta}=\vb*{\theta_0}}\label{eq:derPDET}\,.
\end{align}
Note that, when the errors associated with individual events are small -- as expected for 3G observatories -- the leading contribution to \eq{eq:5terms} comes from the term $\Gamma_{\mathrm{I}}$.
We refer the reader to ref.~\cite{Gair:2022fsj} for a detailed derivation of \eqst{eq:termI}{eq:termV}.
Finally, we note that corrections to \eq{eq:5terms} scale with inverse powers of $N_{\rm det}$. It is therefore essential to assess the validity of these assumptions for the specific scenarios under consideration.

\subparagraph{Implementation. }
The Fisher formalism described above has been implemented as an extension of \texttt{GWFast}~\cite{DeRenzis:2024dvx}. Derivatives with respect to the hyperparameters $\vb*{\lambda}$ are computed using automatic differentiation via the \textsc{jax} library~\cite{jax2018github}. The single-event Fisher matrices are computed through \texttt{GWFast}. 
Equations~\eqref{eq:termI}--\eqref{eq:termV} are then evaluated with re-weighted Monte Carlo integration.
Potential numerical instabilities in $\Gamma_\lambda$ caused by large errors in single-event Fisher matrices 
(e.g., from low-SNR sources or boundary effects), are regularized 
by restricting Monte Carlo sums to the inner 95\% quantile of the target distribution. 
In order to further regularize the computation, $p_{\rm pop}(\vb*{\theta}|\vb*{\lambda})=0$ is set for values below $10^{-12}$. This prevents numerical instabilities when computing derivatives with respect to $\lambda$.

\subsection{Metrics for CBCs}
\label{sec:div9_metrics} 

In this section we describe some key FOMs to assess the accuracy of ET 
observations in constraining the properties of 
compact binary sources. The FOMs we present 
here provide standard metrics for current ET 
configurations, and can be considered as a 
guideline to determine how changes from 
the baseline models could affect the most 
relevant ET science cases. The FOMs presenting number of events are obtained using an SNR detection threshold of 12, unless otherwise specified. 
All FOMs discussed in this 
section can be produced with the open--source tools described previously in this section.

\begin{figure}[t]
    \centering
    \begin{tabular}{c c}
         \includegraphics[width=.48\textwidth]{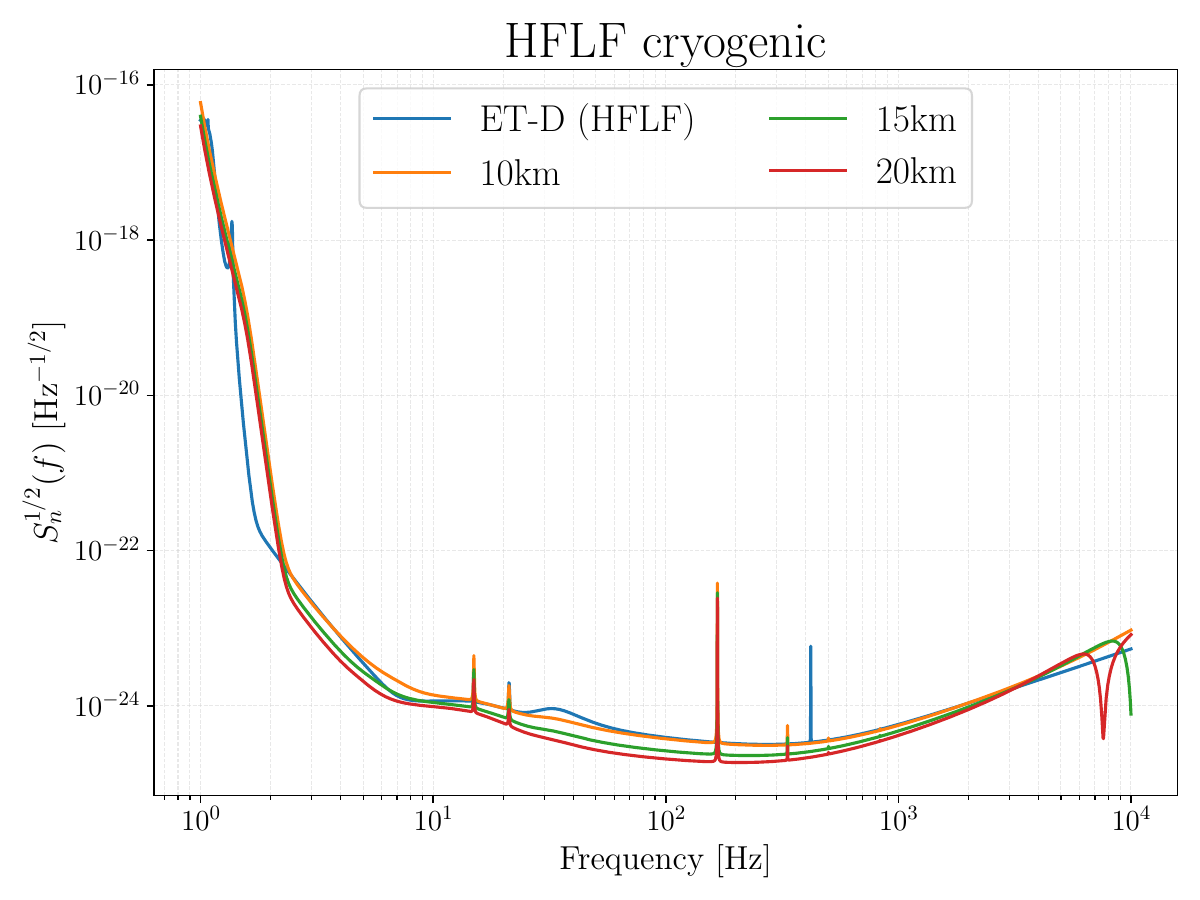}&
         \includegraphics[width=.48\textwidth]{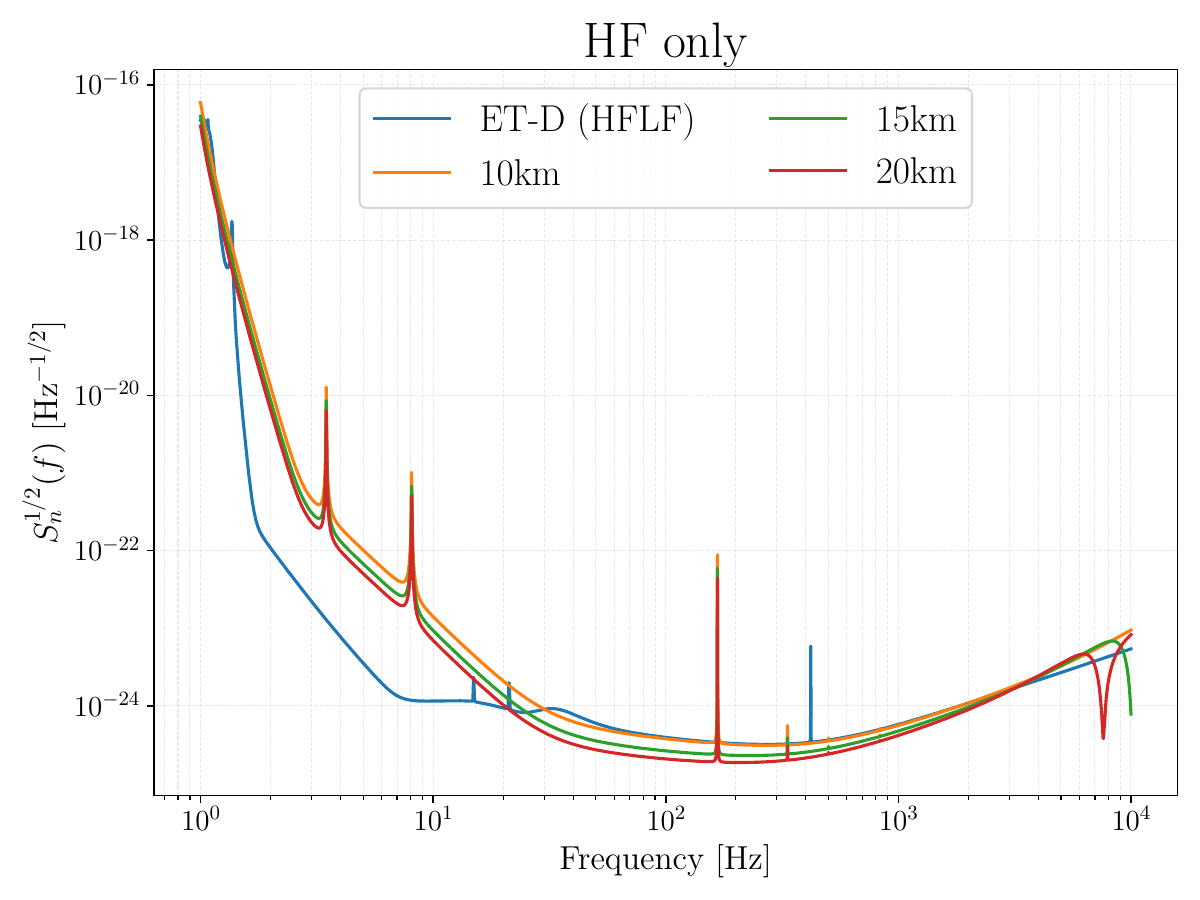}
    \end{tabular}
    \caption{Amplitude spectral densities (ASDs, single-sided) 
    adopted for the 
    FOMs discussed in this section. 
    The left panel shows the sensitivity curves 
    given by the High Frequency and cryogenic Low-Frequency 
    instrument, in a xylophone configuration. The right panel 
    provides the sensitivity resulting from the High 
    Frequency instrument only. We consider ET with 10, 15 
    and 20 km arms, compared with the ASD of the 10 km ET-D. 
    See \cite{Branchesi:2023mws} for further details. \label{fig:ASDsdiv9}} 
    \label{fig:asd}
\end{figure}

The results  are computed  assuming the sensitivity  
curves shown in figure~\ref{fig:asd}, which take 
into account detectors with different arm-length, 
and the possibility of having ET in the full configuration, in which  each detector is made of a high-frequency (HF) and a low-frequency (LF) interferometer, with the latter operating at cryogenic temperatures (curves labeled ``HFLF cryogenic'' or simply ``HFLF-cryo''), or else in a reduced configuration in which only the HF interferometer is operative (labeled ``HF~only''). We refer the 
reader to \cite{Branchesi:2023mws} 
for a detailed investigation on the relevance 
of such configurations for the ET 
science goals, and to \cite{Iacovelli:2024mjy,Maggiore:2024cwf} for studies of further detector network configurations.

We also stress  that the ASDs shown in figure~\ref{fig:ASDsdiv9} always refer to a single L-shaped detector. To obtain the sensitivity of ET in the triangle configuration 
one must take into account that the triangle is made of three nested detectors (a detector being an LF and HF interferometer pair), with an opening angle of $60^{\circ}$. 
One must then  project the GW tensor of the incoming wave onto each of these three  components [see e.g. eqs.~(9)--(11) of \cite{Jaranowski:1998qm} for explicit expressions], and 
combine the results at the level of the SNR  and parameter estimation. This procedure is automatically performed by all PE codes presented above; see also section~\ref{sec:div9_response} for further discussion.

The results for  BBHs and BNSs are obtained using the same catalogs as in \cite{Branchesi:2023mws}  which, for one year of observation, contain about 
 $1.2\times10^5$ BBHs and $7.2\times10^5$ BNSs 
events,\footnote{As already mentioned in footnote~\ref{foot:BNSrate_div2} on page
\pageref{foot:BNSrate_div2}, these correspond to state-of-the art population models tuned to the O3 LVK data. The absence, to date,  of BNS  candidates in the currently ongoing O4 run  implies that the rate used for BNS is now toward the upper range, rather than the mean value.}
and assuming  an independent $85\%$  duty cycle for each L-shaped detector, and in each of the three detectors composing the triangle

\subsubsection{Pattern functions and Earth rotation}
\label{sec:div9_pattern_func_earth_rot} 

As discussed in section~\ref{sec:div9_response}, the response of a GW detector to an incoming signal depends on the position of the source itself and on the detector geometry and location. A useful visualization is provided by a skymap plot of the angular sensitivity of a given network. This is computed as the sum in quadrature of the antenna pattern function of each interferometer in the network. Figure~\ref{fig:pattern_functions_allconf} shows the angular sensitivity for three different ET configurations, namely the triangular one (upper panel), a 2L configuration with misaligned detectors, i.e. with a relative arm orientation of $45^{\circ}$
{(bottom left) and  a 2L configuration with aligned detectors (bottom right).\footnote{The relative angles, $45^{\circ}$ for the misaligned configurartion and $0^{\circ}$ for the aligned  configuration, are  defined with respect to the great circle that connects the two detectors.} In particular, the presence of ``blind spots" with vanishing sensitivity is evident for the latter configuration. 
We note that the angular sensitivity depends on the relative orientation between source and detector, hence on the specific time of the day, which is relevant for sources that remain in the detection band long enough. The explicit time dependence is given in \eqs{eq:patt_func_expr}{eq:pattern:functions:2}. Figure~\ref{fig:pattern_functions_allconf} assumes a fixed GPS time, arbitrarily fixed to Sep. 12, 2030  at 17:19:15 UTC. Figure~\ref{fig:pattern_functions_allconf_6hrs} shows the sensitivity 6 hours later, displaying the different location of the most sensitive areas and of the blind spots.

\begin{figure}[t]
    \centering
    \begin{tabular}{c c}
         \multicolumn{2}{c}{\includegraphics[width=.48\textwidth]{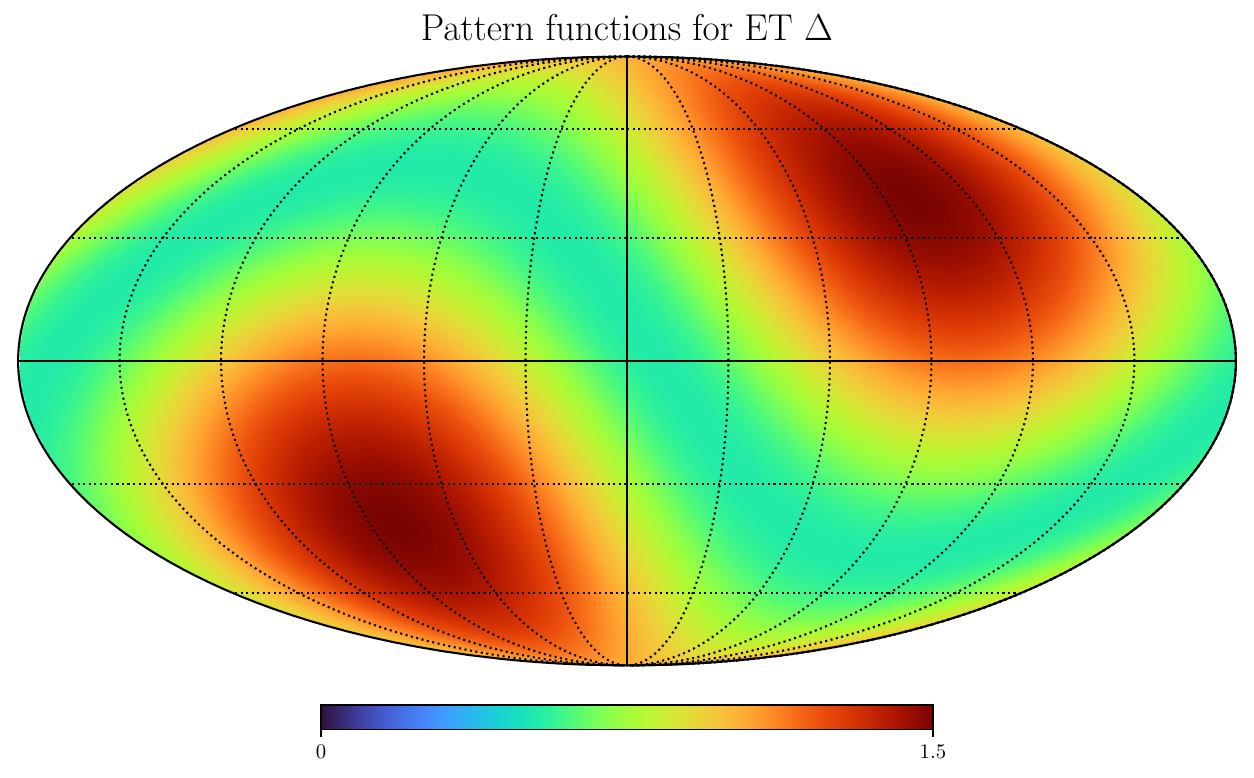}} \\
         \includegraphics[width=.48\textwidth]{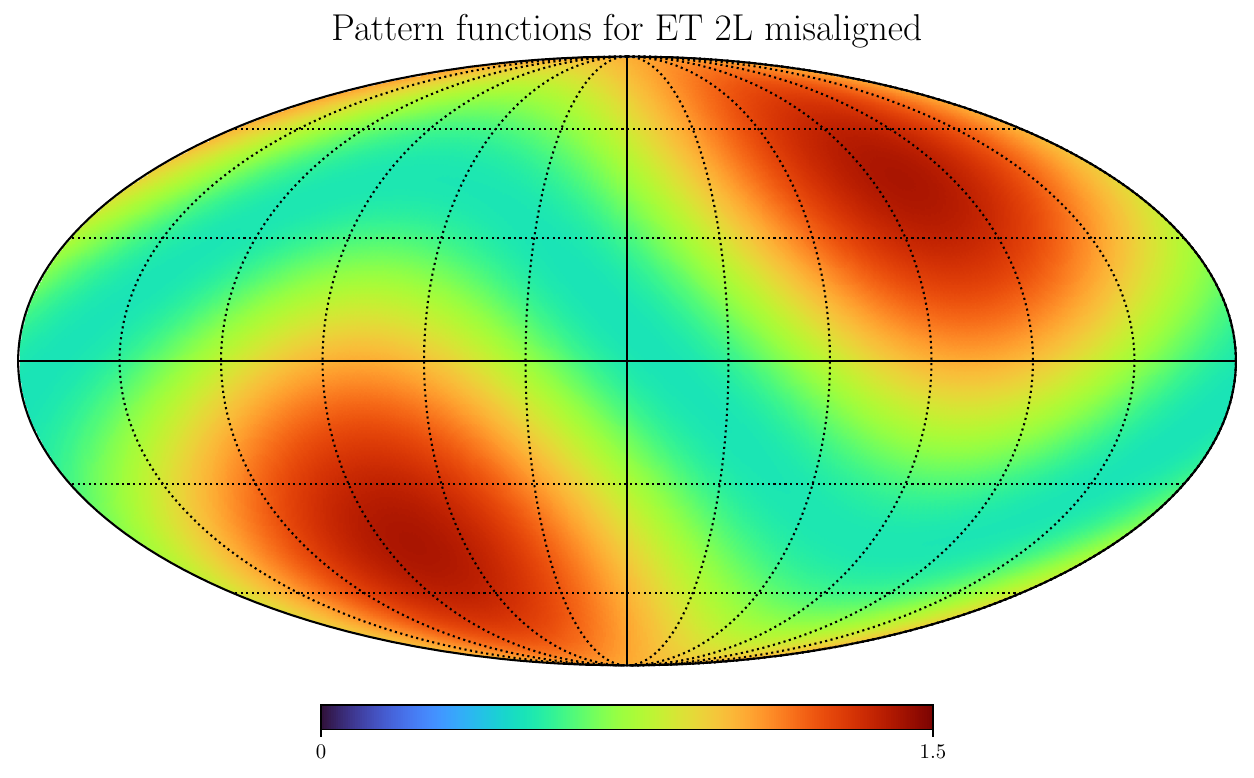} & \includegraphics[width=.48\textwidth]{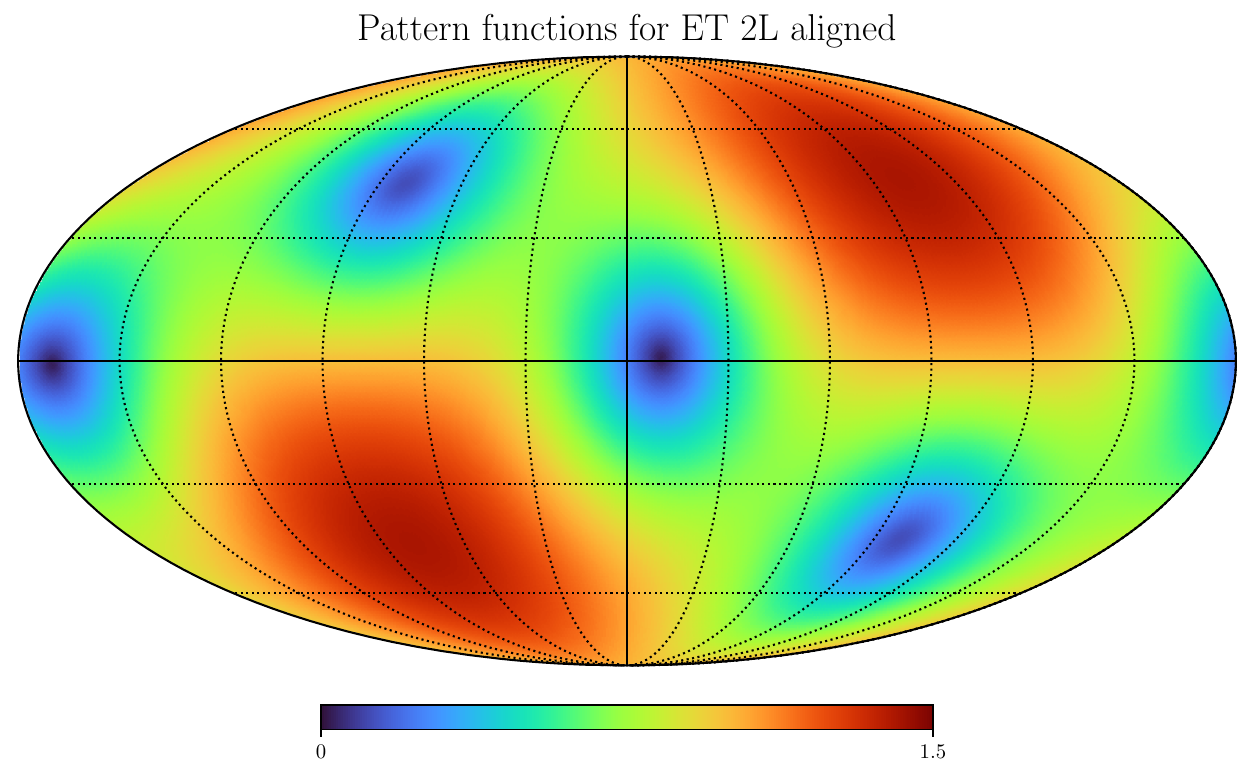}
    \end{tabular}
    \caption{Mollview plots for the angular sensitivity of 
    different ET configurations, at a fixed GPS time. 
    The upper panel shows the antenna pattern functions for a triangular detector, while lower plots 
    refer to two L-shaped detectors in the misaligned (left) and aligned (right) configurations (as defined in \cite{Branchesi:2023mws}). The figures can be reproduced using the 
    notebook available in the 
    \href{https://gitlab.et-gw.eu/div9/tools/-/blob/main/antenna_pattern/notebooks/pattern_functions_Triangle_2L.ipynb?ref_type=heads}{Div.~9 GitLab repository}.}
    \label{fig:pattern_functions_allconf}
\end{figure}
\begin{figure}[t]
    \centering
    \begin{tabular}{c c}
         \multicolumn{2}{c}{\includegraphics[width=.48\textwidth]{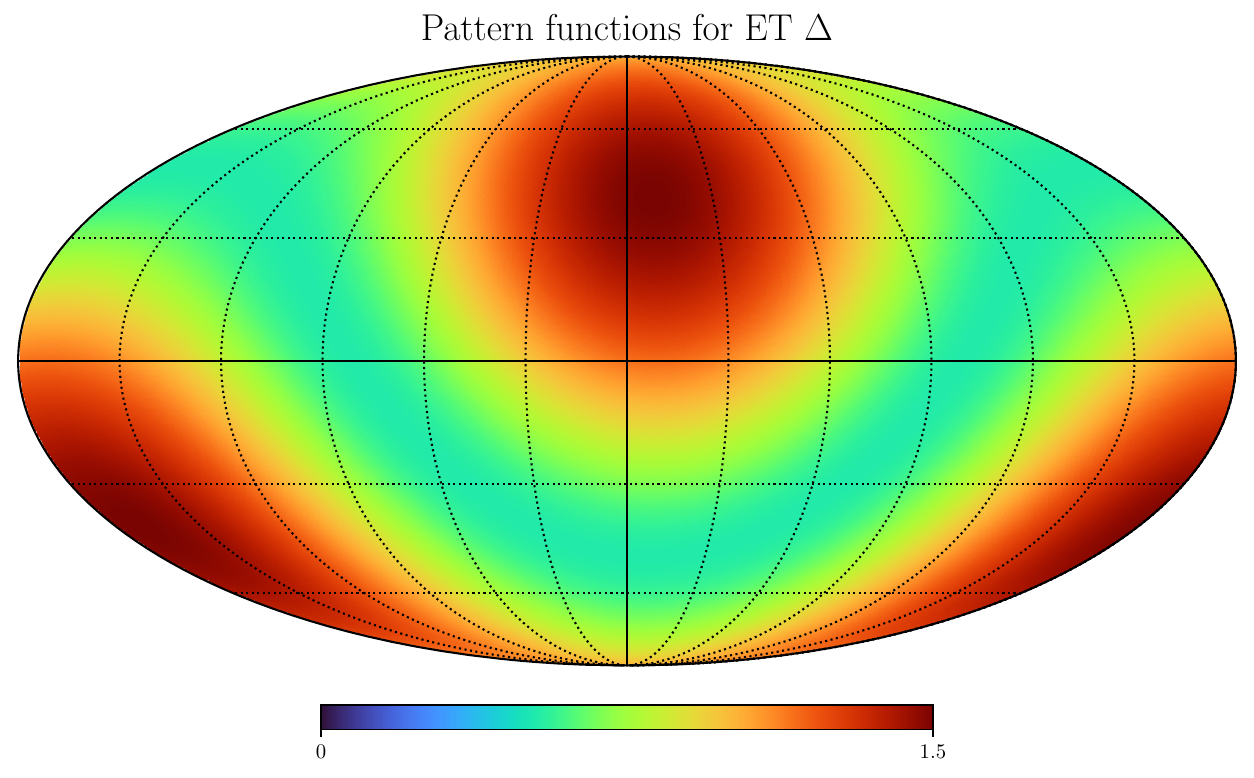}} \\
         \includegraphics[width=.48\textwidth]{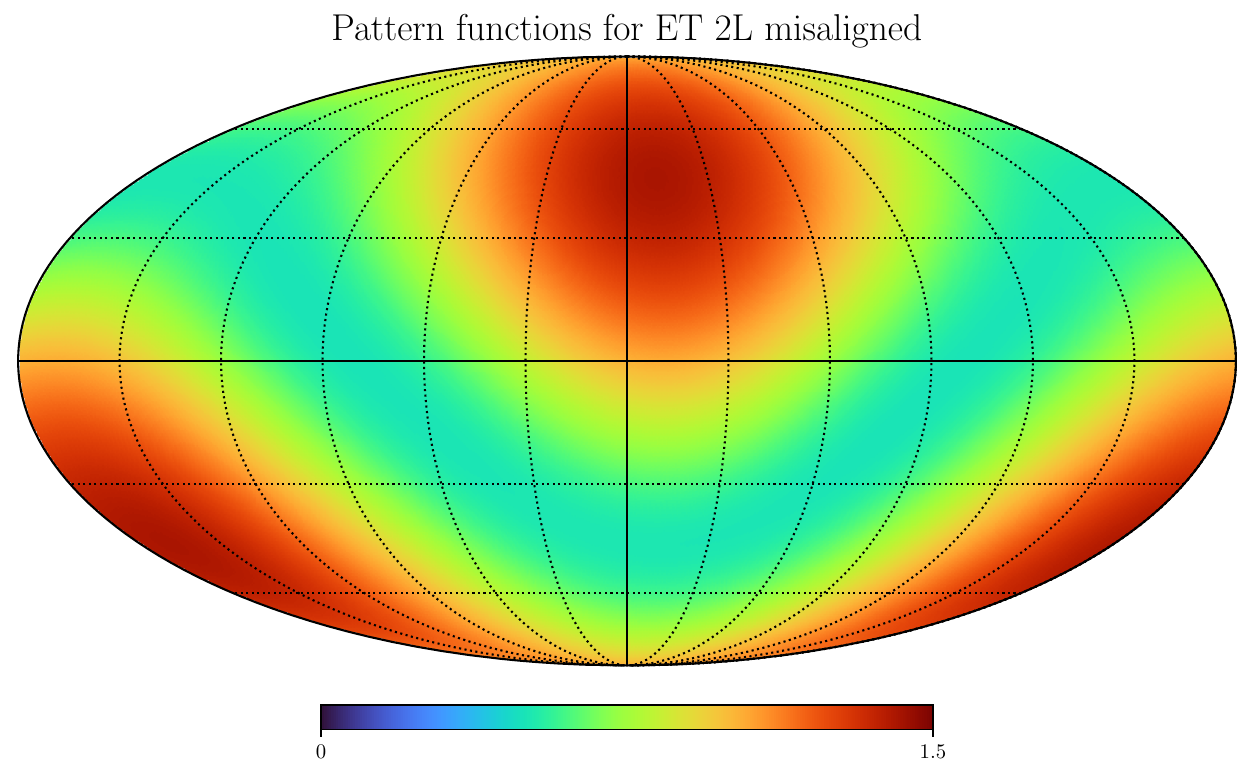} & \includegraphics[width=.48\textwidth]{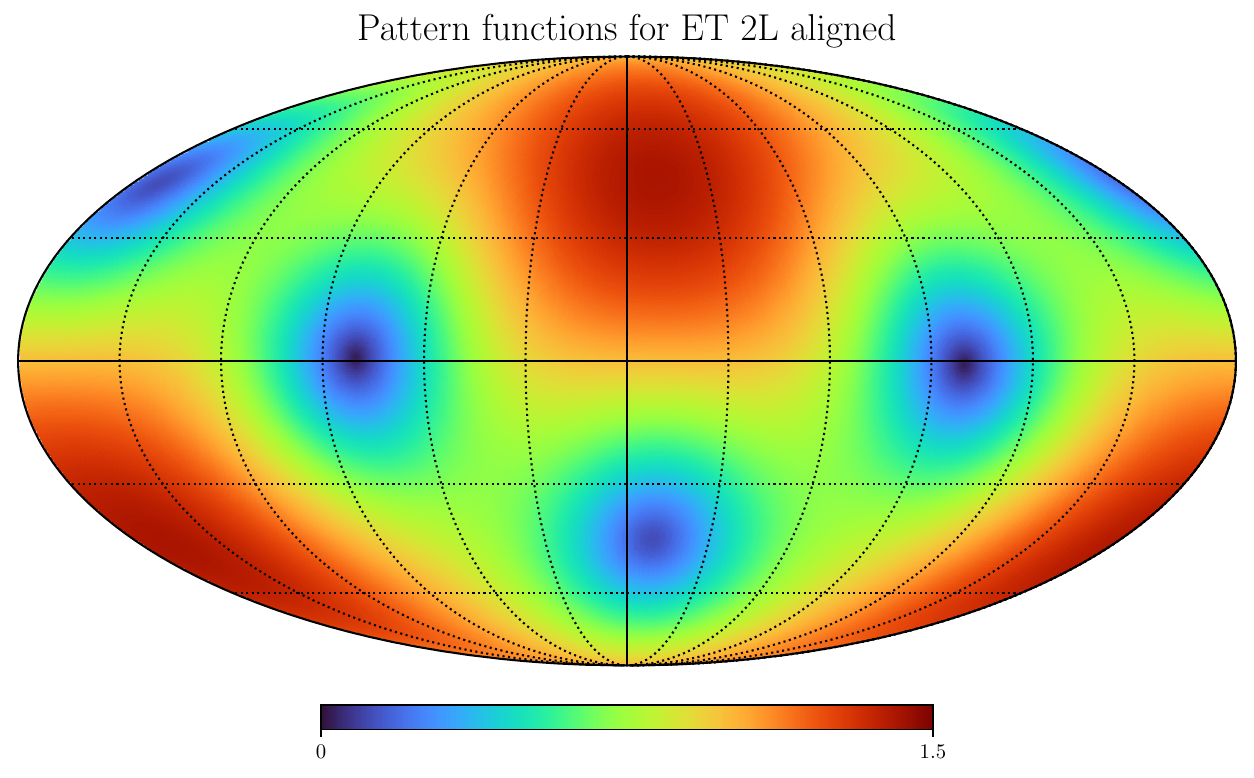}
    \end{tabular}
    \caption{As in figure~\ref{fig:pattern_functions_allconf}, but with a time shift of 6 hrs.}
    \label{fig:pattern_functions_allconf_6hrs}
\end{figure}

\subsubsection{Horizons and Signal to Noise Ratios}
\label{sec:div9_horizons_snr} 

\subparagraph{Horizon.} An important performance indicator for a GW detector is its distance reach. Since the SNR depends on several extrinsic and intrinsic parameters, the maximum distance out to which a binary can be observed  varies from source to source. It is customary to define a FOM that relies on the most optimistic possibility for the extrinsic parameters while tracking the dependence on the intrinsic ones. In particular, the one mostly affecting detectability is the mass of the system. The metric used to quantity the distance reach is the detector horizon, defined as the maximum redshift out of which a binary with optimal sky position and orientation can be observed above a chosen SNR threshold, as a function of the total mass. Equal mass, non--spinning binaries are also usually assumed. Figure~\ref{fig:allconf_hor} shows the horizon plot for different configurations of the ET detector(s) studied in~\cite{Branchesi:2023mws}.

\subparagraph{Inference horizon.} It is important to underline that the concept of detector horizon is an optimistic one. The reason is not only the ideal choice of the source's position and inclination, but more importantly the fact that the definition does not take into account the error on the reconstruction of the source parameters. In simple words, \emph{detecting} a source at some true (but unknown) redshift does not add scientific knowledge unless such detection happens with a precision high enough to \emph{confidently state} that the source is actually located at that redshift. This is particularly relevant for sources at high redshift, which are primary targets of ET inaccessible to current--generation detectors. Since the distance estimate degrades for far away sources, assessing the interplay between true redshift and the best lower bound attainable on it is essential for this specific science case. To this end, ref.~\cite{Mancarella:2023ehn} introduced the notion of an ``inference horizon", defined as the largest
redshift one can possible put a claim on for a given confidence level. This is shown in figure~\ref{fig:inference_hor} as the dashed line for two different ET configurations, choosing a C.L. of $90\%$ as customary in GW astronomy, and denoted by $z_{90\%}$.
Importantly, the sources with best lower bound on the redshift are found not to be those at the horizon. One can define another notion of horizon as the true value of
the redshift where the $z_{90\%}$ is largest. This is shown as the dot--dashed line in figure~\ref{fig:inference_hor} and denoted by $z_{\rm peak}$. In particular, from figure~\ref{fig:inference_hor} one can appreciate that, for systems with total mass between $ \sim [20, 100] \, M_{\odot}$, a 2L configuration might lead to a confident bound of $z \gtrsim 30$, which is usually taken as a threshold for identifying the primordial origin of the system. Such possibility seems not to be guaranteed by the triangular configuration. 


\begin{figure}[tp]
    \centering
    \includegraphics[width=0.8\textwidth]{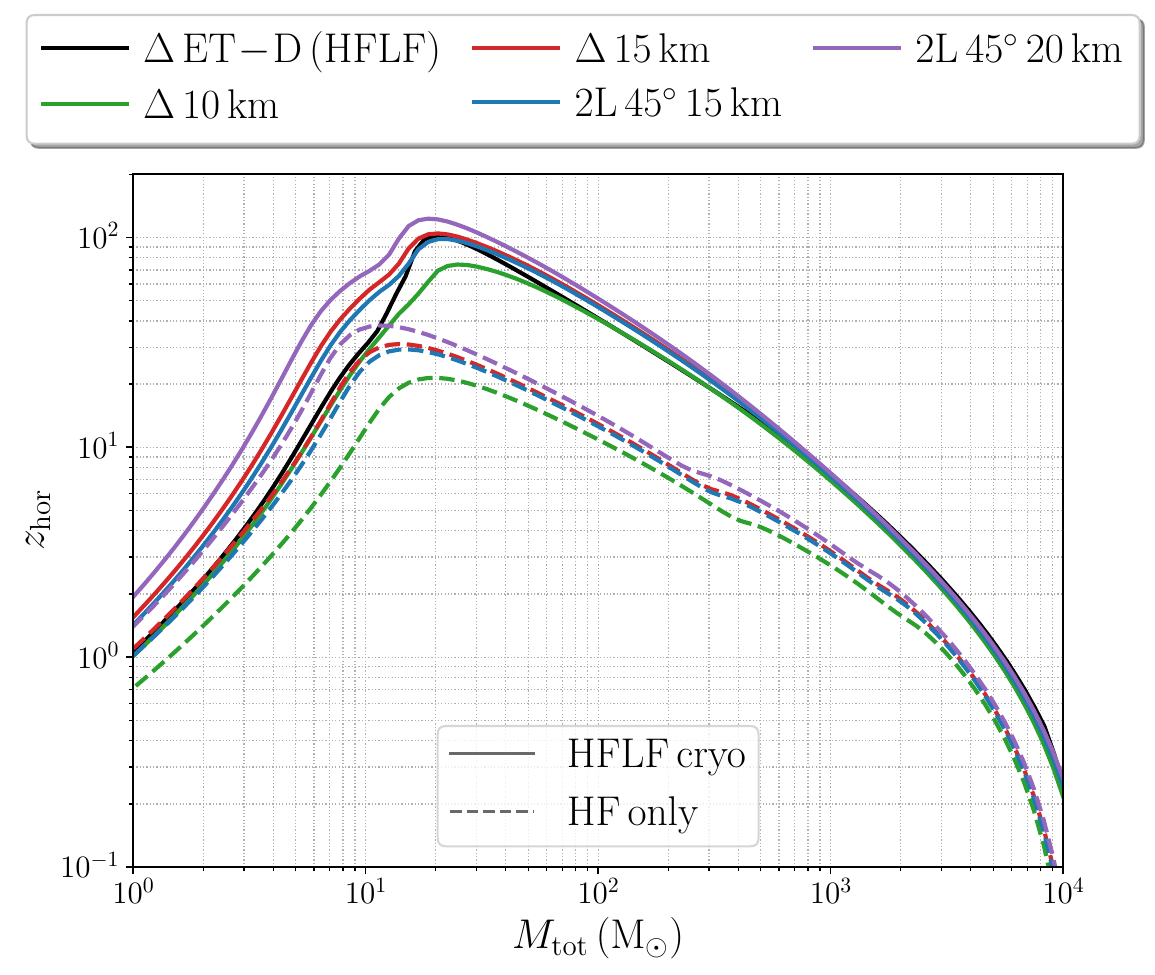}
    \caption{Detection horizons for equal--mass non--spinning binaries as a function of the source--frame total mass for different ET configurations.}
    \label{fig:allconf_hor}
\end{figure}

\begin{figure}[tp]
    \centering
    \includegraphics[width=0.9\textwidth]{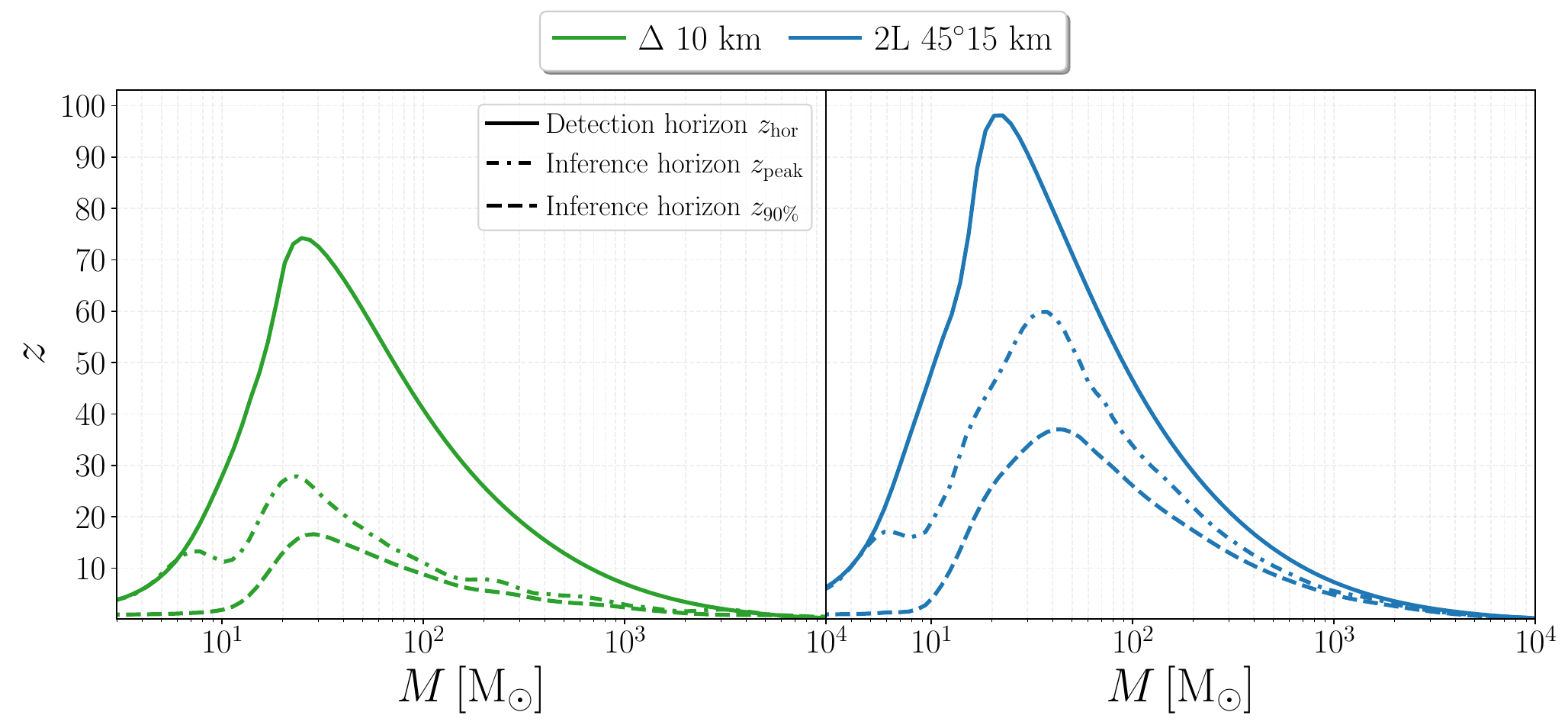}
    \caption{Detection horizon (solid line) compared to the ``inference horizon" $z_{90\%}$ (dashed line), defined as   the highest
redshift  one can confidently put a source beyond (at $90\%$ confidence level), and to the true value of the redshift that maximises the $90\%$ lower bound ($z_{\rm peak}$, dot--dashed line).}
    \label{fig:inference_hor}
\end{figure}

\subparagraph{Early warnings.} 

An important aspect related to the detection of BNS systems is the possibility of detecting the source prior to merger, in order to alert electromagnetic observatories for potential coincident detections. To study this possibility, it is useful to determine  how the SNR accumulates in time, as the frequency of the signal increases. 
In the two panels of figure~\ref{fig:SNRcumul_GW170817_allconf} 
we show the cumulative SNR as a function of 
the frequency, for a GW170817-like event, 
with the top axis of each plot providing 
the time to merger in minutes. Color 
curves refer to different ET configurations. 
The 2L-$45^{\circ}$ design with 20 km  arm 
length yields the largest SNR among the ET 
geometries considered, followed by the triangle 
configuration with 15 km arms. For this 
event, the presence of 
the LF instrument slighlty increases the 
final value of the SNR. However, at 
lower frequencies the difference between 
the HFLF and HF configurations becomes 
significant, with more than an order of 
magnitude improvement at 10 Hz. Of course, besides a sufficiently high SNR, an important aspect of early warnings is the sky localization of the source. This is discussed in detail in section~\ref{sec:div9_skyloc}.

\begin{figure}[t]
    \centering
    \includegraphics[width=\textwidth]{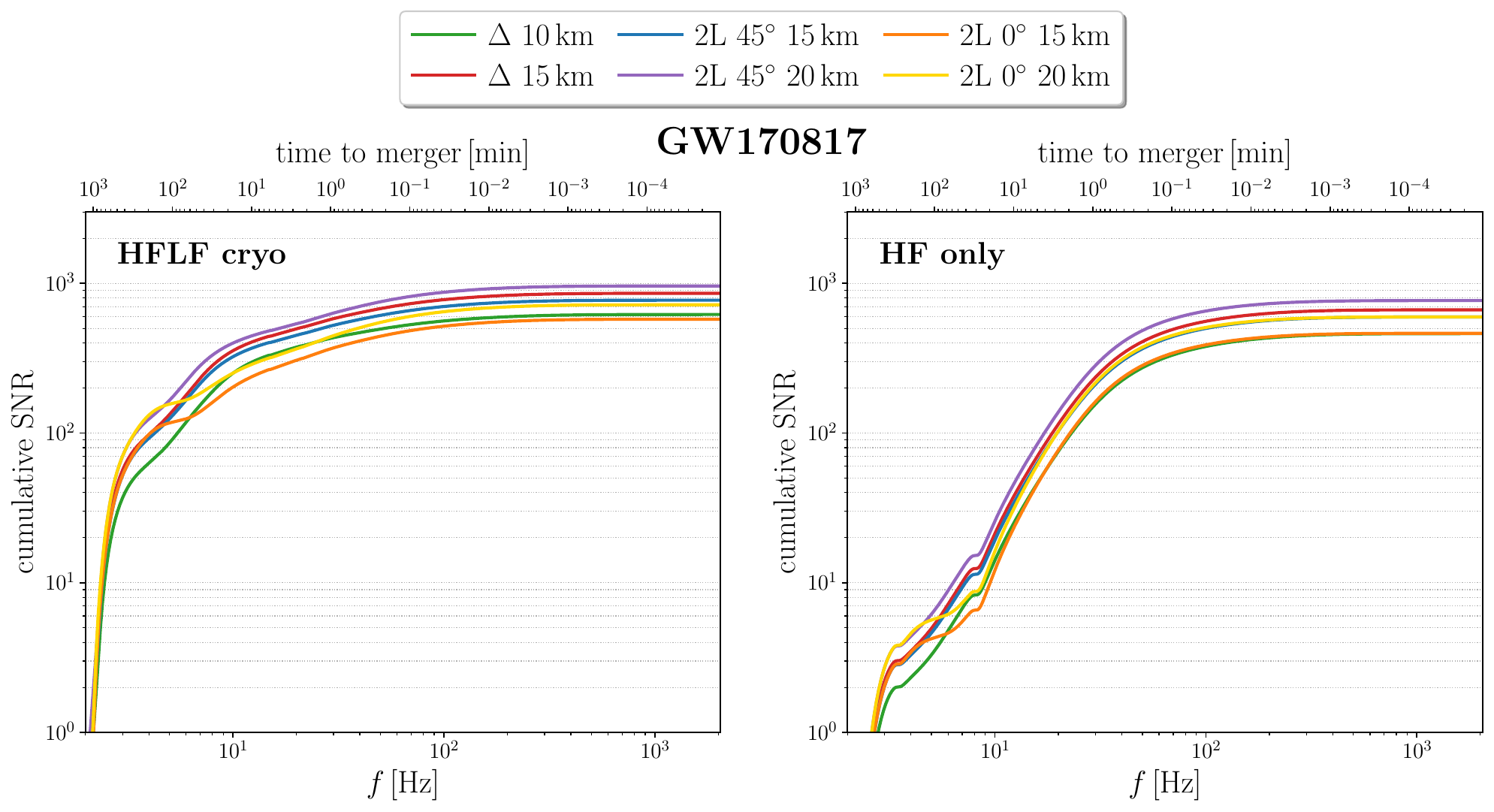}
    \caption{Cumulative SNR for a system with the 
    maximum likelihood parameters of GW170817, 
    observed by different ET configurations, 
    as a function of  GW frequency. The corresponding time to merger is shown in the upper horizontal axis. Left panel: using the ET sensitivity curve which includes both the HF and the LF interferometers, with the latter at cryogenic temperatures. Right panel: using only the  the HF interferometer.}
    \label{fig:SNRcumul_GW170817_allconf}
\end{figure}

\subparagraph{SNR and detection rates.} 
Another obvious metric to determine the potential of a given instrument is the detection rate. 
In figure~\ref{fig:SNRs_cumulatives_allconf_BBH_and_BNS} we show the cumulative number of observed 
events as a function of the SNR and ET design, 
for BBH (top panel) and BNS (bottom panel), using the merger taking place in 1~yr, assuming an independent $85\%$  duty cycle for each L-shaped detector, and in each of the three detectors composing the triangle, and using the same BBH and BNS 
catalogs already used  in \cite{Branchesi:2023mws} (whom we refer the reader for more details).
As a general 
result, we find that the HFLF-cryogenic configuration 
can significantly increase the number of 
detections with SNR between $10^2$ and $10^3$, compared to the HF-only configuration. 
For both BH and NS binaries, the 2L aligned 
design yields the best performance. Note 
however that this picture changes when 
comparing the accuracy reconstruction 
on the intrinsic and extrinsic parameters 
(see discussion afterward), for which 
the 2L-$45^{\circ}$ provides the most 
accurate forecasts.


\begin{figure}[t]
    \centering
    \includegraphics[width=0.49\textwidth]{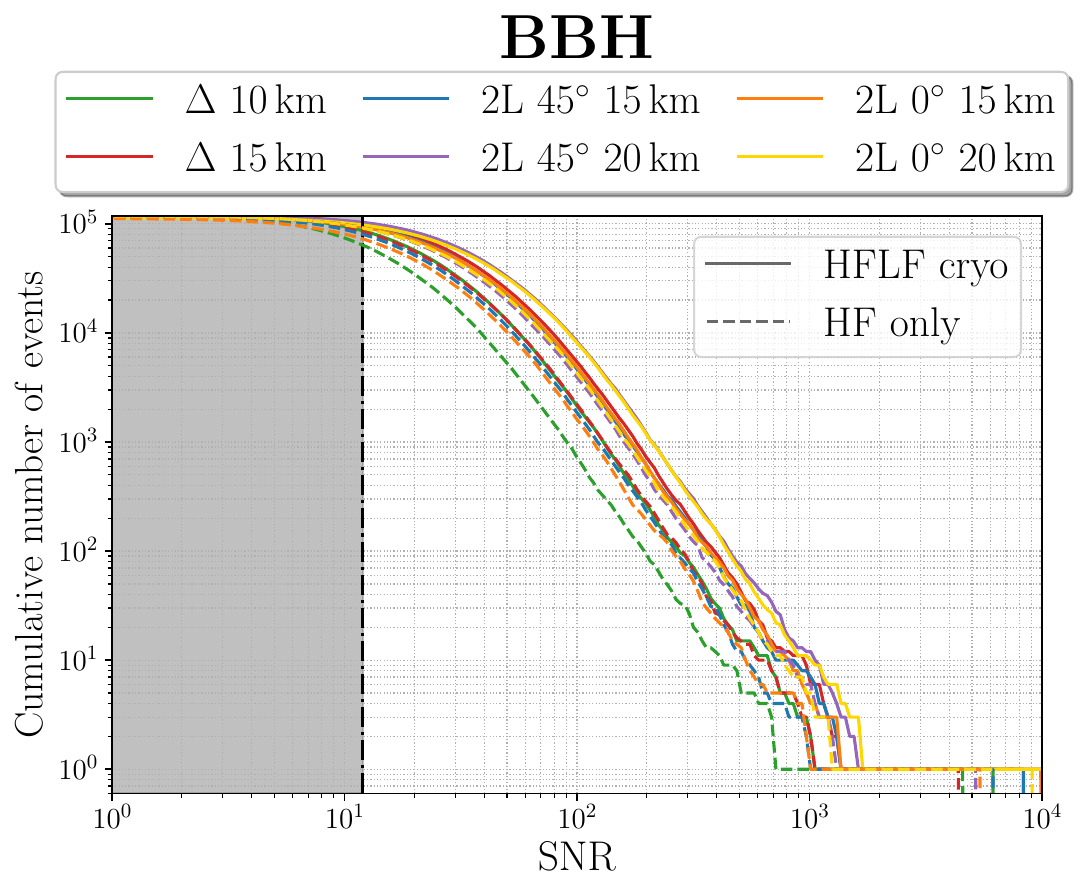}
    \includegraphics[width=0.49\textwidth]{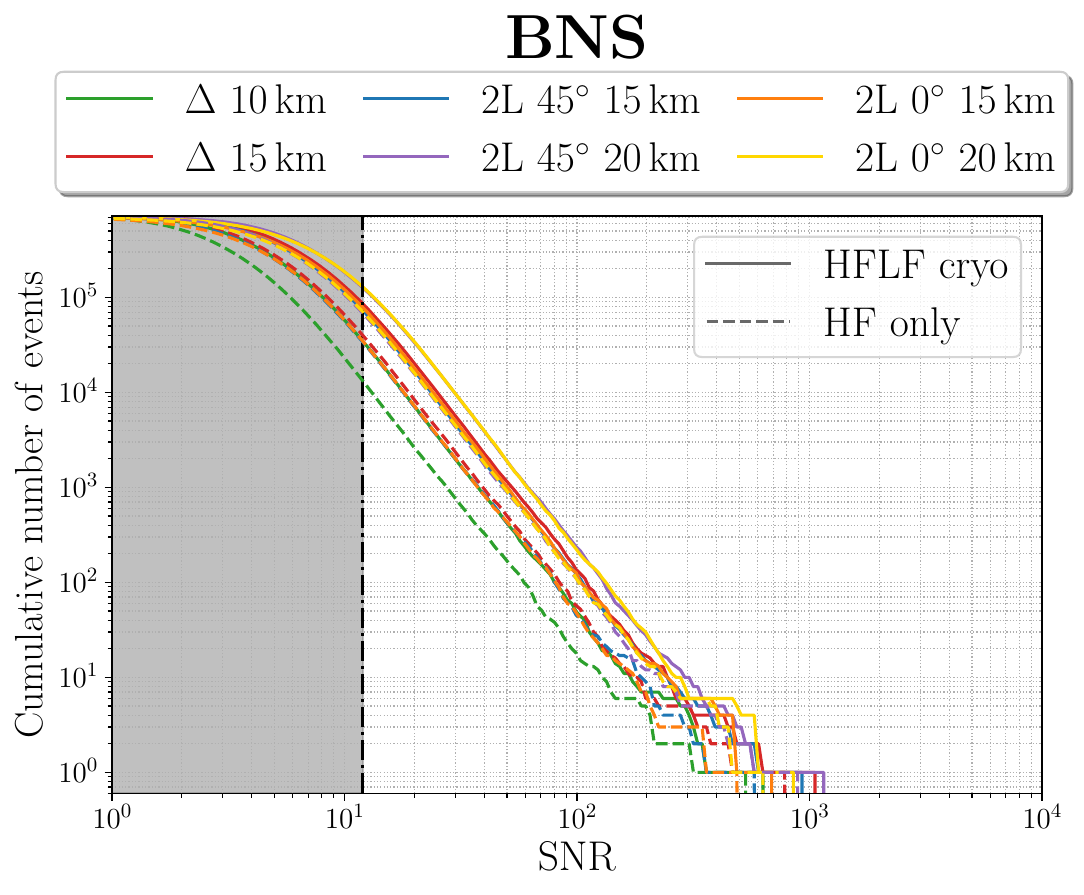}
    \caption{Cumulative distributions of the number of 
    BBH (left panel) and BNS (right panel) detections per year as a function of the SNR, 
    observed by ET for the different configurations  studied 
    in \cite{Branchesi:2023mws}. 
    The shaded area delimited by the dash-dotted line 
    identifies the region with non-detectable events with ${\rm SNR}\leq 12$.}
    \label{fig:SNRs_cumulatives_allconf_BBH_and_BNS}
\end{figure}

\subsubsection{Distance reconstruction and merger and pre-merger sky localization}
\label{sec:div9_skyloc}


The accuracy that ET can reach on  angular localization and luminosity 
distance is shown in figure~\ref{fig:Scumul_extrinsicpars_BBH_allconf_cryoHF} for BBHs and figure~\ref{fig:Scumul_extrinsicpars_BNS_allconf_cryoHF} for BNSs, using again  the same population model as in  \cite{Branchesi:2023mws} and the  same ET configurations as in figures~\ref{fig:SNRcumul_GW170817_allconf} and \ref{fig:SNRs_cumulatives_allconf_BBH_and_BNS}.

\begin{figure}[t]
    \centering
    \includegraphics[width=0.98\textwidth]{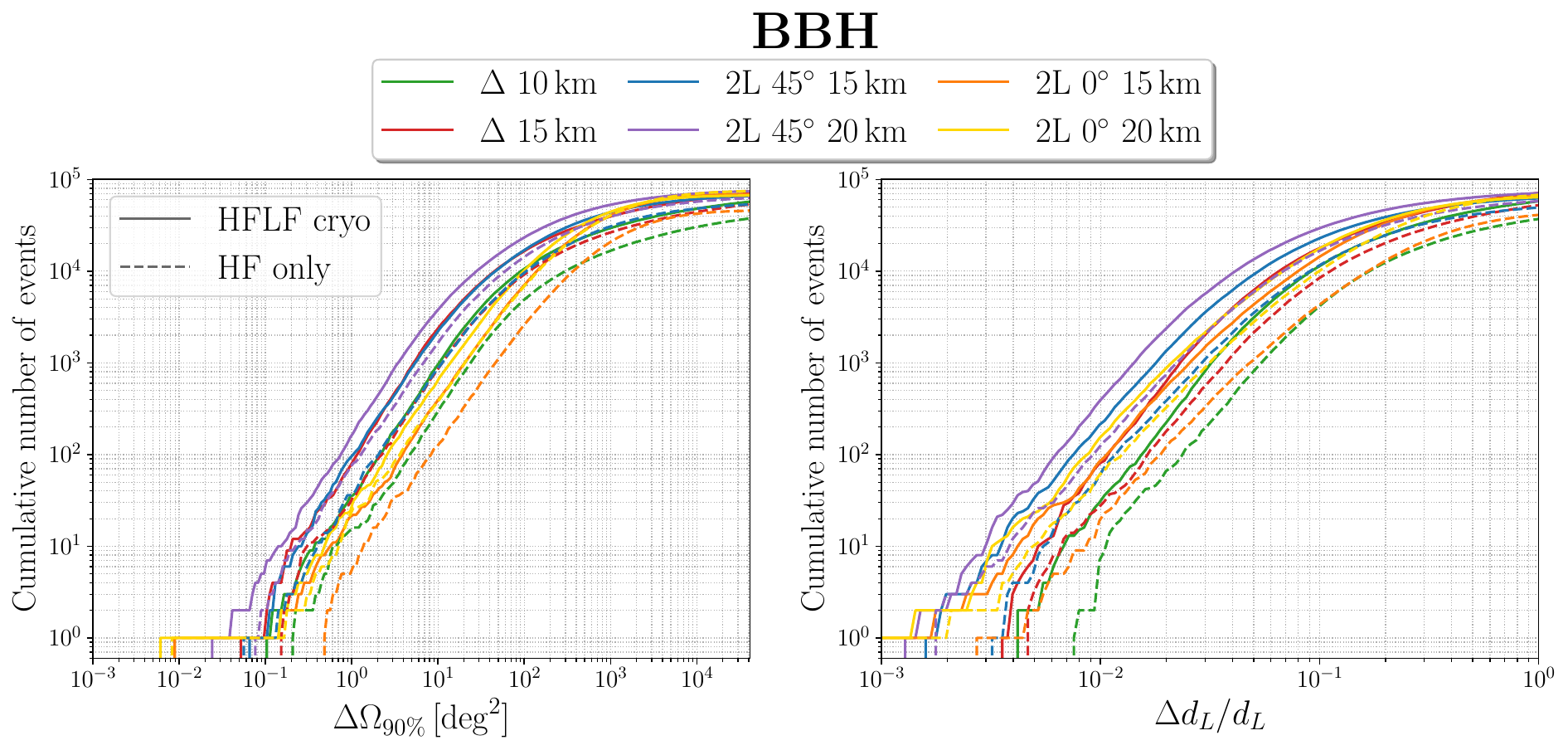}
    \caption{Cumulative distributions of the accuracy on angular localization (left panel) and luminosity distance
     (right panel) for  BBHs
    observed by ET in the different configurations studied 
    in \cite{Branchesi:2023mws}. 
}    \label{fig:Scumul_extrinsicpars_BBH_allconf_cryoHF}
\end{figure}

\begin{figure}[t]
    \centering
    \includegraphics[width=0.98\textwidth]{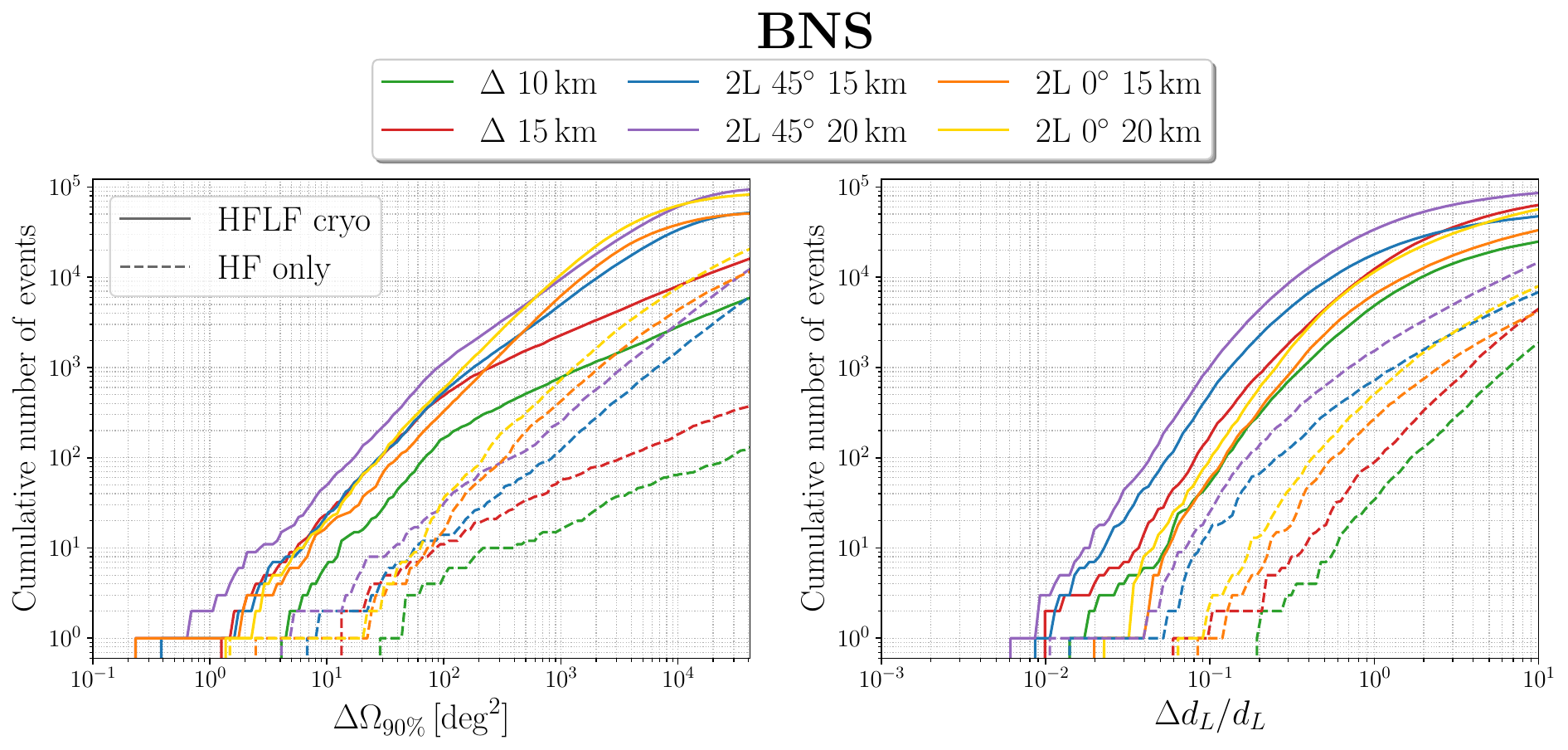}
    \caption{Cumulative distributions of  the accuracy on angular localization (left panel) and luminosity distance
     (right panel) for  BNSs
    observed by ET in the different configurations studied 
    in \cite{Branchesi:2023mws}. 
}   \label{fig:Scumul_extrinsicpars_BNS_allconf_cryoHF}
\end{figure}

\begin{figure}[t]
\centering
\includegraphics[scale=0.9]{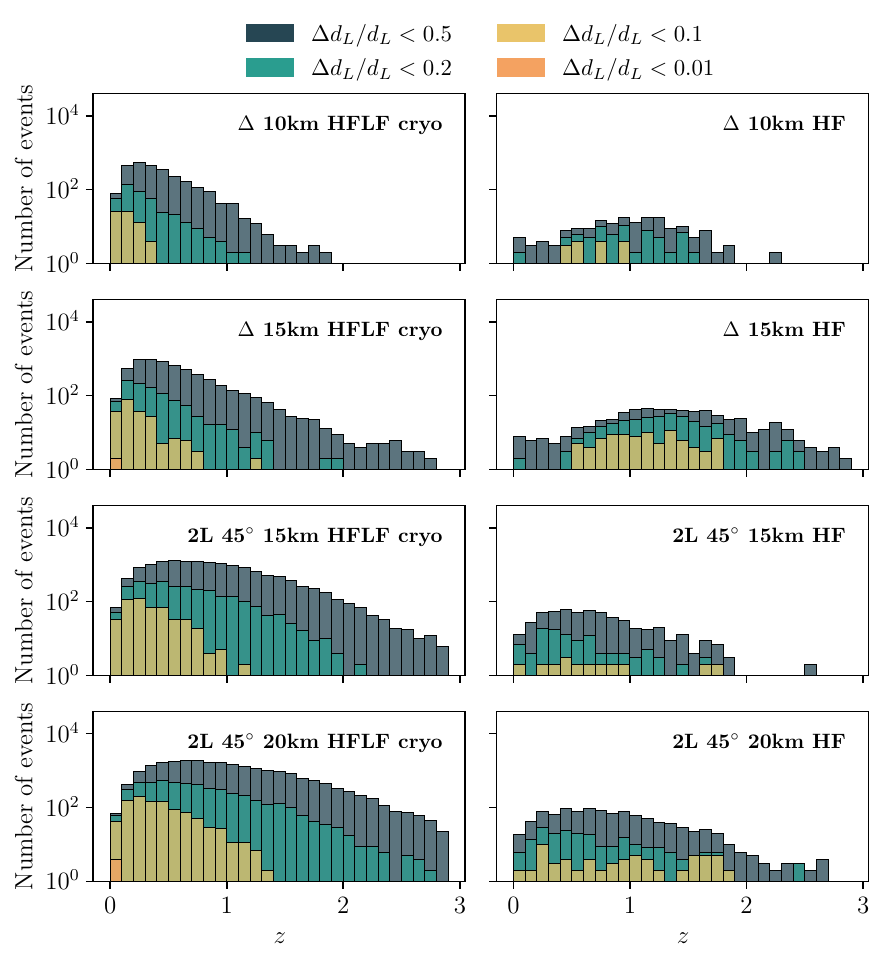}
\caption{Redshift distribution of the relative error on the luminosity distance of BNSs, with cuts at $50\%$, $20\%$, $10\%$ and $1\%$. Each panel represents a different detector configuration: 
the 10 and 15\,km triangle and the 15 and 20\,km 2L geometry. 
The left and right columns adopt full HFLF-cryo and the HF-only  
sensitivity curve, respectively. The total number of 
events corresponds to one year including duty cycle, with the same population of BNS as in \cite{Branchesi:2023mws}.}
\label{fig:dist_redshift}
\end{figure}
\begin{figure}[t]
\centering
\includegraphics[scale=0.9]{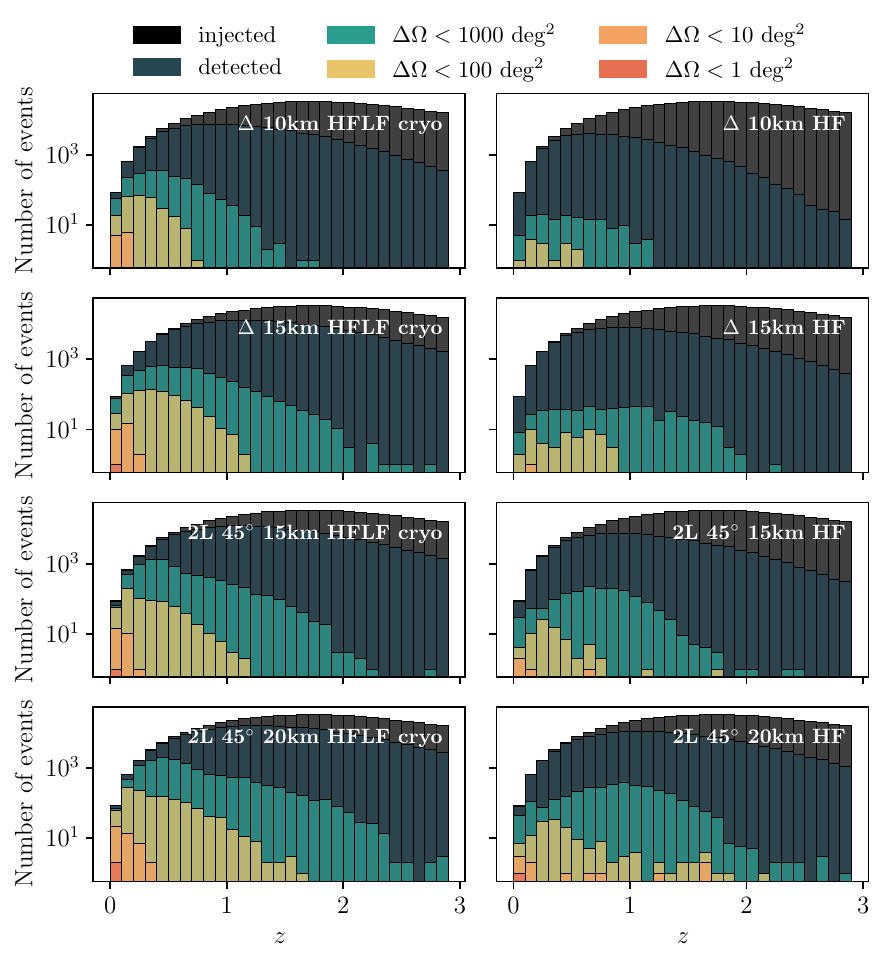}
\caption{Redshift distribution of the uncertainty on 
sky-localization of BNSs at different cuts: $1000$\,deg$^2$, $100$\,deg$^2$, $10$\,deg$^2$ and $1$\,deg$^2$. For comparison, we also 
plot the injected and detected signals with ${\rm SNR} >8$. 
Each panel represents a different detector configuration, considering 
full HFLF-cryo (left plots) and the HF-only detectors (right plots)
as in figure~\ref{fig:dist_redshift}. The total number of events corresponds to one year including duty cycle, with the same population of BNS as in \cite{Branchesi:2023mws}.}
\label{fig:sky_loc}
\end{figure}

Accurate estimates of the source sky-localization and distance play a crucial role in many ET science cases, and are particularly relevant for multi-messenger astrophysics based on detections of electromagnetic (EM) counterparts \cite{Branchesi:2023mws,Ronchini:2022gwk,Banerjee:2023,Loffredo:2024gmx}. 
For example, coincident observations between ET 
and EM observatories would allow us to probe complementary 
aspects of BNS mergers, investigating neutron star physics 
from the GW inspiral to the postmerger kilonovae (KNe) and 
short gamma-ray bursts (GRBs) emission.

Figures~\ref{fig:dist_redshift} and \ref{fig:sky_loc} show the result for BNSs  (and 
the same detector configurations) as a function of redshift (using again the same population as in
\cite{Branchesi:2023mws});
in figure~\ref{fig:dist_redshift} we show the results for luminosity distance.
Different colors 
correspond to cuts on the relative accuracy of 
$50\%$, $20\%$, $10\%$ and $1\%$; 
In figure~\ref{fig:sky_loc} we show the ET sky localization 
performance for BNSs 
as a function of redshift at different cuts of localization precision, $1000$\,deg$^2$, $100$\,deg$^2$, $10$\,deg$^2$ and $1$\,deg$^2$. 

We perform the study on 
different detector setups, the 10 and 15\,km triangle and 
the 15 and 20\,km 2L shape, both using the HFLF-cryogenic 
and the HF-only sensitivity curves. 
Comparison between left and right columns shows that 
losing the low-frequency sensitivity 
provides looser constraints on the source localisation.
ET in a 2L configuration with $20$\,km arms yields 
the best performance, with both the 2L-$15$\,km arm 
and the triangle-$15$\,km leading to similar results. 


A crucial aspect of ET multi-messenger capabilities is 
the localization of the source before the merger. 
Pre-merger alerts can allow EM instruments to narrow 
the binary position in the sky, and to detect prompt and early 
EM emissions that would not be observed otherwise. 
To assess ET pre-merger performances, we compare in  table~\ref{table:premerger_numbers} the 
number of sources that would be detected on a given 
period before the coalescence, with a 
sky-localization of (10, 100, 1000 deg$^{2}$), 
within redshift $z=1.5$. In particular, we distinguish 
between randomly oriented BNS systems and 
on-axis sources with an inclination angle lower 
than $15^{\circ}$, as these events could allow us to 
detect the early/prompt emission associated with 
relativistic jets. For well-localized pre-merger events with sky localization smaller than 100 deg², the 15\,km triangle configuration performs better than both the 10 km triangle and the 2L with 15 km misaligned arms, and its performance is comparable to the 2L with 20 km misaligned arms. Similar results hold for systems with viewing angles below $15^{\circ}$.

\begin{table}
\centering
\caption{Number of BNS mergers per year detected before 
the merger within $z=1.5$ for the different ET 
configurations with HFLF-cryo sensitivity. We 
show three scenarios, with 30, 10, and 1 minute(s) 
before the merger.  For each detector we give the number of 
events with sky-localization ($90\%$CL) within 10, 100, 1000 deg$^{2}$, and the total number of detected sources. The 2L configurations are assumed to be misaligned by $45^{\circ}$ if not otherwise specified.
Observations assume a duty cycle of $0.85$.}
{\small{
\begin{tabular}{ c | c | r | r | r | r | r | r }
\multirow{2}{*}{} &$\Delta\Omega_{90\%}$     & \multicolumn{3}{c|}{All orientation BNSs} & \multicolumn{3}{c}{BNSs with $\Theta_{v} < 15^{\circ}$ } \\
& [deg$^{2}$]    & 30 min  &    10 min                   & 1 min  & 30 min  &    10 min                   & 1 min  \\ \toprule    
\multirow{4}{*}{$\rm \Delta 10 km$} 
&10 &0 &1 &5 &0 &0 &0\\ 
&100 &10 &39 &113 &2 &8 &20\\ 
&1000 &85 &293 &819 &10 &34 &132\\ 
&All detected &905 &4343 &23597 &81 &393 &2312\\
\midrule
\multirow{4}{*}{ $\rm \Delta 15 km$} 
&10 &1 &5 &11 &0 &1 &1\\ 
&100 &41 &109 &281  &6 &14 &36\\
&1000 &279 &806 &2007  &33 &102 &295\\
&All detected &2489 &11303  &48127 &221 &1009 &4024 \\
\midrule
\multirow{4}{*}{ 2L-15 km}
&10 &0 &1 &8 &0 &0 &0\\
&100 &20 &54 &169 &2 &7 &26\\
&1000 &194 &565 &1399 &23 &73 &199\\
&All detected &2172 &9598 &39499 &198 &863 &3432\\
\midrule 
\multirow{3}{*}{ 2L-20 km}
&10 &2 &4 &15 &1 &1 &2\\
&100 &39 &118 &288  &7 &19 &47\\
&1000 &403 &1040 &2427 &47 &128 &346\\
&All detected &4125 &17294 &56611 &363 &1588 &4377 \\
\bottomrule
\end{tabular}
}}
\label{table:premerger_numbers}
\end{table}

\subsubsection{Inference of intrinsic parameters, and golden binaries}
\label{sec:div9_infer_intrinsic_par_golden_binaries} 


The ET exquisite ability to reconstruct the 
source intrinsic parameters, like masses, spins, and eccentricity 
of a binary system is key to study the properties  of the 
astrophysical distributions of detected mergers, see the discussion in section~\ref{section:div3}.
Thanks to the $\sim 10^5$ expected observations of CBCs per year, 
ET would shed new light on the formation channels of compact binaries, 
possibly disentangling sub-populations forming due to distinct 
mechanisms. 

To investigate this aspect of the science case we show in figures~\ref{fig:intrinsicpars_cumulatives_allconf_BBH} and \ref{fig:intrinsicpars_cumulatives_allconf_BNS} the 
cumulative distribution for the errors on some  
intrinsic parameters. Results are shown assuming different 
ET configurations, using again the same catalog as in  \cite{Branchesi:2023mws}.
For BBHs we consider precessing binaries and use as 
GW template the \texttt{IMRPhenomXPHM} waveform 
model~\cite{Pratten:2020ceb}. We restrict the BNS 
analysis to spin--aligned systems and employ the 
\texttt{IMRPhenomD\_NRTidalv2} waveform, which also 
includes tidal effects~\cite{Dietrich:2019kaq}.
Unless otherwise stated, we consider sources detected with an optimal matched--filter SNR larger than 12.

In the top row of both figures we report the $1\sigma$ 
accuracy in the reconstruction of the source--frame 
chirp mass and of the symmetric mass ratio. For BBHs 
we find ET can measure $M_c$ with a relative percentage 
accuracy of ${\cal O}(10^{-3})$, 
while $\eta$ can be inferred with an uncertainty 
as small as $\lesssim{\cal O}(10^{-7})$.  
The 2L-$45^{\circ}$ configuration 
provides the strongest constraints on both mass parameters. 
Indeed this geometry yields the best measurement of the 
luminosity distance, which affect the reconstruction of 
source--frame parameters.

This feature is even more relevant for the reconstruction 
of the BNS chirp mass, with the presence of the LF instrument 
at cryogenic sensitivity also being crucial.
The inference of the symmetric mass ratio is 
also affected in case of loss of the LF instrument, even if in a less dramatic way, with errors on $\eta$ as small as $\lesssim{\cal O}(10^{-5})$. 

In the lower panels of figure~\ref{fig:intrinsicpars_cumulatives_allconf_BBH} we 
show the accuracy measurement on the BBH spins. 
The uncertainties on $\chi_{1,2}$ (where $\chi_1$ refers to the more massive star and $\chi_2$ to the less massive) are not dramatically 
affected by ET geometry. Both spins are reconstructed 
with similar accuracy, 
which can be ${\cal O}(10^{-5})$ in the best case 
scenario, with the
constraints on the spin of the more massive 
binary component being in general stronger. Note however that, for larger errors, 
of the order of ${\cal O}(10^{-2})$, the 
LF instrument play a more significant role. 

The lower left panel of figure~\ref{fig:intrinsicpars_cumulatives_allconf_BNS} 
shows the $1\sigma$ cumulative distribution 
of the effective spin parameter defined in 
\eq{eq:chieff_def} for BNS systems. 
The reconstruction accuracy on this parameter 
improves significantly with the LF instrument, 
allowing for uncertainties as small as 
${\cal O}(10^{-3})$. 
Finally, in the lower right panel of figure~\ref{fig:intrinsicpars_cumulatives_allconf_BNS} 
we present the accuracy distribution on the 
dimensionless tidal parameter $\tilde{\Lambda}$ defined in \eq{eq:tildeLam_def}, whose relevance for  
science cases related to nuclear physics has been discussed in section~\ref{section:div6}. 
The relative errors on $\tilde\Lambda$ depend on the 
detector configuration, with narrower constraints, 
of the order of percent, obtained again by ET in 
HFLF mode.


Figures~\ref{fig:hist_vs_z_allconf_BBH}--\ref{fig:hist_vs_z_allconf_BNS} show the redshift 
distribution of \textit{golden binaries}, i.e. 
BBHs with either SNR larger than 100 
(resp. 30 for BNSs) or with 
reconstruction of the source--frame chirp mass 
smaller than $1\%$ (resp. $5\%$ for BNS). 
Indeed, being able to achieve accurate measurements of 
source parameters up to high--$z$ is pivotal in order 
to disentangle formation channels which might feature 
different redshift evolution. 
For BBH systems, we find that golden binaries can be 
observed up to $z\sim3$ or higher only if the 
LF instrument is present, with the best geometry 
being again the 2L-$45^{\circ}$. Similar results 
hold for BNSs, with the highest redshift being 
$z\sim 1$. 

Finally, in figure~\ref{fig:ecc_errs_allconf_fixedsources} 
we show the relative error on the eccentricity 
in the small $e_0$ limit. 
We focus here on a single-source analysis, 
fixing the binary intrinsic parameters and the 
luminosity distance (see caption of the figure~\ref{fig:ecc_errs_allconf_fixedsources}).
Note that in this case we use as  template the 
\texttt{TaylorF2} waveform approximant~\cite{Buonanno:2009zt, Ajith:2011ec, Mishra:2016whh} augmented with 
the inclusion of the eccentricity~\cite{Moore:2016qxz}. 
Our results show that the error accuracy greatly 
improves for lighter binaries. Indeed, constraints on 
$e_0$ are dominated by the early part of the inspiral 
and improve for systems that stay in band longer, 
performing larger amount of cycles. 
The errors do not change significantly as a function 
of the detector geometry. However, the inclusion 
of the LF instrument allows to constrain  
the eccentricity with a relative error 
smaller than 100\% down to $e_0\sim10^{-4}$.



\begin{figure}[tp]
    \centering
     \includegraphics[width=\textwidth]{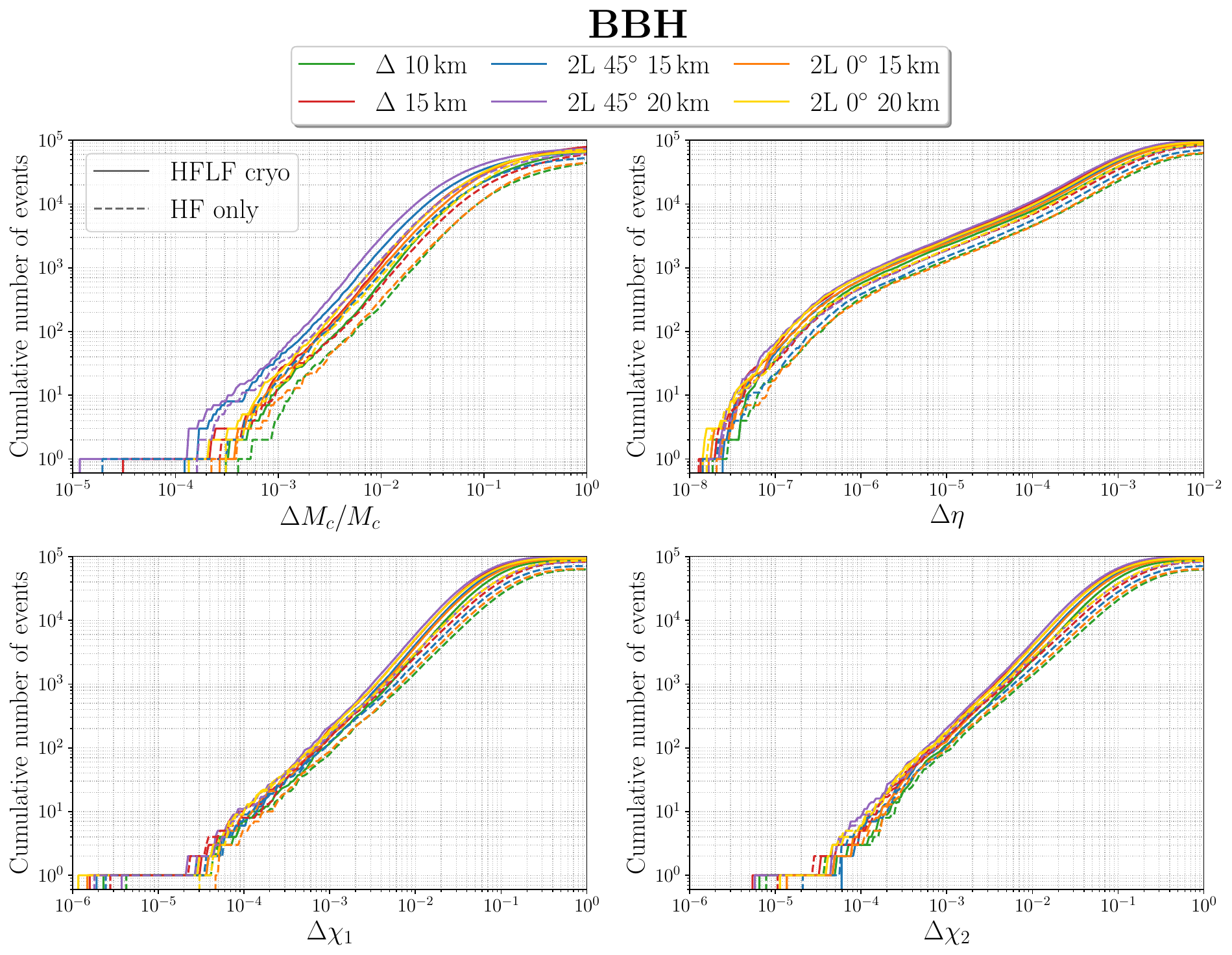} 
    \caption{Cumulative distributions of the number of detections per year for the relative errors on intrinsic parameters of  BBH systems observed by ET in the different configurations studied in \cite{Branchesi:2023mws}. We report the relative uncertainties on the source-frame chirp masses, and the uncertainties on the symmetric mass ratio and spin magnitudes of the two objects.}
    \label{fig:intrinsicpars_cumulatives_allconf_BBH}
\end{figure}

\begin{figure}[tp]
    \centering
     \includegraphics[width=\textwidth]{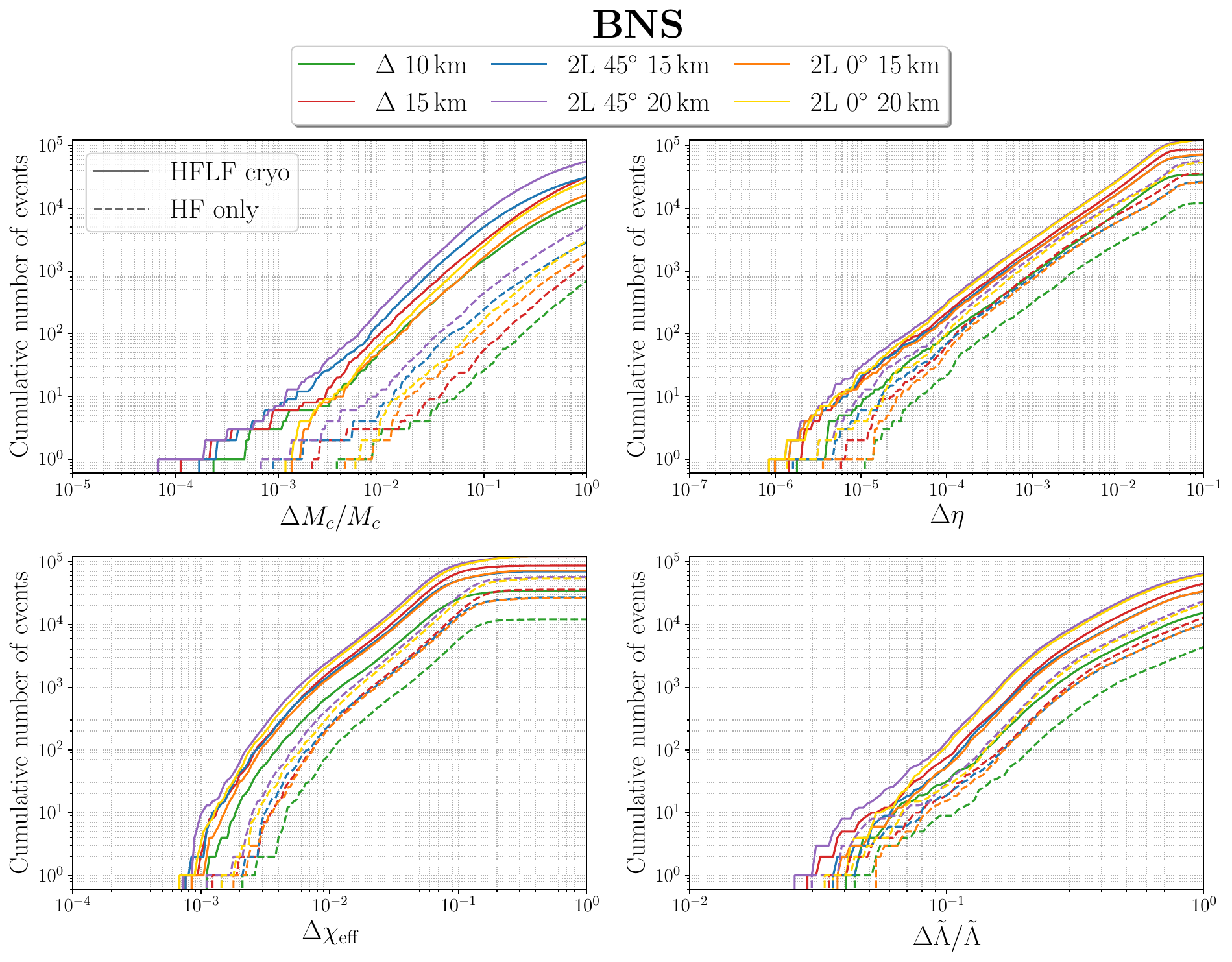}
    \caption{As in figure~\ref{fig:intrinsicpars_cumulatives_allconf_BBH} for BNS sources. We report the relative uncertainties on the source-frame chirp masses and adimensional tidal deformability combinations $\tilde{\Lambda}$, and the uncertainties on the symmetric mass ratio and effective spin parameters.}
    \label{fig:intrinsicpars_cumulatives_allconf_BNS}
\end{figure}

\begin{figure}[tp]
    \centering
    \begin{tabular}{c c}
       \includegraphics[width=.48\textwidth]{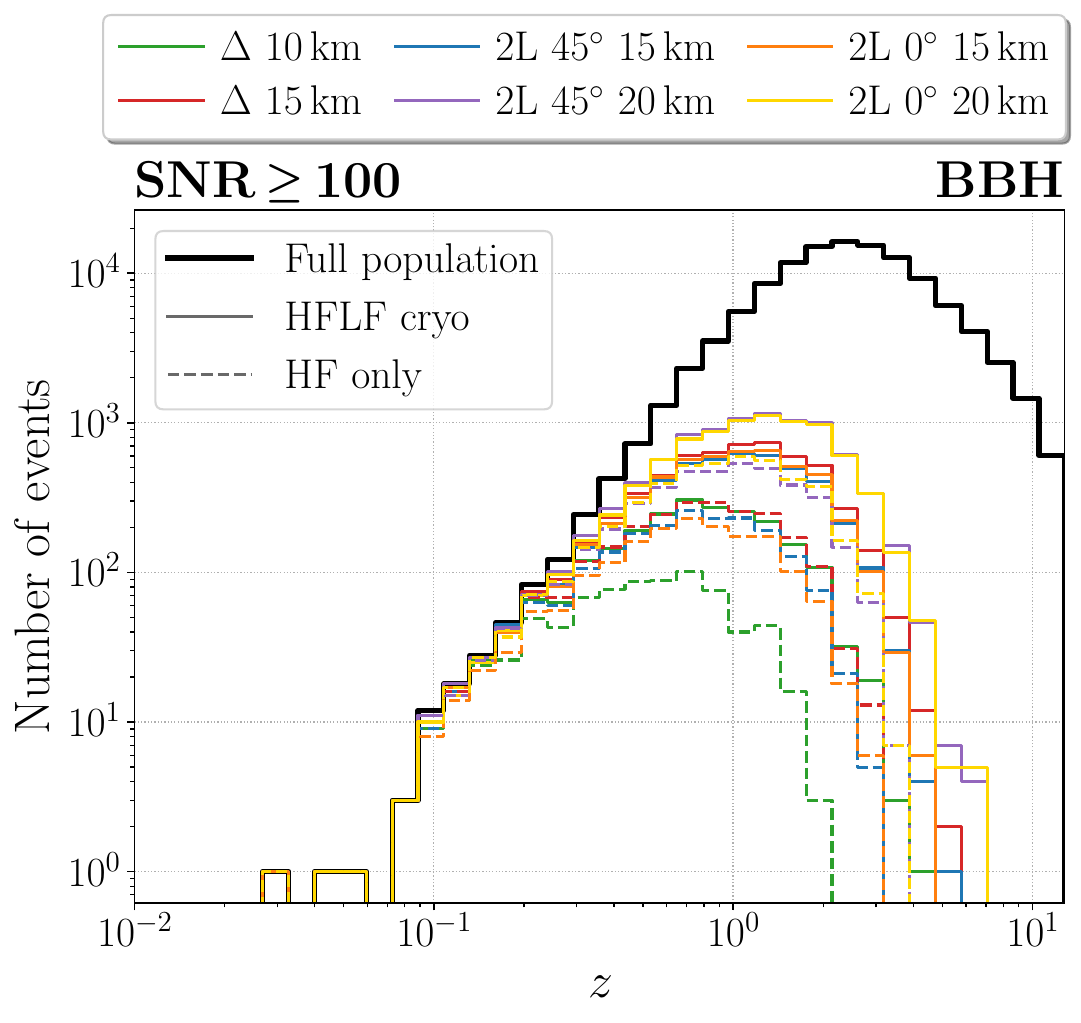}   &  \includegraphics[width=.48\textwidth]{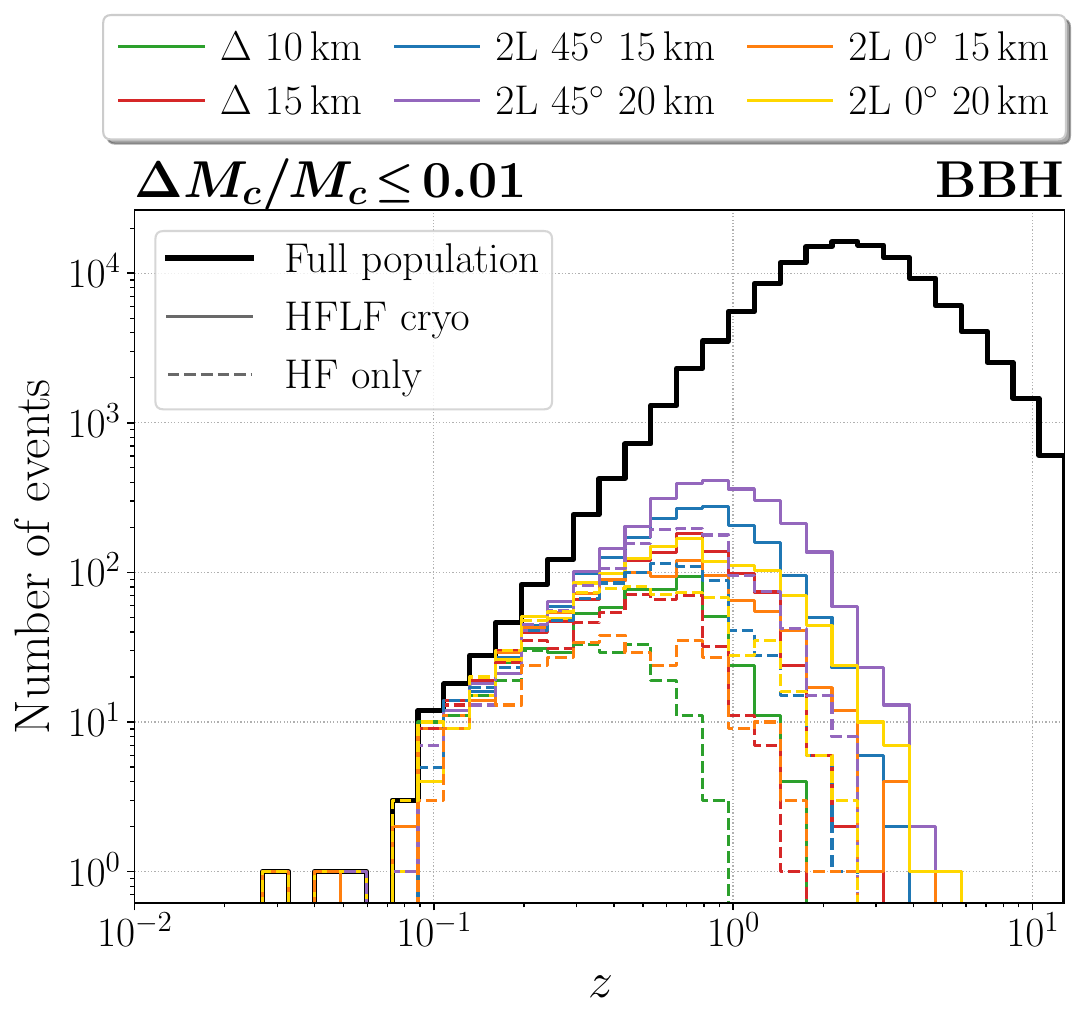}
    \end{tabular}
    \caption{Redshift distributions of detected BBH events at ET in the different configurations studied in \cite{Branchesi:2023mws} having ${\rm SNR}\geq100$ (\emph{left panel}) and relative error on the source-frame chirp mass smaller than $1\%$ (\emph{right panel}).}
    \label{fig:hist_vs_z_allconf_BBH}
\end{figure}

\begin{figure}[tp]
    \centering
    \begin{tabular}{c c}
       \includegraphics[width=.48\textwidth]{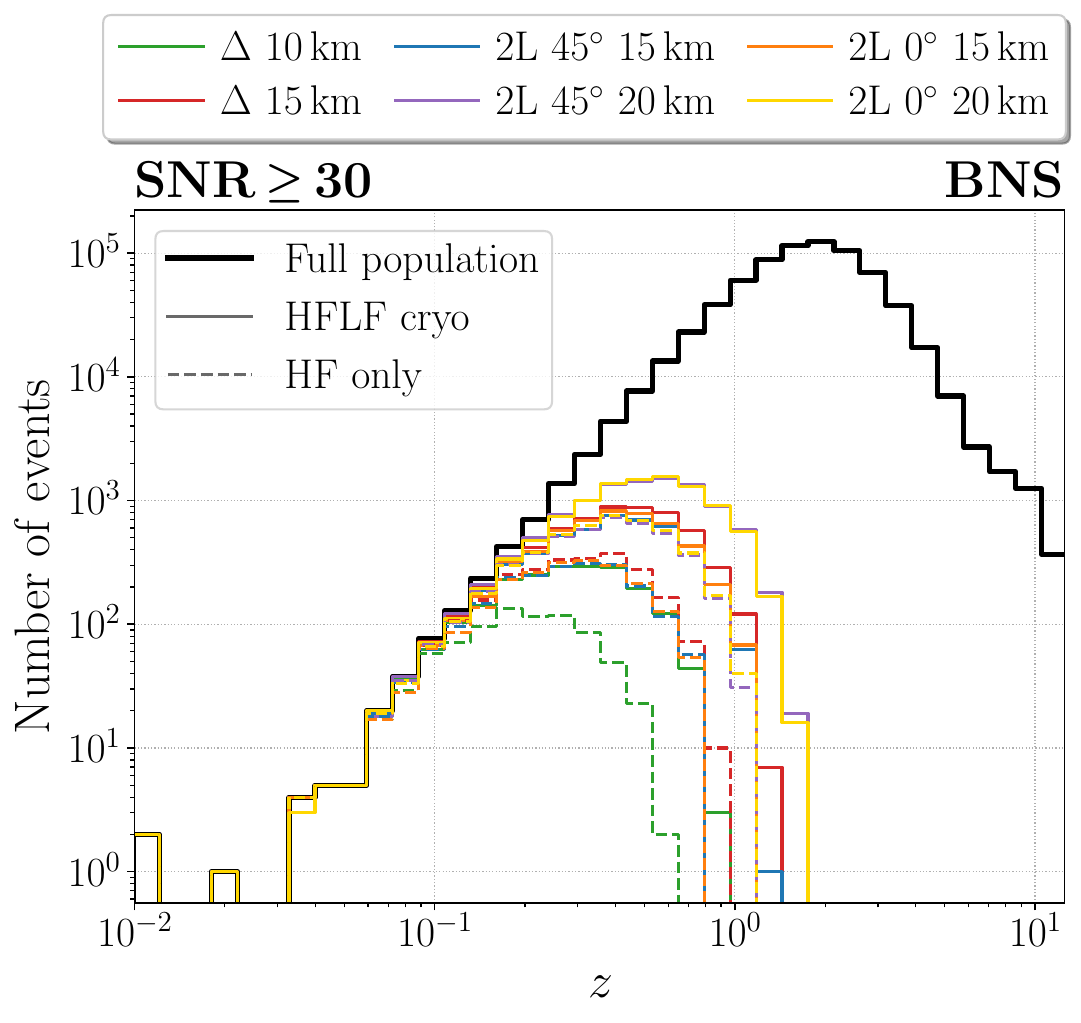}   &  \includegraphics[width=.48\textwidth]{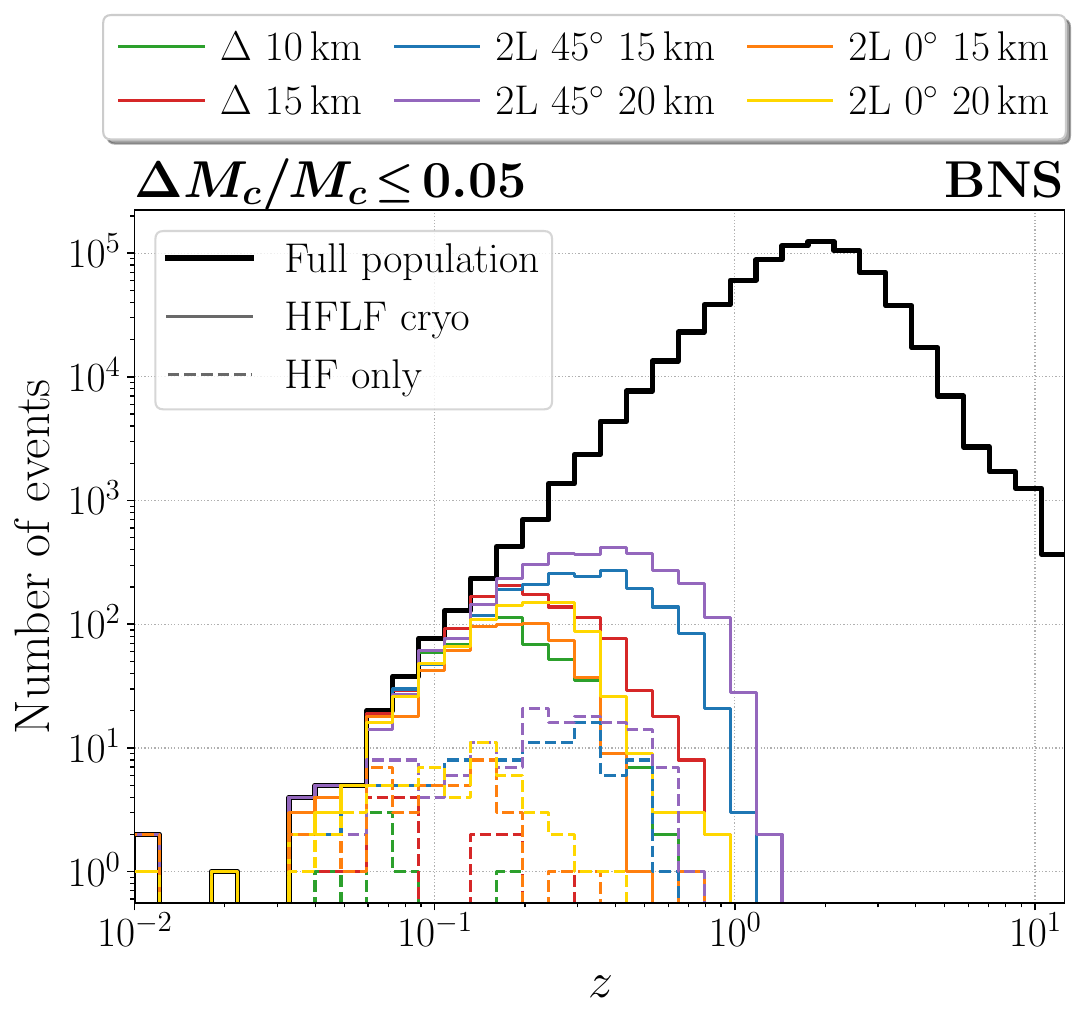}
    \end{tabular}
    \caption{Redshift distributions of detected BNS events at ET in the different configurations studied in \cite{Branchesi:2023mws} having ${\rm SNR}\geq30$ (\emph{left panel}) and relative error on the source-frame chirp mass smaller than $5\%$ (\emph{right panel}).}
    \label{fig:hist_vs_z_allconf_BNS}
\end{figure}

\begin{figure}[tp]
    \centering
    \includegraphics[width=\textwidth]{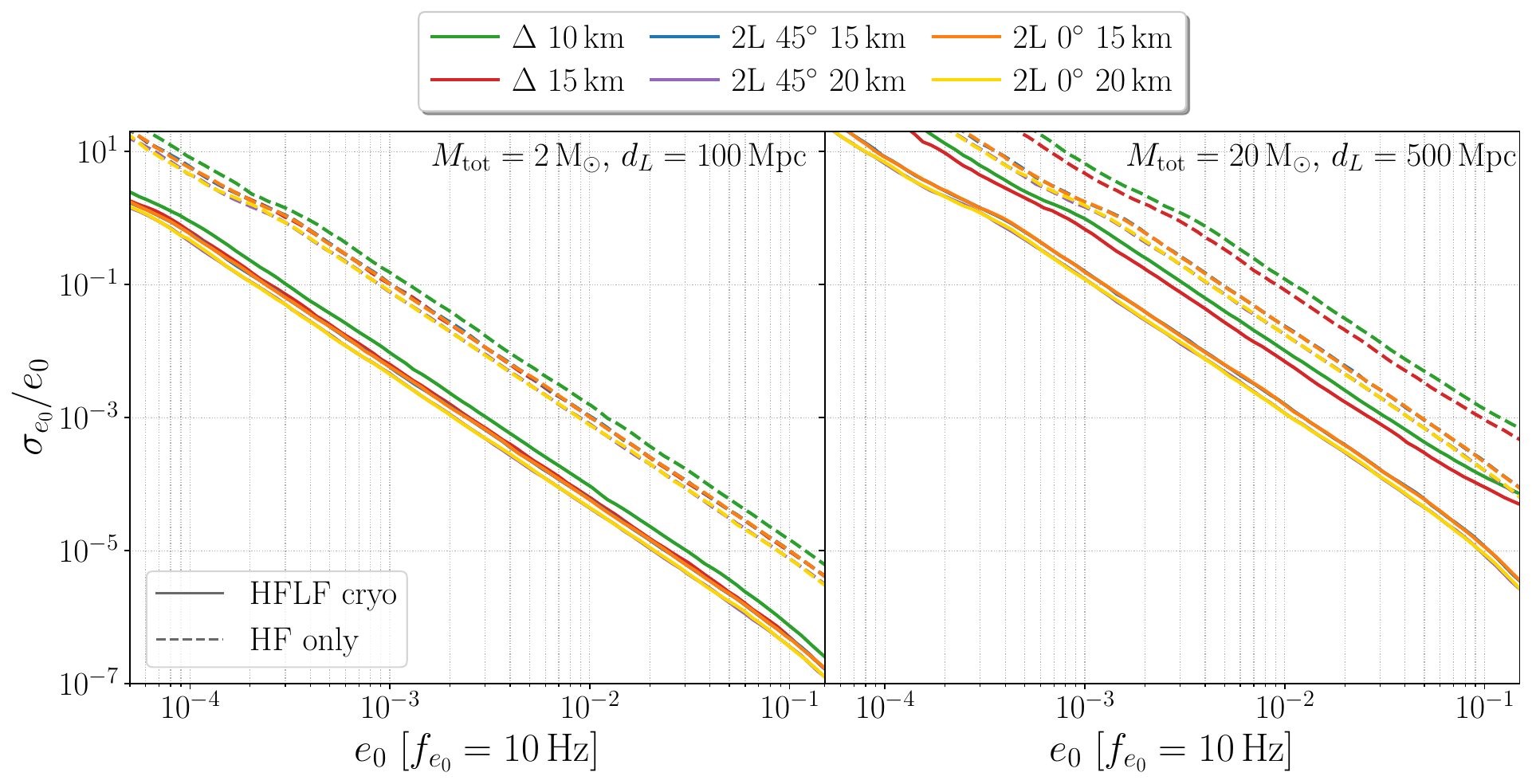}
    \caption{Relative errors on the  eccentricity parameter $e_0$ at $f_{e_{0}} = 10~{\rm Hz}$ attainable at the various ET configurations studied in \cite{Branchesi:2023mws}, in the regime of small eccentricities. We consider equal-mass sources with optimal sky position and orientation for each detector, negligible spins and $M_{\rm tot} = 2~{\rm M}_{\odot}$ at $d_L=100~{\rm Mpc}$ (\emph{left panel}) and $M_{\rm tot} = 20~{\rm M}_{\odot}$ at $d_L=500~{\rm Mpc}$ (\emph{right panel}).}
    \label{fig:ecc_errs_allconf_fixedsources}
\end{figure}

\subsubsection{Detection of population features}
\label{sec:div9_forecast_population_properties} 

Beyond single--event parameters, accurately pinpointing properties and features of the population of compact objects is a central target of 3G observatories. 
For example, the pair-instability supernova process can lead to an accumulation of events, modeled as Gaussian component in the BBH mass spectrum; the tilt angles of the spin vectors are modeled as a mixture of a uniform component, representing dynamically formed binaries with isotropic spin orientations, and a truncated Gaussian component, to account for preferentially aligned spins in isolated binary formation channels; the BBH merger rate is expected to peak at a redshift related to the peak of the star formation, with a subsequent decrease; see section~\ref{section:div5} for extended discussion.
Accurate measurements of those parameters would lead to precision constraints on the physics of compact binaries, and determining ET's capability in constraining them is essential.

As a Figure of Merit, we consider the relative uncertainties for the population hyperparameters as a function of observation time and of the corresponding number of detected events $N_{\rm det}$, quantifying in particular the time needed for achieving a $1\%$ constraint. 
Those are computed with the population-level Fisher matrix software described in section~\ref{sect:inferencepopdiv9}.
 \begin{figure*}[t]
    \includegraphics[width=\textwidth]{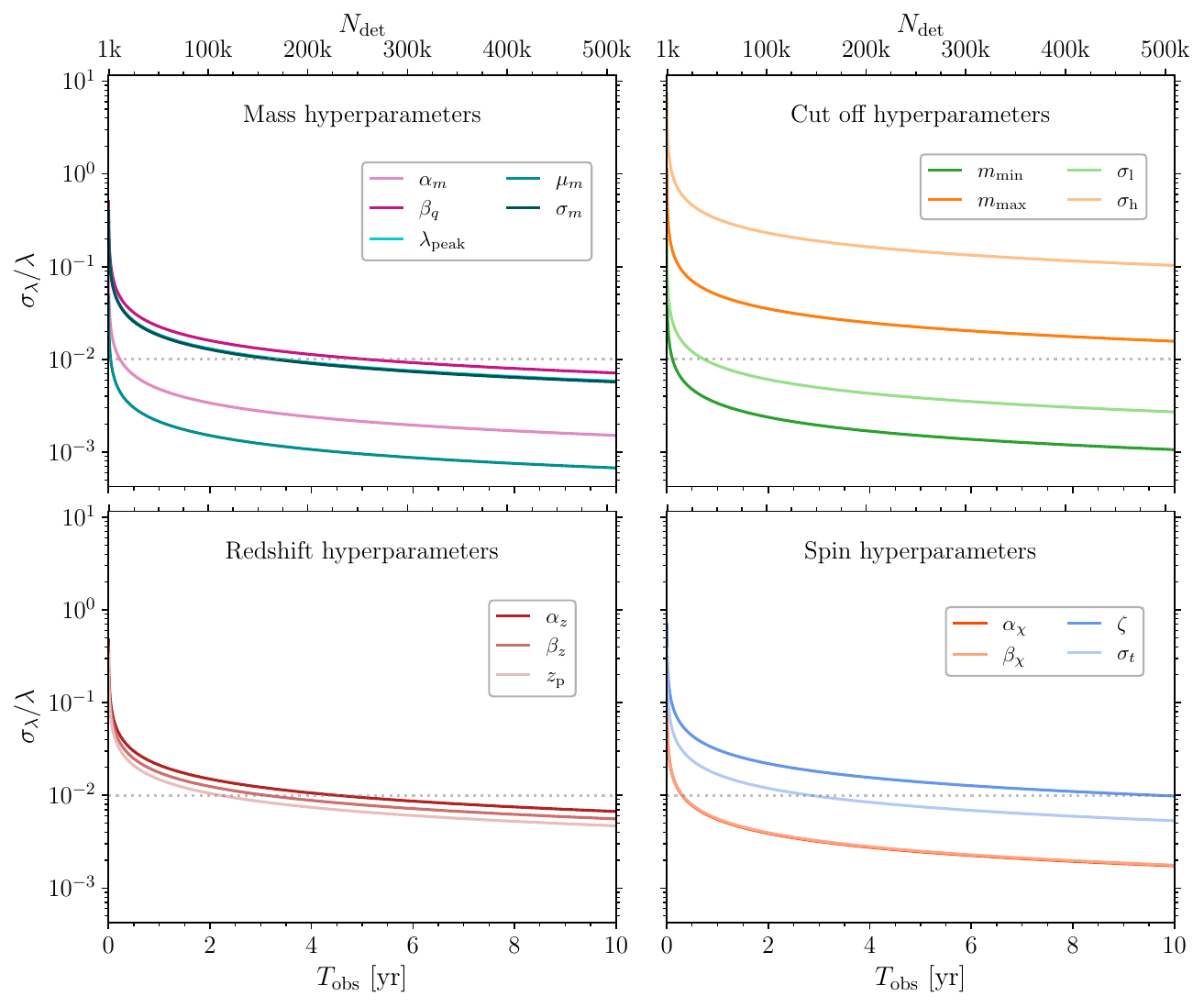}
    \caption{Relative errors on the population hyperparameters $\lambda$ for BBH systems as a function of observation time (bottom $x$-axes) for ET in the triangular configuration, based on the fiducial population model described in the text. The upper $x$-axes display the corresponding number of detected events. Each panel illustrates the hyperparameters associated with broad mass features, mass cutoffs, merger redshift evolution, and BH spins. Dashed horizontal lines indicate the threshold for percent-level accuracy, $\sigma_{\lambda}/\lambda = 1\%$, as a visual reference. Figure adapted from ref.~\cite{DeRenzis:2024dvx}.}\label{fig:relative_errors}
\end{figure*}
Figure~\ref{fig:relative_errors} presents the result for a single, triangular ET detector with 10 km arms located at the Sardinia site.\footnote{We use the 10~km noise curve from ref.~\cite{Branchesi:2023mws}, available at \href{https://apps.et-gw.eu/tds/?r=18213}{apps.et-gw.eu/tds/?r=18213}. The minimum frequency for the analysis is set to 2Hz.
We use the \texttt{IMRPhenomXPHM}~\cite{Pratten:2020fqn,Garcia-Quiros:2020qpx,Pratten:2020ceb} waveform model and the 
single-event Fishers are computed in terms of the 15 source parameters describing a quasi-circular BBH merger, see \eq{CBCparams}.} 
The hyperparameters have values calibrated on the latest LVK observational constraints~\cite{KAGRA:2021duu},\footnote{Additional details on the population models can be found in appendix~A of ref.~\cite{DeRenzis:2024dvx}.} and are categorized into four groups in figure~\ref{fig:relative_errors}, describing respectively: 
\begin{itemize}

\item broad mass features -- $\alpha_m$, $\beta_q$, $\mu_m$, $\sigma_m$, $\lambda_{\rm peak}$. 
The BBH mass spectrum is modeled with the standard {\sc Power--Law + Peak}~\cite{KAGRA:2021duu,Talbot:2018cva},\footnote{Actually, the model slightly differs from the standard LVK implementation for the tapering of the edges of the mass distribution. In particular, here we adopt a differentiable polynomial smoothing function, see Appendix A of Ref~\cite{DeRenzis:2024dvx} for details.} where the primary mass $m_1$ follows a power-law distribution with slope $\alpha_m$, mixed to a Gaussian component (mean $\mu_m$, standard deviation $\sigma_m$, mixing weight $\lambda_{\rm peak}$), and the secondary mass $m_2$ is also described by a power-law distribution (spectral index $\beta_q$). 

\item mass cutoffs -- $m_{\rm max}$, $m_{\rm min}$, $\sigma_{\rm h}$, $\sigma_{\rm l}$. The model supports a minimum mass of $m_{\rm min}-\sigma_{\rm l}$ and a maximum mass of $m_{\rm max}+\sigma_{\rm h}$ with $\sigma_{\rm l}$, $\sigma_{\rm h}$ being the lower and upper tapering scales.

\item redshift evolution -- $z_{\rm p}$, $\alpha_z$, $\beta_z$ parameters of the {\sc Madau--Dickinson} profile~\cite{Madau:2014bja} for the merger rate evolution with redshift [redshift $z_{\rm p}$ at the peak of the rate, power-law indexes governing the increase (decline) at low (high) redshift].

\item spins  -- $\alpha_\chi$, $\beta_\chi$, $\zeta$, $\sigma_t$. Component spin magnitudes follow a Beta distribution with shape parameters $\alpha_\chi$ and $\beta_\chi$.  
The tilt angles of the spin vectors follow a mixture model with a uniform component, and a truncated Gaussian component centered at 1, with standard deviation $\sigma_t$ and weight $\zeta$.

\end{itemize}

\noindent
Figure~\ref{fig:relative_errors} illustrates that certain hyperparameters, such as the minimum mass $m_{\text{min}}$, the spectral index $\alpha_m$, and the Gaussian peak position $\mu_m$ in the primary mass spectrum, are measured with remarkable precision, achieving 1\% accuracy within a few months of 3G observations. These accurate constraints are due to the fact that these hyperparameters influence the mass distribution where the merger rate is higher.
Measuring $m_{\text{min}}$ can help confirm the existence of the hypothesized mass gap between BHs and NSs. The spectral index $\alpha_m$ provides insight into whether the BH mass distribution is steeper or shallower compared to stars, while $\mu_m$ marks the transition between BHs formed via supernovae and those potentially constrained by pair-instability physics.
The secondary-mass spectral index $\beta_q$ is measured with lower precision compared to $\alpha_m$. This disparity arises because the primary mass $m_1$ is inherently easier to constrain than the secondary mass $m_2$, and also due to the functional form of the population model $p_{\rm pop}(\theta|\lambda)$, which separates the two contributions as $p(m_1)p(m_2|m_1)$.
Conversely, hyperparameters like the maximum mass $m_{\text{max}}$ and the smoothing length $\sigma_{\rm h}$ are more challenging to measure. These parameters influence the population where the event rate is lower, and their precision remains above the 1\% threshold even after 10 years of 3G observations.

For our fiducial model, we find that all parameters associated with the redshift evolution achieve 1\% accuracy within a range of 2 to 4.5 years when using ET. Improvements saturate as $T_{\rm obs}$ increases and marginal gains require longer times.

Among the spin hyperparameters, the shape parameters governing the distribution of spin magnitudes are the most tightly constrained. Specifically, $\alpha_\chi$ and $\beta_\chi$ reach percent-level precision within just a few months of observations. Conversely, hyperparameters related to spin orientations require significantly longer observation times to achieve comparable accuracy. Notably, the mixing fraction $\zeta$ demands nearly 10 years of data to obtain 1\% precision.
It is worth noting that, in the case of spins, the population Fisher matrix terms $\Gamma_{\rm II-V}$ become comparable to $\Gamma_{\rm I}$. This indicates that single-event spin measurement uncertainties are no longer negligible in the population analysis.
These findings highlight that accurately identifying the BBH formation channels through spin orientations will remain challenging, even in the 3G detector era. A key factor here will be improving the sensitivity at lower frequencies in order to enable a better resolution of the inspiral phase, where spins are most informative.

\subsection{Tools for the ringdown phase of binary mergers}
\label{sec:div9_ringdown} 
In this section we review the conceptual and implementation details of a Fisher information \texttt{python} code for black hole ringdown spectroscopy, which has been used to assess the capability of ET to perform precision tests  of General Relativity in the strong-field regime. For detailed accounts on the methodology, we refer the reader to \cite{Bhagwat:2021kwv,Bhagwat:2023jwv,Pacilio:2023mvk}. In its present status, it has been used to perform perspective studies of BH spectroscopy with next-generation GW detectors including ET \cite{Bhagwat:2021kwv,Bhagwat:2023jwv,Pacilio:2023mvk} as well as in the recent activities of the Observational Science Board (OSB) \cite{Branchesi:2023mws} and in  section~\ref{section:div1} of the present work. 

\subparagraph{Black-hole spectroscopy.}
A distorted black hole (BH) settles down to equilibrium by emitting GWs with a characteristic set of complex frequencies \cite{Vishveshwara:1970zz,Press:1971wr}, also known as the quasi-normal modes (QNMs) of the BH \cite{Kokkotas:1999bd,Berti:2009kk}. The real parts are the frequencies of oscillation, while the complex parts are the inverse of the damping times of each oscillatory mode. In General Relativity (GR), the no-hair theorems \cite{Israel:1967wq,Carter:1971zc,Hawking:1971vc,Robinson:1975bv} imply that stationary BHs are described by the Kerr metric \cite{Kerr:1963ud,Teukolsky:2014vca}, which is specified solely by the mass $M$ and the dimensionless spin $\chi$ of the BH. This striking consequence of GR holds also at the level of BH perturbations \cite{Teukolsky:1972my,Teukolsky:1973ha},
whose QNMs are expressed as functions of the mass and spin of the perturbed BH at equilibrium. On the other hand, QNMs in theories of gravity beyond GR can be characterized by additional degrees of freedom \cite{Barausse:2008xv,Franchini:2023eda}. It follows that, if one were to measure multiple QNMs from the GWs emitted by a BH, one could put GR to a test by checking that QNMs are consistently described by a two-parameter family \cite{Berti:2018vdi,Gossan:2011ha}. This research program is known as black-hole spectroscopy.
\subparagraph{Ringdown modelling.}
In GW astronomy, one is particularly interested in perturbed BHs generated in the aftermath of BBH coalescence. The corresponding GWs are observed in the so-called ringdown stage of the strain morphology. The degree of excitation of different modes depend on the physical processes that set up the perturbations. While in principle a countably infinite number of modes can be excited, numerical simulations of BBHs show that only a finite set is appreciably excited and one can provide approximate fits for the excitation amplitudes within GR \cite{Kamaretsos:2012bs,Gossan:2011ha,London:2014cma,London:2018gaq,Borhanian:2019kxt,Forteza:2022tgq,Cheung:2023vki}. Moreover, a variety of ready-to-use accurate approximants to the QNMs of Kerr BHs are available, including tabulated data \cite{Berti:2005ys,Berti:webpage}, analytical fits \cite{London:2018nxs} and fast solvers \cite{Stein:2019mop}. Finally, accurate fits from numerical simulations allow us to model the final mass $M_f$ and final spin $\chi_f$ from a BBH merger event as a function of the progenitor masses and spins \cite{Barausse:2012qz,Hofmann:2016yih,Husa:2015iqa,Varma:2018aht,Boschini:2023ryi}. All together, these approximants allow us to model BH ringdowns from Kerr BHs in the aftermath of a BBH coalescence.
\subparagraph{Fisher information formalism.}
Fisher matrix methods can be applied to the case of ringdown GW signals. Due to peculiarities in the ringdown waveform morphology, the resulting formalism is sufficiently distinct from the inspiral-merger-ringdown (IMR) Fisher methods presented in section \ref{sec:div9_FisherIntro}, to justify a separate treatment here. We follow the treatment in \cite{Berti:2005ys} to define approximate notions of SNRs and Fisher-information matrices for ringdown signals in the frequency domain. QNMs are indexed by an angular index $l$, an azimuthal index $m$ and an overtone index $n$. A BH ringdown can be mathematically modelled as a superimposition of damped sinusoids,
    \begin{equation}
        \label{eq:ringdown:strain:1}
        h_+ - i h_\times = \sum_{l>2}\sum_{m=-l}^{l}\sum_{n>0} A_{lmn}e^{-t/\tau_{lmn}}e^{-i(2\pi f_{lmn}+\phi_{lmn})}
    \end{equation}
where $f_{lmn}$ and $\tau_{lmn}$ are respectively the frequency and damping time of the $(l,m,n)$ mode, while $A_{lmn}$ and $\phi_{lmn}$ are the (real) amplitudes and phases of the modes. The above expression is only defined for times $t\geq0$, assuming that the time axis has been shifted such that the ringdown starts at $t=t_{\rm start}=0$. In order to take Fourier transforms, we use the Flanagan-Hughes (FH) method \cite{Flanagan:1997kp} and reflect the waveform about the origin of the time axis; specifically, we assume that the waveform at $t<0$ is identical to the waveform at $t\geq0$ except for the damping factor $e^{-t/\tau_{lmn}}$, which is replaced by $e^{-|t|/\tau_{lmn}}$ in order to ensure that the waveform fall off to zero at $t\to-\infty$. The corresponding doubling of the power spectrum in the signal is compensated by the application of an extra factor of $1/\sqrt{2}$ in the definition of the strain. The FH extension of the waveform allows us to take its Fourier transform and define a notion of SNR via the usual integrated expression in frequency domain, see eqs.~(3.4)--(3.6) in \cite{Berti:2005ys}. Note that the main results in \cite{Berti:2005ys} are based on a $\delta$-function limit when performing integrals in the frequency domain, i.e., they assume that the power spectrum is localized around the QNM frequencies $f_{lmn}$. While this is a good approximation for most cases and it allows us to derive analytical results, our implementation of the integrals is fully numerical and therefore we do not need  this further simplification.
\subparagraph{Extrinsic parameters.}
Bayesian parameter estimations of BH ringdowns usually fix the starting time $t_{\rm start}$ and the sky location $({\rm RA},{\rm dec})$ in order to unambiguously define the data segments at each detector \cite{Isi:2021iql}. Therefore, we exclude $(t_{\rm start},{\rm RA},{\rm dec})$ from the inference parameters $\boldsymbol{\theta}$. Moreover, the typical damping times of stellar-mass BH ringdowns last 1 to 10 ms; therefore, the effect of Earth rotation is negligible during the duration of the signal and we suppress it by fixing $t=t_{\rm start}$ in \eq{eq:pattern:functions:2}. Finally, we keep the polarization angle $\psi$ and the inclination angle $\iota$ fixed at inference time, under the assumptions that they are much more accurately measured during the inspiral phase. This last assumption will be relaxed in a future version of the code.
\subparagraph{Waveform templates.}
We implement three distinct waveform templates. 
\begin{enumerate}
    \item
    \textsc{RingdownDampedSinusoids} parametrizes the strain directly as in \eq{eq:ringdown:strain:1}, in terms of the parameters 
    \begin{equation}
    \boldsymbol{\theta}=\{f_{lmn},\tau_{lmn},A_{lmn},\phi_{lmn}\}\, .
    \end{equation}
    Therefore, there are $4 N_{\rm modes}$ free parameters, where $N_{\rm modes}$ is the number of QNMs included in the model; this model is agnostic about the relation between the QNMs and the Kerr-like nature of the ringing BH.
    \item
    \textsc{RingdownKerr} assumes that the frequencies and damping times are functions of the final mass $M_f$ and final spin $\chi_f$ according to the spectrum of a Kerr BH in GR; it is parametrized by 
    \begin{equation}
    \boldsymbol{\theta}=\{M_f,\chi_f,A_{lmn},\phi_{lmn}\}\, ,
    \end{equation}
    and there are $2+2 N_{\rm modes}$ free parameters.
    \item
    \textsc{RingdownDeltaKerr} places free deviational parameters $\delta f_{lmn}$ and $\delta \tau_{lmn}$ to the frequencies and damping times of a Kerr BH; it is parametrized by 
    \begin{equation}
    \boldsymbol{\theta}=\{M_f,\chi_f,\delta f_{lmn},\delta \tau_{lmn},A_{lmn},\phi_{lmn}\}\, .
    \end{equation}
    As discussed in \cite{Pacilio:2023mvk}, $\delta f_{220}$ and $\delta \tau_{220}$ are degenerate with $M_f$ and $\chi_f$, therefore we fix them to zero both at injection and at inference times, and we are left with $4 N_{\rm modes}$ free parameters. 
\end{enumerate}
The choice of $\log_{10}~A_{lmn}$ in place of $A_{lmn}$ ensures that credible intervals for the amplitudes do not extend to unphysical negative values. All models assume non-precessing binary progenitors, which implies the symmetry 
\be
A_{l-mn}e^{-i\phi_{l-mn}}=(-1)^lA_{lmn}e^{i\phi_{lmn}}\, ,
\ee
together with the reflection symmetry $f_{l-mn}=-f_{lmn}$ and $\tau_{l-mn}=\tau_{lmn}$, generally valid for the QNMs of Kerr BHs in GR, this allows us to recast \eq{eq:ringdown:strain:1} such that the summation runs over positive $m$ only.
\subparagraph{Differentiation, integration and inversion.}
Derivatives of the strain are evaluated with automatic differentiation using the \texttt{JAX} library for \texttt{python}~\cite{jax2018github}. Fisher matrices are computed via numerical integration and inverted via the standard implementation in \texttt{numpy} of the Lower-Upper (LU) triangular decomposition. The code also allows for vectorization, i.e., it can process multiple events simultaneously on a single CPU. We envisage that a relevant future development will be to compute the Fisher information matrix directly in the time domain, e.g.~using the the TiDoFM code described in section \ref{sec:div9_software:tools}, in order for the Fisher approximations to relate directly to well-established time-domain software for BH ringdown analysis \cite{Carullo:2019flw,Isi:2021iql}.

\subparagraph{Applications.}
In figure~\ref{fig:div9:ringdown} we show an example of the forecasts that can be obtained with this code. We focus on a system with final mass $M_f=70~ M_\odot$ and final spin $\chi_f=0.7$, as observed by a single ET detector with a triangular design placed in Sardinia. For illustrative purposes we consider only two ringdown modes, namely the dominant $(l,m,n)=(2,2,0)$ mode and a subdominant mode $(l,m,n)=(3,3,0)$. The relative amplitude is chosen such that $A^R_{330}=A_{330}/A_{220}=0.2$, consistently with estimates from NR simulations of binary black holes --- see for example figure~3 of \cite{Pacilio:2024tdl}. We employ the \textsc{RingdownDeltaKerr} template, which is relevant for the purposes of testing General Relativity. The filled blue contours in figure~\ref{fig:div9:ringdown} (left) display the 68\% and 90\% credible regions on $M_f$ and $\chi_f$ for a signal with amplitudes adjusted to give $\rho_{\rm RD}=50$. For comparison, the empty green contours mark the 68\% and 90\% credible regions for the same signal but with lower SNR $\rho_{\rm RD}=12$. Fig.~\ref{fig:div9:ringdown} (right) displays the corresponding predictions for the deviation parameter $\delta f_{330}$ and for the relative amplitude $A^{R}_{330}$ of the subdominant mode. In order to interpret the constraints on $\delta f_{330}$ physically, it is recommended that the amplitude of the corresponding mode is bounded away from zero to claim mode detection: we see that, while this condition is fully met at $\rho_{\rm RD}=50$, it holds only marginally when $\rho_{\rm RD}=12$. Note that, since we are using a Fisher-matrix formalism in the log-scale amplitudes $\log_10 A_{lmn}$, the posterior is not Gaussian when expressed in terms of the amplitude $A^{R}_{330}$, as it is most evident for the $\rho_{\rm RD}=12$ case.
\begin{figure*}[t]
    \centering
    \includegraphics[width=0.45\textwidth]{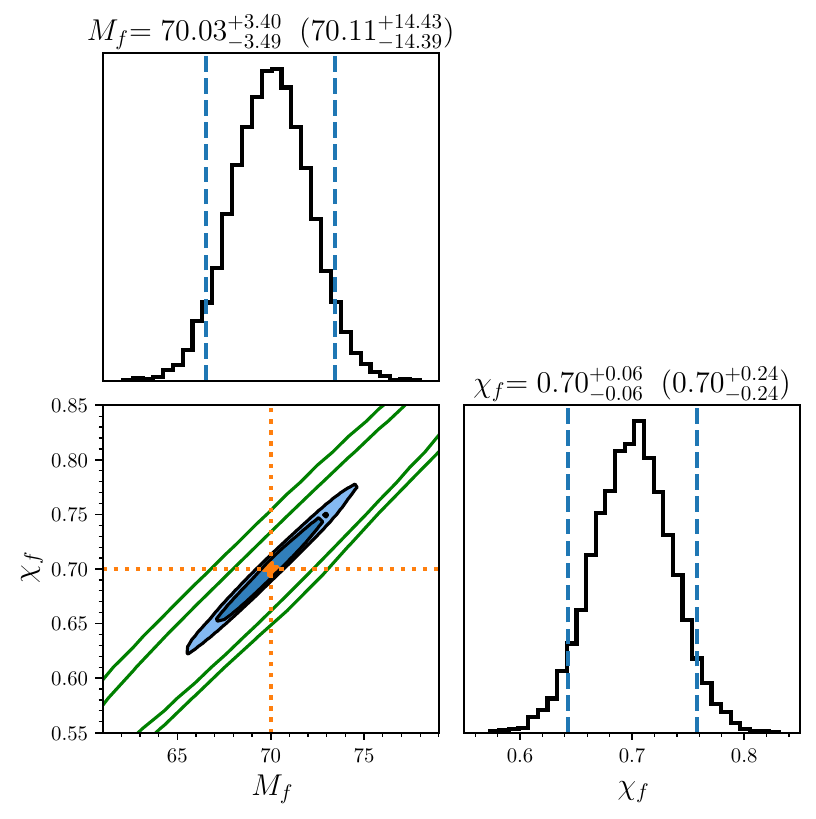}
\includegraphics[width=0.45\textwidth]{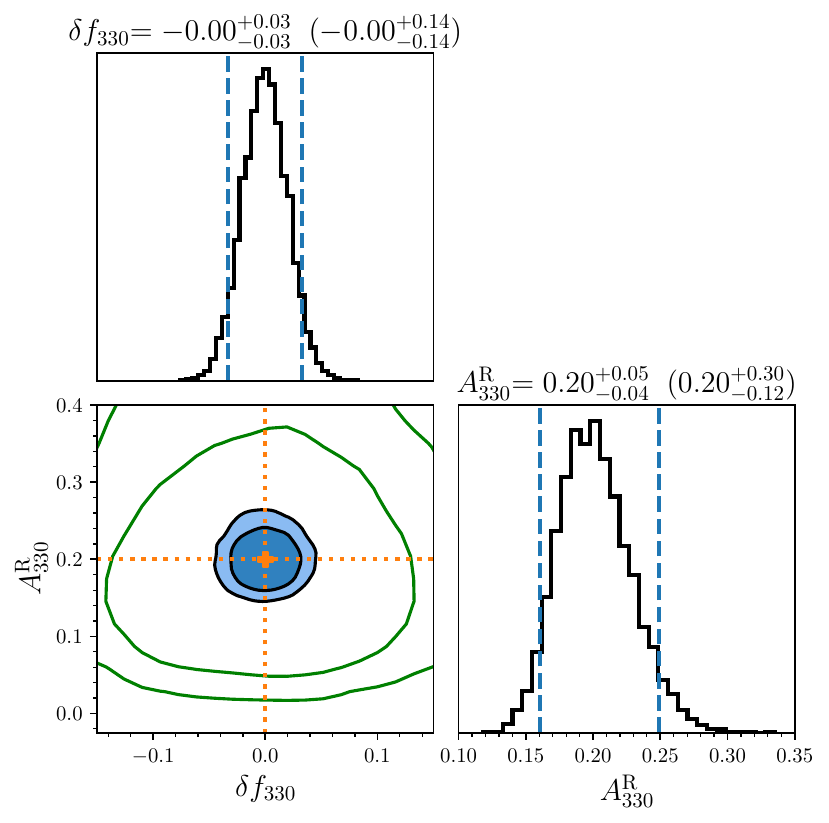}
    \caption{Left: 68\% and 90\% credible regions in the $M_f$-$\chi_f$ plane, for a system with final mass $M_f=70~M_\odot$ and final spin $\chi_f=0.7$. Filled blue contours (resp.~empty green contours) are relative to a system with ringdown SNR $\rho_{\rm RD}=50$ (resp.~$\rho_{\rm RD}=12$). Marginalized 1-d posteriors refer to $\rho_{\rm RD}=50$, with the corresponding 90\% credible intervals indicated by the vertical dashed blue lines and in the plot headers (median and 90\% intervals for the $\rho_{\rm RD}=12$ case are indicated in parentheses for reference).
    Right: Same as left, in the $\delta f_{330}-A_{330}^{\rm R}$ plane. Note that the posterior is not Gaussian because we are mapping samples from $\log_{10} A_{lmn}$ to $A_{330}^{\rm R}$.
    For both $\rho_{\rm RD}=50$ and $\rho_{\rm RD}=12$ we fix the right ascension to $\alpha=1.95$, the declination to $\delta=-1.27$, the polarization to $\phi=0.82$ and the inclination angle to $\iota=\pi/3$, and assume a GPS time $t_{\rm GPS}=1126259462.423$. We change the SNR by a simple rescaling of the amplitudes.
    }
    \label{fig:div9:ringdown}
\end{figure*}

\subsection{Stochastic searches}
\label{sec:div9_stoch_searches}

Astrophysical sources with low SNRs will not be resolved as single events. However, the superposition of such signals will appear in the detector as a Stochastic Gravitational Wave Background (SGWB). Even if 3G detectors resolve a large fraction of the merging binaries in the Universe, still this residual stochastic background is very significant, both for the  astrophysical information that it contains (see  the discussion in section~\ref{sect:Populationbackdiv3}) and for the fact that it could mask a background of cosmological origin.
This is particularly important for BNS; indeed, as we see from the right panel in figure~\ref{fig:SNRs_cumulatives_allconf_BBH_and_BNS}, for a single ET observatory about $(10-30)\%$ (depending on the ET configuration) of the BNS merging in the Universe is resolved, which raises to $(30-50)\%$
and  at a network ET+2CE, see 
figure~\ref{fig:cumul_SNR-Mc-Om-dL_BNS_allconf_ETCE}.
Therefore,  a large fraction of the BNS is still unresolved and contributes to the stochastic background.
For BBHs, the fraction of resolved sources is higher and can be of order $(80-90)\%$ for a single ET (in the full HFLF-cryo configuration, see the left panel of figure~\ref{fig:SNRs_cumulatives_allconf_BBH_and_BNS}) to almost $99\%$ of the population for the best  ET+2CE configuration. Still, in absolute terms even $1\%$ of a population of $10^5$ BBHs means $10^3$ unresolved events per year. On top of this, the reconstruction of the resolved events and their subtraction from the data stream always has an associated error, and the accumulation of the errors from the resolved sources effectively produce  another confusion noise that can degrade the ET sensitivity to  cosmological signals, as we will discuss in detail in section~\ref{sect:subtractastrobkgdi9}. 

Production mechanisms for cosmological stochastic backgrounds have been discussed in detail in section~\ref{sect:SGWBdiv2}.
Here we rather focus on the mathematical formalism for the characterization of the sensitivity of a detector network to a SGWB, the definition, properties and tools for the computation of the power-law integrated sensitivity (PLS) and the formalism and tools that have been developed to address the problem of the subtraction of the astrophysical background to reveal a cosmological background.

\subsubsection{Characterization of stochastic backgrounds}\label{sect:CharStocBackdiv9}

The basic definitions for the characterization of stochastic backgrounds 
have already been given in section~\ref{sec:stochback}, see in
particular \eqst{snrhabdiv2}{avediv2}. For convenience, let us recall here the main formulas.
We write the  superposition of GWs coming from all directions as
\be\label{snrhabdiv9}
h_{ij}(t,{\bf x})
=\int_{-\infty}^{\infty} df \int_{S^2} d^2\hatn\,\sum_{A=+,\times}  \tilde{h}_A(f,\hatn) e^A_{ij}(\hatn) \, \,   e^{-2\pi i f(t-\hatn\cdot{\bf x} /c )}
\, ,
\ee
where $\hatn$ is the unit vector in the propagation direction and $e_{ij}^{A}$ are the polarization tensors, normalized as 
$e_{ij}^{A}(\hatn)e_{ij}^{A'}(\hatn)=2\delta^{AA'}$. The signal observed in a given detector is 
\be\label{hFhFdiv9}
h_a(t)= \int_{-\infty}^{\infty} df \int_{S^2} d^2\hatn\,\sum_{A=+,\times}\tilde{h}_A(f,\hatn) F_a^A(\hatn) \, \,   e^{-2\pi i f(t-\hatn\cdot{\bf x}_a /c )}
\, ,
\ee
where  ${\bf x}_a$ denotes the detector's position, and $F_a^A(\hatn)$ are the detector's pattern functions.
Then, 
\be\label{tildehaFAdiv9}
\tilde{h}_a(f)=\int_{S^2} d^2\hatn \sum_{A=+,\times}\tilde{h}_A(f,\hatn) F_a^A(\hatn) \, \,   
e^{2\pi i f \hatn\cdot{\bf x}_a /c}\, .
\ee
For a stationary,  isotropic and unpolarized stochastic background of GWs, the two-point correlator  is given by
\be\label{avediv9}
\langle \tilde{h}_A^*(f,\hatn)
\tilde{h}_{A'}(f',\hatn')\rangle =
\delta (f-f')\, \frac{\delta(\hatn,\hatn')}{4\pi}\,
 \delta_{AA'}\, \frac{1}{2}S_h(f)\, ,
\ee
which defines the (single-sided) spectral density of the signal. The generalization to polarized backgrounds was given in \eq{eq:gwb_stokes}. 

The GW background can be described by its energy density per unit logarithmic frequency interval, normalized by the critical energy density for closing the Universe $\rho_c$:
\begin{equation}
\label{eq:OmegaGWdiv9}
\Omega_{\rm GW}(f)\equiv\frac{1}{\rho_c}\frac{{\mathrm d}\rho_{\rm GW}}{{\mathrm d}\ln f}=\frac{8\pi G}{3 H_0^2 c^2}\frac{{\mathrm d}\rho_{\rm GW}}{{\mathrm d}\ln f}\,,
\end{equation}
where $\rho_c = 3 c^2 H_0^2/\left(8\pi G\right)$ and $H_0$ is the present-day value of the Hubble parameter. In terms of the spectral density, one has 
\begin{equation}
\label{eq:Shdiv9}
\Omega_{\rm GW}(f) =  \frac{4\pi^2} {3 H_0^2} f^3S_h(f) \, .
\end{equation}
This expression  also holds for polarized backgrounds; indeed, the  Stokes parameters describing linear and circular polarization, see \eq{eq:gwb_stokes}, do not contribute to the energy density~\cite{Belgacem:2024ohp}. To get rid of the uncertainty on $H_0$, is is often convenient to work in terms of the standardized quantity  $h^2\Omega_{\rm GW}(f)$, where
the reduced Hubble constant $h$ is defined from $H_0=100 h~{\rm km} \, {\rm s}^{-1}\, {\rm Mpc}^{-1}$.

The expectation value of the cross--correlation of the outputs of two detectors $a, b$ depends on the overlap reduction function (ORF) of the pair, defined as~\cite{Christensen:1996da,Flanagan:1993ix}
\begin{equation}
\label{eq: norm ORF}
\Gamma_{a b}(f)=\int_{S^2} \frac{\mathrm{d}^2\hat{\bf n}}{4\pi}~e^{2\pi i \frac{f}{c}\hat{\bf n}\cdot{\bf \Delta x}_{a b}}\sum_{A=+,\times}F^A_{a}\left(\hat{\bf n}\right)F^A_{b}\left(\hat{\bf n}\right)\,,
\end{equation}
where ${\bf \Delta x}_{a b}={\bf x}_{b}-{\bf x}_{a}$ is the spatial separation between the two detectors and $F^A_{a}\left(\hat{\bf n}\right)$, $F^A_{b}\left(\hat{\bf n}\right)$ are their antenna pattern functions.
%
In a network  with more than two detectors, for stochastic backgrounds the simplest strategy is to consider separately all possible detector pairs and  sum up in quadrature the signal-to-noise ratio of the various detectors pairs.\footnote{Other possibilities involving multiple detector correlations are however also possible, see refs.~\cite{Allen:1997ad,Zhu:2011bd}.} Then,  considering an optimally filtered cross-correlation search using a network of 
a pair of detectors labeled by indices $a, b$ (with $a\neq b$), the corresponding signal-to-noise ratio $(S/N)_{ab}$, integrated over the observation time $T$ and frequency range $[f_{\rm min}, f_{\rm max}]$,  is given by~\cite{Allen:1997ad, Thrane:2013oya}
\begin{equation}
\label{eq:SNRcross2det}
\(\frac{S}{N}\)_{ab} = \left[ 2T\int_{0}^{\infty} df \,S_h^2(f)\,\ \frac{\Gamma^2_{a b}(f)}{S_n^{(a)}(f) S_n^{(b)}(f)} \right]^{1/2} \, ,
\end{equation}
where $S_n^{(a)}(f)$ and $S_n^{(b)}(f)$ are the noise power spectral densities (PSDs) of the detectors $a$ and $b$, defined by
\be\label{Sn1noT}
\left\langle \tilde{n}_a^*(f)\tilde{n}_b(f') \right\rangle =
\delta(f-f')\frac{1}{2}S_n^{(a)}(f)\delta_{ab}\, ,
\ee
and we assumed that the noise among detectors are uncorrelated (more on this below).
If we have
$\mathcal{N} >2$ detectors with uncorrelated noise, limiting ourselves to two-point correlations, the total signal-to-noise ratio is obtained adding in quadrature the signal-to-noise ratios of the various pairs, so 
\begin{equation}
\label{eq:SNRcross}
\frac{S}{N} = \left[ 2T\int_{0}^{\infty} df \,S_h^2(f)\,\sum_{a=1}^\mathcal{N} \sum_{b>a}^\mathcal{N} \, \frac{\Gamma^2_{a b}(f)}{S_n^{(a)}(f) S_n^{(b)}(f)} \right]^{1/2} \, .
\end{equation}
Eq.~(\ref{eq:SNRcross}) suggests the definition of an effective strain noise power spectral density for the detector network $S_{\rm eff}(f)$,  given by
\begin{equation}
\label{Seff uncorraleted noises}
S_{\rm eff}(f)=\left[\sum_{a=1}^\mathcal{N} \sum_{b>a}^\mathcal{N} \frac{\Gamma_{a b}^2(f)}{S_n^{(a)}(f) S_n^{(b)}(f)}\right]^{-1/2}\, ,
\end{equation}
so that
\be
\frac{S}{N} = \left[ 2T\int_{0}^{\infty} df \,\frac{S_h^2(f)}{S^2_{\rm eff}(f)}\right]^{1/2}\, .
\ee

\subsubsection{Power-Law Sensitivity }
\label{sec:div9_PLSdefinition}


The sensitivity of a network of detectors to an unpolarized and isotropic stochastic GW background can be characterized by the power-law sensitivity (PLS) curve \cite{Thrane:2013oya} (also denoted as the power-law integrated sensitivity curve). Here, we first recall its construction and discuss  some its properties, and we will then show examples of PLS for different ET configurations.

In the PLS construction  one considers a family of power-law GW backgrounds
\begin{equation}
\label{eq: pow law GW}
\Omega_{\rm GW}(f;\beta)=\Omega_{\beta}(f/f_{\rm ref})^\beta \,,
\end{equation}
labeled by  an exponent $\beta$  which can take any real value $\beta\in\left(-\infty,+\infty\right)$; $f_{\rm ref}$ is an arbitrary reference frequency that we  fix once and for all, since changing it   simply  amounts to a redefinition of $\Omega_{\beta}$. A convenient choice could be a value somewhere in the middle of the bandwidth of the detector that we are considering, e.g. 100~Hz for ground based detectors.
For any given value of the exponent $\beta$, one can compute the amplitude $\Omega_\beta$ such that, in a given coincident observation time $T$, the total signal-to-noise ratio in the frequency range $[f_{\rm min}, f_{\rm max}]$ has a given value $\rho$.
%
Explicitly,  
this requirement gives \cite{Thrane:2013oya}
\begin{equation}
\label{eq: Omega_beta}
\Omega_\beta=\rho\left[2T\int_{0}^{\infty} df ~\Omega^{-2}_{\rm eff}(f)\left(\frac{f}{f_{\rm ref}}\right)^{2\beta}\right]^{-1/2}\,,
\end{equation}
where $\Omega_{\rm eff}(f)$ is determined from $S_{\rm eff}(f)$ as in \eq{eq:Shdiv9}, so
\bees
\Omega_{\rm eff}(f)&=&\frac{4\pi^2} {3 H_0^2} f^3S_{\rm eff}(f)\nn\\
&=&\frac{4\pi^2} {3 H_0^2} f^3
\left[\sum_{a=1}^\mathcal{N} \sum_{b>a}^\mathcal{N} \frac{\Gamma_{a b}^2(f)}{S_n^{(a)}(f) S_n^{(b)}(f)}\right]^{-1/2}\, .\label{defOmeffdiv9}
\ees
Inserting \eq{eq: Omega_beta} into \eq{eq: pow law GW}, we see that the family of power-law GW backgrounds of the form (\ref{eq: pow law GW}) with given signal-to-noise ratio $\rho$ is 
given by\footnote{Notice that the arbitrary $f_{\rm ref}$ plays no role, as it cancels in eq.~(\ref{eq: power law family}). We just keep it in the equations so that dimensionless ratios of frequencies appear in the bases of powers.}
\begin{equation}
\label{eq: power law family}
\Omega_{\rm GW}(f;\beta)=\rho\,\left(\frac{f}{f_{\rm ref}}\right)^{\beta}\left[2T\int_{0}^{\infty} d{f^\prime}~\Omega^{-2}_{\rm eff}(f^\prime) \left(\frac{f^\prime}{f_{\rm ref}}\right)^{2\beta}\right]^{-1/2}\,.
\end{equation}
%
Finally, the PLS curve is defined as the envelope of the family (\ref{eq: power law family}), where $\beta$ is the family parameter. The result can be written as~\cite{Thrane:2013oya} (see~\cite{Belgacem:2025oom} for a related conceptual point)
\begin{equation}
\label{eq: PLS basic}
\Omega_{\rm PLS}(f)=\underset{\beta}{\max}~\Omega_{\rm GW}(f;\beta)\,.
\end{equation} 
In a log-log plot, any line tangent to the PLS curve represents a power-law GW background with $S/N$ equal to the value of $\rho$ used in the PLS definition.

The usefulness of the PLS curve is that it allows  studying the detectability (for a chosen $S/N$ threshold $\rho$) of a given power-law GW background~\cite{Thrane:2013oya}: if the background lies everywhere below the PLS curve, then it is not detectable because its $S/N$ is smaller than the threshold $\rho$; if the background rather lies somewhere above the PLS curve, then it will be observed with $S/N$ larger than the threshold $\rho$ (see~\cite{Belgacem:2025oom} for clarifications about this interpretation).

\begin{figure}[t]
    \centering
    \begin{tabular}{c c}
       \includegraphics[width=.48\textwidth]{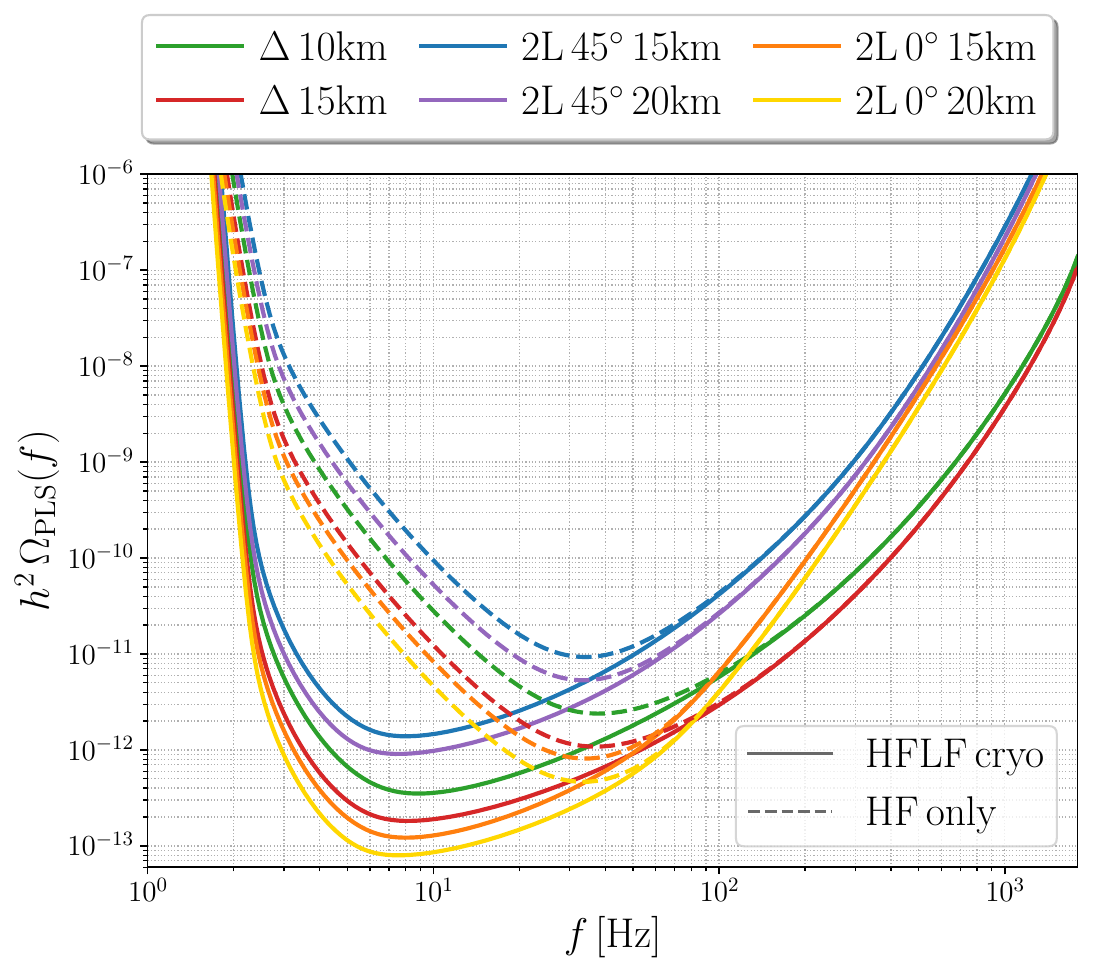}   &  \includegraphics[width=.48\textwidth]{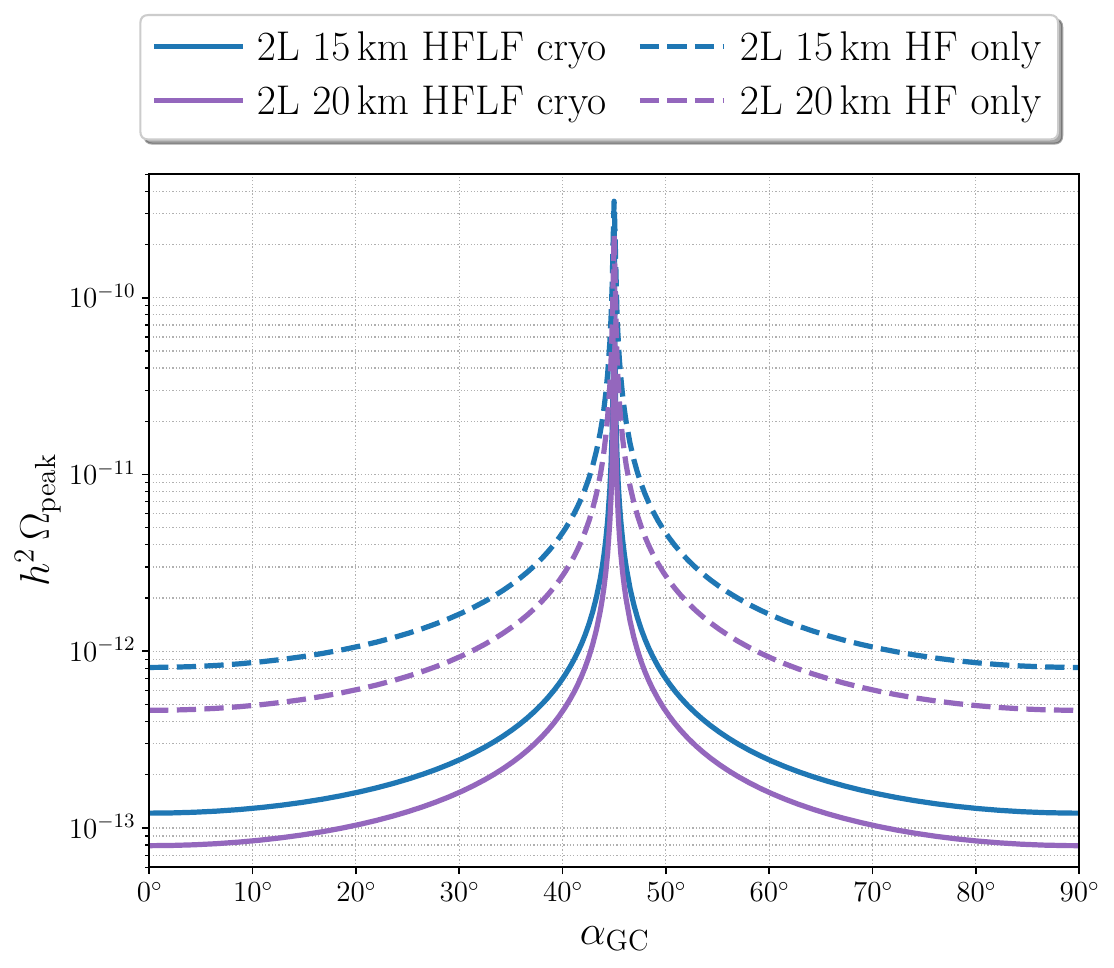}
    \end{tabular}
    \caption{Left panel: the PLS for different ET configurations, colors as in legend. 
    Right panel: the minimum value of the PLS (i.e. the peak sensitivity) for different ET configurations as a function of the relative angle $\alpha_{\rm GC}$ between two interferometers with respect to the great circle that joins them, colors as in legend. In both panels solid  lines refer to the 
    full sensitivity obtained with  the HF interferometer together with  the cryogenic LF interferometer, while dashed curves refer to the sensitivity when only the HF instrument is included. 
    }
    \label{fig:PLS_plots}
\end{figure}

Several properties of the PLS curve, including a parametric equation for it, have been derived analytically in~\cite{Belgacem:2025oom}. Below we mention some of those results.
\begin{itemize}
\item {\bf Exact parametric equation}. The points on the PLS curve are described by the following parametric equation:
\begin{equation}
\label{eq: PLS exact parametric}
\left\{ \begin{aligned} 
  f (\beta)&= f_{\rm ref}\,\exp\left[\frac{\int_{0}^{\infty} d{f^\prime}~\Omega^{-2}_{\rm eff}\left(f^\prime\right)\left(\frac{f^\prime}{f_{\rm ref}}\right)^{2\beta}\ln{\left(\frac{f^\prime}{f_{\rm ref}}\right)}}{\int_{0}^{\infty} d{f^\prime}~\Omega^{-2}_{\rm eff}\left(f^\prime\right)\left(\frac{f^\prime}{f_{\rm ref}}\right)^{2\beta}}\right]\,,\\
  \Omega_{\rm PLS} (\beta)&= \rho\,\frac{\exp\left[\beta\,\frac{\int_{0}^{\infty} d{f^\prime}~\Omega^{-2}_{\rm eff}\left(f^\prime\right)\left(\frac{f^\prime}{f_{\rm ref}}\right)^{2\beta}\ln{\left(\frac{f^\prime}{f_{\rm ref}}\right)}}{\int_{0}^{\infty} d{f^\prime}~\Omega^{-2}_{\rm eff}\left(f^\prime\right)\left(\frac{f^\prime}{f_{\rm ref}}\right)^{2\beta}}\right]}{\left[2T\int_{0}^{\infty} d{f^\prime}~\Omega^{-2}_{\rm eff}\left(f^\prime\right)\left(\frac{f^\prime}{f_{\rm ref}}\right)^{2\beta}\right]^{1/2}}\,.
\end{aligned} \right.
\end{equation}
As the parameter $\beta$ runs over the real numbers, \eq{eq: PLS exact parametric} gives the coordinates of a point that moves on the PLS curve. In a log-log plot, the slope of the PLS curve at that point is equal to the corresponding value of $\beta$. The use of \eq{eq: PLS exact parametric} is the most precise and efficient way to plot the PLS curve.
\item {\bf Convexity}. In a log-log plot the PLS curve is always convex, independently of the details of the network entering in $\Omega_{\rm eff}(f)$, i.e. 
\begin{equation}
\label{eq: loglog convexity PLS}
\frac{{\rm d^2}\ln{\Omega_{\rm PLS}(f)}}{d(\ln{f})^2}>0\,.
\end{equation}
\item {\bf Peak sensitivity}. The peak sensitivity is defined as the minimum value of the PLS curve. We call $f_{\rm peak}$ the frequency at which the minimum PLS value $\Omega_{\rm peak}$ is reached, i.e. 
\begin{equation}
\label{eq: peak sensitivity def}
\Omega_{\rm peak}\equiv\underset{f}{\min}~\Omega_{\rm PLS}(f)=\Omega_{\rm PLS}\left(f_{\rm peak}\right)\,.
\end{equation}
The values of $f_{\rm peak}$ and $\Omega_{\rm peak}$ are given by
\begin{equation}
\label{eq: f_peak}
f_{\rm peak}=f_{\rm ref}\,\exp\left[\frac{\int_{0}^{\infty} d{f^\prime}~\Omega^{-2}_{\rm eff}\left(f^\prime\right)\ln{\left(\frac{f^\prime}{f_{\rm ref}}\right)}}{\int_{0}^{\infty} d{f^\prime}~\Omega^{-2}_{\rm eff}\left(f^\prime\right)}\right]\,,
\end{equation}
\begin{equation}
\label{eq: peak sensitivity result}
\Omega_{\rm peak}=\rho\left[2T\int_{0}^{\infty} d{f^\prime}~\Omega^{-2}_{\rm eff}\left(f^\prime\right)\right]^{-1/2}\,.
\end{equation}

The use of eq.~(\ref{eq: peak sensitivity result}) is convenient for studying the peak sensitivity of a network depending on its configuration ( recall that the information on the detector sensitivities is contained in $\Omega_{\rm eff}(f^\prime)$). It is definitely more efficient than going through the determination of the full PLS curve and then looking numerically for its minimum.
\end{itemize}


%
The left panel of figure~\ref{fig:PLS_plots} shows the PLS curves for different ET configurations, considering a triangle ($\Delta$) or two L-shaped detectors (2L).\footnote{These curves can be obtained with a dedicated public tool  available at 
\href{https://github.com/CosmoStatGW/gwfast}{CosmoStatGW/gwfast}.}
In the right panel of the same figure we plot the peak sensitivity, computed with eq.~(\ref{eq: peak sensitivity result}), as a function of the relative orientation between two L-shaped detectors.
Figure~\ref{fig:PLS_plots} shows in particular the dramatic improvement of the sensitivity below $\sim 10\, \rm Hz$ with a HFLF instrument, as well as the impact of the relative orientation of the detectors for a 2L configuration. In this case, the best sensitivity is reached for aligned detectors, differently from the reconstruction of the source parameters for CBCs, which is enhanced in a misaligned configuration.

It should be stressed that these PLS have been computed assuming that noise among different detectors is uncorrelated, see \eq{Sn1noT}. However, 
correlated instrumental noise will  be present, in particular for ET in its triangular configurations, which features  three colocated detectors; in this case,  correlated Newtonian noise could affect  stochastic searches below ${\cal O}(50-100)$~Hz~\cite{Janssens:2022xmo,Branchesi:2023mws,Janssens:2024jln}, or in any case at least up to 30~Hz assuming an optimistic factor of 10 of Newtonian noise cancellation~\cite{Badaracco:2019vjq}. See sections~\ref{sect:correlated_noise_div2} and \ref{sect:RecCorrNoise_div2} for  detailed discussion.

\subsubsection{Subtraction of the astrophysical background}\label{sect:subtractastrobkgdi9}

As we have discussed, third-generation  detectors such as ET  will  resolve  a huge number of BBH and BNS  signals per year, see figure~\ref{fig:SNRs_cumulatives_allconf_BBH_and_BNS}. 
Nevertheless,  the residual ``confusion noise" due to the unresolved CBCs is still important, and can  affect the search for cosmological backgrounds. This is due to several factors; first, given the large number 
of BBHs and BNSs coalescing per year (about $1\times10^5$~BBH/yr and $7\times10^5$~BNS/yr in the population model used in figure~\ref{fig:SNRs_cumulatives_allconf_BBH_and_BNS}), in absolute terms the number of undetected events is still large. In particular, 
we see from  figure~\ref{fig:SNRs_cumulatives_allconf_BBH_and_BNS} that, setting a detection threshold at ${\rm SNR}=12$, ET alone will detect (depending on the geometry)
between $1\times 10^4$ and $1\times 10^5$ BNS/yr, out of a population of $7\times10^5$~BNS/yr; this means that (in this population model) there will still be of order of $(6-7)\times 10^5$ BNS/yr that are too faint to be resolved; for BBHs a larger fraction of events is resolved (a consequence of the fact that BBHs are more massive than BNSs, and therefore their signal is louder); we see from
figure~\ref{fig:SNRs_cumulatives_allconf_BBH_and_BNS} that only a few percent of the population is missed. Still, missing just a few percent of the events, over a population of ${\cal O}(10^5)$ BBH/yr, means that there will be several thousands of undetected BBH events per year.
The number of unresolved events  decreases further in a ET-2CE network~\cite{Iacovelli:2022bbs,Borhanian:2022czq,Branchesi:2023mws}, but still remain large as absolute numbers.

Even more importantly, the CBCs that are above the detection threshold can only be reconstructed with some approximation, and the accumulation of the errors  in the reconstruction and subtraction on the ${\cal O}(10^5)$ resolved events/yr produces another important source of confusion noise~\cite{Cutler:2005qq,Harms:2008xv,Regimbau:2016ike,Sachdev:2020bkk,Sharma:2020btq,Perigois:2021ovr,Zhou:2022nmt,Pan:2023naq,Zhong:2024dss,Belgacem:2024ntv}. In fact, this can be the dominant source of confusion noise for typical values of the  detection threshold, such as ${\rm SNR}=12$.

Finally, 
this confusion noise must be compared with the PLS of 3G observatories, which significantly improves with respect to 2G networks, and which therefore can also be more easily covered by such a residual background.
As a result, the confusion noise generated by the unresolved sources and by the accumulation on the errors on the resolved sources can be  significant, and could spoil the 3G sensitivity to cosmological backgrounds. It is therefore important to develop the appropriate formalism, as well as the software tools, for accurately estimating this effect, and subtracting the astrophysical confusion noise to reveal possible cosmological backgrounds.

Here we follow the strategy developed in ref.~\cite{Belgacem:2024ntv}, which provides a first-principle approach to the subtraction problem.\footnote{The public code that implements it can be found within the \href{https://github.com/CosmoStatGW/gwfast}{\texttt{GWFast}} package.}
The output of a given GW detector, labeled by an index $a$, in presence of a GW signal, has the form 
$s_a(t) =n_a(t)+h_a(t)$, 
where $n_a(t)$ is the noise in the $a$-th detector and $h_a(t)$ is  a generic GW signal (projected onto the $a$-th detector). 
If the signal is made of a superposition of CBCs plus a cosmological signal, we  have
\be
s_a(t)=n_a(t)+ \sum_{i=1}^{\nev} h^{\rm true}_{a,i}(t) + h_a^{\rm cosmo}(t)  \, ,
\ee
where $h^{\rm true}_{a,i}(t)$  is the projection of the  true $i$-th astrophysical signal onto the $a$-th detector, $\nev$ the number of mergers taking place in the observation time $T$,
and $h_a^{\rm cosmo}(t)$ is the  (possible) cosmological  signal in the $a$-th detector. We are
adding here a label ``true'' to the actual astrophysical signals, since below we will  need to distinguish between the true values and the reconstructed ones.
We assume that an astrophysical signal is detected if the SNR of the full detector network is above a threshold value $\snrth$, and
we denote by $h^{\rm obs}_{a,i}(t)$  the projection of the  reconstructed (i.e., ``observed'')  $i$-th astrophysical signal onto the $a$-th detector. 
Note that whether a signal is above or below threshold, as well as its reconstruction (when it is above threshold)  also depends on the specific realization of the noise in the detector.
Denoting by $\sigma_a(t)$ the output of the $a$-th detector after subtraction of the reconstructed signals, we then have
\be\label{sigmaahconf}
\sigma_a(t)=n_a(t)+h_a^{\rm  conf}(t)+h_a^{\rm cosmo}(t)\, ,
\ee
where
\be\label{defhconf}
h_a^{\rm  conf}(t)=\sum_{i=1}^{\nev}\[  h^{\rm true}_{a,i}(t)- h^{\rm obs}_{a,i}(t) \theta(\snrobs_i - \snrth) \]\, ,
\ee
and $\theta(\snrobs_i - \snrth)$ is a Heaviside theta function,
that expresses the fact that an event is subtracted only if its SNR is above a given threshold $\snrth$.   The notation $h_a^{\rm  conf}(t)$  in \eq{defhconf}  stresses that this quantity is a ``confusion noise'', due to the unresolved sources (the terms in the sum where the theta function is zero), and to the error in the subtraction of resolved sources (the terms in the sum where the theta function is one).
The problem that we face is therefore how to extract a cosmological signal, from the effective noise $n_a^{\rm eff}(t)\equiv n_a(t)+h_a^{\rm  conf}(t)$ or, in Fourier space, 
\be\label{defneff}
\tilde{n}^{\rm eff}_a(f)= 
\tilde{n}_a(f)+  \tilde{h}^{\rm conf}_{a}(f)\, ,
\ee
using the correlation technique between detector pairs. The crucial point  is that, even when we assume that the instrumental noise is uncorrelated among detectors, still   $\tilde{n}^{\rm eff}_a(f)$  is  correlated among the detectors because it depends on   $\tilde{h}^{\rm conf}_{a}(f)$, where the astrophysical signals enter, and the latter  produces a correlated effect in the two detectors. Furthermore, the reconstruction of the signal depends on the noise realization in the whole detector network, and therefore the reconstruction of a signal in one detector is correlated also with the noise in any other detector of the network.

Let us then  discuss, following \cite{Allen:1997ad,Belgacem:2024ntv}, how 
to generalize  the correlation technique discussed in section~\ref{sect:CharStocBackdiv9}, in the presence of correlated noise. 
In general, the signal-to-noise ratio of a two-detector correlation is constructed as follows.
Given two GW detectors, labeled by indices $a, b$ (with $a\neq b$), taking data over a time $T$, one forms the correlator of their outputs, 
\bees\label{defYab}
Y_{ab}&=&\int_{-T/2}^{T/2}dt\int_{-T/2}^{T/2}dt'\, s_a(t)s_b(t')Q_{ab}(t-t')\nonumber\\
&\simeq& \int_{-\infty}^{\infty}df\, \tilde{s}^*_a(f)\tilde{s}_b(f)\tilde{Q}_{ab}(f)\, ,
\ees
where $Q_{ab}(t-t')$ is a filter function.\footnote{The second equality is valid if $fT\gg 1$ over the detector bandwidth, which for 3G  ground-based detectors, operating at $f\,>\, (5-10) $~Hz, is already true for a stretch of data corresponding, e.g., to $T=60$~s.}
One then defines the signal-to-noise ratio $(S/N)_{ab}$ of a pair of detectors $(a,b)$ as
$(S/N)_{ab}=S_{ab}/N_{ab}$, where
$S_{ab}$ is the ensemble average of $Y_{ab}$, and $N_{ab}$ is its root-mean-square  value,
\be\label{defSab}
S_{ab}=\langle Y_{ab} \rangle\, ,\qquad\quad
N_{ab} =  \left[ \langle Y^2_{ab} \rangle - \langle Y_{ab} \rangle^2 \right]^{1/2} \, ,
\ee
and the brackets denotes the ensemble average.
For correlated noise \eq{Sn1noT} generalizes to
\be\label{calNab}
\left\langle \tilde{n}_a^*(f)\tilde{n}_b(f') \right\rangle =
\delta(f-f')\frac{1}{2}{\cal N}_{ab}(f) \, ,
\ee
so that, when $a=b$, ${\cal N}_{aa}(f)=S^{(a)}_n(f)$, while the off-diagonal terms with $a\neq b$ represent the correlated component of the noise among the  two detectors.
Therefore
\be
S_{ab}=\int_{-\infty}^{\infty}df\, \[
\langle\tilde{n}^*_a(f)\tilde{n}_b(f)\rangle + \langle\tilde{h}^*_a(f)\tilde{h}_b(f)\rangle 
\]
\tilde{Q}_{ab}(f)\, ,
\ee
where we assume that the noise in the detector $a$ is uncorrelated with the signal in detector $b$.
Note that, for observations taking place over a finite observation time $T$, the Dirac delta $\delta(f-f')$ in \eqs{avediv9}{calNab} must actually be replaced by the regularized Dirac delta, given by\footnote{See ref.~\cite{Belgacem:2024ohp} for a rigorous derivation.} 
\be
\delta_T(f-f')=\int_{-T/2}^{T/2} dt\, e^{-2\pi i (f-f') t}\, ,
\ee
so in particular 
\be\label{delta0T}
\delta_T(0) =T\, .
\ee
The $\langle hh\rangle$ correlator can be computed using \eqsss{tildehaFAdiv9}{avediv9}{eq: norm ORF}{delta0T}, and one obtains
\be\label{SabcorrQcorrN}
S_{ab}=
\frac{T}{2} \, \int_{-\infty}^{\infty}df\, \[ {\cal N}_{ab}(f) + S_h(f) \Gamma_{ab}(f) \] \tilde{Q}_{ab}(f)\, .
\ee
The computation of the variance of the noise is also standard (e.g. section~7.8.3 of \cite{Maggiore:2007ulw}) and gives (in the limit in which we want to extract a small signal from a much larger noise)
\be\label{Nab2}
N_{ab}=\[ \frac{T}{4}\int_{-\infty}^{\infty}df\,  |\tilde{Q}_{ab}(f)|^2\, S_{n,ab}^2(f) \]^{1/2}\, ,
\ee
where 
\be
S_{n,ab}=\[ S^{(a)}_n(f) S^{(b)}_n(f) \]^{1/2}\, ,
\ee
and $S^{(a)}_n(f), S^{(b)}_n(f)$ are the noise spectral densities of the individual detectors.
In \eq{SabcorrQcorrN} it is convenient to write
$S_{ab}=S_{ab}^{\cal N}+S_{ab}^{\cal H}$, 
where 
\bees
S_{ab}^{\cal N} &=& \frac{T}{2} \, \int_{-\infty}^{\infty}df\,  {\cal N}_{ab}(f) \tilde{Q}_{ab}(f)\, ,\label{defSNdiv9}\\
S_{ab}^{\cal H}&=&\frac{T}{2} \, \int_{-\infty}^{\infty}df\,  
S_h(f) \Gamma_{ab}(f)  \tilde{Q}_{ab}(f)\, .\label{defSHdiv9}
\ees
In the absence of correlated noise, we only need to maximize $S_{ab}^{\cal H}/N_{ab}$. This is obtained choosing \cite{Allen:1997ad}
\be\label{Qabopt}
\tilde{Q}_{ab}(f) = {\rm const.}\, \frac{\Gamma_{ab}(f)S_h(f)}{ S_{n,ab}^2(f)}\, ,
\ee
and leads to \eq{eq:SNRcross2det} (the overall constant in $\tilde{Q}_{ab}$ is irrelevant and cancels in the signal-to-noise ratio). Note that the signal-to-noise ratio in \eq{eq:SNRcross2det}  grows as $T^{1/2}$. This is due to the fact that the signal-signal correlator grows as the observation time $T$ while, for uncorrelated noise, the noise-noise correlator  only grows as $\sqrt{T}$ (as in a random walk), see \eq{Nab2}. Observe that the optimal filter function depends on the stochastic background that we are searching, through its spectral density $S_h(f)$. In the absence of correlated noise, the condition for detectability of a stochastic background is therefore  that, once the optimal filter function has been chosen, $S_{ab}^{\cal H}/N_{ab}>1$ (or whatever confidence level one wishes to impose).

In the presence of correlated noise (and assuming that the correlated component of the noise is smaller compared to the diagonal terms, i.e. 
$|{\cal N}_{12}| \ll {\cal N}_{11}$ and $|{\cal N}_{12}| \ll  {\cal N}_{22}$), the optimal filter function will still be given by \eq{Qabopt}.
However, correlated noise cannot be beaten by integrating for a long time, since $S_{ab}^{\cal N}$ and
$S_{ab}^{\cal H}$ in \eqs{defSNdiv9}{defSHdiv9} both grow as $T$.
Therefore, in order for the signal to beat correlated noise, we must have $S_{ab}^{\cal H}/S_{ab}^{\cal N}> 1$ (or whatever confidence level one wishes to impose). The sensitivity 
of the detector pair to a stochastic background is therefore now determined by two conditions, 
$S_{ab}^{\cal H}/N_{ab}>1$ and 
$S_{ab}^{\cal H}/S_{ab}^{\cal N}> 1$, which can be rewritten as
\be\label{SoverNmax}
\frac{S^{\cal H}_{ab}}{N_{ab}} > {\rm max}\left( 1, \frac{S^{\cal N}_{ab}}{N_{ab}}\right)\, ,
\ee
where $N_{ab}$ is given by \eq{Nab2}, with the filter function chosen as in \eq{Qabopt}. Note that, if one  interprets
$S^{\cal N}_{ab}$ as a signal, correlated among the two detectors, then 
$S^{\cal N}_{ab}/N_{ab}$ is the corresponding signal-to-noise ratio with respect to the uncorrelated noise $N_{ab}$. 
The condition that $S^{\cal H}_{ab}/N_{ab}$ be larger than $S^{\cal N}_{ab}/N_{ab}$ therefore expresses the requirement   that the signal-to-noise ratio of the cosmological signal should be above that of the correlated noise.\footnote{Note that, even when $S^{\cal N}_{ab}/N_{ab}>1$, if the correlated noise is sufficiently well modeled, one could still dig below this new sensitivity curve determined by the correlated noise, by performing   a Bayesian analysis  that determines simultaneously  both the noise and the signal, see also the discussion in section~\ref{sect:RecCorrNoise_div2}. A similar strategy can be applied to separate two cosmological backgrounds, see section~\ref{subsec:source_separability}.}

\begin{figure}[t]
    \centering
    \includegraphics[width=.9\textwidth]{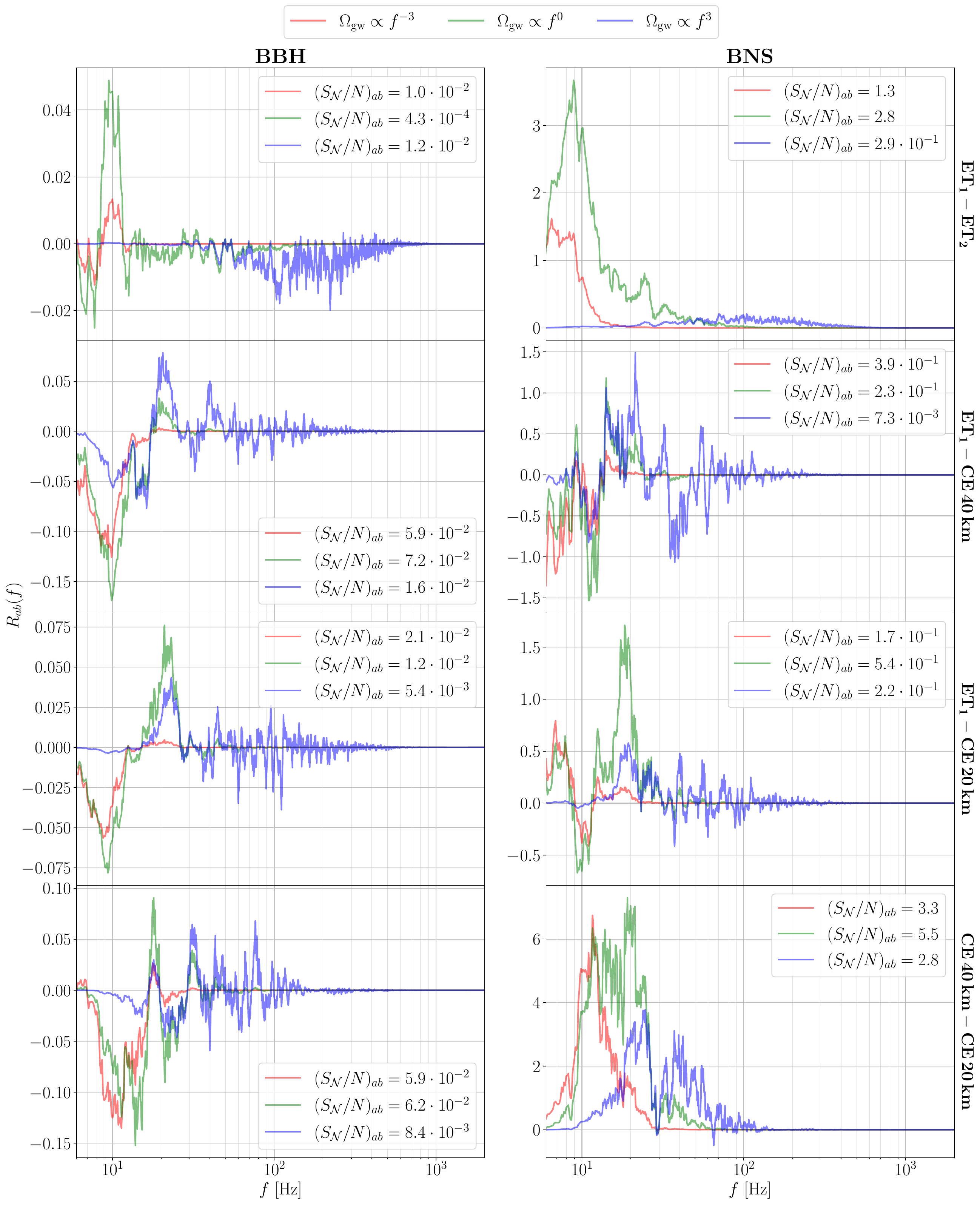}
    \caption{The function $R_{ab}(f)$ that determines the signal-to-noise ratio of the astrophysical confusion noise through \eq{SoverNdShdlogf}, for three different cosmological searches corresponding to $\Omega_{\rm gw}(f)\propto 1/f^3$,
    $\Omega_{\rm gw}(f)\propto {\rm const.}$, and $\Omega_{\rm gw}(f)\propto f^3$. The inset gives the quantity $( S_{\cal N}/N)_{ab}$, defined in \eq{SoverNdShdlogf} as the integral of  $R_{ab}(f)$ over $d\log f$. We use $\snrth=12$ as threshold for the resolved sources. The left column refers to the contribution of BBHs and the right column to BNSs.
    From ref.~\cite{Belgacem:2024ntv}.
    }
    \label{fig:R_ab}
\end{figure}

From the previous expressions, one finds that $S^{\cal N}_{ab}/N_{ab}$ can be rewritten as~\cite{Belgacem:2024ntv}
\be\label{SoverNdShdlogf}
\frac{S^{\cal N}_{ab}}{N_{ab}}=\left| \int_{f=0}^{f=\infty} d\log f\, R_{ab}(f)\right| \, ,
\ee
where
\be\label{defRab}
R_{ab}(f)\equiv\frac{\sqrt{2T}\, f \, {\rm Re} \big[ {\cal N}^{\rm eff}_{ab}(f)  \big] \frac{\Gamma_{ab}(f)S_h(f)}{S_{n,ab}^2(f)}}{
\[ \int_{0}^{\infty}df'\,  \( \frac{\Gamma_{ab}(f')S_h(f')}{ S_{n,ab}(f')} \)^2 \]^{1/2} 
}
\, ,
\ee
so that the function $R_{ab}(f)$ provides  information on the frequency range from which 
$S^{\cal N}_{ab}/N_{ab}$ gets most of its contribution.
Note that $R_{ab}(f)$ also depends on
$S_h(f)$, the spectral density of the cosmological background that we are searching, which enters through the filter function.

The full formalism for the evaluation of the correlator of the effective noise (\ref{defneff})
has been developed in ref.~\cite{Belgacem:2024ntv}, whom we refer the reader for technical details. The main points to be stressed are: (1) the separation between resolved and unresolved sources depends on the specific noise realization. The procedure must therefore be repeated over a large number of noise realizations for all the  BBH and BNS injections, and then  averaged over noise realizations. In ref.~\cite{Belgacem:2024ntv} was considered fo instance the backgrounds produced by 
$10^4$ BBHs and $10^4$ BNS, each reconstructed in $10^3$ different noise realizations. (2) Whether the BBH or BNS background can degrade the sensitivity to a cosmological search does not depend only on the properties of the astrophysical background and of the detector network used, but also on the specific cosmological search performed, which enters through $S_h(f)$ in \eq{defRab}. In particular, the question of whether the stochastic background due to CBCs is detectable at a given detector network is different from the question of whether it can mask a cosmological search: when searching for the CBC stochastic background, the filter function is chosen so to optimize that specific search. Such a filter is very different from the filters used to search for smooth (e.g. power-law) stochastic backgrounds. Therefore, even if, when using the filter tuned to it, the CBC stochastic background could be above detection threshold, still it is quite possible that, in a given cosmological search, it would not affect the  sensitivity of the detector network to that search.

The quality of the reconstruction of the injected signals depends strongly, of course, on the detector network used. As an example, in figure~\ref{fig:R_ab} we show the results obtained resolving the sources with a full network made by ET in the triangle configuration plus two CE detectors of 40~km and 20~km arms, respectively. The plot shows the function $R_{ab}(f)$, where the indices $a,b$ refer to the correlation between different pairs of detectors of this network (here ${\rm ET}_1$ and ${\rm ET}_2$ denotes two independent detectors among the three that form the triangle configuration), and the corresponding value of $S^{\cal N}_{ab}/N_{ab}$ for the given pair are shown in the inset. As mentioned above, the result depends on the shape of the cosmological spectrum that one is searching, and three cases are shown here: a ``red'' spectrum $\Omega_{\rm gw}(f)\propto 1/f^3$, a flat spectrum $\Omega_{\rm gw}(f)\propto {\rm const.}$, and a ``blue'' spectrum $\Omega_{\rm gw}(f)\propto f^3$ (i.e., because of \eq{eq:Shdiv9}, $S_h(f)\propto 1/f^6$, $S_h(f)\propto 1/f^3$ and $S_h(f)\propto {\rm const.}$, respectively). The left column shows the result for the background generated by $10^4$ BBH and the right column for $10^4$ BNS.\footnote{The result for larger number of events, $\nev$, corresponding to longer integration times, can be obtained with approximate rescaling, but the result does not change dramatically since $R_{ab}(f)$  scales as $\nev^{1/2}$, so the result for $10^5$ events are larger only by a factor $\sim \sqrt{10}$.}
We see that for this 3G network made by ET+2CE, that has an excellent resolving power for CBCs, the residual ``confusion noise'' from BBHs can be pushed well below the detector sensitivity (the values of $S^{\cal N}_{ab}/N_{ab}$ in the inset are always well below 1, and the integrand is small at all frequencies), and therefore does not limit the sensitivity to these cosmological searches. The BNS confusion noise is larger, but still  (except for the CE40km-CE20km pair) does not significantly degrades the sensitivity. However, in a network made by ET only (whether in the triangle or 2L configuration) or by only the CE40km--CE20km pair, the resolving power for CBCs will be lower, and the effect of the astrophysical confusion noise will be larger.

\subsection{Tools for the null stream}
\label{sec:div9_null_stream} 

In GW astronomy, a null stream is a signal-free linear combination of inputs from three or more detectors~\cite{Guersel:1989th}. 
It may be formed for one signal at a time with any general network of at least three detectors, and it can also be formed for all signals simultaneously in the case of  ET in the  triangular configuration, because of the closed geometry of its component arms, and the almost negligible light travel times between
detectors.\footnote{Note that each detector is in turn made of two 
interferometers, one optimized for low frequencies and one  for high frequencies. We use here the  terminology from the 2020 ET Design Report Update,
\url{https://apps.et-gw.eu/tds/?r=18715}: the high-frequency (HF) and low-frequency  (LF) interferometers that make the so-called ``xylophone'' configuration are indeed referred to as ``interferometers''. The combination of a HF interferometer  and a LF interferometer (whether in a L-shaped geometry, or with arms at $60^{\circ}$ as in the triangle configuration)
is called a ``detector''.  The whole ensemble of detectors is called an ``observatory''. So, ET in the triangle configuration is made of three detectors for a total of six interferometers, while in the 2L configuration it is made of two detectors, for a total of four interferometers.\label{foot:nomenclaturediv9}}
While not adding any additional information, the null stream reduces the noise modeling uncertainties and enables otherwise-inaccessible tools for data analysis. 
An overview of the benefits and limitations of the null stream of the ET is provided in~\cite{Goncharov:2022dgl} and in ~\cite{Branchesi:2023mws}. 
We briefly outline the known use cases for the ET null stream below:

\begin{enumerate}
    \item It allows direct evaluation of the noise power spectral density (PSD) simultaneously with measuring GW signals, without introducing additional uncertainties associated with distinguishing between the signals and the noise~\cite{Goncharov:2022dgl,Janssens:2022cty}. An example is shown in figure~\ref{fig:nullpsd}. Without the null stream, the PSD is typically evaluated at signal-free segments of data; however, finding such segments would be challenging  for  ET. For instance, with the population model used in \cite{Goncharov:2022dgl},
    around 2 BBHs and 100 BNSs contribute to 1-minute segments above 2 Hz, while 24-hour segments contain approximately 300 BBHs and 2500 BNSs.
    This abundance of signals introduces the so-called ``confusion noise''. The level of confusion noise in ET is shown in figure~5 in~\cite{Goncharov:2022dgl}. Ref.~\cite{Wu:2022pyg} showed the impact of such confusion noise on the loss of matched filtering and estimated the corresponding reduction in the horizon reach of the detector to be the order of a few percent in redshift. However, since Bayesian parameter estimation requires the full background PSD consisting of instrument and confusion noise, the use of the instrument noise PSD still needs to be demonstrated. Additionally, knowledge of the instrument PSD allows, e.g., a straightforward evaluation of the contribution of the stochastic GW background as an alternative method to the correlation based measurements \cite{Regimbau:2012ir}. It should be noted that correlated instrument noise limits the sensitivity of both search methods equally.
    \item It allows a more rapid and computationally efficient gravitational-wave parameter estimation, as shown in~\cite{Wong:2021eun}. The ET null stream allows  decomposing the data into the so-called signal space and the noise space. The signal space is represented, effectively, by only two data streams instead of the standard three data streams from three ET components. 
    \item It allows to eliminate glitches -- non-stationary, non-Gaussian noise artefacts -- down to the Gaussian noise level, as well as to effectively distinguish glitches from signals~\cite{Goncharov:2022dgl}. This becomes possible because glitches and non-stationarities appear as outliers in the Gaussian null stream background, whereas GWs do not appear in the null stream at all. However, since the null stream only contains transients that do not mimick a GW signal consistently in the ET triangle, it is an open question how much the evaluation of the actual search background can be improved with the help of the null stream.
    \item It informs on the relative calibration errors of the ET components~\cite{Schutz:2020hyz,Ajith:2006qk}. A calibration error in a given ET component would make a fraction of a signal leak into the null stream. The procedure would be especially beneficial for the loudest GW signals. Small relative calibration errors would provide indirect evidence that the absolute calibration of the three detectors is accurate as well. Instead, larger than expected relative calibration errors would hint to a problem with detector calibration, which then needs to be addressed by the commissioners.
\end{enumerate}

\begin{figure*}[t]
    \centering
    \begin{subfigure}[b]{0.49\textwidth}
        \includegraphics[width=\textwidth]{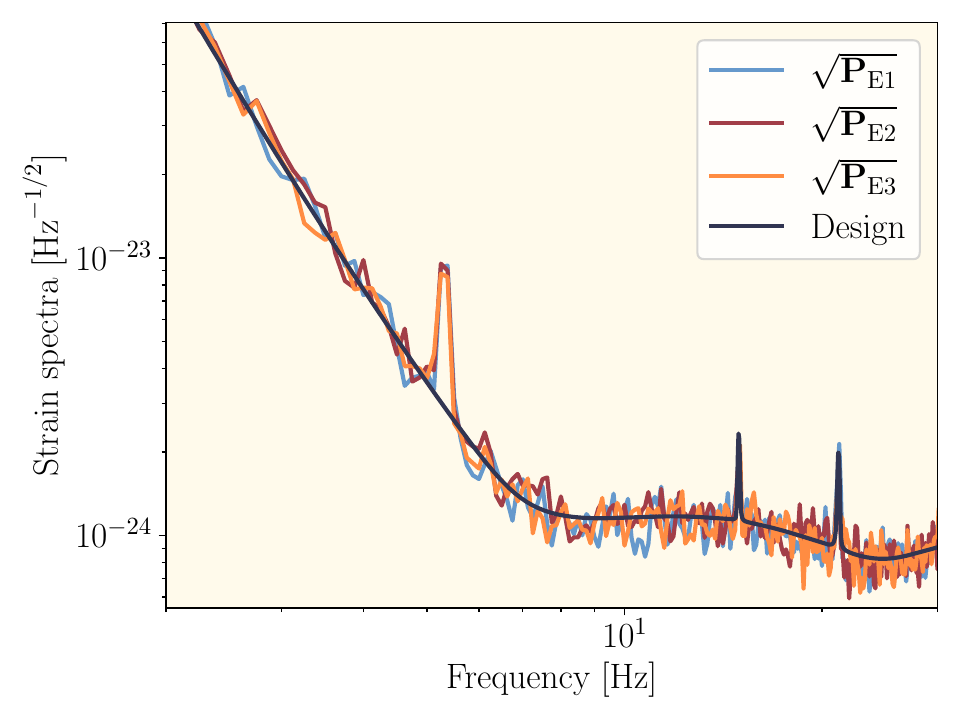}
        \label{fig:asd2-60}
    \end{subfigure}
    \begin{subfigure}[b]{0.49\textwidth}
        \includegraphics[width=\textwidth]{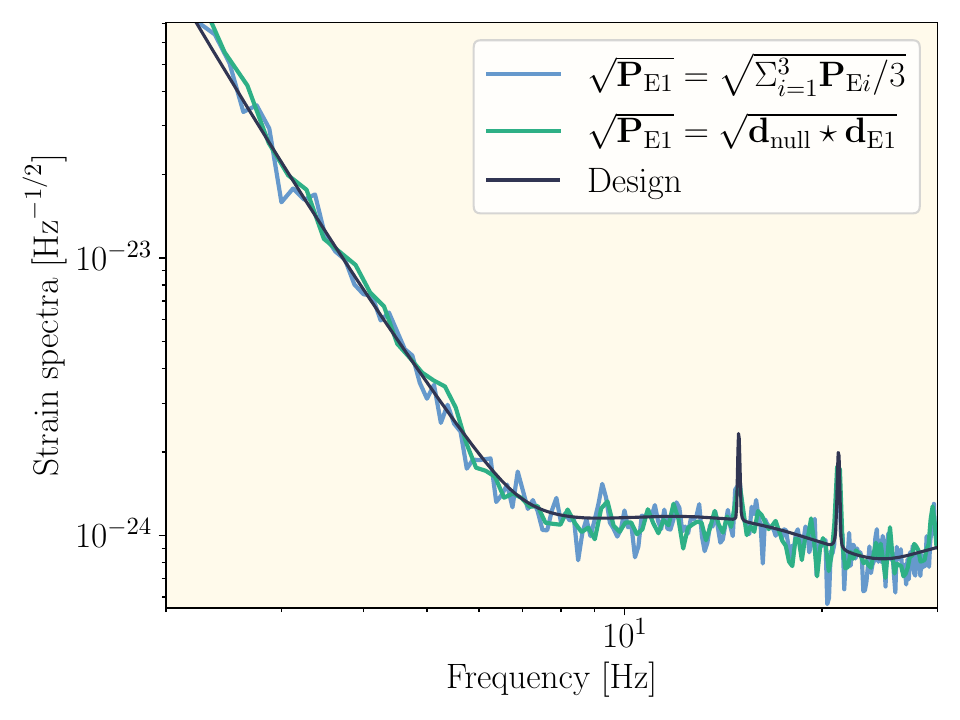}
        \label{fig:asd2-60-60}
    \end{subfigure}
    \vspace{-1.2\baselineskip} 
    \caption{Analysis of the noise PSD in ET based on the null stream. \textbf{Left}: The overall PSD across the ET's components, with the presence of a gravitational-wave signal from a binary neutron star inspiral. It distinctly affects the noise PSD at 5 Hz, deviating from the expected design sensitivity. PSD estimation is based on a 128-second data span, resulting in the BNS signal appearing as a narrowband feature. \textbf{Right}: PSD assessment in one of the ET components, utilizing the cross-power between this component and the null stream, alongside the established linear relationship between the PSD of the null stream and the detector. The impact of the BNS signal on the noise PSD has been effectively eliminated. From ref.~\cite{Goncharov:2022dgl}.
    }
    \label{fig:nullpsd}
\end{figure*}

A null stream \texttt{Python} package based on~\cite{Goncharov:2022dgl} is available in the open-access data repository online.\footnote{\href{https://github.com/bvgoncharov/nullstream}{github.com/bvgoncharov/nullstream}}. 
After an installation, the package may be loaded using the command \texttt{import nullstream}.
In the \texttt{core} module, it currently contains a general class \texttt{NullStreamBaseTD} which is, essentially, a blueprint for the null stream time series. 
In the \texttt{utils} module, there are classes for null stream time series compatible with external packages for GW data analysis, \texttt{Bilby}~\cite{Ashton:2018jfp} and \texttt{PyCBC}~\cite{Usman:2015kfa}. 
Once the null stream time series object is constructed, it can be used for standard analyses such as PSD evaluation as in figure~\ref{fig:nullpsd}, matched filtering of signals and glitches, as well as the background evaluation. 
For \texttt{Bilby}, the package also contains a class to create a null stream noise likelihood object. 
Therefore, the package can also be used for likelihood analyses. 
It is open to further development and contributions.

\subsection{Conclusions}
\label{sec:div9_conclusions} 

ET is expected to have a pivotal 
role among the next generation of GW 
detectors. Its science case is immense. ET observations 
will transform GWs into magnifying lenses of fundamental 
interactions, with the unique opportunity to simultaneously 
bridge the gap between particle physics, cosmology and 
astrophysics. 

To which extent one will achieve such objectives ultimately 
depends on the ET design and geometry. For example, as we showed with the metrics introduced in this section, variations of the sensitivity at low frequencies can have a crucial impact on the distance reach, pre--merger alerts, inference on intrinsic and extrinsic parameters. The relative orientation between the detector arms, as well as the detector geometry, impact the ability to reconstruct the source parameters and the sensitivity to stochastic backgrounds. The ultimate choice about the construction of the instrument has to be informed by those metrics to maximise the science output. In this context, forecasting the detectability 
and measurement accuracy of source parameters based on the 
detector configurations is crucial for determining the 
optimal ET design and maximise its science goals.

Bayesian methods that directly sample the parameter 
distribution of GW signals injected into simulated noise 
are currently computationally too expensive for such tasks. 
This issue becomes even more significant for population-based 
studies, given the expected $\mathcal{O}(10^5)$ 
detections per year by ET. Therefore, it is important to develop approaches that can provide a 
reliable assessment of detector performance while 
maintaining manageable computational costs. This task becomes indispensable when many possible configurations are under study, as was the case for ET in the context of the detailed study~\cite{Branchesi:2023mws}.

In this section we have summarized the efforts performed within the ET Observational Science Board to address 
this problem, developing a variety of numerical 
pipelines, ready-to-use for the ET community. 
The majority of these tools, based on Fisher 
matrix methods, allow forecasting the parameter estimation of 
compact binary signals, and have been extensively used to 
study costs and benefits of ET configurations 
\cite{Branchesi:2023mws}. 
Along with CBC pipelines, however, we have developed different approaches 
that can be used, for example, to forecast 
measurement accuracy of ringdown signals, 
for null-stream analyses, and stochastic background 
studies.

This section presented the essential aspects of the formalism behind the forecasting tools, the different public packages available and their basic features, and a series of metrics and FOM that can be produced with those tools and used to characterize the performance of a network of ground--based interferometers.

While reliable within the limits of validity of the approximations they are based on, the tools presented here, such as FIM codes, can be considered as initial seeds for more sophisticated approaches. As discussed in this section, advances towards refinements in known cases of failure of such approximations  have already been considered. 

Looking forward, it will become necessary to address several aspects of real data analysis situations that cannot be directly captured by the current forecast tools. 
Developing data analysis and simulation tools that progressively include as many realistic elements as possible will become a central task. 
One important aspect is that Fisher matrix forecasts for CBCs assume that each source is resolvable, i.e. that its presence can be detected and isolated from the rest of the data stream. This is the case for current generation observatories, that are most of the time noise--dominated, with a signal detected at a rate of a few per day. On the other hand, future observatories will be signal dominated, in the sense that we should expect overlapping signals in the data stream. This aspect, on which first studies will be presented in section~\ref{section:div10}, will be among the most important future directions of study in the context, 
with the ultimate goal of developing fast data analysis tools for parameter estimation and population studies on real data streams usable by the larger ET community.

\subsection{Executive summary}

This section describes various open--source software tools developed by the ET Observational Science Board.
These tools have been designed to assess how variations in the ET detector configurations, such as sensitivity curves, geometry, and geographical locations, impact its scientific capabilities, through large--scale simulation campaigns. The evaluation of these configurations is based on specific Figures-of-Merit (FoMs), which are presented in the section.
The tools presented here are used to produce scientific results across several other sections of the present work.

\begin{highlightbox}{Software Tools for Compact Binary Coalescence (CBC) Sources}

A suite of Fisher Matrix--based parameter estimation forecast tools for compact binary signals is presented. These have been rigorously tested and validated against each other, enabling robust performance evaluations of ET based on its design, detector geometry, and Earth-based location, and providing simulation schemes used in various other science case studies of ET. 
These tools allow the following:

\begin{itemize}
    \item Estimation of signal-to-noise ratios of CBCs, alongside uncertainties on source parameters, including intrinsic ones (e.g., masses and spin) and sky localization.
    \item Use of analytic, hybrid, and phenomenological waveform models for CBC signals.
    \item Population-level inference analyses for large-scale astrophysical studies.
\end{itemize}

\vspace{0.5em} 
\noindent
\centering
\makebox[\textwidth][c]{%
    \fboxrule=1pt 
    \fboxsep=5pt 
    \fbox{%
        \begin{tabular}{@{\hskip 10pt} c @{\hskip 10pt} c @{\hskip 10pt} c @{\hskip 10pt}} 
            \href{https://gitlab.com/sborhanian/gwbench/-/tree/master}{\textsc{GWBench{}} \faGithub} & 
            \href{https://github.com/CosmoStatGW/gwfast}{\textsc{GWFast{}} \faGithub} & 
            \href{https://github.com/janosch314/GWFish}{\textsc{GWFish{}} \faGithub} \\
            \href{https://github.com/IrisLi135/TiDoFM/tree/main}{\textsc{TiDoFM{}} \faGithub} & 
            \href{https://github.com/andrea-begnoni/GW.jl}{\textsc{GWJulia{}} \faGithub} & 
            \href{https://github.com/jmsdsouzaPhD/GWDALI/}{\textsc{GWDALI{}} \faGithub} \\
        \end{tabular}%
    }%
}

\begin{itemize}
    \item Forecasting constraints on the population properties of merging compact objects.
\end{itemize}

\vspace{0.5em}
\noindent
\centering
\makebox[0pt][c]{%
    \fboxrule=1pt 
    \fboxsep=5pt %
    \fbox{%
        \begin{minipage}{0.4\textwidth} 
           
            \centering 
            \href{https://github.com/CosmoStatGW/gwfast/tree/master/gwfast/population}{\textsc{GWFast/population}{} \faGithub}

        \end{minipage}%
    }%
}

\vspace{1.9em}

\noindent
\centering
Methods and software implementations to address known limitations of Fisher Matrix approaches are also presented, including higher order and exact approximants and inclusion of prior information.

\end{highlightbox}

\begin{highlightbox}{Metrics for CBCs}

Key metrics for assessing the ET science case related to resolved CBC events are introduced and discussed.
These are produced with the software tools discussed in the section, and include:

\begin{itemize}
    \item Detectability and inference horizons, detection rates for compact binaries.
    \item Accuracy assessments for sky localization and source luminosity distance---including pre-merger alerts and multi-messenger analyses---, accounting for the dependency on Earth’s rotation.
    \item Statistical distributions of constraints on intrinsic parameters of compact objects.
    \item Constraints on population--level features depending on observation time. 
\end{itemize}


\end{highlightbox}

\begin{highlightbox}{Tools for Ringdown Analysis of CBCs}

A dedicated tool for assessing the post-merger gravitational wave emission of CBCs is introduced. This tool allows for forecasts of the measurement accuracy on the parameters characterizing the ringdown phase of compact binary mergers, enabling detailed analysis of post-merger signals.

\end{highlightbox}

\begin{highlightbox}{Stochastic Gravitational-Wave Searches}

The tools developed for forecasting the detection of stochastic gravitational wave backgrounds are introduced. Those are designed to:

\begin{itemize}
    \item Characterize the detectability of stochastic backgrounds from astrophysical and cosmological sources via the power--law sensitivity.
    \item Subtract residual "confusion noise" caused by unresolved CBC signals from the astrophysical background in the search for cosmological contributions.
\end{itemize}

\vspace{0.5em} 
\noindent
\centering
\makebox[0pt][c]{%
    \fboxrule=1pt 
    \fboxsep=5pt 
    \fbox{%
        \begin{minipage}{0.4\textwidth} 
           
            \centering 
            \href{https://github.com/CosmoStatGW/gwfast/tree/master/gwfast/stochastic}{\textsc{GWFast/stochastic{}} \faGithub}

        \end{minipage}%
    }%
}

\end{highlightbox}

\begin{highlightbox}{Null Stream Tools for Noise Reduction}

    The implementation of null stream techniques in the triangle configuration to reduce systematic errors is discussed, and a package to assess their impact is presented.

\vspace{0.5em}

\noindent
\centering
\makebox[0pt][c]{%
    \fboxrule=1pt 
    \fboxsep=5pt 
    \fbox{%
        \begin{minipage}{0.4\textwidth} 
           
            \centering 
            \href{https://github.com/bvgoncharov/nullstream}{\textsc{nullstream} \faGithub}

        \end{minipage}%
    }%
}

\end{highlightbox} 
\section{Data Analysis}\label{section:div10}

\subsection{Introduction}
As we have seen in the previous sections, with ET  GW astronomy will take a step further, with the possibility of detecting a large number of compact binary mergers at cosmological distances, and a variety of new types of sources, such as core collapses to neutron stars or black holes, rotating neutron stars, initial instabilities in young neutron stars or stochastic backgrounds. We therefore need to start developing optimal data analysis and parameter estimation methods, to be ready to extract valuable information from the data, in astrophysics, cosmology and fundamental physics.  From a data-analysis perspective, specific problems for ET include long signals from the increase in frequency bandwidth, the presence of  a very large number ($\sim 10^5$)  of signals per year within the sensitivity band~\cite{Borhanian:2022czq,Iacovelli:2022bbs,Branchesi:2023mws}, as well as signals overlapping in the 
detector's band~\cite{Samajdar:2021egv}. All of these involve a steeply rising need for computational resources, or the development of faster and accurate methods of analysing data. With the rising number of signals in the data, characterisation of noise in the detector also becomes challenging since there will be no signal-free stretch of data. In the triangle configuration for ET, the noise characterization will however exploit the existence of the   null stream~\cite{Goncharov:2022dgl,Wong:2021eun,Narola:2024qdh}, both in terms of measuring the noise power spectral density and for characterizing instrumental glitches which can have a significant impact on the quality of parameter estimation (see section~\ref{subsec:null_stream}). A 2L configuration without access to
the null stream may still achieve good noise mitigation but requires more complex methods
that introduce uncertainty into the process (see section~7 of~\cite{Branchesi:2023mws}). One could also envisage fitting directly for the noise model and the signal model to be agnostic of the underlying noise properties~\cite{Littenberg:2014oda}.

In the following, section~\ref{sec:challenges} describes these challenges in detail. In section~\ref{sec:innovative_methods} we briefly review specific innovations particularly in the area of machine learning required for analysing GW data in the ET era. Section~\ref{sec:detection} describes the current status of signal detection, their computational requirements, and possible techniques to address next-generation challenges.
Section~\ref{sec:PE} does the same for parameter estimation.
Section~\ref{sec:pec_ET} outlines peculiarities of the triangle configuration of ET, namely the use of the null stream, and the presence of correlated noise among the three nested detectors forming the triangle. In order to start systematically addressing these challenges we will provide common mock datasets, which can be used for characterisation of algorithms, study their performance and make predictions for required computational cost. The first  Mock Data Challenge  (MDC) dataset is described in section~\ref{sec:MDC}. Section~\ref{sec:common_divs} lists aspects of data analysis that involve synergies with topics discussed in the previous sections; we conclude in section~\ref{sec:conclusiondiv10}.

\subsection{Challenges}\label{sec:challenges}

\subsubsection{Long duration Compact Binary Coalescence signals }
\label{subsec:long_dur}
Current (second generation)  GW detectors have a sensitive frequency band that starts at around 20\,Hz~\cite{KAGRA:2013rdx}, whereas it is expected that ET will have good sensitivity down to $\sim 5$\,Hz. Due to the chirping nature of compact binary coalescences (CBC) signals, where the frequency evolves more slowly at lower frequencies, this means that CBC signals will spend dramatically longer  times in the sensitive frequency band of ET than in previous detectors. For instance, an equal-mass BNS signal with (detector frame) component masses of $1.4 \ \Msun$ will last for $\sim 107$ minutes in the detector's band, while an equal-mass  BBH signal with (detector frame) component  masses of $30 \ \Msun$ will last for $\sim 40$ seconds in the band, assuming that they enter the detector bandwidth at 5 Hz.
In comparison, the same BNS and BBH signals would be $\sim 160$ seconds and $\sim 0.9$ seconds long, respectively, in current generation detectors with a lower frequency sensitivity of 20 Hz.\footnote{Note, however, that the time spent in the bandwidth depends on the detector-frame masses $m_{\rm det}$. These are related to the 
intrinsic (source-frame) masses $m_s$ and the source redshift $z$ by $m_{\rm det}=(1+z)m_s$. Then, for instance, a BBH signal with source-frame component masses of $30\ \Msun +30 \ \Msun$, at $z=2$, will have detector-frame component masses  of $90\ \Msun +90 \ \Msun$, and will stay in the ET bandwidth only about 6~seconds.}
BNS at small distances could enter the ET bandwidth even earlier, and their signal will therefore be corresponding longer. E.g., a source with the properties of GW170817 would stay in the ET bandwidth for about 24 hours, see figure~\ref{fig:GW170817atET}.

\begin{figure}[t]
    \centering
    \includegraphics[width=.8\textwidth]{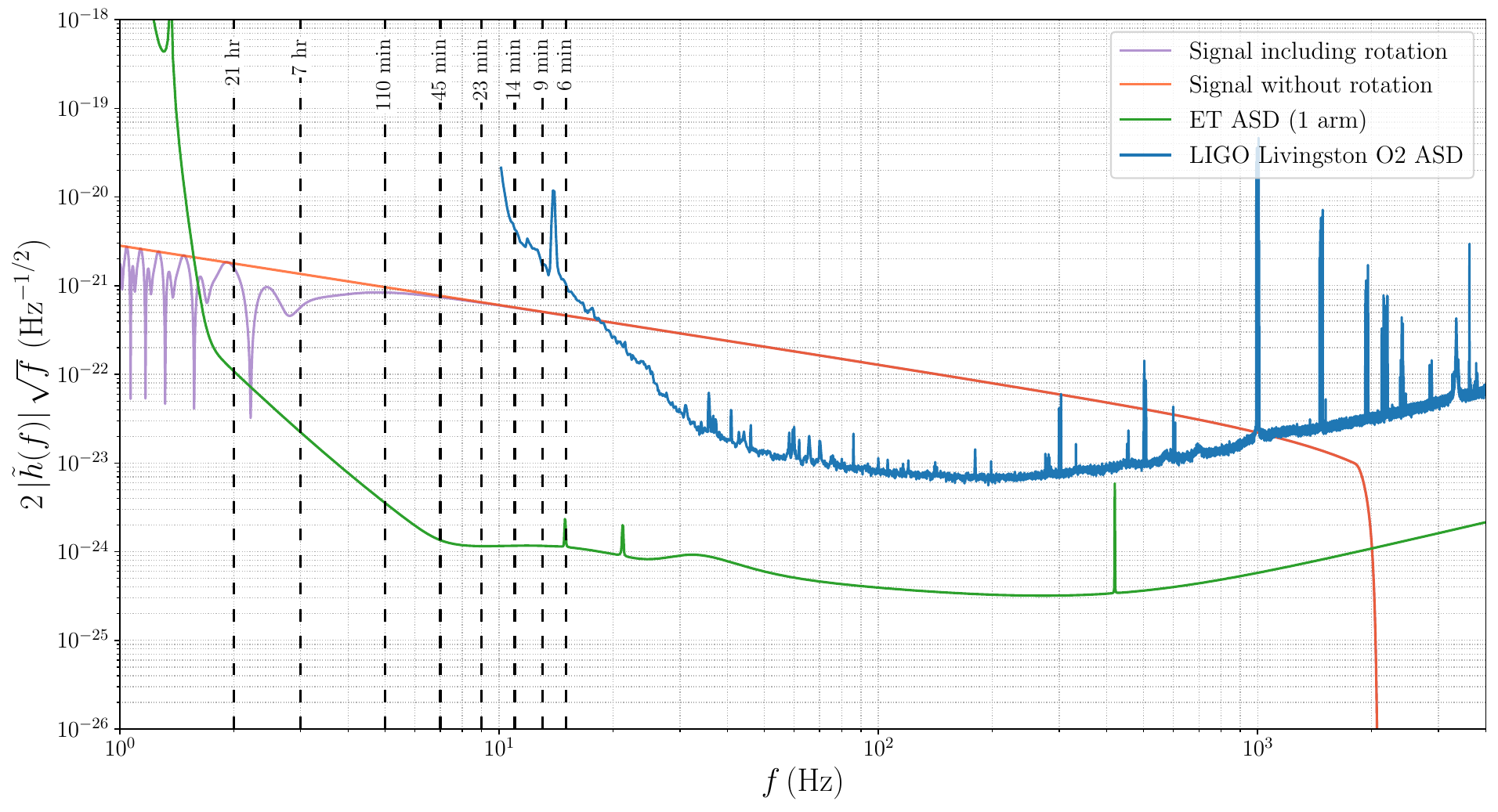}
    \caption{Comparison between the amplitude of  a BNS signal analogous to GW170817, as seen by ET in its  triangular configuration,    accounting for the effect of Earth's rotation (violet curve) and without  accounting for it (orange curve). The dashed vertical lines indicate the amount of time left before coalescence and the amplitude is computed using the waveform model \texttt{IMRPhenomD\_NRTidalv2}. For comparison, we also show a representative LIGO Livingston sensitivity curve during the second observing run.  Figure taken from~\cite{Iacovelli:2022bbs}.}
    \label{fig:GW170817atET}
\end{figure}

These longer signals pose a data analysis problem, as they require an increase in  the number of data points needed to describe them. This translates directly into a linear increase of the computational cost of evaluating the overlap integral or inner product required for standard matched filter-based searches and parameter estimation. This problem, already significant in current generation detectors, becomes prohibitive for low-mass signals in next generation detectors, and requires techniques to accelerate the evaluation of the waveform generation, as already discussed in  section~\ref{sec:acc},  and  the inner product; we will come back to this topic in 
section~\ref{subsec:PEacceleration}.
The measured signal will also be modulated in amplitude and phase by the motion and rotation of the Earth on a 1-day timescale, as visible again in figure~\ref{fig:GW170817atET}. The slow modulation relative to the frequency of the signal has previously been neglected over the minute-long timescale of 2G signals; however, at 3G detectors, for BNS systems in particular this may yield useful information about the localisation of the source, even prior to the merger itself \cite{Magee:2021xdx,Nitz:2020vym, Hu:2023hos}, see also the discussion in section~\ref{sec:div9_pattern_func_earth_rot}. The time and frequency-dependent nature of this modulation has the undesirable effect of making the inner product integrand dependent on the extrinsic sky-location parameters, which normally can be factored out of the problem as a single complex number describing the antenna response. To fully treat this problem, a na\"ive approach would require a template bank to cover not only the intrinsic mass and spin parameters but also the sky location. This problem already exists for continuous wave searches, where the vast number of distinguishable sky locations contributes to the prohibitive cost of an exhaustive all-sky search at sensitivities that should be theoretically possible given the detector performance \cite{KAGRA:2022dwb}. In other words, there is a risk of encountering a computational limit in sensitivity before we exploit the full performance of the detector. While existing methods may perform well across much of the parameter space, combining them with new techniques will be essential for efficient and feasible analysis of signals, particularly given the extreme volumes of data expected from the Einstein Telescope.

In addition to the increase in data quantity, the longer signal duration, particularly for signals lasting many hours, violates several of the assumptions that are made in deriving the standard Whittle likelihood function~\cite{Whittle:1957}. In particular, the detector noise curve may be expected to change over hours-long timescales, due to environmental and technical influences~\cite{Zackay:2019kkv,Edy:2021par,Kumar:2022tto}.
This breaks the assumption of stationarity in the noise-generating process, that results in the possibility of reducing the likelihood computation to a one-dimensional sum over frequencies. If the timescale over which the noise PSD changes is long relative to the autocorrelation time of the noise itself, then the effect of this should be minimal on the actual inference of signals, provided it is modeled sufficiently well. However, techniques to do such modeling are yet to be fully developed into a prototype pipeline for detection or parameter estimation.
A potentially more serious issue comes from the transient noise artefacts known as ``glitches'' that are observed to be present in all detector data to date~\cite{Blackburn:2008ah,LIGOScientific:2016gtq,LIGO:2021ppb}. These glitches have multiple causes, depending on which they may be well modeled or relatively unpredictable in their morphology \cite{Nuttall:2018xhi,Davis:2022dnd}.
As the duration of signals increases, so does the probability that one or more glitches will occur during the time when the signal is in band. This problem has already been observed in 2G detectors, most notably with the BNS signal GW170817~\cite{LIGOScientific:2017vwq}, where the glitch was relatively simple and could be removed from the data prior to analysis of the GW of interest. However, the frequency of this occurrence in 3G data may be expected to overwhelm any human-led method of glitch removal. This motivates simultaneous modeling and analysis of glitches which overlap signals. Such techniques have been demonstrated~\cite{Littenberg:2014oda,Cornish:2014kda,Pankow:2018qpo,Cornish:2020dwh} for 2G and space-based detectors; however, their scaling to the 3G problem is yet to be established. Machine learning approaches based on generative models of detector data are a promising alternative~\cite{Sun:2023vlq,Xiong:2024gpx}, but have yet to be demonstrated for long duration signals.

Finally, the longer signal duration, coupled with the improved sensitivity, means that multiple signals will be present in the sensitive band simultaneously. These overlapping signals may be individually resolvable, or there may be one or more undetectable signal that effectively contributes transiently to the noise process. This problem is described in section~\ref{subsec:overlap} below.

Taken individually, each of the challenges described above has been considered and methods proposed for their solution. The ultimate challenge of 3G analyses lies in developing methods that can address them all simultaneously, whether by making adequate simplifying assumptions to extend traditional methods based on matched filtering or by developing new techniques that change the nature and scaling costs of the analysis problem. A promising avenue for the latter approach lies in machine learning models, in particular simulation-based inference, which relies only on being able to generate examples of the dataset, rather than explicitly writing the likelihood function accounting for all possible physics.

\subsubsection{Overlapping signals}
\label{subsec:overlap}

 Chances of observing \emph{overlapping signals} will increase~\cite{Samajdar:2021egv} in the 3G era. Our current pipelines, designed for single GW signals, may yield biased results when applied to overlapping signals. This is specifically true for parameter estimation where the likelihood-based Bayesian approach assumes a single-signal model to model the data. As detailed in section~\ref{subsec:cbc_det}, previous studies have focused on detection of GWs. An updated search campaign on overlapping signals is proposed by \cite{Wu:2022pyg}, where the authors additionally consider the effects of using the traditional matched filtering and its consequences on estimating the noise properties as well as detection rates of overlapping signals.
 
However, the first parameter inference study~\cite{Samajdar:2021egv} done with simulated overlapping signals in the 3G era and attempted recovery using conventional Bayesian approaches with a single-signal model, did not show significant bias in parameter estimates across most of the parameter space.  Parameter estimates do however show biases when the merger times of two overlapping signals are close to each other and depends furthermore on the ratios of the signals' loudness, with conventional analyses performing worse as signal SNRs become comparable, as well as signal durations. Figure~\ref{fig:overlap-prob} shows an example where the overlapped heavy BBH merger atop a BNS merger with same merger times (top panel) cannot be recovered with a single-signal model. In the same figure, the lighter BBH (lower panel) can be recovered at all times, as well as the heavier BBH when the merger time is before the merger time of the underlying BNS.  The overall conclusions were supported by several other studies~\cite{Pizzati:2021apa, Relton:2021cax, Himemoto:2021ukb, Antonelli:2021vwg, Johnson:2024foj, Wang:2023ldq}. While ref.~\cite{Samajdar:2021egv} was the only one to simulate also binary neutron star sources, ref.~\cite{Relton:2021cax} looked at the effects of spin modulations and if they could mimic effects from overlapping signals, and ref.~\cite{Pizzati:2021apa} looked at the parameter space in more detail and predicted where biases in inference would occur. 
 \begin{figure}
    \centering
    \includegraphics[width=\textwidth]{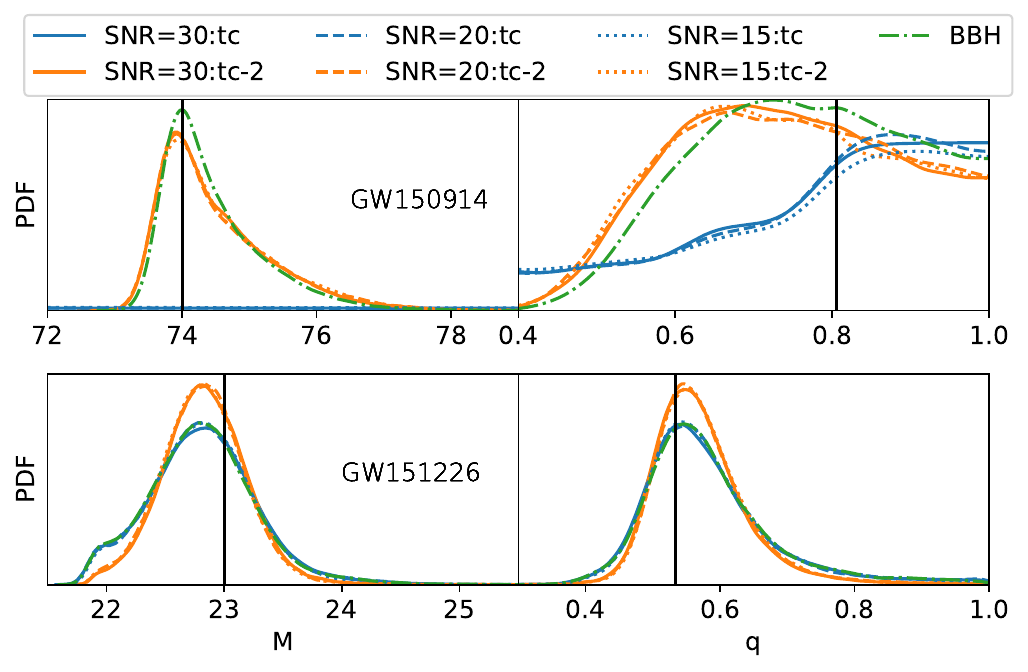}
    \caption{Posterior PDFs for total mass and mass ratio, for the GW150914-like signal 
    (top panel) and the GW151226-like signal (bottom panel) when they are respectively being overlapped with 
    a BNS signal with SNR = 30 (solid lines), SNR = 20 (dashed lines), and  
    SNR = 15 (dotted lines). The overlaps are made so that
    the BBH and the BNS end at the same time (\texttt{tc}), or so that the  
    BBH ends 2 seconds before the BNS (\texttt{tc-2}). 
    Finally, posterior PDFs for the two BBH signals by themselves are shown as 
    green, dashed-dotted lines (\texttt{BBH}). 
    The injected parameter values are indicated by black,  
    vertical lines. Figure taken from~\cite{Samajdar:2021egv}.}
    \label{fig:overlap-prob}
\end{figure}
 
Possible remedies to this problem have been suggested, either from a Fisher Matrix study~\cite{Antonelli:2021vwg} or adapting the signal model accordingly in the Bayesian likelihood~\cite{Janquart:2023hew}. In \cite{Langendorff:2022fzq}, normalizing flows were used as an avenue to deal with the computational burden coming from multiple-signal analyses in case of overlaps. Moreover, \cite{Hu:2022bji} studied the effects of waveform inaccuracy and overlapping signals on tests of GR, concluding that combining signals—i.e., constructing a joint likelihood from multiple data streams to robustly estimate a common parameter like a parameter characterising deviations from GR—can lead to false GR deviations when multiple signal overlaps occur. More recently,~\cite{Dang:2023xkj} extended this to higher post-Newtonian (PN) deformation parameters. Analysing all simulated signals individually for GR deviations and then combining them all, they concluded that a non-negligible number of overlapping signals can lead to false GR violations at the individual event level. However, when the bounds are combined, the biases tend to smooth out, leading to a preference for GR at the population level inference. While overlapping signals will cause biases in estimating parameters in corner cases (depending on ratios of signal loudness, length of overlapped signals), parameter offsets even in a few cases will likely have a large impact on estimating common parameters from a set of signals. Since Bayesian analysis allows a more robust bound by combining several independent data sets, parameters characterising fundamental properties of the Universe like deviations from GR, or Hubble's constant~\cite{Dhani:2024jja} may get systematically biased from offsets in a few analyses. In the case of mass measurement offsets in some signals, the same conclusion will likely affect population properties of compact objects like black holes and neutron stars. Whether these biases average out or accumulate will depend on whether the offsets consistently shift measurements in a particular direction or are effectively random, which can vary depending on the specific problem.

\subsubsection{Noise Background estimation}
\label{subsec:background}

The estimation of the properties of the background noise is an open problem, as ET data will be polluted by a large number of astrophysical signals, making it difficult to evaluate the Power Spectral Density (PSD) of the instruments and the False Alarm Rate even at the level of significance actually used by LIGO and Virgo. 

Early MDC results (2012), where a population of binary neutron stars (BNS) was injected into Gaussian noise, demonstrated that individual PSDs were contaminated by the GW signals and that the use of the null stream (see section~\ref{subsec:null_stream}), in the case of the triangle configuration, would give a better estimate \cite{Regimbau:2012ir}. 
Other studies \cite{Goncharov:2022dgl} indicate that, depending on the length of time segments used to estimate the noise spectra, the presence of a number of simultaneous signals can add a significant contribution to the detector PSD at low frequencies. 

On the other hand, in order to be able to estimate the false alarm rate due to noise alone, current detectors use a very specific technique, which consists of creating time shifts between the data from different detectors. This technique ensures that each coincidence is nothing more than noise.
With ET, we will enter the signal dominated regime and the detection rate will be such that the probability of finding a coincidence between two different sources when applying time shifts will not be negligible anymore. As a result, the false alarm rate estimated with the time-shift method will overestimate the noise background. 
In the case of ET in a triangular configuration, with the help of the null stream, it could be possible to provide a more accurate estimate. Innovative methods should be developed to address this issue and some ideas will be presented below.

\subsubsection{Source subtraction for CBCs}
\label{subsec:source_subtract}

The gravitational wave signal will be dominated by compact binary coalescences that, together, create a foreground that  could  mask a cosmological background from processes that happened in the very early Universe, such as inflation, phase transitions or cosmic strings (see section~\ref{sec:early_universe}). In this context, subtracting individual detected signals correctly is crucial to search for a cosmological background.  Early forecasts of the effect of this confusion noise on stochastic searches  used different heuristic expressions for the residuals errors after CBC subtraction~\cite{Sachdev:2020bkk,Zhou:2022nmt,Pan:2023naq,Song:2024pnk}. A first principle approach to the problem was recently developed in
\cite{Belgacem:2024ntv}, and is discussed in section~\ref{sect:subtractastrobkgdi9}. 

Promising methods, like the projection methods~\cite{Cutler:2005qq,Sharma:2020btq}, the Bayesian method by~\cite{Biscoveanu:2020gds} or the unmodeled notching in the time frequency plan~\cite{Zhong:2022ylh,Zhong:2024dss} will be investigated further using simulated data.

Most of these challenges will also be present in LISA and have been studied for many years. Although data analysis codes may not be directly transferable to ET, some of the methodologies could potentially be adapted.

\subsection{Innovative methods: machine learning applications }
\label{sec:innovative_methods}
The data analysis for the Einstein Telescope detector will present several challenges, as discussed  in section~\ref{sec:challenges}, particularly due to the instrument's enhanced sensitivity overall and its exceptional performance at low frequencies. We must be prepared to process data from astrophysical signals that remain in the band of interest for extended periods, necessitating the analysis of longer time series and transient events. Additionally, in particular for the colocated detectors in the triangle configuration for ET, we must tackle the issue of correlated noise, which could generate correlated transient signals that could be mistaken for astrophysical sources (see section~\ref{subsec:corr_noise} for extended discussion).
The noise may be non-stationary and display non-linearities such as couplings between different sources of noise, a phenomenon observed even in second-generation detectors.
We should aim at efficiently and quickly using signals from noise monitoring (environmental, control, etc.) to enable a rapid alert system, necessary for multi-messenger astrophysics.
These will necessitate the use of alternative and innovative methods compared to the standard data analysis pipelines  currently employed for 2G detectors.

In recent years, significant progress has been made in the development of analysis methods based on machine learning, including applications to gravitational wave analysis \cite{Cuoco:2020ogp}. While we will not delve into the general principles of machine learning in this discussion, readers interested in a more comprehensive overview of the topic can refer to \cite{Theobald:2017, Goodfellow:2016}.

It is possible to identify several domains in which the application of these methods could yield significant benefits in terms of noise characterization, signal detection, early warning strategies, and parameter estimation.
A recent review \cite{Cuoco:2024cdk} outlines the current advancements in machine learning applications for second-generation gravitational wave interferometers. While some of these techniques are still applicable to ET data, their effectiveness may be limited due to the unprecedented sensitivity of  ET, and the new challenges that it poses.

For example, one of the main problems for current generation detectors is the presence of glitches, i.e. transient noise signals which can spoil the detection.
We cannot make predictions on the glitches that will be present in the next generation of detectors, but we could expect many more due to the better sensitivity and, as discussed in section~\ref{subsec:long_dur}, the problem becomes worse for long signals. A key focus of the upcoming MDC preparation will be to incorporate noise glitches resembling those observed in LIGO/Virgo detectors into our datasets. This will allow us to evaluate and refine machine learning-based data cleaning solutions.

The first fundamental approach is to tackle binary classification by attempting to distinguish signals from noise. However, the presence of overlapping signals can significantly complicate this task, making even a seemingly straightforward binary classification highly problematic. When signals overlap, the boundary between what constitutes ``signal" and ``noise" becomes blurred, increasing the challenge of accurate separation.
Machine learning is a promising approach to glitch classification and characterization \cite{Cuoco:2021ngj}. The classification algorithms can be applied to features extracted from glitch time series, e.g. wavelet decomposition, that can then be classified by adopting various approaches such as classification trees \cite{Powell:2015ona,Powell:2016rkl}. 
More complex, deep-learning based approaches can also be  used for glitch classification. Convolutional Neural Networks are well suited for image classification since they can extract and classify features in 2D data, and have therefore been applied to the classification of spectrograms of glitches \cite{Razzano:2018fxb}. This approach is particularly effective since glitch morphologies can be quite rich and complex to analyze with simple image classification algorithms \cite{Zevin:2016qwy}.
The most effective approach to spectrogram classification is based on supervised learning, which requires a large sample of labeled glitches. To this purpose, citizen science projects as GravitySpy \cite{Zevin:2016qwy} and GWitchHunters \cite{Razzano:2022lgg} have been developed, where volunteers across the globe are asked to view and visually analyze and classify glitch spectrograms.

An unsupervised approach has also been applied to the problem of glitch classification, with the goal of overcoming the need for large labeled datasets. To this scope, autoencoders have been shown to be quite effective for finding anomalies among glitch populations, as well as denoising of data \cite{Shen:2019ohi,Laguarta:2023evo}. Another way of tackling the lack of large datasets is to create large synthetic samples of glitches, a task that can be achieved with the application of generative adversarial networks \cite{Lopez:2022lkd,Powell:2022pcg}.

Some of the applications of machine learning techniques for signal detection and parameter estimation will be described in section~\ref{subsec:MLdetection} and section~\ref{subsec:MLPE}.

\subsection{Signal detection method}\label{sec:detection}
\subsubsection{Data analysis methods for Compact Binary Coalescences}\label{subsec:cbc_det}
The detection of GW signals from quasicircular compact binaries coalescence can be performed through the application of matched-filtering \cite{Allen:2005fk}, as the signals are modeled and we know their evolution in amplitude and phase. The method has been successfully applied to find signals in data for the GW interferometers currently in operation and has led to the detection of gravitational waves \cite{Abbott:2016blz}. Therefore, it seems natural to consider using the matched-filter method in ET,  taking however into account the different conditions both in terms of signal characteristics and the nature of the ET noise.

Several pipelines have been designed and are currently used for CBC searches inside the LIGO-Virgo-Kagra (LVK) collaborations \cite{Aubin:2020goo, Ewing:2023qqe, Chu:2020pjv, Usman:2015kfa}; groups external to LVK have implemented as well matched-filter methods and have analysed LVK data \cite{Wadekar:2024zdq}. The pipelines rely on the development of banks of simulated waveforms, called templates, to filter the data; different methods, mainly using geometric or/and stochastic approaches are commonly used to construct the banks (see e.g. \cite{Cokelaer:2007kx, Harry:2009ea, Roy:2017oul}).
As we discussed in section~\ref{subsec:long_dur}, due to the high sensitivity of ET, some CBC signals can be visible for thousands of seconds. This means that long templates are required, which can be computationally expensive. Filtering with such long templates has nevertheless been shown to be doable using approaches already used in the current data analysis platform as decomposition of the analysis in several frequency bands \cite{Aubin:2020goo} or SVD approach \cite{Meacher:2015rex, Ewing:2023qqe}. The computational cost is also linked to the number of templates necessary to filter the data, as this number will grow with the widening of the search frequency band.

A ubiquitous approximation made in today's search algorithms is that the sky location of a compact binary is fixed (from the point of view of an Earth-fixed observer) during the duration of the visible signal in the detector data. This is appropriate because the duration of the inspiral phases, in 2G detectors, goes from fractions of a second for binary black holes to a few minutes for binary neutron stars. This simplification makes the angular response of an interferometer time-independent and degenerate with the luminosity distance of the binary, thus removing the need to explicitly search for the signal over the sky \cite{Allen:2005fk}.  However,  ET 
will extend the useful bandwidth  down to a few Hz, so the inspiral of binary neutron star systems might be visible for thousands of seconds (and up to one day) in the data, and the angular response of the interferometers will
imprint a slow amplitude and phase modulation for at least some signals, especially when a zero of the angular response sweeps close to the
source's sky location (as we see in the example of figure~\ref{fig:GW170817atET}). Neglecting this modulation in the search templates might lead
to a significant loss of signal-to-noise ratio or rejection of a candidate from $\chi^2$ discriminators.  Preliminary studies suggest that this will not be a major concern, because the events missed due to the loss of signal-to-noise ratio will be a small fraction of the overall detections \cite{Meacher:2015rex, Pillas:2023ixj}. However, to our knowledge, a detailed investigation including the effects of signal-based discriminators is not available yet. On the other hand neglecting the modulation is expected to have a stronger impact on parameter estimation, in particular on the localization in the sky.

As already discussed in section~\ref{subsec:overlap}, the presence of overlapping signals will be one of the main features of ET. Several studies have been conducted to check if overlapping coalescing binaries signals could be a problem for detection; the current pipelines have been used in these studies, which mainly focused on the analysis of the first and second ET MDC \cite{Regimbau:2012ir,Meacher:2015rex} or of synthetic data reproducing the  LIGO and Virgo sensitivities \cite{Relton:2022whr}. The results indicate that there should not be any relevant decrease in the detection efficiency; even in the challenging cases of a sub-second difference in the merger times of two signals, an affordable modification of the current methods to cluster the different triggers pertaining to the same event, for example checking for multiple significant triggers in a single templates, should be sufficient to separate the signals \cite{Relton:2022whr}.

\paragraph{Machine learning methods for CBC detection.}
\label{subsec:MLdetection}

In the recent years, the use of machine learning for detecting compact binary coalescence events has emerged as a key research area in the gravitational wave community, especially with advancements in ML technology powered by GPUs.

Typically, this problem is approached as a classification task, where neural networks are trained through supervised learning to distinguish between two main classes: detector noise and detector noise plus a CBC event.
The initial approaches used a deep convolutional neural network (CNN),  providing probabilities for each class (noise or noise plus signal). This method demonstrated that ML could match the sensitivity of traditional matched-filtering techniques \cite{George:2017pmj}.
Subsequent studies confirmed these findings and emphasized comparing ML approaches to matched filtering in realistic scenarios \cite{Gabbard:2017lja}.
Further research sought to enhance the power of ML for CBC detection. Some studies \cite{Gebhard:2019ldz} suggested that while ML approaches, particularly CNNs, could not quantify the statistical significance of GW detections, they are useful for generating real-time triggers for detailed analysis.
Recent advancements have integrated ML with traditional search pipelines~\cite{Gebhard:2019ldz} and improved ML techniques~\cite{Xia:2020vem}, giving better detection sensitivity, especially for low-significance events. Additionally, efforts have been made to improve the reliability of machine learning models and enhance their robustness against adversarial attacks,  i.e. intentional manipulations of input data designed to deceive models by causing incorrect predictions.
In 2021, a mock data challenge for CBC detection using ML was launched, followed by a more realistic challenge aimed at benchmarking ML methods against traditional detection tools~\cite{Schafer:2022dxv}. This challenge highlighted that while traditional methods, like PyCBC, achieved the highest sensitive distances, some ML analyses performed comparably well.
The motivation for using ML in GW searches includes the potential for real-time detection with minimal computational cost, particularly relevant for events involving neutron stars that may produce electromagnetic counterparts \cite{Dax:2024mcn}. Efforts have been made to extend the signal durations considered in ML algorithms, with significant progress in analyzing longer time-series data \cite{Fan:2018vgw,Lin:2019qmd,Krastev:2019koe,Krastev:2020skk}.
Special cases, such as searches for gravitationally lensed signals, have also seen ML applications. Various ML methods, including CNNs and ensemble classifiers, have been developed to identify lensed GW events, demonstrating the versatility and potential of ML in expanding the scope of GW detection \cite{LIGOScientific:2021izm,LIGOScientific:2023bwz}.
XGBoost \cite{Chen:2016btl}, which is a powerful, efficient, and scalable machine learning algorithm based on gradient boosting, is also utilized alongside a version of cWB that targets the detection of BBH events~\cite{Mishra:2021tmu, Mishra:2022ott}.

\paragraph{Computational Requirements.}

Preliminary findings show that searches in the same parameter space presently considered in LIGO-Virgo-Kagra collaborations will not be orders of magnitude more expensive for 3G GW observatories than for current interferometers and will be feasible with present-day computational resources \cite{Lenon:2021zac}. Considering a wider parameter space, i.e. eccentricity, precession, tidal effects etc, will be challenging; improvements in technology may however make these searches a possibility with current methods, but new solutions should also be pursued.

\subsubsection{Data analysis for Gravitational Wave Background}
\label{subsec:sgwb}

Stochastic Gravitational Wave Background (SGWB), both of astrophysical and cosmological origin, have been  discussed in section~\ref{sect:SGWBdiv2}. The background is usually described by the characteristic energy density as a function of the direction of observation $\hatn$ and frequency $f$ as in \eq{eq: OmegaGW_f_n}, that we recall here,
\begin{equation}
\Omega_{\mathrm{gw}} (f, \hatn) = 
\frac{1}{\rho_c} \frac{d\rho_{\rm GW}(f, \hatn)}{d\log f \, d^2 \hatn}\, ,
\end{equation}
where  $\rho_{\rm GW}(f, \hatn)$ the background energy density per unit of logarithmic frequency, and $\rho_c$ is  the critical density  of the Universe today. 
Since the SGWB is described by a random signal, it will look like noise in a single detector. Therefore, the standard search techniques like matched filtering will not work when trying to detect the SGWB. Instead, we have to consider data from multiple detectors and  look for evidence of a common signal. In the case of ET in its triangle configuration we can cross correlate the three  colocated detectors that compose the triangle, which have a $60^{\circ}$ angle between their arms,  while in the 2L configuration we can correlated the two L-shaped detectors, which will not be colocated. To look for SGWB in this way one performs a  cross-correlation, where the output of one detector is used as a template for the data in another detector, see also the discussion in  section~\ref{sec:div9_stoch_searches}. As of today, in the LIGO-Virgo-KAGRA SGWB searches, such cross-correlation techniques are used to look for SGWB. 
The most common prescription is that the SGWB is assumed to be isotropic; one could determine its statistical properties by observing any part of the sky.  However, cosmological and astrophysical SGWB components are expected to be anisotropic due to the nature of spacetime along the line of sight, and for the astrophysical contribution, due to the local distribution of matter and the finiteness of the number of sources \cite{Jenkins:2018nty,Jenkins:2018kxc,Jenkins:2019uzp,Jenkins:2019nks,Cusin:2017fwz,Cusin:2018rsq,Cusin:2019jhg,Bellomo:2021mer,Belgacem:2024ohp}, see the discussion in section~\ref{sect:anisoSGWBdiv2}.  
Two pipelines are tailored to searches for isotropic and anisotropic search. For isotropic SGWB, current LVK  searches use pygwb \cite{Renzini:2024ehg}. This Python package employs cross-correlation spectra of GW detector pairs to construct an optimal estimator of the Gaussian and isotropic SGWB and Bayesian parameter estimation to constrain SGWB models. On the other hand, for anisotropic SGWB current LVK searches use a Python-based pipeline called PyStoch \cite{Ain:2015lea,Ain:2018zvo}. This pipeline employs the GW radiometer technique, which accounts for Earth rotation and time delay. With minimal modifications to these pipelines, one can perform the searches for SGWB even in the ET scenario.

The computational cost of existing stochastic searches is negligible compared to matched filtering for CBC sources. We will need the results of MDCs to estimate better the cost including new developments for ET.

\paragraph{Direction for new methods.}

Though the existing pipelines mentioned above would work for the ET case, there are several scenarios in which many real-life complications will occur. One such complication would be that, at 2G detectors, the noises intrinsic to the detectors are much larger in magnitude than the expected  SGWB signals. This, in general, will no longer be true for 3G detectors. This complication demands new developments in search techniques where the variance from the signal needs to be considered. 

Considering the cross-correlation basics, one can show that the mean and variance associated with SGWB estimators can be written as:
\begin{eqnarray}
    \mu \, &\equiv& \, \langle\hat{C}_{12}\rangle = \langle h^2\rangle = S_h \\
    \sigma^2 \, &\equiv& \, \langle\hat{C}^2_{12}\rangle - \langle\hat{C}_{12}\rangle^2 = (S_{n_1} + S_h)(S_{n_2} + S_h) + S^2 _h
\end{eqnarray}
In the usual cross-correlation searches (employed for the current ground-based detectors), we assume that the noises intrinsic to the detectors are much larger in magnitude than the gravitational strains. This assumption will lead to
\begin{equation}
    \sigma^2 \approx (S_{n_1} + S_h)(S_{n_2} + S_h)
\end{equation}
This assumption is valid for ground-based detectors like LIGO, Virgo, and KAGRA (assuming the current detector sensitivity). However, for 3G detectors like ET, we might  have intrinsic detector noise levels comparable to the level of a real SGWB signal. When the magnitude of the SGWB signal is comparable to the noises intrinsic to the detectors, the calculation of the $\mbox{SNR} = \mu/\sigma$ will become more involved. Although the mean $\mu$ is independent of the relative size of the SGWB and the detector noise, the variance $\sigma$ is not. Thus, we have to change the current way of 'optimal" (which maximizes the SNR) cross-correlations by incorporating the variance term coming from the GW signal. 

Another promising possibility for ET in its triangle configuration is to use the null stream (see section~\ref{subsec:null_stream}). A proof of principle was demonstrated in a mock data challenge that took place in 2012 \cite{Regimbau:2012ir}, where the authors computed the median over the whole dataset of the difference between (one-third of) the null stream PSD and the individual detector PSD. These residuals were shown to be consistent with the median PSD of the injected GW signals in each detector.

\subsubsection{Data analysis for Continuous Wave searches}
Persistent semi-periodic GW signals, called \textit{Continuous Waves} (CW), are produced by various classes of sources, in particular rotating neutron stars asymmetric with respect to their rotation axis, see section~\ref{section:div6}. CW emission is also expected in more exotic sources, including ultra-light boson clouds around spinning black holes (discussed in section~\ref{sect:bosoncloudsdiv1}) and the inspiral of light primordial black hole binaries, with mass $m\ll M_\odot$ (discussed in section~\ref{sec:PBHsdiv3}); see e.g., \cite{Riles:2022wwz} for a recent review of CW sources.

The search for CW is made difficult by the predicted weakness of the signals which, with a few notable exceptions, will limit the search volume to the Milky Way and its neighborhood. This will be also true with 3G GW detectors. Moreover, if the signal parameters are not known in advance, the resulting parameter space can be very large, as a consequence of the long signal duration. This complicates the waveform at the detector due, for instance, to the Doppler effect induced by the Earth motion and the sidereal amplitude and phase modulation due to the detector beam pattern functions. As a result, the search for this kind of signals may be challenging also from the computational point of view. This is true both for current and future detectors.  
No CW signal has been detected so far, although interesting upper limits have been set in several cases, see e.g. \cite{Dergachev:2024knd,DOnofrio:2023amp,LIGOScientific:2022enz,LIGOScientific:2021quq,Steltner:2023cfk,Dergachev:2022lnt,LIGOScientific:2021inr,LIGOScientific:2021hvc,LIGOScientific:2021ozr,LIGOScientific:2021yby,KAGRA:2022osp,KAGRA:2022dwb} for some recent results obtained with the LIGO/Virgo/KAGRA observatory. We refer the reader to section~\ref{sec:Prospects:CW} for more details on current searches of CW from rotating neutron stars and for prospects in the ET era, and to section~\ref{sect:bosoncloudsdiv1}  for prospects about the detection of CW from ultra-light particles.

In order to increase the chance of detection, not only more sensitive detectors are needed, but also improved data analysis methods are crucial. 
In general, search method development should aim to reach three fundamental milestones: a) maximizing sensitivity, b) ensuring robustness against unmodeled signal or noise features, and c) maintaining manageable computational costs. A compromise among these three features must often be found. A standard approach for wide parameter searches is that of developing \textit{semi-coherent} approaches, where the whole dataset is split into a number of smaller segments, which are individually processed and then incoherently combined, that is without requiring signal phase coherence from one segment to the other, see for instance \cite{Astone:2014esa} and reference therein. While this approach introduces a sensitivity loss with respect to optimal searches based on matched filtering (which is used only when the signal parameters are known with high accuracy), it ensures better robustness and a lower computing cost, depending on the segment duration $T_\mathrm{FFT}$. 

It can be shown that for a standard semi-coherent search the sensitivity scales as $T_\mathrm{FFT}^{1/4}$, while the computational load scales at least as $T_\mathrm{FFT}^4$ \cite{Frasca:2005ey}. Finding effective ways to increase the sensitivity of wide parameter searches without making the computational cost explode is a non-trivial and important task for current detectors as well as for future detectors. This target can be pursued at the algorithmic level and/or in the context of a multi-messenger approach. The former case includes optimizations at the level of template placement \cite{Wette:2018bhc,Mukherjee:2022tuc}, efficient exploitation of modern computing architectures, like GPUs \cite{Rosa:2021ptb}, correction of the Doppler effect via hierarchical approaches \cite{DAntonio:2023jxm}, exploration of image processing \cite{Pierini:2023nqf} and Machine Learning techniques \cite{Miller:2019jtp}. The latter consists in exploiting observations in the electromagnetic band to better constrain the portion of parameter space to be investigated, allowing the saved computing resources to be dedicated to making deeper searches, see e.g. \cite{Menon:2023sfa}.    

In the ET era, there will also be other data analysis issues that could play a relevant role and need to be investigated: the impact of the superposition of CBC signals on CW searches, and the superposition of CW signals themselves, which we briefly describe in the following subsections.

It is also worth mentioning that data analysis issues similar to those presented by CW searches in Earth-bound detectors could also affect the search for Extreme Mass Ratio Inspirals (EMRIs) in LISA \cite{Gair:2017ynp}. There could be a mutual benefit from the data analysis developments in these fields.  

\paragraph{Impact of CBC background on CW searches.}
The incoherent superposition of several CBC signals  will constitute an astrophysical background that adds to the detector noise, whose impact on CW searches must be evaluated. This effect would be especially pronounced in the low frequency regime, where the CBC signal density is higher. As a first step, we need to estimate how ET theoretical sensitivity curve compares with such signal background, in order to understand which is the frequency range more affected. Next, we need to study how current CW search procedures are impacted, e.g. in relation to the data power spectrum estimation, and if needed, find suitable solutions. This study will likely be corroborated by a dedicated MDC.  

\paragraph{Superposition of CW signals.}
CW signals emitted by ultra-light boson clouds around rotating black holes, see e.g. \cite{Arvanitaki:2010sy,Arvanitaki:2016qwi,Brito:2017wnc}. As already discussed in section~\ref{sect:bosoncloudsdiv1}, 
see in particular \eq{fgwbosoncloudsdiv1}, they
are characterized by a frequency \cite{LIGOScientific:2021rnv}
\begin{equation}
    f_\mathrm{gw}  \simeq 483\, \text{Hz} \left( \frac{m_\mathrm{b}}{10^{-12}\,\text{eV}} \right) \left[ 1-7\times 10^{-4}\left( \frac{M_\mathrm{bh}}{10M_\odot}\frac{m_\mathrm{b}}{10^{-12} \text{eV}} \right)^2 \right] ,
\end{equation}
where $m_\mathrm{b}$ is the boson mass  and $M_\mathrm{bh}$ is the black hole mass.
Assuming a given boson mass, and a population of boson clouds, the resulting signal frequencies are expected to be concentrated in a range
\begin{equation}
\Delta f_\mathrm{gw}=0.338\, {\rm Hz} \,\left( \frac{m_\mathrm{b}}{10^{-12}\text{eV}} \right)^3\left[\left(\frac{M^{\rm max}_\mathrm{bh}}{10M_\odot}\right)^2-\left(\frac{M^{\rm min}_\mathrm{bh}}{10M_\odot}\right)^2 \right],
\end{equation}
where $M^{\rm min}_\mathrm{bh}$ and $M^{\rm max}_\mathrm{bh}$ are, respectively, the minimum and maximum mass of the black holes that developed a cloud. 
Depending on the black hole mass distribution, such a frequency interval can vary from a fraction of hertz to about 1 Hz, for boson masses in the range $10^{-14} - 10^{-11}$ eV, which produce signals in the sensitivity band of ET. If the number of detectable signals is large, a superposition may take place with a possible impact on search algorithms \cite{Zhu:2020tht}. A preliminary study has been done in \cite{Pierini:2022wgc}, showing that current semi-coherent analysis methods are generally robust concerning this possible issue. A more detailed study, explicitly taking into account the predicted ET sensitivity curve and reasonable models for the boson cloud population, must still be done. It will also be interesting to evaluate if modifications of current semi-coherent schemes, explicitly exploiting the expected signal superposition, can increase the detection probability.   

\subsubsection{Data analysis for burst signals}

Numerous astrophysical processes in the universe are expected to produce transient signals for which we do not have a complete model as in  the case of CBCs. Some notable cases are supernova explosions, gamma-ray bursts, soft-gamma repeaters, fast radio bursts, or possible sources of unknown origin. Given that there is no exhaustive knowledge of the expected signals, alternatives to matched filtering have been developed to search for signals of unknown features but consistent with minimal properties of gravitational waves.

\paragraph{Methods for sub-second long signals.}

Devoted algorithms for short transient signals (bursts), i.e. less than a second, usually share common features. These algorithms identify power excesses from the noises of multiple detectors and combine them in a coherent data stream consistently with the light time difference between detectors. This coherent approach allows to discard of noise features that casually appear in various detectors more effectively than a simple time-coincidence selection. The current pipelines can differ in the way they build the search parameter space; for instance, applying Wavelet or Fourier Time-Frequency transformation, or building an orthogonal basis of Sine-Gaussian functions.
Another difference can be found in the procedure used to differentiate the signal candidates from the noise excesses. For instance, one approach used in the LVK Collaboration for the cWB pipeline~\cite{Drago:2020kic} was to define a subset of noise excesses based on morphology, so as to penalize well known glitches in favor of other signal characteristics, at the cost of the trial factor, i.e. the false alarm rate of a single event is multiplied for the number of total subsets.

In current searches, this approach has been abandoned in favor of machine learning techniques to identify potential signals in the high noise environment, like the application of autoencoder~\cite{Bini:2023gil}, XGBoost~\cite{Szczepanczyk:2022urr} or GMM~\cite{Lopez:2021ikt} in conjunction with coherent WaveBurst~\cite{Drago:2020kic}.
Other algorithms extend some features of these algorithms to restrict the search to a selected class of gravitational waves. Some examples are the use of Convolutional Neural Networks for the recognition of certain patterns in the time-frequency domain, which has been applied to CBC~\cite{Vinciguerra:2017psh} and to supernovae~\cite{LopezPortilla:2020odz,Iess:2023quq}.

Algorithms can be adapted for searches in different conditions, passing from a general all-sky/all-time search, where there is no information of the expected signal, to triggered searches, where information about the arrival time or the sky direction is provided by external information (like neutrino or electromagnetic). Examples of such searches which have been explored are GWs associated with gamma-ray bursts~\cite{LIGOScientific:2021iyk}, fast-radio bursts~\cite{LIGOScientific:2022jpr} or supernova explosions~\cite{LIGOScientific:2019ryq}, where the search for GW is focused around the time of the alerts from non-GW detectors.

The expected detectable signal depends on the shape of the detector PSD; for instance, in the case of the Advanced LVK detectors it is very short for binary black holes (around hundreds of milliseconds). In this case these signals are sufficiently separated in time to not interfere with each other. However, this could be one of the main limiting factors when applying the current pipelines to ET data, given the longer duration of detectable signals, so that we expect more frequently the case that multiple signals would be occurring at the same time. A preliminary study has been performed on overlapping CBC signals~\cite{Relton:2022whr} which demonstrate the feasibility of detection in the case of two signals partially overlapping in time. The considered burst algorithm (cWB) was capable to get most of the overlaps, but as a unique candidate, without distinguishing the separate contributions. This shows that an extension of the algorithm is needed, so to be able to split different signals in case they are happening at the same time.

\paragraph{Data analysis methods for minute-long burst signals.}

Minute-long signals, ranging from 100 to over 1000 seconds, present unique challenges due to the large data volume and the need for high sensitivity over extended periods \cite{KAGRA:2021bhs}. One approach to dealing with these challenges involves creating time-frequency (TF) maps that can cover the entire duration of long signals without fragmenting the waveform across multiple maps. By ``stitching together" multiple TF maps, we effectively create a continuous representation that spans the extended duration of the signal \cite{Macquet:2021ttq}. This technique allows  handling long data segments more efficiently, while maintaining coherence across the signal’s full time span.

The current generation of detection pipelines typically employs a combination of traditional and machine learning methods to enhance detection accuracy and efficiency. One recently developed technique involves using the U-NET architecture for anomaly detection \cite{10020896}. This architecture is particularly effective in distinguishing between normal background noise and potential gravitational wave signals, which appear as anomalies. Given that target signals are relatively well-behaved in the time-frequency domain, the data is analyzed in the form of spectrograms. The resolution of these spectrograms directly impacts the computational requirements, as it determines the size of the 2D array and consequently the batch size used in training.

Training these machine learning models is computationally expensive due to the large data volumes involved. For instance, segments of 512 seconds and 1000 seconds are commonly used in current methodologies \cite{LIGOScientific:2019laf}. The primary goal of these models is to detect gravitational wave signals amidst a background of noise and glitches. Most glitches are shorter than one second, allowing them to be represented as single pixels in the time-frequency array. This simplifies the thresholding process for identifying the loudest background event, which is the most limiting factor for the sensitivity of minute-long searches.

The computational requirements for processing long-duration gravitational wave bursts are significant. High resolution in the time-frequency domain necessitates powerful computational resources to handle the large 2D arrays representing the data. This includes substantial memory and processing power to manage the continuous influx of data and the complex calculations required for anomaly detection and signal processing. The use of advanced machine learning models like U-NET further adds to the computational load, as these models need to be trained on extensive datasets to achieve the necessary pixel-to-pixel accuracy for real-time detection.

In current searches, techniques such as convolutional neural networks (CNNs) and ensemble classifiers, like XGBoost in conjunction with a version of coherent WaveBurst (cWB) \cite{Drago:2020kic} or STAMPAS \cite{Macquet:2021ttq}, are employed to identify potential signals in high-noise environments. These methods enable the processing of large datasets more efficiently and improve the sensitivity and accuracy of gravitational wave searches, particularly for low-significance events.

Moreover, the extraction and isolation of overlapping signals are being actively explored. The approach, detailed in~\cite{Relton:2022whr}, involves advanced methodologies to better separate and identify multiple signals occurring simultaneously.


\subsection{Parameter estimation methods}
\label{sec:PE}

As already discussed in section~\ref{sec:div9_pe}, the parameters of a GW signal are characterised through the use of Bayesian analysis. This requires using Bayes' theorem
\begin{equation}
 p(\vb*{\theta}|\mathcal{H}_s,d) 
 = \frac{p(d|\vb*{\theta},\mathcal{H}_s)\,p(\vb*{\theta}|\mathcal{H}_s)}{p(d|\mathcal{H}_s)},
 \label{eqn:Bayes}
\end{equation}
where $\vb*{\theta}$ is the set of parameter values that describe a GW signal and $\mathcal{H}_s$ is the hypothesis that a GW signal depending on the parameters 
$\vb*{\theta}$ is present in the data $d$. The data is compared to a model by constructing the likelihood $p(d|\vb*{\theta},\mathcal{H}_s)$ under the assumption of Gaussian and stationary noise in GW detectors, while $p(\vb*{\theta}|\mathcal{H}_s)$ is the prior distribution on the parameters.\footnote{Often, the label $\mathcal{H}_s$ is left implicit, and \eq{eqn:Bayes} is rewritten as in \eq{eq:posterior}, where has been used the notation 
$p(\vb*{\theta}|\mathcal{H}_s,d)\equiv p(\vb*{\theta}|d)$, 
$p(d|\vb*{\theta},\mathcal{H}_s)\equiv {\cal L}(d|\vb*{\theta})$ and  
$p(d|\mathcal{H}_s)\equiv {\cal Z}(d)$, and the data has been denoted by $s$ instead of $d$.}
All information about the parameters 
of interest is encoded in the posterior \emph{probability density function} (PDF), $p(\vb*{\theta}|\mathcal{H}_s,d)$.
For parameter estimation purposes, the factor $p(d|\mathcal{H}_s)$, called the \emph{evidence} for the hypothesis  $\mathcal{H}_s$, is set by the requirement that PDFs are normalised. This quantity is particularly important when comparing one model (or hypothesis) to another, using the same data. In a GW data analysis context, under the assumption that the noise is stationary and Gaussian, the likelihood is given by
\begin{equation}
 p(d|\vb*{\theta},\mathcal{H}_s) \propto \exp\left\{ -\frac{1}{2}(d-h(\vb*{\theta})|d-h(\vb*{\theta}))\right\},
 \label{eqn:lhood}
\end{equation}
where the noise-weighted inner product $(\,\cdot\,|\,\cdot\,)$ is defined as
\begin{equation}\label{ascalarbdiv10}
(a|b) = 4\, {\rm Re} \int_{f_{\rm low}}^{f_{\rm high}} df\, \frac{\tilde{a}^\ast(f)\,\tilde{b}(f)}{S_h(f)}\, ,
\end{equation}
see also \eqs{eq:innerprod_def}{eq:gwlik_general}. 
Here a tilde refers to the Fourier transform, and $S_h(f)$ is the PSD. The integral in \eq{ascalarbdiv10} could in principle be taken to run  from $f_{\rm low}=0$ to $f_{\rm high}=\infty$, as in \eq{eq:innerprod_def}, but, in practice, the PSD determines a minimum and a maximum value  of the frequency beyond which the detector is, for all purposes, blind.
A (quasicircular) BBH signal within GR is described by 15 parameters, already given in \eq{CBCparams}, with the parameter space increasing if including eccentricity, or with the presence of matter effects in BNSs, or for exotic compact objects with additional parameters. Stochastic sampling has been traditionally used for sampling this high-dimensional parameter space~\cite{Veitch:2014wba}. With the increased number of detections and improved detector sensitivities, Bayesian pipelines like \texttt{Bilby}~\cite{Ashton:2018jfp} have been developed. 
Improvement of noise sensitivity in the 3G era translates to a lower $f_{\rm low}$, making the signals and therefore each likelihood evaluation extremely expensive for 3G detectors. Considering hours to days long signals in the 3G era, improvements to this pipeline have also been made~\cite{Smith:2019ucc} which, however, relies on heavy computational resources. A further method to enable analysing very long BNS signals like \emph{reduced order quadratures} (ROQ)~\cite{Smith:2021bqc} has also been developed. However, this method suffers from an inherent overhead by the need to first build a reduced basis and moreover construct a new basis each time a new prior may be chosen and further when adding a new parameter. On the one hand, we will be facing extremely long signals requiring very heavy computational resources and on the other, as noted in section~\ref{subsec:overlap}, the data will mostly consist of overlapping signals, making conventional Bayesian analysis potentially extremely computationally expensive by increasing the parameter space dimensionality. In the following, we look at methods developed for solutions of both these problems.

\subsubsection{Analysis of overlapping signals}
\label{subsec:overlap_analysis}

Ref.~\cite{Janquart:2023hew} presented the first methods to analyse overlapping signals effectively. In the presence of overlapping signals, the assumption of the presence of a single signal in the likelihood breaks down. In ref.~\cite{Janquart:2023hew}, simulation sets consisting of two signals in the 3G network at the same time were analysed by expanding the likelihood to consist of two signals, henceforth called \emph{joint likelihood}; however, in reality,  the number of signals present will be  much larger \cite{Samajdar:2021egv} (with the population model used in \cite{Branchesi:2023mws}, at any given moment there are ${\cal O} (10^2)$ overlapping BNS \cite{Belgacem:2024ntv}) and such an analysis may be practically infeasible. Furthermore, the method of \emph{hierarchical subtraction} was also introduced whereby the loudest signals are subtracted from data containing overlapping signals, and signals are analysed one by one. Although computationally more efficient, this method however suffers from the drawback of introducing biases in residuals with each subtraction. Figure~\ref{fig:overlap-soln} shows a comparison of offsets in chirp mass from joint likelihood method compared against the methods of hierarchical subtraction and analysing single signal at a time.
 \begin{figure}
    \centering
    \includegraphics[width=\textwidth]{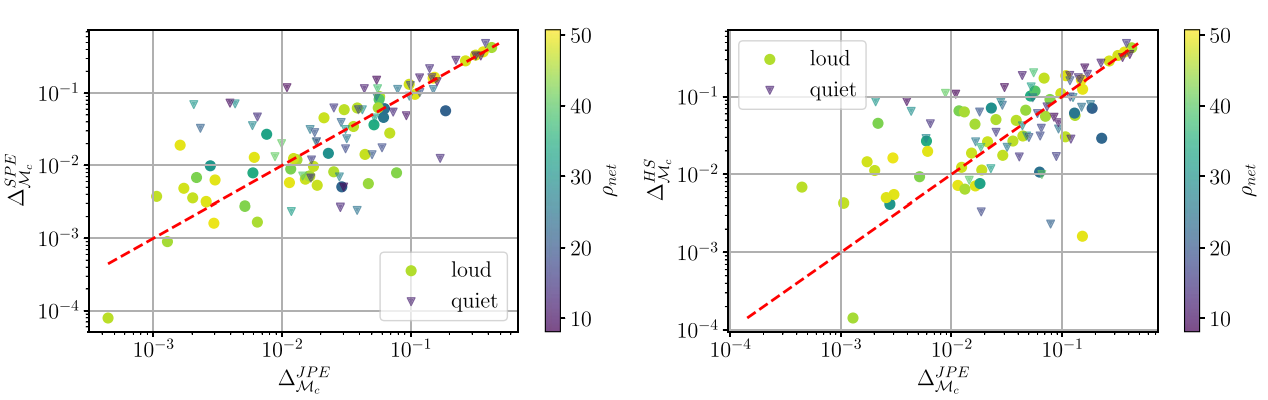}
    \caption{Left: comparison of the offset of the recovered posterior for the chirp mass for joint parameter estimation (JPE) and single parameter estimation (SPE) method. Right: comparison of the
offset of the recovered posteriors for the chirp mass for JPE and hierarchical subtraction (HS). The two plots indicate that the offset is reduced for JPE compared
to HS, due to the better modeling of the noise, while it is still better in the SPE case, where the noise is well modeled and the problem
at hand has a reduced complexity. Figure taken from~\cite{Janquart:2023hew}.}
    \label{fig:overlap-soln}
\end{figure}

\subsubsection{Innovative methods for faster inference}\label{subsec:PEacceleration}

Since PE depends on evaluating the likelihood multiple times (typically $\sim 10^6-10^8$ evaluations for estimating the parameters of a CBC signal),
the speedup can be achieved by constructing reduced-order-quadratures (ROQs)~\cite{Canizares:2014fya, Smith:2016qas}. ROQs were developed for long signals like BNSs relevant for the 3G era in~\cite{Smith:2021bqc}. In addition, reduced-order-methods for waveform generation, discussed in section~\ref{sec:acc}, will facilitate faster likelihood computation. Relative binning~\cite{Dai:2018dca, Zackay:2018qdy} has been proposed as a method to effectively make sparser frequency grids to evaluate long waveforms like those from BNS signals by making use of the frequency-change in the evolution of waveforms. Further advancements in relative binning when using waveforms including higher-order-modes have been made in ~\cite{Leslie:2021ssu, Narola:2023men}. In recent years, \emph{likelihood-free} inference has been gaining popularity due to its immense potential to speed up PE analyses~\cite{Dax:2021tsq, Dax:2022pxd}. Likelihood-free inference, as the name suggests, avoids explicitly computing the likelihood to get parameter estimates. An example of avoiding explicit likelihood computations is by using \emph{machine-learning} methods, eg. \emph{normalising flows} to approximate the likelihood. Simulation-based inference has been applied to PE~\cite{Bhardwaj:2023xph,Alvey:2023naa} and all such methods hold immense promise for longer as well as multi-dimensional analyses necessary in the 3G era due to overlapping signals. Normalising flows have been applied and shown to infer parameters correctly in the case of overlapping signals~\cite{Langendorff:2022fzq}. Testing the theory of gravity or inferring the dense nuclear equation-of-state (EoS) of neutron stars mostly requires inferring additional parameters to characterise their effect, making the PE problem even more complex and computationally challenging. Inference of EoSs using neural networks has been explored in~\cite{McGinn:2024nkd}, while the application of neural networks to testing gravity has been explored in~\cite{Xie:2024ubm}. Bayesian PE using \emph{conditional variational autoencoders} were developed in~\cite{Gabbard:2019rde}. Low-mass BBH sources, with chirp masses as low as $\sim 5 \ \mathrm{M}_\odot$ have also been inferred accurately using neural posterior estimation in~\cite{Kolmus:2024scm}. In the following subsections, we review a few methods for faster inference.

\paragraph{Relative binning method.}\label{subsubsec:relbin}
Relative binning (RB) was developed as a faster alternative to performing Bayesian analyses for the long BNS signal, GW170817~\cite{Dai:2018dca, Zackay:2018qdy}. It has however been extended to other expensive analyses like those with higher-order-modes~\cite{Leslie:2021ssu} and been shown to be effective for 3G detectors in case of lensing analyses~\cite{Narola:2023men}. The principle behind RB is to reduce the computational cost of the  evaluation of the likelihood (\ref{eqn:lhood}) by leveraging summary data calculated on a dense frequency grid for a single \emph{fiducial} waveform, which must resemble the best fit to the data. The key assumption is that the set of parameters contributing non-negligibly to the posterior probability produces similar waveforms, such that their ratio varies smoothly in the frequency domain. Consequently, within each frequency bin $b = [f_{\mathrm{min}}(b), f_{\mathrm{max}}(b)]$, the ratio between the sampled waveforms and the fiducial waveform can be approximated by a linear function of frequency:
\begin{equation}
r(f) = \frac{h(f)}{h_0(f)} = r_0(h, b) + r_1(h, b)[f - f_m(b)] + \cdots,
\label{eqn:RB_ratio}
\end{equation}
where $h_0$ is the fiducial waveform and $f_m(b)$ the central frequency of the frequency bin $b$. This reduces the number of frequencies at which the waveform information must be evaluated, thus reducing the overall cost of the likelihood computation. However, the ratios $r_0(h, b)$ and $r_1(h, b)$ still depend on the source parameters, and so parameter evaluations are necessary. While this method allows the use of computationally expensive samplers like nested sampling~\cite{Skilling:2006gxv}, it requires a fiducial waveform, which assumes an existing set of representative parameter estimates. One may use initial parameter values from detection pipelines, although they are typically associated with larger uncertainties.

\paragraph{Multibanding.}\label{subsubsec:multibanding}
Multibanding is built on the idea that CBC signals evolve from low frequency to high frequency, so the GW data can be downsampled at the early stage as long as the sampling rate is greater than the Nyquist frequency (twice of the highest signal frequency). The downsampling can be applied adaptively to different stages of the signal, resulting in multiple bands. Multibanding has been applied to signal detection in the MBTA pipeline~\cite{Aubin:2020goo}, and its application to PE is also proposed in refs.~\cite{Vinciguerra:2017ngf, Morisaki:2021ngj}. Multibanding accelerates PE in the sense that it reduces the size of data and therefore accelerates waveform and likelihood evaluations. Since multibanding only manipulates the sampling rate, it has the potential to be combined with ROQ (see, e.g. \cite{Morisaki:2021ngj}) and RB (see, e.g. \cite{Dax:2024mcn, Hu:2024oen}). The comparison between ROQ, RB, and multibanding for PE can be found in \cite{Hu:2024mvn}.

\paragraph{Machine learning for parameter estimation.}\label{subsec:MLPE}
\label{subsec:ML_PE}
In addition to the ML-based detection methods laid out in section~\ref{subsec:MLdetection}, 
over the last few years ML has gained traction as an avenue to perform parameter estimation for GW signals since, in principle,  it offers a faster alternative with the same precision as classical methods.  In particular, several approaches, generally based on \emph{normalizing flows} (NFs)~\cite{Papamakarios:2019fms}, have been suggested to speed up parameter inference processes. One can split those into two general trends: (i) the full inference is replaced by a ML framework, i.e. the setup directly outputs posteriors characterizing the event analyzed~\cite{Green:2020hst, Delaunoy:2020zcu, Miller:2021hys, Dax:2021tsq, Dax:2022pxd, Bhardwaj:2023xph, Kolmus:2024scm}, and (ii) part of the process is aided by an ML system~\cite{Williams:2021qyt, Wong:2022xvh, Williams:2023ppp, Wong:2023lgb, Wouters:2024oxj}. In both cases, one observes a significant speed-up in the computation time. Specific ML algorithms developed for parameter estimation are as follows
\begin{itemize}
\item \textit{Neural Posterior Estimation} (NPE) directly outputs posteriors characterizing gravitational waves (GWs) in data~\cite{Green:2020hst, Dax:2021tsq, Dax:2022pxd, Kolmus:2024scm}. It typically consists of two parts: a context network (often a convolutional neural network) that compresses the data, and an inference network that transforms a base distribution (a multivariate Gaussian) into the target posterior. 
Initial NPE methods were PSD-dependent, as they were trained with a specific PSD. To address this, methods have been developed to include the PSD as input and allow for PSD variations during training~\cite{Wildberger:2022agw}, increasing robustness. However, significant detector upgrades may still require re-training. 
Currently, NPE frameworks are reliable for BBHs with component masses above $10 \ \Msun$, though, recently, extension in the neutron star mass regime has been implemented~\cite{Dax:2024mcn, vanStraalen:2024xiq}.
  Adapting these networks for lower mass signals~\cite{Kolmus:2024scm} is challenging due to the need to handle long signal durations and correlations between distant data points. An example posterior obtained with this framework for a system with a $5\,M_{\odot}$ chirp mass injected in an LVK network is shown in figure~\ref{fig:posteriors_5msun}. One approach involves using more effective priors in the training phase---modifying them to have a uniform sample exposure in the mass space, and fine-tuning the network on a per-event basis to focus on relevant regions of the parameter space, significantly improving efficiency while remaining faster than traditional methods. For the Einstein Telescope (ET), which expects high SNR and long-duration signals, further development will be needed. Issues such as overlapping signals~\cite{Samajdar:2021egv, Pizzati:2021apa, Antonelli:2021vwg, Relton:2021cax, Janquart:2023hew} must also be addressed. Initial attempts to analyze overlapping BBH signals using NPEs have shown promise~\cite{Langendorff:2022fzq} but require larger networks and further optimization~\cite{Janquart:2023hew}. Developments are ongoing to adapt NPE for next-generation detectors like ET, indicating a promising direction for future GW data analysis.

\begin{figure}
    \centering
    \includegraphics[keepaspectratio, width=0.9\textwidth]{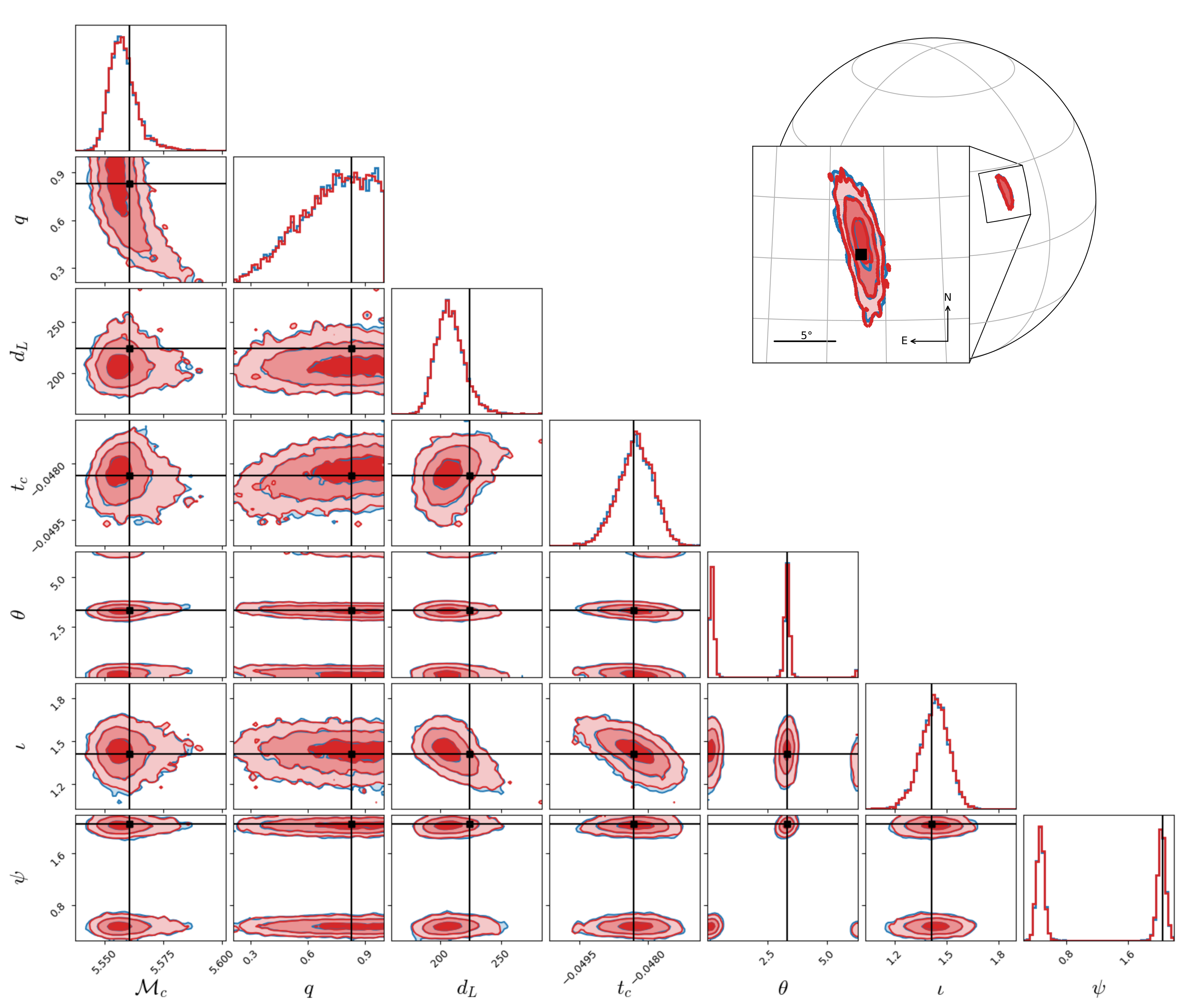}
    \caption{Representation of the posteriors obtained with the approach from~\cite{Kolmus:2024scm} (red) and the posteriors obtained with traditional approaches (blue) for a system with a chirp mass of $5\,M_{\odot}$ injected in an LVK network. A good agreement is obtained between the two posteriors, which required an adapted training procedure due to the relatively low mass of the system. To obtain this agreement, the priors during the training process have been adapted to have an effectively uniform coverage of the mass parameter space.}
    \label{fig:posteriors_5msun}
\end{figure}

\item \textit{Truncated Marginal Neural Ratio Estimator} (TMNRE) generates from the data low-dimensional marginal posterior distributions (1D or 2D)~\cite{Delaunoy:2020zcu, Bhardwaj:2023xph}. Unlike Neural Posterior Estimation (NPE), TMNRE trains a neural network to model the likelihood-to-evidence ratio ($p(d | \vartheta)/p(d)$) for the parameters of interest $\vartheta$. This is done by training a classifier to classify samples between those correlated with the data and those independent from the data. This classifier can then approximate the ratio~\cite{Delaunoy:2020zcu, Bhardwaj:2023xph}. For complex problems like gravitational wave (GW) data analysis, an amortized approach is inefficient as many prior samples do not contribute to the final posterior. Thus, training is done in rounds, shrinking the prior space based on previous outputs until convergence~\cite{Bhardwaj:2023xph}, which is faster than traditional methods. TMNRE adapts well to scenarios like overlapping signals~\cite{Alvey:2023naa}, providing rapid and accurate posteriors. A significant advantage of TMNRE is its scalability for high-dimensional parameter spaces, useful for problems like inferring noise properties in next-generation detectors. However, it does not provide the full set of correlated posteriors, which is necessary for calculating evidence or performing importance sampling. To address this, Neural Ratio Estimation (NRE) can be adapted to be autoregressive (AR), using the probability chain rule to transform the $n$-dimensional joint prior into a product of 1D conditional priors and the likelihood-to-evidence ratio into a product of conditional ratios~\cite{AnauMontel:2023stj}. This approach maintains sample correlations but results can be unstable, depending on the ordering of conditional probabilities~\cite{AnauMontel:2023stj}. TMNRE and autoregressive NRE offer promising machine learning methods for future data analysis in GW research.

\item \textit{Nested sampling with artificial intelligence} ((\textsc{i}-)\textsc{nessai})~\cite{Williams:2021qyt, Williams:2023ppp} integrates ML, specifically NFs, with nested sampling techniques to enhance efficiency in exploring parameter spaces and calculating evidence for GW data. In traditional nested sampling~\cite{Skilling:2006gxv}, samples are drawn from iso-likelihood contours to explore higher likelihood values and build posteriors. \textsc{nessai}~\cite{Williams:2021qyt} uses NFs to represent the sampling space, mapping a multivariate Gaussian distribution to the parameter space. New samples are generated by passing them through the trained NF network and verified via importance sampling, ensuring they remain within prior boundaries. The NF network is retrained periodically based on certain criteria, and this process continues until convergence. This method improves efficiency by reducing the number of likelihood evaluations and computation time compared to usual samplers, making it suitable for the analysis of GW events with future detectors. One bottleneck is the importance sampling step, which can be addressed by using importance nested sampling. This involves approximating the evidence integral with a pseudo-importance sampling density~\cite{Cameron:2013sm, Feroz:2013hea}, using weighted NF distributions to sample directly and bypass the importance sampling step, further reducing errors and rejection sampling time. Comparative studies show that \textsc{i-nessai}~\cite{Williams:2023ppp} significantly reduces sampling time and inference duration, from more than a day to a few hours, as demonstrated with 64 BBH injections in an LVK network, see figure~\ref{fig:nessai_improvement}. Despite its advantages, challenges remain for next-generation detectors with very loud signals and highly peaked posteriors, requiring further testing and development.

\begin{figure}
    \centering
    \includegraphics[keepaspectratio, width=0.6\textwidth]{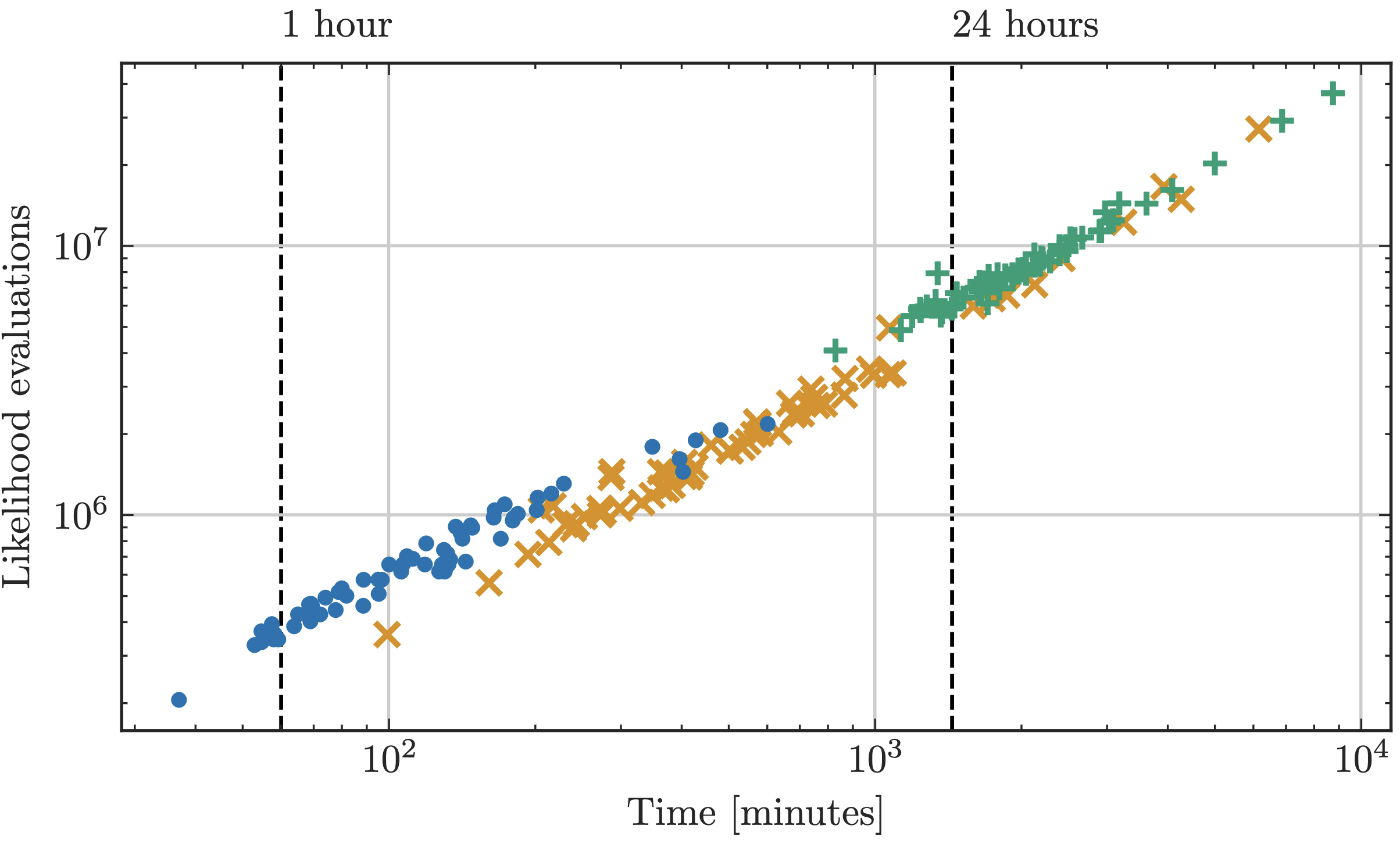}
    \caption{Number of likelihood evaluation and time needed to perform the inference for \textsc{i-nessai}~\cite{Williams:2023ppp} (blue dots), \textsc{nessai}~\cite{Williams:2021qyt} (orange crosses), and \textsc{Dynesty}~\cite{Speagle:2019ivv} (green plusses) for 64 BBH signals injected in an LVK network. One sees that ML-aided nested sampling has a significantly reduced computation time, meaning it is a promising avenue for data analysis in next-generation GW detectors. Figure taken from~\cite{Williams:2023ppp}.}
    \label{fig:nessai_improvement}
\end{figure}

\item \textit{Normalizing Flows for parameter estimation of close encounter signals.} In addition to quasi-circular evolutions of coalescences, dynamical interactions are gaining more and more interest as formation channels for CBCs. In particular, they play a crucial role in the formation of IMBHs in dense stellar environments. Such systems are mainly characterized by having a non-zero eccentricity whose effect is to induce a modulation in the gravitational waveform. In the $e\rightarrow1$ limit the emission is expected to be a series of short-duration bursts peaked around the time of periastron passage, corresponding to the  ``close encounters" case~\cite{Loutrel:2019kky}. Another way of producing a close encounter signal is during capture of two initially unbound objects. In the case of 2-body encounters of stellar mass binaries ($M\lesssim 100 M_\odot$) inside Globular Clusters the peak frequency of the GW emission $f_{\rm GW}$ is expected to be $f_{\rm GW} \lesssim 10 {\,\mathrm{Hz}}$ for eccentricities $-2.5 < \log e < 1.5$ \cite{Samsing:2019dtb, Samsing:2017xmd, Rodriguez:2018pss}, which enhances the importance of the improved low-frequency ET sensitivity.  
Performing accurate and fast PE on these sources is of great importance for many different studies: from fundamental tests of GR in the strong field regime to the investigation of the astrophysical formation channels and even to multimessenger astronomy. In particular, 
electromagnetic counterparts to GW signals are expected to be produced even by BBHs if mergers occur in specific environments like Active Galactic Nuclei (AGNs)~\cite{Graham:2022xxu}. More specifically, the EM signature consists of optical/UV flares induced by shocks in the gas surrounding the merging binary~\cite{McKernan:2019hqs}. 
Due to their repeated burst nature, close encounters are suited for early warning strategies, given that precise localization and time of coalescence can be estimated in advance. Provided that the parameters of each GW burst are well reconstructed and correctly correlated to the previous one(s),   the chances of detecting EM counterparts would increase thanks to the fast and early pointing of telescopes.

Machine learning techniques have proven to be promising for studying this kind of sources, which are characterized by repetitive short burst signals. For example, in \cite{DeSanti:2024oap} a neural network architecture based on Normalizing Flows, \textsc{HYPERION}, obtained successful results in the reconstruction of single-burst parameter posteriors. The proposed network is composed of a convolutional embedding network for extracting the most relevant features and an affine coupling flow to estimate the posterior distribution on the source parameters. The network is trained on simulated close encounter signals, which were realized using the effective fly-by waveform formalism \cite{Loutrel:2019kky}. The computation time for the reconstruction of the posterior discussed in \cite{DeSanti:2024oap} is of 0.5{\,s},  $\sim$5 orders of magnitude faster than that obtained with traditional PE methods.

\item \textit{Markov Chain Monte Carlo aided by Machine Learning}: In Markov Chain Monte Carlo (MCMC) sampling, the goal is to directly assess the posterior distribution without computing the evidence. This is done by evolving chains of samples with a proposal distribution. To improve sample proposals, gradient information can be used to select samples pointing toward higher likelihood values, as seen in the \textsc{MALA}~\cite{Grenander:1994} algorithm. NFs can further enhance MCMC by representing the posterior landscape~\cite{Gabrie:2021tlu}. This combination is implemented in \textsc{JIM}~\cite{Wong:2023lgb, Wouters:2024oxj}, which relies on \textsc{FlowMC}~\cite{Wong:2022xvh, Gabrie:2021tlu} and \textsc{MALA} for sampling and \textsc{Ripple}~\cite{Edwards:2023sak} for \textsc{JAX}-based waveforms. The gradient-based MCMC benefits from \textsc{JAX}'s auto-differentiation and just-in-time compilation for speed. The process involves initial steps of \textsc{MALA} to sample the posterior landscape, using these samples to train an NF network, and iterating between MALA steps and NF training until convergence. Final samples are generated by freezing the network, ensuring the chain's Markovian nature~\cite{Wong:2023lgb}. \textsc{JIM} has shown significant speed-ups~\cite{Wong:2023lgb, Wouters:2024oxj}: BBH analyses in under 3 minutes~\cite{Wong:2023lgb} and BNS analyses in less than 30 minutes~\cite{Wouters:2024oxj}. This method automatically adapts to new noise and signals without pre-training. However, it only returns posteriors, not evidence, which could be addressed with importance sampling~\cite{Williams:2021qyt, Williams:2023ppp}. Efficient implementation requires a fully \textsc{JAX}-based framework~\cite{Edwards:2023sak}, including waveforms.

\item \textit{Deep neural network driven Hamiltonian Monte Carlo} (DeepHMC) is a new approach to parameter estimation for long-duration signals~\cite{porter:2024a}, improving upon standard samplers like MCMC and Nested Sampling, which suffer from high autocorrelations and slow convergence. Hamiltonian Monte Carlo (HMC) avoids these issues by using Hamiltonian trajectories~\cite{Duane:1987de} instead of random jumps, converting the problem into a gravitational one with potential and kinetic energy terms. Properly tuned HMC converges $D$ times faster than MCMC, where $D$ is the parameter space dimensionality. Despite its theoretical advantages, HMC is underused due to tuning difficulties and computational bottlenecks, specifically the need to calculate $D$-dimensional gradients of the log-likelihood at each trajectory step, which is costly for long-duration signals. DeepHMC addresses these issues by combining HMC with deep learning. It generates $\sim\mathcal{O}(10^3)$ Hamiltonian trajectories using numerical differencing to calculate gradients, which are then used to train a deep neural network (DNN) model of the background geometry, called a ``shadow potential". The final inference process involves:
\begin{enumerate}
\item Starting on the true potential and calculating the initial Hamiltonian.
\item Jumping to the shadow potential and evolving the trajectory using DNN gradients.
\item Moving back to the true potential and calculating the final Hamiltonian.
\item Comparing initial and final Hamiltonians using the Metropolis-Hastings ratio.
\end{enumerate}
On a Macbook Pro with an M1 processor, DeepHMC's data-taking phase takes 10 minutes (parallel on ten cores), training/testing the neural network takes 30 minutes, and the final inference phase takes around two hours to produce 5000 independent samples. This reduces Bayesian inference of BNS systems in the LVK band from weeks to hours.

\end{itemize}
While not fully ready for parameter estimation in 3G detectors, the examples above show ML-based methods are a promising avenue for this task in the future, even if more development is needed. Moreover, the applications shown have focused on data resembling the ones we observe in the LIGO-Virgo data. The data will clearly be different for Einstein Telescope, leading to new challenges, and requiring further devlopment.

\subsubsection{Computational Requirements}
Ref.~\cite{Smith:2019ucc} introduced a highly parallelised algorithm, called \emph{parallel bilby}, or \texttt{PBILBY}, to infer properties of a source in an accelerated way. Table~\ref{fig:times} shows a table from their analysis showing the immense improvement in runtimes for a low and asymmetric mass BBH using 640 CPUs which would otherwise be impossible with reduced computational resources. 
\begin{table}
    \centering
    \includegraphics[width=\textwidth]{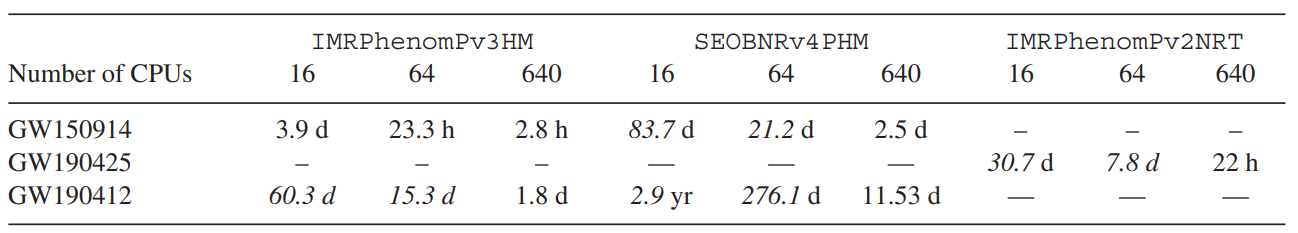}
    \caption{Table showing computational requirements for a variety of waveforms with physical effects including higher-order-modes and tidal deformabilities. For BNS systems as well as highly asymmetric systems, availability of only 16 CPUs make obtaining results infeasible. Table from~\cite{Smith:2019ucc}.}
    \label{fig:times}
\end{table}
\noindent 
As summarised in ref.~\cite{Couvares:2021ajn}, $\sim \mathcal{O}(10)$ events in O2 required about $10^6$ CPU core hours. In the ET era, seeing $\sim 10^3$ CBC signals will therefore scale accordingly. Ref.~\cite{Hu:2024mvn} provides detailed estimates of PE cost in 3G detectors. With one-month observation, PE for BBH sources would cost billions of CPU hours with a vanilla PE setting, but the costs are reduced by at least 4 orders of magnitude when acceleration methods such as ROQ are applied. The costs of BNS and NSBH sources are roughly an order of magnitude higher than BBH when acceleration methods are used. However the advent of faster approaches coupled with ML and use of GPUs and TPUs in addition to CPUs will likely lead to feasible runtimes of longer signals in the ET era. As reported in~\cite{Wouters:2024oxj}, using a single NVIDIA A100-40 GB
GPU can reduce the runtime of a BNS signal to 2.5 minutes.
Machine learning-based methods (such as normalizing flows) can achieve $\mathcal{O}$(second) performance~\cite{Dax:2024mcn, Hu:2024oen} and, according to ref.~\cite{Hu:2024oen}, achieve 1/1000 cost compared with traditional stochastic sampling methods.

\subsection{Peculiarities of a triangular ET}
\label{sec:pec_ET}

\subsubsection{Null stream}
\label{subsec:null_stream}

A null stream can be formed for an arbitrary GW detector network with at least three detectors, but generally only for one GW signal at a time, depending on its sky position~\cite{Guersel:1989th}.
With ET in a triangle configuration,  because of the closed geometry of its component arms and almost negligible light travel times between components, it is possible to construct a null stream  that cancels the gravitational-wave signal,  independently of the sky position of the source. The null stream is obtained by adding together the three detector outputs, 
\begin{align}
	d_{\rm null} (t) = \sum_{i=1}^3 d_i(t)\, .
\end{align}
In the triangular design, it is  then guaranteed that no signal is present in
the null stream~\cite{Goncharov:2022dgl}, so that
\be\label{dnullsumni}
d_{\rm null}(t) = \sum_{i=1}^3 n_i(t)\, .
\ee
The subsections below describe the  opportunities that the sky-independent null stream offers to the data analysis problems described in the sections above.

\paragraph{Noise estimation.}
\label{subsubsec:PSD_est}

Noise estimation is crucial to ensure robust detection \cite{Klimenko:2015ypf,Messick:2016aqy,Usman:2015kfa} and PE \cite{Ashton:2018jfp,Veitch:2014wba}. Furthermore, an accurate noise characterization is essential for understanding the origin of the different noise contributions, and for the development of noise mitigation strategies~\cite{Badaracco:2023vpk}, thereby increasing the overall sensitivity. As explained in section~\ref{subsec:background}, the estimation of the properties of the background noise for ET is still an open problem. The following paragraphs outline a few possibilities to leverage the null stream for this challenge. 

From \eq{dnullsumni} we see that the PSD in each of the three detectors in the triangle configuration\footnote{We use the nomenclature defined in footnote~\ref{foot:nomenclature} on page~\pageref{foot:nomenclature} (see also footnote~\ref{foot:nomenclaturediv9} on page \pageref{foot:nomenclaturediv9}), so ET in its triangle configuration is made of three nested detectors, with each detector composed of two interferometers (one tuned toward low frequencies and one toward high frequencies), for a total of six interferometers.\label{foot:nomendiv10}} can be estimated from the null stream taking the Fourier transform of the correlator
$\langle d^*_{\rm null}(t) d_{i}(0) \rangle$, and assuming that the noise in different detectors are uncorrelated, so that, for each $i$,  $\langle d^*_{\rm null}(t) d_{i}(0) \rangle$ becomes the same as $\langle n^*_i(t) n_{i}(0) \rangle$ (where we used  also the fact that the GW signals are uncorrelated with the noise). Note that this is only valid if $\langle n^*_i(t) n_{j}(0) \rangle\propto \delta_{ij}$, i.e. if the noise in different detectors are uncorrelated.
However, it is highly likely  that noise sources couple coherently to the different ET detectors   of the triangle configuration, particularly at low frequencies~\cite{Janssens:2022cty} (see section~\ref{subsec:corr_noise} for more details). This  would therefore bias the PSD estimates that use the null stream, unless these correlated noise can be identified and characterized, e.g. with witness sensors~\cite{Goncharov:2022dgl}.\footnote{It should also be observed that, as with all applications of the null stream, there is an important dependence on the duty cycle of the detectors, since the null stream can only be formed when all three detectors in the triangle configurations are on-line. Assuming  an uncorrelated 85\% duty cycle in each of the three detectors  composing the triangle, the null stream will be available for $(0.85)^3=61\%$ of the run time, which goes down to $51\%$ of the run time
assuming instead  an independent duty cycle of $80\%$ for each detector \cite{Branchesi:2023mws} (the best duty cycle for the O3b LVK run is $79\%$ \cite{aLIGO:2020wna}). However, the triangle design is considerably more complex than that of 2G detectors, with two interferometers in each detector (an HF interferometer, and a LF interferometer working at cryogenic temperatures), and it is difficult to reliably estimate what the duty cycle will be.}

In the PSD estimate, the null stream can also be useful for removing the  astrophysical confusion noise. This confusion noise is produced by
a large population of unresolvable signals, which forms the stochastic GW background, as well as the foreground population of overlapping signals, which are individually resolvable but behave as an additive noise source given the large number and long duration~\cite{Wu:2022pyg, Coughlin:2018tjc, Coughlin:2016vor}. Such a confusion noise  could affect the detection and bias the PE analysis of individual GW sources.
As outlined in ref.~\cite{Goncharov:2022dgl}, it is possible to remove such background by cross-correlating the null stream with the data stream of the individual detectors. Given the absence of any GW signal in the null stream, this cross-correlation yields the sum of the instrumental Power Spectral Density (PSD) of the detector under consideration and the Cross-power Spectral Density (CSD) of instrumental correlated noise between that detector and the others. Therefore, this method can be used to provide an unbiased estimate of the detector noise, again at frequencies where the correlated noise is negligible.

Another unique opportunity of the null stream is the on-source measurement of the PSD. In GW searches \cite{Klimenko:2015ypf,Messick:2016aqy,Usman:2015kfa} and PE \cite{Ashton:2018jfp,Veitch:2014wba},
detector noise is often modeled as a stationary Gaussian process.
However, actual noise in GW detectors is not stationary \cite{LIGO:2021ppb}.
Figure~\ref{fig:PSD_ratio_distr} illustrates the median and the $1\sigma$ variation of the PSD variation statistic,
akin to the one defined in ref.~\cite{Mozzon:2020gwa}:
\begin{equation}
    v(f;\delta t) = \frac{S_{n}(f;t+\delta t)}{S_{n}(f;t)}\, ,
\end{equation}
for the LIGO Livingston detector during the first half of the third observing run (O3a).
Data segments containing event triggers with a false alarm rate below one per month were excluded from the PSD estimation.
The results indicate significant noise PSD fluctuations over periods of $\mathcal{O}(1)$ hours,
with an estimated fluctuation of $\sim80\%$ in the frequency range of 20-60 Hz.
Extending the time difference to one day results in fluctuations up to $200\%$.

The non-stationarity of noise is a critical issue for  ET,
particularly during long-duration observations of BBH and BNS mergers,
which can span hours to days.
As demonstrated in ref.~\cite{Goncharov:2022dgl},
BNS inspirals can leave a visible imprint on the inferred PSD at $\sim 5\text{ Hz}$.
Inaccurate PSD estimates would lead to incorrect uncertainty estimation in posterior distributions \cite{Edy:2021par},
affecting studies reliant on precise localization volumes,
such as Hubble constant measurements \cite{Kumar:2022tto,Mozzon:2021wam}.
Therefore, PSD estimation using off-source data is not feasible for ET.
The methods discussed above demonstrate the capability of the null stream technique to provide an unbiased noise PSD estimation using on-source data,
even in the presence of GW signals (again, as long as correlated noise can be identified and characterized).

Another option for PSD estimates are
global fitting approaches, such as those used to simultaneously model both GW signals and noise PSD~\cite{Littenberg:2023xpl}. While they offer accurate results,
they substantially increase both the complexity and the computational demands compared to signal-only parameter estimation.
In contrast,
the null stream technique presents a computationally efficient alternative for estimating the on-source noise PSD.
When employing global fitting methods, one must select a parametric model for the PSD, often with a variable number of parameters, which necessitates trans-dimensional sampling, adding further complexity. For example, {\sc BayesWave} employs cubic splines and Lorentzians in this process~\cite{Littenberg:2014oda}. By incorporating null stream information, a non-parametric estimation of the PSD can be directly utilized, streamlining the analysis and bypassing the need for the extensive computational effort required for stochastic sampling in global fitting methods.

\begin{figure}
    \centering
    \begin{subfigure}[b]{0.49\textwidth}
    \includegraphics[width=\textwidth]{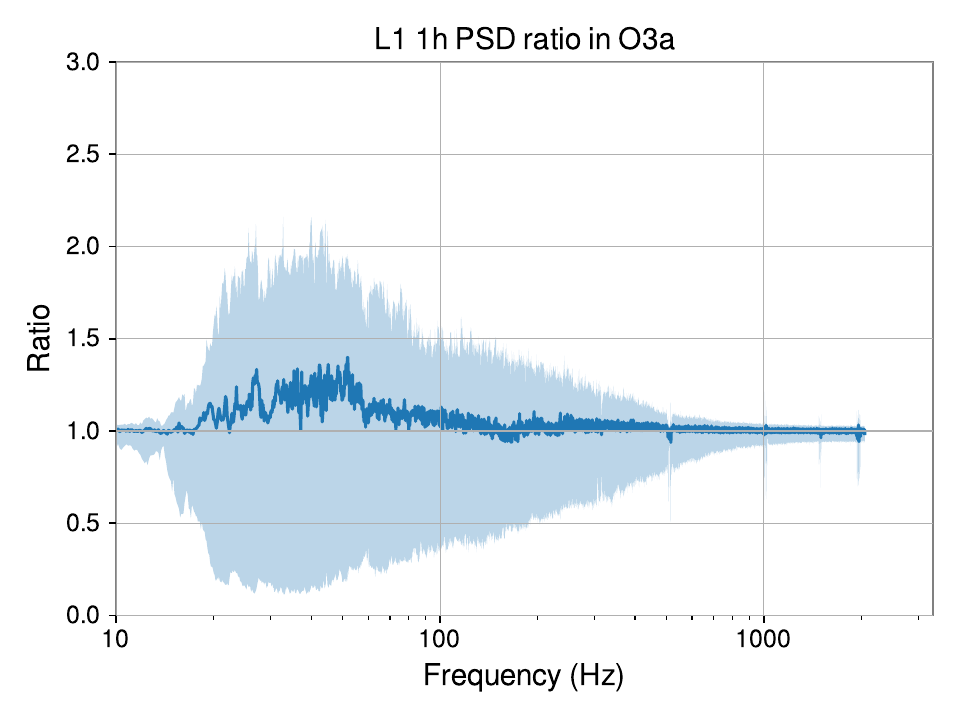}
    \caption{The distribution of the ratio of PSD delayed by 1 hour.}
    \end{subfigure}
    \hfill
    \begin{subfigure}[b]{0.49\textwidth}
        \includegraphics[width=\textwidth]{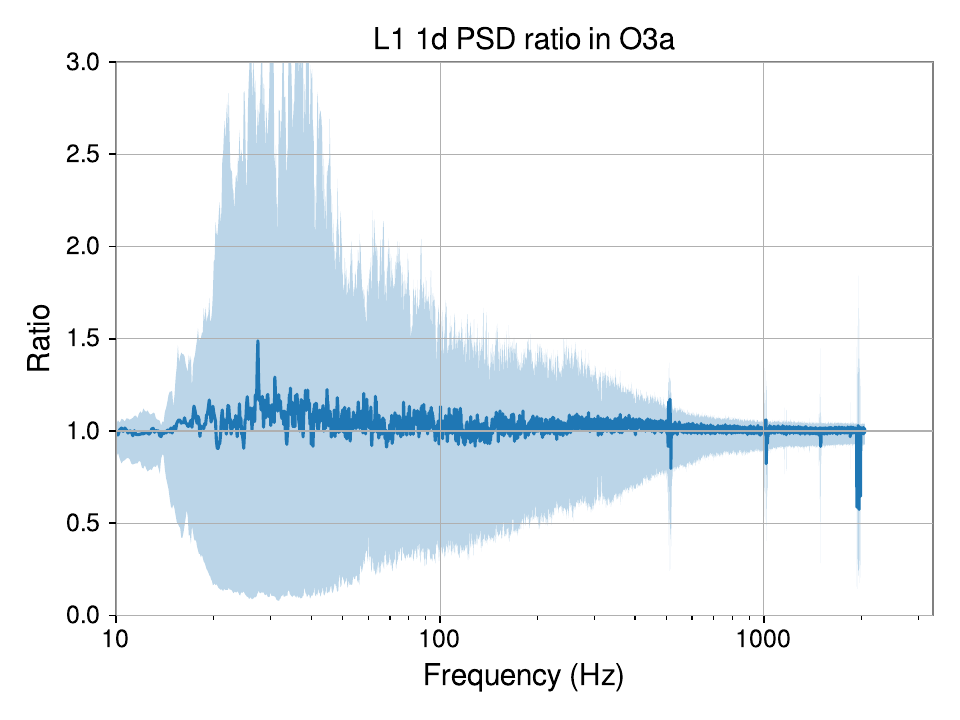}
        \caption{The distribution of the ratio of PSD delayed by 1 day.}
    \end{subfigure}
    \caption{The median and the 1$\sigma$ variation of the PSD variation statistic of the LIGO Livingston detector during the first half of the third observing run.}
    \label{fig:PSD_ratio_distr}
\end{figure}

Another specific advantage of the null stream is the ability to directly measure the spectrum of the confusion noise. As known, the auto-correlation of a single detector provides the sum of the detector noise and confusion noise PSDs. By subtracting the cross-correlation between the null stream and the data stream of that detector from this sum, the instrumental noise PSD is eliminated, leaving only the instrumental correlated and confusion noise components. This method provides an unbiased estimate of the confusion noise PSD at frequencies where the instrumental correlated noise is negligible. Hence, the null-channel method from ref.~\cite{Romano:2016dpx} can be used to improve the measurement of the stochastic GW background described in section~\ref{subsec:sgwb}.

\paragraph{Glitch Identification.}
In addition to non-stationarity, detector noise in GW observatories is also subject to glitches \cite{Virgo:2022fxr,KAGRA:2020agh,LIGO:2021ppb,AdvLIGO:2021oxw}, as outlined in section~\ref{subsec:long_dur} and section~\ref{sec:innovative_methods}.
These glitches can significantly reduce the efficiency of astrophysical searches
\cite{Lee:2024znj} and introduce bias in PE
\cite{Kwok:2021zny,Macas:2022afm,Payne:2022spz,Powell:2018csz}.
Notably, during the O3 run,
the median rate of glitches in the LIGO and Virgo detectors exceeded one per minute for the majority of the run \cite{Davis:2022ird}.
Approximately $20\%$ of the detected signals required some form of mitigation due to glitches \cite{Davis:2022ird}.
The high frequency and significant impact of these glitches highlight the necessity for effective identification and subtraction techniques to ensure the integrity of gravitational wave data and validate transient candidate events \cite{DiRenzo:2022klz}.

The current approach to identifying glitches in GW data primarily involves utilizing witness sensors that are insensitive to GWs to monitor the local environment.
These sensors help detecting non-astrophysical disturbances that might affect data quality.
The data channels from these sensors, known as auxiliary channels, are critical as they provide information for investigating and assessing the quality of the GW data.
Detailed procedures for this process
employed
during O3 are documented in
refs.~\cite{Virgo:2022fxr,LIGO:2021ppb}, outlining the methodologies and tools used to identify and mitigate the impact of glitches on data analysis.
In contrast to GW signals, glitches do not cancel in the null stream.
Therefore, the
null stream offers an additional channel,
apart from the auxiliary channels,
to identify the presence of glitches in strain data.

The null stream has been proposed as an effective tool to discriminate GW signals from noise
\cite{Chatterji:2006nh,Harry:2010fr,Wen:2005ui}.
In the context of a triangular detector network, since the null stream is sky-independent and unique, the presence of glitches in any of the detectors will manifest in the null stream.
Thus, methods utilizing the null stream can directly monitor the ``glitchiness" of strain data.
Ref.~\cite{Goncharov:2022dgl} evaluated the effectiveness of the null stream in detecting glitches in a triangular ET exploring two different approaches: the waveform-dependent null SNR
\cite{Harry:2010fr}
and the waveform-independent likelihood-based approach,
demonstrating that both methods effectively reduce glitches to the Gaussian noise background.
An end-to-end null-stream based glitch-mitigation algorithm for a triangular ET has been developed and implemented recently in~\cite{Narola:2024qdh}. It demonstrates that characterization and subtraction of glitches arbitrarily close to the peak of the signal can be performed using the null stream without any significant effect on the quality of parameter measurements.

Beyond classical statistical tests, the ET null stream could serve as input for machine learning-based methods such as \texttt{Gravity Spy}
\cite{Jarov:2023qpt,Zevin:2016qwy}
and \texttt{iDQ}
\cite{Essick:2020qpo}.
Once a glitch is identified, the specific detector where the glitch is present can be determined by cross-correlating the null stream with the
main GW channels from each detector,
as these channels contain the glitch in a form most similar to the null stream.
Additionally, cross-correlating the null stream with the auxiliary channels can help identify not only the detector where the glitch occurs but also the source of the noise.

\paragraph{Self-calibration.}

In the third gravitational wave transient catalog~\cite{KAGRA:2021vkt}, and previous LVK publications, calibration uncertainty was taken into account by marginalising over the calibration uncertainty estimates provided by the calibration team~\cite{Viets:2017yvy, VIRGO:2021kfv}.

For CBCs, it has been found that the impact of calibration errors is mostly negligible for the current generation of detectors, ~\cite{Szczepanczyk:2023ihe, Vitale:2020gvb, Payne:2020myg}.
However, the localization parameters are the most affected by the inclusion of the calibration error distribution~\cite{Payne:2020myg}; in fact, the area of the 90\% credible interval for the sky position of GW150914 increased by 300\% when calibration uncertainty was taken into account~\cite{LIGOScientific:2016vlm}.
Additionally, calibration errors might lead to biased results for other science cases such as measurements of the Hubble constant~\cite{Huang:2022rdg} or GW stochastic backgrounds~\cite{Yousuf:2023nmz}.
The improved sensitivity of ET will also tighten the requirements on the calibration errors because at high SNR the effect of calibration errors becomes distinguishable from the statistical variation.
More precisely, the calibration error for ET needs to be below 1\% in both amplitude and phase~\cite{Purrer:2019jcp} to avoid biased results in PE for CBC signals.
Therefore, new methods are required to ensure precise calibration that will guarantee the scientific goals set for ET.

One possible method that has been proposed, but never implemented, relies on signal leakage into the null stream~\cite{Schutz:2020hyz}.
First, a null stream is constructed for every detected event. If calibration errors are present, then a small fraction of the GW signal leaks into the null stream, which can be detected with a template bank that consists of convolutions of the detected signal with a parameterised calibration model. Since the SNR of the leaked signal is only a fraction of the SNR of the detected signal, the results of multiple detections have to be combined in a single detection statistic. This approach only works if the calibration error remains constant in between detector upgrades, which is usually a valid assumption. 

A triangular configuration can relax some of the assumptions in the method discussed above because the construction of the null stream is independent of the measured sky position. This can be a major advantage because:
\begin{itemize}
    \item An initial timing error in the calibration model impacts the measured sky-position and hence, if the null stream depends on the sky position, it also impacts the estimated calibration error. This does not happen for the triangle, whose null stream is sky-independent.
    \item The null stream of the triangle  allows us to deal with overlapping signals. Indeed, if the null stream is not sky-position independent, then the SNR of the signal of interest is scaled down by the calibration error ($\sim 1\%$), while the SNR of the secondary signal remains roughly the same. Therefore, it is likely that any signal of interest in the null stream overlaps with several signals of comparable SNR, severely impacting the matched filter performance.
    \item It is possible to abandon the assumptions on the signal altogether by constructing a (Gaussian) likelihood on the null stream because the PSD of the null stream is related to the PSDs of the individual detectors. In this scenario, overlapping signals, unresolvable signals, unmodeled signals and even the SGWB contribute to the self-calibration.
\end{itemize}

Note that we can only improve the relative calibration error because any common mode calibration error across detectors is equivalent to a modified GW signal and therefore does not appear in the null stream. Additionally, for a single event with a fixed polarisation, the response in different detectors is equivalent up to a scaling with a complex factor. Therefore, it is not possible to attribute the calibration error to a specific detector. However, for a collection of events, there is no fixed correspondence between the signals in each detector. Hence, there is a unique set of calibration errors that maximises the likelihood, up to the aforementioned common mode. Finally, the template based method exploits the fixed frequency dependence of the waveform to extrapolate any absolute calibration at a single frequency to the entire spectrum.

\subsubsection{Correlated noise}
\label{subsec:corr_noise}

The triangular configuration of ET with three nested detectors will give rise to correlated noise. So far correlated noise has not been a concern for current detectors, which are situated further apart from one another, 
with the exceptions of the environmental effects like Schumann resonances~\cite{Schumann:1952, Schumann:1952_damping}, 
for which correlations in the magnetic field have been measured between the LIGO and Virgo sites \cite{Thrane:2013npa,Thrane:2014yza,Coughlin:2016vor,Coughlin:2018tjc,Janssens:2022tdj}. Predictions indicate that these could impact the low frequency region of the search for an isotropic gravitational wave background when LVK reaches its design sensitivity \cite{Janssens:2022tdj}. Recent work also reported on correlated magnetic glitches which are predicted to significantly couple to LIGO-Virgo at their A+/AdV+ design sensitivities \cite{Janssens:2022tdj}. The effect of these correlations in magnetic field fluctuations has also been investigated for  ET. In the case in which ET has a similar magnetic coupling function compared to second generation GW interferometric detectors such as LIGO and Virgo, it is predicted that correlated magnetic noise will dominate ET's ASD below 15~Hz and by up to several orders of magnitude for the lowest frequencies \cite{Janssens:2021cta}. The projected impact of correlated magnetic noise remains of similar levels regardless of the choice of a triangular or 2L configuration. Due to the higher sensitivity to correlated noise in the case of the search for a stochastic gravitational wave background, such searches will be more sensitive to these correlated magnetic field fluctuations by about a factor $10$ and will be affected up to $\sim$ 30 Hz \cite{Janssens:2021cta}. Furthermore, lightning strikes would produce about $\mathcal{O}(10^5)$ correlated glitches per week for a third generation detector such as the ET \cite{Janssens:2022tdj}, that is, on average about one correlated glitch every six seconds; however, these are expected to be somewhat clustered in time based on lightning activity.

Due to ET's triangular configuration there is potential for additional noise couplings, which do not occur in the 2L configuration, due to the proximity of sensitive detector parts of different interferometers. One such example is the effect of Newtonian Noise (NN)\footnote{NN is a force directly exerted on GW test-masses caused by density fluctuations in the surrounding medium.} due to correlations in the seismic noise on distance scales of several hundreds of meters \cite{Janssens:2022xmo,Janssens:2024jln}. This noise source is expected to have a limited effect on ET's ASD. However, it is projected to dominate ET's low frequency sensitivity up to 20 Hz-40 Hz for the search for a stochastic gravitational wave background, by many orders of magnitude \cite{Janssens:2022xmo,Janssens:2024jln}, see the discussion in section~\ref{noisecorrseismicdiv2} and figure~\ref{fig:NNBudget_GWB}. In case of such a scenario one could consider only to analyse data starting from 15 Hz--20 Hz for the analysis of a gravitational wave background to eliminate or reduce the effect from correlated NN, as shown in figure~\ref{fig:NNBudget_GWB}. 
However, as shown in section~\ref{sect:RecCorrNoise_div2}, if is it possible to model accurately the noise, a Bayesian analysis can allow to reconstruct the GW signal also at lower frequencies.
Signals with steeper power-laws such as $\alpha=3$ would experience no effect as their key resolvability comes from frequencies which are not affected by the low frequency noise correlations.

Correlated noise can also have an  impact on parameter estimation \cite{Cireddu:2023ssf}. Akin to a network of witness sensors measuring the same physical phenomenon, correlated noise enhances the collective information about the noise processes, thereby reducing the spread of the network likelihood. However, the propagation of this reduced spread to the precision of PE is not trivial and depends on several critical factors, such as the specific structure of the signal gradient. By decomposing the triangular ET observational space into the principal coordinate system \cite{Wong:2021eun}, consisting of a 1-dimensional null space and a 2-dimensional signal space (commonly referred to as the T, A and E channels in the LISA community \cite{Prince:2002hp}), it is observed that the effect of correlated noise varies with the sign of the correlation coefficient between the E1, E2 and E3 detectors \cite{Cireddu:2023ssf}. Specifically, for positive correlation coefficients, the signal space shrinks, resulting in an increase in the SNR compared to the case without correlated noise. Conversely, for negative correlation coefficients, the signal space expands, leading to a reduction in the SNR. Additionally, the impact on the null space must be considered, as an increase or decrease of the noise variance in the null space due to positive or negative correlations, respectively, may indicate a diminished or improved capability to estimate the noise through the null stream.
The sign of the correlations, whether positive or negative, is highly uncertain as it depends on the relative orientation of the detectors and the specific sources of correlated noise. However, due to the triangular geometry, negative correlations may be more common.

Ref.~\cite{Cireddu:2023ssf} showed that in cases of high positive correlations across the entire frequency band, neglecting correlated noise in ET can lead to failures in reconstructing signals that would otherwise be resolvable. Similarly, ignoring negative correlations can lead to an overconfident reconstruction of the posterior distribution, introducing bias into the results. Therefore (for such high values of the noise correlation) accounting for correlated noise becomes important for parameter estimation analysis at low frequencies. This can be achieved by incorporating the estimated CSDs of different detectors directly into the likelihood. 
In particular, the CSDs can be estimated by cross-correlating the data streams from different detectors, using long data stretches and applying the median method to mitigate the impact of confusion noise.

\subsection{Simulations and Mock Data Challenges}
\label{sec:MDC}
ET will have the potential to detect a large number and a large variety of sources, but at the same time, as we have discussed above,  new challenges will appear, such as long duration waveforms or overlapping sources. The purpose of mock data challenges (MDCs) is to produce simulated data that will be used to develop, test and optimize data analysis and parameter estimation techniques and understand the science potential of ET, and also understand the requirements for computing infrastructure.

\subsubsection{Description of the first MDC}

\begin{figure}[t]
    \centering
    \includegraphics[width = 0.8\textwidth]{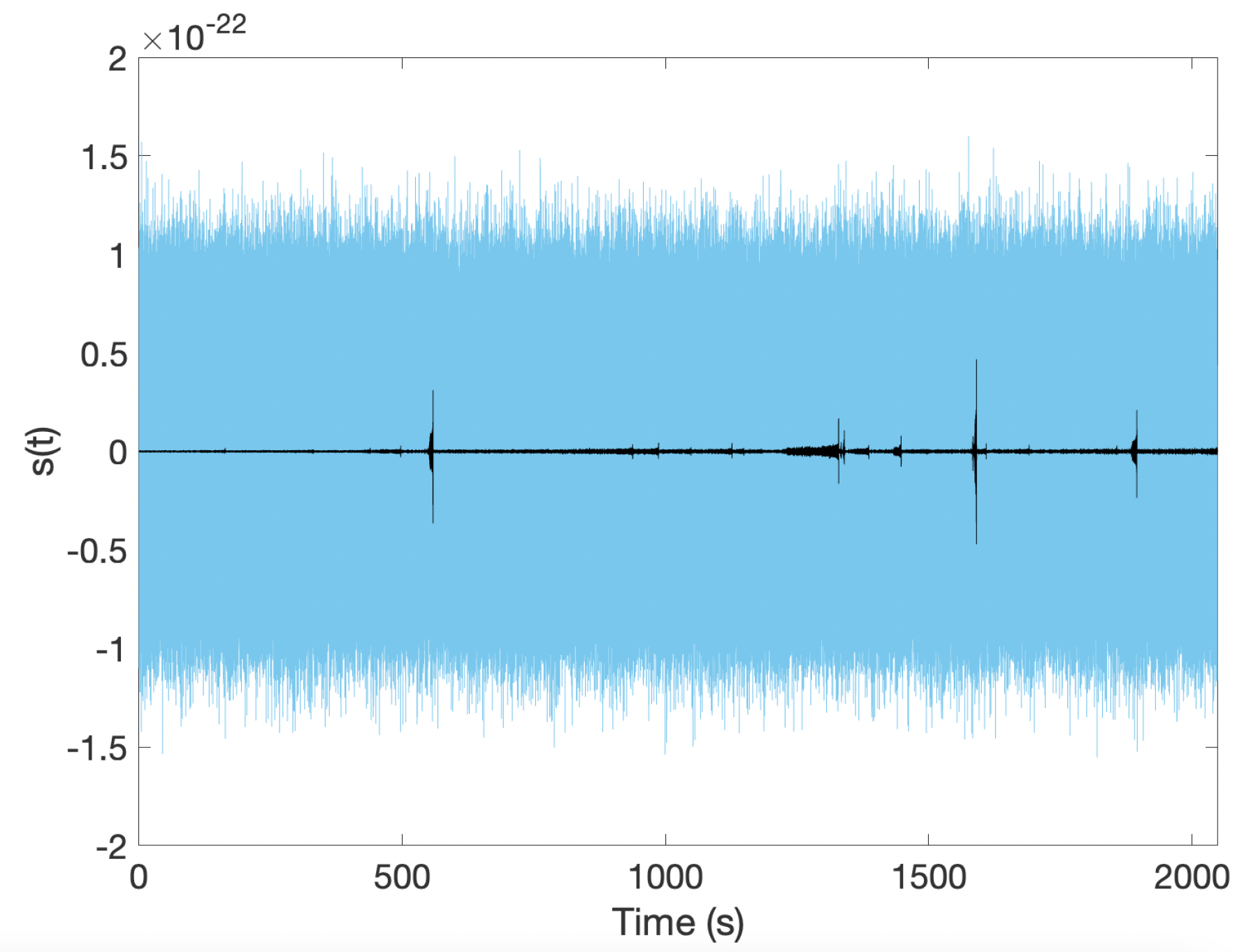}//
     \includegraphics[width = 0.8\textwidth]{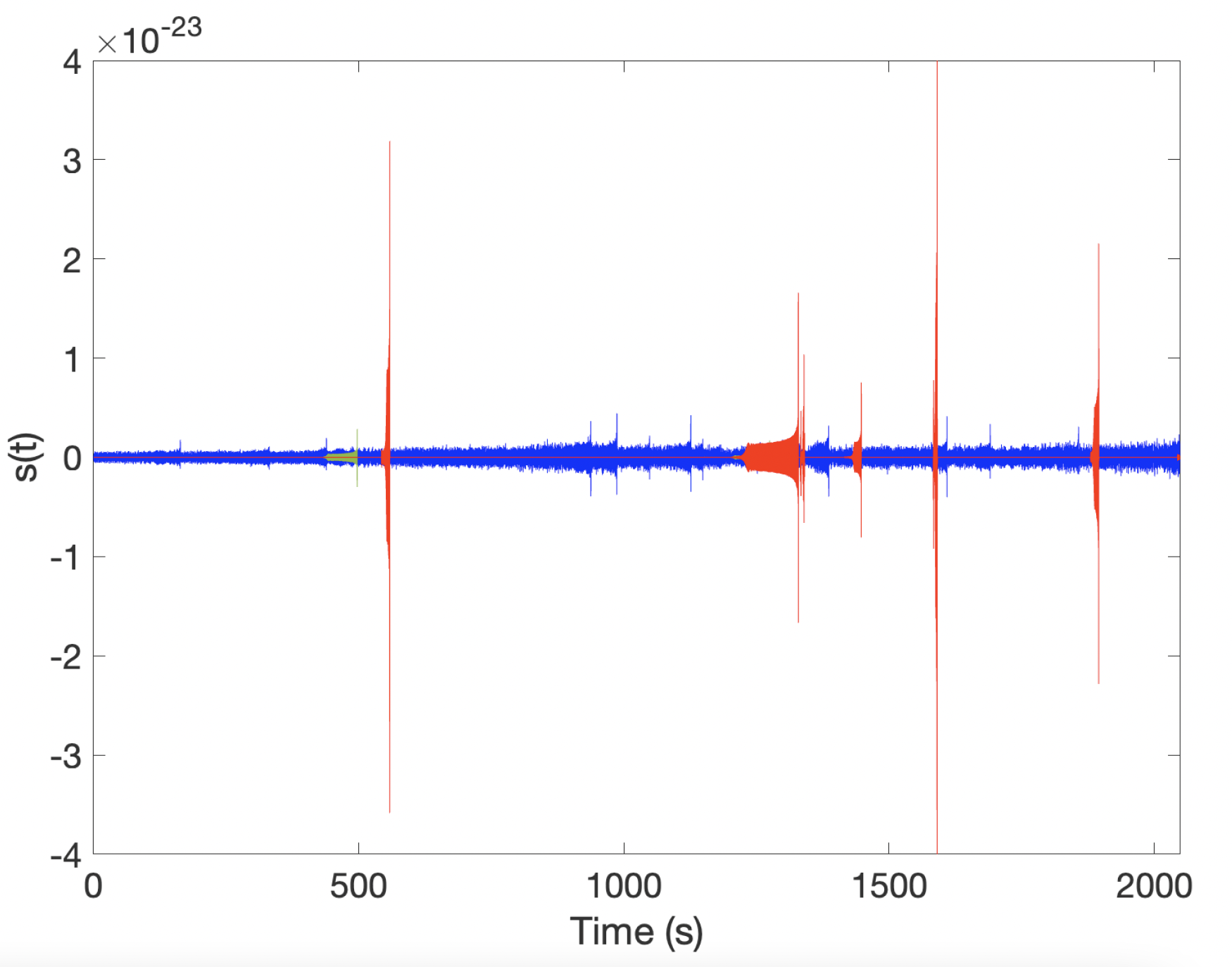}\\
    \caption{Top: Time series of one data segment of length 2048s at a sampling rate of 8192 Hz. The noise realization is in blue and the gravitational-wave signal from compact binary coalescences in black. Bottom: Same as above with the GW signal only, with in blue the  signal from BNSs, in green from BHNSs and in red from BBHs.}
    \label{fig:signal}
\end{figure}

The end product of the MDCs are the time series of the detector outputs, arising both from the instrumental (or environmental) noise and the GW signal (see figure~\ref{fig:signal})

The first ET MDC was generated with a simulation code available on the ET 
gitlab.\footnote{https://gitlab.et-gw.eu/osb/div10/mdc-generation} It consists of one month of data split into 1300 segments of length 2048s with a sampling rate of 8192 Hz. The procedure to generate the simulated data is mostly the same as the one used for the previous ET MDC in 2012-2014 and described in detail in \cite{Regimbau:2012ir}.
The main steps are briefly summarized below.

\subparagraph{Simulation of the Noise.}   
We consider ET in its triangle configuration, which consists of three independent detectors (each made of two interferometers, see footnote~\ref{foot:nomenclature} on page~\pageref{foot:nomenclature} or footnote~\ref{foot:nomenclaturediv9} on page \pageref{foot:nomenclaturediv9})  with 60 degree opening angles and arms of length 10 km, placed underground to reduce the influence of seismic noise. Assuming that there is therefore no instrumental or environmental correlated noise, the noise was simulated independently for each of the three ET detectors E1, E2 and E3, by generating  Gaussian frequency series with a mean of zero and unit variance.\footnote{As discussed in section~\ref{subsec:corr_noise},   the assumption that there is no correlated noise may not be realistic, especially at low frequency,  as the three ET detectors are nearly co-located. A more careful study of the effect of environmental noise will be included in future ET mock data and science challenges.}
These frequency series were then colored with the noise PSD, and inverse Fourier transformed to the time domain. The code generates sequential segments with a stride from one segment to the next, so that the data are continuous over segments. 

\subparagraph{Simulation of the GW signal.} 
The signal contains the gravitational waves from a cosmological population of compact binary coalescences, including binary neutron stars (BNSs), binary black holes (BBHs) and systems of one neutron star and one black hole (BHNSs). The masses, spins and redshift are provided by population synthesis catalogs \cite{Santoliquido:2020axb} for BNS and BHNSs, and \cite{Mapelli:2021syv,Mapelli:2021gyv} for BBHs, and are the same population  described in section~3 of \cite{Branchesi:2023mws}.
To generate a population of compact binaries, we proceeded as follow for each source :

\begin{itemize}
\item Assuming a Poisson process, the time from the previous coalescence was drawn from an exponential distribution $P(\tau) = \exp(- \tau / \lambda)$, where $\lambda=38$~s is the average time interval between successive events (see \cite{Regimbau:2022mdu}), directly calculated from the catalog as the inverse of the ratio between the number of events and the duration of the catalog. 

\item We then selected the type of the binary using the proportion $87\%$ of BNSs, $3\%$ of BHNS and $10\%$ of BBH also provided by the catalogs and drew randomly a source with its masses, spins, distance and redshift from the corresponding catalog. In total the one month dataset contains 59540 BNSs, 6578 BBHs and 1977 BHNSs. 

\item For BNS systems we calculated the tidal parameters of the two neutron stars from their mass given a single equation of state.

\item The inclination was drawn assuming a isotropic distribution, the polarization angle and the phase at coalescence were drawn assuming a uniform distribution.

\item The position in the sky was drawn assuming an isotropic distribution.
 
\item The two polarizations $h_+$ and $h_{\times}$, the antenna pattern functions of the three ET detectors $F^j_+$ and $F^j_{\times}$ ($j=1,2,3$) were calculated, and then the responses $h^j(t)=F^j_+(t) h_+(t)+ F^j_{\times}(t) h_{\times}(t)$ were added to the time series of E1, E2 and E3. In these simulations, we have used the waveforms IMRPhenomPv2 with tidal effects (NRTidalv2) and no spins for BNSs, and IMRPhenomXPHM with higher modes for both BBHs and BHNSs.

\end{itemize}

\subparagraph{The GW background.} 
The combined energy density of the signal from all the sources at all redshift and in all directions, usually referred to as the background or the stochastic background, can be calculated from the list of sources present in the data using the formula given in \cite{Regimbau:2022mdu}:

\begin{equation}
\Omega_{GW}(f) = \frac{f}{c\rho_c} \frac{1}{T} \sum_{k=1}^N \frac{1}{4\pi r^2(z^k)}\frac{dE_{gw}}{df}(\theta^k,f_s) 
\label{eq:omega}
\end{equation}
where  $\rho_{c} =3H_0^2c^2/(8\pi G)$ is the critical energy density required to close the Universe, and $H_0$ is the Hubble constant. The corresponding expressions for BBH, BNS and NSBH are shown in figure~\ref{fig:{monopole_total}}.

\begin{figure}[t]
    \centering
    \includegraphics[width = 0.8\textwidth]{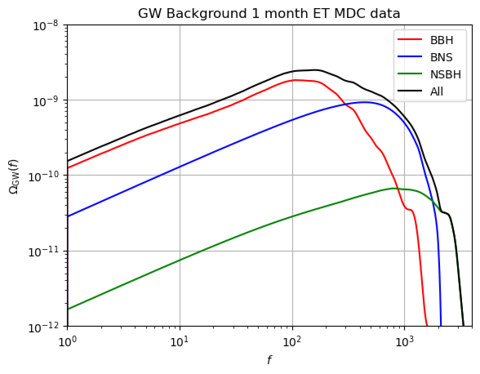}
    \caption{Gravitational energy density from the sources in our one month dataset}
    \label{fig:{monopole_total}}
\end{figure}

\subsection{Challenges}
The first ET MDC is divided into a beginner and a more expert challenge. The beginner challenge consists of the recovery of the six highest signal-to-noise ratio signals for which the time window is given. For this challenge a tutorial is provided to help retrieve the data, calculate the PSD or plot the time series. 
For the expert challenge we ask the participants to run parameter estimation on the largest SNR signals, but also analyse very long BNS or separate overlapping signals.


\subsection{Synergies in data analysis developments}
\label{sec:common_divs}
With the advent of 3G detectors, including ET, there will be access to unprecedented precision of fundamental properties of our Universe. To be able to robustly assess these properties, efficient and feasible methods of data analysis will be of paramount importance. We note below some notable aspects of data analysis to be developed, that will benefit from  synergies with research directions presented in  previous sections of this work.

\begin{itemize}
    \item Synergies with Fundamental Physics (section~\ref{section:div1}). Precision science with ET will require efficient as well as reliable methods of detection and parameter estimation for loud signals. Testing a theory of gravity or probing an equation-of-state of dense nuclear matter introduces additional parameters into the GW data analysis problem. As signals become longer and louder, incorporating these extra parameters will require more refined data analysis methodologies, directly linking advances in fundamental physics to the development of efficient and robust data analysis techniques.
    
    \item Synergies of ET with other GW observatories (section~\ref{section:div5}). In addition to observing CBC signals already in the band of current detectors for much longer durations during the ET era, we will likely detect sources like intermediate-mass black holes (IMBHs), completing the spectrum of CBC GW signals. Furthermore, with pulsar timing arrays (PTAs) already detecting potential signatures of a gravitational wave background from supermassive black hole binaries and the space-based LISA coming online in about a decade, we will be exploring the full GW spectrum from all sources. To maximise the scientific return, data analysis methods developed for other detectors (particularly LISA) could be highly relevant to ET, and vice versa. The synergy between different observatories will require the development of joint analysis frameworks to analyse the full spectrum of observations. This will enable us to gain deeper insights into the populations of sources, but it will also necessitate extensions of the data analysis tools  to handle the combined data from multiple detectors efficiently and robustly.
    
    \item Synergies with developments of waveforms (section~\ref{section:div8}). Most of CBC analyses is done in a modeled framework, making accurate CBC waveforms vital for inferring parameters. As seen  in \eq{eqn:lhood}, the waveform model is directly related to computing the likelihood which is at the heart of Bayesian data analysis. As mentioned  in section~\ref{sec:PE}, waveform acceleration techniques like constructing ROQs will directly affect accelerating likelihood computations. Adding more physical effects like tides, or higher-order-modes or eccentricity increases the parameter space and makes analyses more computationally demanding. In the ET era, it is likely that optimsation of CBC waveforms will be of paramount importance in addition to optimising data analysis methods. While GW detectors have mostly focused on CBC sources and most resources have been spent on developing and refining data analysis for well modeled sources, further development may be necessary for other sources of GWs. 
    
    \item Synergy with the development of tools for assessing the scientific potentials of detector configurations. In section~\ref{section:div9} we have  presented recent development of standardised tools for GW data analysis, including methods like Fisher matrices, which provide rapid estimates of parameter constraints and sensitivities. These tools are especially useful in the early stages of data analysis, where quick, approximate insights are needed to assess signal detectability or to guide more detailed analysis. While Fisher-based methods offer valuable projections, particularly with the projected sensitivities of the ET, they may not fully capture the complexity of real data, such as the effects of glitches, multimodalities, or correlations between parameters. For this reason, there is a crucial synergy with Bayesian inference techniques, which are more robust in handling the detailed structure of the data and addressing challenges that arise from specific waveform models and detector noise. Combining the strengths of both approaches—quick estimates from Fisher matrices and deeper insights from Bayesian analysis—will be key in developing comprehensive data analysis frameworks for ET.
    
\end{itemize}

\subsection{Conclusion}
\label{sec:conclusiondiv10}
Data analyses methods on CBCs, specifically sources targeted by current ground-based detectors, have proven to be effective for detecting and characterising sources in current detectors. However, to fully prepare for ET, we need to develop and test  analyses tailored towards non-CBC sources as well as sources not in the band of current ground-based detectors. For the next MDC we will be extending our dataset in various directions and develop analyses to tackle the following problems:
\begin{itemize}
\item variety of network configurations, the triangle but also 2 L-shaped detectors in Sardinia and the Netherlands, with different orientations (aligned and 45 degrees) and different sizes (10/15/20 km)~\cite{Branchesi:2023mws} (see also \cite{Puecher:2023twf,Bhagwat:2023jwv,Franciolini:2023opt,Iacovelli:2023nbv} for further follow-up studies). To study the effect of network configurations on the estimates of parameters, we will keep the simulated population the same for each specific network.

\item Include transient noise effects, or glitches. While there are ways of simulating blip glitches~\cite{Lopez:2022lkd}, they have been trained on real data from current detectors. Even if it is natural to assume similar shapes  also for the ET data, there could be new kinds of glitches whose effects remain to be seen. For the design of a triangular ET, noise in the 6 interferometers will further be correlated, an additional effect that we need to consider when simulating the next MDC. While mainly studied for the space-based detector LISA, \emph{data gaps} will happen during the long run-times of the interferometers of ET. This may potentially impact parameter estimation of long duration signals like BNSs if present during such a gap event.

\item Include other types of sources, e.g., eccentric binaries, population III binaries, primordial BBHs, IMBHs, bursts, continuous waves and cosmological backgrounds. The detection and characterisation of non-CBC sources as well as signals from non envisaged and possible exotic sources will likely require development of morphology-independent methods.

\item non isotropic model of sky distribution for source populations.

\end{itemize}
Finally, in addition to refinement of data analysis techniques, for example through machine learning, hardware development will be equally important for efficient and robust data analysis. It is likely that detection and parameter estimation of most CBC signals will provide reliable results with existing methods; however, as signal numbers and durations increase, the demand for computational resources will grow significantly, necessitating parallel advancements in hardware development to ensure efficient data processing at the scale required by third-generation detectors. Such hardware advancements include GPUs, Field-Programmable Gate Arrays (FPGAs), advancement and use of quantum computing, and finally need for efficient data-transfer as well as storage of massive volumes of data.

\subsection{Executive summary}

ET will significantly enhance our ability to detect a wide range of GW signals. Its sensitivity and frequency range will allow the observation of compact binary mergers at cosmological distances, stochastic backgrounds, and other exotic sources. In fact a GW signal detected at high redshifts could point to sources of primordial origin. However, the increasing detector bandwidths and the precision that instrumental advancements will introduce bring in substantial challenges for data analysis and parameter estimation of GW signals. Increased frequency bandwidth and sensitivity will lead to increased signal duration, e.g. $\sim$ hours for a low mass binary neutron star (BNS) signals. Additionally this will lead signals to \emph{overlap} in time and frequency. Furthermore, noise properties may change while the signal is being observed. All this leads us to believe that existing data-analysis methods for analysing GW signals will need to be much faster, or automated, and approximations for signal analysis will need to be modified.

\begin{highlightbox}{Key Challenges}
\begin{itemize}
\item Long-Duration Signals: ET's low-frequency sensitivity extends the duration of signals in its band. For example, binary neutron star signals may persist for hours, increasing computational demands for matched filtering and waveform modeling. Additionally, Earth's motion modulates the signal, complicating sky localisation and template bank design.

\item Overlapping Signals: The high detection rate in ET will lead to frequent overlapping signals in the data. Traditional single-signal models may introduce biases, particularly for closely timed or comparable signal amplitudes. New methods to separate and analyze overlapping signals simultaneously are crucial.

\item Noise Background Estimation: The dominance of GW signals complicates noise characterisation, as there will be minimal signal-free data segments. Traditional noise estimation techniques may overestimate the background. For a triangular configuration for ET, the signal-free null stream can be leveraged to produce correct background estimation.
\end{itemize}
\end{highlightbox}

\begin{highlightbox}{Innovative Methods and Tools}
    \begin{itemize}
        \item Machine Learning Applications: Machine learning (ML) methods are emerging as essential tools for ET data analysis. ML-based approaches promise to improve efficiency and scalability for ET's unique data analysis challenges. Applications include:
        \begin{itemize}
            \item Noise characterisation and glitch classification.
    \item Real-time detection of GW signals.
    \item Parameter estimation using simulation-based inference and normalizing flows.
        \end{itemize}
        \item Advanced Signal Detection Techniques: Traditional matched-filtering remains effective for compact binary coalescences (CBCs), but adaptations are needed for ET's longer signal durations and overlapping events. Techniques like relative binning and semi-coherent searches optimize computational resources while maintaining sensitivity.
        \item Parameter Estimation: Parameter estimation in the ET era must account for long signals, high signal-to-noise ratios, and overlapping events. Bayesian methods, accelerated using reduced-order quadratures and ML-based inference, are being developed to handle these challenges efficiently. Novel approaches like hierarchical subtraction and joint likelihood analysis are also being explored for overlapping signals.
        \begin{itemize}
            \item Null Stream Utilisation: For a triangular geometry it is possible to exploit the null stream, which cancels GW signals to provide noise-only data, providing a powerful tool for noise estimation and improving the reliability of detection and parameter inference. Additionally the null stream will let us uniquely characterise and eliminate noise artifacts like glitches to ensure robust parameter estimation.
        \item Correlated noise: In ET, correlated noise acts like a network of sensors, improving noise characterisation and potentially refining parameter estimation when incorporated into likelihood calculations. Depending on the correlation sign, it can either enhance signal loudness (positive correlations) or reduce SNR (negative correlations), affecting detectability. In ET’s triangular configuration, negative correlations may be more common, leading to potential biases if not properly modelled. Additionally, correlated magnetic noise and Newtonian Noise (NN) from seismic fluctuations will dominate ET’s low-frequency spectrum (20–40 Hz), significantly impacting stochastic background searches. The higher rate of correlated glitches also complicates data analysis. While offering benefits for noise modeling, correlated noise introduces low-frequency contamination and must be carefully accounted for to avoid biases.

        \end{itemize}
        \item Simulations and Mock Data Challenges: Mock data challenges (MDCs) play a critical role in testing and validating data analysis pipelines. They provide standardized datasets to benchmark algorithms, evaluate computational requirements, and predict ET's performance in real-world scenarios. Incorporating realistic noise and signal properties will be key to ensuring readiness for ET operations.
    \end{itemize}
\end{highlightbox}

The Einstein Telescope will open new frontiers in GW astronomy, but it demands a paradigm shift in data analysis methodologies. Addressing challenges such as long signal durations, overlapping events, and noise characterization will require innovative techniques, including machine learning and null stream utilization. Continued development and testing through simulations will be essential for maximizing ET's scientific potential. 
\section*{Contributions}

\addcontentsline{toc}{section}{Contributions}

This work has been developed in the context of the activities of the Observational Science Board (OSB) of the Einstein Telescope.

\vspace{2mm}
\noindent
{\em Overall project coordination:}
Marica Branchesi, Archisman Ghosh, Michele Maggiore

\vspace{2mm}
\noindent
Section~\ref{section:div1}.\\
{\em Coordinators:} Chris van den Broeck, Paolo Pani, Rafael Porto\\
{\em Key contributors:} Alessandro Agapito, Michalis Agathos, Conrado Albertus, David Barba-González, Gianfranco Bertone, Swetha Bhagwat, Richard Brito, Tomasz Bulik, Gianluca Calcagni, Roberto Casadio, Philippa Cole, Geoffrey Compère, Francesco Crescimbeni, Valerio De Luca, Daniela Doneva, Gabriele Franciolini, Stefano Foffa, Juan Garcia-Bellido, Boris Goncharov, Leonardo Gualtieri, Maria Haney, Justin Janquart, Gaetano Lambiase, Michele Lenzi, Dicong Liang, Felipe J. Llanes-Estrada, Elisa Maggio, Michele Maggiore, Andrea Maselli, Raissa F. P. Mendes, Pablo Navarro Moreno, Harsh Narola, Alex B. Nielsen, Costantino Pacilio, Cristiano Palomba, Angeles Perez-Garcia, Alessandro Pesci, Adam Pound, Miguel Quartin, Soumen Roy, Nicolas Sanchis-Gual, Lijing Shao, Riccardo Sturani, Giovanni Maria Tomaselli, Elias C. Vagenas, Massimo Vaglio, Stoytcho Yazadjiev

\vspace{2mm}
\noindent
Section~\ref{section:div2}.\\ 
{\em Coordinators:} Archisman Ghosh, Angelo Ricciardone, Mairi Sakellariadou\\
{\em Key contributors:} Raul Abramo, Pierre Auclair, Charles Badger, Nicola Bartolo, Freija Beirnaert, Enis Belgacem, Nicola Bellomo,  Dario Bettoni, Jose J. Blanco-Pillado, Simone Blasi, Nicola Borghi, Gianluca Calcagni, Sofia Canevarolo, Giulia Capurri, Carmelita Carbone, Gergely Dálya, Walter Del Pozzo, Ema Dimastrogiovanni, Hannah Duval, Matteo Fasiello, Stefano Foffa, Gabriele Franciolini, Jacopo Fumagalli,  Ish Gupta, Francesco Iacovelli, Justin Janquart, Kamiel Janssens, Alex C. Jenkins, Sumit Kumar, Sachiko Kuroyanagi, G. Lambiase, Michele Maggiore, Michele Mancarella, Alberto Mariotti, Simone Mastrogiovanni, Isabela Matos, Michele Moresco, Niccol\`o Muttoni, Paolo Pani, Alessandro Pedrotti, Gabriele Perna, Oriol Pujolas, Miguel Quartin, A\"aron Rase,  Alba Romero-Rodriguez, J. Rubio,  G{\'e}raldine Servant, Alexander Sevrin, Peera Simakachorn, Dani\`ele A.~Steer, Riccardo Sturani, Matteo Tagliazucchi, Cezary Turski, Lorenzo Valbusa Dall’Armi, Miguel Vanvlasselaer, Ville Vaskonen

\vspace{2mm}
\noindent
Section~\ref{section:div3}.\\
{\em Coordinators:} Giulia Cusin, Michela Mapelli, Antonio Riotto\\
{\em Key contributors:} John Antoniadis, Fabio Antonini, Manuel Arca Sedda, M. Celeste Artale, Matteo Bachetti, Christopher Berry, Emanuele Berti, Elisa Bortolas, Floor Broekgaarden, Giulia Capurri, Sylvain Chaty, Martyna Chruslinska, Carlo Contaldi,  Valerio De Luca, Irina Dvorkin, Gabriele Franciolini, Anastasios Fragos,  Gianluca Inguglia, Ralf Klessen, Astrid Lamberts, Boyuan Liu, Michele Maggiore, Pablo Marchant, Giorgio Mentasti, Mar Mezcua,  Gijs Nelemans, Paolo Pani, Carole Perigois, Mauro Pieroni,  Filippo Santoliquido, Fabian Schneider, Raffaella Schneider, Jishnu Suresh,  Thomas Tauris, Silvia Toonen, Lieke van Son

\vspace{2mm}
\noindent
Section~\ref{section:div4}.\\ {\em Coordinators:} Giancarlo Ghirlanda, Andrew Levan, Susanna Vergani\\
{\em Key contributors:} Igor Andreoni, Stefano Ascenzi, Maria Grazia Bernardini, Biswajit Banerjee, Sofia Bisero, Marica Branchesi, Enzo Brocato, Tommaso Chiarusi, Martyna Chruslinska, Alberto Colombo, Monica Colpi, Paolo D'Avanzo, Alessio Ludovico De~Santis, Ulyana Dupletsa, Rob Fender, Giuseppe Greco, Andreas Haungs, Nandini Hazra, Alexander van der Horst, Francesco Iacovelli, Giulia Illuminati, Astrid Lamberts, Eleonora Loffredo, Annarita Margiotta, Tista Mukherjee, Lara Nava, Gor Oganesyan, Barbara Patricelli, Céline Péroux, Mathieu Puech, Samuele Ronchini, A. Rossi, Pilar Ruiz-Lapuente, A. Sanna, Om Sharan Salafia, Filippo Santoliquido, Antonio Stamerra, Stephen Smartt, Elena Valenti. 

\vspace{2mm}
\noindent
Section~\ref{section:div5}.\\ {\em Coordinators:} Monica Colpi, Samaya Nissanke, Bangalore Sathyaprakash, Nicola Tamanini\\
{\em Key contributors:} 
Federico Angeloni, M. Celeste Artale, Manuel Arca Sedda, Marica Branchesi, Floor S Broekgaarden, Nazanin Davari, Stephen Fairhurst, Luca Graziani, Ish Gupta, Francesco Iacovelli, David Izquierdo Villalba, Danny Laghi, Michele Maggiore, Michele Mancarella, Alisha Marriott-Best, Sylvain Marsat, Simone Mastrogiovanni, Niccolò Muttoni, Gijs Nelemans,  Peter Pang, Constantino Pacilio, Delphine Perrodin, Arianna Renzini, Lijing Shao, Gianmassimo Tasinato, Rosa Valiante

\vspace{2mm}
\noindent
Section~\ref{section:div6}.\\ 
{\em Coordinators:} Tim Dietrich, Tanja Hinderer, Micaela Oertel\\ 
{\em Key contributors:} Conrado Albertus, Nils Andersson, David Barba-González, Dániel Barta, Andreas Bauswein, Micha{\l}
Bejger, Sebastiano Bernuzzi, Marica Branchesi, Fiorella Burgio, Pablo Cerdá-Durán, Prasanta Char, Sergio Cristallo, Daniela Doneva, Anthea F. Fantina, Tobias Fischer, Bruno Giacomazzo, Fabian Gittins, Stéphane Goriély,
Arus Harutyunyan, Brynmor Haskell, Ian Jones, Kostas Kokkotas, Hao-Jui Kuan, Felipe J. Llanes-Estrada, Eva Lope-Oter, Michele Maggiore, Gabriel Martínez-Pinedo, Philipp Mösta, Chiranjib Mondal, Martin Obergaulinger, Cristiano Palomba, Albino Perego, Maria Angeles Pérez-García, Geraint Pratten, Anna Puecher, Adriana R. Raduta, Violetta Sagun, Patricia Schmidt, Armen Sedrakian, Jan Steinhoff, Nikolaos Stergioulas, Duvier Suárez, Alejandro Torres-Forné, Stoytcho Yazadjiev 

\vspace{2mm}
\noindent
Section~\ref{section:div7}.\\ 
{\em Coordinators:} Ik Siong Heng, Marco Limongi, Cristiano Palomba\\
{\em Key contributors:} Conrado Albertus, Marco Antonelli, Marie Anne Bizouard, Alice Borghese, Maria Teresa Botticella, Marica Branchesi, Enrico Cappellaro, Pablo Cerdá-Durán, Massimo Della Valle, Alessio Ludovico De~Santis, Nancy Elias-Rosa, Ines Francesca Giudice, Brynmor Haskell, Paola Leaci, Michele Maggiore, Simone Mastrogiovanni, Andrea Melandri, Martin Obergaulinger, Francesca Onori, Giulia Pagliaroli, Barbara Patricelli, Alessandro Patruno, Angeles Pérez García, Ornella Juliana Piccinni, Silvia Piranomonte, Jade Powell, Nanda Rea, Dorota Rosinska, Mario Spera, Cristiano Ugolini, Garvin Yim 

\vspace{2mm}
\noindent
Section~\ref{section:div8}.\\ 
{\em Coordinators:} Laura Bernard, Harald P. Pfeiffer, Patricia Schmidt\\
{\em Key contributors:} Simone Albanesi, Angelica Albertini, Tomas Andrade, Dániel Barta,  Juan José Blanco Pillado, Alice Bonino, Matteo Breschi, Alessandra Buonanno, Adam Burrows, Geoffrey Compère, Tim Dietrich, Guillem Domenech, Daniela Doneva, Stefano Foffa, François Foucart, Gabriele Franciolini, Jonathan Gair, Rossella Gamba, Cecilio Garcia-Quiros, László Árpád Gergely, Alejandra Gonzalez, Mark Hannam, Tanja Hinderer, Sascha Husa, Balázs Kacskovics, Sachiko Kuroyanagi, Gaetano Lambiase, François Larrouturou, Georgios Lukes-Gerakopoulos, Sylvain Marsat, Alessandro Nagar, Hector O. Silva,  Lorenzo Pompili, Rafael Porto, Adam Pound, Geraint Pratten, Antoni Ramos-Buades, Piero Rettegno, Diego Rubiera-Garcia,  Stefano Schmidt, Lijing Shao, Thomas Sotiriou, Jan Steinhoff, Riccardo Sturani, Shaun Swain, Jacopo Tissino, Maarten van de Meent, Vijay Varma, Helvi Witek, Stoytcho Yazadjiev

\vspace{2mm}
\noindent
Section~\ref{section:div9}.\\ 
{\em Coordinators:} Micha{\l} Bejger, Michele Mancarella, Andrea Maselli\\
{\em Key contributors:} Ulyana Dupletsa, Francesco Iacovelli, Niccolò Muttoni, Andrea Begnoni, Enis Belgacem, Swetha Bhagwat, Ssohrab Borhanian, Marica Branchesi, Viola De Renzis, Stefano Foffa, Boris Goncharov, Jan Harms, Ik Siong Heng, Yufeng Li, Michele Maggiore, Mendon\c{c}a Soares de Souza, Giorgio Mentasti, Costantino Pacilio, Mauro Pieroni, Angelo Ricciardone, Riccardo Sturani

\vspace{2mm}
\noindent
Section~\ref{section:div10}.\\ 
{\em Coordinators:} Elena Cuoco, Gianluca Guidi, Tania Regimbau, Anuradha Samajdar\\
{\em Key contributors:} Francesco Cireddu, Tito Dal Canton,  Federico De Santi, Marco Drago, Maxime Fays, Jonathan Gair, Justin Janquart, Kamiel Janssens, Tjonnie Li, Cristiano Palomba, Lucia Papalini, Ed Porter, Massimiliano Razzano,  Filippo Santoliquido, John Veitch, Milan Wils, Chun Fung Wong

\section*{Acknowledgments}

\addcontentsline{toc}{section}{Acknowledgments}

This work is partially supported by the ET-PP project (Preparatory Phase for the Einstein Telescope Gravitational Wave Observatory), a Coordination and Support Action funded under the Horizon Europe research programme (Grant Agreement Number 101079696). 

\vspace{2mm}
\noindent
The research of  M.~Maggiore, F.~Iacovelli and N.~Muttoni  is supported by the Swiss National Science Foundation, grant 200020$\_$191957, and by the SwissMap National Center for Competence in Research. E.~Belgacem and M.~Maggiore are supported by the SNSF grant CRSII5$\_$213497. 
The research of F.~Iacovelli is further supported by Miller Fellowship at Johns Hopkins University.
M.~Branchesi, B.~Banerjee, and E.~Loffredo acknowledge financial support by the Italian Ministry of University and Research (MUR) under the PRIN grant METE with contract no. 2020KB33TP.
A.~Ghosh, F.~Beirnaert, and C.~Turski acknowledge support from Fonds Wetenschappelijk Onderzoek (FWO) International Research Infrastructure (IRI) grant ``Essential Technologies for the Einstein Telescope'' (I002123N). Their research is also supported by the Ghent University Special Research Funds (BOF) project BOF/STA/202009/040, the inter-university iBOF project BOF20/IBF/124, and the FWO research project G0A5E24N. 
%
%
R. A. Porto was supported by the ERC Consolidator Grant ``Precision Gravity: From the LHC
to LISA" provided by the European Research Council under the European Union's H2020 research and innovation program (grant agreement No. 817791). R. A. Porto also acknowledges the support of the Deutsche Forschungsgemeinschaft (DFG) under Germany’s Excellence Strategy (EXC 2121) ``Quantum Universe" (390833306).
%
N. Bellomo is supported by PRD/ARPE 2022 ``Cosmology with Gravitational waves and Large Scale Structure - CosmoGraLSS''. 
N.~Bartolo acknowledges financial support from the INFN InDark initiative and from the COSMOS network (www.cosmosnet.it) through the ASI (Italian Space Agency) Grants 2016-24-H.0, 2016-24-H.1-2018 and 2020-9-HH.0. N.~Bartolo's work is partially supported by ICSC - Centro Nazionale di Ricerca in High Performance Computing, Big Data and Quantum Computing, funded by European Union - NextGenerationEU.
G.~Calcagni is supported by grant PID2023-149018NB-C41 funded by the Spanish Ministry of Science, Innovation and Universities MCIN/AEI/10.13039/501100011033.
I. S. Matos is supported by FAPESP grants 2023/02330-0 and 2024/04044-7.
M.~Moresco acknowledges support from the grant PRIN-MUR 2022 2022NY2ZRS 001 “Optimizing the extraction of cosmological information from Large Scale Structure analysis in view of the next large spectroscopic surveys” supported by Next Generation EU and from the grant ASI n. 2024-10-HH.0 “Attività scientifiche per la missione Euclid – fase E”.

J. Rubio is supported by a Ramón y Cajal contract of the Spanish Ministry of Science and Innovation with Ref.~RYC2020-028870-I and the project PID2022-139841NB-I00 of MICIU/AEI/10.13039/501100011033 and FEDER, UE.
H.~Duval,  A.~Mariotti, A.~Rase,  A.~Sevrin and M.~Vanvlasselaer are supported by the Strategic Research Program High-Energy Physics of the Research Council of the Vrije Universiteit Brussel and by the iBOF “Unlocking the Dark Universe with Gravitational Wave Observations: from Quantum Optics to Quantum Gravity” of the VLIR.  A.~Rase and  A.~Sevrin are supported  by the FWO-Vlaanderen through grant number 1152923N and the project G006119N respectively.  
P.~Simakachorn is supported by Generalitat Valenciana Grants:  PROMETEO/2021/083 and CIPROM/2022/69.
P.~Auclair is a Postodoctoral Reseacher of the Fonds de la Recherche Scientifique – FNRS.
Jose J. Blanco-Pillado is supported by the PID2021-123703NB-C21 grant funded by MCIN/ AEI /10.13039/501100011033/ and by ERDF; ``A way of making Europe'', the Basque Government grant (IT-1628-22), and the Basque Foundation for Science (IKERBASQUE).
L. Valbusa Dall'Armi acknowledges financial support from the Supporting TAlent in Re- Search@University of Padova (STARS@UNIPD) for the project “Constraining Cosmology and Astrophysics with Gravitational Waves, Cosmic Microwave Background and Large-Scale Structure cross-correlations”.
S.~Blasi and G.~Servant acknowledge support from the Deutsche Forschungsgemeinschaft under Germany’s Excellence Strategy - EXC 2121 Quantum Universe - 390833306, and S.~Blasi is supported in part by FWO-Vlaanderen through grant number 12B2323N.
R.~Sturani acknowledges support from the FAPESP grants n. 2022/06350-2 and 2021/14335-0
M.~Fasiello acknowledges support from the “Consolidaci\'{o}n Investigadora” grant CNS2022-135590 and the “Ram\'{o}n y Cajal” grant RYC2021-033786-I. M.~Fasiello's work is partially supported by the Spanish Research Agency (Agencia Estatal de Investigaci\'{o}n) through the Grant IFT Centro de Excelencia Severo Ochoa No CEX2020-001007-S, funded by MCIN/AEI/10.13039/501100011033.
S.~Mastrogiovanni is supported by ERC Starting Grant No. 101163912–GravitySirens
The work of M.~Mancarella, A.~Pedrotti and V.~De Renzis received support from the French government under the France 2030 investment plan, as part of the Initiative d’Excellence d’Aix-Marseille Universit\'e – A*MIDEX AMX-22-CEI-02.
S.~Kuroyanagi is supported by the Spanish Attraccion de Talento contract no. 2019-T1/TIC-13177 granted by the Comunidad de Madrid, the I+D grant PID2020-118159GA-C42 funded by MCIN/AEI/10.13039/501100011033, the Consolidaci\'on Investigadora 2022 grant CNS2022-135211, and Japan Society for the Promotion of Science (JSPS) KAKENHI Grant no. 20H01899, 20H05853, and 23H00110.
D.~Bettoni is supported by project  PID2021-
122938NB-I00 funded by the Spanish “Ministerio de Ciencia e Innovación" and FEDER “A way of making Europe” and by project PID2022-139841NB-I00
funded by the Spanish “Ministerio de Ciencia e Innovación”
The work of V. Vaskonen was supported by the European Union's Horizon Europe research and innovation program under the Marie Sk\l{}odowska-Curie grant agreement No. 101065736, and by the Estonian Research Council grants PRG803, RVTT3 and RVTT7 and the Center of Excellence program TK202.
O. Pujolas supported by Grup de Recerca ‘Grup de Fisica Teorica UAB/IFAE’ code: 2021 SGR 00649 funded by the Departament de Recerca i Universitats from Generalitat de Catalunya, projects PID2020-115845GB-I00 funded by MICIU/AEI/10.13039/501100011033 and PID2023-146686NB-C31 funded by 
MICIU/AEI/10.13039/501100011033/ and by FEDER, UE. IFAE is partially funded by the CERCA program of the Generalitat de Catalunya. 
A.~C.~Jenkins was supported by the Gavin Boyle Fellowship at the Kavli Institute for Cosmology, Cambridge, and by the Science and Technology Facilities Council through the UKRI Quantum Technologies for Fundamental Physics Programme (Grant No. ST/T005904/1).
%
A.~Riotto  acknowledges support from the  Swiss National Science Foundation (project number CRSII5\_213497) and from   the Boninchi Foundation for the project ``Pblack holes in the Era of GW Astronomy''.
J.~Antoniadis acknowledges support from the European Commission under project ARGOS-CDS (Grant Agreement number: 101094354).
S.~Toonen acknowledges support from the Netherlands Research
Council NWO (VIDI 203.061 grant). 
This work has received funding from the European Research Council (ERC) under the European Union’s Horizon 2020 research and innovation programme, Grant agreements No.\ 770017 (acronym: DEMOBLACK, PI: M. Mapelli), 945806 (PI: F. Schneider).
F. Schneider, B. Liu, and M. Mapelli acknowledge support from the Deutsche Forschungsgemeinschaft (DFG, German Research Foundation) under Germany’s Excellence Strategy EXC 2181/1-390900948 (the Heidelberg STRUCTURES Excellence Cluster).
R. Klessen acknowledges financial support from the ERC via Synergy Grant ``ECOGAL'' (project ID 855130),  from the German Excellence Strategy via the Heidelberg Cluster ``STRUCTURES'' (EXC 2181 - 390900948), and from the German Ministry for Economic Affairs and Climate Action in project ``MAINN'' (funding ID 50OO2206).  R. Klessen also thanks the 2024/25 Class of Radcliffe Fellows at Harvard for highly interesting and stimulating discussions. 
M.~C.~Artale acknowledges support from FONDECYT Iniciaci\'on 11240540. 
V.~De Luca is supported by funds provided by the Center for Particle Cosmology at the University of Pennsylvania.
M.~Mezcua acknowledges support from the Spanish Ministry of Science and Innovation through the project PID2021-124243NB-C22. This work was partially supported by the program Unidad de Excelencia Mar\'ia de Maeztu CEX2020-001058-M.
M.~Arca Sedda acknowledges funding from the European Union’s Horizon 2020 research and innovation programme under the Marie Skłodowska-Curie grant agreement No. 101025436 (project GRACE-BH), and the MERAC Foundation for the project "The Missing Link".
E.~Berti is supported by NSF Grants No. AST-2307146, PHY-2207502, PHY-090003 and PHY-20043, by NASA Grant No. 21-ATP21-0010, by the John Templeton Foundation Grant 62840, by the Simons Foundation, and by the Italian Ministry of Foreign Affairs and International Cooperation grant No.~PGR01167.
G.~Cusin is supported by the SNSF grant "Gravitational wave propagation in the clustered Universe" and by CNRS. 
T.~Fragos acknowledges support from the Swiss National Science Foundation (project number CRSII5$\_$213497).
P.~Marchant acknowledges support from the FWO senior fellowship number 12ZY523N.
A.~Lamberts is supported by the ANR COSMERGE project, grant ANR-20-CE31-001 of the French Agence Nationale de la Recherche.
F.~Santoliquido acknowledges financial support from the AHEAD2020 project (grant agreement n. 871158).
F.~Antonini is supported by the UK’s Science and Technology Facilities Council grant ST/V005618/1
B.~Liu gratefully acknowledges the support of the Royal Society University Research Fellowship.
E.~Bortolas acknowledges support from the European Union’s Horizon Europe programme under the Marie Sklodowska Curie grant agreement No 101105915 (TESIFA), the European Consortium for Astroparticle Theory in the form of an Exchange Travel Grant, and the European Union's Horizon 2020 Programme under the AHEAD2020 project (grant agreement n. 871158).
%
%
G.~Ghirlanda acknowledges support from the
MUR, PRIN 2017, grant No. 20179ZF5KS. G.~Ghirlanda and O.S.~Salafia acknowledge the project GRB PrOmpt Emission Modular Simulator (POEMS) financed by INAF Grant 1.05.23.06.04 and the funding by the European Union-Next Generation EU, PRIN 2022 RFF M4C21.1 (202298J7KT – PEACE).
The work of M.~Mancarella received support from the French government under the France 2030 investment plan, as part of the Initiative d’Excellence d’Aix-Marseille Université – A*MIDEX AMX-22-CEI-02.
The work of N.~Tamanini received support by the CNRS through the AMORCE funding framework and by the Agence Nationale de la Recherche (ANR) through the project ANR-24-MRS1-0009-01.
S.~Mastrogiovanni is supported by ERC Starting Grant No. 101163912–GravitySirens.
D.~Perrodin acknowledges support from the Italian INAF Large Grant 2023 "Gravitational Wave Detection using Pulsar Timing Arrays" (P.I.~Delphine Perrodin).
We thank Takami Kuroda for providing data shown in Fig.~\ref{fig:GWstrain}.
Work partially supported by the EU under grants 
HORIZON MSCA-2022-PF-01-01, project ProMatExNS (M.A. P\'erez-Garc\'{\i}a, P.Char and others),
824093 STRONG2020 (F.J. Llanes-Estrada and others), ERC SMArt, 101076369, 
NextGenerationEU, ERC Grant HEAVYMETAL No. 101071865 (A.Bauswein),  ERC Consolidator Grant InspiReM-101043372 (S.Bernuzzi); Horizon Project 101131928 -ACME- (N.S.); ERC grant  DynTideEOS, 101151301 (F.Gittins); ERC Advanced Grant KILONOVA No.~885281 (G.Martinez-Pinedo); COST Actions CA16104 and CA16214 (P.Cerda-Duran); Next Generation EU, Mission 4 Component 2 - CUP E53D23002090006, PRIN 2022 Prot. No. 2022KX2Z3B (A.P.); NextGenerationEU RFF M4C2 1.1 PRIN 2022 project "2022RJLWHN URKA" (S.\ Cristallo); PRIN project no. 202275HT58 (M.~Bejger); H2020-MSCA-RISE-2017 -FunFiCO-777740 and HORIZON-MSCA-2021-SE-01 -NewFunFiCO-101086251 (A.Torres-Forné);
German Research Foundation (DFG) Emmy Noether Research Group DO 1771/1-1 (D. Doneva), Project MEMI number BE 6301/2-1 (S. Bernuzzi), project SE 1836/5-3 (A.Sedrakian), Project 279384907 -- SFB 1245 (subproject B07) (A. Bauswein), Project-ID 279384907 - SFB 1245, and MA 4248/3-1 (G.Martinez-Pinedo), Volkswagen Foundation (Hannover, Germany) grant No. 96 839 (A.Haratyunian); Spanish MICINN grants PRX23/00225 (estancias en el extranjero) and PID2022-137003NB-I00 from \\
MCIN/AEI/10.13039/501100011033/ (F.J.Llanes-Estrada), PID2022-137887NB-10 and\\
RED2022-134411-T (M.A. P\'erez-Garc\'{\i}a), Grants No. PGC2018-095984-B-I00 and PID2021-125485NB-C21  (P.Cerda-Duran), Grant PID2021-125485NB-C21 (A.Torres-Forné), Grant PID2021-127495NB-I00 funded by MCIN/AEI/10.13039/501100011033 and by the European Union “NextGenerationEU" and “ESF Investing in your future”, Ramon y Cajal programme (2018-024938-I), Generalitat Valenciana grant PROMETEO/2019/071 (P.Cerda-Duran), PROMETEO grant CIPROM/2022/13 (M.Obergaulinger), Grant CIPROM/2022/49 (A.Torres-Forné), Astrophysics and High Energy Physics programme of the Generalitat Valenciana ASFAE/2022/026 funded by MCIN and the European Union NextGenerationEU (PRTR-C17.I1),  Consejer'ia de Educaci\'on de la Junta de Castilla y Le\'on, European Social Fund, Junta de Castilla y Le\'on SA101P24 and SA091P24; 
French Agence Nationale de la Recherche (ANR) under contract ANR-22-CE31-0001-01 (A.F.Fantina and M.Oertel); 
 National Recovery and Resilience Plan of the Republic of Bulgaria, project No. BG-RRP-2.004-0008-C01 (S. Yazadjiev); FEDER;
Dutch Research Council (NWO) ENW-XL grant (T. Hinderer); 
STFC grant  ST/Y004248/1 FCT (N. Andersson) and ST/V005677/1 (P. Schmidt);
Royal Society: University Research Fellowship URF R1 221500 and RF ERE 221015 (G. Pratten) STFC grant, Research Grant RG R1 241327 (P.Schmidt); 
UKRI-funded ”The next-generation gravitational-wave observatory network” project (Grant No. ST/Y004248/1)  (V. Sagun);
Istituto Nazionale di Fisica Nucleare (INFN) (C.~Palomba);
 INAF 2023 Theory Grant "Understanding R-process and Kilonovae Aspects" (S.~Cristallo);
CNCS/CCCDI–UEFISCDI and PN-IV-P1-PCE-2023-0324 (A. Raduta); 
 Hungarian National Research, Development and Innovation Office (NKFIH) OTKA Grant Agreement No. K138277 (D. Barta);
 Belgian Fonds de la Recherche Scientifique (FNRS)  Grant No IISN 4.4502.19 and EOS Project No O022818F and O000422 co-funded by the Research Foundation Flanders (FWO, Belgium) (C.Mondal);
 Polish National Science Centre (NCN): grant 2020/37/B/ST9/01937 (A.Sedrakian),  grant 2023/49/B/ST9/03941 (T.Fischer), grant 2019/33/B/ST9/00942 (B.Haskell), grant \\ 2021/43/B/ST9/01714 (M.~Bejger);
  The supernova simulations related to the QCD phase transition
  outlined in section~\ref{subsec:GWCCSN} 
  were performed in part at the Wroclaw Centre for Scientific Computing and Networking (WCSS) under `grant328'.
P.~Char has received funding from the European Union’s Horizon 2020 research and innovation programme under the Marie Skłodowska-Curie grant agreement No. 101034371.
%
P.~Cerdá-Durán acknowledges support from the Spanish Agencia Estatal de Investigaci\'on (Grants No. PGC2018-095984-B-I00 and PID2021-125485NB-C21) funded by MCIN/AEI/10.13039/501100011033 and ERDF A way of making Europe, by the Generalitat Valenciana (PROMETEO/2019/071), and by COST Actions CA16104 and CA16214.
C.~Palomba acknowledges support from Istituto Nazionale di Fisica Nucleare (INFN).
M.~Limongi aknowledges support from Kavli IPMU (WPI), UTIAS, The University of Tokyo, Kashiwa, Chiba 277-8583, Japan
N.~E.-R. acknowledges support from the PRIN-INAF 2022, `Shedding light on the nature of gap transients: from the observations to the models'.
M.T.~Botticella acknowledges support from the PRIN-INAF 2022, `Shedding light on the nature of gap transients: from the observations to the models'.
I.F.~Giudice is supported by PNRR Missione 4 - Componente 2 - Investimento 3.1 ETIC (Einstein Telescope Infrastructure Consortium) - IR0000004
N.~Rea is supported by the European Research Council (ERC) via the Consolidator Grant “MAGNESIA” (No. 817661) and the Proof of Concept "DeepSpacePulse" (No. 101189496), by the Catalan grant SGR2021-01269, and by the program Unidad de Excelencia Maria de Maeztu CEX2020-001058-M.
D.~Rosinska is supported by ET-PP HORIZON-INFRA-2021-DEV-02 project (No. 101079696) 
 and by the Ministry of Science and Higher Education project ( W55/HE/2022) 
F.~Onori acknowledges support from MIUR, PRIN 2020 (grant 2020KB33TP) ``Multimessenger astronomy in the Einstein Telescope Era (METE)'' and from INAF-MINIGRANT (2023): "SeaTiDE - Searching for Tidal Disruption Events with ZTF: the Tidal Disruption Event population in the era of wide field surveys".
C.~Albertus and A. Pérez García acknowledge  support by Junta de Castilla y León SA091P24, SA101P24 and Spanish MICIN grant RED2022-134411-T and PID2022-137887NB-I00. 
B.~Haskell acknowledges support by the National Science Centre Poland (NCN) via OPUS-LAP grant 2022/47/I/ST9/01494
M.~Obergaulinger acknowledges support by Grant PID2021-127495NB-I00 funded by\\ MCIN/AEI/10.13039/501100011033 and by the European Union “NextGenerationEU" as well as “ESF Investing in your future”, the Astrophysics and High Energy Physics programme of the Generalitat Valenciana ASFAE/2022/026 funded by MCIN and the European Union NextGenerationEU (PRTR-C17.I1) the Prometeo excellence programme grant CIPROM/2022/13 funded by the Generalitat Valenciana the Ramon y Cajal programme (2018-024938-I) funded by MCIN/AEI.
O.J.~Piccinni is supported by the Australian Research Council Centre of Excellence for Gravitational Wave Discovery (OzGrav), Project Number CE230100016.
M.A.~Bizouard acknowledges support from the axe ondes gravitationnelles of the Observatoire de la Côte d’Azur.
A.~Melandri, S.~Piranomonte, E.~Cappellaro acknowledge support from MIUR, PRIN 2020 (grant 2020KB33TP) ``Multimessenger astronomy in the Einstein Telescope Era (METE)''
Patricia~Schmidt acknowledges support from Science and Technology Facilities Council (STFC) grants ST/V005677/1 and ST/Y00423X/1, and Royal Society Research Grant RG{\textbackslash}R1{\textbackslash}241327. 
Laura~Bernard acknowledges financial support from the ANR PRoGRAM project, grant ANR-21-CE31-0003-001 and the EU Horizon 2020 Research and Innovation Programme under the Marie Sklodowska-Curie Grant Agreement no.101007855.
This research was supported in part by Perimeter Institute for Theoretical Physics. Research at Perimeter Institute is supported in part by the Government of Canada through the Department of Innovation, Science and Economic Development and by the Province of Ontario through the Ministry of Colleges and Universities.
Guillem Domenech acknowledges support from the DFG under the Emmy-Noether program, project No. 496592360, and from the JSPS KAKENHI grant No. JP24K00624.
Vijay~Varma acknowledges support from NSF Grant No. PHY-2309301 and UMass Dartmouth’s Marine and Undersea Technology (MUST) Research Program funded by the Office of Naval Research (ONR) under Grant No. N00014-23-1–2141.
Geraint~Pratten gratefully acknowledges support from a Royal Society University Research Fellowship URF{\textbackslash}R1{\textbackslash}221500 and RF{\textbackslash}ERE{\textbackslash}221015, and STFC grant ST/Y00423X/1.
Shaun~Swain gratefully acknowledges support from a Royal Society University Research Fellowship URF{\textbackslash}R1{\textbackslash}221500 and RF{\textbackslash}ERE{\textbackslash}221015.
Alice~Bonino is supported by STFC, the School of Physics and Astronomy at the University of Birmingham and the Birmingham Institute for Gravitational Wave Astronomy.
Sascha~Husa acknowledges support from the Spanish Agencia Estatal de Investigación grant PID2022-138626NB-I00 funded by MICIU/AEI/10.13039/501100011033 and the ERDF/EU; and the Comunitat Autònoma de les Illes Balears with funds from the European Union - NextGenerationEU/PRTR-C17.I1 (SINCO2022/6719) and from the European Union - European Regional Development Fund (ERDF) (SINCO2022/18146).
Riccardo Sturani acknowledges support from the FAPESP grants n. 2022/06350-2 and 2021/14335-0
Adam Pound acknowledges the support of a Royal Society University Research Fellowship and the ERC Consolidator/UKRI Frontier Research Grant GWModels (selected by the ERC and funded by UKRI [grant number EP/Y008251/1]).
Helvi Witek acknowledges support provided by the National Science Foundation under NSF Award No. OAC-2004879, No.~OAC-2411068 and No.~PHY-2409726.
Stoytcho Yazadjiev is supported by the  European Union-NextGenerationEU, through the National Recovery and Resilience Plan of the Republic of Bulgaria, project No. BG-RRP-2.004-0008C01.
Lijing Shao is supported by the Beijing Natural Science Foundation (1242018), and  the Max Planck Partner Group Program funded by the Max Planck Society.
Daniela Doneva acknowledges financial support via an Emmy Noether Research Group funded by the German Research Foundation (DFG) under Grant No. DO 1771/1-1.
Tim Dietrich acknowledges support from the German Research Foundation (DFG) under project number DI 2553/7, and from the EU Horizon under ERC Starting Grant, no.\ SMArt-101076369. 
Mark Hannam was supported by Science and Technology Facilities Council (STFC) grants ST/V005618/1 and ST/Y004272/1.
Alejandra Gonzalez acknowledges support by the Deutsche Forschungsgemeinschaft (DFG) under Grant No. 406116891 within the Research Training Group RTG 2522/1, and the Comunitat Autònoma de les Illes Balears through the Conselleria d'Educació i Universitats with funds from the European Union - NextGenerationEU/PRTR-C17.I1 (SINCO2022/6719).
Dániel Barta and Balázs Kacskovics acknowledge support from the National Research, Development and Innovation Office (Hungarian abbreviation: NKFIH) under OTKA Grant Agreement No. K138277.
Hector~O.~Silva acknowledges funding from the Deutsche Forschungsgemeinschaft (DFG) - Project~No.:~386119226.
Maarten van de Meent acknowledges financial support by the VILLUM Foundation (grant no. VIL37766) and the DNRF Chair program (grant no. DNRF162) by the Danish National Research Foundation.
Simone~Albanesi acknowledges support from the Deutsche Forschungsgemeinschaft (DFG) project ``GROOVHY'' (BE 6301/5-1 Projektnummer: 523180871).
Cecilio García-Quirós is supported by the Swiss National Science Foundation (SNSF) Ambizione grant PZ00P2\_223711.
Antoni Ramos-Buades is supported by the Veni research programme which is (partly) financed by the Dutch Research Council (NWO) under the grant VI.Veni.222.396; acknowledges support from the Spanish Agencia Estatal de Investigación grant PID2022-138626NB-I00 funded by MICIU/AEI/10.13039/501100011033 and the ERDF/EU; 
is supported by the Spanish Ministerio de Ciencia, Innovación y Universidades (Beatriz Galindo, BG23/00056) and co-financed by UIB.
Rossella Gamba acknowledges support from the NSF Grant PHY-2020275 (Network for Neutrinos, Nuclear Astrophysics, and Symmetries (N3AS)).
%
%
A. Maselli acknowledges financial support from MUR PRIN Grant No. 2022- Z9X4XS, funded by the European Union - Next Generation EU.
S.~Borhanian acknowledges support from the Deutsche Forschungsgemeinschaft (DFG)--project MEMI number BE6301/2-1, ERC Starting Grant No.~945155--GWmining, Cariplo Foundation Grant No.~2021-0555, MUR PRIN Grant No.~2022-Z9X4XS, MUR Grant ``Progetto Dipartimenti di Eccellenza 2023-2027'' (BiCoQ), and the ICSC National Research Centre funded by NextGenerationEU.
G.~Mentasti acknowledges support from the Imperial College London Schrodinger Scholarship scheme.
A.~Begnoni acknowledges support from  ICSC – Centro Nazionale di Ricerca in High Performance Computing, Big Data and Quantum Computing, funded by European Union –
NextGenerationEU.
C.~Pacilio is supported by ERC Starting Grant No.~945155--GWmining, Cariplo Foundation Grant No.~2021-0555, MUR PRIN Grant No.~2022-Z9X4XS, MUR Grant ``Progetto Dipartimenti di Eccellenza 2023-2027'' (BiCoQ), and the ICSC National Research Centre funded by NextGenerationEU.
S.~Bhagwat would like to acknowledge the UKRI Stephen Hawking Fellowship funded by the Engineering and Physical Sciences Research Council (EPSRC) with grant reference number EP/W005727 for support during this work.
A.~Ricciardone, acknowledges financial support the project BIGA - ``Boosting Inference for Gravitational-wave Astrophysics" funded by the MUR Progetti di Ricerca di Rilevante Interesse Nazionale (PRIN) Bando 2022 - grant 20228TLHPE - CUP I53D23000630006
M.~Pieroni acknowledges the hospitality of Imperial College London, which
provided office space during some parts of this project.
The work of J.M.S.~de Souza is supported  by the Brazilian research agency FAPERJ.
%
J. Veitch and Q. Hu were supported by the STFC grant ST/Y004256/1. L. Papalini acknowledges the support of the PhD scholarship in Physics “High Performance Computing and Innovative Data Analysis Methods in Science” (Cycle XXXVIII, Ministerial Decree no. 351/2022) and received funding from the European Union Next-Generation EU—National Recovery and Resilience Plan (NRRP)—MISSION 4 COMPONENT 1, INVESTMENT N.4.1—CUP N.I51J22000630007. M.Wils. is supported by the Research Foundation - Flanders (FWO) through Grant No. 11POK24N. T.G.F. Li, F. Cireddu and I.C.F. Wong are partially supported by the Research Foundation - Flanders (FWO) through Grant No. I002123N.
%
Paolo~Pani acknowledges the MUR PRIN Grant 2020KR4KN2 ``String Theory as a bridge between Gauge Theories and Quantum Gravity'', the FARE programme (GW-NEXT, CUP:~B84I20000100001), and the INFN TEONGRAV initiative. Some numerical simulations have been performed at the Vera cluster supported by the Italian Ministry for Research and by Sapienza University of Rome.
Angeles Perez-Garcia, acknowledges support from Spanish Ministry of Science PID2022-137887NB-10 and RED2022-134411-T projects, Junta de Castilla y León  projects SA101P24, SA091P24.
Harsh~Narola acknowledges the support of the research programme of the Netherlands Organisation for Scientific Research~(NWO). 
Gianluca~Calcagni is supported by grant PID2023-149018NB-C41 funded by the Spanish Ministry of Science, Innovation and Universities MCIN/AEI/10.13039/501100011033.
Geoffrey Comp\`ere is Research Director of the F.R.S.-FNRS. 
Francesco Crescimbeni acknowledges the financial support provided under the “Progetti per Avvio alla Ricerca Tipo 1”, protocol number AR12419073C0A82B.
Nicolas Sanchis-Gual acknowledges support from the Spanish Ministry of Science and Innovation via the Ram\'on y Cajal programme (grants RYC2022-037424-I), funded by \\
MCIN/AEI/10.13039/501100011033 and by ``ESF Investing in your future”, from the Spanish Agencia Estatal de Investigaci\'on (grant PID2021-125485NB-C21) funded by \\ MCIN/AEI/10.13039/501100011033 and ERDF A way of making Europe, and from the European Horizon Europe staff exchange (SE) programme HORIZON-MSCA2021-SE-01 Grant No. NewFunFiCO-101086251.
Valerio~De Luca is supported by funds provided by the Center for Particle Cosmology at the University of Pennsylvania.
Elisa~Maggio is supported by the European Union’s Horizon Europe research and innovation programme under the Marie Sklodowska-Curie grant agreement No. 101107586. E.M. acknowledges funding from the Deutsche Forschungsgemeinschaft (DFG) - project number: 386119226.
Giovanni Maria Tomaselli acknowledges funding provided by the Sivian Fund at the Institute for 
Advanced Study.
Leonardo Gualtieri acknowledges financial support from the EU Horizon 2020 Research and Innovation Programme under the Marie Sklodowska-Curie Grant Agreement No. 101007855.
Stoytcho Yazadjiev 	is supported by the  European Union-NextGenerationEU, through the National Recovery and Resilience Plan of the Republic of Bulgaria, project No. BG-RRP-2.004-0008C01.
Daniela Doneva acknowledges financial support via an Emmy Noether Research Group funded by the German Research Foundation (DFG) under Grant No. DO 1771/1-1.
Juan~Garcia-Bellido acknowledges financial support from the Research Project PID2021-123012NB-C43 [MICINN-FEDER], and the Centro de Excelencia Severo Ochoa Program CEX2020-001007-S at IFT.
Lijing Shao is supported by the Beijing Natural Science Foundation (1242018), the National SKA Program of China (2020SKA0120300), and the Max Planck Partner Group Program funded by the Max Planck Society.
Dicong Liang acknowledges financial support from the National Natural Science Foundation of China (12405065) 
Raissa Mendes acknowledges financial support from the National Council for Scientific and Technological Development (CNPq), and by the Carlos Chagas Filho Research Support Foundation (FAPERJ).
Michele Lenzi is supported by Contracts No. PID2019–106515 GB-I00 and No. PID2022-137674NB-I00 from MCIN/AEI/10.13039/501100011033 (Spanish Ministry of Science and Innovation) and 2017-SGR-1469 and 2021-SGR-01529 (AGAUR, Generalitat de Catalunya). Michele Lenzi is also supported by Juan de la Cierva Contract No. FJC2021-047289-I funded by program MCIN/AEI/10.13039/501100011033 (Spanish Ministry of Science and Innovation) and by NextGenerationEU/PRTR (European Union) and also acknowledge partial support by the program Unidad de Excelencia María de Maeztu CEX2020-001058-M (Spanish Ministry of Science and Innovation).
%
%
Richard Brito acknowledges financial support provided by FCT – Fundação para a Ciência e a Tecnologia, I.P., under the Scientific Employment Stimulus -- Individual Call -- Grant No. 
{2020.00470.CEECIND}. This work is supported by national funds through FCT, under the Project No.  
{2022.01324.PTDC} and under FCT's ERC-Portugal program through the Project ``GravNewFields''.
Michalis Agathos acknowledges support from funds provided by the Kavli Foundation.
Riccardo Sturani acknowledges support from the FAPESP grants n. 2022/06350-2 and 2021/14335-0
Maxime Fays acknowledges the support of the Fonds de la Recherche Scientifique-FNRS, Belgium, under grant No. 4.4501.19.
Philippa Cole is supported by ERC Starting Grant No. 945155–GWmining, Cariplo Foundation Grant No. 2021-0555, MUR PRIN Grant No. 2022-Z9X4XS, MUR Grant “Progetto Dipartimenti di Eccellenza 2023-2027” (BiCoQ), and the ICSC National Research Centre funded by NextGenerationEU.
Tomasz~Bulik was supported by the European Union’s Horizon Europe research and innovation programme under the grant agreement No. 101079696 (HORIZON-INFRA-2021-DEV-02 (ET-PP)) and from the program of the Polish Ministry of Science and Higher Education "Support for the participation of Polish research teams in international research infrastructure projects" under the grant agreement No. W55/HE/2022.
Massimo Vaglio acknowledges support from the PRIN 2022 grant “GUVIRP - Gravity tests in the UltraViolet and InfraRed with Pulsar timing”.
Tanja Hinderer, Maria Haney, Harsh Narola, Peter Pang, Antoni Ramos-Buades, Anuradha Samajdar, Stefano Schmidt, and Chris Van Den Broeck 
are supported by the
research programme of the Netherlands Organisation for
Scientific Research (NWO).
Thomas Sotiriou acknowledges partial support from the STFC Consolidated Grant nos. ST/V005596/1 and ST/X000672/1.
Diego Rubiera-Garcia is supported by the Spanish National Grant PID2022-138607NBI00, funded by MCIN/AEI/10.13039/501100011033 (Spain).
F.~Foucart gratefully acknowledges support from the Department of Energy, Office of Science, Office of Nuclear Physics, under contract number DE-AC02-05CH11231 and from NASA through grant 80NSSC22K0719.
S.~Chaty acknowledges the CNES (Centre National d’Etudes Spatiales) for the continuous funding of multi-wavelength and multi-messenger activities through INTEGRAL/MINE project.
B.~Sathyaprakash acknowledges National Science Foundation (NSF)  awards AST-2307147, PHY-2207638, PHY-2308886 and PHY-2309064 which support I.G. and  B.S.S.
The work of M. Lenzi  and C.F. Sopuerta has been supported  by contracts: PID2019-106515GB-I00/AEI/10.13039/501100011033
(Spanish Ministry of Science and Innovation) and 2017-SGR-1469 (AGAUR, Generalitat de
Catalunya), and partially by the program Unidad de Excelencia María de Maeztu CEX2020-001058-M
(Spanish Ministry of Science and Innovation).
Jacopo Fumagalli is supported by the grant PID2022-136224NB-C22, funded by MCIN/AEI/10.13039
/501100011033/FEDER, UE, and by the grant/ 2021-SGR00872.

\section*{List of Acronyms}

\addcontentsline{toc}{section}{List of Acronyms}

\renewcommand{\arraystretch}{1.2} 

\begin{longtable}{l l}
    \hline
    \textbf{Acronym} & \textbf{Meaning} \\ 
    \hline
    \endfirsthead

    \hline
    \textbf{Acronym} & \textbf{Meaning} \\ 
    \hline
    \endhead

    \hline
    \endfoot

    \hline
    \endlastfoot

        11 HUGS & 11 Mpc H$\alpha$ UV Galaxy Survey \\
        2G & Second Generation \\
        3G & Third Generation \\
        AGB & Asymptotic Giant Branch \\
        AGN & Active Galactic Nucleus \\ 
        AGWB & Astrophysical Gravitational Wave Background \\
        AIC & Accretion-Induced Collapse \\
        ALIGO & Advanced LIGO \\
        ALMA & Atacama Large Millimeter$/$submillimeter Array\\ 
        ALPs & Axion-Like Particles \\
        ASASSN & All Sky Automated Survey for SuperNovae \\
        ASD & Amplitude spectral density \\
        ATLAS & Asteroid Terrestrial Impact Last Alert System \\
        AVirgo & Advanced Virgo \\
        BAO & Baryon Acoustic Oscillation\\
        BBH & Binary Black Hole \\
        BBN & Big Bang Nucleosynthesis \\
        BGS & Bright Galaxy Survey \\
        BH & Black Hole \\
        BHNS & Black Hole -- Neutron Star \\
        BMS & Bondi-Metzner-Sachs\\
        BNS & Binary Neutron Star \\
        BPL & Broken Power Law \\
        BSG & Blue Super Giant \\
        BSM & Beyond Standard Model\\
        CBC & Compact Binary Coalescence \\
        CCSN & Core Collapse Supernova \\
        CCSNR & Core Collapse Supernovae Rate \\
        CE & Cosmic Explorer\\
        CE & Common Envelope {\em (when used in the context of stellar evolution)}\\
        CFS & Chandrasekhar-Friedman-Schutz \\
        CGWB & Cosmological Gravitational Wave Background \\
        CHE & Chemically Homogeneous Evolution \\
        CMB & Cosmic Microwave Background \\
        CNN & Convolutional Neural Network \\
        CO-WD & Carbon-Oxygen White Dwarf \\
        CR & Critical Ratio \\
        CSD & Cross-power Spectral Density \\
        CSM & Circum Stellar Material \\
        CW & Continuous Wave \\
        DC & Direct Collapse \\
        dCS & dynamical Chern-Simons  \\
        DECIGO &Deci-hertz Interferometer Gravitational wave Observatory \\
        DESI & Dark Energy Spectroscopic Instrument \\
        DM & Dark Matter \\
        DNN & Deep Neural Network\\
        DNS & Double Neutron Star \\
        DTD & Delay-Time Distribution \\
        DW & Domain Wall \\
        EC & Electron Capture \\
        ECO & Exotic Compact Object\\
        ECSN & Electron Capture Supernova \\
        EFT & Effective Field Theory \\
        EM & Electromagnetic \\
        EMRI & Extreme Mass-Ratio Inspiral \\
        EOB & Effective One-Body \\
        EOS & Equation of State \\
        EPTA & European Pulsar Timing Array\\
        ESA & European Space Agency \\
        EsGB & Einstein-scalar-Gauss-Bonnet \\
        ET & Einstein Telescope \\
        FAR & False Alarm Rate \\
        FEW & Fast EMRI Waveforms \\
        FIM & Fisher Information Matrix\\
        FIR & Far Infrared \\
        FLRW & Friedmann-Lemaitre-Robertson-Walker \\
        FOM & Figure of Merit \\
        FOPT & First-Order Phase Transition \\
        FoV & Field of View \\
        FRIB & Facility for Rare Isotope Beams \\ 
        FUV & Far Ultraviolet \\
        GC & Globular Cluster \\
        GR & General Relativity \\
        GRB & Gamma Ray Burst \\
        GRMHD & General Relativistic MagnetoHydroDynamics \\
        GSF & Gravitational Self-Force\\
        GW & Gravitational Wave \\
        GWB & Gravitational Wave Background \\
        GWTC & Gravitational Wave Transient Catalog \\
        HF & High Frequency \\
        HFQPO & High-Frequency Quasi-Periodic Oscillations \\
        HLX & Hyper-Luminous X-ray source\\
        HMC &  Hamiltonian Monte Carlo\\
        HMXB & High-Mass X-ray Binary\\ 
        HST & Hubble Space Telescope \\
        IIFSCz & Imperial IRAS Faint Source Catalogue Redshift Database \\
        ILOT & Intermediate Luminosity Optical Transient \\
        ILRT & Intermediate Luminosity Red Transient \\
        IMBHB & Intermediate Mass Black Hole Binary\\
        IMF & Initial Mass Function \\
        IMR & Inspiral-Merger-Ringdown \\
        IMRI & Intermediate Mass-Ratio Inspiral \\
        IR & Infrared \\
        ISB & Instrument Science Board (of ET) \\
        ISCO & Innermost Stable Circular Orbit \\
        ISW & Integrated Sachs-Wolfe \\
        JWST & James Webb Space Telescope \\
        KAGRA & Kamioka Gravitational Wave Detector\\
        KN & Kilonova \\
        LBV & Luminous Blue Variable \\
        LF & Low Frequency \\
        LGRB & Long Gamma Ray Burst \\
        LG & Local Group \\
        LGWA & Lunar Gravitational Wave Antenna\\
        LIGO & Laser Interferometer Gravitational-Wave Observatory \\
        LIRG & Luminous InfraRed Galaxies \\
        LISA & Laser Interferometer Space Antenna \\
        LMC & Large Magellanic Cloud \\
        LMXB & Low-Mass X-ray Binary\\
        LOSS & Lick Observatory Supernova Search \\
        LRN & Luminous Red Novae \\
        LSA & Linearized Signal Approximation\\
        LSS & Large-Scale Structure \\
        LSST & Legacy Survey of Space and Time\\
        LVC & LIGO-Virgo Collaboration \\
        LVK & LIGO-Virgo-KAGRA Collaboration \\
        LW & Lyman-Werner \\
        MACHO & Massive Compact Halo Object \\
        MBHB & Massive Black Hole Binary \\
        MCMC & Markov Chain Monte Carlo \\
        MD & Matter Dominance \\
        MDC & Mock Data Challenge\\
        MHD & MagnetoHydroDynamics\\
        MIR & Mid-infrared \\
        ML & Machine Learning \\
        MSP & Millisecond Pulsar \\
        MW & Milky Way \\
        NEMO & Neutron Star Extreme Matter Observatory \\
        NF & Normalizing Flow \\
        NN & Newtonian Noise \\
        NPE & Neural Posterior Estimation\\
        NR & Numerical Relativity \\
        NS & Neutron Star \\
        NSBH & Neutron Star -- Black Hole \\
        NSC & Nuclear Star Cluster \\
        OSB & Observational Science Board (of ET) \\
        ORF & Overlap Reduction Function \\
        PBH & Primordial Black Hole \\
        PE & Parameter Estimation \\
        PISN & Pair Instability Supernova \\
        PLS & Power-Law Sensitivity \\
        PM & post-Minkowskian \\
        PN & post-Newtonian \\
        PNS & Proto-Neutron Star \\
        PPISN & Pulsation Pair Instability Supernova \\
        PSD & Power Spectral Density \\
        PT & Phase Transition \\
        PTA  & Pulsar Timing Array \\
        QKP & Quasi-Keplerian Parametrization \\
        QRPA & Quasi-particle Random Phase Approximation \\
        QCD & Quantum Chromo-Dynamics \\
        QNM & Quasi-Normal Mode \\
        RB & Relative Binning \\
        RD & Radiation Dominance \\
        RHIC &Relativistic Heavy Ion Collisions \\
        ROM & Reduced-Order Model\\
        ROQ & Reduced-Order Quadrature \\
        RSG & Red Super Giant \\
        RRAT & Rotating Radio Transient \\
        SAGB & Super Asymptotic Giant Branch \\
        SASI & Standing Accretion Shock Instability \\
        SBHB & Stellar-mass Black Hole Binary \\
        SFD & Star Formation Density \\
        SFH & Star Formation History \\
        SFR & Star Formation Rate \\
        SFSR & Single-Field Slow-Roll \\
        SGWB & Stochastic Gravitational Wave Background \\
        SIGW & Scalar-Induced Gravitational Wave \\ 
        SKA & Square Kilometre Array\\
        SLSN & Super Luminous Supernova \\
        SM & Standard Model (of particle physics)\\
        SMBH & Super-Massive Black Hole \\
        SOBBH & Stellar Origin Binary Black Hole \\
        SN & Supernova \\
        SN Ib/c & Type Ib/Ic Supernova \\
        SNIa & Type Ia Supernova \\
        SNIb & Type Ib Supernova \\
        SNIb/c & Type Ib/Ic Supernova \\
        SNIc & Type Ic Supernova \\
        SNIc-BL & Broad-Line Type Ic Supernova \\
        SNII & Type II Supernova \\
        SNII-L & Type II Linear Supernova \\
        SNII-P & Type II Plateau Supernova \\
        SNII-np & Type II nP Supernova \\
        SNIIb & Type IIb Supernova \\
        SNR & Signal-to-Noise Ratio \\
        SPA & Stationary Phase Approximation \\
        SPIRITS & Spitzer InfraRed Intensive Transients Survey \\
        SVD & Singular Value Decomposition \\
        SVOM & Space Variable Objects Monitor\\
        TCW & Transient Continuous Wave \\
        TDE & Tidal Disruption Event \\   
        TF & Time-Frequency\\
        THESEUS & Transient High Energy Sources and Early Universe Surveyor\\ 
        TIR & Total Infrared \\
        TMNRE & Truncated Marginal Neural Ratio Estimator\\ 
        ToO & Target of Opportunity \\
        TP-AGB & Thermally Pulsing Asymptotic Giant Branch \\
        ULX & Ultra-Luminous X-ray source\\
        UV & Ultraviolet \\
        VHE & Very High Energy\\
        VRO & Vera Rubin Observatory \\
        VMS & Very Massive Star \\
        WD & White Dwarf \\
        WR & Wolf-Rayet \\
        WST & Wide-Field Spectroscopic Telescope \\
        XDINS & X-Ray Dim Isolated Neutron Star \\
        YSC & Young Star Cluster\\
        ZAMS & Zero-Age-Main-Sequence \\
        ZTF & Zwicky Transient Facility \\
\hline

\end{longtable}



\clearpage
\bibliographystyle{utphys}
\bibliography{references_merged,refs_temporary}

\end{document}